\documentclass[reqno,10pt,letterpaper]{amsart}

\usepackage{lipsum}
\usepackage{amsmath}
\usepackage{amssymb}
\usepackage{amsthm}
\usepackage{mathrsfs}
\usepackage{accents}
\usepackage{calc}
\usepackage{arydshln}
\usepackage{upgreek}
\usepackage{slashed}
\usepackage{xifthen}
\usepackage{graphicx}
\usepackage{subcaption}
\usepackage{longtable}
\usepackage[inline]{enumitem}

\usepackage{refcount}

\newcommand{\seepages}[4]{%
  \ifnum\getpagerefnumber{#2}=\getpagerefnumber{#4}
    #1 and #3 on page~\pageref{#2}%
  \else
    #1 on page~\pageref{#2} and #3 on page~\pageref{#4}%
  \fi
}

\usepackage{marvosym}
\newcommand{\scissors}{%
  \text{\raisebox{-0.35ex}{\large\Rightscissors}}
}
\newcommand{\textscissors}{%
  \text{\large\Rightscissors}
}

\usepackage{xr}

\newcommand{\citeCD}[1]{\cite[#1]{HintzKerrCD}}
\newcommand{\citeAF}[1]{\cite[#1]{HintzNonstat2}}

\usepackage{xcolor}
\definecolor{myr}{rgb}{0.6,0,0}
\definecolor{myg}{rgb}{0,0.6,0}
\definecolor{myb}{rgb}{0,0,0.7}

\usepackage[pdftex,colorlinks=true,linkcolor=myr,citecolor=myb,urlcolor=myb,breaklinks=true,bookmarksopen=true]{hyperref}

\makeatletter
\newcommand{\myitem}[2]{\item[\rm(#2)]\def\@currentlabel{#2}\label{#1}}
\makeatother

\usepackage{titletoc}

\makeatletter

\def\@tocline#1#2#3#4#5#6#7{
\begingroup
  \par
    \parindent\z@ \leftskip#3 \relax \advance\leftskip\@tempdima\relax
                  \rightskip\@pnumwidth plus 4em \parfillskip-\@pnumwidth
    \ifcase #1 
       \vskip 0.6em \hskip 0em 
       \or
       \or \hskip 0em 
       \or \hskip 1em 
    \fi%
    #6
    \nobreak\relax{\leavevmode\leaders\hbox{\,.}\hfill}
    \hbox to\@pnumwidth {\@tocpagenum{#7}}
  \par
\endgroup
}

 \def\l@section{\@tocline{0}{0pt}{0pc}{}{}}

\renewcommand{\tocsection}[3]{%
  \indentlabel{\@ifnotempty{#2}{ 
    \ignorespaces\bfseries{#2. #3}}}
  \indentlabel{\@ifempty{#2}{\ignorespaces\bfseries{#3}}{}} 
    \vspace{1.5pt}}

\renewcommand{\tocsubsection}[3]{%
  \indentlabel{\@ifnotempty{#2}{
    \ignorespaces#2. #3}}
  \indentlabel{\@ifempty{#2}{\ignorespaces #3}{}}
    \vspace{1.5pt}}

\renewcommand{\tocsubsubsection}[3]{%
  \indentlabel{\@ifnotempty{#2}{
    \ignorespaces#2. #3}}
  \indentlabel{\@ifempty{#2}{\ignorespaces #3}{}}
    \vspace{1.5pt}}

\makeatother

\makeatletter
\def\@nomenstarted{0}
\newlength{\@nomenoldtabcolsep}

\newcommand{\nomenstart}
  {%
    \def\@nomenstarted{1}%
    \setlength{\@nomenoldtabcolsep}{\tabcolsep}%
    \setlength{\tabcolsep}{3.5pt}%
    \begin{longtable}{p{0.11\textwidth} p{0.86\textwidth}}
  }

\newcommand{\nomenitem}[2]{%
    \ifcase\@nomenstarted%
      \or 
      \or \\ 
    \fi%
    #1\,{\leavevmode\leaders\hbox{\,.}\hfill} & #2%
    \def\@nomenstarted{2}%
  }%
\newcommand{\nomenend}
  {\\%
      \end{longtable}%
      \setlength{\tabcolsep}{\@nomenoldtabcolsep}%
      \def\@nomenstarted{0}%
  }
\makeatother

\makeatletter
\newcommand{\BIG}{\bBigg@{3.5}}
\newcommand{\vast}{\bBigg@{4}}
\newcommand{\Vast}{\bBigg@{5}}
\newcommand{\VAST}[1]{\bBigg@{#1}}
\makeatother

\allowdisplaybreaks

\numberwithin{equation}{section}
\numberwithin{figure}{section}
\newtheorem{thm}{Theorem}[section]

\newtheorem{prop}[thm]{Proposition}
\newtheorem{lemma}[thm]{Lemma}
\newtheorem{cor}[thm]{Corollary}
\newtheorem{conj}[thm]{Conjecture}

\newtheorem*{thm*}{Theorem}
\newtheorem*{prop*}{Proposition}
\newtheorem*{cor*}{Corollary}
\newtheorem*{conj*}{Conjecture}

\theoremstyle{definition}
\newtheorem{definition}[thm]{Definition}
\newtheorem{notation}[thm]{Notation}

\theoremstyle{remark}
\newtheorem{rmk}[thm]{Remark}
\newtheorem{example}[thm]{Example}

\makeatletter
\newcommand{\fakephantomsection}{%
  \Hy@MakeCurrentHref{\@currenvir.\the\Hy@linkcounter}
  \Hy@raisedlink{\hyper@anchorstart{\@currentHref}\hyper@anchorend}%
  \Hy@GlobalStepCount\Hy@linkcounter%
}
\makeatother

\newcommand{\mc}{\mathcal}
\newcommand{\cA}{\mc A}

\newcommand{\cC}{\mc C}

\newcommand{\cE}{\mc E}
\newcommand{\cF}{\mc F}
\newcommand{\cG}{\mc G}
\newcommand{\cH}{\mc H}

\newcommand{\cK}{\mc K}
\newcommand{\cL}{\mc L}
\newcommand{\cM}{\mc M}

\newcommand{\cO}{\mc O}

\newcommand{\cR}{\mc R}
\newcommand{\cS}{\mc S}
\newcommand{\cT}{\mc T}

\newcommand{\cV}{\mc V}

\newcommand{\cX}{\mc X}

\newcommand{\ms}{\mathscr}
\newcommand{\sA}{\ms A}

\newcommand{\sC}{\ms C}
\newcommand{\sD}{\ms D}

\newcommand{\sF}{\ms F}
\newcommand{\sG}{\ms G}
\newcommand{\sH}{\ms H}

\newcommand{\scri}{\ms I}
\newcommand{\sscri}{{\!\scri}}

\newcommand{\sR}{\ms R}

\newcommand{\sY}{\ms Y}

\newcommand{\C}{\mathbb{C}}
\newcommand{\N}{\mathbb{N}}
\newcommand{\R}{\mathbb{R}}
\newcommand{\Z}{\mathbb{Z}}

\newcommand{\Sph}{\mathbb{S}}

\newcommand{\sfb}{\mathsf{b}}

\newcommand{\sfe}{\mathsf{e}}
\newcommand{\sff}{\mathsf{f}}

\newcommand{\sfr}{\mathsf{r}}
\newcommand{\sfs}{\mathsf{s}}

\newcommand{\sfG}{\mathsf{G}}

\newcommand{\sfM}{\mathsf{M}}

\newcommand{\sfW}{\mathsf{W}}
\newcommand{\sfX}{\mathsf{X}}

\newcommand{\bfv}{\mathbf{v}}
\newcommand{\bfw}{\mathbf{w}}

\newcommand{\bfB}{\mathbf{B}}

\newcommand{\fa}{\mathfrak{a}}

\newcommand{\fc}{\mathfrak{c}}

\newcommand{\fD}{\mathfrak{D}}

\newcommand{\fk}{\mathfrak{k}}
\newcommand{\fl}{\mathfrak{l}}
\newcommand{\fm}{\mathfrak{m}}

\newcommand{\fP}{\mathfrak{P}}

\newcommand{\fQ}{\mathfrak{Q}}
\newcommand{\fq}{\mathfrak{q}}

\newcommand{\fs}{\mathfrak{s}}
\newcommand{\ft}{\mathfrak{t}}

\newcommand{\sld}{{\slashed{\dd}{}}}
\newcommand{\slg}{{\slashed{g}{}}}

\newcommand{\slsfG}{{\slashed{\sfG}{}}}

\newcommand{\slGamma}{{\slashed{\Gamma}{}}}
\newcommand{\sldelta}{{\slashed{\delta}{}}}
\newcommand{\slDelta}{{\slashed{\Delta}{}}}

\newcommand{\slnabla}{{\slashed{\nabla}{}}}
\newcommand{\slpi}{{\slashed{\pi}{}}}
\newcommand{\slcE}{{\slashed{\cE}{}}}

\newcommand{\slstar}{{\slashed{\star}}}
\newcommand{\slomega}{{\slashed{\omega}{}}}

\newcommand{\sltr}{\operatorname{\slashed\tr}}

\newcommand{\scal}{\mathsf{S}}
\newcommand{\scalspace}{\mathbf{S}}
\newcommand{\vect}{\mathsf{V}}

\newcommand{\vectspace}{\mathbf{V}}

\newcommand{\ran}{\operatorname{ran}}

\newcommand{\End}{\operatorname{End}}

\newcommand{\Hom}{\operatorname{Hom}}
\renewcommand{\Re}{\operatorname{Re}}
\renewcommand{\Im}{\operatorname{Im}}
\newcommand{\Id}{\operatorname{Id}}
\newcommand{\mathspan}{\operatorname{span}}
\newcommand{\supp}{\operatorname{supp}}

\newcommand{\ext}{{\rm ext}}

\newcommand{\tr}{\operatorname{tr}}

\newcommand{\ord}{\operatorname{ord}}

\newcommand{\diag}{\operatorname{diag}}
\newcommand{\Res}{\operatorname{Res}}
\newcommand{\Resspace}{\operatorname{\mathfrak{Res}}}

\newcommand{\ind}{{\rm ind}}

\newcommand{\vecp}{{\!\vec{\;p}}}
\newcommand{\vecq}{{\!\vec{\;q}}}
\newcommand{\vecrho}{{\!\vec{\;\rho}}}

\newcommand{\cd}{\mathfrak{c}}

\newcommand{\Poly}{{\mathrm{Poly}}}

\newcommand{\IVP}{{\mathrm{IVP}}}
\newcommand{\Ups}{\Upsilon}

\newcommand{\eps}{\epsilon}
\newcommand{\ftrans}{\;\!\wh{\ }\;\!}

\newcommand{\hra}{\hookrightarrow}
\newcommand{\la}{\langle}

\newcommand{\extcup}{\mathbin{\ol\cup}}

\newcommand{\ol}{\overline}
\newcommand{\pa}{\partial}
\newcommand{\dd}{{\mathrm d}}
\newcommand{\ra}{\rangle}
\newcommand{\spec}{\operatorname{spec}}
\newcommand{\Spec}{\operatorname{Spec}}
\newcommand{\specb}{\operatorname{spec}_\bop}
\newcommand{\Specb}{\operatorname{Spec}_\bop}
\newcommand{\tot}{\mathrm{tot}}
\newcommand{\aug}{\mathrm{aug}}
\newcommand{\rem}{\mathrm{rem}}

\newcommand{\ul}[1]{\underline{#1}{}}

\newcommand{\wh}{\widehat}
\newcommand{\wt}{\widetilde}
\newcommand{\xra}{\xrightarrow}

\DeclareFontFamily{U}{mathx}{}
\DeclareFontShape{U}{mathx}{m}{n}{<-> mathx10}{}
\DeclareSymbolFont{mathx}{U}{mathx}{m}{n}
\DeclareMathAccent{\widecheck}{0}{mathx}{"71}

\newcommand{\ubar}[1]{\underaccent{\bar}#1}
\newcommand{\pfstep}[1]{$\bullet$\ \underline{\textit{#1}}}
\newcommand{\pfsubstep}[2]{{\bf#1}\ \textit{#2}}

\newcommand{\bop}{{\mathrm{b}}}

\newcommand{\scop}{{\mathrm{sc}}}
\newcommand{\chop}{{\mathrm{c}\semi}}

\newcommand{\cl}{{\mathrm{cl}}}
\newcommand{\scl}{{\mathrm{sc}}}
\newcommand{\ebop}{{\mathrm{eb}}}
\newcommand{\eop}{{\mathrm{e}}}
\newcommand{\tbop}{{3\mathrm{b}}}
\newcommand{\etbop}{{\mathrm{e}3\mathrm{b}}}

\newcommand{\scbtop}{{\mathrm{sc}\text{-}\mathrm{b}}}

\newcommand{\semi}{\hbar}

\newcommand{\cface}{{\mathrm{cf}}}

\newcommand{\scface}{{\mathrm{scf}}}
\newcommand{\sctface}{{\mathrm{sctf}}}
\newcommand{\sface}{{\mathrm{sf}}}
\newcommand{\hbarface}{{\mathrm{\hbar f}}}

\newcommand{\tface}{{\mathrm{tf}}}

\newcommand{\ztface}{{\mathrm{ztf}}}
\newcommand{\zface}{{\mathrm{zf}}}

\newcommand{\rms}{{\mathrm{s}}}
\newcommand{\rmv}{{\mathrm{v}}}
\newcommand{\rmw}{{\mathrm{w}}}
\newcommand{\res}{{\mathrm{res}}}

\newcommand{\cp}{{\mathrm{c}}}

\newcommand{\Diff}{\mathrm{Diff}}

\newcommand{\Vb}{\cV_\bop}

\newcommand{\Diffb}{\Diff_\bop}

\newcommand{\Diffscbt}{\Diff_\scbtop}

\newcommand{\Vtb}{\cV_\tbop}
\newcommand{\Vetb}{\cV_\etbop}
\newcommand{\Vtsc}{\cV_{3\scl}}

\newcommand{\Difftb}{\Diff_\tbop}

\newcommand{\Veb}{\cV_\ebop}

\newcommand{\Diffeb}{\Diff_\ebop}

\newcommand{\Tscbt}{{}^\scbtop T}

\newcommand{\Vsc}{\cV_\scop}

\newcommand{\Tsc}{{}^{\scop}T}

\newcommand{\Ttb}{{}^{\tbop}T}
\newcommand{\Tetb}{{}^{\etbop}T}

\newcommand{\Stb}{{}^{\tbop}S}

\newcommand{\sub}{{\mathrm{sub}}}

\newcommand{\loc}{{\mathrm{loc}}}
\newcommand{\CI}{\cC^\infty}
\newcommand{\CIdot}{\dot\cC^\infty}

\newcommand{\CIc}{\cC^\infty_\cp}

\newcommand{\Hb}{H_{\bop}}

\newcommand{\Htb}{H_\tbop}

\newcommand{\Riem}{\mathrm{Riem}}
\newcommand{\Ric}{\mathrm{Ric}}
\newcommand{\Ein}{\mathrm{Ein}}
\newcommand{\bhm}{\fm}
\newcommand{\bha}{\fa}

\newcommand{\openbigpmatrix}[1]
  {%
    \def\@bigpmatrixsize{#1}%
    \addtolength{\arraycolsep}{-#1}%
    \begin{pmatrix}%
  }
\newcommand{\closebigpmatrix}
  {%
    \end{pmatrix}%
    \addtolength{\arraycolsep}{\@bigpmatrixsize}%
  }

\newlength{\enummargin}

\newcommand{\usref}[1]{{\upshape\ref{#1}}}

\DeclareGraphicsExtensions{.mps}

\makeatletter
\newcommand*{\fwbw}[1]{\expandafter\@fwbw\csname c@#1\endcsname}
\newcommand*{\@fwbw}[1]{\ifcase #1 \or {\rm fw}\or {\rm bw}\fi}
\AddEnumerateCounter{\fwbw}{\@fwbw}
\makeatother

\begin{document}

\title[Nonlinear stability of subextremal Kerr]{Nonlinear stability of subextremal Kerr black holes}

\date{\today}


\author{Peter Hintz}
\address{Department of Mathematics, Pennsylvania State University, 54 McAllister St, State College,\newline PA 16801, United States}
\email{phintz@psu.edu}

\begin{abstract}
  We settle the global nonlinear stability problem for the family of Kerr black holes in the full subextremal range: spacetimes evolving from initial data close to those of a subextremal Kerr black hole as solutions of the Einstein vacuum equation ${\rm Ric}(g)=0$ settle down to a nearby member of the Kerr family at the rate $\mathcal{O}(t_*^{-2-\epsilon_{\mathcal K}})$ in spatially compact regions.

  For the initial data, we require $\mathcal{O}(r^{-1-\epsilon_0})$-decay for $\epsilon_0>0$---more precisely, an arbitrary but finite expansion into terms $r^{-z}(\log r)^k$ where $z>1$, $k\in\mathbb{N}_0$, plus a remainder term with $\mathcal{O}(r^{-3-\epsilon_0})$-decay. We use a generalized wave map gauge modified using gauge source terms that lie in a suitable finite-dimensional space determined by the expansion of the initial data. Like the final black hole parameters (mass and angular momentum) and the gravitational wave tail, the gauge source terms are treated as unknowns in a nonlinear (Nash--Moser) iteration scheme. We work directly with the tensorial equation and in particular do not rely on reductions to scalar equations (except insofar as a reduction to the Teukolsky equation is used in the proof of linear mode stability).

  This paper relies on two companion papers. The first one \cite{HintzKerrCD} introduces a strong form of constraint damping in the full subextremal range, which we use in our formulation of the gauge-fixed Einstein equation as a black box. The second one \cite{HintzNonstat2} provides tame estimates (albeit with weak decay) for forward solutions of a general class of wave-type equations, which we show here to include the linearizations of the gauge-fixed Einstein equation arising in our nonlinear iteration scheme; these estimates are the starting point for our detailed asymptotic analysis.
\end{abstract}

\maketitle
\thispagestyle{empty}

\clearpage

{\pagestyle{empty}
\renewcommand{\contentsname}{Table of Contents}
\setlength{\parskip}{0.00pt}
\tableofcontents
\setlength{\parskip}{0.05in}
\clearpage
}

\pagenumbering{arabic}

\section{Introduction}
\label{SI}

\begin{figure}[!ht]
\centering
\includegraphics{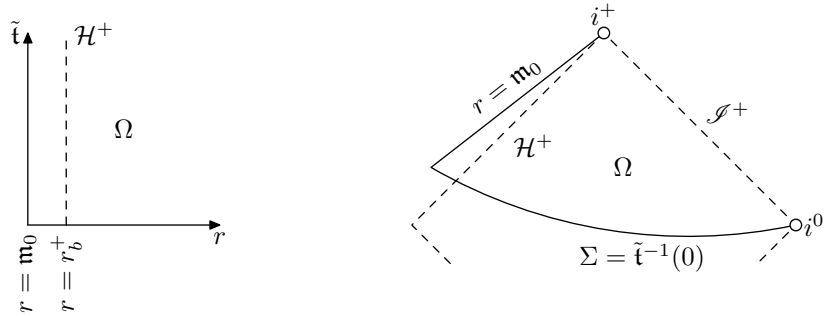}
\caption{\textit{On the left:} the domain $\Omega$, drawn in a product fashion. The dashed line labeled $\cH^+$ is the event horizon for $g_b$. \textit{On the right:} the domain $\Omega$, drawn as a subset of the Penrose diagram of $g_b$.}
\label{FigIOmega}
\end{figure}

The Kerr family \cite{KerrKerr} is an explicit family of Lorentzian metrics, with signature $(-,+,+,+)$, that satisfy the Einstein vacuum equation
\begin{equation}
\label{EqIEin}
  \Ric(g) = 0.
\end{equation}
It is parameterized by $\bhm>0$ and $\bha\in\R^3$, and describes an asymptotically flat black hole with mass $\bhm$ and specific angular momentum $\bha$. The parameters $b=(\bhm,\bha)$ are called \emph{subextremal} when $a:=|\bha|<\bhm$. Introducing $\mu_b(r):=r^2-2\bhm r+a^2$ and $\varrho_b^2(r,\theta):=r^2+a^2\cos^2\theta$, the Kerr metric $g_b$ takes the explicit form
\[
  g_b = -\frac{\mu_b}{\varrho_b^2}(\dd\ft-a\sin^2\theta\,\dd\varphi)^2 + \varrho_b^2\Bigl(\frac{\dd r^2}{\mu_b} + \dd\theta^2\Bigr) + \frac{\sin^2\theta}{\varrho_b^2}\bigl((r^2+a^2)\dd\varphi-a\,\dd\ft\bigr)^2
\]
in Boyer--Lindquist coordinates $\ft\in\R$, $r\in(r^+_b,\infty)$, and $\theta\in(0,\pi)$, $\varphi\in\R/2\pi\Z$ \cite{BoyerLindquistKerr}; here the ``north pole'' $\theta=0$ points in the direction of the axis $\frac{\bha}{|\bha|}$ when $a\neq 0$, and the larger root $r_b^+:=\bhm+\sqrt{\bhm^2-a^2}$ of $\mu_b$ is the radius of the \emph{event horizon}. After a standard coordinate change near $r=r_b^+$---roughly, setting $\dd\tilde\ft\sim\dd\ft+\frac{r^2+a^2}{\mu_b}\,\dd r$ and $\dd\tilde\varphi\sim\dd\varphi+\frac{a}{\mu_b}\,\dd r$---the metric $g_b$ extends analytically to the region $r>r_b^-:=\bhm-\sqrt{\bhm^2-a^2}$ in the coordinates $(\tilde\ft,x)$ where $x\in\R^3$ denotes the Cartesian coordinates corresponding to the polar coordinates $(r,\theta,\tilde\varphi)$. Moreover, the metric $g_b$ is stationary, i.e., the vector field $\pa_{\tilde\ft}$ (which is timelike for large $r$) is Killing. Arranging $\tilde\ft$ to be a time function (i.e., $\dd\tilde\ft$ is everywhere timelike) and $\tilde\ft=\ft$ for large $r$, initial value problems for the scalar wave equation $\Box_{g_b}u=0$ and suitable quasilinear perturbations thereof are then well-posed \cite{DafermosHolzegelRodnianskiTaylorQuasilinear,DafermosHolzegelRodnianskiTaylorQuasilinear2,HintzNonstat2} on the domain
\begin{equation}
\label{EqIOmega}
  \Omega := \{ r\geq\bhm_0,\ \tilde\ft\geq 0 \},\quad\text{with Cauchy hypersurface}\ \Sigma := \tilde\ft^{-1}(0),
\end{equation}
when $b$ is close to the subextremal parameters $b_0=(\bhm_0,\bha_0)$; see Figure~\ref{FigIOmega}.

The Einstein vacuum equation~\eqref{EqIEin} can be cast as a quasilinear wave equation for $g$ after gauge fixing. The initial data are the first and second fundamental form $\gamma$ and $k$ of $\Sigma$ inside of $(\Omega,g)$. The \emph{constraint equations} (or Gauss--Codazzi equations)
\begin{equation}
\label{EqIConstraints}
  {\rm scal}_\gamma - |k|_\gamma^2 + (\tr_\gamma k)^2 = 0,\quad
  \delta_\gamma k + \dd\tr_\gamma k = 0,
\end{equation}
where $(\delta_\gamma k)_\mu:=-\gamma^{\nu\lambda}k_{\mu\nu;\lambda}$, were shown by Choquet-Bruhat \cite{ChoquetBruhatLocalEinstein} to be sufficient (and necessary) for local existence and uniqueness up to isometries. The maximal globally hyperbolic development (MGHD) of an initial data set $(\Sigma,\gamma,k)$, i.e., the maximal spacetime $(M,g)$ with $\Ric(g)=0$ that is uniquely determined by the initial data, was constructed by Choquet-Bruhat--Geroch \cite{ChoquetBruhatGerochMGHD} (see also \cite{SbierskiMGHD}). Denote the initial data of $g_b$ by $(\gamma_b,k_b)$. We can now state a simple version of our main result:

\begin{thm}[Simple version of nonlinear stability: restricted data]
\label{ThmISimple}
  Let $b_0=(\bhm_0,\bha_0)$ be subextremal parameters. Suppose that the initial data $\gamma,k\in\CI(\Sigma;S^2 T^*\Sigma)$ satisfy the constraint equations~\eqref{EqIConstraints}. Suppose moreover that $(\gamma,k)=(\gamma_{b_0},k_{b_0})$ outside of a compact set, and that $\gamma-\gamma_{b_0}$ and $k-k_{b_0}$ are small in the Sobolev space $H^d(\Sigma)$ for some large $d$. Then there exist subextremal parameters $b=(\bhm,\bha)$ close to $b_0$ such that the MGHD of $(\Sigma,\gamma,k)$ contains a region isometric to $(\Omega,g)$, where $g$ is a Lorentzian metric satisfying
  \[
    | g_{\mu\nu}(\tilde\ft,x) - (g_b)_{\mu\nu}(\tilde\ft,x) | \lesssim (1+\tilde\ft)^{-2-\eps_\cK}
  \]
  for some $\eps_\cK>0$ and for all $x$ in a compact set; similarly for finite-order derivatives of $g-g_b$ along $(1+\tilde\ft)\pa_{\tilde\ft}$ and $\pa_x$. Furthermore, $g$ decays at quantitative rates in all asymptotic regions of $\Omega$, namely $r^{-1}(\frac{r}{\tilde\ft})^{2+\eps_\cK}$ for $\frac{r}{\tilde\ft}\leq\frac14$, further $r^{-1}$ for $\frac{r}{\tilde\ft}\geq\frac34$ (including near null infinity), and finally $\tilde t^{-1}\log\tilde t$ for $\frac{r}{\tilde t}\in[\frac14,\frac34]$.
\end{thm}

This settles the Kerr stability conjecture in the full subextremal range. Previous results by Klainerman--Szeftel \cite{KlainermanSzeftelPolarized}, Dafermos--Holzegel--Rodnianski--Taylor \cite{DafermosHolzegelRodnianskiTaylorSchwarzschild}, and Klainerman--Szeftel with Giorgi and Shen \cite{KlainermanSzeftelGCM1,KlainermanSzeftelGCM2,KlainermanSzeftelKerr,GiorgiKlainermanSzeftelStability,ShenGCMKerr} treated the axially symmetric and polarized, ultimately Schwarzschildean, and slowly rotating ($|\bha_0|\ll\bhm_0$) regimes, respectively.\footnote{The smallness and decay conditions on the initial data are different in each of these works. Theorem~\ref{ThmISimple}, or rather its full version (Theorem~\ref{ThmSt}) which applies to initial data of class~\eqref{EqIDataPhg}--\eqref{EqIHbBound}, does not imply the main results of \cite{KlainermanSzeftelPolarized,KlainermanSzeftelKerr} and \cite{DafermosHolzegelRodnianskiTaylorSchwarzschild} which apply to data $(\gamma,r k)$ having ``structureless'' $\cO(r^{-\frac32-\eps_0})$- and $\cO(r^{-\frac52-\eps_0})$-decay, respectively, towards the Kerr data $(\gamma_{b_0},r k_{b_0})$; for structureless data, our result requires $\cO(r^{-3-\eps_0})$-decay. Conversely, these results do not imply Theorem~\ref{ThmSt} in the regimes in which they apply, as we can handle more weakly decaying ($\cO(r^{-1-\eps_0})$), albeit structured, initial data.} Several ingredients of these works (e.g., \cite{KlainermanSzeftelGCM1,KlainermanSzeftelGCM2,ShenGCMKerr}) do not depend on the smallness or vanishing of $|\frac{\bha_0}{\bhm_0}|$; some of the ingredients that do have since been extended to the full subextremal range \cite{ShlapentokhRothmanTeixeiradCTeukolskyI,ShlapentokhRothmanTeixeiradCTeukolskyII,MaSzeftelEnergyKerr,MaSzeftelTeukolsky,HeKlainermanKerr}. See~\S\ref{SsIPrior} for a detailed review of the literature.

\begin{rmk}[Stronger decay: $\tilde\ft^{-3}$]
\label{RmkIt3}
  Our techniques can in fact be used to obtain $\tilde t^{-3}$-decay of $g$ to $g_b$ in spatially compact regions; this is discussed in~\S\ref{SsSt3}. This strengthens the $\tilde t^{-1-\eps}$-decay previously established in the slowly rotating regime.
\end{rmk}

For a precise version of Theorem~\ref{ThmISimple}, see Theorem~\ref{ThmSt}. Let us describe the considerably more permissive class of initial data which, following Theorem~\ref{ThmSt}, we can allow in Theorem~\ref{ThmISimple} (with the same conclusions). We assume that the differences $\tilde\gamma:=\gamma-\gamma_{b_0}$ and $\tilde k:=k-k_{b_0}$ are \emph{partially polyhomogeneous} with $r^{-1-\eps_0}$- and $r^{-2-\eps_0}$-decay, $\eps_0>0$, respectively. More precisely, for a finite subset $\cE_0\subset[1+\eps_0,3]\times\N_0$, we assume that their coefficients in the frame $\pa_{x^i}$, $i=1,2,3$, admit expansions
\begin{equation}
\label{EqIDataPhg}
\begin{split}
  \tilde\gamma_{i j} &= \sum_{(z,m)\in\cE_0} r^{-z}(\log r)^m \tilde\gamma^{(z,m)}_{i j}(\omega) + \tilde\gamma^\flat_{i j}(r,\omega), \\
  \tilde k_{i j} &= \sum_{(z,m)\in\cE_0} r^{-z-1}(\log r)^m \tilde k^{(z+1,m)}_{i j}(\omega) + \tilde k^\flat_{i j}(r,\omega),
\end{split}
\end{equation}
where we write $x=r\omega$, $\omega=\frac{x}{|x|}\in\Sph^2$, and where $\tilde\gamma^{(z,m)}_{i j},\tilde k^{(z+1,m)}_{i j}\in\CI(\Sph^2)$ are small in $H^d(\Sph^2)$ and also
\begin{equation}
\label{EqIHbBound}
  \|\tilde\gamma^\flat_{i j}\|_{\Hb^{d,3+\eps_0}}^2 := \int_{\bhm_0}^\infty \int_{\Sph^2} \bigl| r^{3+\eps_0} (r\pa_x)^{\leq d}\tilde\gamma^\flat_{i j}(r,\omega) \bigr|^2\,\dd\omega\,\frac{\dd r}{r}, \quad
  \|\tilde k^\flat_{i j}\|_{\Hb^{d,4+\eps_0}}^2
\end{equation}
are small.\footnote{By Sobolev embedding, this implies the pointwise bounds $|\tilde\gamma^\flat_{i j}|\lesssim r^{-3-\eps_0}$ and $|\tilde k^\flat_{i j}|\lesssim r^{-4-\eps_0}$, likewise for $(d-2)$-many $r\pa_x$-derivatives; and conversely these pointwise bounds (with $d$ derivatives) imply bounds for the norms~\eqref{EqIHbBound} (with the same $d$) upon reducing $\eps_0$ by an arbitrarily small amount.} A special case of this setup is when $\cE_0=\{(2,m),(3,m)\colon m\leq J\}$ for some $J\in\N_0$; then $\gamma$ and $k$ are smooth in $1/r$ when $J=0$, and log-smooth for general $J\geq 1$.\footnote{For the sake of completeness, we remark that we can also allow for complex exponents $z$, which for reasons of reality must then appear in complex-conjugate pairs $(z,m)$, $(\bar z,m)$.}

\begin{rmk}[Initial data]
\label{RmkIData}
  The existence of nontrivial initial data $(\gamma,k)$ that are exactly Kerrian outside of a compact set was proved using the gluing techniques of Corvino and Schoen \cite{CorvinoScalar,CorvinoSchoenAsymptotics} by Chru\'sciel--Delay \cite{ChruscielDelayMapping}. Initial data that decay at any desired inverse polynomial rate (but no faster) were constructed by Fang--Szeftel--Touati \cite{FangSzeftelTouatiBHData}; data that have partial polyhomogeneous expansions as above can be constructed using a simple modification of their construction (see Example~\ref{ExIDPhg}\eqref{ItIDPhg3}). The question of what constitutes ``physically realistic'' initial data is under active investigation, e.g., by Kehrberger \cite{KehrbergerScri1,KehrbergerScri2,KehrbergerScri3,KehrbergerScri4} with Kadar \cite{KadarKehrbergerPhgScatter}; we remark that the index sets of the polyhomogeneous initial data arising in their works do permit $r^{-1}$-terms, which is slightly too weak compared with our present requirements. (This question is also related to conformal smoothness properties of dynamical metrics at null infinity $\scri^+$; see \cite{FriedrichSmoothScriReview} for an overview, and \cite{NuetziStability} for a recent result of this flavor.) In the present paper, we regard partial polyhomogeneity as a highly convenient condition on initial data which we exploit freely: it plays an important role in our proof, as it is ultimately responsible for our ability to use \emph{finite-dimensional} gauge modifications (see~\S\ref{SssINElim} below). Removing it in favor of ``structureless'' $\cO(r^{-1-\eps_0})$-decay conditions cannot be done using only the ideas introduced in the present paper (see Remark~\ref{RmkINElimNon}). In fact, we expect that the gauge issues become unavoidably infinite-dimensional in nature in this case, and new ideas will be required to treat them. We note that the GCM construction of \cite{KlainermanSzeftelGCM1,KlainermanSzeftelGCM2,ShenGCMKerr} resolves infinite-dimensional gauge issues, albeit in a framework that differs substantially from generalized harmonic gauge. (Also \cite{HintzVasyKdSCosm} encounters infinite-dimensional gauge issues in generalized harmonic gauge, but these are of a different and milder flavor; see Remark~\ref{RmkINInfty}.)
\end{rmk}

We record three consequences of our main result.
\begin{enumerate}
\item The decay obtained in Theorem~\ref{ThmISimple} is strong enough for the main result of Dafermos--Luk \cite{DafermosLukKerrCauchyHorI} on the $\cC^0$-formulation of Penrose's strong cosmic censorship conjecture \cite{PenroseSCC} to apply (see \cite[(4.2)--(4.3)]{DafermosLukKerrCauchyHorI}). We conclude the \emph{unconditional} $\cC^0$-stability of the Cauchy horizon of rotating Kerr black holes.
\item The decay and regularity of the metric proved in Theorem~\ref{ThmSt} allows one to apply the main results of Chen--Klainerman \cite{ChenKlainermanHorizon} and the author \cite{HintzHorizons} concerning dynamical event horizons. Thus, the event horizon of the dynamical spacetime $(\Omega,g)$ is a smooth hypersurface that asymptotes to that of the final black hole metric $g_b$ at a rate $\cO(\tilde\ft^{-2-\eps_\cK})$.
\item One can apply Theorem~\ref{ThmISimple} in conjunction with the main result of \cite{HintzGlueLocIII} to construct black hole merger spacetimes (see \cite[Remark~6.2]{HintzGlueLocIII}). Their initial data are equal to the data $(\gamma_b,k_b)$ of a unit mass Kerr black hole outside of a compact set, and in a compact set are obtained from $(\gamma_b,k_b)$ by gluing in a very light subextremal Kerr black hole with parameters $(\eps\hat\bhm,\eps\hat\bha)$ for some fixed parameters $|\hat\bha|<\hat\bhm$. In evolution, the small black hole can be arranged to cross the event horizon of the unit mass black hole in finite proper time; and the resulting single black hole settles down, at the rates given by Theorems~\ref{ThmISimple} and \ref{ThmSt}, to a final (subextremal) Kerr black hole. It is an interesting problem whether one can estimate the final black hole parameters from the geometry of the setup (i.e., the timelike geodesic that the small black hole follows in the limit $\eps\to 0$).
\end{enumerate}

\subsection{High-level description of the proof}

Delaying a detailed description of the proof of Theorem~\ref{ThmISimple} to~\S\S\ref{SsIGen}--\ref{SsIN} below, we give a broad overview here.

We will construct $g$ in Theorem~\ref{ThmISimple} as a solution of the Einstein equation in a generalized wave map gauge $\Ups(g)-\vartheta=0$, roughly $\Ups(g)=\tr_g(\nabla^g-\nabla^{g^0})$ where $g^0$ is a ``background metric'' interpolating between the initial and final Kerr metrics $g_{b_0}$ and $g_b$, with $b$ and $\vartheta$, as well as the gravitational wave tail $g-g^0$, regarded as unknowns; here $\vartheta$ lies in a suitable finite-dimensional space (determined by the index set $\cE_0$) of 1-forms with appropriate decay properties.\footnote{The identification of a suitable gauge is a central challenge in the analysis of the Einstein equation for long times: while any gauge breaks the diffeomorphism invariance of~\eqref{EqIEin} (i.e., while $\Ric(g)=0$ implies $\Ric(\phi^*g)=\phi^*\Ric(g)$ for all diffeomorphisms $\phi$, it is typically not true that $\Ups(g)-\vartheta=0$ implies $\Ups(\phi^*g)-\vartheta=0$ unless $\phi$ is the identity) and is thus suitable for local-in-time analysis, different gauge choices may lead to different asymptotic behaviors (even decay versus blow-up) of the spacetime metric.} The fact that the gauge issues in the Kerr stability problem, for initial data as in~\eqref{EqIDataPhg}--\eqref{EqIHbBound}, are \emph{finite-dimensional} in nature is one of the key insights of this work. We implement our generalized wave map gauge $\Ups(g)-\vartheta=0$ by solving the gauge-fixed Einstein vacuum equation
\begin{equation}
\label{EqIEinGauged}
  \Ric(g) - \tilde\delta_{g^0}^*\bigl(\Ups(g)-\vartheta\bigr) = 0,
\end{equation}
where $\tilde\delta_{g^0}^*$ is the symmetric gradient plus suitable zeroth-order terms, used to implement (a strong version of) \emph{constraint damping} \cite{BrodbeckFrittelliHubnerReulaSCP,GundlachCalabreseHinderMartinConstraintDamping} via the construction in the companion paper \cite{HintzKerrCD}. We stress that \emph{we work directly with the tensorial equation}~\eqref{EqIEinGauged} (or modifications thereof), which is a quasilinear wave equation for the symmetric 2-tensor $g$. In particular, our analysis does \emph{not} rely on an approximate reduction of the Einstein equation to the Teukolsky equation \cite{TeukolskySeparation} or any other equation governing scalar quantities related to $g$. (The only caveat is that a reduction to the \emph{mode stability} result for the Teukolsky equation---a wave equation for a section of a complex line bundle that governs one component of the linearized Weyl tensor---of Andersson--Ma--Paganini--Whiting \cite{AnderssonMaPaganiniWhitingModeStab} was used by Andersson--H\"afner--Whiting \cite{AnderssonHaefnerWhitingMode} to deduce the mode stability of the subextremal Kerr metric under linearized metric perturbations. This mode stability result, as well as a related result for the 1-form wave operator, is used in the present paper as a black box; see~\S\S\ref{SsWGMode} and \ref{SsWEMode}.)

As in prior work of the author with Vasy on the black hole stability problem for cosmological black holes \cite{HintzVasyKdSStability}, we will solve (a version of) \eqref{EqIEinGauged} using a Nash--Moser iteration in which we solve linearizations thereof, say $L_g h=f$, globally in each step, with precise control on regularity, asymptotics, and decay of the linear solutions $h$. Closing the nonlinear iteration in essence amounts to showing that (after adjusting the gauge condition, if necessary) linear solutions $h$ have sufficient decay for this detailed linear analysis to apply also for the linearization $L_{g+h}$ of~\eqref{EqIEinGauged} around the corrected metric $g+h$. (We will work with function spaces that encode partial expansions of $h$ in several asymptotic regimes. Setting up the nonlinear iteration so that nonlinear forward maps and right inverses of their linearizations are compatible is then a subtle task; see~\S\ref{SssINPhg}.)

The starting point for the linear analysis on dynamical spacetimes $(\Omega,g)$ settling down to Kerr is the general-purpose theory of the companion paper \cite{HintzNonstat2}, which yields high regularity but only weak polynomial bounds for solutions $h$ of $L_g h=f$. (The verification of the assumptions in \cite{HintzNonstat2} on the general wave-type operators studied there in the concrete case of linearizations $L_g$ of~\eqref{EqIEinGauged} is a nontrivial task; see~\S\ref{SssINAdm}. It requires a substantial amount of information on the spectral theory of linearized gauge-fixed Einstein operators on subextremal Kerr spacetimes, including mode stability at nonzero frequencies and a precise description of the resolvent near zero energy.)

We then extract increasingly refined asymptotics for $h$ using asymptotic analysis for the (time-translation-invariant) Kerr model $L_{g_b}$ of $L_g$, $g=g_b+\cO(\tilde\ft^{-2-\eps_\cK})$ (recording only the decay rate in spatially compact sets for this very rough overview), applied to the equation
\[
  L_{g_b}h=f-(L_g-L_{g_b})h.
\]
In particular, one can read off adjustments of the final black hole parameters as well as the required adjustments $\vartheta$ of the gauge condition from these asymptotics: the latter serve to set the linear momentum of the final black hole to zero,\footnote{This \emph{recoil} of the black hole is dealt with here by boosting the initial Cauchy hypersurface in the opposite direction; see already the final part of Theorem~\ref{ThmSt}.} avoid logarithmically divergent or asymptotically constant center-of-mass motions, and remove a plethora of other \emph{pure gauge} contributions to $h$ until one can prove $\tilde\ft^{-2-\eps_\cK}$-decay for $h$ (and appropriate asymptotics and decay also in all other asymptotic regimes of $\Omega$ including null infinity and future timelike infinity). The asymptotic analysis of $L_{g_b}$ relies on a systematic (and, to the author's knowledge, partially novel) general-purpose framework for wave decay on asymptotically flat spacetimes which we explain in~\S\ref{SsIGen}, and which is of independent interest. This framework guides our specific choices involved in the definition of our (linearized) gauge-fixed Einstein operator, as we discuss in~\S\ref{SsIEin}.

The partial polyhomogeneity~\eqref{EqIDataPhg} of the initial data propagates to the partial polyhomogeneity of $h$ as $\tilde\ft\to\infty$. For example, rather than seeing, say, weakly decaying but ``structureless'' $\cO(\tilde\ft^{-\alpha})$, $\alpha\in(0,2+\eps_\cK)$, modulations of the center-of-mass of the black hole, we prove that these modulations are themselves polyhomogeneous, i.e., finite sums (up to sufficiently decaying errors) of motions of the schematic form\footnote{Logarithmic powers of $\tilde\ft$ may appear as well.} $\tilde\ft^{-\alpha}\dot x$ for a \emph{finite} set of $\alpha$ (and associated vectors $\dot x\in\R^3$ describing the direction of motion). We are able to eliminate these \emph{finite-dimensional} pure gauge contributions to $h$ using a matching \emph{finite-dimensional} space of gauge modifications $\vartheta$; see~\S\S\ref{SssIEinElim} and \ref{SssINElim}.

\begin{rmk}[Full subextremal range]
\label{RmkISubex}
  We briefly discuss why we are able to treat the full subextremal range of Kerr black holes in this work.
  \begin{enumerate}
  \item The general linear estimates from \cite{HintzNonstat2} are proved mainly using methods of microlocal (phase space) analysis; these methods translate properties of the lift of the null-geodesic flow to phase space into (microlocal energy) estimates for solutions of wave equations. The main properties of this flow---which hold throughout the subextremal range (but not on extremal Kerr, where $|\bha|=\bhm$)---are:
    \begin{enumerate}
    \item the existence of an $\sfr$-normally hyperbolic trapped set \cite{HirschPughShubInvariantManifolds} (a $\CI$ codimension $3$ submanifold of phase space), as shown by Wunsch--Zworski \cite{WunschZworskiNormHypResolvent} and Dyatlov \cite{DyatlovWaveAsymptotics};
    \item a non-degenerate source at the conormal bundle of the event horizon, as pointed out by Vasy \cite{VasyMicroKerrdS} in a closely related context. (This is related to the non-degenerate red-shift effect \cite{DafermosRodnianskiRedShift}.)
    \end{enumerate}
  \item The mode stability result of Andersson--H\"afner--Whiting \cite{AnderssonHaefnerWhitingMode} for linearized gravity holds in the full subextremal range and is used here as a black box.
  \end{enumerate}
  In particular, our approach treats the full subextremal range of angular momenta in a uniform fashion. Restricting to axisymmetry, ultimately Schwarzschildean spacetimes, or small angular momenta would not lead to simplifications in our arguments.
\end{rmk}

\subsection{Prior work}
\label{SsIPrior}

The black hole stability problem has been the focus of much recent activity. A rough chronology of results on the nonlinear or full linearized Einstein equation on black hole spacetimes is as follows.

\begin{enumerate}[leftmargin=2.5em]
\item Seminal early works on mode stability for the linearized Einstein equation include those by Regge--Wheeler \cite{ReggeWheelerSchwarzschild}, Zerilli \cite{ZerilliPotential}, and Vishveshwara \cite{VishveshwaraSchwarzschild} in the Schwarzschild case, and Teukolsky \cite{TeukolskySeparation}, Wald \cite{WaldKerrPerturbation}, and Whiting \cite{WhitingKerrModeStability} for Kerr.
\item 
  Dafermos--Holzegel--Rodnianski \cite{DafermosHolzegelRodnianskiSchwarzschildStability} proved the linear stability of the Schwarzschild spacetime (in a double null gauge, normalized at the event horizon). This is the statement that solutions of the linearized Einstein vacuum equation $D_{g_\bhm}\Ric(h)=0$ around a Schwarzschild metric $g_\bhm=g_b$, $b:=(\bhm,0)$, decay, at a quantitative (inverse polynomial) rate, to a member
  \begin{equation}
  \label{EqIPriorLin}
    \dot g_b(\dot b):=\frac{\dd}{\dd s}g_{b+s\dot b}\Big|_{s=0},\quad \dot b\in\R^4,
  \end{equation}
  of the linearized Kerr family. The diffeomorphism invariance of the Einstein equation persists in the linearized problem in that $D_{g_b}\Ric(\cL_V g_b)=0$ for all vector fields $V$; here $\cL_V$ is the Lie derivative.
\item 
  Hintz--Vasy \cite{HintzVasyKdSStability} proved the nonlinear stability of slowly rotating Kerr--de~Sitter black holes; these are solutions of the Einstein vacuum equation in the presence of a positive cosmological constant $\Lambda>0$. The approach in the present paper is strongly influenced by \cite{HintzVasyKdSStability}, which introduced the strategy of adjusting generalized harmonic gauges with finite-dimensional gauge modifications designed to eliminate poorly decaying metric perturbations arising in the asymptotics of solutions of the linearized gauge-fixed Einstein equation; we discuss some of the key \emph{differences} between Kerr--de~Sitter and Kerr momentarily. The author subsequently used the same methods to prove the nonlinear stability of (charged and rotating) Kerr--Newman--de~Sitter black holes for subextremal charges and small angular momenta \cite{HintzKNdSStability}.
\item 
  Hung--Keller--Wang \cite{HungKellerWangSchwarzschild} proved the linear stability of the Schwarzschild metric in generalized harmonic gauge (relying on spherically harmonic decompositions and the analysis of gauge-invariant quantities satisfying equations of Regge--Wheeler type \cite{ReggeWheelerSchwarzschild}), extending earlier work by Hung--Keller \cite{HungKellerAxial}; later developments of this approach include \cite{HungSchwarzschildOdd,HungSchwarzschildEven} as well as work by Johnson \cite{JohnsonSchwarzschild}.
\item 
  Klainerman--Szeftel \cite{KlainermanSzeftelPolarized} proved the nonlinear stability of the Schwarzschild spacetime in the class of axially symmetric and polarized spacetimes. These conditions ensure that the final black hole must again be a Schwarzschild black hole, though its center-of-mass may move along the symmetry axis. This work introduces, as a starting point for the foliation of spacetime by null cones, the notion of GCM spheres---topological 2-spheres inside of dynamical spacetimes on which certain geometric quantities attain Schwarzschildean model values---and (spacelike) GCM hypersurfaces foliated by them.
\item 
  Andersson--B\"ackdahl--Blue--Ma \cite{AnderssonBackdahlBlueMaKerr} and shortly after also H\"afner--Hintz--Vasy \cite{HaefnerHintzVasyKerr,HaefnerHintzVasyKerrErratum} proved the linear stability of slowly rotating Kerr spacetimes (i.e., for $|\bha_0|\ll\bhm_0$). The analysis of \cite{AnderssonBackdahlBlueMaKerr} is based on a reduction to the Teukolsky equation and the recovery of the metric from the Teukolsky scalars in the outgoing radiation gauge. The approach of \cite{HaefnerHintzVasyKerr} on the other hand is to analyze the linearized gauge-fixed Einstein equation directly as a tensorial wave-type equation, and use methods of spectral theory (in particular, a detailed description of the resolvent at low energies) to prove asymptotics and decay. In the present paper, we use (technically and conceptually) streamlined versions of arguments introduced in \cite{HaefnerHintzVasyKerr} in order to extract asymptotic expansions for solutions of the linearization of the gauge-fixed Einstein equation around the final (Kerr) black hole metric.
\item 
  Giorgi \cite{GiorgiRNLinear} proved the linear stability of Reissner--Nordstr\"om black holes, which are solutions of the Einstein--Maxwell system describing charged but spherically symmetric black holes, in the full subextremal range of charges (following her earlier work \cite{GiorgiRNLinearSmall} for small charges).
\item 
  Klainerman--Szeftel \cite{KlainermanSzeftelGCM1,KlainermanSzeftelGCM2} began generalizing their approach from \cite{KlainermanSzeftelPolarized} to the Kerr case by constructing GCM spheres on perturbations of subextremal (or extremal) Kerr spacetimes.
\item 
  Dafermos--Holzegel--Rodnianski--Taylor \cite{DafermosHolzegelRodnianskiTaylorSchwarzschild} proved the codimension $3$ stability of the Schwarzschild family of black holes (using a double null gauge, anchored using special spheres). This means that the authors construct a ``full-dimensional'' subset of initial data, near those of Schwarzschild, for which the final black hole has vanishing angular momentum $\bha=0\in\R^3$. (Decay to Schwarzschild is proved at the same time as this subset is constructed.)
\item 
  Klainerman--Szeftel \cite{KlainermanSzeftelKerr} proved the nonlinear stability of the Kerr family of black holes for small angular momenta, i.e., for $|\bha_0|\ll\bhm_0$ in the context of Theorem~\ref{ThmISimple} and for initial data $\gamma,r k=\cO(r^{-\frac32-\eps_0})$. The proof utilizes a formalism developed with Giorgi \cite{GiorgiKlainermanSzeftelFormalism} 
  (see also \cite{BenomioFormalismI}) for the stability problem based on an analysis of the Teukolsky equation and its Chandrasekhar-type transformation (essentially introduced in \cite{ChandrasekharBlackHoles} and used also in \cite{DafermosHolzegelRodnianskiSchwarzschildStability}) to a generalized Regge--Wheeler equation. The decay estimates proved in \cite{GiorgiKlainermanSzeftelStability} 
  and the construction of GCM hypersurfaces on perturbations of Kerr by Shen \cite{ShenGCMKerr} complete the proof.
\item 
  Fang \cite{FangKdSLinear,FangKdS} reworked the proof of the nonlinear stability of slowly rotating Kerr--de~Sitter spacetimes in \cite{HintzVasyKdSStability} from a global (Nash--Moser) nonlinear iteration into a more classical bootstrap argument.
\item 
  Benomio \cite{BenomioSchwarzschildLinear} gave a new proof of the linear stability of the Schwarzschild metric, based on \cite{BenomioFormalismI}.
\item 
  He \cite{HeLinearKerrNewman} proved the linear stability of Kerr--Newman black holes (charged rotating black holes) for small charges and angular momenta, building on the approach of \cite{HaefnerHintzVasyKerr}.
\item 
  H\"afner--Hintz--Vasy \cite{HaefnerHintzVasyKerrLarge} proved the linear stability of Kerr black holes in the full subextremal range; the approach is the same as in \cite{HaefnerHintzVasyKerr}, but it could now be extended beyond the regime of small angular momenta in view of the mode stability result \cite{AnderssonHaefnerWhitingMode} and the verification of certain non-degeneracy properties of the spectral family in \cite{HintzGlueLocIII}.
\item 
  Hintz--Petersen--Vasy \cite{HintzPetersenVasyKdS} proved the nonlinear stability of Kerr--de~Sitter black holes in the full subextremal range, \emph{conditional} on the validity of mode stability for the linearized Einstein equation---which, unlike in the Kerr case, remains an open problem except in the slow-rotation regime (see also \cite{HintzKdSMS} for the small mass regime). The basic strategy is the same as in \cite{HintzVasyKdSStability}, but the implementation of constraint damping is considerably more involved.
\end{enumerate}

We also mention the papers by Fournodavlos--Schlue \cite{FournodavlosSchlueExpanding} and Hintz--Vasy \cite{HintzVasyKdSCosm} on the stability of the cosmological (or expanding) region of Kerr--de~Sitter spacetimes, though in that region (which lies beyond the cosmological horizon of the black hole), the black hole nature of the spacetime plays only a very minor role: it is not any property of the final black hole that is at issue, but rather the behavior of the metric at the conformal boundary, which is de~Sitter-like.

Stability problems in general relativity more broadly have been a driving force behind the development of the modern theory of geometric wave equations. We restrict attention here only to de~Sitter and Minkowski spacetimes. The first proofs of nonlinear stability are Friedrich's proof of the nonlinear stability of de~Sitter space \cite{FriedrichDeSitterPastSimple} (extended and revisited in \cite{AndersonStabilityEvenDS,RingstromEinsteinScalarStability,CicortasScatteringdS,LeimbacherdS}) and, for a restricted class of initial data (later proved to be non-trivial \cite{CorvinoScalar,CorvinoSchoenAsymptotics,ChruscielDelaySimple,ChruscielDelayMapping}), of Minkowski space \cite{FriedrichStability} via a conformal method. Christodoulou--Klainerman \cite{ChristodoulouKlainermanStability} introduced powerful tools to go beyond the (physically unnatural \cite{ChristodoulouNoPeeling,KehrbergerScri1,KehrbergerScri2,KehrbergerScri3,KehrbergerScri4}) conformally smooth case and proved the nonlinear stability of Minkowski space for asymptotically flat data with $\cO(r^{-\frac32-\eps_0})$-decay. There are now many proofs of the nonlinear stability of Minkowski space in various gauges, coupled to various matter models, and under various decay assumptions; these include \cite{KlainermanNicoloEvolution,LindbladRodnianskiGlobalExistence,BieriZipserStability,LindbladRodnianskiGlobalStability,WangThesis,WangRadiation,LeFlochMaEinsteinMassive,WangEinsteinKleinGordon,TaylorEinsteinVlasov,KeirWeak,LindbladTaylorVlasov,HintzVasyMink4,FajmanJoudiouxSmuleviciEinsteinVlasov,ShenMinkExtStab,IonescuPausaderKG,HintzMink4Gauge,ShenMinkBorderline}. The observation of Lindblad--Rodnianski \cite{LindbladRodnianskiWeakNull} that stability can be proved in harmonic gauge is, in a modification of the version of \cite{HintzMink4Gauge}, at the heart of our analysis near $\scri^+$ in the present paper.

\medskip

The Kerr--de~Sitter (KdS) stability problem studied in \cite{HintzVasyKdSStability,FangKdS,HintzPetersenVasyKdS} offers several simplifications compared to the Kerr stability problem, including:
\begin{enumerate}[label=(\roman*),leftmargin=2.5em]
\item Linear waves on KdS decay \emph{exponentially fast} (up to finite-dimensional obstructions which are described by resonances). This was shown for the scalar wave equation on Schwarzschild--de~Sitter (SdS) spacetimes by Bony--H\"afner \cite{BonyHaefnerDecay} and Melrose--S\`a Barreto--Vasy \cite{MelroseSaBarretoVasySdS,MelroseSaBarretoVasyResolvent} using work of S\`a Barreto--Zworski \cite{SaBarretoZworskiResonances} (see also \cite{BachelotSchwarzschildScattering,BachelotMotetBachelotSchwarzschild}), on slowly rotating KdS by Dyatlov \cite{DyatlovQNM,DyatlovQNMExtended,DyatlovAsymptoticDistribution} (see \cite{HintzPsdoInner,HintzVasyKdsFormResonances} for similar results for tensorial wave equations) and Vasy \cite{VasyMicroKerrdS}---with estimates at the trapped set by Wunsch--Zworski \cite{WunschZworskiNormHypResolvent} playing an important role---and in the full subextremal range by Petersen--Vasy \cite{PetersenVasySubextremal}. A corresponding statement for the linearized gauge-fixed Einstein equation is a key ingredient in \cite{HintzVasyKdSStability,HintzPetersenVasyKdS}. On the level of analysis, exponential decay in time at rate $\alpha>0$ relies on the existence of a \emph{meromorphic continuation} of the resolvent $\wh{\Box_g}(\sigma)^{-1}$ of the spectral family of the wave operator $\Box_g$ to a half-space $\Im\sigma\geq-\alpha$. Exponential decay also makes nonlinear effects easy to handle. (We mention here also Mavrogiannis' approach \cite{MavrogiannisSdSMorawetz,MavrogiannisSdSQuasi,MavrogiannisKdSMorawetz,MavrogiannisKdSDecay} to exponential decay which relies more strongly on physical-space energy methods.)
\item The stability of (a neighborhood of) the domain of outer communications of KdS can be proved, by simple domain of dependence considerations, on a manifold $[0,\infty)_{t_*}\times X$ where the spatial manifold $X=[r_-,r_+]\times\Sph^2$ is \emph{compact}; here $r=r_-$, resp.\ $r=r_+$ lies a bit beyond the event, resp.\ cosmological horizon. In particular, there is only \emph{one} asymptotic regime in which asymptotic analysis is necessary, namely $t_*\to\infty$, and spectral theory is the tool of choice there.
\end{enumerate}

\noindent
By contrast:
\begin{enumerate}[leftmargin=2.5em]
\myitem{ItIKerrRate}{i'} Linear waves on Kerr decay only at certain \emph{inverse polynomial} rates; we recall the relevant literature below. On the spectral side, this is related to the fact that the resolvent $\wh{\Box_g}(\sigma)^{-1}$ is no longer meromorphic, but rather has, loosely speaking, an \emph{essential singularity} at $\sigma=0$, and indeed features terms of the type $\sigma^2\log\sigma$ \cite{HintzPrice}. (Away from zero energy, meromorphicity \emph{does} hold, as shown by Stucker \cite{StuckerKerrQNM} and Gajic--Warnick \cite{GajicWarnickKerrQNM}.) Weak decay rates for linear waves---especially in the presence of zero energy bound states or resonances---may render nonlinear interactions potentially disastrous.
\myitem{ItIKerrRegime}{ii'} Kerr has several asymptotic regimes, each with its distinct mechanism for asymptotics and decay. In this paper (adopting the perspective of \cite{HintzPrice,HintzNonstat}), we need to deal with \emph{four} asymptotic regimes of $\Omega$, encoded as ideal boundary hypersurfaces ``at infinity.'' These are: spatial infinity $I^0$ (regarded as a cylinder), where the decay of the initial data simply propagates; future null infinity $\scri^+$, where waves have definite decay rates (typically $\sim r^{-1}$) related to the existence of radiation fields, and where the structure of nonlinear interactions becomes important (cf.\ the null condition \cite{KlainermanNullCondition,ChristodoulouGlobalSolutionsSmallData}); and future timelike infinity, split into two regimes: one corresponding to time going to infinity in spatially compact regions (we call this the \emph{Kerr face} below), where spectral theory is our main tool for proving asymptotics and decay; and another one corresponding to leaving spatially compact regions at positive but subluminal speeds, where certain model problems on the limiting Minkowski spacetime are key for understanding asymptotics and decay.
\end{enumerate}

\medskip

Returning to a review of the literature, we recall that the Teukolsky equation, which plays an important role in many of the aforementioned works on black hole stability, has been studied in its own right, though often, of course, with an eye towards applications to the stability problem. There is one Teukolsky equation for each value $s\in\frac12\Z$ of a spin parameter; for $s=0$, it reduces to the scalar wave equation. For the latter, decay was proved in the full subextremal range by Dafermos--Rodnianski--Shlapentokh-Rothman \cite{DafermosRodnianskiShlapentokhRothmanDecay}, following earlier works in the slowly rotating case by Andersson--Blue \cite{AnderssonBlueHiddenKerr} and Tataru--Tohaneanu \cite{TataruTohaneanuKerrLocalEnergy} as well as by Blue--Soffer \cite{BlueSofferSchwarzschildDecay} in the Schwarzschild case; see also \cite{MarzuolaMetcalfeTataruTohaneanuStrichartz,TohaneanuKerrStrichartz}. A crucial ingredient is the mode stability proved by Whiting~\cite{WhitingKerrModeStability} and Shlapentokh-Rothman \cite{ShlapentokhRothmanModeStability}. Sharp $t^{-3}$-decay for scalar waves (known as Price's law \cite{PriceLawI,PriceLawII,PriceBurkoLaw}) was proved by Tataru \cite{TataruDecayAsympFlat} (see also \cite{MorganDecay,MorganWunschPrice} for related results) in the slowly rotating regime, and (including matching generic lower bounds) by Hintz \cite{HintzPrice} and Angelopoulos--Aretakis--Gajic \cite{AngelopoulosAretakisGajicKerr} in the full subextremal range; earlier work by Donninger--Schlag--Soffer \cite{DonningerSchlagSofferPrice,DonningerSchlagSofferSchwarzschild} (and \cite{AngelopoulosAretakisGajicRNPrice}) treated the Schwarzschild (and related Reissner--Nordstr\"om) case. (We remark that while we do not need sharp decay results here, the \emph{method} introduced in \cite{HintzPrice}, and also in \cite{TataruDecayAsympFlat}, to extract late-time asymptotics of waves from regularity properties of the resolvent at low energies---which goes back to Jensen--Kato \cite{JensenKatoResolvent}---does play a crucial role also in the present paper.) Sharp decay rates on \emph{dynamical} spacetimes are obtained in \cite{MetcalfeTataruTohaneanuPriceNonstationary} by combining integrated local energy decay estimates with properties of solutions to the wave equation on Minkowski space; related results for nonlinear wave equations include \cite{LooiDecayEnergyCritical,LooiDecayCubic,LooiDecayQuasilinearImproved,LukOhTwoTails,LooiXiongSemilinearAsymp}. General results for quasilinear scalar waves on slowly rotating Kerr spacetimes were proved by Lindblad--Tohaneanu \cite{LindbladTohaneanuSchwarzschildQuasi,LindbladTohaneanuKerrQuasi}. Recent developments include \cite{HolzegelKauffmannKerrFirstOrder,ShlapentokhRothmanTohaneanuC1Kerr}, and also the work by Ma--Szeftel \cite{MaSzeftelEnergyKerr} on integrated local energy decay estimates on perturbations of subextremal Kerr that are consistent with their expectations for the dynamical metrics arising in the stability problem; for another novel approach to the study of linear and nonlinear wave equations on subextremal Kerr spacetimes based on a new top order energy estimate, see Dafermos--Holzegel--Rodnianski--Taylor \cite{DafermosHolzegelRodnianskiTaylorQuasilinear,DafermosHolzegelRodnianskiTaylorQuasilinear2}.

The Teukolsky equations with integer spins $s=\pm 2$, resp.\ $s=\pm 1$ arise in the study of linearized gravity around Kerr, resp.\ the Maxwell equations on Kerr. In the case $s=\pm 2$, Blue--Soffer \cite{BlueSofferSchwarzschildSpin2} established the decay estimates in the Schwarzschild case. Sharp decay was proved for \emph{all} $s\in\frac12\Z$ by Millet \cite{MilletTeukolskyDecay} using an adaptation of the methods of \cite{HintzPrice}; prior work includes \cite{MaZhangSchwarzschildDiracSharp,MaZhangTeukolsky}. For nonlinear applications, non-sharp decay results typically suffice, but one may ask for more precise control of regularity (including energy boundedness statements that do not lose derivatives); results of this flavor were proved by Dafermos--Holzegel--Rodnianski \cite{DafermosHolzegelRodnianskiTeukolsky} for $s=\pm 2$ in the slowly rotating regime, and by Shlapentokh-Rothman and Teixeira da Costa \cite{ShlapentokhRothmanTeixeiradCTeukolskyI,ShlapentokhRothmanTeixeiradCTeukolskyII} in the full subextremal range. We also mention the recent work by Ma--Szeftel \cite{MaSzeftelTeukolsky} that extends their previous \cite{MaSzeftelEnergyKerr} to the case of the Teukolsky equation. Work regarding the case $s=\pm 1$ and Maxwell's equations includes \cite{AnderssonBlueMaxwellKerr,SterbenzTataruMaxwellSchwarzschild,MetcalfeTataruTohaneanuMaxwellKerr,MaMaxwellAlmost,MaMaxwellKerr,BenomioTdCMaxwell}.

\medskip

The \emph{extremal} Kerr spacetime offers many unresolved challenges: even decay for solutions of the linear wave equation is presently unknown outside of axisymmetry \cite{AretakisExtremalKerr}. Conditional decay results have been established by Gajic \cite{GajicXKerrInstab}, and mode stability is known by work of Teixeira da Costa \cite{TeixeiradCModes}. The statement of Theorem~\ref{ThmISimple} is widely expected to be \emph{false} for perturbations of the initial data of extremal Kerr black holes, as they are conjectured to lie at the threshold between black hole formation and dispersion. Results of the latter kind in spherically symmetric models have recently been proved by Kehle--Unger \cite{KehleUngerExtremalCritical} (see also \cite{KehleUngerThirdLaw,AngelopoulosKehleUngerModuli}). For a nonlinear stability result for an extremal (Reissner--Nordstr\"om) black hole, see \cite{AngelopoulosKehleUngerXRNStab}.

\subsection{A systematic framework for linear waves on asymptotically flat spacetimes}
\label{SsIGen}

In the remainder of this introduction, we provide an in-depth overview of the methods and ideas behind the proof of Theorem~\ref{ThmISimple}. We begin with a detailed explanation of what determines decay and asymptotics of linear waves on asymptotically flat spacetimes in each of the four asymptotic regimes mentioned in \eqref{ItIKerrRegime} on page~\pageref{ItIKerrRegime}. The application of the framework laid out here to the linearized Einstein equation on subextremal Kerr spacetimes is explained in~\S\ref{SsIEin}; dynamical and nonlinear aspects are the subject of~\S\ref{SsIN}.

First of all, it is highly convenient to make these four regimes (and the transitions between them) tangible and to encode asymptotics in each in a concrete manner by attaching ``ideal'' boundary hypersurfaces ``at infinity'' to the non-compact spacetime manifold $\R_t\times\R^3_x$ (or $\R_t\times(\R^3_x\setminus B_{\bhm_0}(0))$ for Kerr). The utility of compactifications was recognized already by Penrose \cite{PenroseAsymptotics}, although, following Melrose and as in earlier work of the author such as \cite{HintzVasyMink4,HintzNonstat,HintzNonstat2,HintzVasyScrieb,Hintz3b}, it is important to use a more elaborate compactification for analytic (rather than conformal geometric) purposes.

\begin{figure}[!ht]
\centering
\includegraphics{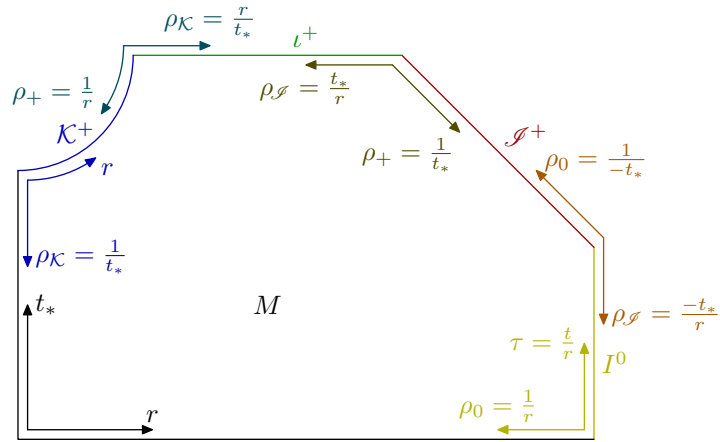}
\caption{Illustration (and construction) of the compactified spacetime manifold $M$ (without the factor of $\Sph^2_\omega$). (In order to include points in $\scri^+$ where $t_*=0$, one simply replaces $\rho_0$ and $\rho_\sscri$ near $I^0\cap\scri^+$ by $\rho_0=\frac{1}{1-t_*}$ and $\rho_\sscri=\frac{1-t_*}{r}$.) If one defines $M$ as a compactification of $\{r\geq\bhm_0,\ t\geq-\frac12 r\}$, it contains the domain $\Omega$ from~\eqref{EqIOmega}.}
\label{FigIMwc}
\end{figure}

Let us split $\R^4=\R_{t_*}\times\R^3_x$. (While we are not yet introducing any metrics, the reader should think of $t_*=t-r$ for $r\gtrsim 1$---smoothly glued to $t_*=t$ for $r\lesssim 1$---on the Minkowski spacetime with metric $\ubar g:=-\dd t^2+\dd r^2+r^2\slg=-\dd t_*^2-2\,\dd t_*\,\dd r+r^2\slg$; and on Kerr $t_*$ would be such that its level sets are transversal to future null infinity and the future event horizon.) We write
\[
  r=|x|,\quad \omega=\frac{x}{|x|},\quad t=t_*+r.
\]
We shall work in $r\geq 1>0$, which (upon replacing $1$ by $\bhm_0$) is the setting relevant for Kerr; the modifications needed to work globally on $\R_{t_*}\times\R^3_x$ are purely notational. We moreover only consider the region $t\geq 0$, or more permissively $t\geq-\frac12 r$. The four regimes we aim to capture are:
\begin{enumerate}
\item \emph{spacelike infinity} $I^0$: $\frac{t}{r}<1$, $r\to\infty$;
\item \emph{future null infinity} $\scri^+$: $t_*\in\R$, $r\to\infty$;
\item \emph{punctured future timelike infinity} $\iota^+$: $\frac{r}{t_*}\in(0,\infty)$ (equivalently: $\frac{r}{t}\in(0,1)$), $t_*\to\infty$;
\item \emph{the Kerr face} $\cK^+$: $1\leq r<\infty$, $t_*\to\infty$.
\end{enumerate}
(As we will explain in detail in~\S\S\ref{SssIGen0}--\ref{SssIGenK}, the influences on decay rates and asymptotic expansions of waves differ for each of these regimes.) To this end, we define a compact manifold $M$ with corners as illustrated in Figure~\ref{FigIMwc}. Thus, $M$ is obtained by taking the disjoint union of $\{(t,r,\omega)\colon r\geq 1,\ t\geq-\frac12 r\}$ and the various coordinate charts, quotiented out by the obvious equivalence relation with finite points (and then by transitivity between different compactified charts); e.g., a point $(t_*,r,\omega)$ with $t_*<0$ is identified with $(\rho_0,\rho_\sscri,\omega)=(\frac{1}{-t_*},\frac{-t_*}{r},\omega)$.\footnote{We will abuse notation throughout and write $\rho_\bullet$ for a local coordinate or globally defined function that defines (i.e., vanishes simply at) the boundary hypersurface corresponding to the subscript ``$\bullet$.'' This is why different choices of $\rho_+$ etc.\ in Figure~\ref{FigIMwc} are inconsistent. We will always make it clear which choice of $\rho_+$ etc.\ we are working with at any given moment.} We can then define $I^0$ to be the boundary hypersurface consisting of all points with $\rho_0=0$, and similarly $\scri^+:=\{\rho_\sscri=0\}$, $\iota^+:=\{\rho_+=0\}$, and $\cK^+:=\{\rho_\cK=0\}$. (The transition from $I^0$ to $\scri^+$ already features in Friedrich's work \cite{FriedrichSpaceConformal}.) Then:
\begin{enumerate}
\item a point in $(I^0)^\circ$ (the interior of $I^0$) is the endpoint of a spacelike geodesic $s\mapsto(t,r,\omega)=(s,s v,\omega)$, $v>1$, $\omega\in\Sph^2$, in Minkowski space;
\item a point in $(\sscri^+)^\circ$ is the endpoint of a null-geodesic $s\mapsto(t_*,r,\omega)=(t_*,s,\omega)$;
\item a point in $(\iota^+)^\circ$ is the endpoint of a timelike geodesic with non-zero spatial velocity, i.e., $s\mapsto(t_*,r,\omega)=(s,s R,\omega)$ where $R\in(0,\infty)$;
\item a point in $(\cK^+)^\circ$ is the endpoint of a timelike geodesic with vanishing spatial velocity, i.e., $s\mapsto(t_*,r,\omega)=(s,r,\omega)$.
\end{enumerate}

Except briefly in~\S\ref{SssINAdm} below, we work with the notion of \emph{b-regularity} on $M$ \cite{MelroseMendozaB,MelroseTransformation,MelroseAPS}, i.e., iterated regularity with respect to smooth vector fields that are tangent to $\pa M$; concretely, in $(t,r,\omega)$-coordinates, this is equivalent to regularity under $\la t-r\ra\pa_t$, $r(\pa_t+\pa_r)$, and $\pa_\omega$ (rotation vector fields), and in $(t_*,r,\omega)$-coordinates and for $t_*\geq 1$ it is equivalent to regularity under $t_*\pa_{t_*}$ and $r\pa_x$ (i.e., $r\pa_r$, $\pa_\omega$). This is the strongest notion of regularity on $M$ that is compatible with the types of expressions (e.g., $r^{-\alpha}$ and $t_*^{-\beta}$) that arise in our asymptotic analysis.

\begin{rmk}[b-regularity]
\label{RmkIb}
  For a detailed discussion of b-regularity and its motivation in the context of waves on asymptotically flat spacetimes, we refer the reader to \citeAF{\S\ref*{SssIMb}}. We only note here that prior stability results for Kerr work with weaker notions of regularity, e.g., regularity under (roughly speaking) $\pa_{t_*}$ and $r\pa_x$ in \cite{KlainermanSzeftelKerr}. Such anisotropic notions (from our compactified perspective) are rather delicate to work with; for example, the decay rates at $(\iota^+)^\circ$ of derivatives $\pa_\mu u$ of regular functions $u$ depend on $\pa_\mu\in\{\pa_{t_*},\pa_{x^1},\pa_{x^2},\pa_{x^3}\}$. On the other hand, b-regularity is very robust, and in any case conceptually easy to recover (as shown in \cite{HintzNonstat2}).
\end{rmk}

In order to describe decay and (partial) asymptotic expansions of functions $u$ defined on (subsets of) $\R_{t_*}\times\R^3_x$, we need to introduce some terminology.
\begin{enumerate}
\item\label{ItIExpDecay}{\rm (Decay.)} We say that $u$ has
  \begin{enumerate}
  \item \emph{(decay) order $\alpha_0$ at $(I^0)^\circ$} if $|u|\lesssim\rho_0^{\alpha_0}$ near compact subsets of $(I^0)^\circ$,
  \item \emph{(decay) orders $\alpha_0$ and $\alpha_\sscri$ at $I^0$ and $\scri^+$} if $|u|\lesssim\rho_0^{\alpha_0}\rho_\sscri^{\alpha_\sscri}$,
  \end{enumerate}
  and so on, with the same bounds holding also for all b-derivatives of $u$. (Such $u$ are often called \emph{conormal} relative to $\rho_0^{\alpha_0}L^\infty$ etc.)
\item{\rm (Partial expansions.)} Let $\cE_0\subset\C\times\N_0$ be an \emph{index set}, which in particular means that $\cE_0\cap\{(z,j)\colon\Re z\leq C_0\}$ is finite for all $C_0$. We then say that $u$ has
  \begin{enumerate}
  \item \emph{order $(\cE_0,\alpha_0)$ at $(I^0)^\circ$} if
    \begin{equation}
    \label{EqIExp0}
      u(\rho_0,\tau,\omega) = \sum_{ \substack{ (z,j)\in\cE_0 \\ \Re z\leq\alpha_0 } } \rho_0^z(\log\rho_0)^j u_{(z,j)}(\tau,\omega) + \tilde u(\rho_0,\tau,\omega)
    \end{equation}
    where $u_{(z,j)}\in\CI([-\frac12,1)_\tau\times\Sph^2_\omega)$, and $\tilde u$ has order $\alpha_0$ at $(I^0)^\circ$,
  \item \emph{orders $(\cE_0,\alpha_0)$ and $\alpha_\sscri$ at $I^0$ and $\scri^+$} if $u=u(\rho_0,\rho_\sscri,\omega)$ has such an expansion, with each coefficient $u_{(z,j)}(\rho_\sscri,\omega)$ (which is a function on $I^0$) having order $\alpha_\sscri$ at $I^0\cap\scri^+=I^0\cap\{\rho_\sscri=0\}$, and with $\tilde u$ having orders $\alpha_0$ and $\alpha_\sscri$.
  \item \emph{orders $(\cE_0,\alpha_0)$ and $(\cE_\sscri,\alpha_\sscri)$ at $I^0$ and $\scri^+$} if
    \[
      u(\rho_0,\rho_\sscri,\omega) = \sum_{ \substack{ (z,j)\in\cE_\sscri \\ \Re z\leq\alpha_\sscri } } \rho_\sscri^z(\log\rho_\sscri)^j u_{(z,j)}(\rho_0,\omega) + \tilde u(\rho_0,\rho_\sscri,\omega)
    \]
    where each $u_{(z,j)}$ (which is a function on $\scri^+$) has order $(\cE_0,\alpha_0)$ at $\scri^+\cap I^0=\scri^+\cap\{\rho_0=0\}$, and $\tilde u$ has orders $(\cE_0,\alpha_0)$ and $\alpha_\sscri$ at $I^0$ and $\scri^+$. Symmetrically, then, $u$ has an expansion also at $I^0$ (i.e., into powers of $\rho_0$ and $\log\rho_0$) with coefficients that have order $(\cE_\sscri,\alpha_\sscri)$ at the other boundary hypersurface $\scri^+$.
  \item{\rm (Full expansions.)} We say that $u$ has \emph{index set $\cE_0$ at $(I^0)^\circ$} if it has an expansion~\eqref{EqIExp0} for all $\alpha_0$; and so on.
  \end{enumerate}
\end{enumerate}
We write $u\in\cA^{\alpha_0}$, $u\in\cA^{\alpha_0,\alpha_\sscri}$, $u\in\cA^{(\cE_0,\alpha_0)}$, $u\in\cA^{(\cE_0,\alpha_0),(\cE_\sscri,\alpha_\sscri)}$, etc.\ in these cases, with the domain on which we work being clear from the context (or made explicit in the notation). See~\S\ref{SssTMExp} for details. We use $L^\infty$-based spaces in this introduction since they are easier to parse. However, we emphasize that in the bulk of the paper we almost exclusively work with $L^2$-based function spaces (as these are compatible with microlocal and spectral theoretic/Fourier-based techniques).

For $\alpha\in\C$ and $j\in\N_0$, we use the short-hand notation
\[
  (\alpha,j) := \{ (\alpha+n,l) \colon n\in\N_0,\ 0\leq l\leq j \}.
\]
For example, consider the solution $u$ of the initial value problem $\ubar\Box u=(-D_t^2+\Delta_x)u=0$ for the wave equation on Minkowski space with initial data $u|_{t=0},\la r\ra\pa_t u|_{t=0}$ of decay order $\alpha_0\in(1,2)$ as $r\to\infty$. Then $u$ has orders $\alpha_0$, $((1,0),\alpha_0)$, $\alpha_0$, and $\alpha_0$ at $I^0$, $\scri^+$, $\iota^+$, and $\cK^+$ (i.e., $u\in\cA^{\alpha_0,\ ((1,0),\alpha_0),\ \alpha_0,\ \alpha_0}$); the leading term in the $\scri^+$-expansion (corresponding to the element $(1,0)$ of the $\scri^+$-index set) encodes the radiation field of $u$. It is easier (and equivalent) to study \emph{forward problems} $\ubar\Box u=f$, i.e., with $u$ supported in the causal future of the support of $f$; the above conclusion for $u$ is then valid when $f$ has order $\alpha_0+2$ at $I^0$ and order $\infty$---weaker orders would suffice, of course---at all other boundary hypersurfaces (i.e., $f\in\cA^{\alpha_0+2,\ \infty,\ \infty,\ \infty}$). We pictorially represent this statement on the compactification in Figure~\ref{FigIMwc} as follows:
\begin{equation}
\label{EqIfuMink}
  \raisebox{-1.5em}{\includegraphics{EqIfuMink}}
\end{equation}
We will use similar pictorial representations in the remainder of the introduction, often showing only parts thereof when discussing asymptotics in a particular region of spacetime.

\begin{figure}[!ht]
\centering
\includegraphics{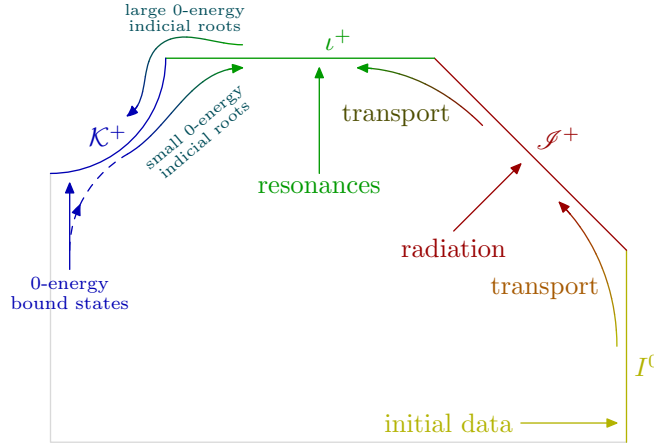}
\caption{This diagram summarizes what influences decay and asymptotics in the four asymptotic regimes of spacetime. The arrow labeled ``initial data'' is discussed in~\S\ref{SssIGen0}, the first ``transport'' arrow and the ``radiation'' arrow in~\S\ref{SssIGenScri}, the second ``transport'' arrow and the ``resonances'' arrow in~\S\ref{SssIGen1}, the ``indicial roots'' arrows in~\S\S\ref{SssIGen2La}--\ref{SssIGen2Sm}, and the ``0-energy bound states'' arrow in~\S\ref{SssIGenK}.}
\label{FigIDecay}
\end{figure}

\bigskip
Our goal is to explain the diagram in Figure~\ref{FigIDecay} that illustrates how the decay of forward solutions of linear wave equations propagates or is influenced by spectral data. The scalar wave operator on $(3+1)$-dimensional Minkowski space $\ubar\Box=:\ubar\Box_{3+1}$ has sufficiently rich structure to illustrate many but not all parts of this diagram; other models to keep in mind are the wave operator $\ubar\Box_{4+1}$ on $(4+1)$-dimensional Minkowski space, couplings $\ubar\Box+V$ with stationary potentials $V$ (e.g., $0\leq V\in\CIc(\R_x^3)$ for maximal simplicity), and the scalar wave operator $\Box_{g_b}$ on a subextremal Kerr spacetime or its tensorial analogues (e.g., the 1-form wave operator). Much of what we discuss applies also for wave-type operators on spacetimes that merely \emph{settle down} at suitable rates to these models.

This diagram will not only explain~\eqref{EqIfuMink} in a conceptual manner, but also more precise and more general versions thereof. For example, if the $I^0$-order of $f$ is $(\cE_0+2,\alpha_0+2)$ where $\alpha_0>1$, one can extract (joint) asymptotic expansions at $\scri^+$, $\iota^+$, and $\cK^+$ with order $\alpha_0-\eps$ errors; and in fact the expansion of $u$ at $\cK^+$ has a very specific structure, e.g., its leading-order term is $c t_*^{-z}$ for some \emph{constant} $c\in\C$ if $\cE_0=(z,0)$ (with appropriate modifications for operators other than $\ubar\Box$). See~\S\ref{SssIGen2La} for (a considerable strengthening of) this.

Some of the features of Figure~\ref{FigIDecay} appear in earlier literature in some form, but their synthesis to a coherent global picture seems to be new; moreover, the precise nature of the communication between $\iota^+$ and $\cK^+$ appears to be noted explicitly and utilized here for the first time.

\subsubsection{Asymptotics at \texorpdfstring{$I^0$}{I0} from initial data}
\label{SssIGen0}

In the coordinates $\tau=\frac{t}{r}$, $\rho_0=\frac{1}{r}$, and $\omega\in\Sph^2$, the schematic form of $L=\ubar\Box$ near $(I^0)^\circ$ is
\begin{equation}
\label{EqIGen0}
  L \sim -\rho_0^2\Bigl( -\pa_\tau^2 + (\rho_0\pa_{\rho_0})^2 + \pa_\omega^2 \Bigr)
\end{equation}
or, more accurately, $-\rho_0^2\bigl(-(1-\tau^2)(\tau\pa_\tau)^2+2\tau\pa_\tau\,\rho_0\pa_{\rho_0}+(\rho_0\pa_{\rho_0})^2+\pa_\omega^2\bigr)$; this is a Lorentzian signature quadratic form in $(\pa_\tau,\rho_0\pa_{\rho_0},\pa_\omega)$ with overall weight $\rho_0^2$, and $\tau$ is a time function for $\tau<1$. (The weight $\rho_0^2$ in~\eqref{EqIGen0} is also the reason for the ``$+2$'' in~\eqref{EqIfuMink}.) The operators $L=\Box_{g_b}$ or $L=\ubar\Box+V$ with $V=\cO(r^{-2})$ (which guarantees the same overall weight $\rho_0^2$) have the same structure. The order of initial data (posed at $\tau=0$, say) as $\rho_0\to 0$ (i.e., decay rates and/or index sets) is thus simply transported in $\tau$. For decay rates, this follows from an energy estimate with weights in $\rho_0$, and for expansions with index sets $\cE_0$ one tests the solution with products of the operators $\rho_0\pa_{\rho_0}-z$ for $(z,j)\in\cE_0$; these commute with $\rho_0^{-2}L$ to leading order at $I^0$.

\subsubsection{Asymptotics at \texorpdfstring{$\scri^+$}{null infinity}: transport from \texorpdfstring{$I^0$}{I0}; radiation field}
\label{SssIGenScri}

Our discussion of the decay rates and asymptotics towards $\scri^+$ is the same as in \cite{HintzMink4Gauge}, which in turn is a streamlined version of \cite{HintzVasyMink4} (see also \cite{KadarKehrbergerPhgScatter} for recent developments); we shall thus be brief. In the coordinates $\rho_0=\frac{1}{-t_*}$, $\rho_\sscri=\frac{-t_*}{r}$, and $\omega\in\Sph^2$ near the corner $I^0\cap\scri^+$ (taken from Figure~\ref{FigIMwc}), the schematic form of $L=\ubar\Box$ (plus an at least inverse quadratic potential) or tensor wave operators on Kerr is
\begin{equation}
\label{EqIGenScriOp}
  L \sim \rho_0^2\rho_\sscri \Bigl(\,\underbrace{\bigl(\rho_\sscri\pa_{\rho_\sscri} - (I+S)\bigr)}_{\text{$\to$\ radiation}} \, \underbrace{( \rho_0\pa_{\rho_0}-\rho_\sscri\pa_{\rho_\sscri} )}_{\text{transport}} + \rho_\sscri\slDelta \Bigr),
\end{equation}
with $S=0$ in these specific cases; see Corollary~\ref{CorExOpBox2} and \eqref{EqExOpLinFormula}. See \eqref{EqExOpLinbNormal} for the linearized gauge-fixed Einstein operator, in which case $S$ is a non-zero endomorphism of the symmetric 2-tensor bundle; this case is discussed further in~\S\ref{SssIEinScri} below. Due to the additional decaying weight $\rho_\sscri$ of the spherical Laplacian $\slDelta$ here, the term $\rho_\sscri\slDelta$ is of lower order for the present asymptotic analysis. (More precisely, note that $\rho_0\pa_{\rho_0}$, $\rho_\sscri\pa_{\rho_\sscri}$, and $\pa_\omega$ are b-vector fields, so $\rho_\sscri\slDelta\sim\rho_\sscri\pa_\omega^2$ has an additional vanishing factor $\rho_\sscri$ at $\scri^+$ as a b-vector field.) The $\scri^+$-order of solutions $u$ of $L u=f$ can be obtained by integrating the two transport operators in~\eqref{EqIGenScriOp}:
\begin{enumerate}
\item{\rm (Transport.)} Integration of $\rho_0\pa_{\rho_0}-\rho_\sscri\pa_{\rho_\sscri}$ transports the decay order of $u$ at $I^0$ (where it is known from~\S\ref{SssIGen0}) to the same decay order of $v=\bigl(\rho_\sscri\pa_{\rho_\sscri}-(I+S)\bigr)u$ at $\scri^+$; see Lemma~\ref{LemmaTIntHyp}.
\item{\rm (Radiation.)} Integrating
  \begin{equation}
  \label{EqIGenScriRad}
    \rho_\sscri\pa_{\rho_\sscri}-(I+S)
  \end{equation}
  towards $\rho_\sscri=0$, we can recover $u$ from $v$. For $S=0$ and $v$ of orders $\alpha_0$ and $\alpha_\sscri\in(1,2)$ at $I^0$ and $\scri^+$, say, this shows that $u$ has orders $\alpha_0$ and $((1,0),\alpha_\sscri)$: the leading-order term $\rho_\sscri=\rho_\sscri^1(\log\rho_\sscri)^0$ arises from the indicial root $1$ of~\eqref{EqIGenScriRad}. See Lemmas~\ref{LemmaTMPhgTest}--\ref{LemmaTMIntFuchs}. (For general $S$ and vector-valued $u$, the projections of $u$ to the various eigenspaces have radiation fields arising at decay orders $z\in 1+\spec S$.)
\end{enumerate}

More accurately, since we are moving $\rho_\sscri\slDelta u$ to the right-hand side of the equation $L u=f$, this integration scheme allows us to extract asymptotics of $u$ one order of $\rho_\sscri$-decay at a time at the cost of some b-derivatives. In particular, to get this procedure started one needs to have control on $u$ in \emph{some} (arbitrarily poorly) \emph{polynomially weighted} function space with (arbitrarily) \emph{high} b-regularity. For finite $t_*$, this is easily accomplished using energy estimates; see \cite{HintzVasyScrieb} for general results and \cite{HintzVasyMink4,HintzMink4Gauge} for the Einstein case.

Taking into account the overall factor of $\rho_0^2\rho_\sscri$ in~\eqref{EqIGenScriOp}, and weakening the $\scri^+$-decay order below that of $I^0$ as a safety margin (which is forced upon us when working with $L^2$-based spaces), these arguments show, in the case $S=0$, that
\[
  \includegraphics{EqIGenScriCon}
\]
More generally, terms in the $I^0$-expansion of $u$, described using an index set $\cE_0$, get transported to $\scri^+$. When $z\notin\spec(I+S)$ for all $(z,0)\in\cE_0$, the subsequent integration of~\eqref{EqIGenScriRad} with right-hand side featuring a term $\rho_\sscri^z$ does not produce logarithmic factors. Under this assumption, and again for $S=0$, we thus have
\[
  \includegraphics{EqIGenScriPhg}
\]
One can, yet more generally, also allow $f$ to have a partial expansion at $\scri^+$ with index set $\cE_\sscri+1$; in that case the $\scri^+$-index set of $u$ gets enlarged to $(1,0)\cup(\cE_0\extcup\cE_\sscri)$ where ``$\extcup$'' denotes the \emph{extended union} that introduces a logarithmic factor when both index sets feature common exponents $z$.

\bigskip

In the context of the Einstein equation, these mechanisms (and appropriate choices of gauge and constraint damping) suffice for settling the \emph{exterior stability problem}, i.e., the problem of constructing a ``piece of $\scri^+$'' from initial data posed at a spacelike hypersurface transversal to $(I^0)^\circ$; this was shown using these techniques in \cite{HintzVasyMink4,HintzMink4Gauge,KadarKehrbergerPhgScatter}. We revisit this problem in the present paper in \S\ref{SEx} and prove the partial polyhomogeneity of metric perturbations for the particular formulation of the gauge-fixed Einstein equation that we introduce in this paper for the global stability problem.

Returning to our general discussion, once one has control on the solution $u$ of a wave equation $L u=f$ up to some hypersurface transversal to $(\scri^+)^\circ$, say, $t_*=1$, one can reduce the problem of understanding the global asymptotics of $u$ to the problem of controlling forward solutions $u'$ of $L u'=f'$, where $f'$ vanishes for $t_*\leq 1$, via a simple cutoff and causality argument (see, e.g., \eqref{EqD2PFwdCutoff}--\eqref{EqD2PFwd}). \emph{From now on, all source terms $f$ will vanish for $t_*\leq 1$, and we will not show $I^0$ in our diagrams anymore.} The order of $u$ at $\scri^+$ is then entirely determined by that of $f$ and the eigenvalues of $I+S$ in~\eqref{EqIGenScriRad}; thus,
\begin{equation}
\label{EqIGenScriFw}
  \raisebox{-1.5em}{\includegraphics{EqIGenScriFw}}
\end{equation}
where, for $S=0$ and index sets $\cE_\sscri$ with $\min\Re\cE_\sscri=\min_{(z,0)\in\cE_\sscri}\Re z>1$, we set $\la\cE_\sscri\ra:=(1,0)\cup\cE_\sscri$, with appropriate modifications for general $S$ which moreover take into account coincidences between $\spec(I+S)$ and $\cE_\sscri$. We moreover require $(2,0)\subset\cE_\sscri$ so that $\cE_\sscri$ can absorb the lower-order corrections from a non-trivial radiation field (which on a technical level arise from integrating the transport ODEs in~\eqref{EqIGenScriOp} with right-hand side given by $\rho_\sscri\slDelta$ applied to the radiation field and its prior corrections). Notice that we have also included the implication ``$\Longleftarrow$'' in~\eqref{EqIGenScriFw}: application of $L$ annihilates the radiation field terms of $u$ (but in general no other terms).

\subsubsection{Asymptotics at \texorpdfstring{$\iota^+$}{i-plus} from transport from \texorpdfstring{$\scri^+$}{null infinity} and resonances}
\label{SssIGen1}

In the coordinates $\rho_\sscri=\frac{t_*}{r}$, $\rho_+=\frac{1}{t_*}$, and $\omega\in\Sph^2$ near the corner $\scri^+\cap\iota^+$ (again taken from Figure~\ref{FigIMwc}), the schematic form of wave-type operators on Minkowski and Kerr is similar to~\eqref{EqIGenScriOp}, namely
\begin{equation}
\label{EqIGen1Op}
  L \sim \rho_\sscri\rho_+^2\Bigl(\,\underbrace{\bigl(\rho_\sscri\pa_{\rho_\sscri}-(I+S)\bigr)}_{\text{$\to$\ radiation}}\,\underbrace{(\rho_\sscri\pa_{\rho_\sscri}-\rho_+\pa_{\rho_+})}_{\text{transport}} + \rho_\sscri\slDelta \Bigr)
\end{equation}
to leading order at $\scri^+$. The second factor transports the $\scri^+$-order of $u$ (without its radiation field) to $\iota^+$. The first factor is the same as in~\eqref{EqIGenScriRad}, and its inversion generates the radiation field as before. In view of the overall factor of $\rho_+^2$ in~\eqref{EqIGen1Op}, this suggests the following picture:
\begin{equation}
\label{EqIGen1Trans}
  \raisebox{-1.5em}{\includegraphics{EqIGen1Trans}}
\end{equation}

This is not correct, however, as the asymptotics as $t_*\to\infty$ cannot (for typical wave-type operators $L$) be determined by arguments that are purely local near $\scri^+$. As a concrete example, consider $\ubar\Box_{4+1}$ (or $\ubar\Box_{3+1}+V_0\la x\ra^{-2}$ for non-zero $V_0>-\frac14$, which is studied in \cite{GajicInverseSquare} and \cite[\S{6.2}]{HintzNonstat}): even for \emph{compactly supported} $f$, the forward solution $u$ has late-time behavior like the fundamental solution, which, for $\ubar\Box_{4+1}$, in the forward cone is given by
\begin{equation}
\label{EqIGen1Pure}
  (t^2-r^2)^{-\frac{4-1}{2}}=t_*^{-3} \Bigl(\frac{v}{v+2}\Bigr)^{\frac32},\quad v:=\frac{t_*}{r}.
\end{equation}
Notice, moreover, that the terms in the $\iota^+$-expansion of $u$ in~\eqref{EqIGen1Trans} cannot be arbitrary, but rather must be annihilated by $L$ to leading order at $\iota^+$ (since $f=L u$ is assumed to have a trivial expansion at $\iota^+$). To make progress, it is thus necessary to capture $L$ \emph{globally} at $\iota^+$ to leading order. On the level of geometry, every asymptotically flat metric (such as Kerr) is equal to the Minkowski metric plus error terms that decay as $r\to\infty$, and thus (noting that $r=\infty$ at $\iota^+$) its $\iota^+$-model is \emph{equal} to the Minkowski metric; similarly then, the $\iota^+$-model of ``geometric'' wave-type operators $L$ is equal to the Minkowskian version $\ubar L$ of $L$. (For example, when $L=\Box_{g_b}$, then $\ubar L=\ubar\Box$.)

To make this concrete, let us work in the coordinates $t_*$, $v=\frac{t_*}{r}$, and $\omega\in\Sph^2$; then the $\iota^+$-normal operator of $\ubar\Box$ is equal to\footnote{We use the convention $\slDelta\geq 0$. The factor $t_*^2$ balances $\rho_+^2$ in~\eqref{EqIGen1Op} and makes this operator homogeneous of degree $0$ with respect to dilations in $t_*$ in the coordinates $(t_*,v,\omega)$.}
\begin{equation}
\label{EqIGen1RescBox}
  t_*^2\ubar\Box = -2 v(v\pa_v+t_*\pa_{t_*})(v\pa_v-1) + v^2\bigl(-(v\pa_v)^2+v\pa_v+\slDelta\bigr).
\end{equation}
The $t_*^{-\lambda}$-coefficient $u_\lambda=u_\lambda(v,\omega)$ in the $\iota^+$-expansion of $u$ must thus lie in the kernel of the operator
\begin{equation}
\label{EqIGen1Normip}
  N_{\iota^+}(\ubar\Box,\lambda) := -2 v(v\pa_v-\lambda)(v\pa_v-1) + v^2\bigl(-(v\pa_v)^2+v\pa_v+\slDelta\bigr)
\end{equation}
on $(\iota^+)^\circ=(0,\infty)_v\times\Sph^2_\omega$, and satisfy certain decay/regularity properties as $v\to\infty$ (i.e., at the other boundary $\iota^+\cap\cK^+$ of $\iota^+$) that we discuss in~\S\ref{SssIGen2La} below. Note that $N_{\iota^+}(\ubar\Box,\lambda)$ is elliptic (owing to the timelike nature of $\pa_{t_*}=\pa_t$), but degenerates as $v\to 0$ or $v\to\infty$. If $(\lambda,0)\in\cE_\sscri$ contributes to the $\scri^+$-expansion of $u$, one therefore expects (given that the transport part of~\eqref{EqIGen1Op} turns $\rho_\sscri^\lambda$ into $\rho_+^\lambda$, i.e., a contribution to $u_\lambda$) that $u_\lambda$ solves an \emph{asymptotic boundary value problem}, with the boundary data being the coefficient of $v^\lambda$ (which is an element of $\CI(\Sph^2)$ that is determined locally via transport from $\scri^+$).

Owing to the origin of~\eqref{EqIGen1Normip} in the asymptotic homogeneity of (massless) wave-type operators under spacetime dilations near $(\iota^+)^\circ$---and, in fact, exact homogeneity in the Minkowski case---it follows that it is related to the spectral family of hyperbolic space via conjugation; the indicial root $1$, resp.\ $\lambda$ of the first term in~\eqref{EqIGen1Normip} is then the outgoing (i.e., relatively decaying when $\Re\lambda\ll 1$), resp.\ incoming (i.e., relatively growing when $\Re\lambda\ll 1$) indicial root. The relationship between $t_*\to\infty$ asymptotics on suitable classes of asymptotically Minkowski spaces and the spectral theory of associated (asymptotically) hyperbolic operators is explained in detail by Vasy \cite[\S{5}]{VasyMicroKerrdS} and Baskin--Vasy--Wunsch \cite{BaskinVasyWunschRadMink,BaskinVasyWunschRadMink2}; see also \cite{VasyMinkDSHypRelation}. Usage of other transversals to the spacetime dilation action (away from $r=0$) yields conjugated versions of~\eqref{EqIGen1Normip}. It is technically advantageous to use the transversal $r=1$, as one then has access to the robust Fredholm framework of \cite{VasyMicroKerrdS} (see also \cite{ZworskiRevisitVasy}, \cite[\S{5}]{DyatlovZworskiBook}, and \cite[Chapter~11]{HintzMicro}) for analysis near $v=0$ (and can thereby circumvent the subtleties of the 0-calculus \cite{MazzeoMelroseHyp,GuillarmouMeromorphic}); the translation to~\eqref{EqIGen1Normip} is then relatively straightforward. See \S\ref{SsipInv}.

Consider again the term~\eqref{EqIGen1Pure} in the $\iota^+$-expansion of waves on $(4+1)$-dimensional Minkowski space; we have $u_3(v,\omega):=(\frac{v}{v+2})^{\frac32}\in\ker N_{\iota^+}(\ubar\Box_{4+1},3)$ in this case (where the latter operator has a form similar to~\eqref{EqIGen1Normip} with some dimensional adjustments, e.g., with $v\pa_v-\frac32$ instead of $v\pa_v-1$). Recall that we must allow for (a generically non-zero scalar multiple of) $t_*^{-3}u_3$ to contribute to the $\iota^+$-expansion of waves even when the source $f$ vanishes near $\scri^+$. Since $u_3$ has \emph{vanishing} incoming data (no $v^3$-term), it is thus what we call a \emph{``pure'' resonant state}.\footnote{The reader may wonder about $(t^2-r^2)^{-\frac{3-1}{2}}=t_*^{-2}\frac{v}{v+2}$ in $3+1$ dimensions; this does solve the wave equation in $t_*>0$, and so $N_{\iota^+}(\ubar\Box_{3+1},2)\frac{v}{v+2}=0$. This term does arise in the $\iota^+$-expansion of waves sourced by $f$ that feature a $\rho_\sscri^3$-term at $\scri^+$. However, we do not regard $\frac{v}{v+2}$ as a pure resonant state since it does not arise for compactly supported $f$ by the sharp Huyghens principle. This is, in any event, an exceptional case; see Remark~\ref{RmkipMeroSpec}.}

\medskip

In summary, one expects the leading-order terms in the $\iota^+$-expansion of $u$ to be ``non-pure'' resonant states arising via transport from $\scri^+$ (and thus at orders determined by the $\scri^+$-index set of $f$) or ``pure'' resonant states (which arise only at a discrete set of exponents $\lambda$ of $t_*^{-\lambda}$ and of which there is only a finite-dimensional space for each such $\lambda$). The details in the Einstein case are described in~\S\ref{SssipP}; see in particular Proposition~\ref{PropipPResPar}. Note also that since, say, $t_*^2\Box_{g_b}$ is equal to $t_*^2\ubar\Box$ only to leading order (as a b-differential operator) at $\iota^+$, a $t_*^{-\lambda}$-term in the $\iota^+$-expansion of a wave typically spawns also terms at orders $t_*^{-\lambda-1}$, $t_*^{-\lambda-2}$, etc. We thus record the schematic picture
\begin{equation}
\label{EqIGen1Full}
  \raisebox{-1.5em}{\includegraphics{EqIGen1Full}}
\end{equation}
where $\cE_+$ is the (extended) union of $\cE_\sscri$ and an index set determined by the pure resonances of the $\iota^+$-model operator. (We discuss the orders of $f$ and $u$ at $\cK^+$ later.) Carefully note, however, that asymptotics at $\iota^+$ cannot be determined locally near $\iota^+$, as they are not decoupled from asymptotics at $\cK^+$---which we discuss only in \S\S\ref{SssIGen2Sm}--\ref{SssIGenK} below.

\subsubsection{Asymptotics at \texorpdfstring{$\cK^+$}{K-plus} coming from \texorpdfstring{$\iota^+$}{i-plus} and large indicial roots at zero energy}
\label{SssIGen2La}

We next discuss how the asymptotic expansion and remainder terms of $u$ at $\iota^+$ influence asymptotics and decay at $\cK^+$. It is thus convenient to switch from $v=\frac{t_*}{r}$ to the coordinate $R:=v^{-1}=\frac{r}{t_*}$, which is a defining function of $\cK^+$ near $\cK^+\cap\iota^+$. (See the top of Figure~\ref{FigIMwc}, where $R$ is denoted by $\rho_\cK$.) Let us proceed systematically and suppose that our wave operator $L$ (such as $L=\Box_{g_b}$) is given by\footnote{This form of the operator is valid for the scalar or tensor wave operator on Schwarzschild when $t_*$ is null. It is important to generalize this structure as in~\eqref{EqWEOpExpl}--\eqref{EqWEOpKerr} in order to accommodate the Kerr case; the additional terms there do not influence any heuristics, however, and thus we drop them here for the sake of brevity.}
\begin{equation}
\label{EqIGen2LaOp}
  L = -2\pa_{t_*}\rho(\rho\pa_\rho-1-S) + \rho^2 \wt L(0)(\rho,\omega,\rho\pa_\rho,\pa_\omega),\quad \rho=r^{-1};
\end{equation}
for example, for $L=\ubar\Box$ we have $S=0$ and $\wt L(0)=-(\rho\pa_\rho)^2+\rho\pa_\rho+\slDelta$. The operator
\begin{equation}
\label{EqIGen2La0}
  \wh L(0) = \rho^2\wt L(0)(\rho,\omega,\rho\pa_\rho,\pa_\omega)
\end{equation}
is the \emph{zero energy operator} of $L$ (so, e.g., $\wh L(0)=\Delta_x$ for $L=\ubar\Box$). The systematic form of~\eqref{EqIGen1Normip} is then
\begin{align}
  N_{\iota^+}(L,\lambda) &= -2 v(v\pa_v-\lambda)(v\pa_v-1-S) + v^2 \wt L(0)(0,\omega,v\pa_v,\pa_\omega) \nonumber\\
\label{EqIGen2LaNorm}
    &= R^{-2}\Bigl( \wt L(0)(0,\omega,-R\pa_R,\pa_\omega) + \cO(R) \Bigr),
\end{align}
where the $\cO(R)$-term is built from the b-vector fields $R\pa_R$ and $\pa_\omega$. (Setting $\rho=0$ in $\wt L(0)$ has the effect of restricting to $\iota^+$.)

\medskip

{\bf Indicial roots and indicial solutions; indicial gap.} Using the Mellin transform in $R$ to analyze (the leading-order term of)~\eqref{EqIGen2LaNorm}, one sees that the asymptotic behavior of $\iota^+$-resonant states at $\iota^+\cap R^{-1}(0)=\iota^+\cap\cK^+$ is determined by the \emph{indicial roots of the zero energy operator}, i.e., those values $\lambda\in\C$ for which there exist non-trivial homogeneous solutions $\ubar u_\pa=\ubar u_\pa(\omega)$ of the equation $\wt L(0)(0,\omega,\rho\pa_\rho,\pa_\omega)(\rho^\lambda\ubar u_\pa(\omega))=0$. (See Proposition~\ref{PropipNAsy}, and Lemma~\ref{LemmaTMSolPhg} for a prototypical result.) For $L=\ubar\Box_{d+1}$, or indeed for the scalar (or tensor) wave operator on \emph{any} asymptotically flat $(d+1)$-dimensional spacetime (including Kerr for $d=3$), these indicial roots are given by $\ldots,-2,-1,0,d-2,d-1,d,\ldots$, and the indicial solutions $\ubar u_\pa$ at the indicial roots $-l$ and $d-2+l$ are degree $l$ spherical harmonics. We call the interval $(0,d-2)$ separating these pairs the \emph{indicial gap}.

The function spaces on which one studies the operator $N_{\iota^+}(L,\lambda)$ (i.e., its Fredholm properties, and then its meromorphic continuation in $\lambda$ from $\Re\lambda\ll 1$ with poles being pure resonances---see \S\ref{SsipInv}) are then those which have an $R$-weight $R^{-\beta}$ where $\beta$ lies in the indicial gap.\footnote{The analytic meaning of the indicial gap is that $N_{\iota^+}(L,\lambda)$ has Fredholm index $0$ on suitable spaces with $R$-weights $-\beta$ when $\beta$ lies in the indicial gap. The possibility of separating into spherical harmonics is thus analytically irrelevant, but of course useful for concrete computations.} (The minus sign arises from the minus sign in front of $R\pa_R$ in~\eqref{EqIGen2LaNorm}.) In particular, $\iota^+$-resonant states have an expansion as $R\to 0$ with leading-order terms of the form $R^{-\lambda}\ubar u_\pa(\omega)$ where $\lambda$ is a zero energy indicial root lying \emph{below} the indicial gap---and thus corresponding to \emph{growing} indicial solutions $\rho^\lambda\ubar u_\pa(\omega)$ of $\hat L(0)$, which is why we call them \emph{large indicial roots} here.

\begin{example}[Indicial roots and asymptotics at $R=0$]
\label{ExIGen2LaAsy}
  Expressing~\eqref{EqIGen1Pure} in terms of $R=v^{-1}$ yields the resonant state $(1+2 R)^{-\frac32}$ of $\ubar\Box_{4+1}$, with leading-order term $R^0$ having coefficient $\CI(\Sph^3)\ni 1\in\ker\wh{\ubar\Box_{4+1}}(0)=\Delta_{\R^4}$. Similarly, the (``non-pure'') $\iota^+$-resonant state $(1+2 R)^{-1}$ of $\ubar\Box_{3+1}$ (and thus also of $\Box_{g_b}$) has the leading-order term $R^0\cdot 1$, $1\in\CI(\Sph^2)$. Moreover, if one attempts to construct a formal solution of $N_{\iota^+}(\ubar\Box_{4+1},3)u_3=0$ with $u_3(R,\omega)=1+o(1)$ as $R\to 0$, one ends up producing the Taylor expansion of $(1+2 R)^{-\frac32}$. --- Somewhat more interestingly, for source terms and operators that are not spherically symmetric, one expects other indicial roots to be activated; for example, in the case of $\ubar\Box_{3+1}$ or $\Box_{g_b}$, the indicial root $-1$ triggers a term $R^{+1}\scal_1$ in the $R\to 0$ asymptotics on $\iota^+$ where $\scal_1\in\mathspan\{Y_{1 m}\colon m=-1,0,1\}$ is a degree $1$ spherical harmonic. (Spherical harmonic decompositions, including in the tensorial setting, are recalled in~\S\ref{SsTY}.)
\end{example}

{\bf Large zero energy states.} Let us return to wave asymptotics \emph{on spacetime}, and suppose that the wave $u$ has a term $t_*^{-\alpha}u_\alpha(R,\omega)$ in its $\iota^+$-expansion, with $u_\alpha$ itself featuring a term $R^{-\lambda}\ubar u_\pa(\omega)$ in its $\iota^+\cap\cK^+$-expansion where $\lambda$ is a large indicial root of the zero energy operator and $\ubar u_\pa$ the corresponding indicial solution. (In the case of the scalar wave operator on Kerr or Minkowski, the reader may consider the case $\lambda=-l$ and $R^{-\lambda}\ubar u_\pa(\omega)=R^l\scal_l$ where $\scal_l$ is a degree $l$ spherical harmonic.) Recalling $R=\frac{r}{t_*}$, this term thus contributes to $u$ via
\begin{equation}
\label{EqIGen2LaTerm}
  t_*^{-\alpha+\lambda}\ubar u^{(\lambda)}(r,\omega),\quad \ubar u^{(\lambda)}(r,\omega):=r^{-\lambda}\ubar u_\pa(\omega).
\end{equation}
As $\lambda$ decreases (as one considers increasingly lower-order terms of $u_\alpha$ as $R\to 0$), the $\cK^+$-order $\alpha-\lambda$ of this increases, while the $\iota^+$-order is of course the constant $\alpha$; in this sense, very large (i.e., very negative) indicial roots contribute to very strongly decaying terms in the $\cK^+$-asymptotics of $u$ (and are thus ultimately uninteresting for the purpose of solving a nonlinear problem, for which some finite decay order suffices).

While $\ubar u^{(\lambda)}(r,\omega)=r^{-\lambda}\ubar u_\pa(\omega)$ in~\eqref{EqIGen2LaTerm} is annihilated by $\wh L(0)$ to leading order at $r=\infty$, it typically does not lie in $\ker\wh L(0)$ when $L=\Box_{g_b}$ or $\ubar\Box+V$ where $0\neq V$; this is to be expected since the analysis of $\iota^+$ only sees the asymptotic structure of $L$ at infinity (here meaning $\iota^+$). The crucial question then becomes whether one can extend $\ubar u^{(\lambda)}(r,\omega)$ to a \emph{large zero energy state}\footnote{This terminology was used already in \cite[\S{4.3}]{HintzNonstat}. In \cite[\S{1.2}]{HintzPrice}, the terminology ``extended bound state'' was used instead.}
\begin{equation}
\label{EqIGen2LaState}
  u^{(\lambda)}(r,\omega) \in \ker \wh L(0),\quad u^{(\lambda)}=\ubar u^{(\lambda)} + \text{(lower order terms)}.
\end{equation}
If this is the case, then one expects to see not~\eqref{EqIGen2LaTerm} but
\begin{equation}
\label{EqIGen2LaStateKerr}
  t_*^{-\alpha+\lambda}u^{(\lambda)}(r,\omega)
\end{equation}
in the late-time asymptotics of $u$.

\begin{example}[Large zero energy states]
\label{ExIGen2LaStates}
  For $L=\ubar\Box$ in $3+1$ dimensions, one simply takes $u^{(\lambda)}=\ubar u^{(\lambda)}$, so large zero energy states are of the form $r^l\scal_l$. For $L=\Box_{g_b}$, one has $u^{(0)}=1$; in the Schwarzschild case, the Minkowskian large zero energy state $\ubar u^{(-1)}_{\rms 1}(\scal_1):=r\scal_1$ can be extended to the large zero energy state $(r-\bhm)\scal_1$ (see \cite[Proposition~6.2]{HaefnerHintzVasyKerr}); see \cite[Remark~5.2]{HintzPrice} for other values of $l$. For $L=\ubar\Box+V$ where $0\leq V\in\CIc(\R^3)$, one can show that every Minkowskian large energy state can be extended to a large zero energy state; for the state $1$, this is discussed in \cite[Theorem~1.9 and the proof of Lemma~2.19]{HintzPrice}. This is true in general when the resolvent $\wh L(\sigma)^{-1}$ (see~\S\ref{SssIGenK}) is regular at $\sigma=0$. When $L$ admits zero energy bound states, this is no longer the case, and there typically do exist Minkowskian large zero energy states $\ubar u^{(\lambda)}$ without extensions~\eqref{EqIGen2LaState}; these require special attention in the Einstein case, as discussed in~\S\ref{SssIEinLa} below.
\end{example}

Similar statements hold also when working with waves that merely have bounds (not expansions) at $\iota^+$, in that one expects terms
\begin{equation}
\label{EqIGen2NoStruct}
  a(t_*)u^{(\lambda)}(r,\omega)
\end{equation}
to arise in the late-time expansion of waves where $a(t_*)=\cO(t_*^{-\alpha+\lambda})$ (and similarly for derivatives of $a$ along $t_*\pa_{t_*}$) is ``structureless'' (i.e., does not have an expansion), when, say, the initial data of $u$ have ``structureless'' $\cO(r^{-\alpha})$-decay.\footnote{While the reader versed in b-normal operator arguments may expect that one proves this via the characterization of spacetime $L^2$-spaces using the Mellin transform in $t_*^{-1}$ in the coordinates $(t_*,R,\omega)$, this is not the route we take in this paper. Instead, we obtain such ``conormal expansions'' via the inverse \emph{Fourier transform} in $t_*$, in the coordinates $(t_*,r,\omega)$, from the output of the low-energy resolvent; cf.\ Proposition~\ref{PropiptfC}.}

\medskip

{\bf Generalized large zero energy states.} The terms spawned by a leading-order term $R^{-\lambda}\ubar u_\pa(\omega)$ in the $R\to 0$ expansion of an element of $\ker N_{\iota^+}(L,\alpha)$, as alluded to in Example~\ref{ExIGen2LaAsy}, which are thus of order $R^{-\lambda+1}$, $R^{-\lambda+2}$, etc.\ (ignoring the possibility of factors of $\log R$), contribute terms of size $t_*^{-\alpha+\lambda-1}\cO(r^{-\lambda+1})$, $t_*^{-\alpha+\lambda-2}\cO(r^{-\lambda+2})$, etc.\ on the spacetime level, which thus have successively higher $\cK^+$-orders (and constant $\iota^+$-order). The question then arises what these contributions are, exactly, globally along $\cK^+$. We can find the answer by determining the failure of the leading-order term~\eqref{EqIGen2LaStateKerr} to lie in the nullspace of the full wave operator, which we assume to be of the form~\eqref{EqIGen2LaOp}. After all, we expect the term~\eqref{EqIGen2LaStateKerr} to arise from, say, source terms $f$ supported near a compact subset of $(\scri^+)^\circ$ with $\rho_\sscri^{\alpha+1}$ in their $\scri^+$-expansion, and thus $L u=f$ vanishes near $\cK^+$. If~\eqref{EqIGen2LaStateKerr} is the term in the $\cK^+$-expansion of $u$ with the least amount of decay, it must therefore be possible to correct it at higher decay orders to increasingly accurate formal solutions of $L$ at $\cK^+$. Now, writing $a(t_*):=t_*^{-\alpha+\lambda}$, we have
\[
  L\bigl(a(t_*)u^{(\lambda)}\bigr) = a(t_*)\underbrace{\wh L(0)u^{(\lambda)}}_{=\,0\ \text{by}\ \eqref{EqIGen2LaState}} + a'(t_*) [L,t_*]u^{(\lambda)}.
\]
This thus has $t_*$-decay $a'=\cO(t_*^{-\alpha+\lambda-1})$ but with spatial dependence $[L,t_*]u^{(\lambda)}=\cO(r^{-\lambda-1})$. If one can write $[L,t_*]u^{(\lambda)}=-\wh L(0)\breve u^{(\lambda),1}$ for some $\breve u^{(\lambda),1}(r,\omega)=\cO(r^{-\lambda+1})$, one thus has
\[
  L\bigl( a(t_*)u^{(\lambda)} + a'(t_*)\breve u^{(\lambda),1} \bigr) = a'(t_*)\bigl(\,\underbrace{[L,t_*]u^{(\lambda)}+\wh L(0)\breve u^{(\lambda),1}}_{=\,0}\,\bigr) + a''(t_*) [L,t_*]\breve u^{(\lambda),1},
\]
which now has improved $\cK^+$-order $\alpha-\lambda+2$. Note that
\begin{subequations}
\begin{equation}
\label{EqIGen2LaGen}
  L\bigl( t_* u^{(\lambda)} + \breve u^{(\lambda),1} \bigr) = 0;
\end{equation}
we thus call
\begin{equation}
\label{EqIGen2LaGen2}
  u^{(\lambda),\leq 1}(t_*) := t_* u^{(\lambda)} + \breve u^{(\lambda),1}
\end{equation}
\end{subequations}
a \emph{generalized (large) zero energy state}. We remark that the existence of $\breve u^{(\lambda),1}$  is again related to the mapping properties of $\wh L(0)$, and thus guaranteed when $\wh L(0)$ is invertible. (See, e.g., Proposition~\ref{PropWG0Large}\eqref{ItWG0Larges01}--\eqref{ItWG0Largevl2}.)

One can push~\eqref{EqIGen2LaGen}--\eqref{EqIGen2LaGen2} further and (attempt to) construct further stationary functions $\breve u^{(\lambda),j}(r,\omega)=\cO(r^{-\lambda+j})$, $j=2,3,4,\ldots$, such that
\begin{equation}
\label{EqIGen2LaGen3}
  u^{(\lambda),\leq k}(t_*,r,\omega) := t_*^k u^{(\lambda)} + \sum_{j=1}^k \pa_{t_*}^j(t_*^k)\,\breve u^{(\lambda),j} \in \ker L.
\end{equation}
A term $a(t_*)u^{(\lambda)}$ in the $\cK^+$-expansion of $u$ then spawns an (arbitrarily long) expansion
\begin{equation}
\label{EqIGen2LaGen4}
  u^{(\lambda),\leq k}(a) := a(t_*) u^{(\lambda)} + \sum_{j=1}^k a^{(j)}(t_*)\breve u^{(\lambda),j};
\end{equation}
and $L\bigl(u^{(\lambda),\leq k}(t_*^{-\alpha+\lambda})\bigr)$ has order $\alpha-\lambda+k+1$ at $\cK^+$ (i.e., more decay as $k$ increases) and constant order $\alpha+2$ at $\iota^+$. (The $t_*^{-\alpha}$-leading-order term of~\eqref{EqIGen2LaGen4} at $\iota^+$ produces the first $k+1$ terms of the Taylor expansion spawned by a leading-order term $R^{-\lambda}\ubar u_\pa(\omega)$, which we already mentioned in Example~\ref{ExIGen2LaAsy}.) The same is true also when $a(t_*)$ is ``structureless'' as explained after~\eqref{EqIGen2NoStruct}.

\begin{example}[Generalized zero energy states]
\label{ExIGen2Gen}
  On Minkowski space and working for a moment in the coordinates $(t,x)$, we have $\wh{\ubar\Box}(0)u^{(0)}=0$ for the large zero energy state $u^{(0)}=1$, and $\ubar\Box(t u^{(0)})=0$; but $\ubar\Box(t^2 u^{(0)})=2=-\wh{\ubar\Box}(\frac16 r^2)$, so $\ubar\Box(t^2 u^{(0)}+\frac16 r^2)=0$. Setting $t_*=t-r$, we have $t u^{(0)}=t_* u^{(0)}+\breve u^{(0),1}$ and $t^2 u^{(0)}+\frac16 r^2=u^{(0),\leq 2}(t_*,r,\omega)=t_*^2 u^{(0)}+2 t_*\breve u^{(0),1}+2\breve u^{(0),2}$ for
  \[
    \breve u^{(0),1} = r,\quad
    \breve u^{(0),2} = \frac{7}{12} r^2.
  \]
\end{example}

Our rough expectation is thus that
\begin{equation}
\label{EqIGen2LaFull}
  \raisebox{-2.2em}{\includegraphics{EqIGen2Full}}
\end{equation}
where $\cE_+$ is as described after~\eqref{EqIGen1Full}, while $\lambda$ runs over the large indicial roots of $\wh L(0)$; and in fact $u$ has an expansion at $\cK^+$ into a sum of expressions of the form~\eqref{EqIGen2LaGen4}, and so does its conormal remainder term, i.e., the order $\alpha_+$-term at $\cK^+$.\footnote{The $\cK^+$-order is $\alpha_+$ when the indicial gap starts at $0$, otherwise it would be $\alpha_+$ minus the infimum of the indicial gap (minus $\eps$ for safety).} If $f$ vanishes to some finite order $\alpha_\cK$ at $\cK^+$, then one might expect the $\cK^+$-order of $u$ to also feature an order $\alpha_\cK$ term at $\cK^+$ that does not have an expansion of any sort (as this is true on Minkowski space). We proceed to discuss how sources at $\cK^+$ affect waves at $\iota^+$ (\S\ref{SssIGen2Sm}) and $\cK^+$ (\S\ref{SssIGenK}).

\subsubsection{Asymptotics at \texorpdfstring{$\iota^+$}{i-plus} coming from \texorpdfstring{$\cK^+$}{K-plus} and small indicial roots at zero energy}
\label{SssIGen2Sm}

Consider sources $f$ supported near a compact subset of $(\cK^+)^\circ$, with $\cK^+$-order $\alpha_\cK$, e.g., $f=\chi_+(t_*)t_*^{-\alpha_\cK}$ times an element of $\CIc(\R^3)$; here $\chi_+$ vanishes for $t_*\leq 1$ and equals $1$ for $t_*\geq 2$. The forward solution $u$ of $\ubar\Box_{d+1}u=f$ can then be seen (upon convolving $f$ with the retarded fundamental solution) to have order $\alpha_\cK$ at $\cK^+$ as well, while its $\iota^+$-order is $\alpha_\cK+(d-2)$. We will argue in~\S\ref{SssIGenK} that the shift $d-2$ is precisely the \emph{supremum} of the indicial gap of the zero energy operator. (See Remark~\ref{RmkIGen2Sm} for a heuristic.) In the case $d=3$ and for the wave operator on Minkowski space or Kerr, we thus expect that
\begin{equation}
\label{EqIGen2SmNaive}
  \raisebox{-1.5em}{\includegraphics{EqIGen2SmNaive}}
\end{equation}
where we ignore all contributions to the $\iota^+$-asymptotics of $u$ from $\iota^+$-resonant states (pure and non-pure ones alike) as well as the $\cK^+$-asymptotics (of the type~\eqref{EqIGen2LaGen4}) that they cause.\footnote{For example, in the case of the scalar wave equation on Kerr,~\eqref{EqIGen2SmNaive} can only hold for $\alpha_\cK<3$ in view of Price's law \cite{HintzPrice,AngelopoulosAretakisGajicKerr}. Moreover, as we discuss in~\S\ref{SssIGenK}, when the wave operator $L$ admits zero energy bound states or resonances, then the $\cK^+$-order of $u$ will be worse than that of $f$.} The bound $\alpha_\cK<\alpha_+$ (discussed already after~\eqref{EqIGen2LaFull}) ensures that terms at $\cK^+$ caused by $\iota^+$-asymptotics have more than $\alpha_\cK$ orders of $\cK^+$-decay, while the bound $\alpha_+<\alpha_\cK+1$ ensures that the limited $\iota^+$-decay order of $u$ caused by the limited $\cK^+$-decay of $f$ is still larger than $\alpha_+$.

\begin{rmk}[Small indicial roots, I]
\label{RmkIGen2Sm}
  While not directly used in the present paper, we make the following observation: when $f=\chi_+(t_*)t_*^{-\alpha_\cK}f_0(x)$ where $f_0\in\CIc(\R^3_x)$, say, then one can attempt to solve $L u=f$ in Taylor series at $\cK^+$ with the initial ansatz $u=t_*^{-\alpha_\cK}u_0(x)+o(t_*^{-\alpha_\cK})$, where $u_0$ must satisfy the zero energy equation $\wh L(0)u_0=f_0$. Given the (weighted) b-differential nature~\eqref{EqIGen2La0} of $\wh L(0)$, and assuming that $\wh L(0)$ is invertible on spaces with $\rho=r^{-1}$-weights lying in the indicial gap, one can then use Mellin transform techniques to show that $u_0$ has a full asymptotic expansion as $\rho\to 0$ into sums of terms of the form $\rho^z u_{(z,0)}(\omega)$ (plus lower-order corrections) where $z$ runs over the \emph{small indicial roots}, i.e., those with $\Re z\geq\sup(\text{indicial gap})$. (See Lemma~\ref{LemmaTMSolPhg}.) Notice then that the term $t_*^{-\alpha_\cK}\rho^z u_{(z,0)}(\omega)=r^{-\alpha_\cK-z}R^\alpha_\cK$ has $\iota^+$-order $\alpha_\cK+z$. This heuristic thus explains how small indicial roots communicate $\cK^+$-asymptotics to $\iota^+$ (much like large indicial roots communicated $\iota^+$-asymptotics to $\cK^+$ in~\S\ref{SssIGen2La}).
\end{rmk}

\begin{rmk}[Small indicial roots, II: bound states]
\label{RmkIGen2Bound}
  Suppose $L$ admits a zero energy bound state $u_{(0)}=u_{(0)}(x)$, by which we mean a solution of $\wh L(0)u_{(0)}=0$ that decays, in the sense that $u_{(0)}=\cO(\rho^\alpha)$ where $\alpha\geq\sup(\text{indicial gap})$. (Examples are given in~\S\S\ref{SssIGenK} and \ref{SssIEinSpec}. In classical self-adjoint spectral theory in dimensions $3$ and $4$, zero energy resonances would fall into this class as well.) Then $u_{(0)}$ admits a expansion as $\rho\to 0$ with leading-order term of the form $\rho^z u_{(0),\pa}(\omega)$ where $z$ is a small indicial root. Consider then a stationary contribution $c u_{(0)}(x)$, $c\in\C$, to the late-time asymptotics of a solution $u$ of $L u=f$ (see the discussion following~\eqref{EqIGenKBdState} below for the origin of such a term) from the perspective of $\iota^+$: it first arises at $\iota^+$-order $(z,0)$ with coefficient $t_*^z\cdot c\rho^z u_\pa(\omega)=c R^{-z} u_\pa(\omega)$. Due to $-z$ being below $-1$ times the indicial gap, this \emph{grows} relative to the weighted spaces on which $N_{\iota^+}(L,z)$ is invertible (or has meromorphic inverse). More dramatically---in the invertible case for simplicity---one can, for \emph{any} value of $c$, construct an element in $\ker N_{\iota^+}(L,z)$ whose $R\to 0$ expansion starts with $c R^{-z}u_\pa(\omega)$. This means that in the presence of zero energy bound states, the asymptotics at $\cK^+$ (encoded by the value $c$ in the example we are considering here) cannot be read off from the terms of the $\iota^+$-expansion: they come from $\cK^+$ proper, and determining them requires the full strength of spectral theory for the Kerr model.
\end{rmk}

\subsubsection{Spectral theory and asymptotics at \texorpdfstring{$\cK^+$}{K-plus} from bound states}
\label{SssIGenK}

In order to understand wave decay at $\cK^+$ (which encompasses decay as $t_*\to\infty$ in spatially compact sets), we use the Fourier transform in $t_*$. (This is only directly effective when the wave-type operator under consideration is stationary, i.e., commutes with $t_*$-translations. On dynamical spacetimes, spectral methods are still effective for extracting precise leading-order asymptotics; we discuss this in the Einstein context in~\S\ref{SssINElim} below.)

\medskip

\begin{figure}[!ht]
\centering
\includegraphics{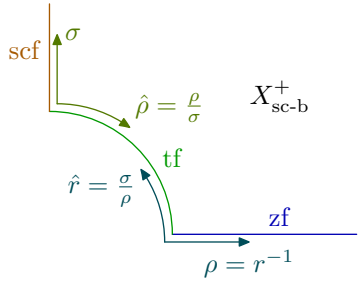}
\caption{The compact manifold $X_\scbtop^+$ with corners suitable for low-energy spectral theory on asymptotically flat spaces $X^\circ\subset\R^3_x$. Here $\rho=r^{-1}=|x|^{-1}$, and $\sigma$ is the spectral parameter (here restricted to $[0,1]$).}
\label{FigIGenKSpec}
\end{figure}

{\bf Fourier transform.} We first recall from \citeAF{\S\S\ref*{SssIPD} and \ref*{SsDFT}}---with polyhomogeneous generalizations stated in Proposition~\ref{PropTFHbphg}---how the Fourier transform
\[
  (\cF u)(\sigma,x) := \int_\R e^{i\sigma t_*}u(t_*,x)\,\dd t_*
\]
relates decay rates on $M$ with those on a suitable compactification of $\pm[0,\infty)_\sigma\times\R^3_x$. We illustrate the part $X_\scbtop^+$ of the compactification of $[0,1]_\sigma\times\R^3_x$ we shall work with in Figure~\ref{FigIGenKSpec}; the set where $\sigma=0$ is a union of two boundary hypersurfaces:
\begin{enumerate}
\item the \emph{zero energy face} $\zface$, where $\frac{\sigma}{\rho}=0$, $\rho:=r^{-1}$;
\item the \emph{transition face} $\tface$, where $\rho+\sigma=0$ (while $\hat r:=\frac{\sigma}{\rho}$ is an affine coordinate on it).
\end{enumerate}
Thus, $\zface$ is a compactification $\ol{\R^3}=\R^3\sqcup\Sph^2$ of $\R^3$, and $\tface=[0,\infty]_{\hat r}\times\Sph^2$ is a cone with a small end ($\hat r=0$) and a large end ($\hat\rho:=\hat r^{-1}=0$). The third boundary hypersurface, $\scface$ (called \emph{scattering face}), is largely irrelevant to our discussion.

\begin{figure}[!ht]
\centering
\includegraphics{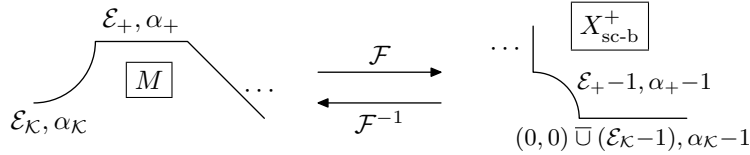}
\caption{Relationships between the $\iota^+$- and $\cK^+$-orders of functions on spacetime $M$ and the $\tface$- and $\zface$-orders of their Fourier transforms on the resolved low-energy space $X_\scbtop^+$. (The conditions on the $\scface$- and $\scri^+$-orders, which are omitted here, are stated precisely in Proposition~\ref{PropTFHbphg}.)}
\label{FigIGenKFT}
\end{figure}

Roughly speaking, the Fourier transform and its inverse have the mapping properties illustrated in Figure~\ref{FigIGenKFT}. See Proposition~\ref{PropTFHbphg} for the complete (and correct) low-frequency statement, which takes space on \emph{$L^2$-based spaces} and keeps track also of matching conditions of Fourier transforms at $\zface$ for $\sigma\in[0,1]$ and $\sigma\in-[0,1]=[-1,0]$. We only explain briefly why we focus on \emph{low frequencies} here: the Fourier transform of a symbol is Schwartz near infinity; similarly then, the Fourier transform of conormal functions on $M$ is Schwartz as $\sigma\to\pm\infty$. Conversely, high $\sigma$-regularity translates into fast $t_*$-decay upon inverse Fourier transforming. For this reason, only a neighborhood of the frequency $\sigma=0$ (where $\sigma$-regularity breaks down and only $\sigma\pa_\sigma$-regularity remains) matters for obtaining polynomial tails. --- There are two caveats:
\begin{enumerate}
\item{\rm (Mode stability.)} We need the wave-type operators under study to satisfy mode stability in $\Im\sigma\geq 0$, $\sigma\neq 0$, so that indeed only low frequencies require special attention. If mode stability failed, waves would not be conormal on $M$ due to oscillations or exponential growth.\footnote{We also do not discuss Klein--Gordon operators $\Box+\mu^2$ here, for which there are two exceptional frequencies $\pm\mu$ (e.g., the resolvent of the Klein--Gordon equation on Minkowski space has branch cuts there). The asymptotic behavior of waves, and thus necessarily also the analysis, is rather different in that setting. For spectral theoretic and microlocal perspectives, see Sussman \cite{SussmanKG} and Baskin--Doll--Gell-Redman \cite{BaskinDollGellRedmanKG}.}
\item{\rm (Trapping.)} Estimates at the (normally hyperbolic) trapped set enter only for control of high frequencies and ensure that inverting a stationary wave operator using its resolvent and the Fourier transform only ever loses a \emph{fixed} finite number of b-derivatives. (See \citeAF{Corollary~\ref*{CorDResHiLoc}} and the discussion preceding it.)
\end{enumerate}

The relationship between $\tface\subset X_\scbtop^+$ and $\iota^+\subset M$ can easily be illustrated by noting that the inverse Fourier transform of a function $w(\sigma,r)=w_0(\hat r)$, where we recall $\hat r=\frac{\sigma}{\rho}=\sigma r$, is
\begin{equation}
\label{EqIGenKIFT}
  (\cF^{-1}w)(t_*,r) = \frac{1}{2\pi}\int e^{-i\sigma t_*}w(\sigma,r)\,\dd\sigma = \frac{1}{2\pi}r^{-1}\int e^{-i\hat r\,t_*/r} w_0(\hat r)\,\dd\hat r = r^{-1} (\cF^{-1}w_0)\Bigl(\frac{t_*}{r}\Bigr),
\end{equation}
with $\frac{t_*}{r}=v$ (in the notation used already in~\eqref{EqIGen1RescBox}) being a projective coordinate on $\iota^+$ that vanishes at $\scri^+$ and equals $+\infty$ at $\cK^+$. Thus, the decay order (which stands in for $\sigma$-regularity) of $w_0$ at $\hat r=0$ translates into decay of $\cF^{-1}w$ at $\cK^+$, consistently with Figure~\ref{FigIGenKFT}.

On the level of operators, consider again $L$ of the form~\eqref{EqIGen2LaOp}, whose spectral family is thus
\[
  \wh L(\sigma) = 2 i\sigma\rho(\rho\pa_\rho-1-S) + \rho^2\wt L(0)(\rho,\omega,\rho\pa_\rho,\pa_\omega).
\]
To leading order at $\tface$, and for $\hat r:=\frac{|\sigma|}{\rho}$, we have
\begin{equation}
\label{EqIGenKNtf}
  |\sigma|^{-2}\wh L(\sigma) \equiv -2 i\hat\sigma\hat r^{-1}(\hat r\pa_{\hat r}+1+S) + \hat r^{-2}\wt L(0)(0,\omega,-\hat r\pa_{\hat r},\pa_\omega) =: N_\tface(L,\hat\sigma),\quad \hat\sigma:=\frac{\sigma}{|\sigma|}.
\end{equation}
The operator $N_\tface(L,\hat\sigma)$ is similar to~\eqref{EqIGen2LaNorm} (and in fact related to the latter via the Fourier transform in $v$, with dual variable $\hat r$ as already suggested by~\eqref{EqIGenKIFT}, and as explained in the arguments leading up to~\eqref{EqipMeroNtf}); and at $\hat r=0$ it is again a b-differential operator with indicial roots given by $-\lambda$ whenever $\lambda$ is an indicial root of the zero energy operator $\wh L(0)$, completely analogously to the behavior of $N_{\iota^+}(L,\lambda)$ at $R=0$. Thus, the considerations in~\S\ref{SssIGen2La} apply \emph{mutatis mutandis} also to the output of the inverse of $N_\tface(L,\hat\sigma)$ (and are essentially a Fourier transformed version thereof); details in the Einstein case are given in~\S\ref{Ssiptf}.

\medskip
{\bf Low-energy resolvent.} We briefly recall how to study the decay of $u$ solving $L u=f$ using Fourier techniques. (This follows \cite[\S{3}]{HintzPrice} and \cite[\S\S{4.2--4.3}]{HintzNonstat}, which in turn were strongly inspired by Guillarmou--Hassell \cite{GuillarmouHassellResI} and Vasy \cite{VasyLowEnergyLag}. For a detailed exposition, see Sussman \cite{SussmanResolventPhg}.) As noted above, only the low-energy part of $u$ matters, and thus we must study the solution of $\wh L(\sigma)\hat u(\sigma)=\hat f(\sigma)$ near $\sigma=0$. We sketch this here for $\sigma\geq 0$, i.e., near $\zface\cup\tface$. Given fairly weak uniform bounds for $\wh L(\sigma)^{-1}$ (for which we will rely on \citeAF{Proposition~\ref*{PropDResLo}}), it suffices to construct an approximate version of $\hat u(\sigma)$ in generalized Taylor series at $\zface$ and $\tface$. Thus, if $\hat f(0)$ (which is a function on $\zface$) has sufficient decay,\footnote{Concretely, the decay order $\alpha+2$ of $\hat f(0)$ must be such that $\alpha$ lies in the indicial gap of $\wh L(0)$, and then the output of $\wh L(0)^{-1}$ has decay order $\alpha$; see~\eqref{EqWEMode0} and Corollary~\ref{CorWE0Solv}.} one can apply the inverse of $\wh L(0)$ (provided it exists) to it to get $\hat u(0)$; if however the decay of $\hat f(0)$ towards $\tface$ is too weak, that means that the decay of $\hat f(\sigma)$ from the perspective of $\tface$ (i.e., in the coordinates $\sigma,\hat r$) towards $\zface$ (i.e., $\hat r\to 0$) is \emph{strong} enough for the applicability of $N_\tface(L,1)^{-1}$ to $\hat f(\sigma)$, which produces the next term in the $\tface$-expansion of $\hat u(\sigma)$. (This algorithm is used frequently in~\S\ref{SD}, e.g., starting on page~\pageref{ItD3Step31} with Step~3.1.) In this manner, one can solve away $\hat f$, and thus construct $\hat u$, in (generalized) Taylor series at $\zface$ and $\tface$.

As in Figure~\ref{FigIGenKFT}, the inverse Fourier transform translates the thereby determined behavior of $\hat u$ at low frequencies to asymptotics and decay of $u$ at $\iota^+$ (consistently with the discussion in~\S\ref{SssIGen1}) as well as at $\cK^+$ (consistently with the expectation in~\eqref{EqIGen2SmNaive} as far as $\cK^+$-orders are concerned). In the bulk of the paper, we use both Fourier-based techniques and $\iota^+$-normal operator arguments in tandem; the latter respect localization near $\iota^+$, but the former are unavoidable near $\cK^+$ (but slightly indirect as far as $\iota^+$-asymptotics are concerned).

\medskip
{\bf Zero energy bound states and resonances.} When $\hat L(0)$ is not invertible,\footnote{but Fredholm of index $0$ when acting between suitably weighted spaces---for which, in the case of wave-type operators on Kerr, \citeAF{Corollary~\ref*{CorSpLoInd0}} provides a general criterion} the above algorithm works after a relatively minor modification, at least in the setting considered in the present paper as well as in \citeAF{\S\ref*{SA2}, equation~(\ref*{EqA2AdmAug})}. Say $0\neq u_{(0)}=u_{(0)}(x)\in\ker\wh L(0)$ (with decay order $\geq\sup(\text{indicial gap})$), then $\wh L(\sigma)(\sigma^{-1}u_{(0)})=\pa_\sigma\wh L(0)u_{(0)}$ is not singular at $\sigma=0$. Thus, one can form the \emph{augmented operator} (or ``Grushin problem'')
\[
  \wt L(\sigma) := \begin{pmatrix} \wh L(\sigma) & \wh L(\sigma)(\sigma^{-1}u_{(0)}) \\ \la\cdot,f^*\ra_{L^2(\R^3_x)} & 0 \end{pmatrix}
\]
acting on the direct sum of a weighted Sobolev space and a copy of $\C$; here $f^*$ is a test function that is \emph{not} orthogonal to $u_{(0)}$. Thus, $\wt L(0)$ is \emph{invertible}. If there are further kernel elements of $\wh L(0)$, or if there is a generalized nullspace, they need to be put into the augmentation as well; see~\eqref{EqAdmLoNaive} for an exemplary version. (The correct construction of $\wt L(\sigma)$ is typically considerably more delicate since the output of the $(1,2)$-entry must be compatible with the weighted function spaces on which the uniform low-energy Fredholm analysis of $\wh L(\sigma)$ takes place; in the Einstein case, this is the subject of \S\ref{SsAdmLo}.) One can now solve $\wh L(\sigma)\hat u(\sigma)=\hat f(\sigma)$ by instead solving the augmented equation
\[
  \wt L(\sigma) \bigl(\hat u_{\rm reg}(\sigma),\,a(\sigma)\bigr) = \bigl(\hat f(\sigma),\,0\bigr)
\]
using the same algorithm as before (where now $a$ is simply a function whose order at $\sigma=0$ matches the $\zface$-order of $\hat u_{\rm reg}$), and then
\begin{equation}
\label{EqIGenKBdState}
  \hat u(\sigma) = \hat u_{\rm reg}(\sigma) + \sigma^{-1}a(\sigma)u_{(0)}.
\end{equation}
Thus, $\hat u$ loses one order at $\zface$ relative to $\hat f$: when the $\zface$-order of $\hat f$ is $((0,0),\alpha_\cK-1)$ (cf.\ Figure~\ref{FigIGenKFT} without $\cE_\cK$), then that of $\hat u$ is $((-1,0),\alpha_\cK-2)$. This has two main consequences.
\begin{enumerate}
\item\label{ItIGenKBdState1} Applying $\cF^{-1}$ to the singular $\sigma^{-1}$-term gives a stationary contribution to $u$ which is a multiple of the zero energy state $u_{(0)}$.\footnote{Since $\sigma^{-1}$ is not locally integrable near $\sigma=0$, some care is required in defining the inverse Fourier transform; we use contours in the upper half plane on $\C_\sigma$ that we shift down to the real axis. See, e.g., the proof of Proposition~\ref{PropD1Alm}.}
\item\label{ItIGenKBdState2} Applying $\cF^{-1}$ to the remainder yields a contribution to $u$ that only has $\cK^+$-order $\alpha_\cK-1$, i.e., one order less than the source term $f$. It is crucial for our purposes (see also \citeAF{\S\ref*{SA2} and Heuristic~I there} for a toy model) to be more precise: the regular part $\cF^{-1}\hat u_{\rm reg}$ contributes at the \emph{lossless order} $\alpha_\cK$, while the less-decaying part $\cF^{-1}(\sigma^{-1}a(\sigma))(t_*)\,u_{(0)}(x)=\cO(t_*^{-\alpha_\cK+1})u_{(0)}(x)$ has spatial dependence given by $u_{(0)}$. The latter is thus another example of a conormal expansion term (similarly to~\eqref{EqIGen2NoStruct}), which now arises from the spectral theory of $L$ (and not from large indicial roots).
\end{enumerate}
When $\wh L(\sigma)^{-1}$ has a \emph{second order} ``pole''\footnote{We put ``pole'' in quotes since the output of $\wh L(\sigma)^{-1}$ can only be properly understood on the resolved space $X_\scbtop^+$ that is \emph{not} a product of a factor for the frequency variable $\sigma$ and a spatial manifold.} (as is the case for the linearized gauge-fixed Einstein operator we shall study), there is also a linear (in $t_*$) contribution to $u$, and the $\cK^+$-decay of the remainder of $u$ loses \emph{two} orders of decay relative to $f$. The pictorial summary of this analysis at $\cK^+$, in the case of a second order pole, is
\begin{equation}
\label{EqIGenKPole2}
  \raisebox{-1.5em}{\includegraphics{EqIGenKPole2}}
\end{equation}
(The $\cK^+$-order $((-1,0),\alpha_\cK-2)$ here arises via the inverse Fourier transform from the $\zface$-order $((-2,0),\alpha_\cK-3)$ of $\hat u(\sigma)=\wh L(\sigma)^{-1}\hat f(\sigma)$.)

\subsection{Linearized gauge-fixed Einstein equation on Kerr backgrounds}
\label{SsIEin}

We now apply the general considerations in~\S\ref{SsIGen} to the case of the linearized gauge-fixed Einstein equations on Kerr.

We work on a fixed subextremal Kerr spacetime with metric $g_b$, $b=(\bhm,\bha)$. Linear stability results for $g_b$ \cite{DafermosHolzegelRodnianskiSchwarzschildStability,AnderssonBackdahlBlueMaKerr,HaefnerHintzVasyKerr,HaefnerHintzVasyKerrLarge} amount to the decay of solutions of initial value problems for the linearized Einstein equation $D_{g_b}\Ric(h)=0$ towards a member $\dot g_b(\dot b)$ of the linearized Kerr family (see~\eqref{EqIPriorLin}). In standard harmonic gauge $\delta_{g_b}\sfG_{g_b}h=0$, where
\[
  (\delta_g h)_\mu:=-g^{\nu\lambda}h_{\mu\nu;\lambda},\quad
  \sfG_g h:=h-\frac12 g\tr_g h,
\]
one can solve $D_{g_b}\Ric(h)=0$ by solving the \emph{gauge-fixed linearized Einstein equation}
\begin{equation}
\label{EqIEinFix}
  (D_{g_b}\Ric + \delta_{g_b}^*\delta_{g_b}\sfG_{g_b})h = 0
\end{equation}
(where $(\delta_g^*\omega)_{\mu\nu}:=\frac12(\omega_{\mu;\nu}+\omega_{\nu;\mu})$) for suitable Cauchy data, followed by an argument involving the second Bianchi identity (recalled in more generality below) for deducing $D_{g_b}\Ric(h)=0$. An important point is that~\eqref{EqIEinFix} is a linear wave equation (on symmetric 2-tensors), with principal part $\frac12\Box_{g_b}$; this structure persists also if one replaces $\delta_{g_b}^*$ and $\delta_{g_b}$ by $0$-th order modifications
\[
  \delta_{g_b,E^\cC}^* := \delta_{g_b}^*+E^\cC,\quad
  \delta_{g_b,E^\Ups} := \delta_{g_b}+E^\Ups.
\]
We will take $E^\cC$, resp.\ $E^\Ups$ to be linear combinations of the operators
\begin{equation}
\label{EqIEinMods}
  \omega\mapsto \fc\otimes_s\omega,\ \ \omega\mapsto g^{-1}(\fc,\omega)g,\quad\text{resp.}\quad h\mapsto \iota_{g^{-1}(\fc)}h,\ \ h\mapsto (\tr_g h)\fc
\end{equation}
for suitable choices of 1-forms $\fc$; we will explain the motivation for our concrete choices as we go on.\footnote{We immediately mention that we will choose $\fc$ to decay like $r^{-1}$ in order to ensure that $L_b$ in~\eqref{EqIEinLb} has the correct approximate homogeneity/decay properties at $I^0$, $\scri^+$, and $\iota^+$ (cf.\ \eqref{EqIGenScriOp} and \eqref{EqIGen1Op}).} The overall aim is to ensure that solutions of
\begin{equation}
\label{EqIEinLb}
  L_b h = 0,\quad L_b := D_{g_b}\Ric + \delta_{g_b,E^\cC}^*\delta_{g_b,E^\Ups}\sfG_{g_b} = \frac12\Box_{g_b} + \text{l.o.t.},
\end{equation}
have as much decay as possible (and as eventually needed for nonlinear purposes), up to terms with ``physical'' meaning (in particular: linearized Kerr metrics) and ``pure gauge'' terms that can be eliminated via modifications of the gauge condition.

We first discuss basic spectral properties of $L_b$ at non-zero and zero frequencies in~\S\ref{SssIEinSpec}; this is essentially a recapitulation of ideas in \cite{HaefnerHintzVasyKerr,HaefnerHintzVasyKerrLarge} and \cite{AnderssonHaefnerWhitingMode}. More refined spectral properties related to the large indicial roots of $\wh{L_b}(0)$ and their effect on $\cK^+$-asymptotics (cf.\ \S\ref{SssIGen2La}) are discussed in~\S\S\ref{SssIEinElim}--\ref{SssIEinLa}; this turns out to be rather delicate and is one of the key novelties of this paper. The discussion of the structure of $L_b$ at $\scri^+$ in~\S\ref{SssIEinScri} is a variation on a theme developed in \cite{HintzVasyMink4,HintzMink4Gauge}.

\subsubsection{Basic spectral theory}
\label{SssIEinSpec}

We first discuss some homogeneous solutions of $L_b h=0$. First, linearizing $\Ric(\phi^*g_b)=\phi^*\Ric(g_b)=0$ around the diffeomorphism $\phi=\Id$ gives $D_{g_b}\Ric\circ\delta_{g_b}^*=0$; thus, for any 1-form $\omega$, we have
\[
  L_b(\delta_{g_b}^*\omega)=0\ \ \text{for}\ \ \omega \in \ker \Box_{g_b,E^\Ups}^\Ups,\ \ \Box_{g_b,E^\Ups}^\Ups:=2\delta_{g_b,E^\Ups}\sfG_{g_b}\delta_{g_b}^*.
\]
We call the operator $\Box_{g_b,E^\Ups}^\Ups=\Box_{g_b}+{\rm l.o.t.}$ the \emph{gauge potential wave operator}, and $\delta_{g_b}^*\omega$ a \emph{pure gauge solution}. The condition $\omega\in\ker\Box_{g_b,E^\Ups}^\Ups$ means that $\delta_{g_b}^*\omega$ satisfies the linear gauge condition $\delta_{g_b,E^\Ups}\sfG_{g_b}(\delta_{g_b}^*\omega)=0$. More generally, if $D_{g_b}\Ric(h)=0$, then upon adding to $h$ a pure gauge term $\delta_{g_b}^*\omega$ (which does not affect the output of $D_{g_b}\Ric$), we can arrange that also the linearized gauge condition $\delta_{g_b,E^\Ups}\sfG_{g_b}(h+\delta_{g_b}^*\omega)=0$ holds; this amounts to solving a linear wave equation for $\omega$ with source $-2\delta_{g_b,E^\Ups}\sfG_{g_b}h$. Then
\[
  L_b(h+\delta_{g_b}^*\omega)=0.
\]
This process ``puts $h$ into the correct (linearized) gauge.''

The first spectral requirement on $L_b$ is that it satisfy mode stability in $\Im\sigma\geq 0$, $\sigma\neq 0$. This was proved by Andersson--H\"afner--Whiting \cite{AnderssonHaefnerWhitingMode} with trivial modifications $E^\Ups=0$, $E^\cC=0$. Let us briefly recall the basic structure of the argument: if $h(t_*,x)=e^{-i\sigma t_*}h_0(x)\in\ker L_b$ is an outgoing mode solution of $L_b$ with $\Im\sigma\geq 0$, then $\delta_{g_b}\sfG_{g_b}L_b h=0$ implies (by the linearized second Bianchi identity)
\[
  \Box_{g_b,E^\cC}^\cC\eta = 0,\quad \Box_{g_b,E^\cC}^\cC:=2\delta_{g_b}\sfG_{g_b}\delta_{g_b,E^\cC}^*=\Box_{g_b}+{\rm l.o.t.},\ \ \eta:=\delta_{g_b,E^\Ups}\sfG_{g_b}h.
\]
If mode stability holds for the \emph{constraint propagation wave operator} $\Box_{g_b,E^\cC}^\cC$---which may or may not hold depending on the choice of $E^\cC$, and which is precisely what \emph{constraint damping} is about---this implies $\eta=0$, i.e., $h$ automatically satisfies the gauge condition $\delta_{g_b,E^\Ups}\sfG_{g_b}h=0$; therefore, $D_{g_b}\Ric(h)=0$. Mode stability for the \emph{ungauged} linearized Einstein equation $D_{g_b}\Ric(h)=0$ (see \cite[Theorem~6.1(1)]{AnderssonHaefnerWhitingMode}) implies that $h$ is the sum of a linearized Kerr metric $\dot g_b(\dot b)$ and a pure gauge term $\delta_{g_b}^*\omega$. In the absence of the former (which can only occur at frequency $\sigma=0$), the gauge condition $\eta=\delta_{g_b,E^\Ups}\sfG_{g_b}h=\delta_{g_b,E^\Ups}\sfG_{g_b}(\delta_{g_b}^*\omega)=0$ implies that $\omega$ is a mode solution of $\Box_{g_b,E^\Ups}^\Ups$, and must therefore vanish (and with it also $h$), provided that mode stability holds for $\Box_{g_b,E^\Ups}^\Ups$.

The zero energy nullspace of $L_b$ is always nontrivial; it is $7$-dimensional when the two 1-form wave operators have no zero modes, and is spanned by
\begin{enumerate}
\myitem{ItIEinSpecKerr}{i} linearized Kerr metrics $\dot g_b^\Ups(\dot b)=\dot g_b(\dot b)+\delta_{g_b}^*\dot\omega_b(\dot b)$, put into the correct gauge (see Proposition~\ref{PropWEMode0Kerr});
\myitem{ItIEinSpecTrans}{ii} deformation tensors $\delta_{g_b}^*\omega_{b,\rms 1}^{(0)}(\scal)$ of asymptotic translations in directions $\scal\in\mathspan\{\frac{x^i}{r}\colon i=1,2,3\}$; the gauge potentials here are $\omega_{b,\rms 1}^{(0)}(\scal)=\dd(r\scal)+\cO(r^{-1})=\dd x^i+\cO(r^{-1})$.
\end{enumerate}
(These are thus zero energy bound states or resonances in the terminology of Remark~\ref{RmkIGen2Bound}.) See Propositions~\ref{PropWG0Symm} and~\ref{PropWEMode0}. There is also a generalized nullspace, spanned by
\begin{enumerate}
\myitem{ItIEinSpecBoost}{iii} deformation tensors $h_{b,\rms 1}^{\leq 1}(\scal)$ of asymptotic Lorentz boosts $\omega_{b,\rms 1}^{(0),\leq 1}(\scal)=t_*\omega_{b,\rms 1}^{(0)}(\scal)+\breve\omega_{b,\rms 1}^{(0),1}(\scal)$; these gauge potentials are equal to $t\,\dd x^i-x^i\,\dd t+{\rm l.o.t}$.
\end{enumerate}
In the terminology of~\S\ref{SssIGenK}, this means that $\wh{L_b}(\sigma)^{-1}$ has a second order pole at $\sigma=0$, and thus forward solutions of $L_b$ lose two powers of $t_*$-decay (at $\cK^+$) and typically have non-decaying contributions given by the tensors in~\eqref{ItIEinSpecKerr}--\eqref{ItIEinSpecBoost}. From the perspective of the nonlinear stability problem, the contribution $\dot g_b^\Ups(\dot b)$ informs the update of the final black hole parameters. On the other hand, we will need a mechanism for ensuring that the final black hole has vanishing linear momentum and is centered (by eliminating contributions from $h_{b,\rms 1}^{\leq 1}$ and $h_{b,\rms 1}$, respectively); see~\S\S\ref{SssIEinElim} and \ref{SssINElim}.

Roughly speaking, this discussion shows that
\[
  \text{\parbox{0.8\textwidth}{\it mode stability for the gauge potential wave operator $\Box_{g_b,E^\Ups}^\Ups$ and the constraint damping wave operator $\Box_{g_b,E^\cC}^\cC$ (both acting on 1-forms) implies mode stability for the gauge-fixed linearized Einstein operator $L_b$, and a simple structure of its (generalized) zero energy nullspace.}}
\]
Mode stability, including at zero frequency, can be arranged perturbatively for $\Box_{g_b,E^\Ups}^\Ups$ (i.e., for small and compactly supported $E^\Ups$); see Proposition~\ref{PropWGMode} (which is a variant of \cite[Proposition~3.7]{HintzGlueLocIII}). The same is true for $\Box_{g_b,E^\cC}^\cC$, as shown in \cite[Proposition~3.7]{HintzGlueLocIII}, and this then suffices for a proof of linear stability. We shall see in~\S\ref{SssIEinLa} though that for stronger decay and nonlinear purposes, we need to place much more stringent requirements on $\Box_{g_b,E^\cC}^\cC$ than mere mode stability, and these require non-perturbative modifications (cf.\ \cite{HintzKerrCD} and Theorem~\ref{ThmWCRec}).

\subsubsection{Basic strategy for eliminating pure gauge terms}
\label{SssIEinElim}

In order to motivate further desiderata for $L_b$, we briefly discuss the basic mechanism for eliminating pure gauge contributions to the late-time asymptotics of solutions $h$ of $L_b h=f$ in two simple cases. For localization near $\cK^+$, fix a cutoff function $\chi_\cK$ that equals $1$ near $\cK^+$ and $0$ near $\scri^+$ and the initial Cauchy hypersurface.\footnote{A possible choice is thus $\chi_\cK=\chi_0(t_*)\chi_1(\frac{r}{t_*})$ where $\chi_0$ equals $1$ near $\infty$ and $\chi_1$ equals $1$ near $0$.}

\begin{figure}[!ht]
\centering
\includegraphics{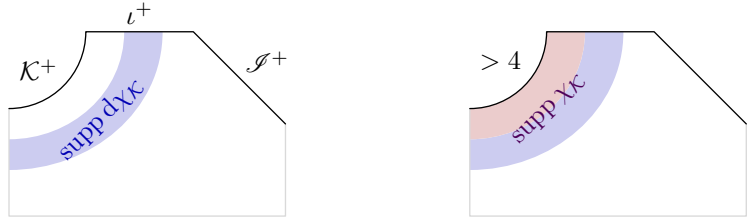}
\caption{Elimination of pure gauge contributions to the late-time asymptotics of metric perturbations using gauge modifications and metric patches. \textit{On the left:} stationary center-of-mass shifts can be eliminated using gauge modifications supported on $\supp\dd\chi_\cK$, with metric perturbation patches supported on the same set. \textit{On the right:} dynamical pure gauge solutions can be eliminated in a similar fashion, except the gauge modification now has full support on $\supp\chi_\cK$ but strong decay ($>4$ orders of decay suffice in this paper) at $\cK^+$.}
\label{FigIEinElim}
\end{figure}

\medskip
{\bf Center-of-mass shifts.} Suppose that the final black hole, in linearized theory, is subject to a stationary center-of-mass motion; that is, $h$ features a contribution $\chi_\cK h_{b,\rms 1}(\scal)=\chi_\cK\delta_{g_b}^*\omega_{b,\rms 1}^{(0)}(\scal)$. Since $h_{b,\rms 1}$ both solves the linearized Einstein equation and satisfies the linearized gauge condition, we then have
\begin{subequations}
\begin{equation}
\label{EqIEinElimRewrite}
\begin{split}
  L_b(\chi_\cK h_{b,\rms 1}) &= D_{g_b}\Ric\bigl(\chi_\cK \delta_{g_b}^*\omega_{b,\rms 1}^{(0)}\bigr) + \delta_{g_b,E^\cC}^* [ \delta_{g_b,E^\Ups}\sfG_{g_b},\chi_\cK]h_{b,\rms 1} \\
    &= D_{g_b}\Ric(h_{\rms 1}^{(0)}) + \delta_{g_b,E^\cC}^*\vartheta_{\rms 1}^{(0)},
\end{split}
\end{equation}
where
\begin{equation}
\label{EqIEinElimRewrite2}
  h_{\rms 1}^{(0)}(\scal) := -[\delta_{g_b}^*,\chi_\cK]\omega_{b,\rms 1}^{(0)}(\scal),\quad
  \vartheta_{\rms 1}^{(0)}(\scal) := [\delta_{g_b,E^\Ups}\sfG_{g_b},\chi_\cK]h_{b,\rms 1}(\scal).
\end{equation}
\end{subequations}
Setting $\vartheta':=\vartheta_{\rms 1}^{(0)}-\delta_{g_b,E^\Ups}\sfG_{g_b}h_{\rms 1}^{(0)}$ (which vanishes near $\cK^+$ and has decay order $2$ at $\iota^+$), this reads\footnote{Later on, we will work directly with~\eqref{EqIEinElimRewrite}, rather than adding and subtracting the gauge term acting on $h_{\rms 1}^{(0)}$; this simplifies some algebra. See~\eqref{EqIDNElimCorrP} and, e.g., ~\eqref{EqD4ParNotildeP2}--\eqref{EqD4ParNoRewrite2}.}
\[
  D_{g_b}\Ric(h_{\rms 1}^{(0)}) + \delta_{g_b,E^\cC}^*\bigl(\delta_{g_b,E^\Ups}\sfG_{g_b}h_{\rms 1}^{(0)} + \vartheta'\bigr) = L_b(\chi_\cK h_{b,\rms 1}).
\]
We interpret this as follows: given the source term $f:=L_b(\chi_\cK h_{b,\rms 1})$---which produces the forward solution $\chi_\cK h_{b,\rms 1}$ that we wish to eliminate---one modifies the gauge condition using $\vartheta'$ to $\delta_{g_b,E^\Ups}\sfG_{g_b}h+\vartheta'=0$, and then the forward solution with this new gauge condition is $h_{\rms 1}^{(0)}$, \emph{which vanishes near $\cK^+$} (and has the acceptable decay order $1$ at $\iota^+$). In other words, changing the gauge condition eliminates the center-of-mass movement at the expense of a patch $h_{\rms 1}^{(0)}$ of the metric perturbation that is, however, localized on $\supp\dd\chi_\cK$; see the left panel of Figure~\ref{FigIEinElim}.

\medskip
{\bf Decaying pure gauge terms.} Suppose now that the $\cK^+$-asymptotics of $h$ feature a pure gauge contribution $\chi_\cK\delta_{g_b}^*(a(t_*)\omega)$, $a(t_*)=t_*^{-\alpha}$,\footnote{We omit a discussion of factors of $\log t_*$ in this introduction since they only cause notational overhead but no conceptual difficulties.} $\alpha>0$, for some stationary 1-form $\omega$.\footnote{We will explain the origin of such terms in~\S\ref{SssIEinLa} below; they will be large zero energy states of $\Box_{g_b,E^\Ups}^\Ups$.} An example to keep in mind is $\omega=\omega_{b,\rms 1}^{(0)}$, in which case this contribution amounts to a decaying center-of-mass shift.\footnote{For $\alpha\leq 2$, this decay is too weak for our purposes, as discussed in~\S\ref{SssINAdm}, and hence we need to eliminate such terms.} Similarly to above, one writes
\begin{equation}
\label{EqIEinElimDec}
  L_b\bigl(\chi_\cK\delta_{g_b}^*(a\omega)\bigr) = -D_{g_b}\Ric\bigl( [\delta_{g_b}^*,\chi_\cK](a\omega) \bigr) + \delta_{g_b,E^\cC}^*\delta_{g_b,E^\Ups}\sfG_{g_b}\bigl(\chi_\cK\delta_{g_b}^*(a\omega)\bigr).
\end{equation}
Since $a$ is not constant, the gauge modification term $\vartheta:=\delta_{g_b,E^\Ups}\sfG_{g_b}(\chi_\cK\delta_{g_b}^*(a\omega))$ requires further study: it equals $\frac12$ times the sum of $[\Box_{g_b,E^\Ups}^\Ups,\chi_\cK](a\omega)$ (which is acceptable as before since it is supported away from $\cK^+$) and
\[
  \chi_\cK[\Box_{g_b,E^\Ups}^\Ups,a]\omega = \chi_\cK a'(t_*)[\Box_{g_b,E^\Ups}^\Ups,t_*]\omega + \cO(a''(t_*)).
\]
Since $a'=\cO(t_*^{-\alpha-1})$, this has order $\alpha+1$ at $\cK^+$, which is higher than that of the term $\delta_{g_b}^*(a\omega)$ we wish to eliminate, but not necessarily sufficiently high: after all, in a nonlinear iteration, this gauge modification would be seen in the next iteration step as a source term of order $\alpha+1$ at $\cK^+$, which turns out to be unacceptable for us unless $\alpha+1>4$ (as discussed in~\S\ref{SssINHeur} below).

To make progress, we recall the discussion around~\eqref{EqIGen2LaGen} and attempt to eliminate not $\chi_\cK\delta_{g_b}^*(a\omega)$ but a more refined (at higher orders of $\cK^+$-decay) tensor, namely
\begin{subequations}
\begin{equation}
\label{EqIEinElimHi}
  \chi_\cK\delta_{g_b}^*\omega^{\leq k}(a(t_*))
\end{equation}
where $\omega^{\leq k}(a(t_*))=a(t_*)\omega+a'(t_*)\breve\omega^1+a''(t_*)\breve\omega^2+\cdots$; here the stationary 1-forms $\breve\omega^j$ are such that $t_*\omega+\breve\omega^1$, $t_*^2\omega+2 t_*\breve\omega^1+2\breve\omega^2$, etc.\ are generalized large zero energy states of the gauge potential wave operator (so elements of $\ker\Box_{g_b,E^\Ups}^\Ups$). Repeating the above computation with $\omega^{\leq k}(a)$ in place of $a\omega$, the previously offensive contribution to the gauge modification term
\begin{equation}
\label{EqIEinElimHiMod}
  \delta_{g_b,E^\Ups}\sfG_{g_b}\bigl(\chi_\cK\delta_{g_b}^*\omega^{\leq k}(a(t_*))\bigr)
\end{equation}
\end{subequations}
is now $\chi_\cK\Box_{g_b,E^\Ups}^\Ups\bigl(\omega^{\leq k}(a(t_*))\bigr)$, so (pretending for brevity that $\Box_{g_b,E^\Ups}^\Ups$ has no $\pa_{t_*}^2$-terms)
\begin{align*}
  &\chi_\cK a'(t_*) \underbrace{\Bigl( [\Box_{g_b,E^\Ups}^\Ups,t_*]\omega + \Box_{g_b,E^\Ups}\breve\omega^1 \Bigr)}_{=\,0} + \chi_\cK a''(t_*)\underbrace{\Bigl( [\Box_{g_b,E^\Ups}^\Ups,t_*]\breve\omega^1 + \Box_{g_b,E^\Ups}^\Ups\breve\omega^2 \Bigr)}_{=\,0} \\
  &\qquad + \cdots + \chi_\cK a^{(k+1)}[\Box_{g_b,E^\Ups}^\Ups,t_*]\breve\omega^k = \cO(t_*^{-\alpha-k-1}).
\end{align*}
This has as much decay as we want if we take $k$ large enough; $\alpha+k+1>4$ suffices for our purposes. (See then~\eqref{EqD4ParNoLog1}--\eqref{EqD4ParNoLog2}.) See the right panel of Figure~\ref{FigIEinElim}. Recalling Example~\ref{ExIGen2LaStates}, the construction of generalized large zero energy states for $\Box_{g_b,E^\Ups}^\Ups$ is straightforward since its zero energy operator is invertible; the ones needed for the stability proof are constructed in Proposition~\ref{PropWG0Large}.

\subsubsection{Large indicial roots and large zero energy states}
\label{SssIEinLa}

The propagation of decay from initial data to $\iota^+$ discussed in~\S\S\ref{SssIGen0}--\ref{SssIGen1} suggests that $r^{-\alpha}$-terms, $\alpha>1$, in the initial data of $h$ solving $L_b h=f$ (corresponding to $r^{-\alpha}$-terms contributing to $\gamma$ or $r k$ in~\eqref{EqIDataPhg}) cause $t_*^{-\alpha}$-terms at $\iota^+$. Sources $f$ arising in the nonlinear iteration furthermore turn out to also cause a $t_*^{-1}$-term at $\iota^+$.\footnote{We refer the reader to Proposition~\ref{PropDScriFormal}. The essence is that on a dynamical background, there is a coupling of the trace-free spherical-spherical part of the metric perturbation, which decays at radiation field order $r^{-1}$, into the $(\dd t_*)^2$-component (which by itself has a radiation field only at the order $r^{-1-2\gamma^\Ups}$ with $0<\gamma^\Ups\ll 1$) of a metric perturbation (cf.\ the $(7,6)$-component of $A_h$ in~\eqref{EqExOpLinAB}); the resulting $r^{-1}$-term in the $(\dd t_*)^2$-component then does propagate from $\scri^+$ to $\iota^+$ and gives the claimed $t_*^{-1}$-term.} Following the observations in~\S\ref{SssIGen2La} concerning solutions of the $\iota^+$-normal operator
\begin{equation}
\label{EqIEinLaNormOp}
  N_{\iota^+}(\lambda)
\end{equation}
of $L_b$ (which is independent of $b$ and only depends on its Minkowskian leading-order terms), we must determine the asymptotics of the coefficients (which are functions on $\iota^+$ valued in a symmetric spacetime-2-tensor bundle) of such terms at $\iota^+\cap\cK^+$; and to this end we need to compute the indicial roots of $\wh{L_b}(0)$. The expectation is then that $t_*^{-\alpha}$-terms of $h$ at $\iota^+$ cause $t_*^{-\alpha+\lambda}$-terms at $\cK^+$ where $\lambda$ is a large indicial root of $\wh{L_b}(0)$. (We will see that this is not quite accurate due to the existence of exceptional indicial roots for which no large zero energy state exists in the usual sense.)

More specifically, we need to
\begin{enumerate}[label=(\alph*)]
\item compute the large indicial roots (after identifying an indicial gap) and the associated indicial solutions (see Lemma~\ref{LemmaWEInd});
\item determine whether the first few of these large Minkowskian zero energy states (namely those that contribute at orders $\leq 4$ at $\cK^+$) can be extended to states on Kerr (cf.\ \eqref{EqIGen2LaState}),
\item and, \emph{crucially}, ensure that the first few of these large zero energy states (namely those that contribute at orders $\leq 2$ at $\cK^+$) on Kerr are \emph{pure gauge}.
\end{enumerate}
The particular $\cK^+$-orders here will be explained in~\S\ref{SssINHeur} below; the idea behind the final requirement is that one can then eliminate these poorly decaying terms in late-time asymptotics via gauge modifications as in~\S\ref{SssIEinElim}, which suggests that one should be able to attain $\cO(t_*^{-2-\eps_\cK})$-decay, $0<\eps_\cK\ll 1$, of the gravitational wave tail in an updated gauge (cf.\ the decay rate stated in Theorem~\ref{ThmISimple}).

The indicial gap of $\wh{L_b}(0)$ will be an interval of the form $(0,\eps_\ind)$ for some $\eps_\ind>0$. (With no modifications $E^\Ups$ and $E^\cC$, it would be $(0,1)$ just as for the scalar wave operator.)

\medskip
{\bf Indicial roots.} Roughly speaking, there are three types of indicial roots of $\wh{L_b}(0)$:
\begin{enumerate}
\item\label{ItIEinLa1} decay rates of physical states (e.g., indicial roots $1$ and $2$ corresponding to the leading-order terms $\rho^1$ and $\rho^2$ as $\rho=r^{-1}\to 0$ of linearized mass and angular momentum changes $\dot g_{\bhm,\bha}^\Ups(\dot\bhm,0)$ and $\dot g_{\bhm,\bha}^\Ups(0,\dot\bha)$);
\item\label{ItIEinLa2} indicial roots of $\wh{\Box_{g_b,E^\Ups}^\Ups}(0)$ (shifted by $1$ to account for the gain in $r$-decay upon applying $\wh{\delta_{g_b}^*}(0)$);
\item\label{ItIEinLa3} indicial roots of $\wh{\Box_{g_b,E^\cC}^\cC}(0)$ (shifted by $-1$, as explained in the proof of\footnote{In short, the indicial roots of $\wh{L_b}(0)^*$ are $1-\bar\lambda$ when $\lambda$ is an indicial root of $\wh{L_b}(0)$. But $\sfG_{g_b}L_b^*\sfG_{g_b}=D_{g_b}\Ric+(\delta_{g_b,E^\Ups})^*(\delta_{g_b,E^\cC}^*)^*\sfG_{g_b}$, so the roles of gauge change $E^\Ups$ and constraint damping $E^\cC$ are reversed for the adjoint. Indicial roots $\mu$ of $\wh{\Box_{g_b,E^\cC}^\cC}(0)^*$ (which could thus be called the ``dual gauge potential wave operator'') thus give rise to indicial roots $\mu+1$ for $\wh{L_b}(0)^*$ (the shift by $1$ accounting for the gain in $r$-decay upon applying $\sfG_{g_b}\wh{\delta_{g_b}^*}(0)$). But $\kappa:=1-\bar\mu$ is an indicial root of $\wh{\Box_{g_b,E^\cC}^\cC}(0)$. Reversing this chain gives $\lambda=\kappa-1$ as an indicial root of $\wh{L_b}(0)$. --- More generally, this reasoning shows how (large) zero energy states of $(\Box_{g_b,E^\cC}^\cC)^*$ give rise to ``dual pure gauge states'' of $L_b^*$, cf.\ Proposition~\ref{PropWC0Dual}\eqref{ItWC0Dualh} and Proposition~\ref{PropWEMode0}\eqref{ItWEMode0Co}.} Lemma~\ref{LemmaWEInd}).
\end{enumerate}
Some of the indicial roots of type~\eqref{ItIEinLa1} are positive, thus not large indicial roots and hence irrelevant for our present discussion; but $-2,-3,\ldots$ turn out to be physical roots as well; they contribute at $\cK^+$-order $t_*^{-1-2}=t_*^{-3}$, $t_*^{-1-3}=t_*^{-4}$, $\ldots$, and are the cause for the limitation of the $\cK^+$-decay rate $3$ (cf.\ Remark~\ref{RmkIt3}). Indicial roots of type~\eqref{ItIEinLa2} are expected to be acceptable because they should correspond to pure gauge terms. Type~\eqref{ItIEinLa3}, however, spells trouble: if $h$ is a large zero energy state corresponding to such an indicial root, then since it must depend on the choice of $E^\cC$ (after all, it satisfies a PDE involving $E^\cC$), it \emph{cannot} solve $D_{g_b}\Ric(h)=0$ and $\delta_{g_b,E^\Ups}\sfG_{g_b}h=0$ separately; rather, the validity $L_b h=0$ must be due to a cancellation of these two terms. Being neither physical nor pure gauge, we have no means to handle or eliminate asymptotic terms of the sort $t_*^{-\alpha}h$.\footnote{In the linear stability problem, this is inconsequential since the linearized constraint equations, and thus the validity of the linearized gauge condition throughout spacetime, guarantee that such terms cannot possibly arise. In a nonlinear context, where we do not have access to the linearized constraints, there is no analogous such argument.} \footnote{More subtly, the indicial roots of $\wh{\Box_{g_b,E^\cC}^\cC}(0)$ affect the decay rate of elements in the cokernel of $\wh{L_b}(0)$ (see Proposition~\ref{PropWC0Dual}\eqref{ItWC0Dualh} and Proposition~\ref{PropWEMode0}\eqref{ItWEMode0Co}) and thus the solvability properties of equations $\wh{L_b}(0)h=f$ that one needs to solve in order to construct large zero energy states with given asymptotic behavior as $r\to\infty$; see the proof of Proposition~\ref{PropWE0}.} \footnote{In the context of the stability of the static patch of de~Sitter space, exponentially growing resonant states of this type were shown to exist \cite[Appendix~C.3]{HintzVasyKdSStability} for poor choices of $E^\cC$.}

What is therefore necessary in our approach to nonlinear stability is to ensure that the indicial roots $\lambda$ of $\wh{\Box_{g_b,E^\cC}^\cC}(0)$ with $\Re\lambda<1$ (so that $\lambda-1$ is a large indicial root of $\wh{L_b}(0)$) in fact satisfy $\Re\lambda<-C_0$ for some finite but reasonably large $C_0$ ($C_0=100$ suffices by a large margin); in other words, we need that
\begin{equation}
\label{EqIEinEnhancedCD}
  \text{\parbox{0.8\textwidth}{\centering\it no indicial root of $\wh{\Box_{g_b,E^\cC}^\cC}(0)$ has real part in $[-C_0,1)$.}}
\end{equation}
Large indicial roots of type~\eqref{ItIEinLa3} then only contribute to the $\cK^+$-asymptotics of solutions $h$ of $L_b h=f$ at very high orders $\gtrsim C_0$ and can thus be treated as remainder terms in our nonlinear analysis. When $E^\cC=0$, then~\eqref{EqIEinEnhancedCD} is true only for $C_0<0$; therefore, arranging~\eqref{EqIEinEnhancedCD} requires a large perturbation $E^\cC$. Since indicial roots are determined by the leading-order (in the sense of decay as $\rho\to 0$) part of $\wh{\Box_{g_b,E^\cC}^\cC}(0)$, the choice of $E^\cC$ must moreover involve 1-forms $\fc$ in~\eqref{EqIEinMods} of size $r^{-1}$. That it is possible to choose $E^\cC$ with the property~\eqref{EqIEinEnhancedCD} is shown in the companion paper \cite{HintzKerrCD}, the main results of which are summarized in Theorem~\ref{ThmWCRec}; we refer the reader to \cite{HintzKerrCD} for a description of the methods used there (which build on \cite{GundlachCalabreseHinderMartinConstraintDamping,HintzVasyKdSStability,HintzPetersenVasyKdS}).

For $\wh{L_b}(0)$ then, the indicial roots $\lambda$ split into two halves separated by an indicial gap $(0,\eps_\ind)$: those with $\Re\lambda\geq\eps_\ind>0$, and those with $\Re\lambda\leq 0$; among the latter, all roots with $\Re\lambda\geq-100$ are of type~\eqref{ItIEinLa1} or (suitably interpreted) of type \eqref{ItIEinLa2}.

\medskip
{\bf Pure gauge indicial roots.} The modification $E^\Ups$ of the linearized gauge condition we work with features small multiples of the terms in~\eqref{EqIEinMods} with $\fc=r^{-1}\,\dd t$; this gives improved decay of certain metric coefficients at $\scri^+$ and shifts indicial roots slightly. We discuss here only three exemplary large indicial roots $\lambda$ of $\wh{L_b}(0)$, and recall that they are expected to yield contributions to the $\cK^+$-asymptotics of $h$ of size $t_*^{-\alpha+\lambda}$ where $\alpha=1$ or $\Re\alpha>1$.
\begin{enumerate}[leftmargin=2.5em]
\item{\rm (A non-exceptional pure gauge indicial root.)} $\wh{L_b}(0)$ has an indicial root $\lambda$ with $-1\ll\lambda<0$ (in fact $2$ of them, denoted $-\lambda^\Ups_{\rms l,1}+1$, $l=0,1$, in Lemma~\ref{LemmaWEInd} using the notation of Lemma~\ref{LemmaWGInd}), which comes from the indicial root $\lambda-1<-1$ of $\wh{\Box_{g_b,E^\Ups}^\Ups}(0)$. Every large indicial root of $\wh{\Box_{g_b,E^\Ups}^\Ups}(0)$ comes with an associated large zero energy state $\omega_b^{(\lambda-1)}=\omega_b^{(\lambda-1)}(x)$ with $r^{-\lambda+1}$-leading-order growth at infinity; and this can be complemented with stationary 1-forms $\breve\omega_b^{(\lambda-1),j}$, $j=1,2,\ldots$, to yield generalized large zero energy states (see Proposition~\ref{PropWG0Large}\eqref{ItWG0Larges01}--\eqref{ItWG0Largeslj}). The corresponding (pure gauge) large zero energy state of $L_b$ is $h_b^{(\lambda)}:=\delta_{g_b}^*\omega_b^{(\lambda-1)}=\cO(r^{-\lambda})$; and its contribution to $\cK^+$-asymptotics is $\chi_\cK t_*^{-\alpha+\lambda}h_b^{(\lambda)}$, or better
  \[
    \chi_\cK \delta_{g_b}^*\omega_b^{(\lambda-1),\leq k}(t_*^{-\alpha+\lambda})
  \]
  (where $k=2$ suffices). This can be eliminated via a gauge modification, as explained after~\eqref{EqIEinElimHi}. (See~\eqref{EqD6CorrNonExc} and Definition~\ref{DefD6Corr}.)
\item{\rm (An exceptional root related to translations.)} $-1$ is always an indicial root, but there exists an indicial solution
  \begin{equation}
  \label{EqIEinhs12}
    \ubar h_{\rms 1}^2(\scal)=\cO(r)
  \end{equation}
  (where $\scal\in\mathspan\{\frac{x^i}{r}\colon i=1,2,3\}$) that \emph{cannot} be extended to a large zero energy state on Kerr. Its origin is more subtle: on Minkowski space, $\ubar\omega_{\rms 1}^{(0)}(\scal):=\dd(r\scal)$---say $\ubar\omega_{\rms 1}^{(0)}=\dd x^i$---is Killing (and thus trivially a large zero energy state of the Minkowskian gauge potential wave operator $\ubar\Box^\Ups_{\ubar E^\Ups}$), as is $t\,\dd x^i-x^i\,\dd t=t_*\ubar\omega_{\rms 1}^{(0)}+\breve{\ubar\omega}_{\rms 1}^{(0),1}$ (which defines the stationary 1-form $\breve{\ubar\omega}_{\rms 1}^{(0),1}$). But as far as generalized large zero energy states are concerned, one need not stop there: there exist $\breve{\ubar\omega}_{\rms 1}^{(0),j}$ for $j=2,3,\ldots$ such that the 1-forms $\ubar\omega_{\rms 1}^{(0),\leq k}(\scal)$ still lie in the kernel of $\ubar\Box^\Ups_{\ubar E^\Ups}$ (even though they are of course no longer Killing for $k\geq 2$). The tensor~\eqref{EqIEinhs12} is then simply the Minkowskian deformation tensor $\ubar\delta^*\ubar\omega_{\rms 1}^{(0),\leq 2}(\scal)$ of the quadratically growing (in $t_*$) 1-form $\ubar\omega_{\rms 1}^{(0),\leq 2}(\scal)$. --- It is now clear how a term $R^1 (r^{-1}\ubar h_{\rms 1}^2(\scal))$ in the expansion at $R=0$ of the $t_*^{-\alpha}$-term of the $\iota^+$-expansion of $h$ must contribute to the $\cK^+$-asymptotics: namely, via
  \begin{equation}
  \label{EqIEinhs12Elim}
  \begin{split}
    &\chi_\cK \delta_{g_b}^* \omega_{b,\rms 1}^{(0),\leq k}(A(t_*)\scal), \\
    &\qquad \omega_{b,\rms 1}^{(0),\leq k}(A(t_*)\scal) = A(t_*)\omega_{b,\rms 1}^{(0)}(\scal) + A'(t_*)\breve\omega_{b,\rms 1}^{(0),1}(\scal) + A''(t_*)\breve\omega_{b,\rms 1}^{(0),2}(\scal) + \cdots,
  \end{split}
  \end{equation}
  where $A''(t_*)=t_*^{-\alpha-1}$ so that the contribution from the $\breve\omega_{b,\rms 1}^{(0),2}$-term matches that of $\ubar h_{\rms 1}^2$ at $\iota^+$; the first two summands are invisible on Minkowski space, but on Kerr they amount to relatively large center-of-mass or linear momentum shifts of the final black hole. The worst case being $\alpha=1$ and thus $A=-\log t_*$, we thus expect to see at worst logarithmically divergent center-of-mass motions; these can again be eliminated as after~\eqref{EqIEinElimHi}. (See~\eqref{EqD4ParRewrite3}--\eqref{EqD4ParNoLog2} and Definition~\ref{DefD4Corr}.)
\item{\rm (An exceptional root related to time re-parameterizations.)} The indicial root $0$ of $\wh{L_b}(0)$ has the peculiar feature that its (Minkowskian, stationary) indicial solution $\approx\dd t_*^2$ is pure gauge, but only with respect to a gauge potential that grows in time (namely, $t_*\pa_{t_*}^\flat$ plus a stationary term, with $\pa_{t_*}^\flat$ being Killing); see Proposition~\ref{PropWG0Large}\eqref{ItWG0Larges00}. This gauge potential can be extended to Kerr as $t_*\omega_{b,\rms 0}^{(0)}+\breve\omega_{b,\rms 0}^{(0),1}$ (with $\omega_{b,\rms 0}^{(0)}=g_b^{-1}(\pa_{t_*},\cdot)$), whose symmetric gradient is stationary and denoted by $h_{b,\rms 0}^{(0)}\approx\dd t_*^2+{\rm l.o.t}$. The term $\chi_\cK t_*^{-\alpha}h_{b,\rms 0}^{(0)}$ in the late-time asymptotics of $h$ can then be rewritten essentially as $\chi_\cK\delta_{g_b}^*(A(t_*)\omega_{b,\rms 0}^{(0),\leq 1})$ where $A'=t_*^{-\alpha}$ and then eliminated as usual. When $\alpha=1$, this is again a logarithmically divergent gauge potential that roughly amounts to a logarithmic re-parameterization of $t_*$.
\end{enumerate}

We remark that there are further indicial solutions for the indicial root $-1$ which correspond to \emph{rotations} of the final black hole instead, in the sense of deformation tensors of generators of rotations (thus corresponding to a rotation of the axis of rotation). These are pure gauge and can again be eliminated in the usual fashion. (By contrast, it is of course \emph{not} possible to eliminate stationary or decaying changes of the magnitude of the angular momentum of the final black hole via gauge modifications; but the latter only arise from spectral theoretic considerations, cf.\ the second terms in columns {\color{myr}$A$} and {\color{myb}$Q$} in Figure~\ref{FigIEinAsy} below.)

\medskip
{\bf Physical indicial roots.} The first large non-pure gauge indicial roots are $-2$ and $-3$, which one thus expects to yield contributions of size $t_*^{-\alpha-2}$ and $t_*^{-\alpha-3}$ to the $\cK^+$-asymptotics, respectively. One can show that there exist corresponding (generalized) large zero energy states on Kerr using normal operator arguments; the states corresponding to the root $-2$ are denoted by
\begin{equation}
\label{EqIEinPhys}
  h_{b,\rms 2}^{(-2)},\quad h_{b,\rmv 2}^{(-2)} = \cO(r^2)
\end{equation}
in Proposition~\ref{PropWE0}.\footnote{What is somewhat delicate, and achieved in Proposition~\ref{PropWE0}, is a construction that depends \emph{smoothly} on the Kerr parameters $b=(\bhm,\bha)$ near $\bha=0$ (since the decay rates of certain elements of the cokernel of $\wh{L_b}(0)$ are discontinuous at $\bha=0$); continuity in $b$ is critical for perturbative purposes (cf.\ \S\ref{SssINElim}), and regularity in $b$ is critical for nonlinear purposes (in order to ensure that the nonlinear forward map taking $b$, the gravitational wave tail, etc.\ into the output of the gauge-fixed Einstein equation depends in a $\cC^2$ fashion on the input data $b$ etc., as needed in \cite{SaintRaymondNashMoser}).}

\medskip
Similarly to~\eqref{EqIGen2NoStruct} and \eqref{EqIGen2LaGen4}, a ``structureless'' $\cO(t_*^{-\alpha})$-term of $h$ at $\iota^+$ spawns analogous terms at $\cK^+$, so $h_{b,\rms 1}^{\leq 1}(\cO(t_*^{-\alpha+1}))$, $h_{b,\rms 0}^{(0)}(\cO(t_*^{-\alpha}))$, $h_{b,\rms 2}^{(-2)}(\cO(t_*^{-\alpha-2}))$, etc. While in linear theory on exact Kerr, the pure gauge terms among them could in principle be eliminated as before, we will not do so when using the present linear asymptotics in the nonlinear iteration scheme, for technical reasons discussed in Remark~\ref{RmkINElimNon}.

We summarize our discussion of $\iota^+$- and $\cK^+$-asymptotics in Figure~\ref{FigIEinAsy}; the term denoted {\color{myb}$Q$} there contains also those conormal terms that arise from the singularity of the resolvent at $\sigma=0$ and have spatial dependence given by a zero energy state, as discussed in point~\eqref{ItIGenKBdState2} on page~\pageref{ItIGenKBdState2}.\footnote{The term $\dot g_b^\Ups(\cO(t_*^{-\alpha_\cK+1}))$ encodes decaying modulations of the mass and angular momentum of the final black hole. That is, our asymptotic analysis automatically tracks these. In particular, it is not necessary to have a special mechanism for identifying the time-dependent angular momentum or the axis of rotation (which in any case would be discontinuous at Schwarzschild).}

\begin{figure}[!ht]
\centering
\includegraphics{FigIEinAsy}
\caption{Schematic description of the decay orders and terms in the $\cK^+$-expansion of the solution $h$ of $L_b h=f$. We will ultimately take $\alpha_\cK>4$ and $\alpha_+>\alpha_\cK-1>3$ close to $4$ and $3$, respectively. We only list terms that correspond to the exemplary large indicial roots discussed above. The index set $\cE_+$ is essentially the union of $\cE_\sscri$, $(1,0)$, and an index set determined by pure resonances (all of which we show to have real part $>1$ in Proposition~\ref{PropipMero}\eqref{ItipMeroInv}) as in~\eqref{EqIGen1Full}. In the second column, the values $z$ are those with $(z,0)\in\cE_+$ (and we do not record logarithmic terms here). The values $k$ in the superscripts ``$\leq k$'' of pure gauge terms are such that the gauge modifications required to eliminate them have more than $4$ orders of $\cK^+$-decay. In the third column, we have dropped the term $h_{b,\rms 2}^{(-2)}(\cO(t_*^{-\alpha_+-2}))$; this can be absorbed into $\tilde h$ when $\alpha_++2>\alpha_\cK$. The terms in {\color{myg}$P$} will be eliminated by gauge modifications, likewise for the spatial translations of {\color{myr}$A$}. The boost and linearized Kerr parameters, and also the gravitational wave tail $\tilde h$ and conormal terms {\color{myb}$Q$}, will be made part of the set of unknowns for the nonlinear stability problem.}
\label{FigIEinAsy}
\end{figure}

\subsubsection{Structure at \texorpdfstring{$\scri^+$}{null infinity}}
\label{SssIEinScri}

We need no fundamentally new insights regarding linear (and nonlinear) analysis at $\scri^+$ compared to \cite{HintzMink4Gauge}: the operator $L_b$ has the form stated in~\eqref{EqIGenScriOp} and \eqref{EqIGen1Op}, with $S$ there being an endomorphism of the symmetric 2-tensor bundle $S^2\cT^*$ that (to leading order at $\scri^+$) is affected by the size-$r^{-1}$-terms of $E^\Ups$ and $E^\cC$; the (small) modification $E^\Ups$ is chosen such that it shifts some eigenvalues of $S$ (which for $E^\Ups=0$ and $E^\cC=0$ are all equal to $0$) to the positive reals, and the (large) modification $E^\cC$ has\footnote{not surprisingly, given that its role is to effect (constraint) \emph{damping}} the same effect. In a certain splitting of $S^2\cT^*$, $S$ has a lower block-triangular form with only positive diagonal entries, except for the entry corresponding to the trace-free spherical-spherical part of $h$. (See~\eqref{EqWEOpMinkS}.) This part thus has a radiation field at order $r^{-1}$ at $\scri^+$ (and its pointwise norm-squared has, albeit in a different gauge, been shown to be the Bondi news function encoding gravitational wave energy \cite[\S{8.2}]{HintzVasyMink4}), while all other metric coefficients decay faster. (As mentioned before, in the nonlinear setting there is a coupling into the $(\dd t_*)^2$-component of $h$ as well, which then also attains a $r^{-1}$-term at $\scri^+$. On the nonlinear level, this is anticipated to encode the Bondi mass.)

\subsection{Aspects of the nonlinear analysis}
\label{SsIN}

We perturb a Kerr black hole with parameters $b_0=(\bhm_0,\bha_0)$. Given initial data as in Theorem~\ref{ThmISimple} or~\eqref{EqIDataPhg}, we prove exterior stability following \cite{HintzMink4Gauge} in Theorem~\ref{ThmEx0} and prove the partial polyhomogeneity of the exterior solution $g_\ext$ of the Einstein vacuum equation following \cite[\S{7}]{HintzVasyMink4} in Theorem~\ref{ThmExPhg}; schematically, we have
\begin{equation}
\label{EqINExt}
  \raisebox{-1.3em}{\includegraphics{EqINExt-r}}
\end{equation}
where $\min\Re\cE_0>1$, and the remainder orders $3+\eps_0>3+\eps_\sscri$ are essentially arbitrary at this point. (They are chosen in this fashion here since they are the minimal requirement for our subsequent black hole stability analysis; see~\S\ref{SssINHeur}.) The notation $\la\cE_\sscri\ra$ indicates that different components of $h_\ext$ have different index sets, all of which include $\cE_\sscri$ ($\supset\cE_0$, with $\min\Re\cE_\sscri>1$) and some additional terms, such as $(1,0)$ for the trace-free spherical-spherical and $(\dd t_*)^2$-components of $h_\ext$. (See Definition~\ref{DefExP}.)

The real task is the extension of such an exterior solution $g_\ext$, which is defined up to a hypersurface $t_*=2$, say, which is transversal to $\scri^+$ and the future event horizon, to all of $\{t_*\geq 1\}$. This can be phrased as a forward problem for a quasilinear wave equation, since the task is to find $g_+$ that is supported in $\{t_*\geq 1\}$ and satisfies
\begin{equation}
\label{EqINFwd}
  \Ric\bigl(\chi_-(t_*)g_\ext+g_+\bigr)=0
\end{equation}
where $\chi_-\in\CI(\R)$ equals $1$ on $(-\infty,1]$ and $0$ on $[2,\infty)$.

Let us start with a minimalist attempt at a (quasilinear wave type) formulation of the gauge-fixed Einstein equation that one may try to use for this purpose. We expect the final black hole to have parameters $b=(\bhm,\bha)$ close to $b_0$; we patch the initial and final metric together via
\begin{equation}
\label{EqINgb0b}
  g_{b_0,b} := (1-\chi_\cK)g_{b_0} + \chi_\cK g_b
\end{equation}
where $\chi_\cK$ is $1$ near $\cK^+$ (as in Figure~\ref{FigIEinElim}). Writing the dynamical metric as $g_{b_0,b}+h$, with $h$ expected to decay in all four asymptotic regimes ($I^0$, $\scri^+$, $\iota^+$, $\cK^+$), we then set
\begin{equation}
\label{EqINPNaive}
  P(b,h) := \Ric(g) - \delta_{g^0,E^\cC}^*\Ups_{E^\Ups}(g,g^0),\quad g:=g_{b_0,b}+h,\ g^0:=g_{b_0,b}.
\end{equation}
The gauge condition is
\[
  \Ups_{E^\Ups}(g,g^0) := \tr_g(\nabla^g-\nabla^{g^0}) - E^\Ups_{g^0}\sfG_{g^0}(g-g^0).
\]
The first term is the familiar generalized harmonic gauge 1-form with components $g_{\kappa\lambda}g^{\mu\nu}(\Gamma(g)_{\mu\nu}^\kappa-\Gamma(g^0)_{\mu\nu}^\kappa)$, and the additional term makes the linearization of $\Ups_{E^\Ups}(g,g^0)$ in the first argument be essentially equal to $-\delta_{g,E^\Ups}\sfG_g$, which is the linear gauge condition we have been working with throughout~\S\ref{SsIEin}.

The linearization of $P$ around $(b_0,0)$ in the second argument is the operator
\[
  D_{(b_0,0)}P(0,\cdot) = D_{g_{b_0}}\Ric + \delta_{g_{b_0},E^\cC}^*\delta_{g_{b_0},E^\Ups}\sfG_{g_{b_0}} = L_{b_0}
\]
introduced already in~\eqref{EqIEinLb}. Linearizing around $(b,0)$ more or less yields $\chi_\cK L_b+(1-\chi_\cK)L_{b_0}$; since $L_b$ and $L_{b_0}$ are well-approximated near $\iota^+$ by their Minkowskian versions, the operator $L_b$ is in fact a good model for this globally at $\cK^+\cup\iota^+$. More generally, finally, for the linearization of $P$ around $(b,h)$ we have, roughly speaking,
\begin{equation}
\label{EqINLinbh}
  \raisebox{-1.5em}{\includegraphics{EqINLinbh-r}}
\end{equation}
where the $\cO$-term is a differential operator built from $r\pa_{t_*}$, $r\pa_x$ with the stated orders. (See Lemma~\ref{Lemma1Lin} for formulas and Proposition~\ref{PropDAdmLin} for precise statements. The shift by $2$ at $\iota^+$ is partly due to the overall weight $\rho_+^2$ in~\eqref{EqIGen1Op}, and partly due to, e.g., the fact that lower-order terms contributing to $D_{(b,h)}P(0,\cdot)-L_b$ involve coordinate derivatives of $h$, each of which gains one power of decay at $\iota^+$, so terms such as $\Gamma(g)\cdot(\Gamma(g)-\Gamma(g_b))$ and $\pa(\Gamma(g)-\Gamma(g_b))$ have $\iota^+$-order $\alpha_++2$.)

\bigskip

The strategy in a Newton-type iteration scheme for solving the nonlinear forward problem $P(b,\chi_-h_\ext+h)=0$ for $b$ and $h$ globally in $t_*\geq 1$ is then as follows (omitting $\chi_-h_\ext$ for notational simplicity). We first only linearize in the second argument, i.e., study the linear wave-type equation
\begin{equation}
\label{EqINLinNaive}
  L_{b,h}u := D_{(b,h)}P(0,u) = -P(b,h)
\end{equation}
on the dynamical spacetime with metric $g=g_{b_0,b}+h$. \emph{If one can show} that the solution $u$ can be written as the linearization of $g_{b_0,B}+H$ (cf.\ $g$ in~\eqref{EqINPNaive}) in $(B,H)$ around $(B,H)=(b,h)$ in the direction $(\dot b,\dot h)$, this morally speaking means that $D_{(b,h)}P(0,u)=D_{(b,h)}P(\dot b,\dot h)$. (See \S\ref{SssINElim} for more precise statements.) More precisely, one ultimately wishes to show that one can solve not~\eqref{EqINLinNaive} but
\begin{equation}
\label{EqINLinNaive2}
  D_{(b,h)}P(\dot b,\dot h)=-P(b,h),
\end{equation}
thereby shuffling the changes of the final black hole parameters from $u$ into $\dot b$; and the remainder $\dot h$ of $u$ now decays. The iteration step is then to set
\[
  (b,h) + (\dot b,\dot h) \rightsquigarrow (b,h).
\]
One repeats the process with these new values of $(b,h)$; and finally $(b,h)$ should converge to a limit, with the limiting metric $g=g_{b_0,b}+h$ solving the nonlinear equation $P(b,h)=0$.

There are many issues with this attempt, starting with the fact that we have not incorporated the boost or spatial translation parameters (see column {\color{myr}A} in Figure~\ref{FigIEinAsy}) into the set of unknowns. Moreover, the linear analysis for the tensorial wave-type equation~\eqref{EqINLinNaive} on a dynamical background is quite involved and subtle; for example, even the polynomial boundedness of solutions $u$ (which we have argued previously to be a necessary starting point for any refined asymptotic analysis) is only expected to hold under fairly strong decay assumptions on $h$---this is discussed in~\S\ref{SssINAdm}. These strong assumptions must then be retrieved for $u$ (or rather $\dot h$) in order to close the nonlinear iteration; but this \emph{fails} unless one can eliminate the poorly decaying terms in column {\color{myg}$P$} in Figure~\ref{FigIEinAsy} (assuming $\alpha_+$ there is large enough so that the terms in column {\color{myb}$Q$} are acceptable). We explain in~\S\ref{SssINElim} how this can be achieved.

\subsubsection{Weak decay and high regularity for linear waves on dynamical spacetimes}
\label{SssINAdm}

We briefly discuss how to obtain \emph{some} (imprecise) polynomial bounds for forward solutions of $L u=f$ where, say, $L=L_{b,h}$ (cf.\ \eqref{EqINLinNaive}); these bounds must also hold for arbitrary b-derivatives of $u$. (This is the first step of a standard two-step procedure, the second step being the proof of improved decay at the cost of giving up regularity.) For this, we rely on the companion paper \cite{HintzNonstat2}, which provides such control for a general class of linear wave-type operators on asymptotically Kerr spacetimes under certain structural and analytic assumptions on the operator and its stationary (i.e., Kerr) limiting model $L_b$. The task in the present paper is thus merely to verify these assumptions. We give a brief overview here.

We begin with the assumptions concerning the stationary model, which is the linearized gauge-fixed Einstein operator $L_b$ from~\eqref{EqIEinLb}.
\begin{enumerate}[leftmargin=2.5em]
\item{\rm (Trapping.)} The subprincipal symbol of $L_b$ at the trapped set (in phase space) must essentially have a sign in order to ensure that high-frequency waves localized near trapping disperse (see also \cite{SbierskiBeams}). The precise condition is \emph{strong trapping admissibility} \citeAF{Definition~\ref*{DefSSTrapAdm}}. This was previously verified in similar contexts, e.g., in \cite[Theorem~4.8]{HintzPsdoInner}, \cite[\S{9.1}]{HintzVasyKdSStability}, \cite[Lemma~3.3]{HintzGlueLocIII}, and \cite[\S{3}]{HintzPetersenVasyKdS}, and we verify it for the present choice of $L_b$ in Proposition~\ref{PropWETr}. This computation is quite short in a tetrad introduced by Marck \cite{MarckParallelNull}.
\item{\rm (Minkowskian model operators.)} The operators $N_\tface(L,\hat\sigma)$ from~\eqref{EqIGenKNtf} governing the transition from zero to nonzero spectral parameters at $r=\infty$ must have trivial kernel and cokernel on appropriate spaces \citeAF{Definition~\ref*{DefSStfAdm}}. This is checked in Proposition~\ref{PropWEtf}. (In the context of~\S\ref{SssIGen2La}, we remark that this also implies the invertibility of the $\iota^+$-normal operators $N_{\iota^+}(L,\lambda)$, and thus the absence of ``pure'' resonances $\lambda$, for $\Re\lambda<1+\eps_\ind$; see Proposition~\ref{PropipMero}\eqref{ItipMeroInv}.)
\item{\rm (2-admissibility.)} The final assumption (see \citeAF{Definition~\ref*{DefSSAlephAdm}(\ref*{ItSSAlephAdm3})}) demands bounds for forward solutions of $L_b u=f$ on certain weighted spacetime Sobolev spaces. These 3b-Sobolev spaces \cite{Hintz3b} (with any desired additional amount of b-regularity) transform in a simple manner under the Fourier transform in $t_*$, and the requisite bounds for $u$ can thus be proved using high- and low-energy resolvent estimates (including holomorphicity considerations in $\Im\sigma>0$); these require fairly detailed control of spectral properties of $\wh{L_b}(\sigma)$ at non-zero and zero frequencies and a careful construction of an augmented spectral problem at low energies. See \S\ref{SAdm} for details. The ``2'' in ``2-admissibility'' is the order of the ``pole'' of $\wh{L_b}(\sigma)^{-1}$ at $\sigma=0$, cf.\ \eqref{EqIGenKPole2}.
\end{enumerate}

The only assumptions on the dynamical wave-type operator $L$ concern the decay of its coefficients towards the model operator $L_b$. Namely, the underlying metric $g$ must decay to $g_b$ at a positive rate $\ell_+$ at $\iota^+$ and at a rate $\ell_\cK>2$ at $\cK^+$, and its coefficients must have suitable decay at $\scri^+$. On the level of operators, one correspondingly requires $L-L_b$ to have coefficients (as an operator built from $r\pa_{t_*}$ and $r\pa_x$) of order $2+\ell_+$ at $\iota^+$ and $\ell_\cK$ at $\cK^+$, and a suitable structure near $\scri^+$ (which is the same that one uses for the exterior stability problem already); the detailed conditions are stated in \citeAF{Definition~\ref*{DefSDWAdm}} and verified for the linearized gauge-fixed Einstein operators arising in our nonlinear iteration scheme in Corollary~\ref{CorDAdm}.

Upon verifying these assumptions, the linear main result of \cite{HintzNonstat2}, which is \citeAF{Theorem~\ref*{ThmF}}, becomes available: in the form \citeAF{(\ref*{EqFTame20})} and using our pictorial representation, it states that forward solutions $u$ of $L u=f$ obey\footnote{The $\cK^+$-order $\frac12$ amounts to $L^2$-membership in $t_*$, which betrays the spectral proof of this result. The $\cK^+$-order of $u$ is $\frac12-2=-\frac32$, the loss of two powers of decay being due to the second order pole of $\wh{L_b}(\sigma)^{-1}$ at $\sigma=0$.}
\begin{equation}
\label{EqINAdm}
  \raisebox{-1.8em}{\includegraphics{EqINAdm}}
\end{equation}
Moreover, $u$ satisfies \emph{tame estimates} in terms of $f$ and the coefficients of $L$; these are a key ingredient for the (eventual) applicability of a Nash--Moser iteration scheme \cite{SaintRaymondNashMoser,HamiltonNashMoser} for the solution of nonlinear equations. See Theorem~\ref{ThmDAdmReg} for details.

\medskip

We briefly comment on the decay requirement at $\cK^+$---which in the context of~\eqref{EqINLinbh} amounts to requiring metric perturbations $h$ to decay at a rate $\ell_\cK=2+\eps_\cK>2$ at $\cK^+$ (cf.\ the conclusion of Theorem~\ref{ThmISimple})---as this is the critical requirement informing our entire setup for nonlinear stability. (See also \citeAF{\S\ref*{SssIML}, item~(\ref*{ItIMLPertDec})}.) As a toy model for an operator whose resolvent has a second order pole at $\sigma=0$ we take $L_0=\pa_{t_*}^2$ (with spectral family $\wh{L_0}(\sigma)=-\sigma^2$). As the dynamical operator, consider
\begin{equation}
\label{EqINODE}
  L=\pa_{t_*}^2-h(t_*)
\end{equation}
where, for the sake of concreteness, we take $h(t_*)=(1+t_*)^{-\ell_\cK}$. For $0<\ell_\cK<2$, the solution of $L u=0$ with initial data $u(0)=1$, $u'(0)=0$, say, grows sub-exponentially ($\sim\exp(t_*^{1-\ell_\cK/2})$ by a Liouville--Green ansatz). For $\ell_\cK>2$ on the other hand, the asymptotics of $u$ match that of solutions of the model problem $L_0 u=0$ to leading order, namely, they are equal to $c t_*+\cO(t_*^{1-(\ell_\cK-2)})=c t_*+o(t_*)$ for some constant $c$. --- The general statement is that in order for the asymptotic behavior of solutions $u$ of dynamical equations $L u=f$ to be equal (to leading order) to that of solutions of $L_0$, the difference $L-L_0$ must decay at a rate $\ell_\cK$ that is faster than the amount of decay that the inversion of $L_0$ loses (here $2$). Otherwise (ignoring the borderline case $\ell_\cK=2$) one should not even expect $u$ to be polynomially bounded.\footnote{This line of reasoning is blind to the structure of the perturbation $L-L_0$. It is an interesting problem, which we do not address here or in \cite{HintzNonstat2}, to determine the structural properties that perturbations with \emph{less} decay would need to have so that one can still describe asymptotics for $L$ using those for $L_0$.}

\subsubsection{Heuristics for nonlinear stability}
\label{SssINHeur}

Extrapolating from the discussion of the ODE model~\eqref{EqINODE} to the linearized gauge-fixed Einstein equation $L u=L_{b,h}u=f$ (using the notation from~\eqref{EqINPNaive} and \eqref{EqINLinNaive}), one expects, for $h$ of $\cK^+$-order $\ell_\cK:=2+\eps_\cK>2$, that $u$ is equal to a deformation tensor of a Lorentz boost $h_{b,\rms 1}^{\leq 1}(\scal)$ in some direction $\scal$ to leading order at $\cK^+$, with a remainder that has $\ell_\cK-2$ orders of decay relative to this linearly growing leading-order term. To prove this, one simply rewrites $L u=f$ (with $f$ having sufficiently strong decay so as to not interfere with the arguments that follow) as
\begin{equation}
\label{EqINHeurRewrite}
  L_b u = f - (L-L_b)u
\end{equation}
and takes advantage of the precise analysis of the stationary problem $L_b$ afforded by the considerations in~\S\S\ref{SsIGen}--\ref{SsIEin}. The point is that the $\cK^+$-order of $(L-L_b)u$ is $\ell_\cK=2+\eps_\cK$ orders better than the decay rate $\alpha_\cK$ originally known for $u$ (so initially $\alpha_\cK=-\frac32$ from~\eqref{EqINAdm}), and the inversion of $L_b$---which loses $2$ orders at $\cK^+$---thus allows one to improve the decay of $u$ by $\ell_\cK-2=\eps_\cK>0$ orders at $\cK^+$, or more precisely to uncover terms in the asymptotic expansion of $u$ at $\cK^+$ that have decay rates between $\alpha_\cK$ and $\alpha_\cK+\eps_\cK$, plus a remainder that decays at the improved rate $\alpha_\cK+\eps_\cK$. Once one has reached the ``decay'' rate $\alpha_\cK=-1-$ corresponding to $\cO(t_*^{1+\eps})$-bounds for all $\eps>0$ (see Proposition~\ref{PropD1Alm}), the next step uncovers the boost term, as claimed (see Proposition~\ref{PropD2Boost}). But if the boost term vanishes, then $u$ has $-1+\eps_\cK$ orders of $\cK^+$-decay, so one can continue this iterative scheme and continue improving the decay of $u$ by $\eps_\cK$, $2\eps_\cK$, $3\eps_\cK$, $\ldots$ orders at $\cK^+$ (see Proposition~\ref{PropD3Alm}) until $\alpha_\cK=0-$, which is next threshold---at which one encounters logarithmically divergent center-of-mass motions as discussed after~\eqref{EqIEinhs12Elim} and stationary center-of-mass shifts and linearized Kerr metrics as discussed in items~\eqref{ItIEinSpecKerr}--\eqref{ItIEinSpecTrans} on page~\pageref{ItIEinSpecKerr}. (See Proposition~\ref{PropD4Par}.)

We defer the detailed explanation of how we will eliminate poorly decaying but polyhomogeneous (in $t_*$) pure gauge contributions (i.e., column {\color{myr}$A$} in Figure~\ref{FigIEinAsy} except for $\dot g_b^\Ups(\dot b)$, and column {\color{myg}$P$}) for solutions of the linearized gauge-fixed Einstein equation on \emph{dynamical} (i.e., asymptotically Kerr) spacetimes to~\S\ref{SssINElim}; for now, we take this for granted. Similarly, we ignore the details of how regarding the final black hole parameters $b=(\bhm,\bha)$ as unknowns allows one to effectively absorb the emergence of linearized Kerr metrics in the late-time asymptotics of $u$.

We can now explain the rough scheme of our nonlinear stability proof as follows. Recall that the task is to extend the exterior solution~\eqref{EqINExt} of the Einstein vacuum equation to $t_*\geq 1$. We claim that this can be done with
\begin{equation}
\label{EqINHeurh}
  \raisebox{-4em}{\includegraphics{EqINHeurh-r}}
\end{equation}
Here $0<\eps_\cK<\eps_+<\eps_\sscri$, further $g_{b_0,b,-\scal}$ is equal to~\eqref{EqINgb0b} except $g_{b_0}$ there is pushed forward by a boost with parameter $\scal$ (this will be discussed further in~\S\ref{SssINElim} below, see~\eqref{EqIDNBoost}), and {\color{myb}$Q$} and {\color{myg}$P$} are taken from Figure~\ref{FigIEinAsy} (with $\alpha_\cK$ and $\alpha_+$ there being $4+\eps_\cK$ and $3+\eps_+$, and with the two $h_{b,\rms 1}^{\leq 1}$-terms combined into one), except the (poorly decaying) pure gauge terms of {\color{myg}$P$} in Figure~\ref{FigIEinAsy} are no longer present here. For now, we pretend that the unknowns are the parameters $b,\scal$, the arguments of the terms comprising {\color{myg}$P$} (which lie in a finite-dimensional space) and {\color{myb}$Q$} (which really lie in weighted Sobolev spaces on the real line), the remainder term $\tilde h$, and a finite-dimensional parameter that parameterizes the gauge modifications $\vartheta$ of the types discussed in~\S\ref{SssIEinElim}. The index set $\mathcal{E}_+$ finally is essentially as in~\eqref{EqIGen2LaFull} and Figure~\ref{FigIEinAsy} (and large enough to account for nonlinear terms), and is comprised of $(1,0)$ and some $(z,k)$ with $\Re z>1$.

First of all, notice that all terms comprising the gravitational wave tail $h=\chi_\cK({\color{myg}P+\color{myb}Q})+\tilde h$ have $\cK^+$-order $2+\eps_\cK>2$ (and the $\iota^+$-order of the terms in {\color{myb}$Q$} is $3+\eps_+$, matching that of the remainder term of $\tilde h$ there).\footnote{For $\dot g_b^\Ups$, say, this uses that $\dot g_b^\Ups(\dot b)=\cO(r^{-1})$ for all $\dot b=(\dot\bhm,\dot\bha)$, so the $\iota^+$-order of $\chi_\cK\dot g_b^\Ups(\cO(t_*^{-3-\eps_\cK}))$ is $4+\eps_\cK>3+\eps_+$.} Therefore, the linearization $L$ of the nonlinear gauge-fixed Einstein operator $P\colon g\mapsto\Ric(g)+\text{(gauge term)}$ around $g$ yields a linear wave-type operator $L$, relative to the metric $g$, for which the basic weak decay/high-regularity result~\eqref{EqINAdm} applies.

Next, the source terms $f$ arising in the nonlinear iteration (cf.\ \eqref{EqINLinNaive}) are of the form $f=-P(g)$. For $g$ as in~\eqref{EqINHeurh}, we would like them to satisfy
\begin{equation}
\label{EqINHeurf}
  \raisebox{-1.5em}{\includegraphics{EqINHeurf}}
\end{equation}
where the $\scri^+$-, $\iota^+$-, and $\cK^+$-orders are shifted by $1$, $2$, and $0$, respectively, relative to those of the tail $\tilde h$ in~\eqref{EqINHeurh}. The membership~\eqref{EqINHeurf} is nontrivial only at $\cK^+$ and $\iota^+$:
\begin{enumerate}[label=(\roman*)]
\item At $\cK^+$, the main point (dropping the gauge term) is that $\Ric(g_b+h)=\Ric(g_b)+D_{g_b}\Ric(h)+\cO(h^2)$; the contribution of $\tilde h$ to the output of $\Ric$ thus has order $4+\eps_\cK$ as desired, and only the term $D_{g_b}\Ric({\color{myb}Q})$ looks like it only has order $2+\eps_\cK$---but in fact it has order $4+\eps_\cK$ as well since $h_{b,\rms 1}^{\leq 1}$ etc.\ lie in the kernel of $D_{g_b}\Ric$. (See Lemma~\ref{LemmaD6Aug5Linb}.) --- This improved decay of the output of the forward map compared to the input (due to the special structure of {\color{myb}$Q$}) already featured prominently in \cite{HintzNonstat2}; see \citeAF{(\ref*{EqIODEDecomp}), (\ref*{EqA2Heur1b})--(\ref*{EqA2Heur1bFwd})}.
\item At $\iota^+$, one expects $f$ to have order $(\cE_++2,5+\eps_+)$. The question is thus why the expansion of $f$ should, in fact, be trivial; this can only happen if the $\iota^+$-expansion of $\tilde h$ has special structure. In linear theory, it is clear from~\S\ref{SssIGen1} what this structure should be: the terms in the $\iota^+$-expansion of $\tilde h$ need to be resonant states of the $\iota^+$-normal operator, plus corrections at higher decay orders to account for the difference between $L$ and its Minkowskian $\iota^+$-model. In the nonlinear problem, these corrections must also take into account nonlinear expressions involving prior expansion terms. We return to this in~\S\ref{SssINPhg} below.
\end{enumerate}

Proceeding with source terms $f$ as in~\eqref{EqINHeurf}, consider now the forward solution $u$ of $L u=f$ that is initially only known to have weak bounds as in~\eqref{EqINAdm}. The orders of $u$ at $\scri^+$ are determined by those of $f$ (cf.\ \S\ref{SssIGenScri}), so the real task is to improve the decay of $u$ at $\iota^+$ and $\cK^+$. To this end, we use the re-writing~\eqref{EqINHeurRewrite} to extract increasingly accurate asymptotics of $u$. The first few steps are as follows.
\begin{enumerate}
\item As mentioned before, the first $\cK^+$-expansion term of $u$ extracted in this fashion is the boost/recoil term $\chi_\cK h_{b,\rms 1}^{\leq 1}(\dot\scal)$; this can be eliminated by adjusting (on the linearized level) the boost parameter $\scal$ of the background metric $g_{b_0,b,-\scal}$. By this we mean subtracting from the original source term $f$ a term given by the linearization of the nonlinear forward map $P$ in $\scal$. (See~\S\S\ref{SssINElim} and \ref{SsD2Boost} for details.)
\item The non-decaying center-of-mass motion terms mentioned above can be eliminated using a gauge modification as explained in~\S\ref{SssIEinElim} (and adapted to the dynamical setting in~\S\ref{SssINElim}). (The detailed implementation is in~\S\ref{SsD4Par}.)
\item The stationary linearized Kerr term $\dot g_b^\Ups(\dot b)$ can be eliminated by adjusting the parameter $b$. (See again~\S\S\ref{SssINElim} and \ref{SsD4Par}.)
\end{enumerate}

This takes care of the terms in column {\color{myr}$A$} in Figure~\ref{FigIEinAsy}.

Further terms in the $\cK^+$-expansion of $u$ arise from $\iota^+$ via the mechanism explained in~\S\ref{SssIGen2La} (see column {\color{myg}$P$} in Figure~\ref{FigIEinAsy}). To describe these, we thus first need to comment on the $\iota^+$-asymptotics of $u$. Since $L-L_b$ almost has $\iota^+$-order $1+2$ (from~\eqref{EqINLinbh} with $\alpha_+=1-$), one expects from~\eqref{EqINHeurRewrite} that one can extract the $\iota^+$-expansion of $u$ almost a full order at a time (until one has to stop at the $\iota^+$-order $3+\eps_+$ determined by the $\iota^+$-order of $f$ in~\eqref{EqINHeurf} minus $2$). Thus, one should be able to extract an expansion of $u$ at $\iota^+$, with index set\footnote{A priori, one might be concerned that the $\iota^+$-index set of $u$ might be larger than $\cE_+$, and then yet larger at the next iteration step, and so on. However, a post-processing of the present analysis using $\iota^+$-normal operator arguments does yield a self-consistent choice for the $\iota^+$-index set $\cE_+$; see~\S\ref{SssINPhg} below.} $\cE_+$, up to a remainder term of decay order $3+\eps_+$.\footnote{In reality, improving asymptotics at $\iota^+$ and $\cK^+$ must more or less be done in turns, though there is considerable flexibility here. See, e.g., \seepages{Step~1.2}{ItD6Step12}{the beginning of Step~2}{ItD6Step2}.} (The restriction $\eps_+<\eps_\sscri$ comes from the transport of the remainder term of $u$ from $\scri^+$.)

Suppose now that $(z,0)\in\cE_+$ (so $z=1$ or $\Re z>1$). In~\S\ref{SssIEinLa}, we discussed the $\cK^+$-expansion terms that the $t_*^{-z}$-term of $u$ at $\iota^+$ might cause. The least-decaying one is a center-of-mass motion $\sim t_*^{-z+1}$ (or $\log t_*$ when $z=1$). This can be eliminated using gauge modifications (see~\S\ref{SsD4Par}, resp.\ \S\ref{SsD5Alm} where this is carried out for $z=1$, resp.\ $\Re z\in(1,2)$); similarly for all other pure gauge terms arising via this mechanism. The first terms we \emph{cannot} eliminate are the physical states $\chi_\cK h_{b,\rms 2/\rmv 2}^{(-2)}$ (see~\eqref{EqIEinPhys}) at order $t_*^{-3}$; crucially, these are acceptable in that they have more than $2$ orders of decay at $\cK^+$.

Finally, consider the ``structureless'' (i.e., not polyhomogeneous) terms in the asymptotics of $u$ (cf.\ the column {\color{myb}$Q$} in Figure~\ref{FigIEinAsy}). As explained in Remark~\ref{RmkINElimNon} below, we do not have a mechanism to eliminate them on dynamical spacetimes (even if they are pure gauge near $\cK^+$ on exact Kerr), so we must ensure that they have sufficient decay---more precisely, that they are as stated in~\eqref{EqINHeurh}. And indeed they do: the second order pole of $\wh{L_b}(\sigma)^{-1}$ creates the first two terms of {\color{myb}$Q$} in~\eqref{EqINHeurh} (encoding $\cO(t_*^{-2-\eps_\cK})$ center-of-mass motions and $\cO(t_*^{-3-\eps_\cK})$ modulations of the mass and angular momentum parameters relative to their current final values), while the large indicial roots $\lambda$, $\Re\lambda\leq 0$, of $\wh{L_b}(0)$, starting with $\lambda=0$ (and the corresponding large zero energy states, which include the previously discussed tensor $h_{b,\rms 0}^{(0)}$), contribute conormal terms at order $t_*^{-3-\eps_++\lambda}$, so also as stated in~\eqref{EqINHeurh}.

We have thus recovered the asymptotics~\eqref{EqINHeurh} for solutions of the linearized problem (up to updating boost and black hole parameters and making finite-dimensional gauge modifications) and therefore closed the iteration scheme. (See Corollary~\ref{CorD6Impr} for the precise statement.) As a by-product, our arguments prove the linear stability of subextremal Kerr with stronger decay rates than previously known (see Corollary~\ref{CorD6LinStab}).

We tie up the loose ends concerning how to eliminate pure gauge terms and how to encode precise asymptotics at $\iota^+$ in~\S\ref{SssINElim} and \ref{SssINPhg}, respectively.

\subsubsection{Improving decay; elimination of pure gauge terms}
\label{SssINElim}

On an exact Kerr spacetime, we argued in~\S\ref{SssIEinElim} how one can eliminate a pure gauge term, say, $h_{b,\rms 1}(\scal)$ (the deformation tensor of asymptotic translations, with gauge potential $\omega_{b,\rms 1}^{(0)}(\scal)=\dd(r\scal)+o(1)\in\ker\Box_{g_b,E^\Ups}^\Ups$), via an algebraic ``trick'' \eqref{EqIEinElimRewrite}; notice that this crucially used the identity $D_{g_b}\Ric\circ\delta_{g_b}^*=0$. For general metrics $g$, however, which do \emph{not} satisfy the Einstein vacuum equation---such metrics arise at intermediate steps of our iteration scheme---the linearization of the identity $\Ric(\phi^*g)=\phi^*\Ric(g)$ around $\phi=\Id$ only yields $D_g\Ric\circ\delta_g^*\omega=\frac12\cL_{\omega^\sharp}\Ric(g)$, which has no reason to vanish for nonzero $\omega$ such as $\omega=\omega_{b,\rms 1}^{(0)}(\scal)$. Moreover, the right-hand side is not of the form $D_g\Ric(\cdots)$ with an argument that has better decay than $\delta_g^*\omega$ itself, and also not of the form $\delta_{g,E^\cC}^*(\cdots)$; therefore, it cannot be re-interpreted as a metric patch or gauge modification.

Given the lack of geometric properties of metrics $g$ arising in the iteration scheme, simple geometric arguments are bound to fail. We instead exploit the closeness of $g$ to a Kerr metric $g_b$ and argue \emph{perturbatively} to eliminate pure gauge terms and to absorb black hole parameter changes. We explain this for the three terms from column {\color{myr}$A$} in Figure~\ref{FigIEinAsy}.

\medskip
{\bf Asymptotic translations.} For the sake of concreteness, let us consider again the linearization $L=L_{b,h}$ of $P$ in~\eqref{EqINPNaive} around $b\approx b_0$ and $h\approx 0$. First, when $(b,h)=(b_0,0)$ and thus $L=L_{b_0}$, we rewrite the identity~\eqref{EqIEinElimRewrite}--\eqref{EqIEinElimRewrite2} in the following way. Define
\begin{equation}
\label{EqIDNElimCorr}
  h_{\rms 1}^{(0)}(\scal) := -[\delta_{g_{b_0}}^*,\chi_\cK]\omega_{b_0,\rms 1}^{(0)}(\scal),\quad
  \vartheta_{\rms 1}^{(0)}(\scal) := [\delta_{g_{b_0},E^\Ups}\sfG_{g_{b_0}},\chi_\cK]h_{b_0,\rms 1}(\scal),
\end{equation}
where we stress that these are defined \emph{with respect to the fixed Kerr parameters $b_0$} of the black hole we are perturbing. Define an augmented version of~\eqref{EqINPNaive} by\footnote{While the argument of $\Ric$ here is the dynamical spacetime metric, we choose the gauge condition not to include the term $h_{\rms 1}^{(0)}$; if we did, it would simply have to be subtracted again via re-defining $\vartheta_{\rms 1}^{(0)}$. The analytic purpose of the gauge condition being the elimination of \emph{infinite-dimensional} gauge freedoms, the important point is that the infinite-dimensional piece $h$ of the metric \emph{is} part of the argument of $\Ups_{E^\Ups}$.}
\begin{equation}
\label{EqIDNElimCorrP}
  P(b,\scal^{(0)},h) := \Ric\bigl( g_{b_0,b} + h_{\rms 1}^{(0)}(\scal^{(0)}) + h \bigr) - \delta_{g_{b_0,b},E^\cC}^* \Bigl( \Ups_{E^\Ups}( g_{b_0,b}+h, g_{b_0,b} ) - \vartheta_{\rms 1}^{(0)}(\scal^{(0)}) \Bigr).
\end{equation}
Thus
\[
  D_{(b_0,0,0)}P(0,\scal,0) = D_{g_{b_0}}\Ric\bigl( h_{\rms 1}^{(0)}(\scal) \bigr) + \delta_{g_{b_0},E^\cC}^*\vartheta_{\rms 1}^{(0)}(\scal)
\]
is exactly the right-hand side of~\eqref{EqIEinElimRewrite}, and thus the full identity reads
\begin{equation}
\label{EqDNElimIdent}
  D_{(b_0,0,0)}P\bigl(0,0,\chi_\cK h_{b_0,\rms 1}(\scal)\bigr) = D_{(b_0,0,0)}P(0,\scal,0).
\end{equation}

For the linearization around $(b,\scal^{(0)},h)$, this identity of course is no longer valid. We instead argue as follows. Suppose for simplicity that $\chi_\cK h_{b,\rms 1}$ always were the first term arising in the $\cK^+$-expansion of general solutions of $L u=f$ where $L:=D_{(b,\scal^{(0)},h)}P(0,0,\cdot)$ (which is equal to our earlier $L_{b,h}$ when $\scal^{(0)}=0$). Then we can define a linear map $S_{b,\scal^{(0)},h}\colon\scal'\mapsto\dot\scal$ on the 3-dimensional vector space $\scalspace_1=\mathspan\{\frac{x^i}{r}\colon i=1,2,3\}$ as follows: $\dot\scal=S_{b,\scal^{(0)},h}(\scal')$ is the argument of the leading-order term $h_{b,\rms 1}(\dot\scal)$ of the $\cK^+$-expansion of the solution of
\[
  L u = D_{(b,\scal^{(0)},h)}P(0,\scal',0).
\]
The crucial point is this: the map $S_{b_0,0,0}$ is the identity map---this is the content of the identity~\eqref{EqDNElimIdent}---and the map $S_{b,\scal^{(0)},h}$ depends continuously on $b,\scal^{(0)},h$. (The norm on $h$ here is some weighted Sobolev norm with sufficiently high differentiability order; one only needs as many derivatives as the extraction of the $\cK^+$-leading-order term---which depends only on a 3-dimensional parameter in $\scalspace_1$---requires.) And
\begin{subequations}
\begin{equation}
\label{EqIDNElimFin}
  \text{\parbox{0.8\textwidth}{\it since $\scalspace_1$ is finite-dimensional, the map $S_{b,\scal^{(0)},h}$ is an isomorphism for all $(b,\scal^{(0)},h)$ close to $(b_0,0,0)$.}}
\end{equation}
But this means that, given a source term $f$, there exists a (unique) $\scal^{(0)\prime}\in\scalspace_1$ with the property that
\begin{equation}
\label{EqIDNElimFin2}
  \text{\parbox{0.8\textwidth}{\it for the solution of the modified equation $L u = f - D_{(b,\scal^{(0)},h)}P(0,\scal^{(0)\prime},0)$, the $\chi_\cK h_{b,\rms 1}$-leading-order term at $\cK^+$ vanishes.}}
\end{equation}
\end{subequations}
Therefore, $u$ now decays at $\cK^+$ (and one can then extract its new, now decaying, leading-order term at $\cK^+$ using~\eqref{EqINHeurRewrite}).\footnote{We are hiding a technical subtlety here: the new source term $f-D_{(b,\scal^{(0)},h)}P(0,\scal',0)$ does not decay at order $5+\eps_+$ at $\iota^+$ anymore when $\scal'\neq 0$, but rather has a partial expansion with remainder of order $5+\eps_+$. Thus, the linear analysis on dynamical spacetimes really requires us to study $L u=f$ for $f$ that admit such partial expansions at $\iota^+$. This is not a conceptual issue, though, as the terms in the $\iota^+$-expansion of $f$ simply spawn the same types of terms in the $\cK^+$-asymptotics of $u$ as those that came from $\scri^+$ via transport already did.} The modified equation reads
\[
  D_{(b,\scal^{(0)},h)}P(0,\scal^{(0)\prime},u) = f;
\]
we have thus shuffled a non-decaying contribution to the solution of the original equation $L u=f$ into a modification of a finite-dimensional parameter $\scal^{(0)}$ (cf.\ the discussion of~\eqref{EqINLinNaive2}). See~\S\ref{SssD4ParNo} for details.

\medskip
{\bf Black hole parameters.} We simplify notation by working again with $P(b,h)$ from~\eqref{EqINPNaive}. Let us now suppose for simplicity that the first term at $\cK^+$ is a linearized Kerr metric $\dot g_b^\Ups(\dot b)$; and to simplify matters further, let us pretend that this is $\dot g_b(\dot b)$ (i.e., that this is already correctly gauged, so lies in $\ker\delta_{g_b,E^\Ups}\sfG_{g_b}$). Then, recalling $g_{b_0,b}=\chi_\cK g_b+(1-\chi_\cK)g_{b_0}$, we have the identity
\[
  D_{(b_0,0)}P(\dot b,0) = D_{g_{b_0}}\Ric\bigl(\chi_\cK\dot g_{b_0}(\dot b)\bigr) = D_{(b_0,0)}P\bigl(0,\chi_\cK\dot g_{b_0}(\dot b)\bigr) - \delta_{g_{b_0},E^\cC}^*[\delta_{g_{b_0},E^\Ups}\sfG_{g_{b_0}},\chi_\cK]\dot g_{b_0}(\dot b).
\]
Defining $\vartheta^{(0)}(\dot b):=[\delta_{g_{b_0},E^\Ups}\sfG_{g_{b_0}},\chi_\cK]\dot g_{b_0}(\dot b)$, this suggests replacing $P$ by
\[
  P(b,h) := \Ric(g_{b_0,b}+h) - \delta_{g_{b_0,b},E^\cC}^*\bigl( \Ups_{E^\Ups}(g_{b_0,b}+h,g_{b_0,b}) - \vartheta^{(0)}(b-b_0) \bigr),
\]
as this new operator $P$ satisfies
\[
  D_{(b_0,0)}P(\dot b,0) = D_{(b_0,0)}P\bigl(0,\chi_\cK\dot g_{b_0}(\dot b)\bigr).
\]
We can now argue as in~\eqref{EqIDNElimFin}--\eqref{EqIDNElimFin2}: the map assigning to $b'$ the parameter $\dot b$ of the leading-order term $\dot g_b(\dot b)$ of the solution $u$ of $L u:=D_{(b,h)}P(0,u)=D_{(b,h)}P(b',0)$ is close to the map for $(b,h)=(b_0,0)$---which is the identity map---and therefore invertible for $(b,h)$ close to $(b_0,0)$. Given $f$, one can thus pick $b'$ such that the solution of the modified equation $L u=f-D_{(b,h)}P(b',0)$, or equivalently
\[
  D_{(b,h)}P(b',u) = f
\]
(cf.\ \eqref{EqINLinNaive2}), has vanishing $\dot g_b$-term at $\cK^+$. See again~\S\ref{SssD4ParNo} for details.

\medskip
{\bf Asymptotic boosts.} The true leading-order term of solutions of the linearized gauge-fixed Einstein equation is $\chi_\cK h_{b,\rms 1}^{\leq 1}(\scal)$. This is a pure gauge term, so one is tempted to try and eliminate it in the usual fashion; however, extrapolating from~\eqref{EqIDNElimCorr}, the metric patch required for this purpose would be $-[\delta_{g_{b_0}}^*,\chi_\cK]\omega_{b_0,\rms 1}^{(0),\leq 1}(\scal)$, which is of order $0$ at $\iota^+\cap\supp\dd\chi_\cK$: it would destroy even the leading-order Minkowskian nature of the spacetime metric there. Thus, a qualitatively different argument is needed, which we proceed to motivate.

Let $\phi_\scal\colon\R^4\to\R^4$ be a Lorentz boost. We do not want the dynamical spacetime metric to look like
\begin{equation}
\label{EqIDNBoosted}
  \chi_\cK\phi_\scal^*g_b + (1-\chi_\cK)g_{b_0} + h,
\end{equation}
as we wish to prove decay to an unboosted Kerr metric.\footnote{Note that $g_b$ and $g_{b_0}$ are equal to the Minkowski metric $\ubar g$ to leading order at $\iota^+$. Since $\phi_\scal$ is an isometry for $\ubar g$, also the patched metric~\eqref{EqIDNBoosted} is equal to the Minkowski metric to leading order at $\iota^+$.} If we push~\eqref{EqIDNBoosted} forward along $\phi_\scal$, one roughly obtains the boosted metric
\[
  g_\scal:=\chi_\cK g_b + (1-\chi_\cK)(\phi_\scal)_*g_{b_0}+(\phi_\scal)_*h.
\]
The initial data $(\gamma,k)$ of~\eqref{EqIDNBoosted} at the Cauchy hypersurface $\Sigma_\IVP$ (so $\tilde\ft=0$ in the notation of Theorem~\ref{ThmISimple}, cf.\ Figure~\ref{FigIOmega}) arise for $g_\scal$ in the form $(\phi_\scal)_*(\gamma,k)$ at the boosted hypersurface $\phi_\scal(\Sigma_\IVP)$. By contrast, at $\Sigma_\IVP$, the data of $g_\scal$ are the push-forwards under $\phi_\scal$ of the data $(\gamma_\scal,k_\scal)$ of $g$ at the boosted hypersurface $\phi_\scal^{-1}(\Sigma_\IVP)$. Unlike $(\gamma,k)$, the data $(\gamma_\scal,k_\scal)$ have non-vanishing ADM linear momentum $\approx\scal$; but the metric $g_\scal$ evolving from them has vanishing final linear momentum (in the sense that it converges to the unboosted metric $g_b$).

Our strategy for avoiding non-zero final linear momenta is thus as follows. Starting with the exterior solution $g_\ext$ of the initial value problem for the Einstein equation, we produce a family of solutions $g_{\ext,-\scal}$ indexed by a (small) boost parameter $\scal$ whose initial data at $\Sigma_\IVP$ are the push-forwards of the initial data of $g_\ext$ at $\phi_\scal^{-1}(\Sigma_\IVP)$. We then regard $\scal\in\scalspace_1$ as an unknown that is to be determined so that the global solution of the gauge-fixed Einstein equation will be of the form
\begin{equation}
\label{EqIDNBoost}
  g_{b_0,b,-\scal} + h,\quad g_{b_0,b,-\scal}:=\chi_\cK g_b + (1-\chi_\cK)(\phi_\scal)_*g_{b_0},
\end{equation}
with $h$ decaying as usual. Defining $h_{-\scal}$ by writing $g_{\ext,-\scal}=(\phi_\scal)_*g_{b_0}+h_{-\scal}$, we thus need to solve
\[
  P(b,\scal,h) := \Ric\bigl( g_{b_0,b,-\scal} + \chi_- h_{-\scal} + h \bigr) - \text{(gauge term)} = 0
\]
for $h$ supported in $t_*\geq 1$ (cf.\ \eqref{EqINFwd}). Now, if the boost term of a solution to the linearized gauge-fixed Einstein equation $L_{b_0}u=f$ is $\chi_\cK h_{b_0,\rms 1}^{\leq 1}(\scal)$, one first uses the trivial identity
\[
  L_{b_0}\bigl(\chi_\cK h_{b_0,\rms 1}^{\leq 1}(\scal)\bigr) = -L_{b_0}\bigl( (1-\chi_\cK)h_{b_0,\rms 1}^{\leq 1}(\scal)\bigr);
\]
but this is the sum of a gauge term $\delta_{g_{b_0},E^\cC}^*\vartheta$ where $\vartheta\sim [\delta_{g_{b_0},E^\Ups}\sfG_{g_{b_0}},\chi_\cK]h_{b_0,\rms 1}^{\leq 1}(\scal)$ has good decay ($2$ orders at $\iota^+$) and a term (up to signs)
\[
  D_{g_{b_0}}\Ric \bigl( (1-\chi_\cK) \cL_{\text{boost}(\scal)}g_{b_0} \bigr),
\]
which is in essence the linearization in $\scal$ of
\[
  \scal \mapsto \Ric(g_{b_0,b_0,-\scal}) - \delta_{g_{b_0,b_0,-\scal},E^\cC}^* \Ups_{E^\Ups}(g_{b_0,b_0,-\scal},g_{b_0,b_0,-\scal}).
\]

Arguing as in~\eqref{EqIDNElimFin}--\eqref{EqIDNElimFin2} (and also inserting the aforementioned gauge term into the definition of $P(b,\scal,h)$), we conclude that, given a source term $f$, one can find a unique $\scal'$ such that the solution of
\[
  D_{(b,\scal,h)}P(0,0,u) = f - D_{(b,\scal,h)}P(0,\scal',0)
\]
has \emph{no} contribution from the boost term $\chi_\cK h_{b,\rms 1}^{\leq 1}$. --- The full computations are carried out in~\S\ref{SssD2BoostNo}. The boost we use is defined in Lemma~\ref{LemmaKBoDef}; it is the identity near $\cK^+$ since in our rewriting above we really only need to ``de-boost'' the initial black hole. The boosted exterior solutions are carefully constructed in~\S\ref{SsExBo}.

\begin{rmk}[Initial boost]
\label{RmkINBoostInit}
  The upshot of~\eqref{EqIDNBoost} is that we will find the solution of the initial value problem for the nonlinear stability of $g_{b_0}$ in the form $g_{b_0,b,-\scal}+h$, with initial data correspondingly given by the push-forward under the boost $\phi_\scal$ of $(\gamma,k)$ at the preimage under $\phi_\scal$ of the original hypersurface (i.e., $\tilde\ft=0$ in the notation of Theorem~\ref{ThmISimple}); see Theorem~\ref{ThmSt}. Thus, the final black hole is boosted relative to the initial one. This is expected to have interesting consequences for the Bondi mass, as indicated in Conjecture~\ref{ConjMass}.
\end{rmk}

\medskip
{\bf Putting the modifications together.} Unlike in linear theory around $g_{b_0}$, the modifications for eliminating boosts, parameter changes, etc.\ are not decoupled. For example, the modification term $D_{(b,\scal^{(0)},h)}P(0,\scal^{(0)\prime},0)$ used in~\eqref{EqIDNElimFin2} may very well trigger a non-trivial boost term in the asymptotics of $u$, \emph{albeit only a small one when $(b,\scal^{(0)},h)$ is close to $(b_0,0,0)$}. Thus, when eliminating asymptotic translations using a parameter $\scal^{(0)\prime}$, one needs to simultaneously use also further (small) modification parameters to eliminate all of the less-decaying terms, such as a (small) parameter $\scal'$ for undoing the (small) boost term triggered by the extra source term $D_{(b,\scal^{(0)},h)}P(0,\scal^{(0)\prime},0)$.

Nonetheless, the elimination of ultimately all pure gauge terms arising in the polyhomogeneous expansion at $\cK^+$ is possible via the same \emph{linear algebra} fact that small perturbations of invertible maps on finite-dimensional vector spaces are invertible.

\begin{rmk}[No elimination of non-polyhomogeneous terms]
\label{RmkINElimNon}
  We can finally explain why we do not eliminate ``structureless'' terms, even if they are pure gauge on exact Kerr---and thus why partial polyhomogeneity is crucial in our approach. For example, consider the term $\chi_\cK h_{b_0,\rms 1}^{\leq 1}(a(t_*))$, $a(t_*)=\cO(t_*^{-2-\eps_\cK})$, in~\eqref{EqINHeurh} (for $b=b_0$). While the (ultimately algebraic) computations around~\eqref{EqIEinElimDec} apply without change, the perturbative arguments sketched above do not work anymore: even though, for a perturbed equation $L_{b,h}u=f$, one can (after various finite-dimensional gauge modifications, as explained above) extract an asymptotic term $\chi_\cK h_{b,\rms 1}^{\leq 1}(a(t_*))$, the map assigning to a \emph{function} $\tilde a(t_*)$, lying in some weighted Sobolev space $\Hb^{k,2+\eps_\cK}$, the term $a(t_*)$ arising in the asymptotic description of the forward solution of
  \[
    L u=\text{(correction term involving $\tilde a(t_*)$)}
  \]
  is, in our analysis that gives away b-derivatives rather freely, a map $\Hb^{\infty,2+\eps_\cK}\to\Hb^{\infty,2+\eps_\cK}$ which is, however, only bounded as a map $\Hb^{k,2+\eps_\cK}\to\Hb^{k-d,2+\eps_\cK}$ for some positive loss of derivatives $d$. Even though for $(b,h)=(b_0,0)$ this map is the identity on $\Hb^{\infty,2+\eps_\cK}$, it is \emph{not} true that general perturbations of this map, in a topology respecting only operator norm continuity in $\cL(\Hb^{k,2+\eps_\cK},\Hb^{k-d,2+\eps_\cK})$, are invertible. --- This also explains the final part of Remark~\ref{RmkIData}: initial data with ``structureless'' $\cO(r^{-1-\delta})$-terms would, in linearized theory, produce ``structureless'' $\cO(t_*^{-\delta})$ center-of-mass motions, which we cannot eliminate as they are parameterized by elements of an infinite-dimensional space like $\Hb^{\infty,\delta}$.
\end{rmk}

\begin{rmk}[Milder infinite-dimensional gauge issues]
\label{RmkINInfty}
  The works \cite{HintzVasyKdSCosm,LeimbacherdS} on the stability of de~Sitter type spacetimes near their conformal boundaries in generalized harmonic gauge also feature infinite-dimensional gauge modifications (see, e.g., \cite[Theorem~3.1]{HintzVasyKdSCosm}). These, however, arise essentially algebraically (when re-interpreting a leading-order term of a metric perturbation at the conformal boundary), \emph{even on the dynamical spacetime}; see \cite[(3.71)--(3.72)]{HintzVasyKdSCosm}.
\end{rmk}

\subsubsection{Nonlinear iteration on spaces encoding partial asymptotic expansions}
\label{SssINPhg}

As mentioned already in the discussion following~\eqref{EqINHeurf}, closing the nonlinear iteration requires us to capture the $\iota^+$-expansion of metric perturbations (i.e., $\tilde h$ in~\eqref{EqINHeurh}) in a very precise fashion in order to ensure that upon applying the non-linear forward map, all polyhomogeneous terms at $\iota^+$ vanish, and only a well-decaying term $f$ as in~\eqref{EqINHeurf} remains.

The basic idea for doing this is simple. Let us consider the problem of constructing the \emph{most general} formal solution (i.e., solution in generalized Taylor series at $\iota^+$) of the nonlinear gauge-fixed Einstein equation, say,~\eqref{EqINPNaive}, that decays towards the Minkowski metric at $\iota^+$ at rate $t_*^{-1}$ (and has decay and asymptotics at $\cK^+$ consistent with~\eqref{EqINHeurh}). The leading-order term $u_1$ (which is $t_*^{-1}$ times a function of $(R,\omega)=(\frac{r}{t_*},\omega)$) must lie in the kernel of the \emph{linear}, \emph{Minkowskian} $\iota^+$-normal operator $N_{\iota^+}(1)$ (see~\eqref{EqIEinLaNormOp}). It is thus uniquely determined by incoming data at $\scri^+$ (specifically, by the $r^{-1}$-term of the $(\dd t_*)^2$-component of $\tilde h$) and also by the final black hole and boost parameters $b$ and $\scal$ (as they are independent quantities not locally determined via transport from $\scri^+$, and since the $t_*^{-1}$-term of $g_{b_0,b,-\scal}$ at $\iota^+$ does depend on $b$ and $\scal$).

Now, $g_{b_0,b,-\scal}+t_*^{-1}u_1$ is an approximate solution of the gauge-fixed Einstein equation at $\iota^+$, but lower-order terms (e.g., the $t_*^{-4}$-term) in its $\iota^+$-expansion need not vanish. Furthermore, one needs to allow for further linear contributions at resonances $\lambda$ with $\Re\lambda>1$. Altogether, one faces error terms at $\iota^+$ (such as the $t_*^{-4}$-term) whose leading-order term at $\iota^+$ (which has more than $t_*^{-3}$-decay) one must eliminate; this can again be done using the inversion of the $\iota^+$-normal operator (see Corollary~\ref{CoripPQhom}), and one then iterates this procedure.

This discussion does not yet address the problem of ensuring strong decay of our formal solutions at $\cK^+$ (i.e., the structure in~\eqref{EqINHeurh}). The gauge modifications required to ensure strong decay at $\cK^+$ arise in such a scheme as follows: the terms arising from the inversion of the $\iota^+$-normal operator typically have non-trivial asymptotic expansions at $\iota^+\cap\cK^+$, i.e., as $R=\frac{r}{t_*}\to 0$. We need to ensure that all of these terms, up to some sufficiently large power of $R$, in fact \emph{vanish}; but this is \emph{precisely} what the gauge modifications and metric patches do (now working entirely on the level of their $\iota^+$-leading-order terms). \emph{There is thus a change in perspective:} the expansion at $\iota^+$ takes center stage, and the asymptotics at $\cK^+$ coming from large indicial roots are determined by it---and therefore so are the gauge modifications required to eliminate the poorly decaying contributions to $\cK^+$-asymptotics that they would otherwise cause.

\bigskip

In summary, we parameterize metric perturbations using the following data: the final Kerr parameters $b$, the boost parameter $\scal$, the coefficients\footnote{For the finite-dimensional space of pure resonant states, this requires only finite-dimensional parameters. The non-pure resonant states are parameterized by their incoming data, which lie in a direct sum of finitely many Sobolev spaces on $\Sph^2$. See Proposition~\ref{PropipPResPar}.} with which $\iota^+$-resonant states contribute---and which thus determine the full $\iota^+$-expansion of $\tilde h$ in~\eqref{EqINHeurh} as well as the gauge modifications required to eliminate terms with $\cK^+$-order below $2+\eps_\cK$---, the terms {\color{myg}$P$} (physical states with $t_*^{-3}$-decay or better) and {\color{myb}$Q$} in~\eqref{EqINHeurh} (conormal terms---involving decaying center-of-mass motions, decaying modulations of the Kerr parameters, and large energy states---that have $t_*^{-2-\eps_\cK}$- but not $t_*^{-4-\eps_\cK}$-decay), and the strongly decaying part of the gravitational tail which is of class
\[
  \includegraphics{EqINPhgTail}
\]
See~\S\ref{SsEfP} and Theorem~\ref{ThmEfP} for details: the data are specified in Definition~\ref{DefEfD}, and the nonlinear gauge-fixed Einstein operator $P$ is given in~\eqref{EqEfPMap}, with Theorem~\ref{ThmEfP} producing the map $\Phi$ that maps data into arguments of $P$.

In order to close the iteration scheme, we must then show that the linear asymptotics, gauge modifications, and parameter changes that arise in the analysis sketched in~\S\ref{SssINHeur} arise equivalently from the linearization of this new parameterization of metric perturbations in the data of Definition~\ref{DefEfD}. But it is clear that this must be so: after all, that parameterization was constructed to produce the \emph{most general} possible metrics compatible with strong decay of the output of the nonlinear forward map (gauge-fixed Einstein), and thus the analogous statement is also true at the linearized level. (The hard input is thus that forward solutions of the linearized problem \emph{do} decay at the rates that are baked into this parameterization ``from infinity;'' recall that this is the content of Corollary~\ref{CorD6Impr}.) The complete argument is given in~\S\ref{SsEfS}. (The reader is encouraged to first study the ODE toy model in~\S\ref{SsEfT}.)

The proof of nonlinear stability in $t_*\geq 1$ now amounts to applying a standard Nash--Moser iteration to the map $P\circ\Phi$, with unknowns given by the aforementioned data. The details are in~\S\ref{SsSt}.

\subsection{Outline}
\label{SsIO}

The paper is structured as follows.

\begin{itemize}
\item \S\ref{ST}: \textit{technical background}. We discuss manifolds with corners and notions of asymptotic analysis on them, including polyhomogeneity, conormality, and b-differential operators and their asymptotic analysis. We also study the Fourier transform and make Figure~\ref{FigIGenKFT} precise; and we recall spherical harmonic decompositions for functions and tensors, as needed in computations for Minkowskian model operators.
\item \S\ref{SK}: \textit{Kerr metrics}. We define the Kerr family, useful (time) coordinates, and the compactification of spacetime (cf.\ Figure~\ref{FigIMwc}) on which our asymptotic analysis will take place.
\item \S\ref{S1}: \textit{gauge-fixed Einstein equation}. We introduce the basic gauge condition and constraint damping modification as well as the gauge-fixed Einstein operator whose linearization around Kerr will subsequently be studied and which will be augmented (with gauge modifications etc.) later on (\S\ref{SD}) in the nonlinear analysis.
\item \S\ref{SEx}: \textit{exterior stability}. We prove the exterior stability and the partial polyhomogeneity of solutions (Theorem~\ref{ThmExPhg}). In preparation for dealing with the recoil problem, we discuss pullbacks by boosts (Theorem~\ref{ThmExBoStab}). We also discuss initial data for Theorem~\ref{ThmISimple} and \eqref{EqIDataPhg} in~\S\ref{SsExID}.
\item \S\ref{SWG}: \textit{gauge potential wave operator on Kerr}. We commence the study of the linearized gauge-fixed Einstein operator on an exact Kerr spacetime. We prove the mode stability of the gauge potential wave operator and moreover construct various (generalized) large zero energy states that arise later on as pure gauge contributions to the late-time asymptotics of gravitational waves.
\item \S\ref{SWC}: \textit{constraint propagation wave operator on Kerr}. We recall the strong version of constraint damping implemented in the companion paper \cite{HintzKerrCD}, and moreover construct dual zero energy states for the linearized gauge-fixed Einstein operator.
\item \S\ref{SWE}: \textit{linearized gauge-fixed Einstein operator on Kerr}. We establish basic spectral properties (mode stability, behavior at zero energy) and verify the subprincipal symbol condition at the trapped set. We also construct the remaining (physical) large zero energy states for our later asymptotic analysis.
\item \S\ref{Sip}: \textit{normal operators at $\iota^+$ and $\tface$}. We describe the mapping properties and asymptotic expansions of solutions of the $\iota^+$- and $\tface$-normal operators. Among other things, we give a full description of the $\cK^+$-asymptotics that are caused by a term in the $\iota^+$-expansion of a metric perturbation (cf.\ \S\ref{SssIEinLa}, and see Proposition~\ref{PropipGr}).
\item \S\ref{SAdm}: \textit{2-admissibility of the linearized gauge-fixed Einstein operator on Kerr.} This is the final section that is concerned with the Kerr model operator $L_b$: we verify the conditions required in the companion paper \cite{HintzNonstat2} (cf.\ \S\ref{SssINAdm}) that make its results on high regularity and weak polynomial bounds for dynamical perturbations of $L_b$ available.
\item \S\ref{SD}: \textit{decay of gravitational waves on dynamical spacetimes.} This is the technical core of our nonlinear stability analysis. Using the strategy outlined in~\S\ref{SssINHeur}, we show how to use carefully designed gauge modifications and metric patches, and how to incorporate the final black hole and boost parameters into augmented versions of the nonlinear gauge-fixed Einstein equation, in order to make solutions of linearized problems have strong decay at $\cK^+$ (and acceptable asymptotics and decay at $\iota^+$).
\item \S\ref{SEf}: \textit{efficient parameterization of metric perturbations.} We implement the ideas outlined in~\S\ref{SssINPhg} on how to capture metric perturbations via their $\iota^+$-asymptotics and some extra data on (the decay towards) the final black hole. The main result is Theorem~\ref{ThmEfS}.
\item \S\ref{SSt}: \textit{proof of nonlinear stability.} The proof of Theorem~\ref{ThmISimple} or its precise version (Theorem~\ref{ThmSt}) is now a standard application of the Nash--Moser iteration scheme based on Theorem~\ref{ThmEfS}. This produces the solution in the forward cone which is then attached to the correctly boosted exterior solution.
\end{itemize}

\subsection*{Acknowledgments}

I am very grateful to Andr\'as Vasy for countless discussions and past collaborations regarding stability problems in general and the Kerr stability problem in particular. The present paper would not have been possible without his deep insights concerning the applicability of spectral theory and microlocal analysis to problems in general relativity. I am also thankful to Mihalis Dafermos, Semyon Dyatlov, Dietrich H\"afner, Stefan Hollands, Gustav Holzegel, Sergiu Klainerman, Rafe Mazzeo, Richard Melrose, Gunther Uhlmann, Jared Wunsch, and Maciej Zworski for discussions about aspects of this problem in the course of the past $\sim 10$ years during which the ideas presented in this paper have been developing.

I am grateful to the U.S.\ National Science Foundation for support under the grants DMS-1955614 and DMS-2554160, and to the Swiss National Science Foundation for support under the SNSF Consolidator Grant TMCG-2\textunderscore{}223167. Part of this research was carried out during the periods I was a Clay Research Fellow and Sloan Research Fellow; I thank the Clay Mathematics Institute and the Sloan Foundation for their support.

Last but not least, I would like to thank my wife Sara for her support that was unwavering and stable under small and large perturbations.

\section{Technical prerequisites}
\label{ST}

\subsection{Manifolds with corners; vector fields; function spaces}
\label{SsTM}

We define an $n$-dimensional manifold with corners $M$ to be locally modeled on $[0,\infty)_k\times\R^{n-k}$, where $k\in\{0,\ldots,n\}$ may vary from chart to chart. Its boundary hypersurfaces are the closures of the connected components of the set of $p\in M$ which lie on the boundary $\{0\}\times\R^{n-1}$ of a chart with $k=1$. Following Melrose \cite{MelroseDiffOnMwc}, we require all boundary hypersurfaces $H$ to be embedded submanifolds. Equivalently, they admit \emph{defining functions} $\rho_H\in\CI(M)$, which are functions $\rho_H\colon M\to[0,1]$ such that $H=\rho_H^{-1}(0)$ and $\dd\rho_H\neq 0$ everywhere on $H$. We say that $M$ is a manifold with boundary if only the cases $k=0$ and $k=1$ occur. A \emph{boundary face} of a manifold with corners is a (non-empty) intersection of boundary hypersurfaces of $M$.

Suppose $H\subset M$ is compact; then for any fixed defining function $\rho_H\in\CI(M)$, there exist $\eps>0$ and a collar neighborhood $[0,\eps)\times H$ of $H$ adapted to $\rho_H$, by which we mean a diffeomorphism from $[0,\eps)\times H$ onto a neighborhood of $H$ such that $\rho_H$ is the coordinate in the first factor (and in particular $\{0\}\times H$ is identified with $H$).

\begin{notation}[Defining functions]
\label{NotTMDef}
  We shall use the same letter $\rho_H$ to denote any smooth function defined on an open subset $U\subset M$ such that $H\cap U=\rho_H^{-1}(0)$ and $\dd\rho_H\neq 0$ everywhere on $H\cap U$; we refer to such $\rho_H$ as a local defining function when $U\neq M$.
\end{notation}

Note that for any two local defining functions $\rho_H,\rho'_H$ defined on $U,U'\subset M$, the quotient $\rho_H/\rho'_H$ is a positive smooth function on $U\cap U'$. For the purpose of measuring decay of functions on $M$ (or $M^\circ$) at $H$ using powers of $\rho_H$, any two choices are thus equivalent on every compact subset of $U\cap U'$. For local coordinate expressions of differential operators however, some choices are more convenient than others.

\begin{example}[Radial compactification]
\label{ExTMRadCpt}
  An important example of a manifold with boundary is the \emph{radial compactification} $\ol{\R^n}$ of $\R^n$, defined as
  \[
    \ol{\R^n} := \Bigl(\R^n \sqcup \bigl( [0,\infty)_\rho\times\Sph^{n-1} \bigr) \Bigr) \Big/ \sim,\quad 0\neq x=r\omega \sim (r^{-1},\omega).
  \]
  Thus $\la x\ra^{-1}=(1+r^2)^{-\frac12}$ is a boundary defining function, and $r^{-1}$ is a local boundary defining function in $\{r>0\}$.
\end{example}

\begin{figure}[!ht]
\centering
\includegraphics{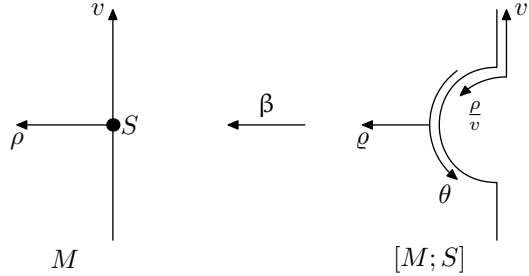}
\caption{The blow-up $[M;S]$ of $M$ at $S$ (here a point) from~\eqref{EqTMBlowup} with some (local) coordinates, and the blow-down map $\upbeta$ mapping $[M;S]$ to $M$.}
\label{FigTMBlowup}
\end{figure}

\begin{example}[Real blow-up]
\label{ExTMBlowup}
  Suppose $M$ is a manifold with corners and $S\subset M$ is a \emph{p-submanifold} (the ``p'' standing for ``product''), which means that for all $p\in S$ there exists a local coordinate chart on $M$ around $p$ such that $S$ is given by the vanishing of some subset (of fixed cardinality, given by the codimension $k$ of $S$ in $M$) of these local coordinates. Suppose also that $S$ is a closed subset of $M$. Then the \emph{real blow-up} $[M;S]$ of $M$ at $S$ is defined to be the manifold with corners obtained by replacing $S$ by the origin of polar coordinates around $S$; the polar coordinate origin is the \emph{front face} (invariantly given by the inward pointing spherical normal bundle of $S$). The \emph{blow-down map} $\upbeta\colon[M;S]\to M$ is the identity map on $M\setminus S$, and the polar coordinate map near $S$. For a set $T\subset M$, we define the \emph{lift} $\upbeta^*T$ of $T$ to $[M;S]$ to be $\upbeta^{-1}(T)$ when $T\subset S$, and the closure of $\upbeta^{-1}(T\setminus S)$ otherwise. We describe $[M;S]$ only in two special cases relevant for this paper. (For a comprehensive discussion, see \cite[\S{5}]{MelroseDiffOnMwc}.)
  \begin{enumerate}
  \item{\rm (Blow-up of a boundary face.)} When $S$ is a compact boundary face, then it has a collar neighborhood $[0,\eps)_{\rho_1}\times\cdots\times[0,\eps)_{\rho_k}\times S$ where $\rho_1,\ldots,\rho_k\in\CI(M)$ are defining functions of the boundary hypersurfaces whose intersection is $S$. Then
    \[
      [M;S] := \Bigl( (M\setminus S) \cup \bigl( [0,\eps)_R \times \Sph^{k-1}_+ \times S \bigr) \Bigr) \Big/ \sim
    \]
    where $R:=(\sum_{j=1}^k\rho_j^2)^{\frac12}$ and $\Sph^{k-1}_+:=\Sph^{k-1}\cap[0,\infty)^k\subset\R^k$; here the equivalence relation $\sim$ identifies a point $(R,\omega,p)$, $R\neq 0$, in the second set with $(R\omega,p)$ in the collar neighborhood. The front face is an $\Sph^{k-1}_+$-bundle over $S$. For example, $[ [0,\infty)^2; \{(0,0)\} ]=[0,\infty)_R\times[0,\frac{\pi}{2}]_\theta$, with the blow-down map to $[0,\infty)^2$ given by $(R,\theta)\mapsto(R\cos\theta,R\sin\theta)$. (Thus the radial coordinate $R$ on $[0,\infty)^2$ lifts to the blow-up as a smooth function, and indeed defining function of the front face.)
  \item{\rm (Blow-up of a boundary submanifold.)} Suppose $S$ is a compact and contained in the interior of a boundary hypersurface $H$ of $M$; for concreteness, suppose $S\subset H^\circ$ has codimension $1$ and orientable normal bundle. Then there exists a collar neighborhood $[0,\eps)_\rho\times(-\eps,\eps)_v\times S$ of $S$, where $\rho\in\CI(M)$ is a defining function of $H$ and $v$ is a smooth function defined in a neighborhood of $S$ in $H$ which vanishes non-degenerately at $S$. Then
    \begin{equation}
    \label{EqTMBlowup}
      [M;S] := \Bigl( (M\setminus S) \cup \bigl( [0,\eps)_\varrho \times \Sph^1_+ \times S \bigr) \Bigr) \Big/ \sim;
    \end{equation}
    identifying $\Sph^1_+=[0,\pi]_\theta$ using polar coordinates, the equivalence relation $\sim$ identifies $(\varrho,\theta,p)$, $\varrho\neq 0$, with the point $(\rho,v,p)=(\varrho\sin\theta,\varrho\cos\theta,p)$. The front face is an $\Sph^1_+$-bundle over $S$ (equivalently, an $\ol\R$-bundle over $S$). See Figure~\ref{FigTMBlowup}.
  \end{enumerate}
\end{example}

\subsubsection{Notions of b-analysis}

The following goes back (at least) to \cite{MelroseTransformation,MelroseMendozaB,MelroseAPS}:

\begin{definition}[Notions of b-analysis]
\label{DefTMb}
  Let $M$ be a manifold with corners.
  \begin{enumerate}
  \item{\rm (b-vector fields.)} The space $\Vb(M)\subset\cV(M)=\CI(M;T M)$ consists of all smooth vector fields on $M$ that are tangent to $\pa M$.
  \item{\rm (b-differential operators.)} Let $m\in\N_0$. The space $\Diffb^m(M)$ consists of all locally finite sums of up to $m$-fold compositions of elements of $\Vb(M)$ (with $0$-fold compositions being multiplication operators by elements of $\CI(M)$).
  \item{\rm (b-densities.)} A \emph{b-density} $\mu$ on $M$ is a smooth 1-density on $M^\circ$ such that $(\prod_H\rho_H)\mu$ is a smooth density on $M$; here $H$ runs over all boundary hypersurfaces of $M$, and $\rho_H\in\CI(M)$ is a defining function of $H$. All densities will be required to be \emph{positive} in this paper.
  \end{enumerate}
\end{definition}

In local coordinates $x^j\in[0,\infty)$, $j=1,\ldots,k$, and $y^l\in\R$, $l=1,\ldots,n-k$, on $M$, b-vector fields are smooth linear combinations of
\[
  x^j\pa_{x^j}\ (j=1,\ldots,k),\quad
  \pa_{y^l}\ (l=1,\ldots,n-k).
\]
Moreover, b-densities can be written as
\[
  a(x^1,\ldots,x^k,y^1,\ldots,y^{n-k}) \Bigl|\frac{\dd x^1}{x^1}\cdots\frac{\dd x^k}{x^k}\,\dd y^1\cdots\dd y^{n-k}\Bigr|,\quad 0<a\in\CI.
\]

Given a b-differential operator $A\in\Diffb^m(M)$ on a manifold with boundary and a collar neighborhood $[0,\eps)_\rho\times\pa M$ of $\pa M$, we can write
\[
  A = \sum_{j=0}^m A_j(\rho)(\rho\pa_\rho)^j,\quad A_j\in\CI\bigl([0,\eps);\Diff^{m-j}(\pa M)\bigr).
\]
\begin{definition}[b-normal operator; indicial roots]
\label{DefTMInd}
  We define the \emph{b-normal operator}, resp.\ the \emph{indicial family} of $A$ at $\pa M$ to be
  \[
    N(A):=\sum_{j=0}^m A_j(0)(\rho\pa_\rho)^j \in \Diff^m_{\bop,{\rm I}}([0,\infty)_\rho\times\pa M),\quad
    N(A,\lambda):=\sum_{j=0}^m A_j(0)\lambda^j \in \Diff^m(\pa M);
  \]
  here the subscript ``${\rm I}$'' refers to the dilation-\emph{i}nvariance of $N(A)$.
  \begin{enumerate}
  \item We say that $\lambda\in\C$ is an \emph{indicial root} of $A$ if there exists a solution $0\neq u\in\CI(\pa M)$ of $N(A,\lambda)u=0$, or equivalently $N(A)(\rho^\lambda u)=0$; we call $u$ or $\rho^\lambda u$ an \emph{indicial solution}.
  \item The \emph{boundary spectrum} $\specb(A)\subset\C$ is the set of all indicial roots of $A$.
  \item We say that $\lambda$ is a \emph{simple} indicial root if there do not exist an indicial solution $u$ and a function $u'\in\CI(\pa M)$ such that $N(A)(\rho^\lambda(\log\rho)u+\rho^\lambda u')=0$.
  \end{enumerate}
  For a weighted operator $A\in\rho^\alpha\Diffb^m(M)$, we define $N(A):=\rho^\alpha N(\rho^{-\alpha}A)$, and its indicial roots as those of $\rho^{-\alpha}A$.
\end{definition}

While $N(A)$ depends on the choice of collar neighborhood, the set of indicial roots (and their simplicity) does not. When the operator $A$ is a b-differential operator that is elliptic as a b-differential operator near $\pa M$,\footnote{Using local coordinates $y$ on $\pa M$ and writing $A=\sum_{j+|\alpha|\leq m} a_{j\alpha}(\rho,y)(\rho\pa_\rho)^j\pa_y^\alpha$, this means that the polynomial $\sum_{j+|\alpha|=m} a_{j\alpha}(\rho,y)\xi^j\eta^\alpha$ has no real roots $(\xi,\eta)\neq(0,0)$.} then $N(A,\lambda)$ is elliptic, and (as a consequence of the large parameter ellipticity of $N(A,\lambda)$ as $|\Re\lambda|\to\infty$ for bounded $|\Im\lambda|$) the set of indicial roots is discrete and has finite intersection with $\{C_1<\Re\lambda<C_2\}$ for all $C_1<C_2$. The simplicity of an indicial root $\lambda_0$ is then moreover equivalent to the order of the pole of $N(A,\lambda)^{-1}$ at $\lambda=\lambda_0$ being equal to $1$.

We next turn to function spaces.

\begin{definition}[Weighted function spaces]
\label{DefTMHb}
  Let $M$ be a compact manifold with corners. Write $\{H_1,\ldots,H_N\}$ for the collection of its boundary hypersurfaces, and $\rho_1,\ldots,\rho_N\in\CI(M)$ for boundary defining functions. Let $\vec\alpha=(\alpha_1,\ldots,\alpha_N)\in\R^N$ and write $\vecrho^{\vec\alpha}:=\prod_{j=1}^N \rho_j^{\alpha_j}$.
  \begin{enumerate}
  \item{\rm (Conormal spaces.)} We write
    \[
      \cA^{\vec\alpha}(M)
    \]
    for the space of all smooth functions $u\colon M^\circ\to\C$ such that for all $m\in\N_0$ and $A\in\Diffb^m(M)$, the function $\vecrho^{-\vec\alpha}A u$ is uniformly bounded.
  \item{\rm (b-Sobolev spaces.)} Fix a b-density $\mu$ on $M$. Let $k\in\N_0$. Then
    \[
      \Hb^{k,\vec\alpha}(M)
    \]
    is the space of all $u\in L^2_\loc(M^\circ)$ such that for all $A\in\Diffb^k(M)$, one has $\vecrho^{-\vec\alpha}A u\in L^2(M,\mu)$. Fixing a finite spanning set $\sA\subset\Diffb^k(M)$ of $\Diffb^k(M)$ over $\CI(M)$, we define the norm
    \[
      \|u\|_{\Hb^{k,\vec\alpha}(M)} := \sum_{A\in\sA} \|A u\|_{\Hb^{0,\vec\alpha}(M)},\quad
      \|u\|_{\Hb^{0,\vec\alpha}(M)} := \|\vecrho^{-\vec\alpha}u\|_{L^2(M,\mu)}.
    \]
  \end{enumerate}
\end{definition}

All b-densities are smooth positive multiples of each other; and when $M$ is compact, the ratio of any two b-densities is uniformly bounded. This implies that these function spaces are well-defined as vector spaces; and different choices (defining functions, b-density, spanning set of $\Diffb^k$) lead to equivalent norms. When $M=[0,\infty)^k\times\R^{n-k}$, or more generally $M=[0,\infty)_x^k\times\R_y^{n-k}\times S$ where $S$ is a closed (i.e., compact, no boundary) smooth manifold, we define the spaces $\cA^{\vec\alpha}(M)$ and $\Hb^{k,\vec\alpha}(M)$ similarly, but for the concrete choices $\mu=|\frac{\dd x^1}{x^1}\cdots\frac{\dd x^k}{x^k}\,\dd y^1\cdots\dd y^{n-k}|\otimes\mu_S$ where $\mu_S$ is a positive smooth density on $S$, and only using as testing operators $A$ those which are compositions of $x^j\pa_{x^j}$, $\pa_{y^l}$, and vector fields on $S$.

\subsubsection{Asymptotic expansions and partial polyhomogeneity}
\label{SssTMExp}

In order to capture asymptotic expansions at boundary hypersurfaces, we first recall:

\begin{definition}[Index sets]
\label{DefTMIndex}
  An \emph{index set} $\cE$ is a subset $\cE\subset\C\times\N_0$ such that
  \begin{equation}
  \label{EqTMIndex}
    (z,k)\in\cE \implies (z+j,k)\in\cE\ \forall\,j\in\N_0,\quad (z,j)\in\cE\ \forall\,j\leq k,
  \end{equation}
  and for all $C\in\R$, there are only finitely many pairwise distinct $(z,k)\in\cE$ with $\Re z\leq C$. We introduce the following further notation:
  \begin{enumerate}
  \item{\rm (Minimal exponents.)} If $\cE\neq\emptyset$, we set $\min\Re\cE:=\inf\{\Re z\colon(z,0)\in\cE\}$. Otherwise, $\min\Re\cE:=\infty$.
  \item\label{ItTMIndexExp}{\rm (Exponents.)} We write $\pi_1\colon\C\times\N_0\to\C$ for the projection, and set $\Re\pi_1\cE:=\{\Re z\colon z\in\pi_1\cE\}\subset\R$.
  \item\label{ItTMIndexMaxLog}{\rm (Maximal logarithmic exponents.)} For $z\in\C$, we set
    \[
      k(\cE,z) := \begin{cases} \max\{ k\in\N_0 \colon (z,k)\in\cE \}, & z\in\pi_1\cE, \\ -1, & \text{otherwise}. \end{cases}
    \]
  \item{\rm (Sums and multiples.)} For two index sets $\cE,\cF$ and for $q\in\C$, we define
    \[
      \cE+\cF := \{ (z+w,k+l) \colon (z,k)\in\cE,\ (w,l)\in\cF \},\quad
      \cE+q := \{ (z+q,k) \colon (z,k)\in\cE \},
    \]
    as well as $2\cE:=\cE+\cE$ and inductively $(j+1)\cE:=j\cE+\cE$ for $j\in\N$. We call $\cE$ with $\min\Re\cE>0$ \emph{nonlinearly closed} if $j\cE\subset\cE$ for all $j\in\N$.
  \item{\rm (Extended union, I.)} For two index sets $\cE,\cF$, we set
    \[
      \cE\extcup\cF := \cE\cup\cF \cup \{(z,k+l+1) \colon (z,k)\in\cE,\ (z,l)\in\cF \}.
    \]
  \item\label{ItTMIndexSingle}{\rm (Extended union, II.)} Given $z\in\C$ and $k\in\N_0$, we write
    \[
      \cE \extcup \{(z,k)\} = \{(z,k)\}\extcup\cE
    \]
    for the smallest index set $\cG\supset\cE$ such that $k(\cG,z)=k(\cE,z)+k+1$.
  \item\label{ItTMIndexSpecial}{\rm (Special index sets.)} Given $z\in\C$ and $k\in\N_0$, we abuse notation and write $(z,k)$ for the index set
    \[
      (z,k) := \{(z+j,l) \colon j\in\N_0,\ 0\leq l\leq k\} \subset \C\times\N_0.
    \]
    (This is the smallest index set containing the pair $(z,k)$.)
  \end{enumerate}
\end{definition}

The operation~\eqref{ItTMIndexSingle} is non-standard. Examples are
\[
  (1,0) \extcup \{(1,0)\} = (1,1),\quad
  \bigl((1,0) \cup (2,1)\bigr) \extcup \{(1,0)\} = (1,1).
\]
The second example shows that $\cE\extcup\{(z,k)\}$ is typically smaller than $\cE\extcup(z,k)$.

\begin{definition}[Partial polyhomogeneity]
\label{DefTMphg}
  Let $M$ be a compact manifold with corners. Write $\sH$ for the set of boundary hypersurfaces, and fix boundary defining functions $\rho_H\in\CI(M)$ for each $H\in\sH$. For $H\in\sH$, write $\sH(H)\subset\sH$ for the subset of $H'\in\sH$ such that $H'\neq H$ and $H\cap H'\neq\emptyset$. Suppose we are given an index set $\cE_H\subset\C\times\N_0$ and a real number $\alpha_H\in\R$ for each $H\in\sH$. Let $k\in\N_0$. If $\min\Re\cE_H>\alpha_H$ for all $H$, set
  \[
    \Hb^{k,(\vec\cE,\vec\alpha)}(M) := \Hb^{k,\vec\alpha}(M),\quad
    \cA^{(\vec\cE,\vec\alpha)}(M) := \cA^{\vec\alpha}(M).
  \]
  Otherwise, we define these by induction on the number of $H\in\sH$ with $\min\Re\cE_H\leq\alpha_H$ as follows: we write $\Hb^{k,(\vec\cE,\vec\alpha)}(M)$ for the space of all $u$ such that for all $H\in\sH$ with $\min\Re\cE_H\leq\alpha_H$ there exist $u_{(z,\ell)}\in\Hb^{k,(\vec\cE(H),\vec\alpha(H))}(H)$, where $(\vec\cE(H),\vec\alpha(H)):=(\cE_{H'},\alpha_{H'})_{H'\in\sH(H)}$, such that, in a collar neighborhood $[0,\eps)_{\rho_H}\times H$ of $H$ and fixing $\chi=\chi(\rho_H)\in\CIc([0,\eps))$, $\chi|_{[0,\frac{\eps}{2}]}=1$,
  \begin{equation}
  \label{EqTMPhgH}
    u - \chi(\rho_H)\sum_{\substack{(z,\ell)\in\cE_H \\ \Re z\leq\alpha_H}} \rho_H^z(\log\rho_H)^\ell u_{(z,\ell)} \in \Hb^{k,(\vec\cE',\vec\alpha)}(M);
  \end{equation}
  here $\vec\cE'=(\cE'_H)_{H\in\sH}$ where $\cE'_H=\emptyset$ and $\cE'_{H'}=\cE_{H'}$ for $H'\neq H$. The definition of $\cA^{(\vec\cE,\vec\alpha)}(M)$ is completely analogous.
\end{definition}

If one uses a different defining function $\rho'_H$ of $H$, then $\rho_H=a\rho'_H$ where $0<a\in\CI([0,\eps)\times H)$. Since therefore $\rho_H^z=a^z(\rho'_H)^z$ and $\log\rho_H=\log\rho'_H+\log a$, with the smooth functions $a,\log a$ admitting Taylor expansions at $\rho'_H=0$, the expansion~\eqref{EqTMPhgH} implies an analogous expansion with $\rho'_H$; we use here the properties~\eqref{EqTMIndex} of $\cE_H$.

\begin{notation}[Alternative notation]
\label{NotTMphg}
  When an ordering of the boundary hypersurfaces $H_1,\ldots,H_N$ of $M$ is fixed, we shall denote the space $\Hb^{k,(\vec\cE,\vec\alpha)}(M)$ by
  \[
    \Hb^{k,\bigl((\cE_1,\alpha_1),\ldots,(\cE_N,\alpha_N)\bigr)}(M) = \Hb^{k,(\cE_1,\alpha_1),\ldots,(\cE_N,\alpha_N)}(M).
  \]
  In the case $N=2$ (for clarity), we moreover write
  \begin{align*}
    \Hb^{k,\alpha_1,(\cE_2,\alpha_2)}(M)=\Hb^{k,(\alpha_1,\ (\cE_2,\alpha_2))}(M) &:= \Hb^{k,\bigl((\emptyset,\alpha_1),\ (\cE_2,\alpha_2)\bigr)}(M), \\
    \Hb^{k,\cE_1,(\cE_2,\alpha_2)}(M)=\Hb^{k,(\cE_1,\ (\cE_2,\alpha_2))}(M) &:= \bigcap_{\alpha_1\in\R} \Hb^{k,\bigl((\cE_1,\alpha_1),\ (\cE_2,\alpha_2)\bigr)}(M),
  \end{align*}
  similarly for obvious variants of these spaces, and similarly also for $\cA$-spaces.
\end{notation}

The space $\cA^{\vec\cE}(M)$ is thus the space of (fully) polyhomogeneous functions on $M$; see also \cite[\S{2A}]{MazzeoEdge} for a brief review. When $M$ is a manifold with boundary, then $\Hb^{\infty,(\cE,\alpha)}(M)=\cA^\cE(M)+\Hb^{\infty,\alpha}(M)$. We caution the reader that the analogous statement on a manifold with corners is not true; the reason is that the terms $u_{(z,\ell)}$ in the generalized Taylor expansion at one boundary hypersurface $H$ of $M$ (see~\eqref{EqTMPhgH}) are not themselves fully polyhomogeneous, but rather only partially polyhomogeneous with a b-Sobolev remainder.

In practice, partial polyhomogeneity is typically proved via one of two methods. The first method is testing with \emph{b-normal vector fields}. To explain this, note first that if we define the vector field $\rho_H\pa_{\rho_H}$ in a collar neighborhood $[0,\eps)_{\rho_H}\times H$ of a boundary hypersurface $H$ of $M$, and similarly define $\rho'_H\pa_{\rho'_H}$ in another collar neighborhood $[0,\eps')_{\rho'_H}\times H$ of $H$ adapted to another (local) defining function $\rho'_H$ of $H$, then $\rho_H\pa_{\rho_H}-\rho'_H\pa_{\rho'_H}\in\rho_H\Vb(U)=\rho'_H\Vb(U)$ on the intersection $U$ of the two collar neighborhoods, as follows from a local coordinate computation; that is, $\rho_H\pa_{\rho_H}$ is well-defined to leading order at $H$ as a b-vector field. We call a vector field $V\in\Vb(M)$ such that $V-\chi(\rho_H)\rho_H\pa_{\rho_H}\in\rho_H\Vb(M)$, with $\chi_H\in\CIc([0,\eps))$ equal to $1$ near $0$, a \emph{b-normal vector field at $H$}.

\begin{lemma}[Normal form]
\label{LemmaTMbNorm}
  Let $V\in\Vb(M)$ be a b-normal vector field at the boundary hypersurface $H\subset M$. Then there exist a defining function $\rho_H\in\CI(M)$ of $H$ and an adapted collar neighborhood $[0,\eps)_{\rho_H}\times H$ of $H$ in $M$ such that $V=\rho_H\pa_{\rho_H}$ in this collar neighborhood.
\end{lemma}
\begin{proof}
  Fix any collar neighborhood $[0,\eps')_x\times H$ of $H$; then $V=x\pa_x+a(x,y)x^2\pa_x+x V_\pa(x)$ where we write $y$ for points in $H$, and $a\in\CI([0,\eps')\times H)$ and $V_\pa\in\CI([0,\eps');\Vb(H))$. For small $s\geq 0$, we can then define the time-$s$-flow of $x^{-1}V=\pa_x+a(x,y)x\pa_x+V_\pa(x)$, i.e., $(s,y)\mapsto e^{s x^{-1}V}y\in[0,\eps')\times H$. (The fact that $V_\pa$ is tangent to the boundary hypersurfaces of $H$ implies that this flow is well-defined.) Since then $s=0$ and $\dd s(\pa_x)=1$ at $H$, this yields a new collar neighborhood $[0,\eps'')_s\times H$ in which $x^{-1}V=\pa_s$ and thus $V=\frac{x}{s}s\pa_s$. To eliminate the factor $\frac{x}{s}=1+s b(s,y)$, we make a further coordinate change in the $s$-variable only, namely
  \[
    \rho_H(s,y) = s(1+c(s,y))
  \]
  where $c\in\CI$ is to be determined so that $V=\rho_H\pa_{\rho_H}$. But
  \[
    V = (1+s b)s\pa_s = (1+s b) s(1+c+s c')\pa_{\rho_H} = \frac{(1+s b)(1+c+s c')}{1+c} \rho_H\pa_{\rho_H},
  \]
  and thus $c$ can be chosen as the solution of $c'=-\frac{b(1+c)}{1+s b}$ with $c(0,y)=0$.
\end{proof}

We then have:

\begin{lemma}[Testing for partial polyhomogeneity]
\label{LemmaTMPhgTest}
  We use the notation of Definition~\usref{DefTMphg}. Let $H\subset M$ be a boundary hypersurface. Assume that $\alpha_H\notin\Re\pi_1\cE_H$. Let $V\in\Vb(M)$ be a b-normal vector field at $H$. Suppose that $u$ is such that $(1-\chi(\rho_H))u\in\Hb^{k,(\vec\cE',\vec\alpha)}(M)$ and
  \begin{equation}
  \label{EqTMPhgTest}
    \Biggl(\;\prod_{\substack{(z,\ell)\in\cE_H \\ \Re z\leq\alpha_H}} (V-z)\Biggr)u \in \Hb^{k,(\vec\cE',\vec\alpha)}(M).
  \end{equation}
  Then $u\in\Hb^{k,(\vec\cE,\vec\alpha)}(M)$. The converse is true if we replace $k$ in~\eqref{EqTMPhgTest} by $k-|\cE_{H,\leq}|$ where $\cE_{H,\leq}=\{(z,\ell)\in\cE_H\colon\Re z\leq\alpha_H\}$.
\end{lemma}
\begin{proof}
  Fix a defining function $\rho_H$ of $H$ and a collar neighborhood as in Lemma~\ref{LemmaTMbNorm}, so $V=\rho_H\pa_{\rho_H}$ near $H$. The ``converse'' part then follows from~\eqref{EqTMPhgH}, as the differential operator in~\eqref{EqTMPhgTest} annihilates the sum. (Terms arising from derivatives falling on $\chi$ vanish near $H$.) To prove the direct statement, it suffices to integrate one vector field $V-z$ at a time and use Lemma~\ref{LemmaTMIntFuchs} below.
\end{proof}

\begin{lemma}[Integration of a Fuchsian ODE]
\label{LemmaTMIntFuchs}
  We work on $[0,1)_x\times H$ where $H$ is a compact manifold with corners, and define b-Sobolev spaces using the vector field $x\pa_x$ and the density $|\frac{\dd x}{x}|$ in the first factor. Let $\chi\in\CIc([0,\frac34))$ be equal to $1$ on $[0,\frac12]$. Let $z\in\C$, $\alpha\neq\Re z$. Suppose $u$ solves
  \[
    (x\pa_x-z)u = f \in \Hb^{k,\ (\cE,\alpha),\ (\vec\cF,\vec\beta)}([0,1)\times H),\quad
    (1-\chi(x))u \in \Hb^{k,\ \infty,\ (\vec\cF,\vec\beta)},
  \]
  where $\cE$ and $\alpha$ are the index set and weight at $\{0\}\times H$, while $\vec\cF$ and $\vec\beta$ are the index sets and weights at $[0,1)$ times the boundary hypersurfaces of $H$. Then $u\in\Hb^{k,\ \bigl(\cE\extcup\{(z,0)\},\,\alpha\bigr),\ (\vec\cF,\vec\beta)}$.
\end{lemma}

\begin{figure}[!ht]
\centering
\includegraphics{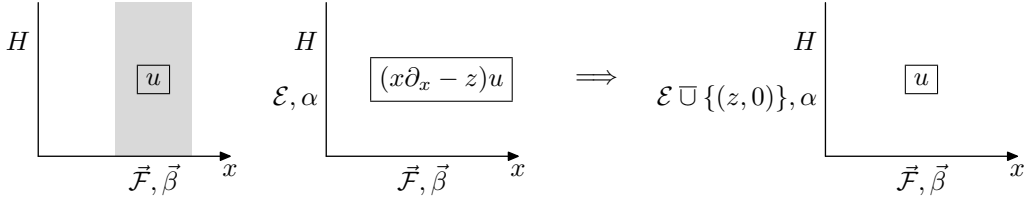}
\caption{Illustration of Lemma~\ref{LemmaTMIntFuchs}: the index sets $\vec\cF$ of $u$ at $\pa H$ get transported from $x\gtrsim 1$ to $x=0$; the index set at $x=0$ increases by $\{(z,0)\}$ relative to $\cE$.}
\label{FigTMIntFuchs}
\end{figure}

See Figure~\ref{FigTMIntFuchs}.

\begin{proof}[Proof of Lemma~\usref{LemmaTMIntFuchs}]
  Note that $x^{-z}u$ satisfies the equation $f=(x\pa_x-z)x^z(x^{-z}u)=x^z x\pa_x(x^{-z}u)$, so $x\pa_x(x^{-z}u)=x^{-z}f$; we may thus reduce to the case $z=0$. Next, denoting points in $H$ by $y$, we compute
  \begin{equation}
  \label{EqTMIntFuchsx}
    x\pa_x(x^w(\log x)^\ell v(y))=w x^w(\log x)^\ell v(y) + \ell x^w(\log x)^{\ell-1}v(y).
  \end{equation}
  Given $w\in\pi_1\cE$ with $w\neq 0$ and $\Re w=\min\Re\cE$, consider the term $\chi(x)x^w(\log x)^\ell v(y)$, $\ell=k(\cE,w)$, in the expansion of $f$ at $x=0$. Then
  \[
    x\pa_x\bigl(u-\chi(x)w^{-1}x^w(\log x)^\ell v(y)\bigr)\in\Hb^{k,\ \bigl(\cE\setminus\{(w,\ell)\},\,\alpha\bigr),\ (\vec\cF,\vec\beta)}.
  \]
  Using this to eliminate all $x^w(\log x)^\ell$-terms of $f$ with $(w,\ell)\in\cE$ and $\Re w\leq\alpha$, we may thus reduce to the case
  \[
    f=f_0+\tilde f,\quad f_0(x,y)=\chi(x)\sum_{j=0}^\ell (\log x)^j v_j(y),\ \tilde f\in\Hb^{k,\ \alpha,\ (\vec\cF,\vec\beta)}.
  \]
  In view of
  \[
    x\pa_x\bigl((\log x)^{j+1}v(y)\bigr) = (j+1)(\log x)^j v(y),
  \]
  we then have
  \[
    x\pa_x\biggl(u - \chi(x)\sum_{j=0}^\ell \frac{1}{j+1}(\log x)^{j+1}v_j(y)\biggr) \in \Hb^{k,\ \alpha,\ (\vec\cF,\vec\beta)}.
  \]

  It remains to study the case $\cE=\emptyset$ (and with $z=0$ still), i.e., $x\pa_x u=f\in\Hb^{k,\ \alpha,\ (\vec\cF,\vec\beta)}$. Using the a priori assumption $(1-\chi(x))u\in\Hb^{k,\ \infty,\ (\vec\cF,\vec\beta)}$, we can subtract $\psi(x)u$ from $u$, where $\psi\in\CIc((\frac34,1])$ equals $1$ near $1$, to reduce to the case that $u=0$ and thus also $f=0$ for $x$ near $1$. Moreover, it suffices to consider the case $k=0$. Since the $y$ variables serve merely as parameters, we drop them. For $x\leq 1$, we then have
  \[
    u(x) = -\int_x^1 f(s)\,\frac{\dd s}{s}.
  \]
  In the case $\alpha<0$, we estimate
  \begin{align*}
    \int_0^1 x^{-2\alpha}|u(x)|^2\,\frac{\dd x}{x} &\leq \int_0^1 x^{-2\alpha}\biggl( \int_x^1 s^{-\alpha}|f(s)|^2\,\frac{\dd s}{s}\biggr)\biggl(\int_x^1 s^\alpha\,\frac{\dd s}{s}\biggr)\,\frac{\dd x}{x} \\
      &\lesssim \int_0^1 x^{-\alpha} \biggl(\int_x^1 s^{-\alpha}|f(s)|^2\,\frac{\dd s}{s}\biggr)\,\frac{\dd x}{x} \\
      &= \int_0^1 s^{-\alpha}|f(s)|^2 \biggl(\int_0^s x^{-\alpha}\,\frac{\dd x}{x}\biggr)\,\frac{\dd s}{s} \\
      &\lesssim \int_0^1 s^{-2\alpha}|f(s)|^2\,\frac{\dd s}{s}.
  \end{align*}
  In the case $\alpha>0$, we note that the integral $u(0)=-\int_0^1 f(s)\,\frac{\dd s}{s}$ converges, and we then similarly estimate the difference $u(x)-u(0)=\int_0^x f(s)\,\frac{\dd s}{s}$ via
  \begin{equation}
  \label{EqTMIntFuchsEst}
  \begin{split}
    \int_0^1 x^{-2\alpha}|u(x)-u(0)|^2\,\frac{\dd x}{x} &\leq \int_0^1 x^{-2\alpha}\biggl(\int_0^x s^{-\alpha}|f(s)|^2\,\frac{\dd s}{s}\biggr)\biggl(\int_0^x s^\alpha\,\frac{\dd s}{s}\biggr)\,\frac{\dd x}{x} \\
      &\lesssim \int_0^1 s^{-2\alpha}|f(s)|^2\,\frac{\dd s}{s}.
  \end{split}
  \end{equation}
  This completes the proof.
\end{proof}

The second method of obtaining polyhomogeneous expansions is via \emph{normal operator arguments}. We describe two typical instances. For both, we use the following setup. Suppose $\pa M$ is a compact manifold without boundary, and put $M:=[0,\rho_0)_\rho\times\pa M$ where $\rho_0>0$. Use the b-density $|\frac{\dd\rho}{\rho}|\otimes\mu$, $0<\mu\in\CI(\pa M;\Omega\pa M)$, to define $\Hb^0(M)=L^2(M)$. For operators $A\in\Diffb^m(M)$ that are elliptic near $\rho=0$, define
\[
  \Specb(A) := \bigl\{ (\lambda,k) \colon N(A,\zeta)^{-1}\ \text{has a pole at $\zeta=\lambda$ of order $k+1$} \bigr\} \subset \C\times\N_0.
\]
(The projection onto the first factor is $\specb(A)$.)

\begin{lemma}[Boundary spectrum and polyhomogeneity]
\label{LemmaTMSolPhg}
  Let $\chi\in\CIc([0,\frac12\rho_0))$ be equal to $1$ on $[0,\frac14\rho_0]$, and let $\tilde\chi\in\CIc([0,\rho_0))$ be $1$ on $[0,\frac12\rho_0]$. Suppose $\tilde\chi u\in\Hb^{\infty,\alpha}(M)$ and $\chi A u\in\Hb^{\infty,(\cF,\beta)}(M)$, with $\beta\neq\Re\lambda$ for all $\lambda\in\specb(A)$. Let $\cE(\alpha)$ denote the smallest index set containing all $(\lambda,k)\in\Specb(A)$ with $\Re\lambda>\alpha$. Then $\tilde\chi u\in\Hb^{\infty,(\cE,\beta)}(M)$ where
  \[
    \cE_0 := \cE(\alpha)\extcup\cF,\quad
    \cE_{k+1} := \cE(\alpha) \extcup \bigl(\cF \cup (\cE_k+1)\bigr),\quad
    \cE := \bigcup_{k\in\N_0} \cE_k.
  \]
\end{lemma}
\begin{proof}
  This is a standard result (see, e.g., \cite[Proposition~5.61]{MelroseAPS}); we sketch the proof for completeness. One rewrites $A(\tilde\chi u)=:f\in\Hb^{\infty,(\cF,\beta)}(M)$ as
  \[
    N(A)(\tilde\chi u) = f - (A-N(A))(\tilde\chi u) =: f' \in \Hb^{\infty,(\cF,\beta)} + \Hb^{\infty,\alpha+1}.
  \]
  One then passes to the Mellin transform
  \[
    \cM(\tilde\chi u)(\lambda,\omega):=\int_0^\infty \rho^{-\lambda}(\tilde\chi u)(\rho,\omega)\,\frac{\dd\rho}{\rho},\quad \omega\in\pa M,
  \]
  initially for $\Re\lambda=\alpha$, and writes $\tilde\chi u=\cM_\alpha^{-1}(N(A,\lambda)^{-1}\cM(f'))$, where
  \[
    \cM_\alpha^{-1}(v)(\rho,\omega):=\frac{1}{2\pi i}\int_{\alpha-i\infty}^{\alpha+i\infty} \rho^\lambda v(\lambda,\omega)\,\dd\lambda.
  \]
  But $N(A,\lambda)^{-1}\cM(f')$ is meromorphic for $\Re\lambda<\min(\beta,\alpha+1)$; the order of its pole at a number $\lambda\in\C$ is the sum of the order of the pole of $N(A,\zeta)^{-1}$ there (which may be $0$) and of $\cM(f')$ (which is at most $k(\cF,\lambda)+1$). Upon shifting the contour to $\Re\lambda=\min(\beta,\alpha+1)$, the residue theorem gives the partial expansion of $\tilde\chi u$ up to an error (via Plancherel's theorem) of class $\Hb^{\infty,\min(\beta,\alpha+1)}$. Iterating this argument yields the claim.
\end{proof}

\begin{rmk}[Mellin transform and polyhomogeneity]
\label{RmkTMMellinPhg}
  The analytic content of Lemma~\ref{LemmaTMSolPhg} is that poles of the Mellin transform of a function give polyhomogeneous expansions of that function. (See \cite[\S{4.14}]{MelroseDiffOnMwc} for a detailed discussion.) Conversely, suppose $u$ is supported in $\{\rho<1\}$ and of class $\Hb^{\infty,(\cE,\beta)}$; then $(\cM u)(\lambda)$ is a holomorphic function on $\{\lambda\colon\Re\lambda<\min(\beta,\min\Re\cE)\}$ that extends meromorphically to $\Re\lambda<\beta$, with poles at most at points $\lambda\in\pi_1\cE$ with order $\leq k(\cE,\lambda)+1$. The key calculation here is that
  \[
    \int_0^1 \rho^{-\lambda} \rho^z(\log\rho)^k\,\frac{\dd\rho}{\rho} = \frac{\dd^k}{\dd w^k} \int_0^1 \rho^{w-\lambda}\,\frac{\dd\rho}{\rho}\biggr|_{w=z} = \frac{(-1)^k k!}{(z-\lambda)^{k+1}}.
  \]
\end{rmk}

The following related result on the existence of formal solutions will be frequently useful:

\begin{lemma}[Formal solutions]
\label{LemmaTMFormal}
  Let $A\in\Diffb^m(M)$ be elliptic. Let $\cF\subset\C\times\N_0$ be an index set, and let $\beta\in\R$. Set
  \[
    \cE := \Biggl\{ (\lambda+j,k+m) \colon (\lambda,k)\in\cF,\ j\in\N_0,\ m\leq\sum_{q=0}^j \ord(\lambda+q) \Biggr\},
  \]
  where $\ord(z)$ denotes the order of the pole of $N(A,\lambda)^{-1}$ at $\lambda=z$ (which equals $0$ when $N(A,\lambda)^{-1}$ is holomorphic at $z$). Then there exists a continuous linear map
  \[
    \cS_{A,\cF,\beta} \colon \cA^\cF(M) \to \cA^\cE(M)
  \]
  with the property that if $f\in\cA^\cF(M)$ and $u=\cS_{A,\cF,\beta}f$, then $f-A u\in\cA^\cG(M)$ where $\cG=\{(\lambda,k)\in\cE\colon\Re\lambda\geq\beta\}$. More generally, we can choose $\cS_{A,\cF,\beta}$ to also map $\Hb^{k,(\cF,\beta)}$ to $\Hb^{k+m,(\cE,\beta)}$ for all $k$, and $f-A u\in\Hb^{k,\beta}$ then. If $A$ depends smoothly on a finite-dimensional parameter $b\in\R^N$ but has $b$-independent normal operator $N(A)$, then one can choose $\cS_{A,\cF,\beta}$ to depend smoothly on $b$ as well. The results on polyhomogeneous spaces remain true also for $\beta=\infty$ if we allow $\cS_{A,\cF,\beta}$ to be non-linear.
\end{lemma}

This is a sharpening of \cite[Lemma~5.44]{MelroseAPS}; and it is largely a special case of \cite[Proposition~3.3]{HintzUnDet}. We give the proof mainly to argue for the smooth parameter dependence.

\begin{proof}[Proof of Lemma~\usref{LemmaTMFormal}]
  Let us consider a single term $\rho^\lambda(\log\rho)^\ell f_0$, $f_0\in H^k(\pa M)$, in the polyhomogeneous expansion of $f$, starting with the term with smallest $\Re\lambda$ (or any one of those when there are several) and $\ell=k(\cF,\lambda)$. Suppose first that $\lambda\notin\specb(A)=\specb(N(A))$. We can then set $u_0:=N(A,\lambda)^{-1}f_0\in H^{k+m}(\pa M)$ and deduce that $\rho^\lambda(\log\rho)^\ell f_0-N(A)(\rho^\lambda(\log\rho)^\ell u_0)$ equals $0$ when $k=0$, or is a sum of terms of class $\rho^\lambda(\log\rho)^j\CI(\pa M)$, $j\leq\ell-1$, when $\ell\geq 1$, which can be solved away iteratively. One thus obtains $u\in\Hb^{k+m,(\lambda,\ell)}(M)$ such that $f-A u\in\Hb^{k,(\cF\setminus\{(\lambda,\ell)\},\beta)}+\Hb^{k,(\lambda+1,\ell)}$. (The second summand, arising from $A-N(A)$ acting on $u$, depends smoothly on $b\in\R^N$ when $A$ does.) One subsequently solves away this remaining error iteratively; one may then also encounter the situation where $\lambda\in\specb(A)$, which we discuss next (using arguments that also apply when $\lambda\not\in\specb(A)$). We have $\rho^\lambda(\log\rho)^\ell=\frac{1}{2\pi i\ell!}\oint_\lambda \rho^\zeta(\zeta-\lambda)^{-\ell-1}\,\dd\zeta$ (counterclockwise contour integral around a small circle surrounding $\lambda$), and thus for
  \begin{equation}
  \label{EqTMFormalu}
    u(\rho,\omega) := \frac{1}{2\pi i\ell!}\oint_\lambda \rho^\zeta(\zeta-\lambda)^{-\ell-1} N(A,\zeta)^{-1}f_0(\omega)\,\dd\zeta
  \end{equation}
  the difference
  \[
    \rho^\lambda(\log\rho)^\ell f_0 - N(A)u = \frac{1}{2\pi i\ell!}\oint_\lambda \rho^\zeta(\zeta-\lambda)^{-\ell-1}f_0(\omega) - N(A) \bigl( \rho^\zeta(\zeta-\lambda)^{-\ell-1}N(A,\zeta)^{-1}f_0\bigr)\,\dd\zeta
  \]
  vanishes. Note that $u\in\Hb^{k+m,(\lambda,\ell+p)}$ if $p$ is the order of the pole of $N(A,\zeta)^{-1}$ at $\zeta=\lambda$. Thus $f-A u\in\Hb^{k,(\cF\setminus\{(\lambda,\ell)\},\beta)}+\Hb^{k,(\lambda+1,\ell+p)}$, and an iterative procedure eliminates this error modulo $\Hb^{k,\beta}$.

  The final statement follows by defining $u$ as an asymptotic sum (which can be done in a continuous, albeit nonlinear, fashion, cf.\ \cite[Proposition~3.17]{HintzMicro}).
\end{proof}

Note that when $\lambda\in\specb(A)$, one can add any multiple of an indicial solution $\rho^\lambda u_0\in\ker N(A)$, $u_0\in\CI(\pa M)$, to~\eqref{EqTMFormalu} without changing the subsequent arguments. From this perspective, the linearity (and regularity in the parameter $b\in\R^N$ when present) of $\cS_{A,\cF,\beta}$ is ensured in the above proof by the choice-free nature of the construction~\eqref{EqTMFormalu}.

\subsubsection{Scattering bundles and operators}

The following notions (originating in \cite{MelroseEuclideanSpectralTheory}) will be useful for the description of asymptotically flat metrics and associated differential operators.

\begin{definition}[Notions of scattering analysis]
\label{DefTMsc}
  Let $n\in\N$.
  \begin{enumerate}
  \item{\rm (Bundles.)} We define the \emph{scattering cotangent bundle} over $\ol{\R^n}$ by $\Tsc^*\ol{\R^n}:=\ol{\R^n}\times\R^n$, and identify a point $(z,\zeta)\in\Tsc^*_{\R^n}\R^n$ with the covector $\zeta_\mu\,\dd z^\mu\in T^*\R^n$ (summation convention). The dual bundle is denoted $\Tsc\ol{\R^n}=\ol{\R^n}\times\R^n$ and carries an identification with $T\R^n$ over $\R^n$, given by $(z,v)\mapsto v^\mu\pa_{z^\mu}$.
  \item{\rm (Vector fields.)} The space of sections of $\Tsc\ol{\R^n}$ is the space
    \[
      \Vsc(\ol{\R^n}) = \CI(\ol{\R^n};\Tsc\ol{\R^n})
    \]
    of \emph{scattering vector fields}; these are thus linear combinations of the coordinate vector fields $\pa_{z^\mu}$, $\mu=0,1,\ldots,n-1$, with coefficients of class $\CI(\ol{\R^n})$.
  \end{enumerate}
\end{definition}

\begin{example}[Minkowski metric]
\label{ExKMfdMink}
  The Minkowski metric
  \begin{equation}
  \label{EqKMfdMink}
    \ubar g := -\dd t^2 + \sum_{j=1}^3 (\dd x^j)^2
  \end{equation}
  defines an element of $\CI(\ol{\R^4};S^2\,\Tsc^*\ol{\R^4})$.
\end{example}

In $\{z^0>0\}$, introduce projective coordinates $\rho_\infty:=\frac{1}{z^0}$ and $\hat z^j:=\frac{z^j}{z^0}$, $j=1,\ldots,n-1$; then
\[
  \pa_{z^0} = -\rho_\infty\biggl(\rho_\infty\pa_{\rho_\infty}+\sum_{j=1}^{n-1}\hat z^j\pa_{\hat z^j}\biggr),\quad
  \pa_{z^j} = \rho_\infty\pa_{\hat z^j}.
\]
Since $\rho_\infty$ is a local boundary defining function, these expressions show that 
\begin{equation}
\label{EqKMfdVscVb}
  \Vsc(\ol{\R^n}) = \rho_n\Vb(\ol{\R^n})
\end{equation}
when $\rho_n\in\CI(\ol{\R^n})$ is a boundary defining function (e.g., $(1+t^2+|x|^2)^{-\frac12}$).

\subsubsection{Notions of 3b-analysis}
\label{SssT3b}

We only recall the basic notions from \cite{Hintz3b} here and refer the reader for more technical aspects (especially those pertaining to variable order operators and spaces) to the companion paper \cite{HintzNonstat2} (which builds on \cite{HintzScaledBddGeo} and is strongly influenced by \cite{Hintz3b}).

\begin{figure}[!ht]
\centering
\includegraphics{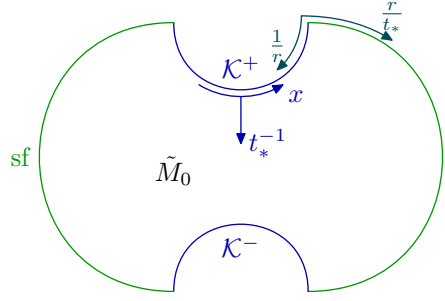}
\caption{The manifold $\tilde M_0$ and its boundary hypersurfaces, and some (partial) local coordinate systems (using polar coordinates $r=|x|$ and dropping spherical variables $\omega=\frac{x}{|x|}$).}
\label{FigT3bMfd}
\end{figure}

Consider the radial compactification $\ol{\R^4}$, and write $\R^4=\R_{t_*}\times\R^3_x$; define the points $\fk^\pm\in\pa\ol{\R^4}$ to be the limits of $(t_*,0)$ as $t_*\to\pm\infty$. Set
\begin{equation}
\label{EqT3btildeM0}
  \tilde M_0 := [\ol{\R^4};\fk^+,\fk^-],\quad\upbeta\colon\tilde M_0\to\ol{\R^4},
\end{equation}
with boundary hypersurfaces denoted
\[
  \sface\ \text{(lift of $\pa\ol{\R^4}$)},\quad
  \cK^\pm\ \text{(lift of $\fk^\pm$)}.
\]
See Figure~\ref{FigT3bMfd}. The space $\Vtsc(\tilde M_0):=\CI(\tilde M_0;\upbeta^*\,\Tsc\ol{\R^4})$ is the $\CI(\tilde M_0)$-span of the space of scattering vector fields on $\ol{\R^4}$ (thus, of coordinate vector fields) on $\R^4$; these are called \emph{3-body-scattering vector fields} by Vasy \cite{VasyThreeBody}.

Thus, a possible choice of defining function of $\sface$ is $\rho_\sface=\la x\ra^{-1}$, while a joint defining function of $\cK^+\cup\cK^-$ is $\rho_\cK=\frac{\la x\ra}{(1+t_*^2+|x|^2)^{\frac12}}$.

\begin{definition}[3b-notions]
\label{DefT3b}
  Recall~\eqref{EqT3btildeM0}.
  \begin{enumerate}
  \item{\rm (Bundles.)} The \emph{3b-cotangent bundle} is defined by $\Ttb^*\tilde M_0:=\tilde M_0\times\R^4$, with a point $(t_*,x;\tau,\xi)$ over $\tilde M_0^\circ=\R^4$ identified with the covector $\tau\,\frac{\dd t_*}{\la x\ra}+\sum_{j=1}^3 \xi_j\frac{\dd x^j}{\la x\ra}$. The dual bundle is denoted $\Ttb\tilde M_0$ and carries an identification with $T\R^4$ over $\R^4$, given by $(t_*,x;v,w)\mapsto v\,\la x\ra\pa_{t_*}+\sum_{j=1}^3 w^j\,\la x\ra\pa_{x^j}$.
  \item{\rm (Vector fields.)} The space of \emph{3b-vector fields} is defined as $\Vtb(\tilde M_0):=\CI(\tilde M_0;\Ttb\tilde M_0)$; it is the $\CI(\tilde M_0)$-span of the vector fields $\la x\ra\pa_{t_*}$ and $\la x\ra\pa_{x^j}$, $j=1,2,3$, or equivalently the set of vector fields of the form $\rho_\sface^{-1}V_0$ where $V_0\in\Vtsc(\tilde M_0)$.
  \item{\rm (Differential operators.)} For $m\in\N_0$, we write $\Difftb^m(\tilde M_0)$ for the space of $m$-th order 3b-differential operators (finite sums of up to $m$-fold compositions of elements of $\Vtb(\tilde M_0)$). For weights $\alpha_\sface,\alpha_\cK\in\R$, we moreover define
    \[
      \Difftb^{m,(\alpha_\sface,\alpha_\cK)}(\tilde M_0) := \rho_\sface^{-\alpha_\sface}\rho_\cK^{-\alpha_\cK}\Difftb^m(\tilde M_0).
    \]
  \end{enumerate}
\end{definition}

One can then define weighted 3b-Sobolev spaces of order $s\in\N_0$ in the usual fashion to consist of functions which remain in $L^2$ under application of up to $s$ many 3b-vector fields. Weighted versions are denoted
\[
  \Htb^{s,(\alpha_\sface,\alpha_\cK)}(\tilde M_0) = \rho_\sface^{\alpha_\sface}\rho_\cK^{\alpha_\cK}\Htb^s(\tilde M_0).
\]
These spaces can also be defined for variable order functions $s$ which are smooth (or merely 3b-regular) functions on $\Stb^*\tilde M_0$, the boundary at fiber infinity of the radially compactified 3b-cotangent bundle $\ol{\Ttb^*}\tilde M_0$; see \citeAF{\S\ref*{SssMUK} and (\ref*{EqMSMixed})}.

Consider now a $t_*$-translation-invariant operator $A\in\Difftb^{m,(\alpha_\sface,0)}(\tilde M_0)$; this means that $A$ is of the form
\begin{equation}
\label{EqT3bOp}
  A = \sum_{j+|\alpha|\leq m} \la x\ra^{\alpha_\sface}a_{j\alpha}(x) (\la x\ra\pa_{t_*})^j (\la x\ra\pa_x)^\alpha,\quad a_{j\alpha} \in\CI(\ol{\R^3}).
\end{equation}
(An example for $m=2$ and $\alpha_\sface=-2$ is the wave operator on Kerr; see also \citeAF{Example~\ref*{ExSSAdmBox}} and the proof of Lemma~\ref{LemmaWGOp} below.) Its spectral family is then defined as the formal conjugation by the Fourier transform in $t_*$, with the convention $(\cF u)(\sigma,x)=\int e^{i\sigma t_*}u(t_*,x)\,\dd t_*$, thus by formally replacing $\pa_{t_*}$ by $-i\sigma$; this gives
\begin{equation}
\label{EqT3bSpec}
  \hat A(\sigma) = \sum_{j+|\alpha|\leq m} \la x\ra^{\alpha_\sface}a_{j\alpha}(x) (-i\sigma\la x\ra)^j (\la x\ra\pa_x)^\alpha.
\end{equation}
When $|\sigma|\leq 1$, the appearance of $\sigma\la x\ra=\frac{\sigma}{\rho}$ where $\rho=\la x\ra^{-1}\in\CI(\ol{\R^3})$ is a defining function of $\pa\ol{\R^3}$ suggests the introduction of:

\begin{definition}[Scattering-b-transition notions]
\label{DefTscbt}
  Set $\tilde X:=\ol{\R^3}$ and define
  \[
    \tilde X_\scbtop := [[0,1]_\varsigma\times\tilde X; \{0\}\times\pa\tilde X].
  \]
  Its boundary hypersurfaces are as follows:
  \begin{enumerate}
  \item the \emph{scattering face} $\scface$ is the lift of $[0,1]\times\pa\tilde X$;
  \item the \emph{transition face} $\tface$ is the front face (i.e., the lift of $\{0\}\times\pa\tilde X$);
  \item the \emph{zero face} $\zface$ is the lift of $\{0\}\times\tilde X$.
  \end{enumerate}
  Write $\rho_\scface$, $\rho_\tface$, and $\rho_\zface$ for defining functions of $\scface$, $\tface$, and $\zface$, respectively.
  \begin{enumerate}
  \item{\rm (Vector fields.)} The space of \emph{scattering-b-transition vector fields} $\cV_\scbtop(\tilde X)$ consists of all $V\in\rho_\scface\Vb(\tilde X_\scbtop)$ with $V\sigma=0$. They are thus the $\CI(\tilde X_\scbtop)$-span of $\rho_\scface\la x\ra\pa_{x^j}$, $j=1,2,3$.
  \item{\rm (Differential operators.)} The corresponding spaces of scattering-b-transition differential operators are denoted by
    \[
      \Diffscbt^{m,(\alpha_\scface,\alpha_\tface,\alpha_\zface)}(\tilde X) := \rho_\scface^{-\alpha_\scface}\rho_\tface^{-\alpha_\tface}\rho_\zface^{-\alpha_\zface}\Diffscbt^m(\tilde X).
    \]
  \item\label{ItTscbtSob}{\rm (Sobolev spaces.)} The space $H_{\scbtop,\varsigma}^{m,(\alpha_\scface,\alpha_\tface,\alpha_\sface)}(\tilde X)$ is defined to be $H_\scop^{m,\alpha_\scface}(\tilde X)$ as a set when $\varsigma>0$ (and $\Hb^{m,\alpha_\tface}(\tilde X)$ when $\varsigma=0$ in the case $\alpha_\zface=0$), but equipped with the $\varsigma$-dependent norm
    \[
      \| u \|_{H_{\scbtop,\varsigma}^{m,(\alpha_\scface,\alpha_\tface,\alpha_\zface)}(\tilde X)} := \sum_{|\alpha|\leq m} \| \rho_\scface^{-\alpha_\scface}\rho_\tface^{-\alpha_\tface}\rho_\zface^{-\alpha_\zface} ( \rho_\scface\la x\ra\pa_x )^\alpha u \|_{L^2(\tilde X,\mu)}
    \]
    with respect to a fixed (weighted) b-density $\mu$ on $\tilde X$. For $k\in\N_0$, we define
    \[
      \|u\|_{H_{(\scbtop,\varsigma);\bop}^{(m;k),(\alpha_\scface,\alpha_\tface,\alpha_\zface)}(\tilde X)} := \sum_{|\beta|\leq k} \|(\la x\ra\pa_x)^\beta u\|_{H_{\scbtop,\varsigma}^{m,(\alpha_\scface,\alpha_\tface,\alpha_\zface)}(\tilde X)}.
    \]
  \end{enumerate}
\end{definition}

These notions were introduced in \cite{HintzKdSMS} based on earlier work by Guillarmou--Hassell \cite{GuillarmouHassellResI}. We also recall that the sc-b-cotangent bundle $\Tscbt^*\tilde X\to\tilde X_\scbtop$ is the dual bundle to the bundle $\Tscbt\tilde X\to\tilde X_\scbtop$ whose smooth sections are precisely the linear combinations of $\rho_\scface\la x\ra\pa_x$ with $\CI(X_\scbtop)$ coefficients. (A frame of $\Tscbt^*X$ is thus given by $\frac{\dd x^j}{\rho_\scface\la x\ra}$, $j=1,2,3$.) One can then define $H_{\scbtop,\varsigma}^{s,(\alpha_\scface,\alpha_\tface,\alpha_\zface)}$-norms also for variable orders $s,r$ (to wit, for $s\in\CI({}^\scbtop S^*X)$ and $r\in\CI(\ol{\Tscbt^*_\scface}X)$).

Setting $\rho:=\la x\ra^{-1}$, possible choices of defining functions are $\rho_\scface=\frac{\rho}{\varsigma+\rho}$, $\rho_\tface=\varsigma+\rho$, and $\rho_\zface=\frac{\varsigma}{\varsigma+\rho}$. In view of
\[
  \varsigma\la x\ra = \frac{\varsigma}{\rho} = \frac{\rho_\zface}{\rho_\scface},\quad
  \la x\ra\pa_{x^j} = \rho_\scface^{-1}\,\rho_\scface\la x\ra\pa_{x^j}\in\rho_\scface^{-1}\cV_\scbtop(\tilde X),
\]
the spectral family~\eqref{EqT3bSpec}, for $\sigma=\hat\sigma|\sigma|$ where $\hat\sigma\in\C$, $|\hat\sigma|=1$, is fixed, satisfies (with $\varsigma:=|\sigma|$)
\[
  \bigl( \hat A(\hat\sigma|\sigma|) \bigr)_{|\sigma|\in[0,1]} \in (\rho_\scface\rho_\tface)^{-\alpha_\sface}\,\rho_\scface^{-m}\Diff_\scbtop^m(\tilde X) = \Diff_\scbtop^{m,(\alpha_\sface+m,\alpha_\sface,0)}(\tilde X).
\]
A consequence of Plancherel's theorem is then:

\begin{lemma}[Fourier transform and 3b-Sobolev spaces]
\label{LemmaT3bFT}
  For $u\in\Htb^{m,(\alpha_\sface,0)}(\tilde M_0,|\dd t_*\,\dd x|)$, we have
  \begin{equation}
  \label{EqT3bFT}
    \int_{-1}^1 \| \hat u(\sigma) \|_{H_{\scbtop,|\sigma|}^{m,(m+\alpha_\sface,\alpha_\sface,0)}(\tilde X)}^2\,\dd\sigma \lesssim \|u\|_{\Htb^{m,(\alpha_\sface,0)}(\tilde M_0,|\dd t_*,\dd x|)}^2.
  \end{equation}
  If $\supp\hat u\subset[-1,1]$, then also the reverse inequality holds. In particular, given a function $\hat u=\hat u(\sigma,x)$ for which the left-hand side is finite, the inverse Fourier transform $u(t_*,x)=\frac{1}{2\pi}\int e^{-i\sigma t_*}\hat u(\sigma,x)\,\dd\sigma$ lies in $\Htb^{m,(\alpha_\sface,0)}$.
\end{lemma} 

A complete spectral characterization of the $\Htb^{m,(\alpha_\sface,0)}$-norm is given in \citeAF{Lemma~\ref*{LemmaMUetbFT}}. (Essentially all the delicate analysis in the present paper takes place at low energies, and hence we will give references to the high-energy notions only at the few places where we directly use them.)

Since every 3b-differential operator $A$ is, a fortiori, also a b-differential operator on $\tilde M_0$, we can define its dilation-invariant b-normal operator at the boundary hypersurface $\sface$ of $\tilde M_0$. (The normal operator at $\cK$ is simply $\hat A(0)$ when $A\in\Difftb^{m,(\alpha_\sface,0)}(\tilde M_0)$.) Let us only discuss this for a stationary operator $A\in\Difftb^{m,(\alpha_\sface,0)}$, expressed as in~\eqref{EqT3bOp}: in this case, we have
\begin{equation}
\label{EqT3bNsf}
  N_\sface(A) = r^{\alpha_\sface} \sum_{j+|\alpha|\leq m} a_{j\alpha}^\infty (r\pa_{t_*})^j (r\pa_x)^\alpha,
\end{equation}
where, upon writing $a_{j\alpha}(x)$ as a function of $(\rho,\omega)$, we define $a_{j\alpha}^\infty=a_{j\alpha}^\infty(\omega)$ as its restriction to $\{\rho=0\}=\pa\ol{\R^3}$. This operator, defined for $r>0$, is homogeneous (of degree $\alpha_\sface$) with respect to spacetime dilations $(t_*,r)\mapsto(\lambda t_*,\lambda r)$, $\lambda>0$, and the difference $A-N_\sface(A)$ is of class $\Difftb^{m,(\alpha_\sface-1,0)}$ (on the set where $N_\sface(A)$ is defined). Passing to inverse spatial polar coordinates $x=\rho^{-1}\omega$, $\rho=|x|^{-1}$, $\omega=\frac{x}{|x|}\in\Sph^2$, write this schematically as
\[
  N_\sface(r^{-\alpha_\sface}A) = \sum_{j+k+|\gamma|\leq m} a_{j k\gamma}(\omega)(\rho^{-1}\pa_{t_*})^j (\rho\pa_\rho)^k \pa_\omega^\gamma,
\]
and subsequently pass to the coordinates $\rho$ and $v=\rho t_*\in\R$ (so $\sface=\ol{\R_v}\times\Sph^2$), then
\[
  N_\sface(r^{-\alpha_\sface}A) = \sum_{j+k+|\gamma|\leq m} a_{j k\gamma}(\omega) \pa_v^j (\rho\pa_\rho+v\pa_v)^k \pa_\omega^\gamma.
\]
Exploiting the invariance under spacetime dilations (which are now dilations in $\rho$, with $(v,\omega)$ held fixed), one can pass to the spectral family via formally Mellin transforming in $\rho$, yielding
\begin{equation}
\label{EqT3bNsfMellin}
  N_\sface(r^{-\alpha_\sface}A, \rho^\lambda) := \sum_{j+k+|\gamma|\leq m} a_{j k\gamma}(\omega)\pa_v^j(\lambda+v\pa_v)^k \pa_\omega^\gamma \in \Diffb^m(\sface).
\end{equation}
The methods of b-analysis are thus available for the study of this operator; a concrete instance is described in~\S\ref{SsipInv}. (It is more common, see \cite[\S{3.3}]{Hintz3b}, to formally Mellin transform with respect to a total boundary defining function of $\sface\cup\cK$, such as $t_*^{-1}$ in the region where $t_*>1$; different conventions lead to the same spectral families up to conjugation.)

\subsubsection{Extendible and supported spaces}
\label{SssTExtSupp}

When solving wave equations, we often encounter the following setup: $M$ is a compact manifold with corners, and $\Omega\subset M$ is a compact submanifold with corners; we require that each of its boundary hypersurfaces $H$ is either contained in a boundary hypersurface of $M$, or else $H^\circ\cap\pa M=\emptyset$ and $H$ is transversal to $\pa M$. We call the boundary hypersurfaces of the second type \emph{interior}. Suppose the set of interior boundary hypersurfaces of $\Omega$ is partitioned into two (possibly empty) sets $\sH_{\rm I}=(H_{{\rm I},1},\ldots,H_{{\rm I},N_{\rm I}})$ and $\sH_{\rm F}=(H_{{\rm F},1},\ldots,H_{{\rm F},N_{\rm F}})$, and that each $H\in\sH_{\rm I}\cup\sH_{\rm F}$ is the zero set of a function $\rho_H\in\CI(U_H)$ defined in an open neighborhood $U_H\subset M$ of $H$ with $\dd\rho_H\neq 0$ everywhere on $H$. For $s\in\N_0$, we then write
\[
  \Hb^s(\Omega)^{\bullet,-}
\]
for a space of b-Sobolev functions with \emph{supported, resp.\ extendible character at each $H_{{\rm I},j}$, resp.\ $H_{{\rm F},l}$}. (The terminology is taken from \cite[Appendix~B]{HormanderAnalysisPDE3}.) We proceed to define this space. When $\sH_{\rm F}=\emptyset$, we write it as
\[
  \Hb^s(\Omega)^\bullet = \dot H_\bop^s(\Omega) := \{ u\in\Hb^s(M) \colon \supp u\subset\Omega \},
\]
with the induced norm; when $\sH_{\rm I}=\emptyset$, we set
\[
  \Hb^s(\Omega)^- = \bar H_\bop^s(\Omega) := \{ u|_\Omega \colon u\in\Hb^s(M) \}
\]
with the quotient norm. To describe the general case, we first arrange that $\rho_H>0$ on $U_H\cap\Omega\setminus H$ (otherwise, replace $\rho_H$ by $-\rho_H$). By compactness, there exists $\delta>0$ such that the domain
\[
  \Omega_\delta := \Omega \cup \bigcup_H U_H \cap \{ \rho_{H'}\geq-\delta\ \forall\,H'\}
\]
(which is a $\delta$-enlargement of $\Omega$) is still a compact manifold with corners. We then define
\[
  \Hb^s(\Omega)^{\bullet,-} := \{ u|_\Omega \colon u\in\dot H_\bop^s(\Omega_\delta),\ u|_{\rho_H<0}=0\ \forall\,H\in\sH_{\rm I} \},
\]
again equipped with the quotient norm. Generalizations of these spaces featuring weights and partial polyhomogeneity are defined in the same fashion.

\begin{notation}[Purely extendible spaces]
\label{NotTMHbbar}
  For better readability, we shall typically omit the bar in the notation for spaces with extendible character at every boundary hypersurface, so $H_\bop^s:=\bar H_\bop^s$.
\end{notation}

\subsection{Integration along certain b-vector fields}
\label{SsTInt}

We record results concerning the integration of the vector fields $x\pa_x$ and $x\pa_x-y\pa_y$ on manifolds with boundary or corners, with $x,y\geq 0$ being boundary defining functions. They are essentially taken from \cite[\S{7}]{HintzVasyMink4}; due to their important role in the present paper (and also their usage in \citeAF{\S\ref*{SsDscri}}), we give detailed proofs. For $d\in\N_0$, $k\in\N_0$, index sets $\cE,\cF\subset\C\times\N_0$, and $\alpha,\beta\in\R$, we write
\[
  \Hb^{k,\ (\cE,\alpha),\ (\cF,\beta)}([0,2)_x\times[0,2)_y\times(-2,2)_q^d)
\]
for the b-Sobolev space (with partial polyhomogeneous expansions) with respect to the density $|\frac{\dd x}{x}\,\frac{\dd y}{y}\,\dd q|$. Fix a cutoff
\begin{equation}
\label{EqTIntCutoff}
  \chi \in \CIc([0,\tfrac34)),\quad \chi|_{[0,\frac12]}=1.
\end{equation}

\begin{lemma}[Solution of a Fuchsian ODE]
\label{LemmaTMIntFuchs2}
  Let $z\in\C$ be such that $\Re z<\min(\alpha,\min\Re\cE)$. Given $f\in\Hb^{k,\ (\cE,\alpha),\ (\cF,\beta)}([0,2)\times[0,2)\times(-2,2)^d)$, there exists a unique solution $u\in\Hb^{k,\ (\cE,\alpha),\ (\cF,\beta)}$ of the equation $(x\pa_x-z)u=f$.
\end{lemma}
\begin{proof}
  We drop the $(y,q)$-variables and reduce to the case $z=0$ (so $\alpha>0$) as in the proof of the Lemma~\ref{LemmaTMIntFuchs}. The terms $x^\ell(\log x)^w v$, $v\in\C$, with $(\ell,w)\in\cE$, $\Re\ell\leq\alpha$, in the expansion of $f$ at $x=0$ can be solved away using~\eqref{EqTMIntFuchsx}; we can thus reduce to the case $\cE=\emptyset$. Defining $u(x)=\int_0^x f(s)\,\frac{\dd s}{s}$ where $f\in\Hb^{k,\alpha}$, we obtain $u\in\Hb^{k,\alpha}$ using the estimate~\eqref{EqTMIntFuchsEst} with $u(x)$ in place of $u(x)-u(0)$.
\end{proof}

\begin{lemma}[Integration of a hyperbolic transport operator]
\label{LemmaTIntHyp}
  Let $z\in\C$. Let $\alpha>\beta$, and suppose that $\Re(w+z)\neq\beta$ for all $(w,0)\in\cE$. Suppose $u$ solves
  \[
    (x\pa_x-y\pa_y+z)u = f \in \Hb^{k,\ (\cE,\alpha),\ (\cF,\beta)}([0,2)\times[0,2)\times(-2,2)^d),\quad
    (1-\chi(y))u \in \Hb^{k,\ (\cE,\alpha),\ \infty},
  \]
  where $\chi$ is as in~\eqref{EqTIntCutoff}. Then $u\in\Hb^{k,\ (\cE,\alpha),\ \bigl((\cE+z)\extcup\cF,\,\beta\bigr)}$.
\end{lemma}

\begin{figure}[!ht]
\centering
\includegraphics{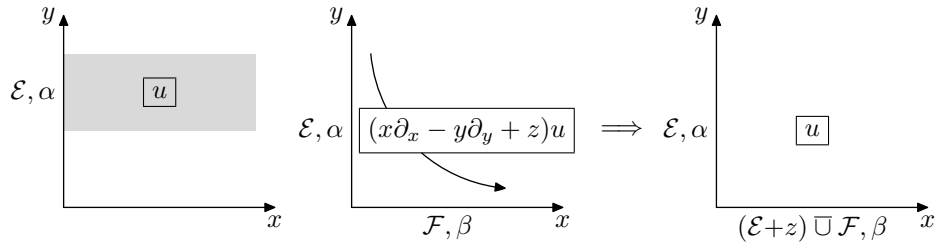}
\caption{Illustration of Lemma~\ref{LemmaTIntHyp}: the index set $\cE$ of $u$ at $x=0$ gets transported from $y\gtrsim 1$ to $y=0$ and interacts with the index set $\cF$ of $(x\pa_x-y\pa_y+z)u$ there.}
\label{FigTIntHyp}
\end{figure}

See Figure~\ref{FigTIntHyp}.

\begin{proof}[Proof of Lemma~\usref{LemmaTIntHyp}]
  Note that $x^z u$ satisfies $(x\pa_x-y\pa_y)(x^z u)=x^z f$, so upon replacing $u$, $f$, $\cE$, and $\alpha$ by $x^z u$, $x^z f$, $\cE+z$, and $\alpha+\Re z$, respectively, we can reduce to the case $z=0$.

  \pfstep{Step~1. Solving away at $x=0$.} Let $w\in\pi_1\cE$ with $\Re w=\min\Re\cE$, and let $\ell=k(\cE,w)$. Consider the term $x^w(\log x)^\ell v(y,q)$ in the expansion of $f$ at $x=0$, with
  \[
    v\in\Hb^{k,(\cF,\beta)}([0,2)_y\times(-2,2)_q^d).
  \]
  Since
  \begin{equation}
  \label{EqTIntHyp1}
    (x\pa_x-y\pa_y)\bigl(x^w(\log x)^\ell a(y,q)\bigr) = x^w(\log x)^\ell(w-y\pa_y)a(y,q) + \ell x^w(\log x)^{\ell-1}a(y,q),
  \end{equation}
  the $x^w(\log x)^\ell$-term in the expansion of $u$ at $x=0$ and for all $y$ with $1-\chi(y)=1$ is given by $x^w(\log x)^\ell a(y,q)$ where $a$ solves the ODE $(w-y\pa_y)a=v$. We can extend $a$ from $\{1-\chi(y)=1\}$ to all of $(0,2)_y\times(-2,2)_q^d$ by solving this ODE; by Lemma~\ref{LemmaTMIntFuchs}, the solution satisfies
  \[
    a\in\Hb^{k,(\cF\extcup\{(w,0)\},\beta)}([0,2)\times(-2,2)^d).
  \]
  For the function
  \[
    \tilde u(x,y,q) := u(x,y,q) - x^w(\log x)^\ell a(y,q),
  \]
  we then have
  \[
    (x\pa_x-y\pa_y)\tilde u =: \tilde f \in \Hb^{k,\ \bigl(\cE\setminus\{(w,\ell)\},\,\alpha\bigr),\ \bigl(\cF\extcup\{(w,0)\},\,\beta\bigr)},\quad
    (1-\chi(y))\tilde u \in \Hb^{k,\ \bigl(\cE\setminus\{(w,\ell)\},\,\alpha\bigr),\ \infty}.
  \]

  We now repeat the same argument for this new equation. Adding all terms $x^w(\log x)^\ell a(y,q)$ obtained in this fashion for all $(w,\ell)\in\cE$ with $\Re w\leq\alpha$, we produce
  \[
    u_0 \in \Hb^{k,\ \cE,\ (\cF\extcup\cE,\beta)}
  \]
  such that for $\tilde u:=u-u_0$, we have
  \begin{equation}
  \label{EqTIntHyp2}
    (x\pa_x-y\pa_y)\tilde u =: \tilde f \in \Hb^{k,\ \alpha,\ (\cE\extcup\cF,\beta)},\quad
    (1-\chi(y))\tilde u\in\Hb^{k,\ \alpha,\ \infty}.
  \end{equation}
  It remains to show that~\eqref{EqTIntHyp2} implies $\tilde u\in\Hb^{k,\ \alpha,\ (\cE\extcup\cF,\beta)}$.

  \pfstep{Step~2. Solving away at $y=0$.} Consider $(w,\ell)\in\cE\extcup\cF$, $\ell=k(\cE\extcup\cF,w)$, with $\Re w\leq\beta<\alpha$. Consider the term $y^w(\log y)^\ell v(x,q)$ in the expansion of $\tilde f$ at $y=0$, with $v\in\Hb^{k,\alpha}([0,2)_x\times(-2,2)_q^d)$. Lemma~\ref{LemmaTMIntFuchs2} allows us to solve $(x\pa_x-w)a(x,q)=v(x,q)$ with $a\in\Hb^{k,\alpha}$. By subtracting $y^w(\log y)^\ell a(x,q)$ from $\tilde u$, and proceeding in this fashion for all such $w$, we produce (upon localizing near $y=0$ using a smooth cutoff)
  \[
    \tilde u_0 \in \Hb^{k,\ \alpha,\ (\cE\extcup\cF,\beta)},\quad
    \tilde f-(x\pa_x-y\pa_y)\tilde u_0 \in \Hb^{k,i\ \alpha,\ \beta},\quad
    (1-\chi(y))\tilde u_0=0.
  \]
  The difference $u_\flat:=\tilde u-\tilde u_0$ therefore satisfies
  \begin{equation}
  \label{EqTIntHypFlat}
    (x\pa_x-y\pa_y)u_\flat =: f_\flat \in \Hb^{k,\ \alpha,\ \beta},\quad
    (1-\chi(y))u_\flat \in \Hb^{k,\ \alpha,\ \infty}.
  \end{equation}

  \pfstep{Step~3. Estimating the remainder.} It remains to prove that $u_\flat\in\Hb^{k,\ \alpha,\ \beta}$; it suffices to do this for $k=0$. Let $\psi\in\CIc([0,1))$ be equal to $1$ on $[0,\frac34]$; then
  \[
    (x\pa_x-y\pa_y) \bigl( \psi(y)u_\flat \bigr) = \psi(y)f_\flat - [y\pa_y,\psi(y)]u_\flat = \psi(y)f_\flat - y\psi'(y)(1-\chi(y))u_\flat
  \]
  still lies in $\Hb^{k,\ \alpha,\ \beta}$, but now $\psi(y)u_\flat$ vanishes near $y=1$. It thus suffices to prove $u_\flat\in\Hb^{k,\ \alpha,\ \beta}$ when $u_\flat$ satisfies~\eqref{EqTIntHypFlat} and in addition $u_\flat=0$ near $y=1$. Since multiplication by powers of $x y$ commutes with $x\pa_x-y\pa_y$, we can moreover reduce to the case $\alpha=0>\beta$. We then estimate the solution
  \[
    u_\flat(x,y) = \int_y^1 f_\flat\Bigl(\frac{x y}{s},s\Bigr)\,\frac{\dd s}{s}
  \]
  of~\eqref{EqTIntHypFlat} via
  \begin{align*}
    &\int_0^1 \int_0^1 y^{-2\beta}|u_\flat(x,y)|^2\,\frac{\dd x}{x}\,\frac{\dd y}{y} \\
    &\qquad \leq \int_0^1 \int_0^1 y^{-2\beta} \biggl(\int_y^1 s^{-\beta}\Bigl|f_\flat\Bigl(\frac{x y}{s},s\Bigr)\Bigr|^2\,\frac{\dd s}{s}\biggr)\biggl(\int_y^1 s^\beta\,\frac{\dd s}{s}\biggr) \,\frac{\dd x}{x}\,\frac{\dd y}{y} \\
    &\qquad \lesssim \int_0^1 \int_0^1 s^{-\beta} \int_0^s y^{-\beta}\Bigl|f_\flat\Bigl(\frac{x y}{s},s\Bigr)\Bigr|^2\,\frac{\dd y}{y}\,\frac{\dd s}{s}\,\frac{\dd x}{x} \\
    &\qquad = \int_0^1 \int_0^1 s^{-\beta} \int_0^x \Bigl(\frac{s t}{x}\Bigr)^{-\beta}|f_\flat(t,s)|^2\,\frac{\dd t}{t}\,\frac{\dd s}{s}\,\frac{\dd x}{x} \\
    &\qquad = \int_0^1 s^{-2\beta} \int_0^1 t^{-\beta} \biggl(\int_t^1 x^\beta\,\frac{\dd x}{x}\biggr) |f_\flat(t,s)|^2\,\frac{\dd t}{t}\,\frac{\dd s}{s} \\
    &\qquad \lesssim \int_0^1\int_0^1 s^{-2\beta}|f_\flat(t,s)|^2\,\frac{\dd t}{t}\,\frac{\dd s}{s}.
  \end{align*}
  This completes the proof.
\end{proof}

\subsection{Fourier transforms}
\label{SsTF}

\begin{figure}[!ht]
\centering
\includegraphics{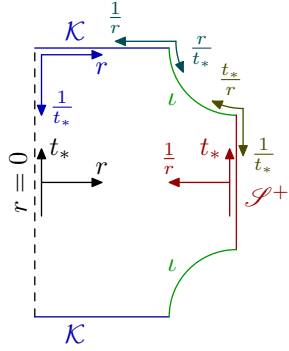}
\caption{Illustration of the subset $\{r=|x|>0\}$ of the manifold $\sfM$ defined in~\eqref{EqTFsfM} and some local coordinate charts; we drop the factor $\Sph^{n-1}_\omega$, $\omega=\frac{x}{|x|}$.}
\label{FigTFM}
\end{figure}

We use the convention
\[
  \hat u(\sigma) := (\cF u)(\sigma) := \int_\R e^{i\sigma t_*}u(t_*)\,\dd t_*,\quad
  (\cF^{-1}v)(t_*) := \frac{1}{2\pi}\int_\R e^{-i\sigma t_*}v(\sigma)\,\dd\sigma
\]
for the Fourier transform and its inverse. We shall prove generalizations of the results in \citeAF{\S\ref*{SsDFT}} on how the (inverse) Fourier transform acts on b-Sobolev spaces on the resolved space
\begin{equation}
\label{EqTFsfM}
  \sfM := [\,\ol{\R_{t_*}} \times \sfX; \{-\infty,\ +\infty\}\times\pa\sfX\,],\quad \sfX:=\ol{\R^n_x}.
\end{equation}
We label the boundary hypersurfaces of $\sfM$ as follows:
\begin{enumerate}
\item $\scri^+$ is the lift of $\ol\R\times\pa\sfX$;
\item $\iota$ is the front face (i.e., the lift of $\pa\ol\R\times\pa\sfX$);
\item $\cK$ is the lift of $\pa\ol\R\times\sfX$.
\end{enumerate}
See Figure~\ref{FigTFM}.

We fix b-densities on $\ol\R$, $\sfX$, and $\sfM$ (e.g., $|\frac{\dd t_*}{\la t_*\ra}|$, $|\frac{\dd x}{\la x\ra^n}|$, and $|\frac{\dd t_*}{\la t_*\ra}\,\frac{\dd x}{\la x\ra^n}|$, respectively) to define $L^2$-spaces. For weights $\beta_\sscri,\beta_\iota,\beta_\cK\in\R$ and index sets $\cE_\sscri,\cE_\iota,\cE_\cK\subset\C\times\N_0$, we shall consider partially homogeneous weighted b-Sobolev spaces
\begin{equation}
\label{EqTFHbM}
  \Hb^{k,\ (\cE_\sscri,\beta_\sscri),\ (\cE_\iota,\beta_\iota),\ (\cE_\cK,\beta_\cK)}(\sfM),
\end{equation}
where the orders and index sets refer to the eponymous boundary hypersurfaces. For the b-differential operators testing for b-regularity, one may take compositions of $\la t_*\ra\pa_{t_*}$ and $\la x\ra\pa_{x^j}$, $j=1,\ldots,n$; one can replace $\la t_*\ra\pa_{t_*}$ by the pair $\pa_{t_*}$, $t_*\pa_{t_*}$ (which has the benefit of transforming in a simpler fashion under the Fourier transform).

On the Fourier transform side, we need two types of function spaces. At high frequencies, we use the spaces
\begin{equation}
\label{EqTFHbplus}
\begin{split}
  H_{\bop^+}^{k,\beta,\gamma}\bigl(\pm[\tfrac12,\infty)\times\sfX\bigr) &:= \bigl\{ v=v(\sigma,x) \colon \sigma^j(\sigma\pa_\sigma)^l v \in |\sigma|^{-\gamma}L^2\bigl(\pm[\tfrac12,\infty),|\tfrac{\dd\sigma}{\sigma}|;\Hb^{k-(j+l),\beta}(\sfX)\bigr) \\
    &\quad\hspace{20em} \forall\,j,l\in\N_0,\ j+l\leq k \bigr\}
\end{split}
\end{equation}
introduced in \citeAF{(\ref*{EqDFTSobHi})}. At low frequencies, we introduce the resolved space
\begin{equation}
\label{EqTFXscbtpm}
  \sfX_\scbtop^\pm := [\,\pm[0,1]_\sigma\times\sfX; \{0\}\times\pa\sfX\,],
\end{equation}
with notation taken from \citeAF{(\ref*{EqDFTXscbtpm})} (and also \S\ref{SssT3b}), whose boundary hypersurfaces we denote by $\scface$ (lift of $\pm[0,1]\times\pa\sfX$), $\tface$ (transition face, lift of $\{0\}\times\pa\sfX$), and $\zface$ (zero face, lift of $\{0\}\times\sfX$). See Figure~\ref{FigIGenKSpec}. For weights $\gamma_\scface,\gamma_\tface,\gamma_\zface\in\R$ and index sets $\cE_\scface,\cE_\tface,\cE_\zface\subset\C\times\N_0$, we then consider partially polyhomogeneous weighted b-Sobolev spaces
\begin{equation}
\label{EqTFHbX}
  \Hb^{k,\ (\cE_\scface,\gamma_\scface),\ (\cE_\tface,\gamma_\tface),\ (\cE_\zface,\gamma_\zface)}(\sfX_\scbtop^\pm).
\end{equation}

As usual, we drop the index sets in~\eqref{EqTFHbM} and \eqref{EqTFHbX} from the notation when they are empty. We then recall from \citeAF{Proposition~\ref*{PropDFTFourier}}:
\begin{subequations}
\begin{align}
\label{EqTFHbLo}
\begin{split}
  \beta_\cK\in(0,\infty)\setminus\N,\ \ 
   \beta_\iota\in\R&,\ \ 
   \beta_\sscri\geq\beta_\iota-\min(\beta_\cK,1-\eps), \\
  \implies (\cF u)|_{\sigma\in\pm[0,1]} &\in \Hb^{k,\ \min(\beta_\sscri,\beta_\iota)-\eps,\ \beta_\iota-1,\ ((0,0),\beta_\cK-1)}(\sfX_\scbtop^\pm);
\end{split} \\
   u\in\Hb^{k+\ell,\ \beta_\sscri,\ \beta_\iota,\ \beta_\cK}(\sfM) \nonumber\\
\label{EqTFHbHi}
  \implies (\cF u)|_{\sigma\in\pm[\frac12,\infty)} &\in H_{\bop^+}^{k,\beta_\sscri,0}(\pm[\tfrac12,\infty)\times\sfX),
\end{align}
\end{subequations}
for suitable $\ell\in\N_0$ (depending only on $\beta_\sscri$, $\beta_\iota$, and $\beta_\cK$), and moreover the restrictions of $\pa_\sigma^j\cF u$ to $\zface$ are equal on $\sfX_\scbtop^+$ and $\sfX_\scbtop^-$ for $j=0,\ldots,\lfloor\beta_\cK-1\rfloor$. Thus, \emph{if the $\scri^+$-decay order $\beta_\sscri$ exceeds an explicit threshold}, the Fourier transform relates the $\iota$- and $\cK$-decay orders of $u$ cleanly to the $\tface$- and $\zface$-decay rates of $\hat u$.

For the inverse Fourier transform, let $\chi=\chi(\sigma)\in\CIc((-1,1))$ be equal to $1$ near $[-\frac12,\frac12]$; we then recall the following consequence of \citeAF{Proposition~\ref*{PropDFTInv}}:
\begin{subequations}
\begin{align}
  &\forall\,\beta_\cK\in(0,\infty)\setminus\N,\ \ \beta_\iota,\,\beta_\scface,\,\beta_\sscri,\,\delta\in\R\ \ \exists\,\ell\in\N_0\ \text{such that:} \nonumber\\
\label{EqTFHbInvLo}
\begin{split}
  &\qquad v_\pm \in \Hb^{k+\ell,\ \beta_\scface,\ \beta_\iota-1,\ ((0,0),\beta_\cK-1)}(\sfX_\scbtop^\pm),\ \pa_\sigma^j v_-=\pa_\sigma^j v_+\ \text{at}\ \zface\ \text{for}\ j=0,\ldots,\lfloor\beta_\cK-1\rfloor \\
  &\qquad \qquad \implies \cF^{-1} v \in \Hb^{k,\ \min(\beta_\scface,\beta_\iota)-\eps,\ \beta_\iota, \ \beta_\cK}(\sfM)\ \forall\,\eps>0,\ \ v:=\chi(\sigma)\sum_\pm H(\pm\sigma)v_\pm;
\end{split} \\
\label{EqTFHbInvHi}
\begin{split}
  &\qquad v_\pm \in H_{\bop^+}^{k+\ell,\beta_\sscri,\delta}(\pm[\tfrac12,\infty)\times\sfX) \\
  &\qquad \qquad \implies \cF^{-1}\bigl((1-\chi(\sigma))v_\pm\bigr) \in \Hb^{k,(\beta_\sscri,\ \beta_\iota,\ \beta_\cK)}(\sfM).
\end{split}
\end{align}
\end{subequations}
Thus, the inverse Fourier transform \emph{always} translates the $\tface$- and $\zface$-decay rates on $\sfX_\scbtop^\pm$ cleanly to $\iota$- and $\cK$-decay rates. The fact that in~\eqref{EqTFHbInvHi} we can obtain any desired $\iota$- and $\cK$-decay rates (upon giving up a sufficient amount $\ell$ of b-regularity in the process) is the reason why the imprecise spaces~\eqref{EqTFHbplus} suffice for our purposes. (In applications, the precise decay at $\scri^+$ of solutions $u$ of wave equations will be obtained using physical space methods.) In order to capture partial expansions at $\iota,\cK$ and $\tface,\zface$, it thus suffices to generalize the low-energy results~\eqref{EqTFHbLo} and \eqref{EqTFHbInvLo}. In order to apply Lemma~\ref{LemmaTMPhgTest}, we first record:

\begin{lemma}[b-normal vector fields]
\label{LemmaTFbNormal}
  Let $\chi\in\CIc((-1,1))$ be equal to $1$ near $0$. Then we have the following b-normal vector fields at the boundary hypersurfaces of $\sfM$ and $\sfX_\scbtop^\pm$:
  \begin{enumerate}
  \item\label{ItTFbNormalK} $-t_*\pa_{t_*}$ (in the coordinates $t_*,x$) at $\cK\subset\sfM$;
  \item\label{ItTFbNormaliota} $-t_*\pa_{t_*}-\chi(r^{-1})r\pa_r$ (in the coordinates $t_*,r=|x|,\omega=\frac{x}{|x|}$) at $\iota\subset\sfM$;\footnote{Near $r=0$, this is defined as $t_*\pa_{t_*}$. The only relevant point is, in any case, that this vector field equals $t_*\pa_{t_*}+r\pa_r$ for large $t_*$ and $r$.}
  \item\label{ItTFbNormalscri} $-\chi(r^{-1})r\pa_r$ (in the coordinates $t_*,r,\omega$) at $\scri\subset\sfM$;
  \item\label{ItTFbNormalzf} $\sigma\pa_\sigma$ (in the coordinates $\sigma,x$) at $\zface\subset\sfX_\scbtop^\pm$;
  \item\label{ItTFbNormaltf} $\sigma\pa_\sigma+\chi(\rho)\rho\pa_\rho$ (in the coordinates $\sigma,\rho=|x|^{-1},\omega=\frac{x}{|x|}$) at $\tface\subset\sfX_\scbtop^\pm$.
  \item\label{ItTFbNormalscf} $\chi(\rho)\rho\pa_\rho$ (in the coordinates $\sigma,\rho,\omega$) at $\scface\subset\sfX_\scbtop^\pm$.
  \end{enumerate}
\end{lemma}
\begin{proof}
  Let us work in $t_*>1$, then $\tau:=t_*^{-1}$ and $x\in\R^n$ are local coordinates near $\{\infty\}\times\R^n\subset\ol\R\times\ol{\R^n}$; this remains true upon blowing up $\{\infty\}\times\pa\ol{\R^n}$, so $\tau$ and $x$ are local coordinates near $\cK^\circ$. Note then that $-t_*\pa_{t_*}=\tau\pa_\tau$. Near $\{\infty\}\times\pa\ol{\R^n}$, we work with inverse polar coordinates $x=\rho^{-1}\omega$, so $\rho=|x|^{-1}$ and $\omega=\frac{x}{|x|}$; then local coordinates near $\iota\cap\cK$ are $\rho_\iota=\rho=\frac{1}{r}$ and $\rho_\cK=\frac{\tau}{\rho}=\frac{r}{t_*}$ (cf.\ Figure~\ref{FigTFM}), and we then compute $-t_*\pa_{t_*}=\rho_\cK\pa_{\rho_\cK}$. This finishes the proof of part~\eqref{ItTFbNormalK}. For part~\eqref{ItTFbNormaliota}, we compute $-t_*\pa_{t_*}-r\pa_r=\rho_\iota\pa_{\rho_\iota}$. One obtains the same expression also in the local coordinates $\rho_\iota=\tau=\frac{1}{t_*}$ and $\rho_\sscri=\frac{\rho}{\tau}=\frac{t_*}{r}$ near $\scri^+\cap\iota$, where we moreover compute $-r\pa_r=\rho_\sscri\pa_{\rho_\sscri}$, proving part~\eqref{ItTFbNormalscri} there.

  For parts~\eqref{ItTFbNormalzf} and \eqref{ItTFbNormaltf}, we only discuss a neighborhood of $\zface\cap\tface$, where we can use the local coordinates $\rho_\zface=\frac{\sigma}{\rho}$, $\rho_\tface=\rho$, and $\omega=\frac{x}{|x|}$, in which $\sigma\pa_\sigma=\rho_\zface\pa_{\rho_\zface}$ and $\sigma\pa_\sigma+\rho\pa_\rho = \rho_\tface\pa_{\rho_\tface}$. For part~\eqref{ItTFbNormalscf}, we use the local coordinates $\rho_\scface=\frac{\rho}{\sigma}$, $\rho_\tface=\sigma$, and $\omega=\frac{x}{|x|}$, in which $\rho\pa_\rho=\rho_\scface\pa_{\rho_\scface}$.
\end{proof}

\begin{prop}[Partial polyhomogeneity]
\label{PropTFHbphg}
  Let $\beta_\cK\in(0,\infty)\setminus\N$, $\beta_\sscri,\beta_\iota\in\R$, and let $\cE_\sscri,\cE_\iota,\cE_\cK\subset\C\times\N_0$ be index sets. Assume that $\beta_\sscri,\beta_\iota-1\notin\Re\pi_1\cE_\sscri$, further $\beta_\iota\notin\Re\pi_1\cE_\iota$, and finally $\beta_\cK\notin\Re\pi_1\cE_\cK$ and $\min\Re\cE_\cK>0$.
  \begin{enumerate}
  \item\label{ItTFHbphgF}{\rm (Fourier transform.)} Suppose that
    \begin{equation}
    \label{EqTFHbphgBetaScri}
      \beta_\sscri \geq \beta_\iota - \beta_\cK,\quad \beta_\sscri > \beta_\iota - \min(1,\min\Re\cE_\cK).
    \end{equation}
    Fix any $\eps>0$ and set $\tilde\beta_\iota:=\min(\beta_\iota,\min\Re\cE_\iota-\eps)$. Then there exists $l\in\N_0$ such that for all $k\in\N_0$ and all $u\in\Hb^{k+l,\ (\cE_\sscri,\beta_\sscri),\ (\cE_\iota,\beta_\iota),\ (\cE_\cK,\beta_\cK)}(\sfM)$, we have
    \begin{equation}
    \label{EqTFHbphgF}
       (\cF u)|_{\sigma\in\pm[0,1]} \in \Hb^{k,\ \bigl(\cE_\sscri,\min(\beta_\sscri,\tilde\beta_\iota)-\eps\bigr),\ \bigl((\cE_\iota-1)\extcup\cE_\sscri,\,\beta_\iota-1\bigr),\ \bigl((0,0)\extcup(\cE_\cK-1),\,\beta_\cK-1\bigr)}(\sfX_\scbtop^\pm).
    \end{equation}
  \item\label{ItTFHbphgFI}{\rm (Inverse Fourier transform.)} Let $\chi\in\CIc((-1,1))$. Let $\beta_\scface\in\R$. Fix any $\eps>0$ and set $\tilde\beta_\iota:=\min(\beta_\iota,\min\Re\cE_\iota-\eps)$. Then there exists $\ell\in\N_0$ such that if $k\in\N_0$ and
    \[
      v_\pm \in \Hb^{k+l,\ \beta_\scface,\ (\cE_\iota-1,\beta_\iota-1),\ \bigl((0,0)\extcup(\cE_\cK-1),\,\beta_\cK-1\bigr)}(\sfX_\scbtop^\pm)
    \]
    satisfy
    \begin{equation}
    \label{EqTFHbphgFIMatch}
      (\sigma\pa_\sigma)^{\ell+1}\pa_\sigma^j v_-=(\sigma\pa_\sigma)^{\ell+1}\pa_\sigma^j v_+,\quad j=0,\ldots,\lfloor\beta_\cK-1\rfloor,\ \ell=k(\cE_\cK,j),
    \end{equation}
    then for $v:=\chi(\sigma)\sum_\pm H(\pm\sigma)v_\pm$, we have
    \begin{equation}
    \label{EqTFHbphgFI}
      \cF^{-1}v \in \Hb^{k,\ \min(\beta_\scface,\tilde\beta_\iota)-\eps,\ (\cE_\iota,\beta_\iota),\ (\cE_\cK,\beta_\cK)}(\sfM).
    \end{equation}
  \end{enumerate}
\end{prop}

This is closely related to \cite[Proposition~2.28]{Hintz3b}, which however deals neither with nontrivial index sets at what in present notation is $\cK$ nor with finite differentiability orders (nor $L^2$-estimates).

\begin{proof}[Proof of Proposition~\usref{PropTFHbphg}]
  \pfstep{Part~\eqref{ItTFHbphgF}, $\cE_\sscri=\emptyset$.} Let $\eps>0$ and set $\tilde\beta_\cK:=\min(\beta_\cK,\min\Re\cE_\cK-\eps)$. Then $u\in\Hb^{k+l,\ \beta_\sscri,\ \tilde\beta_\iota,\ \tilde\beta_\cK}(\sfM)$; and we have $\beta_\sscri\geq\beta_\iota-\tilde\beta_\cK\geq\tilde\beta_\iota-\tilde\beta_\cK$ and $\beta_\sscri\geq\beta_\iota-(1-\eps)\geq\tilde\beta_\iota-(1-\eps)$ when $\eps$ is sufficiently small. For later use, note that if $\tilde\beta_\cK$ increases (which happens when the index set $\cE_\cK$ gets smaller), these inequalities remain valid. Therefore, we can apply~\eqref{EqTFHbLo} to obtain
  \begin{equation}
  \label{EqTFHbphgFu}
    \cF u \in \Hb^{k,\ \min(\beta_\sscri,\tilde\beta_\iota)-\eps,\ \tilde\beta_\iota-1,\ ((0,0),\tilde\beta_\cK-1)}(\sfX_\scbtop^\pm).
  \end{equation}
  Set $\cE_{\iota,\leq}:=\{(z,m)\in\cE_\iota\colon\Re z\leq\beta_\iota\}$ and $\cR_\iota:=\prod_{(z,m)\in\cE_{\iota,\leq}} (-t_*\pa_{t_*}-\chi(r^{-1})r\pa_r-z)$ in the notation of Lemma~\ref{LemmaTFbNormal}; then $\cR_\iota u\in\Hb^{k+l-|\cE_{\iota,\leq}|,\ (\beta_\sscri,\ \beta_\iota,\ \tilde\beta_\cK)}$ and therefore (when $l$ is large enough)
  \[
    \cF(\cR_\iota u) \in \Hb^{k,\ \min(\beta_\sscri,\beta_\iota)-\eps,\ \beta_\iota-1,\ ((0,0),\tilde\beta_\cK-1)}(\sfX_\scbtop^\pm).
  \]
  But since $\cF\circ t_*\pa_{t_*}=-\pa_\sigma\sigma\circ\cF=(-\sigma\pa_\sigma-1)\circ\cF$, we have
  \[
    \Biggl(\;\prod_{(z,m)\in\cE_{\iota,\leq}} (\sigma\pa_\sigma+\chi(\rho)\rho\pa_\rho-(z-1)) \Biggr) \cF u = \cF(\cR_\iota u).
  \]
  Weakening the $\scface$-decay rate of $\cF(\cR_\iota u)$ to $\min(\beta_\sscri,\tilde\beta_\iota)-\eps$ to match~\eqref{EqTFHbphgFu}, we can then apply Lemma~\ref{LemmaTMPhgTest} (and recall Lemma~\ref{LemmaTFbNormal}\eqref{ItTFbNormaltf}) to deduce
  \begin{equation}
  \label{EqTFHbphgFu2}
    \cF u \in \Hb^{k,\ \min(\beta_\sscri,\tilde\beta_\iota)-\eps,\ (\cE_\iota-1,\beta_\iota-1),\ ((0,0),\tilde\beta_\cK-1)}(\sfX_\scbtop^\pm).
  \end{equation}

  To obtain the correct index set at $\zface$, we now test with the vector fields from Lemma~\ref{LemmaTFbNormal}\eqref{ItTFbNormalK} and \eqref{ItTFbNormalzf}; that is, set $\cE_{\cK,\leq}:=\{(z,m)\in\cE_\cK\colon\Re z\leq\beta_\cK\}$ and $\cR_\cK:=\prod_{(z,m)\in\cE_{\cK,\leq}} (-t_*\pa_{t_*}-z)$, then we can apply~\eqref{EqTFHbphgFu2} also to $\cR_\cK u$ and deduce that
  \[
    \Biggl(\;\prod_{(z,m)\in\cE_{\cK,\leq}} (\sigma\pa_\sigma-(z-1))\Biggr) \cF u = \cF(\cR_\cK u) \in \Hb^{k,\ \min(\beta_\sscri,\tilde\beta_\iota)-\eps,\ (\cE_\iota-1,\beta_\iota-1),\ ((0,0),\beta_\cK-1)}(\sfX_\scbtop^\pm).
  \]
  Upon using Lemma~\ref{LemmaTMPhgTest} (or upon, in addition, applying $\prod_{j\leq\lfloor\beta_\cK-1\rfloor} (\sigma\pa_\sigma-j)$ to this equation, so that the output has trivial $\zface$-index set and weight $\beta_\cK-1$, and then applying Lemma~\ref{LemmaTMIntFuchs} repeatedly), we obtain~\eqref{EqTFHbphgF}.

  \pfstep{Part~\eqref{ItTFHbphgF}, general $\cE_\sscri$.} Let $\chi_\cK\in\CI(\sfM)$ be equal to $1$ near $\cK$ and equal to $0$ near $\scri$. Then $\chi_\cK u$ vanishes near $\scri$, so its Fourier transform is already controlled by our arguments thus far. Re-naming $(1-\chi_\cK)u$ as $u$, it thus suffices to consider the case of $u$ vanishing near $\cK$, so $\cE_\cK=\emptyset$ and $\beta_\cK=\infty$ (i.e., arbitrarily large); for present purposes, we fix any $\beta_\cK$ satisfying $\beta_\cK>\beta_\iota$.

  Let now $\tilde\chi\in\CIc(\sfX^\circ)$. Then $\tilde\chi u$ vanishes for large $r$, and thus its Fourier transform lies in $\Hb^{k+l,\ \infty,\ \infty,\ ((0,0),\beta_\cK-1)}(\sfX_\scbtop^\pm)$. On the other hand, setting $\cE_{\sscri,\leq}:=\{(z,m)\in\cE_\sscri\colon\Re z\leq\beta_\sscri\}$ and $\cR_\sscri:=\prod_{(z,m)\in\cE_{\sscri,\leq}} (-\chi(r^{-1})r\pa_r-z)$ in the notation of Lemma~\ref{LemmaTFbNormal}, we have
  \[
    \cR_\sscri u\in\Hb^{k+l-|\cE_{\sscri,\leq}|,\ \beta_\sscri,\ (\cE_\iota,\beta_\iota),\ \beta_\cK}(\sfM)
  \]
  and therefore
  \begin{align*}
    \Biggl(\;\prod_{(z,m)\in\cE_{\sscri,\leq}} (\chi(\rho)\rho\pa_\rho-z)\Biggr)\cF u &= \cF(\cR_\sscri u) \\
      &\in \Hb^{k+l',\ \min(\beta_\sscri,\tilde\beta_\iota)-\eps,\ (\cE_\iota-1,\beta_\iota-1),\ ((0,0),\beta_\cK-1)}(\sfX_\scbtop^\pm)
  \end{align*}
  where $l':=l-|\cE_{\sscri,\leq}|$. We integrate this from $\rho\geq\frac12$ (say) towards $\rho=0$ in two steps. First, near $\zface$, we use the local coordinates $\rho_\zface:=\frac{|\sigma|}{\rho}$ and $\rho_\tface:=\rho$, in which $-\rho\pa_\rho+z=\rho_\zface\pa_{\rho_\zface}-\rho_\tface\pa_{\rho_\tface}+z$ is a transport operator of the type considered in Lemma~\ref{LemmaTIntHyp}, with $(\cE,\alpha)$ and $(\cF,\beta)$ there being $((0,0),\beta_\cK-1)$ and $(\cE_\iota-1,\beta_\iota-1)$ in present notation, respectively; note that our requirement on the (otherwise arbitrary) value of $\beta_\cK$ ensures that $\beta_\cK-1>\beta_\iota-1$. Thus, repeated application of Lemma~\ref{LemmaTIntHyp} implies that the index sets of $\cF u$ at $\zface$ and $\tface$ are $((0,0),\beta_\cK-1)$ and $(\cE_\sscri\extcup(\cE_\iota-1),\beta_\iota-1)$, respectively.

  Secondly, near $\scface$, we use the local coordinates $\rho_\scface:=\frac{\rho}{|\sigma|}$ and $\rho_\tface:=|\sigma|$, in which $\rho\pa_\rho-z=\rho_\scface\pa_{\rho_\scface}-z$. Repeated application of Lemma~\ref{LemmaTMIntFuchs} thus shows that the index set of $\cF u$ at $\scface$ is $(\cE_\sscri,\min(\beta_\sscri,\tilde\beta_\iota)-\eps)$. This finishes the proof of~\eqref{EqTFHbphgF}.

  \pfstep{Part~\eqref{ItTFHbphgFI}, $\cE_\cK=\emptyset$.} We have $v_\pm\in\Hb^{\infty,\ k+l,\ \beta_\scface,\ \tilde\beta_\iota-1,\ ((0,0),\beta_\cK-1)}(\sfX_\scbtop)$, with matching Taylor coefficients at $\zface$ by~\eqref{EqTFHbphgFIMatch}, and therefore~\eqref{EqTFHbInvLo} gives
  \begin{equation}
  \label{EqTFHbphgFI1}
    u := \cF^{-1}v \in \Hb^{k,\ \beta_\sscri,\ \tilde\beta_\iota,\ \beta_\cK}(\sfM),\quad \beta_\sscri:=\min(\beta_\scface,\tilde\beta_\iota)-\eps.
  \end{equation}
  Furthermore, Lemma~\ref{LemmaTFbNormal}\eqref{ItTFbNormaltf} implies that upon setting $\cE_{\tface,\leq}:=\{(z,m)\in\cE_\iota\colon\Re z\leq\beta_\iota\}$ and $\cR_\tface:=\prod_{(z,m)\in\cE_{\tface,\leq}} (\sigma\pa_\sigma+\chi(\rho)\rho\pa_\rho-(z-1))$, we have $\cR_\tface v_\pm\in\Hb^{\infty,\ k+l-|\cE_{\tface,\leq}|,\ \beta_\scface,\ \beta_\iota-1,\ ((0,0),\beta_\cK-1)}(\sfX_\scbtop)$ (still with matching Taylor coefficients at $\zface$). Therefore,
  \[
    \Biggl(\;\prod_{(z,m)\in\cE_{\tface,\leq}} (-t_*\pa_{t_*}-\chi(r^{-1})r\pa_r-z)\Biggr)u = \cF^{-1}(\cR_\tface v) \in \Hb^{k,\ \beta_\sscri,\ \beta_\iota,\ \beta_\cK}(\sfM).
  \]
  Integrating this in a collar neighborhood of $\iota$ and using~\eqref{EqTFHbphgFI1} gives~\eqref{EqTFHbphgFI}.

  \pfstep{Part~\eqref{ItTFHbphgFI}, general $\cE_\cK$.} Set $\tilde\beta_\cK:=\min(\beta_\cK,\min\Re\cE_\cK-\eps)$, then what we have already shown implies
  \[
    u \in \Hb^{k+l-l',\ \beta_\sscri,\ (\cE_\iota,\beta_\iota),\ \tilde\beta_\cK}(\sfM)
  \]
  for $l'=|\cE_{\tface,\leq}|$. Set $\cE_{\cK,\leq}:=\{(z,m)\in\cE_\cK\colon\Re z\leq\beta_\cK\}$ and $\cR_\zface:=\prod_{(z,m)\in\cE_{\cK,\leq}}(\sigma\pa_\sigma-(z-1))$. Then $\cR_\zface v_\pm\in\Hb^{k+l-|\cE_{\cK,\leq}|,\ \beta_\scface,\ (\cE_\iota-1,\beta_\iota-1),\ ((0,0),\beta_\cK)}(\sfX_\scbtop^\pm)$; and~\eqref{EqTFHbphgFIMatch} is precisely the condition that guarantees that the Taylor coefficients of $\cR_\zface v_\pm$ of all orders up to $\lfloor\beta_\cK-1\rfloor$ at $\zface$ agree, so
  \[
    \Biggl(\;\prod_{(z,m)\in\cE_{\cK,\leq}} (-t_*\pa_{t_*}-z)\Biggr)u = \cF^{-1}(\cR_\zface v) \in \Hb^{k,\ \beta_\sscri,\ (\cE_\iota,\beta_\iota),\ \beta_\cK}(\sfM).
  \]
  In view of Lemma~\ref{LemmaTFbNormal}\eqref{ItTFbNormalK}, this implies that the order of $u$ at $\cK$ is, in fact, $(\cE_\cK,\beta_\cK)$, completing the proof.
\end{proof}

\subsection{Spherical harmonic decompositions}
\label{SsTY}

Recalling the discussion of $\iota^+$-normal operators and indicial roots from~\S\ref{SssIGen2La}, the asymptotic behavior at large radii of (large) zero energy states of the linearized gauge-fixed Einstein equation on Kerr and related equations is governed by spectral information of their Minkowskian analogues, which in turn (upon exploiting homogeneity under scaling) can be reduced to spectral information about operators on $\Sph^2$ (a transversal of the scaling action on $\R^3_x\setminus\{0\}$) acting on certain tensor bundles. (We are then mainly interested in low spherical harmonics.) We thus record notation and useful formulas for scalar and vector spherical harmonics, largely following the presentation in \cite[\S{6.1}]{HintzGlueLocI}.

\begin{notation}[Geometric operators]
\label{NotTY}
  We write $\slg$ for the standard metric on $\Sph^2$, so $\slg=\dd\theta^2+\sin^2\theta\,\dd\phi^2$ in polar coordinates. We furthermore write $\sltr=\tr_\slg$ for the trace, $\slstar$ for the Hodge star operator, $\dd\colon\CI(\Sph^2)\to\CI(\Sph^2;T^*\Sph^2)$ for the exterior derivative, $\slDelta=-\sltr\slnabla^2$ for the (non-negative) tensor Laplacian (with $\slnabla$ the Levi-Civita connection), $\sldelta$ for the (negative) divergence (i.e., $\sldelta\omega=-\omega_a{}^{;a}$ and $(\sldelta h)_a=-h_{a b}{}^{;b}$), $\sldelta^*$ for the symmetric gradient (i.e., $(\sldelta^*\omega)_{a b}=\frac12(\omega_{a;b}+\omega_{b;a})$), and
  \[
    \sldelta_0^* := \sldelta^* + \tfrac12\slg\sldelta
  \]
  for the trace-free part of the symmetric gradient.
\end{notation}

On functions, we have $\slDelta=\sldelta\sld$.

\begin{definition}[Spherical harmonics]
\label{DefTY}
  Let $l\in\N_0$.
  \begin{enumerate}
  \item We denote the $(2 l+1)$-dimensional space of spherical harmonics of degree $l$ by
    \[
      \scalspace_l := \ker\bigl(\slDelta - l(l+1)\bigr) \subset \CI(\Sph^2).
    \]
  \item For $l\geq 1$, we write
    \[
      \vectspace_l := \{ \slstar\sld\scal \colon \scal\in\scalspace_l \} \subset \CI(\Sph^2;T^*\Sph^2).
    \]
    We say that $\omega\in\CI(\Sph^2;T^*\Sph^2)$ is of \emph{vector type $l$} if $\omega\in\vectspace_l$, and of \emph{scalar type $l$} if $\omega=\sld\scal$ for some $\scal\in\scalspace_l$.
  \item Let $h\in\CI(\Sph^2;S^2 T^*\Sph^2)$. We then say that $h$ is
    \begin{alignat*}{2}
      \text{of scalar type $l$ if}\ & h = \sldelta_0^*\sld\scal + \scal'\slg,&\quad \scal,\scal'&\in\scalspace_l, \\
      \text{of vector type $l$ if}\ & h = \sldelta^*\vect,&\quad \vect&\in\vectspace_l.
    \end{alignat*}
    The second possibility only exists for $l\geq 1$, and in the first possibility and for $l=0$ we have $h=c\slg$ for a constant $c$.
  \end{enumerate}
  We shall more briefly call a 1-form of scalar type $l$ an ``$\rms l$ 1-form,'' similarly for symmetric 2-tensors and vector types. We moreover say that a function/1-form/symmetric 2-tensor is of \emph{scalar type}, resp.\ of \emph{vector type} if all of its vector type, resp.\ scalar type components vanish.
\end{definition}

We do allow for complex-valued functions and tensors here; we shall not notationally distinguish the subspaces of real functions and tensors.

Note that every vector type 1-form is divergence-free. We recall the spherical harmonic decomposition of functions, 1-forms, and tensors:
\begin{enumerate}
\item Every $\scal\in\CI(\Sph^2)$ can be written as a convergent series $\scal=\sum_{l\in\N_0}\scal_l$ for unique $\scal_l\in\scalspace_l$.
\item Given $\omega\in\CI(\Sph^2;T^*\Sph^2)$, we can write $\sldelta\omega=\sum_{l\geq 1}\scal_l$ and set $\scal:=\sum_{l\geq 1}\frac{1}{l(l+1)}\scal_l$; then $\vect:=\omega-\sld\scal\in\ker\sldelta=\ker\sld\slstar$, so $\sld(\slstar\vect)=0$. But since the first cohomology of $\Sph^2$ is trivial, this gives $\slstar\vect=-\sld\scal'$ for some $\scal'\in\CI(\Sph^2)$, so $\vect=\slstar\sld\scal'$. Decomposing $\scal'$ into spherical harmonics thus shows that $\omega$ can be written as a convergent series of scalar type and vector type 1-forms. (Note that the case $l=0$ does not appear here.)
\item Let $h\in\CI(\Sph^2;S^2 T^*\Sph^2)$, then the tensor $h-\scal'\slg$, $\scal':=\frac12\sltr h$, is trace-free. It remains to consider the case $\sltr h=0$. The operator $\sldelta_0^*$, acting between sections of the rank $2$ bundles $T^*\Sph^2$ and $S^2_0 T^*\Sph^2$ (trace-free symmetric 2-tensors), has principal symbol at $\xi\in T^*\Sph^2\setminus o$ given by $i$ times the map $\omega\mapsto\xi\otimes_s\omega-\frac12\slg\la\xi,\omega\ra$, which is injective and thus invertible. Therefore, we can write $h=\sldelta_0^*\omega+h_0$ where $h_0\in\CI(\Sph^2;S^2_0 T^*\Sph^2)\cap\ker\sldelta$; thus $h_0=0$ by \cite{HiguchiSpherical}. Expanding $\omega$ into scalar and vector type 1-forms yields a decomposition of $h$ into scalar type and vector type tensors. The case $l=0$ is special since there do not exist nontrivial vector type $0$ 1-forms; and also the case $l=1$ is special since
  \begin{equation}
  \label{EqTYTensorl2}
    \sldelta_0^*\sld\scal=0\quad\forall\,\scal\in\scalspace_1.
  \end{equation}
  (It suffices to check this when $\scal$ is the height function, in which case $\slg^{-1}(\sld\scal,\cdot)$ is related, via stereographic projection, to the radial vector field on $\R^2$, which is a conformal Killing vector field; hence, the trace-free part of its symmetric gradient vanishes.)
\end{enumerate}

We moreover record the following identities from \cite[Lemma~6.4]{HintzGlueLocI}, valid for $\scal\in\scalspace_l$ ($l\geq 0$) and $\vect\in\vectspace_l$ ($l\geq 1$):
\begin{equation}
\label{EqTYIdentities}
\begin{alignedat}{3}
  \sldelta\sld\scal&=l(l+1)\scal, & \ \ \sldelta^*\sld\scal&=\sldelta^*_0\sld\scal-\frac{l(l+1)}{2}\slg\scal, & \ \ \slDelta(\sld\scal)&=(l(l+1)-1)\sld\scal, \\
  \slDelta(\slg\scal)&=l(l+1)\slg\scal, & \ \ \sldelta^*(\sldelta^*_0\sld\scal)&=\frac{l(l+1)-2}{2}\sld\scal, & \ \ \slDelta(\sldelta^*_0\sld\scal)&=(l(l+1)-4)\sldelta^*_0\sld\scal, \\
  \slDelta\vect&=(l(l+1)-1)\vect, & \ \ \sldelta\sldelta^*\vect&=\frac{l(l+1)-2}{2}\vect, & \ \ \slDelta\sldelta^*\vect&=(l(l+1)-4)\sldelta^*\vect.
\end{alignedat}
\end{equation}

For the explicit description of the case $l=1$, we introduce:

\begin{definition}[$l=1$ functions and 1-forms]
\label{DefTYs1}
  For $\bha\in\R^3$, we write
  \[
    \scal(\bha) := \bha\cdot\frac{x}{|x|} \in \scalspace_1,\quad
    \vect(\bha) := (\bha\times x)\cdot\dd x \in \vectspace_1.
  \]
  Conversely, given $\scal\in\scalspace_1$ and $\vect\in\vectspace_1$, we write $\bha(\scal)$ and $\bha(\vect)$ for the unique $\bha\in\R^3$ (or $\bha\in\C^3$ when working with complex functions or 1-forms) such that $\scal=\scal(\bha)$ and $\vect=\vect(\bha)$, respectively.
\end{definition}

The weights of $|x|$ in this definition are a matter of convention, and only matter when one regards $\scal(\bha)$ and $\vect(\bha)$ as functions and 1-forms on $\R^3\setminus\{0\}$; our convention is that $\scal(\bha)$ is homogeneous of degree $0$, while $\vect(\bha)$ is dual (with respect to the Euclidean metric) to a rotation vector field.

\subsubsection{Differential operators on Minkowski space}
\label{SssTYMink}

We recall explicit expressions for various geometric differential operators on Minkowski space in polar coordinates from \cite[Lemma~6.6]{HintzGlueLocI}. These will be useful when describing the leading order behavior of wave type operators on Kerr spacetimes at $r=\infty$; thus, we work in the coordinates $t\in\R$, $r=|x|>0$, $\omega=\frac{x}{|x|}\in\Sph^2$, where $0\neq x\in\R^3$. Another useful coordinate system is
\begin{equation}
\label{EqTYMinkStar}
  t_*:=t-r,\ \rho:=r^{-1},\ \omega\in\Sph^2.
\end{equation}
We describe operators on (powers of) the cotangent bundle using the functions
\begin{subequations}
\begin{equation}
\label{EqTYMink01}
  \ubar x^0 := t+r,\quad
  \ubar x^1 := t-r
\end{equation}
(whose level sets are null for the Minkowski metric) and the bundle splittings
\begin{equation}
\label{EqTYMink01Split}
\begin{split}
  T^*\R^4 &= \la\dd\ubar x^0\ra\oplus\la\dd\ubar x^1\ra\oplus r T^*\Sph^2, \\
  S^2 T^*\R^4 &= \la(\dd\ubar x^0)^2\ra \oplus \la 2\,\dd\ubar x^0\,\dd\ubar x^1 \ra \oplus \bigl(2\,\dd\ubar x^0\otimes_s r T^*\Sph^2\bigr) \\
    &\qquad \oplus \la(\dd\ubar x^1)^2\ra \oplus \bigl(2\,\dd\ubar x^1\otimes_s r T^*\Sph^2\bigr) \oplus r^2 S^2 T^*\Sph^2;
\end{split}
\end{equation}
\end{subequations}
the meaning of the factor $r$ in the first splitting is that we identify $(\omega_0,\omega_1,\slomega)$ with the covector $\omega_0\,\dd\ubar x^0+\omega_1\,\dd\ubar x^1+r\slomega$ on $\R_t\times(0,\infty)_r\times\Sph^2$, similarly for the second splitting. When working with sections of $T\R^4$ and $S^2 T\R^4$, we use the dual splittings. With the Minkowski metric being given by
\[
  \ubar g:=-\dd t^2+\dd r^2+r^2\slg=-\dd\ubar x^0\,\dd\ubar x^1 + r^2\slg
\]
and the dual metric therefore being
\begin{equation}
\label{EqTYMinkMetDual}
  \ubar g^{-1}=-4\pa_0\otimes\pa_1 + r^{-2}\slg^{-1} = 2\rho\,\pa_{t_*}\otimes_s \rho\pa_\rho + \rho^2\bigl((\rho\pa_\rho)^2+\slg^{-1}\bigr),
\end{equation}
we thus have
\begin{subequations}
\begin{equation}
\label{EqTYMinkMet}
  \ubar g = \Bigl(0,-\frac12,0,0,0,\slg\Bigr),\quad
  \ul\tr := \tr_{\ubar g} = (0,-4,0,0,0,\sltr).
\end{equation}
For the operator $\sfG_g:=I-\frac12 g\tr_g$, this yields
\begin{equation}
\label{EqTYMinkTrRev}
  \ul\sfG := \sfG_{\ubar g} = I-\frac12\ubar g\,\ul\tr = \begin{pmatrix} 1 & 0 & 0 & 0 & 0 & 0 \\ 0 & 0 & 0 & 0 & 0 & \frac14\sltr \\ 0 & 0 & 1 & 0 & 0 & 0 \\ 0 & 0 & 0 & 1 & 0 & 0 \\ 0 & 0 & 0 & 0 & 1 & 0 \\ 0 & 2\slg & 0 & 0 & 0 & \slsfG \end{pmatrix}
\end{equation}
\end{subequations}

For further computations, we record that the Christoffel symbols of $\ubar g$ in the coordinates $\ubar x^0$, $\ubar x^1$, and coordinates $x^2,x^3$ on $\Sph^2$ vanish except for
\begin{equation}
\label{EqTYMinkChr}
  \ubar\Gamma_{0 b}^c = \ubar\Gamma_{b 0}^c=\frac12 r^{-1}\delta_b^c,\quad
  \ubar\Gamma_{1 b}^c = \ubar\Gamma_{b 1}^c=-\frac12 r^{-1}\delta_b^c,\quad
  \ubar\Gamma_{a b}^0 = -r\slg_{a b},\quad
  \ubar\Gamma_{a b}^1 = r\slg_{a b},\quad
  \ubar\Gamma_{a b}^c = \slGamma_{a b}^c.
\end{equation}
This uses $\pa_0 r=-\pa_1 r=\frac12$. As a simple consequence, the scalar wave operator $\ubar\Box u=-\ubar g^{\mu\nu}u_{;\mu\nu}=-\ubar g^{\mu\nu}(\pa_\mu\pa_\nu u-\ubar\Gamma_{\mu\nu}^\lambda\pa_\lambda u)$ is given by
\[
  \ubar\Box = 4\pa_0\pa_1 + \frac{2}{r}(\pa_1-\pa_0) + r^{-2}\slDelta;
\]
upon passing to the coordinates $t_*=\ubar x^1$, $\rho=r^{-1}=\frac{2}{\ubar x^0-\ubar x^1}$, and using
\[
  \pa_0=-\frac12\rho^2\pa_\rho,\quad
  \pa_1=\pa_{t_*}+\frac12\rho^2\pa_\rho,
\]
this gives
\[
  \ubar\Box = -2\pa_{t_*}\rho(\rho\pa_\rho-1) + \rho^2\bigl(-(\rho\pa_\rho)^2+\rho\pa_\rho+\slDelta \bigr).
\]
(The principal part matches the dual metric~\eqref{EqTYMinkMetDual} upon to an overall minus sign.)

\begin{lemma}[Differential operators]
\label{LemmaTYMinkOp}
  We express differential operators in the coordinates $t_*=t-r$, $\rho=r^{-1}$, and $\omega\in\Sph^2$, and using the bundle splittings~\eqref{EqTYMink01Split}; we furthermore recall Notation~\usref{NotTY}, and write $\slDelta$ for the tensor Laplacian acting component-wise on the summands of~\eqref{EqTYMink01Split}.
  \begin{enumerate}
  \item{\rm (1-forms.)} Acting on 1-forms, we have
    \begin{alignat*}{2}
      \ubar\Box &= -2\pa_{t_*}\rho(\rho\pa_\rho-1) + \wh{\ubar\Box}(0),&\qquad \wh{\ubar\Box}(0)&=\rho^2\left(-(\rho\pa_\rho)^2+\rho\pa_\rho+\slDelta+\begin{pmatrix} 1 & -1 & -\sldelta \\ -1 & 1 & \sldelta \\ -2\sld & 2\sld & 1 \end{pmatrix} \right), \\
      \ubar\delta^*&=\begin{pmatrix} 0 & 0 & 0 \\ \frac12 & 0 & 0 \\ 0 & 0 & 0 \\ 0 & 1 & 0 \\ 0 & 0 & \frac12 \\ 0 & 0 & 0 \end{pmatrix}\pa_{t_*} + \wh{\ubar\delta^*}(0),&\qquad \wh{\ubar\delta^*}(0)&=\rho\begin{pmatrix} -\frac12\rho\pa_\rho & 0 & 0 \\ \frac14\rho\pa_\rho & -\frac14\rho\pa_\rho & 0 \\ \frac12\sld & 0 & -\frac14(\rho\pa_\rho+1) \\ 0 & \frac12\rho\pa_\rho & 0 \\ 0 & \frac12\sld & \frac14(\rho\pa_\rho+1) \\ \slg & -\slg & \sldelta^* \end{pmatrix}.
    \end{alignat*}
  \item{\rm (Symmetric 2-tensors.)} Acting on symmetric 2-tensors, we have $\ubar\Box=-2\pa_{t_*}\rho(\rho\pa_\rho-1)+\wh{\ubar\Box}(0)$ where now
    \begin{equation}
    \label{EqTYMinkOpBox2}
      \wh{\ubar\Box}(0) = \rho^2\left(-(\rho\pa_\rho)^2+\rho\pa_\rho+\slDelta + \begin{pmatrix} 2 & -2 & -2\sldelta & 0 & 0 & -\frac12\sltr \\ -1 & 2 & \sldelta & -1 & -\sldelta & \frac12\sltr \\ -2\sld & 2\sld & 3 & 0 & -2 & -\sldelta \\ 0 & -2 & 0 & 2 & 2\sldelta & -\frac12\sltr \\ 0 & -2\sld & -2 & 2\sld & 3 & \sldelta \\ -2\slg & 4\slg & -4\sldelta^* & -2\slg & 4\sldelta^* & 2 \end{pmatrix}\right).
    \end{equation}
    Furthermore,
    \begin{equation}
    \label{EqTYMinkOpDel}
    \begin{split}
      \ubar\delta &= \begin{pmatrix} 2 & 0 & 0 & 0 & 0 & 0 \\ 0 & 2 & 0 & 0 & 0 & 0 \\ 0 & 0 & 2 & 0 & 0 & 0 \end{pmatrix}\pa_{t_*} + \wh{\ubar\delta}(0), \\
      \wh{\ubar\delta}(0) &= \rho\begin{pmatrix} \rho\pa_\rho-2 & -\rho\pa_\rho+2 & \sldelta & 0 & 0 & \frac12\sltr \\ 0 & \rho\pa_\rho-2 & 0 & -\rho\pa_\rho+2 & \sldelta & -\frac12\sltr \\ 0 & 0 & \rho\pa_\rho-3 & 0 & -\rho\pa_\rho+3 & \sldelta \end{pmatrix}.
    \end{split}
    \end{equation}
  \end{enumerate}
\end{lemma}
\begin{proof}
  Explicit computation; see \cite[\S{6.2}]{HintzGlueLocI} for details.
\end{proof}

Applying the spherical harmonic decompositions described after Definition~\ref{DefTY} to each component of a tensor in the splittings~\eqref{EqTYMink01Split} at each sphere of constant $(t_*,\rho)$ yields the following types of tensors:
\begin{equation}
\label{EqTYSplit1}
\begin{aligned}
  &\text{$\rms 0$ 1-forms}: && (a\scal,b\scal,0), && \scal\in\scalspace_0, \\
  &\text{$\rms l$ 1-forms, $l\geq 1$}: && (a\scal,b\scal,c\,\sld\scal), && \scal\in\scalspace_l, \\
  &\text{$\rmv l$ 1-forms, $l\geq 1$}: && (0,0,c\vect), && \vect\in\vectspace_l;
\end{aligned}
\end{equation}
and symmetric 2-tensors can be decomposed into
\begin{equation}
\label{EqTYSplit2}
\begin{aligned}
  &\text{$\rms 0$ tensors}: && (a\scal,b\scal,0,e\scal,0,p\scal,0), && \scal\in\scalspace_0, \\
  &\text{$\rms 1$ tensors}: && (a\scal,b\scal,c\,\sld\scal,e\scal,f\,\sld\scal,p\scal,0), && \scal\in\scalspace_1, \\
  &\text{$\rms l$ tensors, $l\geq 2$}: && (a\scal,b\scal,c\,\sld\scal,e\scal,f\,\sld\scal,p\scal,q\,\sldelta_0^*\sld\scal), && \scal\in\scalspace_l, \\
  &\text{$\rmv 1$ tensors}: && (0,0,c\vect,0,f\vect,0,0), && \vect\in\vectspace_1, \\
  &\text{$\rmv l$ tensors, $l\geq 2$}: && (0,0,c\vect,0,f\vect,0,q\,\sldelta^*\vect), && \vect\in\vectspace_l.
\end{aligned}
\end{equation}
Here, $a,b,c,e,f,p,q$ are functions of $t_*$ and $\rho$ only. The special nature of low angular frequencies is due to~\eqref{EqTYTensorl2}. The operators in Lemma~\ref{LemmaTYMinkOp} not only preserve the type and spherical harmonic degree, but also the particular spherical harmonic $\scal$ or $\vect$. Thus, for example, on 1-forms, the action of $\wh{\ubar\Box}(0)$ on $\rms l$ 1-forms, $l\geq 1$, is given by $\rho^2$ times the $3\times 3$ matrix
\begin{equation}
\label{EqTYSplitEx}
  -(\rho\pa_\rho)^2+\rho\pa_\rho + \begin{pmatrix} l(l+1) & 0 & 0 \\ 0 & l(l+1) & 0 \\ 0 & 0 & l(l+1)-1 \end{pmatrix} + \begin{pmatrix} 1 & -1 & -l(l+1) \\ -1 & 1 & l(l+1) \\ -2 & 2 & 1 \end{pmatrix}.
\end{equation}
where we used the identities~\eqref{EqTYIdentities}. It is then convenient to introduce a refined classification of indicial roots (Definition~\ref{DefTMInd}):

\begin{definition}[Indicial roots of scalar/vector type]
\label{DefTYIndRoot}
  Consider on $X:=[0,\infty)_\rho\times\Sph^2$ an operator $A\in\rho^\alpha\Diffb^m(X)$. We then say that $\lambda\in\C$ is a \emph{scalar}, resp.\ \emph{vector type $l$ indicial root} of $A$ if there exists a scalar, resp.\ vector type $l$ indicial solution $u\neq 0$ of $N(A)(\rho^\lambda u)=0$.
\end{definition}

For example, upon replacing $\rho\pa_\rho$ in~\eqref{EqTYSplitEx} by $\lambda$, the determinant of the resulting $3\times 3$ matrix vanishes for $\lambda=-l-1,-l,-l+1,l,l+1,l+2$; these $6$ numbers are therefore the scalar type $l$ indicial roots of $\wh{\ubar\Box}(0)$ on 1-forms.

For later use, we record the indicial family of $\wh{\ubar\delta^*}(0)$ for some scalar and vector types in the splittings~\eqref{EqTYSplit1}--\eqref{EqTYSplit2}:
\begin{subequations}
\begin{align}
\label{EqTYdels0}
  N_{\rms 0}\bigl(\wh{\ubar\delta^*}(0),\lambda\bigr) &= \begin{pmatrix} -\frac{\lambda}{2} & 0 \\ \frac{\lambda}{4} & -\frac{\lambda}{4} \\ 0 & \frac{\lambda}{2} \\ 1 & -1 \end{pmatrix}, \\
\label{EqTYdels1}
  N_{\rms 1}\bigl(\wh{\ubar\delta^*}(0),\lambda\bigr) &= \begin{pmatrix} -\frac{\lambda}{2} & 0 & 0 \\ \frac{\lambda}{4} & -\frac{\lambda}{4} & 0 \\ \frac12 & 0 & -\frac14(\lambda+1) \\ 0 & \frac{\lambda}{2} & 0 \\ 0 & \frac12 & \frac14(\lambda+1) \\ 1 & -1 & -1 \end{pmatrix}, \\
\label{EqTYdelv1}
  N_{\rmv 1}\bigl(\wh{\ubar\delta^*}(0),\lambda\bigr) &= \begin{pmatrix} -\frac14(\lambda+1) \\ \frac14(\lambda+1) \end{pmatrix}.
\end{align}
\end{subequations}
The $(6,3)$-entry of~\eqref{EqTYdels1} arises from the second identity in~\eqref{EqTYIdentities}. Note that for all other pure types $\bullet=\rms l,\rmv l$ ($l\geq 2$) and for all $\lambda\in\C$, the operator $N_\bullet(\wh{\ubar\delta^*}(0),\lambda)$ is injective; this follows either from an explicit computation, or from the geometric fact that the only stationary Killing 1-forms on Minkowski space are spacetime translations (scalar type $0$ and $1$) and spatial rotations (vector type $1$). Similarly, $N_{\rms 0}$, $N_{\rms 1}$, and $N_{\rmv 1}$ are injective when $\lambda$ is not equal to $0$, $0$, resp.\ $-1$.

\section{The Kerr family}
\label{SK}

Fix (subextremal) mass and angular momentum parameters
\begin{equation}
\label{EqKParam0}
  \bhm_0 > 0,\quad \bha_0\in\R^3,\ a_0:=|\bha_0|<\bhm_0.
\end{equation}
We will describe the metrics of Kerr black holes with parameters $(\bhm,\bha)$ close to $(\bhm_0,\bha_0)$, as needed for the proof of stability of Kerr with parameters~\eqref{EqKParam0}. We begin by defining the spacetime manifold and a (partial) compactification in~\S\ref{SsKMfd} before putting the Kerr metric on it in~\S\ref{SsKMet}. Pullbacks of Kerr by cut-off Lorentz boosts, which we will eventually utilize to eliminate the asymptotic linear momentum of the black hole upon perturbation, are analyzed in~\S\ref{SsKBo}.

\subsection{Compactified spacetime manifold}
\label{SsKMfd}

We largely follow \citeAF{\S\ref*{SsCM}}, except now spacelike infinity plays a more important role (as we shall solve the Einstein equation with initial data at $t=0$, roughly speaking).

Denote the standard coordinates on $\R^4$ by $t$ and $x=(x^1,x^2,x^3)$, and write $r=|x|$, $\omega=\frac{x}{|x|}\in\Sph^2$ for spatial polar coordinates. Define the following subsets of the radial compactification $\ol{\R^4}$:
\begin{enumerate}
\item $\fk^+\in\pa\ol{\R^4}$ denotes the ``north pole,'' i.e., the point with $\frac{1}{t}=0$, $\frac{x}{t}=0$, lying in the closure of $\{t>0\}$. (Equivalently, this is the limit of $(t,0)$ as $t\to\infty$.) We similarly denote by $\fk^-\in\pa\ol{\R^4}$ the ``south pole'' $\lim_{t\to-\infty}(t,0)\in\pa\ol{\R^4}$.
\item The ``light cone at infinity''
  \begin{equation}
  \label{EqKMfdYplus}
    Y^+\subset\pa\ol{\R^4}
  \end{equation}
  is the 2-sphere defined by $\frac{1}{t}=0$, $\frac{t-r}{t}=0$ in the closure of $\{t>0\}$.
\end{enumerate}

We first define the simplest compactification that carries the Kerr metric and which is suitable for the basic (edge-3b-based) analysis of geometric operators on the exact Kerr spacetime.

\begin{definition}[Minimal compactification]
\label{DefKMfdMin}
  As in~\S\usref{SssT3b}, we define
  \[
    \tilde M_0 := [\ol{\R^4}; \fk^-,\fk^+],
  \]
  with boundary hypersurfaces denoted $\sface$ (lift of $\pa\ol{\R^4}$) and $\cK^\pm$ (lift of $\fk^\pm$). We furthermore define
  \[
    M_0 := \cl_{\tilde M_0} \{ r\geq\bhm_0 \}.
  \]
\end{definition}

Next, we define a compactification on which radiation at null infinity can be captured:

\begin{definition}[Radiative compactification, I]
\label{DefKMfdRad}
  We define\footnote{Compared to \citeAF{Definition~\ref*{DefCMSpacetime}}, we do not blow up anything in $t<0$. This was done in \cite{HintzNonstat2} only to accommodate the global time-translation action on $\tilde M$ more cleanly (which is convenient, for example, when describing 3b-Sobolev spaces in~\citeAF{\S\ref*{SssMUK}}). In the present paper, we can focus on (a neighborhood of) $t\geq 0$. We also remark that what we call $\tilde M$ and $\tilde M_{\frac12}$ here and in Definition~\ref{DefKMfdRad2} below is called $\tilde M_1$ and $\tilde M$ in \cite{HintzNonstat2}, except for differences in blow-ups in $t<0$. The reason is that $\tilde M$ will be the main place of our present analysis, whereas $\tilde M_{\frac12}$ only plays a relatively minor role here in describing the edge-behavior of wave-type operators near null infinity.}
  \[
    \tilde M := [\ol{\R^4}; \fk^+, Y^+],\quad \upbeta\colon\tilde M\to\ol{\R^4}.
  \]
  We label its boundary hypersurfaces as follows:
  \begin{enumerate}
  \item \emph{spacelike infinity}\footnote{The terminology is accurate only in $\frac{t}{r}>-1$, which will be the region in which we shall exclusively work.} $I^0$ is the closure of $\{\frac{t}{r}<1\}$ in $\pa\tilde M$;
  \item \emph{null infinity} $\scri^+$ is the lift of $Y^+$;
  \item \emph{punctured future timelike infinity} $\iota^+$ is the closure of $\{t>0,\ \frac{1}{t}=0,\ \frac{r}{t}\in(0,1)\}$;
  \item the \emph{Kerr face} $\cK^+$ is the lift of $\fk^+$.
  \end{enumerate}
  We write $\rho_0$, $\rho_\sscri$, $\rho_+$, and $\rho_\cK$, respectively, for (local) defining functions of these boundary hypersurfaces. We furthermore define
  \begin{equation}
  \label{EqKMfdRadM}
    M := \cl_{\tilde M}\Bigl\{ r\geq\bhm_0,\ t\geq-\frac12 r \Bigr\},
  \end{equation}
  and denote its boundary hypersurfaces by $I^0$ (by a minor abuse of notation), $\scri^+$, $\cK^+$, $\iota^+$, and
  \[
    \Sigma_{\rm past} := \cl_M\Bigl\{t=-\frac12 r\Bigr\},\quad
    \Sigma_{\rm int} := \cl_M\{r=\bhm_0\}.
  \]
\end{definition}

\begin{figure}[!ht]
\centering
\includegraphics{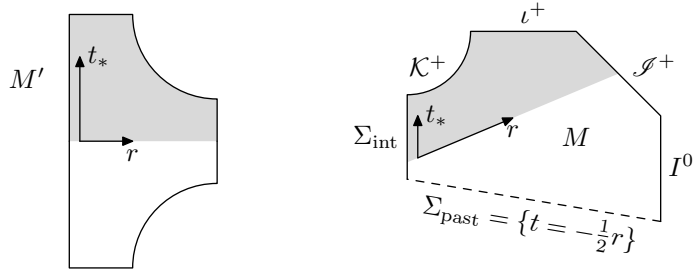}
\caption{Illustration of Lemma~\ref{LemmaKMfdSlice} (dropping the factor $\Sph^2$ of $X=[\bhm_0,\infty]_r\times\Sph^2$), with the domains identified in~\eqref{EqKMfdSlice} highlighted in gray.}
\label{FigKMfdSlice}
\end{figure}

See the right panel of Figure~\ref{FigKMfdSlice}. Concretely, we can cover $M$ with the following coordinate charts. (We shall not state the precise upper bounds on some coordinate functions; they are always to be taken such that the corresponding points in $\R^4$ satisfy the bounds~\eqref{EqKMfdRadM}.)
\begin{subequations}
\begin{enumerate}[label=(\alph*)]
\item\textit{Finite points}: $t,x$ (subject to the bounds~\eqref{EqKMfdRadM}), or $t,r,\omega$, or
  \begin{equation}
  \label{EqKMfdCoordInt}
    t_*=t-r,\quad r,\quad \omega.
  \end{equation}
\item\textit{Near the interior of spacelike infinity}:
  \begin{equation}
  \label{EqKMfdCoordI0}
    \tau=\frac{t}{r}\in[-\tfrac12,1),\quad
    \rho=\frac{1}{r}\in[0,\bhm_0^{-1}),\quad
    \omega=\frac{x}{|x|}\in\Sph^2.
  \end{equation}
\item\textit{Near $I^0\cap\scri^+$}:
  \begin{equation}
  \label{EqKMfdCoordI0Scri}
    \rho_0=\frac{1}{r-t+2}\in[0,\infty),\quad
    \rho_\sscri=\frac{r-t+2}{r}\in[0,1),\quad
    \omega\in\Sph^2.
  \end{equation}
\item\textit{Near the interior of null infinity}:
  \begin{equation}
  \label{EqKMfdCoordScri}
    t_*=t-r\in\R,\quad
    \rho_\sscri=\frac{1}{r}\geq 0,\quad
    \omega\in\Sph^2.
  \end{equation}
\item\textit{Near $\scri^+\cap\iota^+$}:
  \begin{equation}
  \label{EqKMfdCoordScriip}
    \rho_\sscri=\frac{t-r+2}{r}\in[0,\infty),\quad
    \rho_+=\frac{1}{t-r+2}\geq 0,\quad
    \omega\in\Sph^2.
  \end{equation}
\item\textit{Near the interior of punctured future timelike infinity}:
  \begin{equation}
  \label{EqKMfdCoordip}
    \rho_+=\frac{1}{r}\ \text{or}\ \frac{1}{t_*},\quad
    v=\frac{t_*}{r}\in(0,\infty), \quad
    \omega\in\Sph^2.
  \end{equation}
\item\textit{Near $\iota^+\cap\cK^+$}:
  \begin{equation}
  \label{EqKMfdCoordipK}
    \rho_+=\frac{1}{r},\quad
    \rho_\cK=\frac{r}{t_*},\quad
    \omega\in\Sph^2.
  \end{equation}
\item\textit{Near the interior of the Kerr face}:
  \begin{equation}
  \label{EqKMfdCoordK}
    \rho_\cK=t_*^{-1},\quad
    x\ (\text{or}\ r,\omega).
  \end{equation}
\end{enumerate}
\end{subequations}

The boundary hypersurface $\scri^+$ is the total space of a fibration $\upbeta|_{\scri^+}\colon\scri^+\to Y^+\cong\Sph^2$ of $\scri^+$ by copies of $\ol{\R_{t_*}}$; this fibration is given by the projection to the $\omega$-coordinate in the coordinates~\eqref{EqKMfdCoordI0Scri}--\eqref{EqKMfdCoordScriip}. See Figure~\ref{FigIMwc}.

\begin{definition}[Radiative compactification, II: square root blow-up]
\label{DefKMfdRad2}
  We define $\tilde M_{\frac12}:=[\tilde M;\scri^+,\frac12]$ to be the \emph{square root blow-up} of $\tilde M$ at $\scri^+$; that is, $\tilde M_{\frac12}=\tilde M$ as sets, but the smooth structure on $\tilde M_{\frac12}$ is the smallest one for which smooth functions on $\tilde M$ as well as the square root of a defining function of $\scri^+$ are smooth functions. We write $\upbeta_{\frac12}\colon\tilde M_{\frac12}\to\ol{\R^4}$ for the (total) blow-down map. We use the same notation for its boundary hypersurfaces as in Definition~\usref{DefKMfdRad}. We moreover define
  \[
    M_{\frac12} := \cl_{\tilde M_{\frac12}} \Bigl\{r\geq\bhm_0,\ t\geq-\frac12 r \Bigr\}.
  \]
\end{definition}

Essentially the same coordinate charts cover $M_{\frac12}$; the only necessary change is that one needs to replace every $\rho_\sscri$ by its square root
\begin{equation}
\label{EqKMfdCoordHalf}
  x_\sscri = \rho_\sscri^{\frac12},
\end{equation}
which is thus in each case a local defining function of $\scri^+\subset M_{\frac12}$. The boundary hypersurface $\scri^+\subset M_{\frac12}$ is again a $\ol{\R_{t_*}}$-fiber bundle with base $Y^+\cong\Sph^2$.

In the region $t_*\geq 1$, we will frequently use the Fourier transform in $t_*$, with $t_*$-translation-orbits being defined to have constant $x\in\R^3$. From the compactified perspective, this requires the introduction of
\begin{equation}
\label{EqKMfdX}
  X := \cl_{\ol{\R^3}} \{ r\geq\bhm_0 \} \subset \ol{\R^3},\quad \pa X:=X\cap\pa\ol{\R^3},
\end{equation}
and the following result:

\begin{lemma}[Spacetime slicing in $t_*\geq 1$]
\label{LemmaKMfdSlice}
  Define the manifold with corners
  \begin{equation}
  \label{EqKMfdSliceMfd}
    M' := \bigl[\,\ol{\R_{t_*}}\times X; \{\pm\infty\}\times\pa X\,\bigr].
  \end{equation}
  Then the identity map on $\R^4$ extends by continuity and density to a diffeomorphism
  \begin{equation}
  \label{EqKMfdSlice}
    \cl_{M'} \{t_*\geq 1\} \to \cl_M \{t_*\geq 1\}.
  \end{equation}
\end{lemma}

See Figure~\ref{FigKMfdSlice}. Observe that the domain~\eqref{EqKMfdSlice} is a subset of the manifold~\eqref{EqTFsfM} on which we studied the Fourier transform in~\S\ref{SsTF}.

\begin{proof}[Proof of Lemma~\usref{LemmaKMfdSlice}]
  This is essentially proved in \citeAF{\S\ref*{SssCPX}}; we give the proof here for completeness. Away from the front face of $M'$, the functions~\eqref{EqKMfdCoordScri} and~\eqref{EqKMfdCoordK} are local coordinates, so it suffices to consider a neighborhood of the front face. As local coordinates on $\ol\R\times X$ near $\{\infty\}\times X$, we use $\frac{1}{t_*+c}$ for any fixed $c\in\R$, $\rho=\frac{1}{r}$, and $\omega=\frac{x}{|x|}$. Near the lift of $\ol\R\times\pa X$ then, one can use as local coordinates on $M'$ the functions $\frac{t_*+2}{r}$, $\frac{1}{t_*+2}$, and $\omega$, which match~\eqref{EqKMfdCoordScriip}; and near the lift of $\{\infty\}\times X$, one can use $\frac{1}{r}$, $\frac{r}{t_*}$, $\omega$, which match~\eqref{EqKMfdCoordipK}.
\end{proof}

We will capture the asymptotic properties of metrics on $\R^4$ by regarding them as sections of smooth vector bundles over $\tilde M$ based on the scattering bundles defined in Definition~\ref{DefTMsc}:

\begin{notation}[Scattering bundles]
\label{NotKMfdTsc}
  We shall write
  \[
    \cT^*,\ \cT
  \]
  for the bundles $\Tsc^*\ol{\R^4}$ and $\Tsc\ol{\R^4}$, as well as for their pullbacks to $\tilde M$, resp.\ $\tilde M_{\frac12}$ along $\upbeta$, resp.\ $\upbeta_{\frac12}$. We similarly write
  \[
    \cT^*_X,\ \cT_X
  \]
  for the bundles $\Tsc_X^*\ol{\R^3}$ and $\Tsc_X\ol{\R^3}$ over $X$.
\end{notation}

For example, a fortiori, the Minkowski metric discussed in Example~\ref{ExKMfdMink} defines an element also of $\CI(\tilde M;S^2\cT^*)$ and $\CI(\tilde M_{\frac12};S^2\cT^*)$. (The Kerr metric will lie in these spaces over $M$ and $M_{\frac12}$, but not in $\CI(\ol{\R^4}\cap\{r\geq\bhm_0\};S^2\,\Tsc^*\ol{\R^4})$, as discussed in~\S\ref{SsKMet} below.)

\subsection{Kerr metrics; linearizations; initial data}
\label{SsKMet}

We recall \cite{KerrKerr,BoyerLindquistKerr} that in Boyer--Lindquist coordinates
\begin{equation}
\label{EqKMetBLCoord}
  \ft,\,r,\,\theta,\,\varphi,
\end{equation}
the Kerr metric and its dual, with parameters $\bhm>0$ and $a\in\R$, are given in terms of the functions
\[
  \mu_{\bhm,a}(r)=r^2-2\bhm r+a^2,\quad \varrho_a^2(r,\theta)=r^2+a^2\cos^2\theta
\]
by
\begin{align*}
  g_{\bhm,a} &= -\frac{\mu_{\bhm,a}}{\varrho_a^2}(\dd\ft-a\sin^2\theta\,\dd\varphi)^2 + \varrho_a^2\Bigl(\frac{\dd r^2}{\mu_{\bhm,a}}+\dd\theta^2\Bigr) + \frac{\sin^2\theta}{\varrho_a^2}\bigl((r^2+a^2)\dd\varphi-a\,\dd\ft\bigr)^2, \\
  \varrho_a^2 g_{\bhm,a}^{-1} &= -\frac{1}{\mu_{\bhm,a}}\bigl((r^2+a^2)\pa_\ft+a\pa_\varphi\bigr)^2 + \mu_{\bhm,a}\pa_r^2 + \pa_\theta^2 + \frac{1}{\sin^2\theta}(\pa_\varphi+a\sin^2\theta\,\pa_\ft)^2.
\end{align*}
This is well-defined for $r>0$ distinct from the roots $r_{\bhm,a}^\pm=\bhm\pm\sqrt{\bhm^2-a^2}$ of $\mu_{\bhm,a}$; the domain of outer communications is defined by
\begin{equation}
\label{EqKMetBLMfd}
  \cM_{\bhm,a} := \R_\ft \times (r_{\bhm,a}^+,\infty)_r \times \Sph^2_{\theta,\varphi}.
\end{equation}
We shall consider $(\bhm,a)$ close to $(\bhm_0,a_0)$ (see~\eqref{EqKParam0}). We first introduce new coordinates that on the one hand are adapted to null infinity and, on the other hand, in which the metric extends past the event horizon $r=r_{\bhm,a}^+$. (Our presentation is a variation of that in \citeAF{\S\ref*{SsTsK}}; we take care here to make all constructions smooth in $(\bhm,a)$.) Fix cutoff functions
\begin{alignat*}{2}
  \chi_{\cH^+}&\in\CIc([0,4\bhm_0)),&\ \ \chi_{\cH^+}|_{[0,3\bhm_0]}&=1,  \\
  \chi_\infty&\in\CI(\R),&\ \ \chi_\infty|_{[0,4\bhm_0]}&=0,\ \ \chi_\infty|_{[5\bhm_0,\infty)}=1.
\end{alignat*}
with values in $[0,1]$. Let us then set
\begin{equation}
\label{EqKMetTstar}
\begin{alignedat}{2}
  \tilde t_* &:= \ft + T_{\bhm,a}(r),&\quad T_{\bhm,a}(r) &:= \int_{4\bhm_0}^r \bigl(\chi_{\cH^+}(r')-\chi_\infty(r')\bigr)\frac{r'{}^2+a^2}{\mu_{\bhm,a}(r')}\,\dd r', \\
  \phi &:= \varphi + \Phi_{\bhm,a}(r),&\quad \Phi_{\bhm,a}(r) &:= \int_{4\bhm_0}^r \chi_{\cH^+}(r')\frac{a}{\mu_{\bhm,a}(r')}\,\dd r'.
\end{alignedat}
\end{equation}
Note that $\phi=\varphi$ for $r\geq 4\bhm_0$. Moreover, $T_{\bhm,a}(r)+r_*=c+\cO(r^{-1})$ for some $(\bhm,a)$-dependent constant $c$, where we introduce the \emph{tortoise coordinate}
\begin{equation}
\label{EqKMetTortoise}
  r_* := r + 2\bhm\log(r-2\bhm).
\end{equation}
(In fact, $T_{\bhm,a}(r)+r_*$ equals $c$ plus $r^{-1}$ times a smooth function of $r^{-1}$ near $r^{-1}=0$.) We then compute
\begin{align*}
  g_{\bhm,a} &= -\frac{\mu_{\bhm,a}}{\varrho_a^2}(\dd\tilde t_*-a\sin^2\theta\,\dd\phi)^2 + \Bigl[\frac{(1-\chi_{\cH^+}^2)\varrho_a^2}{\mu_{\bhm,a}} - \chi_\infty^2(r^2+a^2)^2\frac{\mu_{\bhm,a}-a^2\sin^2\theta}{\varrho_a^2\mu_{\bhm,a}^2}\Bigr]\dd r^2 + \varrho_a^2\,\dd\theta^2 \\
    &\qquad + 2\chi_{\cH^+}(\dd\tilde t_*-a\sin^2\theta\,\dd\phi)\otimes_s\dd r + \frac{\sin^2\theta}{\varrho_a^2}\bigl((r^2+a^2)\dd\phi-a\,\dd\tilde t_*\bigr)^2 \\
    &\qquad - 2\chi_\infty\Bigl[ \frac{(r^2+a^2)(\mu_{\bhm,a}-a^2\sin^2\theta)}{\varrho_a^2\mu_{\bhm,a}}\,\dd\tilde t_*\,\dd r + \frac{2 a\bhm r(r^2+a^2)\sin^2\theta}{\varrho_a^2\mu_{\bhm,a}}\dd\phi\,\dd r\Bigr], \\
  \varrho_a^2 g_{\bhm,a}^{-1} &= -\frac{1-\chi_{\cH^+}^2}{\mu_{\bhm,a}}\bigl((r^2+a^2)\pa_{\tilde t_*}+a\pa_\phi\bigr)^2 + \mu_{\bhm,a}\pa_r^2 + \pa_\theta^2 + \frac{1}{\sin^2\theta}(\pa_\phi+a\sin^2\theta\,\pa_{\tilde t_*})^2 \\
    &\qquad + 2\chi_{\cH^+}\bigl((r^2+a^2)\pa_{\tilde t_*}+a\pa_\phi\bigr)\otimes_s\pa_r - 2\chi_\infty(r^2+a^2)\pa_{\tilde t_*}\otimes_s\pa_r + \chi_\infty^2\frac{(r^2+a^2)^2}{\mu_{\bhm,a}}\pa_{\tilde t_*}^2.
\end{align*}
These expressions are analytic for $r_{\bhm,a}^+<r<3\bhm_0$, and hence $g_{\bhm,a}$ extends analytically to a (still Ricci-flat) Lorentzian metric in the region $r>r_{\bhm,a}^-$; we only extend it to $r\geq\bhm_0$ for the sake of concreteness.

\begin{lemma}[Time function $t_*$]
\label{LemmaKMetTime}
  There exists a function $\tilde T(r)\in\CI([\bhm_0,\infty))$, which for some constants $C'$ and $C>0$ satisfies $\tilde T(r)=C'-C r^{-1}$ for large $r$, such that the level sets of
  \begin{equation}
  \label{EqKMetTime}
    t_*:=\tilde t_*+\tilde T(r)
  \end{equation}
  are spacelike for the metric $g_{\bhm,a}$ for all $(\bhm,a)$ in a neighborhood of $(\bhm_0,a_0)$ in $(0,\infty)\times[0,\infty)$, and such that moreover $t_*\leq\ft$ everywhere.
\end{lemma}
\begin{proof}
  We first compute, for $r>r_{\bhm_0,a_0}^+$,
  \begin{align*}
    \mu_{\bhm_0,a_0}\varrho_{a_0}^2 g_{\bhm_0,a_0}^{-1}(\dd\ft,\dd\ft) &= -\bigl((r^2+a^2)^2-\mu_{\bhm_0,a_0}a_0^2\sin^2\theta\bigr) \\
      &\leq -\bigl((r^2+a_0^2)^2-(r^2-2\bhm_0 r+a_0^2)a_0^2\bigr) = -(r^4+r^2 a_0^2+2 a_0^2\bhm_0 r) < 0.
  \end{align*}
  Thus, one may take $\tilde T(r)$ to be equal to $-T_{\bhm_0,a_0}(r)$ on any fixed compact subset of $(r_{\bhm_0,a_0}^+,\infty)$, which we choose to contain $[3\bhm_0,5\bhm_0]$. Outside of this interval, we have $\chi_{\cH^+}=1$ and $\chi_\infty=0$ or vice versa. For $r\leq 3\bhm_0$ and using $\dd t_*=\dd\tilde t_*+\tilde T'\,\dd r$, we compute
  \begin{equation}
  \label{EqKMetTimeNorm}
    \varrho_{a_0}^2 g_{\bhm_0,a_0}^{-1}(\dd t_*,\dd t_*) = 2(r^2+a_0^2)\tilde T' + \mu_{\bhm_0,a_0}\tilde T'{}^2 + a_0^2\sin^2\theta.
  \end{equation}
  Where $\tilde T=-T_{\bhm_0,a_0}$, so $\tilde T'=-\frac{r^2+a_0^2}{\mu_{\bhm_0,a_0}}$, this is negative by the previous computation; more generally, when $\mu_{\bhm_0,a_0}>0$, the negativity of~\eqref{EqKMetTimeNorm} holds for a non-empty open interval centered around $-T'_{\bhm_0,a_0}$. For $\mu_{\bhm_0,a_0}\leq 0$ (i.e., for $r\leq r_{\bhm_0,a_0}^+$, and still for $r\geq\bhm_0$), the negativity of~\eqref{EqKMetTimeNorm} holds when $\tilde T'$ is any constant $<-\frac{a_0^2}{2(\bhm_0^2+a_0^2)}$, and then it holds also on a small neighborhood of $[\bhm_0,r_{\bhm,a}^+]$. Altogether, we can thus make a smooth choice of $\tilde T$ interpolating between a constant for $r$ near $[\bhm_0,r_{\bhm_0,a_0}^+]$ and the earlier choice $-T_{\bhm_0,a_0}$ on $[3\bhm_0,5\bhm_0]$ such that~\eqref{EqKMetTimeNorm} is negative for $r\leq 3\bhm_0$. For the same choice of $\tilde T$, and for any fixed radius $R$, the differential $\dd t_*$ is then also timelike for $g_{\bhm,a}$ for $r\leq R$ when $(\bhm,a)$ is close enough (depending on $R$) to $(\bhm_0,a_0)$.

  Turning to $r\geq 5\bhm_0$, we now work directly with $g_{\bhm,a}$ and compute
  \[
    \varrho_a^2 g_{\bhm,a}^{-1}(\dd t_*,\dd t_*) = -2(r^2+a^2)\tilde T' + \mu_{\bhm,a}\tilde T'{}^2 + a^2\sin^2\theta,
  \]
  which is negative provided
  \begin{equation}
  \label{EqKMetTime2Int}
    \Bigl|\tilde T'+\frac{r^2+a^2}{\mu_{\bhm,a}}\Bigr|<\mu_{\bhm,a}^{-1}\sqrt{(r^2+a^2)^2-a^2(r^2-2\bhm r+a^2)}=1+\cO(r^{-1}),\quad r\to\infty.
  \end{equation}
  When $\tilde T'=C r^{-2}$, we have
  \begin{equation}
  \label{EqKMetTime2}
    \varrho_a^2 g_{\bhm,a}^{-1}(\dd t_*,\dd t_*) = -2 C(1+a^2 r^{-2}) + a^2\sin^2\theta + C^2 (r^{-2}-2\bhm r^{-3}+a r^{-4}),
  \end{equation}
  which for sufficiently large $C>0$ is thus negative for all $r\geq 5\bhm_0$ and for all $(\bhm,a)$ near $(\bhm_0,a_0)$. Gluing together this choice of $\tilde T'$ with the already fixed choice of $\tilde T'$ for $r\leq 5\bhm_0$ in a manner compatible with~\eqref{EqKMetTime2Int} finishes the construction. The constant $C'$ in the statement is a constant of integration when integrating $\tilde T'$ from $r=5\bhm_0$ to obtain $\tilde T$; and shifting $t_*$ by a constant does not affect its timelike nature, but can be used to guarantee $t_*\leq\ft$.
\end{proof}

The dual metric in the coordinates $t_*,r,\theta,\phi$ thus reads
\begin{equation}
\label{EqKMetFinal}
\begin{split}
  \varrho_a^2 g_{\bhm,a}^{-1} &= -\frac{1-\chi_{\cH^+}^2}{\mu_{\bhm,a}}\bigl((r^2+a^2)\pa_{t_*}+a\pa_\phi\bigr)^2 + \mu_{\bhm,a}\pa_r^2 + \pa_\theta^2 + \frac{1}{\sin^2\theta}(\pa_\phi+a\sin^2\theta\,\pa_{t_*})^2 \\
    &\qquad + 2\chi_{\cH^+}\bigl( (r^2+a^2)\pa_{t_*}+a\pa_\phi\bigr)\otimes_s(\pa_r+\tilde T'\pa_{t_*}) - 2\chi_\infty(r^2+a^2)\pa_{t_*}\otimes_s(\pa_r+\tilde T'\pa_{t_*}) \\
    &\qquad + \chi_\infty^2\frac{(r^2+a^2)^2}{\mu_{\bhm,a}}\pa_{t_*}^2.
\end{split}
\end{equation}

For the statement of initial value problems, we also need to discuss spacelike hypersurfaces that asymptote to constant $t$ hypersurfaces at large $r$:

\begin{lemma}[Time function $t_\IVP$]
\label{LemmaKMetIVP}
  There exists a function $t_\IVP$ with the following properties:
  \begin{enumerate}
  \item $t_\IVP=t_*$ for $r\leq 5\bhm_0$;
  \item\label{ItKMetIVPShift} $t_\IVP=t+C$, $t:=t_*+r$, for large $r$ and for some constant $C$;
  \item $\dd t_\IVP$ is timelike for $g_{\bhm,a}$ when $(\bhm,a)$ is sufficiently close to $(\bhm_0,a_0)$;
  \item $t_\IVP-t_*$ is a function of $r$ only;
  \item $t_\IVP\leq t$.
  \end{enumerate}
\end{lemma}
\begin{proof}
  Note that
  \begin{equation}
  \label{EqKMetIVPt}
    t=\tilde t_*+\tilde T(r)+r=\ft+T_{\bhm,a}(r)+r+\tilde T(r) = \ft + \cO(\log r)
  \end{equation}
  satisfies $\dd t=\dd\ft+\cO(r^{-1})\dd r$; therefore, since $\dd\ft$ is timelike for large $r$, so is $\dd t$. One can thus set $t=t_*+\breve T(r)$ where $\breve T$ is chosen such that $\breve T(r)=0$ for $r\leq 5\bhm_0$ and $\breve T'(r)=1$ for large $r$ while retaining the negativity of $g_{\bhm,\bha}(\dd t,\dd t)$ (which holds when $\breve T'(r)\in[\delta,2-\delta]$ for $\delta>0$ fixed and $r$ large, cf.\ \eqref{EqKMetTime2Int}).
\end{proof}

\begin{figure}[!ht]
\centering
\includegraphics{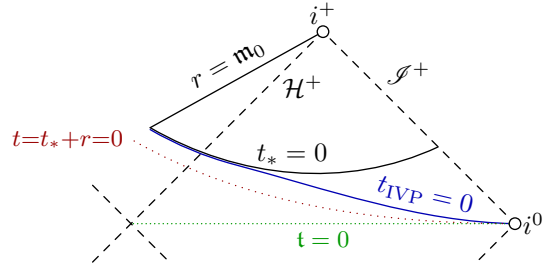}
\caption{Illustration of the functions $t_*$ from Lemma~\ref{LemmaKMetTime} and $t_\IVP$ from Lemma~\ref{LemmaKMetIVP}. We also show the $0$-level sets of $t=t_*+r$ and of the Boyer--Lindquist time coordinate $\ft$.}
\label{FigKMetTime}
\end{figure}

Figure~\ref{FigKMetTime} summarizes the time functions introduced above.

We now transfer the family of Kerr metrics with parameters $(\bhm,\bha)$ where $(\bhm,|\bha|)$ is near $(\bhm_0,|\bha_0|)$ (here $\bha_0\in\R^3$ and $0\leq|\bha_0|<\bhm_0$) to the (fixed) spacetime manifold
\[
  \cM := \R_{t_*}\times[\bhm_0,\infty)\times\Sph^2.
\]
\begin{definition}[Kerr metrics on $\cM$]
\label{DefKMetcM}
  Let $\eps_{\rm param}>0$ be such that the conclusions of Lemmas~\usref{LemmaKMetTime} and \usref{LemmaKMetIVP} and the inequality $r_{\bhm,a}^->\bhm_0$ hold when $|(\bhm,a)-(\bhm_0,a_0)|<\eps_{\rm param}$. Then for all $\bhm>0$ and $\bha\in\R^3$ with $|(\bhm,a)-(\bhm_0,a_0)|<\eps_{\rm param}$ where $a=|\bha|$ and $a_0=|\bha_0|$, we define the Ricci-flat Lorentzian metric $g_{\bhm,\bha}$ on $\cM$ as follows. Fix polar coordinates $\theta\in(0,\pi)$ and $\phi\in(0,2\pi)$ on the unit sphere $\Sph^2\subset\R^3$ such that the ``north pole'' $\theta=0$ is equal to $\bha/|\bha|$ when $\bha\neq 0$. Define then $g_{\bhm,\bha}$ via its dual metric by~\eqref{EqKMetFinal}, i.e., the metric $g_{\bhm,a}$ in the coordinates $t_*$ (from~\eqref{EqKMetTstar} and \eqref{EqKMetTime}), $r$, $\theta$, and $\phi$ (identified with~\eqref{EqKMetTstar}). We shall also use the notation
  \[
    g_b := g_{\bhm,\bha},\quad b=(\bhm,\bha).
  \]
\end{definition}

Thus, $g_{\bhm,\bha}$ is the metric of a Kerr black hole with mass $\bhm$ and specific angular momentum \emph{vector} $\bha\in\R^3$. (When changing only the direction of $\bha$ but not its magnitude, one obtains isometric metrics that are related by a spatial rotation mapping one angular momentum vector to the other.)

We shall prove momentarily that $g_{\bhm,\bha}$ is smooth (which only requires an argument near the poles $\theta=0,\pi$ of $\Sph^2$) and depends smoothly on $(\bhm,\bha)$ (which is only non-trivial near $\bha=0$). The asymptotic behavior of $g_{\bhm,\bha}$ will be described by the Schwarzschild metric modulo $\cO(r^{-2})$, and by the Minkowski metric modulo $\cO(r^{-1})$. More precisely:

\begin{definition}[Reference metrics]
\label{DefKMetRef}
  Recall $\slg=\dd\theta^2+\sin^2\theta\,\dd\phi^2$ and define the \emph{reference Schwarzschild metric} by
  \begin{equation}
  \label{EqKMetRefSchw}
    g_\bhm := -\Bigl(1-\frac{2\bhm}{r}\Bigr)\dd t_*^2 - 2\,\dd t_*\,\dd r + r^2\slg,\quad
    g_\bhm^{-1} = -2\pa_{t_*}\otimes_s\pa_r + \Bigl(1-\frac{2\bhm}{r}\Bigr)\pa_r^2 + r^{-2}\slg^{-1}.
  \end{equation}
  Define the Minkowski metric by $\ubar g=g_0=-\dd t_*^2-2\,\dd t_*\,\dd r+r^2\slg$, with dual metric $\ubar g^{-1}=-2\pa_{t_*}\otimes_s\pa_r+\pa_r^2+r^{-2}\slg^{-1}$.
\end{definition}

\begin{prop}[Kerr metrics on the compactification]
\label{PropKMetCpt}
  Recall $M$ from~\eqref{EqKMfdRadM}. Then for all $b=(\bhm,\bha)$ that are $\eps_{\rm param}$-close to $b_0=(\bhm_0,\bha_0)$, we have
  \begin{equation}
  \label{EqKMetCpt}
    g_b:=g_{\bhm,\bha}\in\CI(M;S^2\cT^*),\quad
    g_b^{-1}\in\CI(M;S^2\cT),
  \end{equation}
  with smooth dependence on $b$. The metrics $g_\bhm$ and $\ubar g$ satisfy~\eqref{EqKMetCpt} as well, and
  \begin{equation}
  \label{EqKMetDiff}
    g_{\bhm,\bha} - g_\bhm \in r^{-2}\CI(M;S^2\cT^*),\quad
    g_\bhm - \ubar g = \frac{2\bhm}{r}\dd t_*^2 \in r^{-1}\CI(M;S^2\cT^*),
  \end{equation}
  so in particular $g_{\bhm,\bha}-\ubar g\in r^{-1}\CI(M;S^2\cT^*)$; similarly for the dual metrics (and then with $S^2\cT$ in place of $S^2\cT^*$).
\end{prop}
\begin{proof}
  We work with the dual metric as in~\eqref{EqKMetFinal}, and exploit the fact that smooth functions of $\rho=r^{-1}$ and $\omega\in\Sph^2$ are smooth on $M$. (This follows either by inspection of the coordinate charts~\eqref{EqKMfdCoordInt}--\eqref{EqKMfdCoordipK}, or more geometrically from the fact that the projection map $\R_{t_*}\times\R^3_x\to\R^3_x$ lifts to a smooth map $M\to X$.) Upon switching from the coordinates $t_*,r,\theta,\phi$ to $t,x$, the vector fields $\pa_{t_*}$ and $\pa_r$ become $\pa_t$ and $\pa_t+\frac{x}{|x|}\cdot\pa_x$, which lie in $\CI(M;\cT)$. Next, at a point $(t_*,x)$ in $\R_{t_*}\times\R^3$ with $|x|\geq\bhm_0$, we have
  \begin{equation}
  \label{EqKMetaExpr}
    a\pa_\phi = \nabla_{\bha \times x},\quad
    a\cos\theta = \bha\cdot \frac{x}{|x|},\quad
    a^2\sin^2\theta = |\bha|^2-\Bigl(\bha\cdot\frac{x}{|x|}\Bigr)^2
  \end{equation}
  where the polar coordinates $\theta,\phi$ are adapted to $\bha\in\R^3$ (with $a=|\bha|$) as in Definition~\ref{DefKMetcM}. Therefore,
  \[
    a\pa_\phi\in r\CI(M;\cT),\quad a\cos\theta,\ a^2\sin^2\theta\in\CI(M),
  \]
  with smooth dependence on $\bha\in\R^3$. This implies that also $\varrho_a^2\in r^2\CI(M)$ depends smoothly on $\bha$; and thus~\eqref{EqKMetCpt} (and their smooth dependence) follows upon writing the last two terms in the first line on the right-hand side of~\eqref{EqKMetFinal} as a sum of three terms,
  \[
    \slg^{-1} + 2 a\pa_\phi\otimes_s\pa_{t_*} + a^2\sin^2\theta\,\pa_{t_*}^2,
  \]
  all of which are smooth.

  In order to compute $g_{\bhm,\bha}^{-1}\bmod r^{-2}\CI(M;S^2\cT)$, we only need to work with large $r$ in~\eqref{EqKMetFinal}, so $\chi_{\cH^+}=0$ and $\chi_\infty=1$; dropping terms in $r^{-2}\CI(M;S^2\cT)$, we thus have
  \begin{align*}
    g_{\bhm,\bha}^{-1} &\equiv r^{-2}\Bigl(-\frac{1}{r^2-2\bhm r}(r^4\pa_{t_*}^2 + 2 r^2\pa_{t_*}\otimes_s a\pa_\phi) + (r^2-2\bhm r)\pa_r^2 + \slg \\
      &\quad \hspace{11em} + 2\pa_{t_*}\otimes_s a\pa_\phi - 2 r^2\pa_{t_*}\otimes_s\pa_r + \frac{r^4}{r^2-2\bhm r}\pa_{t_*}^2\Bigr) \equiv g_\bhm^{-1};
  \end{align*}
  we use here that the $\pa_{t_*}\otimes_s a\pa_\phi$-terms cancel modulo $r^{-2}\CI(M;S^2\cT)$, and the $\pa_{t_*}^2$-terms cancel as well. The claims involving $\ubar g$ follow by inspection of $g_\bhm$.
\end{proof}

Proposition~\ref{PropKMetCpt} allows to define:

\begin{definition}[Linearized Kerr metrics]
\label{DefKMetLin}
  Let $b=(\bhm,\bha)$ be $\eps_{\rm param}$-close to $b_0=(\bhm_0,\bha_0)$. For $\dot b=(\dot\bhm,\dot\bha)\in\R\times\R^3$, we then define the \emph{linearized Kerr metric} by
  \begin{equation}
  \label{EqKMetLin}
    \dot g_{\bhm,\bha}(\dot\bhm,\dot\bha) = \dot g_b(\dot b) := \frac{\dd}{\dd s}g_{b+s\dot b}\Big|_{s=0} \in r^{-1}\CI(M;S^2\cT^*).
  \end{equation}
\end{definition}

We also record, as a consequence of Proposition~\ref{PropKMetCpt}, that
\begin{equation}
\label{EqKMetLina}
  \dot g_{\bhm,\bha}(0,\dot\bha) \in r^{-2}\CI(M;S^2\cT^*).
\end{equation}

Lastly, let us describe the initial data of Kerr metrics. First, note that it is only in the ($\bhm$-dependent) coordinate
\begin{equation}
\label{EqKMetStaticTime}
  \ft:=t_*+r_*,\quad r_*=r+2\bhm\log(r-2\bhm),
\end{equation}
that $g_\bhm$ takes its standard static form $-(1-\frac{2\bhm}{r})\dd\ft^2+(1-\frac{2\bhm}{r})^{-1}\dd r^2+r^2\slg$; this differs from the first coordinate $t=t_*+r=\ft-(r_*-r)$ of $\R^4$ in the notation of~\S\ref{SsKMfd} by a logarithmic term. In the ($\bhm$-independent) coordinates $t,r$ on the other hand, one finds
\begin{equation}
\label{EqKMetSchwtr}
  g_\bhm = -\Bigl(1-\frac{2\bhm}{r}\Bigr)\,\dd t^2 - \frac{4\bhm}{r}\,\dd t\,\dd r + \Bigl(1+\frac{2\bhm}{r}\Bigr)\,\dd r^2 + r^2\slg.
\end{equation}

\begin{definition}[Geometric initial data]
\label{DefKMetData}
  For $b=(\bhm,\bha)$ close to $b_0=(\bhm_0,\bha_0)$, let us define the initial data of $g_b$ at $\Sigma_\IVP:=t_\IVP^{-1}(0)$ (see Lemma~\usref{LemmaKMetIVP}), identified with $X$ (via identifying the Cartesian coordinates $x$ on $X$ and $\Sigma_\IVP$) to be
  \[
    \gamma_b,\,k_b \in \CI(X;S^2\cT^*_X),\quad
    \gamma_b(V,W):=g_b(V,W),\ k_b(V,W):=g_b(\nabla_V\nu_b,W),\ \ V,W\in T X^\circ,
  \]
  using Notation~\usref{NotKMfdTsc}. Here $\nu_b=-\nabla t_\IVP/\sqrt{-g_b(\nabla t_\IVP,\nabla t_\IVP)}$ is the future unit normal at $\Sigma_\IVP$ with respect to $g_b$ (with $\nabla$ being the gradient for $g_b$). We moreover denote by
  \[
    \gamma_\bhm,\ k_\bhm
  \]
  the initial data of $g_\bhm$.
\end{definition}

We (prove and) strengthen the regularity and decay properties of $\gamma_b$ and $k_b$ as follows.

\begin{lemma}[Asymptotics of initial data]
\label{LemmaKMetData}
  We have
  \begin{equation}
  \label{EqKMetData1}
  \begin{split}
    \gamma_\bhm &\equiv \Bigl(1+\frac{2\bhm}{r}\Bigr)\dd r^2+r^2\slg \bmod \CIc(X^\circ;S^2 T^*X^\circ), \\
    \gamma_{\bhm,\bha} &\equiv \gamma_\bhm \bmod r^{-2}\CI(X;S^2\cT^*_X).
  \end{split}
  \end{equation}
  Furthermore,
  \begin{equation}
  \label{EqKMetData2}
  \begin{split}
    k_\bhm &\in r^{-2}\CI(X;S^2\cT^*_X), \\
    k_{\bhm,\bha}-k_\bhm&\in r^{-3}\CI(X;S^2\cT^*_X).
  \end{split}
  \end{equation}
\end{lemma}
\begin{proof}
  The description of $\gamma_\bhm$ follows from~\eqref{EqKMetSchwtr} and the fact that $t_\IVP=t$ for large $r$ (and thus outside of some compact subset of $X^\circ$). The second line of~\eqref{EqKMetData1} follows from $g_{\bhm,\bha}-g_\bhm\in r^{-2}\CI(M;S^2\cT^*)$.

  For the first line of~\eqref{EqKMetData2}, we only need to consider large $r$. Write~\eqref{EqKMetSchwtr} as
  \[
    g_\bhm = -\Bigl(1-\frac{2\bhm}{|x|}\Bigr)\dd t^2 - 4\bhm|x|^{-2}\,\dd t\otimes_s(x\cdot\dd x) + \dd x^2 + 2\bhm|x|^{-3}(x\cdot\dd x)^2.
  \]
  Since the future $g_\bhm$-unit normal at $\Sigma_\IVP$ is $\nu_\bhm=f\pa_t$ for $f\in\CI(X)$ (in fact, $f\equiv 1\bmod r^{-1}\CI(X)$), it suffices to show that the Christoffel symbols of $g_\bhm$ in the coordinates $(z^0,z^1,z^2,z^3)=(t,x^1,x^2,x^3)$ satisfy $\Gamma_{k i 0}\in r^{-2}\CI(X)$, which in turn follows from the membership $\pa_{x^i}g_\bhm(\pa_t,\pa_{x^k})=\pa_{x^i}(-2\bhm|x|^{-2}x^k)\in r^{-2}\CI(X)$. Observe next that~\eqref{EqKMetDiff} implies $\nu_b-\nu_\bhm\in r^{-2}\CI(X;S^2\cT^*)$; since coordinate derivatives map $r^{-2}\CI\to r^{-3}\CI$, this yields the second line of~\eqref{EqKMetData2}.
\end{proof}

\begin{rmk}[Decay of initial data]
\label{RmkKMetDataDecay}
  The initial data for which we prove stability will be equal to $\gamma_\bhm\bmod\cO(r^{-1-\eps})$ and $k_\bhm\bmod\cO(r^{-2-\eps})$ where $\eps>0$. In particular, the contributions of the angular momentum parameter $\bha$ are sub-leading in this sense. In practice, this means that near $I^0\cup\scri^+$, it suffices to describe dynamical spacetime metrics relative to $g_\bhm$; only near $\cK^+$ do we need the Kerr metric as a reference.
\end{rmk}

\subsection{Approximate double null coordinates near \texorpdfstring{$\scri^+$}{null infinity}}
\label{SsKNull}

The analysis of metric perturbations near $\scri^+$ (in generalized harmonic gauge) is most efficiently done in splittings of $S^2\cT^*$ defined using the Schwarzschildean double null coordinates \cite{LindbladRodnianskiGlobalStability,HintzVasyMink4,HintzMink4Gauge}
\begin{equation}
\label{EqKNull}
  x^0 := t_*+2 r_*,\quad
  x^1 := t_* = t-r.
\end{equation}
(In terms of the standard time coordinate $\ft$ in the static region of Schwarzschild from~\eqref{EqKMetStaticTime}, these take on the more familiar form $x^0=\ft+r_*$ and $x^1=\ft-r_*$. These are moreover related to the Minkowskian double null coordinates from~\eqref{EqTYMink01} via $x^0=\ubar x^0+2(r_*-r)$ and $x^1=\ubar x^1$, so $\dd x^0=\dd\ubar x^0+\cO(r^{-1})$ and $\dd x^1=\dd\ubar x^1$.) Note that these coordinates depend on the parameter $\bhm$, although we do not make this explicit in the notation; in practice, $\bhm$ will be fixed to be the ADM mass of the initial data. The mass $\bhm$ Schwarzschild metric~\eqref{EqKMetRefSchw} then reads
\begin{equation}
\label{EqKNullSchw}
  g_\bhm = -\Bigl(1-\frac{2\bhm}{r}\Bigr)\dd x^0\,\dd x^1 + r^2\slg,\quad
  g_\bhm^{-1} = -4\Bigl(1-\frac{2\bhm}{r}\Bigr)^{-1}\pa_0\otimes_s\pa_1 + r^{-2}\slg^{-1},
\end{equation}
where we write $\pa_0=\pa_{x^0}$ and $\pa_1=\pa_{x^1}$. Note that the differentials
\begin{equation}
\label{EqKNullDiff}
  \dd x^0 = \dd t_* + 2\Bigl(1-\frac{2\bhm}{r}\Bigr)^{-1}\,\dd r = \dd t + \frac{1+\frac{2\bhm}{r}}{1-\frac{2\bhm}{r}}\,\dd r,\quad
  \dd x^1 = \dd t_* = \dd t-\dd r \in \CI(M;\cT^*)
\end{equation}
are smooth. For later use, we moreover record:

\begin{lemma}[Computations]
\label{LemmaKNullComp}
  Passing from $t_*,r$ to the coordinates $\rho_0=\frac{1}{2-t_*}$, $\rho_\sscri=\frac{t_*-2}{r}$ from~\eqref{EqKMfdCoordI0Scri} and $\rho_\sscri=\frac{t_*+2}{r}$, $\rho_+=\frac{1}{t_*+2}$ from~\eqref{EqKMfdCoordScriip}, respectively, we have
  \begin{equation}
  \label{EqKNullComp}
  \begin{alignedat}{2}
    \pa_0 &= \frac12\Bigl(1-\frac{2\bhm}{r}\Bigr)\pa_r &&= -\frac12\rho_0\rho_\sscri\Bigl(1-\frac{2\bhm}{r}\Bigr)\rho_\sscri\pa_{\rho_\sscri} \\
      &&&= -\frac12\rho_+\rho_\sscri\Bigl(1-\frac{2\bhm}{r}\Bigr)\rho_\sscri\pa_{\rho_\sscri}; \\
    \pa_1 &= \pa_{t_*}-\frac12\Bigl(1-\frac{2\bhm}{r}\Bigr)\pa_r &&= \rho_0\Bigl[\rho_0\pa_{\rho_0}-\Bigl(1-\frac12\rho_\sscri\Bigl(1-\frac{2\bhm}{r}\Bigr)\Bigr)\rho_\sscri\pa_{\rho_\sscri}\Bigr] \\
      &&&= \rho_+\Bigl[-\rho_+\pa_{\rho_+} + \Bigl(1+\frac12\rho_\sscri\Bigl(1-\frac{2\bhm}{r}\Bigr)\Bigr)\rho_\sscri\pa_{\rho_\sscri}\Bigr].
  \end{alignedat}
  \end{equation}
  In particular,
  \begin{equation}
  \label{EqKNullComp2}
    \pa_0 r=\frac12\Bigl(1-\frac{2\bhm}{r}\Bigr),\quad
    \pa_1 r=-\frac12\Bigl(1-\frac{2\bhm}{r}\Bigr),\quad
    \pa_0 \in \rho_0\rho_\sscri\rho_+\Vb(M),\quad
    \pa_1 \in \rho_0\rho_+\Vb(M).
  \end{equation}
\end{lemma}
\begin{proof}
  Since $\pa_0$ and $\pa_1$ is the dual basis of $\dd x^0$ and $\dd x^1$ (on the 2-dimensional subspace of $\cT^*$ spanned by $\dd x^0$ and $\dd x^1$ and its dual), one obtains the first expressions in~\eqref{EqKNullComp} from~\eqref{EqKNullDiff}; the remaining expressions follow by direct computation.
\end{proof}

In view of the smoothness of~\eqref{EqKNullDiff}, we can introduce the smooth bundle splittings
\begin{align}
\label{EqKNullT}
  \cT^* &= \la\dd x^0\ra \oplus \la\dd x^1 \ra \oplus r T^*\Sph^2, \\
\label{EqKNullS2T}
\begin{split}
  S^2\cT^* &= \la(\dd x^0)^2\ra \oplus \la 2\,\dd x^0\,\dd x^1 \ra \oplus (2\,\dd x^0 \otimes_s r T^*\Sph^2) \\
    &\qquad \oplus \la (\dd x^1)^2\ra \oplus (2\,\dd x^1\otimes_s r T^*\Sph^2) \oplus r^2 S^2 T^*\Sph^2
\end{split}
\end{align}
in the region $r>2\bhm$. As before, the notation means the following: the right-hand side of~\eqref{EqKNullT} is $\ubar\R\oplus\ubar\R\oplus T^*\Sph^2$ as a vector bundle, with a section $(\omega_0,\omega_1,\slomega)$ being identified with the 1-form $\omega_0\,\dd x^0+\omega_1\,\dd x^1+r\slomega$ on $\R_{x^0}\times\R_{x^1}\times\Sph^2$. The meaning of~\eqref{EqKNullS2T} is similar; thus, for example, the Schwarzschild metric is
\begin{equation}
\label{EqKNullSchwComp}
  g_\bhm=\Bigl(0,-\frac12\Bigl(1-\frac{2\bhm}{r}\Bigr),0,0,0,\slg\Bigr).
\end{equation}
We will often split $S^2 T^*\Sph^2$ into the pure trace and traceless parts, thus refining~\eqref{EqKNullS2T} to
\begin{equation}
\label{EqKNullS2TFine}
\begin{split}
  S^2\cT^* &= \la(\dd x^0)^2\ra \oplus \la 2\,\dd x^0\,\dd x^1 \ra \oplus (2\,\dd x^0 \otimes_s r T^*\Sph^2) \\
    &\qquad \oplus \la (\dd x^1)^2\ra \oplus (2\,\dd x^1\otimes_s r T^*\Sph^2) \oplus \la r^2\slg\ra \oplus r^2\ker\sltr.
\end{split}
\end{equation}
To efficiently describe components of tensors in the splittings~\eqref{EqKNullT}--\eqref{EqKNullS2T}, we introduce:

\begin{definition}[Rescaled components]
\label{DefKNullComp}
  For $\mu_1,\ldots,\mu_N\in\{0,1,2,3\}$, we write $\fs(\mu_1,\ldots,\mu_N):=\#\{j\in\{1,\ldots,N\}\colon \mu_j\in\{2,3\}\}$ for the number of spherical indices. For a $(p,q)$-tensor $T$ on $\R^4\cap\{r>2\bhm\}$, and writing $x^2,x^3$ for coordinates on $\Sph^2$, we then write
  \[
    T^{\bar\nu_1\cdots\bar\nu_p}_{\bar\mu_1\cdots\bar\mu_q} := r^{\fs(\nu_1,\ldots,\nu_p)-\fs(\mu_1,\ldots,\mu_q)} T^{\nu_1\cdots\nu_p}_{\mu_1\cdots\mu_q}
  \]
  for its rescaled components.
\end{definition}

For example, for $\omega=(\omega_0,\omega_1,\slomega)$, we have $\omega_{\bar 0}=\omega_0$, $\omega_{\bar 1}=\omega_1$, and $\omega_{\bar a}=\slomega_a$ for $a=2,3$. Furthermore, we have
\[
  (g_{\bhm,\bha})_{\bar\mu\bar\nu} - (g_\bhm)_{\bar\mu\bar\nu} \in r^{-2}\CI(M)
\]
as a consequence of Proposition~\ref{PropKMetCpt}.

\subsection{Pullback along Lorentz boosts}
\label{SsKBo}

A Lorentz boost on Minkowski space in the direction $\vec v\in\R^3$ is given by the time $1$ flow of the vector field $t\vec v\cdot\pa_x+(x\cdot\vec v)\pa_t$. Since $x\cdot\vec v=r\scal$ where $\scal=\frac{x}{|x|}\cdot\vec v\in\scalspace_1$, the 1-form dual to this vector field is given by
\begin{equation}
\label{EqKBo1form}
  \ubar\omega_{\rms 1}^{(0),\leq 1}(\scal) := t\,\dd(r\scal) - r\scal\,\dd t,\quad \scal\in\scalspace_1
\end{equation}
Since Lorentz boosts are isometries for the Minkowski metric $\ubar g$, we have $\delta_{\ubar g}^*\ubar\omega_{\rms 1}^{(0),\leq 1}(\scal)=0$. (Recall here that $\delta_g^*\omega=\frac12\cL_{\omega^\sharp}g$ where $\cL$ denotes the Lie derivative and $\omega^\sharp=g^{-1}(\omega,\cdot)$.) In view of the asymptotic equality of the Kerr and Minkowski metrics (see~\eqref{EqKMetDiff}), one should expect pullbacks along the flow of a vector field dual to $\ubar\omega_{\rms 1}^{(0),\leq 1}(\scal)$ to have only a mild effect on the metric asymptotics as $r\to\infty$. On the other hand, the Lie derivative of the Kerr metric along a boost typically grows like $t$ in spatially compact regions. We shall thus only study the behavior of the Kerr metric under pullbacks by (time $1$ flows of) boosts that are cut off to the exterior of a timelike cone around the black hole. (Recall from~\S\ref{SssINElim} that we shall use pullbacks along boosts to make the final black hole have vanishing linear momentum.)

We first fix a number of cutoff functions.

\begin{definition}[Cutoff functions]
\label{DefKBoCutoff}
  We fix functions $\chi_\cK^\flat,\chi_\cK,\chi_\cK^\sharp\in\CI(M)$ with support in $t>50\bhm_0$ such that
  \[
    \chi_\cK^\flat=1\ \text{for}\ r<\frac{t}{10},\quad
    \supp\chi_\cK^\flat\subset\chi_\cK^{-1}(1),\quad
    \supp\chi_\cK\subset(\chi_\cK^\sharp)^{-1}(1),\quad
    \supp\chi_\cK^\sharp\subset\Bigl\{r<\frac{t}{2}\Bigr\}.
  \]
  Recalling Lemma~\usref{LemmaKMetIVP}, we furthermore fix functions $\chi_\IVP,\chi_\IVP^\sharp\in\CI(M)$ such that
  \[
    \chi_\IVP=1\ \text{for}\ t_\IVP>-\frac{r}{10},\quad
    \supp\chi_\IVP\subset(\chi_\IVP^\sharp)^{-1}(1), \quad
    \supp\chi_\IVP^\sharp \subset \Bigl\{ t_\IVP>-\frac{r}{2}\Bigr\}.
  \]
\end{definition}

See Figure~\ref{FigKBoCutoff}.

\begin{figure}[!ht]
\centering
\includegraphics{FigKBoCutoff}
\caption{Illustration of the cutoff functions from Definition~\ref{DefKBoCutoff}.}
\label{FigKBoCutoff}
\end{figure}

\begin{lemma}[Cut-off Lorentz boosts]
\label{LemmaKBoDef}
  For $\scal\in\scalspace_1$, define the vector field $V(\scal)$ in terms of~\eqref{EqKBo1form} by
  \begin{equation}
  \label{EqKBoVscal}
    V(\scal) := \frac12\chi_\IVP^\sharp(1-\chi_\cK^\flat)\ubar g^{-1}(\ubar\omega_{\rms 1}^{(0),\leq 1}(\scal),\cdot).
  \end{equation}
  Then $V(\scal)\in\Vb(\ol{\R^4})$ is tangent to $Y^+$ (see~\eqref{EqKMfdYplus}), and its lift to $M$ vanishes near $\Sigma_{\rm past}\cup\cK^+\cup\Sigma_{\rm int}$ (Definition~\usref{DefKMfdRad}) and for $r\leq 5\bhm_0$.
\end{lemma}
\begin{proof}
  On the support of $V(\scal)$ in $\ol{\R^4}$, we can work with the local coordinates $\rho:=\frac{1}{r}$, $v:=\frac{t_*}{r}$ (and $\omega=\frac{x}{|x|}\in\Sph^2$). Since
  \[
    \ubar\omega_{\rms 1}^{(0),\leq 1}(\scal)=t_*\,\dd(r\scal) + r\,\dd(r\scal) - r\scal(\dd t_*+\dd r) = t_*\scal\,\dd r+(t_*+r)r\,\dd\scal - r\scal\,\dd t_*
  \]
  we can use~\eqref{EqKMetRefSchw} (with $\bhm=0$) to compute
  \begin{align}
    \ubar g^{-1}(\ubar\omega_{\rms 1}^{(0),\leq 1},\cdot) &= -\scal t_*\pa_{t_*} + \scal\Bigl(1+\frac{t_*}{r}\Bigr)r\pa_r + \Bigl(1+\frac{t_*}{r}\Bigr)\slnabla\scal \nonumber\\
  \label{EqKBoVF}
      &= -\scal v\pa_v - \scal(1+v) (\rho\pa_\rho+v\pa_v) + (1+v)\slnabla\scal.
  \end{align}
  This is a b-vector field and tangent to $Y^+=\{\rho=v=0\}$, as claimed.
\end{proof}

As a consequence of Lemma~\ref{LemmaKBoDef}, the time $1$ flow $\phi_\scal:=\exp(V(\scal))$ of $V(\scal)$ is a diffeomorphism of $\ol{\R^4}$ (depending smoothly on $\scal$) which restricts to a diffeomorphism of $Y^+$; and it lifts to a diffeomorphism
\begin{equation}
\label{EqKBoMap}
  \phi_\scal=\exp(V(\scal)) \colon M\to M
\end{equation}
of $M$ which restricts to a diffeomorphism of $\scri^+$ (and also of $\iota^+$) that is the identity near $\Sigma_{\rm past}\cup\cK^+\cup\Sigma_{\rm int}$ and for $r\leq 5\bhm_0$. Since the pullback of smooth vector fields on $\ol{\R^4}$ by $\phi_\scal$ preserves the property of being tangent to $\pa\ol{\R^4}$, we have $\phi_\scal^*\colon\Vb(\ol{\R^4})\to\Vb(\ol{\R^4})$, and more generally $\phi_\scal^*$ maps any weighted version $(1+|t|^2+|x|^2)^{-\frac{\alpha}{2}}\Vb(\ol{\R^4})$, $\alpha\in\R$, of $\Vb(\ol{\R^4})$ to itself. For $\alpha=1$ and recalling~\eqref{EqKMfdVscVb}, this shows that $\phi_\scal^*\colon\Vsc(\ol{\R^4})\to\Vsc(\ol{\R^4})$, and therefore $\phi_\scal$ induces smooth vector bundle maps
\[
  \phi_\scal^* \colon \cT \to \cT,\quad
  \phi_\scal^* \colon \cT^* \to \cT^*,
\]
covering $\phi_\scal^{-1}$; this holds when regarding $\cT$ and $\cT^*$ as bundles over $\ol{\R^4}$ or over $M$. This then implies
\begin{equation}
\label{EqKBoPullbackT}
  \phi_\scal^* \colon \rho_0^\alpha\rho_\sscri^\beta\rho_+^\gamma\CI(M;S^2\cT^*) \to \rho_0^\alpha\rho_\sscri^\beta\rho_+^\gamma\CI(M;S^2\cT^*)\quad\forall\,\alpha,\beta,\gamma\in\R.
\end{equation}
In particular, we have $\phi_\scal^*g_\bhm\in\CI(M;S^2\cT^*)$. We compute this more precisely:

\begin{lemma}[Pullback by cut-off Lorentz boosts]
\label{LemmaKBo}
  Let $\scal\in\scalspace_1$ be so small that
  \[
    \phi_\scal\bigl(\supp(\chi_\IVP(1-\chi_\cK))\bigr)\subset\supp(\chi_\IVP^\sharp(1-\chi_\cK^\flat)).
  \]
  Set $c=\|\scal\|_{L^\infty(\Sph^2)}$ and
  \begin{equation}
  \label{EqKBoh11}
    h_{\scal,1 1} := 2\bhm\Bigl( \bigl(\cosh(c/2)+\tfrac{\scal}{c}\sinh(c/2)\bigr)^{-3} - 1 \Bigr) \in \CI(\Sph^2);
  \end{equation}
  this depends smoothly on $c^2=\frac{3}{4\pi}\|\scal\|_{L^2(\Sph^2)}^2$ (and thus on $\scal$ itself). Regarding $h_{\scal,1 1}$ as a fiber-constant function on $\scri^+$ and extending it in an arbitrary fashion to a smooth function on $M$, we have (recalling $x^1=t_*$ from~\eqref{EqKNull})
  \begin{equation}
  \label{EqKBo}
    \chi_\IVP(1-\chi_\cK)\Bigl( \phi_\scal^*g_{\bhm,\bha} - (g_{\bhm,\bha}+r^{-1}h_{\scal,1 1}\,(\dd x^1)^2) \Bigr) \in \rho_0\rho_\sscri^2\rho_+\CI(M;S^2\cT^*).
  \end{equation}
\end{lemma}

Recalling that we will describe metric perturbations away from $\cK^+$ relative to $g_\bhm$ (with $\bhm$ the ADM mass of the initial data), this means that a description relative to $\phi_\scal^*g_\bhm$ is equally acceptable; indeed, as we will see already in Definition~\ref{DefExP} and (the proof of) Theorem~\ref{ThmEx0} below (see also \cite[Definition~3.20]{HintzMink4Gauge}), size $r^{-1}$ perturbations of the $(1,1)$-component of the metric tensor arise naturally in our generalized harmonic gauge and thus do not require any further arguments.

\begin{proof}[Proof of Lemma~\usref{LemmaKBo}]
  In view of $g_{\bhm,\bha}-g_\bhm\in r^{-2}\CI(M;S^2\cT^*)$ and~\eqref{EqKBoPullbackT}, it suffices to prove~\eqref{EqKBo} for $g_\bhm$ in place of $g_{\bhm,\bha}$. If we work modulo $\rho_0\rho_\sscri\rho_+\CI(M;S^2\cT^*)$ in~\eqref{EqKBo}, we may moreover replace $g_\bhm$ by $\ubar g$, in which case we note that $\phi_\scal^*\ubar g=\ubar g$ on $\supp(\chi_\IVP(1-\chi_\cK))$ since Lorentz boosts are isometries of $\ubar g$. We thus conclude (again using~\eqref{EqKBoPullbackT}) that the left-hand side of~\eqref{EqKBo} lies in $r^{-1}\CI=\rho_0\rho_\sscri\rho_+\CI$. It thus remains to compute $\phi_\scal^*(g_\bhm-\ubar g)-(g_\bhm-\ubar g)$ near $Y^+$ (and thus, upon lifting the computation to $M$, near $\scri^+$), where we can drop the cutoff functions. We do this in the coordinates $\rho=\frac{1}{r}\geq 0$ and $v=\frac{t_*}{r}\in\R$ using the formula
  \[
    2\vect(\scal) = -(1+v)\scal\rho\pa_\rho - (2+v)\scal v\pa_v + (1+v)\slnabla\scal
  \]
  from~\eqref{EqKBoVF}. For a suitable choice of polar coordinates $\theta,\phi$ on $\Sph^2$, we have
  \[
    \scal = \scal(\theta) = c\cos\theta,\quad c=\|\scal\|_{L^\infty}\geq 0.
  \]
  Since $\slnabla\scal=-c\sin\theta\pa_\theta$, the coordinate $\phi$ is conserved along the $V(\scal)$-flow. We therefore keep track only of the evolution of $\rho,v,\theta$ along the flow; write thus
  \[
    f(\rho,v,\theta;s) = e^{2 s V(\scal)}(\rho,v,\theta) =: \bigl( f_\rho(\rho,v,\theta;s),\ f_v(\rho,v,\theta;s),\ f_\theta(\rho,v,\theta;s) \bigr),
  \]
  so $f(\rho,v,\theta;0)=(\rho,v,\theta)$ and
  \[
    \pa_s f = \Bigl( -(1+f_v)\scal(f_\theta)f_\rho,\ -(2+f_v)\scal(f_\theta)f_v,\ -c(1+f_v)\sin f_\theta \Bigr).
  \]
  Due to the tangency of $V(\scal)$ to the boundary at infinity $\{\rho=0\}$ and to the light cone $\{v=0\}$ (or simply as a consequence of this formula), we have $f_\rho(0,v,\theta;s)=0$ and $f_v(\rho,0,\theta;s)=0$ for all $s$ and $v,\rho$, and moreover $f_\rho(\rho,v,\theta;s)\sim\rho$ for bounded $s$. Direct integration gives
  \begin{equation}
  \label{EqKBoftheta}
    f_\theta(\rho,0,\theta;s) = 2\arctan\Bigl(e^{-c s}\tan\Bigl(\frac{\theta}{2}\Bigr)\Bigr)
  \end{equation}
  for all $\rho$; we shall use this for $\rho=0$.

  Since $g_\bhm-\ubar g=2\bhm\rho(\dd t_*)^2$ and $t_*=\frac{v}{\rho}$, we need to compute $(e^{V(\scal)})^*\rho=f_\rho(\cdot;\frac12)$ modulo terms vanishing quadratically at $Y^+$, and
  \begin{equation}
  \label{EqKBodfvfr}
    (e^{V(\scal)})^*\Bigl(\dd\frac{v}{\rho}\Bigr) = f(\cdot;\tfrac12)^*\Bigl(\dd\frac{v}{\rho}\Bigr) = \dd\Bigl(\frac{f_v(\cdot;\tfrac12)}{f_\rho(\cdot;\tfrac12)}\Bigr).
  \end{equation}
  We must therefore compute $\pa_\rho f_\rho$ and $\pa_v f_\rho$ at $Y^+$; and since $\cT^*$ is locally spanned by $\frac{\dd\rho}{\rho^2}$, $\frac{\dd v}{\rho}$, and $\rho^{-1}$ times 1-forms on $\Sph^2$ (so since we are fixing $\phi$: only $\frac{\dd\theta}{\rho}$), we also need to compute
  \begin{equation}
  \label{EqKBodfvfr2}
  \begin{split}
    \dd\Bigl(\frac{f_v}{f_\rho}\Bigr) &= \Bigl( \frac{\rho}{f_\rho}\rho\pa_\rho f_v - \frac{\rho^2}{f_\rho^2}f_v\,\pa_\rho f_\rho\Bigr)\frac{\dd\rho}{\rho^2} + \frac{\rho}{f_\rho}\Bigl(\pa_v f_v -  \frac{f_v\,\pa_v f_\rho}{f_\rho}\Bigr)\frac{\dd v}{\rho} \\
      &\qquad + \frac{\rho}{f_\rho}\Bigl(\pa_\theta f_v - \frac{f_v\,\pa_\theta f_\rho}{f_\rho}\Bigr)\frac{\dd\theta}{\rho}
  \end{split}
  \end{equation}
  modulo $(\cO(\rho)+\cO(v))\CI([0,1)_\rho\times(-1,1)_v\times(0,\pi)_\theta;\cT^*)$, where we write $\cO(w)$ for the space of functions $a$ defined near $Y^+$ which are of the form $a=w a_0$, with $a_0$ smooth on $\ol{\R^4}$ near $Y^+$. (Thus, $\cO(\rho)+\cO(v)$ is the space of smooth functions near $Y^+$ that vanish at $Y^+$.) Now, writing $z=(\rho,v,\theta)$, we have for $z=(0,0,\theta)\in Y^+$ the following system of ODEs for $\pa_z f_\rho$ and $\pa_z f_v$:
  \begin{equation}
  \label{EqKBopazf}
  \begin{alignedat}{2}
    \pa_z f_\rho(z;0)&=(1,0,0), &\quad \pa_s\pa_z f_\rho(z;s) &= -\scal(f_\theta(z;s))\,\pa_z f_\rho(z;s), \\
    \pa_z f_v(z;0)&=(0,1,0), &\quad \pa_s\pa_z f_v(z;s) &= -2\scal(f_\theta(z;s))\,\pa_z f_\rho(z;s).
  \end{alignedat}
  \end{equation}
  In particular, $\pa_v f_\rho$, $\pa_\theta f_\rho$, $\pa_\rho f_v$, $\pa_\theta f_v=0$ at $Y^+$ for all $s$. In order to determine $\pa_\rho f_\rho$ and $\pa_v f_v$, we use~\eqref{EqKBoftheta} and compute
  \[
    \scal(f_\theta(0,0,\theta;s))=c\,\cos(f_\theta(0,0,\theta;s)) = c\frac{e^{2 c s}-\tan^2\frac{\theta}{2}}{e^{2 c s}+\tan^2\frac{\theta}{2}} = \pa_s \log\Bigl( e^{c s}+e^{-c s}\tan^2\frac{\theta}{2}\Bigr).
  \]
  Plugging this into~\eqref{EqKBopazf} and integrating, we obtain
  \begin{align*}
    \pa_\rho f_\rho(0,0,\theta;s) &= \bigl( \cosh(c s) + \cos\theta\,\sinh(c s) \bigr)^{-1}, \\
    \pa_v f_v(0,0,\theta;s) &= \bigl( \cosh(c s) + \cos\theta\,\sinh(c s) \bigr)^{-2}.
  \end{align*}
  If we define the positive function
  \[
    a(\theta) := \bigl(\cosh(c/2) + \tfrac{\scal(\theta)}{c}\sinh(c/2)\bigr)^{-1}
  \]
  for $\theta\in[0,\pi]$, we therefore have
  \begin{equation}
  \label{EqKBofrfv}
  \begin{split}
    f_\rho(\rho,v,\theta;\tfrac12) &= \rho\bigl( a(\theta) + \cO(\rho) + \cO(v)\bigr), \\
    f_v(\rho,v,\theta;\tfrac12) &= a(\theta)^2 v + \cO(\rho^2) + \cO(\rho v) + \cO(v^2).
  \end{split}
  \end{equation}
  We can now compute~\eqref{EqKBodfvfr} using~\eqref{EqKBodfvfr2} and~\eqref{EqKBofrfv} to be
  \begin{align*}
    (e^{\vect(\scal)})^*\dd t_* &= \Bigl( \bigl(\cO(\rho^2)+\cO(\rho v)\bigr) - \cO(v)\Bigr)\frac{\dd\rho}{\rho^2} \\
    &\quad + (a(\theta)^{-1}+\cO(\rho)+\cO(v))\Bigl(\bigl(a(\theta)^2+\cO(\rho)+\cO(v)\bigr) - \frac{\bigl(\cO(v)+\cO(\rho^2)\bigr)\cdot\cO(\rho)}{\cO(\rho)}\Bigr) \frac{\dd v}{\rho} \\
    &\quad + \bigl( \cO(\rho)+\cO(v) \bigr)\frac{\dd\theta}{\rho}.
  \end{align*}
  The lift of this to $M$ (as a section of $\cT^*$) is given by $a(\theta)\frac{\dd v}{\rho}$ at $\scri^+$. Writing $v=\rho t_*=\frac{t_*}{r}$ and thus $\frac{\dd v}{\rho}=\dd t_*-\frac{t_*}{r}\,\dd r\equiv\dd t_*\bmod\rho_\sscri\CI$, we thus obtain
  \[
    (e^{V(\scal)})^*\bigl(\rho(\dd t_*)^2\bigr) \equiv \rho a(\theta)^3\,(\dd t_*)^2 \bmod \rho_0\rho_\sscri^2\rho_+\CI.
  \]

  Altogether, we thus compute near $Y^+$ that
  \[
    \phi_\scal^*g_\bhm - g_\bhm = \phi_\scal^*(g_\bhm-\ubar g) - (g_\bhm-\ubar g) = 2\bhm\rho\bigl( a(\theta)^3 - 1 \bigr)\,(\dd t_*)^2,
  \]
  which proves~\eqref{EqKBo}.
\end{proof}

\section{A first version of the gauge-fixed Einstein equation}
\label{S1}

We will study nonlinear perturbations of a subextremal Kerr metric $g_b$ (Definition~\ref{DefKMetcM}) using a carefully chosen gauge-fixed version of the Einstein vacuum equation $\Ric(g)=0$. In this section, we discuss a basic version of the gauge-fixed operator we shall use. Only when studying the detailed asymptotic behavior of its forward solutions in~\S\ref{SD} will we devise more refined versions, which incorporate different background metrics and modifications of the gauge condition.

The gauge in which we will find the perturbed spacetime metric $g$ will be a generalized harmonic gauge relative to some background metric $g^0$. (For now the reader can think of $g^0$ as being equal to $g_b$.)

\begin{definition}[Gauge 1-form]
\label{Def1Gauge}
  Denote by $g$ and $g^0$ two metrics on $M^\circ$ (see Definition~\usref{DefKMfdRad}). Let $\cd^\Ups\in r^{-1}\CI(X;\cT_X^*)$ and $\cd^\Ups_{\cH^+}\in\CIc(\{r<5\bhm_0\};\cT_X^*)$ be stationary 1-forms (chosen later), with $\cd^\Ups=r^{-1}\,\dd t$ for large $r$. Given $E^\Ups=(\cd^\Ups,e^\Ups,\gamma^\Ups;\cd^\Ups_{\cH^+},\gamma^\Ups_{\cH^+})$ with $e^\Ups,\gamma^\Ups,\gamma^\Ups_{\cH^+}\in\R$,\footnote{We will ultimately choose $e^\Ups,\gamma^\Ups,\gamma^\Ups_{\cH^+}>0$ small.} define the bundle map
  \begin{alignat}{2}
    &&&\hspace{-6em}E^\Ups_{g^0} = E^\Ups_{g^0,\infty} + E^\Ups_{g^0,\cH^+}, \nonumber\\
  \label{Eq1GaugeInfty}
    &\quad E^\Ups_{g^0,\infty} &\colon h &\mapsto -\gamma^\Ups\bigl(2 \iota_{(g^0)^{-1}(\cd^\Ups)}h - (1-e^\Ups)\cd^\Ups\tr_{g^0}h \bigr), \\
  \label{Eq1GaugeHor}
    &\quad E^\Ups_{g^0,\cH^+} &\colon h &\mapsto \gamma^\Ups_{\cH^+}\bigl(2 \iota_{(g^0)^{-1}(\cd^\Ups_{\cH^+})}h - \cd^\Ups_{\cH^+}\tr_{g^0}h\bigr)
  \end{alignat}
  from $S^2 T^*M^\circ$ to $T^*M^\circ$. Define then the \emph{gauge 1-form} $\Ups_{E^\Ups}(g,g^0)$ by
  \begin{equation}
  \label{Eq1UpsDef}
    \Ups_0(g,g^0) := g(g^0)^{-1}\delta_g\sfG_g g^0,\quad
    \Ups_{E^\Ups}(g,g^0) := \Ups_0(g,g^0) - E^\Ups_{g^0}\sfG_{g^0}(g-g^0),
  \end{equation}
  where $\sfG_g h:=h-\frac12 g\tr_g h$ is the \emph{trace reversal}, as well as the \emph{modified divergence} $\delta_{g^0,E^\Ups}:=\delta_{g^0}+E^\Ups_{g^0}$.
\end{definition}

In local coordinates, we have
\begin{equation}
\label{Eq1Ups0}
  \Ups_0(g,g^0)_\mu = g_{\mu\nu}g^{\kappa\lambda}\bigl( \Gamma(g)_{\kappa\lambda}^\nu - \Gamma(g^0)_{\kappa\lambda}^\nu\bigr).
\end{equation}
The vanishing of this is equivalent to the statement that the identity map $(M^\circ,g)\to(M^\circ,g^0)$ is a wave map. When $g^0$ is the Minkowski metric in standard coordinates on $\R^4\supset M^\circ$, this is moreover equivalent to the harmonicity of the coordinate functions with respect to $g$, i.e., the standard wave coordinate gauge. When perturbing Kerr, it is however more natural to take $g^0$ to be (related to) the Kerr metric. One may then consider the gauge-fixed Einstein equation
\begin{equation}
\label{Eq1EinNoCD}
  \Ric(g) - \delta_g^* \Ups_{E^\Ups}(g,g^0) = 0;
\end{equation}
solving this with suitable (correctly gauged) initial data yields a solution of $\Ric(g)=0$ in the gauge $\Ups_{E^\Ups}(g,g^0)=0$. The utility of modifying the gauge condition $\Ups_0(g,g^0)=0$ in the present context was discussed in prior works:
\begin{enumerate}
\item The modification $E^\Ups_{g^0,\infty}$ (with $e^\Ups=0$ and $\gamma^\Ups>0$) was introduced in \cite{HintzMink4Gauge}. In the gauge $\Ups_{E^\Ups}(g,g_\bhm)=0$, certain coefficients of the metric perturbation $g-g_\bhm$ have stronger decay at $\scri^+$ than when $\gamma^\Ups=0$; roughly speaking, a positive value of $\gamma^\Ups$ shifts some indicial roots of the linearized gauge-fixed Einstein operator at $\scri^+$ from $1$ to $1+\cO(\gamma^\Ups)>1$ (cf.\ the diagonal entries of $A_h$ in~\eqref{EqExOpLinAB}). The flexibility afforded by allowing for nonzero $e^\Ups$ simplifies some bookkeeping: it is used in the present paper to reduce the multiplicity of these indicial roots as well as to remove integer coincidences in indicial roots of zero energy operators, which leads to fewer logarithmic terms (at low exponents) in asymptotic expansions.\footnote{A concrete example is that upon perturbing away the $\rms 1$ indicial root $0$---which is absent indeed in Lemma~\ref{LemmaWEInd}---mildly simplifies the construction of improved zero energy states corresponding to asymptotic boosts; cf.\ the discussion of~\eqref{EqAdmLoImBoostConstr}.}
\item The possible utility of a modification $E^\Ups_{g^0,\cH^+}$, which here will be supported near the horizon, was first mentioned in \cite[Remark~10.14]{HaefnerHintzVasyKerr} for the purpose of ensuring mode stability of the \emph{gauge potential wave operator}
  \begin{equation}
  \label{Eq1BoxUps}
    \Box^\Ups_{g,E^\Ups} := 2\delta_{g,E^\Ups}\sfG_g\delta_g^*
  \end{equation}
  for $g=g_b$, including at zero frequency. We recall that if $\omega\in\ker\Box^\Ups_{g,E^\Ups}$, then the pure gauge tensor $\delta_g^*\omega$ (which always lies in the kernel of $D_g\Ric$ since $g=g_b$ is Ricci-flat) satisfies the linearized gauge condition
  \[
    D_1|_{(g,g)}\Ups_{E^\Ups}(\delta_g^*\omega) := D_{(g,g)}\Ups_{E^\Ups}(\delta_g^*\omega,0) = (-\delta_g\sfG_g-E^\Ups_g\sfG_g)(\delta_g^*\omega)=-\delta_{g,E^\Ups}\sfG_g\delta_g^*\omega=0.
  \]
  Mode stability at zero frequency is convenient when constructing and controlling gauge potentials $\omega$ and their associated pure gauge states that appear in the late-time expansion of linearized metric perturbations. The modification $E^\Ups_{g^0,\cH^+}$ was subsequently implemented in the full subextremal range in \cite[Proposition~3.12]{HintzGlueLocIII}; we recall the argument in~\S\ref{SsWGMode} below.
\end{enumerate}

It is important for us to modify the operator $\delta_g^*$ in~\eqref{Eq1EinNoCD} by a zeroth order term; this procedure is called \emph{constraint damping} \cite{BrodbeckFrittelliHubnerReulaSCP,GundlachCalabreseHinderMartinConstraintDamping}. On a practical level, this amounts to adding linear expressions in the components of $\Ups_{E^\Ups}(g,g^0)$ to the gauge-fixed equation. While this does not change the gauge condition itself, it does modify the total gauge-fixed equation and thus the asymptotic behavior of general solutions (i.e., those not necessarily arising from initial data satisfying the constraint equations), similarly for the linearized problem; this will be crucial for our nonlinear iteration scheme, much as in \cite{HintzVasyKdSStability,HintzVasyMink4,HintzMink4Gauge}, and has been used before in several places in the literature including, e.g., \cite[\S{1.5}]{RingstromEinsteinScalarStability}.

\begin{definition}[Modified symmetric gradient]
\label{Def1Symm}
  Let $g^0$ denote a metric on $M^\circ$. Fix a stationary 1-form $\cd^\cC\in r^{-1}\CI(X;\cT^*)$ that for large $r$ is equal to $r^{-1}(\dd t-v^\cC\,\dd r)$ for some fixed $v^\cC\in(0,1)$. Given $E^\cC=(\cd^\cC,e^\cC,\gamma^\cC)$ with $e^\cC,\gamma^\cC\in\R$,\footnote{We will later choose $e^\cC>0$ small and $\gamma^\cC>0$ large.}, define the bundle map
  \begin{equation}
  \label{Eq1Symm}
    E^\cC_{g^0}\omega := \gamma^\cC\bigl(2 \cd^\cC\otimes_s \omega - (1-e^\cC)g^0 \iota_{(g^0)^{-1}(\cd^\cC)}\omega \bigr)
  \end{equation}
  from $T^*M^\circ$ to $S^2 T^*M^\circ$. Define moreover the \emph{modified symmetric gradient}
  \[
    \delta_{g^0,E^\cC}^* := \delta_{g^0}^* + E^\cC_{g^0}.
  \]
\end{definition}

The precise choice of $E^\cC$ will be taken from \cite{HintzKerrCD}; we recall this in~\S\ref{SWC}. We can then define the gauge-fixed Einstein operator
\begin{equation}
\label{Eq1Ein}
  P(g,g^0) := \Ric(g) - \delta_{g^0,E^\cC}^* \Ups_{E^\Ups}(g,g^0).
\end{equation}
Applying $\delta_g\sfG_g$ to $P(g,g^0)$ annihilates the term $\Ric(g)$ by the second Bianchi identity; the operator acting on $\Ups_{E^\Ups}(g,g^0)$ is then the \emph{constraint propagation wave operator}
\begin{equation}
\label{Eq1BoxC}
  \Box^\cC_{g,E^\cC} := 2\delta_g\sfG_g\delta_{g,E^\cC}^*
\end{equation}
when $g=g^0$. Roughly speaking, constraint damping amounts to ensuring the strong decay of homogeneous solutions of $\Box^\cC_{g,E^\cC}\Ups=0$ (given appropriately decaying initial data).

\begin{rmk}[Background metrics]
\label{Rmk1BgMet}
  There is no a priori reason to use the same metric $g^0$ both for the symmetric gradient and for the gauge condition. It turns out, however, that for our purposes we do not need the additional flexibility afforded by decoupling the two.
\end{rmk}

\subsection{Linearizations}
\label{Ss1Lin}

We denote the linearization of~\eqref{Eq1Ein} in the first argument by
\begin{equation}
\label{Eq1EinLin}
  L_{g,g^0} := D_1|_{(g,g^0)}P := D_{(g,g^0)}P(\cdot,0),\quad L_g:=L_{g,g}.
\end{equation}
Thus, $L_{g,g^0}\dot g:=\frac{\dd}{\dd s}P(g+s\dot g,g^0)|_{s=0}$. We record the following expressions (see, e.g., \cite{GrahamLeeConformalEinstein}):

\begin{lemma}[Linearizations]
\fakephantomsection
\label{Lemma1Lin}
  \begin{enumerate}
  \item Given metrics $g$ and $g^0$, define
    \[
      C(g,g^0)_{\mu\nu}^\lambda := \Gamma(g)_{\mu\nu}^\lambda - \Gamma(g^0)_{\mu\nu}^\lambda,\quad
      \sC_{g,g^0}(\dot g)_\kappa := g_{\kappa\lambda}C(g,g^0)_{\mu\nu}^\lambda\dot g^{\mu\nu},\quad
      \sY_{g,g^0}(\dot g)_\kappa := \Ups_0(g,g^0)^\lambda\dot g_{\kappa\lambda},
    \]
    where we raise and lower indices using $g$. Then
    \begin{equation}
    \label{Eq1LinUps}
    \begin{split}
      D_1|_{(g,g^0)}\Ups &= -\delta_g\sfG_g - \sC_{g,g^0} + \sY_{g,g^0}, \\
      D_1|_{(g,g^0)}\Ups_{E^\Ups} &= -\delta_g\sfG_g - \sC_{g,g^0} + \sY_{g,g^0} - E^\Ups_{g^0}\sfG_{g^0}.
    \end{split}
    \end{equation}
  \item Upon setting
    \[
      (\sR_g\dot g)_{\mu\nu} := \Riem(g)^\kappa{}_{\mu\nu}{}^\lambda\dot g_{\kappa\lambda} + \frac12\bigl(\Ric(g)_\mu{}^\kappa\dot g_{\kappa\nu} + \Ric(g)^\kappa{}_\nu\dot g_{\mu\kappa}\bigr),
    \]
    and defining $(\Box_g\dot g)_{\mu\nu}=-g^{\kappa\lambda}\dot g_{\mu\nu;\kappa\lambda}$ to be the tensor wave operator, we have
    \begin{equation}
    \label{Eq1LinEin}
      D_g\Ric = \frac12\Box_g - \delta_g^*\delta_g\sfG + \sR_g.
    \end{equation}
  \item We have
    \begin{equation}
    \label{Eq1LinEinGauged}
      L_{g,g^0} = \frac12\Box_g + E^\cC_{g^0}\delta_g\sfG_g + (\delta_{g^0}^*-\delta_g^*)\delta_g\sfG_g + \sR_g + \delta_{g^0,E^\cC}^*\bigl(\sC_{g,g^0}-\sY_{g,g^0}+E_{g^0}^\Ups\sfG_{g^0}\bigr).
    \end{equation}
  \end{enumerate}
\end{lemma}
\begin{proof}
  We give a proof for completeness. Given~\eqref{Eq1Ups0} (which is most easily checked at the origin in a $g$-normal coordinate system), the first formula in~\eqref{Eq1LinUps} follows from
  \[
    D_1|_{(g,g^0)}\Ups(\dot g)_\mu = \dot g_{\mu\nu} \Ups_0(g,g^0)^\nu - g_{\mu\nu}\dot g^{\kappa\lambda}C_{\kappa\lambda}^\nu + \tfrac12 g_{\mu\nu}g^{\kappa\lambda} (\dot g_\kappa{}^\nu{}_{;\lambda} + \dot g_\lambda{}^\nu{}_{;\kappa} - \dot g_{\kappa\lambda}{}^{;\nu} );
  \]
  the first term is $\sY_{g,g^0}(\dot g)_\mu$, the second term equals $\sC_{g,g^0}(\dot g)_\mu$, and the final term equals $\dot g_{\mu\kappa}{}^{;\kappa}-\frac12\dot g_\kappa{}^\kappa{}_{;\mu}=-(\delta_g\sfG_g\dot g)_\mu$. For the second formula, we note that the linearization of the final term in~\eqref{Eq1UpsDef} in $g$ is $-E_{g^0}^\Ups\sfG_{g^0}$. For~\eqref{Eq1LinEin}, we first note that
  \[
    \bigl(D_g\Riem(\dot g)\bigr)^\kappa{}_{\lambda\mu\nu} = \bigl(D_g\Gamma(\dot g)_{\lambda\nu}^\kappa\bigr)_{;\mu} - \bigl(D_g\Gamma(\dot g)_{\lambda\mu}^\kappa\bigr)_{;\nu},\quad
    D_g\Gamma(\dot g)_{\mu\nu}^\lambda = \frac12(\dot g_\mu{}^\lambda{}_{;\nu} + \dot g_\nu{}^\lambda{}_{;\mu} - \dot g_{\mu\nu}{}^{;\lambda}),
  \]
  and thus
  \[
    2\bigl(D_g\Ric(\dot g)\bigr)_{\lambda\nu} = 2\bigl(D_g\Riem(\dot g)\bigr)^\kappa{}_{\lambda\kappa\nu} = (\dot g_\lambda{}^\kappa{}_{;\nu\kappa} - \dot g_\lambda{}^\kappa{}_{;\kappa\nu}) - \dot g_{\lambda\nu}{}^{;\kappa}{}_\kappa + \dot g_\nu{}^\kappa{}_{;\lambda\kappa} - \dot g_\kappa{}^\kappa{}_{;\lambda\nu} + \dot g_{\lambda\kappa}{}^{;\kappa}{}_\nu.
  \]
  The first term on the right equals
  \begin{equation}
  \label{Eq1LinRic1}
    \Riem(g)^\rho{}_{\lambda\nu\kappa}\dot g_\rho{}^\kappa + \Riem(g)_\rho{}^\kappa{}_{\nu\kappa}\dot g_\lambda{}^\rho = \Riem(g)^\rho{}_{\lambda\nu}{}^\kappa\dot g_{\rho\kappa} + \Ric(g)^\rho{}_\nu\dot g_{\lambda\rho}.
  \end{equation}
  The second term is $(\Box_g\dot g)_{\lambda\nu}$. The sum of the remaining terms equals
  \[
    (-2\delta_g^*\delta_g\sfG_g\dot g)_{\lambda\nu} = \dot g_\nu{}^\kappa{}_{;\kappa\lambda} - \dot g_\kappa{}^\kappa{}_{;\lambda\nu} + \dot g_{\lambda\kappa}{}^{;\kappa}{}_\nu
  \]
  plus
  \[
    \dot g_\nu{}^\kappa{}_{;\lambda\kappa} - \dot g_\nu{}^\kappa{}_{;\kappa\lambda} = \Riem(g)^\rho{}_{\nu\lambda}{}^\kappa\dot g_{\rho\kappa} + \Ric(g)^\rho{}_\lambda\dot g_{\nu\rho}.
  \]
  The sum of this and~\eqref{Eq1LinRic1} equals $(2\sR_g\dot g)_{\lambda\nu}$. The formula~\eqref{Eq1LinEinGauged} is now immediate.
\end{proof}

The expression~\eqref{Eq1LinEinGauged} will be useful for computing the form of $L_{g,g^0}$ near null infinity. Near the Kerr face, a more geometric expression is more useful: since $\sC_{g,g}=0=\sY_{g,g}$, we have
\begin{equation}
\label{Eq1Lin}
  L_g = D_g\Ric + \delta_{g,E^\cC}^*\delta_{g,E^\Ups}\sfG_g.
\end{equation}

\section{Exterior stability with partial polyhomogeneity}
\label{SEx}

In this section, we will work in the subset of $M$ (see Definition~\ref{DefKMfdRad}) bounded to the future by $t_*^{-1}=T$ (as well as by the hypersurface $\Sigma_{\rm int}=r^{-1}(\bhm_0)$ in the black hole interior), i.e.,
\begin{equation}
\label{EqExDomain}
  \Omega_\ext := \cl_M \{ t_*\leq T \},
\end{equation}
where $T$ is arbitrary but fixed; upon redefining $t_*$ by a constant shift, we may take $T=1$, and do so in the sequel for notational simplicity. We will prove the nonlinear stability of the subextremal Kerr metric $g_b$ in the subset of this region bounded in the past by $\Sigma_\IVP=t_\IVP^{-1}(0)$ (see Lemma~\ref{LemmaKMetIVP} and Definition~\ref{DefKMetData}), and in fact in a slightly bigger region which contains the images of $\Sigma_\IVP$ under small Lorentz boosts.

The setup, estimates, and arguments are very similar to those in~\cite{HintzMink4Gauge} (which in turn is strongly inspired by \cite{HintzVasyMink4} and \cite{LindbladRodnianskiGlobalExistence,LindbladRodnianskiGlobalStability}), except for minor (essentially only notational) adjustments arising from the fact that the modifications in Definitions~\ref{Def1Gauge} and \ref{Def1Symm} differ slightly from those in \cite{HintzMink4Gauge}. There are two novelties, however:
\begin{enumerate}
\item We will prove the partial polyhomogeneity of metric perturbations in the gauge $\Ups_{E^\Ups}=0$ (using arguments that are more robust than those in \cite[\S7]{HintzVasyMink4} in that they apply also to non-Ricci-flat metrics), and explicitly describe partially polyhomogeneous function spaces between which direct operators and their forward solution operators act; variants of these considerations will be particularly important for our later analysis in $t_*\geq 1$.
\item We show that pullbacks along Lorentz boosts (cf.\ \S\ref{SsKBo}) are compatible with our exterior stability setup. In particular, accommodating boosted data does not necessitate any substantial changes.
\end{enumerate}

We begin in~\S\ref{SsExP} to describe the class of (partially polyhomogeneous) metric perturbations. In~\S\ref{SsExOp}, we compute the form of the linearized gauge-fixed Einstein operator $L_{g,g^0}$ (see~\eqref{Eq1EinLin}) when $g$ equals $g_\bhm$ plus such a metric perturbation, and $g^0$ is a smooth asymptotically Minkowskian background metric. The forward mapping properties of the linearized and nonlinear gauge-fixed Einstein operators are then deduced in~\S\ref{SsExFw}. The basic exterior stability result, with only basic asymptotic control as in \cite{HintzMink4Gauge} (thus capturing metric perturbations modulo $o(r^{-1})$ remainders), is given in~\S\ref{SsEx0}. In~\S\ref{SsExPhg}, we prove the partial polyhomogeneity of the metrics thus obtained. In~\S\ref{SsExBo}, we discuss the initial value problem for boosted data. Finally, \S\ref{SsExID} relates geometric initial data and gauged Cauchy data.

\bigskip

We fix some notation for the remainder of this section. Since we work away from $\iota^+$ and $\cK^+$, we only record weights and index sets at $I^0$ and $\scri^+$ in our notation for function spaces. Thus, we write
\begin{equation}
\label{EqExHb}
  \Hb^{k,\alpha_0,\alpha_\sscri}(\Omega_\ext) := \rho_0^{\alpha_0}\rho_\sscri^{\alpha_\sscri}\Hb^k(\Omega_\ext)
\end{equation}
for weighted b-Sobolev spaces on $\Omega_\ext$ relative to a b-density. (As usual, the b-nature here refers only to the boundary hypersurfaces of $\Omega_\ext$ at infinity, i.e., those contained in $I^0\cup\scri^+$. A more precise notation would be $\bar H_\bop^k$; see however Notation~\ref{NotTMHbbar}.) Recall here from Definition~\ref{DefKMfdRad} that $\rho_0$ and $\rho_\sscri$ are defining functions of $I^0$ and $\scri^+$, respectively. On $\Omega_\ext$, one can use the expressions in~\eqref{EqKMfdCoordI0Scri}. Recalling Definition~\ref{DefKMfdRad2}, we moreover define
\begin{equation}
\label{EqExOmega12}
  \Omega_{\ext,\frac12} := \cl_{M_{\frac12}} \{ t_*\leq 1 \} \subset M_{\frac12}.
\end{equation}

\begin{definition}[edge-b-notions]
\label{DefExeb}
  The space $\Veb(\Omega_{\ext,\frac12})$ of \emph{edge-b-vector fields} (or \emph{eb-vector fields} for short) on $\Omega_{\ext,\frac12}$ consists of all $V\in\Vb(\Omega_{\ext,\frac12})$ that are tangent to the fibers of $\scri^+$ (cf.\ the discussion after~\eqref{EqKMfdCoordHalf}). We write $\Diffeb^m(\Omega_{\ext,\frac12})$ for the associated space of $m$-th order \emph{edge-b-differential operators}.
\end{definition}

In local coordinates as in~\eqref{EqKMfdCoordI0Scri} but with $x_\sscri=\rho_\sscri^{\frac12}$, a frame of $\Veb(\Omega_{\ext,\frac12})$ is given by
\[
  \rho_0\pa_{\rho_0},\quad
  x_\sscri\pa_{x_\sscri}=2\rho_\sscri\pa_{\rho_\sscri},\quad
  x_\sscri\pa_\omega=\rho_\sscri^{\frac12}\pa_\omega,
\]
where we use the schematic notation $\pa_\omega$ to denote smooth vector fields on $\Sph^2_\omega$. A more convenient frame is given by $\rho_0^{-1}x_\sscri^{-2}\pa_0 = r\pa_0$, $\rho_0^{-1}\pa_1$, and $x_\sscri\pa_\omega$. (That this is indeed a frame follows from the expressions~\eqref{EqKNullComp} and $\rho_\sscri\pa_{\rho_\sscri}=2 x_\sscri\pa_{x_\sscri}$.) In particular,
\begin{equation}
\label{EqExebpa01}
  \pa_0 \in r^{-1}\Veb = \rho_0 x_\sscri^2\Veb,\quad
  \pa_1 \in \rho_0\Veb,\quad
  r^{-1}\pa_\omega \in \rho_0 x_\sscri\Veb.
\end{equation}
We use powers of $x_\sscri$ for weights at $\scri^+\subset\Omega_{\ext,\frac12}$, so for example
\[
  \Hb^{k,(\alpha_0,2\alpha_\sscri)}(\Omega_{\ext,\frac12}) := \rho_0^{\alpha_0}x_\sscri^{2\alpha_\sscri}\Hb^k(\Omega_{\ext,\frac12}).
\]
(This particular space is the same space as~\eqref{EqExHb}.)

\subsection{Class of metric perturbations}
\label{SsExP}

As will follow from Proposition~\ref{PropExOpLin} below, when $g=g^0=g_\bhm$ is the Schwarzschild metric~\eqref{EqKMetRefSchw}, then the operator $L_{g_\bhm}$, expressed in the coordinates $t_*$, $\rho=r^{-1}$, $\omega\in\Sph^2$ near $(\scri^+)^\circ$, takes the form
\[
  \rho^{-1}L_{g_\bhm} \equiv -\bigl(\rho\pa_\rho - (1+A_0)\bigr)\pa_{t_*} + B_0
\]
modulo $\rho\Diffb^2$, where, in a certain permutation of the summands of~\eqref{EqKNullS2TFine}, $A_0$ has diagonal entries that are equal to $0$ or positive multiples of $\gamma^\Ups$ and $\gamma^\cC$ (when $e^\cC,e^\Ups>0$ are small), and $B_0$ is lower triangular. Ignoring couplings between the metric coefficients, the diagonal entries of $A_0$ dictate the decay rates of the coefficients of a solution of $L_{g_\bhm}\dot g=0$; we introduce notation to describe this:

\begin{definition}[Metric perturbation components]
\label{DefExPComp}
  Denote by $\ubar\R:=M\times\R\to M$ the trivial real rank $1$ vector bundle. In terms of the splitting~\eqref{EqKNullS2TFine} and recalling the notation in Definition~\usref{DefKNullComp}, we then define the bundle projections
  \begin{alignat*}{2}
    \pi^\cC &\colon S^2\cT^* \to \ul\R \oplus T^*\Sph^2 \oplus \ul\R,
      & \qquad h &\mapsto \Bigl(h_{0 0}, h_{0\bar b}, \frac12\sltr h\Bigr), \quad \sltr h:=\slg^{a b}h_{\bar a\bar b}, \\
    \pi_{0 1} &\colon S^2\cT^* \to \ul\R,
      &\qquad h &\mapsto h_{0 1}, \\
    \pi_{1 /} & \colon S^2\cT^* \to T^*\Sph^2,
      &\qquad h &\mapsto h_{1\bar b}, \\
    \slpi_0 &\colon S^2\cT^* \to \ker\sltr,
      &\qquad h &\mapsto h_{\bar a\bar b}-\frac12(\slg^{c d}h_{\bar c\bar d})\slg_{a b}, \\
    \pi_{1 1} &\colon S^2\cT^* \to \ul\R,
      &\qquad h &\mapsto h_{1 1}.
  \end{alignat*}
\end{definition}

Then, roughly speaking, the piece $\pi^\cC h$ will have strong decay at $\scri^+$ when $\gamma^\cC$ is large, while $\pi_{0 1}h$ and $\pi_{1 /}h$ have $r^{-1-(1-e^\Ups)\gamma^\Ups}$- and $r^{-1-\gamma^\Ups}$-decay, respectively; and $\slpi_0 h$ has a $r^{-1}$ radiation field, which couples into $\pi_{1 1}h$ (which without this coupling would have $r^{-1-2\gamma^\Ups}$-decay); also $\pi_{0 1}h$ and $\pi_{1 /}h$ couple into $\pi_{1 1}h$. This discussion serves as partial motivation for the following definition. (The full justification of this definition is that we will show that it is compatible with the structure of metric perturbations in the gauge $\Ups_{E^\Ups}=0$.)

\begin{definition}[Metric perturbations in the exterior region]
\label{DefExP}
  Fix $\gamma^\Ups\in(0,\frac12)$ and $e^\Ups\in(0,1)$.\footnote{These will be the same parameters as in Definition~\ref{Def1Gauge}.} Let
  \[
    \cE_0,\ \cE_\sscri^\cC \subset \C\times\N_0
  \]
  be index sets such that:
  \begin{enumerate}
  \item\label{ItExPE0} $\min\Re\cE_0>0$, and $j\cE_0\subset\cE_0$ for all $j\in\N$;
  \item\label{ItExPEscri} $\min\Re\cE_\sscri^\cC>1+2\gamma^\Ups$;
  \item\label{ItExPTotNL} setting
    \begin{equation}
    \label{EqEinPertTot}
      \cE_\sscri^\tot := \cE_\sscri^\cC \cup (1,0) \cup (1+(1-e^\Ups)\gamma^\Ups,0) \cup (1+\gamma^\Ups,0) \cup (1+2\gamma^\Ups,0),
    \end{equation}
    we have $j\cE_\sscri^\tot\subset\cE_\sscri^\cC$ for all $j\geq 2$, and moreover $\cE_\sscri^\cC+(\cE_\sscri^\tot-1)\subset\cE_\sscri^\cC$.
  \end{enumerate}
  Define further
  \begin{equation}
  \label{EqExPOther}
  \begin{alignedat}{2}
    \cE_{\sscri,0 1} &:= \cE_\sscri^\cC \cup (1+(1-e^\Ups)\gamma^\Ups,0), &\qquad
    \cE_{\sscri,1 /} &:= \cE_\sscri^\cC \cup (1+\gamma^\Ups,0), \\
    \slcE_{\sscri,0} &:= \cE_\sscri^\cC \cup (1,0), &\qquad
    \cE_{\sscri,1 1} &:= \cE_\sscri^\cC \cup (1,0) \cup (1+(1-e^\Ups)\gamma^\Ups,0) \cup (1+2\gamma^\Ups,0).
  \end{alignedat}
  \end{equation}
  Let $k\geq 5$; let $\ell_0>0$ and $\ell_\sscri>1$. Then, recalling Definition~\usref{DefTMphg} and Notation~\usref{NotTMphg}, the space\footnote{The dependence on $\gamma^\Ups$ and $e^\Ups$ is not made explicit in the notation. The notation $\la\cE_\sscri^\cC\ra$ is meant to indicate that the $\scri^+$-index sets of the different components of $h$ differ, but have explicit expressions in terms of $\cE_\sscri^\cC$.}
  \begin{equation}
  \label{EqExPSpace}
    \Hb^{k,\ (\cE_0,\ell_0),\ \bigl(\la\cE_\sscri^\cC\ra,\ell_\sscri\bigr)}(\Omega_\ext) \subset H_\bop^{k,0,0}(\Omega_\ext;S^2\cT^*)
  \end{equation}
  consists of all real\footnote{If a term with, say, $\rho_0^z$ asymptotics, $(z,0)\in\cE_0$, appears in the asymptotic expansion of $h$ at $I^0$, with $z\notin\R$, then the reality of $h$ forces also the complex conjugate term to appear, yielding $\rho_0^{\Re z}\cos(\log\rho_0)$- and $\rho_0^{\Re z}\sin(\log\rho_0)$-terms.}  symmetric 2-tensors $h$ such that
  \begin{equation}
  \label{EqExPSpaceProj}
  \begin{alignedat}{2}
    \pi^\cC h &\in H_\bop^{k,(\cE_0,\ell_0),(\cE_\sscri^\cC,\ell_\sscri)}(\Omega_\ext;\ul\R\oplus T^*\Sph^2\oplus\ul\R), \\
    \pi_{0 1}h &\in H_\bop^{k,(\cE_0,\ell_0),(\cE_{\sscri,0 1},\ell_\sscri)}(\Omega_\ext), &\qquad
    \pi_{1 /}h &\in H_\bop^{k,(\cE_0,\ell_0),(\cE_{\sscri,1 /},\ell_\sscri)}(\Omega_\ext;T^*\Sph^2), \\
    \slpi_0 h &\in H_\bop^{k,(\cE_0,\ell_0),(\slcE_{\sscri,0},\ell_\sscri)}(\Omega_\ext;\ker\sltr), &\qquad
    \pi_{1 1}h &\in H_\bop^{k,(\cE_0,\ell_0),(\cE_{\sscri,1 1},\ell_\sscri)}(\Omega_\ext).
  \end{alignedat}
  \end{equation}
  If $0<\ell_0<\min\Re\cE_0$, we write $\ell_0$ in~\eqref{EqExPSpace} instead of $(\cE_0,\ell_0)$, and if\footnote{We have $(2,0)\subset\cE_\sscri^\cC$, see~\eqref{EqExPEC} below.} $1<\ell_\sscri<\min\Re\cE_\sscri^\cC\leq 2$, we write $(-,\ell_\sscri)$ instead of $(\la\cE_\sscri^\cC\ra,\ell_\sscri)$.
\end{definition}

We explain some aspects of this definition:
\begin{enumerate}[label=(\roman*)]
\item The restriction $\gamma^\Ups<\frac12$ is useful for bookkeeping: it guarantees that $1+2\gamma^\Ups<2$. (To facilitate our analysis of the Kerr model in~\S\ref{SWE}, we will take $0<\gamma^\Ups\ll 1$.)
\item Given any index set $\tilde\cE_\sscri^\cC$ with $\min\Re\tilde\cE_\sscri^\cC>1+2\gamma^\Ups$, there exists a smallest index set $\cE_\sscri^\cC\supseteq\tilde\cE_\sscri^\cC$ satisfying conditions~\eqref{ItExPEscri}--\eqref{ItExPTotNL}. To see this, note that the map $\Phi$ taking $\tilde\cE_\sscri^\cC$ to the union of $\tilde\cE_\sscri^\cC$ with $j\tilde\cE_\sscri^\tot$ (where $\tilde\cE_\sscri^\tot$ is defined by~\eqref{EqEinPertTot} with $\tilde\cE_\sscri^\cC$ in place of $\cE_\sscri^\cC$), $j\geq 2$, and $\tilde\cE_\sscri^\cC+(\tilde\cE_\sscri^\tot-1)$ has the property that the iterates $\Phi^n(\tilde\cE_\sscri^\cC)=\Phi(\cdots(\Phi(\tilde\cE_\sscri^\cC))\cdots)$ stabilize when intersected with $\{\Re z<C\}$ for any fixed $C$, i.e., these intersections are independent of $n$ for $n\geq n(C)$.
\item Condition~\eqref{ItExPTotNL} implies, in particular, that
  \begin{equation}
  \label{EqExPEC}
    (2,0)\cup(2+(1-e^\Ups)\gamma^\Ups,0)\cup(2+\gamma^\Ups,0)\cup(2+2\gamma^\Ups,0)\subset\cE_\sscri^\cC.
  \end{equation}
  Furthermore, the intersections of $\cE_{\sscri,0 1}$ etc.\ with $\{\Re z>1+2\gamma^\Ups\}$ are all equal.
\item The restriction $e^\Ups\in(0,1)$ implies that all $\scri^+$-index sets other than $\slcE_{\sscri,0}$ and $\cE_{\sscri,1 1}$ have $\min\Re>1$. The lower bound $1+2\gamma^\Ups$ for $\Re\cE_\sscri^\cC$ in~\eqref{ItExPEscri} is the largest indicial root for which we will need to keep track of associated radiation fields in our stability proof.
\item The conditions on $j\cE_0$ and $j\cE_\sscri^\tot$ in parts~\eqref{ItExPE0} are made such that the index sets of nonlinear expressions arising in the computation of Christoffel symbols etc.\ of metrics $g_b+h$ do not get bigger than those of $h$ itself.
\item The final condition in part~\eqref{ItExPTotNL} will essentially be forced on us when controlling the asymptotic behavior of solutions of the gauge-fixed Einstein equation from initial data (cf.\ \eqref{EqExPhgEscri} below, which indeed implies $\cE_\sscri^\cC+(\cE_\sscri^\tot-1)\subset\cE_\sscri^\cC$)
\item The lower bound $k\geq 5$ ensures that $H_\bop^{k-2,0,0}(\Omega_\ext)$, with $k-2>\frac{\dim\Omega_\ext}{2}=2$, is an algebra under pointwise multiplication.
\end{enumerate}

The following class of metric perturbations is useful for capturing only the very leading order behavior at $\scri^+$:

\begin{definition}[Metric perturbations in the exterior region: leading order]
\label{DefExPLead}
  Let $\cE_0\subset\C\times\N_0$ be an index set satisfying $\min\Re\cE_0>0$ and $j\cE_0\subset\cE_0$ for all $j\in\N$. Let $k\geq 5$ and $\ell_0>0$, $\ell_\sscri>1$. Then the space
  \[
    \Hb^{k,(\cE_0,\ell_0),\ \bigl(\la(2,0)\ra_0,\ell_\sscri\bigr)}(\Omega_\ext)
  \]
  consists of all $h$ such that $\pi^\cC h$, $\pi_{0 1}h$, $\pi_{1 /}h\in H_\bop^{k,(\cE_0,\ell_0),((2,0),\ell_\sscri)}$ and $\slpi_0 h$, $\pi_{1 1} h\in H_\bop^{k,(\cE_0,\ell_0),((1,0),\ell_\sscri)}$.
\end{definition}

Note that $\Hb^{k,(\cE_0,\ell_0),(\la(2,0)\ra_0,\ell_\sscri)}\subset\Hb^{k,(\cE_0,\ell_0),(\la\cE_\sscri^\cC\ra,\ell_\sscri)}$ for all index sets $\cE_\sscri^\cC$ satisfying the conditions of Definition~\ref{DefExP}.

\begin{rmk}[Pullback by boosts]
\label{RmkExPBo}
  Lemma~\ref{LemmaKBo} implies $\phi_\scal^*g_b-g_b\in\Hb^{\infty,\ \bigl((1,0),\infty\bigr),\ \bigl(\la(2,0)\ra_0,\infty\bigr)}(\Omega_\ext)$, so $\phi_\scal^*g_b$ belongs to the class of metric perturbations of $g_b$ that Definitions~\ref{DefExPLead} (and \ref{DefExP}) allow for. We already point out here that the dynamical metric perturbations we will consider in this paper will have $I^0$-decay $o(r^{-1})$, i.e., $\min\Re\cE_0>1$ and $\ell_0>1$, and we allow for weaker decay at $I^0$ in these definitions only to explicitly accommodate $\phi_\scal^*g_b$ as well.
\end{rmk}

\subsection{Structure of the linearized gauge-fixed Einstein operator}
\label{SsExOp}

In the notation of Definition~\ref{DefExP}, let us consider a Lorentzian metric
\begin{equation}
\label{EqExOpMet}
  g=g_b+h,\quad h\in\Hb^{k,(\cE_0,\ell_0),(\la\cE_\sscri^\cC\ra,\ell_\sscri)}(\Omega_\ext).
\end{equation}
We wish to compute the precise form of the linearized gauge-fixed Einstein operator $L_{g,g^0}$ (for appropriate $g^0$, specified below) using the expression~\eqref{Eq1LinEinGauged}. The first step is the computation of the metric coefficients, Christoffel symbols, and components of the curvature tensor. We work in the coordinates $x^0,x^1$ from~\eqref{EqKNull} and write $x^2,x^3$ for local coordinates on $\Sph^2$; we correspondingly use spacetime indices $0,1,2,3$, and we use Latin letters $a,b,c,\ldots$ to denote spherical indices ($2,3$).

\begin{lemma}[Metric coefficients]
\label{LemmaExOpMet}
  The expressions for $g_{\mu\nu}\bmod r^{\fs(\mu,\nu)}(r^{-2}\CI+H_\bop^{k,(\cE_0,\ell_0),(\cE_\sscri^\cC,\ell_\sscri)})$ (recalling Definition~\usref{DefKNullComp}) are
  \begin{alignat*}{3}
    g_{0 0} &\equiv 0, &\qquad g_{0 1} &\equiv -\tfrac12(1-2\bhm r^{-1}) + h_{0 1}, &\qquad g_{0 b}&\equiv 0, \\
    g_{1 1} &\equiv h_{1 1}, &\qquad g_{1 b} &\equiv r h_{1\bar b}, &\qquad g_{a b}&\equiv r^2\slg_{a b}+r^2 h_{\bar a\bar b}.
  \end{alignat*}
  The expressions for $g^{\mu\nu}\bmod r^{-\fs(\mu,\nu)}(r^{-2}\CI+H_\bop^{k,(\cE_0,\ell_0),(\cE_\sscri^\cC,\ell_\sscri)})$ are
  \begin{equation}
  \label{EqExOpMetDual}
  \begin{alignedat}{3}
    g^{0 0} &\equiv -4 h_{1 1}, &\qquad g^{0 1} &\equiv -2(1+2\bhm r^{-1}) - 4 h_{0 1}, &\qquad g^{0 b} &\equiv -2 r^{-1}h_1{}^{\bar b}, \\
    g^{1 1} &\equiv 0, &\qquad g^{1 b} &\equiv 0, &\qquad g^{a b} &\equiv r^{-2}\slg^{a b}-r^{-2}h^{\bar a\bar b}.
  \end{alignedat}
  \end{equation}
  Here, spherical indices of $h$ are raised using the standard metric $\slg$ on $\Sph^2$.
\end{lemma}
\begin{proof}
  Proposition~\ref{PropKMetCpt} allows us to replace $g_b$ in $g=g_b+h$ by $g_\bhm$. The expressions for $g_{\mu\nu}$ then follow from~\eqref{EqKNullSchw}. Note that the coefficients $h_{0 0}$ and $h_{0\bar b}$ only contribute error terms (as does $\sltr h$, though we do not separate the pure trace and trace-free parts of $h_{a b}$ here).

  For the computation of the dual metric, we write, for small $r^{-1}$,
  \[
    g^{-1} = (g_b+h)^{-1} = g_b^{-1} (I + h g_b^{-1})^{-1} = g_b^{-1} - g_b^{-1}h g_b^{-1} + \sum_{j=2}^\infty (-1)^j g_b^{-1}(h g_b^{-1})^j.
  \]
  From $g_b^{-1}\equiv -2(1+2\bhm r^{-1})\cdot 2\pa_0\otimes_s\pa_1 + r^{-2}\slg^{-1}\bmod r^{-2}\CI(\Omega_\ext;S^2\cT^*)$ (see Proposition~\ref{PropKMetCpt} and~\eqref{EqKNullSchw}), we obtain the stated expressions for $g^{\mu\nu}$ from the first two terms on the right. For the nonlinear (in $h$) remainder, we note that $h g_b^{-1}\in H_\bop^{k,(\cE_0,\ell_0),(\cE_\sscri^\tot,\ell_\sscri)}$ as a section of $\cT^*\otimes\cT=\End(\cT)$, and thus its $j$-th power satisfies (a fortiori)
  \[
    (h g_b^{-1})^j \in H_\bop^{k,(j\cE_0,\ell_0),(j\cE_\sscri^\tot,\ell_\sscri)}.
  \]
  But $j\cE_0\subset\cE_0$ and $j\cE_\sscri^\tot\subset\cE_\sscri^\cC$ for $j\geq 2$ by Definition~\ref{DefExP}\eqref{ItExPE0}, \eqref{ItExPTotNL}.
\end{proof}

\begin{lemma}[Christoffel symbols]
\label{LemmaExOpChr}
  The expressions for $\Gamma_{\kappa\mu\nu}=\frac12(\pa_\mu g_{\kappa\nu} + \pa_\nu g_{\kappa\mu} - \pa_\kappa g_{\mu\nu})$, modulo $r^{\fs(\mu,\nu,\kappa)}(\rho_0^3\rho_\sscri^2\CI+H_\bop^{k-1,(\cE_0+1,\ell_0+1),(\cE_\sscri^\cC,\ell_\sscri)})$, are
  \begin{alignat*}{2}
    \Gamma_{0 0 0} &\equiv 0, &\qquad \Gamma_{0 0 1} &\equiv 0, \\
    \Gamma_{1 0 0} &\equiv -\tfrac12\bhm r^{-2}, &\qquad \Gamma_{1 0 1} &\equiv 0, \\
    \Gamma_{c 0 0} &\equiv 0; &\qquad \Gamma_{c 0 1} &\equiv 0; \\
    \Gamma_{0 0 b} &\equiv 0, &\qquad \Gamma_{0 1 1} &\equiv \tfrac12\bhm r^{-2} + \pa_1 h_{0 1}, \\
    \Gamma_{1 0 b} &\equiv 0, &\qquad \Gamma_{1 1 1} &\equiv \tfrac12\pa_1 h_{1 1}, \\
    \Gamma_{c 0 b} &\equiv \tfrac12(r-2\bhm)\slg_{b c}; &\qquad \Gamma_{c 1 1} &\equiv r\pa_1 h_{1\bar c}; \\
    \Gamma_{0 1 b} &\equiv 0, &\qquad \Gamma_{0 a b} &\equiv -\tfrac12(r-2\bhm)\slg_{a b}, \\
    \Gamma_{1 1 b} &\equiv 0, &\qquad \Gamma_{1 a b} &\equiv \tfrac12(r-2\bhm)\slg_{a b} - \tfrac12 r^2\pa_1 h_{\bar a\bar b}, \\
    \Gamma_{c 1 b} &\equiv -\tfrac12(r-2\bhm)\slg_{b c} + \tfrac12 r^2\pa_1 h_{\bar b\bar c}; &\qquad \Gamma_{c a b} &\equiv r^2\slGamma_{c a b}.
  \end{alignat*}
  The expressions for $\Gamma_{\mu\nu}^\kappa=g^{\kappa\lambda}\Gamma_{\lambda\mu\nu}$, modulo $r^{\fs(\mu,\nu)-\fs(\kappa)}(\rho_0^3\rho_\sscri^2\CI+H_\bop^{k-1,(\cE_0+1,\ell_0+1),(\cE_\sscri^\cC,\ell_\sscri)})$, are
  \begin{alignat*}{2}
    \Gamma_{0 0}^0 &\equiv \bhm r^{-2}, &\qquad \Gamma_{0 1}^0 &\equiv 0, \\
    \Gamma_{0 0}^1 &\equiv 0, &\qquad \Gamma_{0 1}^1 &\equiv 0, \\
    \Gamma_{0 0}^c &\equiv 0; &\qquad \Gamma_{0 1}^c &\equiv 0; \\
    \Gamma_{0 b}^0 &\equiv 0, &\qquad \Gamma_{1 1}^0 &\equiv -\pa_1 h_{1 1}, \\
    \Gamma_{0 b}^1 &\equiv 0, &\qquad \Gamma_{1 1}^1 &\equiv -\bhm r^{-2}-2\pa_1 h_{0 1}, \\
    \Gamma_{0 b}^c &\equiv \tfrac12 r^{-1}(1-2\bhm r^{-1})\delta_b^c; &\qquad \Gamma_{1 1}^c &\equiv r^{-1}\pa_1 h_1{}^{\bar c}; \\
    \Gamma_{1 b}^0 &\equiv 0, &\qquad \Gamma_{a b}^0 &\equiv -r\slg_{a b} + r^2\pa_1 h_{\bar a\bar b}, \\
    \Gamma_{1 b}^1 &\equiv 0, &\qquad \Gamma_{a b}^1 &\equiv r\slg_{a b}, \\
    \Gamma_{1 b}^c &\equiv -\tfrac12 r^{-1}(1-2\bhm r^{-1})\delta_b^c + \tfrac12\pa_1 h_{\bar b}{}^{\bar c}; &\qquad \Gamma_{a b}^c &\equiv \slGamma_{a b}^c.
  \end{alignat*}
\end{lemma}

We regard as error terms those which have one more power of $\rho_0$ compared to the remainder terms in Lemma~\ref{LemmaExOpMet}. The reason for this is that all derivatives $\pa_0\in\rho_0\rho_\sscri\Vb(M)$, $\pa_1\in\rho_0\Vb(M)$, $r^{-1}\pa_c\in\rho_0\rho_\sscri\Vb(M)$ gain one power of $\rho_0$. Note also that $\pa_1$ does not gain a power of $\rho_\sscri$, while $\pa_0$ and $r^{-1}\pa_c$ do.

\begin{proof}[Proof of Lemma~\usref{LemmaExOpChr}]
  In the computation of $\Gamma_{\kappa\mu\nu}$, derivatives of the coefficients of $g_\bhm$ are computed using~\eqref{EqKNullComp2}. For derivatives of the coefficients of $h$---which at $\scri^+$ have index sets contained in $\cE_\sscri^\tot$ with $\min\Re\cE_\sscri^\tot\geq 1$---we note that derivatives along $\pa_0$ and $r^{-1}\pa_c$ yield an additional order of decay at $\scri^+$ and thus produce error terms; only derivatives along $\pa_1$ can produce leading order contributions. For $\Gamma_{c 1 1}$, we note that $\pa_1 h_{1 c}=\pa_1(r h_{1\bar c})=r\pa_1 h_{1\bar c}-\frac12(1-2\bhm r^{-1})h_{1\bar c}$, with the second term being an error term; similarly for $\Gamma_{c 1 b}$, $\Gamma_{1 a b}$. (The point is that $\pa_1$- and also $\pa_0$-derivatives falling on powers of $r$ gain a power of $r^{-1}$.)

  We compute $\Gamma_{\mu\nu}^\kappa$ using~\eqref{EqExOpMetDual}. Note that an error term $\rho_0^3\rho_\sscri^2\CI+H_\bop^{k-1,(\cE_0+1,\ell_0+1),(\cE_\sscri^\cC,\ell_\sscri)}$ contributing to $\Gamma_{\bar\lambda\bar\mu\bar\nu}:=r^{-\fs(\mu,\nu,\lambda)}\Gamma_{\lambda\mu\nu}$ produces, upon raising the first index using $g^{\bar\kappa\bar\lambda}\in\CI+H_\bop^{k,(\cE_0,\ell_0),(\cE_\sscri^\tot,\ell_\sscri)}$, a term in
  \[
    \rho_0^3\rho_\sscri^2\CI + H_\bop^{k-1,(\cE_0+1,\ell_0+1),(\cE_\sscri^\cC,\ell_\sscri)} + H_\bop^{k,(\cE_0+3,\ell_0+3),(\cE_\sscri^\tot+2,\ell_\sscri+2)} + H_\bop^{k-1,(2\cE_0+1,\ell_0+1),(\cE_\sscri^\cC+\cE_\sscri^\tot,\ell_\sscri)};
  \]
  for the final summand, arising from the product of the $H_\bop$-terms, we use only that $\min\Re\cE_0>0$. Since \eqref{EqEinPertTot} and~\eqref{EqExPEC} imply $\cE_\sscri^\tot+2\subset\cE_\sscri^\tot+1\subset\cE_\sscri^\cC$ and $\cE_\sscri^\cC+\cE_\sscri^\tot\subset 2\cE_\sscri^\tot\subset\cE_\sscri^\cC$, this lies in $\rho_0^3\rho_\sscri^2\CI+H_\bop^{k-1,(\cE_0+1,\ell_0+1),(\cE_\sscri^\cC,\ell_\sscri)}$, i.e., it is an error term. Thus, only the leading-order terms of $\Gamma_{\lambda\mu\nu}$ contribute leading-order terms to $\Gamma_{\mu\nu}^\kappa$.

  Products of terms in $g^{\bar\kappa\bar\lambda}$ involving $h$, which are of class $H_\bop^{k,(\cE_0,\ell_0),(\cE_\sscri^\tot,\ell_\sscri)}$, with terms in $\Gamma_{\bar\lambda\bar\mu\bar\nu}$ involving $h$, which are of class $H_\bop^{k-1,(\cE_0+1,\ell_0+1),(\cE_\sscri^\tot,\ell_\sscri)}$, are error terms as well since $2\cE_\sscri^\tot\subset\cE_\sscri^\cC$. The stated expressions for $\Gamma_{\mu\nu}^\kappa$ now follow easily.
\end{proof}

For bookkeeping, it is convenient to introduce
\[
  \Gamma_{\bar\mu\bar\nu}^{\bar\kappa} := r^{\fs(\kappa)-\fs(\mu,\nu)}\Gamma_{\mu\nu}^\kappa,
\]
which is formally an instance of Definition~\ref{DefKNullComp} (except the Christoffel symbols are not components of a tensor). We note that
\begin{equation}
\label{EqExOpChrMem}
  \Gamma_{\bar\mu\bar\nu}^{\bar\kappa}\in r^{-1}\CI + H_\bop^{k-1,(\cE_0+1,\ell_0+1),(\cE_\sscri^\tot,\ell_\sscri)}.
\end{equation}

\begin{lemma}[Curvature tensor]
\label{LemmaExOpRiem}
  The expressions for
  \[
    \Riem(g)^\kappa{}_{\lambda\mu\nu}=\pa_\mu\Gamma_{\nu\lambda}^\kappa+\Gamma_{\mu\rho}^\kappa\Gamma_{\lambda\nu}^\rho-\pa_\nu\Gamma_{\mu\lambda}^\kappa-\Gamma_{\nu\rho}^\kappa\Gamma_{\lambda\mu}^\rho,
  \]
  modulo $r^{\fs(\mu,\nu,\lambda)-\fs(\kappa)}(\rho_0^3\rho_\sscri^2\CI+H_\bop^{k-2,(\cE_0+2,\ell_0+2),(\cE_\sscri^\cC,\ell_\sscri)})$, are $\Riem(g)^\kappa{}_{\mu\nu\lambda}=-\Riem(g)^\kappa{}_{\mu\lambda\nu}\equiv 0$ for all $\kappa,\mu,\nu,\lambda$ with $\nu\leq\lambda$ except for
  \begin{equation}
  \label{EqExOpRiem}
    \Riem(g)^d{}_{1 1 b} \equiv \frac12 \pa_1^2 h_{\bar b}{}^{\bar d},\quad
    \Riem(g)^0{}_{c 1 b} \equiv r^2\pa_1^2 h_{\bar b\bar c}.
  \end{equation}
  Moreover, we have
  \[
    \Ric(g)_{\mu\nu} \equiv 0 \bmod r^{\fs(\mu,\nu)}(\rho_0^3\rho_\sscri^2\CI+H_\bop^{k-2,(\cE_0+2,\ell_0+2),(\cE_\sscri^\cC,\ell_\sscri)}).
  \]
\end{lemma}

This result does not quite follow from \cite[Corollary~3.26]{HintzMink4Gauge} since the present error space is smaller. It is consistent with the expressions \cite[(A.7)]{HintzVasyMink4} where the metric coefficients are allowed to have even less decay at $\scri^+$ than in the present paper. We also note that if one were to reduce the error space from $\rho_0^3\rho_\sscri^2\CI$ to $\rho_0^4\rho_\sscri^2\CI$, one would obtain additional nonzero components coming from the long-range mass term of the Schwarzschild metric.

\begin{proof}[Proof of Lemma~\usref{LemmaExOpRiem}]
  Error terms of the Christoffel symbols $\Gamma_{\mu\nu}^\kappa$ contribute error terms also to the components of $\Riem(g)$:
  \begin{enumerate}
  \item For the terms linear in Christoffel symbols, this follows from~\eqref{EqKNullComp2} and $r^{-1}\pa_a\in\rho_0\rho_\sscri\Vb(M)$.
  \item For the quadratic terms, \emph{all} contributions of $h$ to the Christoffel symbols (which lie in $H_\bop^{k-1,(\cE_0+1,\ell_0+1),(\cE_\sscri^\tot,\ell_\sscri)}$) yield error terms, as follows from~\eqref{EqExOpChrMem} and the fact that
    \[
      (r^{-1}\CI+H_\bop^{k-1,(\cE_0+1,\ell_0+1),(\cE_\sscri^\tot,\ell_\sscri)}) \cdot H_\bop^{k-1,(\cE_0+1,\ell_0+1),(\cE_\sscri^\tot,\ell_\sscri)} \subset H_\bop^{k-1,(\cE_0+2,\ell_0+2),(\cE_\sscri^\cC,\ell_\sscri)};
    \]
    here we use $\cE_\sscri^\tot+1$, $2\cE_\sscri^\tot\subset\cE_\sscri^\cC$. Thus, only the Schwarzschildean contributions to the Christoffel symbols matter here. 
  \end{enumerate}
  For the linear terms, we observe moreover that derivatives along $\pa_0$ and $r^{-1}\pa_c$ falling on the terms involving $h$ contribute error terms, since $\pa_0$ and $r^{-1}\pa_c$ map the space $H_\bop^{k-1,(\cE_0+1,\ell_0+1),(\cE_\sscri^\tot,\ell_\sscri)}$ into the error space $H_\bop^{k-2,(\cE_0+2,\ell_0+2),(\cE_\sscri^\cC,\ell_\sscri)}$ (again using $\cE_\sscri^\tot+1\subset\cE_\sscri^\cC$). Since the Riemann curvature tensor of Schwarzschild is of class $r^{-3}\CI(M;\cT\otimes\cT^*\otimes\cT^*\otimes\cT^*)$, it only contributes an error term. The only nontrivial contributions thus arise from derivatives of components of $h$ (other than $\pi^\cC h$) along $\pa_1^2$. The claim regarding $\Riem(g)$ now follows by a simple computation.

  For the Ricci tensor, only the $(1,1)$ component can be affected by the nontrivial components~\eqref{EqExOpRiem} of $\Riem(g)$; but $\Riem(g)^d{}_{1 d 1}\equiv-\frac12\pa_1^2(\slg^{c d}h_{\bar c\bar d})\equiv 0$ in view of the membership of $\pi^\cC h$ (specifically, its component $\sltr h$) in~\eqref{EqExPSpaceProj}.
\end{proof}

Proceeding to differential operators, we first relate covariant derivatives to ordinary derivatives. Concretely, set
\[
  \cT^{p,q}:=\cT^{\otimes p}\otimes(\cT^*)^{\otimes q},
\]
and write $\pa_0$, $\pa_1$ for derivatives of the components of $(p,q)$-tensors in the splitting of $\cT^{p,q}$ induced by~\eqref{EqKNullT}. We record two types of memberships: one in spaces of b-differential operators on $M$, which are most convenient for extracting asymptotic expansions; and one in spaces of edge-b-differential operators on $M_{\frac12}$ (see Definition~\ref{DefKMfdRad2}), which (following \cite{HintzVasyScrieb,HintzMink4Gauge}) are better suited for basic (energy or microlocal) estimates since they capture the hyperbolic nature of linear wave operators non-degenerately. (In~\S\ref{SD}, edge-b-notions are only used to verify the structural properties of dynamical wave-type operators near $\scri^+$ as needed for the application of the results of \cite{HintzNonstat2}.)

\begin{cor}[Covariant derivatives]
\label{CorExCov}
  Denote by $\nabla$ the Levi-Civita connection for the metric $g$ from~\eqref{EqExOpMet}. Write $\nabla_0$ for the map $T\mapsto\nabla_0 T=\nabla_{\pa_0}T$ where $T$ is a section of $\cT^{p,q}$; similarly for $\nabla_1$. Then, as maps on sections of $\cT^{p,q}$,
  \begin{subequations}
  \begin{align}
  \label{EqExCov0}
    \nabla_0 &\equiv \pa_0 \bmod (r^{-2}\CI + H_\bop^{k-1,(\cE_0+1,\ell_0+1),(\cE_\sscri^\cC,\ell_\sscri)}), \\
  \label{EqExCov1}
    \nabla_1 &\equiv \pa_1 \bmod (r^{-1}\CI + H_\bop^{k-1,(\cE_0+1,\ell_0+1),(\cE_\sscri^\tot,\ell_\sscri)}), \\
  \label{EqExCov2}
    r^{-1}\nabla_a &\equiv r^{-1}\slnabla_a \bmod (r^{-1}\CI+\Hb^{k-1,(\cE_0+1,\ell_0+1),(\cE_\sscri^\tot,\ell_\sscri)}).
  \end{align}
  \end{subequations}
  Furthermore,
  \begin{subequations}
  \begin{align}
  \label{EqExCov02}
    \nabla_0 &\in (r^{-1}\CI+H_\bop^{k-1,(\cE_0+1,\ell_0+1),(\cE_\sscri^\cC,\ell_\sscri)})\Diffb^1(\Omega_\ext;\cT^{p,q}), \\
  \label{EqExCov12}
    \nabla_1 &\in (\rho_0\CI+H_\bop^{k-1,(\cE_0+1,\ell_0+1),(\cE_\sscri^\tot,\ell_\sscri)})\Diffb^1(\Omega_\ext;\cT^{p,q}), \\
  \label{EqExCovc2}
    r^{-1}\nabla_a &\in (r^{-1}\CI+H_\bop^{k-1,(\cE_0+1,\ell_0+1),(\cE_\sscri^\tot,\ell_\sscri)})\Diffb^1(\Omega_\ext;\cT^{p,q}).
  \end{align}
  \end{subequations}
  The memberships~\eqref{EqExCov02}--\eqref{EqExCov12} also hold when replacing $\Diffb^1(\Omega_\ext)$ by $\Diffeb^1(\Omega_{\ext,\frac12})$ upon also replacing $\cE_\sscri^\cC,\cE_\sscri^\tot,\ell_\sscri$ by $2\cE_\sscri^\cC,2\cE_\sscri^\tot,2\ell_\sscri$, whereas
  \begin{equation}
  \label{EqExCovc2eb}
    r^{-1}\nabla_a \in (\rho_0 x_\sscri\CI+H_\bop^{k-1,(\cE_0+1,\ell_0+1),(2\cE_\sscri^\tot,2\ell_\sscri)})\Diffeb^1(\Omega_{\ext,\frac12};\cT^{p,q}).
  \end{equation}
\end{cor}
\begin{proof}
  We only consider the case $p=0$, $q=1$, the general case being only notationally more burdensome. By Lemma~\ref{LemmaExOpChr}, we have $\Gamma^{\bar\sigma}_{0\bar\mu}\equiv 0$ modulo $r^{-2}\CI+H_\bop^{k-1,(\cE_0+1,\ell_0+1),(\cE_\sscri^\cC,\ell_\sscri)}$ for all $\sigma$ and $\mu$ except
  \begin{equation}
  \label{EqExCovChr0}
    \Gamma^{\bar c}_{0\bar b} \equiv \tfrac12 r^{-1}(1-2\bhm r^{-1})\delta_b^c.
  \end{equation}
  Thus, using~\eqref{EqKNullComp2},
  \[
    (\nabla_0 T)_{\bar\mu} = r^{-\fs(\mu)}(\nabla_0 T)_\mu = r^{-\fs(\mu)}\pa_0(r^{\fs(\mu)}T_{\bar\mu}) - \Gamma_{0\bar\mu}^{\bar\sigma}T_{\bar\sigma} = \pa_0 T_{\bar\mu} + \fs(\mu)\tfrac12(1-2\bhm r^{-1})T_{\bar\mu} - \Gamma_{0\bar\mu}^{\bar\sigma}T_{\bar\sigma}.
  \]
  The contribution of $\Gamma_{0\bar\mu}^{\bar\sigma}$ cancels the second summand, yielding~\eqref{EqExCov0}. The proof of~\eqref{EqExCov1} is analogous, now using that $\Gamma^{\bar\sigma}_{1\bar\mu}\equiv 0$ modulo $r^{-2}\CI+H_\bop^{k-1,(\cE_0+1,\ell_0+1),(\cE_\sscri^\tot,\ell_\sscri)}$ for all $\sigma$ and $\mu$ except
  \begin{equation}
  \label{EqExCovChr1}
    \Gamma_{1\bar b}^{\bar c} \equiv -\tfrac12 r^{-1}(1-2\bhm r^{-1})\delta_b^c.
  \end{equation}
  The memberships~\eqref{EqExCov02} and \eqref{EqExCov12} now follow from~\eqref{EqKNullComp2}. Finally, to prove~\eqref{EqExCov2} and~\eqref{EqExCovc2}, we note that $\Gamma_{\bar a\bar\mu}^{\bar\sigma} \in r^{-1}\CI+H_\bop^{k-1,(\cE_0+1,\ell_0+1),(\cE_\sscri^\tot,\ell_\sscri)}$ for all $\mu$ and $\sigma$, which is a special case of~\eqref{EqExOpChrMem}.

  The statements in the edge-b-setting follow from~\eqref{EqExebpa01}.
\end{proof}

\begin{cor}[Tensor wave operator]
\label{CorExOpBox2}
  Fix
  \begin{equation}
  \label{EqExOpBox2Flat}
    \ell_0^\flat\in(0,\ell_0],\ \ell_0^\flat<\min\Re\cE_0;\qquad \ell_\sscri^\flat\in(1,\ell_\sscri],\ \ell_\sscri^\flat<\min\Re\cE_\sscri^\cC,\ 2\ell_\sscri^\flat\leq 3.
  \end{equation}
  Then the tensor wave operator $(\Box_g\dot g)_{\mu\nu}=-g^{\kappa\lambda}\dot g_{\mu\nu;\kappa\lambda}$ satisfies
  \begin{subequations}
  \begin{align}
  \label{EqExOpBox2b}
    \Box_g &\equiv 4\pa_1 r^{-1}\pa_0 r \bmod (\rho_0^2\rho_\sscri^2\CI+H_\bop^{k-2,(\cE_0+2,\ell_0+2),(\cE_\sscri^\cC,\ell_\sscri)})\Diffb^2(\Omega_\ext;S^2\cT^*), \\
  \label{EqExOpBox2eb}
    \Box_g &\equiv 4\pa_1 r^{-1}\pa_0 r + r^{-2}\slDelta \bmod \rho_0^2 x_\sscri^2(x_\sscri\CI + H_\bop^{k-2,\ell_0^\flat,2(\ell_\sscri^\flat-1)})\Diffeb^2(\Omega_{\ext,\frac12},S^2\cT^*),
  \end{align}
  \end{subequations}
  where the spherical tensor Laplacian $\slDelta$ acts componentwise in the splitting~\eqref{EqKNullS2T}. The leading-order terms here satisfy
  \[
    4\pa_1 r^{-1}\pa_0 r \in \rho_0^2\rho_\sscri\Diffb^2(M),\quad
    4\pa_1 r^{-1}\pa_0 r+r^{-2}\slDelta \in \rho_0^2 x_\sscri^2\Diffeb^2(M_{\frac12}),
  \]
  and the remainders have stronger decay at $\scri^+$ than these.
\end{cor}

Since we will use the edge-b-perspective only for obtaining basic energy estimates (without precise asymptotics at $I^0$ and $\scri^+$), the unstructured form of the remainder in~\eqref{EqExOpBox2eb} will be sufficient. On the other hand, for the extraction of polyhomogeneous expansions with reasonably precise control on index sets, we shall use the b-perspective; for this purpose, the (partially) polyhomogeneous form~\eqref{EqExOpBox2b} of the remainder will be important.

\begin{proof}[Proof of Corollary~\usref{CorExOpBox2}]
  We need to compute
  \begin{equation}
  \label{EqExOpBox2}
    (\Box_g\dot g)_{\bar\mu\bar\nu} = \underbrace{-g^{\bar\kappa\bar\lambda} r^{-\fs(\mu,\nu,\kappa,\lambda)}\pa_\lambda(r^{\fs(\mu,\nu,\kappa)}\dot g_{\bar\mu\bar\nu;\bar\kappa})}_{{\rm I}}{}+ \underbrace{g^{\bar\kappa\bar\lambda}\Gamma_{\bar\mu\bar\lambda}^{\bar\sigma}\dot g_{\bar\sigma\bar\nu;\bar\kappa}}_{{\rm II}}{} + \underbrace{g^{\bar\kappa\bar\lambda}\Gamma_{\bar\nu\bar\lambda}^{\bar\sigma}\dot g_{\bar\mu\bar\sigma;\bar\kappa}}_{{\rm III}}{} + \underbrace{g^{\bar\kappa\bar\lambda}\Gamma^{\bar\sigma}_{\bar\kappa\bar\lambda}\dot g_{\bar\mu\bar\nu;\bar\sigma}}_{{\rm IV}}.
  \end{equation}
  We start with the first term. Consider first the error terms of $g^{\bar\kappa\bar\lambda}$ in~\eqref{EqExOpMetDual}, i.e., the contributions of class $r^{-2}\CI+H_\bop^{k,(\cE_0,\ell_0),(\cE_\sscri^\cC,\ell_\sscri)}$: using Corollary~\ref{CorExCov}, their contributions to the term ${\rm I}$ are of class
  \[
    (r^{-2}\CI+H_\bop^{k,(\cE_0,\ell_0),(\cE_\sscri^\cC,\ell_\sscri)})\circ\rho_0\Diffb^1\circ(\rho_0\CI+H_\bop^{k-1,(\cE_0+1,\ell_0+1),(\cE_\sscri^\tot,\ell_\sscri)})\Diffb^1
  \]
  acting on $\dot g$ (with the largest possible contribution at $\scri^+$ coming from $\pa_1^2$); this is contained in the error space $(r^{-2}\CI+H_\bop^{k-2,(\cE_0+2,\ell_0+2),(\cE_\sscri^\cC,\ell_\sscri)})\Diffb^2$. (On the edge-b-scale, these contributions lie in the error space $(\rho_0^2 x_\sscri^4\CI+H_\bop^{k-2,(\cE_0+2,\ell_0+2),(2\cE_\sscri^\cC,2\ell_\sscri)})\Diffeb^2(\Omega_{\ext,\frac12})$.) Next, the contributions of class $H_\bop^{k,(\cE_0,\ell_0),(\cE_\sscri^\tot,\ell_\sscri)}$ to $g^{0 0}$, $g^{0 1}$, $g^{0\bar b}$, and $g^{\bar a\bar b}$ (i.e., all terms involving $h$) in~\eqref{EqExOpMetDual} to term ${\rm I}$ are of class
  \begin{align*}
    &H_\bop^{k,(\cE_0,\ell_0),(\cE_\sscri^\tot,\ell_\sscri)}\circ\rho_0\Diffb^1\circ(r^{-1}\CI+H_\bop^{k-1,(\cE_0+1,\ell_0+1),(\cE_\sscri^\tot,\ell_\sscri)})\Diffb^1, \\
    &H_\bop^{k,(\cE_0,\ell_0),(\cE_\sscri^\tot,\ell_\sscri)}\circ r^{-1}\Diffb^1\circ(\rho_0\CI+H_\bop^{k-1,(\cE_0+1,\ell_0+1),(\cE_\sscri^\tot,\ell_\sscri)})\Diffb^1
  \end{align*}
  (with the largest possible contribution arising from derivatives along $\pa_1$ and $r^{-1}\nabla_a$); this is also contained in the error space $H_\bop^{k-2,(\cE_0+2,\ell_0+2),(\cE_\sscri^\cC,\ell_\sscri)}$, using again that $\cE_\sscri^\tot+1\subset\cE_\sscri^\cC$. (To analyze these terms on the eb-scale, one replaces the weight $r^{-1}$ by $\rho_0 x_\sscri$, cf.\ the difference between~\eqref{EqExCovc2} and \eqref{EqExCovc2eb}, and thus the total $x_\sscri$-weight at $\scri^+\subset M_{\frac12}$ is at least $3$; this is the reason for imposing the final upper bound in~\eqref{EqExOpBox2Flat}.)

  The contribution of the term $\slg^{a b}$ of $g^{\bar a\bar b}$ to term ${\rm I}$ in~\eqref{EqExOpBox2} is of class
  \[
    r^{-1}\Diffb^1\circ(r^{-1}\CI+H_\bop^{k-1,(\cE_0+1,\ell_0+1),(\cE_\sscri^\tot,\ell_\sscri)})\Diffb^1
  \]
  by~\eqref{EqExCovc2}, which is contained in the error space. (On the edge-b-scale, $-r^{-2}\slg^{a b}\pa_a\pa_b$ is of leading order. We may replace $\pa_a$ by $\slnabla_a$ upon committing an error of class $\rho_0^2 x_\sscri^3\Diffeb^2(\Omega_{\ext,\frac12})$, since in any local coordinate system on $\Sph^2$, the difference $x_\sscri\nabla_a-x_\sscri\pa_a$ is a multiplication operator by an element of $x_\sscri\CI$.) Likewise for the contributions of the term $-4\bhm r^{-1}$ of $g^{1 0}=g^{0 1}$ to the term ${\rm I}$, which is given by error terms of class
  \begin{align*}
    r^{-1}\CI\circ r^{-1}\Diffb^1 \circ (\rho_0\CI+H_\bop^{k-1,(\cE_0+1,\ell_0+1),(\cE_\sscri^\tot,\ell_\sscri)})\Diffb^1, \\
    r^{-1}\CI\circ \rho_0\Diffb^1 \circ (r^{-1}\CI+H_\bop^{k-1,(\cE_0+1,\ell_0+1),(\cE_\sscri^\cC,\ell_\sscri)})\Diffb^1
  \end{align*}
  by~\eqref{EqExCov12} and \eqref{EqExCov02}, respectively. (On the edge-b-scale, the weights are the same, except for the usual factor of $2$ in the $\scri^+$-weight.) Modulo an element of the error space $(\rho_0^2\rho_\sscri^2\CI+H_\bop^{k-2,(\cE_0+2,\ell_0+2),(\cE_\sscri^\cC,\ell_\sscri)})\Diffb^2(\Omega_\ext)$ acting on $\dot g$, we thus have
  \begin{equation}
  \label{EqExOpBox21}
    {\rm I}{} \equiv 2 r^{-\fs(\mu,\nu)}\pa_1(r^{\fs(\mu,\nu)}\dot g_{\bar\mu\bar\nu;0})+2 r^{-\fs(\mu,\nu)}\pa_0(r^{\fs(\mu,\nu)}\dot g_{\bar\mu\bar\nu;1}) \equiv 4\pa_0\pa_1\dot g_{\bar\mu\bar\nu} + r^{-1}\fs(\mu,\nu)\dot g_{\bar\mu\bar\nu;1}.
  \end{equation}
  by~\eqref{EqKNullComp2} and~\eqref{EqExCov0}--\eqref{EqExCov1}; and we can further replace $\dot g_{\bar\mu\bar\nu;1}$ by $\pa_1\dot g_{\bar\mu\bar\nu}$ in view of~\eqref{EqExCov1}. (On the edge-b-scale, term ${\rm I}$ is given, modulo $(\rho_0^2 x_\sscri^3\CI+\Hb^{k-2,(\cE_0+1,\ell_0+1),(2\cE_\sscri^\cC,2\ell_\sscri)})\Diffeb^2(\Omega_{\ext,\frac12})$, by the sum of this expression and $r^{-2}\slDelta$.)

  We proceed to evaluate the term ${\rm II}$ in~\eqref{EqExOpBox2}. Since $\Gamma^{\bar\sigma}_{\bar\mu\bar\lambda}\in r^{-1}\CI+H_\bop^{k-1,(\cE_0+1,\ell_0+1),(\cE_\sscri^\tot,\ell_\sscri)}$ by~\eqref{EqExOpChrMem}, we conclude from~\eqref{EqExCov02}--\eqref{EqExCovc2} that terms with $\kappa\neq 1$ contribute error terms. For $\kappa=1$ then, terms with $\lambda\neq 0$ contribute only error terms by~\eqref{EqExOpMetDual}. For $\lambda=0$, we get (modulo error terms) $-2\Gamma_{\bar\mu 0}^{\bar\sigma}\pa_1\dot g_{\bar\sigma\bar\nu}\equiv -r^{-1}\fs(\mu)\pa_1\dot g_{\bar\mu\bar\nu}$ by~\eqref{EqExCovChr0}. The term ${\rm III}$ in~\eqref{EqExOpBox2} is the same but with $\mu$ and $\nu$ exchanged; thus
  \begin{equation}
  \label{EqExOpBox223}
    {\rm II}{} + {\rm III}{} \equiv -r^{-1}\fs(\mu,\nu)\pa_1\dot g_{\bar\mu\bar\nu}.
  \end{equation}

  In the term ${\rm IV}$ of~\eqref{EqExOpBox2}, again only $\sigma=1$ contributes, and we then use
  \[
    g^{\bar\kappa\bar\lambda}\Gamma^1_{\bar\kappa\bar\lambda} \equiv 2 r^{-1} \bmod (r^{-2}\CI+H_\bop^{k-1,(\cE_0+1,\ell_0+1),(\cE_\sscri^\cC,\ell_\sscri)})
  \]
  to conclude that
  \begin{equation}
  \label{EqExOpBox24}
    {\rm IV}{} \equiv 2 r^{-1}\pa_1\dot g_{\bar\mu\bar\nu}.
  \end{equation}
  Combining~\eqref{EqExOpBox21}, \eqref{EqExOpBox223}, and \eqref{EqExOpBox24} gives
  \[
    \Box_g \equiv 2(2\pa_0+r^{-1})\pa_1 \equiv 4 \pa_1 r^{-1} \pa_0 r.
  \]
  (On the edge-b-scale, we have the same expression plus $-r^{-2}\slg^{a b}\slnabla_a\slnabla_b$, modulo the error space in~\eqref{EqExOpBox2eb}.)
\end{proof}

We are now ready to prove the following variant of \cite[Lemma~3.8]{HintzVasyMink4} and \cite[Proposition~3.29]{HintzMink4Gauge}:

\begin{prop}[Structure of $L_{g,g^0}$: exterior region]
\label{PropExOpLin}
  Consider a Lorentzian metric $g=g_b+h$, $h\in\Hb^{k,(\cE_0,\ell_0),(\la\cE_\sscri^\cC\ra,\ell_\sscri)}(\Omega_\ext)$ as in~\eqref{EqExOpMet} and Definition~\usref{DefExP}. Let $g^0$ denote a metric which (in the notation of Definition~\usref{DefKMetRef}) satisfies\footnote{We use the notation of Definition~\ref{DefExPLead} here. An equivalent formulation of this condition is that $g^0-g_\bhm\in r^{-1}\CI(M;S^2\cT^*)$ and $\pi(g^0-g_\bhm)\in\rho_0\rho_\sscri^2\CI$ for $\pi\in\{\pi^\cC,\,\pi_{0 1},\,\pi_{1 /}\}$.} $g^0-g_\bhm\in\Hb^{\infty,\ \bigl((1,0),\infty\bigr),\ \bigl(\la(2,0)\ra_0,\infty\bigr)}(\Omega_\ext)$. Fix (as in~\eqref{EqExOpBox2Flat}) quantities
  \begin{equation}
  \label{EqExOpLinFlat}
    \ell_0^\flat\in(0,\ell_0],\ \ell_0^\flat<\min\Re\cE_0;\qquad \ell_\sscri^\flat\in(1,\ell_\sscri],\ \ell_\sscri^\flat<\min\Re\cE_\sscri^\cC,\ 2\ell_\sscri^\flat\leq 3.
  \end{equation}
  Then the operator $L_{g,g^0}=D_1|_{(g,g^0)}P$ from~\eqref{Eq1EinLin} satisfies
  \begin{subequations}
  \begin{align}
  \label{EqExOpLinb}
    \begin{split}
      L_{g,g^0} &\equiv (2 r^{-1}\pa_0 r + r^{-1}A_h)\pa_1 + r^{-1} B_h \\
        &\quad\qquad \bmod (\rho_0^2\rho_\sscri^2\CI+H_\bop^{k-2,(\cE_0+2,\ell_0+2),(\cE_\sscri^\cC,\ell_\sscri)})\Diffb^2(\Omega_\ext;S^2\cT^*),
    \end{split} \\
  \label{EqExOpLineb}
    \begin{split}
      L_{g,g^0} &\equiv (2 r^{-1}\pa_0 r + r^{-1}A_h)\pa_1 + r^{-2}\slDelta + r^{-1} B_h \\
        &\quad\qquad \bmod \rho_0^2 x_\sscri^2(x_\sscri\CI+H_\bop^{k-2,\ell_0^\flat,2(\ell_\sscri^\flat-1)})\Diffeb^2(\Omega_{\ext,\frac12};S^2\cT^*),
    \end{split}
  \end{align}
  \end{subequations}
  where, in the bundle splitting
  \begin{equation}
  \label{EqExOpLinSplit}
  \begin{split}
    S^2\cT^* &= \la(\dd x^0)^2\ra \oplus \bigl(2\,\dd x^0\otimes_s r T^*\Sph^2\bigr) \oplus \la r^2\slg\ra \oplus \la 2\,\dd x^0\,\dd x^1\ra \\
      &\qquad \oplus \bigl(2\,\dd x^1\otimes_s r T^*\Sph^2\bigr) \oplus r^2\ker\sltr{}\oplus \la(\dd x^1)^2\ra,
  \end{split}
  \end{equation}
  the operator $A_h$ is lower triangular and $B_h$ is strictly lower triangular; explicitly,
  \begin{equation}
  \label{EqExOpLinAB}
  \begin{split}
    A_h&=
      \scalebox{0.8}{$\openbigpmatrix{3pt}
        2(1{-}v^\cC)\gamma^\cC & 0 & 0 & 0 & 0 & 0 & 0 \\
        0 & (1{-}v^\cC)\gamma^\cC & 0 & 0 & 0 & 0 & 0 \\
        2(1{-}e^\cC)(1{+}v^\cC)\gamma^\cC & 0 & (1{-}e^\cC)(1{-}v^\cC)\gamma^\cC & 0 & 0 & 0 & 0 \\
        e^\cC(1{+}v^\cC)\gamma^\cC{+}\gamma^{\!\Ups}{+}2\pa_1(r h_{0 1}) & 0 & \frac12(e^\cC(1{-}v^\cC)\gamma^\cC{+}e^{\!\Ups}\gamma^{\!\Ups}) & (1{-}e^{\!\Ups})\gamma^{\!\Ups} & 0 & 0 & 0 \\
        2\pa_1(r h_{1\bar a}) & (1{+}v^\cC)\gamma^\cC{+}\gamma^{\!\Ups} & 0 & 0 & \gamma^{\!\Ups} & 0 & 0 \\
        2\pa_1(r h_{\bar a\bar b}) & 0 & 0 & 0 & 0 & 0 & 0 \\
        0 & 2\pa_1(r h_1{}^{\bar a}) & (1{+}v^\cC)\gamma^\cC{+}e^{\!\Ups}\gamma^{\!\Ups}{-}2\pa_1(r h_{0 1}) & 2(1{-}e^{\!\Ups})\gamma^{\!\Ups} & 0 & {-}\frac12\pa_1(r h^{\bar a\bar b}) & 2\gamma^{\!\Ups}
      \closebigpmatrix$}, \\
    B_h&=
      \begin{pmatrix}
        0 & 0 & 0 & 0 & 0 & 0 & 0 \\
        0 & 0 & 0 & 0 & 0 & 0 & 0 \\
        0 & 0 & 0 & 0 & 0 & 0 & 0 \\
        2\pa_1^2(r h_{0 1}) & 0 & 0 & 0 & 0 & 0 & 0 \\
        2\pa_1^2(r h_{1\bar a}) & 0 & 0 & 0 & 0 & 0 & 0 \\
        2\pa_1^2(r h_{\bar a\bar b}) & 0 & 0 & 0 & 0 & 0 & 0 \\
        2\pa_1^2(r h_{1 1}) & 0 & 0 & 0 & 0 & 0 & 0
      \end{pmatrix},
  \end{split}
  \end{equation}
  The components of $A_h$ are of class $\CI+\Hb^{k-1,(\cE_0,\ell_0),(\cE_\sscri^\tot-1,\ell_\sscri-1)}$ on $\Omega_\ext$, and those of $B_h$ are of class $\Hb^{k-2,(\cE_0+1,\ell_0+1),(\cE_\sscri^\tot-1,\ell_\sscri-1)}$.
\end{prop}

The lower triangular structure of $A_h$ and $B_h$ was noted in a similar setting in \cite[\S{3.3}]{HintzVasyMink4}.

\begin{proof}[Proof of Proposition~\usref{PropExOpLin}]
  We work in the bundle splittings~\eqref{EqKNullT} and \eqref{EqKNullS2TFine} until the final step of the proof. We use the formula~\eqref{Eq1LinEinGauged} for $L_g$. Corollary~\ref{CorExOpBox2} gives the form of the first term of~\eqref{Eq1LinEinGauged}. For the second term, we note that in the splitting $\cT=\la\pa_0\ra\oplus\la\pa_1\ra\oplus r^{-1}T\Sph^2$ dual to~\eqref{EqKNullS2T}, we have (recalling Definition~\ref{Def1Symm},~\eqref{EqKNullDiff}, and \eqref{EqKNullSchwComp} for $\bhm=0$)
  \begin{align*}
    \cd^\cC &\equiv r^{-1}(\dd t-v^\cC\,\dd r) \equiv r^{-1}\bigl(\tfrac12(1-v^\cC), \tfrac12(1+v^\cC), 0\bigr) \bmod r^{-2}\CI(M;\cT^*), \\
    g^0 &\equiv (0, -\tfrac12,0,0,0,1,0) \bmod r^{-1}\CI(\Omega_\ext;S^2\cT^*), \\
    (g^0)^{-1}(\cd^\cC,\cdot) &\equiv -r^{-1}(1+v^\cC,1-v^\cC,0) \bmod r^{-2}\CI(\Omega_\ext;\cT).
  \end{align*}
  Therefore, 
  \begin{equation}
  \begin{split}
  \label{EqExOpLinEC0}
    E^\cC_{g^0} &\equiv \gamma^\cC r^{-1}
      \begin{pmatrix}
        1-v^\cC & 0 & 0 \\
        \frac12 e^\cC(1+v^\cC) & \frac12 e^\cC(1-v^\cC) & 0 \\
        0 & 0 & \frac12(1-v^\cC) \\
        0 & 1+v^\cC & 0 \\
        0 & 0 & \frac12(1+v^\cC) \\
        (1-e^\cC)(1+v^\cC) & (1-e^\cC)(1-v^\cC) & 0 \\
        0 & 0 & 0
      \end{pmatrix} \\
    &\quad\qquad \bmod r^{-2}\CI(\Omega_\ext;\Hom(\cT^*,S^2\cT^*)).
  \end{split}
  \end{equation}
  Consider next the form of $(\delta_g\dot g)_{\bar\mu}=-g^{\bar\kappa\bar\lambda}\dot g_{\bar\mu\bar\kappa;\bar\lambda}$. We shall regard as error terms all b-operators whose coefficients are of class $r^{-1}\CI(\Omega_\ext)+H_\bop^{k-2,(\cE_0+1,\ell_0+1),(\cE_\sscri^\cC-1,\ell_\sscri-1)}(\Omega_\ext)$ and all eb-operators whose coefficients are of class $\rho_0 x_\sscri\CI(\Omega_{\ext,\frac12})+H_\bop^{k-2,\ell_0^\flat+1,2(\ell_\sscri^\flat-1)}(\Omega_{\ext,\frac12})$. By Corollary~\ref{CorExCov} and recalling that $\cE_\sscri^\tot\subset\cE_\sscri^\cC-1$, the terms with $\lambda\neq 1$ contribute error terms, and those terms with $\lambda=1$ but $\kappa\neq 0$ do as well by~\eqref{EqExOpMetDual}. The $(\kappa,\lambda)=(0,1)$ term contributes $2\pa_1\dot g_{\bar\mu 0}$ modulo error terms. Therefore,
  \[
    \delta_g \equiv \begin{pmatrix} 2\pa_1 & 0 & 0 & 0 & 0 & 0 & 0 \\ 0 & 2\pa_1 & 0 & 0 & 0 & 0 & 0 \\ 0 & 0 & 2\pa_1 & 0 & 0 & 0 & 0 \end{pmatrix}
  \]
  modulo error terms. Lastly, Lemma~\ref{LemmaExOpMet} gives $g\equiv(0,-\frac12,0,0,0,1,0)\bmod r^{-1}\CI+H_\bop^{k,(\cE_0,\ell_0),(\cE_\sscri^\tot,\ell_\sscri)}$ and $\tr_g\equiv(0,-4,0,0,0,2,0)$, and therefore
  \begin{equation}
  \label{EqExOpLinGg}
    \sfG_g \equiv
      \begin{pmatrix}
        1 & 0 & 0 & 0 & 0 & 0 & 0 \\
        0 & 0 & 0 & 0 & 0 & \frac12 & 0 \\
        0 & 0 & 1 & 0 & 0 & 0 & 0 \\
        0 & 0 & 0 & 1 & 0 & 0 & 0 \\
        0 & 0 & 0 & 0 & 1 & 0 & 0 \\
        0 & 2 & 0 & 0 & 0 & 0 & 0 \\
        0 & 0 & 0 & 0 & 0 & 0 & 1
      \end{pmatrix} \bmod (r^{-1}\CI+H_\bop^{k,(\cE_0,\ell_0),(\cE_\sscri^\tot,\ell_\sscri)})(\Omega_\ext;\End(S^2\cT^*)).
  \end{equation}
  Therefore,
  \begin{equation}
  \label{EqExOpLinEC}
    E^\cC_{g^0}\delta_g\sfG_g \equiv
      \gamma^\cC r^{-1}\begin{pmatrix}
        2(1-v^\cC) & 0 & 0 & 0 & 0 & 0 & 0 \\
        e^\cC(1+v^\cC) & 0 & 0 & 0 & 0 & \frac12 e^\cC(1-v^\cC) & 0 \\
        0 & 0 & 1-v^\cC & 0 & 0 & 0 & 0 \\
        0 & 0 & 0 & 0 & 0 & 1+v^\cC & 0 \\
        0 & 0 & 1+v^\cC & 0 & 0 & 0 & 0 \\
        2(1-e^\cC)(1+v^\cC) & 0 & 0 & 0 & 0 & (1-e^\cC)(1-v^\cC) & 0 \\
        0 & 0 & 0 & 0 & 0 & 0 & 0
      \end{pmatrix}
      \pa_1
  \end{equation}
  modulo errors of the classes stated in~\eqref{EqExOpLinb}--\eqref{EqExOpLineb}.

  Consider next the term $\sR_g$ in~\eqref{Eq1LinEinGauged}: using Lemmas~\ref{LemmaExOpMet} and~\ref{LemmaExOpRiem}, the only nontrivial curvature components are $\Riem(g)^{\bar d}{}_{1 1}{}^{\bar b}\equiv\frac12\pa_1^2 h^{\bar b\bar d}$, $\Riem(g)^{\bar d}{}_{1\bar b}{}^0\equiv\pa_1^2 h_{\bar b}{}^{\bar d}$, $\Riem(g)^0{}_{\bar c 1}{}^{\bar b}\equiv\pa_1^2 h_{\bar c}{}^{\bar b}$, and $\Riem(g)^0{}_{\bar c\bar b}{}^0\equiv 2\pa_1^2 h_{\bar b\bar c}$. Therefore,
  \begin{equation}
  \label{EqExOpLinRiemOp}
  \begin{split}
    \sR_g &\equiv
      \begin{pmatrix}
        0 & 0 & 0 & 0 & 0 & 0 & 0 \\
        0 & 0 & 0 & 0 & 0 & 0 & 0 \\
        0 & 0 & 0 & 0 & 0 & 0 & 0 \\
        0 & 0 & 0 & 0 & 0 & 0 & \frac12\pa_1^2 h^{\bar a\bar b} \\
        0 & 0 & \pa_1^2 h_{\bar a}{}^{\bar b} & 0 & 0 & 0 & 0 \\
        0 & 0 & 0 & 0 & 0 & 0 & 0 \\
        2\pa_1^2 h_{\bar a\bar b} & 0 & 0 & 0 & 0 & 0 & 0
      \end{pmatrix} \\
      &\quad\qquad \bmod (r^{-2}\CI+\Hb^{k-2,(\cE_0+2,\ell_0+2),(\cE_\sscri^\cC,\ell_\sscri)})(\Omega_\ext;\End(S^2\cT^*)).
  \end{split}
  \end{equation}
  We use here that the pure trace part $\pa_1^2(\slg_{a b}h^{\bar a\bar b})$ only contributes a term in the error space.

  We proceed with the computation of~\eqref{Eq1LinEinGauged} by noting that, modulo $r^{-1}\Diffb^1$ or $\rho_0 x_\sscri\Diffeb^1$,
  \begin{equation}
  \label{EqExOpLindelgstar}
    \delta_{g^0,E^\cC}^* \equiv \delta_{g^0}^* \equiv
      \begin{pmatrix}
        0 & 0 & 0 \\
        \frac12 & 0 & 0 \\
        0 & 0 & 0 \\
        0 & 1 & 0 \\
        0 & 0 & \frac12 \\
        0 & 0 & 0 \\
        0 & 0 & 0
      \end{pmatrix}
      \pa_1
  \end{equation}
  by Corollary~\ref{CorExCov} for $g=g^0$.\footnote{This is where the required structure of $g^0$ is useful, as otherwise Corollary~\ref{CorExCov} would not be applicable and we would need to do further computations for the Christoffel symbols of $g^0$.} We use Lemma~\ref{LemmaExOpChr} to deduce that $C(g,g^0)_{\bar\kappa\bar\lambda}^{\bar\nu}\equiv 0$ modulo $r^{-2}\CI+H_\bop^{k-1,(\cE_0+1,\ell_0+1),(\cE_\sscri^\cC,\ell_\sscri)}$ except for
  \begin{equation}
  \label{EqExOpLinCgg0}
  \begin{split}
    C(g,g^0)_{1 1}^0&\equiv-\pa_1 h_{1 1}, \\
    C(g,g^0)_{1 1}^1&\equiv -2\pa_1 h_{0 1}, \\
    C(g,g^0)_{1 1}^{\bar c}&\equiv \pa_1 h_1{}^{\bar c}, \\
    C(g,g^0)_{\bar a\bar b}^0&\equiv \pa_1 h_{\bar a\bar b}, \\
    C(g,g^0)_{1\bar b}^{\bar c}=C(g,g^0)_{\bar b 1}^{\bar c}&\equiv\frac12\pa_1 h_{\bar b}{}^{\bar c}.
  \end{split}
  \end{equation}
  Therefore,
  \begin{align*}
    \sC_{g,g^0} &\equiv
      \begin{pmatrix}
        4\pa_1 h_{0 1} & 0 & 0 & 0 & 0 & 0 & 0 \\
        2\pa_1 h_{1 1} & 0 & 0 & 0 & 0 & 0 & -\frac12\pa_1 h^{\bar a\bar b} \\
        4\pa_1 h_{1\bar a} & 0 & -2\pa_1 h_{\bar a}{}^{\bar b} & 0 & 0 & 0 & 0
      \end{pmatrix} \\
    &\quad\qquad \bmod (r^{-2}\CI+H_\bop^{k-1,(\cE_0+1,\ell_0+1),(\cE_\sscri^\cC,\ell_\sscri)})(\Omega_\ext;\Hom(S^2\cT^*,\cT^*)),
  \end{align*}
  and thus
  \begin{equation}
  \label{EqExOpLinCg}
  \begin{split}
    \delta_{g^0,E^\cC}^*\circ\sC_{g,g^0} &\equiv
      \begin{pmatrix}
        0 & 0 & 0 & 0 & 0 & 0 & 0 \\
        2\pa_1^2 h_{0 1} & 0 & 0 & 0 & 0 & 0 & 0 \\
        0 & 0 & 0 & 0 & 0 & 0 & 0 \\
        2\pa_1^2 h_{1 1} & 0 & 0 & 0 & 0 & 0 & -\frac12\pa_1^2 h^{\bar a\bar b} \\
        2\pa_1^2 h_{1\bar a} & 0 & -\pa_1^2 h_{\bar a}{}^{\bar b} & 0 & 0 & 0 & 0 \\
        0 & 0 & 0 & 0 & 0 & 0 & 0 \\
        0 & 0 & 0 & 0 & 0 & 0 & 0
      \end{pmatrix} \\
    &\quad +
      \begin{pmatrix}
        0 & 0 & 0 & 0 & 0 & 0 & 0 \\
        2\pa_1 h_{0 1} & 0 & 0 & 0 & 0 & 0 & 0 \\
        0 & 0 & 0 & 0 & 0 & 0 & 0 \\
        2\pa_1 h_{1 1} & 0 & 0 & 0 & 0 & 0 & -\frac12\pa_1 h^{\bar a\bar b} \\
        2\pa_1 h_{1\bar a} & 0 & -\pa_1 h_{\bar a}{}^{\bar b} & 0 & 0 & 0 & 0 \\
        0 & 0 & 0 & 0 & 0 & 0 & 0 \\
        0 & 0 & 0 & 0 & 0 & 0 & 0
      \end{pmatrix}\pa_1.
  \end{split}
  \end{equation}
  Since $((\delta_{g^0}^*-\delta_g^*)\omega)_{\bar\mu\bar\nu}=C(g,g^0)_{\bar\mu\bar\nu}^{\bar\kappa}\omega_{\bar\kappa}$, we also deduce from~\eqref{EqExOpLinCgg0} that
  \begin{align*}
    \delta_{g^0}^*-\delta_g^* &\equiv
      \begin{pmatrix}
        0 & 0 & 0 \\
        0 & 0 & 0 \\
        0 & 0 & 0 \\
        -\pa_1 h_{1 1} & -2\pa_1 h_{0 1} & \pa_1 h_1{}^{\bar a} \\
        0 & 0 & \frac12\pa_1 h_{\bar a}{}^{\bar b} \\
        0 & 0 & 0 \\
        \pa_1 h_{\bar a\bar b} & 0 & 0
      \end{pmatrix} \\
      &\quad\qquad \bmod (r^{-2}\CI+H_\bop^{k-1,(\cE_0+1,\ell_0+1),(\cE_\sscri^\cC,\ell_\sscri)})(\Omega_\ext;\Hom(\cT^*,S^2\cT^*)),
  \end{align*}
  and hence, modulo errors of the classes stated in~\eqref{EqExOpLinb}--\eqref{EqExOpLineb},
  \begin{equation}
  \label{EqExOpLinDiffdelg}
    (\delta_{g^0}^*-\delta_g^*)\delta_g\sfG_g \equiv
      \begin{pmatrix}
        0 & 0 & 0 & 0 & 0 & 0 & 0 \\
        0 & 0 & 0 & 0 & 0 & 0 & 0 \\
        0 & 0 & 0 & 0 & 0 & 0 & 0 \\
        -2\pa_1 h_{1 1} & 0 & 2\pa_1 h_1{}^{\bar a} & 0 & 0 & -2\pa_1 h_{0 1} & 0 \\
        0 & 0 & \pa_1 h_{\bar a}{}^{\bar b} & 0 & 0 & 0 & 0 \\
        0 & 0 & 0 & 0 & 0 & 0 & 0 \\
        2\pa_1 h_{\bar a\bar b} & 0 & 0 & 0 & 0 & 0 & 0
      \end{pmatrix}\pa_1.
  \end{equation}

  From~\eqref{EqExOpLinCgg0} and the expression~\eqref{Eq1Ups0} (and recalling the stronger decay of the spherical pure trace part of $h$ at $\scri^+$), we deduce that $\Ups_0(g,g^0)_{\bar\mu}\equiv 0$ modulo $r^{-2}\CI+H_\bop^{k-1,(\cE_0+1,\ell_0+1),(\cE_\sscri^\cC,\ell_\sscri)}$, and hence the term $\delta_{g^0,E^\cC}^*\circ\sY_{g,g^0}$ in~\eqref{Eq1LinEinGauged} is an error term.

  For the final term in~\eqref{Eq1LinEinGauged}, we first note that
  \[
    \cd^\Ups = r^{-1}\,\dd t \equiv r^{-1}(\tfrac12,\tfrac12,0) \bmod r^{-2}\CI(M;\cT^*),
  \]
  therefore $(g^0)^{-1}(\cd^\Ups,\cdot)\equiv r^{-1}(-1,-1,0)$, and thus
  \begin{equation}
  \label{EqExOpLinEUps}
    E^\Ups_{g^0} \equiv
      \gamma^\Ups r^{-1}
      \begin{pmatrix}
        2 & 2 e^\Ups & 0 & 0 & 0 & 1-e^\Ups & 0 \\
        0 & 2 e^\Ups & 0 & 2 & 0 & 1-e^\Ups & 0 \\
        0 & 0 & 2 & 0 & 2 & 0 & 0
      \end{pmatrix}
      \bmod r^{-2}\CI(M;\Hom(S^2\cT^*,\cT^*)).
  \end{equation}
  Together with~\eqref{EqExOpLinGg} (for $g=g^0$) and~\eqref{EqExOpLindelgstar}, this gives
  \begin{equation}
  \label{EqExOpLinUpsComp}
    \delta_{g^0,E^\cC}^*E^\Ups_{g^0}\sfG_{g^0} \equiv
      \gamma^\Ups r^{-1}\begin{pmatrix}
        0 & 0 & 0 & 0 & 0 & 0 & 0 \\
        1 & 1-e^\Ups & 0 & 0 & 0 & \frac12 e^\Ups & 0 \\
        0 & 0 & 0 & 0 & 0 & 0 & 0 \\
        0 & 2(1-e^\Ups) & 0 & 2 & 0 & e^\Ups & 0 \\
        0 & 0 & 1 & 0 & 1 & 0 & 0 \\
        0 & 0 & 0 & 0 & 0 & 0 & 0 \\
        0 & 0 & 0 & 0 & 0 & 0 & 0
      \end{pmatrix}\pa_1 \bmod r^{-2}\Diffb^1(M;S^2\cT^*).
  \end{equation}

  Adding~\eqref{EqExOpBox2b}--\eqref{EqExOpBox2eb} to~\eqref{EqExOpLinEC}, \eqref{EqExOpLinRiemOp}, \eqref{EqExOpLinCg}, \eqref{EqExOpLinDiffdelg}, and \eqref{EqExOpLinUpsComp} yields~\eqref{EqExOpLinb}--\eqref{EqExOpLineb}, with the bundle endomorphisms $A_h$ and $B_h$ given by
  \begin{align}
  \label{EqExOpLinAhBh}
    A_h&=
      \scalebox{0.8}{$\openbigpmatrix{3pt}
        2(1{-}v^\cC)\gamma^\cC & 0 & 0 & 0 & 0 & 0 & 0 \\
        e^\cC(1{+}v^\cC)\gamma^\cC{+}\gamma^{\!\Ups}{+}2\pa_1(r h_{0 1}) & (1{-}e^{\!\Ups})\gamma^{\!\Ups} & 0 & 0 & 0 & \frac12(e^\cC(1{-}v^\cC)\gamma^\cC{+}e^{\!\Ups}\gamma^{\!\Ups}) & 0 \\
        0 & 0 & (1{-}v^\cC)\gamma^\cC & 0 & 0 & 0 & 0 \\
        0 & 2(1{-}e^{\!\Ups})\gamma^{\!\Ups} & 2\pa_1(r h_1{}^{\bar a}) & 2\gamma^{\!\Ups} & 0 & (1{+}v^\cC)\gamma^\cC{+}e^{\!\Ups}\gamma^{\!\Ups}{-}2\pa_1(r h_{0 1}) & \ {-}\frac12\pa_1(r h^{\bar a\bar b}) \\
        2\pa_1(r h_{1\bar a}) & 0 & (1{+}v^\cC)\gamma^\cC{+}\gamma^{\!\Ups} & 0 & \gamma^{\!\Ups} & 0 & 0 \\
        2(1{-}e^\cC)(1{+}v^\cC)\gamma^\cC & 0 & 0 & 0 & 0 & (1{-}e^\cC)(1{-}v^\cC)\gamma^\cC & 0 \\
        2\pa_1(r h_{\bar a\bar b}) & 0 & 0 & 0 & 0 & 0 & 0
      \closebigpmatrix$}, \\
    B_h&=
      \begin{pmatrix}
        0 & 0 & 0 & 0 & 0 & 0 & 0 \\
        2\pa_1^2(r h_{0 1}) & 0 & 0 & 0 & 0 & 0 & 0 \\
        0 & 0 & 0 & 0 & 0 & 0 & 0 \\
        2\pa_1^2(r h_{1 1}) & 0 & 0 & 0 & 0 & 0 & 0 \\
        2\pa_1^2(r h_{1\bar a}) & 0 & 0 & 0 & 0 & 0 & 0 \\
        0 & 0 & 0 & 0 & 0 & 0 & 0 \\
        2\pa_1^2(r h_{\bar a\bar b}) & 0 & 0 & 0 & 0 & 0 & 0
      \end{pmatrix} \nonumber
  \end{align}
  in the bundle splitting~\eqref{EqKNullS2TFine}. Upon passing to the splitting~\eqref{EqExOpLinSplit}, one obtains the expressions~\eqref{EqExOpLinAB}, except with $2\pa_1 r^{-1}\pa_0 r$ in place of $2 r^{-1}\pa_0 r\pa_1$. But modulo $\rho_0^2\rho_\sscri^2\Diffb^1$ or $\rho_0^2 x_\sscri^4\Diff_\ebop^1$, we have
  \[
    \pa_1 r^{-1}\pa_0 r\equiv r^{-1}\pa_1\pa_0 r = r^{-1}\pa_0\pa_1 r \equiv r^{-1}\pa_0 r\pa_1.
  \]
  This finishes the proof.
\end{proof}

As a consequence of the formulas~\eqref{EqKNullComp}, we have
\begin{equation}
\label{EqExOpLinFormula}
  2r^{-1}\pa_0 r\pa_1 \equiv -\rho_0^2\rho_\sscri(\rho_\sscri\pa_{\rho_\sscri}-1)(\rho_0\pa_{\rho_0}-\rho_\sscri\pa_{\rho_\sscri})
\end{equation}
modulo $\rho_0^2\rho_\sscri^2\Diffb^2$ and also modulo $\rho_0^2 x_\sscri^4\Diffeb^2$. Therefore,~\eqref{EqExOpLinb} and \eqref{EqExOpLineb} imply
\begin{subequations}
\begin{align}
\label{EqExOpLinbNormal}
\begin{split}
  L_{g,g^0} &\equiv -\rho_0^2\rho_\sscri\Bigl[(\rho_\sscri\pa_{\rho_\sscri}-(I+A_h))(\rho_0\pa_{\rho_0}-\rho_\sscri\pa_{\rho_\sscri}) - \rho_0^{-1}B_h\Bigr] \\
        &\quad\qquad \bmod (\rho_0^2\rho_\sscri^2\CI+H_\bop^{k-2,(\cE_0+2,\ell_0+2),(\cE_\sscri^\cC,\ell_\sscri)})\Diffb^2(\Omega_\ext;S^2\cT^*),
\end{split} \\
\label{EqExOpLinebNormal}
\begin{split}
  2 L_{g,g^0} &\equiv -\rho_0^2 x_\sscri^2\frac12\Bigl[(x_\sscri\pa_{x_\sscri}-2(I+A_h))(2\rho_0\pa_{\rho_0}-x_\sscri\pa_{x_\sscri}) - 4\rho_0^{-1}B_h\Bigr] + r^{-2}\slDelta \\
        &\quad\qquad \bmod \rho_0^2 x_\sscri^2(x_\sscri\CI+H_\bop^{k-2,\ell_0^\flat,2(\ell_\sscri^\flat-1)})\Diffeb^2(\Omega_{\ext,\frac12};S^2\cT^*).
\end{split}
\end{align}
\end{subequations}
Note that since $\pa_1\in\rho_0\Vb(M)$, the entries of $\rho_0^{-1}B_h$ are of class $H_\bop^{k-1,(\cE_0,\ell_0),(\cE_\sscri^\tot-1,\ell_\sscri-1)}$: they do not blow up at $\scri^+$ (and they decay as $\rho_0\to 0$). The entries of $A_h$ are of class $\CI+H_\bop^{k-1,(\cE_0,\ell_0),(\cE_\sscri^\tot-1,\ell_\sscri-1)}$, and thus are bounded down to $\scri^+$ as well.

We now revisit the discussion at the beginning of~\S\ref{SsExP}. Given~\eqref{EqExOpLinbNormal}, one expects that if $L_g\dot g=0$ (with suitably decaying initial data for $\dot g$), then various components of $\dot g$ decay like $\cO(\rho_\sscri^{1+z})$ where $z$ is an eigenvalue of $A_h$. Conditions ensuring at least $\cO(\rho_\sscri)$ decay are, therefore,
\begin{equation}
\label{EqExOpParam}
  v^\cC<1,\ \ 0<e^\cC<1,\ \ \gamma^\cC>0,\ \ 0<e^\Ups<1,\ \ \gamma^\Ups>0.
\end{equation}
(For an open set of such parameters, the diagonal entries of $A_h$ are pairwise distinct; this will be convenient for bookkeeping.) Since $\pi^\cC h$ (see~\eqref{EqExPSpaceProj}), i.e., the first three components of $h$ in the splitting~\eqref{EqExOpLinSplit}, are required to have stronger decay (as far as its polyhomogeneous part is concerned) than the other components of $h$, it is moreover natural to require that the smallest one of the corresponding eigenvalues $2(1-v^\cC)\gamma^\cC$, $(1-v^\cC)\gamma^\cC$, $(1-e^\cC)(1-v^\cC)\gamma^\cC$ exceeds the largest remaining eigenvalue (so $(1-e^\cC)(1-v^\cC)\gamma^\cC>2\gamma^\Ups$). We will require, more strongly, a lower bound by the conormal remainder decay rate $\ell_\sscri$ (which will amount to not having to record radiation fields for the components of $\pi^\cC h$ even when keeping track of polyhomogeneous expansions up to terms with $\cO(r^{-\ell_\sscri}$)-decay); this leads to the condition
\[
  1+(1-e^\cC)(1-v^\cC)\gamma^\cC > \ell_\sscri,
\]
which implies the former when $\ell_\sscri>2>1+2\gamma^\Ups$. In our eventual application, $v^\cC\in(0,1)$ will be fixed, $e^\cC>0$ will be small, $e^\Ups>0$ will be fixed, $\gamma^\Ups>0$ will be small, $\ell_\sscri>3$, and $\gamma^\cC$ can be taken to be arbitrarily large.

\begin{rmk}[Admissibility]
\label{RmkExOpAdm}
  As remarked in \cite[Example~3.12]{HintzVasyScrieb} in the context of the closely related \cite{HintzMink4Gauge}, the operator $2 L_{g,g^0}$ is an admissible operator in the sense of \cite[Definition~3.7]{HintzVasyScrieb}, with $p_1=A_h$ and $p_0=B_h$ (provided $k=\infty$ and using Sobolev embedding to control $h$ in $L^\infty$-based conormal spaces). The admissibility of the metric $g=g_b+h$ follows easily from the characterization~\cite[(3.12)]{HintzVasyScrieb}.
\end{rmk}

\subsection{Forward mapping properties}
\label{SsExFw}

We can use Proposition~\ref{PropExOpLin} to obtain a precise (as far as index sets are concerned) description of the forward mapping properties of $h\mapsto P(g_b+h,g^0)$ and $\dot g\mapsto L_{g_b+h,g^0}\dot g$ in the notation of~\eqref{Eq1Ein} and~\eqref{Eq1EinLin}. The spaces in which $h,\dot g$ here lie are the spaces $\Hb^{k,(\cE_0,\ell_0),(\la\cE_\sscri^\cC\ra,\ell_\sscri)}$ from Definition~\ref{DefExP} (in which we will, after all, seek the metric perturbations that solve linearized or nonlinear gauge-fixed Einstein equation).

\begin{definition}[Source terms near $I^0\cap\scri^+$]
\label{DefExFwSrc}
  Let $\cE_0,\cE_\sscri^\cC\subset\C\times\N_0$ be as in Definition~\usref{DefExP}. Let $k\geq 3$ and $\ell_0>0$, $\ell_\sscri>1$.
  \begin{enumerate}
  \item\label{ItExFwSrc1} Set
    \begin{equation}
    \label{EqExFwSrc11}
      \cF_{\sscri,1 1} := 1 + \bigl( \cE_\sscri^\cC \cup (1,0) \cup (1+(1-e^\Ups)\gamma^\Ups,0) \bigr).
    \end{equation}
    Then the space\footnote{We use the notation $\la\cE_\sscri^\cC+1\ra'$ to distinguish this space more clearly from~\eqref{EqExPSpace}.}
    \begin{equation}
    \label{EqExFwSrc}
      \Hb^{k,\ (\cE_0+2,\ell_0+2),\ \bigl(\la\cE_\sscri^\cC+1\ra',\ell_\sscri+1\bigr)}(\Omega_\ext) \subset H_\bop^{k,2,0}(\Omega_\ext;S^2\cT^*)
    \end{equation}
    consists of all real symmetric 2-tensors $f$ such that
    \begin{align*}
      \pi^\cC f,\ \pi_{0 1}f,\ \pi_{1 /}f,\ \slpi_0 f &\in H_\bop^{k,\ (\cE_0+2,\ell_0+2),\ (\cE_\sscri^\cC+1,\ell_\sscri+1)}(\Omega_\ext), \\
      \pi_{1 1}f &\in H_\bop^{k,\ (\cE_0+2,\ell_0+2),\ (\cF_{\sscri,1 1},\ell_\sscri+1)}(\Omega_\ext).
    \end{align*}
  \item If $\ell_0<\min\Re\cE_0$ (in particular, when $\cE_0=\emptyset$), we write $\ell_0+2$ in~\eqref{EqExFwSrc} instead of $(\cE_0+2,\ell_0+2)$, and if $\ell_\sscri<\min\Re\cE_\sscri^\cC$, we write $(-',\ell_\sscri+1)$ instead of $(\la\cE_\sscri^\cC+1\ra',\ell_\sscri+1)$.
  \item Given any index set $\cE_0\subset\C\times\N_0$, we write
    \[
      \Hb^{k,\ (\cE_0+2,\ell_0+2),\ \bigl(\la(3,0)\ra'_0,\ell_\sscri+1\bigr)}(\Omega_\ext)
    \]
    for the space of all real $f$ such that $\pi^\cC f$, $\pi_{0 1}f$, $\pi_{1 /}f$, $\slpi_0 f\in H_\bop^{k,(\cE_0+2,\ell_0+2),((3,0),\ell_\sscri+1)}$ and $\pi_{1 1}f\in H_\bop^{k,(\cE_0+2,\ell_0+2),((2,0),\ell_\sscri+1)}$.
  \end{enumerate}
\end{definition}

\begin{prop}[Forward mapping properties of $L_{g,g^0}$]
\label{PropExFwL}
  Consider a Lorentzian metric $g=g_b+h$, $h\in\Hb^{k,\ (\cE_0,\ell_0),\ (\la\cE_\sscri^\cC\ra,\ell_\sscri)}(\Omega_\ext)$. Let $g^0$ be a metric with $g^0-g_\bhm\in\Hb^{\infty,\ ((1,0),\infty),\ \bigl(\la(2,0)\ra_0,\infty\bigr)}(\Omega_\ext)$. Then
  \begin{equation}
  \label{EqExFwL}
    L_{g,g^0} \colon \Hb^{k,\ (\cE_0,\ell_0),\ (\la\cE_\sscri^\cC\ra,\ell_\sscri)}(\Omega_\ext) \to \Hb^{k-2,\ (\cE_0+2,\ell_0+2),\ \bigl(\la\cE_\sscri^\cC+1\ra',\ell_\sscri+1\bigr)}(\Omega_\ext).
  \end{equation}
  Moreover, if $h\in\Hb^{k,\ (\cE_0,\ell_0),\ \bigl(\la(2,0)\ra_0,\ell_\sscri\bigr)}(\Omega_\ext)$, then
  \[
    L_{g,g^0} \colon \Hb^{k,\ (\cE_0,\ell_0),\ \bigl(\la(2,0)\ra_0,\ell_\sscri\bigr)}(\Omega_\ext) \to \Hb^{k-2,\ (\cE_0+2,\ell_0+2),\ \bigl(\la(3,0)\ra'_0,\ell_\sscri+1\bigr)}(\Omega_\ext).
  \]
\end{prop}
\begin{proof}
  The error term in~\eqref{EqExOpLinbNormal} maps $\dot g\in H_\bop^{k,\ (\cE_0,\ell_0),\ (\cE_\sscri^\tot,\ell_\sscri)}(\Omega_\ext;S^2\cT^*)$ into the space
  \[
    H_\bop^{k-2,\ (\cE_0+2,\ell_0+2),\ (\cE_\sscri^\tot+2,\ell_\sscri+2)} + H_\bop^{k-2,\ (2\cE_0+2,\ell_0+2),\ (\cE_\sscri^\cC+\cE_\sscri^\tot,\ell_\sscri+1)};
  \]
  here we use that $\min\Re(\cE_\sscri^\tot\setminus\{(1,0)\})>1$. Since $\cE_\sscri^\cC+\cE_\sscri^\tot\subset\cE_\sscri^\cC+1$ by Definition~\ref{DefExP}\eqref{ItExPTotNL}, this space is indeed contained in the target space $\Hb^{k-2,(\cE_0+2,\ell_0+2),(\la\cE_\sscri^\cC+1\ra',\ell_\sscri+1)}$ in~\eqref{EqExFwL}.

  It remains to consider the effect of the explicit leading-order term in~\eqref{EqExOpLinbNormal}. Note that $\rho_0\rho_\sscri B_h\in H_\bop^{k-1,(\cE_0+2,\ell_0+2),(\cE_\sscri^\tot,\ell_\sscri)}$ only has nonzero entries in the first column, which thus multiply $\dot g_{0 0}\in H_\bop^{k,(\cE_0,\ell_0),(\cE_\sscri^\cC,\ell_\sscri)}$ and produce an output in the target space of~\eqref{EqExFwL}. It remains to prove
  \[
    \rho_0^2\rho_\sscri(\rho_\sscri\pa_{\rho_\sscri}-(I+A_h)) \colon \Hb^{k-1,\ (\cE_0,\ell_0),\ (\la\cE_\sscri^\cC\ra,\ell_\sscri)} \to \Hb^{k-2,\ (\cE_0+2,\ell_0+2),\ (\la\cE_\sscri^\cC+1\ra',\ell_\sscri+1)},
  \]
  which we do using the expression~\eqref{EqExOpLinAB}: the first three columns of $A_h$ act on the components of $\pi^\cC\dot g\in H_\bop^{k,(\cE_0+2,\ell_0+2),(\cE_\sscri^\cC,\ell_\sscri)}$ and thus produce an output in the desired target space. The $(4,4)$-entry $1+(1-e^\Ups)\gamma^\Ups$ of $I+A_h$ annihilates the $\rho_\sscri^{1+(1-e^\Ups)\gamma^\Ups}$ leading-order term of $\pi_{0 1}\dot g$ (whence the $01$-component of the output has $\scri^+$-index set equal to $\cE_{\sscri,0 1}\setminus\{(1+(1-e^\Ups)\gamma^\Ups,0)\}=\cE_\sscri^\cC$); similarly, the $(5,5)$-, resp.\ $(6,6)$-entry $1+\gamma^\Ups$, resp.\ $1$ annihilates the $\rho_\sscri^{1+\gamma^\Ups}$, resp.\ $\rho_\sscri^1$ leading-order term of $\pi_{1 /}\dot g$, resp.\ $\slpi_0\dot g$. The $(7,4)$ and $(7,6)$ entries of $I+A_h$ couple $\pi_{0 1}\dot g$ and $\slpi_0\dot g$ into the $1 1$-component of the output, while the $(7,7)$ entry of $I+A_h$ annihilates the $\rho_\sscri^{1+2\gamma^\Ups}$-term in the expansion of $\dot g_{1 1}$; hence the definition of $\cF_{\sscri,1 1}$ in~\eqref{EqExFwSrc11}.

  The final claim follows from~\eqref{EqExFwL} and the observation that only integer powers appear in the asymptotic expansions at $\scri^+$ of all coefficients of $L_{g,g^0}$ and in the evaluation of $L_{g,g^0}\dot g$ when the asymptotic expansions of $h$ and $\dot g$ themselves only involve integer powers.
\end{proof}

\begin{cor}[Forward mapping properties of $P$]
\label{CorExFwN}
  Let $g=g_b+h$, $h\in\Hb^{k,\ (\cE_0,\ell_0),\ (\la\cE_\sscri^\cC\ra,\ell_\sscri)}(\Omega_\ext)$, be a Lorentzian metric, and let $g^0$ be a metric with $g^0-g_\bhm\in\Hb^{\infty,\ ((1,0),\infty),\ \bigl(\la(2,0)\ra_0,\infty\bigr)}(\Omega_\ext)$, and with $\Ric(g^0)=0$. Recall $P(g,g^0)$ from~\eqref{Eq1Ein}. Then:
  \begin{enumerate}
  \item\label{ItExFwN1}{\rm (Nonlinear operator, I.)} We have
    \begin{equation}
    \label{EqExFwNMem}
      P(g,g^0) \in \Hb^{k-2,\ (\cE_0+2,\ell_0+2),\ \bigl(\la\cE_\sscri^\cC+1\ra',\ell_\sscri+1\bigr)}(\Omega_\ext).
    \end{equation}
  \item\label{ItExFwN2}{\rm (Nonlinear operator, II.)} For $h\in\Hb^{k,\ (\cE_0,\ell_0),\ \bigl(\la(2,0)\ra_0,\ell_\sscri\bigr)}(\Omega_\ext)$, we have
    \begin{equation}
    \label{EqExFwNMemSm}
      P(g,g^0)\in\Hb^{k-2,\ (\cE_0+2,\ell_0+2),\ \bigl(\la(3,0)\ra'_0,\,\ell_\sscri+1\bigr)}(\Omega_\ext).
    \end{equation}
  \end{enumerate}
\end{cor}
\begin{proof}
  Since $\Ric(g^0)=0$ (by assumption on $g^0$) and $\Ups_{E^\Ups}(g^0,g^0)=0$ (which holds for any metric $g^0$), we have $P(g^0,g^0)=0$. Therefore, upon setting $\tilde h:=h-(g^0-g_b)$, we have
  \begin{equation}
  \label{EqExFwNLin}
    P(g_b+h,g^0) = P(g^0+\tilde h,g^0) = \int_0^1 L_{g^0+s\tilde h,g^0}\tilde h\,\dd s = \int_0^1 L_{g_b+(1-s)(g^0-g_b)+s h,g^0}\tilde h\,\dd s.
  \end{equation}
  The membership~\eqref{EqExFwNMem} now follows from Proposition~\ref{PropExFwL}. The proof of~\eqref{EqExFwNMemSm} is completely analogous.
\end{proof}

\subsection{The basic exterior stability result}
\label{SsEx0}

We now prove the exterior stability of Kerr, and indeed the solvability of the gauge-fixed Einstein equation for general small Cauchy data, using the arguments of \cite{HintzMink4Gauge}. These arguments require only minor modifications in the present formulation~\eqref{Eq1Ein} of the gauge-fixed Einstein equation $P(g,g^0)=0$.

Concretely, let us work with the parameters $b_0=(\bhm_0,\bha_0)$ of a subextremal Kerr metric, and recall the time functions $t_*=:t-r$ and $t_\IVP$ (equal to $t+C$ for large $r$) from Lemmas~\ref{LemmaKMetTime} and~\ref{LemmaKMetIVP}. Since $\Ric(g_{b_0})=0$ and $\Ups_{E^\Ups}(g_{b_0},g_{b_0})=0$ in the notation of Definition~\ref{Def1Gauge}, we have
\[
  P(g_{b_0},g_{b_0})=0.
\]
In the definition of $P$, we fix the parameters $v^\cC,e^\cC,\gamma^\cC$ and $e^\Ups,\gamma^\Ups$ according to~\eqref{EqExOpParam}. (Further conditions will only be placed on these parameters starting in~\S\ref{SsExPhg}.) Consider moreover weights
\begin{equation}
\label{EqEx0ells}
  \ell_0^\flat > 1, \quad
  1<\ell_\sscri^\flat<\min(\ell_0^\flat,1+(1-e^\Ups)\gamma^\Ups,\tfrac32).
\end{equation}
(These quantities satisfy the conditions~\eqref{EqExOpLinFlat} for $\ell_0=\ell_0^\flat$, $\ell_\sscri=\ell_\sscri^\flat$.) Recalling the final part of Definition~\ref{DefExP}, the space $\Hb^{k,\ell_0^\flat,(-,\ell_\sscri^\flat)}(\Omega_\ext)$ thus consists of all symmetric 2-tensors $h$ such that $\pi^\cC h$, $\pi_{0 1}h$, $\pi_{1 /}h\in H_\bop^{k,\ell_0^\flat,\ell_\sscri^\flat}(\Omega_\ext)$, while $\pi_{1 1}h$, $\slpi_0 h\in H_\bop^{k,\ell_0^\flat,((1,0),\ell_\sscri^\flat)}(\Omega_\ext)$ are permitted to have $r^{-1}$-terms at $\scri^+$; this will be the simplest space in which our nonlinear iteration scheme for solving initial value problems for the quasilinear wave equation $P(g_{b_0}+h,g_{b_0})=0$ closes, with the sought-after $h$ then lying in $\Hb^{k,\ell_0^\flat,(-,\ell_\sscri^\flat)}(\Omega_\ext)$. We first clarify the relevant spacetime domains via a variant of \cite[Definition~3.20]{HintzMink4Gauge} (see also \cite[Lemma~3.6 and \S{6}]{HintzVasyScrieb} and \citeAF{Lemma~\ref*{LemmaSDGTime}}, the latter of which includes a shift of $\ell_\sscri^\flat$ by $1$).

\begin{lemma}[Spacetime domain]
\label{LemmaEx0Dom}
  Set
  \[
    \Omega_{\ext,r_0} := \Bigl\{ -\frac{r-r_0}{4} \leq t_\IVP \leq 0 \Bigr\} \cup \Bigl\{ 0 \leq t_\IVP \leq r + 2 r^{-\ell_\sscri^\flat+1} - (r_0+2 r_0^{-\ell_\sscri^\flat+1}) \Bigr\},
  \]
  which is contained in $M$ (see Definition~\usref{DefKMfdRad}) for large enough $r_0$. Denote the boundary hypersurfaces of $\Omega_{\ext,r_0}$ by
  \begin{align*}
    \Sigma_{{\rm past},r_0}&=\Omega_{\ext,r_0}\cap\Bigl\{t_\IVP=-\frac{r-r_0}{4}\Bigr\}, \\
    \Sigma_{{\rm fut},r_0}&=\Omega_{\ext,r_0}\cap\bigl\{ t_\IVP=r+2 r^{-\ell_\sscri^\flat+1} - (r_0+2 r_0^{-\ell_\sscri^\flat+1}) \bigr\},
  \end{align*}
  and define moreover (in the notation of Definition~\usref{DefKMetData})
  \begin{equation}
  \label{EqEx0DomSigmaIVP}
    \Sigma_{\IVP,r_0} := \Sigma_\IVP\cap\Omega_{\ext,r_0} = \Omega_{\ext,r_0} \cap \{t_\IVP=0\}.
  \end{equation}
  There exists $R_0>3\bhm_0$ such that the following holds: for $r_0\geq R_0$ and for all $h\in\Hb^{3,\ell_0^\flat,(-,\ell_\sscri^\flat)}(\Omega_{\ext,r_0})$ (Definition~\usref{DefExP}) with $\|h\|_{\Hb^{3,\ell_0^\flat,(-,\ell_\sscri^\flat)}}<1$, the hypersurfaces $\Sigma_{{\rm past},r_0}$, $\Sigma_{\IVP,r_0}$, and $\Sigma_{{\rm fut},r_0}$ are spacelike for $g_{b_0}+h$.
\end{lemma}

\begin{figure}[!ht]
\centering
\includegraphics{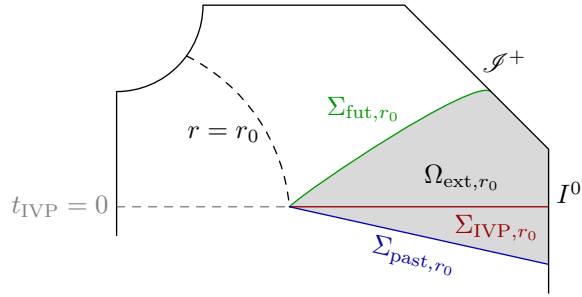}
\caption{Illustration of the domain $\Omega_{\ext,r_0}$ and the hypersurfaces $\Sigma_{{\rm past},r_0}$, $\Sigma_{\IVP,r_0}$, and $\Sigma_{{\rm fut},r_0}$.}
\label{FigEx0Dom}
\end{figure}

See Figure~\ref{FigEx0Dom}.

\begin{proof}[Proof of Lemma~\usref{LemmaEx0Dom}]
  Since $\dd t_\IVP=\dd t$ (for large $r$) is timelike for $g_{b_0}$, the same is true also for all sufficiently small (in $L^\infty$, as sections of $S^2\cT^*$) perturbations of $g_{b_0}$. Similarly, $\zeta:=\dd(t_\IVP+\frac{r-r_0}{4})=\dd t+\frac14\,\dd r=\dd t_*+\frac54\,\dd r$ is timelike for the Minkowski metric (cf.\ \eqref{EqKMetRefSchw} for $\bhm=0$) since $\frac54\in(1,2)$, with $\ubar g^{-1}(\zeta,\zeta)=-2\frac{5}{4}+(\frac{5}{4})^2=-\frac{15}{16}<0$, and therefore the same is true for $g_{b_0}$ when $r$ is large, and then also for small (in $L^\infty$, as sections of $S^2\cT^*$) perturbations thereof. This proves the claim for $\Sigma_{{\rm past},r_0}$ and $\Sigma_{\IVP,r_0}$.

  The argument for $\Sigma_{{\rm fut},r_0}$ is more subtle since $t_\IVP-r$ is null for the Minkowski (and Schwarzschild) metric~\eqref{EqKMetRefSchw} for large $r$. We use $t-r=x^1$ and $\dd r=\frac12(1-\frac{2\bhm_0}{r})(\dd x^0-\dd x^1)$ (from~\eqref{EqKNullDiff}) to compute
  \begin{align*}
    \zeta &:= \dd\bigl(t_\IVP-r-2 r^{-\ell_\sscri^\flat+1}\bigr) \\
      &= \dd x^1 + (\ell_\sscri^\flat-1)r^{-\ell_\sscri^\flat}(1-2\bhm_0 r^{-1})(\dd x^0-\dd x^1) \\
      &= (\ell_\sscri^\flat-1)(1-2\bhm_0 r^{-1})r^{-\ell_\sscri^\flat}\dd x^0 + \bigl(1-(\ell_\sscri^\flat-1)(1-2\bhm_0 r^{-1})r^{-\ell_\sscri^\flat}\bigr)\,\dd x^1.
  \end{align*}
  By Lemma~\ref{LemmaExOpMet} and Sobolev embedding, we have
  \[
    |g^{0 0}|\leq C(\rho_0^2\rho_\sscri^2+\rho_0^{\ell_0^\flat}\rho_\sscri), \quad
    |g^{0 1}+2| \leq C(\rho_0\rho_\sscri+\rho_0^{\ell_0^\flat}\rho_\sscri), \quad
    |g^{1 1}|\leq C(\rho_0^2\rho_\sscri^2+\rho_0^{\ell_0^\flat}\rho_\sscri^{\ell_\sscri^\flat}).
  \]
  Using moreover that $r^{-1}=\rho_0\rho_\sscri$ and writing $f:=1-2\bhm_0 r^{-1}=1-2\bhm_0\rho_0\rho_\sscri$, we therefore have
  \begin{align*}
    g^{-1}(\zeta,\zeta) &\leq \bigl(-4 + C(\rho_0\rho_\sscri+\rho_0^{\ell_0^\flat}\rho_\sscri)\bigr) (\ell_\sscri^\flat-1)f \bigl( 1-(\ell_\sscri^\flat-1)r^{-\ell_\sscri^\flat}f\bigr) \rho_0^{\ell_\sscri^\flat}\rho_\sscri^{\ell_\sscri^\flat} \\
      &\quad + C(\rho_0^2\rho_\sscri^2+\rho_0^{\ell_0^\flat}\rho_\sscri) \rho_0^{2\ell_\sscri^\flat}\rho_\sscri^{2\ell_\sscri^\flat} + C(\rho_0^2\rho_\sscri^2+\rho_0^{\ell_0^\flat}\rho_\sscri^{\ell_\sscri^\flat}).
  \end{align*}
  Since $\rho_0^{\ell_0^\flat} / \rho_0^{\ell_\sscri^\flat} \to 0$ as $\rho_0\to 0$ in view of~\eqref{EqEx0ells}, and thus on $\Omega_{\ext,r_0}$ when $r_0\to\infty$, the first term on the right dominates the third (and also the second) term on $\Omega_{\ext,r_0}$ for sufficiently large $r_0$, and thus $g^{-1}(\zeta,\zeta)<0$.
\end{proof}

We then have the following result, which is very similar to \cite[Corollary~3.36]{HintzMink4Gauge}:

\begin{thm}[Exterior solution]
\label{ThmEx0}
  There exists $k_0\in\N$ such that the following holds. Let $\ell_0^\flat,\ell_\sscri^\flat$ be as in~\eqref{EqEx0ells}, and let $r_0\geq R_0$ where $R_0$ is as in Lemma~\usref{LemmaEx0Dom}. Then there exists $\eps_0>0$ such that for all
  \[
    h_0,\ h_1 \in H_\bop^{\infty,\ell_0^\flat}(\Sigma_{\IVP,r_0};S^2\cT^*)
  \]
  with $\|h_j\|_{H_\bop^{k_0,\ell_0^\flat}}<\eps_0$ for $j=0,1$, the initial value problem
  \begin{equation}
  \label{EqEx0IVP}
    \left\{
    \begin{aligned}
      P(g_{b_0}+h,g_{b_0})&=0 && \text{in}\ \Omega_{\ext,r_0}, \\
      h&=h_0 && \text{on}\ \Sigma_{\IVP,r_0}, \\
      r\cL_{\pa_{t_\IVP}}h&=h_1 && \text{on}\ \Sigma_{\IVP,r_0}
    \end{aligned}
    \right.
  \end{equation}
  has a unique solution $h\in\Hb^{\infty,\ell_0^\flat,(-,\ell_\sscri^\flat)}(\Omega_{\ext,r_0})$ in the notation of Definition~\usref{DefExP}. Moreover, for all $\eps>0$ and $k\in\N$, we can choose $k_0\in\N_0$ and $\eps_0>0$ such that we have the smallness property $\|h\|_{\Hb^{k,\ell_0^\flat,(-,\ell_\sscri^\flat)}}<\eps$.
\end{thm}

\begin{rmk}[Stability of the exterior region]
\label{RmkEx0IVPEin}
  If the geometric initial data determined by $h_0,h_1$---i.e., the first and second fundamental form of $\hat g:=g_{b_0}+h_0+t_\IVP\cdot r h_1$, where $h_0,h_1$ are regarded as $t_\IVP$-independent tensors on $\R_{t_\IVP}\times\Sigma_{\IVP,r_0}$---satisfy the vacuum constraint equations on $\Sigma_{\IVP,r_0}$, and if moreover $\hat g$ satisfies the gauge condition $\Ups_{E^\Ups}(\hat g,g_{b_0})=0$ pointwise at $\Sigma_{\IVP,r_0}$ (see~\S\ref{SsExID} regarding the latter), then standard arguments involving the second Bianchi identity show that $\Ups(g_{b_0}+h,g_{b_0})=0$ and $\Ric(g_{b_0}+h)=0$ in $\Omega_{r_0}$.
\end{rmk}

\begin{rmk}[Weaker decay at $I^0$]
\label{RmkEx0I0Decay}
  If one were to allow for the $I^0$-decay rate $\ell_0^\flat$ to lie in $(0,1]$, corresponding to $r^{-1}$ (or weaker) decay of metric perturbations at spacelike infinity, one would need to revisit the exterior stability argument: one would have to contend with $\rho_\sscri^1$ (or weaker) bounds for metric components at $\scri^+$. Work by Kadar--Kehrberger \cite{KadarKehrbergerPhgScatter} suggests that this can be done (by using a metric other than the Minkowski metric as the background metric off of which one perturbs) in generalized harmonic gauges. There also exist more geometric techniques to deal with weakly decaying initial data \cite{BieriZipserStability,ShenMinkBorderline}. Our nonlinear stability proof in $\{t_*\geq 1\}$ frequently takes advantage of $r^{-1-\eps}$-decay of initial data, however, and thus we do not discuss weaker decay rates further here.
\end{rmk}

\begin{proof}[Proof of Theorem~\usref{ThmEx0}]
  The proof is the same (with minor, only notational, modifications) as that of \cite[Corollary~3.36]{HintzMink4Gauge}, and hence we shall be very brief. The key is to show that for $h\in\Hb^{\infty,\ell_0^\flat,(-,\ell_\sscri^\flat)}(\Omega_{\ext,r_0})$ with small $\Hb^{d,\ell_0^\flat,(-,\ell_\sscri^\flat)}$-norm (for sufficiently large $d$), one has (tame) decay estimates for solutions of the linearized problem
  \[
    L_{g_{b_0}+h,g_{b_0}} \dot g = f \in \Hb^{\infty,\ell_0^\flat+2,(-',\ell_\sscri^\flat+1)}(\Omega_{\ext,r_0});
  \]
  here $f$ is the error from the previous iteration (cf.\ Corollary~\ref{CorExFwN}). These estimates take the form
  \begin{equation}
  \label{EqEx0Tame}
  \begin{split}
    \| \dot g \|_{\Hb^{k,\ell_0^\flat,(-,\ell_\sscri^\flat)}(\Omega_{\ext,r_0})} &\leq C_k\Bigl( \|f\|_{\Hb^{k+d,\ell_0^\flat+2,(-',\ell_\sscri^\flat+1)}(\Omega_{\ext,r_0})} \\
      &\quad\qquad + \|h\|_{\Hb^{k+d,\ell_0^\flat,(-,\ell_\sscri^\flat)}(\Omega_{\ext,r_0})}\|f\|_{\Hb^{2 d,\ell_0^\flat+2,(-',\ell_\sscri^\flat+1)}(\Omega_{\ext,r_0})}\,\Bigr)
  \end{split}
  \end{equation}
  and are proved (for $d=11$) in \cite[Theorem~3.35]{HintzMink4Gauge}; for the basic, first order, energy estimate (from which~\eqref{EqEx0Tame} follows via commutation with $\rho_0\pa_{\rho_0}$, $\rho_\sscri\pa_{\rho_\sscri}$, and spherical vector fields), one uses the structure~\eqref{EqExOpLinebNormal} (which matches \cite[Proposition~3.29]{HintzMink4Gauge}, keeping in mind also the weights in \cite[Definition~3.27]{HintzMink4Gauge}). An application of the Nash--Moser iteration scheme \cite{SaintRaymondNashMoser} then implies Theorem~\ref{ThmEx0}. --- The proof of the tame estimate~\eqref{EqEx0Tame} in turn proceeds by first proving a tame energy estimate for $\dot g$ in terms of $f$ on weighted b-Sobolev spaces (with weight $\ell_0^\flat$ at $I^0$ and with any fixed weight $<1$ at $\scri^+$) as in \cite[Proposition~3.32]{HintzMink4Gauge}. In a second step, the partial asymptotic expansion of $\dot g$ as encoded in the space $\Hb^{k,\ell_0^\flat,(-,\ell_\sscri^\flat)}$ (i.e., the membership in $\Hb^{k,\ell_0^\flat,\ell_\sscri^\flat}$ of all components of $\dot g$ except for $\pi_{1 1}\dot g$ and $\slpi_0 \dot g$, which have a $\rho_\sscri^1\Hb^{k,\ell_0^\flat}(\Omega_{\ext,r_0}\cap\scri^+)$-radiation field plus an $\Hb^{k,\ell_0^\flat,\ell_\sscri^\flat}$-remainder) is obtained by integrating the leading-order term in~\eqref{EqExOpLinbNormal} while putting the error term (acting on $\dot g$) on the right-hand side; this second step is performed in detail (and with longer polyhomogeneous expansions) in the proof of Theorem~\ref{ThmExPhg} below.

  Unlike in \cite{HintzMink4Gauge}, the domain $\Omega_{\ext,r_0}$ also includes a region $-\frac{r}{4}\leq t_\IVP\leq 0$ in the past of $\Sigma_{\IVP,r_0}$; this region is however disjoint from past null infinity, and indeed is treated in the same way as the region $0\leq t_\IVP\leq\frac{r}{4}$ which is disjoint from future null infinity.
\end{proof}
  
Below, we will not state tame estimates explicitly anymore; we will thus content ourselves with the following variant of~\eqref{EqEx0Tame}: for any fixed $h\in\Hb^{\infty,\ell_0^\flat,(-,\ell_\sscri^\flat)}(\Omega_{\ext,r_0})$ with small norm in $\Hb^{d,\ell_0^\flat,(-,\ell_\sscri^\flat)}$ (where $d$ is large but fixed)---so, in particular, $g_{b_0}+h$ is a Lorentzian metric on $\Omega_{\ext,r_0}$---we can estimate the solution $\dot g$ of the initial value problem
\[
  \left\{
    \begin{aligned}
      L_{g_{b_0}+h,g_{b_0}}\dot g &= f\in\Hb^{k+d,\alpha_0+2,(-',\ell_\sscri^\flat+1)}(\Omega_{\ext,r_0}) && \text{in}\ \Omega_{\ext,r_0}, \\
      \dot g &= \dot g_0 \in H_\bop^{k+d,\alpha_0}(\Sigma_{\IVP,r_0}) && \text{on}\ \Sigma_{\IVP,r_0}, \\
      r\cL_{\pa_{t_\IVP}}\dot g &= \dot g_1 \in H_\bop^{k+d,\alpha_0}(\Sigma_{\IVP,r_0}) && \text{on}\ \Sigma_{\IVP,r_0}
    \end{aligned}
  \right.
\]
by
\begin{equation}
\label{EqEx0Est}
  \| \dot g \|_{\Hb^{k,\alpha_0,(-,\ell_\sscri^\flat)}} \leq C\Bigl( \| f \|_{\Hb^{k+d,\alpha_0+2,(-',\ell_\sscri^\flat+1)}} + \|\dot g_0\|_{H_\bop^{k+d,\alpha_0}} + \|\dot g_1\|_{H_\bop^{k+d,\alpha_0}}\Bigr)
\end{equation}
for all $\alpha_0>\ell_\sscri^\flat$ (with $C$ depending on $k,\alpha_0,\ell_\sscri^\flat,h$, with $\ell_\sscri^\flat$ as in~\eqref{EqEx0ells}); this follows in the same manner as \cite[Proposition~3.32 and Theorem~3.35]{HintzMink4Gauge}.

\begin{rmk}[Other background metrics]
\label{RmkEx0BgMetric}
  Note that the structure of the linearized operator $L_{g,g^0}$ recorded in Proposition~\ref{PropExOpLin} is insensitive to the particular choice of $g^0$ (as long as it satisfies the conditions in Proposition~\ref{PropExOpLin} for $\bhm=\bhm_0$). The mapping properties of the nonlinear operator in Corollary~\ref{CorExFwN}\eqref{ItExFwN1} only require $\Ric(g^0)=0$. Therefore, we can more generally solve the initial value problem for
  \[
    P(g^0+h,g^0)=0
  \]
  provided $\Ric(g^0)=0$, with the initial data of $h$ being $h_0,h_1$ as in~\eqref{EqEx0IVP}; we only need to ensure that the boundary and Cauchy hypersurfaces of $\Omega_{\ext,r_0}$ are spacelike for $g^0$, which holds if $g^0-g_{b_0}$ is small enough in the sense of Lemma~\ref{LemmaEx0Dom}. As an important special case, this allows for $g^0=\phi_\scal^*g_{b_0}$ when $\scal\in\scalspace_1$ is small (see Lemma~\ref{LemmaKBo}).
\end{rmk}

\subsection{Partial polyhomogeneity}
\label{SsExPhg}

We now show that if the initial data $h_0,h_1$ in~\eqref{EqEx0IVP} are partially polyhomogeneous, then so is the solution $h$. We do this by adapting (and streamlining) the arguments of \cite{HintzVasyMink4} to the present setting.

\begin{thm}[Partial polyhomogeneity: exterior region]
\label{ThmExPhg}
  Let $\cE_0\subset\C\times\N_0$ be an index set with $\min\Re\cE_0>1+2\gamma^\Ups$ and $j\cE_0\subset\cE_0$ for all $j\in\N$. Fix $\ell_0,\ell_\sscri>1$ with $\ell_0\notin\Re\pi_1\cE_0$ and
  \[
    \ell_\sscri<\ell_0,\quad
    \ell_\sscri<1+(1-e^\cC)(1-v^\cC)\gamma^\cC.
  \]
  Then there exists an index set $\cE_\sscri^\cC\subset\C\times\N_0$ for which the conditions in Definition~\usref{DefExP} are satisfied and such that the following holds. Let $\ell_0^\flat=1+\frac12(1-e^\Ups)\gamma^\Ups$ and $\ell_\sscri^\flat:=1+\frac14(1-e^\Ups)\gamma^\Ups$.\footnote{These quantities satisfy the conditions~\eqref{EqEx0ells}.} Suppose, in the notation of Lemma~\usref{LemmaEx0Dom}, that
  \begin{equation}
  \label{EqExPhgData}
    h_0,\,h_1 \in H_\bop^{\infty,(\cE_0,\ell_0)}(\Sigma_{\IVP,r_0};S^2\cT^*)
  \end{equation}
  satisfy the smallness condition in Theorem~\usref{ThmEx0} (i.e., $\|h_j\|_{H_\bop^{k_0,\ell_0^\flat}}<\eps_0$ for $j=1,2$), and denote by $h\in\Hb^{\infty,\ell_0^\flat,(-,\ell_\sscri^\flat)}(\Omega_{\ext,r_0})$ the unique solution of the initial value problem~\eqref{EqEx0IVP}. Then, in fact,
  \begin{equation}
  \label{EqExPhgh}
    h \in \Hb^{\infty,\ (\cE_0,\ell_0),\ \bigl(\la\cE_\sscri^\cC\ra,\ell_\sscri\bigr)}(\Omega_{\ext,r_0}).
  \end{equation}
  Moreover, given any $k\in\N$ and $\eps>0$, there exist $k_0\in\N$ and $\eps_0>0$ such that
  \begin{equation}
  \label{EqExPhgSmall}
    \|h_j\|_{\Hb^{k_0,(\cE_0,\ell_0)}}<\eps_0,\ j=1,2 \implies \|h\|_{\Hb^{k,\ (\cE_0,\ell_0),\ (\la\cE_\sscri^\cC\ra,\ell_\sscri)}}<\eps.
  \end{equation}
\end{thm}

This differs slightly from \cite[Theorem~7.1]{HintzVasyMink4} in that our present gauge condition leads to different index sets at $\scri^+$. (Moreover, in the present paper we do not need $g_{b_0}+h$ to satisfy the Einstein equation.) The method of proof is quite similar, however. At $I^0$ (Step~1 below), we use the approximate commutation of the linearized operator with $\rho_0\pa_{\rho_0}-z$ for $(z,k)\in\cE_0$ to extract the expansion (cf.\ Lemma~\ref{LemmaTMIntFuchs} and~\S\ref{SssIGen0}). At $\scri^+$ (Step~2 below), we replace the linearized operator by its normal operator and regard lower order terms (in the sense of decay) as errors that we put on the right-hand side.

\begin{rmk}[Polyhomogeneity and smallness]
\label{RmkExPhgSmall}
  The smallness of $\|h_j\|_{\Hb^{k_0,\ell_0^\flat}}$, $j=1,2$, and the polyhomogeneity~\eqref{EqExPhgData} required in Theorem~\ref{ThmExPhg} do \emph{not} imply that the coefficients of the partial expansions of the $h_j$ (or of $h$) are themselves small; this is why we must state~\eqref{EqExPhgSmall} separately. To illustrate this, note that if $\chi\in\CIc([0,1);[0,1])$ equals $1$ on $[0,\frac12]$, then the functions $u_{C,C'}(x):=C x\chi(C' x)$, where $C>0$ and $C'\geq 1$, are polyhomogeneous at $x=0$ with index set $(1,0)$, but not small in $\Hb^{0,((1,0),\ell)}$ for any $\ell>1$ unless the $x^1$-coefficient $C$ is small. On the other hand, we have
  \[
    \|u_{C,C'}\|_{L^2}^2 := \int_0^1 |u_{C,C'}(x)|^2\,\frac{\dd x}{x} \leq \int_0^{1/C'} |C x|^2\,\frac{\dd x}{x} = \frac12\Bigl(\frac{C}{C'}\Bigr)^2,
  \]
  which tends to zero for $C=k$ (in which case the polyhomogeneous norm blows up) and $C'=k^2$.
\end{rmk}

\begin{proof}[Proof of Theorem~\usref{ThmExPhg}]
  \pfstep{Step~1. Partial polyhomogeneity at $I^0$.} Assume that for some $\alpha_0\in[\ell_0^\flat,\ell_0)$, we have already shown
  \begin{equation}
  \label{EqExPhgAssmI0}
    h\in\Hb^{\infty,(\cE_0,\alpha_0),(-,\ell_\sscri^\flat)}.
  \end{equation}
  This holds for $\alpha_0=\ell_0^\flat$ by Theorem~\ref{ThmEx0}.

  Since $P(g_{b_0}+h,g_{b_0})=0$, we can rewrite the gauge-fixed Einstein equation following~\eqref{EqExFwNLin} as
  \begin{equation}
  \label{EqExPhgRewrite}
    L_{g_{b_0}}h = L_{g_{b_0},g_{b_0}}h = -\int_0^1 (L_{g_{b_0}+s h,g_{b_0}}-L_{g_{b_0},g_{b_0}})h\,\dd s.
  \end{equation}
  Now, Proposition~\ref{PropExOpLin} implies
  \begin{equation}
  \label{EqExPhgI0}
    L_{g_{b_0}+s h,g_{b_0}}-L_{g_{b_0},g_{b_0}} \in H_\bop^{\infty,(\cE_0+2,\alpha_0+2),((1,0),\ell_\sscri^\flat)}\Diffb^2(\Omega_{\ext,r_0};S^2\cT^*).
  \end{equation}
  Indeed, the smooth terms of $L_{g_{b_0}+s h,g_{b_0}}$ and $L_{g_{b_0},g_{b_0}}$ cancel, and what remains are those terms from $r^{-1}A_h\pa_1$ and $r^{-1}B_h$ that involve $h$. These are of class $H_\bop^{\infty,(\cE_0+2,\alpha_0+2),((1,0),\ell_\sscri^\flat)}$; and when computing the action of~\eqref{EqExPhgI0} on $h$, they act on $\pi^\cC h\in H_\bop^{\infty,(\cE_0,\alpha_0),\ell_\sscri^\flat}$, except for the $(7,6)$-component of $A_h$ in~\eqref{EqExOpLinAB} which acts on $\slpi_0 h$ and contributes only to the $(\dd x^1)^2$ component of the output. Recalling that $\min\Re\cE_0>1$, we also note that the b-Sobolev remainder in~\eqref{EqExOpLinb} acts on~\eqref{EqExPhgAssmI0} to produce an element of $H_\bop^{\infty,(\cE_0+3,\ell_0+3),\ell_\sscri^\flat+1}$. Altogether then, the action of~\eqref{EqExPhgI0} on $h$ produces an element of
  \[
    \Hb^{\infty,(\cE_0+3,\alpha_0+3),(-',\ell_\sscri^\flat+1)},
  \]
  which is one order better at $I^0$ than what Proposition~\ref{PropExFwL} alone would give for $L_{g_{b_0}}h$, given~\eqref{EqExPhgAssmI0}. Therefore,~\eqref{EqExPhgRewrite} implies that $h$ solves the initial value problem
  \begin{equation}
  \label{EqExPhgI0Comm}
    \left\{
      \begin{aligned}
        L_{g_{b_0}}h&\in\Hb^{\infty,\ (\cE_0+3,\alpha_0+3),\ (-',\ell_\sscri^\flat+1)} && \text{in}\ \Omega_{\ext,r_0}, \\
        h&=h_0\in H_\bop^{\infty,(\cE_0,\ell_0)} && \text{on}\ \Sigma_{\IVP,r_0}, \\
        r\cL_{\pa_t}h&=h_1\in H_\bop^{\infty,(\cE_0,\ell_0)} && \text{on}\ \Sigma_{\IVP,r_0}.
      \end{aligned}
    \right.
  \end{equation}
  Reducing $\alpha_0$ by an arbitrarily small amount (if necessary) to ensure $\min(\alpha_0+1,\ell_0)\notin\Re\pi_1\cE_0$, we claim that this implies
  \begin{equation}
  \label{EqExPhgConcI0}
    h \in \Hb^{\infty,\ \bigl(\cE_0,\min(\alpha_0+1,\ell_0)\bigr),\ (-,\ell_\sscri^\flat)},
  \end{equation}
  which would thus uncover one full power more of the expansion of $h$ at $I^0$ compared to~\eqref{EqExPhgAssmI0}. To this end, define the b-differential operator $\cR:=\prod_{(z,\ell)\in\cE_{0,\leq}}(\rho_0\pa_{\rho_0}-z)$ where $\cE_{0,\leq}:=\{(z,\ell)\in\cE_0\colon\Re z\leq\min(\alpha_0+1,\ell_0)\}$; the number of these $(z,\ell)$ is a finite number $N:=|\cE_{0,\leq}|$, so $\cR\in\Diffb^N$. Since $L_{g_{b_0}}\in\rho_0^2\rho_\sscri\Diffb^2$, we obtain for $\cR h$, which by~\eqref{EqExPhgAssmI0} lies in $\Hb^{\infty,\alpha_0,(-,\ell_\sscri^\flat)}$, the equation
  \[
    L_{g_{b_0}}(\cR h) = \rho_0^2\rho_0^{-2}L_{g_{b_0}}(\cR h) = \rho_0^2\cR(\rho_0^{-2}L_{g_{b_0}}h) + \rho_0^2[\rho_0^{-2}L_{g_{b_0}},\cR]h.
  \]
  The first term lies in $\Hb^{\infty,\ \min(\alpha_0+1,\ell_0)+2,\ (-',\ell_\sscri^\flat+1)}$ by~\eqref{EqExPhgI0Comm}. For the second term, we use that for the particular operator $\cR$ defined above, we have $[\cR,\rho_\sscri^m\Diffb^k]\subset\rho_\sscri^m\rho_0\Diffb^{k+N-1}$ for all $m,k$, which implies that $\rho_0^2[\rho_0^{-2}L_{g_{b_0}},\cR]\in\rho_0^3\rho_\sscri\Diffb^{N+1}$ maps~\eqref{EqExPhgAssmI0} into $\Hb^{\infty,\ (\cE_0+3,\alpha_0+3),\ (-',\ell_\sscri^\flat+1)}$. The initial data for $\cR h$ are of class $H_\bop^{\infty,\min(\alpha_0+1,\ell_0)}(\Sigma_{\IVP,r_0})$. The estimate~\eqref{EqEx0Est} thus shows that
  \[
    \cR h \in \Hb^{\infty,\ \min(\alpha_0+1,\ell_0),\ (-,\ell_\sscri^\flat)}.
  \]
  Integrating $\cR$ (see Lemma~\ref{LemmaTMPhgTest}) gives~\eqref{EqExPhgConcI0}.

  Iterating this argument finitely many times yields~\eqref{EqExPhgAssmI0} for $\alpha_0=\ell_0$.

  \pfstep{Step~2. Partial polyhomogeneity at $\scri^+$.} Suppose now that for some $\alpha_\sscri\in[\ell_\sscri^\flat,\ell_\sscri)$ (so in particular $\alpha_\sscri>1$), we have
  \begin{equation}
  \label{EqExPhgScri0}
    h \in \Hb^{\infty,\ (\cE_0,\ell_0),\ (\la\cE_\sscri^\cC\ra,\alpha_\sscri)}
  \end{equation}
  where $\cE_\sscri^\cC$ satisfies the conditions of Definition~\ref{DefExP}. The first step of the proof establishes this for $\alpha_\sscri=\ell_\sscri^\flat$ and any $\cE_\sscri^\cC$ that is allowed in Definition~\ref{DefExP}. Since the components $\pi_{1 1}h$ and $\slpi_0 h$ contribute to the $\scri^+$-normal operator of $L_{g_{b_0}+s h,g_{b_0}}$ (via $A_h$ and $B_h$ in~\eqref{EqExOpLinb}), we cannot use~\eqref{EqExPhgRewrite} to extract the expansion of $h$ at $\scri^+$; rather, we must work relative to a metric capturing also the leading-order terms of $\pi_{1 1}h,\slpi_0 h$ at $\scri^+$. For this purpose, let us introduce
  \[
    g_0 := g_{b_0} + h_0,
  \]
  where $h_0\in\Hb^{\infty,\ (\cE_0,\ell_0),\ \bigl(\la(2,0)\ra_0,\infty\bigr)}(\Omega_{\ext,r_0})$ is chosen such that $(r h_0)_{1 1}=(r h)_{1 1}$ and $\slpi_0(r h_0)=\slpi_0(r h)$ at $\scri^+$. Thus, the polyhomogeneous expansion of
  \[
    \tilde h:=h-h_0
  \]
  does not contain any $\rho_\sscri^1$ terms: the index sets of its components are, using the notation~\eqref{EqExPOther},
  \begin{equation}
  \label{EqExPhgScriStart}
    \pi^\cC\tilde h:\ \cE_\sscri^\cC,\quad
    \pi_{0 1}\tilde h:\ \cE_{\sscri,0 1},\quad
    \pi_{1 /}\tilde h:\ \cE_{\sscri,1 /},\quad
    \slpi_0\tilde h:\ \slcE_{\sscri,0}\setminus\{(1,0)\}=\cE_\sscri^\cC,\quad
    \pi_{1 1}\tilde h:\ \cE_{\sscri,1 1}\setminus\{(1,0)\}.
  \end{equation}
  Moreover, Corollary~\ref{CorExFwN} and the fact that only integer powers arise at $\scri^+$ yield
  \begin{equation}
  \label{EqExPhgf0}
    P(g_0,g_{b_0}) \in \Hb^{\infty,\ (\cE_0+2,\ell_0+2),\ \bigl(\la(3,0)\ra'_0,\infty\bigr)}.
  \end{equation}

  \pfsubstep{Step~2.1.}{Rewriting the PDE.} We write $P(g_{b_0}+h,g_{b_0})=P(g_0+\tilde h,g_{b_0})=0$ as
  \begin{equation}
  \label{EqExPhgRewrite0}
    L_{g_0,g_{b_0}}\tilde h = -P(g_0,g_{b_0}) -\int_0^1 (L_{g_0+s\tilde h,g_{b_0}}-L_{g_0,g_{b_0}})\tilde h\,\dd s.
  \end{equation}
  Recalling~\eqref{EqExOpLinbNormal}, we can write
  \begin{equation}
  \label{EqExPhgScriKerr}
  \begin{split}
    &\hspace{-3em}\rho_0^{-2}\rho_\sscri^{-1}L_{g_0,g_{b_0}} = L^0 + L^1 + \tilde L, \\
    L^0 &= -(\rho_\sscri\pa_{\rho_\sscri}-(I+A_0))(\rho_0\pa_{\rho_0}-\rho_\sscri\pa_{\rho_\sscri}), \\
    L^1 &= (A_{h_0}-A_0)(\rho_0\pa_{\rho_0}-\rho_\sscri\pa_{\rho_\sscri}) + \rho_0^{-1}B_{h_0}, \\
    \tilde L &\in \bigl(\rho_\sscri\CI+H_\bop^{\infty,\ (\cE_0,\ell_0),\ \bigl((1,0),\infty\bigr)}\bigr)\Diffb^2,
  \end{split}
  \end{equation}
  and moreover
  \begin{equation}
  \label{EqExPhgScriRem}
  \begin{split}
    &\hspace{-3em}\rho_0^{-2}\rho_\sscri^{-1}\bigl(L_{g_0+s\tilde h,g_{b_0}}-L_{g_0,g_{b_0}}\bigr) = R_{s\tilde h} + \tilde R_{s\tilde h}, \\
    R_{s\tilde h} &= (A_{s\tilde h}-A_0)(\rho_0\pa_{\rho_0}-\rho_\sscri\pa_{\rho_\sscri}) + \rho_0^{-1}B_{s\tilde h}, \\
    \tilde R_{s\tilde h} &\in H_\bop^{\infty,(\cE_0,\ell_0),(\cE_\sscri^\cC-1,\alpha_\sscri-1)}\Diffb^2,
  \end{split}
  \end{equation}
  where we used~\eqref{EqExPhgScri0} for the membership of $\tilde R_{s\tilde h}$. In order to make the arguments below more transparent, we note that in the splitting~\eqref{EqExOpLinSplit}, and writing $*$ for a constant nonzero entry, we have
  \[
    A_0 =
    \begin{pmatrix}
      2(1-v^\cC)\gamma^\cC & 0 & 0 & 0 & 0 & 0 & 0 \\
      0 & (1-v^\cC)\gamma^\cC & 0 & 0 & 0 & 0 & 0 \\
      * & 0 & (1-e^\cC)(1-v^\cC)\gamma^\cC & 0 & 0 & 0 & 0 \\
      * & 0 & * & (1-e^\Ups)\gamma^\Ups & 0 & 0 & 0 \\
      0 & * & 0 & 0 & \gamma^\Ups & 0 & 0 \\
      0 & 0 & 0 & 0 & 0 & 0 & 0 \\
      0 & 0 & * & * & 0 & 0 & 2\gamma^\Ups
    \end{pmatrix}
  \]
  and
  \[
    A_h-A_0 =
    \begin{pmatrix}
      0 & 0 & 0 & 0 & 0 & 0 & 0 \\
      0 & 0 & 0 & 0 & 0 & 0 & 0 \\
      0 & 0 & 0 & 0 & 0 & 0 & 0 \\
      2\pa_1(r h_{0 1}) & 0 & 0 & 0 & 0 & 0 & 0 \\
      2\pa_1(r h_{1\bar a}) & 0 & 0 & 0 & 0 & 0 & 0 \\
      2\pa_1(r h_{\bar a\bar b}) & 0 & 0 & 0 & 0 & 0 & 0 \\
      0 & 2\pa_1(r h_1{}^{\bar a}) & -2\pa_1(r h_{0 1}) & 0 & 0 & -\frac12\pa_1(r h^{\bar a\bar b}) & 0
    \end{pmatrix}.
  \]
  We can then rewrite~\eqref{EqExPhgRewrite0} further as
  \begin{equation}
  \label{EqExPhgRewrite2}
  \begin{split}
    &L^0\tilde h = -L^1\tilde h - \int_0^1 R_{s\tilde h}\tilde h\,\dd s + f_0 + \tilde f, \\
    &\qquad f_0:=-\rho_0^{-2}\rho_\sscri^{-1}P(g_0,g_{b_0}),\quad \tilde f :=  -\tilde L\tilde h - \int_0^1 \tilde R_{s\tilde h}\tilde h\,\dd s.
  \end{split}
  \end{equation}
  We shall use that
  \[
    \tilde h\in H_\bop^{\infty,(\cE_0,\ell_0),(\tilde\cE_\sscri^\tot,\alpha_\sscri)},\quad \tilde\cE_\sscri^\tot:=\cE_\sscri^\tot\setminus\{(1,0)\};
  \]
  here $\Re\tilde\cE_\sscri^\tot\geq 1+(1-e^\Ups)\gamma^\Ups>1$. Since $\cE_\sscri^\cC-1\supset(1,0)$, this gives
  \[
    \tilde L\tilde h,\ \tilde R_{s\tilde h}\tilde h \in H_\bop^{\infty,(\cE_0,\ell_0),(\cE_\sscri^\cC+\tilde\cE_\sscri^\tot-1,\alpha_\sscri+\eta)}
  \]
  for all $\eta<\min(\alpha_\sscri-1,(1-e^\Ups)\gamma^\Ups)\in(0,1)$; we demand that $\alpha_\sscri+\eta\leq\ell_\sscri$. Therefore,
  \[
    \tilde f \in H_\bop^{\infty,(\cE_0,\ell_0),(\cE_\sscri^\cC+\tilde\cE_\sscri^\tot-1,\alpha_\sscri+\eta)}.
  \]

  \pfsubstep{Step~2.2.}{Improving $\tilde h_{0 0}$, $\tilde h_{0\bar a}$, $\sltr\tilde h$.} The first component of~\eqref{EqExPhgRewrite2} reads
  \[
    -\bigl(\rho_\sscri\pa_{\rho_\sscri}-(1+2(1-v^\cC)\gamma^\cC)\bigr)(\rho_0\pa_{\rho_0}-\rho_\sscri\pa_{\rho_\sscri})\tilde h_{0 0} \in H_\bop^{\infty,(\cE_0,\ell_0),\bigl((2,0)\cup(\cE_\sscri^\cC+\tilde\cE_\sscri^\tot-1),\alpha_\sscri+\eta\bigr)},
  \]
  with the $\scri^+$-index set $(2,0)$ coming from $f_0$ (via~\eqref{EqExPhgf0} times $\rho_0^{-2}\rho_\sscri^{-1}$). Since $1+2(1-v^\cC)\gamma^\cC>\ell_\sscri$, Lemma~\ref{LemmaTMIntFuchs} implies
  \begin{equation}
  \label{EqExPhgDiffh00}
    (\rho_0\pa_{\rho_0}-\rho_\sscri\pa_{\rho_\sscri})\tilde h_{0 0} \in H_\bop^{\infty,(\cE_0,\ell_0),\ \bigl((2,0)\cup(\cE_\sscri^\cC+\tilde\cE_\sscri^\tot-1),\alpha_\sscri+\eta\bigr)},
  \end{equation}
  and then Lemma~\ref{LemmaTIntHyp} and the condition
  \begin{equation}
  \label{EqExPhgEscri}
     \cE_0\extcup\Bigl( (2,0) \cup \bigl[ \cE_\sscri^\cC+(\tilde\cE_\sscri^\tot-1) \bigr] \Bigr)\subset\cE_\sscri^\cC,
  \end{equation}
  which we impose now on $\cE_\sscri^\cC$, give $\tilde h_{0 0} \in H_\bop^{\infty,(\cE_0,\ell_0),(\cE_\sscri^\cC,\alpha_\sscri+\eta)}$. This proves the polyhomogeneity of $\tilde h_{0 0}$ modulo conormal remainders with $\eta$ more orders of decay than initially assumed in~\eqref{EqExPhgScri0}.

  For the second, i.e., $0\bar a$-, component, of~\eqref{EqExPhgRewrite2}, the same argument applies, giving the same membership for $\pi_{0 /}\tilde h$. Similarly for the third component of~\eqref{EqExPhgRewrite2}, except this now couples to the first component due to the nonzero lower triangular $(3,1)$-entry $(A_0)_{3,1}$; putting the corresponding term $(A_0)_{3,1}(\rho_0\pa_{\rho_0}-\rho_\sscri\pa_{\rho_\sscri})\tilde h_{0 0}\in H_\bop^{\infty,(\cE_0,\ell_0),\ \bigl((2,0)\cup(\cE_\sscri^\cC+\tilde\cE_\sscri^\tot-1),\alpha_\sscri+\eta\bigr)}$ on the right-hand side, we can again integrate using Lemmas~\ref{LemmaTMIntFuchs} and \ref{LemmaTIntHyp}. We have thus far shown
  \begin{subequations}
  \begin{align}
  \label{EqExPhgh00}
    * \in \{\tilde h_{0 0},\,\tilde h_{0\bar a},\,\sltr\tilde h\} \implies \hspace{8em} * &\in H_\bop^{\infty,(\cE_0,\ell_0),(\cE_\sscri^\cC,\alpha_\sscri+\eta)}, \\
  \label{EqExPhgh00Diff}
    (\rho_0\pa_{\rho_0}-\rho_\sscri\pa_{\rho_\sscri})*&\in H_\bop^{\infty,(\cE_0,\ell_0),\ \bigl((2,0)\cup(\cE_\sscri^\cC+\tilde\cE_\sscri^\tot-1),\alpha_\sscri+\eta\bigr)}.
  \end{align}
  \end{subequations}
  The second membership implies membership in $\Hb^{\infty,(\cE_0,\ell_0),(\cE_\sscri^\cC,\alpha_\sscri+\eta)}$ by~\eqref{EqExPhgEscri}.

  \pfsubstep{Step~2.3.}{Improving $\tilde h_{0 1}$, $\pi_{1 /}\tilde h$.} To proceed, we reduce $\eta$ by an arbitrary small amount, if necessary, so as to ensure that $\alpha_\sscri+\eta\notin\{1+(1-e^\Ups)\gamma^\Ups,\,1+\gamma^\Ups,\,1+2\gamma^\Ups\}$. (No such reduction is needed when $\alpha_\sscri+\eta=\ell_\sscri$ since $\ell_\sscri>1+2\gamma^\Ups$.) Consider then the fourth component of~\eqref{EqExPhgRewrite2}. We again put the lower triangular contributions on the right-hand side; these are:
  \begin{enumerate}[label=(\roman*)]
  \item\label{ItExPhg231} from $L^0$ via $(A_0)_{4,1}$ and $(A_0)_{4,3}$: $H_\bop^{\infty,(\cE_0,\ell_0),\ \bigl((2,0)\cup(\cE_\sscri^\cC+\tilde\cE_\sscri^\tot-1),\alpha_\sscri+\eta\bigr)}$ (from~\eqref{EqExPhgh00Diff});
  \item\label{ItExPhg232} from $L^1$ via $(A_{h_0}-A_0)_{4,1}\in\Hb^{\infty,(\cE_0,\ell_0),((2,0),\infty)}$, the action of which on~\eqref{EqExPhgDiffh00} also produces an element of this space. For later use, we point out that for this conclusion it suffices that the coefficient $(A_{h_0}-A_0)_{4,1}$ lies in $H_\bop^{\infty,(\cE_0,\ell_0),(\cE_\sscri^\tot-1,\alpha_\sscri-1)}$, since
    \begin{align*}
      &(\cE_\sscri^\tot-1) + \bigl( (2,0) \ \cup\  (\cE_\sscri^\cC+\tilde\cE_\sscri^\tot-1) \bigr) \\
      &\qquad \subset (2,0) \cup (\tilde\cE_\sscri^\tot+1) \ \cup\ \bigl(\cE_\sscri^\cC+(\cE_\sscri^\tot-1) + \tilde\cE_\sscri^\tot-1 \bigr) \\
      &\qquad \subset (2,0) \cup (\cE_\sscri^\cC+\tilde\cE_\sscri^\tot-1),
    \end{align*}
    where in the passage to the third line we use Definition~\ref{DefExP}\eqref{ItExPTotNL} as well as $(2,0)\subset\cE_\sscri^\cC$;
  \item\label{ItExPhg233} from $L^1$ via $\rho_0^{-1}(B_{h_0})_{4,1}\in H_\bop^{\infty,(\cE_0,\ell_0),(\tilde\cE_\sscri^\tot-1,\alpha_\sscri-1)}$, the action of which on~\eqref{EqExPhgh00} also produces an element of this space;
  \item\label{ItExPhg234} from $R_{s\tilde h}$ via $(A_{s\tilde h}-A_0)_{4,1}$, $\rho_0^{-1}(B_{s\tilde h})_{4,1}\in H_\bop^{\infty,(\cE_0,\ell_0),(\tilde\cE_\sscri^\tot-1,\alpha_\sscri-1)}$: likewise.
  \end{enumerate}
  Altogether, we thus obtain
  \[
    -\bigl(\rho_\sscri\pa_{\rho_\sscri}-(1+(1-e^\Ups)\gamma^\Ups)\bigr)(\rho_0\pa_{\rho_0}-\rho_\sscri\pa_{\rho_\sscri})\tilde h_{0 1} \in H_\bop^{\infty,(\cE_0,\ell_0),\bigl((2,0)\cup(\cE_\sscri^\cC+\tilde\cE_\sscri^\tot-1),\alpha_\sscri+\eta\bigr)}.
  \]
  Lemma~\ref{LemmaTMIntFuchs} gives
  \begin{equation}
  \label{EqExPhgh01}
    (\rho_0\pa_{\rho_0}-\rho_\sscri\pa_{\rho_\sscri})\tilde h_{0 1} \in H_\bop^{\infty,(\cE_0,\ell_0),\ \bigl((1+(1-e^\Ups)\gamma^\Ups,0)\cup(2,0)\cup(\cE_\sscri^\cC+\tilde\cE_\sscri^\tot-1),\alpha_\sscri+\eta\bigr)},
  \end{equation}
  which we integrate using Lemma~\ref{LemmaTIntHyp} to obtain
  \[
    \tilde h_{0 1} \in H_\bop^{\infty,(\cE_0,\ell_0),(\cE_{\sscri,0 1},\alpha_\sscri+\eta)};
  \]
  here we use~\eqref{EqExPhgEscri} again as well as the definition~\eqref{EqExPOther} of $\cE_{\sscri,0 1}$, and we exploit the fact that $\min\Re\cE_0>1+2\gamma^\Ups>1+(1-e^\Ups)\gamma^\Ups$. The same arguments apply for the fifth component of~\eqref{EqExPhgRewrite2} and imply
  \[
    \pi_{1 /}\tilde h \in H_\bop^{\infty,(\cE_0,\ell_0),(\cE_{\sscri,1 /},\alpha_\sscri+\eta)}.
  \]

  \pfsubstep{Step~2.4.}{Improving $\slpi_0\tilde h$.} Consider now the sixth component of~\eqref{EqExPhgRewrite2}. The contributions from $(A_{s\tilde h}-A_0)_{6,1}$ and $\rho_0^{-1}(B_{s\tilde h})_{6,1}$ again lie in $H_\bop^{\infty,(\cE_0,\ell_0),((2,0)\cup(\cE_\sscri^\cC+\tilde\cE_\sscri^\tot-1),\alpha_\sscri+\eta)}$ (cf.\ item~\ref{ItExPhg234} in Step~2.3), as does the contribution of $(A_{h_0}-A_0)_{6,1}$ from $L^1$ (cf.\ item~\ref{ItExPhg232} in Step~2.3). The term $\rho_0^{-1}(B_{h_0})_{6,1}\in H_\bop^{\infty,(\cE_0,\ell_0),((0,0),\infty)}$ acting on $\tilde h_{0 0}$, for which we have already shown~\eqref{EqExPhgh00}, yields an element of $H_\bop^{\infty,(\cE_0,\ell_0),(\cE_\sscri^\cC,\alpha_\sscri+\eta)}$. Altogether,
  \[
    -(\rho_\sscri\pa_{\rho_\sscri}-1)(\rho_0\pa_{\rho_0}-\rho_\sscri\pa_{\rho_\sscri})\slpi_0\tilde h \in H_\bop^{\infty,(\cE_0,\ell_0),(\cE_\sscri^\cC,\alpha_\sscri+\eta)}.
  \]
  Integrating this equation using Lemmas~\ref{LemmaTMIntFuchs} and \ref{LemmaTIntHyp} gives
  \begin{align}
  \label{EqExPhgslpi0h}
    (\rho_0\pa_{\rho_0}-\rho_\sscri\pa_{\rho_\sscri})\slpi_0\tilde h &\in H_\bop^{\infty,(\cE_0,\ell_0),((1,0)\cup\cE_\sscri^\cC,\alpha_\sscri+\eta)}, \\
  \nonumber
    \slpi_0\tilde h&\in H_\bop^{\infty,(\cE_0,\ell_0),(\slcE_{\sscri,0}^\sharp,\alpha_\sscri+\eta)},
  \end{align}
  where, recalling $\slcE_{\sscri,0}=\cE_\sscri^\cC\cup(1,0)$, we set
  \[
    \slcE_{\sscri,0}^\sharp := \slcE_{\sscri,0} \extcup(\cE_0 \cap \{\Re z\geq\alpha_\sscri\}).
  \]
  (The index set in $\{\Re z<\alpha_\sscri\}$ is of course unchanged relative to the starting point~\eqref{EqExPhgScriStart}.)

  \pfsubstep{Step~2.5.}{Improving $\tilde h_{1 1}$.} Consider finally the seventh component of~\eqref{EqExPhgRewrite2}. The contributions of $(A_0)_{7,3}$, $(A_{h_0}-A_0)_{7,2}$, $(A_{s\tilde h}-A_0)_{7,2}$, $(A_{s\tilde h}-A_0)_{7,3}$, and $\rho_0^{-1}(B_{s\tilde h})_{7,1}$ can be put on the right-hand side as before; these lie in $H_\bop^{\infty,(\cE_0,\ell_0),((2,0)\cup(\cE_\sscri+\tilde\cE_\sscri^\tot-1),\alpha_\sscri+\eta)}$. The contribution of $\rho_0^{-1}(B_{h_0})_{7,1}$ lies in $H_\bop^{\infty,(\cE_0,\ell_0),(\cE_\sscri^\cC,\alpha_\sscri+\eta)}$. The remaining contributions are twofold:
  \begin{enumerate}[label=(\roman*)]
  \item from $(A_0)_{7,4}$: this acts on~\eqref{EqExPhgh01} and produces an element of the space in~\eqref{EqExPhgh01};
  \item from $(A_{h_0}-A_0)_{7,6}\in H_\bop^{\infty,(\cE_0,\ell_0),((0,0),\infty)}$ and $(A_{s\tilde h}-A_0)_{7,6}\in H_\bop^{\infty,(\cE_0,\ell_0),(\cE_\sscri^\cC-1,\alpha_\sscri)}$: these act on~\eqref{EqExPhgslpi0h} and produce elements of $H_\bop^{\infty,(\cE_0,\ell_0),\ ((1,0)\cup\cE_\sscri^\cC,\alpha_\sscri+\eta)}$, where we use $\cE_\sscri^\cC+(\cE_\sscri^\cC-1)\subset\cE_\sscri^\cC$.
  \end{enumerate}
  Therefore,
  \begin{equation}
  \label{EqExPhgh11}
    -(\rho_\sscri\pa_{\rho_\sscri}-(1+2\gamma^\Ups))(\rho_0\pa_{\rho_0}-\rho_\sscri\pa_{\rho_\sscri})\tilde h_{1 1} \in H_\bop^{\infty,(\cE_0,\ell_0),\bigl((1,0)\cup(1+(1-e^\Ups)\gamma^\Ups,0)\cup\cE_\sscri^\cC,\alpha_\sscri+\eta\bigr)}.
  \end{equation}
  Integration yields
  \[
    \tilde h_{1 1} \in H_\bop^{\infty,(\cE_0,\ell_0),(\cE_{\sscri,1 1}^\sharp,\alpha_\sscri+\eta)},
  \]
  where, recalling $\cE_{\sscri,1 1}=\cE_\sscri^\cC\cup(1,0)\cup(1+(1-e^\Ups)\gamma^\Ups,0)\cup(1+2\gamma^\Ups,0)$, we set
  \[
    \cE_{\sscri,1 1}^\sharp := \cE_{\sscri,1 1} \extcup(\cE_0\cap\{\Re z\geq\alpha_\sscri\}).
  \]
  (Again, the index set in $\{\Re z<\alpha_\sscri\}$ is unchanged relative to~\eqref{EqExPhgScriStart}.)

  \pfsubstep{Step~2.6.}{Conclusion.} From the membership $h\in\Hb^{\infty,(\cE_0,\ell_0),(\la\cE_\sscri^\cC\ra,\alpha_\sscri)}$ in~\eqref{EqExPhgScri0}, we have now deduced that
  \[
    h \in \Hb^{\infty,\ (\cE_0,\ell_0),\ \bigl(\la\cE_\sscri^{\cC,{\rm new}}\ra,\min(\alpha_\sscri+\eta,\ell_\sscri)\bigr)}
  \]
  where $\eta$ can be taken arbitrarily close to $\min(\alpha_\sscri-1,(1-e^\Ups)\gamma^\Ups)$, and $\cE_\sscri^{\cC,{\rm new}}$ is the smallest index set containing
  \[
    \cE_\sscri^{\cC,\sharp} := \cE_\sscri^\cC \extcup(\cE_0\cap\{\Re z\geq\alpha_\sscri\})
  \]
  and satisfying the conditions of Definition~\ref{DefExP} as well as~\eqref{EqExPhgEscri}. If we always choose $\eta>\frac12\min(\alpha_\sscri-1,(1-e^\Ups)\gamma^\Ups)$ (as we may), then after finitely many iterations we obtain the desired partial polyhomogeneity of $h$ at $\scri^+$.

  The smallness statement~\eqref{EqExPhgSmall} follows from the proof, as the terms in the expansion of $h$ are constructed from those of the initial data via integration.
\end{proof}

\subsection{Pullback by boosts}
\label{SsExBo}

Continuing Remark~\ref{RmkEx0BgMetric}, we observe that Theorem~\ref{ThmExPhg} remains valid (with the same proof) in the more general setting of writing metrics as perturbations of $g^0$ satisfying $\Ric(g^0)=0$ and $g^0-g_{\bhm_0}\in\Hb^{\infty,\ ((1,0),\infty),\ \bigl(\la(2,0)\ra_0,\infty\bigr)}(\Omega_{\ext,r_0})$. By Lemma~\ref{LemmaKBo}, this includes the special case $g^0=\phi_\scal^*g_{b_0}$ in the notation of~\eqref{EqKBoMap}.

To motivate further adjustments, suppose we apply Theorem~\ref{ThmEx0} with initial data satisfying the constraint equations; then $\Ric(g_{b_0}+h)=0$ and $\Ups_{E^\Ups}(g_{b_0}+h,g_{b_0})=0$. Pulling back the solution along $\phi_\scal$, we note that
\[
  \Ric(\phi_\scal^*g_{b_0} + \phi_\scal^*h) = \phi_\scal^*\Ric(g_{b_0}+h) = 0
\]
still; but we do not have $\Ups_{E^\Ups}(\phi_\scal^*g_{b_0}+\phi_\scal^*h,\phi_\scal^*g_{b_0})=0$ unless $\gamma^\Ups=0$ in Definition~\ref{Def1Gauge}, in which case we do have
\begin{equation}
\label{EqExBoUps0}
  \phi_\scal^*\Ups_0(g_{b_0}+h,g_{b_0}) = \Ups_0(\phi_\scal^*g_{b_0}+\phi_\scal^*h,\phi_\scal^*g_{b_0}).
\end{equation}
The issue in the case $\gamma^\Ups\neq 0$ is that one would also need to pull back $E^\Ups_{g^0}$, which however would lead to a different structure of the gauge condition (in particular to leading order at $\scri^+$). Rather than tracking this change, we instead note that the stationary nature of the gauge modification $E^\Ups_{g^0}$ in Definition~\ref{Def1Gauge} played no role in the proofs of Theorems~\ref{ThmEx0} and \ref{ThmExPhg}, and neither did the particular choice of $\cd^\Ups$ away from $\scri^+$ (or more precisely: modulo $\rho_0\rho_\sscri^2\CI(\Omega_\ext;\cT^*)$). We may thus transition from the gauge 1-form $\Ups_0$ near $\Sigma_\IVP$ to $\Ups_{E^\Ups}$ near $\scri^+$ as follows. Fix a cutoff function
\[
  \check\chi_0\in\CIc((-\tfrac14,\tfrac14)),\quad \check\chi_0|_{[-\frac18,-\frac18]}=1,
\]
and define
\[
  \chi_0 := 1-\check\chi_0\Bigl(\frac{t_\IVP}{r}\Bigr) \in \CI(M);
\]
this equals $1$ near $\scri^+$ and $0$ near $\Sigma_\IVP$. We can then work with the gauge 1-form
\begin{equation}
\label{EqExBoGauge}
  \Ups_{\chi_0 E^\Ups}(g,g^0) := \Ups_0(g,g^0) - \chi_0 E^\Ups_{g^0}\sfG_{g_0}(g-g^0),
\end{equation}
which equals $\Ups_0(g,g^0)$ near $\Sigma_\IVP$ and $\Ups_{E^\Ups}(g,g^0)$ near $\scri^+$. (The only difference to~\eqref{Eq1UpsDef} is the presence of the cutoff $\chi_0$.) We then note that if $\scal\in\scalspace_1$ is small enough so that $\phi_\scal(\Sigma_\IVP)\subset\{\chi_0=1\}$, we do have
\[
  \Ups_{\chi_0 E^\Ups}(\phi_\scal^*g_{b_0}+\phi_\scal^*h,\phi_\scal^*g_{b_0}) = 0\quad\text{near}\ \Sigma_{\IVP,r_1}
\]
in the notation of~\eqref{EqEx0DomSigmaIVP}, as a consequence of~\eqref{EqExBoUps0}; here we fix $r_1>r_0$ (say, $r_1=r_0+1$) to ensure that $\phi_\scal^*h$ is defined on $\Sigma_{\IVP,r_1}$ (for small enough $\scal$). Thus, in a neighborhood of $\Sigma_{\IVP,r_1}$, $\phi_\scal^*h$ is equal to the (unique) solution $h_\scal$ of the initial value problem
\begin{subequations}
\begin{equation}
\label{EqExBoIVP}
  \left\{
  \begin{aligned}
    \Ric(\phi_\scal^*g_{b_0}+h_\scal) - \delta_{\phi_\scal^*g_{b_0},E^\cC}^*\Ups_{\chi_0 E^\Ups}(\phi_\scal^*g_{b_0}+h_\scal,\phi_\scal^*g_{b_0})&=0 && \text{in}\ \Omega_{\ext,r_1}, \\
    h_\scal &=h_{\scal,0} && \text{on}\ \Sigma_{\IVP,r_1}, \\
    r\cL_{\pa_{t_\IVP}}h_\scal&=h_{\scal,1} && \text{on}\ \Sigma_{\IVP,r_1},
  \end{aligned}
  \right.
\end{equation}
if we use the initial data
\begin{equation}
\label{EqExBoIVPData}
  h_{\scal,0} := (\phi_\scal^*h)|_{\Sigma_{\IVP,r_1}},\quad
  h_{\scal,1} := \bigl(r\cL_{\pa_{t_\IVP}}(\phi_\scal^*h)\bigr)\big|_{\Sigma_{\IVP,r_1}}.
\end{equation}
\end{subequations}
Equivalently, upon using $\phi_\scal$ to identify $\Sigma_{\IVP,r_1}$ and $\Sigma_{\IVP,r_1,\scal}:=\phi_\scal(\Sigma_{\IVP,r_1})$, we have $h_{\scal,0}=\phi_\scal^*(h|_{\Sigma_{\IVP,r_1,\scal}})$ and $h_{\scal,1}:=\phi_\scal^*( (\cL_{(\phi_\scal)_*(r\pa_{t_\IVP})}h)|_{\Sigma_{\IVP,r_1,\scal}})$. Since $h\in\Hb^{\infty,\ell_0^\flat}$, resp.\ $\Hb^{\infty,(\cE_0,\ell_0)}$ away from $\Omega_{\ext,r_1}\cap\scri^+$ in the setting of Theorem~\ref{ThmEx0}, resp.\ \ref{ThmExPhg}, the initial data $h_{\scal,j}$, $j=0,1$, have the same weight and partial polyhomogeneity as the original data $h_j$ (and the same degree of b-regularity up to a finite loss, which using standard hyperbolic theory one can show to be $0$, but which in any case we do not need to track for the purposes of the present paper). If one uses the latter description of $h_{\scal,j}$, one also needs to observe that $(\phi_\scal)_*(r\pa_{t_\IVP})$ is still a b-vector field near $(I^0)^\circ$ (since $\phi_\scal$ is the time $1$ flow along a b-vector field, and $r\pa_{t_\IVP}$ is a b-vector field); if, instead, one uses the former description, one needs to use that pullback by $\phi_\scal$ maps elements of (weighted and partially polyhomogeneous) b-Sobolev spaces near compact subsets of $(I^0)^\circ$ to elements of b-Sobolev spaces (with the same weights and index sets).

We summarize our discussion as follows (only in the partially polyhomogeneous setting):

\begin{thm}[Stability with boosted background and data]
\label{ThmExBoStab}
  Let small initial data $h_0,h_1$ on $\Sigma_{\IVP,r_0}$ be given as in Theorem~\usref{ThmExPhg}, and denote by $h\in\Hb^{\infty,(\cE_0,\ell_0),(\la\cE_\sscri^\cC\ra,\ell_\sscri)}(\Omega_{\ext,r_0})$ the solution of the initial value problem for
  \begin{equation}
  \label{EqExBoStabPDE}
    \Ric(g_{b_0}+h) - \delta_{g_{b_0},E^\cC}^*\Ups_{\chi_0 E^\Ups}(g_{b_0}+h,g_{b_0}) = 0
  \end{equation}
  from~\eqref{EqExPhgh}. For all $\scal\in\scalspace_1$ with sufficiently small norm (depending only on $b_0$ and the parameters $\ell_0^\flat,\ell_\sscri^\flat$ in Theorem~\usref{ThmEx0}), define the data $h_{\scal,0},h_{\scal,1}$ by~\eqref{EqExBoIVPData}. Let $h_\scal\in\Hb^{\infty,\ (\cE_0,\ell_0),\ (\la\cE_\sscri^\cC\ra,\ell_\sscri)}(\Omega_{\ext,r_1})$ be the solution of~\eqref{EqExBoIVP} on $\Omega_{\ext,r_1}$, $r_1:=r_0+1$. Then:
  \begin{enumerate}
  \item $h_\scal=\phi_\scal^*h$ in a neighborhood of $\Sigma_{\IVP,r_1}$ (and $h_0=h$ on all of $\Omega_{\ext,r_1}$);
  \item $h_\scal$ depends smoothly on $\scal$;\footnote{The regularity is $\cC^l$ for any fixed $l$ when measuring $h$ in a finite-regularity space $\Hb^{k,(\cE_0,\ell_0),(\la\cE_\sscri^\cC\ra,\ell_\sscri)}$ provided the initial data have sufficiently high regularity depending on $k$ and $l$.}
  \item\label{ItExBoStabEinstein} if the first and second fundamental form induced by $h_0,h_1$ satisfy the constraint equations, and if $\Ups_0(g_{b_0}+h,g_{b_0})=0$ at $\Sigma_{\IVP,r_0}$ (and hence $\Ric(g_{b_0}+h)=0$ and $\Ups_{\chi_0 E^\Ups}(g_{b_0}+h,g_{b_0})=0$), then
    \[
      \Ric(\phi_\scal^*g_{b_0}+h_\scal)=0,\quad
      \Ups_{\chi_0 E^\Ups}(\phi_\scal^*g_{b_0}+h_\scal,\phi_\scal^*g_{b_0})=0.
    \]
  \end{enumerate}
\end{thm}
\begin{proof}
  Only the regularity of $h_\scal$ in $\scal$ has not yet been discussed. First of all, the initial data $h_{\scal,j}$, $j=0,1$, depend smoothly on $\scal$, as follows by direct differentiation of~\eqref{EqExBoIVPData}. Consider next the linearization of~\eqref{EqExBoIVP} in $\scal$ around $\scal_0\in\scalspace_1$; let $\scal\colon(-1,1)\to\scalspace_1$ be smooth with $\scal(0)=\scal_0$ and $\dot\scal:=\scal'(0)$. Writing $L_{g,g^0}^{\chi_0}$ for the linearization of $P^{\chi_0}(g,g^0):=\Ric(g)-\delta_{g^0,E^\cC}^*\Ups_{\chi_0 E^\Ups}(g,g^0)$ in $g$, we then find that, formally, $\dot h:=D_\scal h_\cdot(\dot\scal)=\frac{\dd}{\dd s}h_{\scal(s)}|_{s=0}$ satisfies the equation
  \[
    L_{\phi_\scal^*g_{b_0}+h_\scal,\phi_\scal^*g_{b_0}}^{\chi_0}\dot h = -\frac{\dd}{\dd s}P^{\chi_0}(\phi_{\scal(s)}^*g_{b_0}+h_{\scal_0},\phi_{\scal(s)}^*g_{b_0})\Big|_{s=0}.
  \]
  The right-hand side lies in the space~\eqref{EqExFwNMem}, and thus the linear estimates for forward solutions of $L_{g,g^0}^{\chi_0}$ recorded in~\eqref{EqEx0Tame} (see also Remark~\ref{RmkEx0BgMetric}) apply; the partial polyhomogeneity of $\dot h$ then follows from (the proof of) Theorem~\ref{ThmExPhg}. That $\dot h$ is, indeed, the derivative of $h_\scal$ in $\scal$ follows by considering finite difference quotients and taking limits. Higher regularity is proved similarly.
\end{proof}

\begin{rmk}[Trivial initial data]
\label{RmkExBoCpt}
  If $h_0$ and $h_1$ vanish on $\Sigma_{\IVP,r_0}$, then the unique solution in Theorem~\ref{ThmExBoStab} is $h_\scal=0$ on $\Omega_{\ext,r_0}$ by finite speed of propagation, so the spacetime metric is $\phi_\scal^*g_{b_0}$ simply. (But even in this case, the material developed in this section concerning the structure of the linearized gauge-fixed Einstein operator at $\scri^+$ is indispensable for our later analysis.) On all of $\Sigma_\IVP$, such data amount to compactly supported perturbations of the Kerr data $\gamma_{b_0}$ and $k_{b_0}$, as constructed in \cite{CorvinoSchoenAsymptotics,ChruscielDelayMapping,ChruscielDelaySimple}; see~\S\ref{SsExID} for more on initial data.
\end{rmk}

\subsection{Initial data for dynamical perturbations of Kerr}
\label{SsExID}

Recalling the time function $t_\IVP$ from Lemma~\ref{LemmaKMetIVP} and
\[
  \Sigma_\IVP = t_\IVP^{-1}(0) \cong X = \cl_{\ol{\R^3}}\{r\geq\bhm_0\},
\]
we describe the passage from geometric initial data (first and second fundamental forms) to Cauchy data for the gauge-fixed Einstein equation in the partially polyhomogeneous category.

\begin{prop}[Gauged Cauchy data]
\label{PropExID}
  Let $\cE_0$ be an index set with $\min\Re\cE_0>0$ and $j\cE_0\subset\cE_0$ for all $j\in\N$, and let $\ell_0>0$. Let $\gamma=\gamma_{b_0}+\tilde\gamma$ and $k=k_{b_0}+\tilde k$ where
  \begin{align*}
    \tilde\gamma &\in \Hb^{\infty,(\cE_0,\ell_0)}(\Sigma_\IVP;S^2 T^*\Sigma_\IVP), \\
    \tilde k &\in \Hb^{\infty,(\cE_0+1,\ell_0+1)}(\Sigma_\IVP;S^2 T^*\Sigma_\IVP),
  \end{align*}
  with $\tilde\gamma$ small in $\Hb^3$ (so that $\gamma$ is a Riemannian metric). Then there exist
  \begin{equation}
  \label{EqExIDhj}
    h_j \in \Hb^{\infty,(\cE_0,\ell_0)}(\Sigma_\IVP;S^2\cT^*_{\Sigma_\IVP}),\quad j=0,1,
  \end{equation}
  depending continuously on $\tilde\gamma$ and $\tilde k$ and equal to $0$ when $(\tilde\gamma,\tilde k)=(0,0)$, such that the following holds for any spacetime metric $g$ with $(g-g_{b_0})|_{\Sigma_\IVP}=h_0$ and $(r\cL_{\pa_{t_\IVP}}(g-g_{b_0}))|_{\Sigma_\IVP}=h_1$:
  \begin{enumerate}
  \item{\rm (Initial data.)} The first and second fundamental form of $\Sigma_\IVP$ with respect to $g$ are equal to $\gamma$ and $k$, respectively.
  \item{\rm (Gauge condition.)} The gauge condition $\Ups_0(g,g_{b_0})=\tr_g(\nabla^g-\nabla^{g_{b_0}})=0$ (see~\eqref{Eq1UpsDef} and \eqref{Eq1Ups0}) holds at $\Sigma_\IVP$.
  \end{enumerate}
\end{prop}
\begin{proof}
  This is very similar to \cite[Proposition~3.10]{HintzVasyKdSStability} and \cite[Theorem~4.2]{HintzVasyKdSCosm}; the only additional feature here is the need to keep track of index sets and decay as $r\to\infty$. Working in a product neighborhood $(-1,1)_{t_\IVP}\times\Sigma_\IVP$ of $\Sigma_\IVP$, we can write
  \[
    g_{b_0} = a\,\dd t_\IVP^2 + 2\,\dd t_\IVP\otimes_s b + \gamma_{b_0},\quad a\in\CI(\Sigma_\IVP),\ b\in\CI(\Sigma_\IVP;T^*\Sigma_\IVP)
  \]
  Setting $h_0:=\tilde\gamma$, we then conclude that $g_0:=g_{b_0}+h_0=a\,\dd t_\IVP^2+2\,\dd t_\IVP\otimes_s b+\gamma$ has first fundamental form $\gamma$ at $\Sigma_\IVP$.

  Write now $\nu_{b_0}$ and $\nu_0$ for the future unit normals of $\Sigma_\IVP$ for $g_{b_0}$ and $g_0$, respectively, so $\nu_0=\nu_{b_0}+\tilde\nu$ for some $\tilde\nu\in\Hb^{\infty,(\cE_0,\ell_0)}(\Sigma_\IVP;\cT_{\Sigma_\IVP})$. Our task is now to construct $h_1=r\tilde h_1$ of class~\eqref{EqExIDhj} such that for $g:=g_0+r^{-1}t_\IVP h_1=g_0+t_\IVP\tilde h_1$ (which still has unit normal $\nu_0$) we have, at $\Sigma_\IVP$,
  \[
    k(X,Y) = g(\nabla^g_X Y,\nu_0) = g_{b_0}(\nabla^{g_{b_0}}_X Y,\nu_{b_0}) + h_0(\nabla^{g_{b_0}}_X Y,\nu_{b_0}) + g_0(\nabla^{g_{b_0}}_X Y,\tilde\nu) + g_0\bigl((\nabla^g_X-\nabla^{g_{b_0}}_X)Y,\nu\bigr),
  \]
  or equivalently (since the first term on the right equals $k_{b_0}(X,Y)$)
  \begin{equation}
  \label{EqExIDk}
    g_0\bigl((\nabla_X^{g_0+t_\IVP\tilde h_1}-\nabla_X^{g_0})Y,\nu\bigr) = \tilde k(X,Y) - h_0(\nabla^{g_{b_0}}_X Y,\nu_{b_0}) - g_0(\nabla^{g_{b_0}}_X Y,\tilde\nu) - g_0\bigl((\nabla^{g_0}_X-\nabla^{g_{b_0}}_X)Y,\nu\bigr).
  \end{equation}
  In view of the b-regularity of $h_0$ (so each $x$-derivative, arising in the computation of Christoffel symbols, gains one power of $r^{-1}$) and the assumption that $\cE_0$ is nonlinearly closed, the right-hand side here is the evaluation at $(X,Y)$ of an element of $\Hb^{\infty,(\cE_0+1,\ell_0+1)}(\Sigma_\IVP;S^2 T^*\Sigma_\IVP)$. The gauge condition reads
  \begin{equation}
  \label{EqExIDGauge}
    \tr_{g_0}(\nabla^{g_0+t_\IVP\tilde h_1}-\nabla^{g_0}) = -\tr_{g_0}(\nabla^{g_0}-\nabla^{g_{b_0}});
  \end{equation}
  the right-hand side of this lies in $\Hb^{\infty,(\cE_0+1,\ell_0+1)}$.

  Let us use a frame of $\cT$ (the spacetime tangent bundle) over $\Sigma_\IVP$ with $e_0=\nu$, while $e_1,e_2,e_3$ are unit vectors for $g_0$ that are $g^0$-orthogonal to $e_0$ (and thus span $T\Sigma_\IVP$); in particular, $e_0 t_\IVP>0$ is of class $\CI+\Hb^{\infty,(\cE_0,\ell_0)}$ and bounded away from $0$, while $e_i t_\IVP=0$ for $i=1,2,3$. Christoffel symbols in this frame satisfy, at $t_\IVP=0$,
  \[
    2\bigl(\Gamma(g_0+t_\IVP\tilde h_1)_{\lambda\mu\nu} - \Gamma(g_0)_{\lambda\mu\nu}\bigr) = (t_\IVP)_{;\mu}(\tilde h_1)_{\nu\lambda} + (t_\IVP)_{;\nu}(\tilde h_1)_{\mu\lambda} - (t_\IVP)_{;\lambda}(\tilde h_1)_{\mu\nu}.
  \]
  The left-hand side of~\eqref{EqExIDk} for $(X,Y)=(e_i,e_j)$ thus equals $-\frac12(e_0 t_\IVP)(\tilde h_1)_{i j}$, which thus uniquely determines $(\tilde h_1)_{i j}\in\Hb^{\infty,(\cE_0+1,\ell_0+1)}$. Write $e^\mu$, $\mu=0,1,2,3$, for the dual frame; then the $e^\lambda$-component of~\eqref{EqExIDGauge} (with indices lowered using $g^0$) reads
  \[
    (g^0)^{0\mu} (e_0 t_\IVP) (\tilde h_1)_{\mu\lambda} - \frac12 (e_\lambda t_\IVP) (g^0)^{\mu\nu} (\tilde h_1)_{\mu\nu} = f_\lambda \in \Hb^{\infty,(\cE_0+1,\ell_0+1)}
  \]
  where $f_\lambda$ is the $e^\lambda$-component of $-\tr_{g_0}(\nabla^{g_0}-\nabla^{g_{b_0}})$. For $\lambda=j=1,2,3$, this uniquely determines $(\tilde h_1)_{0 j}\in\Hb^{\infty,(\cE_0+1,\ell_0+1)}$; and for $\lambda=0$ one obtains $-\frac12(e_0 t_\IVP)(\tilde h_1)_{0 0}=f_0+\frac12(e_0 t_\IVP)(g^0)^{i j}(\tilde h_1)_{i j}$, which thus determines $(\tilde h_1)_{0 0}\in\Hb^{\infty,(\cE_0+1,\ell_0+1)}$.
\end{proof}

The Cauchy data $h_0$ and $h_1$ produced by Proposition~\ref{PropExID} can be used in Theorem~\ref{ThmExPhg}, and thus in Theorem~\ref{ThmExBoStab}; and if $(\gamma,k)$ satisfy the constraint equations, these theorems thus produce solutions of the Einstein vacuum equation (see Remark~\ref{RmkEx0IVPEin} and Theorem~\ref{ThmExBoStab}\eqref{ItExBoStabEinstein}).

\subsubsection{Initial data at Boyer--Lindquist time \texorpdfstring{$0$}{0}}
\label{SssExIBL}

Recall that $t_\IVP$-level sets are equal to $t$-level sets near spatial infinity, where $t$ is a $\cO(\log r)$-modification of the Boyer--Lindquist time coordinate $\ft$ (see~\eqref{EqKMetIVPt}). If one instead wishes to specify initial data on a Cauchy hypersurface that is equal to a $\ft$-level set---say, $\ft^{-1}(0)$---near spatial infinity, one can easily do so, as we proceed to show for the sake of completeness. By the finite speed of propagation, it suffices to work in regions of large $r$. Denote by $\gamma_{b_0}^{\rm BL}$ and $k_{b_0}^{\rm BL}$ the first and second fundamental form, respectively, of $g_{b_0}$ at $\Sigma_\IVP^{\rm BL}:=\ft^{-1}(0)$. Let $\gamma^{\rm BL}=\gamma_{b_0}^{\rm BL}+\tilde\gamma^{\rm BL}$ and $k^{\rm BL}=k_{b_0}^{\rm BL}+\tilde k^{\rm BL}$ where
\begin{equation}
\label{EqExIDBL}
\begin{split}
  \tilde\gamma^{\rm BL} &\in \Hb^{\infty,(\cE_0,\ell_0)}(\Sigma_\IVP^{\rm BL};S^2 T^*\Sigma_\IVP^{\rm BL}), \\
  \tilde k^{\rm BL} &\in \Hb^{\infty,(\cE_0+1,\ell_0+1)}(\Sigma_\IVP^{\rm BL};S^2 T^*\Sigma_\IVP^{\rm BL}),
\end{split}
\end{equation}
are small in $\Hb^{d,(\cE_0,\ell_0)}$ and $\Hb^{d,(\cE_0+1,\ell_0+1)}$, respectively, where $d\in\N$ is a fixed large number. Suppose $\gamma^{\rm BL}$ and $k^{\rm BL}$ satisfy the constraint equations. Denote by
\[
  h_j^{\rm BL}\in\Hb^{\infty,(\cE_0,\ell_0)}(\Sigma_\IVP^{\rm BL};S^2\cT^*_{\Sigma_\IVP^{\rm BL}}),\quad j=0,1,
\]
small tensors such that $\hat g:=g_{b_0}+h_0^{\rm BL}+r^{-1}\ft h_1^{\rm BL}$ has initial data $\gamma^{\rm BL}$ and $k^{\rm BL}$ at $\Sigma_\IVP^{\rm BL}$ and satisfies the gauge condition $\Ups_0(\hat g,g_{b_0})=0$ pointwise at $\Sigma_\IVP^{\rm BL}$; such $h_j^{\rm BL}$ can be constructed by repeating the proof of Proposition~\ref{PropExID}. We can then solve the gauge-fixed Einstein equation
\begin{align*}
  \left\{
  \begin{aligned}
    \Ric(g_{b_0} + h) - \delta_{g_{b_0},E^\cC}^*\Ups_0(g_{b_0}+h,g_{b_0}) &= 0, \\
    h&=h_0^{\rm BL} && \text{on}\ \Sigma_\IVP^{\rm BL}, \\
    r\cL_{\pa_\ft}h&=h_1^{\rm BL} && \text{on}\ \Sigma_\IVP^{\rm BL}
  \end{aligned}
  \right.
\end{align*}
in a small conic neighborhood $U_\delta:=\{|\frac{\ft}{r}|<\delta,\ r>\delta^{-1}\}$ of spatial infinity. (This is an instance of the ``boost theorem'' \cite{ChristodoulouOMurchadhaBoost} and is in any case easily proved using the arguments involved in the proof of Theorem~\ref{ThmEx0}, except one now works in $[-\delta,\delta]_\tau\times[0,\delta]_\rho\times\Sph^2$ where $\tau:=\frac{\ft}{r}$.) Again, the polyhomogeneity of the initial data propagates, so
\[
  h \in \Hb^{\infty,(\cE_0,\ell_0)}(U_\delta;S^2\cT^*).
\]
Note then that for a sufficiently large value of $r_0$, the piece $\{t_\IVP=0,\ r>r_0\}$ of the Cauchy hypersurface $\Sigma_\IVP$ considered before is contained in $U_\delta$. We can therefore read off from $g=g_{b_0}+h$ the data
\begin{subequations}
\begin{equation}
\label{EqExIDBLData}
  (h_0,h_1) := \bigl(h|_{\Sigma_\IVP},\ \bigl(r\cL_{\pa_{t_\IVP}}h\bigr)\big|_{\Sigma_\IVP}\bigr),
\end{equation}
or the geometric initial data $(\gamma,k)=(\gamma_{b_0}+\tilde\gamma,k_{b_0}+\tilde k)$ of $\Sigma_\IVP$ with respect to $g$. The only subtlety is the following: $\Sigma_\IVP$ is \emph{not} a smooth submanifold of $[-\delta,\delta]_\tau\times[0,\delta]_\rho\times\Sph^2$ due to the logarithmic discrepancy of $\ft$ and $t_\IVP$. Thus, while the $r$-decay order $\ell_0$ is unchanged when restricting to the hypersurface $\Sigma_\IVP$, the index sets of $h_0$ and $h_1$ may attain additional factors of $\log r$ relative to $h_0^{\rm BL}$ and $h_1^{\rm BL}$; this was already noted in \cite[(2.41)]{HintzVasyMink4}, so
\begin{equation}
\label{EqExIDBLLog}
  h_j \in \Hb^{\infty,(\cE_0+\cE_{\rm log},\ell_0)}(\Sigma_\IVP\cap\{r>r_0\};S^2\cT_{\Sigma_\IVP}^*),\quad \cE_{\rm log}:=\{ (j,k) \colon j,k\in\N_0,\ k\leq j \}.
\end{equation}
\end{subequations}
The proof of this membership relies on the following observation. Consider a term $a(\tau)\rho^z(\log\rho)^k$, $a\in\CI(\R_\tau)$, in the polyhomogeneous expansion of a metric coefficient of $h$ at $\rho=r^{-1}=0$. Passing to the smooth structure on $M$ amounts to passing from $\tau=\frac{\ft}{r}$ to $\tau_\IVP:=\frac{t_\IVP}{r}$; since adding to $t_\IVP$ a smooth function of $r^{-1}$ changes $\tau_\IVP$ by a smooth term, we may replace $t_\IVP$ by $\ft+c\log r$ for present purposes, so $\tau_\IVP=\tau+c\frac{\log r}{r}=\tau-c\rho\log\rho$ and thus
\[
  a(\tau)\rho^z(\log\rho)^k = a(\tau_\IVP-c\rho\log\rho) \rho^z(\log\rho)^k.
\]
Taylor expanding $a$ around $\tau_\IVP$ shows that this is polyhomogeneous on $\R_{\tau_\IVP}\times[0,\infty)_\rho$ with index set $(z,k)+\cE_{\rm log}$; this gives~\eqref{EqExIDBLLog}.

\begin{rmk}[Logarithms]
\label{RmkExIDLog}
  When $\cE_0=(2,0)$, so $\gamma^{\rm BL}$ and $r k^{\rm BL}$ are smooth in $r^{-1}$ (up to order $r^{-\ell_0}$ remainders), the Cauchy data~\eqref{EqExIDBLLog} induced at $\Sigma_\IVP$ have index set $(2,0)+\cE_{\rm log}=(2,0)\cup(3,1)\cup(4,2)\cup\cdots$, so are in general only log-smooth. This is one reason why in the analysis in the forward causal cone in the bulk of the paper, we do allow for logarithmic terms in partial asymptotic expansions.
\end{rmk}

\begin{example}[Initial data with strong decay]
\label{ExIDPhg}
  There are several results in the literature that produce initial data with arbitrary decay towards Kerr data.
  \begin{enumerate}
  \item{\rm (Exactly Kerr near spatial infinity.)} Gluing techniques introduced by Corvino and Schoen \cite{CorvinoScalar,CorvinoSchoenAsymptotics} and further developed and used by Chru\'sciel and Delay \cite{ChruscielDelayMapping,ChruscielDelaySimple} allow one to take any asymptotically flat data set and glue it across a large annulus to the initial data of a suitable Kerr metric $g_{b_0}$. The resulting initial data are thus \emph{equal} to $(\gamma_{b_0}^{\rm BL},k_{b_0}^{\rm BL})$ outside of a compact set; and the initial data induced on $\Sigma_\IVP$ are correspondingly equal to $(\gamma_{b_0},k_{b_0})$ outside of a compact set. Such data satisfy the asymptotic conditions of Proposition~\ref{PropExID} for $\cE_0=\emptyset$ and all $\ell_0$.
  \item{\rm (Kerr plus non-trivial polynomial tails.)} Using a variant of the conformal method, Fang--Szeftel--Touati \cite{FangSzeftelTouatiBHData} construct initial data that are close to Kerr data in the weighted norms we are using, and with any desired decay rate (but which typically do not decay faster than that rate to Kerr data). The differences with exactly Kerr data thus satisfy~\eqref{EqExIDBL} for $\cE_0=\emptyset$ and any desired (finite) value of $\ell_0$. Another flexible method for the construction of initial data sets with arbitrary decay was introduced by Chen--Klainerman \cite{ChenKlainermanConstraints}.
  \item\label{ItIDPhg3}{\rm (Kerr plus partially polyhomogeneous tails.)} There do not seem to exist explicit statements in the literature regarding the construction of initial data close to Kerr which feature non-trivial polyhomogeneous expansions. However, in any construction based on the conformal method---say in the context of \cite{FangSzeftelTouatiBHData}---we can make the following observation. Suppose one takes the seed data to be partially polyhomogeneous, i.e., $\widecheck{g}$ and $\widecheck{\pi}$ in \cite[(1.5)]{FangSzeftelTouatiBHData} with $q=1$ and $\delta\in(0,1)$ fixed, and with $(\widecheck{g},r\widecheck{\pi})$ being polyhomogeneous with index set $\cE_0$ (which must then satisfy $\min\Re\cE_0>1+\delta$) up to a conormal remainder term with any desired decay rate $\ell_0>1+\delta$. Then one notes that the inversion of the relevant elliptic (Laplacian-type) operators on $\R^3$ (i.e., $D\Phi[g_{m,a},\pi_{m,a}]$ in \cite[(1.6)]{FangSzeftelTouatiBHData}) preserves polyhomogeneity (and non-linear terms act on polyhomogeneous expansions by producing additional lower-order expansion terms); this is due to the fact that these operators are weighted b-differential operators near $\pa\ol{\R^3}$, and thus general results on asymptotic expansions of their solutions apply (such as Lemma~\ref{LemmaTMSolPhg} and \cite[Proposition~5.61]{MelroseAPS}). One can then close the Picard iteration on partially polyhomogeneous spaces by making purely notational modifications to the arguments in \cite{FangSzeftelTouatiBHData}.
  \end{enumerate}
\end{example}

\section{Gauge potential wave operator}
\label{SWG}

We begin our study of the stability problem in the forward cone $\{t_*>1\}$ by describing the (spectral) properties of the gauge potential wave operator
\[
  \Box_{g_b,E^\Ups}^\Ups := 2\delta_{g_b,E^\Ups}\sfG_{g_b}\delta_{g_b}^*.
\]
introduced in~\eqref{Eq1BoxUps}. Here, $b=(\bhm,\bha)$ is close to the subextremal parameter set $b_0=(\bhm_0,\bha_0)$; and we write $a:=|\bha|<\bhm$. We recall $\delta_{g_b,E^\Ups}=\delta_{g_b}+E^\Ups_{g_b}$ from Definition~\ref{Def1Gauge}, with $E^\Ups_{g_b}=E^\Ups_{g_b,\infty}+E^\Ups_{g_b,\cH^+}$ being the sum of two bundle maps which are supported near $r=\infty$ and near $r=r_b^+=\bhm+\sqrt{\bhm^2-|\bha|^2}$, respectively; \emph{we demand throughout that}
\begin{equation}
\label{EqWGEUpsSupp}
  \text{\parbox{0.8\textwidth}{\it\centering $E^\Ups_{g_b}=0$ near the projection of the trapped set of $g_b$.}}
\end{equation}
(The trapped set lies over $r^{-1}(I)$, for some fixed compact interval $I\subset(r_{b_0}^+,\infty)$, for all $b$ near $b_0$.) It is convenient to write $\Box_{g_b,E^\Ups}^\Ups$ as
\begin{equation}
\label{EqWGOp}
  \Box_{g_b,E^\Ups}^\Ups = \Box_{g_b} + 2 E^\Ups_{g_b}\sfG_{g_b}\delta_{g_b}^*,
\end{equation}
where $\Box_{g_b}$ is the (Hodge or tensor, both being the same since $\Ric(g_b)=0$) 1-form wave operator.

In~\S\ref{SsWGTr}, we study the skew-adjoint part of $\Box_{g_b,E^\Ups}^\Ups$ at the trapped set. In~\S\ref{SsWGInd}, we compute the indicial roots of the zero energy operator $\wh{\Box_{g_b,E^\Ups}^\Ups}(0)$. In~\S\ref{SsWGtf}, we recall and study the \emph{transition face normal operator} $N_\tface(\Box_{g_b,E^\Ups}^\Ups,\hat\sigma)$, $\hat\sigma\in e^{i[0,\pi]}$, governing the low-energy behavior. In~\S\ref{SsWGMode}, we prove (largely following \cite[\S\S{3.2}--{3.3}]{HintzGlueLocIII}) that for a suitable choice of $E^\Ups$, mode stability (including at zero energy) holds for $\Box_{g_b,E^\Ups}^\Ups$; this concerns the spectral family of $\Box_{g_b,E^\Ups}^\Ups$, which is defined by
\begin{equation}
\label{EqWGBoxSpecFam}
  \wh{\Box_{g_b,E^\Ups}^\Ups}(\sigma) := e^{i\sigma t_*} \Box_{g_b,E^\Ups}^\Ups e^{-i\sigma t_*} \in \Diff^2(X^\circ;\cT^*_X)
\end{equation}
acting on stationary 1-forms; here $t_*$ is the time function from Lemma~\ref{LemmaKMetTime}, and $X^\circ=\{r\geq\bhm_0\}\subset\R^3$ (cf.\ \eqref{EqKMfdX}). The detailed late-time asymptotics of solutions of the (linearized) gauge-fixed Einstein equation will feature a large number of pure gauge zero energy states; these are constructed in~\S\ref{SsWG0Kerr}.

We will strongly rely on notions and results from \citeAF{\S\S\ref*{STs}, \ref*{SS}, \ref*{SSp}}. First of all, recalling $M$ and $X$ from~\eqref{EqKMfdRadM} and \eqref{EqKMfdX}, we note:

\begin{lemma}[Stationary wave-type operator]
\label{LemmaWGOp}
  The operator $\Box_{g_b,E^\Ups}^\Ups$ is a \emph{stationary wave-type operator} on $(M,g_b)$ in the sense of \citeAF{Definition~\ref*{DefSSAdm}}. More precisely:
  \begin{enumerate}
  \item\label{ItWGOpStruct}{\rm (Structure.)} In the coordinates $t_*$ (from Lemma~\usref{LemmaKMetTime}) and $\rho=r^{-1}$, we have
    \begin{subequations}
    \begin{equation}
    \label{EqWGOp1}
      \Box_{g_b,E^\Ups}^\Ups = -2\pa_{t_*}\rho(\rho\pa_\rho-1-S_{E^\Ups}) + \wh{\Box_{g_b,E^\Ups}^\Ups}(0) + Q\pa_{t_*} - g^{0 0}\pa_{t_*}^2
    \end{equation}
    where $g^{0 0}=g_b^{-1}(\dd t_*,\dd t_*)\in\rho^2\CI(X)$ and
    \begin{equation}
    \label{EqWGOp2}
      \wh{\Box_{g_b,E^\Ups}^\Ups}(0) \in \rho^2\Diffb^2(X;\cT^*_X),\quad
      S_{E^\Ups}\in\CI(X;\End(\cT^*_X)),\quad
      Q\in\rho^3\Diffb^1(X;\cT^*_X).
    \end{equation}
    \end{subequations}
  \item\label{ItWGOpMink}{\rm (Minkowskian large-scale asymptotics.)} Use underbars for operators defined relative to the Minkowski metric $\ubar g$ from Definition~\usref{DefKMetRef}. In $\{r>0\}$, define
    \begin{equation}
    \label{EqWGOpMinkE}
      \ubar E^\Ups \colon h \mapsto -\gamma^\Ups\bigl( 2\iota_{\ubar g^{-1}(\ubar\cd^\Ups)}h - (1-e^\Ups)\ubar\cd^\Ups\ul\tr\,h\bigr),\quad \ubar\cd^\Ups:=r^{-1}\,\dd t,\ t=t_*+r,
    \end{equation}
    as well as $\ubar\delta_{\ubar E^\Ups}:=\ubar\delta+\ubar E^\Ups$ and
    \[
      \ubar\Box_{\ubar E^\Ups}^\Ups := 2\ubar\delta_{\ubar E^\Ups}\ul\sfG\,\ubar\delta^*.
    \]
    Upon writing (analogously to~\eqref{EqWGOp1}--\eqref{EqWGOp2}, with the analogues of $Q$ and $g^{0 0}$ vanishing)
    \begin{equation}
    \label{EqWGBoxMink}
      \ubar\Box_{\ubar E^\Ups}^\Ups = -2\pa_{t_*}\rho(\rho\pa_\rho-1-\ubar S_{\ubar E^\Ups}) + \wh{\ubar\Box_{\ubar E^\Ups}^\Ups}(0)
    \end{equation}
    in $r>0$, we have
    \begin{equation}
    \label{EqWGOpDiff}
      S_{E^\Ups} - \ubar S_{\ubar E^\Ups} \in \rho\CI(X;\End(\cT^*_X)),\quad
      \wh{\Box_{g_b,E^\Ups}^\Ups}(0) - \wh{\ubar\Box_{\ubar E^\Ups}^\Ups}(0) \in \rho^3\Diffb^2(X;\cT^*_X).
    \end{equation}
  \end{enumerate}
\end{lemma}
\begin{proof}
  For the 1-form operators $\Box_{g_b}$ and $\ubar\Box$, this is the content of \citeAF{Example~\ref*{ExSSAdmBox}}; in this case, $S_{E^\Ups}\in\rho\CI$ and $\ubar S_{\ubar E^\Ups}=0$.\footnote{We caution the reader that $\ubar S$ in the reference has a different meaning than in the present paper. (In the reference, it is the bottom of the spectrum of $S$.)} One can prove~\eqref{EqWGOpDiff} in this case by using~\eqref{EqKMetDiff} and the consequence $\Gamma(g_b)_{\mu\nu}^\lambda-\Gamma(\ubar g)_{\mu\nu}^\lambda\in\rho^2\CI(X)$ for the ($t_*$-independent) Christoffel symbols in the coordinates $t_*\in\R$ and $x\in\R^3$. We record an invariant consequence of this, following \cite{HintzNonstat2}: recall from Definition~\ref{DefT3b} that $\Difftb^m(M_0)$ (with $M_0=[\ol{\R^4};\fk^-,\fk^+]\cap\{r\geq\bhm_0\}$ from Definition~\ref{DefKMfdMin}) is the space of $m$-th order differential operators constructed from $r\pa_{t_*}=\rho^{-1}\pa_{t_*}$, $r\pa_r=-\rho\pa_\rho$, and $\pa_\omega$ with coefficients of class $\CI(M_0)$. Then
  \begin{equation}
  \label{EqWGOpNabla}
    \nabla^{g_b} \in \rho\Difftb^1(M_0;\cT^{p,q},\cT^{p,q+1}),\quad
    \nabla^{g_b}-\ubar\nabla \in \rho^2\CI(M_0;\Hom(\cT^{p,q},\cT^{p,q+1})).
  \end{equation}
  From the membership $\Box_{g_b}\in\rho^2\Difftb^2(M_0;\cT^{p,q})$ and principal symbol considerations (cf.\ the dual metric in~\eqref{EqKMetRefSchw} for $\bhm=0$), one can then read off~\eqref{EqWGOp1}--\eqref{EqWGOp2}.

  Smooth and compactly supported perturbations of $\Box_{g_b}$ only affect the terms~\eqref{EqWGOp2} in compact subsets of $X^\circ$, and hence they do not affect the validity of~\eqref{EqWGOpDiff}. Note next that $E^\Ups_{g_b,\infty}\in\rho\CI(X;\Hom(S^2\cT^*_X,\cT^*_X))$, which follows from Definition~\ref{Def1Gauge};\footnote{This uses that we use a 1-form $\cd^\Ups\in\rho\CI(X;\cT^*_X)$ to define the gauge modifications in Definition~\ref{Def1Gauge}.} and this equals $E^\Ups_{\ubar g,\infty}$ modulo $\rho^2\CI$ in view of Proposition~\ref{PropKMetCpt}, and this further equals $\ubar E^\Ups$ near $\pa X$. Since $\delta_{g_b}\in\rho\Difftb^1(M_0;S^2\cT^*,\cT^*)$ (by~\eqref{EqWGOpNabla}) and $\sfG_{g_b}\in\CI(M_0;\End(S^2\cT^*))$, which are moreover equal to their Minkowskian counterparts to leading order, we obtain
  \[
    \delta_{g_b}^*\sfG_{g_b}E^\Ups_{g_b} \in \rho^2\Difftb^1(M_0;\cT^*),\quad
    \delta_{g_b}^*\sfG_{g_b}E^\Ups_{g_b}-\ubar\delta^*\ul\sfG\ubar E^\Ups \in \rho^3\Difftb^1(M_0;\cT^*);
  \]
  these operators are $t_*$-translation-invariant. It then remains to note that the first operator here is the sum of terms of the schematic form $\CI(X;\End(\cT^*_X))\rho\pa_{t_*}$ (contributing to $S_{E^\Ups}$) as well as $\CI\rho^2\,\rho\pa_\rho$ and $\CI\rho^2\,\pa_\omega$ (contributing to $\wh{\Box_{g_b,E^\Ups}^\Ups}(0)$).
\end{proof}

In terms of the formula~\eqref{EqWGOp1}, the spectral family of $\Box_{g_b,E^\Ups}^\Ups$ takes the form
\begin{equation}
\label{EqWGOpSpecFam}
  \wh{\Box_{g_b,E^\Ups}^\Ups}(\sigma) = 2 i\sigma\rho(\rho\pa_\rho-1-S_{E^\Ups}) + \wh{\Box_{g_b,E^\Ups}^\Ups}(0) - i\sigma Q + g^{0 0}\sigma^2.
\end{equation}

\begin{rmk}[Consequences for the spectral family]
\label{RmkWGOpCons}
  Lemma~\ref{LemmaWGOp}\eqref{ItWGOpStruct} and the fact that the principal symbol of $\Box_{g_b,E^\Ups}^\Ups$ is given by the dual metric function of the subextremal Kerr metric $g_b$ imply that the general-purpose results in \citeAF{\S\ref*{SSp}} apply to its spectral family (with the high-energy estimates requiring subprincipal symbol information at the trapped set, which is discussed in~\S\ref{SsWGTr} below); we describe this in more detail in subsequent sections.
\end{rmk}

Next, we compute $\ubar\Box_{\ubar E^\Ups}^\Ups$ explicitly:

\begin{lemma}[Minkowskian operators]
\label{LemmaWGOpMink}
  In the bundle splittings~\eqref{EqTYMink01Split}, we have
  \begin{equation}
  \label{EqWGOpMinkE2}
    \ubar\delta_{\ubar E^\Ups} - \ubar\delta = \ubar E^\Ups = \gamma^\Ups\rho\begin{pmatrix} 2 & 2 e^\Ups & 0 & 0 & 0 & \frac12(1-e^\Ups)\sltr \\ 0 & 2 e^\Ups & 0 & 2 & 0 & \frac12(1-e^\Ups)\sltr \\ 0 & 0 & 2 & 0 & 2 & 0 \end{pmatrix}.
  \end{equation}
  Moreover, in the expression~\eqref{EqWGBoxMink}, we have
  \begin{equation}
  \label{EqWGOpMinkBox}
  \begin{split}
    \ubar S_{\ubar E^\Ups} &= \gamma^\Ups\begin{pmatrix} 1-e^\Ups & 0 & 0 \\ 1-e^\Ups & 2 & 0 \\ 0 & 0 & 1 \end{pmatrix}, \\
    \wh{\ubar\Box_{\ubar E^\Ups}^\Ups}(0) &= \wh{\ubar\Box}(0) + \gamma^\Ups\rho^2\begin{pmatrix} -(1+e^\Ups)\rho\pa_\rho+2 e^\Ups & -(1-e^\Ups)\rho\pa_\rho-2 e^\Ups & -e^\Ups\sldelta \\ (1-e^\Ups)\rho\pa_\rho+2 e^\Ups & (1+e^\Ups)\rho\pa_\rho-2 e^\Ups & -e^\Ups\sldelta \\ 2\sld & 2\sld & 0 \end{pmatrix}.
  \end{split}
  \end{equation}
\end{lemma}
\begin{proof}
  Direct computation using~\eqref{EqTYMinkMet}--\eqref{EqTYMinkTrRev} and Lemma~\ref{LemmaTYMinkOp}; note that $\ubar\cd^\Ups=\rho(\frac12,\frac12,0)$ and $\ubar g^{-1}(\ubar\cd^\Ups)=\rho(-1,-1,0)$, and $\ubar\Box_{\ubar E^\Ups}^\Ups=\ubar\Box+2\ubar E^\Ups\ul\sfG\,\ubar\delta^*$. The expression~\eqref{EqWGOpMinkE2} matches~\eqref{EqExOpLinEUps}.
\end{proof}

\subsection{Strong trapping admissibility}
\label{SsWGTr}

Recall the domain of outer communications $\cM_b=\cM_{\bhm,a}$ from~\eqref{EqKMetBLMfd}. For a principally scalar differential operator $A\in\Diff^m(\cM_b;\cE)$ acting on sections of a vector bundle $\cE\to\cM_b$, we recall from \cite{HintzPsdoInner,DenckerPolarization} that in a local trivialization of $\cE$, its \emph{subprincipal operator} is given by
\begin{equation}
\label{EqWGSsub}
  S_\sub(A) := -i H_a + \upsigma_\sub(A),
\end{equation}
where $a:=\upsigma^m(A)$ is the principal symbol of $A$ and $\upsigma_\sub(A)$ is the matrix of subprincipal symbols of $A$ relative to a choice of trivialization of the 1-density bundle, in the present section given by the metric volume density $|\dd g_b|$ (cf.\ \cite[\S{5.2}]{DuistermaatHormanderFIO2}); this operator does not depend on the choice of trivialization and defines an element of $\Diff^1(T^*\cM_b\setminus o;\pi^*\cE)$ where $\pi\colon T^*\cM_b\setminus o\to\cM_b$ is the base projection. When $A=\Box_g^{(p,q)}$ is the tensor wave operator for a Lorentzian metric $g$, acting on sections of the $(p,q)$-tensor bundle $T^{p,q}$, then \cite[Proposition~4.1]{HintzPsdoInner} shows that
\begin{equation}
\label{EqWGTrSub}
  S_\sub(\Box^{(p,q)}) = -i\nabla_g^{\pi^*T^{p,q}}
\end{equation}
where $\nabla_g^{\pi^*T^{p,q}}$ denotes the pullback connection. We equip $T^*\cM_b$ with its symplectic volume form. Our goal is to prove:

\begin{lemma}[Strong trapping admissibility]
\label{LemmaWGTr}
  The operator $\Box_{g_b,E^\Ups}^\Ups$ is \emph{strongly trapping admissible} in the sense of \citeAF{Definition~\ref*{DefSSTrapAdm}}. That is, for all $\eps>0$ there exists a time-translation-invariant positive definite fiber inner product on $\pi^*(\cT^*)$, homogeneous of degree $0$ with respect to fiber-dilations in $T^*\cM_b\setminus o$, such that
  \begin{equation}
  \label{EqWGTr}
    \sigma^{-1}\frac{1}{2 i}\bigl(S_\sub(\Box_{g_b,E^\Ups}^\Ups)-S_\sub(\Box_{g_b,E^\Ups}^\Ups)^*\bigr) < \eps\quad\text{at}\ \Gamma_0,
  \end{equation}
  where $\sigma=-\la\cdot,\pa_\ft\ra$ is the (negative) time momentum in Boyer--Lindquist coordinates~\eqref{EqKMetBLCoord}, and the trapped set $\Gamma_0\subset T^*\cM_b^\circ$ (contained in the set of future null covectors) is defined in \citeAF{Definition~\ref*{DefTs3bOTrap0}}.
\end{lemma}

For the proof, we need the following two pieces of information about $\Gamma_0$. First, its projection $\pi(\Gamma_0)$ is the product of $\R_\ft$ with a \emph{compact} subset of $\{r_b^+<r<\infty\}$. Second, the \emph{Carter constant}
\[
  \sC := \eta_\theta^2 + \frac{1}{\sin^2\theta} (-a\sin^2\theta\,\sigma + \eta_\varphi )^2
\]
does not vanish on $\Gamma_0$; here we use coordinates on $T^*\cM_b^\circ$ defined by writing covectors in the form $-\sigma\,\dd\ft+\xi\,\dd r+\eta_\theta\,\dd\theta+\eta_\varphi\,\dd\varphi$. (See~\citeAF{equation~(\ref*{EqTs3bOMetHam})}.) We recall from \cite{CarterHamiltonJacobiEinstein} and \citeAF{\S\ref*{SssTs3bO}} that this quantity is conserved along the Hamiltonian flow of $G_b$ in the characteristic set $G_b^{-1}(0)$, where $G_b(\zeta):=g_b^{-1}(\zeta,\zeta)$ is the dual metric function. Given~\eqref{EqWGEUpsSupp} and~\eqref{EqWGOp}, we only need to prove the strong trapping admissibility of the 1-form wave operator $\Box_{g_b}$.

The proof of Lemma~\ref{LemmaWGTr} will be based on the computation of $S_\sub(\Box_{g_b})$ in a special frame of $\pi^*T^*\cM_b$, following \cite{MarckParallelNull} and \cite[\S{3.2.3}]{HintzGlueLocII}. To describe this, we first define the tetrad
\begin{equation}
\label{EqWGTrTetrad}
\begin{alignedat}{2}
  \omega^{(0)} &:= \frac{\sqrt{\mu_b}}{\varrho_a}(\dd\ft-a\,\sin^2\theta\,\dd\varphi), &\qquad \omega^{(1)} &:= \frac{\varrho_a}{\sqrt{\mu_b}}\dd r, \\
  \omega^{(2)} &:= \varrho_a\,\dd\theta, &\qquad \omega^{(3)} &:= \varrho_a^{-1}\sin\theta\,(a\,\dd\ft-(r^2+a^2)\,\dd\varphi);
\end{alignedat}
\end{equation}
write $\omega_{(\mu)}$ for the dual tetrad. (One can then check that $\sC=\varrho_a^2(\omega_{(2)}^2+\omega_{(3)}^2)$.) At a null covector $(z,\zeta)\in T^*\cM_b\setminus o$, with $z=(\ft,r,\theta,\varphi)$, we then define a frame $e^\mu$ of $(\pi^*T^*\cM_b)_{(z,\zeta)}=T^*_z\cM_b$ by
\begin{alignat*}{2}
  e^0 &:= (\omega_{(0)},\omega_{(1)},\omega_{(2)},\omega_{(3)}), &\qquad e^1 &:= (r\omega_{(1)},r\omega_{(0)},-a\cos\theta\,\omega_{(3)}, a\cos\theta\,\omega_{(2)}), \\
  e^2 &:= \frac{\varrho_a^2}{2}(\omega_{(0)},\omega_{(1)},-\omega_{(2)},-\omega_{(3)}), &\qquad e^3 &:= (a\cos\theta\,\omega_{(1)}, a\cos\theta\,\omega_{(0)}, r\omega_{(3)}, -r\omega_{(2)});
\end{alignat*}
that is, these are the coefficients in the tetrad~\eqref{EqWGTrTetrad}. (In particular, $e^0=\zeta$.) From \cite[Lemma~3.20]{HintzGlueLocII}, we then recall that $e^\mu$ is a smooth stationary section of $\pi^*T^*\cM_b$ which forms a quasi-orthonormal frame of $\pi^*T^*\cM_b$ at all null covectors $\zeta\in T^*_z\cM_b$ for which $\sC\neq 0$; in fact,
\begin{subequations}
\begin{equation}
\label{EqWGTrMarckInner}
  \bigl(g_b^{-1}(e^\mu,e^\nu)\bigr)_{\mu,\nu=0,\ldots,3} = \sC \begin{pmatrix} 0 & 0 & -1 & 0 \\ 0 & 1 & 0 & 0 \\ -1 & 0 & 0 & 0 \\ 0 & 0 & 0 & 1 \end{pmatrix}.
\end{equation}
Moreover, we recall from \cite[Proposition~3.21]{HintzGlueLocII} that in the smooth bundle splitting
\begin{equation}
\label{EqWGTrMarckSplit}
  \la\sfe^0\ra\oplus\la\sfe^1\ra\oplus\la\sfe^2\ra\oplus\la\sfe^3\ra,\quad \sfe^\mu:=\frac{1}{\sqrt\sC}e^\mu,
\end{equation}
of $\pi^*T^*\cM_b$ over the set $\{\sC>0\}\subset T^*\cM_b\setminus o$ (which contains $\Gamma_0$)---with each $\sfe^\mu$ being homogeneous of degree $0$ with respect to dilations in $T^*\cM_b\setminus o$---we have
\begin{equation}
\label{EqWGTrMarckNabla}
  \nabla^{\pi^*T^*\cM_b}_{H_{G_b}} = H_{G_b} + 2\sigma\begin{pmatrix} 0 & 1 & 0 & 0 \\ 0 & 0 & 1 & 0 \\ 0 & 0 & 0 & 0 \\ 0 & 0 & 0 & 0 \end{pmatrix}.
\end{equation}
\end{subequations}

\begin{proof}[Proof of Lemma~\usref{LemmaWGTr}]
  In view of~\eqref{EqWGTrSub} and \eqref{EqWGTrMarckNabla}, it suffices to observe that with respect to the inner product $\diag(1,\eta^{-1},\eta^{-2},\eta^{-3})$ in the splitting~\eqref{EqWGTrMarckSplit}, where $\eta\in(0,1)$, the matrix in~\eqref{EqWGTrMarckNabla} has operator norm of size $\cO(\eta)$, and thus also real part of size $\cO(\eta)$. Choosing $\eta$ to be a small multiple of $\eps$ yields~\eqref{EqWGTr}.
\end{proof}

Lemma~\ref{LemmaWGTr} provides the necessary ingredient for the validity of high-energy estimates for the spectral family $\wh{\Box_{g_b,E^\Ups}^\Ups}(\sigma)$, i.e., when $|\sigma|\to\infty$, $\Im\sigma\geq 0$, as stated in \citeAF{Theorem~\ref*{ThmSpHi}}. In particular:

\begin{prop}[Mode stability at high energies]
\label{PropWGTrHi}
  There exists a constant $C>0$ such that when the modification parameters $e^\Ups,\gamma^\Ups,\gamma^\Ups_{\cH^+}$ in Definition~\usref{Def1Gauge} vary in some fixed compact set (with the 1-forms $\cd^\Ups$ and $\cd^\Ups_{\cH^+}$ held fixed), the operator $\wh{\Box_{g_b,E^\Ups}^\Ups}(\sigma)$ is invertible, as a map between the spaces in \citeAF{equation~(\ref*{EqSpBMap})}, when $|\sigma|\geq C$ and $\Im\sigma\geq 0$.
\end{prop}

\subsection{Indicial roots}
\label{SsWGInd}

By Lemma~\ref{LemmaWGOp}, the normal operator of $\wh{\Box_{g_b,E^\Ups}^\Ups}(0)$ at $\pa X$ is equal to the Minkowskian operator $\wh{\ubar\Box_{\ubar E^\Ups}^\Ups}(0)$ (which is homogeneous with respect to dilations in $\rho$, and hence equal to its normal operator).

\begin{lemma}[Indicial roots]
\label{LemmaWGInd}
  All indicial roots of $\wh{\Box_{g_b,E^\Ups}^\Ups}(0)$ (or equivalently: of $\wh{\ubar\Box_{\ubar E^\Ups}^\Ups}(0)$) are simple. Using the terminology of Definition~\usref{DefTYIndRoot}, they are as follows:
  \begin{align*}
    \text{$\rms 0$ roots:}\quad & {-}\lambda^\Ups_{\rms 0,1},\ 0,\ \lambda^\Ups_{\rms 0,1},\ 2, \\
    \text{$\rms l$ roots, $l\geq 1$:}\quad & {-}\lambda^\Ups_{\rms l,l+1},\ -\lambda^\Ups_{\rms l,l},\ -l+1,\ \lambda^\Ups_{\rms l,l},\ \lambda^\Ups_{\rms l,l+1},\ l+2, \\
    \text{$\rmv l$ roots, $l\geq 1$:}\quad & {-}l,\ l+1,
  \end{align*}
  where
  \begin{align*}
    \lambda^\Ups_{\rms l,l} &= \biggl( l(l+1)+\tfrac12+2 e^\Ups(\gamma^\Ups)^2 - \sqrt{l(l+1)+\bigl(\tfrac12+2 e^\Ups(\gamma^\Ups)^2\bigr)^2} \biggr)^{\frac12}, \\
    \lambda^\Ups_{\rms l,l+1} &= \biggl( l(l+1)+\tfrac12+2 e^\Ups(\gamma^\Ups)^2 + \sqrt{l(l+1)+\bigl(\tfrac12+2 e^\Ups(\gamma^\Ups)^2\bigr)^2} \biggr)^{\frac12}.
  \end{align*}
  In particular, $\lambda^\Ups_{\rms 0,1} = \sqrt{1+4 e^\Ups(\gamma^\Ups)^2}$. The space of $\rms 0$, resp.\ $\rms 1$ indicial solutions for the root $0$ is spanned by $\dd t$, resp.\ $\dd(r\scal)$, $\scal\in\scalspace_1$.
\end{lemma}
\begin{proof}
  We use the explicit expressions~\eqref{EqWGBoxMink} and~\eqref{EqWGOpMinkBox} and those from Lemma~\ref{LemmaTYMinkOp}; and we use the description~\eqref{EqTYSplit1} of forms of pure type. Acting on forms of pure type, the indicial family of $\Box_{g_b,E^\Ups}^\Ups$ is given by
  \begin{align*}
    N_{\rms 0}(\Box_{g_b,E^\Ups}^\Ups,\lambda) &= -\lambda^2+\lambda+\begin{pmatrix} 1 & -1 \\ -1 & 1 \end{pmatrix} + \gamma^\Ups\begin{pmatrix} -(1+e^\Ups)\lambda+2 e^\Ups & -(1-e^\Ups)\lambda-2 e^\Ups \\ (1-e^\Ups)\lambda+2 e^\Ups & (1+e^\Ups)\lambda-2 e^\Ups \end{pmatrix}, \\
    N_{\rms l}(\Box_{g_b,E^\Ups}^\Ups,\lambda) &= -\lambda^2+\lambda + \begin{pmatrix} l(l+1) & 0 & 0 \\ 0 & l(l+1) & 0 \\ 0 & 0 & l(l+1)-1 \end{pmatrix} + \begin{pmatrix} 1 & -1 & -l(l+1) \\ -1 & 1 & l(l+1) \\ -2 & 2 & 1 \end{pmatrix} \\
      &\qquad + \gamma^\Ups\begin{pmatrix} -(1+e^\Ups)\lambda+2 e^\Ups & -(1-e^\Ups)\lambda-2 e^\Ups & -e^\Ups l(l+1) \\ (1-e^\Ups)\lambda+2 e^\Ups & (1+e^\Ups)\lambda-2 e^\Ups & -e^\Ups l(l+1) \\ 2 & 2 & 0 \end{pmatrix}, \\
    N_{\rmv l}(\Box_{g_b,E^\Ups}^\Ups,\lambda) &= -\lambda^2+\lambda + l(l+1);
  \end{align*}
  here $l\geq 1$. One then finds by explicit calculation that $\det N_{\rms 0}(\Box_{g_b,E^\Ups}^\Ups,\lambda)$ factors, with the stated $\rms 0$ roots; likewise for the $\rms 1$ and $\rmv 1$ roots.

  That $\dd t$ and $\dd(r\scal)$ are indicial solutions follows from the fact that they are dual to time and spatial translations, which are Killing vector fields for the Minkowski metric.
\end{proof}

For small $e^\Ups$ and $\gamma^\Ups$, we Taylor expand $\lambda^\Ups_{\rms l,l+j}$, $j=0,$, in the parameter $z:=e^\Ups(\gamma^\Ups)^2$ and obtain
\begin{align*}
  \lambda_{\rms l,l}^\Ups &= l + \tfrac{2}{2 l+1}z - \tfrac{8 l^2+12 l+2}{(2 l+1)^3 l}z^2 + \cO((z/l)^3), \\
  \lambda_{\rms l,l+1}^\Ups &= (l+1) + \tfrac{2}{2 l+1}z + \tfrac{8 l^2+4 l-2}{(2 l+1)^3(l+1)}z^2 + \cO((z/l)^3).
\end{align*}
Thus, when $0<e^\Ups\ll 1$ and $\gamma^\Ups\neq 0$, and hence $z$ is small and positive, we have
\begin{subequations}
\begin{equation}
\label{EqWGIndOrder1}
  \lambda^\Ups_{\rms l,l+j}\in(l+j,l+j+\tfrac14),\quad l\in\N,\ j=0,1.
\end{equation}
Similarly, $\lambda^\Ups_{\rms 0,1}\in(1,\tfrac54)$. The index $l+j$ thus refers to the integer $l+j$ closest to $\lambda^\Ups_{\rms l,l+j}$. Furthermore, when $z$ is sufficiently small, then the indicial roots are ordered according to
\begin{equation}
\label{EqWGIndOrder2}
  \lambda^\Ups_{\rms l+1,l+1}<1+\lambda^\Ups_{\rms l,l}<\lambda^\Ups_{\rms l,l+1},\ \ l\geq 1;\qquad \lambda^\Ups_{\rms 1,1} < \lambda^\Ups_{\rms 0,1}.
\end{equation}
\end{subequations}
Note, in particular, that no indicial root lies in the interval $(0,\lambda_{\rms 1,1}^\Ups)$, which is thus an indicial gap for $\Box_{g_b,E^\Ups}^\Ups$. By \citeAF{Theorem~\ref*{ThmSp0}}, the operator $\wh{\Box_{g_b,E^\Ups}^\Ups}(0)$ is thus Fredholm on weighted b-Sobolev spaces $\bar H_\bop^{s,\alpha}(X)$ (defined with respect to a Euclidean density on $X$) when the shifted weight $\alpha+\frac32$ lies in this indicial gap, for suitable\footnote{Concretely, $s$ needs to be larger than a threshold quantity, denoted $\frac12+\vartheta_{\cH^+}$ in \citeAF{Definition~\ref*{DefSSOrderAdm}}, arising from a generalized radial point estimate at the conormal bundle of the event horizon. In \cite{HintzNonstat2}, the order $s$ is induced by a order function on spacetime, as described in \citeAF{\S\ref*{SsSpTs}}.} regularity orders $s$; we continue the analysis of $\wh{\Box_{g_b,E^\Ups}^\Ups}(0)$ in~\S\ref{SsWGMode}.

\begin{rmk}[Interpretation]
\label{RmkWGIndInt}
  The indicial roots below the indicial gap are the exponents of $\rho=r^{-1}$ that describe the leading-order asymptotics of stationary solutions of $\Box_{g_b,E^\Ups}^\Ups\omega=0$ (which we call \emph{large zero energy states}). These, in turn, give rise to pure gauge solutions $\delta_{g_b}^*\omega$ of the linearized gauge-fixed Einstein equation, which then appear in the late-time asymptotic expansion of general solutions of these, with the indicial root roughly (i.e., up to constant shifts) corresponding to the power of $t_*^{-1}$ at which they contribute. Lemma~\ref{LemmaWGInd} thus says that gauge modifications near $r=\infty$ (encoded by $\cd^\Ups$, $\gamma^\Ups$, and $e^\Ups$) can lead to slightly improved decay rates of some of these pure gauge contributions. Moreover, since the $\lambda^\Ups_{\rms l,l+j}$ are not integers, we can avoid some integer coincidences, which will simplify the bookkeeping of logarithmic terms arising in the construction of large zero energy states. For further discussion, see~\S\ref{SsWG0Kerr}.
\end{rmk}

\subsection{\texorpdfstring{$\tface$-}{tf-}admissibility}
\label{SsWGtf}

The transition from zero and nonzero energies is mediated at large $r$ (more precisely, for $r\sim|\sigma|^{-1}$, i.e., $\rho\sim|\sigma|$) by the \emph{transition face normal operator} of the spectral family of $\Box_{g_b,E^\Ups}^\Ups$. This can be defined as follows: fixing $\hat\sigma\in e^{i[0,\pi]}$, consider the operator $|\sigma|^{-2}\wh{\Box_{g_b,E^\Ups}^\Ups}(\sigma)$ where $\sigma=\varsigma\hat\sigma$, $\varsigma:=|\sigma|\in(0,1)$; expressed in terms of the coordinates $\hat\rho:=\frac{\rho}{\varsigma}$, $\omega\in\Sph^2$, it has a well-defined limit as $\varsigma\searrow 0$ as a differential operator on
\[
  (0,\infty)_{\hat\rho}\times\Sph^2 \subset \tface := [0,\infty]_{\hat\rho}\times\Sph^2
\]
acting on sections of $\pi_\tface^*(\cT^*_X)$ where $\pi_\tface\colon(0,\infty)\times\Sph^2\to\Sph^2=\pa X$ is the projection; this is the $\tface$-normal operator. We can describe this explicitly using~\eqref{EqWGOpSpecFam}: only the Minkowskian terms from~\eqref{EqWGBoxMink} survive in the limit, and thus
\begin{subequations}
\begin{equation}
\label{EqWGtf1}
  N_\tface(\Box_{g_b,E^\Ups}^\Ups,\hat\sigma) = 2 i\hat\sigma\hat\rho(\hat\rho\pa_{\hat\rho}-1-\ubar S_{\ubar E^\Ups}) + \hat\rho^2\wt{\ubar\Box_{\ubar E^\Ups}^\Ups}(0)(\omega,\hat\rho\pa_{\hat\rho},\pa_\omega),
\end{equation}
where we define the dilation-invariant operator $\wt{\ubar\Box_{\ubar E^\Ups}^\Ups}(0)\in\Diffb^2([0,\infty)_\rho\times\Sph^2;\pi_\tface^*(\cT^*_X))$ via
\begin{equation}
\label{EqWGtf2}
  \wh{\ubar\Box_{\ubar E^\Ups}^\Ups}(0) =: \rho^2 \wt{\ubar\Box_{\ubar E^\Ups}^\Ups}(0)(\omega,\rho\pa_\rho,\pa_\omega).
\end{equation}
\end{subequations}
(Note that $\rho\pa_\rho=\hat\rho\pa_{\hat\rho}$.) The estimates and the characterization of kernel and cokernel elements in \citeAF{Theorem~\ref*{ThmSptf}} apply to $N_\tface(\Box_{g_b,E^\Ups}^\Ups,\hat\sigma)$, with the indicial gap there being $(0,\lambda^\Ups_{\rms 1,1})$.

\begin{lemma}[$\tface$-admissibility]
\label{LemmaWGtf}
  For all sufficiently small $\gamma^\Ups$ and $e^\Ups\geq 0$ and all $\gamma^\Ups_{\cH^+}$ in Definition~\usref{Def1Gauge}, the operator $\Box_{g_b,E^\Ups}^\Ups$ is \emph{$\tface$-admissible with weight $\beta\in(0,\lambda^\Ups_{\rms 1,1})$} in the sense of \citeAF{Definition~\ref*{DefSStfAdm}}. That is, for all $\alpha\in\R$, the nullspace of $N_\tface(\Box_{g_b,E^\Ups}^\Ups,\hat\sigma)$ on
  \begin{equation}
  \label{EqWTGtfSpace}
    \cA^{\alpha,-\beta}(\tface;\pi_\tface^*(\cT^*_X)) = \rho_\sctface^\alpha\rho_\ztface^{-\beta}\cA^{0,0}(\tface;\pi_\tface^*(\cT^*_X)),\quad \rho_\sctface:=\frac{\hat\rho}{1+\hat\rho},\ \ \rho_\ztface:=\frac{1}{1+\hat\rho}
  \end{equation}
  is trivial for all $\hat\sigma\in e^{i[0,\pi]}$, and so is the nullspace of $N_\tface(\Box_{g_b,E^\Ups}^\Ups,1)^*$ (with the adjoint defined using the Euclidean density $\hat\rho^{-3}|\frac{\dd\hat\rho}{\hat\rho}\,\dd\slg|$ and any smooth fiber inner product on $\cT^*_X$) on
  \[
    e^{-2 i/\rho_\sctface}\cA^{\alpha,-1+\beta}(\tface;\pi_\tface^*(\cT^*_X)).
  \]
\end{lemma}
\begin{proof}
  We shall deduce this perturbatively from the corresponding statement for the unmodified operator $\Box_{g_b}$, which was proved in \citeAF{Proposition~\ref*{PropSptfAdm}}.

  The Fredholm estimates \citeAF{(\ref*{EqSptfEst})} apply uniformly to $N_\tface(\Box_{g_b,E^\Ups}^\Ups,\hat\sigma)$ for all $\hat\sigma\in e^{i[0,\pi]}$ and for all sufficiently small $\gamma^\Ups$ and $e^\Ups$; here one takes the scattering-b-regularity order $\sfs$ and scattering decay order $\sfr$ to be induced (in the sense explained in~\citeAF{(\ref*{SssSpOrderscbt})}) by a stationary-$\Box_{g_b}$-admissible order function (see \citeAF{Definition~\ref*{DefSSOrderAdm}}) with weights $0,0$;\footnote{This amounts to taking $\alpha_+=0$ in~\citeAF{(\ref*{EqSpOrderStart})}. The choice of $\alpha_+$ is inconsequential for the estimates for the spectral family, as only the sum of $\alpha_+$ and the scattering decay order arises in them; the threshold conditions for this \emph{sum} are stated, e.g., in \citeAF{(\ref*{EqSpBThrIn})--(\ref*{EqSpBThrOut})}.} and the decay order $q$ is fixed such that $-q+\frac32\in(0,1)$, with $(0,1)$ being an indicial gap of the unmodified operator $\Box_{g_b}$; let us take $-q+\frac32=\frac12$ for the sake of concreteness. (We remark that in the present context, the quantities denoted $\vartheta_{\pa\cK^+,{\rm in}}$, $\vartheta_{\pa\cK^+,{\rm out}}$, and $\ubar S$ in the reference vanish since the Minkowskian 1-form wave operator $\ubar\Box$ is symmetric if one uses the fiber inner product for which the standard coordinate differentials are orthonormal.) Since by \citeAF{Proposition~\ref*{PropSptfAdm}} the operator $\Box_{g_b}$ is tf-admissible for the weight $\beta=\frac12$ (and in fact for all weights $\beta\in(0,1)$), a simple functional analytic argument (see, e.g., \cite[\S{2.7}]{VasyMicroKerrdS} or \citeAF{\S\ref*{SssSpBPf}}) implies that the error terms in the estimates in \citeAF{Theorem~\ref*{ThmSptf}(\ref*{ItSptfReal})--(\ref*{ItSptfImag})} can be dropped when $e^\Ups$ and $\gamma^\Ups$ are sufficiently small; this relies on the compactness of the inclusion of the error spaces $H_{\scop,\bop}^{-N,(-N,-N)}$ in~\citeAF{(\ref*{EqSptfEst})} into the spaces of the left-hand sides. By \citeAF{Theorem~\ref*{ThmSptf}(\ref*{ItSptfKer})}, and recalling that $(0,\lambda^\Ups_{\rms 1,1})$ is an indicial gap for $\Box_{g_b,E^\Ups}^\Ups$, this proves the claim. (The condition $e^\Ups\geq 0$ is only used to ensure that $\lambda^\Ups_{\rms 1,1}$ is indeed the smallest indicial root larger than $0$.)
\end{proof}

\subsection{Mode stability}
\label{SsWGMode}

By~\cite[Theorem~5.1]{AnderssonHaefnerWhitingMode}, mode stability holds for the 1-form wave operator $\Box_{g_b}$ for all frequencies $\sigma\in\C$ with $\Im\sigma\geq 0$ and $\sigma\neq 0$, in the precise sense that $\wh{\Box_{g_b}}(\sigma)$ is invertible between appropriate b-Sobolev spaces. As shown in \cite[Theorem~3.5]{HintzGlueLocIII}, this invertibility then also holds when considering the spectral family as acting between certain weighted scattering Sobolev spaces---which are the spaces used in \citeAF{Theorem~\ref*{ThmSpB}}. That is, $\wh{\Box_{g_b,E^\Ups}^\Ups}(\sigma)$ is invertible between the spaces in~\citeAF{equation~(\ref*{EqSpBMap})} for all such $\sigma$. (One can avoid reference to \cite{HintzGlueLocIII} here by noting that \cite[Theorem~5.1]{AnderssonHaefnerWhitingMode} implies the absence of nontrivial kernel elements of $\wh{\Box_{g_b,E^\Ups}^\Ups}(\sigma)$ of class $\cA^{1-\eps}(X;\cT^*_X)$ for all $\eps\in(0,1)$, and thus by \citeAF{Theorem~\ref*{ThmSpB}(\ref*{ItSpBNull})} the operator $\wh{\Box_{g_b,E^\Ups}^\Ups}(\sigma)$ is injective as a map~\citeAF{(\ref*{EqSpBMap})}; but since it has index $0$, it is invertible.)

\begin{prop}[Mode stability away from zero]
\label{PropWGModeNon0}
  Let $c_0>0$. Then, for fixed 1-forms $\cd^\Ups$ and $\cd^\Ups_{\cH^+}$, there exists $\eps>0$ such that if $|e^\Ups|$, $|\gamma^\Ups|$, $|\gamma^\Ups_{\cH^+}|<\eps$, then mode stability\footnote{as defined above; see also the statement of Proposition~\ref{PropWGMode} below} holds for $\Box_{g_b,E^\Ups}^\Ups$ at all $\sigma\in\C$ with $\Im\sigma\geq 0$ and $|\sigma|\geq c_0$.
\end{prop}
\begin{proof}
  In view of the mode stability at high energies $|\sigma|$ recorded in Proposition~\ref{PropWGTrHi}, we only need to consider bounded frequencies $c_0\leq|\sigma|\leq C$. Since the Fredholm estimates in \citeAF{Theorem~\ref*{ThmSpB}} hold uniformly for small perturbations of $\Box_{g_b}$ (within the class of stationary wave-type operators, to which $\Box_{g_b,E^\Ups}^\Ups$ belongs by Lemma~\ref{LemmaWGOp}), it is again the same standard functional analytic argument as mentioned in the proof of Lemma~\ref{LemmaWGtf} that implies the invertibility of $\wh{\Box_{g_b,E^\Ups}^\Ups}(\sigma)$ for small $e^\Ups,\gamma^\Ups,\gamma^\Ups_{\cH^+}$.
\end{proof}

In the remainder of this section, we show that for a suitable choice of $\cd^\Ups_{\cH^+}$ and $\gamma^\Ups$, mode stability holds also for small frequencies $|\sigma|\leq c_0$. First of all, as shown in \citeAF{Corollary~\ref*{CorSpLoInd0}}, a consequence of Lemma~\ref{LemmaWGtf} is that the Fredholm index of the map
\begin{equation}
\label{EqWGMode0Map}
  \wh{\Box_{g_b,E^\Ups}^\Ups}(0) \colon \bigl\{ u\in\bar H_\bop^{s,\alpha}(X;\cT^*_X) \colon \wh{\Box_{g_b,E^\Ups}^\Ups}(0)u\in\bar H_\bop^{s-1,\alpha+2}(X;\cT^*_X) \bigr\} \to \bar H_\bop^{s-1,\alpha+2}(X;\cT^*_X)
\end{equation}
(where use a Euclidean density to define the underlying $L^2$-spaces) equals $0$ for $\alpha\in\R$ with $\alpha+\frac32\in(0,\lambda^\Ups_{\rms 1,1})$, and for all regularity orders $s$ obeying a threshold condition discussed in~\S\ref{SsWGInd}; this threshold condition depends on the subprincipal symbol of $\wh{\Box_{g_b,E^\Ups}^\Ups}(0)$ at the conormal bundle of the event horizon and thus on the modification $E^\Ups_{\cH^+}$, but as we shall take the latter modification to be small, one can work with any fixed regularity order suitable for the unmodified operator $\wh{\Box_{g_b}}(0)$.\footnote{We recall that the proof proceeds by proving the constancy of the Fredholm index of the spectral family as the spectral parameter goes from real infinity (where the spectral family is invertible by Proposition~\ref{PropWGTrHi} and thus has index $0$) down to $0$ (with uniform low-energy estimates---which strongly rely on the $\tface$-admissibility---for a certain Grushin problem giving the constancy of the index of the spectral family when passing from non-zero frequencies to zero frequency).} \footnote{This can also be deduced perturbatively from the corresponding statement for the 1-form wave operator $\Box_{g_b}$, which is noted in \cite[Theorem~5.1]{AnderssonHaefnerWhitingMode}, or deduced from a deformation argument as in the proof of \cite[Theorem~4.3]{HaefnerHintzVasyKerr}.} For the unmodified 1-form wave operator, we recall:

\begin{lemma}[Kernel and cokernel of $\Box_{g_b}$ at zero energy]
\label{LemmaWGMode0}
  In terms of the Boyer--Lindquist coordinates~\eqref{EqKMetBLCoord}, define $t_0=\ft+\int_{4\bhm}^r\frac{r^2+a^2}{\mu_b}\,\dd r$ and $\varphi_0=\varphi+\int_{4\bhm}^r\frac{a}{\mu_b}\,\dd r$.\footnote{These functions thus equal $\tilde t_*$ and $\phi$ from~\eqref{EqKMetTstar} for $r\leq 3\bhm_0$.} The kernel of $\wh{\Box_{g_b}}(0)$ acting between the spaces~\eqref{EqWGMode0Map} (with $\wh{\Box_{g_b,E^\Ups}^\Ups}(0)$ replaced by $\wh{\Box_{g_b}}(0)$) is then spanned by the (divergence-free) 1-form
  \[
    \omega_{(0)} := \frac{r}{\varrho_a^2}(\dd t_0-a\sin^2\theta\,\dd\varphi_0) + \frac{r_b^+-r}{\mu_b}\,\dd r \in r^{-1}\CI(X;\cT^*_X),
  \]
  and the cokernel is spanned by
  \[
    \omega_{(0)}^* := \delta(r-r_b^+)\,\dd r.
  \]
\end{lemma}
\begin{proof}
  This is the content of \cite[Theorem~5.1(2)]{AnderssonHaefnerWhitingMode}.
\end{proof}

Largely following \cite[Propositions~3.7 and 3.12]{HintzGlueLocIII}, we will now use the term $E^\Ups_{g_b,\cH^+}$ in Definition~\ref{Def1Gauge} to eliminate the zero energy nullspace of $\Box_{g_b,E^\Ups}^\Ups$; and we do this in such a way as to not destroy the mode stability that holds for $\Box_{g_b}$ for (small) nonzero frequencies. The main difference to \cite{HintzGlueLocIII} is that we now need to deal with the gauge modification near $r=\infty$. We need the following auxiliary result:

\begin{lemma}[Pairings]
\label{LemmaWG0Pair}
  Using the spatial volume density $|\dd g_b|_X|=\varrho_a^2\sin\theta\,|\dd r\,\dd\theta\,\dd\varphi_0|$ of the Kerr metric, and the (indefinite) fiber inner product on $\cT^*_X$ induced by $g_b$, we have
  \begin{equation}
  \label{EqWG0Pair1}
    \la [\Box_{g_b},t_*]\omega_{(0)}, \omega_{(0)}^*\ra_{L^2(X;\cT^*_X)} = 4\pi.
  \end{equation}
  Furthermore, there exists a 1-form $\cd^\Ups_{\cH^+}\in\CIc(X^\circ;\cT^*_X)$ that is independent of $b$ and has support arbitrarily close to $r=r_{b_0}^+$ such that for the map
  \begin{equation}
  \label{EqWG0EMod}
    E^\Ups_{g_b,\cH^+,1} \colon h \mapsto 2\iota_{g_b^{-1}(\cd^\Ups_{\cH^+})}h - \cd_{\cH^+}^\Ups \tr_{g_b}h
  \end{equation}
  (which is the same as~\eqref{Eq1GaugeHor} for $\gamma^\Ups_{\cH^+}=1$), we have
  \begin{equation}
  \label{EqWG0Pair2}
    \la 2 E^\Ups_{g_b,\cH^+,1}\sfG_{g_b}\delta_{g_b}^*\omega_{(0)}, \omega_{(0)}^* \ra_{L^2(X;\cT^*_X)} \in (3\pi,5\pi)
  \end{equation}
  for all $b$ near $b_0$.
\end{lemma}
\begin{proof}
  The identity~\eqref{EqWG0Pair1} was proved in \cite[Lemma~3.8]{HintzGlueLocIII} using a ``boundary pairing'' computation. We briefly recall the argument for~\eqref{EqWG0Pair2} from \cite[Proposition~3.12]{HintzGlueLocIII}: first of all, we have $E_{g_b,\cH^+,1}^\Ups\sfG_{g_b}=2\iota_{g_b^{-1}(\cd_{\cH^+}^\Ups)}$, and thus~\eqref{EqWG0Pair2} equals twice $\la\delta_{g_b}^*\omega_{(0)},2\cd_{\cH^+}^\Ups\otimes_s\omega_{(0)}^*\ra=\la\eta,\cd^\Ups_{\cH^+}\ra$ where $\eta:=\iota_{g_b^{-1}(\omega_{(0)}^*)}\delta_{g_b}^*\omega_{(0)}$. Consider the case $b=b_0$: since $\eta$ is supported at $r=r_{b_0}^+$, we can choose $\cd^\Ups_{\cH^+}$ as required provided $\eta=\delta(r-r_{b_0}^+)\iota_{\nabla r}\delta_{g_{b_0}}^*\omega_{(0)}\neq 0$; this in turn follows from an explicit computation. Fixing thus $\cd_{\cH^+}^\Ups$ so that~\eqref{EqWG0Pair2} equals $4\pi$ for $b=b_0$, it remains to note that the value of~\eqref{EqWG0Pair2} depends continuously on $b$ and thus remains close to its value $4\pi$ for $b=b_0$.
\end{proof}

\begin{prop}[Mode stability of the gauge potential wave operator]
\label{PropWGMode}
  Fix $\cd^\Ups_{\cH^+}$ as in Lemma~\usref{LemmaWG0Pair}, and fix $\cd^\Ups$ to be equal to $r^{-1}\,\dd t$ for large $r$, and with $\supp\cd^\Ups_{\cH^+}$ and $\supp\cd^\Ups$ disjoint from the trapped set. Then there exists $\eps_0>0$ such that for all $\gamma^\Ups_{\cH^+}\in(0,\eps_0]$ there exists $\eps_1>0$ such that for all $e^\Ups\in[0,\eps_1]$ and $|\gamma^\Ups|<\eps_1$, mode stability holds for the operator $\Box_{g_b,E^\Ups}^\Ups$ (given by~\eqref{Eq1BoxUps} and Definition~\usref{Def1Gauge}) for all frequencies $\sigma\in\C$, $\Im\sigma\geq 0$. More precisely, $\wh{\Box_{g_b,E^\Ups}^\Ups}(\sigma)$ is invertible as a map~\citeAF{(\ref*{EqSpBMap})} (and thus has trivial nullspace on $\cA^\alpha(X;\cT^*_X)$ for all $\alpha\in\R$) when $\sigma\neq 0$ and as a map~\eqref{EqWGMode0Map} (which is the same as~\citeAF{(\ref*{EqSp0Est})}) for all $\alpha$ with $\alpha+\frac32\in(0,\lambda^\Ups_{\rms 1,1})$ when $\sigma=0$ (and thus has trivial nullspace on $\cA^\eps(X;\cT^*_X)$ for all $\eps>0$).
\end{prop}

The order of the choice of parameters reflects the proof below: we first prove mode stability for the (compactly supported) perturbation of $\Box_{g_b}$ encoded by $\cd_{\cH^+}^\Ups$ and $E^\Ups_{g_b,\cH^+}$; and then mode stability for the perturbation near infinity encoded by $E^\Ups_{g_b,\infty}$ holds by a perturbative argument.

\begin{proof}[Proof of Proposition~\usref{PropWGMode}]
  In view of Proposition~\ref{PropWGModeNon0}, we only need to show that there exist $c_0,\eps_0>0$ such that for $\gamma^\Ups_{\cH^+}\in(0,\eps_0]$ and all sufficiently small $e^\Ups$ and $\gamma^\Ups$, mode stability holds for $|\sigma|\leq c_0$; this requires perturbative arguments at low energies. For this purpose, we use notions from scattering-b-transition analysis, recalled in Definition~\ref{DefTscbt}, and results and notation from \cite{HintzNonstat2}.

  \pfstep{Step~1. Mode stability for compactly supported perturbations.} The proof is very similar to that of \cite[Proposition~3.7]{HintzGlueLocIII}; since there are a number of (largely notational) differences, mainly due to our usage of the hyperboloidal time function $t_*$ (as opposed to $t$), we nonetheless spell out the proof in some detail.

  Let us omit the metric $g_b$ from the notation, so in particular $\Box=\Box_{g_b}$, and further abbreviate
  \[
    \Box_\gamma := \Box^\Ups_{g_b,(\cd^\Ups,0,0;\cd^\Ups_{\cH^+},\gamma)} = 2(\delta+\gamma E_{\cH^+})\sfG\delta^* = \Box + 2\gamma E_{\cH^+}\sfG\delta^*,\quad E_{\cH^+}:=E_{g_b,\cH^+,1}^\Ups,
  \]
  where we used the notation~\eqref{EqWG0EMod}; this is thus the operator $\Box^\Ups_{g_b,E^\Ups}$ when only the modification~\eqref{Eq1GaugeInfty} near the event horizon is used. We claim that when $\eps_0>0$ is small enough, then mode stability holds for $\Box_\gamma$ when $\gamma\in(0,\eps_0]$ and $|\sigma|\leq\eps_0$. To do this, we consider the joint parameter set $(\sigma,\gamma)\neq(0,0)$ and introduce
  \[
    \lambda := |(\sigma,\gamma)|,\quad (\hat\sigma,\hat\gamma):=\lambda^{-1}(\sigma,\gamma)
  \]
  as well as the augmented operator\footnote{Since we are working in a hyperboloidal foliation, i.e., with the spectral family defined by its action on $e^{-i\sigma t_*}=e^{-i\sigma(t-r)}$, no oscillatory factor $e^{i\sigma r}$ is needed here in order to effect a leading-order cancellation in the term $\pa_\sigma\wh\Box(0)\omega_{(0)}$ below, unlike in \cite{HintzGlueLocIII}.}
  \begin{equation}
  \label{EqWGModeAug}
    \wt\Box(\lambda,\hat\sigma,\hat\gamma) := \begin{pmatrix} \wh{\Box_\gamma}(\sigma) & f(\lambda,\hat\sigma,\hat\gamma) \\ \la\cdot,f^*\ra_{L^2} & 0 \end{pmatrix},\qquad
    f(\lambda,\hat\sigma,\hat\gamma) := \wh{\Box_\gamma}(\sigma)(\lambda^{-1}\omega_{(0)}),
  \end{equation}
  where we fix $f^*\in\CIc(X^\circ;\cT^*_X)$ with $\la\omega_{(0)},f^*\ra_{L^2}=1$. Note that since $\wh\Box(0)\omega_{(0)}=0$, we have
  \[
    f(\lambda,\hat\sigma,\hat\gamma) = \hat\sigma\pa_\sigma\wh\Box(0)\omega_{(0)} + 2\hat\gamma E_{\cH^+}\sfG\wh{\delta^*}(0)\omega_{(0)} + \frac12\lambda\hat\sigma^2\pa_\sigma^2\wh\Box(0)\omega_{(0)} + 2\lambda\hat\sigma\hat\gamma E_{\cH^+}\sfG\,\pa_\sigma\wh{\delta^*}(0)\omega_{(0)}.
  \]
  Since $\pa_\sigma\wh\Box(0)\equiv 2 i\rho(\rho\pa_\rho-1)\bmod\rho^2\Diffb^1(X;\cT^*_X)$ by~\eqref{EqWGOpSpecFam} and~\eqref{EqWGOpDiff}, \eqref{EqWGOpMinkBox}, with $\rho\pa_\rho-1$ annihilating the $r^{-1}$ leading-order term of $\omega_{(0)}$, the first summand is of class $\hat\sigma\rho^3\CI(X;\cT^*_X)$.\footnote{It is here that not yet introducing the modification of $\Box$ at infinity is important: otherwise $\pa_\sigma\wh{\Box_{g_b,E^\Ups}^\Ups}(0)$ would be equal to $2 i\rho(\rho\pa_\rho-1-\ubar S_{\ubar E^\Ups})$ to leading order, with $\rho\pa_\rho-1-\ubar S_{\ubar E^\Ups}$ (with $\ubar S_{\ubar E^\Ups}$ as in~\eqref{EqWGOpMinkBox} and with $\gamma^\Ups\neq 0$) not annihilating the $r^{-1}$ leading-order term of $\omega_{(0)}$ anymore. (One would instead need to correct $\omega_{(0)}$ at $\tface\subset X_\scbtop$ by solving away the resulting $r^{-2}$ error using the inverse of $N_\tface(\Box,\hat\sigma)$.)} The second summand is of class $\hat\gamma\CIc(X^\circ;\cT^*_X)$. For the third summand, we use that $\pa_\sigma^2\wh\Box(0)\in\rho^2\CI(X)$, so this lies in $\lambda\hat\sigma^2\rho^3\CI(X;\cT^*_X)$; and the fourth summand finally lies in $\lambda\hat\sigma\hat\gamma\CIc(X^\circ;\cT^*_X)$, so
  \begin{equation}
  \label{EqWGModefMem}
    f \in (\hat\sigma\rho^3\CI+\hat\gamma\CIc) + \lambda\hat\sigma(\hat\sigma\rho^3\CI + \hat\gamma\CIc).
  \end{equation}
  This can be restricted to $\lambda=0$; since $\pa_\sigma\wh\Box(0)=-i[\Box,t_*]\ftrans(0)$, Lemma~\ref{LemmaWG0Pair} gives
  \begin{equation}
  \label{EqWGModeInner}
    \la f(0,\hat\sigma,\hat\gamma),\omega_{(0)}^*\ra_{L^2} = 4\pi(-i\hat\sigma+c_b\hat\gamma) \neq 0
  \end{equation}
  whenever $\Im\sigma\geq 0$, $\gamma\geq 0$ (and $(\sigma,\gamma)\neq(0,0)$ still); here $c_b\in(\frac34,\frac54)$.

  \pfsubstep{(1.1)}{Statement of the main estimate; consequences.} Fix a stationary-$\Box_{g_b}$-admissible order with weights $0,0$, and denote by $\sfs$ and $\sfr_\pm$ the induced scattering-b-transition regularity and scattering decay orders as in \citeAF{(\ref*{EqSpOrderscbt})}. We recall that the requirements on these orders are as follows: they are monotonically non-increasing along the future-directed null-bicharacteristic flow; $\sfs$ is larger than a threshold quantity at the conormal bundle of the event horizon; at the zero section $\cR_{\rm out}$ of ${}^\scbtop T^*_\scface X$ (in the notation introduced around~\citeAF{(\ref*{EqMUscbtTstar})}), we have $\sfr_\pm<-\frac12$, and at the incoming radial set $\cR_{\pm 1,{\rm in}}\subset{}^\scbtop T^*_\scface X$ (see \citeAF{(\ref*{EqSpOrderRad})}), we have $\sfr_\pm>-\frac12$. (We use here that $\vartheta_{\pa\cK^+,{\rm in}}=\vartheta_{\pa\cK^+,{\rm out}}=0$ for $\Box=\Box_{g_b}$, as noted already in the proof of Lemma~\ref{LemmaWGtf}.) Define the norm
  \[
    \|(u,c)\|_{\tilde H_{\scbtop,\varsigma}^{s,(r,l,0)}} := \|u\|_{H_{\scbtop,\varsigma}^{s,(r,l,0)}} + |c|,
  \]
  where we use a Euclidean density to define the norm of $u$ (cf.\ Definition~\ref{DefTscbt}\eqref{ItTscbtSob}). We claim that, for fixed $\alpha$ with $\alpha+\frac32\in(0,1)$ (the indicial gap for $\Box$),
  \begin{equation}
  \label{EqWGModeEst}
    \|(u,c)\|_{H_{\scbtop,|\sigma|}^{\sfs,(\sfr,\alpha,0)}} \leq C\| \wt\Box(\lambda,\hat\sigma,\hat\gamma)(u,c) \|_{H_{\scbtop,|\sigma|}^{\sfs-1,(\sfr+1,\alpha+2,0)}},
  \end{equation}
  for all $\hat\sigma\in\C$, $\hat\gamma\in\R$, and $\lambda\geq 0$ with $\Im\hat\sigma\geq 0$, $\hat\gamma\geq 0$, $|(\hat\sigma,\hat\gamma)|=1$, and $\lambda\leq\lambda_0$ when $\lambda_0>0$ is sufficiently small. (This is analogous to the estimate~\citeAF{(\ref*{EqSpLoEst})}, except now for an augmented operator.) The estimate~\eqref{EqWGModeEst} and the index $0$ property of $\wt\Box(\lambda,\hat\sigma,\hat\gamma)$ between the direct sum of the b-spaces in~\citeAF{(\ref*{EqSp0Map})} for $\sigma=\lambda\hat\sigma=0$ and the scattering spaces in~\citeAF{(\ref*{EqSpBMap})} for $\sigma\neq 0$ implies the invertibility of $\wt\Box(\lambda,\hat\sigma,\hat\gamma)$ between these spaces, and thus by inspection of the first row of $\wt\Box(\lambda,\hat\sigma,\hat\gamma)$ for $\lambda\neq 0$ the surjectivity (and thus invertibility) of $\wh{\Box_\gamma}(\sigma)$ between the original spaces \citeAF{(\ref*{EqSp0Map}), (\ref*{EqSpBMap})}. In particular, for every $\gamma\in(0,\frac12\lambda_0)$, this yields the mode stability for $\Box_\gamma$ for all $\sigma\in\C$ with $\Im\sigma\geq 0$ and $|\sigma|\leq\frac12\lambda_0$, as desired.

  \pfsubstep{(1.2)}{Proof of the main estimate.} As far as notation is concerned, we follow the proof of \citeAF{Theorem~\ref*{ThmSpLo}}, while the particularities of our augmented operator are handled similarly to the proof of \cite[Proposition~3.7]{HintzGlueLocIII}.\footnote{There are some minor differences: staying close to \cite{HintzNonstat2}, we first invert the zero parameter operator here and only later the $\tface$-normal operator; this is the reverse order compared to \cite{HintzGlueLocIII}. Moreover, we use an observation from \citeAF{\S\ref*{SsSpLo}} which allows us to avoid the need to start with a symbolic estimate (even if this is a simple consequence of the results in \citeAF{\S\ref*{SSp}}).}

  \pfsubstep{(1.2.1)}{Consistency.} We first show that the augmentation terms of $\wt\Box(\lambda,\hat\sigma,\hat\gamma)$ have mapping properties consistent with~\eqref{EqWGModeEst}. First, since $f^*$ is compactly supported in $X^\circ$, the operator $u\mapsto\la u,f^*\ra_{L^2}$ is uniformly bounded as a map $H_{\scbtop,|\sigma|}^{s,(r,\alpha,0)}\to\C$. Second, we need to show that $f(\lambda,\hat\sigma,\hat\gamma)$ has uniformly bounded $H_{\scbtop,|\sigma|}^{\sfs-1,(\sfr+1,\alpha+2,0)}$-norm. This is clear for the $\CIc(X^\circ;\cT^*_X)$-pieces in~\eqref{EqWGModefMem}, and follows for the $\rho^3\CI(X;\cT^*_X)$-pieces in~\eqref{EqWGModefMem} from the uniform boundedness of the inclusions
  \[
    \rho^3\CI(X;\cT^*_X) \hra \bar H_\bop^{\infty,\frac32-\eps}(X;\cT^*_X) \hra H_{\scbtop,|\sigma|}^{\infty,(\sfr+1,\frac32-\eps,0)}(X)
  \]
  for any fixed $\eps\in(0,1)$; here we use that infinite b-regularity implies infinite scattering decay away from the zero section of $\Tscbt^*X$ over $\scface$ (cf.\ \citeAF{Remark~\ref*{RmkSpBNob}}), and $\frac32-\eps>\sfr+1$ near the zero section when $\eps<1$.

  \pfsubstep{(1.2.2)}{Inversion of $\wt\Box(0,\hat\sigma,\hat\gamma)$.} Next, we use that $\wt\Box(0,\hat\sigma,\hat\gamma)$ is invertible, with
  \[
    \|(u,c)\|_{\tilde H_\bop^{\sfs,\alpha}} := \|u\|_{\bar H_\bop^{\sfs,\alpha}(X;\cT^*_X)}+|c| \leq C\|\wt\Box(0,\hat\sigma,\hat\gamma)(u,c)\|_{\tilde H_\bop^{\sfs-1,\alpha+2}}.
  \]
  This follows from the validity of a Fredholm estimate, which features an additional relatively compact error term on the right-hand side, and the injectivity of $\wt\Box(0,\hat\sigma,\hat\gamma)$ on $\bar H_\bop^{\sfs,\alpha}(X;\cT^*_X)\oplus\C$, which follows from the fact that pairing $\wh{\Box_\gamma}(0)u+c f(0,\hat\sigma,\hat\gamma)=0$ with $\omega_{(0)}^*$ implies $c=0$ by~\eqref{EqWGModeInner}, but then $u=c'\omega_{(0)}$ and $0=\la u,f^*\ra=c'$, so also $u=0$. Fix the defining function $\rho_\zface:=\frac{|\sigma|}{\rho+|\sigma|}$ of $\zface\subset X_\scbtop$. Let $\chi_\zface=\chi(\rho_\zface)$ be equal to $1$ for $\rho_\zface\in[0,\frac14]$ and $0$ for $\rho_\zface\geq\frac12$; then we have uniform bounds
  \begin{align*}
    \|(\chi_\zface u,c)\|_{\tilde H_{\scbtop,|\sigma|}^{\sfs,(\sfr,\alpha,0)}} &\sim \|(\chi_\zface u,c)\|_{\tilde H_\bop^{\sfs,\alpha}} \\
      &\lesssim \|\wt\Box(0,\hat\sigma,\hat\gamma)(\chi_\zface u,c)\|_{\bar H_\bop^{\sfs-1,\alpha+2}} \sim \|\wt\Box(0,\hat\sigma,\hat\gamma)(\chi_\zface u,c)\|_{\tilde H_{\scbtop,|\sigma|}^{\sfs-1,(\sfr+1,\alpha+2,0)}}.
  \end{align*}
  (Here we use the first norm equivalence in~\citeAF{(\ref*{EqMUscbtNormEquiv})}. Note that the scattering decay order is arbitrary since $\scface\cap\supp\chi_\zface=\emptyset$.) Replacing $\wt\Box(0,\hat\sigma,\hat\gamma)$ here by $\wt\Box(\lambda,\hat\sigma,\hat\gamma)$ creates two error terms (cf.\ the first row in~\eqref{EqWGModeAug}): one from $f(\lambda,\hat\sigma,\hat\gamma)-f(0,\hat\sigma,\hat\gamma)\in\sigma(\hat\sigma\rho^3\CI+\hat\gamma\CIc)$ (see~\eqref{EqWGModefMem}), whose norm is thus of size $|\sigma||c|$; and one from $\wh{\Box_{\lambda\hat\gamma}}(\lambda\hat\sigma)-\wh\Box(0)$, which can be written as the sum of two errors, namely, from $\wh{\Box_\gamma}(\sigma)-\wh{\Box_\gamma}(0)$ and $\wh{\Box_{\lambda\hat\gamma}}(0)-\wh\Box(0)$, both acting on $\chi_\zface u$. Since the former operator is of class $\sigma\rho\Diffb^1(X)+\sigma^2\rho^2\Diffb^0(X)$ (which vanishes to first, resp.\ second order at $\zface$, resp.\ $\tface\subset X_\scbtop$), its contribution has norm bounded by $\|\chi_\zface u\|_{H_{\scbtop,|\sigma|}^{\sfs,(\sfr,\alpha,-\eta)}}$ for $\eta=1$, and thus a fortiori for all $\eta\leq 1$. The latter operator is $2\lambda\hat\gamma E_{\cH^+}\sfG\wh{\delta^*}(0)$ and thus has compactly supported coefficients; its contribution has norm bounded by $\lambda\|\chi_\zface u\|_{H_{\scbtop,|\sigma|}^{\sfs,(\sfr,\alpha,0)}}$. We proceed to commute $\wt\Box(\lambda,\hat\sigma,\hat\gamma)$ through the cutoff $\chi_\zface$; here,
  \[
    [\wt\Box(\lambda,\hat\sigma,\hat\gamma),\chi_\zface\times\Id_\C] = \begin{pmatrix} [\wh{\Box_\gamma}(\sigma),\chi_\zface] & (1-\chi_\zface)f(\lambda,\hat\sigma,\hat\gamma) \\ \la(\chi_\zface-1)\cdot,f^*\ra_{L^2} & 0 \end{pmatrix},
  \]
  with the second row vanishing for small $|\sigma|$ since $f^*$ has compact support. The $(1,1)$-entry is supported where $\rho\sim|\sigma|$ and vanishes quadratically as $\rho\to 0$ as a b-differential (and thus sc-b-transition-differential) operator; for the $(1,2)$-entry on the other hand, we note that the $H_{\scbtop,|\sigma|}^{\sfs-1,(\sfr+1,\alpha+2,0)}$-norm of $|\sigma|^{-\delta}(1-\chi_\zface)f$ is uniformly bounded provided the inequality $\frac32-\delta>\alpha+2$ of $\tface$-decay orders holds. (Note here that $|\sigma|^{-\delta}(1-\chi_\zface)\in\cA^{0,-\delta,\infty}(X_\scbtop)$.) Altogether, we can now conclude the uniform estimate
  \begin{equation}
  \label{EqWGModeEst0}
    \|(u,c)\|_{\tilde H_{\scbtop,|\sigma|}^{\sfs,(\sfr,\alpha,0)}} \lesssim \|\wt\Box(\lambda,\hat\sigma,\hat\gamma)(u,c)\|_{\tilde H_{\scbtop,|\sigma|}^{\sfs-1,(\sfr+1,\alpha+2,0)}} + \|u\|_{H_{\scbtop,|\sigma|}^{\sfs,(\sfr,\alpha,-\eta)}} + \lambda\|u\|_{H_{\scbtop,|\sigma|}^{\sfs,(\sfr,\alpha,0)}} + |\sigma|^\delta |c|.
  \end{equation}
  For small $\lambda$ (and thus small $|\sigma|$), we can absorb the final two terms into the left-hand side.

  \pfsubstep{(1.2.3)}{Inversion of the $\tface$-normal operators.} Next, we estimate the second term on the right of~\eqref{EqWGModeEst0} using the $\tface$-admissibility of $\Box_\gamma$ (Lemma~\ref{LemmaWGtf}): this provides us with the estimate~\citeAF{(\ref*{EqSptfAdm})} (with $k=0$, and with $\frac{\sigma}{|\sigma|}$ in place of $e^{i\theta}$), with uniform constants for small $\gamma$. Using the second norm equivalence in \citeAF{(\ref*{EqMUscbtNormEquiv})}, this gives
  \begin{align*}
    \|\chi_\tface u\|_{H_{\scbtop,|\sigma|}^{\sfs,(\sfr,\alpha,-\eta)}} &\sim |\sigma|^{\frac32-\alpha} \|\chi_\tface u\|_{H_{\scop,\bop}^{\sfs,\sfr,-\alpha-\eta}(\tface)} \\
      &\lesssim |\sigma|^{\frac32-\alpha} \| N_\tface(\Box_\gamma,\tfrac{\sigma}{|\sigma|})(\chi_\tface u) \|_{H_{\scop,\bop}^{\sfs-1,\sfr+1,-\alpha-\eta-2}(\tface)} \\
      &\sim \|N_\tface(\Box_\gamma,\tfrac{\sigma}{|\sigma|})(\chi_\tface u)\|_{H_{\scbtop,|\sigma|}^{\sfs-1,(\sfr+1,\alpha,-\eta-2)}} \sim \| |\sigma|^{-2}N_\tface(\Box_\gamma,\tfrac{\sigma}{|\sigma|})(\chi_\tface u)\|_{H_{\scbtop,|\sigma|}^{\sfs-1,(\sfr+1,\alpha+2,-\eta)}}
  \end{align*}
  for all sufficiently small $\eta>0$ (namely, those for which the $\ztface$-decay order $-\alpha-\eta$ satisfies $-(-\alpha-\eta)+\frac32\in(0,1)$); here $\chi_\tface\in\CI(X_\scbtop)$ is a cutoff localizing to a neighborhood of $\tface$. The difference between $|\sigma|^{-2}N_\tface(\Box_\gamma,\frac{\sigma}{|\sigma|})\chi_\tface$ and $\wh{\Box_\gamma}(\sigma)\chi_\tface$ is of class $\rho_\scface\rho_\tface^3\Diff_\scbtop^2(X;\cT^*_X)$ (see \citeAF{(\ref*{EqSpLoNtfDiff})}), as is the commutator of $\wh{\Box_\gamma}(\sigma)$ with $\chi_\tface$, and thus we obtain
  \[
    \|u\|_{H_{\scbtop,|\sigma|}^{\sfs,(\sfr,\alpha,-\eta)}} \lesssim \|\wh{\Box_\gamma}(\sigma)u\|_{H_{\scbtop,|\sigma|}^{\sfs-1,(\sfr+1,\alpha+2,-\eta)}} + \|u\|_{H_{\scbtop,|\sigma|}^{\sfs,(\sfr,\alpha-1,-\eta)}}.
  \]
  (Note that the norm of $(1-\chi_\tface)u$ is trivially controlled by the second term on the right.) We have now improved~\eqref{EqWGModeEst0} (with the final two terms already absorbed) to
  \begin{align*}
    &\|(u,c)\|_{\tilde H_{\scbtop,|\sigma|}^{\sfs,(\sfr,\alpha,0)}} \\
    &\qquad \lesssim \|\wt\Box(\lambda,\hat\sigma,\hat\gamma)(u,c)\|_{\tilde H_{\scbtop,|\sigma|}^{\sfs-1,(\sfr+1,\alpha+2,0)}} + \|\wt\Box(\lambda,\hat\sigma,\hat\gamma)(u,0)\|_{\tilde H_{\scbtop,|\sigma|}^{\sfs-1,(\sfr+1,\alpha+2,-\eta)}} + \|u\|_{H_{\scbtop,|\sigma|}^{\sfs,(\sfr,\alpha-1,-\eta)}}.
  \end{align*}
  The final term on the right is bounded by $|\sigma|^\eta\|u\|_{H_{\scbtop,|\sigma|}^{\sfs,(\sfr,\alpha,0)}}$ (since $\eta\leq 1$) and can thus be absorbed. In the second term on the right, we wish to replace $0$ by $c$; this creates an error
  \[
    |c| \| f(\lambda,\hat\sigma,\hat\gamma) \|_{H_{\scbtop,|\sigma|}^{\sfs-1,(\sfr+1,\alpha+2,-\eta)}} = |c| |\sigma|^\eta \| f(\lambda,\hat\sigma,\hat\gamma) \|_{H_{\scbtop,|\sigma|}^{\sfs-1,(\sfr+1,\alpha+2+\eta,0)}} \lesssim |c| |\sigma|^\eta
  \]
  when $\eta$ is small enough (namely, such that $\alpha+2+\eta<\frac32$ still); this can be absorbed, too.

  This completes the proof of~\eqref{EqWGModeEst}.

  \pfstep{Step~2. Mode stability for perturbations near infinity.} The zero energy estimates in \citeAF{(\ref*{EqSp0Est})} hold for $\Box_\gamma$, $\gamma\in(0,\eps_0]$, without the error term: for the direct problem and in the notation of the reference, they read $\|u\|_{\bar H_\bop^{\sfs_0,\alpha}(X;\cT^*_X)}\leq C\|\wh{\Box_\gamma}(0)u\|_{\bar H_\bop^{\sfs_0-1,\alpha+2}(X;\cT^*_X)}$; similarly for the adjoint estimate. But since the Fredholm estimates \citeAF{(\ref*{EqSp0Est})} hold uniformly also for small perturbations of $\Box_\gamma$ within the class of stationary wave-type operators, the usual functional analytic argument (now using the compactness of the inclusion $\bar H_\bop^{-N,-N}\hra\bar H_\bop^{\sfs_0,\alpha}$ and, for the adjoint estimate, of $\dot H_\bop^{-N,-N}\hra\dot H_\bop^{-\sfs_0+1,-\alpha-2}$) implies the invertibility also of such small perturbations. The $\tface$-normal operator invertibility similarly follows for small perturbations (as already discussed in the proof of Lemma~\ref{LemmaWGtf}). Therefore, the low-energy estimate in \citeAF{Theorem~\ref*{ThmSpLo}} holds uniformly as well (as this estimate follows from a combination of the zero energy and $\tface$-estimates), for a fixed neighborhood $\{|\sigma|\leq c_0,\ \Im\sigma\geq 0\}$, $c_0>0$, of $\sigma=0$ in the closed upper half plane, for small perturbations of $\Box_\gamma=\Box^\Ups_{g_b,(\cd^\Ups,0,0;\cd^\Ups_{\cH^+},\gamma)}$ (with the differential and decay orders fixed relative to the operator $\Box_\gamma$, and with $c_0>0$ being as in \citeAF{Theorem~\ref*{ThmSpLo}}), so in particular for $\Box^\Ups_{g_b,E^\Ups}$ when $E^\Ups=(\cd^\Ups,e^\Ups,\gamma^\Ups;\cd_{\cH^+}^\Ups,\gamma_{\cH^+}^\Ups)$ where $\gamma_{\cH^+}^\Ups=\gamma\in(0,\eps_0]$, and $e^\Ups,\gamma^\Ups$ are small (depending on $\gamma$). This implies the triviality of the kernel of $\wh{\Box^\Ups_{g_b,E^\Ups}}(\sigma)$ for such $\sigma$, and thus mode stability for such $\sigma$. Carefully note that $c_0>0$ is independent of the parameters $\gamma_{\cH^+}^\Ups,e^\Ups$, and $\gamma^\Ups$ (provided they are small in the sense just explained). In view of the mode stability for $|\sigma|\geq c_0$ noted in Proposition~\ref{PropWGModeNon0}, the proof is complete.
\end{proof}

\subsection{Large (generalized) zero energy states}
\label{SsWG0Kerr}

\emph{We fix $E^\Ups$ according to Proposition~\usref{PropWGMode}.} If $\omega\in\ker\Box_{g_b,E^\Ups}^\Ups$, then $\delta_{g_b}^*\omega$ lies both in $\ker D_{g_b}\Ric$ and $\ker\delta_{g_b,E^\Ups}\sfG_{g_b}$, and therefore in the kernel of the linearized gauge-fixed Einstein operator $L_b:=L_{g_b}$ in the notation of~\eqref{Eq1EinLin} (see also~\eqref{Eq1Lin}). Such 2-tensors will feature in the late-time asymptotics of solutions of $L_b$. In preparation for the zero energy analysis of $L_b$, we thus construct 1-forms $\omega\in\ker\Box_{g_b,E^\Ups}^\Ups$ for later use. The guiding principle is that for every indicial root $\leq 0$ in Lemma~\ref{LemmaWGInd}, there exists a corresponding stationary 1-form in the kernel of $\Box_{g_b,E^\Ups}^\Ups$ with leading-order term as $r\to\infty$ given by the corresponding indicial solution. This is an instance of Melrose's relative index theorem \cite[Theorem~6.5]{MelroseAPS}; we prove this by hand here.

It is convenient to re-state (and slightly expand on) the invertibility of~\eqref{EqWGMode0Map}:
\begin{lemma}[Invertibility of the zero energy operator]
\label{LemmaWG0Inv}
  Define $E^\Ups$ according to Proposition~\usref{PropWGMode}. Use an unweighted b-density (such as $|\frac{\dd\rho}{\rho}\,\dd\slg|$) to define $L^2(X)$. For all sufficiently large\footnote{It suffices that $s$ exceeds a certain threshold quantity, which using the form of $\nabla_{H_G}^{\pi^*T^*M^\circ}$ in \cite[Appendix~A]{HaefnerHintzVasyKerrLarge} one can compute to be equal to $\frac32$ when $\gamma^\Ups_{\cH^+}=0$, and thus close to $\frac32$ for small $\gamma^\Ups_{\cH^+}$.} $s\in\R$ and for all $\alpha\in(0,\lambda^\Ups_{\rms 1,1})$, the operator
  \[
    \wh{\Box_{g_b,E^\Ups}^\Ups}(0) \colon \bigl\{ u\in\bar H_\bop^{s,\alpha}(X;\cT^*_X) \colon \wh{\Box_{g_b,E^\Ups}^\Ups}(0)\omega \in \bar H_\bop^{s-1,\alpha+2}(X;\cT^*_X) \bigr\} \to \bar H_\bop^{s-1,\alpha+2}(X;\cT^*_X)
  \]
  is invertible. Moreover, if $f\in\cA^{2+\cE}(X;\cT^*_X)$ where $\min\Re\cE>0$, then $\wh{\Box_{g_b,E^\Ups}^\Ups}(0)^{-1}f\in\cA^{\cE'}(X;\cT^*_X)$ where $\min\Re\cE'>0$.
\end{lemma}
\begin{proof}
  The final part is a special case of Lemma~\ref{LemmaTMSolPhg} (which moreover gives a concrete description of the index set $\cE'$).
\end{proof}

\emph{We will work with an unweighted b-density on $X$ for the remainder of this section.}

\subsubsection{Gauge potentials corresponding to asymptotic symmetries}

We begin with constructing 1-forms $\omega$ for which $\delta_{g_b}^*\omega$ decays (as a section of $\cT^*_X)$ as $r\to\infty$; since $\delta_{g_b}^*$ equals $\ubar\delta^*:=\delta_{\ubar g}^*$ to leading order as $r\to\infty$, these are related to Minkowskian symmetries. \emph{We denote geometric operators on Minkowski space, with the metric $\ubar g$ from Definition~\usref{DefKMetRef}, using underbars.} We will frequently use:

\begin{lemma}[Kerr and Minkowskian operators]
\label{LemmaWG0KerrMink}
  Let $b=(\bhm,\bha)$ be close to $b_0=(\bhm_0,\bha_0)$. Write $b_S:=(\bhm,0)$. Then
  \[
    \wh{\delta_{g_b}^*}(0) \in \rho\Diffb^1(X;\cT^*_X,S^2\cT^*_X),\quad
    \wh{\delta_{g_b}^*}(0)-\wh{\ubar\delta^*}(0) \in \rho^2\Diffb^0,\quad
    \wh{\delta_{g_b}^*}(0)-\wh{\delta_{g_{b_S}}^*}(0) \in \rho^3\Diffb^0.
  \]
  Similarly,
  \[
    \wh{\Box_{g_b,E^\Ups}^\Ups}(0) \in \rho^2\Diffb^2(X;\cT^*_X),\quad
    \wh{\Box_{g_b,E^\Ups}^\Ups}(0)-\wh{\ubar\Box_{\ubar E^\Ups}^\Ups}(0) \in \rho^3\Diffb^2,\quad
    \wh{\Box_{g_b,E^\Ups}^\Ups}(0)-\wh{\Box_{g_{b_S},E^\Ups}^\Ups}(0) \in \rho^4\Diffb^2.
  \]
\end{lemma}
\begin{proof}
  This follows directly from~\eqref{EqKMetDiff} and its consequence~\eqref{EqWGOpNabla} (for comparing $g_b$ and $\ubar g$) and its analogue
  \[
    \nabla^{g_b} - \nabla^{g_{b_S}} \in \rho^3\CI(M;\Hom(\cT^{p,q},\cT^{p,q+1}))
  \]
  (for comparing $g_b$ and $g_{b_S}\equiv g_b\bmod\rho^2\CI(M;S^2\cT^*)$).
\end{proof}

In the result below, $\eps_\ind>0$ denotes any fixed number less than $d(\Re\lambda,\Z)$ for all indicial roots $\lambda$ in Lemma~\ref{LemmaWGInd} with $\Re\lambda\neq\Z$ and $|\Re\lambda|\leq 10$.

\begin{prop}[Zero energy states, I: asymptotic symmetries]
\label{PropWG0Symm}
  On $\R_t\times(0,\infty)_r\times\Sph^2\subset\R^4$ and for $\scal\in\scalspace_1$ and\footnote{As in Definition~\ref{DefTYs1}, we use the convention that $\vect$ is dual to a rotation vector field with respect to the Euclidean metric; thus, up to scaling and a choice of polar coordinates, non-zero $\vect$ take the form $\vect=\pa_\phi^\flat=r^2\sin^2\theta\,\dd\phi$. In Cartesian coordinates, this is further equal to $x\,\dd y-y\,\dd x$, and thus defines an element of $\rho^{-1}\CI(X;\cT^*_X)$.} $\vect\in\vectspace_1$, define\footnote{Thus, $\ubar\omega_{\rms 1}^{(0)}(\scal)$ is dual to a spatial translation on Minkowski space, $t_*\ubar\omega_{\rms 1}^{(0)}(\scal)+\breve{\ubar\omega}_{\rms 1}^{(0),1}(\scal)=t\,\dd(r\scal)-r\scal\,\dd t$ is dual to a Lorentz boost, and $\ubar\omega_{\rmv 1}^{(-1)}(\vect)$ is dual to a spatial rotation.}
  \begin{equation}
  \label{EqWG0SymmMink}
  \begin{split}
    \ubar\omega_{\rms 0}^{(0)} &:= \ubar g(\pa_{t_*},\cdot) = -\dd t = -\dd t_*-\dd r, \\
    \ubar\omega_{\rms 1}^{(0)}(\scal) &:= \dd(r\scal), \\
    \breve{\ubar\omega}_{\rms 1}^{(0),1}(\scal) &:= r\,\dd(r\scal) - r\scal\,\dd t, \\
    \ubar\omega_{\rmv 1}^{(-1)}(\vect) &:= \vect.
  \end{split}
  \end{equation}
  There exists an index set $\cE_\ind\subset\C\times\N_0$ with $\min\Re\cE_\ind\geq\eps_\ind$ such that the following holds.
  \begin{enumerate}
  \item\label{ItWG0SymmStates}{\rm (States.)} For Kerr parameters $b=(\bhm,\bha)$ close to $b_0=(\bhm_0,\bha_0)$, there exist stationary 1-forms
    \begin{equation}
    \label{EqWG0Symmomega}
    \begin{split}
      \omega_{b,\rms 0}^{(0)} := g_b^{-1}(\pa_{t_*},\cdot) &\in \cA^{(0,0)}(X;\cT^*_X), \\
      \omega_{b,\rms 1}^{(0)}(\scal) &\in \cA^{(0,0)\cup(1+\cE_\ind)}(X;\cT^*_X), \\
      \breve\omega_{b,\rms 1}^{(0),1}(\scal) &\in \cA^{(-1,0)\cup(0,1)\cup\cE_\ind}(X;\cT^*_X), \\
      \omega_{b,\rmv 1}^{(-1)}(\vect) &\in \cA^{(-1,0)\cup(1+\cE_\ind)}(X;\cT^*_X), \\
    \end{split}
    \end{equation}
    which depend linearly on $\scal$ and $\vect$ and smoothly on $b$, have leading-order terms $\ubar\omega_{\rms 0}^{(0)}$, $\ubar\omega_{\rms 1}^{(0)}(\scal)$, $\breve{\ubar\omega}_{\rms 1}^{(0),1}(\scal)$, and $\ubar\omega_{\rmv 1}^{(-1)}(\vect)$ (in the precise sense that $\omega_{b,\rms 1}^{(0)}(\scal)-\ubar\omega_{\rms 1}^{(0)}(\scal)\in\cA^{1+\cE_\ind}(X;\cT^*_X)$, $\breve\omega_{b,\rms 1}^{(0),1}(\scal)-\breve{\ubar\omega}_{\rms 1}^{(0),1}(\scal)\in\cA^{(0,1)\cup\cE_\ind}(X;\cT^*_X)$, and $\omega_{b,\rmv 1}^{(-1)}(\vect)-\ubar\omega_{\rmv 1}^{(-1)}(\vect)\in\cA^{(1,0)\cup(1+\cE_\ind)}(X;\cT^*_X)$), and satisfy
    \begin{align*}
      \Box_{g_b,E^\Ups}^\Ups\omega_{b,\rms 0}^{(0)} &= 0, \\
      \Box_{g_b,E^\Ups}^\Ups\omega_{b,\rms 1}^{(0)}(\scal) &= 0, \\
      \Box_{g_b,E^\Ups}^\Ups\bigl(t_*\omega_{b,\rms 1}^{(0)}(\scal)+\breve\omega_{b,\rms 1}^{(0),1}(\scal)\bigr) &= 0, \\
      \Box_{g_b,E^\Ups}^\Ups\omega_{b,\rmv 1}^{(-1)}(\vect) &= 0.
    \end{align*}
    When $\vect$ and $\vect(\bha)$ (in the notation of Definition~\usref{DefTYs1}) are linearly dependent, then $\omega_{b,\rmv 1}^{(-1)}(\vect)=g_b\ubar g^{-1}\vect$ (i.e., this is dual to a rotation vector field).
  \item{\rm (Logarithmic leading-order term.)} We have
    \begin{equation}
    \label{EgWG0SymmLog}
      \breve\omega_{b,\rms 1}^{(0),1}(\scal)-\breve{\ubar\omega}_{\rms 1}^{(0),1}(\scal)-(\log\rho)c_{b,\rms 1}\ubar\omega_{\rms 1}^{(0)}(\scal)\in\cA^{(0,0)\cup\cE_\ind}(X;\cT^*_X)
    \end{equation}
    for some $c_{b,\rms 1}$ depending smoothly on $b$ (in fact, on $\bhm$ only).
  \item\label{ItWG0SymmGrad}{\rm (Symmetric gradients: pure gauge states.)} We have $\delta_{g_b}^*\omega_{b,\rms 0}^{(0)}=0$ and
    \begin{subequations}
    \begin{align}
    \label{EqWG0hs1}
      h_{b,\rms 1}(\scal) &:= \delta_{g_b}^*\omega_{b,\rms 1}^{(0)}(\scal) \in \cA^{(2,0)\cup(2+\cE_\ind)}(X;S^2\cT^*_X), \\
    \label{EqWG0hs1breve}
      \delta_{g_b}^*\bigl(t_*\omega_{b,\rms 1}^{(0)}(\scal)+\breve\omega_{b,\rms 1}^{(0),1}(\scal)\bigr) &=: t_* h_{b,\rms 1}(\scal) + \breve h_{b,\rms 1}^1(\scal),\quad \breve h_{b,\rms 1}^1(\scal) \in \cA^{(1,0)\cup(1+\cE_\ind)}(X;S^2\cT^*_X), \\
    \label{EqWG0hv1}
      h_{b,\rmv 1}(\vect) &:= \delta_{g_b}^*\omega_{b,\rmv 1}^{(-1)}(\vect) \in \cA^{(2,0)\cup(1+\cE_\ind)}(X;S^2\cT^*_X),
    \end{align}
    \end{subequations}
    and the $\rho^1$ leading-order term of $\breve h_{b,\rms 1}^1(\scal)$ is of scalar type $1$. The tensors $h_{b,\rms 1}(\scal)$, $t_* h_{b,\rms 1}(\scal)+\breve h_{b,\rms 1}^1(\scal)$, and $h_{b,\rmv 1}(\vect)$ lie in the kernels of $D_{g_b}\Ric$ and $\delta_{g_b,E^\Ups}\sfG_{g_b}$.
  \end{enumerate}
\end{prop}

\begin{rmk}[Notation]
\label{RmkWG0Not}
  The subscript records the Kerr parameters and the scalar/vector type. The number in parentheses in the superscript encodes the decay rate, i.e., power of $\rho$.
\end{rmk}

The reason for the symmetric gradient of $\omega_{b,\rms 1}^{(0)}$ to gain two powers of $r$-decay over $\omega_{b,\rms 1}^{(0)}$ (rather than only one power, as expected from the weight $\rho$ of $\wh{\delta_{g_b}^*}(0)\in\rho\Diffb^1$) is that $g_b$ is, to leading order as $r\to\infty$, translation-invariant (being equal to the Minkowski metric to leading order). Similarly, the reason why $\omega_{b,\rmv 1}^{(-1)}$ gains three powers of decay over $\omega_{b,\rmv 1}^{(-1)}$ itself is that the Kerr metric is spherically symmetric modulo sub-sub-leading terms (see~\eqref{EqKMetDiff}).

\begin{rmk}[Killing 1-forms]
\label{RmkWG0Killing}
  Note that $\delta_{g_b}^*\omega_{b,\rms 0}^{(0)}=0$. Similarly, when $\bha(\vect)$ and $\bha=\bha(b)$ are linearly dependent (so for all $\vect$ when $\bha=0$, and for all $\vect$ such that $\bha(\vect)$ is parallel to $\bha$ otherwise), then $\vect_b:=g_b\ubar g^{-1}\vect$ is a Killing vector field, and correspondingly $\omega_{b,\rmv 1}^{(-1)}(\vect)=\vect_b$ and $h_{b,\rmv 1}(\vect)=0$ in this case.
\end{rmk}

\begin{proof}[Proof of Proposition~\usref{PropWG0Symm}]
  We fix $\scal\in\scalspace_1$ and $\vect\in\vectspace_1$ and omit them from the notation. We shall moreover write $\cE_\ind$ for any index set with $\Re\cE_\ind>0$; it may get larger from line to line.

  Since $\pa_{t_*}$ is Killing, we have $\delta_{g_b}^*\omega_{b,\rms 0}^{(0)}=0$ and thus also $\Box_{g_b,E^\Ups}^\Ups\omega_{b,\rms 0}^{(0)}=0$.

  \pfstep{Asymptotic translations.} Since $\ubar\delta^*\ubar\omega_{\rms 1}^{(0)}=0$, we have
  \[
    \delta_{g_b}^*\ubar\omega_{\rms 1}^{(0)} = \bigl(\wh{\delta_{g_b}^*}(0)-\wh{\ubar\delta^*}(0)\bigr)\ubar\omega_{\rms 1}^{(0)} \in \rho^2\Diff^0(X;\cT^*_X,S^2\cT^*_X)\bigl(\CI(X;\cT^*_X)\bigr) \subset \rho^2\CI(X;S^2\cT^*_X)
  \]
  by Lemma~\ref{LemmaWG0KerrMink}; therefore,
  \[
    f_b := \Box_{g_b,E^\Ups}^\Ups\ubar\omega_{\rms 1}^{(0)} \in \rho^3\CI(X;\cT^*_X).
  \]
  By Lemma~\ref{LemmaWG0Inv}, we have $\omega'_b:=-\wh{\Box_{g_b,E^\Ups}^\Ups}(0)^{-1}f_b\in\bar H_\bop^{\infty,\alpha}(X;\cT^*_X)$ for all $\alpha<1$, and in fact $\omega'_b$ is polyhomogeneous. Since $1$ is not an indicial root of $\wh{\Box_{g_b,E^\Ups}^\Ups}(0)$, we have $\omega'_b\in\cA^{(1,0)\cup(1+\cE_\ind)}$ where (an upper bound for) the index set $\cE_\ind=(\lambda^\Ups_{\rms 1,1}-1,0)\cup\cdots$ can be determined from the indicial roots in Lemma~\ref{LemmaWGInd} and the index set $(3,0)$ of $f_b$. We then set
  \begin{equation}
  \label{EqWG0s1Omega}
    \omega_{b,\rms 1}^{(0)} := \ubar\omega_{\rms 1}^{(0)} + \omega'_b,\quad \text{with}\ \omega'_b\in\cA^{(1,0)\cup(1+\cE_\ind)}(X;\cT^*_X).
  \end{equation}
  By construction, $\Box_{g_b,E^\Ups}^\Ups\omega_{b,\rms 1}^{(0)}=0$; and, again by Lemma~\ref{LemmaWG0KerrMink},
  \begin{align*}
    \delta_{g_b}^*\omega_{b,\rms 1}^{(0)} = \wh{\delta_{g_b}^*}(0)\omega_{b,\rms 1}^{(0)} &= \wh{\delta_{g_b}^*}(0)\bigl(\omega_{b,\rms 1}^{(0)}-\ubar\omega_{\rms 1}^{(0)}\bigr) + \bigl(\wh{\delta_{g_b}^*}(0)-\wh{\ubar\delta^*}(0)\bigr)\ubar\omega_{\rms 1}^{(0)} \\
      &\in \rho\Diffb^1\bigl(\cA^{(1,0)\cup(1+\cE_\ind)}\bigr) + \rho^2\Diffb^0\bigl(\CI\bigr) \subset \cA^{(2,0)\cup(2+\cE_\ind)}.
  \end{align*}

  For the regularity in $b$, consider a smooth function $b(s)$ with $b(0)=b$; then formally differentiating the equation $\wh{\Box_{g_{b(s)},E^\Ups}^\Ups}(0)\omega_{b(s),\rms 1}^{(0)}=0$ at $s=0$ yields for $\dot\omega'_b:=\frac{\dd}{\dd s}\omega'_{b(s)}|_{s=0}$ the equation
  \begin{equation}
  \label{EqWG0s1Der}
    \wh{\Box_{g_b,E^\Ups}^\Ups}(0) \dot\omega'_b = -\Bigl(\frac{\dd}{\dd s}\wh{\Box_{g_{b(s)},E^\Ups}^\Ups}(0)\Bigr)\Big|_{s=0} ( \ubar\omega_{\rms 1}^{(0)} + \omega'_b ).
  \end{equation}
  Since $\wh{\Box_{g_{b(s)},E^\Ups}^\Ups}(0)$ differs from the $s$-independent operator $\wh{\ubar\Box_{\ubar E^\Ups}^\Ups}(0)$ by $\rho^3\Diffb^2$ (see~\eqref{EqWGOpDiff}), the right-hand side here lies in $\rho^3(\CI+\cA^{(1,0)\cup(1+\cE_\ind)})=\cA^{(3,0)\cup(4+\cE_\ind)}(X;\cT^*_X)$; thus we can control $\dot\omega_b'$ in $\bar H_\bop^{\infty,\alpha}$ for all $\alpha<1$ using Lemma~\ref{LemmaWG0Inv}. Approximating the $s$-derivative by finite difference quotients and taking limits, one finds that $\omega'_{b(s)}$ is indeed differentiable at $s=0$, with derivative equal to the solution $\dot\omega_b'$ of~\eqref{EqWG0s1Der}. (The index set of $\dot\omega'_b$ must, of course, be a subset of that of $\omega'_b$.) Higher regularity is proved similarly.

  \pfstep{Asymptotic boosts.} Using~\eqref{EqWG0s1Omega}, we write
  \begin{align*}
    \breve f_b := \Box_{g_b,E^\Ups}^\Ups\bigl( t_* \omega_{b,\rms 1}^{(0)} + \breve{\ubar\omega}_{\rms 1}^{(0),1} \bigr) &= [\Box_{g_b,E^\Ups}^\Ups,t_*]\omega_{b,\rms 1}^{(0)} + \Box_{g_b,E^\Ups}^\Ups\breve{\ubar\omega}_{\rms 1}^{(0),1} \\
      &= \Bigl( [\ubar\Box_{\ubar E^\Ups}^\Ups,t_*]\ubar\omega_{\rms 1}^{(0)} + \ubar\Box_{\ubar E^\Ups}^\Ups\breve{\ubar\omega}_{\rms 1}^{(0),1}\Bigr) + [\Box_{g_b,E^\Ups}^\Ups-\ubar\Box^\Ups_{\ubar E^\Ups},t_*]\ubar\omega_{\rms 1}^{(0)} \\
      &\quad \qquad + [\Box_{g_b,E^\Ups}^\Ups,t_*]\omega'_b + \Bigl(\wh{\Box_{g_b,E^\Ups}^\Ups}(0)-\wh{\ubar\Box_{\ubar E^\Ups}^\Ups}(0)\Bigr)\breve{\ubar\omega}_{\rms 1}^{(0),1}.
  \end{align*}
  The first term vanishes. The second term is given by an element of $\rho^2\Diffb^1$ (by Lemma~\ref{LemmaWGOp}) acting on $\CI$, and thus of class $\rho^2\CI$. The third term is the action of an element of $\rho\Diffb^1$ on $\omega'_b$, and hence of class $\cA^{(2,0)\cup(2+\cE_\ind)}$; and the final term is the action of an element of $\rho^3\Diffb^2$ on an element of $\rho^{-1}\CI$, so of class $\rho^2\CI$. Altogether, we thus have $\breve f_b\in\cA^{(2,0)\cup(2+\cE_\ind)}(X;\cT^*_X)$. We can then apply Lemma~\ref{LemmaTMFormal} with $\beta=4$, say, to find $\breve\omega_b\in\cA^{(0,1)\cup\cE_\ind}(X;\cT^*_X)$ (with smooth dependence on $b$) such that $\breve f'_b:=\breve f_b+\wh{\Box_{g_b,E^\Ups}^\Ups}(0)\breve\omega_b\in\cA^\cG$ where $\cG\subset\{(\lambda,k)\colon\Re\lambda\geq 4\}$. Finally, we put $\breve\omega'_b:=-\wh{\Box_{g_b,E^\Ups}^\Ups}(0)^{-1}\breve f'_b\in\cA^{1+\cE_\ind}$ and
  \[
    \breve\omega_{b,\rms 1}^{(0),1} := \breve{\ubar\omega}_{\rms 1}^{(0),1} + \breve\omega_b + \breve\omega_b'.
  \]
  The logarithmic leading-order term of $\breve\omega_b$ arises from the indicial root $0$ of $\wh{\Box_{g_b,E^\Ups}^\Ups}(0)$ in Lemma~\ref{LemmaWGInd}, whose associated indicial solutions are $\dd t$ (for the $\rms 0$ root $0$) and $\dd(r\scal)$ (for the $\rms 1$ root $0$). We can be more precise still: since the departure from spherical symmetry arises at \emph{sub}-sub-leading order, $\breve f_b$ is of scalar type $1$ to leading order, i.e., the $\rho^2$ leading-order term of $\breve f_b$ is of scalar type $1$, and moreover is described by the same $\scal\in\scalspace_1$ as the one we fixed from the outset. Therefore, only the scalar type $1$ indicial solution (with the same $\scal$) can appear as the logarithmic leading-order term of $\breve\omega_b$.

  The smoothness of $\breve\omega_b$ and $\breve\omega'_b$ (and thus of $\breve\omega_{b,\rms 1}^{(0),1}$) in $b$ is a consequence of the regularity in $b$ of solutions of the equations involving $\wh{\Box_{g_b,E^\Ups}^\Ups}(0)$ on the spaces in Lemma~\ref{LemmaWG0Inv} in the above construction (for $\breve\omega'_b$), and of the smooth parameter dependence recorded in Lemma~\ref{LemmaTMFormal} (for $\breve\omega_b$).

  The membership~\eqref{EqWG0hs1breve} of $\breve h_{b,\rms 1}^1$ follows from
  \begin{align*}
    \breve h_{b,\rms 1}^1 = [\delta_{g_b}^*,t_*]\omega_{b,\rms 1}^{(0)} + \delta_{g_b}^*\breve\omega_{b,\rms 1}^{(0),1} &= \Bigl([\ubar\delta^*,t_*]\ubar\omega_{\rms 1}^{(0)} + \ubar\delta^*\breve{\ubar\omega}_{\rms 1}^{(0),1}\Bigr) + [\delta_{g_b}^*-\ubar\delta^*,t_*]\ubar\omega_{\rms 1}^{(0)} + [\delta_{g_b}^*,t_*]\omega'_b \\
      &\quad\qquad + \bigl(\wh{\delta_{g_b}^*}(0)-\wh{\ubar\delta^*}(0)\bigr)\breve{\ubar\omega}_{\rms 1}^{(0),1} + \wh{\delta_{g_b}^*}(0)(\breve\omega_b+\breve\omega'_b).
  \end{align*}
  The first term vanishes since Lorentz boosts are isometries of Minkowski space. The second term vanishes as well since $[\delta_g^*,t_*]=\dd t_*\otimes_s(\cdot)$ for any metric $g$. The third term is of class $\cA^{(1,0)\cup(1+\cE_\ind)}$ by~\eqref{EqWG0s1Omega}. The first term in the second line is of class $\rho^2\Diffb^0(\rho^{-1}\CI)\subset\rho\CI$ by Lemma~\ref{LemmaWG0KerrMink}, and the final term lies in $\rho\Diffb^1(\cA^{(0,1)\cup\cE_\ind})\subset\cA^{(1,1)\cup(1+\cE_\ind)}$; but its logarithmic leading-order term is the symmetric gradient of $c_{b,\rms 1}\,\dd(r\scal)$ (from~\eqref{EgWG0SymmLog}) relative to the Minkowski metric and thus vanishes. Finally, the scalar type $1$ nature of the leading-order term of $\breve h_{b,\rms 1}^1$ is clear in the spherically symmetric setting $b=b_S$, and follows for general $b$ again by the sub-sub-leading nature of the departure from spherical symmetry.

  \pfstep{Asymptotic rotations.} Writing $b=(\bhm,\bha)$ and $b_S=(\bhm,0)$, we use Lemma~\ref{LemmaWG0KerrMink} and the spherical symmetry of $g_{b_S}$ to conclude that
  \begin{equation}
  \label{EqWG0SymmRot}
    \delta_{g_b}^*\ubar\omega_{\rmv 1}^{(-1)} = \bigl(\wh{\delta_{g_b}^*}(0)-\wh{\delta_{g_{b_S}}^*}(0)\bigr) \ubar\omega_{\rmv 1}^{(-1)} \in \rho^3\Diffb^0(\rho^{-1}\CI) \subset \rho^2\CI,
  \end{equation}
  and therefore $f_b:=\wh{\Box_{g_b,E^\Ups}^\Ups}(0)\ubar\omega_{\rmv 1}^{(-1)}\in\rho^3\CI$. This can be solved away using Lemma~\ref{LemmaWG0Inv}, and thus we obtain
  \[
    \omega_{b,\rmv 1}^{(-1)} := \ubar\omega_{\rmv 1}^{(-1)} + \omega'_b,\quad \omega'_b:=-\wh{\Box_{g_b,E^\Ups}^\Ups}(0)^{-1}f_b \in \cA^{(1,0)\cup(1+\cE_\ind)}(X;\cT^*_X);
  \]
  note here that $1$ is not an indicial root by Lemma~\ref{LemmaWGInd}. Therefore,
  \[
    \delta_{g_b}^*\omega_{b,\rmv 1}^{(-1)} = \delta_{g_b}^*\ubar\omega_{\rmv 1}^{(-1)} + \wh{\delta_{g_b}^*}(0)\omega'_b = \delta_{g_{b_S}}^*\ubar\omega_{\rmv 1}^{(-1)} + \bigl(\wh{\delta_{g_b}^*}(0)-\wh{\delta_{g_{b_S}}^*}(0)\bigr)\ubar\omega_{\rmv 1}^{(-1)} + \wh{\delta_{g_b}^*}(0)\omega_b'.
  \]
  The first term vanishes, and by Lemma~\ref{LemmaWG0KerrMink}, the second term lies in $\rho^3\Diffb^1(\rho^{-1}\CI)\subset\rho^2\CI$ by~\eqref{EqWG0SymmRot} and the final term in $\cA^{(2,0)\cup(2+\cE_\ind)}(X;S^2\cT^*_X)$.
\end{proof}

\subsubsection{Gauge potentials for pure gauge large zero energy states}
\label{SssWG0L}

Looking ahead at the indicial roots of the linearized gauge-fixed Einstein operator at zero energy, $\wh{L_b}(0)$, in Lemma~\ref{LemmaWEInd}, we next construct gauge potentials whose symmetric gradients have decay rates given by those indicial roots $\lambda$ (and leading-order behavior given by the corresponding indicial solutions) of $\wh{L_b}(0)$ with $\Re\lambda\in[-3,0]$. They appear, roughly speaking, for the first time at order $t_*^{-1+\lambda}$ in the late-time asymptotics of metric perturbations. This means that we will capture all pure gauge state contributions up to and including $t_*^{-4}$-decay. (Contributions with stronger decay can be  put into the ``structureless'' gravitational wave tail for our nonlinear stability result.) Moreover, if, say, a term $h_{b,\rms 1}(\scal(t_*))$ arises in the late-time asymptotics, where $\scal(t_*)=\log t_*$ (for the sake of example), then also $\breve h_{b,\rms 1}^1(\scal'(t_*))=\cO(t_*^{-1})$ appears as well as further terms modulated by higher derivatives of $\scal(t_*)$; the spatial parts will be higher-order versions of $\breve h_{b,\rms 1}^1(\scal)$ in~\eqref{EqWG0hs1breve}, arising from symmetric gradients of higher-degree polynomials in $t_*$ (starting with $t_*^j\omega_{b,\rms 1}^{(0)}+j t_*^{j-1}\breve\omega_{b,\rms 1}^{(0),1}+\cdots$) in the kernel of $\Box_{g_b,E^\Ups}^\Ups$---recall that we call these \emph{generalized (large) zero energy states}. This motivates the scope of Proposition~\ref{PropWG0Large} below. To aid the reader in parsing it, we use the following conventions.
\begin{enumerate}
\item We list the zero energy states according to the $\rho$-decay rate of their symmetric gradients; that is, they correspond to the indicial roots of the linearized gauge-fixed Einstein operator at zero energy, $\wh{L_b}(0)$, in Lemma~\ref{LemmaWEInd} (\emph{not} of $\wh{\Box_{g_b,E^\Ups}^\Ups}(0)$).
\item Once we have constructed states $\omega_{b,\bullet}^{(\lambda)}\in\ker\wh{\Box_{g_b,E^\Ups}^\Ups}(0)$ and stationary 1-forms $\breve\omega_{b,\bullet}^{(\lambda),i}$, we write
  \begin{subequations}
  \begin{equation}
  \label{EqWG0GenState}
    \omega_{b,\bullet}^{(\lambda),\leq k} := t_*^k\omega_{b,\bullet}^{(\lambda)} + \sum_{i=1}^k t_*^{k-i}\frac{k!}{(k-i)!}\breve\omega_{b,\bullet}^{(\lambda),i}
  \end{equation}
  for the $k$-th order generalized zero energy state. (The 1-forms $\breve\omega_{b,\bullet}^{(\lambda),i}$ are constructed so that this lies in $\ker\Box_{g_b,E^\Ups}^\Ups$.) We use analogous notation for symmetric 2-tensors, thus writing
  \begin{equation}
  \label{EqWG0GenStateh}
    h_{b,\bullet}^{(\lambda),\leq k} := t_*^k h_b^{(\lambda)} + \sum_{i=1}^k t_*^{k-i}\frac{k!}{(k-i)!}\breve h_b^{(\lambda),i}.
  \end{equation}
  \end{subequations}
  For generalized zero energy states whose leading $t_*$-coefficient is $h_{b,\rms 1}$ or $h_{b,\rmv 1}$, we use similar notation, so for example equation~\eqref{EqWG0hs1breve} reads
  \begin{equation}
  \label{EqWG0GenStateNot}
    \delta_{g_b}^*\omega_{b,\rms 1}^{(0),\leq 1}(\scal) = h_{b,\rms 1}^{\leq 1}(\scal) := t_* h_{b,\rms 1}(\scal) + \breve h_{b,\rms 1}^1(\scal).
  \end{equation}
  We also recall that $\Box_{g_b,E^\Ups}^\Ups\omega=0$ is, by definition of $\Box_{g_b,E^\Ups}^\Ups$, the same as the second identity of
  \[
    D_{g_b}\Ric(h) = 0,\ \ 
    \delta_{g_b,E^\Ups}\sfG_{g_b}h = 0,\quad h:=\delta_{g_b}^*\omega,
  \]
  while the first identity follows from $D_{g_b}\Ric\circ\delta_{g_b}^*=0$. We shall only mention this once in the statement of Proposition~\ref{PropWG0Large} below to avoid excessive repetition.
\item The leading-order terms of many of the states constructed below are indicial solutions of the Minkowskian model $\ubar L$ of $L_b$; this is given by
  \[
    \ubar L=D_{\ubar g}\Ric + \ubar\delta_{\ubar E^\cC}^*\ubar\delta_{\ubar E^\Ups}\ul\sfG,
  \]
  where the terms with underbars are defined like their Kerr counterparts, but only using the Minkowski metric and the leading-order terms at $r=\infty$; see~\eqref{EqWGOpMinkE} and \eqref{EqWCRecE}, and also Lemmas~\ref{LemmaWEOp} and \ref{LemmaWEOpMink} below for the explicit formulas.
\end{enumerate}

\begin{prop}[Zero energy states, II: generalized zero energy states]
\label{PropWG0Large}
  We use the notation of Proposition~\usref{PropWG0Symm}. There exists an index set $\cE_\ind\subset\C\times\N_0$ with $\min\Re\cE_\ind\geq\eps_\ind$ such that the following statements hold for all $b$ near $b_0$.
  \begin{enumerate}
  \item\label{ItWG0Larges00}{\rm (Exceptional $\rms 0$ root $0$.)} There exists $\breve\omega_{b,\rms 0}^{(0),1} \in \cA^{(-1,0)\cup(-1+\cE_\ind)}(X;\cT^*_X)$, with scalar type $0$ leading-order term, such that $\omega_{b,\rms 0}^{(0),\leq 1}=t_*\omega_{b,\rms 0}^{(0)}+\breve\omega_{b,\rms 0}^{(0),1}\in\ker\Box_{g_b,E^\Ups}^\Ups$ and
    \begin{equation}
    \label{EqWG0Larges00}
      h_{b,\rms 0}^{(0)} := \delta_{g_b}^*\omega_{b,\rms 0}^{(0),\leq 1} \in \cA^{(0,0)\cup\cE_\ind}(X;S^2\cT^*_X).
    \end{equation}
    The leading-order term of $h_{b,\rms 0}^{(0)}$ is equal to an $\rms 0$ tensor $\ubar h_{\rms 0}^{(0)}$ which spans the space of indicial solutions of $N_{\rms 0}(\wh{\ubar L}(0),0)$. Furthermore, there exist $\breve\omega_{b,\rms 0}^{(0),k} \in \cA^{(-k,0)\cup(-k+\cE_\ind)}(X;\cT^*_X)$, $k=2,3,4$, such that $\omega_{b,\rms 0}^{(0),\leq k}\in\ker\Box_{g_b,E^\Ups}^\Ups$; and for the tensors $\breve h_{b,\rms 0}^{(0),k}$, $k=1,2,3$, for which
    \begin{subequations}
    \begin{equation}
    \label{EqWG0Larges002}
      h_{b,\rms 0}^{(0),\leq k} = t_*^k h_{b,\rms 0}^{(0)} + \sum_{i=1}^k t_*^{k-i}\frac{k!}{(k-i)!}\breve h_{b,\rms 0}^{(0),i} := \delta_{g_b}^*\omega_{b,\rms 0}^{(0),\leq k+1},
    \end{equation}
    we have\footnote{These are given explicitly in terms of the $\breve\omega_{b,\rms 0}^{(0),j}$; see~\eqref{EqWG0Larges00Pf}.}
    \begin{equation}
    \label{EqWG0Larges003}
      \breve h_{b,\rms 0}^{(0),k} \in \cA^{(-k,0)\cup(-k+\cE_\ind)}(X;S^2\cT^*_X).
    \end{equation}
    \end{subequations} 
    We have $h_{b,\rms 0}^{(0),\leq k}\in\ker D_{g_b}\Ric\cap\ker\delta_{g_b,E^\Ups}\sfG_{g_b}$ for $k=0,1,2,3$.
  \item\label{ItWG0Larges1m1}{\rm (Exceptional $\rms 1$ root $-1$.)} There exist $\breve\omega_{b,\rms 1}^{(0),k}(\scal)\in\cA^{(-k,0)\cup(-k+\cE_\ind)}(X;\cT^*_X)$, $k=2,3,4$, depending linearly on $\scal\in\scalspace_1$, with scalar type $1$ leading-order terms, such that $\omega_{b,\rms 1}^{(0),\leq k}(\scal)\in\ker\Box_{g_b,E^\Ups}^\Ups$, and the constant coefficient of
      \[
        h_{b,\rms 1}^{\leq k}(\scal) := \delta_{g_b}^*\omega_{b,\rms 1}^{\leq k}(\scal)
      \]
      (which is a degree $k$ polynomial in $t_*$) for $k=2,3,4$ is $k!$ times
      \[
        \breve h_{b,\rms 1}^k(\scal) \in \cA^{(-k+1,0)\cup(-k+1+\cE_\ind)}(X;S^2\cT^*_X).
      \]
      The leading-order terms $\ubar h_{\rms 1}^2(\scal)$ of the tensors $\breve h_{b,\rms 1}^2(\scal)$ comprise the space $\ker N_{\rms 1}(\wh{\ubar L}(0),-1)$ (and are $\rms 1$ tensors with respect to $\scal$).
  \item\label{ItWG0Largev1m1}{\rm (Exceptional $\rmv 1$ root $-1$.)} There exist $\breve\omega_{b,\rmv 1}^{(-1),k}(\vect)\in\cA^{(-k-1,0)\cup(-k-1+\cE_\ind)}(X;\cT^*_X)$, $k=1,2,3$, depending linearly on $\vect\in\vectspace_1$ and with vector type $1$ leading-order terms, such that $\omega_{b,\rmv 1}^{(-1),\leq k}(\vect)\in\ker\Box_{g_b,E^\Ups}^\Ups$, and the constant coefficient of
    \[
      h_{b,\rmv 1}^{\leq k}(\vect) := \delta_{g_b}^*\omega_{b,\rmv 1}^{(-1),\leq k}(\vect)
    \]
    for $k=1,2,3$ is $k!$ times
    \[
      \breve h_{b,\rmv 1}^k(\vect) \in \cA^{(-k,0)\cup(-k+\cE_\ind)}(X;S^2\cT^*_X).
    \]
    The leading-order terms $\ubar h_{\rmv 1}^1(\vect)$ of the tensors $\breve h_{b,\rmv 1}^1(\vect)$ comprise the space $\ker N_{\rmv 1}(\wh{\ubar L}(0),-1)$ (and are $\rmv 1$ tensors with respect to $\vect$); explicitly, in the splitting~\eqref{EqTYMink01Split},
    \begin{equation}
    \label{EqWG0Largev1m1hbar}
      \ubar h_{\rmv 1}^1(\vect) = \frac14\Bigl(0,0,\Bigl(1+\frac{\gamma^\Ups}{2}\Bigr)\vect,0,\Bigl(1-\frac{\gamma^\Ups}{2}\Bigr)\vect,0\Bigr).
    \end{equation}
  \end{enumerate}
  The remaining four classes of pure gauge solutions we record are not ``special'' in that the corresponding gauge potentials do not involve (approximate) symmetries;\footnote{and thus the (generalized zero energy state) gauge potentials and corresponding pure gauge states are polynomials in $t_*$ of the same degree} these are:
  \begin{enumerate}
  \setcounter{enumi}{3}
  \item\label{ItWG0Larges01}{\rm ($\rms 0$ root $-\lambda_{\rms 0,1}^\Ups+1$.)} Set $\lambda:=-\lambda_{\rms 0,1}^\Ups+1$. There exist $\omega_{b,\rms 0}^{(\lambda-1)}\in\cA^{(\lambda-1,0)\cup(\lambda-1+\cE_\ind)}(X;\cT^*_X)$ and $\breve\omega_{b,\rms 0}^{(\lambda-1),k}\in\cA^{(\lambda-1-k,0)\cup(\lambda-1-k+\cE_\ind)}(X;\cT^*_X)$, $k=1,2$, with scalar type $0$ leading-order terms, such that $\omega_{b,\rms 0}^{(\lambda-1),\leq k}\in\ker\Box_{g_b,E^\Ups}^\Ups$, and with the leading-order term of
    \begin{equation}
    \label{EqWG0Larges01h}
      h_{b,\rms 0}^{(\lambda)} := \delta_{g_b}^*\omega_{b,\rms 0}^{(\lambda-1)} \in \cA^{(\lambda,0)\cup(\lambda+\cE_\ind)}(X;S^2\cT^*_X)
    \end{equation}
    equal to an $\rms 0$ tensor $\ubar h_{\rms 0}^{(\lambda)}$ which spans $\ker N_{\rms 0}(\wh{\ubar L}(0),\lambda)$. The constant coefficient of
    \[
      h_{b,\rms 0}^{(\lambda),\leq k}:=\delta_{g_b}^*\omega_{b,\rms 0}^{(\lambda-1),\leq k}
    \]
    for $k=1,2$ is $k!$ times
    \begin{equation}
    \label{EqWG0Larges01hbreve}
      \breve h_{b,\rms 0}^{(\lambda),k} \in \cA^{(\lambda-k,0)\cup(\lambda-k+\cE_\ind)}(X;S^2\cT^*_X).
    \end{equation}
  \item\label{ItWG0Largeslj}{\rm ($\rms l$ roots $-\lambda^\Ups_{\rms l,l+j}+1$, $l\geq 1$.)} For $l=1,2,3$, $j=0,1$, set $\lambda:=-\lambda_{\rms l,l+j}+1$.\footnote{Thus $\lambda<-l-j+1$ is close to $-l-j+1$; recall here~\eqref{EqWGIndOrder1}.} Then there exist $\omega_{b,\rms l}^{(\lambda-1)}(\scal)\in\cA^{(\lambda-1,0)\cup(\lambda-1+\cE_\ind)}(X;\cT^*_X)$, $\breve\omega_{b,\rms l}^{(\lambda-1),k}(\scal)\in \cA^{(\lambda-1-k,0)\cup(\lambda-1-k+\cE_\ind)}(X;\cT^*_X)$, $k=0,1,2$, depending linearly on $\scal\in\scalspace_l$ and having scalar type $l$ leading-order terms, such that $\omega_{b,\rms l}^{(\lambda-1),\leq k}\in\ker\Box_{g_b,E^\Ups}^\Ups$, and with the leading-order term of
    \[
      h_{b,\rms l}^{(\lambda)}(\scal) := \delta_{g_b}^*\omega_{b,\rms l}^{(\lambda-1)} \in \cA^{(\lambda,0)\cup(\lambda+\cE_\ind)}(X;S^2\cT^*_X)
    \]
    equal to an $\rms l$ tensor $\ubar h_{\rms l}^{(\lambda)}(\scal)$, with $\ubar h_{\rms l}^{(\lambda)}(\scalspace_l)=\ker N_{\rms l}(\wh{\ubar L}(0),\lambda)$. The constant coefficient of
    \[
      h_{b,\rms l}^{(\lambda),\leq k}:=\delta_{g_b}^*\omega_{b,\rms l}^{(\lambda-1),\leq k}
    \]
    for $k=1,2$ is $k!$ times
    \[
      \breve h_{b,\rms l}^{(\lambda),k} \in \cA^{(\lambda-k,0)\cup(\lambda-k+\cE_\ind)}(X;S^2\cT^*_X).
    \]
  \item\label{ItWG0Largesl2}{\rm ($\rms l$ roots $-l+2$, $l\geq 2$.)} For $2\leq l\leq 5$, there exist $\omega_{b,\rms l}^{(-l+1)}(\scal)\in\cA^{(-l+1,0)\cup(-l+1+\cE_\ind)}(X;\cT^*_X)$ and $\breve\omega_{b,\rms l}^{(-l+1),k}(\scal)\in\cA^{(-l+1-k,0)\cup(-l+1-k+\cE_\ind)}(X;\cT^*_X)$, $k=1,\ldots,5-l$, depending linearly on $\scal\in\scalspace_l$ and with scalar type $l$ leading-order terms, such that $\omega_{b,\rms l}^{(-l+1),\leq k}\in\ker\Box_{g_b,E^\Ups}^\Ups$, and with the leading-order terms $\ubar h_{\rms l}^{(-l+2)}(\scal)$ of
    \[
      h_{b,\rms l}^{(-l+2)}(\scal) := \delta_{g_b}^*\omega_{b,\rms l}^{(-l+1)}(\scal) \in \cA^{(-l+2,0)\cup(-l+2+\cE_\ind)}(X;S^2\cT^*_X)
    \]
    spanning $\ker N_{\rms l}(\wh{\ubar L}(0),-l+2)$. The constant coefficient of
    \[
      h_{b,\rms l}^{(-l+2),\leq k}(\scal) := \delta_{g_b}^*\omega_{b,\rms l}^{(-l+1),\leq k}(\scal)
    \]
    for $k=1,\ldots,5-l$ is $k!$ times
    \[
      \breve h_{b,\rms l}^{(-l+2),k}(\scal) \in \cA^{(-l+2-k,0)\cup(-l+2-k+\cE_\ind)}(X;S^2\cT^*_X).
    \]
  \item\label{ItWG0Largevl2}{\rm ($\rmv l$ roots $-l+1$, $l\geq 2$.)} For $l=2,3,4$, there exist $\omega_{b,\rmv l}^{(-l)}(\vect)\in\cA^{(-l,0)\cup(-l+\cE_\ind)}(X;\cT^*_X)$ and $\breve\omega_{b,\rmv l}^{(-l),k}(\vect)\in\cA^{(-l-k,0)\cup(-l-k+\cE_\ind)}(X;\cT^*_X)$, $k=1,\ldots,4-l$, depending linearly on $\vect\in\vectspace_l$ and having vector type $l$ leading-order terms, such that $\omega_{b,\rmv l}^{(-l),\leq k}(\vect)\in\ker\Box_{g_b,E^\Ups}^\Ups$, and with the leading-order terms $\ubar h_{\rmv l}^{(-l+1)}(\vect)$ of
    \[
      h_{b,\rmv l}^{(-l+1)}(\vect) := \delta_{g_b}^*\omega_{b,\rmv l}^{(-l)}(\vect) \in \cA^{(-l,0)\cup(-l+\cE_\ind)}(X;S^2\cT^*_X)
    \]
    spanning $\ker N_{\rmv l}(\wh{\ubar L}(0),-l+1)$. The constant coefficient of
    \[
      h_{b,\rmv l}^{(-l+1),\leq k}(\vect) := \delta_{g_b}^*\omega_{b,\rmv l}^{(-l),\leq k}(\vect)
    \]
    for $k=1,\ldots,4-l$ is $k!$ times
    \[
      \breve h_{b,\rmv l}^{(-l+1),k}(\vect) \in \cA^{(-l+1-k,0)\cup(-l+1-k+\cE_\ind)}(X;S^2\cT^*_X).
    \]
  \end{enumerate}
  All 1-forms in this statement can be chosen to depend smoothly on $b$ near $b_0$.
\end{prop}

For each indicial root $\lambda$ in Lemma~\ref{LemmaWGInd} (all of which are real) with $\lambda<-1$, there exists an element in $\ker\wh{\Box_{g_b,E^\Ups}^\Ups}(0)$ with $\rho^\lambda$-growth and leading-order term given by an indicial solution corresponding to the indicial root $\lambda$. The symmetric gradient thereof is a pure gauge state with nontrivial $\rho^{\lambda+1}$ leading-order term. This is how the stationary states in parts~\eqref{ItWG0Larges01}--\eqref{ItWG0Largevl2} arise (as explained in detail in the proof below). The indicial roots $\lambda=-1,0$ are special:
\begin{enumerate}
\item The scalar type $0$ zero energy state $\omega_{b,\rms 0}^{(0)}$ corresponding to $\lambda=0$ is a Killing 1-form, so its symmetric gradient vanishes.
\item The scalar type $1$ state $\omega_{b,\rms 1}^{(0)}$ (corresponding to $\lambda=0$) and the vector type $1$ state $\omega_{b,\rmv 1}^{(-1)}$ (with $\lambda=-1$) are asymptotically Killing, so their symmetric gradients have better than $\rho^{\lambda+1}$-decay (as discussed in Proposition~\ref{PropWG0Symm}).
\end{enumerate}

Nonetheless, looking ahead at Lemma~\ref{LemmaWEInd}, the linearized gauge-fixed Einstein operator has an $\rms 0$ indicial root $0$ (even though $-1$ is not an indicial root of $\wh{\Box_{g_b,E^\Ups}^\Ups}(0)$), and an $\rms 1$ and $\rmv 1$ indicial root $-1$ (even though $-2$ is not an $\rms 1$ or $\rmv 1$ indicial root). The former corresponds to the stationary large zero energy state $h_{b,\rms 0}^{(0)}$ constructed in part~\eqref{ItWG0Larges00}. The latter two do \emph{not} correspond to stationary states on Kerr. Rather, the indicial solution for the $\rms 1$ root $-1$ is a stationary tensor $\ubar h_{\rms 1}^2(\scal)=\ubar\delta^*(t_*^2\ubar\omega_{\rms 1}^{(0)}(\scal)+2 t_*\breve{\ubar\omega}_{\rms 1}^{(0),1}(\scal)+2\breve{\ubar\omega}_{\rms 1}^{(0),2}(\scal))\in\ker\ubar\Box_{\ubar E^\Ups}^\Ups$ for an appropriate $\breve{\ubar\omega}_{\rms 1}^{(0),2}$ (homogeneous of degree $-2$); this stationarity is due to $t_*\ubar\omega_{\rms 1}^{(0)}(\scal)+\breve{\ubar\omega}_{\rms 1}^{(0),1}(\scal)$ being a Killing 1-form (namely, a Lorentz boost). The corresponding Kerr state, by contrast, is quadratic in $t_*$, with $t_*^2$- and $t_*^1$-terms that decay in space, while the $t_*^0$-term is of size $\cO(\rho^{-1})$ with leading-order term $\ubar h_{\rms 1}^2(\scal)$ at $\rho=0$. (This was already discussed around~\eqref{EqIEinhs12}.) The $\rmv 1$ root $-1$ arises similarly as the $t_*^0$-term of a typically linear-in-$t_*$ generalized zero energy state related to $\omega_{b,\rmv 1}^{(-1)}$.

\begin{rmk}[More on the $\rmv 1$ root $-1$]
\label{RmkWG0Largev1}
  The situation in the $\rmv 1$ case is in fact more subtle. Write $b=(\bhm,\bha)$ and note that $\omega_{b,\rmv 1}^{(-1)}(\vect)$ is a Killing 1-form when $\bha=0$; this means that in this case
  \begin{equation}
  \label{EqWG0Largev1}
    h_{b,\rmv 1}^{\leq 1}(\vect) = \breve h_{b,\rmv 1}^1(\vect) = \dd t_*\otimes_s\omega_{b,\rmv 1}^{(-1)}(\vect) + \wh{\delta_{g_b}^*}(0)\breve\omega_{b,\rmv 1}^{(-1),1}(\vect) \in \cA^{(-1,0)\cup(-1+\cE_\ind)}(X;S^2\cT^*_X)
  \end{equation}
  is, in fact, a large zero energy state. When $\bha\neq 0$, then $\omega_{b,\rmv 1}^{(-1)}(\vect)$ is Killing only when $\vect$ is a multiple of $\vect(\bha)$ (in the notation of Definition~\ref{DefTYs1}), since in this case $\omega_{b,\rmv 1}^{(-1)}(\vect)$ is dual to the generator of rotations around the axis of rotation $\frac{\bha}{|\bha|}$ of the black hole (cf.\ Remark~\ref{RmkWG0Killing}), and then~\eqref{EqWG0Largev1} again holds. For all other $\vect$ however, $h_{b,\rmv 1}^{\leq 1}(\vect)$ has a non-zero $t_*$-coefficient. The difference in the dimensions of the spaces of large zero energy states of size $\cO(\rho^{-1})$ between the cases $\bha=0$ and $\bha\neq 0$ will be matched by a difference in the dimensions of the spaces of elements in the cokernel of $\wh{L_b}(0)$ with $o(\rho^2)$ decay; see the discussion of~\eqref{EqWC0hv1} below.
\end{rmk}

\begin{proof}[Proof of Proposition~\usref{PropWG0Large}]
  We write
  \[
    \Box_b:=\Box_{g_b,E^\Ups}^\Ups.
  \]
  for better readability. The smoothness of all constructions in $b$ follows as in the proof of Proposition~\ref{PropWG0Symm}; we shall not comment on this further. It is also convenient to decompose
  \[
    \Box_b = A_0 + A_1\pa_{t_*} + A_2\pa_{t_*}^2,\quad A_0=\wh{\Box_b}(0)\in\rho^2\Diffb^2,\ A_1\in\rho\Diffb^1,\ A_2=-g^{0 0}\in\rho^2\CI;
  \]
  the memberships follow from~\eqref{EqWGOp1}. Similarly, we shall frequently use
  \begin{equation}
  \label{EqWG0SymmGrad}
    \delta_{g_b}^* = \underbrace{\wh{\delta_{g_b}^*}(0)}_{\in\rho\Diffb^1} + \underbrace{\dd t_*\otimes_s(\cdot)}_{\in\CI}.
  \end{equation}

  \pfstep{Part~\eqref{ItWG0Larges00}.} Since $\delta_{g_b}^*\omega_{b,\rms 0}^{(0)}=0$, we have $\delta_{g_b}^*(t_*\omega_{b,\rms 0}^{(0)})=\dd t_*\otimes_s\omega_{b,\rms 0}^{(0)}\in\CI(X;S^2\cT^*_X)$, and therefore $\Box_b(t_*\omega_{b,\rms 0}^{(0)})\in\rho\CI=\cA^{(1,0)}$, with $\rms 0$ leading- and sub-leading-order term. We can write this as $-\wh{\Box_b}(0)\breve\omega_{b,\rms 0}^{(0),1}$ for a 1-form $\breve\omega_{b,\rms 0}^{(0),1}\in\cA^{(-1,0)\cup(0,1)\cup(1+\cE_\ind)}$ whose $\rho^{-1}$-, $\log\rho$-, and $\rho^0$-terms are determined using Lemma~\ref{LemmaTMFormal}, while the $\cA^{1+\cE_\ind}$-remainder of $\breve\omega_{b,\rms 0}^{(0),1}$ arises from Lemma~\ref{LemmaWG0Inv}. We simplify the index set of $\breve\omega_{b,\rms 0}^{(0),1}$ to $(-1,0)\cup(-1+\cE_\ind)$. Next, we compute
  \[
    \delta_{g_b}^*\bigl(t_*\omega_{b,\rms 0}^{(0)}+\breve\omega_{b,\rms 1}^{(0),1}\bigr) = \dd t_*\otimes_s\omega_{b,\rms 0}^{(0)} + \wh{\delta_{g_b}^*}(0)\breve\omega_{b,\rms 1}^{(0),1} \in (\CI + \cA^{(0,0)\cup\cE_\ind})(X;S^2\cT^*_X).
  \]
  Finally, note that $\wh{L_b}(0)h_{b,\rms 0}^{(0)}=0$ implies that the leading-order ($\rho^0$) term $\ubar h_{\rms 0}^{(0)}$ of $h_{b,\rms 0}^{(0)}$ lies in $\ker N(\wh{L_b}(0),0)=\ker N(\wh{\ubar L}(0),0)$, which is a $1$-dimensional space. Moreover, we have $\ubar h_{\rms 0}^{(0)}\neq 0$ since the leading-order term $\dd t_*\otimes_s\ubar\omega_{\rms 0}^{(0)}=-\dd t_*\otimes_s\dd t$, which under the identification~\eqref{EqTYSplit2} of $\rms 0$-tensors with elements of $\R^4$ is equal to $(0,-\frac14,-\frac12,0)$, does not lie in the range of $N_{\rms 0}(\wh{\ubar\delta^*}(0),-1)$ in~\eqref{EqTYdels0}.

  Suppose we have constructed $\breve\omega_{b,\rms 0}^{(0),j}$ for $j<k\in\{1,2,3,4\}$, with $\Box_b\omega_{b,\rms 0}^{(0),\leq j}=0$. Writing this out in powers of $t_*$, the $t_*^j$-coefficient reads $A_0\omega_{b,\rms 0}^{(0)}=0$, the $t_*^{j-1}$-coefficient reads $A_0\breve\omega_{b,\rms 0}^{(0),1}+A_1\omega_{b,\rms 0}^{(0)}=0$, and the $t_*^{j-l}$-coefficient for $l\geq 2$ reads
  \[
    A_0\breve\omega_{b,\rms 0}^{(0),l} + A_1\breve\omega_{b,\rms 0}^{(0),l-1} + A_0\breve\omega_{b,\rms 0}^{(0),l-2} = 0,
  \]
  where we set $\breve\omega_{b,\rms 0}^{(0),0}:=\omega_{b,\rms 0}^{(0)}$. This equation holds for all $l\geq 0$ if we additionally use the convention that $\breve\omega_{b,\rms 0}^{(0),m}=0$ for $m\leq -1$. We conclude that the construction of $\breve\omega_{b,\rms 0}^{(0),k}$ requires solving
  \begin{align*}
    A_0\breve\omega_{b,\rms 0}^{(0),k} = -A_1\breve\omega_{b,\rms 0}^{(0),k-1} - A_2\breve\omega_{b,\rms 0}^{(0),k-2} &\in \rho\cA^{(-k+1,0)\cup(-k+1+\cE_\ind)} + \rho^2\cA^{(-k+2,0)\cup(-k+2+\cE_\ind)} \\
      &\subset \cA^{(-k+2,0)\cup(-k+2+\cE_\ind)}(X;\cT^*_X).
  \end{align*}
  Since $-k$ is not an $\rms 0$ indicial root of $A_0$, we can indeed find $\breve\omega_{b,\rms 0}^{(0),k}$, with scalar type $0$ leading-order term, by combining Lemmas~\ref{LemmaTMFormal} and~\ref{LemmaWG0Inv}. (At each step of this construction, the set $\cE_\ind$ may need to be enlarged, though the property $\min\Re\cE_\ind\geq\eps_\ind$ will remain valid.) We moreover compute the $t_*^0$-coefficient of $\delta_{g_b}^*\omega_{b,\rms 0}^{(0),\leq k}$ (which is a degree $(k-1)$ polynomial in $t_*$) to be equal to
  \begin{equation}
  \label{EqWG0Larges00Pf}
    \frac{k!}{0!} \wh{\delta_{g_b}^*}(0)\breve\omega_{b,\rms 0}^{(0),k} + \frac{k!}{1!} \dd t_*\otimes_s\breve\omega_{b,\rms 0}^{(0),k-1} =: \frac{(k-1)!}{0!} \breve h_{b,\rms 0}^{(0),k-1}.
  \end{equation}
  This gives~\eqref{EqWG0Larges003}.

  \pfstep{Part~\eqref{ItWG0Larges1m1}.} With $\omega_{b,\rms 1}^{(0)}(\scal)$ and $\breve\omega_{b,\rms 1}^{(0),1}(\scal)$ already constructed in Proposition~\ref{PropWG0Symm}, we can iteratively construct $\breve\omega_{b,\rms 1}^{(0),k}(\scal)$ for $k\geq 2$ by solving
  \begin{equation}
  \label{EqWG0Larges1}
    A_0\breve\omega_{b,\rms 1}^{(0),k}(\scal) = -A_1\breve\omega_{b,\rms 1}^{(0),k-1}(\scal) - A_2\breve\omega_{b,\rms 1}^{(0),k-2}(\scal) \in \cA^{(-k+2,0)\cup(-k+2+\cE_\ind)}(X;\cT^*_X)
  \end{equation}
  (with linear dependence on $\scal\in\scalspace_1$). Since the right-hand side has a scalar type $1$ leading-order term (as follows inductively) and since $-k$ is not a scalar type $1$ indicial root of $A_0=\wh{\Box_b}(0)$, this admits a solution $\breve\omega_{b,\rms 1}^{(0),k}\in\cA^{(-k,0)\cup(-k+\cE_\ind)}(X;\cT^*_X)$ by Lemmas~\ref{LemmaTMFormal} and \ref{LemmaWG0Inv}, with $\rms 1$ leading-order term. For $k=2$, we compute
  \begin{align*}
    \delta_{g_b}^*\omega_{b,\rms 1}^{(0),\leq 2} &= t_*^2 \wh{\delta_{g_b}^*}(0)\omega_{b,\rms 1}^{(0)} + 2 t_*\bigl( \dd t_*\otimes_s\omega_{b,\rms 1}^{(0),1}+\wh{\delta_{g_b}^*}(0)\breve\omega_{b,\rms 1}^{(0),1} \bigr) + 2 \bigl( \dd t_*\otimes_s\breve\omega_{b,\rms 1}^{(0),1} + \wh{\delta_{g_b}^*}(0)\breve\omega_{b,\rms 1}^{(0),2}\bigr) \\
      &= t_*^2 h_{b,\rms 1} + 2 t_*\breve h_{b,\rms 1}^1 + \breve h_{b,\rms 1}^2.
  \end{align*}
  Note then that the $\rho^{-1}$ leading-order term of $\breve\omega_{b,\rms 1}^{(0),1}$ is $\breve{\ubar\omega}_{\rms 1}^{(0),1}(\scal)=r(-\scal\,\dd(t-r)+r\,\sld\scal)$ (see~\eqref{EqWG0SymmMink}), which in the splitting~\eqref{EqTYMink01Split} and in terms of~\eqref{EqTYSplit1} equals $r$ times $(0,-1,1)$, and thus the leading-order term of $\dd t_*\otimes_s\breve\omega_{b,\rms 1}^{(0),1}$ in terms of~\eqref{EqTYSplit2} is $(0,0,0,-1,\frac12,0)$. Using the formula~\eqref{EqTYdels1} with $\lambda=-2$, the range of $N_{\rms 1}(\rho^{-1}\wh{\delta_{g_b}^*}(0),-2)$ is easily checked to not contain $(0,0,0,-1,\frac12,0)$, and therefore the leading-order term $\ubar\delta^*\ubar\omega_{\rms 1}^{(0),\leq 2}(\scal)$ of $\breve h_{b,\rms 1}^2(\scal)$ is non-zero when $\scal\neq 0$. --- A more conceptual proof using material from~\S\S\ref{SWC}--\ref{SWE} is that if $\breve h_{b,\rms 1}^2(\scal)$, $\scal\neq 0$, were of size $\cO(\rho^{-1+\eps})$ for some $\eps>0$, then taking the $L^2$-inner product of $L_b(t_*^2 h_{b,\rms 1}+2 t_*\breve h_{b,\rms 1}^1)=-\wh{L_b}(0)\breve h_{b,\rms 1}^2$ with the dual states $h_{b,\rms 1}^*(\scal)=\cO(\rho^2)$ constructed in~\eqref{EqWC0Dualh} would vanish (as follows by an integration by parts), which contradicts Lemma~\ref{LemmaWEPair}.

  For $k\geq 3$, the claimed membership of $\breve h_{b,\rms 1}^k(\scal)$ follows from
  \[
    \breve h_{b,\rms 1}^k(\scal) = \wh{\delta_{g_b}^*}(0)\breve\omega_{b,\rms 1}^{(0),k}(\scal) + \dd t_*\otimes_s\breve\omega_{b,\rms 1}^{(0),k-1}(\scal).
  \]

  \pfstep{Part~\eqref{ItWG0Largev1m1}.} With $\omega_{b,\rmv 1}^{(-1)}(\vect)$ already constructed in Proposition~\ref{PropWG0Symm}, we first construct $\breve\omega_{b,\rmv 1}^{(-1),1}(\vect)$ by solving
  \[
    A_0\breve\omega_{b,\rmv 1}^{(-1),1}(\vect) = -A_1\omega_{b,\rmv 1}^{(-1)}(\vect) \in \cA^{(0,0)\cup\cE_\ind}(X;\cT^*_X);
  \]
  the $\rho^0$ leading-order term of the right-hand side is of vector type $1$. Since $-2$ is not a $\rmv 1$ indicial root of $A_0=\wh{\Box_b}(0)$, this can be solved (using Lemmas~\ref{LemmaTMFormal} and \ref{LemmaWG0Inv} as usual) for $\breve\omega_{b,\rmv 1}^{(-1),1}(\vect)\in\cA^{(-2,0)\cup(-2+\cE_\ind)}$ with $\rmv 1$ leading-order term. The construction of $\breve\omega_{b,\rmv 1}^{(-1),k}(\vect)$ for $k\geq 2$ is then completely analogous (with only shifts in decay rates) to the arguments around~\eqref{EqWG0Larges1}.

  We next compute
  \[
    \delta_{g_b}^*\omega_{b,\rmv 1}^{(-1),\leq 1} = t_*\wh{\delta_{g_b}^*}(0)\omega_{b,\rmv 1}^{(-1)} + \bigl(\dd t_*\otimes_s\omega_{b,\rmv 1}^{(-1)} + \wh{\delta_{g_b}^*}(0)\breve\omega_{b,\rmv 1}^{(-1),1}\bigr) = t_* h_{b,\rmv 1} + \breve h_{b,\rmv 1}^1.
  \]
  We can compute the leading-order term $\breve{\ubar\omega}_{\rmv 1}^{(-1),1}(\vect)$ of $\breve\omega_{b,\rmv 1}^{(-1),1}(\vect)$ explicitly as follows. In the splitting~\eqref{EqTYMink01Split} and using the description~\eqref{EqTYSplit1} in the vector type $1$ case relative to the fixed element $0\neq\vect\in\vectspace_1$, we have $\rho\ubar\omega_{\rmv 1}^{(-1)}(\vect)=1$ simply, while $N_{\rmv 1}(\rho^{-1}A_1,\lambda)=-2(\lambda-1-\gamma^\Ups)$ (the indicial family of the $(3,3)$-component of $-2\rho(\rho\pa_{\rho}-1-\ubar S_{\ubar E^\Ups})$ from~\eqref{EqWGBoxMink}, determined using \eqref{EqWGOpMinkBox}), and $N_{\rmv 1}(\rho^{-2}A_0,\lambda)=N_{\rmv 1}(\rho^{-2}\wh{\ubar\Box}(0),\lambda)=-\lambda^2+\lambda+2$ (from Lemma~\ref{LemmaTYMinkOp}), so in view of $-N_{\rmv 1}(\rho^{-2}A_0,-2)^{-1} \rho^{-1}N_{\rmv 1}(\rho^{-1}A_1,-1) = 1+\frac{\gamma^\Ups}{2}$ we get
  \[
    \breve{\ubar\omega}_{\rmv 1}^{(-1),1}(\vect) = \rho^{-2}\Bigl(0,0,\Bigl(1+\frac{\gamma^\Ups}{2}\Bigr)\rho\vect\Bigr).
  \]
  In the $\rmv 1$ case of the description~\eqref{EqTYSplit2}, we thus compute $\ubar h_{\rmv 1}^1(\vect)=\dd t_*\otimes_s\ubar\omega_{\rmv 1}^{(-1)}+\wh{\ubar\delta^*}(0)\breve{\ubar\omega}_{\rmv 1}^{(-1),1}$ using $\dd t_*=\dd\ubar x^1$ and~\eqref{EqTYdelv1} to be given by $(0,\frac12)+N_{\rmv 1}(\rho^{-1}\wh{\ubar\delta^*}(0),-2)(1+\frac{\gamma^\Ups}{2})=\frac14(1+\frac{\gamma^\Ups}{2},1-\frac{\gamma^\Ups}{2})$, which is~\eqref{EqWG0Largev1m1hbar}.

  \pfstep{Part~\eqref{ItWG0Larges01}.} Take a nontrivial indicial solution $\ubar\omega_{\rms 0}^{(\lambda-1)}$ of $\wh{\Box_b}(0)$ corresponding to the indicial root $\lambda-1=-\lambda^\Ups_{\rms 0,1}$ (which lies in $(-2,-1)$ and is close to $-1$), and solve the equation
  \[
    \wh{\Box_b}(0)\omega' = -\wh{\Box_b}(0)\ubar\omega_{\rms 0}^{(\lambda-1)} = -\underbrace{\bigl(\wh{\Box_b}(0)-\wh{\ubar\Box}(0)\bigr)}_{\in\rho^3\Diffb^2}\ubar\omega_{\rms 0}^{(\lambda-1)} \in \rho^{\lambda+2}\CI(X;\cT^*_X).
  \]
  We can solve this as usual for $\omega'\in\cA^{(\lambda,0)\cup(1+\cE_\ind)}(X;\cT^*_X)$ (with smooth dependence on $b$), and can then set $\omega_{b,\rms 0}^{(\lambda-1)}=\ubar\omega_{\rms 0}^{(\lambda-1)}+\omega'$. Its symmetric gradient $h_{b,\rms 0}^{(\lambda)}\in\cA^{(\lambda,0)\cup(\lambda+\cE_\ind)}$ has non-zero leading-order term since $N(\wh{\ubar\delta^*}(0),\lambda-1)$ is injective (see the discussion following~\eqref{EqTYdelv1}). We then solve
  \[
    A_0\breve\omega_{b,\rms 0}^{(\lambda-1),1} = -A_1\omega_{\rms 0}^{(\lambda-1)} \in \cA^{(\lambda,0)\cup(\lambda+\cE_\ind)}(X;\cT^*_X)
  \]
  for $\breve\omega_{b,\rms 0}^{(\lambda-1),1}\in\cA^{(\lambda-2,0)\cup(\lambda-2+\cE_\ind)}$ using Lemmas~\ref{LemmaTMFormal} and \ref{LemmaWG0Inv} as usual. Similarly then, we solve
  \[
    A_0\breve\omega_{b,\rms 0}^{(\lambda-1),2} = -A_1\breve\omega_{\rms 0}^{(\lambda-1),1}-A_2\omega_{\rms 0}^{(\lambda-1)} \in \cA^{(\lambda-1,0)\cup(\lambda-1+\cE_\ind)}
  \]
  for $\breve\omega_{b,\rms 0}^{(\lambda-1),2}\in\cA^{(\lambda-3,0)\cup(\lambda-3+\cE_\ind)}$. The statement~\eqref{EqWG0Larges01hbreve} follows from~\eqref{EqWG0SymmGrad} as usual.

  \pfstep{Part~\eqref{ItWG0Largeslj}.} This is proved using exactly the same arguments as for part~\eqref{ItWG0Larges01}.

  \pfstep{Part~\eqref{ItWG0Largesl2}.} We again start with a nontrivial indicial solution $\ubar\omega_{\rms l}^{(-l+1)}(\scal)$, $\scal\in\scalspace_l$, of $\wh{\ubar L}(0)$, and construct $\omega_{b,\rms l}^{(-l+1)}(\scal)$ as the sum of $\ubar\omega_{\rms l}^{(-l+1)}(\scal)$ and a solution of $\wh{\Box_b}(0)\omega'=-\wh{\Box_b}(0)\ubar\omega_{\rms l}^{(-l+1)}(\scal)\in\rho^{-l+3}\CI(X;\cT^*_X)$; potential logarithmic terms of the solution (obtained via Lemmas~\ref{LemmaTMFormal} and \ref{LemmaWG0Inv}) are subsumed into the index set $-l+1+\cE_\ind$, and then $h_{b,\rms l}^{(-l+2)}(\scal)=\wh{\delta_{g_b}^*}(0)\omega_{b,\rms l}^{(-l+1)}(\scal)$. The construction of $\breve\omega_{b,\rms l}^{(-l+1),j}(\scal)$ proceeds as before.

  \pfstep{Part~\eqref{ItWG0Largevl2}.} The same arguments apply, \emph{mutatis mutandis}.
\end{proof}

\section{Constraint propagation wave operator}
\label{SWC}

The modification
\begin{equation}
\label{EqWCDelta}
  \delta_{g_b,E^\cC}^*=\delta_{g_b}^*+E^\cC_{g_b},\quad
  E^\cC_{g_b} \colon \omega \mapsto \gamma^\cC\bigl(2\cd^\cC\otimes_s\omega - (1-e^\cC)g_b \iota_{g_b^{-1}(\cd^\cC)}\omega \bigr),
\end{equation}
of the symmetric gradient in Definition~\ref{Def1Symm} (and consequently in the gauge-fixed Einstein operator in~\eqref{Eq1Ein}) we use in this paper is not perturbative. While in \cite[Proposition~10.12]{HaefnerHintzVasyKerr} (small angular momenta) and \cite[Proposition~3.7]{HintzGlueLocIII} (full subextremal range), perturbative modifications suffice for eliminating the zero energy mode of the 1-form wave operator $\Box_{g_b}$, the requirements for our stability proof are considerably stronger: we need to widen the indicial gap of $\Box_{g_b}$ from $(0,1)$ to an indicial gap $[-C_0,1)$ for
\begin{equation}
\label{EqWCBox}
  \Box_{g_b,E^\cC}^\cC = 2\delta_{g_b}\sfG_{g_b}\delta_{g_b,E^\cC}^*,
\end{equation}
for a sufficiently large constant $C_0$ (with $C_0=100$ being sufficient); this is called ``enhanced mode stability at zero energy'' in \cite{HintzKerrCD}. That this (enhanced) mode stability can be arranged is shown in the companion paper \cite{HintzKerrCD}; we recall precise statements in~\S\ref{SsWCRec}. In~\S\ref{SsWC0}, we record important consequences regarding large dual states of $\Box_{g_b,E^\cC}^\cC$, whose trace-reversed symmetric gradients will be shown to span the cokernel of the linearized gauge-fixed Einstein operator in~\S\ref{SsWEMode}.

\subsection{Indicial roots, tf-admissibility, mode stability}
\label{SsWCRec}

We begin with preliminary observations regarding the structure of $\Box_{g_b,E^\cC}^\cC$ that mirror those in~\S\ref{SWG}. First of all, since the 1-form $\cd^\cC$ in Definition~\ref{Def1Symm} is of class $\rho\CI(X;\cT^*_X)$, the operator $\Box_{g_b,E^\cC}^\cC$ is a stationary wave-type operator in the sense of \citeAF{Definition~\ref*{DefSSAdm}}; its spectral family is thus of the form
\[
  \wh{\Box_{g_b,E^\cC}^\cC}(\sigma) = 2 i\sigma\rho(\rho\pa_\rho-1-S_{E^\cC}) + \wh{\Box_{g_b,E^\cC}^\cC}(0) - i\sigma Q + g^{0 0}\sigma^2
\]
(analogously to~\eqref{EqWGOpSpecFam}). To leading order at $r=\infty$ (in the analogous sense to Lemma~\ref{LemmaWGOp}\eqref{ItWGOpMink}), the operator $\Box_{g_b,E^\cC}^\cC$ is given by the Minkowskian operator
\begin{align}
  &\ubar\Box_{E^\cC}^\cC := 2\ubar\delta\ul\sfG\,\ubar\delta_{\ubar E^\cC}^*, \nonumber\\
\label{EqWCRecE}
  &\quad \ubar\delta_{\ubar E^\cC}^* := \ubar\delta^*+\ubar E^\cC,\quad \ubar E^\cC \colon h \mapsto \gamma^\cC \bigl( 2 \ubar\cd^\cC\otimes_s\omega - (1-e^\cC)\ubar g \iota_{\ubar g^{-1}(\ubar\cd^\cC)}\omega \bigr),\quad \ubar\cd^\cC := r^{-1}(\dd t-v^\cC\,\dd r).
\end{align}
Its $\tface$-normal operator is defined in the same fashion as in~\S\ref{SsWGtf}: writing
\[
  \wh{\ubar\Box_{\ubar E^\cC}^\cC}(0) = \rho^2\wt{\ubar\Box_{\ubar E^\cC}^\cC}(0)(\omega,\rho\pa_\rho,\pa_\omega)
\]
for the zero energy operator of $\ubar\Box_{E^\cC}^\cC$, we have
\begin{equation}
\label{EqWCRecNtf}
  N_\tface(\Box_{g_b,E^\cC}^\cC,\hat\sigma) = 2 i\hat\sigma\hat\rho(\hat\rho\pa_{\hat\rho}-1-\ubar S_{\ubar E^\cC}) + \hat\rho^2\wt{\ubar\Box_{\ubar E^\cC}^\cC}(0)(\omega,\hat\rho\pa_{\hat\rho},\pa_\omega).
\end{equation}
where $\ubar S_{\ubar E^\cC}$ is given by \citeCD{(\ref*{EqM00S})} (with the roles of $v,e,h$ in the reference played by $v^\cC,e^\cC,\frac{1}{\gamma^\cC}$ in present notation); for later use, we record the explicit expressions in the bundle splitting~\eqref{EqTYMink01Split} to be
\begin{align}
\label{EqWCRecEC}
  \ubar E^\cC &= \gamma^\cC\rho\begin{pmatrix} 1-v^\cC & 0 & 0 \\ \frac12 e^\cC(1+v^\cC) & \frac12 e^\cC(1-v^\cC) & 0 \\ 0 & 0 & \frac12(1-v^\cC) \\ 0 & 1+v^\cC & 0 \\ 0 & 0 & \frac12(1+v^\cC) \\ (1-e^\cC)(1+v^\cC)\slg & (1-e^\cC)(1-v^\cC)\slg & 0 \end{pmatrix}, \\
  \ubar S_{\ubar E^\cC} &= \gamma^\cC\begin{pmatrix} 2(1-v^\cC) & 0 & 0 \\ (1-e^\cC)(1+v^\cC) & (1-e^\cC)(1-v^\cC) & 0 \\ 0 & 0 & 1-v^\cC \end{pmatrix}, \nonumber\\
\label{EqWCRecBox0}
  \wt{\ubar\Box_{\ubar E^\cC}^\cC}(0) &= \rho^{-2}\wh{\ubar\Box}(0) + \gamma^\cC\begin{pmatrix} q_{0 0} & q_{0 1} & q_{0 /} \\ q_{1 0} & q_{1 1} & q_{1 /} \\ q_{/ 0} & q_{/ 1} & q_{/ /} \end{pmatrix}
\end{align}
where the zero energy operator $\wh{\ubar\Box}(0)$ of the 1-form wave operator on the Minkowski spacetime is given in Lemma~\ref{LemmaTYMinkOp}, and the operators $q_{\bullet\bullet}=q_{\bullet\bullet}(\omega,\rho\pa_\rho,\pa_\omega)$ are given by
\begin{align*}
  q_{0 0} &= (1-3 v^\cC+e^\cC+e^\cC v^\cC)\rho\pa_\rho + (-1+3 v^\cC + e^\cC + e^\cC v^\cC) , \\
  q_{1 0} &= (1-e^\cC)(1+v^\cC)\rho\pa_\rho - (1+e^\cC)(1+v^\cC), \\
  q_{/ 0} &= -2 e^\cC(1+v^\cC)\sld, \\
  q_{0 1} &= -(1-e^\cC)(1-v^\cC)\rho\pa_\rho + (1+e^\cC)(1-v^\cC), \\
  q_{1 1} &= (-1-3 v^\cC-e^\cC+e^\cC v^\cC)\rho\pa_\rho + (1+3 v^\cC-e^\cC+e^\cC v^\cC), \\
  q_{/ 1} &= -2 e^\cC(1-v^\cC)\sld, \\
  q_{0 /} &= (1-v^\cC)\sldelta, \\
  q_{1 /} &= (1+v^\cC)\sldelta, \\
  q_{/ /} &= -2 v^\cC\rho\pa_\rho+4 v^\cC.
\end{align*}
Note that the smallest eigenvalue of $\ubar S_{\ubar E^\cC}$ equals $\gamma^\cC(1-e^\cC)(1-v^\cC)$. Moreover, the expression for $\ubar E^\cC$ already appeared in~\eqref{EqExOpLinEC0}. We summarize the main results from \cite{HintzKerrCD} (in the slightly weakened form sufficient for our purposes) as follows.

\begin{thm}[Constraint propagation wave operator]
\label{ThmWCRec}
  Fix $v^\cC\in(0,1)$ and $C_0>0$. Then there exist $e^\cC\in(0,1)$ and a stationary past timelike 1-form $\cd^\cC\in\rho\CI(X;\cT^*_X)$, with $\cd^\cC=r^{-1}(\dd t-v^\cC\,\dd r)$ for all sufficiently large $r$, such that for all sufficiently large $\gamma^\cC>1$, the following statements hold for the operator $\Box_{g_b,E^\cC}^\cC$ defined by~\eqref{EqWCDelta}--\eqref{EqWCBox}, provided $b=(\bhm,\bha)$ sufficiently close to $b_0=(\bhm_0,\bha_0)$.
  \begin{enumerate}
  \item\label{ItWCRecNon0}{\rm (Mode stability, $\sigma\neq 0$: \citeCD{Theorem~\ref*{ThmT}(\ref*{ItTNonzero})}.)} For $\sigma\in\C$, $\sigma\neq 0$, $\Im\sigma\geq 0$, the operator $\wh{\Box_{g_b,E^\cC}^\cC}(\sigma)$ has trivial nullspace on $\cA^\gamma(X;\cT^*_X)$ for all $\gamma\in\R$.
  \item\label{ItWCRec0}{\rm (Enhanced mode stability, $\sigma=0$: \citeCD{Theorem~\ref*{ThmT}(\ref*{ItTZero})}.)} For all $\alpha\in[-C_0-\frac32,-\frac12)$ and $s\geq -C_0$, the operator
    \begin{equation}
    \label{EqWCRec0}
      \wh{\Box_{g_b,E^\cC}^\cC}(0) \colon \bigl\{ \omega\in\bar H_\bop^{s,\alpha}(X;\cT^*_X) \colon \wh{\Box_{g_b,E^\cC}^\cC}(0)\omega \in \bar H_\bop^{s-1,\alpha+2}(X;\cT^*_X) \bigr\} \to \bar H_\bop^{s-1,\alpha+2}(X;\cT^*_X)
    \end{equation}
    is invertible; here, we use a Euclidean volume density to define $L^2(X)$.
  \item\label{ItWCRecInd}{\rm (Indicial roots: \citeCD{Theorem~\ref*{ThmM0}}.)} The indicial roots $\lambda\in\C$ of $\wh{\Box_{g_b,E^\cC}^\cC}(0)$ satisfy $\Re\lambda\geq 1$ or $\Re\lambda<-C_0$ and accumulate only at $\pm\infty$. The only indicial root with $\Re\lambda=1$ is $\lambda=1$, and this is a simple scalar type $0$ and a simple scalar type $1$ root in the terminology of Definition~\usref{DefTYIndRoot}.
  \item\label{ItWCRectf}{\rm (tf-admissibility: \citeCD{Theorem~\ref*{ThmMiInv}}.)} The operator $\Box_{g_b,E^\cC}^\cC$ is $\tface$-admissible for every weight $\beta\in[-C_0,1)$ in the sense of \citeAF{Definition~\ref*{DefSStfAdm}}.
  \end{enumerate}
\end{thm}

One can also verify that $\Box_{g_b,E^\cC}^\cC$ is strongly trapping admissible in the sense of \citeAF{Definition~\ref{DefSSTrapAdm}}, though we do not need this information here. (One could then upgrade Theorem~\ref{ThmWCRec}\eqref{ItWCRecNon0} to the invertibility of $\wh{\Box_{g_b,E^\cC}^\cC}(\sigma)$ between the spaces in \citeAF{(\ref*{EqSpBMap})}.) We organize the indicial roots of $\wh{\Box_{g_b,E^\cC}^\cC}(0)$ (or equivalently of $\wh{\ubar\Box_{\ubar E^\cC}^\cC}(0)$) according to their type by
\begin{equation}
\label{EqWCIndRoots}
\begin{split}
    \text{$\rms 0$ roots:}\quad & \lambda^\cC_{\rms 0,\ll,1},\ \lambda^\cC_{\rms 0,\ll,2},\ 1,\ \lambda_{\rms 0,+}^\cC, \\
    \text{$\rms 1$ roots:}\quad & \lambda^\cC_{\rms 1,\ll, 1},\ \lambda^\cC_{\rms 1,\ll, 2},\ \lambda^\cC_{\rms 1,\ll, 3},\ 1,\ \lambda^\cC_{\rms 1,+,1},\ \lambda^\cC_{\rms 1,+,2}, \\
    \text{$\rms l$ roots, $l\geq 2$:}\quad & \lambda^\cC_{\rms l,\ll, 1},\ \lambda^\cC_{\rms l,\ll,2},\ \lambda^\cC_{\rms l,\ll,3},\ \lambda^\cC_{\rms l,+,1},\ \lambda^\cC_{\rms l,+,2},\ \lambda^\cC_{\rms l,+,3}, \\
    \text{$\rmv l$ roots, $l\geq 1$:}\quad & \lambda^\cC_{\rmv l,\ll},\ \lambda^\cC_{\rmv l,+}\quad (\text{$l=1$:}\ {-}1-2 v^\cC\gamma^\cC,\ 2).
\end{split}
\end{equation}
Here the roots with subscript ``$\ll$'' have real parts $<-C_0$, and those with subscript ``$+$'' have real parts $>1$.

\begin{proof}[Proof of Theorem~\usref{ThmWCRec}]
  We shall apply the results of \cite{HintzKerrCD} for the parameters $b=b_0$. Parts~\eqref{ItWCRecInd} and \eqref{ItWCRectf} only see the Minkowskian model and are thus independent of $b$. The statements of parts~\eqref{ItWCRecNon0} and \eqref{ItWCRec0} do depend on $b$, but once they have been established for $b=b_0$, they follow for nearby $b$ by the usual Fredholm perturbation arguments.

  For part~\eqref{ItWCRecNon0}, we note that \citeCD{Theorem~\ref*{ThmT}(\ref*{ItTNonzero})} asserts the triviality of all solutions $\omega$ of $\Box_{g_b,E^\cC}^\cC\omega=0$ which are of the form $\omega=e^{-i\sigma(\ft-r_*)}\omega_1$ where $\omega_1\in\cA^\gamma(X;\cT^*_X)$; see \citeCD{Definition~\ref*{DefTOutgoing}}, and note that the coordinate $t$ used there is the Boyer--Lindquist coordinate $\ft$ from~\eqref{EqKMetBLCoord} in present notation. But $e^{-i\sigma(\ft-r_*)}=e^{-i\sigma t_*}e^{-i\sigma F}$ where $F(r)=\ft-r_*-t_*-c\in\rho\CI(X)$ for some constant $c\in\R$ (as follows from~\eqref{EqKMetTstar} and \eqref{EqKMetTime}), so $e^{-i\sigma F}$ is smooth on $X$. Thus, \citeCD{Theorem~\ref*{ThmT}(\ref*{ItTNonzero})} shows that if $\omega=e^{-i\sigma t_*}\omega_2$ with $\omega_2\in\cA^\gamma(X;\cT^*_X)$ and $\Box_{g_b,E^\cC}^\cC\omega=0$, or equivalently $\wh{\Box_{g_b,E^\cC}^\cC}(\sigma)\omega_2=0$, then $\omega_2=0$.

  Parts~\eqref{ItWCRec0} and \eqref{ItWCRecInd} are re-statements of \citeCD{Theorem~\ref*{ThmT}(\ref*{ItTZero})} (see also \citeCD{Theorem~\ref*{ThmM0}}).

  We claim that part~\eqref{ItWCRectf} follows from \citeCD{Theorem~\ref*{ThmMiInv}}, in which the notation $\wh{\Box_g^{\cC_h}}(\sigma)$ (see~\citeCD{(\ref*{EqMiOpB})}) is used for what is $N_\tface(\Box_{g_b,E^\cC}^\cC,\hat\sigma)$ in present notation (with $\hat\sigma=\sigma\in e^{i[0,\pi]}$). To prove this, consider first an element $\omega\in\cA^{\alpha,-\beta}(\tface;\pi_\tface^*(\cT^*_X))$ in the kernel of $N_\tface(\Box_{g_b,E^\cC}^\cC,\sigma)$ (with $\alpha\in\R$ arbitrary), then in fact $\omega\in\cA^{1+\gamma^\cC(1-e^\cC)(1-v^\cC)-\eps,-\beta}$ for all $\eps>0$ by a normal operator argument (cf.\ the discussion of the spectrum of $\ubar S_{\ubar E^\cC}$ after~\eqref{EqWCRecNtf}, and see also \citeAF{Theorem~\ref*{ThmSptf}(\ref*{ItSptfKer})} and its proof). This is then further contained in the space $H_{\scop,\bop}^{s,(\ell,\beta')}(\tface;\pi_\tface^*(\cT^*_X))$ for all $s\in\R$ and $\ell<-\frac12+\gamma^\cC(1-e^\cC)(1-v^\cC)$ and for $\beta':=-\beta+\frac32-\eps$ if we use the density $\hat\rho^{-3}|\frac{\dd\hat\rho}{\hat\rho}\,\dd\slg|$ on $\tface$, the shift by $\frac32$ in the weights being due to the weight $\hat\rho^{-3}$ of this density. But it is the triviality of $\ker N_\tface(\Box_{g_b,E^\cC}^\cC,\hat\sigma)$ on this space that is asserted in \citeCD{Theorem~\ref*{ThmMiInv}}.

  The triviality of the kernel of $N_\tface(\Box_{g_b,E^\cC}^\cC,1)^*$ on $e^{-2 i/\rho_\sctface}\cA^{\alpha,-1+\beta}(\tface;\pi_\tface^*(\cT^*_X))$ follows in a similar fashion. Indeed, if $\omega\in e^{-2 i/\rho_\sctface}\cA^{\alpha,-1+\beta}(\tface)$ lies in $\ker N_\tface(\Box_{g_b,E^\cC}^\cC,1)^*$ for some $\alpha\in\R$, then one can improve $\alpha$ to $1+\gamma^\cC(1-e^\cC)(1+v^\cC)-\eps$ for all $\eps>0$ (cf.\ the arguments following \citeCD{(\ref*{EqMiFredZero})}, which amount to computing the analogue of $\ubar S_{\ubar E^\cC}$ for the operator $e^{2 i/\hat\rho}N_\tface(\Box_{g_b,E^\cC}^\cC,1)^*e^{-2 i/\hat\rho}$ and showing that the bottom of its spectrum is $\gamma^\cC(1-e^\cC)(1+v^\cC)$); and then one concludes that $\omega\in H_{\scop,\bop}^{-s+2,(-\ell-1,-\beta'+2)}(\tface;\pi_\tface^*(\cT^*_X))$ for all $s\in\R$, for $\beta'=-\beta+\frac32-\eps$ as before, and for all (variable scattering decay orders) $\ell$ such that $-\ell-1<-\frac12+\gamma^\cC(1-e^\cC)(1+v^\cC)$ at the set $\cR_{\rm in}$ (the graph over $\hat\rho=0$ of $\dd(-2 r)$) in the notation of \citeCD{Theorem~\ref*{ThmMiInv}}); the latter bound precisely matches the lower bound in~\citeCD{(\ref*{EqMiInvRange})}.
\end{proof}

\subsection{Large zero energy states of the adjoint operator}
\label{SsWC0}

\emph{We fix $E^\cC$ according to Theorem~\usref{ThmWCRec} for fixed choices of $v^\cC\in(0,1)$ and $C_0>0$.} We record a dual statement of~\eqref{EqWCRec0}:

\begin{lemma}[Zero energy operator adjoint]
\fakephantomsection
\label{LemmaWC0Dual}
  \begin{enumerate}
  \item{\rm (Invertibility.)} Use an unweighted b-density (such as $|\frac{\dd\rho}{\rho}\,\dd\slg|$) to define $L^2(X)$, but use the spatial volume density $|\dd g_b|_X|$ (or any smooth positive multiple of the Euclidean density) to define $L^2$-adjoints. Then for all $\alpha\in(0,C_0+1]$ and $s\leq C_0+1$, the operator
    \begin{equation}
    \label{EqWC0Dual}
      \wh{\Box_{g_b,E^\cC}^\cC}(0)^* \colon \bigl\{ \omega \in \dot H_\bop^{s,\alpha}(X;\cT^*_X) \colon \wh{\Box_{g_b,E^\cC}^\cC}(0)^*\omega \in \dot H_\bop^{s-1,\alpha+2}(X;\cT^*_X) \bigr\} \to \dot H_\bop^{s-1,\alpha+2}(X;\cT^*_X)
    \end{equation}
    is invertible.
  \item{\rm (Indicial roots.)} The indicial roots of $\wh{\Box_{g_b,E^\cC}^\cC}(0)^*$ are equal to $1-\bar\lambda$ where $\lambda$ runs over the indicial roots of $\wh{\Box_{g_b,E^\cC}^\cC}(0)$ (given in~\eqref{EqWCIndRoots}). In particular, the indicial roots of $\wh{\Box_{g_b,E^\cC}^\cC}(0)^*$ are $0$ (scalar type $0$ or $1$), have negative real part, or have real part $>1+C_0$.
  \end{enumerate}
\end{lemma}
\begin{proof}
  The dual space (with respect to $|\dd g_b|_X|$) of the target space of~\eqref{EqWCRec0} is $\dot H_\bop^{-s+1,-\alpha-2}$. Switching to an unweighted b-density, this is $\dot H_\bop^{-s+1,-\alpha-\frac12}$, with $-\alpha-\frac12\in(0,C_0+1]$ and $-s+1\leq C_0+1$.

  For the second statement, write $\Box=\wh{\Box_{g_b,E^\cC}^\cC}(0)$; writing $*$ for adjoints with respect to $\rho^{-3}\,|\frac{\dd\rho}{\rho}\,\dd\slg|$ or $\dd\slg$ and $*_\bop$ for adjoints with respect to the unweighted b-density $|\frac{\dd\rho}{\rho}\,\dd\slg|$, we have
  \begin{equation}
  \label{EqWC0DualIndRoot}
  \begin{split}
    N(\rho^{-2}\Box,\lambda)^* &= (\rho^{-\lambda}N(\rho^{-2}\Box)\rho^\lambda)^* = \rho^{\bar\lambda} N(\rho^{-2}\Box)^{*_\bop} \rho^{-\bar\lambda} = \rho^{\bar\lambda}\rho^{-3} N(\rho^{-2}\Box)^*\rho^3\rho^{-\bar\lambda} \\
      &= \rho^{\bar\lambda}\rho^{-3} \rho^2 N(\rho^{-2}\Box^*) \rho^{-2} \rho^3\rho^{-\bar\lambda} = N(\rho^{-2}\Box^*,1-\bar\lambda).
  \end{split}
  \end{equation}
  This implies the claim.
\end{proof}

Note that since
\begin{equation}
\label{EqWC0DualOp}
  (\Box_{g_b,E^\cC}^\cC)^* = 2 (\delta_{g_b,E^\cC}^*)^*\sfG_{g_b}\delta_{g_b}^*,
\end{equation}
Killing 1-forms on Minkowski space will give rise to large zero energy states of $(\Box_{g_b,E^\cC}^\cC)^*$, and their trace-reversed symmetric gradients decay as $r\to\infty$ and lie in the kernel of $L_{g_b}^*$ (see~\eqref{EqWEModeAdj} below for the explicit expression); we discuss the relevance of these dual pure gauge states in~\S\ref{SsWEMode}. We have the following analogue of Proposition~\ref{PropWG0Symm}:

\begin{prop}[Zero energy dual states]
\label{PropWC0Dual}
  Recall $\ubar\omega_{\rms 0}^{(0)}$, $\ubar\omega_{\rms 1}^{(0)}(\scal)$, $\breve{\ubar\omega}_{\rms 1}^{(0),1}(\scal)$, and $\ubar\omega_{\rmv 1}^{(-1)}(\vect)$ from equation~\eqref{EqWG0SymmMink}.
  \begin{enumerate}
  \item{\rm (States.)} For Kerr parameters $b=(\bhm,\bha)$ close to $b_0=(\bhm_0,\bha_0)$, there exist stationary 1-forms
    \begin{equation}
    \label{EqWC0Dualomega}
    \begin{split}
      \omega_{b,\rms 0}^{*(0)},\ \omega_{b,\rms 1}^{*(0)}(\scal) &\in \dot H_\bop^{1,((0,0),C_0+1)}(X;\cT^*_X), \\
      \omega_{b,\rmv 1}^{*(-1)}(\vect) &\in \dot H_\bop^{1,((-1,0),C_0+1)}(X;\cT^*_X), 
    \end{split}
    \end{equation}
    which depend linearly on $\scal$ and $\vect$, and in a $\cC^k$ fashion on $b$ in $\dot H_\bop^{1-k,((0,0),C_0+1)}$ and $\dot H_\bop^{1-k,((-1,0),C_0+1)}$, respectively, have leading-order terms $\ubar\omega_{\rms 0}^{(0)}$, $\ubar\omega_{\rms 1}^{(0)}(\scal)$, and $\ubar\omega_{\rmv 1}^{(-1)}(\vect)$, and indeed with
    \[
      \omega_{b,\rms 0}^{*(0)} - \chi g_b(\pa_{t_*},\cdot) \in \dot H_\bop^{1,C_0+1}(X;\cT^*_X)
    \]
    where $\chi\in\CI(X)$ equals $1$ near $\pa X$ and $0$ near $r=\bhm$, and which satisfy
    \begin{equation}
    \label{EqWC0DualEq}
      \omega_{b,\rms 0}^{*(0)},\ \omega_{b,\rms 1}^{*(0)}(\scal),\ \omega_{b,\rmv 1}^{*(-1)}(\vect) \in (\Box_{g_b,E^\cC}^\cC)^*.
    \end{equation}
    When $\vect$ and $\vect(\bha)$ (in the notation of Definition~\usref{DefTYs1}) are linearly dependent, then
    \begin{equation}
    \label{EqWC0omegav1}
      \omega_{b,\rmv 1}^{*(-1)}(\vect) - \chi g_b\ubar g^{-1}\vect \in \dot H_\bop^{1,C_0+1}(X;\cT^*_X).
    \end{equation}
  \item\label{ItWC0Dualh}{\rm (Symmetric gradients: dual pure gauge states.)} We have
    \begin{equation}
    \label{EqWC0Dualh}
    \begin{alignedat}{2}
      h_{b,\rms 0}^* &:= \sfG_{g_b}\delta_{g_b}^*\omega_{b,\rms 0}^{*(0)} &&\in \dot H_\bop^{0,C_0+2}(X;S^2\cT^*_X), \\
      h_{b,\rms 1}^*(\scal) &:= \sfG_{g_b}\delta_{g_b}^*\omega_{b,\rms 1}^{*(0)}(\scal) &&\in \dot H_\bop^{0,((2,0),C_0+2)}(X;S^2\cT^*_X), \\
      h_{b,\rmv 1}^*(\vect) &:= \sfG_{g_b}\delta_{g_b}^*\omega_{b,\rmv 1}^{*(-1)}(\vect) &&\in \dot H_\bop^{0,((2,0),C_0+2)}(X;S^2\cT^*_X),
    \end{alignedat}
    \end{equation}
    with $\cC^k$-regularity in $b$ when measured in $\dot H_\bop^{-k,C_0+2}$ and $\dot H_\bop^{-k,((2,0),C_0+2)}$, respectively.
  \end{enumerate}
\end{prop}

Note that $\wh{\Box_{g_b,E^\cC}^\cC}(0)^*\omega_{b,\rms 0}^{*(0)}=0$ is a wave equation in the region $r<r_b^+$, so since $\omega_{b,\rms 0}^{*(0)}$, lying in a Sobolev space with supported character at $r=\bhm_0$, vanishes for $r\leq\bhm_0$, it must in fact vanish for $r<r_b^+$; similarly for the other dual states.

We also remark that the memberships in~\eqref{EqWC0Dualomega} (and thus~\eqref{EqWC0Dualh}) can be improved to full polyhomogeneity near $\pa X$. Note however that the output of the inverse of~\eqref{EqWC0Dual} has limited regularity (albeit at the conormal bundle of the event horizon only). In any case, for our purposes the high order $\approx C_0$ of decay of the non-polyhomogeneous remainder terms will be sufficient.

\begin{proof}[Proof of Proposition~\usref{PropWC0Dual}]
  \pfstep{Asymptotic time translations.} Since $\pa_{t_*}$ is a Killing vector field, we have
  \[
    \delta_{g_b}^* (\chi g_b(\pa_{t_*},\cdot)) = \dd\chi\otimes_s g_b(\pa_{t_*},\cdot) \in \CIc(X^\circ;\cT^*_X),
  \]
  and hence $(\Box_{g_b,E^\cC}^\cC)^*(\chi g_b(\pa_{t_*},\cdot))=-\wh{\Box_{g_b,E^\cC}^\cC}(0)^*\omega'_b$ for a unique $\omega'_b\in\dot H_\bop^{1,C_0+1}(X;\cT^*_X)$ by Lemma~\ref{LemmaWC0Dual}. We then set
  \[
    \omega_{b,\rms 0}^{*(0)} := \chi g_b(\pa_{t_*},\cdot) + \omega'_b \in \ker (\Box_{g_b,E^\cC}^\cC)^*.
  \]
  Using Lemma~\ref{LemmaWG0KerrMink}, this satisfies
  \[
    \delta_{g_b}^*\omega_{b,\rms 0}^{(0)} = \dd\chi\otimes_s g_b(\pa_{t_*},\cdot) + \wh{\delta_{g_b}^*}(0)\omega'_b \in \dot H_\bop^{0,C_0+2}(X;S^2\cT^*_X).
  \]

  As for the regularity of $\omega'_b$ in $b$, consider a smooth function $b(s)$ with $b(0)=b$, then $\dot\omega'_b:=\frac{\dd}{\dd s}\omega_{b(s)}|_{s=0}$ formally satisfies
  \[
    \wh{\Box_{g_b,E^\cC}^\cC}(0)^*\dot\omega_b' = -\frac{\dd}{\dd s} \wh{\Box_{g_{b(s)},E^\cC}^\cC}(0)^*\Big|_{s=0}\omega'_b - \frac{\dd}{\dd s} \bigl[ \wh{\Box_{g_{b(s)},E^\cC}^\cC}(0)^*, \chi \bigr] g_{b(s)}(\pa_{t_*},\cdot)\Big|_{s=0}.
  \]
  The first term on the right lies in $\rho^3\Diffb^2(X)(\dot H_\bop^{1,C_0+1}(X;\cT^*_X))\subset\dot H_\bop^{-1,C_0+4}(X;\cT^*_X)$, and the second term lies in $\CIc(X^\circ;\cT^*_X)$. Lemma~\ref{LemmaWC0Dual} thus gives an estimate for $\dot\omega_b'$ in $\dot H_\bop^{0,C_0+1}$. (Notice the loss of one derivative here.) An iteration of this argument shows that $\omega_b'$ is $k$-times differentiable with values in $\dot H_\bop^{1-k,C_0+1}$.

  \pfstep{Asymptotic translations.} Analogously to the proof of Proposition~\ref{PropWG0Symm}, but inserting the cutoff $\chi$ to localize near $r=\infty$, one can set
  \[
    \omega_{b,\rms 1}^{*(0)}(\scal) = \chi\ubar\omega_{\rms 1}^{(0)} - \bigl(\wh{\Box_{g_b,E^\cC}^\cC}(0)^*\bigr)^{-1} f_b,\quad f_b:=(\Box_{g_b,E^\cC}^\cC)^*(\chi\ubar\omega_{\rms 1}^{(0)}) \in \rho^3\CI(X;\cT^*_X).
  \]
  The description of the indicial roots after Lemma~\ref{LemmaWC0Dual} implies that the second summand lies in $\dot H_\bop^{1,((1,0),C_0+1)}(X;\cT^*_X)$. (One can also obtain this by first solving away the expansion terms of $f_b$ using Lemma~\ref{LemmaTMFormal} before applying Lemma~\ref{LemmaWC0Dual} to the fast decaying remainder.) Regularity of $b$ is proved as before; and the membership of $h_{b,\rms 1}^*$ in~\eqref{EqWC0Dualh} follows similarly to the proof of Proposition~\ref{PropWG0Symm}.

  \pfstep{Asymptotic rotations.} Write $b=(\bhm,\bha)$ and $b_S=(\bhm,0)$. By~\eqref{EqWG0SymmRot}, we have
  \[
    f_b:=\wh{\Box_{g_b,E^\cC}^\cC}(0)^*(\chi\ubar\omega_{\rmv 1}^{(-1)})\in\rho^3\CI
  \]
  (with support disjoint from $r=\bhm$). This can be solved away using Lemma~\ref{LemmaWC0Dual}, and thus
  \[
    \omega_{b,\rmv 1}^{*(-1)} := \chi\ubar\omega_{\rmv 1}^{(-1)} + \omega'_b,\quad \omega'_b:=-\bigl(\wh{\Box_{g_b,E^\cC}^\cC}(0)^*\bigr)^{-1}f_b \in \dot H_\bop^{1,((1,0),C_0+1)}(X;\cT^*_X).
  \]
  Therefore,
  \[
    \delta_{g_b}^*\omega_{b,\rmv 1}^{*(-1)} = [\delta_{g_b}^*,\chi]\ubar\omega_{\rmv 1}^{(-1)} + \chi\delta_{g_b}^*\ubar\omega_{\rmv 1}^{(-1)} + \wh{\delta_{g_b}^*}(0)\omega'_b.
  \]
  The first term lies in $\CIc(X^\circ;\cT^*_X)$, the second term in $\rho^2\CI$ by~\eqref{EqWG0SymmRot}, and the final term in $\dot H_\bop^{0,((2,0),C_0+2)}(X;S^2\cT^*_X)$ by Lemma~\ref{LemmaWG0KerrMink}.

  To prove~\eqref{EqWC0omegav1}, we instead construct $\omega_{b,\rmv 1}^{*(-1)}(\vect)$ as $\chi g_b\ubar g^{-1}\vect+\omega''_b$ where
  \[
    \omega''_b = -\bigl(\wh{\Box_{g_b,E^\cC}^\cC}(0)^*\bigr)^{-1} \Bigl( 2 \wh{\delta_{g_b,E^\cC}^*}(0)^*\sfG_{g_b}\wh{\delta_{g_b}^*}(0)(\chi g_b\ubar g^{-1}\vect)\Bigr) \in \dot H_\bop^{C_0+1}(X;\cT^*_X)
  \]
  when $\vect$ and $\vect(\bha)$ are linearly dependent, since then $g_b\ubar g^{-1}\vect$ is Killing. (Since $\chi\ubar\omega_{\rmv 1}^{(-1)}\equiv\chi g_b\ubar g^{-1}\vect\bmod\rho\CI$, this produces the same 1-form indeed.)
\end{proof}

\subsubsection{Refined description of vector type 1 dual states}
\label{SssWC0v1}

Remark~\ref{RmkWG0Largev1} has an analogue for the dual states $h_{b,\rmv 1}^*(\vect)$ defined by~\eqref{EqWC0Dualh}: write $b=(\bhm,\bha)$, then~\eqref{EqWC0omegav1} gives
\begin{equation}
\label{EqWC0hv1}
  h_{b,\rmv 1}^*(\vect) \in \dot H_\bop^{0,C_0+2}(X;S^2\cT^*_X)
\end{equation}
for all $\vect\in\vectspace_1$ when $\bha=0$; for $\bha\neq 0$, this only holds when $\vect$ is a multiple of $\vect(\bha)$, as otherwise $g_b\ubar g^{-1}\vect$ is not Killing. (In fact, in this case $h_{b,\rmv 1}^*(\vect)$ has a non-zero $\rho^2$ leading-order term, as follows from Proposition~\ref{PropWC0hbv1} below.) The space of $\vect$ for which~\eqref{EqWC0hv1} holds thus depends in a discontinuous fashion on $\bha$ at $\bha=0$. (Since inner products with the dual states~\eqref{EqWC0Dualh} determine the solvability of $\wh{L_b}(0)h=f$ where $L_b=L_{g_b}$ is the linearized gauge-fixed Einstein operator from~\eqref{Eq1Lin}, this will present a technical challenge---which we will overcome using the material developed below---for guaranteeing the smoothness in $b$ of certain large zero energy states in Proposition~\ref{PropWE0} below.)

Notice now that the vector $\bha\times\bha(\vect)\in\R^3$ quantifies how far $\bha(\vect)$ deviates from the axis of rotation. We proceed to show that the $\rho^2$ leading-order term of $h_{b,\rmv 1}^*(\vect)$ depends on $\vect$ only through $\bha\times\bha(\vect)$. More precisely:

\begin{prop}[Structure of $\rmv 1$ dual states]
\label{PropWC0hbv1}
  Write $\bha(b):=\bha$ when $b=(\bhm,\bha)$. The dual states $h_{b,\rmv 1}^*(\vect)$ constructed in Proposition~\usref{PropWC0Dual} can be written in the form
  \begin{equation}
  \label{EqWC0hbv1}
    h_{b,\rmv 1}^*(\vect) = h_{b,\rmv 1}^{*\sharp}(\vect) + h_{b,\rmv 1}^{*\flat}(\vect),\quad\vect\in\vectspace_1,
  \end{equation}
  where the summands $h_{b,\rmv 1}^{*\sharp}(\vect)$ and $h_{b,\rmv 1}^{*\flat}(\vect)$ have the following properties.
  \begin{enumerate}
  \item There exists $h_{b,\rmv 1}^\sharp(\bfw)\in\rho^2\CI(X;S^2\cT^*_X)$, depending smoothly on $b$ and linearly on $\bfw\in\R^3$, and vanishing near $r=\bhm_0$, such that
    \[
      h_{b,\rmv 1}^{*\sharp}(\vect) = h_{b,\rmv 1}^\sharp(\bha(b)\times\bha(\vect)).
    \]
    Moreover, the $\rho^2$ leading-order term $\ubar h_{\rmv 1}^\sharp(\bfw)$ of $h_{b,\rmv 1}^\sharp(\bfw)$ (and thus also of $h_{b,\rmv 1}^{*\sharp}(\vect)$) is of vector type $1$ and explicitly given by
    \begin{equation}
    \label{EqWC0hbv1Lead}
      \ubar h_{\rmv 1}^\sharp(\bfw) = -\frac{\bhm}{1+2 v^\cC\gamma^\cC}\rho^2 \bigl( (1+2(v^\cC-1)\gamma^\cC)\,\dd x^0 + (1+2(v^\cC+1)\gamma^\cC)\,\dd x^1\bigr) \otimes_s \rho\vect(\bfw).
    \end{equation}
    Lastly, for all $\bfw\in\R^3$, we have
    \begin{equation}
    \label{EqWC0hbv1SharpSol}
      (\delta_{g_b,E^\cC}^*)^* h_{b,\rmv 1}^\sharp(\bfw),\ \wh{D_{g_b}\Ric}(0)^* h_{b,\rmv 1}^\sharp(\bfw) \in \rho^{\lfloor C_0\rfloor+2}\CI(X;\cT^*_X).
    \end{equation}
  \item $h_{b,\rmv 1}^{*\flat}(\vect)\in\dot H_\bop^{0,C_0+1}(X;\cT^*_X)$ depends linearly on $\vect$ and has $\cC^k$-regularity in $b$ when measured in $\dot H_\bop^{-k,C_0+1}$.
  \end{enumerate}
\end{prop}

This will be used in~\S\ref{SsWE0}. The key computation for the proof is the following.

\begin{lemma}[Lie derivative of $g_b$ along rotation vector fields]
\label{LemmaWC0Rot}
  Let $b=(\bhm,\bha)$. Let $\vect\in\vectspace_1$ and set $\vect_b:=g_b\ubar g^{-1}\vect$; set $\scal':=\scal(\bha(b)\times\bha(\vect))\in\scalspace_1$. Then
  \begin{equation}
  \label{EqWC0Rot}
    \delta_{g_b}^*\vect_b = \rho^2\bigl(-2\bhm(\dd t_*+\dd r)\otimes_s\rho\vect(\scal')\bigr) - \delta^*_{g_b}\bigl(\rho\scal(\bha)\scal'\,\dd r+\tfrac12\sld(\scal(\bha)\scal')\bigr) + \widetilde h_b(\scal')
  \end{equation}
  where $\widetilde h_b\colon\scalspace_1\to\rho^3\CI(X;S^2\cT^*_X)$ is linear and depends smoothly on $b$.
\end{lemma}
\begin{proof}
  Define the rotation vector field $R=R(\vect):=\bfv\times(\cdot)$ where $\bfv:=\bha(\vect)\in\R^3$. (Equivalently, $R(\vect)=\ubar g^{-1}\vect$.) We then have
  \[
    \delta_{g_b}^*\vect_b=\frac12\cL_R g_b = -\frac12 g_b(\cL_R g_b^{-1})g_b.
  \]

  \pfstep{Evaluation of $\cL_R g_b^{-1}$.} Since $\cL_R$ annihilates spherically symmetric functions, 1-forms, and tensors, and thus in particular the Schwarzschild dual metric $g_{b_S}^{-1}$ where $b_S=(\bhm,0)$, it is convenient to write
  \begin{equation}
  \label{EqWC0RotSum}
    g_b^{-1} = g_{b_S}^{-1} + (\varrho_a^{-2}-r^{-2})r^2 g_{b_S}^{-1} + \varrho_a^{-2}\bigl(\varrho_a^2 g_b^{-1} - r^2 g_{b_S}^{-1}\bigr).
  \end{equation}

  The Lie derivative of $g_{b_S}^{-1}$ along $R$ vanishes. Using the expressions from Definition~\ref{DefTYs1} and~\eqref{EqKMetaExpr}, we can write
  \[
    (\varrho_a^{-2}-r^{-2})r^2 = -r^{-2}\scal(\bha)^2 + r^{-4}f(\scal(\bha)^2,r),\quad \scal(\bha)=\bha\cdot\frac{x}{|x|},
  \]
  where $f\in\CI(\R^3\times[\bhm_0,\infty])$; note then that
  \[
    \cL_R\scal(\bha) = \cL_R\Bigl(\bha\cdot\frac{x}{|x|}\Bigr) = \Big\la \bha, \bfv \times \frac{x}{|x|}\Big\ra_{\R^3} = \Big\la \bha\times\bfv, \frac{x}{|x|}\Big\ra = \scal',
  \]
  and therefore the second term of~\eqref{EqWC0RotSum} is of the form
  \begin{equation}
  \label{EqWC0Rot2}
    \cL_R((\varrho_a^{-2}-r^{-2})r^2 g_{b_S}^{-1}) = -2 \rho^2 \scal(\bha)\scal' \bigl( -2\pa_{t_*}\otimes_s\pa_r + \pa_r^2 + r^{-2}\slg^{-1}\bigr) + \widetilde h_b(\scal').
  \end{equation}
  \emph{Here, and also in the remainder of the proof, we write $\widetilde h_b(\scal')$ for an element of $\rho^3\CI(X;S^2\cT^*_X)$ or $\rho^3\CI(X;S^2\cT_X)$ that depends linearly on $\scal'\in\scalspace_1$ and smoothly on $b$, but may change from line to line.}

  Consider next the third term in~\eqref{EqWC0RotSum}. Since $\varrho_a^2 g_b^{-1}-r^2 g_{b_S}^{-1}\in\CI(X;S^2\cT^*_X)$ and $\varrho_a^{-2}\in\rho^2\CI(X)$, the contribution from differentiating $\varrho_a^{-2}=(r^2+\scal(\bha)^2)^{-1}$ along $\cL_R$ depends linearly on $\scal'$ and is of class $r^{-4}\cdot\CI=\rho^4\CI(X;S^2\cT^*_X)$, and hence can be put into $\widetilde h_b(\scal')$. We next consider the coefficients of $h_3:=\varrho_a^2 g_b^{-1}-r^2 g_{b_S}^{-1}$ and $f_3:=\cL_R h_3$ one by one. We recall the expression~\eqref{EqKMetFinal} (which relates to $g_{\bhm,\bha}$ as described in Definition~\ref{DefKMetcM}), where we note that $\chi_{\cH^+}$ and $\chi_\infty$ are functions of $r$ only. 
  \begin{enumerate}
  \item The $\pa_{t_*}^2$-coefficient of $h_3$ is
    \begin{align*}
      -(1-\chi_{\cH^+}^2-\chi_\infty^2)\Bigl(\frac{(r^2+a^2)^2}{\mu_{\bhm,a}}-\frac{r^4}{\mu_{\bhm,0}}\Bigr) + \Bigl[|\bha|^2-\Bigl(\bha\cdot\frac{x}{|x|}\Bigr)^2\Bigr] + 2(\chi_{\cH^+}-\chi_\infty)|\bha|^2\tilde T'(r).
    \end{align*}
    The only term that is not spherically symmetric is $-(\bha\cdot\frac{x}{|x|})^2$. Thus, the contribution to $f_3$ is $-2\scal(\bha)\scal'\pa_{t_*}^2$, the product of which with $\varrho_a^{-2}$ is $-2\rho^2\scal(\bha)\scal'\pa_{t_*}^2$ plus contributions to $\widetilde h_b(\scal')$.
  \item The $\pa_r^2$-coefficient of $h_3$ is $\mu_{\bhm,a}-\mu_{\bhm,0}=|\bha|^2$ and thus spherically symmetric.
  \item The $2\pa_{t_*}\otimes_s\pa_r$-coefficient of $h_3$ is $(\chi_{\cH^+}-\chi_\infty)|\bha|^2$ and thus spherically symmetric as well.
  \item The terms in $h_3$ involving the product of $\pa_{t_*}$ with a spherical vector field are
    \[
      2\Bigl[-\Bigl((1-\chi_{\cH^+}^2)\frac{r^2+a^2}{\mu_{\bhm,a}} - 1\Bigr) + \chi_{\cH^+}\tilde T'(r) \Bigr] \pa_{t_*}\otimes_s\nabla_{\bha\times(\cdot)}.
    \]
    The prefactor here is spherically symmetric and equal to $-\frac{2\bhm}{r}+r^{-2}\CI([\bhm_0,\infty])$. Since
    \begin{equation}
    \label{EqWC0RotLie}
      \cL_R\nabla_{\bha\times(\cdot)} = [\nabla_{\bfv\times(\cdot)},\nabla_{\bha\times(\cdot)}] = \nabla_{(\bha\times\bfv)\times(\cdot)},
    \end{equation}
    the contribution to $f_3$ is $-\frac{4\bhm}{r}\pa_{t_*}\otimes_s\nabla_{(\bha\times\bfv)\times x}$. Its product with $\varrho_a^{-2}$ is therefore $-\frac{4\bhm}{r^3}\pa_{t_*}\otimes_s\nabla_{(\bha\times\bfv)\times(\cdot)}$ plus further terms that we can put into $\widetilde h_b(\scal')$.
  \item The only term in $h_3$ involving the product of $\pa_r$ with a spherical vector field is $2\chi_{\cH^+}\pa_r\otimes_s\nabla_{\bha\times(\cdot)}$; it has compact support, and ultimately (using again~\eqref{EqWC0RotLie}) contributes only terms to $\widetilde h_b(\scal')$.
  \item Finally, the spherical 2-tensor component of $r^2 g_{b_S}$ is $\pa_\theta^2+\frac{1}{\sin^2\theta}\pa_\phi^2$, and thus that of $h_3$ is
    \begin{equation}
    \label{EqWC0RotSph}
      -\frac{1-\chi_{\cH^+}^2}{\mu_{\bhm,a}}\nabla_{\bha\times(\cdot)}\otimes_s\nabla_{\bha\times(\cdot)}.
    \end{equation}
    Its Lie derivative along $\cL_R$ can again be computed using~\eqref{EqWC0RotLie}, and the leading-order contribution to $\cL_R g_b^{-1}$ is $-2 r^{-4}\nabla_{\bha\times(\cdot)}\otimes_s\nabla_{(\bha\times\bfv)\times(\cdot)}$.
  \end{enumerate}
  Altogether, recalling~\eqref{EqWC0Rot2}, we have now shown that
  \begin{equation}
  \label{EqWC0RotIM}
  \begin{split}
    \rho^{-2}\cL_R g_b^{-1} &= -2\scal(\bha)\scal'\bigl((\pa_{t_*}-\pa_r)^2 + r^{-2}\slg^{-1}\bigr) \\
      &\qquad - \frac{4\bhm}{r}\pa_{t_*}\otimes_s\nabla_{(\bha\times\bfv)\times(\cdot)} - 2 r^{-2}\nabla_{\bha\times(\cdot)}\otimes_s\nabla_{(\bha\times\bfv)\times(\cdot)} + \widetilde h_b(\scal').
  \end{split}
  \end{equation}

  \pfstep{Re-writing of $\delta_{g_b}^*\vect_b$.} Since $g_b\equiv\ubar g=-\dd t_*^2-2\,\dd t_*\otimes_s\dd r+r^2\slg\bmod\rho\CI(X;S^2\cT^*_X)$, we obtain from~\eqref{EqWC0RotIM} the expression
  \begin{equation}
  \label{EqWC0RotIM2}
    \rho^{-2}\delta_{g_b}^*\vect_b = - 2\bhm(\dd t_*+\dd r)\otimes_s\rho\vect(\scal') + \scal(\bha)\scal'(\dd r^2+r^2\slg) + \rho\vect(\scal(\bha))\otimes_s\rho\vect(\scal') + \rho^{-2}\widetilde h_b(\scal').
  \end{equation}

  Note now that $\scal(\bha)\scal'\in\scalspace_2$ (since the two factors are orthogonal, corresponding to rotations around orthogonal axes), and also $\vect(\scal(\bha))\otimes_s\vect(\scal')$ is of scalar type $2$ (by the representation theory of $O(3)$, see, e.g., \cite[(6.1)]{HintzGlueLocI}, and again using the orthogonality of $\scal(\bha)$ and $\scal'$); but, ultimately, the zero energy state $h_{b,\rmv 1}(\vect)$ (as constructed in~\eqref{EqWG0hv1} via~\eqref{EqWG0SymmRot}, where one can equivalently start with $\vect_b$ instead of $\ubar\omega_{\rmv 1}^{(-1)}$) cannot have any scalar type $2$ contributions at the leading-order level since $2$ is not an $\rms 2$ indicial root by Lemma~\ref{LemmaWEInd} below. It is thus natural to expect that one can express the sum of the second and third terms in~\eqref{EqWC0RotIM2} as a symmetric gradient to leading order; we proceed to do this.

  First, we claim for $\vect=\vect(\scal)\in\vectspace_1$ and on the unit 2-sphere that
  \begin{equation}
  \label{EqWC0VV}
    \vect\otimes_s\vect = \biggl(\frac{1}{4\pi}\int_{\Sph^2} \scal^2\,\dd\slg\biggr)\slg + \bigl(Y(\scal)\slg + \sldelta_0^*\sld Y(\scal)\bigr),\quad
    Y(\scal) := \frac12\biggl(\frac{1}{4\pi}\int_{\Sph^2}\scal^2\,\dd\slg - \scal^2\biggr) \in \scalspace_2.
  \end{equation}
  (This is thus the decomposition of $\vect\otimes_s\vect$ into its scalar type $0$ and $2$ pieces.) By scaling and pullback along rotations, it suffices to check this in the special case $\scal=\cos\theta$, in which case $\vect=\pa_\phi^\flat=\sin^2\theta\,\dd\phi$; in this case $\frac{1}{4\pi}\int_{\Sph^2}\scal^2\,\dd\slg=\frac13$, so $Y(\scal)=\frac12\sin^2\theta-\frac13$, and together with $\sldelta_0^*\sld Y(\scal)=\frac12\sin^2\theta(-\dd\theta^2+\sin^2\theta\,\dd\phi^2)$, this gives~\eqref{EqWC0VV}. For $L^2(\Sph^2)$-orthogonal $\scal,\scal'\in\scalspace_1$, one can then polarize the identity~\eqref{EqWC0VV} to obtain, in light of $\frac14(Y(\scal+\scal')-Y(\scal-\scal'))=-\frac12\scal\scal'$,
  \begin{equation}
  \label{EqWC0RotVV2}
    \vect(\scal)\otimes_s\vect(\scal') = -\frac12\scal\scal'\slg - \frac12\sldelta_0^*\sld(\scal\scal').
  \end{equation}
  For orthogonal $\scal,\scal'\in\scalspace_1$ in the splitting~\eqref{EqTYMink01Split} and in the description~\eqref{EqTYSplit2} of elements of scalar type $2$ tensors on Minkowski space (now relative to $\scal\scal'\in\scalspace_2$ instead of $\scal$), the expression~\eqref{EqWC0RotVV2} equals $(0,0,0,0,0,-\frac12,-\frac12)$.

  Next, since $\dd r^2=\frac14((\dd\ubar x^0)^2-2\,\dd\ubar x^0\otimes_s\dd\ubar x^1+(\dd\ubar x^1)^2)$ in the notation of~\eqref{EqTYMink01}, the expression for $\scal\scal'(\dd r^2+r^2\slg)+\rho\vect(\scal)\otimes_s\rho\vect(\scal')$ in terms of~\eqref{EqTYSplit2} is
  \[
    (\tfrac14,-\tfrac14,0,\tfrac14,0,\tfrac12,-\tfrac12).
  \]
  This in turn is equal to
  \[
    N_{\rms 2}(\rho^{-1}\wh{\ubar\delta^*}(0),1) (-\tfrac12,\tfrac12,-\tfrac12),
  \]
  where we use~\eqref{EqTYSplit1}--\eqref{EqTYSplit2} in the $\rms 2$ sector and use the form of
  \[
    N_{\rms 2}(\rho^{-1}\wh{\ubar\delta^*}(0),\lambda) = \begin{pmatrix} -\frac{\lambda}{2} & 0 & 0 \\ \frac{\lambda}{4} & -\frac{\lambda}{4} & 0 \\ \frac12 & 0 & -\frac14(\lambda+1) \\ 0 & \frac{\lambda}{2} & 0 \\ 0 & \frac12 & \frac14(\lambda+1) \\ 1 & -1 & -3 \\ 0 & 0 & 1 \end{pmatrix}
  \]
  which one obtains from Lemma~\ref{LemmaTYMinkOp} (and upon using the second identity in~\eqref{EqTYIdentities} for the $(6,3)$- and $(7,3)$-entries). Writing this out, we have thus proved that
  \[
    \rho^2\bigl(\scal\scal'(\dd r^2+r^2\slg) + \rho\vect(\scal)\otimes_s\rho\vect(\scal')\bigr) = -\wh{\ubar\delta^*}(0)\Bigl(\rho\bigl( \scal\scal'\,\dd r+\tfrac12 r\sld(\scal\scal')\bigr)\Bigr),\quad \scal,\scal'\in\scalspace_1,\ \scal\perp\scal'.
  \]
  Plugging this into~\eqref{EqWC0RotIM2}, replacing $\ubar\delta^*$ by $\delta_{g_b}^*$, and absorbing the resulting error terms of class $\rho^3\CI$ into $\widetilde h_b$ finishes the proof.
\end{proof}

\begin{proof}[Proof of Proposition~\usref{PropWC0hbv1}]
  Write $\Box_b^*:=(\Box_{g_b,E^\cC}^\cC)^*=2(\delta_{g_b,E^\cC}^*)^*\sfG_{g_b}\delta_{g_b}^*$. Let $\chi\in\CI(X)$ be equal to $1$ near $\pa X$ and $0$ near $r=\bhm$. We revisit the construction of $\omega_{b,\rmv 1}^*(\vect)$ in the proof of Proposition~\ref{PropWC0Dual}; motivated by~\eqref{EqWC0Rot}, we make the rather precise ansatz
  \[
    \omega_{b,\rmv 1}^{*(-1)}(\vect) = \chi\Bigl(\vect_b + \rho\scal(\bha)\scal' + \frac12\sld\bigl(\scal(\bha)\scal'\bigr) + \chi\rho\omega_0\Bigr) + \omega', \quad \scal':=\scal(\bha(b)\times\bha(\vect))
  \]
  where we shall determine $\omega_0\in\CI(\pa X;\cT^*_X|_{\pa X})$ and $\omega'\in\dot H_\bop^{0,((2,0),C_0+1)}(X;\cT^*_X)$ so that this lies in the kernel of $\Box_b^*$. Thus,
  \begin{equation}
  \label{EqWC0hbv1Dels}
    \delta_{g_b}^*\omega_{b,\rmv 1}^{*(-1)}(\vect) \equiv -2\bhm\rho^2(\dd t_*+\dd r)\otimes_s\rho\vect(\scal') + \rho^2 N(\rho^{-1}\wh{\ubar\delta^*}(0),1)\omega_0 \bmod \dot H_\bop^{0,((3,0),C_0+2)}(X;\cT^*_X).
  \end{equation}

  \pfstep{Determination of $\omega_0$.} In the bundle splitting~\eqref{EqTYSplit2}, abbreviating $\vect':=\vect(\scal')$, and noting that $\dd t_*+\dd r=\frac12(\dd x^0+\dd x^1)$, we have $2\bhm\rho^2(\dd t_*+\dd r)\otimes_s\rho\vect'=\frac{\bhm}{2}\rho^2(0,0,\vect',0,\vect',0)$. Thus, $\omega_0$ is uniquely determined by the equation
  \begin{equation}
  \label{EqWC0hbv1Omega0}
    N\bigl(\rho^{-2}\wh{\Box_b^*}(0),1\bigr)\omega_0 = 2 N\bigl(\rho^{-1}\wh{\ubar\delta^*_{\ubar E^\cC}}(0)^*,2\bigr)\ubar\sfG\tfrac{\bhm}{2}(0,0,\vect',0,\vect',0) = -4\bhm\gamma^\cC(0,0,\vect');
  \end{equation}
  here we used $(\ubar\delta_{\ubar E^\cC}^*)^*=\ubar\delta+(\ubar E^\cC)^*$, the expression for $N(\rho^{-1}\wh{\ubar\delta}(0),2)$ arising from Lemma~\ref{LemmaTYMinkOp} (which annihilates $\ubar\sfG(0,0,\vect',0,\vect',0)=(0,0,\vect',0,\vect',0)$), and $(\ubar E^\cC)^*=\gamma^\cC(2\iota_{\ubar g^{-1}(\ubar\cd^\cC)}(\cdot)-(1-e^\cC)\ubar\cd^\cC\,\ul\tr)$ (from~\eqref{EqWCRecE}), which in light of $\ubar\cd^\cC=\rho(\frac12(1-v^\cC),\frac12(1+v^\cC),0)$ (so $\ubar g^{-1}(\ubar\cd^\cC)=-\rho(1+v^\cC,1-v^\cC,0)$) gives
  \[
    \rho^{-1}(\ubar E^\cC)^*=\gamma^\cC
      \openbigpmatrix{3pt}
        {-}2(1{+}v^\cC) & {-}2 e^\cC(1{-}v^\cC) & 0 & 0 & 0 & {-}\frac12(1{-}e^\cC)(1{-}v^\cC)\sltr \\
        0 & {-}2 e^\cC(1{+}v^\cC) & 0 & {-}2(1{-}v^\cC) & 0 & {-}\frac12(1{-}e^\cC)(1{+}v^\cC)\sltr \\
        0 & 0 & {-}2(1{+}v^\cC) & 0 & {-}2(1{-}v^\cC) & 0
      \closebigpmatrix
  \]
  which acts on $\bhm(0,0,\vect',0,\vect',0)$ to give~\eqref{EqWC0hbv1Omega0}. Now, since the right-hand side of~\eqref{EqWC0hbv1Omega0} is of vector type $1$, we then note that since $\ubar\sfG$ is the identity on $\rmv 1$ 2-tensors, we have
  \begin{align*}
    N_{\rmv 1}\bigl(\rho^{-2}\wh{\Box_b^*}(0),1\bigr) &= 2 N_{\rmv 1}\bigl(\rho^{-1}\wh{\ubar\delta^*_{\ubar E^\cC}}(0)^*,2\bigr)N_{\rmv 1}\bigl(\rho^{-1}\wh{\ubar\delta^*}(0),1\bigr) \\
      &= 2 \Bigl( (-1,1) + \gamma^\cC\bigl(-2(1+v^\cC),-2(1-v^\cC)\bigr)\Bigr) \begin{pmatrix} -\frac12 \\ \frac12 \end{pmatrix} = 2+4\gamma^\cC v^\cC
  \end{align*}
  (which is a $1\times 1$ matrix, cf.\ \eqref{EqTYSplit1} in the vector type $1$ sector). Therefore,
  \begin{equation}
  \label{EqWC0hbv1Omega0Sol}
    \omega_0 = -\frac{2\bhm\gamma^\cC}{1+2 v^\cC\gamma^\cC}\rho\vect'.
  \end{equation}

  We record here already that the $\rho^2$ leading-order term of $h_{b,\rmv 1}^*(\vect)=\sfG_{g_b}\delta_{g_b}^*\omega_{b,\rmv 1}^{*(-1)}(\vect)$ will be equal to $\ubar h_{b,\rmv 1}^\sharp(\bha(\vect'))$ (thus, crucially, only depends on $\vect$ through $\bha(\vect')=\bha(b)\times\bha(\vect)$) where, for general $\bfw\in\R^3$, we put
  \[
    \ubar h_{b,\rmv 1}^\sharp(\bfw) := \rho^2\Bigl( -2\bhm(\dd t_*+\dd r) + \frac{4\bhm\gamma^\cC}{1+2 v^\cC\gamma^\cC}\,\dd r\Bigr) \otimes_s (\rho\vect(\bfw)) \in \rho^2\CI(\pa X;S^2\cT^*_X|_{\pa X}).
  \]
  (We use here that $\ubar\delta^*(-2\bhm\gamma^\cC\,r^{-2}\vect')=4\bhm\gamma^\cC r^{-3}\,\dd r\otimes_s\vect'$.) This is equal to~\eqref{EqWC0hbv1Lead} upon writing $\dd t_*+\dd r=\frac12(\dd x^0+\dd x^1)$ and $\dd r=\frac12(\dd x^0-\dd x^1)$.

  \pfstep{Conclusion of the proof.} Set
  \[
    \omega_{b,\rmv 1}^{\sharp(-1)}(\vect) := \vect_b + \rho\scal(\bha)\scal'\,\dd r + \tfrac12\sld\bigl(\scal(\bha)\scal'\bigr) - \frac{2\bhm\gamma^\cC}{1+2 v^\cC\gamma^\cC}\rho\vect',
  \]
  where $\scal'=\scal(\bha(b)\times\bha(\vect))$ and $\vect'=\vect(\scal')$. By Lemma~\ref{LemmaWC0Rot}, we have
  \begin{equation}
  \label{EqWC0hbv1widetilde}
    \sfG_{g_b}\delta_{g_b}^*\omega_{b,\rmv 1}^{\sharp(-1)}(\vect) =: \ubar h_{b,\rmv 1}^\sharp(\bha(\vect')) + \widetilde h_b(\scal')
  \end{equation}
  for a linear function $\widetilde h_b\colon\scalspace_1\to\rho^3\CI(X;S^2\cT^*_X)$ that depends smoothly on $b$. Thus, $\omega'$ in the ansatz $\omega_{b,\rmv 1}^{*(-1)}(\vect)=\chi\omega_{b,\rmv 1}^{\sharp(-1)}(\vect)+\omega'$ must satisfy
  \begin{equation}
  \label{EqWC0hbv1Omegap}
    \wh{\Box_b^*}(0)\omega' = -2(\delta_{g_b,E^\cC}^*)^*\bigl(\chi\ubar h_{b,\rmv 1}^\sharp(\bha(\vect'))\bigr) - 2 (\delta_{g_b,E^\cC}^*)^*\bigl(\chi \widetilde h_b(\scal')\bigr) - 2(\delta_{g_b,E^\cC}^*)^*\sfG_{g_b}\bigl(\dd\chi\otimes_s\omega_{b,\rmv 1}^{\sharp(-1)}(\vect)\bigr).
  \end{equation}
  By construction, the first summand lies not merely in $\rho^3\CI(X;\cT^*_X)$, but in $\rho^4\CI$. The second summand also lies in $\rho^4\CI$. We can compute the output of $\wh{\Box_b^*}(0)^{-1}$ acting on the sum of the first two summands by first constructing a formal solution $\chi\omega^{\prime\sharp}_1(\scal')\in\rho^2\CI(X;\cT^*_X)$ up to decay order $C_0+1$ using Lemma~\ref{LemmaTMFormal} (which depends smoothly on $b$) and then using Lemma~\ref{LemmaWC0Dual} to solve away the remaining error by means of an element $\omega^{\prime\flat}_1(\scal')\in\dot H_\bop^{1,C_0+1}(X;\cT^*_X)$. The second summand on the other hand lies in $\CIc(X^\circ;\cT^*_X)$, so upon applying $\wh{\Box_b^*}(0)^{-1}$ to it we obtain an element $\omega'_2(\vect)\in\dot H_\bop^{1,C_0+1}(X;\cT^*_X)$. Altogether, this gives
  \begin{equation}
  \label{EqWC0hbv1Dels2}
    \omega' = \chi\omega^{\prime\sharp}_1(\scal') + \omega^{\prime\flat}_1(\scal') + \omega'_2(\vect)
  \end{equation}
  We emphasize that this construction defines the 1-form $\omega_1^{\prime\sharp}$ (and also the 1-form $\omega_1^{\prime\flat}$) for \emph{all} $\scal'\in\scalspace_1$.

  We can now write
  \[
    h_{b,\rmv 1}^*(\vect) = \sfG_{g_b}\delta_{g_b}^*\bigl(\chi\omega_{b,\rmv 1}^{\sharp(-1)}(\vect) + \chi\omega_1^{\prime\sharp}(\scal') + \omega_1^{\prime\flat}(\scal') + \omega'_2(\vect)\bigr) = h_{b,\rmv 1}^\sharp(\bha(b)\times\bha(\vect)) + h_{b,\rmv 1}^{*\flat}(\vect),
  \]
  where in view of~\eqref{EqWC0hbv1Dels2} we can write
  \begin{align*}
    h_{b,\rmv 1}^\sharp(\bfw) &= \chi\ubar h_{b,\rmv 1}^\sharp(\bfw) + \chi\widetilde h_b(\scal(\bfw)) + \underbrace{\sfG_{g_b}\delta_{g_b}^*\bigl(\chi\omega_1^{\prime\sharp}(\scal(\bfw))\bigr)}_{\in\rho^3\CI(X;S^2\cT^*_X)} \in \rho^2\CI(X;S^2\cT^*_X), \\
    h_{b,\rmv 1}^{*\flat}(\vect) &=\sfG_{g_b}\bigl(\dd\chi\otimes_s\omega_{b,\rmv 1}^{\sharp(-1)}(\vect)\bigr) + \sfG_g\delta_{g_b}^*\omega'_2(\vect) \\
      &\quad\hspace{6em} + \sfG_{g_b}\delta_{g_b}^*\omega_1^{\prime\flat}\bigl(\scal(\bha(b)\times\bha(\vect))\bigr) \in \dot H_\bop^{0,C_0+2}(X;S^2\cT^*_X),
  \end{align*}
  which is the desired splitting~\eqref{EqWC0hbv1}.

  Finally, in order to prove that $h_{b,\rmv 1}^\sharp$ is a formal solution of the adjoint linearized gauge and Einstein equation as asserted in~\eqref{EqWC0hbv1SharpSol}, note that modulo elements of $\dot H_\bop^{0,C_0+2}(X;\cT^*_X)$, the above construction gives
  \begin{equation}
  \label{EqWC0hbv1SharpSolPf}
    h_{b,\rmv 1}^\sharp(\bfw) \equiv \ubar h_{b,\rmv 1}^\sharp(\bfw) + \widetilde h_b(\scal(\bfw)) - 2\sfG_{g_b}\delta_{g_b}^*\wh{\Box_b^*}(0)^{-1}\Bigl( (\delta_{g_b,E^\cC}^*)^*\bigl(\ubar h_{b,\rmv 1}^\sharp(\bfw) + \widetilde h_b(\scal(\bfw))\bigr) \Bigr).
  \end{equation}
  (The second summand here arises from the term $\omega_1^{\prime\sharp}$, which in turn arises from the action of $\wh{\Box_b^*}(0)^{-1}$ on the first two summands on the right in~\eqref{EqWC0hbv1Omegap}.) Since $\ubar h^\sharp_{b,\rmv 1}(\bfw)+\widetilde h_b(\scal(\bfw))\in\ran\sfG_{g_b}\delta_{g_b}^*$ by definition (see~\eqref{EqWC0hbv1widetilde}), this lies in $\ran\sfG_{g_b}\delta_{g_b}^*\subset\ker D_{g_b}\Ric\circ\sfG_{g_b}$ (modulo error terms with $\cO(\rho^{C_0+2})$-decay). Moreover, applying $(\delta_{g_b,E^\cC}^*)^*$ to~\eqref{EqWC0hbv1SharpSolPf}, the two terms on the right-hand side cancel since $2(\delta_{g_b,E^\cC}^*)^*\sfG_{g_b}\delta_{g_b}^*\wh{\Box_b^*}(0)^{-1}$ is the identity operator.
\end{proof}

\section{Linearized gauge-fixed Einstein operator: basic spectral properties}
\label{SWE}

We now turn to the study of the linearization of the gauge-fixed Einstein operator around a subextremal Kerr metric $g_b$. Recalling Definition~\ref{DefKMetcM}, we consider Kerr parameters $b=(\bhm,\bha)$ that are close to subextremal parameters $b_0=(\bhm_0,\bha_0)$. We fix the following data.
\begin{enumerate}
\item We fix 1-forms $\cd^\Ups_{\cH^+}\in\CIc(X^\circ;\cT^*_X)$ and $\cd^\Ups\equiv r^{-1}\,\dd t\bmod\CIc(X^\circ;\cT^*_X)$ (with supports disjoint from the trapped set) as well as parameters $\gamma^\Ups_{\cH^+}>0$ and $0<e^\Ups\ll 1$ and $0<\gamma^\Ups\ll 1$ according to Proposition~\ref{PropWGMode}; recall that one can choose $\gamma^\Ups$ arbitrarily small. Thus, mode stability holds for the gauge potential wave operator $\Box_{g_b,E^\Ups}^\Ups=2\delta_{g_b,E^\Ups}\sfG_{g_b}\delta^*_{g_b}$ where $\delta^*_{g_b,E^\Ups}$ is given by Definition~\ref{Def1Gauge}, and its indicial roots are given by Lemma~\ref{LemmaWGInd} and satisfy the properties~\eqref{EqWGIndOrder1}--\eqref{EqWGIndOrder2}.
\item We fix $v^\cC\in(0,1)$ (say, $v^\cC=\frac12$) and $C_0=100$ and apply Theorem~\ref{ThmWCRec} to obtain a past timelike 1-form $\cd^\cC\equiv r^{-1}(\dd t-v^\cC\,\dd r)\bmod\CIc(X^\circ;\cT^*_X)$ as well as parameters $e^\cC\in(0,1)$ and $\gamma^\cC>0$ (which can be taken to be arbitrarily large) such that mode stability (including enhanced mode stability at zero energy) holds for the constraint propagation wave operator $\Box_{g_b,E^\cC}^\cC=2\delta_{g_b}\sfG_{g_b}\delta_{g_b,E^\cC}^*$, where $\delta_{g_b,E^\cC}^*$ is given by Definition~\ref{Def1Symm}. Its indicial roots are given by~\eqref{EqWCIndRoots}. We moreover require $\gamma^\cC$ to be so large that $(1-e^\cC)(1-v^\cC)\gamma^\cC>100$.
\end{enumerate}
For these choices, we then define the linearized gauge-fixed Einstein operator
\begin{equation}
\label{EqWEOp}
  L_b = D_{g_b}\Ric + \delta_{g_b,E^\cC}^*\delta_{g_b,E^\Ups}\sfG_{g_b}.
\end{equation}
(This equals $L_{g_b}$ in the notation of~\eqref{Eq1EinLin} and~\eqref{Eq1Lin}.)

In this section and~\S\ref{SAdm} below, we show that $L_b$ satisfies all assumptions for stationary wave-type operators required for the applicability of \cite{HintzNonstat2}. In~\S\ref{SsWETr}, we verify a sign condition for the subprincipal part of $L_b$ at the trapped set. In~\S\ref{SsWEInd}, we compute the indicial roots of the zero energy operator $\wh{L_b}(0)$. The invertibility of the transition face normal operator $N_\tface(L_b,\hat\sigma)$, $\hat\sigma\in e^{i[0,\pi]}$, is verified in~\S\ref{SsWEtf}. The mode stability of $L_b$ away from zero energy and a first description of the zero energy behavior of $L_b$ are discussed in~\S\ref{SsWEMode}. Further large/generalized zero energy states are constructed in~\S\ref{SsWE0}; these play a crucial role in the precise description of the late-time behavior of solutions of $L_b$ (as discussed in~\S\S\ref{SssIEinLa} and~\ref{SD}). Rough late-time bounds for solutions of $L_b$ will be discussed in~\S\ref{SAdm}.

We begin by describing the structure of $L_b$ similarly to Lemma~\ref{LemmaWGOp}:

\begin{lemma}[Stationary wave-type operator]
\label{LemmaWEOp}
  The operator $2 L_b$ is a stationary wave-type operator on $(M,g_b)$ in the sense of \citeAF{Definition~\ref*{DefSSAdm}}. More precisely:
  \begin{enumerate}
  \item{\rm (Structure.)} In the coordinates $t_*$ (from Lemma~\usref{LemmaKMetTime}) and $\rho=r^{-1}$, we have
    \begin{subequations}
    \begin{equation}
    \label{EqWEOpExpl}
      2 L_b = -2\pa_{t_*}\rho(\rho\pa_\rho-1-S_{E^\Ups,E^\cC}) + \wh{L_b}(0) + Q\pa_{t_*} - g^{0 0}\pa_{t_*}^2
    \end{equation}
    where $g^{0 0}=g_b^{-1}(\dd t_*,\dd t_*)\in\rho^2\CI(X)$ and
    \begin{equation}
    \label{EqWEOpKerr}
      \wh{L_b}(0)\in\rho^2\Diffb^2(X;S^2\cT^*_X),\quad
      S_{E^\Ups}\in\CI(X;\End(S^2\cT^*_X)),\quad
      Q\in\rho^3\Diffb^1(X;S^2\cT^*_X).
    \end{equation}
    \end{subequations}
  \item{\rm (Minkowskian large-scale asymptotics.)} Denote operators of Minkowski space by underbars. Define $\ubar E^\Ups$ and $\ubar E^\cC$ as in~\eqref{EqWGOpMinkE} and~\eqref{EqWCRecE}, respectively; set $\ubar\delta_{\ubar E^\Ups}:=\ubar\delta+\ubar E^\Ups$ and $\ubar\delta_{\ubar E^\cC}^*:=\ubar\delta^*+\ubar E^\cC$ as well as
    \begin{subequations}
    \begin{equation}
    \label{EqWEOpMink}
      \ubar L := D_{\ubar g}\Ric + \ubar\delta_{\ubar E^\cC}^*\ubar\delta_{\ubar E^\Ups}\ul\sfG.
    \end{equation}
    Upon writing
    \begin{equation}
    \label{EqWEOpMink2}
      2\ubar L = -2\pa_{t_*}\rho(\rho\pa_\rho-1-\ubar S_{\ubar E^\Ups,\ubar E^\cC}) + \wh{\ubar L}(0)
    \end{equation}
    \end{subequations}
    in $r>0$, we have
    \begin{equation}
    \label{EqWEOpKerrMink}
      S_{E^\Ups,E^\cC}-\ubar S_{\ubar E^\Ups,\ubar E^\cC} \in \rho\CI(X;\End(S^2\cT^*_X)),\quad
      \wh{L_b}(0) - \wh{\ubar L}(0) \in \rho^3\Diffb^2(X;S^2\cT^*_X).
    \end{equation}
  \end{enumerate}
\end{lemma}
\begin{proof}
  Equation~\eqref{Eq1LinEinGauged} in Lemma~\ref{Lemma1Lin} reads
  \begin{equation}
  \label{EqWEOpLb}
    2 L_b = \Box_{g_b} + 2 E_{g_b}^\cC\delta_{g_b}\sfG_{g_b} + 2\delta_{g_b,E^\cC}^* E^\Ups_{g_b}\sfG_{g_b} + 2\sR_{g_b}.
  \end{equation}
  Here $\Box_{g_b}$ is the tensor wave operator on symmetric 2-tensors, which is described in \citeAF{Example~\ref*{ExSSAdmBox}}. Since $\sR_{\ubar g}=0$ and the Christoffel symbols of $g_b$ in the coordinates $t_*,x$ are of class $\rho^2\CI(X)$, we have $\sR_{g_b}\in\rho^3\CI(X;\End(S^2\cT^*_X))$. Moreover, as in the proof of Lemma~\ref{LemmaWGOp}, the memberships $E^\Ups_{g_b},E^\cC_{g_b}\in\rho\CI$ (as maps between $\cT^*_X$ and $S^2\cT^*_X$) imply that the second and third terms are of the required form. The second part follows as before from~\eqref{EqKMetDiff} and~\eqref{EqWGOpNabla}.
\end{proof}

The explicit Minkowskian expressions are quite lengthy; we only need them for the purpose of verifying claims about indicial roots below in a purely computational fashion.

\begin{lemma}[Minkowskian operators]
\label{LemmaWEOpMink}
  In the expression~\eqref{EqWEOpMink2} for the operator $\ubar L$ in~\eqref{EqWEOpMink} and in the bundle splitting~\eqref{EqTYMink01Split}, we have
  \[
    \ubar S_{\ubar E^\Ups,\ubar E^\cC} =
      \scalebox{0.8}{$\openbigpmatrix{3pt}
        2(1-v^\cC)\gamma^\cC & 0 & 0 & 0 & 0 & 0 \\
        e^\cC(1+v^\cC)\gamma^\cC+\gamma^\Ups & (1-e^\Ups)\gamma^\Ups & 0 & 0 & 0 & \frac14(e^\cC(1-v^\cC)\gamma^\cC+e^\Ups\gamma^\Ups)\sltr \\
        0 & 0 & (1-v^\cC)\gamma^\cC & 0 & 0 & 0 \\
        0 & 2(1-e^\Ups)\gamma^\Ups & 0 & 2\gamma^\Ups & 0 & \frac12((1+v^\cC)\gamma^\cC+e^\Ups\gamma^\Ups)\sltr \\
        0 & 0 & (1+v^\cC)\gamma^\cC+\gamma^\Ups & 0 & \gamma^\Ups & 0 \\
        2(1-e^\cC)(1+v^\cC)\gamma^\cC & 0 & 0 & 0 & 0 & \tfrac12(1-e^\cC)(1-v^\cC)\gamma^\cC \\
        0 & 0 & 0 & 0 & 0 & 0
      \closebigpmatrix$},
  \]
  while $2\wh{\ubar L}(0)$ is given by
  \begin{equation}
  \label{EqWEOpMink0}
  \begin{split}
      2\wh{\ubar L}(0) = \wh{\ubar\Box}(0)
        &+ \gamma^\cC
          \scalebox{0.6}{$\openbigpmatrix{3pt}
            2(1{-}v^\cC)(\rho\pa_\rho{-}2) & 4(1{-}v^\cC) & 2(1{-}v^\cC)\sldelta & 0 & 0 & {-}\frac12(1{-}v^\cC)(\rho\pa_\rho{-}2) \\
            e^\cC(1{+}v^\cC)(\rho\pa_\rho{-}2) & 4 e^\cC v^\cC & e^\cC(1{+}v^\cC)\sldelta & {-}e^\cC(1{-}v^\cC)(\rho\pa_\rho{-}2) & e^\cC(1{-}v^\cC)\sldelta & {-}\frac12 e^\cC v^\cC(\rho\pa_\rho{-}2)\sltr \\
            0 & {-}2(1{-}v^\cC)\sld & (1{-}v^\cC)(\rho\pa_\rho{-}3) & 0 & {-}(1{-}v^\cC)(\rho\pa_\rho{-}3) & (1{-}v^\cC)\sldelta\slsfG \\
            0 & {-}4(1{+}v^\cC) & 0 & {-}2(1{+}v^\cC)(\rho\pa_\rho{-}2) & 2(1{+}v^\cC)\sldelta & \frac12(1{+}v^\cC)(\rho\pa_\rho{-}2)\sltr \\
            0 & {-}2(1{+}v^\cC)\sld & (1{+}v^\cC)(\rho\pa_\rho{-}3) & 0 & {-}(1{+}v^\cC)(\rho\pa_\rho{-}3) & (1{+}v^\cC)\sldelta\slsfG \\
            2(1{-}e^\cC)(1{+}v^\cC)(\rho\pa_\rho{-}2)\slg & 8(1{-}e^\cC)v^\cC\slg & 2(1{-}e^\cC)(1{+}v^\cC)\slg\sldelta & {-}2(1{-}e^\cC)(1{-}v^\cC)(\rho\pa_\rho{-}2)\slg & 2(1{-}e^\cC)(1{-}v^\cC)\slg\sldelta & {-}(1{-}e^\cC)v^\cC(\rho\pa_\rho{-}2)\slg\sltr
          \closebigpmatrix$} \\
       &+ \gamma^\Ups
         \scalebox{0.6}{$\openbigpmatrix{0pt}
           -2(\rho\pa_\rho+1) & -2(1-e^\Ups)(\rho\pa_\rho+1) & 0 & 0 & 0 & -\frac12 e^\Ups(\rho\pa_\rho+1)\sltr \\
           \rho\pa_\rho+1 & 0 & 0 & -(\rho\pa_\rho+1) & 0 & 0 \\
           2\sld & 2(1-e^\Ups)\sld & -(\rho\pa_\rho+2) & 0 & -(\rho\pa_\rho+2) & \frac12 e^\Ups\sld\sltr \\
           0 & 2(1-e^\Ups)(\rho\pa_\rho+1) & 0 & 2(\rho\pa_\rho+1) & 0 & \frac12 e^\Ups(\rho\pa_\rho+1)\sltr \\
           0 & 2(1-e^\Ups)\sld & \rho\pa_\rho+2 & 2\sld & \rho\pa_\rho+2 & \frac12 e^\Ups\sld\sltr \\
           4\slg & 0 & 4\sldelta^* & -4\slg & 4\sldelta^* & 0
         \closebigpmatrix$} \\
       &+ \gamma^\cC\gamma^\Ups
         \scalebox{0.6}{$\openbigpmatrix{0pt}
           4(1-v^\cC) & 4(1-e^\Ups)(1-v^\cC) & 0 & 0 & 0 & e^\Ups(1-v^\cC)\sltr \\
           2 e^\cC(1+v^\cC) & 4 e^\cC(1-e^\Ups) & 0 & 2 e^\cC(1-v^\cC) & 0 & e^\cC e^\Ups\sltr \\
           0 & 0 & 2(1-v^\cC) & 0 & 2(1-v^\cC) & 0 \\
           0 & 4(1-e^\Ups)(1+v^\cC) & 0 & 4(1+v^\cC) & 0 & e^\Ups(1+v^\cC)\sltr \\
           0 & 0 & 2(1+v^\cC) & 0 & 2(1+v^\cC) & 0 \\
           4(1-e^\cC)(1+v^\cC)\slg & 8(1-e^\cC)(1-e^\Ups)\slg & 0 & 4(1-e^\cC)(1-v^\cC)\slg & 0 & 2(1-e^\cC)e^\Ups\slg\sltr
         \closebigpmatrix$},
  \end{split}
  \end{equation}
  with $\wh{\ubar\Box}(0)$ given by~\eqref{EqTYMinkOpBox2}. The block matrix $\ubar S_{\ubar E^\Ups,\ubar E^\cC}$ is lower triangular in the splitting~\eqref{EqExOpLinSplit}, and
  \begin{equation}
  \label{EqWEOpMinkSpec}
    \min\spec\ubar S_{\ubar E^\Ups,\ubar E^\cC} = 0.
  \end{equation}
\end{lemma}

Note that the operator $\ubar S_{\ubar E^\Ups,\ubar E^\cC}$ already appeared in~\eqref{EqExOpLinAhBh} (with $h=0$ there). In the splitting~\eqref{EqExOpLinSplit}, it takes the schematic form
\begin{equation}
\label{EqWEOpMinkS}
  \ubar S_{\ubar E^\Ups,\ubar E^\cC}=
    \begin{pmatrix}
      2(1-v^\cC)\gamma^\cC & 0 & 0 & 0 & 0 & 0 & 0 \\
      0 & (1-v^\cC)\gamma^\cC & 0 & 0 & 0 & 0 & 0 \\
      * & 0 & (1-e^\cC)(1-v^\cC)\gamma^\cC & 0 & 0 & 0 & 0 \\
      0 & 0 & * & (1-e^\Ups)\gamma^\Ups & 0 & 0 & 0 \\
      0 & * & 0 & 0 & \gamma^\Ups & 0 & 0 \\
      0 & 0 & 0 & 0 & 0 & 0 & 0 \\
      0 & 0 & * & * & 0 & 0 & 2\gamma^\Ups
    \end{pmatrix},
\end{equation}
where ``$*$'' indicates a constant non-zero entry; this is the same as $A_0$ in the notation of~\eqref{EqExOpLinAB}.

\begin{proof}[Proof of Lemma~\usref{LemmaWEOpMink}]
  Write
  \[
    2\ubar L = \ubar\Box + 2\ubar E^\cC\ubar\delta\ul\sfG + 2(\ubar\delta^*+\ubar E^\cC)\ubar E^\Ups\ul\sfG.
  \]
  By Lemma~\ref{LemmaTYMinkOp}, the bundle map $\ubar S_{\ubar E^\Ups,\ubar E^\cC}$ is equal to $\frac12\rho^{-1}$ times the sum of $\pa_{t_*}$-coefficients of the second and third terms; it then remains to combine the expressions in~\eqref{EqTYMinkTrRev} and Lemma~\ref{LemmaTYMinkOp} with~\eqref{EqWGOpMinkE2} (for $\ubar E^\Ups$) and~\eqref{EqWCRecEC} (for $\ubar E^\cC$). In the bundle splitting~\eqref{EqExOpLinSplit} (with $x^0$ and $x^1$ replaced by $\ubar x^0$ and $\ubar x^1$, respectively, in order to match~\eqref{EqTYMink01Split}), the $7\times 7$ matrix for $\ubar S_{\ubar E^\Ups,\ubar E^\cC}$ is lower triangular, and given by the expression in~\eqref{EqExOpLinAB} with $h=0$; all eigenvalues (diagonal entries) are $\geq 0$, which gives~\eqref{EqWEOpMinkSpec}. The expression for $2\wh{\ubar L}(0)-\wh{\ubar\Box}(0)=2\ubar E^\cC\wh{\ubar\delta}(0)\ul\sfG+2(\wh{\ubar\delta^*}(0)+\ubar E^\cC)\ubar E^\Ups\ul\sfG$ recorded in~\eqref{EqWEOpMink0} is similarly obtained by a direct computation.
\end{proof}

\subsection{Strong trapping admissibility}
\label{SsWETr}

We prove the analogue of Lemma~\ref{LemmaWGTr} for the linearized gauge-fixed Einstein operator.

\begin{prop}[Strong trapping admissibility]
\label{PropWETr}
  The operator $2 L_b$ is strongly trapping admissible in the sense of \citeAF{Definition~\ref*{DefSSTrapAdm}}. That is, for all $\eps>0$ there exists a time-translation-invariant positive definite fiber inner product on $\pi^*(S^2\cT^*)$, homogeneous of degree $0$ with respect to fiber-dilations in $T^*\cM_b\setminus o$, such that
  \begin{equation}
  \label{EqWETr}
    \sigma^{-1}\frac{1}{2 i}\bigl(S_\sub(2 L_b) - S_\sub(2 L_b)^*\bigr) < \eps\quad\text{at}\ \Gamma_0,
  \end{equation}
  where we use the notation of Lemma~\usref{LemmaWGTr}.
\end{prop}
\begin{proof}
  In the subextremal Kerr--de~Sitter setting in \cite[\S{3.2}]{HintzPetersenVasyKdS}, a result was proved that is similar but has a critical deficit for our purposes: for fixed $\eps>0$ it was shown that there exists a fiber inner product such that~\eqref{EqWETr} holds when $e^\cC>0$ is sufficiently small \emph{depending on $\eps$}. We need a stronger result here, namely, that for \emph{fixed} modification parameters in Definitions~\ref{Def1Gauge} and \ref{Def1Symm} and for \emph{all} $\eps>0$ there exists a fiber inner product with~\eqref{EqWETr}.\footnote{This will be crucial for the proof of b-regularity of solutions $u$ of $L_b u=f$ with fixed regularity loss relative to $f$; see the discussions after \citeAF{(\ref*{EqA1AdmHiFinal})} and~\eqref{EqAdmPfHiRes} below.} We prove this simply by doing a more comprehensive computation, which takes the full form of $E^\cC_{g_b}$ into account.

  Since the gauge modification (encoded by $E_\Ups$) is supported away from the base projection of the trapped set $\Gamma_0$, only the constraint damping modifications enter in this computation. (Without constraint damping, the conclusion would follow from the fact, which is a consequence of~\eqref{EqWGTrMarckNabla}, that $\nabla_{H_{g_b}}^{\pi^*S^2 T^*\cM_b}$ has a nilpotent structure.) The main task is thus the evaluation of
  \begin{equation}
  \label{EqWETrSsub}
    i S_\sub(2 L_b) = i S_\sub(\Box_{g_b}) + 2 E_{g_b}^\cC\upsigma^1(i\delta_{g_b})\sfG_{g_b}
  \end{equation}
  (cf.\ \eqref{EqWGSsub} and \eqref{EqWEOpLb}). Recall the smooth frame $\sfe^\mu$ of $\pi^*T^*\cM_b$ over the superset $\{\sC>0\}\subset T^*\cM_b\setminus o$ of $\Gamma_0$ from~\eqref{EqWGTrMarckSplit}. Denote by $\sfe_\mu$, $\mu=0,\ldots,3$, the dual basis to $\{\sfe^\mu\}$. Let us work with the smooth bundle splitting
  \begin{equation}
  \label{EqWETrSplit}
  \begin{split}
    &\la(\sfe^0)^2\ra \oplus \la 2\sfe^0\otimes_s\sfe^1\ra \oplus \la 2\sfe^0\otimes_s\sfe^2\ra \oplus \la 2\sfe^0\otimes_s\sfe^3\ra \\
    &\quad \oplus \la(\sfe^1)^2\ra \oplus \la 2\sfe^1\otimes_s\sfe^2\ra \oplus \la 2\sfe^1\otimes_s\sfe^3\ra \oplus \la(\sfe^2)^2\ra \oplus \la 2\sfe^2\otimes_s\sfe^3\ra \oplus \la(\sfe^3)^2\ra
  \end{split}
  \end{equation}
  of $\pi^*S^2 T^*\cM_b$ over $\{\sC>0\}$. Then~\eqref{EqWGTrMarckNabla} and the Leibniz rule $\nabla_{H_{G_b}}^{\pi^*S^2 T^*\cM_b}(\sfe\otimes\sff)=(\nabla_{H_{G_b}}^{\pi^*T^*\cM_b}\sfe)\otimes\sff+\sfe\otimes(\nabla_{H_{G_b}}^{\pi^*T^*\cM_b}\sff)$ imply
  \[
    i S_\sub(\Box_{g_b}) = \nabla_{H_{G_b}}^{\pi^*S^2 T^*\cM_b} = H_{G_b} + 2\sigma
      \begin{pmatrix}
        0 & 2 & 0 & 0 & 0 & 0 & 0 & 0 & 0 & 0 \\
        0 & 0 & 1 & 0 & 1 & 0 & 0 & 0 & 0 & 0 \\
        0 & 0 & 0 & 0 & 0 & 1 & 0 & 0 & 0 & 0 \\
        0 & 0 & 0 & 0 & 0 & 0 & 1 & 0 & 0 & 0 \\
        0 & 0 & 0 & 0 & 0 & 2 & 0 & 0 & 0 & 0 \\
        0 & 0 & 0 & 0 & 0 & 0 & 0 & 1 & 0 & 0 \\
        0 & 0 & 0 & 0 & 0 & 0 & 0 & 0 & 1 & 0 \\
        0 & 0 & 0 & 0 & 0 & 0 & 0 & 0 & 0 & 0 \\
        0 & 0 & 0 & 0 & 0 & 0 & 0 & 0 & 0 & 0 \\
        0 & 0 & 0 & 0 & 0 & 0 & 0 & 0 & 0 & 0
      \end{pmatrix}.
  \]

  Turning to the second term of~\eqref{EqWETrSsub}, note first that $g_b^{-1}=-2\sfe_0\otimes_s\sfe_2+\sfe_1^2+\sfe_3^2$ and $g_b=-2\sfe^0\otimes_s\sfe^2+(\sfe^1)^2+(\sfe^3)^2$ by~\eqref{EqWGTrMarckInner}, so in the splitting~\eqref{EqWETrSplit} and its dual,
  \[
    g_b = (0, 0, -1, 0, 1, 0, 0, 0, 0, 1),\quad
    \tr_{g_b} = (0,0,-2,0,1,0,0,0,0,1),
  \]
  and therefore
  \[
    \sfG_{g_b} = I-\frac12 g_b\tr_{g_b} =
      \begin{pmatrix}
        1 & 0 & 0 & 0 & 0 & 0 & 0 & 0 & 0 & 0 \\
        0 & 1 & 0 & 0 & 0 & 0 & 0 & 0 & 0 & 0 \\
        0 & 0 & 0 & 0 & \frac12 & 0 & 0 & 0 & 0 & \frac12 \\
        0 & 0 & 0 & 1 & 0 & 0 & 0 & 0 & 0 & 0 \\
        0 & 0 & 1 & 0 & \frac12 & 0 & 0 & 0 & 0 & -\frac12 \\
        0 & 0 & 0 & 0 & 0 & 1 & 0 & 0 & 0 & 0 \\
        0 & 0 & 0 & 0 & 0 & 0 & 1 & 0 & 0 & 0 \\
        0 & 0 & 0 & 0 & 0 & 0 & 0 & 1 & 0 & 0 \\
        0 & 0 & 0 & 0 & 0 & 0 & 0 & 0 & 1 & 0 \\
        0 & 0 & 1 & 0 & -\frac12 & 0 & 0 & 0 & 0 & \frac12
      \end{pmatrix}.
  \]
  Recall moreover that $\sfe^0=\frac{1}{\sqrt\sC}\zeta$ at $(z,\zeta)\in T^*\cM_b$; therefore, in the splittings~\eqref{EqWETrSplit} and~\eqref{EqWGTrMarckSplit},
  \begin{align*}
    \upsigma^1(i\delta_g)(z,\zeta) &= \iota_{g_b^{-1}(\zeta)} = \sqrt\sC \iota_{g_b^{-1}(\sfe^0)} = -\sqrt\sC \iota_{\sfe_2} \\
    &= \sqrt\sC
       \begin{pmatrix}
         0 & 0 & -1 & 0 & 0 & 0  & 0 & 0  &  0 & 0 \\
         0 & 0 & 0  & 0 & 0 & -1 & 0 & 0  &  0 & 0 \\
         0 & 0 & 0  & 0 & 0 & 0  & 0 & -1 &  0 & 0 \\
         0 & 0 & 0  & 0 & 0 & 0  & 0 & 0  & -1 & 0
       \end{pmatrix}.
  \end{align*}
  Finally, writing $\cd^\cC=\cd_\mu\sfe^\mu$, we have $g_b^{-1}(\cd^\cC)=-\cd_0\sfe_2+\cd_1\sfe_1-\cd_2\sfe_0+\cd_3\sfe_3=(-\cd_2,\cd_1,-\cd_0,\cd_3)$. The map $E^\cC_{g_b}$ defined in~\eqref{Eq1Symm} is thus equal to
  \[
    E^\cC_{g_b} = \gamma^\cC
      \begin{pmatrix}
        2\cd_0 & 0 & 0 & 0 \\
        \cd_1 & \cd_0 & 0 & 0 \\
        \cd_2 & 0 & \cd_0 & 0 \\
        \cd_3 & 0 & 0 & \cd_0 \\
        0 & 2\cd_1 & 0 & 0 \\
        0 & \cd_2 & \cd_1 & 0 \\
        0 & \cd_3 & 0 & \cd_1 \\
        0 & 0 & 2\cd_2 & 0 \\
        0 & 0 & \cd_3 & \cd_2 \\
        0 & 0 & 0 & 2\cd_3
      \end{pmatrix}
      - \gamma^\cC(1-e^\cC)
      \begin{pmatrix}
        0 & 0 & 0 & 0 \\
        0 & 0 & 0 & 0 \\
        \cd_2 & -\cd_1 & \cd_0 & -\cd_3 \\
        0 & 0 & 0 & 0 \\
        -\cd_2 & \cd_1 & -\cd_0 & \cd_3 \\
        0 & 0 & 0 & 0 \\
        0 & 0 & 0 & 0 \\
        0 & 0 & 0 & 0 \\
        0 & 0 & 0 & 0 \\
        -\cd_2 & \cd_1 & -\cd_0 & \cd_3
      \end{pmatrix}.
  \]
  Altogether, then, upon writing $\gamma:=\frac{\sqrt\sC}{\sigma}\gamma^\cC$, which is homogeneous of degree $0$ and positive on $\Gamma_0$ (since $\sigma>0$ there by \citeAF{(\ref*{EqTs3bOTrapSign})}), we have
  \[
    i S_\sub(2 L_b) = H_{G_b} +
    \sigma A,\quad A:=\scalebox{0.78}{$\openbigpmatrix{3pt}
      0 & 4 & 0 & 0 & {-}2\gamma\cd_0 & 0 & 0 & 0 & 0 & {-}2\gamma\cd_0 \\
      0 & 0 & 2 & 0 & 2{-}\gamma\cd_1 & {-}2\gamma\cd_0 & 0 & 0 & 0 & {-}\gamma\cd_1 \\
      0 & 0 & 0 & 0 & {-}e^\cC\gamma\cd_2 & 2{-}2(1{-}e^\cC)\gamma\cd_1 & 0 & {-}2 e^\cC\gamma\cd_0 & {-}2(1{-}e^\cC)\gamma\cd_3 & {-}e^\cC\gamma\cd_2 \\
      0 & 0 & 0 & 0 & {-}\gamma\cd_3 & 0 & 2 & 0 & {-}2\gamma\cd_0 & {-}\gamma\cd_3 \\
      0 & 0 & 0 & 0 & {-}(1{-}e^\cC)\gamma\cd_2 & 4{-}2(1{+}e^\cC)\gamma\cd_1 & 0 & {-}2(1{-}e^\cC)\gamma\cd_0 & 2(1{-}e^\cC)\gamma\cd_3 & {-}(1{-}e^\cC)\gamma\cd_2 \\
      0 & 0 & 0 & 0 & 0 & {-}2\gamma\cd_2 & 0 & 2{-}2\gamma\cd_1 & 0 & 0 \\
      0 & 0 & 0 & 0 & 0 & {-}2\gamma\cd_3 & 0 & 0 & 2{-}2\gamma\cd_1 & 0 \\
      0 & 0 & 0 & 0 & 0 & 0 & 0 & {-}4\gamma\cd_2 & 0 & 0 \\
      0 & 0 & 0 & 0 & 0 & 0 & 0 & {-}2\gamma\cd_3 & {-}2\gamma\cd_2 & 0 \\
      0 & 0 & 0 & 0 & {-}(1{-}e^\cC)\gamma\cd_2 & 2(1{-}e^\cC)\gamma\cd_1 & 0 & {-}2(1{-}e^\cC)\gamma\cd_0 & {-}2(1{+}e^\cC)\gamma\cd_3 & {-}(1{-}e^\cC)\gamma\cd_2
    \closebigpmatrix$}.
  \]
  The eigenvalues of the $A$ are (with multiplicity)
  \[
    0\ (6\times),\quad
    -2(1-e^\cC)\gamma\cd_2\ (1\times),\quad
    -2\gamma\cd_2\ (2\times),\quad
    -4\gamma\cd_2\ (1\times).
  \]
  This follows from the observation that $A$ is upper triangular in the basis
  \begin{equation}
  \label{EqWGTrSplit2}
    (\sfe^0)^2,\ 2\sfe^0\otimes_s\sfe^3,\ 2\sfe^0\otimes_s\sfe^1,\ 2\sfe^1\otimes_s\sfe^3,\ 2\sfe^0\otimes_s\sfe^2,\ (\sfe^1)^2-(\sfe^3)^2,\ (\sfe^3)^2,\ 2\sfe^1\otimes_s\sfe^2,\ 2\sfe^2\otimes_s\sfe^3,\ (\sfe^2)^2,
  \end{equation}
  with these diagonal entries (in the stated order). Recall now that $\zeta$ and thus $\sfe^0$ are \emph{future} null at points $(z,\zeta)\in\Gamma_0$ in the trapped set inside of the future characteristic set. Since $\cd^\cC$ is \emph{past} timelike (see Theorem~\ref{ThmWCRec}), we thus have
  \[
    \cd_2=-g_b^{-1}(\cd^\cC,\sfe^0)<0.
  \]
  Therefore, all eigenvalues of $A$ are $\geq 0$. If in the basis~\eqref{EqWGTrSplit2} we define an inner product by $\diag(1,\eta^{-1},\ldots,\eta^{-9})$ where $\eta>0$, then in the orthonormal basis given by rescaling~\eqref{EqWGTrSplit2}, the off-diagonal entries of $A$ are of size $\cO(\eta)$ as $\eta\to 0$, and thus $A+A^*\geq-C\eta$ for some $\eta$-independent $C$. Since the inner product is constant and $H_{G_b}$ acts component-wise, we have $H_{G_b}^*=-H_{G_b}$, and therefore,
  \[
    \sigma^{-1}\frac{1}{2}\Bigl(i S_\sub(2 L_b) + \bigl(i S_\sub(2 L_b)\bigr)^*\Bigr) > -\eps
  \]
  when $\eta$ is small enough, as desired.
\end{proof}

\subsection{Indicial roots}
\label{SsWEInd}

We can evaluate the indicial roots of $\wh{L_b}(0)$ using the explicit expression for its normal operator $\wh{\ubar L}(0)$ in~\eqref{EqWEOpMink0} and \eqref{EqTYMinkOpBox2}. In the terminology of Definition~\ref{DefTYIndRoot}, the result is:

\begin{lemma}[Indicial roots]
\label{LemmaWEInd}
  All indicial roots of $\wh{L_b}(0)$ (or equivalently: of $\wh{\ubar L}(0)$) are simple. In terms of the numbers in Lemma~\usref{LemmaWGInd} and~\eqref{EqWCIndRoots}, they are given as follows:
  \begin{align*}
    \text{$\rms 0$ roots:}\quad & \lambda^\cC_{\rms 0,\ll,1}{-}1,\ \lambda^\cC_{\rms 0,\ll 2}{-}1,\ -\lambda^\Ups_{\rms 0,1}{+}1,\ 0,\ \lambda^\cC_{\rms 0,+}{-}1,\ 1,\ \lambda^\Ups_{\rms 0,1}{+}1,\ 3; \\
    \text{$\rms 1$ roots:}\quad & \lambda^\cC_{\rms 1,\ll,1}{-}1,\ \lambda^\cC_{\rms 1,\ll,2}{-}1,\ \lambda^\cC_{\rms 1,\ll,3}{-}1,\ -\lambda^\Ups_{\rms 1,2}{+}1,\ -1,\ -\lambda^\Ups_{\rms 1,1}{+}1, \\
      & \quad \lambda^\cC_{\rms 1,+,1}{-}1,\ \lambda^\cC_{\rms 1,+,2}{-}1,\ 2,\ \lambda^\Ups_{\rms 1,1}{+}1,\ \lambda^\Ups_{\rms 1,2}{+}1,\ 4; \\
    \text{$\rms l$ roots, $l\geq 2$:}\quad & \lambda^\cC_{\rms l,\ll,1}{-}1,\ \lambda^\cC_{\rms l,\ll,2}{-}1,\ \lambda^\cC_{\rms l,\ll,3}{-}1,\ -\lambda^\Ups_{\rms l,l+1}{+}1,\ -l,\ -\lambda^\Ups_{\rms l,l}{+}1,\ -l{+}2, \\
      & \quad \lambda^\cC_{\rms l,+,1}{-}1,\ \lambda^\cC_{\rms l,+,2}{-}1,\ \lambda^\cC_{\rms l,+,3}{-}1,\ l+1,\ \lambda^\Ups_{\rms l,l}{+}1,\ \lambda^\Ups_{\rms l,l+1}{+}1,\ l{+}3; \\
    \text{$\rmv 1$ roots:}\quad & \lambda^\cC_{\rmv 1,\ll}{-}1,\ -1,\ 2,\ 3; \\
    \text{$\rmv l$ roots, $l\geq 2$:}\quad & \lambda^\cC_{\rmv l,\ll}{-}1,\ -l,\ -l{+}1,\ \lambda^\cC_{\rmv l,+}{-}1,\ l{+}1,\ l{+}2.
  \end{align*}
  In particular, there is a positive number
  \[
    \eps_\ind>0
  \]
  such that no indicial root has real part in the interval $(0,\eps_\ind)$. Moreover, the number of roots of a fixed type with real parts $\leq 0$, resp.\ $\geq\eps_\ind$ are equal.
\end{lemma}

Thus, $(0,\eps_\ind)$ is contained in the indicial gap of $\wh{L_b}(0)$. It is convenient for the remainder of the paper to reduce $\eps_\ind$ further so that
\begin{equation}
\label{EqWEIndEps}
  \parbox{0.86\textwidth}{\it\centering $d(\Re\lambda,\Z)\geq\eps_\ind$ for all non-integer indicial roots $\lambda$ with $|\lambda|\leq 50$.}
\end{equation}
(The number $50$ here is arbitrary; we only really use this for $|\lambda|\leq 4$.) Recall also that the enhanced mode stability of \cite{HintzKerrCD} implies that among the indicial roots in Lemma~\ref{LemmaWEInd} with non-positive real parts, those with superscript ``$\cC$'' (for ``constraint damping'') have real parts $<-100$.

\begin{proof}[Proof of Lemma~\usref{LemmaWEInd}]
  Let us briefly explain some of the relationships between the stated roots and those in Lemma~\ref{LemmaWGInd} and \eqref{EqWCIndRoots}; this expands on the discussion in~\S\ref{SssIEinLa}. First of all, if $\lambda$ is an indicial root of $\wh{\ubar\Box_{\ubar E^\Ups}^\Ups}(0)$ with indicial solution $\omega_0$, so $\ubar\Box_{\ubar E^\Ups}^\Ups(\rho^\lambda\omega)=0$, then $h:=\ubar\delta^*(\rho^\lambda\omega_0)=\rho^{\lambda+1}N(\rho^{-1}\wh{\ubar\delta^*}(0),\lambda)\omega_0$ lies in the kernel of $\wh{\ubar L}(0)$, so $\lambda+1$ is an indicial root of $\wh{\ubar L}(0)$---unless $h=0$, which happens if and only if $\rho^\lambda\omega_0$ is a Killing 1-form on Minkowski space (spacetime translations or spatial rotations), and thus only for $\lambda=-1,0$.

  Similarly, an indicial root $\mu$ of $\wh{\ubar\Box_{\ubar E^\cC}^\cC}(0)^*$ (which is equal to $\mu=1-\bar\lambda$ where $\lambda$ is an indicial root of $\wh{\ubar\Box_{\ubar E^\cC}^\cC}(0)$, cf.\ the computation~\eqref{EqWC0DualIndRoot}) gives rise, via application of $\ul\sfG\,\ubar\delta^*$ to an indicial solution, to an indicial root $\mu+1$ (and corresponding indicial solution) of $\wh{\ubar L}(0)^*$. Therefore, barring possible cancellations when $\mu=-1,0$, the number $1-\ol{\mu+1}=-\bar\mu=\lambda-1$ is an indicial root of $\wh{\ubar L}(0)$.

  Conversely, suppose $\lambda$ is an indicial root of $\wh{\ubar L}(0)$, with indicial solution $u_0\in\CI(\pa X;S^2\cT^*_X)$, so $\ubar L(\rho^\lambda u_0)=0$. Applying $\ubar\delta\,\ul\sfG$ and using the linearized second Bianchi identity $\ubar\delta\,\ul\sfG\,D_{\ubar g}\Ric=0$, one finds that $N(\wh{\ubar\Box_{\ubar E^\cC}^\cC}(0),\lambda+1)\eta_0=0$ where $\eta=\rho^{\lambda+1}\eta_0:=\ubar\delta_{\ubar E^\Ups}\,\ul\sfG(\rho^\lambda u_0)$. If $\lambda+1$ is not an indicial root of $\wh{\ubar\Box_{\ubar E^\cC}^\cC}(0)$, one can conclude that $\eta_0=0$ (so $\rho^\lambda u_0$ satisfies the linearized gauge condition), and thus in fact $N(\wh{D_{\ubar g}\Ric}(0),\lambda)u_0=0$. The space of solutions of this equation was described in \cite[Proposition~7.6]{HintzGlueLocI}; except for exceptional values of $\lambda$ and scalar/vector types, $u_0$ must be pure gauge, so $u_0=N(\wh{\ubar\delta^*}(0),\lambda-1)\omega_0$. But then the linearized gauge condition gives $N(\wh{\ubar\Box_{\ubar E^\Ups}^\Ups}(0),\lambda-1)\omega_0=0$. If $\lambda-1$ is not an indicial root of $\wh{\ubar\Box_{\ubar E^\Ups}^\Ups}(0)$, we thus obtain $\omega_0=0$, and thus $\lambda$ is not an indicial root of $\wh{\ubar L}(0)$.

  We shall, however, not carry out the analysis of the exceptional cases in detail here. Instead, we simply compute the expressions for $N(\wh{\ubar L}(0),\lambda)$ acting on the various pure types, which are explicit square matrices (obtained using the formulas~\eqref{EqTYIdentities}), and relate their determinants to the determinants of the matrices of normal operator families of $\wh{\ubar\Box_{\ubar E^\Ups}^\Ups}(0)$ and $\wh{\ubar\Box_{\ubar E^\cC}^\cC}(0)$. The full computations are lengthy and were performed with \texttt{Wolfram Mathematica}; the results are as follows.
  \begin{enumerate}
  \item{\rm ($\rms 0$ roots.)} Identifying the space of $\rms 0$ tensors on $\pa X$ for fixed $0\neq\scal\in\scalspace_0$ with $\C^4$ via~\eqref{EqTYSplit2}, $N_{\rms 0}(\wh{\ubar L}(0),\lambda)$ is a $4\times 4$ matrix (with entries which are quadratic polynomials in $\lambda$); and using~\eqref{EqTYSplit1} to express also the normal operators of the gauge potential and constraint propagation wave operators as $2\times 2$ matrices, one has
    \[
      \det\Bigl(N_{\rms 0}\bigl(\rho^{-2}\wh{\ubar L}(0),\lambda\bigr)\Bigr) = \det\Bigl(N_{\rms 0}\bigl(\wh{\ubar\Box_{\ubar E^\Ups}^\Ups}(0),\lambda-1\bigr)\Bigr) \det\Bigl(N_{\rms 0}\bigl(\wh{\ubar\Box_{\ubar E^\cC}^\cC}(0),\lambda+1\bigr)\Bigr).
    \]
    This expresses the indicial roots of the linearized gauge-fixed Einstein operator in terms of the indicial roots of the gauge potential and constraint damping wave operators.
  \item{\rm ($\rms 1$ roots.)} $N_{\rms 1}(\wh{\ubar L}(0),\lambda)$ is a $6\times 6$ matrix, while $N_{\rms 1}(\wh{\ubar\Box_{\ubar E^\bullet}^\bullet}(0),\lambda)$, $\bullet=\Ups,\cC$, are $3\times 3$-matrices. One has
    \begin{align*}
      \det\Bigl(N_{\rms 1}\bigl(\wh{\ubar L}(0),\lambda\bigr)\Bigr) &= (\lambda+1)(\lambda-2) \\
        &\qquad \times (\lambda-1)^{-1}\det\Bigl(N_{\rms 1}\bigl(\wh{\ubar\Box_{\ubar E^\Ups}^\Ups}(0),\lambda-1\bigr)\Bigr) \\
        &\qquad \times \lambda^{-1}\det\Bigl(N_{\rms 1}\bigl(\wh{\ubar\Box_{\ubar E^\cC}^\cC}(0),\lambda+1\bigr)\Bigr).
    \end{align*}
    Note that the expressions in the second and third line are polynomials (i.e., regular), seeing as $0$ and $1$ are $\rms 1$ indicial roots of $\wh{\ubar\Box_{\ubar E^\Ups}^\Ups}(0)$ and $\wh{\ubar\Box_{\ubar E^\cC}^\cC}(0)$, respectively. The $\rms 1$ roots of $\wh{\ubar L}(0)$ can thus be read off from this.
  \item{\rm ($\rms l$ roots, $l\geq 2$.)} The indicial family of $\wh{\ubar L}(0)$ is now a $7\times 7$ matrix, while the indicial families of the gauge potential and constraint damping wave operators are still $3\times 3$-matrices. One finds
    \begin{align*}
      -\det\Bigl(N_{\rms l}\bigl(\wh{\ubar L}(0),\lambda\bigr)\Bigr) &= (\lambda+l)(\lambda-l-1) \\
        &\qquad \times \det\Bigl(N_{\rms l}\bigl(\wh{\ubar\Box_{\ubar E^\Ups}^\Ups}(0),\lambda-1\bigr)\Bigr) \det\Bigl(N_{\rms l}\bigl(\wh{\ubar\Box_{\ubar E^\cC}^\cC}(0),\lambda+1\bigr)\Bigr).
    \end{align*}
  \item{\rm ($\rmv 1$ roots.)} Now, $N_{\rmv 1}(\wh{\ubar L}(0),\lambda)$ is a $2\times 2$ matrix, while the indicial families of the 1-form wave operators are $1\times 1$ matrices, i.e., quadratic polynomials in $\lambda$; then
    \begin{align*}
      \det\Bigl(N_{\rmv 1}\bigl(\wh{\ubar L}(0),\lambda\bigr)\Bigr) &= (\lambda+1)(\lambda-2) \\
        &\qquad \times \lambda^{-1} N_{\rmv 1}\bigl(\wh{\ubar\Box_{\ubar E^\Ups}^\Ups}(0),\lambda-1\bigr) \times (\lambda-1)^{-1} N_{\rmv 1}\bigl(\wh{\ubar\Box_{\ubar E^\cC}^\cC}(0),\lambda+1\bigr).
    \end{align*}
    Note that $-1$ and $2$ are $\rmv 1$ indicial roots of $\wh{\ubar\Box_{\ubar E^\Ups}^\Ups}(0)$ and $\wh{\ubar\Box_{\ubar E^\cC}^\cC}(0)$ by Lemma~\ref{LemmaWGInd} and~\eqref{EqWCIndRoots}, respectively.
  \item Finally, in the $\rmv l$ sector, $l\geq 2$, $N_{\rmv l}(\wh{\ubar L}(0),\lambda)$ is a $3\times 3$ matrix with
    \begin{align*}
      &-\det\Bigl(N_{\rmv l}\bigl(\wh{\ubar L}(0),\lambda\bigr)\Bigr) \\
      &\quad = (\lambda+l)(\lambda-l-1) \times N_{\rmv l}\bigl(\wh{\ubar\Box_{\ubar E^\Ups}^\Ups}(0),\lambda-1\bigr) \times N_{\rmv l}\bigl(\wh{\ubar\Box_{\ubar E^\cC}^\cC}(0),\lambda+1\bigr),
    \end{align*}
    from which we can read off the indicial roots.
  \end{enumerate}
  The final statement follows from the fact that $\Re\lambda^\cC_{\rms 0,+}>1$ and $\Re\lambda^\cC_{\rms l,+,j}>1$ for all $l,j$, as noted after~\eqref{EqWCIndRoots}.
\end{proof}

For later use, we record:
\begin{lemma}[Indicial solutions at $\lambda=0$ satisfy the linearized gauge condition]
\label{LemmaWEInds00}
  We have
  \[
    \ker N(\wh{\ubar L}(0),0)\subset\ker N(\wh{\ubar\delta_{\ubar E^\Ups}}(0)\ul\sfG,0).
  \]
\end{lemma}
\begin{proof}
  Since $0$ is only an indicial root of scalar types $0$ and $2$, we can discuss these two pure types separately.

  Given the 1-dimensional nature of $\ker N_{\rms 0}(\wh{\ubar L}(0),0)$ established in Lemma~\ref{LemmaWEInd}, note that $\ubar h_{\rms 0}^{(0)}$, defined in Proposition~\ref{PropWG0Large}\eqref{ItWG0Larges00}, is the leading-order term of $h_{b,\rms 0}^{(0)}$---which, in particular, satisfies $\delta_{g_b,E^\Ups}\sfG_{g_b}h_{b,\rms 0}^{(0)}=0$. The $\rho^1$ leading-order term of this equation reads
  \begin{equation}
  \label{EqWEInds00Pf}
    \wh{\ubar\delta_{\ubar E^\Ups}}(0)\ul\sfG\,\ubar h_{\rms 0}^{(0)}=0,
  \end{equation}
  as desired. One can avoid the somewhat unnatural passage to Kerr states by instead constructing $\ubar h_{\rms 0}^{(0)}=\ubar\delta^*\ubar\omega_{\rms 0}^{(0),\leq 1}$, $\ubar\omega_{\rms 0}^{(0),\leq 1}:=t_*\ubar\omega_{\rms 0}^{(0)}+\breve{\ubar\omega}_{\rms 0}^{(0),1}$, where $\ubar\omega_{\rms 0}^{(0)}=-\dd t$ is a Killing 1-form (from~\eqref{EqWG0SymmMink}) and $\breve{\ubar\omega}_{\rms 0}^{(0),1}:=-N(\wh{\ubar\Box_{\ubar E^\Ups}^\Ups}(0),-1)^{-1}[\ubar\Box_{\ubar E^\Ups}^\Ups,t_*]\ubar\omega_{\rms 0}^{(0)}$ (which is well-defined since $-1$ is not an $\rms 0$ indicial root); since $\ubar\omega_{\rms 0}^{(0),\leq 1}\in\ker\ubar\Box_{\ubar E^\Ups}^\Ups$, \eqref{EqWEInds00Pf} follows.

  The arguments for the scalar type $2$ nullspace of $N(\wh{\ubar L}(0),0)$ are similar (and, in fact, simpler, since it is spanned by $\ubar h_{\rms 2}^{(0)}=\ubar\delta^*\ubar\omega_{\rms 2}^{(-1)}$ for $\ubar\omega_{\rms 2}^{(-1)}\in\ker N_{\rms 2}(\wh{\ubar\Box_{\ubar E^\Ups}^\Ups}(0),-1)$).
\end{proof}

\subsection{tf-admissibility}
\label{SsWEtf}

Similarly to~\S\ref{SsWGtf}, we define the $\tface$-normal operator
\begin{equation}
\label{EqWEtfOp}
\begin{split}
  N_\tface(L_b,\hat\sigma) = N_\tface(\ubar L,\hat\sigma) &= 2 i\hat\sigma\hat\rho(\hat\rho\pa_{\hat\rho}-1-\ubar S_{\ubar E^\Ups,\ubar E^\cC}) + \hat\rho^2\wt{\ubar L}(0)(\omega,\hat\rho\pa_{\hat\rho},\pa_\omega) \\
    &= -2 i\hat\sigma\hat r^{-1}(\hat r\pa_{\hat r}+1+\ubar S_{\ubar E^\Ups,\ubar E^\cC}) + \hat r^{-2}\wt{\ubar L}(0)(\omega,-\hat r\pa_{\hat r},\pa_\omega),\quad \hat r:=\hat\rho^{-1}, \\
  &\hspace{-4em} \wh{\ubar L}(0) =: \rho^2\wt{\ubar L}(0)(\omega,\rho\pa_\rho,\pa_\omega)
\end{split}
\end{equation}
as a differential operator on $\tface=[0,\infty]_{\hat\rho}\times\Sph^2$ acting on sections of $\pi_\tface^*(S^2\cT^*_X)$. Here $\hat\rho=\frac{\rho}{|\sigma|}$. (See Lemma~\ref{LemmaWEOpMink} for the explicit expressions for $\ubar S_{\ubar E^\Ups,\ubar E^\cC}$ and $\wh{\ubar L}(0)$.)

\begin{prop}[$\tface$-admissibility]
\label{PropWEtf}
  For all sufficiently small values of $\gamma^\Ups>0$ in Definition~\usref{Def1Gauge} (while keeping the other parameters and all 1-forms fixed), the operator $L_b$ is $\tface$-admissible with weight $\beta\in(0,\eps_\ind)$ in the sense of \citeAF{Definition~\ref*{DefSStfAdm}}. That is, for all $\alpha\in\R$, the nullspace of $N_\tface(L_b,\hat\sigma)$ on
  \[
    \cA^{\alpha,-\beta}(\tface;\pi_\tface^*(S^2\cT^*_X)) = \rho_\sctface^\alpha\rho_\ztface^{-\beta}\cA^{0,0}(\tface;\pi_\tface^*(S^2\cT^*_X)),
  \]
  with $\rho_\sctface=\frac{\hat\rho}{1+\hat\rho}$ and $\rho_\ztface=\frac{1}{1+\hat\rho}$ as in~\eqref{EqWTGtfSpace}, is trivial for all $\hat\sigma\in e^{i[0,\pi]}$, and so is the nullspace of $N_\tface(L_b,1)^*$ (with the adjoint defined using the Euclidean density $\hat\rho^{-3}|\frac{\dd\hat\rho}{\hat\rho}\,\dd\slg|$ and any smooth fiber inner product on $S^2\cT^*_X$) on $e^{-2 i/\rho_\sctface}\cA^{\alpha,-1+\beta}(\tface;\pi_\tface^*(S^2\cT^*_X))$.
\end{prop}
\begin{proof}
  \pfstep{Triviality of the kernel.} Suppose $u_0=u_0(\hat\rho,\omega)\in\cA^{\alpha,-\beta}(\tface;\pi_\tface^*(S^2\cT^*_X))$ satisfies
  \[
    N_\tface(L_b,\hat\sigma)u_0=0.
  \]
  We first use a normal operator argument at $\hat r=0$ to obtain more information on $u_0$ there. The indicial roots of $N_\tface(L_b,\hat\sigma)$ at $\hat r=0$ are given by $-\lambda$ where $\lambda$ runs through the roots in Lemma~\ref{LemmaWEInd}; thus the first indicial root larger than $-\beta$ is the $\rms 0$ and $\rms 2$ root $0$, with indicial solutions given by $\ubar h_{\rms 0}^{(0)}$ and $\ubar h_{\rms 2}^{(0)}$ from Proposition~\ref{PropWG0Large}\eqref{ItWG0Larges00} and \eqref{ItWG0Largesl2}. Fix a cutoff function $\chi=\chi(\hat r)\in\CIc([0,2))$ that equals $1$ on $[0,1]$; we then conclude that, for some $\scal_0\in\scalspace_0$ and $\scal_2\in\scalspace_2$,
  \[
    u_0 = \chi(\hat r)\bigl(\scal_0\ubar h_{\rms 0}^{(0)} + \ubar h_{\rms 2}^{(0)}(\scal_2)\bigr) + \tilde u_0 \in \cA^{\alpha,\;((0,0),\delta)},\quad \tilde u_0 \in \cA^{\alpha,\delta}(\tface;\pi_\tface^*(S^2\cT^*_X)),
  \]
  where $\delta\in(0,1)$ is such that $\Re\lambda<-\delta$ for all indicial roots in Lemma~\ref{LemmaWEInd} whose real part is negative.

  We wish to apply the linearized second Bianchi identity to get information about the linearized gauge 1-form of $u_0$. This can be implemented using the multiplicativity of the $\tface$-normal operator map. We instead present a direct treatment by first passing to an equation on spacetime. The scaling-invariance of $\wh{\ubar L}(\sigma)$ in $(\sigma,\rho)$ implies that $\wh{\ubar L}(\sigma)u_0(\rho/|\sigma|,\omega)=0$ for $\sigma=|\sigma|\hat\sigma$ and any $|\sigma|>0$, or equivalently $\ubar L(e^{-i\sigma t_*}u_0(\rho/|\sigma|,\omega))=0$. For $|\sigma|=1$, this reads $\ubar L u=0$ where
  \[
    u(t_*,\rho,\omega):=e^{-i\hat\sigma t_*}u_0(\rho,\omega).
  \]
  Apply $\ubar\delta\,\ul\sfG$ and use the linearized second Bianchi identity $\ubar\delta\,\ul\sfG\,D_{\ubar g}\Ric=0$ to conclude that
  \[
    0 = \ubar\delta\,\ul\sfG\,\ubar\delta_{\ubar E^\cC}^* \eta = \frac12\ubar\Box_{\ubar E^\cC}^\cC\eta,\quad \eta := \ubar\delta_{\ubar E^\Ups}\ul\sfG u = e^{-i\hat\sigma t_*}\eta_0,\quad \eta_0=\wh{\ubar\delta_{\ubar E^\Ups}\ul\sfG}(\hat\sigma)u_0,
  \]
  so $\wh{\ubar\Box_{\ubar E^\cC}}(\hat\sigma)\eta_0(\rho,\omega)=0$. Now, $\wh{\ubar\delta_{\ubar E^\Ups}}(\hat\sigma)=\rho A_1+\hat\sigma A_0$ where $A_1:=\rho^{-1}\wh{\ubar\delta_{\ubar E^\Ups}}(0)=A_1(\omega,\rho\pa_\rho,\pa_\omega)$ (cf.\ the weights $\rho$ in~\eqref{EqTYMinkOpDel} and~\eqref{EqWGOpMinkE2}) and $A_0=\dd t_*\otimes_s(\cdot)$ are dilation-invariant (as bundle maps from $\cT^*_X|_{\pa X}$ to $S^2\cT^*_X|_{\pa X}$), so since $\rho=\hat\rho=\frac{\rho}{|\sigma|}$ for $|\sigma|=1$, we have $\eta_0(\hat\rho,\omega)=\hat\rho A_1(\omega,\hat\rho\pa_{\hat\rho},\pa_\omega)u_0(\hat\rho,\omega)+\hat\sigma A_0 u_0(\hat\rho,\omega)$, which in view of $\hat\rho=\rho_\sctface\rho_\ztface^{-1}$ implies
  \begin{equation}
  \label{EqWEtfeta0}
    \eta_0 \in \cA^{\alpha+1,\;((-1,0),\delta-1)} + \cA^{\alpha,\;((0,0),\delta)} = \cA^{\alpha,\;((-1,0),\delta-1)}(\tface;\pi_\tface^*(\cT^*_X)),\quad
    N_\tface\bigl(\wh{\ubar\Box_{\ubar E^\cC}^\cC}(0),\hat\sigma\bigr)\eta_0 = 0.
  \end{equation}
  The $\hat r^{-1}$ leading-order term of $\eta_0$ is given in terms of the leading-order term $\scal_0\ubar h_{\rms 0}^{(0)}+\ubar h_{\rms 2}^{(0)}(\scal_2)$ of $u_0$ by $N(\rho^{-1}\wh{\ubar\delta_{\ubar E^\Ups}}(0)\ul\sfG,0)(\scal_0\ubar h_{\rms 0}^{(0)}+\ubar h_{\rms 2}^{(0)}(\scal_2))$; this vanishes by Lemma~\ref{LemmaWEInds00}. Therefore, we in fact have
  \[
    \eta_0 \in \cA^{\alpha,\;\delta-1}(\tface;\pi_\tface^*(\cT^*_X)).
  \]
  In light of the equation~\eqref{EqWEtfeta0} and since $-(\delta-1)=1-\delta<1$, we can now appeal to Theorem~\ref{ThmWCRec}\eqref{ItWCRectf} to conclude that $\eta_0=0$. But this implies that $u$ satisfies the linearized gauge-fixed Einstein equation regardless of the choice of modified symmetric gradient, so in particular for the unmodified symmetric gradient, i.e.,
  \[
    \ubar L_{\gamma^\Ups}u = 0,\quad \ubar L_{\gamma^\Ups} := D_{\ubar g}\Ric + \ubar\delta^* \ubar\delta_{\ubar E^\Ups}\ul\sfG,
  \]
  and thus $N_\tface(\ubar L_{\gamma^\Ups},\hat\sigma)u_0=0$; we keep here the parameter $e^\Ups$ and the 1-form $\ubar\cd^\Ups$ fixed but keep $\gamma^\Ups$ variable.

  Now, for $\gamma^\Ups=0$, i.e., for $\ubar\delta$ in place of $\ubar\delta_{\ubar E^\Ups}$, the operator $2\ubar L_0=\ubar\Box$ is the Minkowskian wave operator acting on symmetric 2-tensors, which is $\tface$-admissible with weight $(0,1)$ by \citeAF{Proposition~\ref*{PropSptfAdm}}. Since the Fredholm estimates \citeAF{(\ref*{EqSptfEst})} hold uniformly for small $\gamma^\Ups$, a standard functional analytic argument implies the invertibility of $N_\tface(\ubar L_{\gamma^\Ups},\hat\sigma)$ on the spaces in \citeAF{(\ref*{EqSptfOp})} for all sufficiently small $\gamma^\Ups$ and all $\hat\sigma\in e^{i[0,\pi]}$. For small enough $\gamma^\Ups$ then, we can thus conclude that $u_0=0$. (One first improves the $\sctface=\scface\cap\tface$-decay rate $\alpha$ to $1-\eps$ for every $\eps>0$ using~\eqref{EqWEOpMinkSpec}, and then notes that $\cA^{1-\eps,-\beta}(\tface)$ is contained in the domain of~\citeAF{(\ref*{EqSptfOp})} in view of the threshold condition $\sfr+\alpha_+<-\frac12$ at the outgoing radial set in the notation of the reference, analogously to the proof of Theorem~\ref{ThmWCRec}\eqref{ItWCRectf}; see also the parenthetical remark in \citeAF{Theorem~\ref*{ThmSptf}(\ref*{ItSptfKer})}.)

  \pfstep{Triviality of the cokernel.} We use an abstract argument based on Fredholm index considerations. By \citeAF{Theorem~\ref*{ThmSptf}(\ref*{ItSptfReal}), (\ref*{ItSptfKer})}, we need to show that the index of $N_\tface(L_b,1)$ as a map~\citeAF{(\ref*{EqSptfOp})} is $0$, i.e., taking $\alpha_+=0$ in~\citeAF{(\ref*{EqSpOrderStart})} (and working with the induced orders) and $q\in\R$ with $-q+\frac32\in(0,\eps_\ind)$, that the index of the map
  \begin{equation}
  \label{EqWEtfNtf1}
    N_\tface(L_b,1) \colon \bigl\{ u\in H_{\scop,\bop}^{\sfs,(\sfr,q)}(\tface,\hat r^2\,|\dd\hat r\,\dd\slg|;\pi^*(S^2\cT^*_X)) \colon N_\tface(L_b,1)u \in H_{\scop,\bop}^{\sfs-2,(\sfr+1,q-2)} \bigr\} \to H_{\scop,\bop}^{\sfs-2,(\sfr+1,q-2)},
  \end{equation}
  is $0$. It suffices to show this for the restriction of $N_\tface(L_b,1)$ to each spherical harmonic type and degree, with $N_\tface(L_b,1)$ being a family of ordinary differential operators. Working with a fixed type, say scalar type $l\geq 2$ for concreteness, and indeed with scalar type $l$ tensors relative to a fixed $1$-dimensional subspace of $\scalspace_l$, increase then $q$ to exceed $\frac32-\Re\lambda$ for all scalar type $l$ indicial roots $\lambda$ of $\wh{L_b}(0)$. By doing so, one crosses the real parts of $3$ indicial roots of $N_\tface(L_b,1)$ at $\ztface$ (which are $-1$ times the indicial roots of $\wh{L_b}(0)$ listed in Lemma~\ref{LemmaWEInd}), and hence the index jumps by $-3$, and remains such when one continuously shifts the modification parameters $\gamma^\Ups$ and $\gamma^\cC$ in Definitions~\ref{Def1Gauge} and \ref{Def1Symm} to $0$ (provided $q$ is large enough so that $q$ exceeds $\frac32-\Re\lambda$ for all $\rms l$ indicial roots arising during this deformation). Once one has reached $\gamma^\Ups=0=\gamma^\cC$, in which case $N_\tface(L_b,1)=N_\tface(\frac12\ubar\Box,1)$, one shifts $q$ back to the central interval, i.e., to a value with $-q+\frac32\in(0,1)$; by doing so, one crosses three indicial roots (cf.\ the example after Definition~\ref{DefTYIndRoot}) and thus increases the Fredholm index by $3$. But for such $q$, the invertibility (and in particular Fredholm index $0$ property) of $N_\tface(\ubar\Box,1)$ is the content of \citeAF{Proposition~\ref*{PropSptfAdm}}. Tracing the index back to that of~\eqref{EqWEtfNtf1} shows that the index of~\eqref{EqWEtfNtf1} is $0$, as claimed.
\end{proof}

As a consequence, the quantitative estimates for $N_\tface(L_b,\hat\sigma)$ on scattering-b-Sobolev spaces on $\tface$ (including with additional global b-regularity) stated in \citeAF{Corollary~\ref*{CorSptfAdm}} hold. We only state a version here that takes place entirely on b-Sobolev spaces on $\tface$ and consider only $\hat\sigma=\pm 1$.

\begin{cor}[$\tface$-normal operator inversion on b-Sobolev spaces]
\label{CorWEtfb}
  We use the unweighted b-density $|\frac{\dd\hat\rho}{\hat\rho}\,\dd\slg|=|\frac{\dd\hat r}{\hat r}\,\dd\slg|$ on $\tface$. Fix
  \begin{subequations}
  \begin{equation}
  \label{EqWEtfb1}
    \beta_\sctface < 1+\min\Re\ubar S_{\ubar E^\Ups,\ubar E^\cC} = 1,\quad
    \beta_\ztface \in (-\eps_\ind,0).
  \end{equation}
  Then there exists $l\in\N_0$ such that for all $k\in\N_0$, we have
  \begin{equation}
  \label{EqWEtfb2}
    N_\tface(L_b,\pm 1)^{-1} \colon \Hb^{k+l,\ \beta_\sctface+1,\ \beta_\ztface-2}(\tface) \to \bigcap_{\eps>0}\Hb^{k,\ \beta_\sctface-\eps,\ \beta_\ztface}(\tface).
  \end{equation}
  \end{subequations}
\end{cor}
\begin{proof}
  The (lossy) translation between the variable scattering-b-Sobolev spaces (with additional integer amounts of b-regularity) used in \citeAF{(\ref*{EqSptfAdm})} and pure b-Sobolev spaces is explained in \citeAF{Lemma~\ref*{LemmaDRes}} on $X$ near $\pa X$, which is the same as on $\tface$ near $\sctface=\tface\cap\scface$.
\end{proof}

\subsection{Mode stability and basic zero energy analysis}
\label{SsWEMode}

\emph{We fix the modification parameters as explained before~\eqref{EqWEOp}, with $\gamma^\Ups$ in addition small enough so that the conclusions of Proposition~\usref{PropWEtf} hold.}

\begin{prop}[Mode stability for non-zero frequencies]
\label{PropWEMode}
  Mode stability holds for $L_b$ at all $0\neq\sigma\in\C$ with $\Im\sigma\geq 0$: the kernel of $\wh{L_b}(\sigma)$ on $\cA^\alpha(X;S^2\cT^*_X)$ is trivial for all $\alpha\in\R$, and $\wh{L_b}(\sigma)$ is invertible as a map~\citeAF{(\ref*{EqSpBMap})}.
\end{prop}

This is the same statement as \cite[Proposition~3.15]{HintzGlueLocIII}, the only difference being that our present definition of $L_b$ implements large parameter constraint damping. 

\begin{proof}[Proof of Proposition~\usref{PropWEMode}]
  By the Fredholm index $0$ property of $\wh{L_b}(\sigma)$ as a map \citeAF{(\ref*{EqSpBMap})} (which follows from Proposition~\ref{PropWETr}), the invertibility statement follows from the injectivity (see also \citeAF{Theorem~\ref*{ThmSpB}(\ref*{ItSpBNull})}). Given $u\in\cA^\alpha(X;S^2\cT^*_X)$ with $\wh{L_b}(\sigma)u=0$, we have $L_b(e^{-i\sigma t_*}u)=0$. Applying $\delta_{g_b}\sfG_{g_b}$ to this equation and using the second Bianchi identity $\delta_{g_b}\sfG_{g_b}D_{g_b}\Ric=0$ yields
  \[
    \wh{\Box_{g_b,E^\cC}^\cC}(\sigma)\eta=0,\quad \eta:=\wh{\delta_{g_b,E^\Ups}}(\sigma)\sfG_{g_b}u \in \cA^{\alpha+1}(X;\cT^*_X).
  \]
  By Theorem~\ref{ThmWCRec}\eqref{ItWCRecNon0}, this implies $\eta=0$. Therefore, $D_{g_b}\Ric(e^{-i\sigma t_*}u)=0$. By \cite[Theorem~6.1(1)]{AnderssonHaefnerWhitingMode}, we obtain $u=\wh{\delta_{g_b}^*}(\sigma)\omega$ where $\omega\in\cA^{\alpha'}(X;\cT^*_X)$ for some $\alpha'\in\R$. But then $\eta=0$ implies
  \[
    \wh{\Box_{g_b,E^\Ups}^\Ups}(\sigma)\omega = 0.
  \]
  Mode stability for $\Box_{g_b,E^\Ups}^\Ups$ (Proposition~\ref{PropWGMode}) implies that $\omega=0$, and hence $u=0$.
\end{proof}

At zero frequency, we shall prove an analogue of \cite[Proposition~3.16]{HintzGlueLocIII}. First, we need to bring the linearized Kerr metrics $\dot g_b(\dot b)$ from Definition~\ref{DefKMetLin} into the desired linearized gauge:

\begin{prop}[Gauged linearized Kerr metrics]
\label{PropWEMode0Kerr}
  There exist an index set $\cE_\ind\subset\C\times\N_0$ with $\min\Re\cE_\ind\geq\eps_\ind$ and stationary 1-forms
  \begin{align}
  \label{EqWEMode0KerrOmega}
    \dot\omega_b(1,0) &= c_b(\log\rho)\ubar\omega_{\rms 0}^{(0)} + \dot\omega_b'(1,0),\quad \dot\omega_b'(1,0) \in \cA^{(0,0)\cup\cE_\ind}(X;\cT^*_X), \\
    \dot\omega_b(0,\dot\bha) &\in \cA^{(1,0)\cup(1+\cE_\ind)}(X;\cT^*_X), \nonumber
  \end{align}
  depending smoothly on $b$ near $b_0$ and linearly on $\dot\bha\in\R^3$, such that, with $\dot\omega_b(\dot b)=\dot\omega_b(\dot\bhm,\dot\bha):=\dot\bhm\dot\omega_b(1,0)+\dot\omega_b(0,\dot\bha)$, we have
  \[
    \dot g_b^\Ups(\dot b) := \dot g_b(\dot b) + \delta_{g_b}^*\dot\omega_b(\dot b) \in \ker \wh{L_b}(0).
  \]
  Furthermore, we have
  \begin{equation}
  \label{EqWEMode0KerrMem}
  \begin{split}
    \dot g_b^\Ups(1,0) &\in \cA^{(1,0)\cup(1+\cE_\ind)}(X;S^2\cT^*_X), \\
    \dot g_b^\Ups(0,\dot\bha) &\in \cA^{(2,0)\cup(2+\cE_\ind)}(X;S^2\cT^*_X).
  \end{split}
  \end{equation}
\end{prop}
\begin{proof}
  \pfstep{Changes of the mass.} We have $\dot g_b(1,0)\in\rho\CI(X;S^2\cT^*_X)$ by~\eqref{EqKMetLin}; its $\rho^1$ leading-order term is spherically symmetric (since the Kerr metric is spherically symmetric modulo $r^{-2}\CI=\rho^2\CI$, cf.\ \eqref{EqKMetDiff}), i.e., of scalar type $0$. Therefore,
  \[
    \eta := 2\delta_{g_b,E^\Ups}\sfG_{g_b}\dot g_b(1,0) \in \rho^2\CI(X;\cT^*_X),
  \]
  with $\rms 0$ leading-order term. We can use (the proof of) Lemma~\ref{LemmaTMFormal} to solve $\wh{\Box_{g_b,E^\Ups}^\Ups}(0)\omega=-\eta$ to leading order, namely with
  \[
    \omega = c(\log\rho)\ubar\omega_{\rms 0}^{(0)} + \omega_{(0)}
  \]
  for some constant $c\in\C$ and a scalar type $0$ 1-form $\omega_{(0)}\in\CI(\pa X;\cT^*_X|_{\pa X})$ which both depend smoothly on $b$ (in fact, on $\bhm$ only). Thus,
  \[
    \eta + \wh{\Box_{g_b,E^\Ups}^\Ups}(0)\bigl(c(\log\rho)\ubar\omega_{\rms 0}^{(0)}+\omega_{(0)}\bigr) \in \cA^{(3,1)}(X;\cT^*_X)
  \]
  can be written as $-\wh{\Box_{g_b,E^\Ups}^\Ups}(0)\omega'$ where $\omega'\in\cA^{\cE_\ind}(X;\cT^*_X)$ (using Lemma~\ref{LemmaWG0Inv}). We then set $\dot\omega_b(1,0):=c(\log\rho)\ubar\omega_{\rms 0}^{(0)}+\omega_{(0)}+\omega'$ and compute
  \begin{align*}
    \dot g_b^\Ups(1,0) &= \dot g_b(1,0) + \wh{\delta_{g_b}^*}(0) \bigl( c(\log\rho)\ubar\omega_{\rms 0}^{(0)} + \omega_{(0)} + \omega' \bigr) \\
      &= \dot g_b(1,0) + \wh{\delta_{g_b}^*}(0)(\omega_{(0)}+\omega') + c[\wh{\ubar\delta^*}(0),\log\rho]\ubar\omega_{\rms 0}^{(0)} + \bigl(\wh{\delta_{g_b}^*}(0)-\wh{\ubar\delta^*}(0)\bigr)\ubar\omega_{\rms 0}^{(0)} \\
      &\in \rho\CI + \cA^{(1,0)\cup(1+\cE_\ind)} + \rho\CI + \rho^2\CI,
  \end{align*}
  where we used Lemma~\ref{LemmaWG0KerrMink} and $[\ubar\delta^*,\log\rho]=\frac{\dd\rho}{\rho}\otimes_s(\cdot)=-\rho\,\dd r\otimes_s(\cdot)$. This gives~\eqref{EqWEMode0KerrMem}. The smoothness in $b$ is proved as in the proof of Proposition~\ref{PropWG0Symm}.

  \pfstep{Changes of the specific angular momentum.} We have $\dot g_b(0,\dot\bha)\in\rho^2\CI(X;S^2\cT^*_X)$, and therefore
  \[
    2\delta_{g_b,E^\Ups}\sfG_{g_b}\dot g_b(0,\dot\bha)\in\rho^3\CI.
  \]
  This is equal to $-\wh{\Box_{g_b,E^\Ups}^\Ups}(0)\dot\omega_b(0,\dot\bha)$ for some $\dot\omega_b(0,\dot\bha)\in\cA^{(1,0)\cup(1+\cE_\ind)}(X;\cT^*_X)$ by Lemma~\ref{LemmaWG0Inv} and the fact that $1$ is not an indicial root. Then
  \[
    \dot g_b^\Ups(0,\dot\bha) = \dot g_b(0,\dot\bha) + \wh{\delta_{g_b}^*}(0)\dot\omega_b(0,\dot\bha) \in \rho^2\CI + \cA^{(2,0)\cup(2+\cE_\ind)},
  \]
  as claimed.
\end{proof}

\begin{rmk}[Angular momentum changes that are pure gauge]
\label{RmkWEMode0Ang}
  Consider a rotation vector field $R=\bfv\times(\cdot)$ and its time $s$-flow $e^{s R}\colon\Sph^2\to\Sph^2$; then $g_{\bhm,\bha(s)}=(e^{-s R})^*g_{\bhm,\bha}$ where $\bha(s)=e^{s R}\bha$; differentiating this relation at $s=0$ gives
  \[
    \dot g_{\bhm,\bha}(0,\bfv\times\bha) = -\cL_R g_{\bhm,\bha} = -2\delta_{g_{\bhm,\bha}}^*\vect_b(\bfv),\quad \vect_b(\bfv):=g_b\ubar g^{-1}\vect(\bfv).
  \]
  When $\bha\neq 0$, then given $\dot\bha\perp\bha$, there exists $\bfv\in\R^3$ with $\bfv\perp\bha$ and $|\bfv|=|\dot\bha||\bha|^{-1}$ such that $\bfv\times\bha=\dot\bha$; and we can then conclude that
  \begin{equation}
  \label{EqWEMode0Ang}
    \dot g_{\bhm,\bha}^\Ups(0,\dot\bha) = -2 h_{b,\rmv 1}(\vect(\bfv))
  \end{equation}
  in the notation of~\eqref{EqWG0hv1}. While such infinitesimal angular momentum changes are thus pure gauge, the corresponding gauge potential $\omega_{b,\rmv 1}^{(-1)}(\vect(\bfv))$ has size $2|\bfv|=2|\dot\bha||\bha|^{-1}$, which for $|\dot\bha|=1$, say, is unbounded as $b$ approaches the parameters of a Schwarzschild, i.e., non-rotating, black hole. For the determination of the final black hole parameters in the stability problem, it is thus necessary to encode the final angular momentum as a vector in $\R^3$ (unless one studied only perturbations of \emph{nontrivially} rotating black holes, in which case one could keep the axis of rotation fixed).
\end{rmk}

We can now characterize the mapping properties of $\wh{L_b}(0)$:

\begin{prop}[Zero frequency (co)kernel]
\label{PropWEMode0}
  For appropriate\footnote{For small $\gamma^\Ups_{\cH^+}$ in Definition~\ref{Def1Gauge}---which is what we do take in Proposition~\ref{PropWGModeNon0}---it suffices to take $s\geq 3$: $s>\frac52$ works when $\gamma^\Ups_{\cH^+}=0$, and thus $s\geq 3$ works also for all small $\gamma^\Ups_{\cH^+}$. One can also take variable orders $s$ provided they are monotonically non-increasing along the future-directed null-bicharacteristic flow. Such order functions arise naturally in~\S\ref{SAdm}, following \citeAF{\S\ref*{SSp}}.} $s\in\R$ and for all $\alpha\in(0,\eps_\ind)$, the Fredholm index of the map
  \begin{equation}
  \label{EqWEMode0}
    \wh{L_b}(0) \colon \bigl\{ u\in\bar H_\bop^{s,\alpha}(X;S^2\cT^*_X) \colon \wh{L_b}(0)u\in\bar H_\bop^{s-1,\alpha+2}(X;S^2\cT^*_X) \bigr\} \to \bar H_\bop^{s-1,\alpha+2}(X;S^2\cT^*_X),
  \end{equation}
  with the underlying $L^2$-spaces defined using an unweighted b-density $\mu_\bop$ (such as $\frac{|\dd x|}{|x|^3}$), is equal to $0$.
  \begin{enumerate}
  \item{\rm (Kernel.)} The nullspace of~\eqref{EqWEMode0} is $7$-dimensional and spanned by the symmetric 2-tensors
    \begin{equation}
    \label{EqWEMode0Ker}
    \begin{alignedat}{2}
      \dot g_b^\Ups(1,0) &\in \cA^{(1,0)\cup(1+\cE_\ind)}(X;S^2\cT^*_X), \\
      \dot g_b^\Ups(0,\dot\bha) &\in \cA^{(2,0)\cup(2+\cE_\ind)}(X;S^2\cT^*_X), &\quad \dot\bha&\in\R^3, \\
      h_{b,\rms 1}(\scal) &\in \cA^{(2,0)\cup(2+\cE_\ind)}(X;S^2\cT^*_X), &\quad \scal&\in\scalspace_1,
    \end{alignedat}
    \end{equation}
    in the notation of~\eqref{EqWG0hs1} and Proposition~\usref{PropWEMode0Kerr}. (Here $\cE_\ind$ is an index set with $\min\Re\cE_\ind\geq\eps_\ind$ that is independent of $b$.)
  \item\label{ItWEMode0Co}{\rm (Cokernel.)} The cokernel of~\eqref{EqWEMode0}, i.e., the kernel of $\wh{L_b}(0)^*$ (with adjoint defined using the spatial volume density $|\dd g_b|_X|$ of Kerr and the indefinite fiber inner product on $S^2\cT^*_X$ induced by the Kerr metric) on the space\footnote{Making the densities explicit in the notation, we have $\bar H_\bop^{s-1,\alpha+2}(X,\mu_\bop)=\bar H_\bop^{s-1,\alpha+\frac12}(X,|\dd g_b|_X|)$, the $L^2(X,|\dd g_b|_X|)$-dual of which is $\dot H_\bop^{-s+1,-\alpha-\frac12}(X,|\dd g_b|_X)=\dot H_\bop^{-s+1,-\alpha+1}(X,\mu_\bop)$.} $\dot H_\bop^{-s+1,-\alpha+1}(X;S^2\cT^*_X)$, is the $7$-dimensional space spanned by
    \begin{equation}
    \label{EqWEMode0Coker}
    \begin{alignedat}{2}
      h_{b,\rms 0}^* &\in \dot H_\bop^{0,C_0+2}(X;S^2\cT^*_X), \\
      h_{b,\rms 1}^*(\scal) &\in \dot H_\bop^{0,((2,0),C_0+2)}(X;S^2\cT^*_X),&\quad \scal&\in\scalspace_1, \\
      h_{b,\rmv 1}^*(\vect) &\in \dot H_\bop^{0,((2,0),C_0+2)}(X;S^2\cT^*_X),&\quad \vect&\in\vectspace_1,
    \end{alignedat}
    \end{equation}
    in the notation of~\eqref{EqWC0Dualh}.
  \end{enumerate}
\end{prop}
\begin{proof}
  \pfstep{Kernel.} By a normal operator argument (cf.\ \citeAF{Theorem~\ref*{ThmSpB}(\ref*{ItSpBNull})}), every element $u$ of the nullspace of~\eqref{EqWEMode0} lies in $\cA^{\eps_\ind-\eps}(X;S^2\cT^*_X)$ for all $\eps>0$. Arguing similarly to the proof of Proposition~\ref{PropWEMode}, we then find that $\wh{\Box_{g_b,E^\cC}^\cC}(0)\eta=0$ where now $\eta:=\wh{\delta_{g_b,E^\Ups}}(0)\sfG_{g_b}u\in\cA^{1+\eps_\ind-\eps}(X;\cT^*_X)$. Theorem~\ref{ThmWCRec}\eqref{ItWCRec0} implies that $\eta=0$. Therefore, $\wh{D_{g_b}\Ric}(0)(u)=0$. By \cite[Theorem~6.1(2)]{AnderssonHaefnerWhitingMode}, with $q=\eps_\ind-\frac32-\eps$ for an arbitrarily small $\eps>0$ so that $q\in(-\frac32,-\frac12)$, there exist $\dot b\in\R^4$ and a stationary 1-form $\omega\in\bar H_\bop^{\infty,q-1}(X;\cT^*_X)\subset\cA^{-1+\eps_\ind-\eps}$ such that $u = \dot g_b(\dot b) + \delta_{g_b}^*\omega$. By Proposition~\ref{PropWEMode0Kerr}, we may replace $\dot g_b(\dot b)$ by $\dot g_b^\Ups(\dot b)$ upon adjusting $\omega$ appropriately, so
  \[
    u = \dot g_b^\Ups(\dot b) + \wh{\delta_{g_b}^*}(0)\omega.
  \]
  Since $u,\dot g_b^\Ups(\dot b)\in\ker\wh{L_b}(0)\cap\ker\wh{\delta_{g_b,E^\Ups}}(0)\sfG_{g_b}$, we must also have $\wh{\delta_{g_b}^*}(0)\omega\in\ker\wh{\delta_{g_b,E^\Ups}}(0)\sfG_{g_b}$, i.e.,
  \[
    \omega \in \cA^{-1+\eps_\ind-\eps}(X;\cT^*_X) \cap \ker \wh{\Box_{g_b,E^\Ups}^\Ups}(0).
  \]
  Since by Lemma~\ref{LemmaWGInd} the smallest indicial root of $\wh{\Box_{g_b,E^\Ups}^\Ups}(0)$ with real part $>-1$ is $0$, which is a scalar type $0$ and $1$ indicial root, the final part of Lemma~\ref{LemmaWGInd} implies that $\omega$ has a bounded leading-order term equal to a linear combination of $\dd t$ and $\ubar\omega_{\rms 1}^{(0)}(\scal)=\dd(r\scal)$ (in the notation of~\eqref{EqWG0SymmMink}). Upon subtracting from $\omega$ a suitable multiple of $\omega_{b,\rms 0}^{(0)}=g_b(\pa_{t_*},\cdot)\in\ker\wh{\delta_{g_b}^*}(0)\subset\ker\wh{\Box_{g_b,E^\Ups}^\Ups}(0)$ and $\omega_{b,\rms 1}^{(0)}(\scal)\in\ker\wh{\Box_{g_b,E^\Ups}^\Ups}(0)$ for a suitable $\scal\in\scalspace_1$, we have $\omega\in\cA^\eps(X;\cT^*_X)$ for $\eps>0$ (in fact, one can take $\eps$ to be any number $<\lambda^\Ups_{\rms 1,1}$ by Lemma~\ref{LemmaWGInd}), and $\wh{\Box_{g_b,E^\Ups}^\Ups}(0)\omega=0$ still. By Proposition~\ref{PropWGMode}, this implies $\omega=0$. Therefore, the original $\omega$ is a linear combination of $\omega_{b,\rms 0}^{(0)}$ and $\omega_{b,\rms 1}^{(0)}(\scal)$, and hence $\wh{\delta_{g_b}^*}(0)\omega=h_{b,\rms 1}(\scal)$.

  We have now shown that $\ker\wh{L_b}(0)$ is contained in the span of~\eqref{EqWEMode0Ker}. Conversely, each tensor in~\eqref{EqWEMode0Ker} lies in $\ker\wh{L_b}(0)$. That $\ker\wh{L_b}(0)$ is $7$-dimensional follows from Lemma~\ref{LemmaWEPair} below: if $u=\dot g_b^\Ups(\dot b)+h_{b,\rms 1}(\scal)=0$ where $\dot b=(\dot\bhm,\dot\bha)$, we first observe that $0=\la[L_b,t_*]u,h_{b,\rms 0}^*\ra=-16\pi\dot\bhm$ implies $\dot\bhm=0$; then $0=\la[L_b,t_*]u,h_{b,\rmv 1}^*(\vect(\fq))\ra=-16\pi\bhm(\fq\cdot\dot\bha)$ for all $\fq\in\R^3$ implies $\dot\bha=0$. Finally, $h_{b,\rms 1}(\scal)=0$ if and only if $\scal=0$, as follows from its construction and the fact that $\dd(r\scal)$ is not a Killing vector field for the Schwarzschild (and thus also not for the Kerr) metric.

  \pfstep{Cokernel.} The map~\eqref{EqWEMode0} has Fredholm index $0$, as follows from Propositions~\ref{PropWETr} and~\ref{PropWEtf} together with \citeAF{Corollary~\ref*{CorSpLoInd0}}. Since the adjoints of $L_b$ and $\Box_{g_b,E^\cC}^\cC$ are given by
  \begin{equation}
  \label{EqWEModeAdj}
    L_{g_b}^* = \sfG_{g_b}\circ D_{g_b}\Ric\circ\sfG_{g_b} - \sfG_{g_b}(\delta_{g_b,E^\Ups})^*(\delta_{g_b,E^\cC}^*)^*,\quad
    (\Box_{g_b,E^\cC}^\cC)^* = 2(\delta_{g_b,E^\cC}^*)^*\sfG_{g_b}\delta_{g_b}^*,
  \end{equation}
  we have $\sfG_{g_b}\delta_{g_b}^*\ker(\Box_{g_b,E^\cC}^\cC)^*\subset\ker L_{g_b}^*$, and therefore~\eqref{EqWC0DualEq} and \eqref{EqWC0Dualh} imply that the tensors in~\eqref{EqWEMode0Coker} indeed lie in $\ker\wh{L_b}(0)^*$. To conclude, it suffices to argue that the space spanned by~\eqref{EqWEMode0Coker} is $7$-dimensional, which again follows by considering the pairings in Lemma~\ref{LemmaWEPair} below.
\end{proof}

To complete the proof, we need the following result from \cite{HintzGlueLocIII}, which in turn builds on \cite[Theorem~9.6]{HintzGlueLocI}:

\begin{lemma}[$L^2$-pairings]
\label{LemmaWEPair}
  We use the notation $\vect(\fq)\in\vectspace_1$ and $\scal(\fq)\in\scalspace_1$ from Definition~\usref{DefTYs1}. Recall $b=(\bhm,\bha)\in\R\times\R^3$. Computing $L^2(X;S^2\cT^*_X)$-pairings using the spatial volume density $|\dd g_b|_X|$ and the (indefinite) fiber inner product induced by $g_b$, we have:
  \begin{alignat*}{2}
    \la[L_b,t_*]\dot g_b^\Ups(\dot\bhm,0),&\ h_{b,\rms 0}^*\ra &&= -16\pi\dot\bhm, \\
    \la[L_b,t_*]\dot g_b^\Ups(\dot\bhm,0),&\ h_{b,\rmv 1}^*(\vect(\fq))\ra &&= -16 \pi(\fq\cdot\bha)\dot\bhm, \\
    \la[L_b,t_*]\dot g_b^\Ups(\dot\bhm,0),&\ h_{b,\rms 1}^*(\scal(\fq))\ra &&= 0,
  \end{alignat*}
  further
  \begin{alignat*}{2}
    \la[L_b,t_*]\dot g_b^\Ups(0,\dot\bha),&\ h_{b,\rms 0}^*\ra &&= 0, \\
    \la[L_b,t_*]\dot g_b^\Ups(0,\dot\bha),&\ h_{b,\rmv 1}^*(\vect(\fq))\ra &&= -16 \pi\bhm(\fq\cdot\dot\bha), \\
    \la[L_b,t_*]\dot g_b^\Ups(0,\dot\bha),&\ h_{b,\rms 1}^*(\scal(\fq))\ra &&= 0,
  \end{alignat*}
  and finally
  \begin{equation}
  \label{EqWEPairs1}
  \begin{alignedat}{2}
    \Big\la \frac12[[L_b,t_*],t_*]h_{b,\rms 1}(\scal(c))+[L_b,t_*]\breve h_{b,\rms 1}^1(\scal(c)),&\ h_{b,\rms 0}^*\Big\ra &&= 0, \\
    \Big\la \frac12[[L_b,t_*],t_*]h_{b,\rms 1}(\scal(c))+[L_b,t_*]\breve h_{b,\rms 1}^1(\scal(c)),&\ h_{b,\rmv 1}^*(\vect(\fq))\Big\ra &&= 0, \\
    \Big\la \frac12[[L_b,t_*],t_*]h_{b,\rms 1}(\scal(c))+[L_b,t_*]\breve h_{b,\rms 1}^1(\scal(c)),&\ h_{b,\rms}^*(\scal(\fq))\Big\ra &&= 8\pi\bhm(\fq\cdot c).
  \end{alignedat}
  \end{equation}
\end{lemma}
\begin{proof}
  The proof of \cite[Lemma~3.17]{HintzGlueLocIII} applies verbatim.
\end{proof}

Since solutions of $\wh{L_b}(0)u=f$ in the space~\eqref{EqWEMode0} are not unique even when they exist, we note the following construction:

\begin{cor}[Linear right inverse]
\label{CorWE0Solv}
  Fix $f_j^*,f_j^\sharp\in\CIc(X^\circ;S^2\cT^*_X)$, $j=1,\ldots,7$, such that $u\mapsto(\la u,f_j^*\ra_{L^2})_{j=1,\ldots,7}$ and $u^*\mapsto(\la f_j^\sharp,u^*\ra_{L^2})_{j=1,\ldots,7}$ restrict to isomorphisms $\ker\wh{L_{b_0}}(0)\to\C^7$ and $\ker\wh{L_b}(0)^*\to\C^7$, respectively (which map into $\R^7$ when restricted to real 2-tensors). Define
  \[
    \wt L_b := \begin{pmatrix} \wh{L_b}(0) & (f_j^\sharp)_{j=1,\ldots,7} \\ (\la\cdot,f_j^*\ra_{L^2})_{j=1,\ldots,7} & 0 \end{pmatrix}.
  \]
  \begin{enumerate}
  \item\label{ItWE0Solv1}{\rm (Invertibility.)} For all $b$ near $b_0$, the operator $\wt L_b$ is invertible as a map
  \[
    \bigl\{u\in\bar H_\bop^{s,\alpha}\colon\wh{L_b}(0)u\in\bar H_\bop^{s-1,\alpha+2}\bigr\}\oplus\C^7\to\bar H_\bop^{s-1,\alpha+2}\oplus\C^7,
  \]
  and the inverse depends smoothly on $b$ as a map $\cA^{\alpha+2}(X;S^2\cT^*_X)\oplus\C^7\to\cA^{\alpha-\eps}(X;S^2\cT^*_X)\oplus\C^7$ for all $\alpha\in(0,\eps_\ind)$ and $\eps>0$.
  \item\label{ItWE0Solv2}{\rm (Inversion of $\wh{L_b}(0)$.)} If $f\in\cA^{\alpha+2}(X;S^2\cT^*_X)$ is $L^2$-orthogonal to all dual states~\eqref{EqWEMode0Coker} and $(u,c)=\wt L_b^{-1}(f,0)$, then $\wh{L_b}(0)u=f$. If $f$ depends smoothly on $b$ and satisfies the orthogonality condition for all $b$, then $u$ depends smoothly on $b$ as well.
  \end{enumerate}
\end{cor}
\begin{proof}
  The requirements on $f_j^*$ and $f_j^\sharp$ are satisfied also for the (co)kernel of $\wh{L_b}(0)$ when $b$ is close to $b_0$. Thus, if $\wt L_b(u,c)=(0,0)$, then $\wh{L_b}(0)u+\sum_{j=1}^7 c_j f_j^\sharp=0$. Taking the inner product with all $h^*\in\ker\wh{L_b}(0)^*$ in~\eqref{EqWEMode0Coker} implies that $c_1=\cdots=c_7=0$, so $u\in\ker\wh{L_b}(0)$; and then $\la u,f_j^*\ra_{L^2}=0$ for all $j$ implies that $u=0$. Since $\wt L_b$ is Fredholm of index $0$ (since $\wh{L_b}(0)$ is), this proves the first half of part~\eqref{ItWE0Solv1}. Arguments as in \cite[\S{2.7}]{VasyMicroKerrdS} or \citeAF{\S\ref*{SssSpBPf}, Part~(3)} then imply the continuous dependence of $\wt L_b^{-1}\colon\bar H_\bop^{s-1,\alpha+2}\to H_\bop^{s-\eps,\alpha-\eps}$ on $b$ in the operator norm for any $\eps>0$. By direct differentiation, this implies the $\cC^k$-dependence on $b$ of $\wt L_b^{-1}$ as a map $\bar H_\bop^{s-1+k,\alpha+2}\to H_\bop^{s-\eps,\alpha-\eps}$, and thus the second half of part~\eqref{ItWE0Solv1}.

  For part~\eqref{ItWE0Solv2}, we only need to note that the inner product of $\wh{L_b}(0)u+\sum_{j=1}^7 c_j f_j^\sharp=f$ with $h^*\in\ker\wh{L_b}(0)^*$ implies $\sum_{j=1}^7 c_j\la f_j^\sharp,h^*\ra_{L^2}$, and thus $c_1=\cdots=c_7=0$ by assumption on the $f_j^\sharp$.
\end{proof}

\subsection{Large (generalized) zero energy states: physical states}
\label{SsWE0}

Consider the indicial roots $\lambda$ of $\wh{L_b}(0)$ in Lemma~\ref{LemmaWEInd} with $\Re\lambda\in[-3,0]$ (which thus excludes the indicial roots with superscript ``$\cC$''). For most of them, Proposition~\ref{PropWG0Large} produces corresponding large zero energy states of $L_b$ with leading-order behavior at $\pa X$ given by an indicial solution of size $\rho^\lambda$ (with appropriate modifications for the $\rms 1$ and $\rmv 1$ indicial root $-1$). The only exceptional indicial roots are the $\rms l$ and $\rmv l$ roots $-l$ for $l\geq 2$. They are \emph{not} related to pure gauge states; we shall instead construct them using properties of $L_b$ directly. This is more delicate than the arguments in~\S\ref{SsWG0Kerr} since, unlike $\wh{\Box_{g_b,E^\Ups}^\Ups}(0)$ (see Lemma~\ref{LemmaWG0Inv}), the operator $\wh{L_b}(0)$ has a nontrivial cokernel by Proposition~\ref{PropWEMode0}; and considerable care must be exercised to produce zero energy states \emph{with smooth dependence on $b=(\bhm,\bha)$} when $\bha$ is near $0$ (since the cokernel of $\wh{L_b}(0)$ on the relevant weighted spaces depends discontinuously on $b$, cf.\ \eqref{EqWC0hv1}).

\begin{prop}[Zero energy states, III: physical states]
\label{PropWE0}
  There exists an index set $\cE_\ind\subset\C\times\N_0$ with $\min\Re\cE_\ind\geq\eps_\ind$ such that the following statements hold for all $b$ near $b_0$.
  \begin{enumerate}
  \item\label{ItWE0s2}{($\rms 2$ and $\rmv 2$ root $-2$.)} There exist
    \begin{equation}
    \label{EqWE0s2}
    \begin{alignedat}{3}
      &h_{b,\rms 2}^{(-2)}(\scal),\ &&h_{b,\rmv 2}^{(-2)}(\vect) &&\in \cA^{(-2,0)\cup(-2+\cE_\ind)}(X;S^2\cT^*_X), \\
      &\breve h_{b,\rms 2}^{(-2),1}(\scal),\ &&\breve h_{b,\rmv 2}^{(-2),1}(\vect) &&\in \cA^{(-3,0)\cup(-3+\cE_\ind)}(X;S^2\cT^*_X),
    \end{alignedat}
    \end{equation}
    depending linearly on $\scal\in\scalspace_2$ and $\vect\in\vectspace_2$ and smoothly on $b$ near $b_0$, such that\footnote{We recall here from~\eqref{EqWG0GenStateh} notation such as $h_{b,\rms 2}^{(-2),\leq 1}(\scal)=t_* h_{b,\rms 2}^{(-2)}(\scal)+\breve h_{b,\rms 2}^{(-2),1}(\scal)$.}
    \[
      h_{b,\rms 2}^{(-2)}(\scal),\ h_{b,\rmv 2}^{(-2)}(\vect),\ h_{b,\rms 2}^{(-2),\leq 1}(\scal),\ h_{b,\rmv 2}^{(-2),\leq 1}(\vect) \in \ker D_{g_b}\Ric \cap \ker \delta_{g_b,E^\Ups}\sfG_{g_b},
    \]
    and with the $\rho^{-2}$ leading-order terms $\ubar h_{\rms 2}^{(-2)}(\scal)$ and $\ubar h_{\rmv 2}^{(-2)}(\vect)$ of $h_{b,\rms 2}^{(-2)}(\scal)$ and $h_{b,\rmv 2}^{(-2)}(\vect)$ comprising the spaces $N_{\rms 2}(\wh{\ubar L}(0),-2)$ and $N_{\rmv 2}(\wh{\ubar L}(0),-2)$, respectively.
  \item\label{ItWE0s3}{($\rms 3$ and $\rmv 3$ root $-3$.)} There exist
    \begin{equation}
    \label{EqWE0s3}
      h_{b,\rms 3}^{(-3)}(\scal),\ h_{b,\rmv 3}^{(-3)}(\vect) \in \cA^{(-3,0)\cup(-3+\cE_\ind)}(X;S^2\cT^*_X),
    \end{equation}
    depending linearly on $\scal\in\scalspace_3$ and $\vect\in\vectspace_3$ and smoothly on $b$ near $b_0$, such that
    \[
      h_{b,\rms 3}^{(-3)}(\scal),\ h_{b,\rmv 3}^{(-3)}(\vect) \in \ker D_{g_b}\Ric \cap \ker \delta_{g_b,E^\Ups}\sfG_{g_b},
    \]
    and with the $\rho^{-3}$ leading-order terms $\ubar h_{\rms 3}^{(-3)}(\scal)$ and $\ubar h_{\rmv 3}^{(-3)}(\vect)$ of $h_{b,\rms 3}^{(-3)}(\scal)$ and $h_{b,\rmv 3}^{(-3)}(\vect)$ comprising the spaces $N_{\rms 3}(\wh{\ubar L}(0),-3)$ and $N_{\rmv 3}(\wh{\ubar L}(0),-3)$, respectively.
  \end{enumerate}
\end{prop}

In order to deal with the cokernel of $\wh{L_b}(0)$, we record:

\begin{lemma}[$L^2$-pairings, II]
\label{LemmaWE0Pair}
  Recall $\breve h_{b,\rms 1}^2(\scal)$ and $\breve h_{b,\rmv 1}^1(\vect)\in\cA^{(-1,0)\cup(-1+\cE_\ind)}(X;S^2\cT^*_X)$ from Proposition~\usref{PropWG0Large}\eqref{ItWG0Larges1m1}. We use the spatial volume density $|\dd g_b|_X|$ and the (non-definite) fiber inner product induced by $g_b$ to define the $L^2(X;S^2\cT^*_X)$-inner product.
  \begin{enumerate}
  \item\label{ItWE0Pair2s1}{\rm ($\rms 1$ pairings.)} We have $\la\wh{L_b}(0)\breve h_{b,\rms 1}^2(\scal),h^*\ra_{L^2}=0$ for $h^*=h_{b,\rms 0}^*$ and $h_{b,\rmv 1}^*(\vect)$ for all $\vect\in\vectspace_1$. On the other hand, the linear map
    \[
      \scalspace_1 \ni \scal \mapsto \la \wh{L_b}(0)\breve h_{b,\rms 1}^2(\scal), h_{b,\rms 1}^*(\cdot)\ra_{L^2} \in \scalspace_1^*
    \]
    is an isomorphism that depends smoothly on $b$ near $b_0$.
  \item\label{ItWE0Pair2v1}{\rm ($\rmv 1$ pairings.)} We have $\la\wh{L_b}(0)\breve h_{b,\rmv 1}^1(\vect),h^*\ra_{L^2}=0$ for $h^*=h_{b,\rms 0}^*$ and $h_{b,\rms 1}^*(\scal)$ for all $\scal\in\scalspace_1$. On the other hand,
    \begin{equation}
    \label{EqWE0Pairv1}
      \la \wh{L_b}(0)\breve h_{b,\rmv 1}^1(\vect), h_{b,\rmv 1}^*(\vect')\ra_{L^2} = -8\pi\bhm \la \bha(\vect), \bha(b)\times\bha(\vect')\ra_{\R^3},\quad b=(\bhm,\bha).
    \end{equation}
  \end{enumerate}
\end{lemma}
\begin{proof}
  \pfstep{Part~\eqref{ItWE0Pair2s1}.} We only need to note that the equation
  \begin{align*}
    0 = L_b h_{b,\rms 1}^{\leq 2} &= t_*^2 L_b h_{b,\rms 1} + 2 t_*\bigl([L_b,t_*]h_{b,\rms 1}+L_b\breve h_{b,\rms 1}^1\bigr) \\
      &\quad\qquad + 2\Bigl(\frac12[[L_b,t_*],t_*]h_{b,\rms 1} + [L_b,t_*]\breve h_{b,\rms 1}^1 + L_b\breve h_{b,\rms 1}^2\Bigr)
  \end{align*}
  implies that $L_b\breve h_{b,\rms 1}^2=-\frac12[[L_b,t_*],t_*]h_{b,\rms 1}-[L_b,t_*]\breve h_{b,\rms 1}^1$; the claim then follows from~\eqref{EqWEPairs1}.

  \pfstep{Part~\eqref{ItWE0Pair2v1}.} We present two arguments, one conceptual based on Remark~\ref{RmkWEMode0Ang}, and one via direct computation. The identity~\eqref{EqWEMode0Ang} with $\dot\bha=\bfv\times\bha$ implies
  \begin{align*}
    \la \wh{L_b}(0)\breve h_{b,\rmv 1}^1(\vect(\bfv)), h_{b,\rmv 1}^*(\vect') \ra &= -\la [L_b,t_*]h_{b,\rmv 1}(\vect(\bfv)), h_{b,\rmv 1}^*(\vect')\ra \\
      &= \frac12\la [L_b,t_*] \dot g_b^\Ups(0,\bfv\times\bha), h_{b,\rmv 1}^*(\vect')\ra \\
      &= -8\pi\bhm\la\bha(\vect'),\bfv\times\bha\ra_{\R^3}
       = -8\pi\bhm\la\bha(\vect'),\bha(\vect)\times\bha(b)\ra_{\R^3},
  \end{align*}
  where we used Lemma~\ref{LemmaWEPair} in the passage to the final line. This equals~\eqref{EqWE0Pairv1} by the cyclicity of the scalar triple product.

  The direct computation is as follows. Let $\chi_\eps=\chi_1(\rho/\eps)$ where $\chi_1\in\CI(\R)$ vanishes on $[0,1]$ and equals $1$ on $[2,\infty)$. Using the decomposition~\eqref{EqWC0hbv1}, we then write the left-hand side of~\eqref{EqWE0Pairv1} as
  \begin{align*}
    &\lim_{\eps\searrow 0} \la \breve h_{b,\rmv 1}^1(\vect), \wh{L_b}(0)^*\chi_\eps h_{b,\rmv 1}^*(\vect')\ra \\
    &\qquad = \lim_{\eps\searrow 0} \la \breve h_{b,\rmv 1}^1(\vect), [\wh{L_b}(0)^*,\chi_\eps]h_{b,\rmv 1}^*(\vect') \ra \\
    &\qquad = \lim_{\eps\searrow 0} \la \breve h_{b,\rmv 1}^1(\vect), [\wh{L_b}(0)^*,\chi_\eps](h_{b,\rmv 1}^\sharp(\bfw) + h_{b,\rmv 1}^{*\flat}(\vect')) \ra,\quad \bfw := \bha(b)\times\bha(\vect').
  \end{align*}
  Now, $\breve h_{b,\rmv 1}^1(\vect)\in\cA^{-1}(X;S^2\cT^*_X)$, so since the density $|\dd g_b|_X|$ is $\rho^{-3}$ times an unweighted b-density, all contributions to $h_{b,\rmv 1}^\sharp(\bfw)$ of class $\cA^{2+\delta}$ for any $\delta>0$ vanish in the limit $\eps\searrow 0$, so only the leading-order term $\ubar h_{\rmv 1}^\sharp(\bfw)$ matters; similarly, the contribution of $h_{b,\rmv 1}^{*\flat}(\vect')$ vanishes in this limit. By the same token, we may replace $\wh{L_b}(0)^*$ by its normal operator $\wh{\ubar L}(0)^*$ and $\breve h_{b,\rmv 1}^1(\vect)$ by its leading-order term $\ubar h_{\rmv 1}^1(\vect)$ (see Proposition~\ref{PropWG0Large}\eqref{ItWG0Largev1m1}). Thus, we have reduced our task to the evaluation of
  \begin{equation}
  \label{EqWE0Pair2v12}
    \lim_{\eps\searrow 0} \la \ubar h_{\rmv 1}^1(\vect), [\wh{\ubar L}(0)^*,\chi_\eps]\ubar h_{\rmv 1}^\sharp(\bfw) \ra = -\lim_{\eps\searrow 0} \la[\wh{\ubar L}(0),\chi_\eps]\ubar h_{\rmv 1}^1(\vect), \ubar h_{\rmv 1}^\sharp(\bfw)\ra.
  \end{equation}
  Now, using the splitting~\eqref{EqTYSplit2} in the vector type $1$ sector, the expression~\eqref{EqWEOpMink0} (together with~\eqref{EqTYMinkOpBox2}) becomes a $2\times 2$-matrix of operators, whose indicial operator can be computed (using~\eqref{EqTYIdentities}) to be
  \[
    2 N_{\rmv 1}(\rho^{-2}\wh{\ubar L}(0),\lambda) \equiv -\lambda^2+\lambda + \lambda\gamma^\cC\begin{pmatrix} 1-v^\cC & -(1-v^\cC) \\ 1+v^\cC & -(1+v^\cC) \end{pmatrix} + \lambda\gamma^\Ups\begin{pmatrix} -1 & -1 \\ 1 & 1 \end{pmatrix},
  \]
  modulo $\lambda$-independent terms. In these splittings, we also recall from~\eqref{EqWG0Largev1m1hbar} and~\eqref{EqWC0hbv1Lead} that
  \begin{align*}
    \rho\ubar h_{\rmv 1}^1(\vect) &= \frac14\Bigl(1+\frac{\gamma^\Ups}{2},1-\frac{\gamma^\Ups}{2}\Bigr)\ [\text{w.r.t.}\ \vect], \\
    \rho^{-2}\ubar h_{\rmv 1}^\sharp(\bfw) &= -\frac{\bhm}{1+2 v^\cC\gamma^\cC}\bigl( 1+2(v^\cC-1)\gamma^\cC,\ 1+2(v^\cC+1)\gamma^\cC \bigr)\ [\text{w.r.t.}\ \vect(\bfw)].
  \end{align*}
  Since for $\eta,\eta'\in T^*\Sph^2$ the Minkowskian inner product of $2\,\dd\ubar x^0\otimes_s\eta$ and $2\,\dd\ubar x^1\otimes_s\eta'$ is $-4\la\eta,\eta'\ra_{\slg^{-1}}$, the fiber inner product for $\rmv 1$ tensors is
  \[
    \begin{pmatrix} 0 & -4\slg^{-1} \\ -4\slg^{-1} & 0 \end{pmatrix};
  \]
  note also that the $L^2(\Sph^2;T^*\Sph^2)$-inner product of $\vect=-\slstar\sld\scal(\bha(\vect))$ and $\vect(\bfw)=-\slstar\sld\scal(\bfw)$ is
  \[
    \int_{\Sph^2} \la\sld\scal(\bha(\vect)),\sld\scal(\bfw)\ra_{\slg^{-1}}\,\dd\slg = \la\sldelta\sld\scal(\bha(\vect)),\scal(\bfw)\ra_{L^2(\Sph^2)} = \frac{8\pi}{3}\la\bha(\vect),\bfw\ra_{\R^3} = \frac{8\pi}{3}\la\bha(\vect),\bha(b)\times\bha(\vect')\ra_{\R^3}.
  \]
  Altogether, then,~\eqref{EqWE0Pair2v12} is equal to
  \begin{align*}
    &-\frac{8\pi}{3}\la\bha(\vect),\bha(b)\times\bha(\vect')\ra_{\R^3}\cdot\frac14\cdot\Bigl(-\frac{\bhm}{1+2 v^\cC\gamma^\cC}\Bigr) \\
    &\qquad \cdot\pa_\lambda N_{\rmv 1}(\rho^{-2}\wh{\ubar L}(0),\lambda)|_{\lambda=-1} \Bigl(1+\frac{\gamma^\Ups}{2},1-\frac{\gamma^\Ups}{2}\Bigr) \begin{pmatrix} 0 & -4 \\ -4 & 0 \end{pmatrix} \begin{pmatrix} 1+(2 v^\cC-1)\gamma^\cC \\ 1+(2 v^\cC+1)\gamma^\cC \end{pmatrix}.
  \end{align*}
  (The form of this expression is a special case of \cite[Lemma~2.5]{HintzGlueLocI}.) After a short computation, this gives~\eqref{EqWE0Pairv1}.
\end{proof}

\begin{proof}[Proof of Proposition~\usref{PropWE0}]
  We only discuss the scalar type cases, the vector type cases being completely analogous. We write $b=(\bhm,\bha)$ as usual.

  \pfstep{Part~\eqref{ItWE0s2}, scalar type $2$.} Fix scalar type $2$ indicial solutions $\ubar h_{\rms 2}^{(-2)}(\scal)$ (homogeneous of degree $-2$ in $\rho$) of $\wh{\ubar L}(0)$ that depend linearly on $\scal\in\scalspace_2$ and span $\ker N_{\rms 2}(\wh{\ubar L}(0),-2)$. We then have
  \[
    \wh{L_b}(0)\ubar h_{\rms 2}^{(-2)}(\scal) =: f_b \in \rho\CI(X;S^2\cT^*_X).
  \]

  \pfsubstep{Step~0.}{A first choice of the zero energy state.}\label{ItWE0s2Step0} We first show that there exists a solution $h'_b(\scal)\in\cA^{-2+\cE_\ind}(X;S^2\cT^*_X)$, depending linearly on $\scal$ and smoothly on $b$, of the equation
  \begin{equation}
  \label{EqWE0s2Pf1}
    \wh{L_b}(0)h'_b = -f_b,
  \end{equation}
  which would yield the kernel element $\ubar h_{\rms 2}^{(-2)}(\scal)+h'_b\in\ker\wh{L_b}(0)$. It suffices to do this for finitely many $\scal$ that form a basis of $\scalspace_2$, and construct $h'_b(\scal)$ for all remaining $\scal\in\scalspace_2$ by linearity; we thus henceforth suppress $\scal$ from the notation.

  Now, for $\alpha\in(-1-\eps_\ind,-1)$, the operator $\wh{L_b}(0)$ is Fredholm as a map~\eqref{EqWEMode0}. (This is a special case of \citeAF{Theorem~\ref*{ThmSp0}} and uses that such $\alpha$ are not the real part of any indicial root of $\wh{L_b}(0)$.) The kernel of its adjoint consists of those tensors~\eqref{EqWEMode0Coker} which are of class $\dot H_\bop^{0,1-\alpha}(X;S^2\cT^*_X)$; here $1-\alpha\in(2,2+\eps_\ind)$ exceeds $2$. As a consequence of the discussion of~\eqref{EqWC0hv1}, this cokernel is thus $4$-dimensional when $\bha=0$ (it is spanned by $h_{b,\rms 0}^*$ and $h_{b,\rmv 1}^*(\vectspace_1)$), whereas for $\bha\neq 0$ it is $2$-dimensional (and spanned by $h_{b,\rms 0}^*$ and $h_{b,\rmv 1}^*(\vect(\bha))$).\footnote{This is consistent with the relative index theorem \cite[Theorem~6.5]{MelroseAPS} since the dimension of the kernel of $\wh{L_b}(0)$ increases by $2$ more when decreasing the weight $\alpha$ from $-1+\eps$ to $-1-\eps$, $0<\eps\ll 1$, in the case $\bha=0$ as compared to the case $\bha\neq 0$; cf.\ \eqref{EqWG0Largev1}.} In either case, the relevant cokernel of $\wh{L_b}(0)^*$ consists of distributions $h^*$ with $\rho^{C_0+2}$-decay by~\eqref{EqWC0Dualh}, which is more than enough to integrate by parts to obtain
  \[
    \la f_b,h^*\ra_{L^2}=\la\ubar h_{\rms 2}^{(-2)},\wh{L_b}(0)^*h^*\ra_{L^2}=0;
  \]
  this guarantees the existence of \emph{some} solution $h'_b\in\Hb^{\infty,-1-\eps}(X;S^2\cT^*_X)$ for all $\eps>0$, which is then in fact polyhomogeneous by Lemma~\ref{LemmaTMSolPhg}. \emph{However}, due to the discontinuous nature of $\ker_{\dot H_\bop^{0,1-\alpha}}\wh{L_b}(0)^*$ in $b$, it is not clear from this argument how to ensure the \emph{smoothness} of $h'_b$ in $b$. (When $\bha$ is away from $0$, a simple Grushin type argument as in the proof of Corollary~\ref{CorWE0Solv} does produce a smooth right inverse of $\wh{L_b}(0)$ on the required spaces; the delicate situation is when $\bha_0=0$ and thus $\bha$ lies in a neighborhood of $0$.)

  \pfsubstep{Step~1.}{Formal solution and relationship with the cokernel.} In order to produce \emph{smooth} (in $b$) solutions of~\eqref{EqWE0s2Pf1}, we proceed as follows. First, Lemma~\ref{LemmaTMFormal} produces $h'_{b,1}\in\cA^{-2+\cE_\ind}(X;S^2\cT^*_X)$ with smooth dependence on $b$ such that
  \[
    \wh{L_b}(0)h_{b,\rms 2}^{(-2),\approx} \in \cA^{2+\cE_\ind}(X;S^2\cT^*_X),\quad h_{b,\rms 2}^{(-2),\approx} := \ubar h_{\rms 2}^{(-2)} + h'_{b,1}
  \]
  (or any other desired rate of decay). While this has sufficient decay to be paired in $L^2$ with all elements of the cokernel of $\wh{L_b}(0)^*$ described in Proposition~\ref{PropWEMode0}, these pairings need not vanish, except we always have
  \begin{subequations}
  \begin{equation}
  \label{EqWE0s2s0}
    \la\wh{L_b}(0)h_{b,\rms 2}^{(-2),\approx}, h_{b,\rms 0}^*\ra = 0.
  \end{equation}
  The linear functional
  \begin{equation}
  \label{EqWE0s2s1}
    \lambda_{b,\rms 1} \colon \scalspace_1 \ni \scal_1' \mapsto \la\wh{L_b}(0)h_{b,\rms 2}^{(2),\approx},h_{b,\rms 1}^*(\scal_1')\ra \in \R
  \end{equation}
  (which is smooth in $b$) captures part of the obstruction for solving away $\wh{L_b}(0)h_{b,\rms 2}^{(-2),\approx}$ using some $h'_{b,2}\in\cA^{\cE_\ind}$. For the dual states $h_{b,\rmv 1}^*(\vect'_1)$, $\vect'_1\in\vectspace_1$, on the other hand, we use Proposition~\ref{PropWC0hbv1} to compute, with $\chi_\eps=\chi_1(\rho/\eps)$ where $\chi_1\in\CI(\R)$ vanishes on $[0,1]$ and equals $1$ on $[2,\infty)$, that
  \begin{align}
  \label{EqWE0s2v10}
    \lambda_{b,\rmv 1}(\vect'_1) &:= \la\wh{L_b}(0)h_{b,\rms 2}^{(-2),\approx}, h_{b,\rmv 1}^*(\vect'_1)\ra \\
      &= \lim_{\eps\searrow 0} \la\wh{L_b}(0)h_{b,\rms 2}^{(-2),\approx}, \chi_\eps h_{b,\rmv 1}^*(\vect'_1)\ra \nonumber\\
      &= \lim_{\eps\searrow 0} \la h_{b,\rms 2}^{(-2),\approx}, \wh{L_b}(0)^*\bigl(\chi_\eps h_{b,\rmv 1}^*(\vect'_1) \ra \nonumber\\
      &= \lim_{\eps\searrow 0} \Big\la h_{b,\rms 2}^{(-2),\approx}, [\wh{L_b}(0)^*,\chi_\eps] \bigl( h_{b,\rmv 1}^\sharp(\bha(b)\times\bha(\vect'_1)) + h_{b,\rmv 1}^{*\flat}(\vect'_1)\bigr) \Big\ra. \nonumber
  \end{align}
  \end{subequations}
  Due to the fast decay of $h_{b,\rmv 1}^{*\flat}(\vect'_1)$, this term only gives a vanishing contribution in the limit $\eps\searrow 0$, so we are left with
  \begin{align*}
    &\lim_{\eps\searrow 0} \la h_{b,\rms 2}^{(-2),\approx}, [\wh{L_b}(0)^*,\chi_\eps]h_{b,\rmv 1}^\sharp(\bha(b)\times\bha(\vect_1'))\ra \\
    &\qquad = \la \wh{L_b}(0)h_{b,\rms 2}^{(-2),\approx}, h_{b,\rmv 1}^\sharp(\bha(b)\times\bha(\vect'_1)) \ra - \la h_{b,\rms 2}^{(-2),\approx}, \wh{L_b}(0)^* h_{b,\rmv 1}^\sharp(\bha(b)\times\bha(\vect'_1))\ra.
  \end{align*}
  We use here that $\wh{L_b}(0)^*h_{b,\rmv 1}^\sharp\in\rho^{\lfloor C_0\rfloor+2}\CI$ decays more than fast enough for the second pairing to be well-defined. Crucially (for smoothness in $b$), we can define a more general linear functional
  \[
    \lambda^\sharp_{b,\rmv 1}(\bfw) := \la\wh{L_b}(0)h_{b,\rms 2}^{(-2),\approx}, h_{b,\rmv 1}^\sharp(\bfw)\ra - \la h_{b,\rms 2}^{(-2),\approx}, \wh{L_b}(0)h_{b,\rmv 1}^\sharp(\bfw)\ra
  \]
  on $\R^3$; this is given by the Euclidean inner product with some\footnote{The subtle but crucial point here is that while $\lambda_{b,\rmv 1}$ only induces a functional on the subspace $\{\bha(b)\times\bha(\vect'_1)\colon\vect'_1\subset\vectspace_1\}\subset\R^3$---which depends in a discontinuous fashion on $\bha$, and thus so does its dual---, the functional $\lambda_{b,\rmv 1}^\sharp$ is defined on \emph{all} of $\R^3$.}
  \[
    \bfv_b\in\R^3
  \]
  depending smoothly on $b$, so
  \begin{equation}
  \label{EqWE0s2v1}
    \lambda_{b,\rmv 1}(\vect'_1) = \lambda_{b,\rmv 1}^\sharp(\bha(b)\times\bha(\vect'_1)) = \la \bfv_b, \bha(b)\times\bha(\vect_1')\ra_{\R^3}.
  \end{equation}

  \pfsubstep{Step~2.}{Eliminating the obstructions.} We now claim that there exist $\scal_1\in\scalspace_1$ and $\vect_1\in\vectspace_1$, with smooth dependence on $b$, such that
  \begin{alignat}{2}
  \label{EqWE0s2s1Solv}
    \big\la\wh{L_b}(0)\bigl(\breve h_{b,\rms 1}^2(\scal_1) + \breve h_{b,\rmv 1}^1(\vect_1)\bigr), h_{b,\rms 1}^*(\scal_1') \big\ra &= -\lambda_{b,\rms 1}(\scal'_1)&\quad&\forall\,\scal_1'\in\scalspace_1, \\
  \label{EqWE0s2v1Solv}
    \big\la\wh{L_b}(0)\bigl(\breve h_{b,\rms 1}^2(\scal_1) + \breve h_{b,\rmv 1}^1(\vect_1)\bigr), h_{b,\rmv 1}^*(\vect_1') \big\ra &= -\lambda_{b,\rms 1}(\vect'_1)&\quad&\forall\,\vect_1'\in\vectspace_1.
  \end{alignat}
  By Lemma~\ref{LemmaWE0Pair}, the left-hand side of~\eqref{EqWE0s2s1Solv} is independent of the choice of $\vect_1$, and thus we obtain a unique solution $\scal_1$ (with smooth dependence on $b$). Similarly, the left-hand side of~\eqref{EqWE0s2v1Solv} is independent of $\scal_1$, and equals $-8\pi\bhm\la\bha(\vect_1),\bha(b)\times\bha(\vect_1')\ra_{\R^3}$; by comparison with~\eqref{EqWE0s2v1}, we can thus take $\vect_1\in\vectspace_1$ to be the unique element with $-8\pi\bhm\bha(\vect_1)=\bfv_b$.

  \pfsubstep{Step~3.}{A smooth construction of the zero energy state.} Having arranged for~\eqref{EqWE0s2s1Solv}--\eqref{EqWE0s2v1Solv}, we now conclude from~\eqref{EqWE0s2s0}, \eqref{EqWE0s2s1}, and~\eqref{EqWE0s2v10} that
  \[
    h_{b,\rms 2}^{(-2),\approx,2}:=h_{b,\rms 2}^{(-2),\approx} + \breve h_{b,\rms 1}^2(\scal_1) + \breve h_{b,\rmv 1}^1(\vect_1) \implies
    \la\wh{L_b}(0)h_{b,\rms 2}^{(-2),\approx,2}, h^*\big\ra = 0
  \]
  for all $h^*=h_{b,\rms 0}^*,h_{b,\rms 1}^*(\scal'_1),h_{b,\rmv 1}^*(\vect'_1)$. Note then that $\wh{L_b}(0)\in\rho^2\Diffb^2(X;S^2\cT^*_X)$ maps $\breve h_{b,\rms 1}^2$ and $\breve h_{b,\rmv 1}^1\in\rho^{-1}\CI$ into $\rho^2\CI$ (since $N(\wh{\ubar L}(0),-1)$ annihilates their leading-order terms $\ubar h_{\rms 1}^2$ and $\ubar h_{\rmv 1}^1$, cf.\ Proposition~\ref{PropWG0Large}\eqref{ItWG0Larges1m1}--\eqref{ItWG0Largev1m1}), so
  \[
    \wh{L_b}(0)h_{b,\rms 2}^{(-2),\approx,2}\in\cA^{(2,0)\cup(2+\cE_\ind)}(X;S^2\cT^*_X),
  \]
  with smooth dependence on $b$. Using Lemma~\ref{LemmaTMFormal}, we can construct $h_{b,(0)}\in\cA^{(0,k)}(X;S^2\cT^*_X)$ (for some $k\in\N_0$) such that
  \[
    f_b := -\wh{L_b}(0)\bigl(h_{b,\rms 2}^{(-2),\approx,2} + h_{b,(0)}\bigr) \in \cA^{2+\cE_\ind}(X;S^2\cT^*_X).
  \]
  The $\cO(\rho^2)$-decay of the dual states $h^*$ of $\wh{L_b}(0)$ justifies the integration by parts in $\la\wh{L_b}(0)h_{b,(0)},h^*\ra=\la h_{b,(0)},\wh{L_b}(0)^*h^*\ra=0$, and hence we conclude that $f_b$ is still orthogonal to the cokernel of $\wh{L_b}(0)$. Therefore, Corollary~\ref{CorWE0Solv}\eqref{ItWE0Solv2} produces $h_{b,\flat}\in\cA^{\cE_\ind}(X;S^2\cT^*_X)$, with smooth dependence on $b$, such that
  \begin{equation}
  \label{EqWE0s2State}
    h_{b,\rms 2}^{(-2),=} := h_{b,\rms 2}^{(-2),\approx,2} + h_{b,\flat} \implies \wh{L_b}(0)h_{b,\rms 2}^{(-2),=} = 0.
  \end{equation}
  This finishes the construction of a large zero energy state with leading-order term $\ubar h_{\rms 2}^{(-2)}$ and smooth dependence on $b$.

  \pfsubstep{Step~4.}{Improved construction; generalized zero energy state.} While~\eqref{EqWE0s2State} is an acceptable large zero energy state by itself, it turns out that it may not be the $t_*$-coefficient of a linear (in $t_*$) generalized zero energy state due to the failure of $[L_b,t_*]h_{b,\rms 2}^{(-2),=}\in\cA^{(-1,0)\cup(-1+\cE_\ind)}$ (with scalar type $2$ leading-order term) to be orthogonal to some of the dual zero energy states~\eqref{EqWEMode0Coker}. The flexibility which we have not yet exploited is to add to $h_{b,\rms 2}^{(-2),=}$ elements of $\ker\wh{L_b}(0)$ with $o(\rho^{-2})$-decay; since pure gauge zero energy states do come with associated generalized states by Proposition~\ref{PropWG0Large}, only the linearized Kerr metrics will be useful here.

  Since $-3$ is not an $\rms 2$ root by Lemma~\ref{LemmaWEInd}, we can first use Lemma~\ref{LemmaTMFormal} to produce a formal solution $\breve h_{b,\rms 2}^{(-2),1,\approx}\in\cA^{(-3,0)\cup(-3+\cE_\ind)}(X;S^2\cT^*_X)$ of
  \[
    [L_b,t_*]h_{b,\rms 2}^{(-2),=} + \wh{L_b}(0)\breve h_{b,\rms 2}^{(-2),1,\approx} \in \cA^{2+\cE_\ind}(X;S^2\cT^*_X).
  \]
  Our task is then to determine $\dot b=(\dot\bhm,\dot\bha)$ and $\scal_1\in\scalspace_1$ (possibly different from the one in Steps~2 and 3 above), with smooth dependence on $b$, such that
  \begin{equation}
  \label{EqWE0s2Gen}
    \Big\la [L_b,t_*]\bigl(h_{b,\rms 2}^{(-2),=}+\dot g_b^\Ups(\dot b)\bigr) + \wh{L_b}(0)\bigl(\breve h_{b,\rms 2}^{(-2),1,\approx}+\breve h_{b,\rms 1}^2(\scal_1)\bigr), h^*\Big\ra = 0
  \end{equation}
  for $h^*=h_{b,\rms 0}^*,h_{b,\rms 1}^*(\scal'_1),h_{b,\rmv 1}^*(\vect'_1)$. (We do not need the flexibility of adding $\breve h_{b,\rmv 1}^1(\vect_1)$ here anymore, since the contribution from $\dot g_b^\Ups(\dot b)$ will already suffice to tweak the inner products with $h_{b,\rmv 1}^*(\vect'_1)$.) For $\bullet=\rms 0,\rms 1,\rmv 1$, define
  \begin{align*}
    \lambda_\bullet^{(-2),=} &:= \la[L_b,t_*]h_{b,\rms 2}^{(-2),=},h_{b,\bullet}^*\ra, \\
    \breve\lambda_\bullet^{(-2),1,\approx} &:= \la\wh{L_b}(0)\breve h_{b,\rms 2}^{(-2),1,\approx},h_{b,\bullet}^*\ra,
  \end{align*}
  so $\lambda^{(-2),=}_{\rms 0}\in\R$, while $\lambda^{(-2),=}_{\rms 1}\in\scalspace_1^*$ and $\lambda^{(-2),=}_{\rmv 1}\in\vectspace_1^*$; similarly for $\breve\lambda_\bullet^{(-2),1,\approx}$. We now evaluate the left-hand side of~\eqref{EqWE0s2Gen} for the three different classes of dual states using Lemma~\ref{LemmaWEPair}:
  \begin{enumerate}
  \item For $h^*=h_{b,\rms 0}^*$, we need $\lambda_{\rms 0}^{(-2),=}-16\pi\dot\bhm=0$, which forces us to take $\dot\bhm=\lambda_{\rms 0}^{(-2),=}/16\pi$.
  \item For $h^*=h_{b,\rms 1}^*(\scal'_1)$, we need $\lambda_{\rms 1}^{(-2),=}(\scal'_1)+\breve\lambda_\bullet^{(-2),1,\approx}(\scal'_1)+\la\wh{L_b}(0)\breve h_{b,\rms 1}^2(\scal_1),h_{b,\rms 1}^*(\scal'_1)\ra=0$ for all $\scal'_1\in\scalspace_1$. By Lemma~\ref{LemmaWE0Pair}, this uniquely determines $\scal_1$.
  \item For $h^*=h_{b,\rmv 1}^*(\vect'_1)$, we need
    \[
      \lambda_{\rmv 1}^{(-2),=}(\vect'_1) - 16\pi\dot\bhm\la\bha(b),\bha(\vect'_1)\ra_{\R^3} - 16\pi\bhm\la\dot\bha,\bha(\vect'_1)\ra_{\R^3} + \breve\lambda_{\rmv 1}^{(-2),1,\approx}(\vect'_1) = 0
    \]
    for all $\vect'_1\in\vectspace_1$, which uniquely determines $\dot\bha$.
  \end{enumerate}
  Having arranged~\eqref{EqWE0s2Gen}, we can then find $\breve h_{b,\flat}\in\cA^{(0,k)\cup\cE_\ind}$ (using Lemma~\ref{LemmaTMFormal} to solve away the $\rho^2$ leading-order term of $[L_b,t_*]\dot g_b^\Ups(\dot b)+\wh{L_b}(0)\breve h_{b,\rms 1}^2(\scal_1)$, followed by an application of Corollary~\ref{CorWE0Solv}\eqref{ItWE0Solv2}), with smooth dependence on $b$, such that
  \[
    [L_b,t_*]\bigl(h_{b,\rms 2}^{(-2),=}+\dot g_b^\Ups(\dot b)\bigr) + \wh{L_b}(0)\bigl(\breve h_{b,\rms 2}^{(-2),1,\approx}+\breve h_{b,\rms 1}^2(\scal_1)+\breve h_{b,\flat}\bigr) = 0.
  \]
  This completes the construction of
  \[
    h_{b,\rms 2}^{(-2)} := h_{b,\rms 2}^{(-2),=} + \dot g_b^\Ups(\dot b),\quad
    \breve h_{b,\rms 2}^{(-2),1} := \breve h_{b,\rms 2}^{(-2),1,\approx}+\breve h_{b,\rms 1}^2(\scal_1) + \breve h_{b,\flat}.
  \]

  \pfsubstep{Step~5.}{Linearized gauge condition and Einstein equation.} Given any solution $h$ of $\wh{L_b}(0)h=0$ with $h\in\cA^\mu(X;S^2\cT^*_X)$ where $\mu\geq -C_0-1$, the second Bianchi identity implies
  \[
    0 = \delta_{g_b}\sfG_{g_b}L_b h = \frac12\Box_{g_b,E^\cC}^\cC \eta,\quad \eta := \delta_{g_b,E^\Ups}\sfG_{g_b}h \in \cA^{\mu+1}(X;\cT^*_X),
  \]
  so $\wh{\Box_{g_b,E^\cC}^\cC}(0)\eta=0$ since $\eta$ is stationary. Since $\mu+1\geq-C_0$, the enhanced mode stability recorded in Theorem~\ref{ThmWCRec}\eqref{ItWCRec0} implies that $\eta=0$, and therefore $0=L_b h=D_{g_b}\Ric(h)=0$. This applies, in particular, to $h=h_{b,\rms 2}^{(-2)}(\scal)$, $\scal\in\scalspace_2$.

  For generalized mode solutions $t_* h+\breve h$ where $h\in\cA^{\mu+1}\cap\ker L_b$ (and thus $h\in\ker D_{g_b}\Ric\cap\ker\delta_{g_b,E^\Ups}\sfG_{g_b}$) and $\breve h\in\cA^\mu$, $\mu\geq-C_0-1$, we similarly have
  \begin{align*}
    0 = \delta_{g_b}\sfG_{g_b}L_b(t_* h+\breve h) &= \frac12\Box_{g_b,E^\cC}^\cC\bigl(\delta_{g_b,E^\Ups}\sfG_{g_b}(t_* h+\breve h)\bigr) \\
      &= \frac12\wh{\Box_{g_b,E^\cC}^\cC}(0)\bigl( [\delta_{g_b,E^\Ups}\sfG_{g_b},t_*]h + \delta_{g_b,E^\Ups}\sfG_{g_b}\breve h\bigr).
  \end{align*}
  The argument of $\wh{\Box_{g_b,E^\cC}^\cC}(0)$ lies in the space $\cA^{\mu+1}$ and thus again vanishes (by enhanced mode stability); and this then also implies $D_{g_b}\Ric(t_* h+\breve h)=0$.

  \pfstep{Part~\eqref{ItWE0s2}, vector type $2$.} The arguments are completely analogous to the scalar type $2$ case.

  \pfstep{Part~\eqref{ItWE0s3}.} For the construction of $h_{b,\rms 3}^{(-3)}$ and $h_{b,\rmv 3}^{(-3)}$, we can use completely analogous arguments; the only difference is that the formal solution at the beginning of Step~1 above will now lie in $\cA^{-3+\cE_\ind}(X;S^2\cT^*_X)$ (in fact, it is log-smooth). Since we do not need any corresponding generalized zero energy states, Step~4 can be omitted.
\end{proof}

\section{Spectral theory at transition faces}
\label{Sip}

\emph{We fix the modifications $E^\cC$ and $E^\Ups$ as at the beginning of~\S\usref{SWE}, with the parameter $\gamma^\Ups$ moreover chosen small enough such that the conclusions of Proposition~\usref{PropWEtf} hold.} The operator
\[
  \ubar L = D_{\ubar g}\Ric + \ubar\delta_{\ubar E^\cC}^*\ubar\delta_{\ubar E^\Ups}\ul\sfG
\]
defined in~\eqref{EqWEOpMink} is homogeneous of degree $-2$ with respect to dilations in the base (which we define to act trivially in the fibers of $S^2 T^*\R^4$ split via~\eqref{EqTYMink01Split} or, equivalently, trivialized via differentials of linear coordinates). In view of Lemma~\ref{LemmaWEOp}, it is thus the $\sface$-normal operator of $L_b\in\rho^2\Difftb^2(M_0)$ in the terminology of Definition~\ref{DefKMfdMin} and \S\ref{SssT3b}. We recall that
\[
  2\ubar L = -2\rho\pa_{t_*}(\rho\pa_\rho-1-\ubar S) + \rho^2\wt{\ubar L}(0)(\omega,\rho\pa_\rho,\pa_\omega)
\]
where we drop the subscript ``$\ubar E^\Ups,\ubar E^\cC$'' for brevity, and $\wt{\ubar L}(0)=\rho^{-2}\wh{\ubar L}(0)\in\Diffb^2(\ol{\R^3}\setminus o)$ (introduced already in~\eqref{EqWEtfOp}, with principal part $(\rho D_\rho)^2+\slDelta$) is dilation-invariant. We recall the important fact
\begin{equation}
\label{EqSipBd}
  \min\spec\ubar S = 0.
\end{equation}
Following the discussion leading to \eqref{EqT3bNsfMellin}, let us pass to the coordinates
\[
  \rho,\quad v:=\rho t_*,\quad\omega\in\Sph^2,
\]
thus
\[
  2\rho^{-2}\ubar L = -2\pa_v(\rho\pa_\rho+v\pa_v-1-\ubar S) + \wt{\ubar L}(0)(\omega,\rho\pa_\rho+v\pa_v,\pa_\omega),
\]
and correspondingly define its spectral family by
\begin{equation}
\label{EqipSpecFam}
  N_\sface(\lambda) := N(2\rho^{-2}\ubar L,\rho^\lambda) := -2\pa_v(v\pa_v+\lambda-1-\ubar S) + \wt{\ubar L}(0)(\omega,v\pa_v+\lambda,\pa_\omega).
\end{equation}
In this section, we shall prove several results concerning the inversion of $N_\sface(\lambda)$ (\S\ref{SsipInv}) as well as related results on the spectral side for $N_\tface(\ubar L,\pm 1)$ (\S\ref{Ssiptf}) which explain how source terms near $\iota^+\subset M$, or on the spectral side near $\tface\subset X_\scbtop^\pm=[\pm[0,1]_\sigma\times X;\{0\}\times\pa X]$, produce waves whose asymptotics at $\iota^+$ and $\cK^+$ can be described very precisely; see~\S\S\ref{SssIGen2La} and \ref{SssIEinLa} for heuristics, and Proposition~\ref{PropipGr} for a concrete result of this flavor.

\subsection{Inversion of the spectral family}
\label{SsipInv}

Notice that the operator $N_\sface(\lambda)$ is elliptic for $v>0$, and indeed elliptic as a b-differential operator also at
\[
  R:=v^{-1}=\frac{r}{t_*}=0;
\]
but it is hyperbolic for $v\in(-2,0)$. (The set $\{v=-2\}$ corresponds to the light cone at \emph{past} null infinity.) It is, in fact, a prototypical example of an operator that fits into the non-elliptic Fredholm framework introduced in \cite{VasyMicroKerrdS}; apart from the fact that $N(\rho^{-2}\ubar L,\rho^\lambda)$ does not extend smoothly across the ``north pole'' $R=0$ as a differential operator on $\pa\ol{\R^4}$, this can be regarded as a special case of the computations in \cite[\S{3.3}]{BaskinVasyWunschRadMink}. Since we are studying \emph{forward} solutions of $L_b$, we are interested in values of $\lambda$ for which \emph{forward} solutions of $N(\rho^{-2}\ubar L,\rho^\lambda)u=0$ exist, i.e., $u=u(v,\omega)$ with support in $\{v\geq 0\}$; this thus corresponds to studying $N_\sface(\lambda)$ on function spaces with supported character at a (spacelike) hypersurface $v=v_0\leq 0$, say $v=-1$ (not $v=0$, as this is the delicate place where $N_\sface(\lambda)$ changes type), cf.\ \cite[\S{5}]{BaskinVasyWunschRadMink} (see also \cite{VasyMinkDSHypRelation}). Since $\{v=0\}$, which over $\sface$ can be regarded as a subset of either $\pa\ol{\R^4}$ or $\sface$, is the light cone at infinity $Y^+\subset\sface$ (see~\eqref{EqKMfdYplus}), it will be important for us to keep track of conormal regularity (and partial polyhomogeneity) there. We thus work with the following function spaces:

\begin{definition}[Weighted Sobolev spaces on $\sface$]
\label{DefipHb}
  For $s,\mu\in\R$, we define
  \[
    \dot H_\bop^{s,\mu}([-1,\infty]_v\times\Sph^2)
  \]
  to be the space of elements of $\Hb^{s,\mu}(\sface)=\la v\ra^{-\mu}\Hb^s(\sface)$, $\sface=\ol{\R_v}\times\Sph^2$, with support in $\{v\geq -1\}$. For $k\in\N_0$, we moreover define
  \[
    \dot H_{\bop;\bop}^{(s;k),\mu}([-1,\infty]\times\Sph^2)
  \]
  to consist of all $u\in\dot H_\bop^{s,\mu}$ such that $(v\pa_v)^j\pa_\omega^\gamma u\in\dot H_\bop^{s,\mu}$ for all $j+|\gamma|\leq k$.
\end{definition}

The operator $\rho^{-2}\ubar L$, as the $\sface$-normal operator of $\rho^{-2}L_b$, acts on sections of the bundle $S^2\cT^*|_\sface$, which (lying over $\la x\ra^{-1}=0$) is naturally isomorphic to $S^2\cT^*_X|_{\pa X}$; we denote this by $S^2\cT^*$ simply for brevity.

\begin{prop}[Meromorphic continuation]
\label{PropipMero}
  Write
  \[
    \sface_{\geq -1} := [-1,\infty]_v\times\Sph^2 \subset \sface.
  \]
  Fix the volume density $|\frac{\dd v}{2+v}\,\dd\slg|$ on $\sface_{\geq 1}$ to define $L^2$-space underlying the Sobolev spaces in Definition~\usref{DefipHb}.\footnote{This density is an unweighted b-density at $R=0$, and it is smooth for $v\geq -1$.} Recall the small number $\eps_\ind>0$ from~\eqref{EqWEIndEps}.
  \begin{enumerate}
  \item\label{ItipMeroFred}{\rm (Fredholm property.)} Let $\lambda\in\C$. For $\mu\in(\Re\lambda-\eps_\ind,\Re\lambda)$ and $s<\frac32+\min\spec\ubar S-\Re\lambda=\frac32-\Re\lambda$, the operator
    \begin{equation}
    \label{EqipMeroMap}
    \begin{split}
      N_\sface(\lambda) &\colon \bigl\{ u\in\dot H_\bop^{s,\mu}(\sface_{\geq -1};S^2\cT^*) \colon N_\sface(\lambda)u \in \dot H_\bop^{s-1,\mu}(\sface_{\geq -1};S^2\cT^*) \bigr\} \\
        &\qquad \to \dot H_\bop^{s-1,\mu}(\sface_{\geq -1};S^2\cT^*)
    \end{split}
    \end{equation}
    is Fredholm of index $0$.
  \item\label{ItipMeroInv}{\rm (Invertibility.)} For all $\lambda\in\C$ with $\Re\lambda<1+\eps_\ind$, the map~\eqref{EqipMeroMap} is invertible. Furthermore, for every $C$, the map~\eqref{EqipMeroMap} is invertible for all but finitely many $\lambda\in\C$ with $\Re\lambda\leq C$, and the inverse $N_\sface(\lambda)^{-1}$ is finite-meromorphic.\footnote{That is, the Schwartz kernel extends meromorphically from $\Re\lambda<1+\eps_\ind$ to the entire complex plane, with finite rank principal parts at each pole. Alternatively, one can work with fixed $s,\mu$ and deduce the meromorphicity of $N_\sface(\lambda)^{-1}$ for all $\lambda$ that lie in a strip for which $\mu$ and $s$ are admissible orders in part~\eqref{ItipMeroFred}; and one can cover the complex plane in $\lambda$ by such strips.}
  \item\label{ItipMeroHi}{\rm (Higher b-regularity at $v=0$.)} If $u\in\dot H_\bop^{s,\mu}$ satisfies $N_\sface(\lambda)u\in\dot H_{\bop;\bop}^{(s-1;k),\mu}$ for some $k\in\N_0$, then in fact $u\in\dot H_{\bop;\bop}^{(s;k),\mu}$.
  \end{enumerate}
\end{prop}

While closely related to \citeAF{Proposition~\ref*{PropNip}}, this is a different result, mainly as we work here across $v=0$: this allows us to use the very robust methods of \cite{VasyMicroKerrdS} for the analysis of $N_\sface(\lambda)$, whereas the 0-analysis in \citeAF{\S\ref*{SsNip}} becomes very delicate as soon as $\Re\lambda$ (in the notation of \cite{HintzNonstat2}: $-\Im\sigma$, under the identification $\sigma=-i\lambda$) exceeds $\min\Re\ubar S$ (as this would correspond to studying the meromorphic continuation of a resolvent in the zero calculus, cf.\ \cite{MazzeoMelroseHyp,GuillarmouMeromorphic,SaBarretoWangResolvent}). In addition, part~\eqref{ItipMeroInv} is stronger than the analogous statement in \citeAF{Proposition~\ref*{PropNip}(\ref*{ItNipB})}, which only covers the range $\Re\lambda<1$. We also remark that due to the particular (singular) structure of $N_\sface(\lambda)$ near $R=0$ in the high-frequency regime $|\Im\lambda|\to\infty$, Proposition~\ref{PropipMero} is \emph{not} a consequence of the results in \cite[\S{5}]{BaskinVasyWunschRadMink}.

\begin{proof}[Proof of Proposition~\usref{PropipMero}]
  \pfstep{Semi-Fredholm estimate.} For fixed $\lambda\in\C$, the principal part of $N_\sface(\lambda)$ is $2D_v v D_v+(v D_v)^2+\slDelta$. We claim that we have the semi-Fredholm estimate
  \[
    \|u\|_{\dot H_\bop^{s,\mu}} \leq C\Bigl(\|N_\sface(\lambda)u\|_{\dot H_\bop^{s-1,\mu}} + \|u\|_{\dot H_\bop^{-N,\mu}} \Bigr)
  \]
  for any fixed $N$, provided $s<\frac32+\min\spec\ubar S-\Re\lambda$. This is a by now standard consequence of radial point estimates at the radial set
  \[
    \cR:=N^*\{v=0\}\setminus o
  \]
  and real principal type propagation estimates (on the characteristic set over $v\leq 0$ away from $\cR$) as well as microlocal elliptic estimates (and elliptic estimates in the b-setting near $R=0$); see \cite[\S{2}]{VasyMicroKerrdS} and also \cite{ZworskiRevisitVasy}, \cite[Chapter~5.6]{DyatlovZworskiBook}, and \cite[Chapter~11.3]{HintzMicro}. The upper bound on the regularity order $s$ here arises without computation as follows: the characteristic exponents of the b-normal operator of $v N_\sface(\lambda)$ at $v=0$ are $0$ (from $v\pa_v$) and $1+\alpha-\lambda$ where $\alpha\in\spec\ubar S$ (from $v\pa_v+\lambda-1-\ubar S$ in~\eqref{EqipSpecFam}); the below-threshold radial point estimate \cite[Theorem~10.7(2)]{HintzMicro} (which is relevant here) must \emph{allow} for all of latter indicial roots, which in view of $\chi_+^{1+\alpha-\lambda}(v)\in H_\loc^{\frac32+\alpha-\Re\lambda}$ (where $\chi_+^z(v)=v_+^z/\Gamma(1+z)$ as usual) leads to the stated condition on $s$. (We leave it to the reader to check that a computation of this threshold regularity following \cite[Chapter~10.2]{HintzMicro}, for a positive definite fiber inner product on $S^2\cT^*$ in which the summands of a bundle splitting in which $\ubar S$ is diagonal are orthogonal, yields the same result.)

  Furthermore, since the b-normal operator of $N_\sface(\lambda)$ at $R=0$ is given by $\wt{\ubar L}(0)(\omega,-R\pa_R+\lambda,\pa_\omega)$, its indicial roots are given by all $\mu\in\C$ for which $-\mu+\lambda$ is an indicial root of $\wt{\ubar L}(0)$ or equivalently of $\wh{\ubar L}(0)$, i.e., one of the numbers recorded in Lemma~\ref{LemmaWEInd}. If $\Re(-\mu+\lambda)$ is not equal to $\Re\lambda'$ for all indicial roots $\lambda'$ of $\wh{\ubar L}(0)$, a normal operator argument (see, e.g., \citeAF{(\ref*{EqMUbEstNearInfty})}) allows us to weaken the error term in this estimate. This applies in particular to $\mu\in(\Re\lambda-\eps_\ind,\Re\lambda)$ and yields the semi-Fredholm estimate
  \begin{equation}
  \label{EqipMeroSemi}
    \|u\|_{\dot H_\bop^{s,\mu}} \leq C\Bigl( \|N_\sface(\lambda)u\|_{\dot H_\bop^{s-1,\mu}} + \|u\|_{\dot H_\bop^{-N,\mu_0}}\Bigr)
  \end{equation}
  for all $\mu_0<\mu$. (If $\mu_0\in(\Re\lambda-\eps_\ind,\Re\lambda)$, as we shall assume from now on, then this estimate holds in the strong sense that if the right-hand side is finite, then in fact $u\in\dot H_\bop^{s,\mu}$ and the estimate holds.)

  \pfstep{Higher regularity: proof of part~\eqref{ItipMeroHi}.} Away from $v=0$ (i.e., upon multiplication by a cutoff function which equals $0$ near $v=0$ and $1$ near $v=\pm\infty$), membership in $\dot H_{\bop;\bop}^{(s;k),\mu}$ is equivalent to membership in $\dot H_\bop^{s+k,\mu}$ (and thus to membership in $\dot H_\loc^{s+k}$ on $[-1,\infty)_v\times\Sph^2$). By b-elliptic regularity at $v=\infty$, we thus only need to prove higher regularity in $v\leq 0$. But in $v<0$, we merely need to use the standard real principal type propagation of regularity on $H_\loc^{s+k}$, and it remains to propagate $k$ orders of b-regularity (relative to $H_\loc^s$) into the characteristic set of $N_\sface(\lambda)$ over $v=0$, which is given by $\cR$. The required higher regularity result at the radial set $\cR$ is then an instance of a general result of Haber--Vasy, specifically, the first part of \cite[Theorem~6.3]{HaberVasyPropagation} (which is a module regularity type result in the spirit of \cite{HassellMelroseVasySymbolicOrderZero}, with additional microlocalization near individual points of $\cR$ which we do not need in the present paper however).

  \pfstep{Injectivity for $\Re\lambda<1+\eps_\ind$.} If $u\in\dot H_\bop^{s,\mu}$, with $\mu\in(\Re\lambda-\eps_\ind,\Re\lambda)$, satisfies $N_\sface(\lambda)u=0$, then $u\in\dot H_{\bop;\bop}^{(s;\infty),\mu}$ by part~\eqref{ItipMeroHi}. Since $\Re\lambda<2+\min\spec\ubar S=2$, we can take $s>-\frac12$ here (while still $s<\frac32+\min\spec\ubar S-\Re\lambda$). But then the identification
  \begin{equation}
  \label{EqipMerodotH}
    \dot H^s([0,\infty),|\dd v|)=H_\bop^{s,s}([0,\infty),|\dd v|)=H_\bop^{s,s-\frac12}([0,\infty),|\tfrac{\dd v}{v}|),\quad s>-\frac12,
  \end{equation}
  (where the weights are powers of $v$) recorded in \citeAF{Lemma~\ref*{LemmaDFTLb}} (see also \cite[Lemma~3.12]{LiInternalCorners} for $s\in(-\frac12,\frac12)$) is available, yielding (in view of the infinite b-regularity at $v=0$ and $v=\infty$)
  \begin{equation}
  \label{EqipMeroRestrMem}
    u|_{v>0} \in H_\bop^{\infty,(1+\min\spec\ubar S-\Re\lambda-\eps,\mu)}([0,\infty]_v\times\Sph^2,|\tfrac{\dd v}{v}|) \subset \cA^{1+\min\spec\ubar S-\Re\lambda-\eps,\mu}\quad\forall\,\eps>0,
  \end{equation}
  where we write $H_\bop^{s,(\alpha,\beta)}([0,\infty]_v\times\Sph^2):=\rho_0^\alpha\rho_\infty^\beta H_\bop^s$, with $\rho_0=\frac{v}{v+1}$ and $\rho_\infty=\frac{1}{v+1}$ being defining functions of $\{v=0\}$ and $\{v^{-1}=0\}$, respectively. Given the membership~\eqref{EqipMeroRestrMem}, the extension of $u$ to $\{v<0\}$ by $0$ lies in $L^1_\loc$. The subsequent arguments are very similar to those in the proof of \citeAF{Lemma~\ref*{LemmaNipBker}}, except that the a priori information that $N_\sface(\lambda)u=0$ holds \emph{for all} $v\geq -1$ (so including near $v=0$) allows us to treat a larger range of $\lambda$. To wit, we Fourier transform the equation $N_\sface(\lambda)u=0$ in $v$, with the convention
  \[
    \hat u(\hat r,\omega) = \int_\R e^{i\hat r v}u(v,\omega)\,\dd v.
  \]
  We then have
  \[
    w_\pm(\hat r,\omega) := \hat u(\pm\hat r,\omega) \in \cA^{((0,0),\mu-1-\eps),\ 1+\min\spec\ubar S+1-\Re\lambda-\eps}([0,\infty]_{\hat r}\times\Sph^2)
  \]
  for all $\eps>0$ by \cite[Lemma~2.25]{Hintz3b}, with the index set and weights referring to the boundary hypersurfaces $\hat r^{-1}(0)$ and $\hat r^{-1}(\infty)$ (in this order). It suffices to show that $w_\pm=0$, since then $\supp\hat u\subset\{\hat r=0\}$, so $\hat u$ is a sum of differentiated $\delta$-distributions at $\hat r=0$, and hence $u$ is a polynomial in $v$, which must therefore vanish since $u=0$ for $v<-1$. Now, upon conjugating the operator~\eqref{EqipSpecFam} by the Fourier transform, one finds that $w_\pm$ satisfies
  \[
    \Bigl(\pm 2 i\hat r(-\pa_{\hat r}\hat r+\lambda-1-\ubar S) + \wt{\ubar L}(0)(\omega,-\pa_{\hat r}\hat r+\lambda,\pa_\omega)\Bigr) w_\pm = 0,
  \]
  and therefore
  \[
    w_\pm^\flat(\hat r,\omega):=\hat r^{-\lambda+1}w_\pm(\hat r,\omega) \in \cA^{((-\lambda+1,0), \mu-\Re\lambda-\eps),\ 1+\min\spec\ubar S-\eps}([0,\infty]\times\Sph^2)
  \]
  satisfies
  \begin{equation}
  \label{EqipMeroNtf}
    0 = \Bigl(\mp 2 i\hat r(\hat r\pa_{\hat r}+1+\ubar S) + \wt{\ubar L}(0)(\omega,-\hat r\pa_{\hat r},\pa_\omega)\Bigr) w_\pm^\flat = \hat r^{-2}N_\tface(\ubar L,\pm 1)w_\pm^\flat,
  \end{equation}
  where we recall~\eqref{EqWEtfOp}. But since $\Re(-\lambda+1)>-\eps_\ind$, Proposition~\ref{PropWEtf} applies and shows that $w_\pm^\flat=0$, as desired.

  \pfstep{Index $0$ property.} Dually to~\eqref{EqipMeroSemi}, we have $\|u\|_{\bar H_\bop^{-s+1,-\mu}}\lesssim\|N_\sface(\lambda)^*u\|_{\bar H_\bop^{-s,-\mu}}+\|u\|_{\bar H_\bop^{-N,-\mu_1}}$ for all fixed $N,\mu_1$, (and in the strong sense if $\mu_1\in(\Re\lambda-\eps_\ind,\Re\lambda)$), where the adjoint is defined using the same density $|\frac{\dd v}{2+v}\,\dd\slg|$ as the one used to define $L^2$; this implies the Fredholm property of the map~\eqref{EqipMeroMap}. However, the Fourier transform cannot be utilized directly as in the previous step to prove the injectivity of $N_\sface(\lambda)^*$ on $\bar H_\bop^{-s+1,-\mu}$ since the latter space consists of extendible distributions.

  We get around this impasse in an indirect way, as follows: fix $\lambda_0\in\C$, and fix $\mu\in(\Re\lambda_0-\eps_\ind,\Re\lambda_0)$ and $s<\frac32+\min\spec\ubar S-\Re\lambda_0$ as in part~\eqref{ItipMeroFred}. Consider $\lambda=\lambda_0+i\varsigma$ where $\varsigma\geq 0$; then $N_\sface(\lambda)$ is Fredholm as a map~\eqref{EqipMeroMap} for all such $\lambda$, and its index is independent of $\varsigma$. We then consider the semiclassical limit $\varsigma=h^{-1}\to+\infty$ and prove that $N_\sface(\lambda)=N_\sface(\lambda_0+i h^{-1})$ is invertible for all sufficiently small $h>0$ and hence has index $0$. We do this by proving semiclassical (high-energy) estimates on \emph{semiclassical cone Sobolev spaces} $\dot H_\chop^{s,(\gamma_\cface,\gamma_\tface,\sfb)}(\sface_{\geq -1};S^2\cT^*)$ which were introduced in \cite{HintzConicPowers,HintzConicProp} and used for essentially the same purpose (except for the aforementioned difference in treating the behavior near $v=0$) in \citeAF{\S\ref*{SssNipHi}}. We recall the norm on these spaces for integer regularity ($s$) and semiclassical ($\sfb$) orders and refer to the references for the case of variable $\sfb$ (which we use below). Upon writing $\rho_\infty:=\la v\ra^{-1}$ for a defining function of $\pa\ol{\R_v}\times\Sph^2$ and setting
  \[
    \rho_\cface := \frac{\rho_\infty}{\rho_\infty+h},\quad
    \rho_\tface := \rho_\infty+h,\quad
    \rho_\hbarface := \frac{h}{\rho_\infty+h}
  \]
  (which are defining functions of the boundary hypersurfaces $\cface$, $\tface$, and\footnote{Unlike after \citeAF{(\ref*{EqMUipSingle})}, we do not use the notation $\sface$ for the semiclassical face, i.e., the lift of $\{0\}\times\sface$, here, to avoid a clash of notation.} $\hbarface$ of $\sface_\chop:=[[0,1)_h\times\sface;\{0\}\times\pa\sface]$), we define, for $s\in\N_0$,
  \[
    \|u\|_{H_\chop^s(\sface)} := \sum_{j+|\alpha|\leq s} \| (\rho_\hbarface \la v\ra\pa_v)^j (\rho_\hbarface\pa_\omega)^\alpha u \|_{L^2(\sface)},\quad L^2(\sface):=L^2(\sface,|\tfrac{\dd v}{\la v\ra}\,\dd\slg|),
  \]
  and $\|u\|_{H_\chop^{s,(\gamma_\cface,\gamma_\tface,\sfb)}}:=\|\rho_\cface^{-\gamma_\cface}\rho_\tface^{-\gamma_\tface}\rho_\hbarface^{-\sfb}u\|_{H_\chop^s}$. To bring ourselves closer to the setting considered in~\citeAF{\S\ref*{SsNip}} as far as the analysis near the conic singularity at $R=0$ is concerned, we first conjugate $R^2 N_\sface(\lambda)$ by $(t_*\rho)^\lambda=R^{-\lambda}$ near $R=0$, which amounts to considering the indicial family of $R^{-2}\rho^2\ubar L=t_*^2\ubar L$ in terms of its action on functions of the form $t_*^{-\lambda}u(R,\omega)$ (cf.\ \citeAF{Definition~\ref*{DefNipOp}}). This conjugation, if done globally, would destroy the smooth extendability of $N_\sface(\lambda)$ across $v=0$, and thus we must localize it: fix
  \[
    \tilde R(R) \in \CI(\R),\quad
    \tilde R(R) \begin{cases} =R, & R<\frac12, \\ \geq\frac12, & \frac12\leq R\leq 1, \\ =1, & R\geq 1, \end{cases}
  \]
  and define the operator
  \[
    N_h := \tilde R^{-\lambda}\tilde R^{-2}N_\sface(\lambda)\tilde R^\lambda,\quad \lambda=\lambda_0+i h^{-1}.
  \]
  For $R\geq 1$ (so $v\leq 1$), this equals $N_\sface(\lambda)$. For $R\leq\frac12$, we compute using~\eqref{EqipSpecFam} that
  \begin{equation}
  \label{EqipMeroHi}
    N_h = -2 R^{-1}(R\pa_R+\lambda)(R\pa_R+1+\ubar S) + R^{-2}\wt{\ubar L}(0)(\omega,-R\pa_R,\pa_\omega),
  \end{equation}
  which is equal to the operator $N_{\iota^+}^0(P_0,-i\lambda)$ in the notation of \citeAF{(\ref*{EqNipOpMT})}. For $R\in[\frac12,1]$, where $h^2 N_h$ is simply a semiclassical differential operator, conjugation by $\tilde R^\lambda=\exp((i+h\lambda_0)\log(\tilde R)/h)$ amounts to a pullback in phase space of the principal symbol (and thus of the null-bicharacteristic flow) along the translations by the (smooth and bounded) 1-form $\dd\log\tilde R=\frac{\dd\tilde R}{\tilde R}$. For $\gamma_\cface\in(-\eps_\ind,0)$ and suitable (variable) semiclassical orders $\sfb$, we then have
  \begin{equation}
  \label{EqipMeroHiEst}
    \| u \|_{\dot H_\chop^{s,(\gamma_\cface,\gamma_\cface,\sfb)}(\sface_{\geq -1})} \leq C \| h^2 N_h u \|_{\dot H_\chop^{s-1,(\gamma_\cface-2,\gamma_\cface,\sfb+1)}(\sface_{\geq -1})}
  \end{equation}
  for all sufficiently small $h>0$. This is proved in essentially the standard fashion as in \cite[Chapter~11.4]{HintzMicro}, with modifications as in \citeAF{\S\ref*{SssNipHi}} (and going back to \cite{HintzConicProp}) near $R=0$: starting with semiclassical energy estimates near $v=-1$, one uses the semiclassical propagation of regularity across $v=0$ towards $R=0$. Near $R=0$, one uses radial point estimates in the semiclassical cone calculus as in \citeAF{Step~(1.3) in \S\ref*{SssNipHi}} to propagate semiclassical control through $R=0$ (over the semiclassical face $\hbarface$ of $\sface_\chop$), from where, finally, real principal type propagation yields control in a punctured neighborhood of the conormal bundle of $v=0$ at fiber infinity, where a semiclassical (below-threshold) radial point estimate applies. This gives the estimate~\eqref{EqipMeroHiEst} but with an error term $\|u\|_{\dot H_\chop^{s_0,(\gamma_\cface,\gamma_\cface,\sfb_\flat)}}$ where $s_0<s$, $\sfb_\flat<\sfb$. Inversion of the $\tface$-normal operator of $h^2 N_h$---which is equal to $N_\tface(\ubar L,1)$---as in \citeAF{Step~2 in \S\ref*{SssNipHi}}---allows one to improve the second ($\tface$-)order to a number $<\gamma_\cface$, and thus this new error term is \emph{small} compared to the left-hand side of~\eqref{EqipMeroHiEst} when $h>0$ is sufficiently small, and can thus be absorbed.\footnote{This argument works for \emph{any} value of $\lambda_0$. Indeed, the only place where a lower bound on $\Im\sigma=-\Re\lambda_0$ (with the identification $\sigma=-i\lambda$) arises in \citeAF{\S\ref*{SssNipHi}} is in the estimates near the ``0-end'' $x_\sscri=0$ (in the notation of the reference). But presently we study $N_\sface(\lambda)$ on different function spaces where the delicate issues of 0-analysis do not arise. (That is, the present analysis is related to that in \citeAF{\S\ref*{SssNipHi}} in a manner similar to how \cite{VasyMicroKerrdS} relates to \cite{MazzeoMelroseHyp}.)} Note that this, together with the analogous result for $\lambda=\lambda_0-i h^{-1}$, also implies the finiteness statement of part~\eqref{ItipMeroInv}.

  The adjoint version of the estimate of~\eqref{EqipMeroHiEst} is proved similarly (using now the estimates for $N_\tface(\ubar L,1)^*$ implied by the $\tface$-admissibility referred to in \citeAF{Corollary~\ref*{CorSptfAdm}}, i.e., \citeAF{(\ref*{EqSptfEst})} without the second term on the right). Since for fixed $h>0$ the $\dot H_\chop^{s,(\gamma_\cface,\gamma_\cface,\sfb)}$-norm is equivalent to the $\dot H_\bop^{s,\gamma_\cface}$-norm, this then implies the invertibility of $N_h$ and thus of~\eqref{EqipMeroMap} for such $h$ and $\lambda=\lambda_0+i h^{-1}$.
\end{proof}

We now translate parts of Proposition~\ref{PropipMero} into a different functional setting appropriate for the study of the spectral family
\begin{equation}
\label{EqipSpecFam2}
\begin{split}
  N_{\iota^+}(\ubar L,\lambda) &:= t_*^\lambda\,2 t_*^2\ubar L\, t_*^{-\lambda} = (t_*\rho)^\lambda (t_*\rho)^2 \,\bigl(\rho^{-\lambda}\,2\rho^{-2}\ubar L\,\rho^\lambda\bigr)\, (t_*\rho)^{-\lambda} \\
    &= v^{2+\lambda} N_\sface(\lambda) v^{-\lambda} = R^{-2-\lambda} N_\sface(\lambda) R^\lambda
\end{split}
\end{equation}
of $\ubar L$ defined via formally Mellin transforming in $t_*$ (in the splitting $(0,\infty)_{t_*}\times(0,\infty)_v\times\Sph^2$ of $(0,\infty)_{t_*}\times(\R^3_x\setminus\{0\})$); this was already computed in~\eqref{EqipMeroHi}, which we also record in the coordinates $v,\omega$ to read
\begin{equation}
\label{EqipSpecFam2Expr}
\begin{split}
  N_{\iota^+}(\ubar L,\lambda) &= -2 R^{-1}(R\pa_R+\lambda)(R\pa_R+1+\ubar S) + R^{-2}\wt{\ubar L}(0)(\omega,-R\pa_R,\pa_\omega) \\
     &=-2 v(v\pa_v-\lambda)(v\pa_v-1-\ubar S) + v^2\wt{\ubar L}(0)(\omega,v\pa_v,\pa_\omega).
\end{split}
\end{equation}
At $v=\infty$, the indicial roots of $\wt{\ubar L}(0)(\omega,-R\pa_R,\pa_\omega)$, i.e., the negatives of the indicial roots recorded in Lemma~\ref{LemmaWEInd}, dictate the asymptotic behavior of solutions $h$ of equations of the form $N_{\iota^+}(\ubar L,\lambda)h=f$. At $v=0$ on the other hand, the indicial roots of $v\pa_v-1-\ubar S$ take on this role; these are $1$ plus the eigenvalues of $\ubar S=\ubar S_{\ubar E^\Ups,\ubar E^\cC}$, and thus we are led to essentially the same index sets of the various components of $h$ as in Definition~\ref{DefExP}. Thus, let us define:

\begin{definition}[Index sets at $\scri^+\cap\iota^+$ and polyhomogeneity]
\label{DefipIndex}
  Recall that we have fixed the parameters $0<\gamma^\Ups\ll 1$, $e^\Ups\in(0,1)$ and $v^\cC\in(0,1)$, $e^\cC\in(0,1)$, and $\gamma^\cC>0$.
  \begin{enumerate}
  \item{\rm (Index sets.)} We define
    \begin{equation}
    \label{EqipIndex0}
      \cE_{\iota^+,\sscri}^\tot := (1,0) \cup (1{+}(1{-}e^\Ups)\gamma^\Ups,0) \cup (1{+}\gamma^\Ups,0) \cup (1{+}2\gamma^\Ups,0), \quad
      \cE_{\iota^+,\sscri}^\cC := \cE_{\iota^+,\sscri}^\tot + 1,
    \end{equation}
    as well as
    \begin{equation}
    \label{EqipIndex1}
    \begin{alignedat}{2}
      \cE_{\iota^+,\sscri,0 1} &:= \cE_{\iota^+,\sscri}^\cC{\cup}(1{+}(1{-}e^\Ups)\gamma^\Ups,0), &\quad
      \cE_{\iota^+,\sscri,1 /} &:= \cE_{\iota^+,\sscri}^\cC{\cup}(1{+}\gamma^\Ups,0), \\
      \slcE_{\iota^+,\sscri,0} &:= \cE_{\iota^+,\sscri}^\cC{\cup}(1,0), &\quad
      \cE_{\iota^+,\sscri,1 1} &:= \cE_{\iota^+,\sscri}^\cC{\cup}(1{+}(1{-}e^\Ups)\gamma^\Ups,0){\cup}(1{+}2\gamma^\Ups,0), \\
    \end{alignedat}
    \end{equation}
  \item{\rm (Function spaces.)} For an index set $\cE_\cK\subset\C\times\N_0$, we denote by
    \[
      \cA^{\la\cE_{\iota^+,\sscri}^\cC\ra,\ \cE_\cK}(\iota^+)
    \]
    the space of real sections $h$ of $S^2\cT^*\to\iota^+$ such that, in the notation of Definition~\usref{DefExPComp},
    \begin{equation}
    \label{EqipIndexComp}
      \pi^\cC h \in \cA^{\cE_{\iota^+,\sscri}^\cC,\ \cE_\cK}(\iota^+;S^2\cT^*),\quad
      \pi_\bullet h \in \cA^{\cE_{\iota^+,\sscri,\bullet},\ \cE_\cK},\ \bullet \in \{ 0 1,\,1 /,\,1 1 \}, \quad
      \slpi_0 h \in \cA^{\slcE_{\iota^+,\sscri,0},\ \cE_\cK}.
    \end{equation}
    The index sets refer to $\iota^+\cap\scri^+$ and $\iota^+\cap\cK^+$ (in this order). Spaces of partially polyhomogeneous tensors, resp.\ tensors with finite b-regularity are denoted by $\cA^{(\la\cE_{\iota^+,\sscri}^\cC\ra,\ell_\sscri),\ (\cE_\cK,\ell_\cK)}$, resp.\ $\Hb^{k,\ (\la\cE_{\iota^+,\sscri}^\cC\ra,\ell_\sscri),\ (\cE_\cK,\ell_\cK)}$.
  \end{enumerate}
\end{definition}

Besides not requiring any nonlinear properties (or the propagation of polyhomogeneity from $I^0$), another (minor) difference is that we do not need to include the index set $(1,0)$ in $\cE_{\iota^+,\sscri,1 1}$, cf.\ \eqref{EqExPOther}. (This arose from the coupling of the $r^2\ker\sltr$-component with the $(\dd x^1)^2$-component via the $(7,6)$-component of $A_h$ in~\eqref{EqExOpLinAB}; but this coupling is absent when $h=0$.)

\begin{prop}[Inversion of $N_{\iota^+}(\ubar L,\lambda)$]
\label{PropipNInv}
  Suppose $\lambda\in\C$ is such that $N_\sface(\lambda)$ is invertible. Given $f\in\CIdot(\iota^+;S^2\cT^*)$ (i.e., $f$ vanishes to infinite order at $\pa\iota^+=\iota^+\cap(\scri^+\cup\cK^+)$), set
  \begin{equation}
  \label{EqipNInvhDef}
    h := N_{\iota^+}(\ubar L,\lambda)^{-1}f := \bigl(R^{-\lambda} N_\sface(\lambda)^{-1}(R^{2+\lambda}f)\bigr)\big|_{(0,\infty)_R\times\Sph^2}.
  \end{equation}
  Then $N_{\iota^+}(\ubar L,\lambda)h=f$. Set $\cE_\cK:=(0,0)\cup\bigcup_{ \substack{ 0\leq l\leq 3,\ j=0,1, \\ 1\leq l+j\leq 3 } } (\lambda^\Ups_{\rms l,l+j}-1,0)$ (so in particular $\min\Re(\cE_\cK\setminus\{(0,0)\})>0$). Then:
  \begin{enumerate}
  \item\label{ItipNInvCon}{\rm (Conormality.)} We have\footnote{In fact, $h$ has a full polyhomogeneous expansion at $R=0$, with an index set that includes elements $(z,k)$ with $\Re z>3$ beyond those contained in $\cE_\cK$. This index set can be computed from the indicial roots of $\wh{\ubar L}(0)$ with negative real parts.}
    \begin{equation}
    \label{EqipNInvh}
      h \in \cA^{\bigl(\la\cE_{\iota^+,\sscri}^\cC\ra,\ell_\sscri\bigr),\ (\cE_\cK,\,3+\eps_\ind-\eps)}(\iota^+)
    \end{equation}
    for all $\ell_\sscri<1+(1-e^\cC)(1-v^\cC)\gamma^\cC$. Furthermore, in $v<1$ we have
    \begin{equation}
    \label{EqipNInvhAsy}
      (v\pa_v-1-\ubar S)h \in \cA^{\cE_{\iota^+,\sscri}^\tot+1}([0,1)_v\times\Sph^2;S^2\cT^*).
    \end{equation}
  \item\label{ItipNInvb}{\rm (b-regularity.)} Suppose $\ell_\sscri\notin\Re\pi_1\cE_{\iota^+,\sscri}^\tot$ and $\ell_\sscri<1+(1-e^\cC)(1-v^\cC)\gamma^\cC$, and let $\ell_\cK\in(-\eps_\ind,3+\eps_\ind)$ with $\ell_\cK\notin\Re\pi_1\cE_\cK$. Then the map $f\mapsto h$ extends to a continuous linear map
    \begin{equation}
    \label{EqipNInvb}
      N_{\iota^+}(\ubar L,\lambda)^{-1} \colon \Hb^{k+d,\ \ell_\sscri+1,\ \ell_\cK-2}(\iota^+;S^2\cT^*) \ni f \mapsto h \in \Hb^{k,\ \bigl(\la\cE_{\iota^+,\sscri}^\cC\ra,\ell_\sscri\bigr),\ (\cE_\cK,\ell_\cK)}(\iota^+)
    \end{equation}
    for all $k\in\N_0$ depending finite-meromorphically (i.e., meromorphically with finite-rank principal parts) on $\lambda\in\C$ for
    \[
      \Re\lambda<\ell_\sscri,
    \]
    with the principal parts at poles being independent of $k$, $\ell_\sscri$, and $\ell_\cK$ and having ranges in the space~\eqref{EqipNInvh}. In~\eqref{EqipNInvb}, $d\in\N_0$ depends only on $\Re\lambda$, and we use the unweighted b-density $|\frac{\dd v}{v}\,\dd\slg|=|\frac{\dd R}{R}\,\dd\slg|$ to define $L^2(\iota^+)$. Moreover,
    \[
      (v\pa_v-1-\ubar S)h\in\Hb^{k-1,\ (\cE_{\iota^+,\sscri}^\tot+1,\ell_\sscri)}([0,1)_v\times\Sph^2;S^2\cT^*).
    \]
  \end{enumerate}
\end{prop}

Note that~\eqref{EqipNInvhAsy} improves over the trivial membership in $\cA^{\cE_{\iota^+,\sscri}^\tot}$ by one order of decay at $\scri^+\cap\iota^+$; it reflects the fact that the asymptotics of $h$ at $\scri^+$, i.e., as $v\to 0$, are of radiation field type (with characteristic exponents given by the spectrum of $1+\ubar S$). In Proposition~\ref{PropipNAsy} below, we describe the asymptotic expansion of $h$ at $R=0$.

\begin{rmk}[Relationship to resolvents on hyperbolic space]
\label{RmkipNInvHyp}
  If $N_{\iota^+}(\ubar L,\lambda)$ were the scalar operator given by~\eqref{EqipSpecFam2Expr} with $\ubar S=0$ and $\wt{\ubar L}(0)(\omega,\rho\pa_\rho,\pa_\omega)=-(\rho\pa_\rho)^2+\rho\pa_\rho+\slDelta$ (the Laplacian on $\R^3$), then it would be the spectral family (with respect to spacetime dilations) of the Minkowski wave operator and thus a conjugation of the spectral family on hyperbolic 3-space. The inverse picked out by Proposition~\ref{PropipNInv} would then produce, for $\Re\lambda<0$ and upon undoing the conjugation, elements of $L^2$, i.e., the resolvent in the sense of $L^2$-spectral theory. The analytic continuation of this resolvent always produces the ``outgoing'' indicial root, which in the present context really is a set $\spec(1+\ubar S)$ of indicial roots, each associated with the corresponding eigenbundle of $\ubar S$. See also \cite[\S{7}]{BaskinVasyWunschRadMink}.
\end{rmk}

\begin{proof}[Proof of Proposition~\usref{PropipNInv}]
  \pfstep{Conormal solution; polyhomogeneity at $\iota^+\cap\cK^+$.} Proposition~\ref{PropipMero} gives
  \begin{equation}
  \label{EqipNInvtildehDef}
    \tilde h:=N_\sface(\lambda)^{-1}(R^{2+\lambda}f)\in\dot H_{\bop;\bop}^{(\frac32-\Re\lambda-\eps;\infty),\Re\lambda-\eps}\quad\forall\,\eps>0.
  \end{equation}
  When $\Re\lambda<2$, then the Sobolev regularity here is $>-\frac12$, so we have
  \begin{equation}
  \label{EqipNInvtildeh}
    \tilde h \in \Hb^{\infty,\frac32-\Re\lambda-\eps,\Re\lambda-\eps}([0,\infty]_v\times\Sph^2,|\tfrac{\dd v}{2+v}\,\dd\slg|)
  \end{equation}
  by~\eqref{EqipMerodotH}; here we recall the notation $\Hb^{s,\alpha_\sscri,\alpha_\cK}:=\rho_\sscri^{\alpha_\sscri}\rho_\cK^{\alpha_\cK}\Hb^s$ for weighted b-Sobolev spaces on $\iota^+=[0,\infty]\times\Sph^2$, with $\rho_\sscri=\frac{v}{1+v}$ and $\rho_\cK=\frac{R}{1+R}$ being defining functions of $\iota^+\cap\scri^+$ and $\iota^+\cap\cK^+$, respectively, on $\iota^+$. In the (overlapping) case $\Re\lambda>1$, the Sobolev regularity of $\tilde h$ is $<\frac12$, and thus the restriction of $\tilde h$ to $(0,\infty)_v\times\Sph^2$, which (by definition of extendible spaces) lies in $\bar H_{\bop;\bop}^{(\frac32-\Re\lambda-\eps;\infty);\Re\lambda-\eps}([0,\infty]\times\Sph^2)$, satisfies~\eqref{EqipNInvtildeh} as well, this time by using the isomorphism
  \begin{equation}
  \label{EqipNInvHbar}
    \bar H^s([0,\infty),|\dd v|)=\Hb^{s,s-\frac12}([0,\infty),|\tfrac{\dd v}{v}|),\quad s<\frac12,
  \end{equation}
  obtained by dualizing~\eqref{EqipMerodotH} (with respect to the density $|\dd v|$).

  In both cases (and thus for \emph{all} $\lambda$ for which $N_\sface(\lambda)$ is invertible), upon passing to the b-density $|\frac{\dd v}{v}\,\dd\slg|$, we obtain
  \[
    h := v^\lambda\tilde h \in \Hb^{\infty,1-\eps,-\eps}(\iota^+,|\tfrac{\dd v}{v}\,\dd\slg|)\subset\cA^{1-\eps,-\eps}([0,\infty]_v\times\Sph^2)\quad\forall\,\eps>0.
  \]
  Since $N_{\iota^+}(\ubar L,\lambda)$ is b-elliptic at $R=v^{-1}=0$, this can be improved to full polyhomogeneity at $\iota^+\cap\cK^+=\{R=0\}$, where an upper bound for the resulting index set $\cE_\cK$ can be determined from $-1$ times the indicial roots in Lemma~\ref{LemmaWEInd} with non-positive real parts. The precise description of this index set up to decay rate $3+\eps_\ind$ as in~\eqref{EqipNInvh} can be read off from~\eqref{EqipNAsyu0} below.

  \pfstep{Expansion at $\iota^+\cap\scri^+$ when $\Re\lambda<1$.} At $\iota^+\cap\scri^+=\{v=0\}$ and working in $v<1$ (thus omitting the index set at $v=\infty$), we have
  \begin{equation}
  \label{EqipNInvRegSing}
    (v\pa_v-\lambda)(v\pa_v-1-\ubar S)h = -\frac{1}{2 v}f + \frac{v}{2}\wt{\ubar L}(0)(\omega,v\pa_v,\pa_\omega)h \in \cA^{2-\eps}([0,1)_v\times\Sph^2).
  \end{equation}
  The a priori bound $|h|\lesssim v^{1-\eps}$ implies that if $\Re\lambda<1$, then only the indicial roots $\spec(1+\ubar S)$ of this regular-singular ODE contribute to the asymptotics of $h$; recalling the form of $\ubar S$ from~\eqref{EqWEOpMinkS} and solving component by component (which is a simple special case of the \emph{nonlinear} arguments in the proof of Theorem~\ref{ThmExPhg}), this establishes $h\in\cA^{(\la\cE_{\iota^+,\sscri}^\cC\ra,2-\eps),\ \cE_\cK}(\iota^+)$. Plugging this back into the right-hand side of~\eqref{EqipNInvRegSing}, all indicial roots already used now get shifted by $1$, and thus subsequent integrations of~\eqref{EqipNInvRegSing} produce (at most) the index set $\cE_{\iota^+,\sscri}^\cC$ for the lower-order terms of all components of $h$. We stop the expansion at the order $v^{\ell_\sscri}$, beyond which the large eigenvalues of $\ubar S$ would enter (which we do not keep track of here, much as in~\S\ref{SEx}).

  For $\Re\lambda<1$, we can moreover prove~\eqref{EqipNInvhAsy} by noting that integrating $v\pa_v-\lambda$ in~\eqref{EqipNInvRegSing} and again using the a priori bound $|h|\lesssim v^{1-\eps}$, we get $(v\pa_v-1-\ubar S)h\in\cA^{2-\eps}$. Using the already established polyhomogeneity of $h$ with index set $\cE_{\iota^+,\sscri}^\tot$, this implies that $(v\pa_v-1-\ubar S)h$ has $\scri^+$-index set $\cE_{\iota^+,\sscri}^\tot\cap\{(z,k)\colon\Re z\geq 2\}=\cE_{\iota^+,\sscri}^\tot+1$.

  \pfstep{No additional contributions from $v\pa_v-\lambda$.} For arbitrary $\lambda$, one can still integrate~\eqref{EqipNInvRegSing}, except we now need to argue that the indicial root $\lambda$ of $v\pa_v-\lambda$ does not lead to additional contributions not captured by~\eqref{EqipNInvh}. In line with Remark~\ref{RmkipNInvHyp}, we use an analytic continuation argument for this purpose. Define $h(\mu):=N_{\iota^+}(\ubar L,\mu)^{-1}f$ for all $\mu\in\C$ for which $N_\sface(\mu)$ is invertible; thus $h(\mu)$ is, for each $\mu$, polyhomogeneous at $v=0$. The coefficient of $v^{\kappa_0}$, $\kappa_0\in\C$, in its expansion can be computed as follows. Let $\chi\in\CIc([0,2))$ be equal to $1$ on $[0,1]$; then for $\Re\kappa<\Re\kappa_0$ we have
  \begin{align*}
    \cM(\chi v^{\kappa_0})(\kappa) &:= \int_0^\infty v^{-\kappa}\,\chi(v)v^{\kappa_0}\,\frac{\dd v}{v} = \frac{1}{\kappa-\kappa_0} \int_0^\infty v^{\kappa_0-\kappa} v\pa_v\chi(v)\,\frac{\dd v}{v} \\
      &= \frac{1}{\kappa-\kappa_0}\bigl(1+(\kappa-\kappa_0){\rm hol.}(\kappa)\bigr),
  \end{align*}
  where ${\rm hol.}(\kappa)$ is a holomorphic function of $\kappa$. Applying this with $\kappa_0=\mu$, we conclude that the $v^\mu$-coefficient of $h(\mu)$ is equal to the residue
  \begin{equation}
  \label{EqipNInvResidue}
    \res_{\kappa=\mu}\,\cM(\chi h(\mu))(\kappa)=\frac{1}{2\pi i}\oint_\mu \cM(\chi h(\mu))(\kappa)\,\dd\kappa,
  \end{equation}
  where we integrate along the boundary of any disc around $\mu$ that does not contain any $\kappa'\in\C$ (other than $\mu$) for which $(\kappa',0)$ lies in the index set of $h(\mu)$.

  Recall the notation $\pi_1 \colon \C\times\N_0 \to \C$ for the projection. From what we have already shown,~\eqref{EqipNInvResidue} vanishes when $\Re\mu<1$; therefore, it continues to vanish for all $\mu\notin\pi_1\cE_{\iota^+,\sscri}^\tot$, since such $\mu$ can be connected to points with real part $<1$ along a path $\tilde\mu$ that avoids $\pi_1\cE_{\iota^+,\sscri}^\tot\subset\C$, and the expression on the right in~\eqref{EqipNInvResidue} is holomorphic in $\mu$ near such a path. (See Figure~\ref{FigipNInvRes}.) This proves the absence of a $v^\lambda$-term in the expansion of $h$ when $\lambda\notin\pi_1\cE_{\iota^+,\sscri}^\tot$; more finely, the same arguments imply that $\pi^\cC h$ has no $v^\lambda$-term when $\lambda\notin\pi_1\cE_\sscri^\cC$, similarly for the other components in~\eqref{EqipIndexComp}.

  \begin{figure}[!ht]
  \centering
  \includegraphics{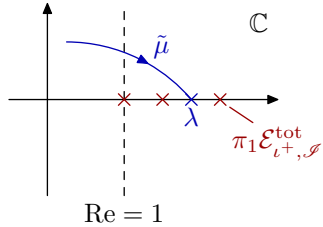}
  \caption{Illustration of the argument after~\eqref{EqipNInvResidue} concerning the vanishing of the $v^\lambda$-coefficient of $h$. The crosses mark points in the projection to the first factor of the index set $\cE_{\iota^+,\sscri}^\tot$. The curve labeled $\tilde\mu$ starts in $\Re<1$ and ends at $\lambda$ while not passing through any of these points.}
  \label{FigipNInvRes}
  \end{figure}

  It remains to consider the case $\lambda\in\pi_1\cE_{\iota^+,\sscri}^\tot$; we must argue for the absence of a nontrivial $v^\lambda\log v$-term in the expansion of $h$. For concreteness, we consider $\lambda=1$ (and thus really only the component $\slpi_0 h$). The $v^\lambda\log\lambda$-coefficient of $h(\mu)$ is given by $\res_{\kappa=\lambda}\bigl( (\kappa-\lambda)\cM(\chi h(\mu))(\kappa) \bigr)$; for $\mu$ close to $\lambda=1$, let us thus study the function
  \begin{equation}
  \label{EqipNInvResidueLog}
    \mu \mapsto \frac{1}{2\pi i}\int_\gamma (\kappa-\mu) \cM(\chi h(\mu))(\kappa)\,\dd\kappa,
  \end{equation}
  where $\gamma$ is the boundary of a disc containing $1$ and $\mu$ but no other points in $\pi_1\cE_{\iota^+,\sscri}^\tot$. For $\Re\mu<1$, this is equal to $(1-\mu)\res_{\kappa=1} \cM(\chi h(\mu))(\kappa)\,\dd\kappa$: this is $1-\mu$ times a $\CI(\Sph^2;S^2\cT^*)$-valued holomorphic function of $\mu$, and computes $1-\mu$ times the $v\log v$-coefficient of $h(\mu)$. Analytically continuing to $\mu=1$ shows that the $v^1\log v$-coefficient of $h(1)$ (given by~\eqref{EqipNInvResidueLog} for $\mu=1$) vanishes, as claimed.

  \pfstep{Proof of~\eqref{EqipNInvhAsy} for general $\lambda$.} The claim~\eqref{EqipNInvhAsy} is equivalent to the vanishing of all $v^z(\log v)^k$-terms in the expansion of $(v\pa_v-1-\ubar S)h(\mu)$ for $\Re z<2$ and $\mu=\lambda$, where again $h(\mu)=N_{\iota^+}(\ubar L,\mu)^{-1}f$. It thus follows by unique continuation from its validity for $\Re\mu<1$.

  \pfstep{Finite decay and finite b-regularity.} The proof of~\eqref{EqipNInvb} follows from the same arguments, with the quantity $d$ accounting for b-regularity losses incurred in the passage from standard to weighted b-Sobolev spaces. Only the first step~\eqref{EqipNInvtildehDef} of the proof requires us to observe that the interpretation of $R^{2+\lambda}f\in\Hb^{k+d,\ \ell_\sscri-1-\Re\lambda,\ \ell_\cK+\Re\lambda}(\iota^+)$ as a supported distribution on $\sface_{\geq -1}$ via extension by $0$ is immediate when $\ell_\sscri-1-\Re\lambda>-1$ (since then $R^{2+\lambda}f$ is in $L^1_\loc$ near $v=0$); this is the reason for the restriction on $\Re\lambda$. The rest of the argument is as before.

  The principal part at a pole of $N_{\iota^+}(\ubar L,\lambda)^{-1}$ is equal to that of $R^{-\lambda}N_\sface(\lambda)^{-1}R^{2+\lambda}$. The principal part of the inverse $N_\sface(\lambda)^{-1}$ of~\eqref{EqipMeroMap} is independent of $s,\mu$ (satisfying the assumptions there) and maps into infinitely b-regular functions by Proposition~\ref{PropipMero}\eqref{ItipMeroHi}; and therefore so does the principal part of $N_{\iota^+}(\ubar L,\lambda)^{-1}$.

  The final assertion of part~\eqref{ItipNInvb} concerns the vanishing of certain coefficients in the expansion of $(v\pa_v-1-\ubar S)h$ at $v=0$, which follows by density from the corresponding statement~\eqref{EqipNInvhAsy} for $f\in\dot\cC^\infty$ proved above.
\end{proof}

The following definition is suggested by Propositions~\ref{PropipMero} and \ref{PropipNInv}:

\begin{definition}[Boundary spectrum]
\label{DefipMeroSpec}
  We denote by
  \[
    \spec_{\iota^+}(\ubar L) \subset \{z \in \C \colon \Re z\geq 1+\eps_\ind \}
  \]
  the set of poles of $N_{\iota^+}(\ubar L,\lambda)^{-1}$ acting on $\CIdot(\iota^+;S^2\cT^*)$, and by
  \[
    \Spec_{\iota^+}(\ubar L) \subset \spec_{\iota^+}(\ubar L)\times\N_0 \subset \C\times\N_0
  \]
  the set of pairs $(\lambda,k)$ where $\lambda$ is a pole of $N_{\iota^+}(\ubar L,\lambda)^{-1}$ of order equal to $k+1$.
\end{definition}

Elements of $\spec_{\iota^+}(\ubar L)$ were called ``pure'' resonances in~\S\ref{SssIGen1}. By Proposition~\ref{PropipNInv}\eqref{ItipNInvb}, the intersection of $\Spec_{\iota^+}(\ubar L)$ with the half-space $\{\Re\lambda<\ell_\sscri\}$ is equal to the divisor of $N_{\iota^+}(\ubar L,\lambda)^{-1}$ acting on $\Hb^{\infty,\ \ell_\sscri+1,\ \ell_\cK-2}$ as in~\eqref{EqipNInvb}.

\begin{rmk}[$N_\sface(\lambda)^{-1}$ vs.\ $N_{\iota^+}(\ubar L,\lambda)^{-1}$]
\label{RmkipMeroSpec}
  Poles of $N_\sface(\lambda)^{-1}$ which disappear upon restriction to $\{v>0\}$ can only arise when $\ker N_\sface(\lambda)$ contains distributions supported at $v=0$, i.e., sums of (differentiated) $\delta$-distributions. These therefore do not correspond to terms in the $\iota^+$-expansion of solutions of $L_b u=f$, but rather relate to the asymptotics at the radiation field; see, e.g., \cite[Remarks~7.5 and 8.2]{BaskinVasyWunschRadMink}. Working with $N_{\iota^+}(\ubar L,\lambda)^{-1}$ in Definition~\ref{DefipMeroSpec} ensures that we only record poles that do correspond to terms in the $\iota^+$-expansion of waves.
\end{rmk}

\subsubsection{Precise asymptotics at \texorpdfstring{$\iota^+\cap\cK^+$}{the Kerr face}}
\label{SssipAsy}

The asymptotics of the solutions $h$ produced by Proposition~\ref{PropipNInv} at $\iota^+\cap\cK^+=\{R=0\}\subset\iota^+$ (and thus the index set $\cE_\cK$) can be described precisely.\footnote{The number of terms we need in the $R\to 0$ expansion depends on the context. We will stop at the order $R^3$ since we can treat terms with more decay as error terms in our nonlinear stability arguments later on (cf.\ the discussion in~\S\ref{SssIEinLa}). Only starting at intermediate stages of the precise asymptotic description of forward solutions on dynamical spacetimes do we need terms beyond the $R^0$-term.} As discussed in~\S\ref{SssipGr}, this will give detailed information on the near-field asymptotics (i.e., asymptotics at $\cK^+$) of waves that sources in the far field (i.e., roughly speaking, sources supported near $(\iota^+)^\circ$) with $t_*^{-\lambda-2}$-decay produce.

Before stating the precise result, we need to recall/introduce some notation. We write
\begin{subequations}
\begin{equation}
\label{EqipAsys0}
  \ubar h_{\rms 0}^{(0)}\in\CI(\pa X;S^2\cT^*_X|_{\pa X}),\quad \breve{\ubar h}_{\rms 0}^{(0),k} \in \rho^{-k}\CI(\pa X;S^2\cT^*_X|_{\pa X})
\end{equation}
for the (homogeneous in $\rho$ of the stated degree) leading-order terms of $h_{b,\rms 0}^{(0)}$ and $\breve h_{b,\rms 0}^{(0),k}$ in~\eqref{EqWG0Larges00} and \eqref{EqWG0Larges003}, respectively. The equation $L_b h_{b,\rms 0}^{(0)}=0$ implies $\ubar L \ubar h_{\rms 0}^{(0)}=0$, and similarly $L_b h_{b,\rms 0}^{(0),\leq 1}=L_b(t_* h_{b,\rms 0}^{(0)}+\breve h_{b,\rms 0}^{(0),1})=0$ implies
\begin{equation}
\label{EqipAsys0Eq}
  \ubar L \ubar h_{\rms 0}^{(0),\leq k} = \ubar L\biggl( t_*^k\ubar h_{\rms 0}^{(0)} + \sum_{i=1}^k t_*^{k-i}\frac{k!}{(k-i)!}\breve{\ubar h}_{\rms 0}^{(0),i}\biggr) = 0
\end{equation}
\end{subequations}
for $k=1$, similarly for $k=2,3$. We similarly recall/introduce
\begin{subequations}
\begin{equation}
\label{EqipAsys1}
  \ubar h_{\rms 1}^2(\scal),\ \breve{\ubar h}_{\rms 1}^k(\scal)\ (k=3,4),\quad
  \ubar h_{\rmv 1}^1(\vect),\ \breve{\ubar h}_{\rmv 1}^k(\vect)\ (k=2,3)
\end{equation}
from Proposition~\ref{PropWG0Large}\eqref{ItWG0Larges1m1}--\eqref{ItWG0Largev1m1}, with $\breve{\ubar h}_{\rms 1}^k$, resp.\ $\breve{\ubar h}_{\rmv 1}^k$ being the (homogeneous of degree $-k-1$, resp.\ $-k$ in $\rho$) leading-order term of $\breve h_{b,\rms 1}^k$, resp.\ $\breve h_{b,\rmv 1}^k$; thus
\begin{equation}
\label{EqipAsys1Eq}
\begin{alignedat}{3}
  \ubar L\ubar h_{\rms 1}^2&=0,&\quad
  \ubar L(t_*\ubar h_{\rms 1}^2+\breve{\ubar h}_{\rms 1}^3)&=0,&\quad
  \ubar L\bigl(t_*^2\ubar h_{\rms 1}^2+2 t_*\breve{\ubar h}_{\rms 1}^3+2\breve{\ubar h}_{\rms 1}^4\bigr)&=0, \\
  \ubar L\ubar h_{\rmv 1}^1&=0,&\quad
  \ubar L(t_*\ubar h_{\rmv 1}^1+\breve{\ubar h}_{\rmv 1}^2)&=0,&\quad
  \ubar L\bigl(t_*^2\ubar h_{\rmv 1}^1+2 t_*\breve{\ubar h}_{\rmv 1}^2+2\breve{\ubar h}_{\rmv 1}^3\bigr)&=0,
\end{alignedat}
\end{equation}
\end{subequations}
as follows by computing the $\rho^1$, $\rho^0$, and $\rho^{-1}$ term at $\rho=0$ of $L_b h_{b,\rms 1}^{\leq k}=0$ for $k=2$, $3$, and $4$, respectively, and from $L_b h_{b,\rmv 1}^{\leq k}=0$ for $k=1,2,3$. The discussion of the remaining (generalized) large energy states from Proposition~\ref{PropWG0Large}\eqref{ItWG0Larges01}--\eqref{ItWG0Largevl2} can be streamlined; we denote the leading-order terms of the tensors constructed there by
\begin{subequations}
\begin{equation}
\label{EqipAsyRem}
\begin{alignedat}{2}
  &\ubar h_{\rms 0}^{(\lambda)},\ \breve{\ubar h}_{\rms 0}^{(\lambda),k},&\quad &\lambda=-\lambda_{\rms 0,1}^\Ups+1,\ k=1,2; \\
  &\ubar h_{\rms l}^{(\lambda)},\ \breve{\ubar h}_{\rms l}^{(\lambda),k},&\quad &\lambda=-\lambda_{\rms l,l+j}^\Ups+1,\ l=1,2,3,\ j=0,1,\ k=1,2; \\
  &\ubar h_{\rms l}^{(-l+2)},\ \breve{\ubar h}_{\rms l}^{(-l+2),k},&\quad &l=2,3,4,5,\ k=1,\ldots,5-l; \\
  &\ubar h_{\rmv l}^{(-l+1)},\ \breve{\ubar h}_{\rmv l}^{(-l+1),k},&\quad &l=2,3,4,\ k=1,\ldots,4-l.
\end{alignedat}
\end{equation}
We similarly denote the leading-order terms of the states constructed in Proposition~\ref{PropWE0} by
\begin{equation}
\label{EqipAsyPhys}
\begin{alignedat}{2}
  \ubar h_\bullet^{(-2)},\ \breve{\ubar h}_\bullet^{(-2),1},&\quad \bullet&=\rms 2,\rmv 2; \\
  \ubar h_\bullet^{(-3)}, &\quad \bullet&=\rms 3,\rmv 3.
\end{alignedat}
\end{equation}
\end{subequations}

Suppose now that the late-time expansion of a solution of $\ubar L h=0$ on Minkowski space (without $\{r=0\}$) involves a term of the form $a(t_*)\ubar h_{\rms 0}^{(0)}$, say, where $a(t_*)\in\cA^\lambda([1,\infty]_{t_*})$ is conormal at infinity with $t_*^{-\lambda}$-decay. Then
\begin{align*}
  \ubar L\bigl(a(t_*)\ubar h_{\rms 0}^{(0)}\bigr)=[\ubar L,a(t_*)]\ubar h_{\rms 0}^{(0)}&=a'(t_*)[\ubar L,t_*]\ubar h_{\rms 0}^{(0)} \\
    &=-a'(t_*)\ubar L\breve{\ubar h}_{\rms 0}^{(0),1}=-\ubar L\bigl(a'(t_*)\breve{\ubar h}_{\rms 0}^{(0),1}\bigr) + [\ubar L,a'(t_*)]\breve{\ubar h}_{\rms 0}^{(0),1}.
\end{align*}
The first line shows that $\ubar L(a(t_*)\ubar h_{\rms 0}^{(0)})=\cO(t_*^{-\lambda-1}r^{-1})=\cO(\rho_+^{\lambda+2}\rho_\cK^{\lambda+1})$. The re-writing in the second line allows us to conclude that
\[
  \ubar L\bigl( a(t_*)\ubar h_{\rms 0}^{(0)} + a'(t_*)\breve{\ubar h}_{\rms 0}^{(0),1} \bigr) = a''(t_*)[\ubar L,t_*]\breve{\ubar h}_{\rms 0}^{(0),1} = \cO(t_*^{-\lambda-2}r^0) = \cO(\rho_+^{\lambda+2}\rho_\cK^{\lambda+2});
\]
and, by continuing this re-writing, more generally
\begin{equation}
\label{EqipAsyCompGen}
\begin{split}
  &\ubar h_{\rms 0}^{(0),\leq k}(a(t_*)) := a(t_*)\ubar h_{\rms 0}^{(0)} + \sum_{j=1}^k a^{(j)}(t_*)\breve{\ubar h}_{\rms 0}^{(0),j} \\
  &\quad \implies \ubar L\bigl(\ubar h_{\rms 0}^{(0),\leq k}(a(t_*))\bigr) = a^{(k+1)}(t_*)[\ubar L,t_*]\breve{\ubar h}_{\rms 0}^{(0),k} = \cO(\rho_+^{\lambda+2}\rho_\cK^{\lambda+k+1}).
\end{split}
\end{equation}
(This expands on the discussion around~\eqref{EqIGen2LaGen4}.) We can thus produce increasingly accurate formal solutions at $\cK^+$ using generalized zero energy states; and we will use the same computation, \emph{mutatis mutandis}, on the Kerr spacetime, e.g., in~\S\ref{SssipGr}. (An improvement of the decay of the error term at $\iota^+$ requires the inversion of the $\iota^+$-normal operator, which is based on the dilation-homogeneity rather than $t_*$-translation-invariance of $\ubar L$.) In the case $a(t_*)=t_*^{-\lambda}$, one can restrict $t_*^{\lambda+2}\ubar L\ubar h_{\rms 0}^{(0),\leq k}(t_*^{-\lambda})$ to $\iota^+$ as an element
\begin{equation}
\label{EqipAsyCompGen2}
  \bigl(t_*^\lambda t_*^2\ubar L\ubar h_{\rms 0}^{(0),\leq k}(t_*^{-\lambda})\bigr)\big|_{\iota^+} \in \rho_\cK^{k-1}\CI([0,1)_{\rho_\cK}\times\Sph^2),
\end{equation}
where one can locally work with $\rho_\cK=R=\frac{r}{t_*}$.\footnote{It is, in fact, not necessary to spell out the restriction to $\iota^+$ if one works in the coordinates $t_*$, $R=\frac{r}{t_*}=\frac{1}{\rho t_*}$, and $\omega\in\Sph^2$ since $t_*^\lambda t_*^2\ubar L\ubar h_{\rms 0}^{(0),\leq k}(t_*^{-\lambda})$ is independent of $t_*$.} This means that $(t_*^\lambda\ubar h_{\rms 0}^{(0),\leq k}(t_*^{-\lambda}))|_{\iota^+}$ is a good formal solution of $N_{\iota^+}(\ubar L,\lambda)=t_*^2 t_*^\lambda\ubar L t_*^{-\lambda}$ on $\iota^+$ at $\iota^+\cap\cK^+=\iota^+\cap R^{-1}(0)$ with leading-order term $\ubar h_{\rms 0}^{(0)}$ at $R=0$. In general, let us introduce:

\begin{definition}[Formal solutions at $\cK^+$: Minkowskian version]
\label{DefipAsyForm}
  For a function $a=a(t_*)$, define $\ubar h_{\rms 0}^{(0),\leq k}(a(t_*))$ by~\eqref{EqipAsyCompGen}; and define $\ubar h_{\rms l}^{(-l+2),\leq k}(a(t_*))$ etc.\ analogously, now with $a=a(t_*)$ valued in $\scalspace_l$ etc. In the exceptional scalar, resp.\ vector type $1$ cases, we write, for $\scalspace_1$-, resp.\ $\vectspace_1$-valued $a=a(t_*)$,
  \begin{alignat*}{2}
    \ubar h_{\rms 1}^{\leq k}(a(t_*)) &:= \ubar h_{\rms 1}^2(a(t_*)) + \breve{\ubar h}_{\rms 1}^3(a'(t_*)) + \cdots + \breve{\ubar h}_{\rms 1}^k(a^{(k-2)}(t_*)),&\quad k&=2,3,4, \\
    \ubar h_{\rmv 1}^{\leq k}(a(t_*)) &:= \ubar h_{\rmv 1}^1(a(t_*)) + \breve{\ubar h}_{\rmv 1}^2(a'(t_*)) + \cdots + \breve{\ubar h}_{\rmv 1}^k(a^{(k-1)}(t_*)),&\quad k&=1,2,3.
  \end{alignat*}
\end{definition}

The generalization of~\eqref{EqipAsyCompGen2} for large zero energy states $\ubar h^{(\mu)}\in\rho^\mu\CI(\pa X;S^2\cT^*X|_{\pa X})$ which are homogeneous of degree $\mu$ (e.g., $\ubar h_{\rms l}^{(-l+2)}$, $l\geq 2$, with $\mu=-l+2$) is then that
\begin{equation}
\label{EqipAsyCompGen3}
  \bigl(t_*^\lambda t_*^2\ubar L \ubar h^{(\mu),\leq k}(t_*^{-\lambda+\mu})\bigr)\big|_{\iota^+} \in \rho_\cK^{-\mu+k-1}\CI([0,1)_R\times\Sph^2).
\end{equation}
Note moreover that
\begin{equation}
\label{EqipAsyCompGen4}
  t_*^\lambda\ubar h^{(\mu),\leq k}(t_*^{-\lambda+\mu}) = (\rho t_*)^\mu \underbrace{(\rho^{-\mu}\ubar h^{(\mu)})}_{\in\CI(\pa X)} + \sum_{j=1}^k \underbrace{\rho^{\mu-j}t_*^\lambda \Bigl(\frac{\dd^j}{\dd t_*^j}t_*^{-\lambda+\mu}\Bigr)}_{={\rm const.}\times(\rho t_*)^{\mu-j}} \underbrace{(\rho^{-\mu+j}\breve{\ubar h}^{(\mu),j})}_{\in\CI(\pa X)},
\end{equation}
as a function of $(t_*,R,\omega)$, is, in fact, $t_*$-independent, and can thus be regarded as a function on $\iota^+$. The observation~\eqref{EqipAsyCompGen3} then suggests the form of the following result:

\begin{prop}[Asymptotic expansion at $R=0$]
\label{PropipNAsy}
  Suppose $N_{\iota^+}(\ubar L,\lambda)u=f$ on $[0,1)_R\times\Sph^2\subset\iota^+$; here $f\in\Hb^{k,\ell_\cK-2}([0,1)_R\times\Sph^2;|\frac{\dd R}{R}\,\dd\slg|)$ where $\ell_\cK<3+\eps_\ind$ is not equal to $\Re(-\mu)$ for any indicial root $\mu$ of $\wh{\ubar L}(0)$, and $u\in\Hb^{k+2,\alpha}([0,1)\times\Sph^2)$ for some $\alpha\in(-\eps_\ind,0)$. Then $u=u_0+\tilde u$ where $\tilde u\in\Hb^{k+2,\ell_\cK}$ and, recalling that the tensors~\eqref{EqipAsyCompGen4} are functions of $(R,\omega)$ only,
  \begin{equation}
  \label{EqipNAsyu0}
  \begin{split}
    u_0(R,\omega) &= t_*^\lambda\ubar h_{\rms 1}^{\leq 4}(t_*^{-\lambda-1}\scal_1^{(-1)}) + t_*^\lambda\ubar h_{\rmv 1}^{\leq 3}(t_*^{-\lambda-1}\vect_1^{(-1)}) + \scal_0^{(0)} t_*^\lambda\ubar h_{\rms 0}^{(0),\leq 3}(t_*^{-\lambda}) \\
      &\quad + \sum_{\substack{\mu=-\lambda_{\rms l,l+j}^\Ups+1 \\ 0\leq l\leq 3,\ j=0,1, \\ 1\leq l+j\leq 3}} t_*^\lambda\ubar h_{\rms l}^{(\mu),\leq 2}(t_*^{-\lambda+\mu}\scal_l^{(\mu)}) \\
      &\quad + \sum_{l=2}^5 t_*^\lambda\ubar h_{\rms l}^{(-l+2),\leq 5-l}(t_*^{-\lambda-l+2}\scal_l^{(-l+2)}) + \sum_{l=2}^4 t_*^\lambda\ubar h_{\rmv l}^{(-l+1),\leq 4-l}(t_*^{-\lambda-l+1}\vect_l^{(-l+1)}) \\
      &\quad + t_*^\lambda\ubar h_{\rms 2}^{(-2),\leq 1}(t_*^{-\lambda-2}\scal_2^{(-2)}) + t_*^\lambda\ubar h_{\rmv 2}^{(-2),\leq 1}(t_*^{-\lambda-2}\vect_2^{(-2)}) \\
      &\quad + t_*^\lambda\ubar h_{\rms 3}^{(-3)}(t_*^{-\lambda-3}\scal_3^{(-3)}) + t_*^\lambda\ubar h_{\rmv 3}^{(-3)}(t_*^{-\lambda-3}\vect_3^{(-3)})
  \end{split}
  \end{equation}
  for suitable $\scal_l^{(\mu)}\in\scalspace_l$ and $\vect_l^{(\mu)}\in\vectspace_l$.
\end{prop}

Note that all terms have the same structure; thus, this is in fact a very systematic expansion. Viewed from the perspective of the Kerr spacetime, it has a rather simple interpretation; see~\S\ref{SssipGr}.

\begin{proof}[Proof of Proposition~\usref{PropipNAsy}]
  We recall from~\eqref{EqipSpecFam2Expr} that
  \[
    R^2 N_{\iota^+}(\ubar L,\lambda)=A_0+R A_1,\quad
    A_0:=\wt{\ubar L}(0)(\omega,-R\pa_R,\pa_\omega),\ A_1:=-2(R\pa_R+\lambda)(R\pa_R+1+\ubar S).
  \]
  The equation for $u$ thus reads
  \begin{equation}
  \label{EqipNAsy}
    A_0 u=R^2 f-R A_1 u\in\Hb^{k,\min(\ell_\cK,\alpha+1)}.
  \end{equation}
  The standard Mellin transform and normal operator argument as in Lemma~\ref{LemmaTMSolPhg} thus implies that $u\in\Hb^{k,\ell_\cK}$ when $\ell_\cK<0$, while in the case $\ell_\cK>0$ we get a partial expansion into indicial solutions corresponding to the indicial roots in Lemma~\ref{LemmaWEInd} with real parts in $(\min(\ell_\cK,\alpha+1),0]$. The first two terms, arising from the $\rms 0$ and $\rms 2$ indicial root $0$, are thus
  \begin{equation}
  \label{EqipNAsyPf1}
    u = R^0\bigl(\scal_0^{(0)}\ubar h_{\rms 0}^{(0)} + \ubar h_{\rms 2}^{(0)}(\scal_2^{(0)})\bigr) + \tilde u,\quad\tilde u\in\Hb^{k,\min(\ell_\cK,\eps_\ind)}.
  \end{equation}
  Instead of immediately plugging this back into~\eqref{EqipNAsy}, we first promote the two leading-order terms to formal solutions of $N_{\iota^+}(\ubar L,\lambda)$ to a rather high order. To this end, we rewrite the equation satisfied by $u$ as
  \begin{equation}
  \label{EqipNAsyPfEq}
    t_*^\lambda t_*^2\ubar L(t_*^{-\lambda}u)=f
  \end{equation}
  (with $t_*^2\ubar L$ being dilation-invariant and thus its own $\iota^+$-normal operator). By~\eqref{EqipAsyCompGen2}, we have
  \[
    t_*^\lambda t_*^2\ubar L\bigl(t_*^{-\lambda} \cdot t_*^\lambda\ubar h_{\rms 0}^{(0),\leq 3}(t_*^{-\lambda})\bigr) \in R^2\CI([0,1)_R\times\Sph^2).
  \]
  Therefore, if we subtract $\scal_0^{(0)}\cdot(t_*^\lambda\ubar h_{\rms 0}^{(0),\leq 3}(t_*^{-\lambda}))|_{\iota^+}$ from $u$, the new $u$ thus obtained has $\scal_0^{(0)}=0$ and satisfies the same equation as the original $u$, except for $f$ modified by an element of $R^2\CI([0,1)_R\times\Sph^2)$; the new $f$ thus still lies in $\Hb^{k,\ell_\cK-2}$ since $\ell_\cK-2<2$. We similarly eliminate the term $\ubar h_{\rms 2}^{(0)}(\scal_2^{(0)})$ by subtracting $(t_*^\lambda\ubar h_{\rms 2}^{(0),\leq 3}(t_*^{-\lambda}))|_{\iota^+}$ from $u$ and updating $f$ accordingly by an element of $R^2\CI$.

  This produces the first two terms in the expansion of $u_0$ in the statement of the Proposition. It then remains to study the equation $N_{\iota^+}(\ubar L,\lambda)u=f$ under the assumption $u\in\Hb^{k,\alpha}$ where now $0<\alpha\leq\ell_\cK$ in the notation of~\eqref{EqipNAsyPf1}. We plug this into the right-hand side of~\eqref{EqipNAsy} and read off the next terms in the expansion of $u$ at $R=0$. Let us only describe two types of indicial roots that arise as one iterates this argument.

  First, consider, say, the root $\mu=-\lambda_{\rms l,l+j}^\Ups+1$ where $0\leq l\leq 3$, $j=0,1$, and $1\leq l+j\leq 3$. This yields a contribution to $u$ of the form
  \begin{equation}
  \label{EqipNAsyPfLot}
    R^{-\mu}\cdot \underbrace{\rho^{-\mu}\ubar h_{\rms l}^{(\mu)}(\scal)}_{\in\CI(\pa X)} = t_*^\mu \ubar h_{\rms l}^{(\mu)}(\scal),\quad\scal\in\scalspace_l,
  \end{equation}
  and thus to the argument $t_*^{-\lambda}u$ of~\eqref{EqipNAsyPfEq} the contribution $t_*^{-\lambda+\mu}\ubar h_{\rms l}^{(\mu)}(\scal)$. The formal solution of the $\iota^+$-normal operator with $R^{-\mu}\cdot\rho^{-\mu}\ubar h_{\rms l}^{(\mu)}(\scal)$ as its leading-order term at $R=0$ and with overall weight $t_*^{-\lambda}$ is thus $\ubar h_{\rms l}^{(\mu),\leq 2}(t_*^{-\lambda+\mu}\scal)$ by~\eqref{EqipAsyCompGen3}. Multiplying this by $t_*^\lambda$, one obtains a tensor that is $t_*$-independent in the coordinates $(t_*,R,\omega)$; subtracting it from $u$ amounts to eliminating the contribution~\eqref{EqipNAsyPfLot}, while modifying $f$ by an element of $R^{-\mu+1}\CI\subset\Hb^{\infty,\ell_\cK-2}$ (using~\eqref{EqipAsyCompGen3} and recalling that $3-\mu\geq 3+\eps_\ind>\ell_\cK$). --- The roots $-l+2$ corresponding to the states~\eqref{EqipAsyRem} are handled in the same fashion, as are the scalar and vector type roots $-l$.

  Finally, let us consider the scalar type $1$ indicial root $-1$ of $\wh{\ubar L}(0)$. This yields a contribution to $u$ of the form
  \[
    R^1\cdot \rho^1\ubar h_{\rms 1}^2(\scal) = t_*^{-1}\ubar h_{\rms 1}^2(\scal),\quad \scal\in\scalspace_1,
  \]
  and thus to $t_*^{-\lambda}u$ a contribution $t_*^{-\lambda-1}\ubar h_{\rms 1}^2(\scal)$. We are thus led to subtract from $t_*^{-\lambda}u$ the term $\ubar h_{\rms 1}^{\leq 4}(t_*^{-\lambda-1}\scal)$, or equivalently from $u$ the term $t_*^\lambda\ubar h_{\rms 1}^{\leq 4}(t_*^{-\lambda-1}\scal)$, which leads to a modification of $f$ by a term of class $\rho^2\CI$ again. --- The vector type $1$ root $-1$ is handled similarly.

  Once the contributions of all indicial roots of $\wh{\ubar L}(0)$ with real parts in $[-3,0]$ to the $R\to 0$ asymptotics have thus been accounted for, the remainder term of $u$ is of class $\Hb^{k+2,\ell_\cK}$; recall here that the definition of $\eps_\ind$ implies that no indicial roots have real parts in $(3,3+\eps_\ind)$.
\end{proof}

\subsubsection{Grafting solutions at \texorpdfstring{$\iota^+$}{future timelike infinity} into spacetime}
\label{SssipGr}

Proposition~\ref{PropipNAsy} means, roughly speaking, that source terms for $\ubar L$ of the form $f=t_*^{-\lambda-2}f_0(R,\omega)$ (where we recall $R=\frac{r}{t_*}$) produce waves which at $\iota^+\cap\cK^+=\iota^+\cap R^{-1}(0)$ include terms asymptotic to $t_*^{-\lambda+\mu}\ubar h^{(\mu)}$, where $\mu$ is an indicial root (see Lemma~\ref{LemmaWEInd}) of the zero energy operator $\wh{L_b}(0)$ (or equivalently of $\wh{\ubar L}(0)$) with $\Re\mu\leq 0$ and $\ubar h^{(\mu)}$ is the corresponding indicial solution. When $\ubar h^{(\mu)}$ is the leading-order term of a large zero energy state $h_b^{(\mu)}$ on the Kerr spacetime, this means that the source $f$ for the Kerr operator $L_b$ produces a contribution $t_*^{-\lambda+\mu}h_b^{(\mu)}$ in the late-time asymptotics at the Kerr face $\cK^+$ (cf.\ \S\S\ref{SssIGen2La} and \ref{SssIEinLa}).

Thus, while the asymptotic expansion at $R=0$ in Proposition~\ref{PropipNAsy} looks somewhat intimidating, it is fairly harmless (and not only because all terms have the same structure): each term in this expansion can be extended to an approximate solution (at $\cK^+$) of the linearized gauge-fixed Einstein equation on Kerr, which is moreover (up to a fast-decaying remainder) a sum of pure gauge states from Propositions~\ref{PropWG0Symm} and \ref{PropWG0Large} as well as of the physical states from Proposition~\ref{PropWE0}. We state this precisely using the following variant of Definition~\ref{DefipAsyForm}:

\begin{definition}[Formal solutions at $\cK^+$: Kerr version]
\label{DefipGrKerr}
  For a function $a=a(t_*)$, define
  \[
    h_{b,\rms 0}^{(0),\leq k}(a(t_*)) := a(t_*)h_{b,\rms 0}^{(0)} + \sum_{j=1}^k a^{(j)}(t_*)\breve h_{b,\rms 0}^{(0),j};
  \]
  define $h_{b,\rms l}^{(-l+2),\leq k}(a(t_*))$ etc.\ analogously, now with $a=a(t_*)$ valued in $\scalspace_l$ etc. We analogously define the 1-forms $\omega_{b,\rms l}^{(-l+1),\leq k}(a(t_*))$ etc. In the exceptional scalar, resp.\ vector type $1$ cases, we write, for $\scalspace_1$-, resp.\ $\vectspace_1$-valued $a=a(t_*)$,
  \begin{alignat*}{2}
    h_{b,\rms 1}^{\leq k}(a(t_*)) &:= h_{b,\rms 1}(a(t_*)) + \sum_{j=1}^k \breve h_{b,\rms 1}^j(a^{(j)}(t_*)), &\quad k&=2,3,4, \\
    h_{b,\rmv 1}^{\leq k}(a(t_*)) &:= h_{b,\rmv 1}(a(t_*)) + \sum_{j=1}^k \breve h_{b,\rmv 1}^j(a^{(j)}(t_*)), &\quad k&=1,2,3.
  \end{alignat*}
\end{definition}
Analogously to~\eqref{EqipAsyCompGen}, we then have
\begin{equation}
\label{EqipGrKerrComp}
\begin{split}
  L_b\bigl(h_{b,\rms 0}^{(0),\leq k}(a(t_*))\bigr) &= a^{(k+1)}(t_*)\Bigl([L_b,t_*]\ftrans(0)\breve h_{b,\rms 0}^{(0),k} + \frac12[[L_b,t_*],t_*]\breve h_{b,\rms 0}^{(0),k-1}\Bigr) \\
    &\qquad + a^{(k+2)}(t_*)\frac12[[L_b,t_*],t_*]\breve h_{b,\rms 0}^{(0),k},
\end{split}
\end{equation}
with the conventions $\breve h_{b,\rms 0}^{(0),-1}=0$ and $\breve h_{b,\rms 0}^{(0),0}=h_{b,\rms 0}^{(0)}$. For example, when $a(t_*)=t_*^{-\lambda}$, this is of class $t_*^{-\lambda-k-1}(\rho\cdot\rho^{-k}+\rho^2\cdot\rho^{-(k-1)})+t_*^{-\lambda-k-2}\rho^2\cdot\rho^{-k}=\cO(\rho_+^{\lambda+2}\rho_\cK^{\lambda+k+1})$ and thus vanishes to increasingly high orders at $\cK^+$ when $k$ increases, while the order of decay at $\iota^+$ is the usual $2$ more than the $\iota^+$-order $\lambda$ of the input $h_{b,\rms 0}^{(0),\leq k}(t_*^{-\lambda})$.

Recalling the compactified spacetime manifold $M$ from~\eqref{EqKMfdRadM}, we set
\begin{equation}
\label{EqipGrOmegaStar}
  \Omega_* := \cl_M \{ \ft_*\geq 1 \} \subset M,
\end{equation}
where $\ft_*$ is a time function such as $t_*-C$ for any (arbitrarily large) constant $C$ or the one discussed in~\eqref{EqDftstar} below. The only relevant properties of $\Omega_*$ for the time being are that it contains neighborhoods of $\iota^+$ and $\cK^+$ and is disjoint from $I^0$; thus, it intersects only the ideal boundary hypersurfaces $\scri^+$, $\iota^+$, and $\cK^+$ of $M$, and spaces of functions defined on $\Omega_*$ correspondingly have three orders (or index sets), which we always list in the order $\scri^+,\iota^+,\cK^+$. When we use the notation $\la\cE_{\iota^+,\sscri}^\cC\ra$ for the $\scri^+$-index set of a section $h$ of $S^2\cT^*$, this means that the components of $h$ have $\scri^+$-index sets analogously to~\eqref{EqipIndexComp}. We fix a cutoff function $\chi_\cK\in\CI(M)$ that equals $1$ near $\cK^+$ and $0$ near $\ft_*^{-1}(1)\cup\scri^+$.

\begin{prop}[Grafting]
\label{PropipGr}
  Let $\lambda\notin\spec_{\iota^+}(\ubar L)$ (see Definition~\usref{DefipMeroSpec}) and with $\Re\lambda>0$, and let $\ell_\cK<3+\eps_\ind$, $\ell_\cK\neq\Re(-\lambda)$ for all indicial roots $\lambda$ of $\wh{L_b}(0)$ with $\Re\lambda\in[-3,0]$. Let $\ell_\sscri<1+(1-e^\cC)(1-v^\cC)\gamma^\cC$ with $\ell_\sscri\notin\Re\pi_1\cE_{\iota^+,\sscri}^\tot$. Let $f_+\in\Hb^{\infty,\ \ell_\sscri+1,\ \ell_\cK-2}(\iota^+;S^2\cT^*)$ and write $u_+:=N_{\iota^+}(\ubar L,\lambda)^{-1}f_+$ in the form $u_+=\chi_\cK|_{\iota^+}u_0+\tilde u$ where (using Propositions~\usref{PropipNInv} and \usref{PropipNAsy}) $u_0$ is given by~\eqref{EqipNAsyu0} and
  \[
    \tilde u=\tilde u(R,\omega)\in\Hb^{\infty,\ \bigl(\la\cE_{\iota^+,\sscri}^\cC\ra,\ell_\sscri\bigr),\ \ell_\cK}(\iota^+)
  \]
  in the notation of Definition~\usref{DefipIndex}. Define
  \begin{subequations}
  \begin{equation}
  \label{EqipGra}
    a_{\rms 1}(t_*)
      :=\begin{cases}
         \frac{1}{\lambda(\lambda-1)}t_*^{-\lambda+1}, & \lambda\neq 1, \\
         -\log t_*, & \lambda=1,
       \end{cases}
     \qquad\qquad
    a_{\rmv 1}(t_*) := -\frac{1}{\lambda}t_*^{-\lambda},
  \end{equation}
  and $u_{b,0}=u_{b,0}(t_*,r,\omega)$ by
  \begin{equation}
  \label{EqipGr}
  \begin{split}
    u_{b,0} &:= h_{b,\rms 1}^{\leq 4}(a_{\rms 1}(t_*)\scal_1^{(-1)}) + h_{b,\rmv 1}^{\leq 3}(a_{\rmv 1}(t_*)\vect_1^{(-1)}) + \scal_0^{(0)} h_{b,\rms 0}^{(0),\leq 3}(t_*^{-\lambda}) \\
      &\quad + \sum_{\substack{\mu=-\lambda_{\rms l,l+j}^\Ups+1 \\ 0\leq l\leq 3,\ j=0,1, \\ 1\leq l+j\leq 3}} h_{b,\rms l}^{(\mu),\leq 2}(t_*^{-\lambda+\mu}\scal_l^{(\mu)}) \\
      &\quad + \sum_{l=2}^5 h_{b,\rms l}^{(-l+2),\leq 5-l}(t_*^{-\lambda-l+2}\scal_l^{(-l+2)}) + \sum_{l=2}^4 h_{b,\rmv l}^{(-l+1),\leq 4-l}(t_*^{-\lambda-l+1}\vect_l^{(-l+1)}) \\
      &\quad + h_{b,\rms 2}^{(-2),\leq 1}(t_*^{-\lambda-2}\scal_2^{(-2)}) + h_{b,\rmv 2}^{(-2),\leq 1}(t_*^{-\lambda-2}\vect_2^{(-2)}) \\
      &\quad + h_{b,\rms 3}^{(-3)}(t_*^{-\lambda-3}\scal_3^{(-3)}) + h_{b,\rmv 3}^{(-3)}(t_*^{-\lambda-3}\vect_3^{(-3)}).
  \end{split}
  \end{equation}
  \end{subequations}
  Let $\chi_\iota\in\CI(M)$ be equal to $1$ near $\iota^+$ and supported in $\Omega_*$, and set
  \begin{equation}
  \label{EqipGru}
    u := \chi_\cK u_{b,0} + \chi_\iota t_*^{-\lambda}\tilde u \in \Hb^{\infty,\ \bigl(\la\cE_{\iota^+,\sscri}^\cC\ra,\ell_\sscri\bigr),\ (\lambda,0)\cup(\lambda+\cE_\ind),\ (\cE_\cK^{\rm gr}(\lambda),\ell_\cK)}(\Omega_*)^{\bullet,-},
  \end{equation}
  where $\cE_\cK^{\rm gr}(\lambda)$ is an index set containing the index sets of $a_{\rms 1}$, $a_{\rmv 1}$, and $(\kappa,0)$ for all exponents $\kappa$ of $t_*^{-1}$ appearing in~\eqref{EqipGr}, and $\cE_\ind$ is an index set with $\min\Re\cE_\ind\geq\eps_\ind$ depending only on $\ubar L$. Then we have
  \begin{equation}
  \label{EqipGrLb}
    \chi_\iota t_*^{-\lambda-2}f_+ - L_b u \in \Hb^{\infty,\ \bigl(\cE_{\iota^+,\sscri}^\tot+2,\,\ell_\sscri+1\bigr),\ \lambda+2+\cE_\ind,\ \lambda+\ell_\cK}(\Omega_*;S^2\cT^*)^{\bullet,-}.
  \end{equation}
\end{prop}

Thus, $u$ is an approximate solution of $L_b u=t_*^{-\lambda-2}f_+$ at $\iota^+$ (note that the remainder term~\eqref{EqipGrLb} has more than $\lambda+2$ orders of decay at $\iota^+$) which is moreover rather accurate also at $\cK^+$. The smallest $\lambda$ for which we will apply this result is $\lambda=1$, in which case the remainder in~\eqref{EqipGrLb} has more than $4$ orders of decay at $\cK^+$---enough to be regarded as a remainder term.

The only subtlety in this result concerns the first two terms in~\eqref{EqipGr}; this was already discussed around~\eqref{EqIEinhs12Elim}. While $\ubar h_{\rms 1}^{\leq 4}(A(t_*))$ is the sum of three terms $\ubar h_{\rms 1}^{\leq 2}(A(t_*))+\breve{\ubar h}_{\rms 1}^{\leq 3}(A'(t_*))+\breve{\ubar h}_{\rms 1}^{\leq 4}(A''(t_*))$, we have $h_{b,\rms 1}^{\leq 4}(a(t_*))=h_{b,\rms 1}(a(t_*))+\breve h_{b,\rms 1}^1(a'(t_*))+\sum_{j=2}^4\breve h_{b,\rms 1}^j(a^{(j)}(t_*))$; the definition of $a_{\rms 1}$ in~\eqref{EqipGra} is such that the coefficient $a''_{\rms 1}\scal_1^{(-1)}$ of $\breve h_{b,\rms 1}^2$ is $t_*^{-\lambda-1}\scal_1^{(-1)}$ and thus matches the coefficient of $\ubar h_{\rms 1}^2$ encoded in the first term of~\eqref{EqipNAsyu0}. Thus, a $\cO(t_*^{-\lambda-2})$ source term $f$ can lead to a center-of-mass movement of the black hole of size $\cO(t_*^{-\lambda+1})$ (or $\cO(\log t_*)$ when $\lambda=1$). Similarly for the $h_{b,\rmv 1}^{\leq 3}$ term, which includes the term $h_{b,\rmv 1}(a_{\rmv 1}(t_*)\vect_1^{(-1)})$ that is not visible at order $t_*^{-\lambda}$ at $\iota^+$, and is only non-zero when $\bha=\bha(b)\neq 0$, in which case it corresponds to a $\cO(|\bha|)$ change of the axis of rotation (cf.\ Remark~\ref{RmkWEMode0Ang}).

\begin{proof}[Proof of Proposition~\usref{PropipGr}]
  The definition of $u$ ensures that $(t_*^\lambda u)|_{\iota^+}=u_+$. We use here that the leading-order terms at $\rho=0$ of $h_{b,\rms 2}^{(-2)}$ etc.\ are given by $\ubar h_{b,\rms 2}^{(-2)}$ etc., while the special terms $h_{b,\rms 1}(a_{\rms 1}(t_*)\scal_1^{(-1)})$, $\breve h_{b,\rms 1}^1(a'_{\rms 1}(t_*)\scal_1^{(-1)})$, and $h_{b,\rmv 1}(a_{\rmv 1}(t_*)\vect_1^{(-1)})$ are of size $\rho^2 t_*^{-\lambda+1}$ (or $\rho\log t_*$ when $\lambda=-1$), $\rho t_*^{-\lambda}$, and $\rho t_*^{-\lambda}$ and thus of size $\cO(t_*^{-\lambda-1})=o(t_*^{-\lambda})$ at $(\iota^+)^\circ$ (where $t_*^{-1}\sim\rho$). Thus, they do not contribute to $(t_*^\lambda u)|_{\iota^+}$.

  Let us first work away from $\cK^+$; we drop the $\cK^+$-order from the notation. Then we have $u\in\Hb^{\infty,\ (\la\cE_{\iota^+,\sscri}^\cC\ra,\ell_\sscri),\ (\lambda,0)\cup(\lambda+\cE_\ind)}$, which $L_b\in\rho_\sscri\rho_+^2\Diffb^2$ maps into the space
  \[
    \Hb^{\infty,\ (\cE_{\iota^+,\sscri}^\tot+1,\,\ell_\sscri+1),\ (\lambda+2,0)\cup(\lambda+2+\cE_\ind)}.
  \]
  We first argue that the $\scri^+$-index here is, in fact, contained in $\cE_{\iota^+,\sscri}^\tot+2$: by Lemma~\ref{LemmaWEOp}, it suffices to consider the action of the first term in~\eqref{EqWEOpMink2} on $t_*^{-\lambda}\tilde u(R,\omega)$. The conjugation of $-t_*^2\cdot 2\rho\pa_{t_*}(\rho\pa_\rho-1-\ubar S)$ by $t_*^\lambda$ is $-2 t_*^2\rho(\pa_{t_*}-\lambda t_*^{-1})(\rho\pa_\rho-1-\ubar S)$, so upon passing to $t_*$ and $v=\rho t_*$, this acts on $\tilde u(R,\omega)=\tilde u(v^{-1},\omega)$ as $-2 v(v\pa_v-\lambda)(v\pa_v-1-\ubar S)\tilde u$ (cf.\ \eqref{EqipSpecFam2Expr}); but $(v\pa_v-1-\ubar S)\tilde u\in\cA^{\cE_{\iota^+,\sscri}^\tot+1}([0,1)_v\times\Sph^2)$ by Proposition~\ref{PropipNInv}, and acting on this with $-2 v(v\pa_v-\lambda)$ produces the claimed index set $\cE_{\iota^+,\sscri}^\tot+2$. Note, moreover, that the restriction of $t_*^{\lambda+2}L_b u$ to $\iota^+$ is given by $t_*^\lambda t_*^2\ubar L t_*^{-\lambda}u_+=f_+$. Altogether, we thus conclude that
  \[
    t_*^{-\lambda-2}f_+-L_b u\in\Hb^{\infty, (\cE_{\iota^+,\sscri}^\tot+2,\,\ell_\sscri+1),\ \lambda+2+\cE_\ind}
  \]
  away from $\cK^+$.

  Near $\cK^+$, and dropping the $\scri^+$-order from the notation, we note that $L_b$ maps $\chi_\iota t_*^{-\lambda}\tilde u\in\Hb^{\infty,\ (\lambda,0),\ \lambda+\ell_\cK}$ into $\Hb^{\infty,\ (\lambda+2,0),\ \lambda+\ell_\cK}$; and similarly $[L_b,\chi_\cK]u_{b,0}\in\Hb^{\infty,\ (\lambda+2,0)\cup(\lambda+2+\cE_\ind),\ \infty}$. It remains to show that, on $\supp\chi_\cK$,
  \[
    L_b u_{b,0} \in \Hb^{\infty,\ (\lambda+2,0)\cup(\lambda+2+\cE_\ind),\ \lambda+\ell_\cK}.
  \]
  (Once this is shown, the same arguments as before apply as regards the index set of~\eqref{EqipGrLb} at $\iota^+$.) This holds, in fact, with $\ell_\cK$ replaced by any number $<3+\eps_\ind$. Consider the third term in the first line of~\eqref{EqipGr}. Then the identity~\eqref{EqipGrKerrComp}, with $a(t_*)=\scal_0^{(0)}t_*^{-\lambda}$ and $k=3$, implies that
  \begin{align*}
    L_b\bigl( \scal_0^{(0)}h_{b,\rms 0}^{(0),\leq 3}(t_*^{-\lambda})\bigr) &\in t_*^{-\lambda-4}\Bigl(\rho\Diffb^1(X)\cA^{(-3,0)\cup(-3+\cE_\ind)}(X) + \rho^2\CI(X)\cA^{(-2,0)\cup(-2+\cE_\ind)}(X)\Bigr) \\
      &\quad + t_*^{-\lambda-5}\rho^2\CI(X)\cA^{(-3,0)\cup(-3+\cE_\ind)}(X) \\
      &\in \cA^{(\lambda+2,0)\cup(\lambda+2+\cE_\ind),\ \lambda+4} \subset \Hb^{\infty,\ (\lambda+2,0)\cup(\lambda+2+\cE_\ind),\ \lambda+\ell_\cK};
  \end{align*}
  here we use~\eqref{EqWG0Larges003}. Similar arguments apply to all other terms in~\eqref{EqipGr}; the $\cK^+$-order of $L_b(h_b^{(\mu),\leq k}(t_*^{-\lambda+\mu}))$ is $\lambda-\mu+k+1$, and our choices of $k$ ensure that $-\mu+k+1\geq 3+\eps_\ind$ in all cases.
\end{proof}

We next solve away source terms at $\iota^+$ that feature logarithmic factors, and at orders $\lambda_0\in\C$ that may lie in $\spec_{\iota^+}(\ubar L)$. (We only need this for $\Re\lambda_0>1$.) We first note that, for fixed $\eps>0$, Proposition~\ref{PropipGr} produces a map
\begin{subequations}
\begin{equation}
\label{EqipGrPhi}
\begin{split}
  \Phi_{\iota,b}(\lambda) &\colon \Hb^{\infty,\ \ell_\sscri+1,\ \ell_\cK-(\lambda_0+2)-\eps}(\iota^+;S^2\cT^*) \ni f_+ \\
  &\qquad \mapsto u \in \Hb^{\infty,\ \bigl(\la\cE_{\iota^+,\sscri}^\cC\ra,\ell_\sscri\bigr),\ (\lambda,0)\cup(\lambda+\cE_\ind),\ (\cE_\cK^{\rm gr}(\lambda),\ell_\cK)}(\Omega_*)^{\bullet,-}
\end{split}
\end{equation}
which is holomorphic for $\lambda$ near $\lambda_0$ with $\lambda\notin\spec_{\iota^+}(\ubar L)$ when regarded as taking values in the fixed space $\Hb^{\infty,\ (\la\cE_{\iota^+,\sscri}^\cC\ra,\ell_\sscri),\ \lambda_0-\eps,\ \lambda_0-1-\eps}(\Omega_*)^{\bullet,-}$, and such that
\begin{equation}
\label{EqipGrF}
  F_{\iota,b}(\lambda)f_+ := \chi_\iota t_*^{-\lambda-2}f_+ - L_b\bigl(\Phi_{\iota,b}(\lambda)f_+\bigr) \in \Hb^{\infty,\ (\cE_{\iota,\sscri}^\tot+2,\ell_\sscri+1),\ \lambda+2+\cE_\ind,\ \lambda+\ell_\cK}(\Omega_*;S^2\cT^*)^{\bullet,-}
\end{equation}
\end{subequations}
is holomorphic when regarded as taking values in $\Hb^{\infty,\ (\cE_{\iota,\sscri}^\tot+2,\ell_\sscri+1),\ \lambda_0+2+\eps_\ind-\eps,\ \lambda_0+\ell_\cK-\eps}(\Omega_*)^{\bullet,-}$.

\begin{cor}[Solving away quasi-homogeneous sources at $\iota^+$]
\label{CoripGrQhom}
  Let $\lambda_0\in\C$ be arbitrary with $\Re\lambda_0>1$, and define $l_0\in\N_0$ by $(\lambda_0,l_0-1)\in\Spec_{\iota^+}(\ubar L)$ when $\lambda_0\in\spec_{\iota^+}(\ubar L)$ (see Definition~\usref{DefipMeroSpec}) and $l_0=0$ otherwise. Let $\ell_\cK<3+\eps_\ind$, with $\ell_\cK\neq\Re(-\lambda)$ for all indicial roots $\lambda$ of $\wh{L_b}(0)$ with $\Re\lambda\in[-3,0]$. Let $\ell_\sscri<1+(1-e^\cC)(1-v^\cC)\gamma^\cC$ with $\ell_\sscri\notin\Re\pi_1\cE_{\iota^+,\sscri}^\tot$, and suppose that $\Re\lambda_0<\ell_\sscri$. Let $k\in\N_0$ and
  \[
    f \in \Hb^{\infty,\ \ell_\sscri+1,\ (\lambda_0+2,k),\ \lambda_0+\ell_\cK-\eps}(\Omega_*;S^2\cT^*)^{\bullet,-}\quad\forall\,\eps>0.
  \]
  Then there exist $u_{b,0}^{(j)}$, $j=0,\ldots,k+l_0$, of the form~\eqref{EqipGr} with $\lambda_0$ in place of $\lambda$ and with a factor of $(\log t_*)^j$ in each argument (so $a_{\rms 1}(t_*)(\log t_*)^j$, $a_{\rmv 1}(t_*)(\log t_*)^j$, $t_*^{-\lambda_0}(\log t_*)^j$, etc.), and $\tilde u^{(j)}\in\bigcap_{\eps>0}\Hb^{\infty,\ (\la\cE_{\iota^+,\sscri}^\cC\ra,\ell_\sscri),\ \ell_\cK-\eps}(\iota^+;S^2\cT^*)$, $j=0,\ldots,k+l_0$, such that
  \begin{equation}
  \label{EqipGrQhomu}
  \begin{split}
    u &:= \sum_{j=0}^{k+l_0} \chi_\cK u_{b,0}^{(j)} + \chi_\iota t_*^{-\lambda_0}(\log t_*)^j\tilde u^{(j)} \\
      &\ \in \bigcap_{\eps>0}\Hb^{\infty,\ \bigl(\la\cE_{\iota^+,\sscri}^\cC\ra,\ell_\sscri\bigr),\ (\lambda_0,k+l_0)\cup(\lambda_0+\cE_\ind),\ \bigl(\cE_\cK^{\rm gr}(\lambda_0)+(0,k+l_0),\,\ell_\cK-\eps\bigr)}(\Omega_*)^{\bullet,-}
  \end{split}
  \end{equation}
  satisfies
  \[
    f - L_b u \in \bigcap_{\eps>0}\Hb^{\infty,\ (\cE_{\iota^+,\sscri}^\tot+2,\,\ell_\sscri+1),\ \lambda_0+2+\cE_\ind,\ \lambda_0+\ell_\cK-\eps}(\Omega_*;S^2\cT^*)^{\bullet,-}
  \]
  where $\cE_\ind$ is an index set with $\min\Re\cE_\ind\geq\eps_\ind$.
\end{cor}
\begin{proof}
  The membership of $f$ implies that we can write
  \begin{align*}
    f(t_*,R,\omega) &= \sum_{j=0}^k \chi_\iota t_*^{-\lambda_0-2}(\log t_*)^j f_+^{(j)}(R,\omega), \\
    &\qquad f_+^{(j)}=f_+^{(j)}(R,\omega)\in\bigcap_{\eps>0}\Hb^{\infty,\ \ell_\sscri+1,\ \ell_\cK-2-\eps}(\iota^+;S^2\cT^*),\ j=0,\ldots,k,
  \end{align*}
  up to a remainder in $\Hb^{\infty,\ \ell_\sscri+1,\ \infty,\ \lambda_0+\ell_\cK-\eps}$ (which we may drop).  We can express $f$ as
  \[
    f = \frac{1}{2\pi i}\oint_{\lambda_0} \sum_{j=0}^k (-1)^j j!(\lambda-\lambda_0)^{-j-1}\chi_\iota t_*^{-\lambda-2} f_+^{(j)}\,\dd\lambda,
  \]
  where we integrate along a small circle around $\lambda_0$. Therefore, for
  \[
    u := \frac{1}{2\pi i}\oint_{\lambda_0} \sum_{j=0}^k (-1)^j j!(\lambda-\lambda_0)^{-j-1}\Phi_{\iota,b}(\lambda)f_+^{(j)}\,\dd\lambda,
  \]
  which by (the lines after)~\eqref{EqipGrPhi} lies in $\Hb^{\infty,\ (\la\cE_{\iota^+,\sscri}^\cC\ra,\ell_\sscri),\ \lambda_0-\eps,\ \lambda_0-1-\eps}$, we have
  \[
    f - L_b u = \frac{1}{2\pi i}\oint_{\lambda_0} \sum_{j=0}^k (-1)^j j! (\lambda-\lambda_0)^{-j-1}F_{\iota,b}(\lambda)f_+^{(j)}\,\dd\lambda.
  \]
  By (the line after)~\eqref{EqipGrF}, this lies in $\Hb^{\infty,\ (\cE_{\iota^+,\sscri}^\tot+2,\ell_\sscri+1),\ \lambda_0+2+\eps_\ind-\eps,\ \lambda_0+\ell_\cK-\eps}$. The polyhomogeneity of $u$ at $\iota^+$ and $\cK^+$ and of $f-L_b u$ at $\iota^+$ follows by expanding $\Phi_{\iota,b}(\lambda)f_+^{(j)}$ and $F_{\iota,b}(\lambda)$ in $t_*$ and noting that the maximal power of $\log t_*$ is bounded by the order of the pole of the integrands at $\lambda=\lambda_0$ minus $1$.
\end{proof}

\subsubsection{Further properties of \texorpdfstring{$N_{\iota^+}(\underbar{L},\lambda)$}{the ip-normal operator family}}
\label{SssipP}

--- \emph{The following material will be used only in~\S\usref{SEf} below.} In preparation for providing precise constraints on the possible asymptotic behavior of solutions of the linearized gauge-fixed Einstein equation on dynamical Kerr spacetimes at $\iota^+$ in~\S\ref{SEf}, we collect further results on the meromorphic and mapping properties of $N_{\iota^+}(\ubar L,\lambda)$. 

The $\scri^+$-order of source terms $f$ on spacetime will be of the form $(\la\cE_\sscri^\cC+1\ra',\ell_\sscri+1)$ where
\begin{equation}
\label{EqipPellscri}
  1<\ell_\sscri<1+(1-e^\cC)(1-v^\cC)\gamma^\cC
\end{equation}
is fixed; the notation here is taken from Definition~\ref{DefExFwSrc}. (In~\S\S\ref{SD}--\ref{SEf}, we shall work with $\ell_\sscri=3+\eps_\sscri$ where $0<\eps_\sscri\ll 1$.) Terms in the $\iota^+$-expansion of such source terms (provided they are partially polyhomogeneous at $\iota^+$) thus typically have the same index sets at $\iota^+\cap\scri^+$; and inverting $N_{\iota^+}(\ubar L,\lambda)$ on such terms then should produce $\scri^+$-orders $(\la\cE_\sscri^\cC\ra,\ell_\sscri)$. Analogously to Definitions~\ref{DefExP}, \ref{DefipIndex}, and \ref{DefExFwSrc}, we must therefore introduce spaces
\begin{equation}
\label{EqipPSpaces}
  \Hb^{k,\ \bigl(\la\cE_\sscri^\cC\ra,\ell_\sscri\bigr),\ (\cE_\cK,\ell_\cK)}(\iota^+),\quad
  \Hb^{k,\ \bigl(\la\cE_\sscri^\cC+1\ra',\ell_\sscri+1\bigr),\ \ell_\cK-2}(\iota^+)
\end{equation}
of sections of $S^2\cT^*\to\iota^+$: recalling~\eqref{EqExPOther}, membership of $h$ in the first space means that $\pi^\cC h\in\Hb^{k,(\cE_\sscri^\cC,\ell_\sscri),(\cE_\cK,\ell_\cK)}$, $\pi_\bullet h\in\Hb^{k,(\cE_{\sscri,\bullet},\ell_\sscri),(\cE_\cK,\ell_\cK)}$ for $\bullet\in\{0 1,\ 1 /,\ 1 1\}$, and $\slpi_0 h\in\Hb^{k,(\slcE_{\sscri,0},\ell_\sscri),(\cE_\cK,\ell_\cK)}$, while membership of $f$ in the second space means that $\pi^\cC f$, $\pi_{0 1}f$, $\pi_{1 /}f$, $\slpi_0 f\in\Hb^{k,(\cE_\sscri^\cC+1,\ell_\sscri+1),\ell_\cK-2}$ and $\pi_{1 1}f\in\Hb^{k,(\cF_{\sscri,1 1},\ell_\sscri+1),\ell_\cK-2}$ where $\cF_{\sscri,1 1}$ is given by~\eqref{EqExFwSrc11}.

Our first task is then to study $N_{\iota^+}(\ubar L,\lambda)^{-1}$ on partially polyhomogeneous inputs. An important and elementary toy model is the operator family $x\pa_x-\lambda$, which we analyze on partially polyhomogeneous spaces $\bar H_\bop^{k,(\cE,\alpha)}([0,1)_x;|\frac{\dd x}{x}|)$ (with extendible character at $x=1$).

\begin{lemma}[Inversion of $x\pa_x-\lambda$]
\label{LemmaipPxpax}
  Let $\cE\subset\C\times\N_0$ be an index set and $\alpha\in\R$. Define the operator $\cR(\lambda)$ by
  \[
    \bigl(\cR(\lambda)f\bigr)(x) := x^\lambda \int_0^x t^{-\lambda}f(t)\,\frac{\dd t}{t},\quad f\in\CIdot([0,1]),\ \lambda\in\C;
  \]
  this is thus a right inverse of $x\pa_x-\lambda$. Then $\cR(\lambda)$ extends from $\Re\lambda\ll -1$ to a finite-meromorphic family of operators
  \[
    \cR(\lambda) \colon \bar H_\bop^{k,(\cE,\alpha)}([0,1)) \to \bar H_\bop^{k+1,(\cE,\alpha)}([0,1)),\quad \Re\lambda<\alpha,
  \]
  with divisor equal to $\{(z,k(\cE,z))\colon z\in\pi_1\cE,\ \Re z<\alpha\}$ in the notation of Definition~\usref{DefTMIndex}.
\end{lemma}
\begin{proof}
  Consider first the case $\cE=\emptyset$. Then for $\Re\lambda<\alpha$ and $f\in\CI([0,1))$ vanishing to infinite order at $0$, we can estimate, for $\eps:=\frac12(\alpha-\Re\lambda)>0$,
  \begin{align*}
    \|\cR(\lambda)f\|_{\bar H_\bop^{0,\alpha}([0,1))}^2 &= \int_0^1 \biggl| x^{\lambda-\alpha}\int_0^x t^{-\lambda+\alpha} t^{-\alpha}f(t)\,\frac{\dd t}{t}\biggr|^2\,\frac{\dd x}{x} \\
      &\leq \int_0^1 x^{2(\lambda-\alpha)}\biggl(\int_0^x t^{2(-\lambda+\alpha)-\eps}\,\frac{\dd t}{t}\biggr) \biggl(\int_0^x t^\eps|t^{-\alpha}f(t)|^2\,\frac{\dd t}{t}\biggr)\,\frac{\dd x}{x} \\
      &= C_\eps \int_0^1 t^\eps|t^{-\alpha}f(t)|^2\,\biggl(\int_t^1 x^{-\eps}\,\frac{\dd x}{x}\biggr)\,\frac{\dd t}{t} \\
      &\leq C'_\eps \|f\|_{\bar H_\bop^{0,\alpha}([0,1))}^2.
  \end{align*}
  An estimate on $\bar H_\bop^{1,\alpha}$ follows from $x\pa_x\cR(\lambda)f=\lambda f$, and higher regularity is then clear.

  To treat the case of general $\cE$, it suffices to study $\cR(\lambda)f$ for $f(x)=x^z(\log x)^k$. Since $\cR(\lambda)x^z=(z-\lambda)^{-1}x^z$ (initially for $\Re\lambda<\Re z$, and then extended by meromorphicity to $\lambda\in\C$), we have
  \begin{equation}
  \label{EqipPxpaxSing}
    \cR(\lambda)f = \frac{\dd^k}{\dd z^k}\Bigl(\frac{1}{z-\lambda}x^z\Bigr).
  \end{equation}
  This indeed lies in $\Hb^{\infty,(z,k)}$ and has a pole of order $k+1$ at $z=\lambda$.
\end{proof}

We use this to construct formal solutions of $N_{\iota^+}(\ubar L,\lambda)$ on $\iota^+$ near $\scri^+$ (so on the set $[0,1)_v\times\Sph^2$); we thus only record orders at $\scri^+$.

\begin{lemma}[Formal solutions near $\scri^+$]
\label{LemmaipPScri}
  Let $\ell_\sscri$ be as in~\eqref{EqipPellscri}. Set $l:=2\lceil\ell_\sscri\rceil+10$, and let $f\in\Hb^{k+l,\,(\la\cE_\sscri^\cC+1\ra',\ell_\sscri+1)}([0,1)_v\times\Sph^2)$. Then there exists $u(\lambda)\in\Hb^{k+2,\,(\la\cE_\sscri^\cC\ra,\ell_\sscri)}([0,1)_v\times\Sph^2)$ such that
  \begin{enumerate}
  \item $u(\lambda)$ is meromorphic in $\lambda\in\C$ for $\Re\lambda<\ell_\sscri$, with divisor contained in $\{(z,k(\cE_{\sscri,\sharp}^\cC))\colon (z,k)\in\cE_{\sscri,\sharp}^\cC\}$ where
    \begin{equation}
    \label{EqipPScriEsharp}
      \cE_{\sscri,\sharp}^\cC:=(1,0)\cup(1+(1-e^\Ups)\gamma^\Ups,0)\cup\cE_\sscri^\cC \subset \cE_\sscri^\tot;
    \end{equation}
  \item away from the poles of $u(\lambda)$, we have $f-N_{\iota^+}(\ubar L,\lambda)u(\lambda)\in\Hb^{k,\ell_\sscri+1}([0,1)_v\times\Sph^2)$.
  \end{enumerate}
\end{lemma}
\begin{proof}
  For $f\in\Hb^{k+l,\ell_\sscri+1}$, we may simply take $u(\lambda)=0$. The task is thus to solve away the partial expansion of $f$ at $v=0$. Recall from~\eqref{EqipSpecFam2Expr} that
  \[
    -\frac{1}{2 v}N_{\iota^+}(\ubar L,\lambda) = (v\pa_v-\lambda)(v\pa_v-1-\ubar S) + \tilde N,\quad \tilde N:=-\frac{1}{2 v}v^2\wt{\ubar L}(0)(\omega,v\pa_v,\pa_\omega)\in v\Diffb^2,
  \]
  with $\ubar S$ given by~\eqref{EqWEOpMinkS}. Consider first the exceptional terms in the $(\dd x^1)^2$-component of $-\frac{1}{2 v}f$ arising from the terms other than $\cE_\sscri^\cC$ in~\eqref{EqExFwSrc11}; these are thus $f_{\rm exc}(v,\omega):=v f_1(\omega)+v^{1+(1-e^\Ups)\gamma^\Ups}f_2(\omega)$ times $(\dd x^1)^2$. Restricted to $\la(\dd x^1)^2\ra$, the endomorphism $1+\ubar S$ is multiplication by $1+2\gamma^\Ups$, and thus
  \[
    (v\pa_v-1-\ubar S)w^\flat_{\rm exc} = f_{\rm exc}
  \]
  if we set
  \[
    w^\flat_{\rm exc}(v,\omega) := \bigl(1-(1+2\gamma^\Ups)\bigr)^{-1} v f_1(\omega) + \bigl(1+(1-e^\Ups)\gamma^\Ups-(1+2\gamma^\Ups)\bigr)^{-1} v^{1+(1-e^\Ups)\gamma^\Ups}f_2(\omega)
  \]
  Defining $w_{\rm exc}(\lambda,v,\omega):=(v\pa_v-\lambda)^{-1}w^\flat_{\rm exc}$ using Lemma~\ref{LemmaipPxpax} (which is thus meromorphic with simple poles at $\lambda=1$ and $\lambda=1+(1-e^\Ups)\gamma^\Ups$), we thus have
  \[
    f - N_{\iota^+}(\ubar L,\lambda)w_{\rm exc}(\lambda) \in \Hb^{k+l-2,\ (\cE_\sscri^\cC+1,\ell_\sscri+1)},
  \]
  with meromorphic dependence on $\lambda$.

  It remains to solve away terms $v^z(\log v)^k f_{(z,k)}(\omega)$, in the $\scri^+$-expansion of $-\frac{1}{2 v}f$; here $(z,k)\in\cE_\sscri^\cC$ and $\Re z\leq\ell_\sscri$. For such $z$, we have $\Re z>1+2\gamma^\Ups$ (as required in Definition~\ref{DefExP}\eqref{ItExPEscri}) and $\Re z<1+(1-e^\cC)(1-v^\cC)\gamma^\cC$ by~\eqref{EqipPellscri}, and hence the endomorphism $z-1-\ubar S$ of $S^2\cT^*$ is invertible. We can thus define
  \[
    w^\flat(v,\omega) := \frac{\dd^k}{\dd z^k} \Bigl( v^z (z-1-\ubar S)^{-1} f_{(z,k)} \Bigr),
  \]
  which solves $(v\pa_v-1-\ubar S)w^\flat=v^z(\log v)^k f_{(z,k)}$, and then define $w(\lambda):=(v\pa_v-\lambda)^{-1}v^\flat$ using Lemma~\ref{LemmaipPxpax} (which is thus meromorphic with divisor at most $\{(z,k)\}$). Moreover,
  \[
    v^z(\log v)^k f_{(z,k)} - N_{\iota^+}(\ubar L,\lambda)\Bigl(-\frac{1}{2 v}w(\lambda)\Bigr) \in \Hb^{k+l-4,\ ((z+1,k),\ell_\sscri+1)}.
  \]
  We can solve this away iteratively, until after finitely many steps the remainder has decay order $\ell_\sscri+1$.
\end{proof}

We can now invert $N_{\iota^+}(\ubar L,\lambda)$ globally on $\iota^+$ on spaces with partial expansions at $\iota^+\cap\scri^+$. Recall from Proposition~\ref{PropipNInv}\eqref{ItipNInvb} that $N_{\iota^+}(\ubar L,\lambda)^{-1}$ is well-defined on inputs with $\scri^+$-order $2-\eps$ provided $\Re\lambda<1-\eps$; this thus includes inputs with $\scri^+$-order $(\la\cE_\sscri^\cC+1\ra',\ell_\sscri+1)$.

\begin{prop}[Meromorphicity of $N_{\iota^+}(\ubar L,\lambda)^{-1}$ on partially polyhomogeneous spaces]
\label{PropipPMeroPhg}
  We use the notation of Lemma~\usref{LemmaipPScri}. Let $\ell_\cK\in(-\eps_\ind,3+\eps_\ind)$ with $\ell_\cK\notin\Re\pi_1\cE_\cK$, where $\cE_\cK$ was defined in Proposition~\usref{PropipNInv}. Then the map $N_{\iota^+}(\ubar L,\lambda)^{-1}$ extends from $\Re\lambda<1$ to a meromorphic map
  \begin{equation}
  \label{EqipPMeroPhg}
    N_{\iota^+}(\ubar L,\lambda)^{-1} \colon \Hb^{k+l,\ \bigl(\la\cE_\sscri^\cC+1\ra',\ell_\sscri+1\bigr),\ \ell_\cK-2}(\iota^+) \to \Hb^{k,\ \bigl(\la\cE_\sscri^\cC\ra,\ell_\sscri\bigr),\ (\cE_\cK,\ell_\cK)}(\iota^+),\quad \Re\lambda<\ell_\sscri.
  \end{equation}
  Its poles lie in the union of $\spec_{\iota^+}(\ubar L)$ (from Definition~\usref{DefipMeroSpec}) and $\pi_1\cE_{\sscri,\sharp}^\cC$ (from~\eqref{EqipPScriEsharp}), and the order of a pole $\lambda$ is no greater than the sum of the order of $\lambda$ recorded in $\Spec_{\iota^+}(\ubar L)$ and $k(\cE_{\sscri,\sharp}^\cC,\lambda)+1$.
\end{prop}
\begin{proof}
  Fix a cutoff function $\chi_\sscri\in\CIc([0,1)_v)$ which equals $1$ on $[0,\frac12]$. Given $f$ in the domain of~\eqref{EqipPMeroPhg}, define $u(\lambda)$ using Lemma~\ref{LemmaipPScri}; then
  \[
    f'(\lambda) := f - N_{\iota^+}(\ubar L,\lambda)\bigl(\chi_\sscri u(\lambda)\bigr) \in \Hb^{k,\ \ell_\sscri+1,\ \ell_\cK-2}(\iota^+;S^2\cT^*),
  \]
  with meromorphic dependence on $\lambda$ for $\Re\lambda<\ell_\sscri$. We can then apply the inverse of $N_{\iota^+}(\ubar L,\lambda)$ from Proposition~\ref{PropipNInv}\eqref{ItipNInvb} to $f'(\lambda)$, giving a meromorphic $u'(\lambda)\in\Hb^{k,\ (\la\cE_{\iota^+,\sscri}^\cC\ra,\ell_\sscri),\ (\cE_\cK,\ell_\cK)}$ such that
  \[
    N_{\iota^+}(\ubar L,\lambda)\bigl( \chi_\sscri u(\lambda) + u'(\lambda) \bigr) = f.
  \]
  A fortiori, we have $u'(\lambda)\in\Hb^{k,\ (\la\cE_\sscri^\cC\ra,\ell_\sscri),\ (\cE_\cK,\ell_\cK)}$, and thus~\eqref{EqipPMeroPhg} follows.
\end{proof}

Since the terms in the $\scri^+$-expansion of $f$ are arbitrary sections of $S^2\cT^*_X|_{\pa X}\to\pa X\cong\Sph^2$ and thus lie in infinite-dimensional spaces, the singular coefficients of $N_{\iota^+}(\ubar L,\lambda)^{-1}$ in~\eqref{EqipPMeroPhg} are \emph{not} finite rank operators, so while $N_{\iota^+}(\ubar L,\lambda)^{-1}$ is meromorphic, it is \emph{not} finite-meromorphic. Analogously to Definition~\ref{DefipMeroSpec}, we introduce:

\begin{definition}[Boundary spectrum]
\label{DefipPSpec}
  In the setting of Proposition~\ref{PropipPMeroPhg}, we write
  \[
    \spec_{\iota^+}^{<\ell_\sscri}(\ubar L,\cE_\sscri^\cC) \subset \{\Re\lambda<\ell_\sscri\}
  \]
  for the set of poles of the map~\eqref{EqipPMeroPhg}, and
  \[
    \Spec_{\iota^+}^{<\ell_\sscri}(\ubar L,\cE_\sscri^\cC) \subset \spec_{\iota^+}^{<\ell_\sscri}(\ubar L,\cE_\sscri^\cC) \times \N_0
  \]
  for the set of $(\lambda,k)$ such that the order of the pole of~\eqref{EqipPMeroPhg} at $\lambda$ is equal to $k+1$.
\end{definition}

Proposition~\ref{PropipPMeroPhg} and Lemma~\ref{LemmaipPScri} imply that
\begin{equation}
\label{EqipPSpecEx}
  (1,0) \in \Spec_{\iota^+}^{<\ell_\sscri}(\ubar L,\cE_\sscri^\cC);\quad
  1\neq\lambda\in\spec_{\iota^+}(\ubar L,\cE_\sscri^\cC) \implies \Re\lambda>1.
\end{equation}

\begin{rmk}[Poles depending on $\cE_\sscri^\cC$]
\label{RmkipPMeroPhg}
  Expanding on the discussion in~\S\ref{SssIGen2La}, we can interpret the poles of~\eqref{EqipPMeroPhg} that depend on $\cE_\sscri^\cC$ in the following manner. For the sake of concreteness, consider some $(\alpha,0)\in\cE_\sscri^\cC$ with $1+2\gamma^\Ups<\alpha<\ell_\sscri$ such that $\alpha\notin\spec_{\iota^+}(\ubar L)$. Set $f(v,\omega):=v^{\alpha+1}f_0(\omega)$ where $0\neq f_0\in\CI(\pa X;S^2\cT^*_X)$. Then the proof of Proposition~\ref{PropipPMeroPhg} shows that $N_{\iota^+}(\ubar L,\lambda)^{-1}f$ typically has a simple pole at $\lambda=\alpha$. The residue $u(v,\omega)$ can be described explicitly: it has a non-zero $v^\alpha$-term (cf.\ \eqref{EqipPxpaxSing}), say $v^\alpha u_0(\omega)$, which using (the arguments in) Lemma~\ref{LemmaipPScri} can be upgraded via addition of suitable $v^{\alpha+j}$-terms, $j=1,2,\ldots$, to an element of the approximate nullspace of $N_{\iota^+}(L,\alpha)$; and subtracting the output of $N_{\iota^+}(L,\alpha)^{-1}$ acting on the remaining error (which is of class $\Hb^{\infty,\ell_\sscri+1}$ near $v=0$) produces $u$. From the perspective of scattering on asymptotically hyperbolic spaces, this means that $u$ is a solution of the scattering problem at frequency $\lambda=\alpha$, with ``incoming'' data $u_0(\omega)$ (specified at the ``incoming'' indicial root $\alpha$). The ``outgoing'' data, which are produced by the (global) inverse $N_{\iota^+}(\ubar L,\lambda)^{-1}$ (suitably interpreted at poles) are the radiation fields of $u$, corresponding to the ``outgoing'' indicial roots $\spec(1+\ubar S)$. --- The index set $\cE_\sscri^\cC$ thus specifies at which frequencies $\lambda$ one allows for nontrivial ``incoming'' data.
\end{rmk}

When $\lambda$ is not a pole of $u(\lambda)$, then from the equation $N_{\iota^+}(L,\lambda)u(\lambda)\in\Hb^{k+l,\ell_\cK-2}([0,1)_R\times\Sph^2)$ one can use Proposition~\ref{PropipNAsy} to extract an expansion of $u(\lambda)$ at $R=0$. The same applies to the most singular term in the Laurent expansion of $u(\lambda)$ at a pole, and also (by an iterative argument) to the less singular terms. We shall not give the details here, and will instead only develop the spacetime version of such an expansion below.

We construct exact (quasi-)homogeneous solutions of $\ubar L u=f$ when $f=f(t_*,R,\omega)$ is \mbox{(quasi-)}ho\-mo\-ge\-neous in $t_*^{-1}$. There are two main differences to Corollary~\ref{CoripGrQhom}: first, we allow for $f$ with nontrivial expansions at $\scri^+$; secondly, we do not graft the solutions into the Kerr spacetime. Moreover, we do allow for $\lambda_0=1$. The proof is however very similar.

\begin{cor}[Quasihomogeneous solutions at $\iota^+$]
\label{CoripPQhom}
  Let $\lambda_0\in\C$ with $\Re\lambda_0<\ell_\sscri$, where $\ell_\sscri$ is as in~\eqref{EqipPellscri}; let $l=2\lceil\ell_\sscri\rceil+10$. Let $\ell_\cK\in(-\eps_\ind,3+\eps_\ind)$ with $\ell_\cK\notin\Re\pi_1\cE_\cK$ where $\cE_\cK$ is as in Proposition~\usref{PropipNInv}. Let $J\in\N_0$, and suppose we are given
  \begin{equation}
  \label{EqipPQhomf}
    f = f(t_*,R,\omega) = \sum_{j=0}^J t_*^{-\lambda_0-2}(\log t_*)^j f^{(\lambda_0+2,j)}(R,\omega),\quad f^{(\lambda_0+2,j)}\in\Hb^{k+l,\ \bigl(\la\cE_\sscri^\cC+1\ra',\ell_\sscri+1\bigr),\ \ell_\cK-2}(\iota^+).
  \end{equation}
  Let $K$ be equal to the order of the pole of~\eqref{EqipPMeroPhg} at $\lambda=\lambda_0$. Then:
  \begin{enumerate}
  \item{\rm (Existence.)} There exists a solution of $\ubar L u=f$, depending linearly on $f$,\footnote{We state the linear dependence on $f$ explicitly here since, as we shall discuss momentarily, the relevant nullspace of $\ubar L$ is infinite-dimensional.} of the form
    \begin{equation}
    \label{EqipPQhomu}
      u = u(t_*,R,\omega) = \sum_{j=0}^{J+K} t_*^{-\lambda_0}(\log t_*)^j u^{(\lambda_0,j)}(R,\omega),\quad u^{(\lambda_0,j)}\in\Hb^{k,\ \bigl(\la\cE_\sscri^\cC\ra,\ell_\sscri\bigr),\ (\cE_\cK,\ell_\cK)}(\iota^+).
    \end{equation}
  \item\label{ItipPQhomExp}{\rm (Expansion at $R=0$.)} We can write $u=u_0+\tilde u$ where $u_0$ is a sum of terms of the form~\eqref{EqipNAsyu0} but with $t_*^{-\lambda}$ replaced by $t_*^{-\lambda_0}(\log t_*)^j$, $j=0,\ldots,J+K$, and $\tilde u$ is a sum of terms of the form $t_*^{-\lambda_0}(\log t_*)^j\Hb^{k,\ (\la\cE_\sscri^\cC\ra,\ell_\sscri),\ \ell_\cK}(\iota^+)$.
  \end{enumerate}
\end{cor}
\begin{proof}
  \pfstep{Existence.} Upon multiplying the $f^{(\lambda_0+2,j)}$ by suitable scalar factors, we have
  \[
    f = \frac{1}{2\pi i}\oint_{\lambda_0} t_*^{-\lambda-2} \sum_{j=0}^J (\lambda-\lambda_0)^{-j-1}f^{(\lambda_0+2,j)}\,\dd\lambda.
  \]
  We then set
  \[
    u = \frac{1}{2\pi i}\oint_{\lambda_0} t_*^{-\lambda} \sum_{j=0}^J (\lambda-\lambda_0)^{-j-1}N_{\iota^+}(\ubar L,\lambda)^{-1}f^{(\lambda_0+2,j)}\,\dd\lambda,
  \]
  which is well-defined by Proposition~\ref{PropipPMeroPhg}, with the order of the pole of the integrand at $\lambda=\lambda_0$ being at most $J+1+K$. Finally, since $\ubar L(t_*^{-\lambda}\tilde u(R,\omega))=t_*^{-\lambda-2}N_{\iota^+}(\ubar L,\lambda)\tilde u$, we have
  \[
    \ubar L u = \frac{1}{2\pi i}\oint_{\lambda_0} t_*^{-\lambda-2} N_{\iota^+}(\ubar L,\lambda)\Biggl(\;\sum_{j=0}^J (\lambda-\lambda_0)^{-j-1}N_{\iota^+}(\ubar L,\lambda)^{-1}f^{(\lambda_0+2,j)}\Biggr)\,\dd\lambda = f.
  \]

  \pfstep{Expansion at $R=0$.} We prove part~\eqref{ItipPQhomExp} more generally for $u$ of the form~\eqref{EqipPQhomu} such that $\ubar L u=f$ where $f$ is as in~\eqref{EqipPQhomf} but may now have up to $J+K$ many powers of $\log t_*$. Relabeling $J+K$ as $J$, let thus $u$ be as in~\eqref{EqipPQhomu} (with $K=0$ now). We work near $R=0$ and thus omit the $\scri^+$-order from the notation. Furthermore, we do not keep track of b-regularity orders anymore and work with $k=\infty$ for brevity. The term with the largest power of $\log t_*$ satisfies
  \[
    N_{\iota^+}(\ubar L,\lambda_0) u^{(\lambda_0,J)} \in \Hb^{\infty,\ell_\cK-2}([0,1)_R\times\Sph^2)
  \]
  and thus, by Proposition~\ref{PropipNAsy}, admits an expansion modulo $\Hb^{\infty,\ell_\cK}$ as in~\eqref{EqipNAsyu0}. When $J=0$, we are done, so let us assume that $J\geq 1$. Suppose initially that there is only one term in this expansion, namely $\scal_0^{(0),(J)}t_*^{\lambda_0}\ubar h_{\rms 0}^{(0),\leq 3}(t_*^{-\lambda_0})$. Define
  \[
    h = h(t_*,R,\omega) := \scal_0^{(0),(J)}\ubar h_{\rms 0}^{(0),\leq 3}\bigl(t_*^{-\lambda_0}(\log t_*)^{J}\bigr),
  \]
  which is thus also of the form~\eqref{EqipPQhomu}, with $t_*^{-\lambda_0}(\log t_*)^J$-coefficient equal to $\scal_0^{(0),(J)}\ubar h_{\rms 0}^{(0)}$ to leading order at $R=0$. Furthermore, recalling from~\eqref{EqipAsyCompGen2} that $\ubar L\ubar h_{\rms 0}^{(0),\leq 3}(t_*^{-\lambda_0})\in t_*^{-\lambda_0-2}R^2\CI([0,1)_R\times\Sph^2)$, we conclude that
  \[
    \ubar L h = (-1)^J\scal_0^{(0),(J)}\frac{\dd^{J}}{\dd\lambda_0^{J}} \ubar L \ubar h_{\rms 0}^{(0),\leq 3}(t_*^{-\lambda_0}) \in \sum_{j=0}^J t_*^{-\lambda_0-2}(\log t_*)^j R^2\CI([0,1)_R\times\Sph^2).
  \]
  In view of $R^2\CI\subset\Hb^{\infty,\ell_\cK-2}$, this is (a fortiori) of the same class as $f$. Upon subtracting $h$ from $u$, we thus obtain a new $u$, still of the form~\eqref{EqipPQhomu} (with $K=0$) and with $\ubar L u=f\in\sum_{j=0}^J t_*^{-\lambda_0-2}(\log t_*)^j\Hb^{\infty,\ell_\cK-2}$ still, \emph{but} with $u^{(\lambda_0,J)}$ not having an $R^0$-coefficient anymore. --- When $u^{(\lambda_0,J)}$ has a full expansion~\eqref{EqipNAsyu0}, we can eliminate each term in the same fashion; we have thus reduced to the case that $u^{(\lambda_0,J)}\in\Hb^{k,\ell_\cK}$.

  We can now set up an inductive argument: assuming that for some $j_0\in\N_0$, $j_0<J$, we have $u^{(\lambda_0,j)}\in\Hb^{\infty,\ell_\cK}$ for all $j=j_0+1,\ldots,J$, the equation satisfied by $u^{(\lambda_0,j_0)}$ is
  \[
    N_{\iota^+}(\ubar L,\lambda_0)u^{(\lambda_0,j_0)} - (j_0+1)\pa_\lambda N_{\iota^+}(\ubar L,\lambda_0)u^{(\lambda_0,j_0+1)} \in \Hb^{\infty,\ell_\cK-2}.
  \]
  But since $\pa_\lambda N_{\iota^+}(\ubar L,\lambda_0)\in R^{-1}\Diffb^1([0,1)_R\times\Sph^2)$, the second term here is of class $\Hb^{\infty,\ell_\cK-1}$ by the inductive hypothesis. Thus, the same arguments as before can be used to eliminate the expansion of $u^{(\lambda_0,j_0)}$. Once $j_0=0$, the proof is complete.
\end{proof}

The solution $u$ of $\ubar L u=f$ constructed in Corollary~\ref{CoripPQhom} is typically not unique. The freedom is captured by:

\begin{definition}[Resonant states]
\label{DefipPRes}
  Let $\lambda\in\C$ with $\Re\lambda<\ell_\sscri$; here $\ell_\sscri$ is as in~\eqref{EqipPellscri}. Fix an index set $\cE_\sscri^\cC$ as in Definition~\usref{DefExP}. Let $\ell_\cK\in(-\eps_\ind,3+\eps_\ind)$ with $\ell_\cK\notin\Re\pi_1\cE_\cK$, where $\cE_\cK$ was defined in Proposition~\ref{PropipNInv}. We then define, for $k\in\N_0\cup\{\infty\}$,
  \begin{equation}
  \label{EqipPResDef}
  \begin{split}
    \Resspace_{\iota^+}^k(\ubar L,\cE_\sscri^\cC,\lambda) &:= \Biggl\{ u(t_*,v,\omega)=\sum_{j=0}^K t_*^{-\lambda}(\log t_*)^j u^{(\lambda,j)}(v,\omega) \colon \\
      &\quad\hspace{5em} \ubar L u=0,\ u^{(\lambda,j)}\in\Hb^{k,\ \bigl(\la\cE_\sscri^\cC\ra,\ell_\sscri\bigr),\ (\cE_\cK,\ell_\cK)}(\iota^+),\ K\in\N_0 \Biggr\}.
  \end{split}
  \end{equation}
\end{definition}

That is, the solution $u$ in~\eqref{EqipPQhomu} is unique modulo $\Resspace_{\iota^+}^k(\ubar L,\cE_\sscri^\cC,\lambda)$.

\begin{rmk}[Asymptotic expansion of elements of $\Resspace_{\iota^+}^k(\ubar L,\cE_\sscri^\cC,\lambda)$]
\label{RmkipPResAsy}
  Note that (the proof of) Corollary~\ref{CoripPQhom}\eqref{ItipPQhomExp} applies also to elements of $\Resspace_{\iota^+}^k(\ubar L,\cE_\sscri^\cC,\lambda)$ (with $\ell_\cK$ there arbitrarily close to $3+\eps_\ind$), thus giving a complete description of the asymptotics of $u$ as $R=v^{-1}\to 0$ modulo $\cO(R^{3+\eps_\ind-\eps})$-errors.
\end{rmk}

Remark~\ref{RmkipPMeroPhg} explains why the space~\eqref{EqipPResDef} does depend on $\cE_\sscri^\cC$; it also implies that $\Resspace_{\iota^+}^k(\ubar L,\cE_\sscri^\cC,\lambda)$ is \emph{infinite-dimensional}, since the ``incoming'' data are parameterized by arbitrary tensors over $\Sph^2$ (namely, sections of $S^2\cT^*_X|_{\pa X}\to\pa X\cong\Sph^2$) with appropriate regularity; we make this precise in Proposition~\ref{PropipPResPar} below.\footnote{This is consistent with Proposition~\ref{PropipMero}\eqref{ItipMeroHi}: the b-regularity statement there only applies to elements in the (approximate) nullspace of $N_\sface(\ubar L,\lambda)$ that have \emph{vanishing} incoming data, as encoded in the choice of function space (cf.\ \cite{VasyMicroKerrdS}).} This is why we include a regularity parameter $k$ in Definition~\ref{DefipPRes}. We first give a characterization of~\eqref{EqipPResDef} on the spectral (i.e., Mellin transform) side. (See \cite[Proposition~11.20]{HintzMicro} for a simple version of this.)

\begin{lemma}[Resonant states via residues]
\label{LemmaipPRes}
  The space $\Resspace^k_{\iota^+}(\ubar L,\cE_\sscri^\cC,\lambda)$ is independent of the choice of $\ell_\cK$. For $k=\infty$, it is equal to the space of all tensors on $\iota^+$ of the form
  \begin{equation}
  \label{EqipPResres}
    \res_{z=\lambda}\,\Bigl(t_*^{-z}N_{\iota^+}(\ubar L,z)^{-1}f(z)\Bigr)
  \end{equation}
  where $f$ is a polynomial in $z\in\C$ with values in $\Hb^{\infty,\ \bigl(\la\cE_\sscri^\cC+1\ra',\ell_\sscri+1\bigr),\ \beta-2}(\iota^+)$; here $\beta\in(-\eps_\ind,0)$ is arbitrary. For finite $k$, it is contained in (resp.\ contains) the space of all such tensors where $f$ has b-regularity $k-l$ (resp.\ $k+l$) for some $l=l(\ell_\sscri)$.
\end{lemma}
\begin{proof}
  The independence of~\eqref{EqipPResDef} of the choice of $\ell_\cK$ follows from the fact that its elements have an expansion as in Corollary~\ref{CoripPQhom}\eqref{ItipPQhomExp}.

  Applying $t_*^2\ubar L$ to~\eqref{EqipPResres} and writing $\res_{z=\lambda}(\cdots)=\frac{1}{2\pi i}\oint_\lambda(\cdots)\,\dd z$, one obtains the expression $\frac{1}{2\pi i}\oint_\lambda t_*^{-z}f(z)\,\dd z=0$ since $f$ is holomorphic. Thus, the tensor~\eqref{EqipPResres} lies in the nullspace of $\ubar L$; and in view of~\eqref{EqipPMeroPhg}, it has the form required for membership in~\eqref{EqipPResDef}.

  Conversely, suppose that $u=\sum_{j=0}^K t_*^{-\lambda}(\log t_*)^j u^{(\lambda,j)}\in\Resspace^k_{\iota^+}(\ubar L,\cE_\sscri^\cC,\lambda)$, then (upon replacing $u^{(\lambda,j)}$ by an appropriate scalar multiple) $u=\frac{1}{2\pi i}\oint_\lambda t_*^{-z}\tilde u(z)\,\dd z$ where $\tilde u(z):=\sum_{j=0}^K (z-\lambda)^{-j-1}u^{(\lambda,j)}$. Since $\ubar L u=0$, we have $\res_{z=\lambda}\bigl(t_*^{-z-2}N_{\iota^+}(\ubar L,z)\tilde u(z)\bigr)=0$ for all $t_*$, which implies that $N_{\iota^+}(\ubar L,z)\tilde u(z)$ is holomorphic. If $K$ is the order of the pole of $N_{\iota^+}(\ubar L,z)^{-1}$ at $z=\lambda$, define $f(z)$ to be the Taylor expansion of $N_{\iota^+}(\ubar L,z)\tilde u(z)$ around $z=\lambda$ up to order $K+1$; then $u$ is equal to~\eqref{EqipPResres}. It remains to show that $f(z)$ takes values in $\Hb^{k-2,\ (\la\cE_\sscri^\cC+1\ra',\ell_\sscri+1),\ \beta-2}$. This is clear near $R=0$ since $N_{\iota^+}(\ubar L,z)\in R^{-2}\Diffb^2$ and $u\in\Hb^{k-2,-\eps}$ for all $\eps>0$ there. Near $v=0$ on the other hand, we only need to observe that the term $-2 v(v\pa_v-z)(v\pa_v-1-\ubar S)$ of $N_{\iota^+}(\ubar L,z)$ maps the index set $\la\cE_\sscri^\cC\ra$ (in the sense of~\eqref{EqipPSpaces}) to $\la\cE_\sscri^\cC+1\ra'$ (as follows from Proposition~\ref{PropExFwL} or, more directly, by an inspection of~\eqref{EqWEOpMinkS}), and the other term $v^2\wt{\ubar L}(0)$ maps $\la\cE_\sscri^\cC\ra$, or indeed $\cE_\sscri^\tot$, into $\cE_\sscri^\tot+2\subset\cE_\sscri^\cC+1$ (cf.\ Definition~\ref{DefExP}\eqref{ItExPTotNL}).
\end{proof}

Note that if in~\eqref{EqipPResres} we restrict to $f\in\Hb^{k,\,\ell_\sscri+1,\,\beta-2}$ (i.e., they have a trivial partial expansion at $\scri^+$), then by the \emph{finite}-meromorphic nature of $N_{\iota^+}(\ubar L,z)^{-1}$ asserted in Proposition~\ref{PropipNInv}\eqref{ItipNInvb}, we obtain a \emph{finite}-dimensional subspace
\begin{equation}
\label{EqipPResFin}
\begin{split}
  \Resspace_{\iota^+}(\ubar L,\lambda) &:= \Bigl\{ \res_{z=\lambda}\Bigl(t_*^{-z}N_{\iota^+}(\ubar L,z)^{-1}f(z)\Bigr) \colon \\
    &\quad\hspace{3em} f\ \text{is a polynomial in $z$ with values in $\Hb^{k,\,\ell_\sscri+1,\,\beta-2}$} \Bigr\},\quad \Re\lambda<\ell_\sscri,
\end{split}
\end{equation}
of $\Resspace_{\iota^+}^\infty(\ubar L,\cE_\sscri^\cC,\lambda)$ consisting of infinitely b-regular tensors that is independent of the b-regularity order $k$. The quotient space $\Hb^{k,\ (\la\cE_\sscri^\cC+1\ra',\ell_\sscri+1),\ \beta-2}/\Hb^{k,\ \ell_\sscri+1,\ \beta-2}$ can be identified with the space
\begin{equation}
\label{EqipPResFinHb}
\begin{split}
  &\Hb^{k,\,\bigl[\la\cE_\sscri^\cC+1\ra',\leq\ell_\sscri+1\bigr]} \\
  &\quad := \Biggl\{ \chi_\sscri\Biggl( \bigl( f_{1 1}^{(2,0)}v^2 + f_{1 1}^{(2+(1-e^\Ups)\gamma^\Ups,0)}v^{2+(1-e^\Ups)\gamma^\Ups} \bigr)(\dd x^1)^2 + \sum_{\substack{(\lambda,j)\in\cE_\sscri^\cC+1 \\ \Re\lambda\leq\ell_\sscri+1}} v^\lambda(\log v)^j f_{(\lambda,j)}(\omega)\Biggr) \colon \\
  &\quad \hspace{14em} f_{1 1}^{(2,0)},\,f_{1 1}^{(2+(1-e^\Ups)\gamma^\Ups,0)}\in H^k(\pa X),\ f_{(\lambda,j)}\in H^k(\pa X;S^2\cT^*_X|_{\pa X}) \Biggr\}
\end{split}
\end{equation}
of elements of $\Hb^{k,\,(\la\cE_\sscri^\cC+1\ra',\ell_\sscri+1),\,\infty}(\iota^+)$ that have a finite expansion without any remainder terms; here $\chi_\sscri=\chi_\sscri\in\CIc([0,1)_v)$ is any fixed cutoff function that equals $1$ near $0$. (This space is thus isomorphic to a finite number of copies of $H^k(\pa X)$.) In combination, we have thus proved most of the following result.

\begin{prop}[Parameterization of spaces of resonant states]
\label{PropipPResPar}
  We use the notation of Definition~\usref{DefipPRes}. Let $\lambda\in\C$ with $\Re\lambda<\ell_\sscri$. Let $K(\lambda)$ denote the order of the pole of $N_{\iota^+}(\ubar L,z)^{-1}$, as a map~\eqref{EqipPMeroPhg}, at $z=\lambda$. Then there exists an $l=l(\ell_\sscri)\in\N_0$ such that for all $k\geq l$, the map
  \begin{equation}
  \label{EqipPResPar}
  \begin{split}
    \Res_{\iota^+}(\ubar L,\cE_\sscri^\cC,\lambda) &\colon \Resspace_{\iota^+}(\ubar L,\lambda) \oplus \Poly^{K(\lambda)}\Bigl(\C_z;\Hb^{k,\ \bigl[\la\cE_\sscri^\cC+1\ra',\leq\ell_\sscri+1\bigr]}\Bigr) \to \Resspace_{\iota^+}^{k-l}(\ubar L,\cE_\sscri^\cC,\lambda), \\
      &\quad (u^0,f) \mapsto u^0 + \res_{z=\lambda} \Bigl( t_*^{-z}N_{\iota^+}(\ubar L,z)^{-1}f(z) \Bigr),
  \end{split}
  \end{equation}
  which is defined in a $k$-independent manner, is well-defined and continuous; let us denote it by $\Res_{\iota^+}^k(\ubar L,\cE_\sscri^\cC,\lambda)$ to record the regularity order $k$ of its domain. Moreover, the space $\Resspace_{\iota^+}^k(\ubar L,\cE_\sscri^\cC,\lambda)$ is contained in the range of $\Res_{\iota^+}^{k-l}(\ubar L,\cE_\sscri^\cC,\lambda)$; more precisely, there exists a right inverse of~\eqref{EqipPResPar} for $k=\infty$ which is continuous also as a map
  \[
    \Resspace^k(\ubar L,\cE_\sscri^\cC,\lambda)\to\Resspace_{\iota^+}(\ubar L,\lambda)\oplus\Poly^{K(\lambda)}\Bigl(\C;\Hb^{k-l,\ \bigl[\la\cE_\sscri^\cC+1\ra',\leq\ell_\sscri+1\bigr]}\Bigr)
  \]
  for all $k$.
\end{prop}
\begin{proof}
  The proof of Lemma~\ref{LemmaipPRes} produces an explicit map
  \[
    \Resspace^k(\ubar L,\cE_\sscri^\cC,\lambda)\ni u\mapsto f\in\Poly^{K(\lambda)}\Bigl(\C_z;\Hb^{k-l,\ \bigl(\la\cE_\sscri^\cC+1\ra',\ell_\sscri+1\bigr),\ \beta-2}(\iota^+)\Bigr)
  \]
  with $u=\res_{z=\lambda}(t_*^{-z}N_{\iota^+}(\ubar L,z)^{-1}f(z))$. Here, we fix $\beta\in(-\eps_\ind,0)$; the definition of this map does not depend on $k$ (so it is the continuous extension of this map restricted to the space $\Resspace^\infty(\ubar L,\cE_\sscri^\cC,\lambda)$). We can furthermore uniquely write $f=f_1+f_2$ where $f_1$ and $f_2$ take values in $\Hb^{k-l,\,\ell_\sscri+1,\,\beta-2}$ and $\Hb^{k-l,\ \bigl[\la\cE_\sscri^\cC+1\ra',\leq\ell_\sscri+1\bigr]}$, respectively. The desired right inverse then maps
  \[
    u\mapsto\bigl(\res_{z=\lambda}(t_*^{-z}N_{\iota^+}(\ubar L,z)^{-1}f_1(z)),\ f_2\bigr).\qedhere
  \]
\end{proof}

The map~\eqref{EqipPResPar} parameterizes the quasi-homogeneous nullspace $\Resspace_{\iota^+}^k(\ubar L,\cE_\sscri,\lambda)$ of $\ubar L$ using two pieces of data: one (finite-dimensional) piece $\Resspace_{\iota^+}(\ubar L,\lambda)$ capturing all ``pure'' resonant states (with trivial ``incoming'' data), and another piece (which has a finite description by~\eqref{EqipPResFinHb} but is infinite-dimensional) that encodes nontrivial ``incoming'' data, albeit indirectly (namely, on the level of sources $f$ rather than incoming coefficients such as $u_0$ in Remark~\ref{RmkipPMeroPhg}).

\begin{rmk}[Over-parameterization]
\label{RmkipPResOver}
  The map~\eqref{EqipPResPar} is an over-parameterization of the space $\Resspace_{\iota^+}(\ubar L,\cE_\sscri^\cC,\lambda)$. For example, when $\lambda=1$ (in which case $\Resspace_{\iota^+}(\ubar L,1)=\{0\}$ is trivial by Proposition~\ref{PropipMero}\eqref{ItipMeroInv}), the tensor $N_{\iota^+}(\ubar L,z)^{-1}f$ has no pole at $z=1$ \emph{unless} $f_{1 1}^{(2,0)}\neq 0$ in~\eqref{EqipPResFinHb}.
\end{rmk}

\begin{rmk}[Elements of $\Resspace_{\iota^+}^\infty(\ubar L,\cE_\sscri^\cC,\lambda)$ with trivial expansions]
\label{RmkipPResTriv}
  Suppose $\lambda\in\spec_{\iota^+}^{<\ell_\sscri}(\ubar L,\cE_\sscri^\cC)\setminus\spec_{\iota^+}(\ubar L)$ (e.g., $\lambda=1$), and consider $f$ in~\eqref{EqipPResPar} for which $u:=\Res_{\iota^+}(\ubar L,\cE_\sscri^\cC,\lambda)(0,f)$ is non-zero. Since $\ubar L$ diagonalizes upon decomposing into spherical harmonics, we conclude that if $f$ is supported in sufficiently high spherical harmonic degrees $l$, then the expansion of $u$ at $R=0$ is trivial modulo $\cO(R^{\ell_\cK})$-errors for all $\ell_\cK<3+\eps_\ind$. (Concretely, $l\geq 6$ suffices since all terms in~\eqref{EqipNAsyu0} have spherical harmonic degree $\leq 5$.) Thus, even if one imposes an arbitrarily strong but finite order of vanishing of ``non-pure'' resonant states at $R=0$, there is still an infinite-dimensional space of states satisfying this vanishing condition.
\end{rmk}

\subsection{Inversion of the \texorpdfstring{$\tface$-}{Transition face }normal operator and grafting}
\label{Ssiptf}

The (inverse) Fourier transform in $t_*$ relates asymptotic expansions and decay at $\iota^+\subset M$ and $\tface\subset X_\scbtop^\pm=[\pm[0,1]_\sigma\times X;\{0\}\times\pa X]$ (defined in~\eqref{EqTFXscbtpm})---see, e.g., Proposition~\ref{PropTFHbphg}---and it also relates the $\iota^+$-normal operator of $L_b$ (or equivalently of $\ubar L$) and the $\tface$-normal operator of the spectral family $\wh{L_b}(\sigma)$ (or equivalently of $\wh{\ubar L}(\sigma)$) for real $\sigma$. (This relationship was already exploited around~\eqref{EqipMeroNtf}.) For the purposes of controlling aspects of the low-energy asymptotics for $\wh{L_b}(\sigma)^{-1}$, we now prove analogues of Propositions~\ref{PropipNAsy}--\ref{PropipGr} and Corollary~\ref{CoripGrQhom} on the spectral side for the operators $N_\tface(\ubar L,\pm 1)$ (defined in~\eqref{EqWEtfOp}) and $\wh{L_b}(\sigma)$. The results proved here will be used from~\S\ref{SsD3Alm} below onwards. \emph{For the sake of brevity, we only state results with infinite b-regularity;} we merely point out that results at finite b-regularity orders lose only fixed amounts of b-regularity.

\subsubsection{Precise asymptotics at \texorpdfstring{$\zface$}{the zero face}}
\label{Sssiptf0}

Recall that $\tface=[0,\infty]_{\hat r}\times\pa X$, $\pa X=\Sph^2$, where $\hat r:=\frac{|\sigma|}{\rho}=\hat\rho^{-1}$ is a local defining function of $\ztface=\{0\}\times\pa X$. Recall moreover that $N_\tface(\ubar L,\pm 1)$ is equal to $|\sigma|^{-2}\wh{\ubar L}(\sigma)$, $\pm\sigma>0$, under the identification $\hat r=r|\sigma|$. Since the Fourier transform in $t_*$ intertwines $\pa_{t_*}$ and $-i\sigma$, the computation~\eqref{EqipAsyCompGen} and $\sigma=\pm|\sigma|=\pm\hat r\rho$ suggest that
\begin{equation}
\label{EqiptfIdent}
  N_\tface(\ubar L,\pm 1)\biggl( \ubar h_{\rms 0}^{(0)} + \sum_{j=1}^k (\mp i\hat r)^j \underbrace{(\rho^j\breve{\ubar h}_{\rms 0}^{(0),j})|_{\pa X}}_{\in\CI(\pa X;S^2\cT^*)}\biggr) \in \hat r^{-1+k}\CI([0,1)_{\hat r}\times\pa X)
\end{equation}
for $k=0,1,2,3$. (Recall here that $\breve{\ubar h}_{\rms 0}^{(0),j}$ is homogeneous of degree $-j$ in $\rho$.) This of course also follows by direct computation: writing $\ubar L=\rho^2\wt{\ubar L}(0)(\omega,\rho\pa_\rho,\pa_\omega)+\rho\wt{\ubar L_1}(\rho\pa_\rho)\pa_{t_*}$ where $\wt{\ubar L}(0)$ is as in~\eqref{EqWEtfOp} and $\wt{\ubar L_1}(\rho\pa_\rho)=-2(\rho\pa_\rho-1-\ubar S)$ (cf.\ \eqref{EqWEOpMink2}), we have
\begin{equation}
\label{EqiptfNtf}
\begin{split}
  N_\tface(\ubar L,\pm 1) &= \hat\rho^2\wt{\ubar L}(0)(\omega,\hat\rho\pa_{\hat\rho},\pa_\omega) \mp i\hat\rho\wt{\ubar L_1}(\hat\rho\pa_{\hat\rho}) \\
    &= \hat r^{-2}\wt{\ubar L}(0)(\omega,-\hat r\pa_{\hat r},\pa_\omega) \mp i\hat r^{-1}\wt{\ubar L_1}(-\hat r\pa_{\hat r}).
\end{split}
\end{equation}
Therefore, the identity $\ubar L(t_*\ubar h_{\rms 0}^{(0)}+\breve{\ubar h}_{\rms 0}^{(0),1})$, which is equivalent to the pair of identities
\begin{align*}
  \wh{\ubar L}(0)(\omega,0,\pa_\omega)(\ubar h_{\rms 0}^{(0)}|_{\pa X})&=0, \\
  \wt{\ubar L_1}(0)(\ubar h_{\rms 0}^{(0)}|_{\pa X}) + \wt{\ubar L}(0)(\omega,-1,\pa_\omega)\bigl((\rho\breve{\ubar h}_{\rms 0}^{(0),1})|_{\pa X}\bigr)&=0,
\end{align*}
implies~\eqref{EqiptfIdent} for $k=0,1$; similarly for higher $k$. In order to state the analogue of Proposition~\ref{PropipNAsy}, we are thus led to introduce the following notation (which uses the notation from~\eqref{EqipAsys0}, \eqref{EqipAsys1}, \eqref{EqipAsyRem}, and \eqref{EqipAsyPhys}):

\begin{definition}[Formal solutions of $N_\tface(\ubar L,\pm 1)$ at $\ztface$]
\label{DefiptfFormal}
  For $\lambda=0$ and $\lambda=-\lambda^\Ups_{\rms 0,1}+1$, define
  \begin{equation}
  \label{EqiptfFormal}
    \ubar h_{\rms 0}^{(\lambda),\leq k}(\pm\hat r) := \hat r^{-\lambda}(\rho^{-\lambda}\ubar h_{\rms 0}^{(\lambda)})|_{\pa X} + \sum_{j=1}^k (\mp i\hat r)^j \hat r^{-\lambda} (\rho^{-\lambda-j}\ubar h_{\rms 0}^{(\lambda),j})|_{\pa X},
  \end{equation}
  which is a function of $(\hat r,\omega)$ valued in $\pi_\tface^*(S^2\cT^*_X|_{\pa X})\to\pa X$ (where $\pi_\tface\colon\tface\to\pa X$ is the projection). Define $\ubar h_{\rms l}^{(\lambda),\leq k}(\pm\hat r,\scal)$, $\ubar h_{\rms l}^{(-l+2),\leq k}(\pm\hat r,\scal)$, etc.\ analogously; and finally
  \begin{align*}
    \ubar h_{\rms 1}^{\leq k}(\pm\hat r,\scal) &:= \sum_{j=0}^{k-2} (\mp i\hat r)^j \hat r \bigl(\rho^{1+j}\ubar h_{\rms 1}^{2+j}(\scal)\bigr)\big|_{\pa X}, \\
    \ubar h_{\rmv 1}^{\leq k}(\pm\hat r,\vect) &:= \sum_{j=0}^{k-1} (\mp i\hat r)^j \hat r \bigl(\rho^{1+j}\ubar h_{\rmv 1}^{1+j}(\vect)\bigr)\big|_{\pa X}
  \end{align*}
  for $k=2,3,4$ and $k=1,2,3$, respectively.
\end{definition}

The factor $\hat r^{-\lambda}\rho^{-\lambda}=|\sigma|^{-\lambda}$ in~\eqref{EqiptfFormal} is natural as it commutes with the spectral family. Moreover, $\hat r^{-\lambda}(\rho^{-\lambda}\ubar h_{\rms 0}^{(\lambda)})$ has order $-\lambda$ at $\ztface$, which is consistent with $-\lambda$ being an indicial root of $N_\tface(\ubar L,\pm 1)$ at $\hat r=0$. More to the point, we have
\begin{equation}
\label{EqiptfApprox}
  N_\tface(\ubar L,\pm 1)\ubar h_{\rms 0}^{(\lambda),\leq k}(\pm\hat r) \in \hat r^{-\lambda+k-1}\CI([0,1)_{\hat r}\times\pa X)
\end{equation}
(and similarly for the action on the other tensors defined here), which, given that $N_\tface(\ubar L,\pm 1)\in\hat r^{-2}\Diffb^2$, is $k+1$ powers of $\hat r$ better than what the action on a generic element of $\hat r^{-\lambda}\CI([0,1)\times\pa X)$ would produce.

We will use the objects from Definition~\ref{DefiptfFormal} to describe the $\hat r\to 0$ asymptotics of solutions of $N_\tface(\ubar L,\pm 1)$. We introduce the function
\begin{equation}
\label{EqiptfPowerReg}
  \fl^\beta(\hat r) := \begin{cases} \frac{\hat r^\beta-1}{\beta}, &\beta\in\C\setminus\{0\}, \\ \log\hat r, & \beta=0. \end{cases}
\end{equation}
For any $C_0\in\R$, observe that for $\Re\beta>C_0$, the function $\fl^\beta(\hat r)\in\Hb^{\infty,C_0}([0,1)_{\hat r};|\frac{\dd\hat r}{\hat r}|)$ depends holomorphically on $\beta$ (including at $\beta=0$ when $C_0<0$). On the other hand, the coefficients $\pm\beta^{-1}$ of the terms $\hat r^\beta$ and $\hat r^0$ in the polyhomogeneous expansion of $\fl^\beta(\hat r)$ at $\hat r=0$ are \emph{not} holomorphic at $\beta=0$, but rather meromorphic with simple poles at $\beta=0$. (This subtle difference between polyhomogeneity and conormality is important for understanding the statement of Lemma~\ref{LemmaiptfAsy} below.) --- The somewhat peculiar formulation of the following result is needed for the efficient treatment of source terms with logarithmic terms at $\tface$ in (the proof of) Proposition~\ref{PropiptfGr} below.

\begin{lemma}[Asymptotic expansion at $\hat r=0$]
\label{LemmaiptfAsy}
  Fix $\alpha_0\in\C$ and let $\ell_\scface,\ell_\zface\in\R$ be such that
  \[
    \Re\alpha_0\in(1,2+\eps_\ind),\quad
    \ell_\scface<1,\quad
    -\eps_\ind<\ell_\zface-\Re\alpha_0+2<3+\eps_\ind,
  \]
  and such that $\ell_\zface-\Re\alpha_0+2\neq\Re(-\lambda)$ for all indicial roots $\lambda$ of $\wh{\ubar L}(0)$.\footnote{In particular, $\ell_\zface-\Re\alpha_0+2\notin\N_0$.} Let
  \[
    f_0\in\Hb^{\infty,\ \ell_\scface+1,\ \bigl((0,0),\ell_\zface\bigr)}(\tface;|\tfrac{\dd\hat r}{\hat r}\,\dd\slg|).
  \]
  Fix a defining function $\rho_\ztface\in\CI(\tface)$ of $\ztface=\tface\cap\zface$, e.g., $\rho_\ztface=\frac{\hat r}{\hat r+1}$. For $\alpha\in\C$ close to $\alpha_0$ with $\Re\alpha\in(1,2+\eps_\ind)$, denote by $u_\alpha=u_\alpha(\hat r,\omega)$ the solution of
  \[
    N_\tface(\ubar L,\pm 1)u_\alpha = \rho_\ztface^{-\alpha}f_0(\hat r,\omega) \in \Hb^{\infty,\ \ell_\scface+1,\ \bigl((-\alpha,0),\ell_\zface-\alpha\bigr)}(\tface).
  \]
  produced by Corollary~\usref{CorWEtfb}.\footnote{Fixing $\eps_0<\eps_\ind$, then for $\Re\alpha<2+\eps_0$, so $\Re(-\alpha)>-2-\eps_0>-2-\eps_\ind$ and $\Re(\ell_\zface-\alpha)>-2-\eps_\ind$, this depends holomorphically on $\alpha$ near $\alpha_0$ in the space $\Hb^{\infty,\ \ell_\scface-\eps,\ \beta_\ztface}(\tface)$ for all $\eps>0$ and $-\eps_\ind<\beta_\ztface<\min(0,-\eps_0,\ell_\zface-\Re\alpha_0+2)$.} Fix a cutoff function $\chi_\ztface\in\CI(\tface)$ which equals $1$ near $\ztface=\tface\cap\hat r^{-1}(0)$ and $0$ near $\sctface=\tface\cap\hat r^{-1}(\infty)$. Then we can write
  \begin{subequations}
  \begin{equation}
  \label{EqiptfAsy0}
    u_\alpha(\hat r,\omega)=\chi_\ztface(u_{\alpha,0}+\hat r^{-\alpha+2}u_{\alpha,1}) + (1-\chi_\ztface)u_{\alpha,\infty}
  \end{equation}
  where, for $\alpha$ that are close to $\alpha_0$,
  \begin{equation}
  \label{EqiptfAsy1}
    u_{\alpha,1}\in\Hb^{\infty,\ 0,\ \bigl((0,0),\ell_\zface\bigr)}(\tface),\quad
    u_{\alpha,\infty}\in\Hb^{\infty,\ \ell_\scface-\eps,\ 0}(\tface)
  \end{equation}
  depend holomorphically on $\alpha$ close to $\alpha_0$, and
  \begin{equation}
  \label{EqiptfAsy2}
  \begin{split}
    u_{\alpha,0}(\hat r,\omega) &= \fl^{-\alpha+2}(\hat r)\ubar h_{\rms 1}^{\leq 4}(\pm\hat r,\scal_1^{(-1)}) + \fl^{-\alpha+2}(\hat r)\ubar h_{\rmv 1}^{\leq 3}(\pm\hat r,\vect_1^{(-1)}) + \fl^{-\alpha+2}(\hat r)\scal_0^{(0)}\ubar h_{\rms 0}^{(0),\leq 3}(\pm\hat r) \\
      &\quad + \sum_{\substack{\mu=-\lambda_{\rms l,l+j}^\Ups+1 \\ 0\leq l\leq 3,\ j=0,1, \\ 1\leq l+j\leq 3}} \fl^{-\alpha+2-\{\lambda^\Ups_{\rms l,l+j}\}}(\hat r)\ubar h_{\rms l}^{(\mu),\leq 2}(\pm\hat r,\scal_l^{(\mu)}) \\
      &\quad + \sum_{l=2}^5 \fl^{-\alpha+2}(\hat r)\ubar h_{\rms l}^{(-l+2),\leq 5-l}(\pm\hat r,\scal_l^{(-l+2)}) + \sum_{l=2}^4 \fl^{-\alpha+2}(\hat r)\ubar h_{\rmv l}^{(-l+1),\leq 4-l}(\pm\hat r,\vect_l^{(-l+1)}) \\
      &\quad + \fl^{-\alpha+2}(\hat r)\ubar h_{\rms 2}^{(-2),\leq 1}(\pm\hat r,\scal_2^{(-2)}) + \fl^{-\alpha+2}(\hat r)\ubar h_{\rmv 2}^{(-2),\leq 1}(\pm\hat r,\vect_2^{(-2)}) \\
      &\quad + \fl^{-\alpha+2}(\hat r)\ubar h_{\rms 3}^{(-3)}(\pm\hat r,\scal_3^{(-3)}) + \fl^{-\alpha+2}(\hat r)\ubar h_{\rmv 3}^{(-3)}(\pm\hat r,\vect_3^{(-3)})
  \end{split}
  \end{equation}
  \end{subequations}
  for some $\scal_l^{(\mu)}\in\scalspace_l$ and $\vect_l^{(\mu)}\in\vectspace_l$ which depend holomorphically on $\alpha$ (not made explicit in the notation). Here $\{x\}:=x-\lfloor x\rfloor$ denotes the fractional part of a real number.
\end{lemma}

\begin{rmk}[Form of $u_{\alpha,0}$]
\label{RmkiptfAsyInt}
  The description~\eqref{EqiptfAsy2} is designed to be regular when some of the exponents of $\fl$ tend to $0$:
  \begin{enumerate}
  \item The coefficients of the polyhomogeneous term $u_{\alpha,0}$ are meromorphic, with simple poles when $-\alpha+2$ (which has real part in $(-\eps_\ind,1)$) equals $0$, $\lambda^\Ups_{\rms 0,1}-1$, or $\lambda^\Ups_{\rms l,l+j}-(l+j)$ for some values of $l=1,2,3$, $j=0,1$, and $1\leq l+j\leq 3$; these are thus precisely the values of $-\alpha+2$ which have integer coincidences with the indicial roots of $N_\tface(\ubar L,\pm 1)$ at $\hat r=0$ (which are $-1$ times the indicial roots from Lemma~\ref{LemmaWEInd}). For these exceptional values of $\alpha$, $u_{\alpha,0}$ features logarithmic terms in $\hat r$.
  \item\label{ItiptfAsyInt2} For all \emph{other} fixed values of $\alpha$, the description~\eqref{EqiptfAsy2} can be simplified in that one can replace $\fl^\beta(\hat r)$ by $1$ simply, provided one replaces $\scal_1^{(-1)}$ etc.\ by $-\frac{1}{-\alpha+2}\scal_1^{(-1)}$ etc., since the additional terms $\frac{1}{-\alpha+2}\hat r^{-\alpha+2}\scal_0^{(0)}\ubar h_{\rms 0}^{(0),\leq 3}$ etc.\ can be absorbed into $\hat r^{-\alpha+2}u_{\alpha,1}$ in~\eqref{EqiptfAsy0}. However, this re-writing is not regular in the parameter $\alpha$ when $\alpha$ tends to an exceptional value (e.g., $\frac{1}{-\alpha+2}$ blows up as $\alpha\to 2$).
  \item\label{ItiptfAsyIntReg} If we consider $\chi_\ztface u_{\alpha,0}$ as an element of the space $\Hb^{\infty,\ \beta_\zface,\ 0}(\tface)$ for any fixed weight $\beta_\zface<\min(0,-\Re\alpha_0+2)$, it depends holomorphically on $\alpha$. This either follows from the explicit expression, or from the holomorphicity of $u_\alpha$, $u_{\alpha,1}$, and $u_{\alpha,\infty}$ in~\eqref{EqiptfAsy0} in this sense, which implies that of $u_{\alpha,0}$.
  \end{enumerate}
\end{rmk}

\begin{rmk}[Smooth source terms]
\label{RmkiptfAsySmooth}
  In the case $\alpha=\alpha_0=2$, a subtle difference to Proposition~\ref{PropipNAsy} is that the source term $f_0$ has a nontrivial, \emph{but smooth}, expansion at $\hat r=0$ (whereas $f$ in Proposition~\ref{PropipNAsy} did not have an expansion at $R=0$). Some source terms $f_0$ to which we will ultimately apply Lemma~\ref{LemmaiptfAsy} (via Proposition~\ref{PropiptfGr} below) will arise as (restrictions to $\tface$ of) Fourier transforms of functions with good decay properties at $\cK^+$, which translates into high regularity, but not vanishing, at $\zface$, as discussed after~\eqref{EqTFHbLo}. (In a similar vein, the smooth contributions to $u_{\alpha,1}$ for $\alpha=2$ at $\hat r=0$ are essentially acceptable: if the $(\pm\hat r)^j$-terms, $j=0,1,\ldots$, of solutions of $\tface$-model problems on $X_\scbtop^+$ and $X_\scbtop^-$ match, they contribute only Schwartz terms (at $\cK^+$) to the inverse Fourier transform.)
\end{rmk}

\begin{proof}[Proof of Lemma~\usref{LemmaiptfAsy}]
  Given the identity $N_\tface(\ubar L,\pm 1)(\chi_\ztface u_\alpha)=\chi_\ztface\rho_\ztface^{-\alpha}f_0+[N_\tface(\ubar L,\pm 1),\chi_\ztface]u_\alpha$ and the membership $u_\alpha\in\Hb^{\infty,\ \ell_\scface-\eps,\ -\eps_\ind}(\tface)$, we can reduce to the case that $f_0$ and $u_\alpha$ are supported near $\hat r=0$, say, in $\hat r^{-1}([0,\frac12])$. On $\supp u_\alpha$, we may then without loss take $\rho_\ztface=\hat r$. Strictly speaking, the latter choice can be made only if we allow $f_0$ to depend holomorphically on $\alpha$ in order to absorb the factor $(\rho_\ztface\hat r^{-1})^\alpha$, which we can (and shall) indeed allow for in all of our arguments below. From now on, we fix $\chi_\ztface\in\CIc([0,1))$ to equal $1$ on $[0,\frac12]$.

  Using the notation from~\eqref{EqiptfNtf}, we then study the equation
  \begin{align}
  \label{EqiptfAsyRewrite}
    &(N_0+\hat r N_1)u_\alpha = \hat r^{-\alpha+2}f_0\in\Hb^{\infty,\ \bigl((-\alpha+2,0),\,\ell_\zface-\alpha+2\bigr)}([0,1)_{\hat r}\times\Sph^2), \\
    &\qquad N_0 := \wt{\ubar L}(0)(\omega,-\hat r\pa_{\hat r},\pa_\omega),\quad N_1 := \mp i\wt{\ubar L_1}(-\hat r\pa_{\hat r}) \nonumber
  \end{align}
  using normal operator arguments. (We drop the bundle $\pi_\tface^*(S^2\cT^*_X|_{\pa X})$ from the notation.) We shall extract the leading-order asymptotics of $u_\alpha$ from this and subtract them, appropriately extended (e.g., as in~\eqref{EqiptfAsy2}), from $u_\alpha$; this will lead to modifications of $f_0$ along the way, and in particular we will eliminate the $\hat r^0$, $\hat r^1$, etc.\ terms of $f_0$ successively.

  Suppose then that $u_\alpha\in\Hb^{\infty,\beta}([0,1)_{\hat r}\times\Sph^2)$ for some $\beta>-\eps_\ind$, with holomorphic dependence on $\alpha$ near $\alpha_0$; this indeed holds for $\beta>-\eps_\ind$ close enough to $-\eps_\ind$. We then have
  \begin{equation}
  \label{EqiptfAsyN0}
    N_0 u_\alpha = \hat r^{-\alpha+2}f_0 - \hat r N_1 u_\alpha \in \Hb^{\infty,\ \bigl((-\alpha+2+j,0),\,\tilde\beta\bigr)}([0,1)_{\hat r}\times\Sph^2),\quad \tilde\beta:=\min(\ell_\zface-\Re\alpha+2,\beta+1),
  \end{equation}
  where $j\in\N_0$ is such that $-\Re\alpha+2+j\in(\beta,\tilde\beta]$; or if there is no such $j$, then $N_0 u_\alpha\in\Hb^{\infty,\tilde\beta}$ simply. (We again stress that $f_0$ will be modified in the course of our argument, so at the present stage we are implicitly assuming that the $\hat r^0$, $\ldots$, $\hat r^{j-1}$-terms of $f_0$ have already been solved away, i.e., $f_0\in\Hb^{\infty,((j,0),\ell_\zface)}([0,1)_{\hat r}\times\Sph^2)$. For $j=0$, we are working with the original $f_0$.)

  We pass to the Mellin transform in $\hat r$, with the convention $\hat u(\lambda,\cdot):=\int_0^1 \hat r^{-\lambda}u(\hat r,\cdot)\,\frac{\dd\hat r}{\hat r}$. Write $N_0(\lambda):=\wt{\ubar L}(0)(\omega,-\lambda,\pa_\omega)\in\Diff^2(\Sph^2)$ and $N_1(\lambda)=\mp i\wt{\ubar L_1}(-\lambda)$ for the indicial families of $N_0$ and $N_1$; then
  \begin{equation}
  \label{EqiptfAsyMT}
    \wh{u_\alpha}(\lambda) = N_0(\lambda)^{-1}\wh{f_0}(\lambda+\alpha-2) - N_0(\lambda)^{-1}N_1(\lambda-1)\wh{u_\alpha}(\lambda-1),
  \end{equation}
  initially for $\Re\lambda<\beta$ (with $L^2_{\Im\lambda}$-restriction to $\Re\lambda=\beta$ by Plancherel), but the right-hand side provides a meromorphic continuation to $\Re\lambda<\tilde\beta$, with poles only at indicial roots of $N_0$ or for $\lambda+\alpha-2\in\N_0$ (or both, in which case one gets a double pole); recall here Remark~\ref{RmkTMMellinPhg}, which explains why smooth $f_0$ have Mellin transforms $\wh{f_0}(\mu)$ with poles at $\mu\in\N_0$. By reducing $\tilde\beta$ by an arbitrarily small amount if necessary, we may assume that no $\lambda$ with $\Re\lambda=\tilde\beta$ is an indicial root or lies in $-\alpha+2+\N_0$. (By our assumptions on $\ell_\zface$, no reduction of $\tilde\beta$ is necessary when $\tilde\beta=\ell_\zface-\Re\alpha+2$.)

  \pfstep{Case~1. No integer coincidences.} Suppose that $\alpha_0$ is such that for $\Re\lambda\in(\beta,\tilde\beta)$ we either have $\lambda+\alpha_0-2\notin\N_0$ or $\lambda$ is not an indicial root of $N_0$; this remains valid also for $\alpha$ sufficiently close to $\alpha_0$. Taking the inverse Mellin transform of~\eqref{EqiptfAsyMT} at $\Re\lambda=\beta$ and shifting the contour to $\Re\lambda=\tilde\beta$, we obtain
  \begin{equation}
  \label{EqiptfAsyNoCoinSol}
    u_\alpha(\hat r,\omega) = \chi_\ztface\Bigl(\hat r^{-\alpha+2}\hat r^j u_\alpha^{(j)}(\omega) + u_{\alpha,{\rm ind}}(\hat r,\omega) + \tilde u_\alpha(\hat r,\omega)\Bigr),
  \end{equation}
  and then $u_\alpha^{(j)}(\omega)=N_0(-\alpha+2+j)^{-1}\res_{z=j}\wh{f_0}(z)$ depends holomorphically on $\alpha$ and smoothly on $\omega\in\Sph^2$; or the term $u_\alpha^{(j)}$ is absent when no $j\in\N_0$ with $\Re(-\alpha+2+j)\in(\beta,\tilde\beta)$ exists. Furthermore, $u_{\alpha,{\rm ind}}(\hat r,\omega)$ is the sum of residues at the poles of $N_0(\lambda)^{-1}$ (discussed momentarily), and $\tilde u_\alpha\in\Hb^{\infty,\tilde\beta}$ depends holomorphically on $\alpha$.

  Let us describe $u_{\alpha,{\rm ind}}$ in more detail. The contribution of the indicial root $\lambda=0$ (present only if $0\in(\beta,\tilde\beta)$) is equal to $\scal_0^{(0)}\ubar h_{\rms 0}^{(0)} + \ubar h_{\rms 2}^{(0)}(\scal_2^{(0)})$. We can thus eliminate it by subtracting from $u_\alpha$ the tensor
  \begin{equation}
  \label{EqiptfAsyNoCoin}
    \chi_\ztface\Bigl(\scal_0^{(0)}\ubar h_{\rms 0}^{(0),\leq 3}(\pm\hat r) + \ubar h_{\rms 2}^{(0),\leq 3}(\pm\hat r,\scal_2^{(0)})\Bigr);
  \end{equation}
  note that $\hat r^2 N_\tface(\ubar L,\pm 1)$ maps this into $\hat r^4\CI([0,1)_{\hat r}\times\Sph^2)$, which thus (with some margin) lies in the conormal remainder space $\Hb^{\infty,\ell_\zface-\alpha+2}$ for $\hat r^{-\alpha+2}f_0$ in~\eqref{EqiptfAsyRewrite} since $\ell_\zface-\Re\alpha+2<4$. As already observed in Remark~\ref{RmkiptfAsyInt}\eqref{ItiptfAsyInt2}, we can equally well subtract from $u_\alpha$ the tensor $(\alpha-2)\scal_0^{(0)}\frac{\hat r^{-\alpha+2}-1}{-\alpha+2}\ubar h_{\rms 0}^{(0),\leq 3}(\pm\hat r)=(\alpha-2)\fl^{-\alpha+2}(\hat r)\scal_0^{(0)}\ubar h_{\rms 0}^{(0),\leq 3}(\pm 3)$ and its scalar type $2$ analogue, which differs from~\eqref{EqiptfAsyNoCoin} by a term in $\hat r^{-\alpha+2}\CI([0,1)_{\hat r}\times\Sph^2)$; the $\hat r^{-\alpha+2}\CI(\Sph^2)$ leading-order term of this can absorbed into $u_\alpha^{(j)}$ by re-defining the latter, while the sub-leading terms of class $\hat r^{-\alpha+3}\CI$ can be absorbed into $\tilde u_\alpha$.

  The other indicial roots are treated in the same fashion; for example, the indicial root $\lambda=-\lambda^\Ups_{\rms 0,1}+1$ of $\wh{\ubar L}(0)$ contributes via $\scal_0^{(\lambda)}\ubar h_{\rms 0}^{(\lambda)}(\pm\hat r)$; we can eliminate this by subtracting from $u_\alpha$ the tensor $\chi_\ztface\scal_0^{(\lambda)}\ubar h_{\rms 0}^{(\lambda),\leq 2}(\pm\hat r)\in\hat r^{-\lambda}\CI([0,1)_{\hat r}\times\Sph^2)$. If we subtract not $1$ times this but $(\alpha-3+\lambda^\Ups_{\rms 0,1})\fl^{-\alpha+3-\lambda^\Ups_{\rms 0,1}}(\hat r)=(\alpha-2+\lambda)\fl^{-\alpha+2-\lambda}(\hat r)$ times this from $u_\alpha$, we produce extra contributions of class $\hat r^{-\alpha+2+\lambda}\cdot\hat r^{-\lambda}\CI=\hat r^{-\alpha+2}\CI$, which can be absorbed into $u_\alpha^{(j)}$ and $\tilde u_\alpha$ as before.

  In summary, if we subtract from $u_\alpha$ the terms~\eqref{EqiptfAsyNoCoin} etc.\ (which thus contribute to $u_{\alpha,0}$ in~\eqref{EqiptfAsy2}), the equation~\eqref{EqiptfAsyN0} remains valid but with a different $f_0\in\Hb^{\infty,\ ((0,0),\ell_\zface)}$ (now depending holomorphically on $\alpha$). If, moreover, we subtract from $u_\alpha$ the leading-order term $\hat r^{-\alpha+2}\hat r^j u_\alpha^{(j)}(\omega)$ (which solves away the $\hat r^{-\alpha+2+j}$ leading-order term of $\hat r^{-\alpha+2}f_0$), then the new $u_\alpha$ is of class $\Hb^{\infty,\tilde\beta}$ (like $\tilde u_\alpha$ in~\eqref{EqiptfAsyNoCoinSol}), and the new $f_0$ is of class $\Hb^{\infty,((j+1,0),\ell_\zface)}([0,1)_{\hat r}\times\Sph^2)$, i.e., vanishes to one order more at $\hat r=0$.

  \pfstep{Case~2. Integer coincidences.} For the sake of clarity, let us consider the case that $0\in(\beta,\tilde\beta)$ (so $j=0$ in~\eqref{EqiptfAsyN0}), and $\alpha_0-2=0$, so the first term in~\eqref{EqiptfAsyMT} has a double pole at $0$ when $\alpha=\alpha_0=2$. Let us write
  \[
    \wh{f_0}(\mu) = \mu^{-1}f^{(0)} + \tilde f(\mu)
  \]
  where $f^{(0)}\in\CI(\Sph^2)$, while $\tilde f(\mu)$ is holomorphic in $\mu\in\C$ near $0$ with values in $\CI(\Sph^2)$. We moreover write
  \[
    N_0(\lambda)^{-1} = \lambda^{-1}\Pi^{(0)} + \tilde I_0(\lambda)
  \]
  where $\Pi^{(0)}=\res_{\lambda=0}N_0(\lambda)^{-1}$ is the rank $6=\dim\scalspace_0+\dim\scalspace_2$ operator with range spanned by $\ubar h_{\rms 0}^{(0)}$ and $\ubar h_{\rms 2}^{(0)}(\scalspace_2)\subset\CI(\Sph^2)$, while $\tilde I_0(\lambda)$ is holomorphic near $\lambda=0$ with values in the space of bounded operators $H^k(\Sph^2)\to H^{k+2}(\Sph^2)$ (for all $k$). Writing $\lambda^{-1}\mu^{-1}=\frac{1}{\mu-\lambda}(\lambda^{-1}-\mu^{-1})$, we then have
  \begin{equation}
  \label{EqiptfAsyPoles}
  \begin{split}
    N_0(\lambda)^{-1}\wh{f_0}(\lambda+\alpha-2) &= \frac{1}{\alpha-2}\Bigl(\frac{1}{\lambda}-\frac{1}{\lambda+\alpha-2}\Bigr)\Pi^{(0)}f^{(0)} + \lambda^{-1}\Pi^{(0)}\tilde f(\lambda+\alpha-2) \\
      &\quad + (\lambda+\alpha-2)^{-1}\tilde I_0(\lambda)f^{(0)} + \tilde I_0(\lambda)\tilde f(\lambda+\alpha-2).
  \end{split}
  \end{equation}
  The poles of $N_0(\lambda)^{-1}$ and $\wh{f_0}(\lambda+\alpha-2)$ other than $\lambda=0$ are a positive distance apart for all $\alpha$ near $\alpha_0=2$, so they contribute to $u_\alpha$ like $u_{\alpha,{\rm ind}}$ in~\eqref{EqiptfAsyNoCoinSol}; we denote this contribution by $u_{\alpha,{\rm ind},{\rm rest}}$ below. Upon passing to the inverse Mellin transform in~\eqref{EqiptfAsyMT} at $\Re\lambda=\beta$ and shifting the contour to $\Re\lambda=\tilde\beta$, we thus obtain
  \begin{align*}
    u_\alpha(\hat r,\omega) &= \chi_\ztface\Bigl( -\frac{1}{\alpha-2}(1-\hat r^{-\alpha+2})(\Pi^{(0)}f^{(0)})(\omega) + u_{\alpha,{\rm ind},0}(\omega) + \hat r^{-\alpha+2}u_\alpha^{(0)}(\omega) \\
      &\quad\hspace{18em} + u_{\alpha,{\rm ind},{\rm rest}}(\hat r,\omega) + \tilde u_\alpha(\hat r,\omega) \Bigr),
  \end{align*}
  with $u_{\alpha,{\rm ind},0}$ and $\hat r^{-\alpha+2}u_\alpha^{(0)}$ capturing the contributions of the second and third terms in~\eqref{EqiptfAsyPoles}. We can eliminate $u_{\alpha,{\rm ind},{\rm rest}}$ as before. Note that $u_{\alpha,{\rm ind},0}$ is a linear combination of $\ubar h_{\rms 0}^{(0)}$ and $\ubar h_{\rms 2}^{(0)}(\scal)$ for some $\scal\in\scalspace_2$; but writing $\ubar h_{\rms 0}^{(0)}=\frac{1}{\alpha-2}(1-\hat r^{-\alpha+2})(\alpha-2)\ubar h_{\rms 0}^{(0)}+\hat r^{-\alpha+2}\ubar h_{\rms 0}^{(0)}$, we can absorb the second term into $u_\alpha^{(0)}$ and the first term into $\Pi^{(0)}f^{(0)}$ at the (harmless) expense of making the latter $\alpha$-dependent (in a holomorphic fashion); in this manner, we can eliminate the term $u_{\alpha,{\rm ind},0}$ and are left to deal with a contribution
  \begin{equation}
  \label{EqiptfAsyPoleContr}
    \chi_\ztface \fl^{-\alpha+2}(\hat r)\Bigl(\scal_0^{(0)}(\alpha)\ubar h_{\rms 0}^{(0)} + \ubar h_{\rms 2}^{(0)}\bigl(\scal_2^{(0)}(\alpha)\bigr)\Bigr)
  \end{equation}
  to $u_\alpha$, where $\scal_l^{(0)}(\alpha)\in\scalspace_l$, $l=0,2$, is holomorphic in $\alpha$ near $\alpha_0$.

  We shall eliminate~\eqref{EqiptfAsyPoleContr} by subtracting from $u_\alpha$ the term
  \begin{equation}
  \label{EqiptfAsyPoleContr2}
    \chi_\ztface\fl^{-\alpha+2}(\hat r)\Bigl(\scal_0^{(0)}(\alpha)\ubar h_{\rms 0}^{(0),\leq 3}(\pm\hat r) + \ubar h_{\rms 2}^{(0),\leq 3}\bigl(\pm\hat r,\scal_2^{(0)}(\alpha)\bigr)\Bigr).
  \end{equation}
  We need to evaluate the action of the operator $\hat r^2 N_\tface(\ubar L,\pm 1)$ on this term. Commuting this operator through $\chi_\ztface$ yields an error that can be absorbed into the remainder term of $f_0$ in~\eqref{EqiptfAsyN0}, so it remains to consider
  \[
    \hat r^2 N_\tface(\ubar L,\pm 1)\Bigl(\fl^{-\alpha+2}(\hat r)\ubar h_{\rms 0}^{(0),\leq 3}(\pm\hat r)\Bigr)
  \]
  (and similarly the scalar type $2$ term). This is equal to
  \[
    \fl^{-\alpha+2}(\hat r) \underbrace{\hat r^2 N_\tface(\ubar L,\pm 1)\ubar h_{\rms 0}^{(0),\leq 3}(\hat r)}_{\in\hat r^4\CI([0,1)_{\hat r}\times\Sph^2)}{} + [N_0+\hat r N_1,\fl^{-\alpha+2}(\hat r)]\ubar h_{\rms 0}^{(0),\leq 3}.
  \]
  The first summand depends holomorphically on $\alpha$ near $2$ as an element of $\Hb^{\infty,\gamma}$ for all fixed $\gamma<4$, and can thus be absorbed into the conormal remainder of $\hat r^{-\alpha+2}f_0$ in~\eqref{EqiptfAsyN0}. For the second summand, we note that $[\hat r\pa_{\hat r},\fl^\beta(\hat r)]=[\hat r\pa_{\hat r},\beta^{-1}(\hat r^\beta-1)]=\hat r^\beta$, so $[N_0+\hat r N_1,\fl^{-\alpha+2}(\hat r)]\in\hat r^{-\alpha+2}\Diffb^1([0,1)_{\hat r}\times\Sph^2)$ maps $\ubar h_{\rms 0}^{(0),\leq 3}$ into $\hat r^{-\alpha+2}\CI([0,1)_{\hat r}\times\Sph^2)$ (with holomorphic dependence of the $\CI$-factor on $\alpha$); as in the arguments in Case~1, the $\hat r^{-\alpha+2}$ leading-order term can be absorbed into $u_\alpha^{(0)}$, while the $\hat r^{-\alpha+3}\CI$ remainder can be put into $\tilde u_\alpha$.

  Upon subtracting from $u_\alpha$ the terms~\eqref{EqiptfAsyPoleContr2}, $\hat r^{-\alpha+2}u_\alpha^{(0)}$, and also $u_{\alpha,{\rm ind},{\rm rest}}$ using the arguments from Case~1, we obtain a new $u_\alpha$ which is now of class $\Hb^{\infty,\tilde\beta}$ (with holomorphic dependence on $\alpha$) and a new $f_0$ that is of class $\Hb^{\infty,((1,0),\ell_\zface)}$.

  The arguments at all other exceptional values of $\alpha$ are completely analogous. After finitely many iterations of these arguments, we have reduced to the case that $u_\alpha$ is of class $\Hb^{\infty,\ell_\zface-\alpha+2}$, which is captured by the term involving $u_{\alpha,1}$ in~\eqref{EqiptfAsy0}.
\end{proof}

\subsubsection{Grafting into the low-energy resolvent space}
\label{SssiptfGr}

With $|\sigma|^2 N_\tface(\ubar L,\pm 1)$ being the model operator for $\wh{L_b}(\sigma)$ at $\tface\subset X_\scbtop^\pm$ for $\pm\sigma\in[0,1]$, we now extend solutions of $\tface$-model problems to a full neighborhood of $\zface\cup\tface$ without incurring additional index sets of the remaining error at $\zface$ (which one might naively expect from the terms in~\eqref{EqiptfAsy2}); the key, as in Proposition~\ref{PropipGr}, is to realize that the terms in~\eqref{EqiptfAsy2} are the leading-order terms of large generalized zero energy states on Kerr (now interpreted on the spectral side). Note that $\mp i\hat r\rho=-i\sigma$ for $\hat r=\frac{|\sigma|}{\rho}=|\sigma|r$, $\sigma=\pm|\sigma|$. We thus introduce:

\begin{definition}[Formal solutions of $\wh{L_b}(\sigma)$ at $\zface$]
\label{DefiptfFormalKerr}
  For $\lambda=0$ and $\lambda=-\lambda_{\rms 0,1}^\Ups+1$, define
  \[
    h_{b,\rms 0}^{(\lambda),\leq k}(\sigma) := h_{b,\rms 0}^{(\lambda)} + \sum_{j=1}^k (-i\sigma)^j \breve h_{b,\rms 0}^{(\lambda),j},
  \]
  which is a function of $(\sigma,r,\omega)$ valued in $S^2\cT^*X$. Define $h_{b,\rms l}^{(\lambda),\leq k}(\sigma,\scal)$, $h_{b,\rms l}^{(-l+2),\leq k}(\sigma,\scal)$, etc.\ analogously. Finally, we set
  \begin{align}
  \label{EqiptfFormalKerrs1}
    \breve h_{b,\rms 1}^{\leq k}(\sigma,\scal) &:= \sum_{j=1}^k (-i\sigma)^{j-1}\breve h_{b,\rms 1}^j(\scal), \\
    h_{b,\rmv 1}^{\leq k}(\sigma,\vect) &:= h_{b,\rmv 1}(\vect) + \sum_{j=1}^k (-i\sigma)^j\breve h_{b,\rmv 1}^j(\vect). \nonumber
  \end{align}
\end{definition}

Thus, for example, $\wh{L_b}(\sigma)h_{b,\rms 0}^{(\lambda),\leq k}(\sigma)$ vanishes to order $\cO(|\sigma|^{k+1})$ as $|\sigma|\to 0$. Carefully note that~\eqref{EqiptfFormalKerrs1} is exceptional in that the term $h_{b,\rms 1}$ is absent; it will be inserted in a particular manner (see~\eqref{EqiptfGrKerr} and Remark~\ref{RmkiptfGrs1}). We do have
\begin{equation}
\label{EqiptfFormals1}
  \wh{L_b}(\sigma) \bigl( h_{b,\rms 1}(\scal) - i\sigma\breve h_{b,\rms 1}^{\leq 4}(\scal) \bigr) \in |\sigma|^5\CI([0,1)_{|\sigma|}\times X^\circ),\quad \pm\sigma\geq 0.
\end{equation}

\begin{prop}[Solving away leading-order terms at $\tface$, I: polyhomogeneous sources]
\label{PropiptfGr}
  Let $\alpha_0\in\C$ with $\Re\alpha_0\in(1,2+\eps_\ind)$, and let $k\in\N_0$. Let $\ell_\scface<1$ and $-\eps_\ind<\ell_\zface-\Re\alpha_0+2$, $\ell_\zface-\min(\Re\alpha_0-2,0)<3+\eps_\ind$, with $\ell_\zface-\Re\alpha_0+2\notin\Re(-\lambda)$ for all indicial roots $\lambda$ of $\wh{L_b}(0)$. Let
  \[
    f = f(\sigma,r,\omega) \in \Hb^{\infty,\ \ell_\scface+1,\ (\alpha_0,k),\ ((0,0),\ell_\zface)}(X_\scbtop^\pm;S^2\cT^*_X).
  \]
  \begin{enumerate}
  \item\label{ItiptfGrRough}{\rm (Rough description.)} There exists
    \begin{equation}
    \label{EqiptfGru}
      u = u(\sigma,r,\omega) \in \bigcap_{\eps>0}\Hb^{\infty,\ \ell_\scface-\eps,\ \alpha_0-2-\eps,\ \min(-1,\alpha_0-3)-\eps}(X_\scbtop^\pm;S^2\cT^*_X),
    \end{equation}
    depending smoothly on the Kerr parameters $b$ near $b_0$, such that
    \begin{equation}
    \label{EqiptfGrErr}
      \bigl(f(\sigma) - \wh{L_b}(\sigma)u(\sigma)\bigr)\big|_{\sigma\in\pm[0,1]} \in \Hb^{\infty,\ \ell_\scface+1,\ (\alpha_0,k)+\cE_\ind,\ ((0,0),\ell_\zface)}(X_\scbtop^\pm;S^2\cT^*_X),
    \end{equation}
    where $\cE_\ind$ is an index set (independent of $b$, $\alpha_0$, and $k$) with $\min\Re\cE_\ind\geq\eps_\ind$.
  \item\label{ItiptfGrPrec}{\rm (Precise description.)} Let $\chi_\tface\in\CI(X_\scbtop)$ be equal to $1$ near $\tface$, with support in a small neighborhood thereof, and let $\chi_\zface\in\CI(X_\scbtop)$ be equal to $1$ near $\zface$ and $0$ near $\scface$. Recall $\hat r=|\sigma|r=\frac{|\sigma|}{\rho}$. For $k=0$, we can then take $u$ (with the properties~\eqref{EqiptfGru}--\eqref{EqiptfGrErr}) to be of the form
    \begin{subequations}
    \begin{equation}
    \label{EqiptfGruPrec}
      u(\sigma,r,\omega) = \chi_\zface|\sigma|^{\alpha_0-2}u_b(\sigma,r,\omega) + \chi_\tface \rho_\tface^{\alpha_0-2}u_\tface(\hat r,\omega)
    \end{equation}
    where $u_\tface\in\Hb^{\infty,\ \ell_\scface-\eps,\ ((0,0),\ell_\zface)}(\tface)$ and (recalling $\fl^\beta$ from~\eqref{EqiptfPowerReg})
    \begin{equation}
    \label{EqiptfGruPrec2}
    \begin{split}
      u_b &= -\fl^{-\alpha_0+2}(|\sigma|)|\sigma|\sigma^{-2}h_{b,\rms 1}(\scal_1^{(-1)}) + \fl^{-\alpha_0+2}(\hat r)|\sigma|i\sigma^{-1}\breve h_{b,\rms 1}^{\leq 4}(\sigma,\scal_1^{(-1)}) \\
        &\qquad + \fl^{-\alpha_0+2}(\hat r)|\sigma|i\sigma^{-1}h_{b,\rmv 1}^{\leq 3}(\sigma,\vect_1^{(-1)}) \\
        &\qquad + \fl^{-\alpha_0+2}(\hat r)\scal_0^{(0)} h_{b,\rms 0}^{(0),\leq 3}(\sigma) \\
        &\qquad + \sum_{\substack{\mu=-\lambda^\Ups_{\rms l,l+j}+1 \\ 0\leq l\leq 3,\ j=0,1, \\ 1\leq l+j\leq 3}} \fl^{-\alpha_0+2-\{\lambda^\Ups_{\rms l,l+j}\}}(\hat r)|\sigma|^{-\mu}h_{b,\rms l}^{(\mu),\leq 2}(\sigma,\scal_l^{(\mu)}) \\
        &\qquad + \sum_{l=2}^5 \fl^{-\alpha_0+2}(\hat r)|\sigma|^{l-2}h_{b,\rms l}^{(-l+2),\leq 5-l}(\sigma,\scal_l^{(-l+2)}) \\
        &\qquad + \sum_{l=2}^4 \fl^{-\alpha_0+2}(\hat r)|\sigma|^{l-1}h_{b,\rmv l}^{(-l+1),\leq 4-l}(\sigma,\vect_l^{(-l+1)}) \\
        &\qquad + \fl^{-\alpha_0+2}(\hat r)|\sigma|^2 h_{b,\rms 2}^{(-2),\leq 1}(\sigma,\scal_2^{(-2)}) + \fl^{-\alpha_0+2}(\hat r)|\sigma|^2 h_{b,\rmv 2}^{(-2),\leq 1}(\sigma,\vect_2^{(-2)}) \\
        &\qquad + \fl^{-\alpha_0+2}(\hat r)|\sigma|^3 h_{b,\rms 3}^{(-3)}(\scal_3^{(-3)}) + \fl^{-\alpha_0+2}(\hat r)|\sigma|^3 h_{b,\rmv 3}^{(-3)}(\vect_3^{(-3)})
    \end{split}
    \end{equation}
    \end{subequations}
    for some $\scal_l^{(\mu)}\in\scalspace_l$ and $\vect_l^{(\mu)}\in\vectspace_l$. For general values of $k\in\N_0$, we can take $u$ to be a sum of terms of the form~\eqref{EqiptfGruPrec} and up to $k$-fold derivatives of such terms in the parameter $\alpha_0$.
  \end{enumerate}
\end{prop}

Mirroring Remark~\ref{RmkiptfAsyInt}\eqref{ItiptfAsyInt2}, one can, for $\alpha_0\neq 2$, replace the contribution
\[
  \chi_\zface|\sigma|^{\alpha_0-2}\fl^{-\alpha_0+2}(\hat r)\scal_0^{(0)}h_{b,\rms 0}^{(0),\leq 3}(\sigma)
\]
to $\chi_\zface|\sigma|^{\alpha_0-2}u_b$ by $-\frac{|\sigma|^{\alpha_0-2}}{-\alpha_0+2}\scal_0^{(0)}h_{b,\rms 0}^{(0),\leq 3}(\sigma)$; the other term $\chi_\zface\frac{\rho^{\alpha_0-2}}{-\alpha_0+2}\scal_0^{(0)}h_{b,\rms 0}^{(0),\leq 3}(\sigma)$ is smooth at $\sigma=0$ and could thus be combined with $\chi_\zface\rho_\tface^{\alpha_0-2}u_\tface\in\Hb^{\infty,\ \ell_\scface-\eps,\ (\alpha_0-2,0)\cup(\alpha_0-2+\cE_\ind),\ ((0,0),\ell_\zface)}(X_\scbtop^\pm)$; similarly for the other terms. The formulation~\eqref{EqiptfGruPrec2} has the advantage of being valid across $\alpha_0=2$ (and other exceptional values of $\alpha_0$) and differentiable at all $\alpha_0$ (as needed for the final statement of Proposition~\ref{PropiptfGr}).

\begin{rmk}[Size of the source term]
\label{RmkiptfGrSize}
  Proposition~\ref{PropiptfGr} asserts that $u$ solves away $f$ to leading order at $\tface$ without creating any new error terms at $\zface$ that have worse index sets than what $f$ itself has. Note also that for $\Re\alpha_0\leq 2$, the restriction of $f$ to $\zface$ is of class $\cA^{\alpha_0}(X)$ (when $k=0$), which fails to lie in the domain of $\wh{L_b}(0)^{-1}$ (or an augmentation thereof) as far as weights are concerned (and thus ignoring the failure of invertibility of the latter), cf.\ Corollary~\ref{CorWE0Solv}. In the computation of the low-energy resolvent output, the inversion of the $\tface$-model operator thus becomes necessary when the zero energy operator inverse cannot be applied. (This was stated explicitly already in \cite{HintzPrice} and recalled in~\S\ref{SssIGenK}.)
\end{rmk}

\begin{rmk}[Size of the approximate solution]
\label{RmkiptfGrSizeSol}
  The weaker the $\tface$-order of $f$ is, the more singular the approximate solution $u$ constructed in Proposition~\ref{PropiptfGr} is at $\zface$. This phenomenon already occurs for the spectral family for the Laplacian on Euclidean $\R^3$, with $(\Delta-\sigma^2)^{-1}\la x\ra^{-\alpha}$, $\alpha\in(1,2]$, having a $|\sigma|^{-2+\alpha}$-singularity ($\log|\sigma|$ when $\alpha=2$) as $|\sigma|\to 0$. (A uniform description as in~\eqref{EqiptfGruPrec}--\eqref{EqiptfGruPrec2} is $|\sigma|^{\alpha-2}\fl^{-\alpha+2}(|\sigma|)$.) The singularity of~\eqref{EqiptfGru} is one order stronger than this still; the origin of this singularity is the exceptional $h_{b,\rms 1}$ term in~\eqref{EqiptfGruPrec2}. Note also that the term $h_{b,\rms 0}^{(0),\leq 3}$ in~\eqref{EqiptfGruPrec2} contributes to~\eqref{EqiptfGruPrec} via \emph{two} terms when $\alpha_0\neq 2$, namely via the singular term $|\sigma|^{\alpha_0-2}h_{b,\rms 0}^{(0),\leq 3}$ and the smooth (in $\sigma$) term $h_{b,\rms 0}^{(0),\leq 3}$; similarly, say, $h_{b,\rms 2}^{(-2),\leq 1}$ contributes via the singular term $|\sigma|^{\alpha_0-2}|\sigma|^2 h_{b,\rms 2}^{(-2),\leq 1}$ and the smooth term $|\sigma|^2 h_{b,\rms 2}^{(-2),\leq 1}$. When $\alpha_0\to 2$, the two terms combine to give logarithmic singularities, i.e., $(\log|\sigma|)h_{b,\rms 0}^{(0),\leq 3}$ and $(\log|\sigma|)|\sigma|^2 h_{b,\rms 2}^{(-2),\leq 1}$.
\end{rmk}

\begin{rmk}[Relationship to approximate solutions at $\iota^+$]
\label{RmkiptfGrCompip}
  For $\alpha_0\neq 2$, say, the $\cO(\rho_\tface^{\alpha_0})$-size of the source term at $\tface$ roughly corresponds (cf.\ \eqref{EqTFHbLo} and \eqref{EqTFHbInvLo}) to a $\cO(\rho_+^{\alpha_0+1})$ source at $\iota^+$; in the context of Proposition~\ref{PropipGr}, this creates, among other things, a term $t_*^{-\alpha_0+1}h_{b,\rms 0}^{(0)}$, which on the Fourier transform side is essentially $|\sigma|^{\alpha_0-2}h_{b,\rms 0}^{(0)}$, which we do have in~\eqref{EqiptfGruPrec}--\eqref{EqiptfGruPrec2} as well (keeping in mind the observation in Remark~\ref{RmkiptfGrSizeSol}).
\end{rmk}

\begin{proof}[Proof of Proposition~\usref{PropiptfGr}]
  Write $\rho_\tface\in\CI(X_\scbtop^\pm)$ for a defining function of $\tface$ (e.g., $\rho+|\sigma|$), then $\rho_\tface=|\sigma|\rho_\zface^{-1}$ where $\rho_\zface\in\CI(X_\scbtop^\pm)$ is a defining function of $\zface$. It suffices to solve away a single term
  \begin{equation}
  \label{EqiptfGrTerm}
    \chi_\tface\rho_\tface^{\alpha_0}(\log\rho_\tface)^k f_0(\hat r,\omega),\quad
    f_0 \in \Hb^{\infty,\ \ell_\scface+1,\ ((0,0),\ell_\zface)}(\tface),
  \end{equation}
  in the expansion of $f$ at $\tface$.

  \pfstep{Step~1. No logarithms at $\tface$.} When $k=0$, we write $\rho_\tface^{\alpha_0}f_0=|\sigma|^{\alpha_0}\rho_\zface^{-\alpha_0}f_0$; with $N_\tface(\ubar L,\pm 1)$ being the $\tface$-normal operator of $|\sigma|^{-2}\wh{L_b}(\sigma)$, the strategy is to solve away $\chi_\tface\rho_\tface^{\alpha_0}f_0$ at $\tface$ using $|\sigma|^{\alpha_0-2}$ times a suitable grafting of
  \[
    u_{\tface,\alpha_0}(\hat r,\omega) := N_\tface(\ubar L,\pm 1)^{-1}(\rho_\zface^{-\alpha_0}f_0)
  \]
  into $X_\scbtop^\pm$. In order to prepare for the treatment of the case $k\geq 1$ via differentiation in $\alpha_0$, we do this grafting here in a fashion that is regular also across exceptional values of $\alpha_0$. Thus, with $\chi_\ztface:=\chi_\zface|_\tface$, we use Lemma~\ref{LemmaiptfAsy} to write
  \begin{equation}
  \label{EqiptfGrutf}
  \begin{split}
    u_{\tface,\alpha_0} &= \chi_\ztface ( u_{\tface,\alpha_0,0} + \hat r^{-\alpha_0+2}u_{\tface,\alpha_0,1} ) + (1-\chi_\ztface)u_{\tface,\alpha_0,\infty}, \\
    &\quad u_{\tface,\alpha_0,1}\in\Hb^{\infty,\ 0,\ ((0,0),\ell_\zface)}(\tface),\quad
    u_{\tface,\alpha_0,\infty}\in\Hb^{\infty,\ \ell_\scface-\eps,\ 0}(\tface).
  \end{split}
  \end{equation}
  For the sake of brevity, let us only discuss the grafting of two of the summands in~\eqref{EqiptfAsy2} (the remaining cases being completely analogous); so let us suppose that
  \begin{equation}
  \label{EqiptfGru0}
    u_{\tface,\alpha_0,0} = \fl^{-\alpha_0+2}(\hat r)\ubar h_{\rms 1}^{\leq 4}(\pm\hat r,\scal_1^{(-1)}) + \fl^{-\alpha_0+2}(\hat r)\ubar h_{\rms 2}^{(-2),\leq 1}(\pm\hat r,\scal_2^{(-2)}).
  \end{equation}
  We then define $u_{\alpha_0}=u_{\alpha_0}(\sigma,r,\omega)$ by
  \begin{subequations}
  \begin{equation}
  \label{EqiptfGrDef}
    u_{\alpha_0} := |\sigma|^{\alpha_0-2}\chi_\zface u_{b,\alpha_0} + |\sigma|^{\alpha_0-2}\chi_\tface\bigl( \chi_\zface\hat r^{-\alpha_0+2}u_{\tface,\alpha_0,1}(\hat r,\omega) + (1-\chi_\zface)u_{\tface,\alpha_0,\infty}(\hat r,\omega) \bigr),\quad \hat r=\frac{|\sigma|}{\rho},
  \end{equation}
  where we set
  \begin{equation}
  \label{EqiptfGrKerr}
  \begin{split}
    u_{b,\alpha_0} &:= -\fl^{-\alpha_0+2}(|\sigma|)|\sigma|\sigma^{-2}h_{b,\rms 1}(\scal_1^{(-1)}) + \fl^{-\alpha_0+2}(\hat r)|\sigma|i\sigma^{-1}\breve h_{b,\rms 1}^{\leq 4}(\sigma,\scal_1^{(-1)}) \\
      &\qquad + \fl^{-\alpha_0+2}(\hat r) |\sigma|^2 h_{b,\rms 2}^{(-2),\leq 1}(\sigma,\scal_2^{(-2)}).
  \end{split}
  \end{equation}
  \end{subequations}
  (We explain the exceptional form of the first summand below; see also Remark~\ref{RmkiptfGrs1}.) We note that since the terms comprising $u_{\tface,\alpha_0}$ are holomorphic in appropriate senses (see also Remark~\ref{RmkiptfAsyInt}\eqref{ItiptfAsyIntReg}), also $(u_\alpha(\sigma))|_{\sigma\in\pm[0,1]}$ is holomorphic in $\alpha$ near $\alpha_0$ when regarded as an element
  \begin{equation}
  \label{EqiptfGrHol}
    u_\alpha \in \Hb^{\infty,\ \ell_\scface-\eps,\ \alpha_0-2-\eps,\ \min(-1,\alpha_0-3)-\eps}(X_\scbtop^\pm),
  \end{equation}
  for any fixed $\eps>0$. (The dominant contribution at $\zface$ arises from $|\sigma|^{\alpha_0-2}$ times the first term of~\eqref{EqiptfGrKerr}.) For $\alpha=\alpha_0$, the arguments in Step~1.1 below give
  \[
    u_{\alpha_0} \in \Hb^{\infty,\ \ell_\scface-\eps,\ (\alpha_0-2,0)\cup(\alpha_0-2+\cE_\ind),\ \min(-1,\alpha_0-3)-\eps}(X_\scbtop^\pm).
  \]

  In the context of the description~\eqref{EqiptfGruPrec}, we note that the second summand of~\eqref{EqiptfGrDef} can be written as $\chi_\tface$ times $\rho^{\alpha_0-2}\chi_\zface u_{\tface,\alpha_0,1}+|\sigma|^{\alpha_0-2}(1-\chi_\zface)u_{\tface,\alpha_0,\infty}$; but near $\supp\chi_\zface$, resp.\ $\supp(1-\chi_\zface)$, the function $\rho$, resp.\ $|\sigma|$ is a local defining function of $\tface$, so we can combine both terms into $\rho_\tface^{\alpha_0-2}u_\tface$ in~\eqref{EqiptfGruPrec}.

  \pfsubstep{Step~1.1.}{Verification that $|\sigma|^{-\alpha_0+2}u_{\alpha_0}$ extends $u_{\tface,\alpha_0}$ off $\tface.$} Comparing~\eqref{EqiptfGrDef} with~\eqref{EqiptfGrutf}, we need to show that the restriction of $u_{b,\alpha_0}$ to $\tface$ is equal to $\chi_\ztface u_{\tface,\alpha_0,0}$. The restriction of $|\sigma|^2 h_{b,\rms 2}^{(-2),\leq 1}(\sigma,\scal_2^{(-2)})=\hat r^2\rho^2 h_{b,\rms 2}^{(-2),\leq 1}(\sigma,\scal_2^{(-2)})$ to $\tface$ is equal to $\ubar h_{\rms 2}^{(-2),\leq 1}(\pm\hat r,\scal_2^{(-2)})$, and, by the same token, the restrictions of $|\sigma|i\sigma^{-1}(-i\sigma)^{j-1}\breve h_{b,\rms 1}^j(\scal_1^{(-1)})=|\sigma|(-i\sigma)^{j-2}\breve h_{b,\rms 1}^j(\scal_1^{(-1)})$ for $j=2,3,4$ to $\tface$ add up to $\ubar h_{\rms 1}^{\leq 4}(\pm\hat r,\scal_1^{(-1)})$ (cf.\ Definition~\ref{DefiptfFormal}). On the other hand, the terms
  \begin{align*}
    \chi_\zface\fl^{-\alpha_0+2}(|\sigma|)|\sigma|\sigma^{-2} h_{b,\rms 1}(\scal_1^{(-1)}) &\in \cA^{\infty,\ \min(1,3-\Re\alpha_0),\ \min(-1,-\Re\alpha_0+1)}(X_\scbtop^\pm), \\
    \chi_\zface\fl^{-\alpha_0+2}(\hat r)|\sigma|i\sigma^{-1}\breve h_{b,\rms 1}^1(\scal_1^{(-1)}) &\in \cA^{\infty,\ 1,\ \min(0,-\Re\alpha_0+1)}(X_\scbtop^\pm),
  \end{align*}
  which have no analogue in~\eqref{EqiptfGru0}, vanish at $\tface$. A consequence of this leading-order behavior at $\tface$ is that $\chi_\tface\rho_\tface^{\alpha_0}f_0 - \wh{L_b}(\sigma)u_{\alpha_0}$ is of size $o(\rho_\tface^{\alpha_0})$ at $\tface$, since the restriction of $|\sigma|^{-\alpha_0}$ times this to $\tface$ is given by $\rho_\zface^{-\alpha_0}f_0-N_\tface(\ubar L,\pm 1)u_{\tface,\alpha_0}$.

  \pfsubstep{Step~1.2.}{Decay of the remaining error term.} We claim that
  \begin{equation}
  \label{EqiptfGrError}
    \chi_\tface\rho_\tface^{\alpha_0}f_0 - \wh{L_b}(\sigma)u_{\alpha_0} \in \Hb^{\infty,\ \ell_\scface+1,\ \alpha_0+\cE_\ind,\ ((0,0),\ell_\zface)}(X_\scbtop^\pm).
  \end{equation}
  The fact that the $\tface$-index set is $\alpha_0+\cE_\ind$ (and thus $\min\Re\cE_\ind\geq\eps_\ind$ better than the index set $(\alpha_0,0)$ of $\rho_\tface^{\alpha_0}f_0$) follows from the arguments in Step~1.1 and the polyhomogeneity of $h_{b,\rms 1}$ etc.\ appearing in~\eqref{EqiptfGrKerr} (as constructed in Propositions~\ref{PropWG0Symm}, \ref{PropWG0Large}, and \ref{PropWE0}). That the $\scface$-decay order is, in fact, $\ell_\scface+2-\eps$ (and we only record $\ell_\scface+1$ in the statement of the Proposition) follows from the observation that $\wh{L_b}(\sigma)-|\sigma|^2 N_\tface(\ubar L,\pm 1)\in\rho_\scface^2\rho_\tface^3\Diffb^2(X_\scbtop^\pm)$ (as follows from~\eqref{EqWEOpKerr}, \eqref{EqWEOpKerrMink}, and~\eqref{EqWEtfOp}), the point being the order $2$ vanishing at $\scface$.

  It remains to discuss the $\zface$-order, which we only need to determine at $\zface^\circ$. To this end, note that the $\zface$-order of the second summand in~\eqref{EqiptfGrDef} is $((0,0),\ell_\zface)$ (since $|\sigma|^{\alpha_0-2}\hat r^{-\alpha_0+2}$ is smooth down to $\zface^\circ$). Since $(\wh{L_b}(\sigma))|_{\sigma\in\pm[0,1]}\in\rho_\scface\rho_\tface^2\Diffb^2(X_\scbtop^\pm)$, we thus only need to show that the $\zface$-order of $\wh{L_b}(\sigma)(|\sigma|^{\alpha_0-2}u_{b,\alpha_0})$ is, in fact, $((0,0),\ell_\zface)$ as well, which we proceed to demonstrate.

  We begin with the second line of~\eqref{EqiptfGrKerr} and write
  \begin{align*}
    &\wh{L_b}(\sigma)\bigl(|\sigma|^{\alpha_0-2}\fl^{-\alpha_0+2}(\hat r)|\sigma|^2 h_{b,\rms 2}^{(-2),\leq 1}(\sigma,\scal_2^{(-2)})\bigr) \\
    &\qquad = |\sigma|^{\alpha_0-2} \fl^{-\alpha_0+2}(\hat r) \wh{L_b}(\sigma)\bigl(|\sigma|^2 h_{b,\rms 2}^{(-2),\leq 1}(\sigma,\scal_2^{(-2)})\bigr) \\
    &\qquad \qquad + |\sigma|^{\alpha_0-2}[\wh{L_b}(\sigma),\fl^{-\alpha_0+2}(\hat r)] \cdot|\sigma|^2 h_{b,\rms 2}^{(-2),\leq 1}(\sigma,\scal_2^{(-2)}).
  \end{align*}
  Since $\wh{L_b}(\sigma)h_{b,\rms 2}^{(-2),\leq 1}(\sigma,\scal_2^{(-2)})\in|\sigma|^2\CI([0,1)_\sigma\times X^\circ)$ and $\fl^{-\alpha_0+2}(\hat r)\in\cA^{\min(0,-\Re\alpha_0+2)-\eps}([0,1)_\sigma\times X^\circ)$ for all $\eps>0$ (regardless of whether $\alpha_0=2$ or not), the first term on the right is of class $\cA^{\min(\Re\alpha_0+2,4)-\eps}([0,1)_\sigma\times X^\circ)$ and thus (a fortiori) has $\zface^\circ$-order $\ell_\zface$, as required. For the second term, we note that
  \begin{equation}
  \label{EqiptfGrCommW}
    \Bigl[\pa_x,\fl^\beta\Bigl(\frac{|\sigma|}{\rho(x)}\Bigr)\Bigr]=-\frac{\pa_x\rho}{\rho} \Bigl(\frac{|\sigma|}{\rho}\Bigr)^\beta;
  \end{equation}
  applying this with $\rho(x)=|x|^{-1}$ (so $\frac{|\sigma|}{\rho}=\hat r$) and $\beta=-\alpha_0+2$, the second term thus lies in $|\sigma|^{\alpha_0-2}\hat r^{-\alpha_0+2}|\sigma|^2\CI([0,1)_\sigma\times X^\circ)=|\sigma|^2\CI$.\footnote{For grafts of other terms from~\eqref{EqiptfAsy2}, the analogue of this term has other integer powers of $|\sigma|$. For example, for the term with $\mu=-\lambda^\Ups_{\rms l,l+j}+1$, the power of $|\sigma|$ is $(\alpha_0-2)+(-\alpha_0+2-\{\lambda^\Ups_{\rms l,l+j}\})+(-\mu)=l+j-1$.}

  We finally discuss the $\zface$-order of the contribution of the first line of~\eqref{EqiptfGrKerr}, i.e., of
  \begin{align}
    &|\sigma|^{\alpha_0-2}\wh{L_b}(\sigma)\Bigl(-\fl^{-\alpha_0+2}(|\sigma|)|\sigma|\sigma^{-2}h_{b,\rms 1} + \fl^{-\alpha_0+2}(\hat r)|\sigma|i\sigma^{-1}\breve h_{b,\rms 1}^{\leq 4}(\sigma)\Bigr) \nonumber\\
    &\qquad = |\sigma|^{\alpha_0-2}\fl^{-\alpha_0+2}(|\sigma|)|\sigma| \wh{L_b}(\sigma)\bigl( -\sigma^{-2}h_{b,\rms 1} + i\sigma^{-1}\breve h_{b,\rms 1}^{\leq 4} \bigr) \nonumber\\
    &\qquad \qquad + |\sigma|^{\alpha_0-2}\bigl( \fl^{-\alpha_0+2}(\hat r)-\fl^{-\alpha_0+2}(|\sigma|) \bigr) |\sigma| \wh{L_b}(\sigma)\bigl(i\sigma^{-1}h_{b,\rms 1}^{\leq 4}(\sigma)\bigr) \nonumber\\
  \label{EqiptfGrs1Sing}
    &\qquad \qquad + |\sigma|^{\alpha_0-2} [\wh{L_b}(\sigma),\fl^{-\alpha_0+2}(\hat r)] |\sigma| i\sigma^{-1}\breve h_{b,\rms 1}^{\leq 4}(\sigma).
  \end{align}
  The first line is of class $\cA^{\min(0,\Re\alpha_0-2)-\eps}([0,1)_{|\sigma|})\cdot|\sigma|\cdot|\sigma|^3\CI(X)=\cA^{\min(\Re\alpha_0+2,4)-\eps}([0,1))\CI(X)$ (cf.\ \eqref{EqiptfFormals1}), and thus has $\zface^\circ$-order $\ell_\zface$. We next note that
  \[
    \fl^\beta(\hat r) - \fl^\beta(|\sigma|) = \frac{(|\sigma|/\rho)^\beta-1}{\beta} - \frac{|\sigma|^\beta-1}{\beta} = |\sigma|^\beta \fl^\beta(\rho^{-1})
  \]
  for $\beta=-\alpha_0+2$; the second line is thus smooth down to $\sigma=0$. In the third line, we again use~\eqref{EqiptfGrCommW} to conclude that it is of class $|\sigma|\cdot|\sigma|^{-1}\CI=\CI$ in $\sigma$.

  \pfstep{Step~2. Logarithms at $\tface$.} Consider now a general term~\eqref{EqiptfGrTerm}; we write this as
  \[
    \frac{\dd^k}{\dd\alpha^k}\bigl(\chi_\tface\rho_\tface^\alpha f_0(\hat r,\omega)\bigr)\Big|_{\alpha=\alpha_0}.
  \]
  Now, $u_\alpha$, defined by~\eqref{EqiptfGrDef} and (in a reduced setting for notational brevity) \eqref{EqiptfGrKerr}, is holomorphic near $\alpha=\alpha_0$, as observed around~\eqref{EqiptfGrHol}. Using~\eqref{EqiptfGrError}, we thus obtain
  \begin{equation}
  \label{EqiptfGrLogErr}
    u := \frac{\dd^k}{\dd\alpha^k}u_\alpha\Big|_{\alpha=\alpha_0} \implies \chi_\tface\rho_\tface^{\alpha_0}(\log\rho_\tface)^k f_0 - \wh{L_b}(\sigma)u \in \Hb^{\infty,\ \ell_\scface+1,\ \alpha_0+\eps_\ind-\eps,\ ((0,0),\ell_\zface)}(X_\scbtop^\pm)
  \end{equation}
  for all $\eps>0$. But direct differentiation of~\eqref{EqiptfGrDef}--\eqref{EqiptfGrKerr} in the parameter (there denoted $\alpha_0$) gives an explicit formula for $u$ which, in particular, shows that it is polyhomogeneous at $\tface$ and thus, in fact, the error term~\eqref{EqiptfGrLogErr} has $\tface$-index set $(\alpha_0,k)+\cE_\ind$.
\end{proof}

\begin{rmk}[Exceptional grafting of the $h_{b,\rms 1}$-term]
\label{RmkiptfGrs1}
  If instead of $\fl^{-\alpha_0+2}(|\sigma|)$ we used $\fl^{-\alpha_0+2}(\hat r)$ for the first term in~\eqref{EqiptfGrKerr}, then in~\eqref{EqiptfGrs1Sing} we would get an additional term
  \[
    -|\sigma|^{\alpha_0-2}[\wh{L_b}(\sigma),\fl^{-\alpha_0+2}(\hat r)] |\sigma|\sigma^{-2}h_{b,\rms 1}\in|\sigma|^{-1}\CI,
  \]
  which is singular at $\sigma=0$.
\end{rmk}

\subsubsection{Conormal sources at \texorpdfstring{$\tface$}{the transition face}}
\label{SssiptfC}

We shall need a variant of Proposition~\ref{PropiptfGr} for the purpose of solving away terms at $\tface$ that are conormal rather than polyhomogeneous. The approximate solutions are correspondingly also only conormal at $\tface$ and $\zface$. We only need this for $\alpha<2$.

\begin{prop}[Solving away leading-order terms at $\tface$, II: conormal sources]
\label{PropiptfC}
  Let $\alpha\in(1,2)$, $\ell_\scface<1$, and $-\eps_\ind<\ell_\zface-\alpha+2<3+\eps_\ind$, with $-\alpha+2+j$, $\ell_\zface-\alpha+2\neq\Re(-\lambda)$ for all $j\in\N_0$ and all indicial roots $\lambda$ of $\wh{L_b}(0)$ with $\Re\lambda\in[-3,0]$. Let $k_0\in\N_0$. Then for all
  \begin{equation}
  \label{EqiptfCf}
    f = f(\sigma,r,\omega) \in \Hb^{\infty,\ \ell_\scface+1,\ \alpha,\ \bigl((k_0,0),\ell_\zface\bigr)}(X_\scbtop^\pm;S^2\cT_X^*)
  \end{equation}
  there exists
  \begin{equation}
  \label{EqiptfCu}
    u = u(\sigma,r,\omega) \in \bigcap_{\eps>0} \Hb^{\infty,\ \ell_\scface-\eps,\ \alpha-2,\ \alpha-3}(X_\scbtop^\pm;S^2\cT^*_X)
  \end{equation}
  such that
  \begin{equation}
  \label{EqiptfCErr}
    \bigl(f(\sigma) - \wh{L_b}(\sigma)u(\sigma)\bigr)\big|_{\sigma\in\pm[0,1]} \in \Hb^{\infty,\ \ell_\scface+1,\ \alpha+\eps_\ind-\eps,\ \bigl((k_0,0),\ell_\zface\bigr)}(X_\scbtop^\pm;S^2\cT^*_X).
  \end{equation}
  More precisely, one can take $u=u(\sigma,r,\omega)$ to be of the form
  \begin{subequations}
  \begin{equation}
  \label{EqiptfCuPrec}
    u(\sigma,r,\omega) = \chi_\zface u_b(\sigma,r,\omega) + \chi_\tface\tilde u(\sigma,r,\omega)
  \end{equation}
  where $\tilde u\in\Hb^{\infty,\ \ell_\scface-\eps,\ \alpha-2,\ \bigl((k_0,0),\ell_\zface\bigr)}(X_\scbtop^\pm)$ and
  \begin{equation}
  \label{EqiptfCuPrec2}
  \begin{split}
    u_b &= h_{b,\rms 1}^{\leq 4}\bigl(\sigma,|\sigma|^{\alpha-3}\scal_1^{(-1)}(\sigma)\bigr) + h_{b,\rmv 1}^{\leq 3}\bigl(\sigma,|\sigma|^{\alpha-2}\vect_1^{(-1)}) + h_{b,\rms 0}^{(0),\leq 3}\bigl(\sigma,|\sigma|^{\alpha-2}\scal_0^{(0)}(\sigma)\bigr) \\
      &\quad + \sum_{ \substack{\mu=-\lambda^\Ups_{\rms l,l+j}+1 \\ 0\leq l\leq 3,\ j=0,1, \\ 1\leq l+j\leq 3} } h_{b,\rms l}^{(\mu),\leq 2}\bigl(\sigma,|\sigma|^{\alpha-3+\lambda^\Ups_{\rms l,l+j}}\scal_l^{(-\lambda^\Ups_{\rms l,l+j}+1)}(\sigma)\bigr) \\
      &\quad + \sum_{l=2}^5 h_{b,\rms l}^{(-l+2),\leq 5-l}\bigl(\sigma,|\sigma|^{\alpha-2+(l-2)}\scal_l^{(-l+2)}(\sigma)\bigr) + \sum_{l=2}^4 h_{b,\rmv l}^{(-l+1),\leq 4-l}\bigl(\sigma,|\sigma|^{\alpha-2+(l-1)}\vect_l^{(-l+1)}(\sigma)\bigr) \\
      &\quad + h_{b,\rms 2}^{(-2),\leq 1}\bigl(\sigma,|\sigma|^\alpha\scal_2^{(-2)}(\sigma)\bigr) + h_{b,\rmv 2}^{(-2),\leq 1}\bigl(\sigma,|\sigma|^\alpha\vect_2^{(-2)}(\sigma)\bigr) \\
      &\quad + h_{b,\rms 3}^{(-3)}\bigl(\sigma,|\sigma|^{\alpha+1}\scal_3^{(-3)}(\sigma)\bigr) + h_{b,\rmv 3}^{(-3)}\bigl(\sigma,|\sigma|^{\alpha+1}\vect_3^{(-3)}(\sigma)\bigr)
  \end{split}
  \end{equation}
  \end{subequations}
  for some $\scal_l^{(\mu)}\in\Hb^{\infty,0}(\pm[0,1)_\sigma,|\frac{\dd\sigma}{\sigma}|;\scalspace_l)$ and $\vect_l^{(\mu)}\in\Hb^{\infty,0}(\pm[0,1)_\sigma,|\frac{\dd\sigma}{\sigma}|;\vectspace_l)$.
\end{prop}

The point is that the error term~\eqref{EqiptfCErr} vanishes to higher order at $\tface$ than the original source term~\eqref{EqiptfCf}. If $f$ is in fact polyhomogeneous at $\tface$, \eqref{EqiptfCu} is consistent with~\eqref{EqiptfGru}--\eqref{EqiptfGrErr}, and~\eqref{EqiptfCuPrec2} is consistent with~\eqref{EqiptfGruPrec2}.

\begin{proof}[Proof of Proposition~\usref{PropiptfC}]
  Write $f=|\sigma|^{\alpha-2}|\sigma|^2 f_0$ where
  \[
    f_0\in\Hb^{\infty,\ \ell_\scface+1,\ 0,\ \bigl((-\alpha+k_0,0),\,\ell_\zface-\alpha\bigr)}(X_\scbtop^\pm).
  \]
  Roughly speaking, since $|\sigma|^2 N_\tface(\ubar L,\pm 1)$ is the $\tface$-model of $\wh{L_b}(\sigma)$, we need to control $N_\tface(\ubar L,\pm 1)^{-1}$ acting on $\chi_\tface f_0(\sigma,\cdot)$ (parametrically in $\sigma$) where $\chi_\tface\in\CI(X_\scbtop^\pm)$ is $1$ in a small neighborhood of $\tface$ and supported nearby. (The output must then be multiplied by $|\sigma|^{\alpha-2}$ and grafted into $X_\scbtop^\pm$.) However, since $[0,1]_{\pm\sigma}\times\tface$ does \emph{not} map smoothly into a collar neighborhood of $\tface$ in $X_\scbtop^\pm$,\footnote{Rather, a neighborhood of the lift of $\{0\}\times\tface$ to ${\rm C}\tface:=[[0,1]_{\pm\sigma}\times\tface;\{0\}\times\ztface]$ is naturally diffeomorphic to a neighborhood of $\tface$ in $X_\scbtop^\pm$; it is the front face of ${\rm C}\tface$ (\emph{not} the boundary hypersurface $[0,1]\times\ztface$ of $[0,1]\times\tface$) that carries the partial expansion of $f_0$. The reader is invited to interpret the proof below from this geometric singular perspective.}  some care is required to carry this out.

  \pfstep{Step~1. Solving away part of the $\zface$-expansion of $f$.} Consider the case $k_0=0$. Denote the $\hat r^{-\alpha}$-coefficient of $f_0$ at $\zface$ by $f_0^{(0)}=f_0^{(0)}(\rho,\omega)\in\Hb^{\infty,0}([0,\rho_0)\times\Sph^2)$, $\rho_0:=\frac{1}{100\bhm}$. (We require $\supp\chi_\tface\subset\{\rho<\rho_0\}$.) Inverting the relationship $\hat r=\frac{|\sigma|}{\rho}$ gives $\rho=\frac{|\sigma|}{\hat r}$; recalling that
  \[
    N_\tface(\ubar L,\pm 1) = \hat r^{-2}\wt{\ubar L}(0)(\omega,-\hat r\pa_{\hat r},\pa_\omega) + \hat r^{-1}N_1(\hat r\pa_{\hat r})
  \]
  from~\eqref{EqiptfNtf} where we write $N_1:=\mp i\wt{\ubar L_1}(-\hat r\pa_{\hat r})\in\Diffb^1$, let us thus define $u_0^{(0)}(\rho,\omega)$ parametrically in $\rho$ as
  \[
    u_0^{(0)}(\rho,\cdot) := \wt{\ubar L}(0)(\omega,\alpha-2,\pa_\omega)^{-1} f_0^{(0)}(\rho,\cdot).
  \]
  (We use here that $\alpha-2$ is not an indicial root of $\wh{\ubar L}(0)$.) We then compute
  \begin{align*}
    &\hat r^{-\alpha}f_0^{(0)} - N_\tface(\ubar L,\pm 1)\Bigl( \hat r^{-\alpha+2} u_0^{(0)}\Bigl(\frac{|\sigma|}{\hat r},\omega\Bigr) \Bigr) \\
    &\qquad = -\hat r^{-2}\hat r^{-\alpha+2}\tilde f\Bigl(\frac{|\sigma|}{\hat r},\omega\Bigr) -\hat r^{-1}N_1(\hat r\pa_{\hat r}) \Bigl( \hat r^{-\alpha+2} u_0^{(0)}\Bigl(\frac{|\sigma|}{\hat r},\omega\Bigr) \Bigr),
  \end{align*}
  where $\tilde f\in\rho\Hb^{\infty,0}([0,\rho_0)_\rho\times\Sph^2)$ collects all terms that arise when at least one $\hat r\pa_{\hat r}$-derivative of $\wt{\ubar L}(0)$ falls on $u_0^{(0)}(\frac{|\sigma|}{\hat r},\omega)$; this produces a factor of $\frac{|\sigma|}{\hat r}=\rho$. The second summand is of class $\hat r^{-\alpha+1}\Hb^{\infty,0}([0,\rho_0)_\rho\times\Sph^2)$ and thus has an additional factor of $\hat r$.

  Multiplying through by $|\sigma|^{\alpha-2}$ and using that $|\sigma|^{\alpha-2}\hat r^{-\alpha+2}=\rho^{\alpha-2}$, we conclude that if $\chi_\zface\in\CI(X_\scbtop^\pm)$ is $1$ near $\zface$ and $0$ near $\scface$, then
  \[
    f - \wh{L_b}(\sigma)\bigl( \chi_\zface\chi_\tface\rho^{\alpha-2}u_0^{(0)}(\rho,\omega) \bigr) \in \Hb^{\infty,\ \ell_\scface+1,\ \alpha+1,\ ((0,0),\ell_\zface)} + \Hb^{\infty,\ \ell_\scface+1,\ \alpha,\ ((1,0),\ell_\zface)}
  \]
  The first summand here is already an acceptable error term for~\eqref{EqiptfCErr}, and hence it remains to treat the second summand; note that we have thus reduced our task to the case $k_0=1$. Now, for $f\in\Hb^{\infty,\ \ell_\scface+1,\ \alpha,\ ((1,0),\ell_\zface)}$, we repeat the above argument for the $\hat r^{-\alpha+1}$-term $\hat r^{-\alpha+1}f_0^{(1)}(\rho,\omega)$, $f_0^{(1)}\in\Hb^{\infty,0}$, of $f_0=|\sigma|^{-\alpha}f$. Using now that $-\alpha+3$ is not an indicial root, this produces $u_0^{(1)}\in\Hb^{\infty,0}$ such that
  \[
    \hat r^{-\alpha+1}f_0^{(1)} - N_\tface(\ubar L,\pm 1)\Bigl(\hat r^{-\alpha+3}u_0^{(1)}\Bigl(\frac{|\sigma|}{\hat r},\omega\Bigr)\Bigr) \in \hat r^{-\alpha+1}\rho\Hb^{\infty,0} + \hat r^{-\alpha+2}\Hb^{\infty,0}.
  \]
  Using $\chi_\zface\chi_\tface |\sigma|^{\alpha-2}\hat r^{-\alpha+3}u_0^{(1)}(\rho,\omega)=\hat r\chi_\zface\chi_\tface\rho^{\alpha-2}u_0^{(1)}(\rho,\omega)$, we can thus eliminate also the $\hat r^1$-term of $f$ at $\zface$.

  Repeating this procedure $\lfloor\ell_\zface\rfloor$ many times produces
  \begin{equation}
  \label{EqiptfCu0}
    u_0=u_0(\sigma,\rho,\omega) \in \Hb^{\infty,\ \infty,\ \alpha-2,\ (0,0)}(X_\scbtop^\pm)
  \end{equation}
  with support near $\tface\cap\zface$ such that
  \begin{equation}
  \label{EqiptfCf12}
    f - \wh{L_b}(\sigma)u_0 = f_1 + f_2,\quad f_1 \in \Hb^{\infty,\ \ell_\scface+1,\ \alpha,\ \ell_\zface},\quad f_2 \in \Hb^{\infty,\ \ell_\scface+1,\ \alpha+1,\ ((0,0),\ell_\zface)},
  \end{equation}
  with $f_2$ being an acceptable error term.

  \pfstep{Step~2. Solving away $f_1$.} Consider now
  \[
    \chi_\tface|\sigma|^{-\alpha}f_1 \in \Hb^{\infty,\ \ell_\scface+1,\ 0,\ \ell_\zface-\alpha}(X_\scbtop^\pm).
  \]
  Using the defining functions $\rho_\scface=\frac{\rho}{\rho+|\sigma|}=\frac{1}{1+\hat r}$ and $\rho_\zface=\frac{|\sigma|}{\rho+|\sigma|}=\frac{\hat r}{\hat r+1}$ on $\tface$, this lies in
  \begin{align*}
    &(1+\hat r)^{-(\ell_\scface+1)} \Bigl(\frac{\hat r}{\hat r+1}\Bigr)^{\ell_\zface-\alpha} \Hb^{\infty,0}\bigl( \pm[0,1]_\sigma; \Hb^{\infty,0,0}(\tface)\bigr) \\
    &\qquad = \Hb^{\infty,0}\bigl(\pm[0,1]_\sigma;\Hb^{\infty,\ \ell_\scface+1,\ \ell_\zface-\alpha}(\tface)\bigr)
  \end{align*}
  where we use unweighted b-densities on $[0,1]$ and $\tface=[0,\infty]_{\hat r}\times\Sph^2$, i.e., $|\frac{\dd\sigma}{\sigma}|$ and $|\frac{\dd\hat r}{\hat r}\,\dd\slg|$. Since $\ell_\zface-\alpha+2>-\eps_\ind$, we can use Corollary~\ref{CorWEtfb} to apply $N_\tface(\ubar L,\pm 1)^{-1}$ to $\chi_\tface|\sigma|^{-\alpha}f_1$, with the output inheriting the regularity in the parameter $\sigma$, so
  \[
    u_1(\sigma,\hat r,\omega) := N_\tface(\ubar L,\pm 1)^{-1}\bigl( \chi_\tface|\sigma|^{-\alpha}f_1|_{\{\sigma\}\times\tface} \bigr) \implies u_1\in\Hb^{\infty,0}\bigl(\pm[0,1]_\sigma;\Hb^{\infty,\ \ell_\scface-\eps,\ \beta_\ztface}(\tface)\bigr)
  \]
  where $-\eps_\ind<\beta_\ztface<\min(0,\ell_\zface-\alpha+2)$.

  The usual normal operator argument for $N_\tface(\ubar L,\pm 1)$ at $\ztface=\hat r^{-1}(0)$, applied with parametric dependence on $\sigma$, now implies that $u_1$ in fact has a partial expansion at $\ztface$ featuring $-\lambda$ where $\lambda$, with $\Re\lambda\in[0,\ell_\zface-\alpha+2)$, is an indicial root of $\wh{L_b}(0)$. For the sake of brevity, we only discuss two terms that arise in this fashion (analogously to~\eqref{EqiptfGru0}); so let us suppose that
  \begin{align}
  \label{EqiptfCu1}
    &u_1(\sigma,\hat r,\omega) = \chi_\ztface\Bigl( \ubar h_{\rms 1}^{\leq 4}\bigl(\pm\hat r,\scal_1^{(-1)}(\sigma)\bigr) + \ubar h_{\rms 2}^{(-2),\leq 1}\bigl(\pm\hat r,\scal_2^{(-2)}(\sigma)\bigr)\Bigr) + \tilde u_1(\sigma,\hat r,\omega), \\
    &\qquad \scal_1^{(-1)}\in\Hb^{\infty,0}(\pm[0,1];\scalspace_1),\ \ \scal_2^{(-2)}\in\Hb^{\infty,0}(\pm[0,1];\scalspace_2), \nonumber\\
    &\qquad \tilde u_1\in\Hb^{\infty,0}\bigl(\pm[0,1];\Hb^{\infty,\ \ell_\scface-\eps,\ \ell_\zface-\alpha+2}(\tface)\bigr). \nonumber
  \end{align}
  We then claim that
  \begin{align}
  \label{EqipCFinalErr}
  \begin{split}
    f_1 - {}&\wh{L_b}(\sigma)\biggl[ |\sigma|^{\alpha-2}\chi_\zface\Bigl( -\sigma^{-2}|\sigma|h_{b,\rms 1}\bigl(\scal_1^{(-1)}(\sigma)\bigr) + i\sigma^{-1}|\sigma|\breve h_{b,\rms 1}^{\leq 4}\bigl(\sigma,\scal_1^{(-1)}(\sigma)\bigr) \\
      &\hspace{19em} + |\sigma|^2 h_{b,\rms 2}^{(-2),\leq 1}\bigl(\sigma,\scal_2^{(-2)}(\sigma)\bigr)\Bigr) \\
      &\hspace{4em} + |\sigma|^{\alpha-2}\chi_\tface\tilde u_1\Bigl(\sigma,\frac{|\sigma|}{\rho},\omega\Bigr) \biggr]
  \end{split} \\
      & \in \Hb^{\infty,\ \ell_\scface+1,\ \alpha+\eps_\ind-\eps,\ \ell_\zface}(X_\scbtop^\pm), \nonumber
  \end{align}
  which is an improvement over $f_1$ in~\eqref{EqiptfCf12} by $\eps_\ind-\eps$ orders at $\tface$ (where $\eps>0$ is arbitrary). That the output of $\wh{L_b}(\sigma)$ is of class $\Hb^{\infty,\ \ell_\scface+1-\eps,\ \alpha,\ \ell_\zface}$ is clear for the contribution of $\tilde u_1$ and follows for the contribution of $h_{b,\rms 1}$ and $\breve h_{b,\rms 1}^{\leq 4}$ from~\eqref{EqiptfFormals1}, similarly for the contribution of $h_{b,\rms 2}^{(-2),\leq 1}$ (see the discussion after Definition~\ref{DefiptfFormalKerr}). Working modulo $\Hb^{\infty,\ \ell_\scface+1-\eps,\ \alpha+\eps_\ind-\eps,\ \ell_\zface}$, the membership~\eqref{EqipCFinalErr} then follows from $|\sigma|^\alpha u_1$ solving the $\tface$-model problem of $|\sigma|^{-2}\wh{L_b}(\sigma)$ with right-hand side $f_1$. That the $\scface$-order is $\ell_\scface+2-\eps$ (and thus a fortiori $\ell_\scface+1$) follows from the equality of the $\scface$-normal operators of $\wh{L_b}(\sigma)$ and $|\sigma|^2 N_\tface(\ubar L,\pm 1)$ noted already after~\eqref{EqiptfGrError}.

  The precise description~\eqref{EqiptfCuPrec}--\eqref{EqiptfCuPrec2} is an immediate consequence of the above construction; the term $\tilde u$ is the sum of $|\sigma|^{\alpha-2}\chi_\tface\tilde u_1$ from~\eqref{EqipCFinalErr} and $u_0$ from~\eqref{EqiptfCu0}, while the term $u_b$ collects the first two lines of the argument of $\wh{L_b}(\sigma)$ in~\eqref{EqipCFinalErr}.

  The term in square brackets in~\eqref{EqipCFinalErr} is a sum of terms of class $\chi_\zface\Hb^{\infty,\alpha-3}([0,1]_{\pm\sigma})\cA^2(X)\subset\Hb^{\infty,\ \infty,\ \alpha-1-\eps,\ \alpha-3}(X_\scbtop^\pm)$, further $\chi_\zface\Hb^{\infty,\alpha-2}\cA^1$, $\chi_\zface\Hb^{\infty,\alpha-2+\mu}\cA^{-\mu}$ (for $-\mu\in[-3,0]$ being the real part of an indicial root of $\wh{L_b}(0)$, so $\mu=1$ and $2$ for the second and third terms of~\eqref{EqipCFinalErr}), all of which lie in $\Hb^{\infty,\ \infty,\ \alpha-2,\ \alpha-2-\eps}(X_\scbtop^\pm)$, and finally $\Hb^{\infty,\ \ell_\scface-\eps,\ \alpha-2,\ \ell_\zface}(X_\scbtop^\pm)$. The sum with~\eqref{EqiptfCu0} defines $u$ satisfying~\eqref{EqiptfCu}--\eqref{EqiptfCErr}.
\end{proof}

\section{Analysis of the stationary problem: 2-admissibility}
\label{SAdm}

We continue writing $L_b$ for the linearized gauge-fixed Einstein operator~\eqref{EqWEOp}, with the modifications $E^\cC$ and $E^\Ups$ fixed as at the beginning of~\S\ref{SWE}, and with the parameter $\gamma^\Ups$ moreover chosen small enough so that the conclusions of Proposition~\ref{PropWEtf} hold. Recall then that $L_b$ has an indicial gap containing the interval $(0,\eps_\ind)$ (see Lemma~\ref{LemmaWEInd}). The main result of this section is:

\begin{thm}[2-admissibility]
\label{ThmAdm}
  Let $\alpha_\sface\in\R$ be such that $\alpha_\sface+\frac32\in(0,\eps_\ind)$, and let $\delta>\alpha_\sface+\frac32$. Then the operator $L_b$ is \emph{2-admissible with $\sface$-weight $\alpha_\sface$, $\sface$-loss $\delta$, and margin $1$} in the terminology of \citeAF{Definition~\ref*{DefSSAlephAdm} and Remark~\ref*{RmkSSAlephAdmRelax}}.
\end{thm}

We recall that in addition to the (strong) trapping admissibility (verified in Proposition~\usref{PropWETr}) and $\tface$-ad\-miss\-i\-bil\-i\-ty with weight $\alpha_\sface+\frac32$ (verified in Proposition~\usref{PropWEtf}), this demands a quantitative estimate for forward solutions of $L_b u=f$. We recall the notation and function spaces for the precise statement. Recall $\tilde M_0:=[\ol{\R^4};\fk^+,\fk^-]$ from Definition~\ref{DefKMfdMin}, with boundary hypersurfaces $\sface$ and $\cK^\pm$. The Kerr metrics $g_b$, with $b=(\bhm,\bha)$ near $b_0=(\bhm_0,\bha_0)$, are defined on the subset
\begin{equation}
\label{EqAdmM0}
  M_0:=\cl_{\tilde M_0}\{r\geq\bhm_0\}
\end{equation}
of $\tilde M_0$ (and indeed smooth sections of the pullback $S^2\cT^*$ of $S^2\,\Tsc^*\ol{\R^4}$, as follows from the proof of Proposition~\ref{PropKMetCpt} as well as from \citeAF{Lemma~\ref*{LemmaTsKLMetric}(\ref*{ItTsKLMetric3b})}). On $M_0$ then, a defining function of $\sface$ is $\rho_\sface=r^{-1}$; and on the domain
\begin{equation}
\label{EqAdmOmega}
  \Omega := \cl_{M_0}\{t_*\geq 1\} = \cl_{\tilde M_0} \{ t_*\geq 1,\ r\geq\bhm_0 \} \subset M_0,
\end{equation}
a defining function of $\cK^+$ is $\rho_\cK=\frac{r}{t_*+r}$. Over $M_0$, one can take $r\pa_{t_*}$, $r\pa_r$, and $\pa_\omega$ as a spanning set of $\Vtb(M_0)$. (Cf.\ the discussion preceding~\eqref{EqWGOpNabla}.) Recall from~\S\ref{SssT3b} that the $m$-th order 3b-Sobolev space $\Htb^m(\tilde M_0)$ consists of all functions on $\R^4$ which remain in $L^2$ upon application of up to $m$-fold compositions of 3b-vector fields. On the domain $\Omega$, with initial, resp.\ final hypersurface $t_*^{-1}(1)$, resp.\ $r^{-1}(\bhm_0)$, we thus have supported/extendible spaces
\[
  H_\tbop^m(\Omega)^{\bullet,-}
\]
by the general construction explained in~\S\ref{SssTExtSupp}. Weighted versions are denoted by
\[
  H_\tbop^{m,(\alpha_\sface,\alpha_\cK)}(\Omega)^{\bullet,-} := \rho_\sface^{\alpha_\sface}\rho_\cK^{\alpha_\cK}H_\tbop^m(\Omega)^{\bullet,-}.
\]
One can define more general spaces which encode also additional $k\in\N_0$ degrees of b-regularity on $\tilde M_0$; on $\Omega$, this is equivalent to regularity with respect to $t_*\pa_{t_*}$ and $r\pa_x$ (i.e., $r\pa_r,\pa_\omega$). These spaces are denoted
\[
  H_{\tbop;\bop}^{(m;k),(\alpha_\sface,\alpha_\cK)}(\Omega)^{\bullet,-}.
\]
Finally, one can define versions of these spaces where the regularity order $m$ is variable (namely, a smooth function on the 3b-cosphere bundle); this is a special case of the construction \citeAF{(\ref*{EqMSMixed})} originating in \cite{HintzScaledBddGeo}. To prove Theorem~\ref{ThmAdm}, we must then show:
\begin{equation}
\label{EqAdm}
  \parbox{0.86\textwidth}{
    \it Let $\sfs$ be a stationary-$L_b$-admissible order function with weights $\alpha_\sface,-2$ and margin $1$ (see \citeAF{Definition~\ref*{DefSSOrderAdm}}). Then for all $k\in\N_0$ and
    \[
      f \in H_{\tbop;\bop}^{(\sfs;k),(\alpha_\sface+2,0)}(\Omega,|\dd g_b|;S^2\cT^*)^{\bullet,-},
    \]
    the unique forward solution $u$ of $L_b u=f$ satisfies
    \[
      u \in H_{\tbop;\bop}^{(\sfs-1;k),(\alpha_\sface-\delta,-2)}(\Omega,|\dd g_b|;S^2\cT^*)^{\bullet,-},
    \]
    with norm bounded by a ($k$-dependent) constant times that of $f$.
  }
\end{equation}

\begin{rmk}[Upper bound on $\sfs$ in outgoing directions]
\label{RmkAdmUpper}
  In the threshold condition at the outgoing radial set in \citeAF{Definition~\ref*{DefSSOrderAdm}(\ref*{ItSSOrderAdm5})}, the quantity $\vartheta_{\pa\cK^+,{\rm out}}$ is equal to $\min\spec S_{E^\Ups,E^\cC}=0$ (in the notation of Lemma~\ref{LemmaWEOp}) by \citeAF{Lemma~\ref{LemmaSSOrderThr}}, so the requirement is that $\sfs+\alpha_\sface<-\frac12-1$ ($1$ being the margin) at the outgoing radial set. (There is no need for an $\eps$-loss since $S_{E^\Ups,E^\cC}$ is diagonalizable.)
\end{rmk}

\begin{figure}[!ht]
\centering
\includegraphics{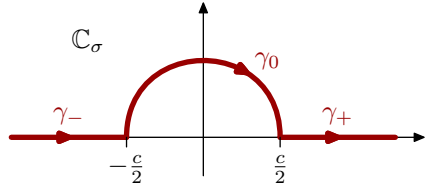}
\caption{The contours used in~\eqref{EqAdmIFT}.}
\label{FigAdmIFT}
\end{figure}

Following essentially the same strategy as in \citeAF{\S\ref*{SsA2Adm}} (and in a simpler setting, namely without zero energy bound states, in~\citeAF{\S\ref*{SsA1Adm}}), we shall prove~\eqref{EqAdm} as follows: using the large $\Im\sigma$ estimates of \citeAF{Theorem~\ref*{ThmSpHi}}, the Paley--Wiener argument in the proof of \citeAF{Theorem~\ref*{ThmA1Adm}} applies to yield the representation
\[
  u(t_*,\cdot) = \frac{1}{2\pi} \int_{\Im\sigma=\gamma} e^{-i\sigma t_*}\wh{L_b}(\sigma)^{-1}\hat f(\sigma,\cdot)\,\dd\sigma
\]
of the forward solution in~\eqref{EqAdm} when $\gamma>0$ is sufficiently large. Using high-energy (large $|\Re\sigma|$) estimates for $\wh{L_b}(\sigma)^{-1}$ and the absence of poles of $\wh{L_b}(\sigma)^{-1}$ (to handle regions of bounded $|\Re\sigma|$)---which is a consequence of mode stability (Proposition~\ref{PropWEMode}) via \citeAF{Theorem~\ref*{ThmSpB}}---one can shift the contour down to any positive value $\gamma>0$. Using the continuity of $\wh{L_b}(\sigma)^{-1}$ down to $\R\setminus\{0\}$ (see \citeAF{Theorem~\ref*{ThmSpB}(\ref*{ItSpBInvReg})} and \citeAF{Theorem~\ref*{ThmSpHi}(\ref*{ItSpHiInvReg})}), one can shift the contour down to $\R$ away from $0$, so
\begin{equation}
\label{EqAdmIFT}
\begin{split}
  u(t_*,\cdot) &= \frac{1}{2\pi}\int_{\gamma_-\cup\gamma_0\cup\gamma_+} e^{-i\sigma t_*}\wh{L_b}(\sigma)^{-1}\hat f(\sigma)\,\dd\sigma, \\
  &\qquad \gamma_-=(-\infty,-\tfrac{c}{2}],\ \ \gamma_0=\tfrac{c}{2}e^{i[\pi,0]},\ \ \gamma_+=[\tfrac{c}{2},\infty),
\end{split}
\end{equation}
for any fixed $c>0$; see Figure~\ref{FigAdmIFT}. The high-energy pieces $\gamma_\pm$ will be easily handled using high-energy estimates for $\wh{L_b}(\sigma)$ (which, as shown in \citeAF{\S\ref*{SsSpHi}}, follow in particular from the strong trapping admissibility of $L_b$ recorded in Proposition~\ref{PropWETr}) exactly as in \cite{HintzNonstat2}; we recall the argument in~\S\ref{SssAdmPfHi}. The key difficulty is to control $\wh{L_b}(\sigma)^{-1}$ near $\sigma=0$, where the presence of a nontrivial nullspace of $\wh{L_b}(0)$ (Proposition~\ref{PropWEMode0}) leads to the blow-up of the resolvent. To capture this precisely, we proceed similarly to \citeAF{\S\ref*{SsA2Adm}} by introducing a carefully designed \emph{finite rank augmentation} (i.e., Grushin problem) $\wt{L_b}(\sigma)$ of $\wh{L_b}(\sigma)$, which \emph{is} invertible at $\sigma=0$ and for which thus uniform low-energy resolvent estimates are available; certain components of $\wt{L_b}(\sigma)$ feature $\wh{L_b}(\sigma)$ acting on singular (as $\sigma\to 0$) terms. See~\S\ref{SsAdmLo} for details. Upon carefully estimating the contributions of these terms in~\eqref{EqAdmIFT}, we shall obtain Theorem~\ref{ThmAdm}. The details are given in~\S\ref{SssAdmPfLo}.

\subsection{Preparation of the low-energy analysis}
\label{SsAdmLo}

Given the description of the nullspace of $\wh{L_b}(0)$ in Proposition~\ref{PropWEMode0}, a naive attempt (analogous to \citeAF{(\ref*{EqA2AdmLo})}) at defining an augmentation of $\wh{L_b}(0)$ with invertible zero energy operator is
\begin{equation}
\label{EqAdmLoNaive}
  \begin{pmatrix}
    \wh{L_b}(\sigma) & \wh{L_b}(\sigma)(i\sigma^{-1}\dot g_b^\Ups(\cdot)) & \wh{L_b}(\sigma)\bigl(-\sigma^{-2}h_{b,\rms 1}(\cdot)+i\sigma^{-1}\breve h_{b,\rms 1}^1(\cdot)\bigr) \\
    \la\cdot,f^*_1\ra_{L^2(X;S^2\cT^*_X)} & 0 & 0 \\
    \la\cdot,f^*_2\ra_{L^2(X;S^2\cT^*_X)} & 0 & 0
  \end{pmatrix},
\end{equation}
acting on $(u,\dot b,\dot\scal)$ where $u$ is a section of $S^2\cT^*_X\to X^\circ$ and $\dot b\in\R^4$, $\dot\scal\in\scalspace_1$; here we fix
\begin{subequations}
\begin{equation}
\label{EqAdmLofstar}
  f^*_1 \in \CIc(X^\circ;S^2\cT^*_X)\otimes\R^4,\quad
  f^*_2 \in \CIc(X^\circ;S^2\cT^*_X)\otimes\scalspace_1,
\end{equation}
such that the map
\begin{equation}
\label{EqAdmLofstar2}
  \R^4 \oplus \scalspace_1\ni(\dot b,\dot\scal)\mapsto \Bigl(\la\dot g_b^\Ups(\dot b),f^*_1\ra_{L^2},\ \la h_{b,\rms 1}(\dot\scal),f^*_2\ra_{L^2}\Bigr) \in \R^4 \oplus \scalspace_1
\end{equation}
\end{subequations}
is an isomorphism. The factors of $i$ in the first row in~\eqref{EqAdmLoNaive} are motivated by the fact that the Fourier transforms of $H(t_*)$ and $t_* H(t_*)$ are $\frac{i}{\sigma+i 0}$ and $-\frac{1}{(\sigma+i 0)^2}$, respectively. Note moreover that the $(1,2)$-entry
\begin{equation}
\label{EqAdmLoKerr}
\begin{split}
  \wh{L_b}(\sigma)\bigl(i\sigma^{-1}\dot g_b^\Ups(\dot b)\bigr) &= i\sigma^{-1}\underbrace{\wh{L_b}(0)\dot g_b^\Ups(\dot b)}_{=0} + i\pa_\sigma\wh{L_b}(0)\dot g_b^\Ups(\dot b) + \frac{i\sigma}{2}\pa_\sigma^2\wh{L_b}(0)\dot g_b^\Ups(\dot b) \\
    &= [L_b,t_*]\ftrans(0)\dot g_b^\Ups(\dot b) - \frac{i\sigma}{2}[[L_b,t_*],t_*]\dot g_b^\Ups(\dot b)
\end{split}
\end{equation}
of~\eqref{EqAdmLoNaive} is regular at $\sigma=0$, as is the $(1,3)$-entry
\begin{equation}
\label{EqAdmLos1}
\begin{split}
  &\wh{L_b}(\sigma)\bigl(-\sigma^{-2}h_{b,\rms 1}(\dot\scal)+i\sigma^{-1}\breve h_{b,\rms 1}^1(\dot\scal)\bigr) \\
  &\quad = -\sigma^{-2}\underbrace{\wh{L_b}(0)h_{b,\rms 1}(\dot\scal)}_{=\,0} + i\sigma^{-1} \underbrace{\bigl( [L_b,t_*]\ftrans(0)h_{b,\rms 1}(\dot\scal) + \wh{L_b}(0)\breve h_{b,\rms 1}^1(\dot\scal) \bigr)}_{=\,0\ \text{by Proposition~\ref{PropWG0Symm}\eqref{ItWG0SymmGrad}}} \\
  &\quad\qquad + [[L_b,t_*],t_*]h_{b,\rms 1}(\dot\scal) + [L_b,t_*]\ftrans(0)\breve h_{b,\rms 1}^1(\dot\scal) - \frac{i\sigma}{2}[[L_b,t_*],t_*]\breve h_{b,\rms 1}^1(\dot\scal) \\
  &\quad = [L_b,t_*]\ftrans(0)\breve h_{b,\rms 1}^1(\dot\scal) + \frac12[[L_b,t_*],t_*]\bigl(h_{b,\rms 1}(\dot\scal) - i\sigma\breve h_{b,\rms 1}^1(\dot\scal)\bigr).
\end{split}
\end{equation}

The issue, however, is that these entries only have $r^{-2}$ decay; indeed, we record here
\begin{subequations}
\begin{align}
\label{EqAdmLo12}
  [L_b,t_*]\ftrans(0)\dot g_b^\Ups(1,0) &\in \cA^{(2,0)\cup(2+\cE_\ind)}(X;S^2\cT^*_X), \\
\label{EqAdmLo13}
  [L_b,t_*]\ftrans(0)\breve h_{b,\rms 1}^1(\dot\scal) &\in \cA^{(2,0)\cup(2+\cE_\ind)}(X;S^2\cT^*_X),
\end{align}
\end{subequations}
with leading-order terms of scalar type $0$, resp.\ scalar type $1$, as consequences of~\eqref{EqWEMode0KerrMem} and \eqref{EqWG0hs1breve}. (Only terms arising from angular momentum changes,
\[
  [L_b,t_*]\ftrans(0)\dot g_b^\Ups(0,\dot\bha) \in \cA^{(3,0)\cup(3+\cE_\ind)},
\]
have acceptable $o(r^{-2})$-decay.) Therefore, they do not lie in the target space of $\wh{L_b}(0)$ in~\eqref{EqWEMode0}---which we are forced to work with for uniform low-energy resolvent estimates. (Acting on weaker function spaces, even the free resolvent on $\R^3$ becomes singular at $\sigma=0$; see Remark~\ref{RmkiptfGrSizeSol}.) Essentially the same issue already arose \cite[(3.32)]{HintzGlueLocIII}.\footnote{In the reference it only affected the $h_{b,\rms 1}$ sector, since the $\rho^1$ leading-order terms of $\dot g_b^\Ups(1,0)$ and $h_{b,\rms 1}$ were annihilated by the commutator of the linearized gauge-fixed Einstein operator used there (which has vanishing $\ubar S$ in the notation of~\eqref{EqWEOpMink2}) with $t_*$ (an analogous cancellation occurs in the 1-form wave operator case discussed in \citeAF{Lemma~\ref*{LemmaA212}}), whereas in the present paper the lack of decay is present in more components of~\eqref{EqAdmLoNaive}.}

The strategy for overcoming this issue developed in \cite[Lemmas~3.18--3.20 and Proposition~3.21]{HintzGlueLocIII} is to modify the naive attempt~\eqref{EqAdmLoNaive} using corrections obtained by inverting the $\tface$-normal operator $N_\tface(L_b,\hat\sigma)$; this leads to augmentation terms having acceptable $o(\rho^2)$-decay at $\tface\subset X_\scbtop=[[0,1]_\varsigma\times X;\{0\}\times\pa X]$ (for $\varsigma=|\sigma|$ and for all $\hat\sigma:=\frac{\sigma}{|\sigma|}\in e^{i[0,\pi]}$) in the notation of Definition~\ref{DefTscbt}. We present a conceptually cleaner strategy (which in particular avoids the somewhat delicate holomorphicity considerations in $\Im\sigma>0$ of \cite[\S3]{HintzGlueLocIII} and replaces them by simple Paley--Wiener type observations) and instead construct the augmentation terms as Fourier transforms of certain tensors \emph{on spacetime} which, roughly speaking, have large-$t_*$-behavior given by $\dot g_b^\Ups(\dot b)$ and $t_* h_{b,\rms 1}(\dot\scal)+\breve h_{b,\rms 1}^1(\dot\scal)$.

Upon taking the inverse Fourier transform, the $\tface$-decay order $2$ of the term\footnote{with boundary hypersurface ordered as $\scface,\tface,\zface$, this membership follows from~\eqref{EqAdmLo12} and $[[L_b,t_*],t_*]\in\rho^2\CI(X)$}
\[
  \wh{L_b}(\sigma)(i \sigma^{-1}\dot g_b^\Ups(1,0))\in\cA^{2,2,(0,0)}(X_\scbtop)
\]
corresponds to $L_b\dot g_b^\Ups(1,0)$ having $\iota^+$-decay order $3$ (see~\eqref{EqTFHbInvLo}), which we must thus consider to be (borderline) unacceptable. Note also that the index set of $\dot g_b^\Ups(1,0)$ at $\rho^{-1}(0)$ arises from the boundary spectrum of $\wh{L_b}(0)$ and thus is unrelated to the asymptotics of waves at $\scri^+$ which arise from the spectrum of $\ubar S_{\ubar E^\Ups,\ubar E^\cC}$ in the notation of Lemma~\ref{LemmaWEOp}. It is via inverting the $\iota^+$-normal operator of $L_b$ (on the error term $L_b\dot g_b^\Ups(1,0)$, suitably localized), which we do using material from~\S\ref{SsipInv}, that one can both improve the $\iota^+$-decay rate and obtain the correct asymptotics at $\scri^+$. We present the detailed construction of the resulting ``improved zero energy states''  and describe their Fourier transforms in~\S\ref{SssAdmLoIm}. A correctly augmented operator is then defined and analyzed in~\S\ref{SssAdmLoDef}. An augmentation on the spectral side \emph{but only for real $\sigma$} is constructed in~\S\ref{SssAdmLoSpec}; this will be useful when holomorphicity considerations in $\Im\sigma>0$ are not needed.

\subsubsection{Improved zero energy states}
\label{SssAdmLoIm}

We first work on the resolved Kerr spacetime $M$ from Definition~\ref{DefKMfdRad} in the subset
\[
  \Omega := \cl_M \{ t_*\geq 1 \}.
\]
(We use the same symbol as in~\eqref{EqAdmOmega}; it will always be clear from the context whether we regard $\Omega$ as a subset of $M$ or $M_0$.) We write $\cA^{\alpha_\sscri,\alpha_+,\alpha_\cK}(\Omega)$ for ($L^\infty$-based) conormal functions on $M$ with support in $\Omega$ and weight $\rho_\sscri^{\alpha_\sscri}\rho_+^{\alpha_+}\rho_\cK^{\alpha_\cK}$; similarly for polyhomogeneous spaces. We use the notation $\la\cE_{\iota^+,\sscri}^\cC\ra$ for the $\scri^+$-index sets of sections of $S^2\cT^*$ as introduced already after Proposition~\ref{PropipGr} based on Definition~\ref{DefipIndex}.

\begin{lemma}[Construction of improved zero energy states]
\label{LemmaAdmLoIm}
  There exist index sets $\cE_\ind$ and $\cE_\cK$ with $\min\Re\cE_\ind,\min\Re\cE_\cK\geq\eps_\ind$ such that, for all $b=(\bhm,\bha)$ near $b_0=(\bhm_0,\bha_0)$, there exist tensors
  \begin{align}
  \label{EqAdmLoImKerr}
    \dot g_b^{\Ups,\aug}(1,0) &\in \cA^{\la\cE_{\iota^+,\sscri}^\cC\ra,\ (1,0)\cup(1+\cE_\ind),\ \bigl((0,0)\cup(1+\cE_\cK),4+\eps_\ind-\eps\bigr)}(\Omega), \\
  \label{EqAdmLoImBoost}
    h_{b,\rms 1}^{\leq 1,\aug}(\dot\scal) &\in \cA^{\la\cE_{\iota^+,\sscri}^\cC\ra,\ (1,0)\cup(1+\cE_\ind),\ \bigl((-1,0)\cup(0,1)\cup(1+\cE_\cK),4+\eps_\ind-\eps\bigr)}(\Omega),\quad \dot\scal\in\scalspace_1,\ \eps>0,
  \end{align}
  that depend smoothly on $b$ (and linearly on $\dot\scal\in\scalspace_1$ in the case of~\eqref{EqAdmLoImBoost}), have leading-order terms $\dot g_b^\Ups(1,0)$ and $h_{b,\rms 1}^{\leq 1}(\dot\scal)$ at $\cK^+$ (see~\eqref{EqAdmLoImLot} for a more precise statement), and satisfy
  \begin{equation}
  \label{EqAdmLoImLb}
    L_b\dot g_b^{\Ups,\aug}(1,0),\ L_b h_{b,\rms 1}^{\leq 1,\aug}(\dot\scal) \in \cA^{\cE_{\iota^+,\sscri}^\tot+2,\ 3+\cE_\ind,\ 4+\eps_\ind-\eps} \subset \cA^{3,\ 3+\eps_\ind-\eps,\ 4+\eps_\ind-\eps}(\Omega)
  \end{equation}
  for all $\eps>0$.
\end{lemma}

The (constructive) proof will show that if we fix $\chi_\cK\in\CI(M)$ to be equal $1$ near $\cK^+$ and to vanish near $\scri^+$ and for $t_*\leq 1$, then we have (in the notation of~\eqref{EqWG0GenStateNot}, \eqref{EqWG0Larges00}, and \eqref{EqWEMode0KerrMem})
\begin{equation}
\label{EqAdmLoImLot}
\begin{alignedat}{2}
  \dot g_b^{\Ups,\aug}(1,0) &- \chi_\cK \bigl( \dot g_b^\Ups(1,0) + c_b t_*^{-1} h_{b,\rms 0}^{(0)} \bigr) &&\in \cA^{\la\cE_{\iota^+,\sscri}^\cC\ra,\ (1,0)\cup(1+\cE_\ind),\ (1+\cE_\cK,\,4+\eps_\ind-\eps)}, \\
  h_{b,\rms 1}^{\leq 1,\aug}(\dot\scal) &- \chi_\cK\bigl(h_{b,\rms 1}^{\leq 1}(\dot\scal) + c'_b h_{b,\rms 1}^{\leq 1}(\dot\scal\log t_*)\bigr) &&\in \cA^{\la\cE_{\iota^+,\sscri}^\cC\ra,\ (1,0)\cup(1+\cE_\ind),\ (1+\cE_\cK,\,4+\eps_\ind-\eps)},
\end{alignedat}
\end{equation}
for some constants $c_b$ and $c'_b$ (depending smoothly on $b$).

\begin{proof}[Proof of Lemma~\usref{LemmaAdmLoIm}]
  \pfstep{Linearized mass changes.} With $\chi_\cK$ as above, note that
  \[
    \chi_\cK\dot g_b^\Ups(1,0)\in\cA^{\infty,\ (1,0)\cup(1+\cE_\ind),\ (0,0)}(\Omega).
  \]
  (This follows from the fact that $\rho=r^{-1}$ is a joint defining function of $\scri^+\cup\iota^+$.) Therefore,
  \begin{equation}
  \label{EqAdmLoImLbgb}
    L_b(\chi_\cK\dot g_b^\Ups(1,0)) = [L_b,\chi_\cK]\dot g_b^\Ups(1,0) \in \rho_+^2\Diffb^2\bigl(\cA^{\infty,\ (1,0)\cup(1+\cE_\ind),\ \infty}\bigr) \subset \cA^{\infty,\ (3,0)\cup(3+\cE_\ind),\ \infty}(\Omega).
  \end{equation}
  We capture the leading-order term of this at $\iota^+$ using
  \begin{equation}
  \label{EqAdmLoImfb}
    f_b := -\bigl(t_*^3 L_b(\chi_\cK\dot g_b^\Ups(1,0))\bigr)\big|_{\iota^+} \in \CIc((\iota^+)^\circ;S^2\cT^*).
  \end{equation}
  This is of scalar type $0$. Applying Proposition~\ref{PropipGr} to $f_b$ with $\lambda=1$ (which does not lie in $\spec_{\iota^+}(\ubar L)$ by Proposition~\ref{PropipMero}\eqref{ItipMeroInv}), arbitrary $\ell_\cK<3+\eps_\ind$, and $\ell_\sscri=3+\eps_\sscri$ produces $u_b=\chi_\cK u_{b,0}+\chi_\iota t_*^{-1}\tilde u$ as in~\eqref{EqipGru} such that $\chi_\iota t_*^{-3}f_b-L_b u_b\in\Hb^{\infty,\ (\cE_{\iota^+,\sscri}^\tot+2,\ 4+\eps_\sscri),\ 3+\cE_\ind,\ 4+\eps_\ind-\eps}(\Omega_*;S^2\cT^*)^{\bullet,-}$ for all $\eps>0$. We then set
  \[
    \dot g_b^{\Ups,\aug}(1,0) := \chi_\cK\dot g_b^\Ups(1,0) + u_b.
  \]
  This arranges~\eqref{EqAdmLoImKerr} and the first membership of~\eqref{EqAdmLoImLb}, and the construction gives also the first line of~\eqref{EqAdmLoImLot}; note that only the scalar type $0$ terms of~\eqref{EqipGr} contribute to $u_b$, so the term with the least amount of $t_*$-decay is $\chi_\cK t_*^{-1} h_{b,\rms 0}^{(0)}$.

  \pfstep{Boosts.} We have $L_b h_{b,\rms 1}^{\leq 1}(\dot\scal)=0$ and $\chi_\cK h_{b,\rms 1}^{\leq 1}(\dot\scal)\in\cA^{\infty,\ (1,0)\cup(1+\cE_\ind),\ (-1,0)}(\Omega)$, and therefore
  \[
    f_b(\dot\scal) := -\Bigl(t_*^3 L_b\bigl(\chi_\cK h_{b,\rms 1}^{\leq 1}(\dot\scal)\bigr)\Bigr)\Big|_{\iota^+} \in \CIc((\iota^+)^\circ;S^2\cT^*).
  \]
  Note that $f_b(\dot\scal)$ is of scalar type $1$ (cf.\ the discussion after~\eqref{EqAdmLo13}). We now define $u_b(\dot\scal)$ using Proposition~\ref{PropipGr} and set
  \begin{equation}
  \label{EqAdmLoImBoostConstr}
    h_{b,\rms 1}^{\leq 1,\aug}(\dot\scal) := \chi_\cK h_{b,\rms 1}^{\leq 1}(\dot\scal) + u_b(\dot\scal).
  \end{equation}
  This satisfies all requirements including, by construction,~\eqref{EqAdmLoImLot} since only the scalar type $1$ terms of~\eqref{EqipGr} feature in $u_b(\dot\scal)$.
\end{proof}

We next prove estimates for the Fourier transforms of (tensors related to) these improved states. We begin with estimates on the scattering-b-transition Sobolev spaces used in the low-energy analysis of \citeAF{\S\ref*{SsSpLo}}:

\begin{lemma}[Fourier transforms of source terms]
\label{LemmaAdmLoImFT}
  We use the notation of Lemma~\usref{LemmaAdmLoIm}. For $\sigma\in\C$, $\Im\sigma\geq 0$, define $\hat f(\sigma,\cdot)$ to be the Fourier transform in $t_*$ (on $\R_{t_*}\times X$) of $f=f(t_*,x)=L_b\dot g_b^{\Ups,\aug}(1,0)$, resp.\ $L_b h_{b,\rms 1}^{\leq 1,\aug}(\dot\scal)$, $\dot\scal\in\scalspace_1$. For $\sigma\neq 0$, we introduce $\hat\sigma:=\frac{\sigma}{|\sigma|}\in e^{i[0,\pi]}$. Let $m\in\N_0$ be an arbitrary scattering-b-transition order, $\alpha_\sface\in(-\frac32,-\frac32+\eps_\ind)$, and let $\sfr\in\CI(\ol{\Tscbt^*_\scface}X)$ be a scattering decay order such that $\sfr+\alpha_\sface<-\frac12$ at the zero section of $\Tscbt^*_\scface X$. We use the (asymptotically Euclidean) density $|\dd g_b|_X|$ to define $L^2(X)$.
  \begin{enumerate}
  \item\label{ItAdmLoImFT0}{\rm (Zero frequency.)} We have $\hat f(0)\in\cA^{2+\eps_\ind-\eps}(X;S^2\cT^*_X)$ for all $\eps>0$, and thus $\hat f(0)\in\Hb^{\infty,\frac12+\eps_\ind-\eps}(X;S^2\cT^*_X)$ for all $\eps>0$. Furthermore,
    \begin{equation}
    \label{EqItAdmLoImFT0}
      \hat f(0) \equiv [L_b,t_*]\dot g_b^\Ups(1,0),\ \text{resp.}\ \frac12[[L_b,t_*],t_*]h_{b,\rms 1}(\dot\scal) + [L_b,t_*]\breve h_{b,\rms 1}^1(\dot\scal)
    \end{equation}
    modulo $\wh{L_b}(0)(\bigcap_{\eps>0}\cA^{-\eps}(X;S^2\cT^*_X))$.
  \item\label{ItAdmLoImFTUnif}{\rm (Uniform bounds.)} Let $k\in\N_0$. Then $\hat f(|\sigma|\hat\sigma,\cdot)$ is uniformly bounded in the norm of $H_{(\scbtop,|\sigma|);\bop}^{(m;k),(\sfr+\alpha_\sface+1,\alpha_\sface+2,0)}(X;S^2\cT^*_X)$ (see Definition~\usref{DefTscbt}) for all $\hat\sigma\in e^{i[0,\pi]}$ and $|\sigma|\leq 1$. The same holds for derivatives, of any finite order, of $\hat f(\sigma,\cdot)$ along $\sigma\pa_\sigma$.
  \item\label{ItAdmLoImFTLo}{\rm (Regularity at $0$.)} Let $\chi_\zface\in\CI(X_\scbtop)$ be equal to $1$ near $\zface$ and $0$ near $\scface$. Regarding functions on $X^\circ$ as $\varsigma$-independent functions on $(0,1)_\varsigma\times X^\circ$, the difference
    \[
      \hat f(\sigma) - \chi_\zface\hat f(0)
    \]
    is uniformly (in $\hat\sigma\in e^{i[0,\pi]}$ and $|\sigma|\leq 1$) bounded in $H_{(\scbtop,|\sigma|);\bop}^{(m;k),(\sfr+\alpha_\sface+1,\alpha_\sface+2,\eps_\ind-\eps)}(X;S^2\cT^*_X)$ for every fixed $k\in\N_0$ and $\eps>0$.
  \item\label{ItAdmLoImFTHolo}{\rm (Holomorphicity.)} $\hat f(\sigma,\cdot)$ is holomorphic in $\Im\sigma>0$ with values in $H_\scop^{m,\sfr+\alpha_\sface+1}(X;S^2\cT^*_X)$.
  \end{enumerate}
\end{lemma}
\begin{proof}
  By~\eqref{EqAdmLoImLb}, $f(t_*,x)$ is integrable in $t_*$ for all $x\in X^\circ$ (as its $\cK^+$-order is $>1$), and hence $\hat f(\sigma,x)$ is continuous in $\sigma$.

  \pfstep{Part~\eqref{ItAdmLoImFT0}.} Note that $\hat f(0,\cdot)=\int f(t_*,\cdot)\,\dd t_*$. We use~\eqref{EqAdmLoImLot}, so for fixed $\dot\scal\in\scalspace_1$, we write
  \begin{align*}
    \dot g_b^{\Ups,\aug}(1,0) &= \chi_\cK \dot g_b^\Ups(1,0) + \chi_\cK c_b t_*^{-1}h_{b,\rms 0}^{(0)} + h', \\
    h_{b,\rms 1}^{\leq 1,\aug} &= \chi_\cK h_{b,\rms 1}^{\leq 1}(\dot\scal) + \chi_\cK c'_b h_{b,\rms 1}^{\leq 1}(\dot\scal\log t_*) + h'',
  \end{align*}
  where
  \[
    h',h''\in\cA^{1,1,1+\eps_\ind}(\Omega).
  \]

  We first control the contributions of $h'$, resp.\ $h''$; both are of the same class, so we only consider $h'$. Note that, a fortiori,
  \[
    h'\in\bigcap_{\eps>0}\cA^{-\eps,1,1+\eps}=\bigcap_{\eps>0} \dot\cA^{1+\eps}([1,\infty]_{t_*};\cA^{-\eps}(X)),
  \]
  so the $t_*$-integral of $h'$ satisfies $\wh{h'}(0)\in\cA^{-\eps}(X)$, while the total $t_*$-integral of $\pa_{t_*}^j h'$ vanishes when $j\geq 1$. Writing
  \[
    L_b h' = \wh{L_b}(0) h' + [L_b,t_*]\ftrans(0)\pa_{t_*}h' + \frac12[[L_b,t_*],t_*]\pa_{t_*}^2 h'
  \]
  and integrating in $t_*$, we thus find that
  \[
    \int L_b h'\,\dd t_* = \wh{L_b}(0) \wh{h'}(0) \in \wh{L_b}(0)\Biggl(\bigcap_{\eps>0}\cA^{-\eps}(X)\Biggr)
  \]
  is an error term. More generally, we have shown that terms of class $L_b(\bigcap_{\eps>0}\cA^{-\eps,1,1+\eps})$ only contribute error terms.

  Next, let $\chi_+=\chi_+(t_*)$ be equal to $0$ for $t_*\leq 1$ and equal to $1$ for $t_*\geq 2$. Then the action of $L_b$ on $(\chi_\cK-\chi_+)\dot g_b^\Ups(1,0)\in\cA^{1,1,\infty}(X)$ contributes an error term, while
  \begin{align*}
    \int L_b(\chi_+\dot g_b^\Ups(1,0))\,\dd t_* &= \biggl(\int \chi_+'(t_*)\,\dd t_*\biggr) [L_b,t_*]\dot g_b^\Ups(1,0) + \biggl(\int \chi_+''(t_*)\,\dd t_*\biggr) \frac12[[L_b,t_*],t_*]\dot g_b^\Ups(1,0) \\
      &= [L_b,t_*]\dot g_b^\Ups(1,0).
  \end{align*}
  Finally, the action of $L_b$ on $(\chi_\cK-\chi_+)t_*^{-1}h_{b,\rms 0}^{(0)}\in\cA^{0,1,\infty}$ contributes an error term, while
  \begin{align*}
    &\int L_b\bigl(\chi_+ t_*^{-1}h_{b,\rms 0}^{(0)}\bigr)\,\dd t_* \\
      &\qquad = \underbrace{\biggl(\int \pa_{t_*}(\chi_+(t_*)t_*^{-1})\,\dd t_*\biggr)}_{=0} [L_b,t_*]h_{b,\rms 0}^{(0)} + \underbrace{\biggl(\int \pa_{t_*}^2(\chi_+(t_*)t_*^{-1})\,\dd t_*\biggr)}_{=0} \frac12[[L_b,t_*],t_*]h_{b,\rms 0}^{(0)} = 0.
  \end{align*}
  This proves the first part of~\eqref{EqItAdmLoImFT0}.

  Turning to $h_{b,\rms 1}^{\leq 1,\aug}(\dot\scal)$, we similarly note that the action of $L_b$ on $(\chi_\cK-\chi_+)h_{b,\rms 1}^{\leq 1}(\dot\scal)\in\cA^{1,1,\infty}$ and $(\chi_\cK-\chi_+)h_{b,\rms 1}^{\leq 1}(\dot\scal\log t_*)\in\cA^{1,2-\eps,\infty}$ contributes error terms. We next compute
  \begin{align*}
    &\int L_b\bigl(\chi_+(t_*)(t_* h_{b,\rms 1}+\breve h_{b,\rms 1}^1)\bigr)\,\dd t_* \\
    &\qquad = \int (t_*\chi_+)'[L_b,t_*]h_{b,\rms 1} + \chi_+\wh{L_b}(0)\breve h_{b,\rms 1}^1 + \frac12(t_*\chi_+)''[[L_b,t_*],t_*]h_{b,\rms 1} + \chi_+'[L_b,t_*]\breve h_{b,\rms 1}^1 \\
    &\qquad \hspace{26em} + \frac12\chi_+''[[L_b,t_*],t_*]\breve h_{b,\rms 1}^1\,\dd t_*.
  \end{align*}
  The first two terms in the integrand add up to the sum of $\chi_+ L_b h_{b,\rms 1}^{\leq 1}=0$ and $t_*\chi_+'[L_b,t_*]h_{b,\rms 1}=-t_*\chi_+'\wh{L_b}(0)\breve h_{b,\rms 1}^1$; the $t_*$-integral of the latter is thus a constant multiple of $\wh{L_b}(0)\breve h_{b,\rms 1}^1$, and hence in view of $\breve h_{b,\rms 1}^1\in\cA^1$ an error term. The third and fourth terms integrate to $\frac12[[L_b,t_*],t_*]h_{b,\rms 1}+[L_b,t_*]\breve h_{b,\rms 1}^1$. The final term integrates to $0$. Similarly, writing $\chi_+(t_*)t_*^{-1}=(\chi_+\log t_*)'-\chi_+'\log t_*$, we have
  \begin{align*}
    &\int L_b\bigl(\chi_+(t_*)((\log t_*)h_{b,\rms 1}+t_*^{-1}\breve h_{b,\rms 1}^1)\bigr)\,\dd t_* \\
    &\qquad = \int (\chi_+\log t_*)'' \Bigl( \frac12[[L_b,t_*],t_*]h_{b,\rms 1}+[L_b,t_*]\breve h_{b,\rms 1}^1\Bigr) + (\chi_+\log t_*)''' \frac12[[L_b,t_*],t_*]\breve h_{b,\rms 1}^1 \\
    &\qquad\hspace{26em} - L_b\bigl( \chi'_+(t_*)(\log t_*)\breve h_{b,\rms 1}^1\bigr)\,\dd t_*.
  \end{align*}
  The first two summands integrate to $0$ since $(\chi_+\log t_*)'$ and $(\chi_+\log t_*)''$ vanish at infinity. We rewrite the third summand as the sum of three terms: $-(\chi_+'\log t_*)'[L_b,t_*]\breve h_{b,\rms 1}^1$ (which integrates to $0$), further $-(\chi_+'\log t_*)''\frac12[[L_b,t_*],t_*]\breve h_{b,\rms 1}^1$ (which also integrates to $0$), and $-\chi'_+(t_*)(\log t_*)\wh{L_b}(0)\breve h_{b,\rms 1}^1$ (which integrates to a multiple of $\wh{L_b}(0)\breve h_{b,\rms 1}^1$, which is thus an error term). This proves the second part of~\eqref{EqItAdmLoImFT0}.

  \pfstep{Part~\eqref{ItAdmLoImFTUnif}, almost-real $\sigma$.} From~\eqref{EqAdmLoImLb} we obtain $f\in\rho_\sscri^{3-\eps}\rho_+^{3+\eps_\ind-\eps}\rho_\cK^{4+\eps_\ind-\eps}\Hb^\infty(\Omega)$ for all $\eps>0$ (with respect to an unweighted b-density). By~\eqref{EqTFHbLo}, we thus have
  \begin{equation}
  \label{EqAdmLoImfHb}
    (\hat f(\pm\varsigma,\cdot)|_{\varsigma\in[0,1]} \in \Hb^{\infty,\ 3-\eps,\ 2+\eps_\ind-\eps,\ ((0,0),3+\eps_\ind-\eps)}(X_\scbtop),\quad\eps>0,
  \end{equation}
  where we use an unweighted b-density on $X_\scbtop$ to define $L^2(X_\scbtop)$; a fortiori, using the notation of Definition~\ref{DefTscbt},
  \[
    \rho_\scface^{-(3-\eps)}\rho_\tface^{-(2+\eps_\ind-\eps)}\hat f(\pm\varsigma,\cdot) \in \cA^0\bigl([0,1)_\sigma;\Hb^{\infty,0}(X,\mu_\bop)\bigr),
  \]
  where $\mu_\bop$ is an unweighted b-density on $X$. This in turn implies uniform bounds for the restrictions $\hat f(\pm\varsigma,\cdot)$ to $\varsigma$-level sets with respect to the norm
  \begin{equation}
  \label{EqAdmLoImfHb2}
    \|\hat f(\pm\varsigma,\cdot)\|_{H_{\bop,\varsigma}^{k,(3-\eps,\ 2+\eps_\ind-\eps,\ 0)}(X,\mu_\bop)} := \| \rho_\scface^{-(3-\eps)}\rho_\tface^{-(2+\eps_\ind-\eps)}\hat f(\pm\varsigma,\cdot)\|_{\Hb^k(X,\mu_\bop)}
  \end{equation}
  for all $k\in\N_0$ and $\eps>0$. Finally, we use~\citeAF{(\ref*{EqDResscbt1})} to deduce uniform bounds for $\hat f(\hat\sigma|\sigma|,\cdot)$, $\hat\sigma=\pm 1$, $|\sigma|\in[0,1]$, in the space $H_{(\scbtop,|\sigma|);\bop}^{(m;k),(\sfr+\alpha_\sface+1,\alpha_\sface+2,0)}$ for all sc-b-transition regularity orders $m$, b-regularity orders $k$, and for all $\alpha_\sface<(2+\eps_\ind-\eps)-\frac72$, i.e., $\alpha_\sface<-\frac32+\eps_\ind$ since $\eps>0$ is arbitrary, under the sole assumption that at the zero section of $\Tscbt^*X$ over $\scface$, we have $\sfr+\alpha_\sface+1<\frac32-\eps$, i.e., $\sfr+\alpha_\sface<\frac12$ since $\eps>0$ is arbitrary. (For use after~\eqref{EqAdmLoImFExp} below, we remark that if we only had the decay order $2-\eps$ at $\scface$, the same conclusions would remain valid except one would need $\sfr+\alpha_\sface<-\frac12$.) Since the Fourier transform intertwines $-\pa_{t_*}t_*$ and $\sigma\pa_\sigma$, the higher regularity statement in part~\eqref{ItAdmLoImFTUnif} follows from the fact that application of $\sigma\pa_\sigma$ preserves the space~\eqref{EqAdmLoImfHb}.

  We can treat the case of $\hat\sigma\in e^{i[0,\frac{\pi}{4}]}\cup e^{i[\frac{\pi}{4},\frac{3\pi}{4}]}$ similarly: write
  \begin{equation}
  \label{EqAdmLoImfHb3}
    \hat f(\sigma,\cdot) = \int e^{-(\Im\sigma)t_*} e^{i(\Re\sigma)t_*}f(t_*,\cdot)\,\dd t_* = \cF\bigl( e^{-(\Im\hat\sigma)|\sigma|t_*}f(t_*,\cdot)\bigr) ( |\sigma|\Re\hat\sigma ).
  \end{equation}
  Since $|\sigma|\Re\hat\sigma\sim\pm|\sigma|$, we then only need to observe that, in view of $\supp f\subset\{t_*\geq 1\}$, the family $\{e^{-\delta t_*}f(t_*,\cdot)\colon 0\leq\delta\leq 1\}$ is uniformly bounded in the space $\rho_\sscri^{3-\eps}\rho_+^{3+\eps_\ind-\eps}\rho_\cK^{4+\eps_\ind-\eps}\Hb^\infty(\Omega)$ for every fixed $\eps>0$, just like $f$ itself; this follows from the fact that $e^{-\delta t_*}$ is uniformly bounded in the space $\cA^{0,0,0}(\Omega)$ (which here amounts to the uniform, in $\delta$ and for $t_*\geq 1$, boundedness of any finite-order derivative of $e^{-\delta t_*}$ along $t_*\pa_{t_*}$).

  \pfstep{Part~\eqref{ItAdmLoImFTLo}, almost real $\sigma$.} From~\eqref{EqAdmLoImfHb}, we obtain
  \[
    \bigl(\hat f(\pm\varsigma,\cdot) - \chi_\zface\hat f(0,\cdot)\bigr)\big|_{\varsigma\in[0,1]} \in \Hb^{\infty,\ 3-\eps,\ 2+\eps_\ind-\eps,\ 1-\eps}(X_\scbtop),\quad\eps>0.
  \]
  This implies uniform bounds for the $H_{\bop,\varsigma}^{k,(3-\eps,\ 2+\eps_\ind-\eps,\ 1-\eps)}(X,\mu_\bop)$-norms of restrictions to $\varsigma$-level sets; here $k\in\N_0$ is arbitrary. Analogously to the considerations following~\eqref{EqAdmLoImfHb2}, this implies uniform bounds in $H_{(\scbtop,\varsigma);\bop}^{(m;k),(\sfr+\alpha_\sface+1,\alpha_\sface+2,1-\eps)}$. This treats the case $\hat\sigma=\pm 1$; the case $\hat\sigma\in e^{i[0,\frac{\pi}{4}]}\cup e^{i[\frac{3\pi}{4},\pi]}$ follows similarly using the observation~\eqref{EqAdmLoImfHb3}.

  \pfstep{Part~\eqref{ItAdmLoImFTUnif}, almost-imaginary $\sigma$.} Consider $\hat\sigma\in e^{i[\frac{\pi}{4},\frac{3\pi}{4}]}$. Since $|\Re\hat\sigma|\leq\Im\hat\sigma$, we have
  \[
    e^{i\sigma t_*} = e^{-\frac12|\sigma|(\Im\hat\sigma)t_*} e^{|\sigma| (i\Re\hat\sigma - \frac12\Im\hat\sigma)t_*}.
  \]
  The family $e^{|\sigma|(i\Re\hat\sigma-\frac12\Im\hat\sigma)t_*}f(t_*,\cdot)$ is uniformly bounded in $\rho_\sscri^{3-\eps}\rho_+^{3+\eps_\ind-\eps}\rho_\cK^{4+\eps_\ind-\eps}\Hb^\infty(\Omega)$ for every fixed $\eps>0$, and thus it suffices to consider the case $\hat\sigma=e^{i\frac{\pi}{2}}=i$. We only need to use that
  \begin{equation}
  \label{EqAdmLoImFProd}
    f \in \dot H_\bop^{\infty,1+\delta}\bigl([1,\infty]_{t_*}; \Hb^{\infty,2+\eps_\ind-\eps-\delta}(X,\mu_\bop)\bigr)
  \end{equation}
  for all $\eps>0$ and $\delta\in(0,\eps_\ind-\eps)$, and note that $e^{-|\sigma|t_*}f$ is uniformly bounded in this space for $|\sigma|\in[0,1]$; since $\la t_*\ra^{-1-\delta}$ is integrable in $t_*$, we conclude that
  \[
    \int e^{-|\sigma|t_*}f(t_*,\cdot)\,\dd t_* \in \Hb^{\infty,2+\eps_\ind-\eps-\delta}(X,\mu_\bop)
  \]
  is uniformly bounded for any fixed $\eps,\delta>0$. Therefore, a fortiori,
  \begin{equation}
  \label{EqAdmLoImFExp}
    \|\hat f(i|\sigma|,\cdot)\|_{H_{\bop,|\sigma|}^{k,(2-\eps,\ 2+\eps_\ind-\eps,\ 0)}(X,\mu_\bop)}
  \end{equation}
  is uniformly bounded (where we relaxed the $\scface$-order) for $|\sigma|\in[0,1]$ and for any fixed $\eps>0$, which by \citeAF{(\ref*{EqDResscbt1})} implies the claim (as mentioned before). Higher $\sigma\pa_\sigma$-regularity follows as before from the fact that applications of $t_*\pa_{t_*}$ preserve the space~\eqref{EqAdmLoImFProd}.

  \pfstep{Part~\eqref{ItAdmLoImFTLo}, almost-imaginary $\sigma$.} We only use the information~\eqref{EqAdmLoImFProd}. Fix any $\delta'\in(0,\delta)$ and write
  \begin{equation}
  \label{EqAdmLoImFDiff}
    \int e^{-|\sigma|t_*}f(t_*,\cdot)\,\dd t_* - \int f(t_*,\cdot)\,\dd t_* = |\sigma|^{\delta'}\int \tilde e(|\sigma|t_*) t_*^{\delta'}f(t_*,\cdot)\,\dd t_*,
  \end{equation}
  where the function
  \[
    \tilde e(x) := \frac{e^{-x}-1}{x^{\delta'}},\quad x>0,
  \]
  is uniformly bounded, as is $(x\pa_x)^j\tilde e$ for all $j\in\N$. Since
  \[
    t_*^{\delta'}f(t_*,\cdot)\in\dot H_\bop^{\infty,1+\delta-\delta'}\bigl([1,\infty]_{t_*};\Hb^{\infty,2+\eps_\ind-\eps-\delta}(X,\mu_\bop)\bigr)
  \]
  is still integrable in $t_*$, we infer that~\eqref{EqAdmLoImFDiff} is uniformly (for $|\sigma|\in[0,1]$) bounded in the norm of $|\sigma|^{\delta'}\Hb^{k,2+\eps_\ind-\eps-\delta}(X,\mu_\bop)$ for every fixed $k\in\N_0$. Since $|\sigma|=\rho_\zface\rho_\tface$, this norm is equal to the $H_{\bop,|\sigma|}^{k,(2+\eps_\ind-\eps-\delta,\ 2+\eps_\ind-\eps+\delta-\delta',\ \delta')}(X,\mu_\bop)$-norm. Let us relax the $\scface$-order to $2-\eps$. Taking $\eps\searrow 0$ and $\delta',\delta\nearrow\eps_\ind$ then yields the desired result.

  \pfstep{Part~\eqref{ItAdmLoImFTHolo}.} This Paley--Wiener type statement follows immediately from the forward support property of $f$, with the membership in $H_\scop^{m,\sfr+\alpha_\sface+1}$ being a consequence of part~\eqref{ItAdmLoImFTUnif}.
\end{proof}

The zero energy states $\dot g_b^\Ups(0,\dot\bha)$, $\dot\bha\in\R^3$, corresponding to linearized changes of the angular momentum parameter have stronger (namely, $\cO(r^{-2})$- instead of $\cO(r^{-1})$-) decay as $r\to\infty$. For full consistency with Lemma~\ref{LemmaAdmLoIm}, we nonetheless define ``improved'' zero energy states also in this case, to wit,
\begin{subequations}
\begin{equation}
\label{EqAdmLoImKerr2}
  \dot g_b^{\Ups,\aug}(0,\dot\bha) := \chi_\cK\dot g_b^\Ups(0,\dot\bha),\quad\dot\bha\in\R^3.
\end{equation}
We then have
\begin{equation}
\label{EqAdmLoImLb2}
  L_b\dot g_b^{\Ups,\aug}(0,\dot\bha) \in \cA^{\cE_{\iota^+,\sscri}^\tot+2,\ 3+\cE_\ind,\ 4+\eps_\ind-\eps} \subset \cA^{3,\ 3+\eps_\ind-\eps,\ 4+\eps_\ind-\eps}(\Omega)
\end{equation}
exactly as in~\eqref{EqAdmLoImLb}. (In fact, the index sets at $\scri^+$ and $\cK^+$ are empty, and the index set at $\iota^+$ is $(4,0)\cup(4+\cE_\ind)$.) Lemma~\ref{LemmaAdmLoImFT} generalizes similarly (with the same proof) and gives
\begin{equation}
\label{EqAdmLoImF2}
  \cF\bigl(L_b\dot g_b^{\Ups,\aug}(0,\dot\bha)\bigr)(0) \equiv [L_b,t_*]\dot g_b^\Ups(0,\dot\bha) \bmod\wh{L_b}(0)\Biggl(\bigcap_{\eps>0}\cA^{-\eps}(X)\Biggr);
\end{equation}
\end{subequations}
and also Lemma~\ref{LemmaAdmLoImFT}\eqref{ItAdmLoImFTUnif}--\eqref{ItAdmLoImFTHolo} are valid for the Fourier transform of $L_b\dot g_b^{\Ups,\aug}(0,\dot\bha)$.

For use in rather precise low-energy resolvent expansions (starting in~\S\ref{SsD3Alm}), we record:

\begin{lemma}[Partial polyhomogeneity of Fourier transforms]
\label{LemmaAdmLob}
  Let $\dot\scal\in\scalspace_1$ and $\dot b=(\dot\bhm,\dot\bha)\in\R\times\R^3$, and consider $h\in\{\dot g_b^{\Ups,\aug}(\dot b), h_{b,\rms 1}^{\leq 1,\aug}(\dot\scal)\}$. Then
  \begin{equation}
  \label{EqAdmLob}
    \bigl(\cF(L_b h)(\pm\varsigma)\bigr)\big|_{\varsigma\in[0,1]} \in \Hb^{\infty,\ 3-\eps,\ \cE_\ind+2,\ \bigl((0,0),3+\eps_\ind-\eps\bigr)}(X_\scbtop;S^2\cT^*_X)\quad\forall\,\eps>0,
  \end{equation}
  where $\cE_\ind$ is an index set with $\min\Re\cE_\ind\geq\eps_\ind$.
\end{lemma}
\begin{proof}
  This is a special case of Proposition~\ref{PropTFHbphg}\eqref{ItTFHbphgF} with $\beta_\cK:=4+\eps_\ind-\eps$, $\cE_\sscri:=\cE_{\iota^+,\sscri}^\tot+2$, and $\cE_\iota:=3+\cE_0^\Ups$, where we write $\cE_0^\Ups$ for the index set denoted $\cE_\ind$ in Lemma~\ref{LemmaAdmLoIm}, while $\beta_\sscri$ and $\beta_\iota$ there are arbitrary (but subject to~\eqref{EqTFHbphgBetaScri}) and $\cE_\cK=\emptyset$. The $\tface$-index set of $\cF(L_b h)$ is thus $(\cE_0^\Ups+2)\extcup(\cE_{\iota^+,\sscri}^\tot+2):=\cE_\ind+2$ for $\cE_\ind:=\cE_0^\Ups\extcup\cE_{\iota^+,\sscri}^\tot$; recall here that $\min\Re\cE_{\iota^+,\sscri}^\tot\geq 1$. (We relax the $\scface$-order in~\eqref{EqAdmLob} to $3-\eps$ since we will not have any use for the slightly more precise consequence of~\eqref{EqTFHbphgF}.)
\end{proof}

\subsubsection{Augmented operator: definition and uniform resolvent bounds}
\label{SssAdmLoDef}

--- \emph{Unless otherwise noted, we use the asymptotically Euclidean density $|\dd g_b|_X|$ to define $L^2$-spaces on $X$.} Using the notation of Lemma~\ref{LemmaAdmLoIm} as well as~\eqref{EqAdmLoImKerr2}, we now define
\begin{equation}
\label{EqAdmLoDef}
  \wt{L_b}(\sigma):=
    \begin{pmatrix}
      \wh{L_b}(\sigma) & \wh{L_b}(\sigma)\bigl(\cF(\dot g_b^{\Ups,\aug}(\cdot))(\sigma)\bigr) & \wh{L_b}(\sigma)\bigl(\cF(h_{b,\rms 1}^{\leq 1,\aug}(\cdot)(\sigma))\bigr) \\
      \la\cdot,f_1^*\ra_{L^2} & 0 & 0 \\
      \la\cdot,f_2^*\ra_{L^2} & 0 & 0
    \end{pmatrix}
\end{equation}
acting on the direct sum of $\sD'(X^\circ;S^2\cT^*_X)$, $\C^4$, and $\scalspace_1$; here $f_1^*$ and $f_2^*$ are as in~\eqref{EqAdmLofstar}--\eqref{EqAdmLofstar2}. To shorten notation, for a function space $H$ on $X^\circ$, we shall write
\[
  \tilde H := H \oplus \C^4 \oplus \scalspace_1,\quad
  \|(u,\dot b,\dot\scal)\|_{\tilde H} := \|u\|_H + |\dot b| + |\dot\scal|,
\]
where we fix any linear isomorphism $\scalspace_1\cong\C^3$ for the final term. We first prove an analogue of \citeAF{Lemma~\ref*{LemmaA2Adm0Op}}:

\begin{prop}[Zero energy operator]
\label{PropAdmLoDef0}
  For appropriate b-regularity orders\footnote{such as the order induced by $\sfs$ in~\eqref{EqAdm} on the spectral side at zero frequency, as explained in \citeAF{\S\ref*{SsSpTs}}, or for all sufficiently large $s_0$} $s_0$ and for all $\alpha_\sface\in(-\frac32,-\frac32+\eps_\ind)$, the map
  \begin{equation}
  \label{EqAdmLoDef0}
    \wt{L_b}(0) \colon \bigl\{ (u,\dot b,\dot\scal)\in\tilde H_\bop^{s_0,\alpha_\sface}(X,|\dd g_b|_X|;S^2\cT^*_X) \colon \wt{L_b}(0)(u,\dot b,\dot\scal)\in\tilde H_\bop^{s_0-1,\alpha_\sface+2} \bigr\} \to \tilde H_\bop^{s_0-1,\alpha_\sface+2}
  \end{equation}
  is invertible. (When working with an unweighted b-density instead, this holds for all $\alpha_\sface\in(0,\eps_\ind)$.)
\end{prop}

This is a variant of the construction in Corollary~\ref{CorWE0Solv}, albeit a more delicate one since our present demands (concerning uniform analysis of $\wh{L_b}(\sigma)^{-1}$ as $|\sigma|\to 0$, rather than merely the right inversion of $\wh{L_b}(0)$) are considerably stronger.

\begin{proof}[Proof of Proposition~\usref{PropAdmLoDef0}]
  By Proposition~\ref{PropWEMode0}, the map~\eqref{EqAdmLoDef0} is Fredholm of index $0$. Therefore, we only need to prove its injectivity. Consider a kernel element $(u,(\dot\bhm,\dot\bha),\dot\scal)$. By Lemma~\ref{LemmaAdmLoImFT}, we then have
  \[
    \wh{L_b}(0)u + [L_b,t_*]\dot g_b^\Ups(\dot\bhm,\dot\bha) + \Bigl(\frac12[[L_b,t_*],t_*]h_{b,\rms 1}(\dot\scal) + [L_b,t_*]\breve h_{b,\rms 1}^1(\dot\scal)\Bigr) + \wh{L_b}(0)h' = 0
  \]
  where $h'\in\bigcap_{\eps>0}\cA^{-\eps}(X;S^2\cT^*_X)$ captures the error terms of~\eqref{EqItAdmLoImFT0} and thus depends linearly on $(\dot\bhm,\dot\bha,\dot\scal)$. Let us take the $L^2$-inner product, defined as in Lemma~\ref{LemmaWEPair}, of this with $h_{b,\rms 0}^*$. For the first term, we have $\la\wh{L_b}(0)u,h_{b,\rms 0}^*\ra=\la u,\wh{L_b}(0)^*h_{b,\rms 0}^*\ra=0$. Recall moreover that $h_{b,\rms 0}^*$ has $\cO(r^{-2})$ decay (as do all other dual states in~\eqref{EqWC0Dualh}), so integration by parts is justified also in the computation
  \[
    \la\wh{L_b}(0)h',h_{b,\rms 0}^*\ra=\la h',\wh{L_b}(0)^*h_{b,\rms 0}^*\ra=0.
  \]
  (Given the almost boundedness of $h'$, this in fact only requires $\cO(r^{-1-\eta})$-decay of $h_{b,\rms 0}^*$ for any $\eta>0$.) By Lemma~\ref{LemmaWEPair}, we thus obtain $-16\pi\dot\bhm=0$. Pairing with $h_{b,\rmv 1}^*(\vect)$ for all $\vect\in\vectspace_1$ then yields $\dot\bha=0$, again by Lemma~\ref{LemmaWEPair}. Finally, we pair with $h_{b,\rms 1}^*(\scal)$ for all $\scal\in\scalspace_1$ to deduce that $\dot\scal=0$. With $(\dot\bhm,\dot\bha,\dot\scal)=(0,0,0)$ and thus $h'=0$, we now conclude that $u\in\ker\wh{L_b}(0)$ is a linear combination of $\dot g_b^\Ups(b')$ and $h_{b,\rms 1}(\scal')$ for some $b'\in\R^4$ and $\scal'\in\scalspace_1$. But then the vanishing of the inner products of $u$ with $f_1^*$ and $f_2^*$ implies that $b'=0$ and $\scal'=0$, so $u=0$.
\end{proof}

We shall now prove the following analogue of \citeAF{Theorem~\ref*{ThmSpLo}}. We denote by $\sfs$ and $\sfr_\pm$ the regularity and decay orders (for positive, resp.\ negative frequencies) induced by the order function in~\eqref{EqAdm} as described in \citeAF{\S\ref*{SssSpOrderscbt}}.

\begin{prop}[Uniform low-energy estimates]
\label{PropAdmLoUnif}
  Let $\alpha_\sface\in(-\frac32,-\frac32+\eps_\ind)$. Then there exists $c>0$ such that for all $0\neq\sigma\in\C$ with $\Im\sigma\geq 0$ and $|\sigma|\leq c$, the operator $\wt{L_b}(\sigma)$ is invertible as a map
  \[
    \wt{L_b}(\sigma) \colon \bigl\{ (u,\dot b,\dot\scal) \in \tilde H_\scop^{\sfs,\sfr+\alpha_\sface}(X;S^2\cT^*_X) \colon \wt{L_b}(\sigma)(u,\dot b,\dot\scal) \in \tilde H_\scop^{\sfs-1,\sfr+\alpha_\sface+1} \bigr\} \to \tilde H_\scop^{\sfs-1,\sfr+\alpha_\sface+1}
  \]
  where $\sfr=\sfr_+$, resp.\ $\sfr_-$ when $\arg\sigma\in[0,\frac{\pi}{4}]$, resp.\ $[\frac{3\pi}{4},\pi]$, and $\sfr$ is arbitrary when $\arg\sigma\in[\frac{\pi}{4},\frac{3\pi}{4}]$ except for satisfying $\sfr+\alpha_\sface<-\frac12$ at the outgoing radial set over $\pa X$. Moreover, for such $\sigma$, and for all $k\in\N_0$, we have the uniform estimate
  \begin{equation}
  \label{EqAdmLoUnifEst}
    \|(u,\dot b,\dot\scal)\|_{\tilde H_{(\scbtop,|\sigma|);\bop}^{(\sfs;k),(\sfr+\alpha_\sface,\alpha_\sface,0)}(X;S^2\cT^*_X)} \leq C\|\wt{L_b}(\sigma)(u,\dot b,\dot\scal)\|_{\tilde H_{(\scbtop,|\sigma|);\bop}^{(\sfs-1;k),(\sfr+\alpha_\sface+1,\alpha_\sface+2,0)}(X;S^2\cT^*_X)}.
  \end{equation}
  Finally, $j$-fold conormal derivatives of the resolvent in $\sigma$ are uniformly bounded as maps
  \begin{equation}
  \label{EqAdmLoUnifInvReg}
    (\sigma\pa_\sigma)^j\wt{L_b}(\sigma)^{-1} \colon \tilde H_{(\scbtop,|\sigma|);\bop}^{(\sfs-1;k+j),(\sfr+\alpha_\sface+1,\alpha_\sface+2,0)} \to \tilde H_{(\scbtop,|\sigma|);\bop}^{(\sfs+j;k),(\sfr+\alpha_\sface,\alpha_\sface,0)}.
  \end{equation}
\end{prop}

The only novelty compared to \citeAF{Theorem~\ref*{ThmSpLo}} is our handling of the augmentation terms, which are more general than those considered in \citeAF{Remark~\ref*{RmkSpLoAug}} (but merely require a more attentive bookkeeping).

\begin{proof}[Proof of Proposition~\usref{PropAdmLoUnif}]
  We first discuss the case $k=0$. We write $\sigma=|\sigma|\hat\sigma$. We moreover fix $\chi_\zface,\chi_\tface\in\CI(X_\scbtop)$, with $\chi_\zface$ equal to $1$ near $\zface$ and $0$ near $\scface$, and $\chi_\tface$ equal to $1$ near $\tface$ and supported nearby.

  \pfstep{Estimate near $\zface$.} Using the equivalence of $\scbtop$- and $\bop$-Sobolev norms near $\zface$ (see~\citeAF{(\ref*{EqMUscbtNormEquiv})}), we first use Proposition~\ref{PropAdmLoDef0} to estimate, for any fixed $N$,
  \begin{equation}
  \label{EqAdmLoUnif0}
  \begin{split}
    \|(\chi_\zface u,\dot b,\dot\scal)\|_{\tilde H_{\scbtop,|\sigma|}^{\sfs,(\sfr+\alpha_\sface,\alpha_\sface,0)}} &\sim \|(\chi_\zface u,\dot b,\dot\scal)\|_{\tilde H_\bop^{\sfs,\alpha_\sface}} \\
      &\lesssim \|\wt{L_b}(0)(\chi_\zface u,\dot b,\dot\scal)\|_{\tilde H_\bop^{\sfs-1,\alpha_\sface+2}} \sim \|\wt{L_b}(0)(\chi_\zface u,\dot b,\dot\scal)\|_{\tilde H_{\scbtop,|\sigma|}^{\sfs-1,(-N,\alpha_\sface+2,0)}}.
  \end{split}
  \end{equation}

  We wish to replace $\wt{L_b}(0)$ on the right by $\wt{L_b}(\sigma)$ and need to control the resulting error term. Note then that the difference $\wt{L_b}(\sigma)-\wt{L_b}(0)$ is (uniformly in $|\sigma|$) of class
  \begin{equation}
  \label{EqAdmLoUnifDiff}
    \begin{pmatrix}
      \rho_\tface^2\rho_\zface\Diff_\scbtop^1 & H_{\scbtop,|\sigma|}^{m,(\sfr_0+\alpha_0+1,\alpha_0+2,\eps_0)} & H_{\scbtop,|\sigma|}^{m,(\sfr_0+\alpha_0+1,\alpha_0+2,\eps_0)} \\ 0 & 0 & 0 \\ 0 & 0 & 0
    \end{pmatrix}
  \end{equation}
  for any fixed $m\in\N$, $\alpha_0\in(-\frac32,-\frac32+\eps_\ind)$, and $\eps_0\in(0,\eps_\ind)$, and with $\sfr_0$ chosen to satisfy the requirements of Lemma~\ref{LemmaAdmLoImFT}. Here, for the $(1,1)$-entry, we use the notation from Definition~\ref{DefTscbt} and an inspection of
  \begin{equation}
  \label{EqAdmSpecFam}
    2\wh{L_b}(\sigma) = 2 i\sigma\rho(\rho\pa_\rho-1-\underbrace{S_{E^\Ups,E^\cC}}_{\in\CI(X)})\ \ + \underbrace{\wh{L_b}(0)}_{\in\rho^2\Diffb^2(X)} -\  i\sigma\underbrace{Q}_{\in\rho^3\Diffb^1(X)} + \underbrace{g^{0 0}}_{\in\rho^2\CI(X)}\sigma^2
  \end{equation}
  (from Lemma~\ref{LemmaWEOp}), while for the $(1,2)$- and $(1,3)$-entries we use Lemma~\ref{LemmaAdmLoImFT}\eqref{ItAdmLoImFTLo}. Now, a function that is uniformly bounded in $H_{\scbtop,|\sigma|}^{m,(\sfr_0+\alpha_0+1,\alpha_0+2,\eps_0)}$ has $H_{\scbtop,|\sigma|}^{\sfs-1,(-N,\alpha_\sface+2,0)}$-norm bounded uniformly by $C|\sigma|^{\eps_0}$ provided $-N\leq\sfr_0+\alpha_0+1$ and $\alpha_0+2\geq\alpha_\sface+2+\eps_0$; indeed, multiplication by $|\sigma|^{-\eps_0}=\rho_\zface^{-\eps_0}\rho_\tface^{-\eps_0}$ then maps the former space in a uniformly bounded fashion to the latter space. We can arrange these conditions by first choosing $\alpha_0$ close enough to $-\frac32+\eps_\ind$ so that it exceeds $\alpha_\sface$, and then choosing $\eps_0<\eps_\ind$ with $\eps_0\leq\alpha_0-\alpha_\sface$. We thus obtain the estimate
  \[
    \bigl\|\bigl(\wt{L_b}(\sigma)-\wt{L_b}(0)\bigr)(\chi_\zface u,\dot b,\dot\scal)\bigr\|_{\tilde H_{\scbtop,|\sigma|}^{\sfs-1,(-N,\alpha_\sface+2,0)}} \lesssim \|\chi_\zface u\|_{H_{\scbtop,|\sigma|}^{\sfs,(-N,\alpha_\sface,-1)}} + |\sigma|^{\eps_0}(|\dot b|+|\dot\scal|)
  \]
  for any fixed $\eps_0<-\frac32+\eps_\ind-\alpha_\sface\in(0,\eps_\ind)$. The second term on the right can be absorbed into the left-hand side of~\eqref{EqAdmLoUnif0} when $|\sigma|$ is small enough, so we have now proved
  \begin{equation}
  \label{EqAdmLoUnif1}
    \|(u,\dot b,\dot\scal)\|_{\tilde H_{\scbtop,|\sigma|}^{\sfs,(\sfr+\alpha_\sface,\alpha_\sface,0)}} \lesssim \|\wt{L_b}(\sigma)(\chi_\zface u,\dot b,\dot\scal)\|_{\tilde H_{\scbtop,|\sigma|}^{\sfs-1,(-N,\alpha_\sface+2,0)}} + \|u\|_{H_{\scbtop,|\sigma|}^{\sfs,(\sfr+\alpha_\sface,\alpha_\sface,-\delta)}}
  \end{equation}
  for any fixed $\delta\leq 1$; here we estimated the norm of $(1-\chi_\zface)u$ (which vanishes near $\zface$) by the right-most term.

  We next commute $\chi_\zface\oplus I\oplus I$ through $\wt{L_b}(\sigma)$. We have the uniform (in $|\sigma|$) description
  \begin{align*}
    &\bigl[\wt{L_b}(\sigma),\chi_\zface\oplus I\oplus I\bigr] \\
    &\qquad =\openbigpmatrix{3pt}
         [\wh{L_b}(\sigma),\chi_\zface] & (1-\chi_\zface)H_{\scbtop,|\sigma|}^{m,(\sfr_0+\alpha_0+1,\alpha_0+2,0)} & (1-\chi_\zface)H_{\scbtop,|\sigma|}^{m,(\sfr_0+\alpha_0+1,\alpha_0+2,0)} \\
         \la(\chi_\zface-1)\cdot,f_1^*\ra_{L^2} & 0 & 0 \\
         \la(\chi_\zface-1)\cdot,f_2^*\ra_{L^2} & 0 & 0
       \closebigpmatrix.
  \end{align*}
  Note that $[\wh{L_b}(\sigma),\chi_\zface]\in\rho_\scface^\infty\rho_\tface^2\rho_\zface^\infty\Diffscbt^1$. Since $1-\chi_\zface=0$ near $\zface$, and again using $\alpha_0+2\geq\alpha_\sface+2+\eps_0$, we thus find that the first component of the action of this commutator on $(u,\dot b,\dot\scal)$ is bounded in $H_{\scbtop,|\sigma|}^{\sfs-1,(-N,\alpha_\sface+2,0)}$ by $\|u\|_{H_{\scbtop,|\sigma|}^{\sfs,(-N,\alpha_\sface,-N)}}+|\sigma|^{\eps_0}(|\dot b|+|\dot\scal|)$ for any fixed $N$; the second term here can again be absorbed. The second and third components vanish since by the compact support property of $f_j^*$, $j=1,2$, on $X^\circ$ we have $(\chi_\zface-1)f_j^*=0$ on $X_\scbtop$ for all sufficiently small $|\sigma|$. For small enough $|\sigma|$, then, we now deduce from~\eqref{EqAdmLoUnif1} the estimate
  \begin{equation}
  \label{EqAdmLoUnif2}
    \|(u,\dot b,\dot\scal)\|_{\tilde H_{\scbtop,|\sigma|}^{\sfs,(\sfr+\alpha_\sface,\alpha_\sface,0)}} \lesssim \|\wt{L_b}(\sigma)(u,\dot b,\dot\scal)\|_{\tilde H_{\scbtop,|\sigma|}^{\sfs-1,(\sfr+\alpha_\sface+1,\alpha_\sface+2,0)}} + \|u\|_{H_{\scbtop,|\sigma|}^{\sfs,(\sfr+\alpha_\sface,\alpha_\sface,-\delta)}}.
  \end{equation}
  (This holds with $\scface$-order of the first term on the right equal to $-N$, but we take $\sfr+\alpha_\sface+1$ for compatibility with the $\tface$-estimate that follows next.)

  \pfstep{Estimate near $\tface$; conclusion.} We proceed to estimate the final term in~\eqref{EqAdmLoUnif2}; this is completely analogous to the arguments around \citeAF{(\ref*{EqSpLoNtfDiff})}. Let us fix $\delta>0$ in~\eqref{EqAdmLoUnif2} such that $\alpha_\sface+\delta<-\frac32+\eps_\ind$ still. We first use the uniform norm equivalence \citeAF{(\ref*{EqMUscbtNormEquiv})}) to obtain
  \[
    \|\chi_\tface u\|_{H_{\scbtop,|\sigma|}^{\sfs,(\sfr+\alpha_\sface,\alpha_\sface,-\delta)}(X)} \sim |\sigma|^{\frac32-\alpha_\sface} \|\chi_\tface u\|_{H_{\scop,\bop}^{\sfs,(\sfr+\alpha_\sface,-\alpha_\sface-\delta)}(\tface)},
  \]
  where we use the Euclidean density $\hat r^2\,|\dd\hat r\,\dd\slg|$ on $\tface=[0,\infty]_{\hat r}\times\Sph^2$, $\hat r:=r|\sigma|$. The $\tface$-admissibility of $L_b$ (Proposition~\ref{PropWEtf}) implies, via \citeAF{Corollary~\ref*{CorSptfAdm}}, that we can estimate this further by
  \[
    |\sigma|^{\frac32-\alpha_\sface} \| N_\tface(L_b,\hat\sigma)(\chi_\tface u) \|_{H_{\scop,\bop}^{\sfs-2,(\sfr+\alpha_\sface+1,-\alpha_\sface-\delta-2)}(\tface)} \sim \| |\sigma|^2 N_\tface(L_b,\hat\sigma)(\chi_\tface u)\|_{H_{\scbtop,|\sigma|}^{\sfs-2,(\sfr+\alpha_\sface+1,\alpha_\sface+2,-\delta)}(X)}.
  \]
  We replace $|\sigma|^2 N_\tface(L_b,\hat\sigma)\chi_\tface$ by $\wh{L_b}(\sigma)\chi_\tface$; the difference of these two operators lies in $\rho_\scface\rho_\tface^3\Diffscbt^2$. We then commute $\chi_\tface$ through $\wh{L_b}(\sigma)$, and finally obtain the uniform estimate
  \begin{equation}
  \label{EqAdmLoUnif3}
    \|u\|_{H_{\scbtop,|\sigma|}^{\sfs,(\sfr+\alpha_\sface,\alpha_\sface,-\delta)}} \lesssim \| \wh{L_b}(\sigma)u \|_{H_{\scbtop,|\sigma|}^{\sfs-2,(\sfr+\alpha_\sface+1,\alpha_\sface+2,-\delta)}} + \|u\|_{H_{\scbtop,|\sigma|}^{\sfs,(\sfr+\alpha_\sface,\alpha_\sface-1,-\delta)}}.
  \end{equation}
  (The norm of $(1-\chi_\tface)u$, which is supported away from $\tface$, is trivially bounded by the right-most term.) The final term here is $\lesssim |\sigma|^\delta \|u\|_{H_{\scbtop,|\sigma|}^{\sfs,(\sfr+\alpha_\sface,\alpha_\scface,0)}}$ and can thus be absorbed into the left-hand side of~\eqref{EqAdmLoUnif2} for small enough $|\sigma|$. We moreover need to replace $\wh{L_b}(\sigma)u$ in~\eqref{EqAdmLoUnif3} by $\wt{L_b}(\sigma)(u,\dot b,\dot\scal)$ and thus estimate the difference
  \begin{align*}
    &\bigl\| |\sigma|^\delta \bigl(\wh{L_b}(\sigma)u - \wt{L_b}(\sigma)(u,\dot b,\dot\scal)\bigr) \bigr\|_{\tilde H_{\scbtop,|\sigma|}^{\sfs-2,(\sfr+\alpha_\sface+1,\alpha_\sface+2+\delta,0)}} \\
    &\qquad = |\sigma|^\delta \sum_{j=1}^2 \| \hat f_j(\sigma) \|_{H_{\scbtop,|\sigma|}^{\sfs-2,(\sfr+\alpha_\sface+1,\alpha_\sface+2+\delta,0)}} + |\sigma|^\delta|\la u,f_1^*\ra_{L^2}| + |\sigma|^\delta|\la u,f_2^*\ra_{L^2}|
  \end{align*}
  where $\hat f_1(\sigma):=\wh{L_b}(\sigma)(\cF(\dot g_b^{\Ups,\aug}(\dot b))(\sigma))$ and $\hat f_2(\sigma):=\wh{L_b}(\sigma)(\cF(h_{b,\rms 1}^{\leq 1,\aug}(\dot\scal))(\sigma))$. The first term is bounded by $C|\sigma|^\delta(|\dot b|+|\dot\scal|)$ if we fix $\delta\in(0,-\frac32+\eps_\ind-\alpha_\sface)$, as before. The final terms are bounded by $|\sigma|^\delta\|u\|_{H_{\scbtop,|\sigma|}^{\sfs,(\sfr+\alpha_\sface,\alpha_\sface,0)}}$ (it only matters that the $\zface$-order of the norm used here is $0$). All of these terms can thus be absorbed for small $|\sigma|$. This finishes the proof of the estimate~\eqref{EqAdmLoUnifEst}.

  \pfstep{Higher spatial b-regularity.} The case of general b-regularity orders $k\in\N$ is handled in the same way and gives~\eqref{EqAdmLoUnifEst} for an a priori $k$-dependent neighborhood of $0$. But in the original neighborhood (for $k=0$) and away from $\sigma=0$, higher b-regularity follows from the case $k=0$ via \citeAF{Theorem~\ref*{ThmSpB}(\ref*{ItSpBb})}, so we in fact obtain~\eqref{EqAdmLoUnifEst} in a $k$-independent neighborhood of $0$.

  \pfstep{Conormal regularity in $\sigma$.} We indicate the proof of~\eqref{EqAdmLoUnifInvReg} for $j=1$; higher regularity is proved similarly. Using the resolvent identity $\sigma\pa_\sigma\wt{L_b}(\sigma)^{-1}=-\wt{L_b}(\sigma)^{-1}\circ\sigma\pa_\sigma\wt{L_b}(\sigma)\circ\wt{L_b}(\sigma)^{-1}$, we claim that each arrow in
  \begin{alignat}{2}
    \tilde H_{(\scbtop,|\sigma|);\bop}^{(\sfs-1;k+1),(\sfr+\alpha_\sface+1,\alpha_\sface+2,0)} &\xra{\wt{L_b}(\sigma)^{-1}}&\ & \tilde H_{(\scbtop,|\sigma|);\bop}^{(\sfs;k+1),(\sfr+\alpha_\sface,\alpha_\sface,0)} \nonumber\\
  \label{EqAdmLoUnifDer}
      &\xra{\sigma\pa_\sigma\wt{L_b}(\sigma)}&\ & \tilde H_{(\scbtop,|\sigma|);\bop}^{(\sfs;k),(\sfr+\alpha_\sface+1,\alpha_\sface+2,0)} \\
      &\xra{\wt{L_b}(\sigma)^{-1}}&\ & \tilde H_{(\scbtop,|\sigma|);\bop}^{(\sfs;k),(\sfr+\alpha_\sface,\alpha_\sface,0)} \nonumber
  \end{alignat}
  is uniformly bounded.\footnote{The differentiability of $\wt{L_b}(\sigma)^{-1}$ in $\sigma\neq 0$, and thus the justification of this formal computation, is the content of \citeAF{Theorem~\ref*{ThmSpB}(\ref*{ItSpBInvReg})} and proved in \citeAF{\S\ref*{SssSpBPf}}.} This only requires an argument for the middle arrow. Only the first row of $\sigma\pa_\sigma\wt{L_b}(\sigma)$ is non-zero. Consider its $(1,1)$-entry: by~\eqref{EqAdmSpecFam}, we have $\sigma\pa_\sigma\wh{L_b}(\sigma)\in \sigma\rho\Diffb^1+\sigma\rho^3\Diffb^1+\sigma^2\rho^2\CI$, which thus indeed gains one power of $\sigma\rho=\rho_\scface\rho_\tface^2$ at the expense of one b-derivative. The $(1,2)$- and $(1,3)$-entries of $\sigma\pa_\sigma\wt{L_b}(\sigma)$ lie in the desired target space by Lemma~\ref{LemmaAdmLoImFT}\eqref{ItAdmLoImFTUnif}.
\end{proof}

\subsubsection{An augmentation constructed on the spectral side}
\label{SssAdmLoSpec}

--- \emph{The following material will be used only in~\S\usref{SsD6Better} below.} When holomorphicity considerations in $\{\Im\sigma>0\}$ are not needed, it will be slightly more convenient to use an augmentation that is defined directly on the spectral side, with the lack of decay of the terms~\eqref{EqAdmLo12}--\eqref{EqAdmLo13} resolved via the inversion of the $\tface$-normal operators $N_\tface(L_b,\pm 1)$. Thus, recalling~\eqref{EqAdmLoKerr}, consider first
\begin{equation}
\label{EqAdmLoErrtf1}
\begin{split}
  \wh{L_b}(\sigma)\bigl(\chi_\zface i\sigma^{-1}\dot g_b^\Ups(\dot b)\bigr) &= i\sigma^{-1}[\wh{L_b}(\sigma),\chi_\zface]\dot g_b^\Ups(\dot b) \\
    &\qquad + \chi_\zface [L_b,t_*]\ftrans(0)\dot g_b^\Ups(\dot b) - \chi_\zface\frac{i\sigma}{2}[[L_b,t_*],t_*]\dot g_b^\Ups(\dot b).
\end{split}
\end{equation}
Since $[\wh{L_b}(\sigma),\chi_\zface]\in\rho_\scface^\infty\rho_\tface^2\rho_\zface^\infty\Diff_\scbtop^1(X_\scbtop^\pm)$, the first two terms on the right lie in the space $\cA^{\infty,\ (2,0)\cup(2+\cE_\ind),\ \infty}(X_\scbtop^\pm)$ for $\sigma\in\pm[0,1]$, with scalar type $0$ leading-order term at $\tface$ which only depends on the component $\dot\bhm$ of $\dot b=(\dot\bhm,\dot\bha)$. (The final term in~\eqref{EqAdmLoErrtf1} is of class $\cA^{\infty,\ (4,0)\cup(4+\cE_\ind),\ \infty}$ and thus vanishes to higher order at $\tface$.) We then apply Proposition~\ref{PropiptfGr} to~\eqref{EqAdmLoErrtf1} for $\dot b=(1,0)$, with $(\alpha_0,k)=(2,0)$ and $\ell_\zface=3+\eps_\ind-\eps$; this produces
\begin{subequations}
\begin{equation}
\label{EqAdmLoErrtf1Corr}
\begin{split}
  &\dot g_b^{\Ups,{\rm corr}}(1,0) = \chi_\zface\Bigl(\scal_0^{(0)}\log(\hat r)h_{b,\rms 0}^{(0),\leq 3}(\sigma) + |\sigma|^{\lambda^\Ups_{\rms 0,1}-1} h_{b,\rms 0}^{(-\lambda^\Ups_{\rms 0,1}+1),\leq 2}\bigl(\sigma,\scal_0^{(-\lambda^\Ups_{\rms 0,1}+1)}\bigr) \Bigr) \\
  &\hspace{7em} + \chi_\tface\dot g_{b,\tface}^{\Ups,{\rm corr}}(1,0), \\
  &\qquad \dot g_{b,\tface}^{\Ups,{\rm corr}}(1,0) \in \Hb^{\infty,\ 1-\eps,\ (0,0),\ ((0,0),3+\eps_\ind-\eps)}(X_\scbtop^\pm;S^2\cT^*_X),
\end{split}
\end{equation}
(depending on the choice of sign, not made explicit in the notation) such that
\begin{equation}
\label{EqAdmLoErrtf1CorrErr}
\begin{split}
  &\dot g_b^{\Ups,{\rm sing}}(\sigma)(\dot b) := \chi_\zface i\sigma^{-1}\dot g_b^\Ups(\dot b) + \dot\bhm\dot g_b^{\Ups,{\rm corr}}(1,0) \\
  &\qquad \implies \wh{L_b}(\sigma)\bigl(\dot g_b^{\Ups,{\rm sing}}(\sigma)(\dot b)\bigr) \in \Hb^{\infty,\ 2-\eps,\ 2+\cE_\ind,\ ((0,0),3+\eps_\ind-\eps)}(X_\scbtop^\pm).
\end{split}
\end{equation}
\end{subequations}
Similarly, recalling~\eqref{EqAdmLo12}, the application of Proposition~\ref{PropiptfGr} to the (scalar type $1$, size $\rho_\tface^2$) leading-order term of $\wh{L_b}(\sigma)\bigl(\chi_\zface(-\sigma^{-2}h_{b,\rms 1}(\dot\scal)+i\sigma^{-1}\breve h_{b,\rms 1}^1(\dot\scal))\bigr)$ produces
\begin{subequations}
\begin{equation}
\label{EqAdmLoErrtf2Corr}
\begin{split}
  h_{b,\rms 1}^{\leq 1,{\rm corr}}(\dot\scal) &= \chi_\zface\Bigl( -\sigma^{-1}(\log|\sigma|)h_{b,\rms 1}(\scal_1^{(-1)}) + (\log\hat r)i\breve h_{b,\rms 1}^{\leq 4}(\sigma,\scal_1^{(-1)}) \\
    &\quad\hspace{7em} + |\sigma|^{\lambda^\Ups_{\rms 1,1}-1}h_{b,\rms 1}^{(-\lambda^\Ups_{\rms 1,1}+1),\leq 2}\bigl(\sigma,\scal_1^{(-\lambda^\Ups_{\rms 1,1}+1)}\bigr) \Bigr) + \chi_\tface h_{b,\rms 1,\tface}^{\leq 1,{\rm corr}}(\dot\scal), \\
    &\qquad h_{b,\rms 1,\tface}^{\leq 1,{\rm corr}}(\dot\scal) \in \Hb^{\infty,\ 1-\eps,\ (0,0),\ ((0,0),3+\eps_\ind-\eps)}(X_\scbtop^\pm;S^2\cT^*_X),
\end{split}
\end{equation}
such that
\begin{equation}
\label{EqAdmLoErrtf2CorrErr}
\begin{split}
  &h_{b,\rms 1}^{\leq 1,{\rm sing}}(\sigma)(\dot\scal) := \chi_\zface\bigl(-\sigma^{-2}h_{b,\rms 1}(\dot\scal)+i\sigma^{-1}\breve h_{b,\rms 1}^1(\dot\scal)\bigr) + h_{b,\rms 1}^{\leq 1,{\rm corr}}(\dot\scal) \\
  &\qquad \implies \wh{L_b}(\sigma)\bigl(h_{b,\rms 1}^{\leq 1,{\rm sing}}(\sigma)\bigr) \in \Hb^{\infty,\ 2-\eps,\ 2+\cE_\ind,\ ((0,0),3+\eps_\ind-\eps)}(X_\scbtop^\pm).
\end{split}
\end{equation}
\end{subequations}

\begin{rmk}[Memberships]
\label{RmkAdmLoMem}
  By inspection of~\eqref{EqAdmLoErrtf1Corr}--\eqref{EqAdmLoErrtf2CorrErr}, we have
  \begin{equation}
  \label{EqAdmLoMem}
  \begin{split}
    \bigl(\sigma\dot g_b^{\Ups,{\rm sing}}(\sigma)\bigr)\big|_{\sigma\in\pm[0,1]} &\in \Hb^{\infty,\ 1-\eps,\ (1,0)\cup(1+\cE_\ind),\ \bigl((0,0)\cup(1,1),\,1+\eps_\ind-\eps\bigr)}, \\
    \bigl(\sigma^2 h_{b,\rms 1}^{\leq 1,{\rm sing}}(\sigma)\bigr)\big|_{\sigma\in\pm[0,1]} &\in \Hb^{\infty,\ 1-\eps,\ (2,0)\cup(2+\cE_\ind),\ \bigl((0,0)\cup(1,1),\,2+\eps_\ind-\eps\bigr)}.
  \end{split}
  \end{equation}
  The logarithmic terms at $\zface$ (and dropping terms with $\zface$-orders $\geq 1+\eps_\ind$ and $\geq 2+\eps_\ind$, respectively) are $\sigma(\log|\sigma|)h_{b,\rms 0}^{(0)}(\sigma)$ and $-\sigma(\log|\sigma|)h_{b,\rms 1}^{\leq 1}(\scal_1^{(-1)})$, respectively. (These memberships are also what one can show for $\cF(-i\pa_{t_*}\dot g_b^{\Ups,\aug})$ and $\cF(-\pa_{t_*}^2 h_{b,\rms 1}^{\leq 1,\aug})$ in~\eqref{EqAdmPfLoDerF1} and \eqref{EqAdmPfLoDerF2} below, with the logarithmic terms here corresponding to the $t_*^{-1}$- and $\log t_*$-terms in~\eqref{EqAdmLoImLot}.) For later use, we note that replacing $\log(\hat r)$ in~\eqref{EqAdmLoErrtf1Corr} by $\log|\sigma|=\log(\hat r)+\log\rho$ produces an error of class $\Hb^{\infty,\ \infty,\ (0,1)\cup\cE_\ind,\ (0,0)}$ there and thus of class $\Hb^{\infty,\ \infty,\ (1,1)\cup(1+\cE_\ind),\ ((0,0),1+\eps_\ind-\eps)}$ in the first line of~\eqref{EqAdmLoMem}; similarly in~\eqref{EqAdmLoErrtf2Corr}, where this replacement causes an extra $\tface$-index set $(2,1)$ in the second line of~\eqref{EqAdmLoMem}.
\end{rmk}

Let us then define, analogously to~\eqref{EqAdmLoDef},
\begin{equation}
\label{EqAdmLoSpec}
  \wt{L_b'}(\sigma) := \begin{pmatrix} \wh{L_b}(\sigma) & \wh{L_b}(\sigma)\dot g_b^{\Ups,{\rm sing}}(\sigma) & \wh{L_b}(\sigma)h_{b,\rms 1}^{\leq 1,{\rm sing}}(\sigma) \\ \la\cdot,f_1^*\ra_{L^2} & 0 & 0 \\ \la\cdot,f_2^*\ra_{L^2} & 0 & 0 \end{pmatrix}.
\end{equation}
The outputs of $\wh{L_b}(\sigma)$ in~\eqref{EqAdmLoErrtf1CorrErr} and \eqref{EqAdmLoErrtf2CorrErr} have order $>2$ at $\tface$ and satisfy the conclusions of Lemma~\ref{LemmaAdmLoImFT}\eqref{ItAdmLoImFT0}--\eqref{ItAdmLoImFTLo}; thus, also the conclusions of Proposition~\ref{PropAdmLoUnif} apply to it.

\subsection{Proof of 2-admissibility}
\label{SsAdmPf}

We now adapt the arguments used in \citeAF{\S\ref*{SsA2Adm}} for the proof of the 1-admissibility of the 1-form wave operator on subextremal Kerr to prove the 2-admissibility of our linearized gauge-fixed Einstein operator $L_b$; that is, we will prove~\eqref{EqAdm} starting with the expression~\eqref{EqAdmIFT} for the forward solution. We recall that by a density argument we may (and do) assume $f\in\CIc(\Omega^\circ;S^2\cT^*)$. With $c>0$ given by Proposition~\ref{PropAdmLoUnif} and shrunk, if necessary, so that $c\leq 1$, fix a cutoff function
\begin{equation}
\label{EqAdmPfCutoff}
  \chi=\chi(|\sigma|)\in\CIc([0,c)),\quad \chi|_{[0,\tfrac{c}{2}]}=1,
\end{equation}
to separate high and low frequencies, which we discuss in~\S\S\ref{SssAdmPfHi} and \ref{SssAdmPfLo}, respectively.

\subsubsection{High-frequency estimate}
\label{SssAdmPfHi}

For the part of the integral~\eqref{EqAdmIFT} that excludes $\gamma_0$, we prove:

\begin{lemma}[High-frequency estimate]
\label{LemmaAdmPfHi}
  Using the notation~\eqref{EqAdmIFT} and~\eqref{EqAdmPfCutoff}, set
  \begin{equation}
  \label{EqAdmPfHiDef}
    u_{\rm hi}(t_*,\cdot) := \frac{1}{2\pi} \int_{\R\setminus[-\frac{c}{2},\frac{c}{2}]} e^{-i\sigma t_*}(1-\chi(|\sigma|))\wh{L_b}(\sigma)^{-1}\hat f(\sigma)\,\dd\sigma.
  \end{equation}
  Then for all $k\in\N_0$ we have\footnote{Note that $u_{\rm hi}$ is (typically) not supported in $\Omega$, which is why we use the 3b-Sobolev norm on all of $M_0$ here.}
  \begin{equation}
  \label{EqAdmPfHi}
    \|u_{\rm hi}\|_{H_{\tbop;\bop}^{(\sfs;k),(\alpha_\sface,0)}(M_0,|\dd g_b|;S^2\cT^*)} \leq C\|f\|_{H_{\tbop;\bop}^{(\sfs;k),(\alpha_\sface+1,0)}(\Omega,|\dd g_b|;S^2\cT^*)^{\bullet,-}}.
  \end{equation}
\end{lemma}

This estimate is lossless as far as the $\cK^+$-decay order $0$ is concerned; losses will only arise from the low-energy behavior of $\wh{L_b}(\sigma)^{-1}$ (specifically, its singularity at $\sigma=0$). Similarly, the order $2$ (or rather $2+\delta$) difference of the $\sface$-decay orders in~\eqref{EqAdm} only arises in the low-energy theory. (Of course, the estimate~\eqref{EqAdmPfHi} is stronger than, i.e., immediately implies, the estimate $\|u_{\rm hi}\|_{H_{\tbop;\bop}^{(\sfs;k),(\alpha_\sface-\delta,-2)}}\leq C\|f\|_{H_{\tbop;\bop}^{(\sfs;k),(\alpha_\sface+2,0)}}$ matching~\eqref{EqAdm}.)

\begin{proof}[Proof of Lemma~\usref{LemmaAdmPfHi}]
  This uses the same arguments as in the proofs of \citeAF{Theorems~\ref*{ThmA1Adm} and \ref*{ThmA2Adm}}. The key are the high-energy estimates (including with b-regularity) of \citeAF{Theorem~\ref*{ThmSpHi}(\ref*{ItSpHiInvReg})}; this involves the semiclassical loss function $\delta_\Gamma(\eta)$, which was defined in \citeAF{(\ref*{EqSpHiTrLoss})}, and where in view of the strong trapping admissibility proved in Proposition~\ref{PropWETr} one can take $\gamma_+$ (see \citeAF{(\ref*{EqSpHiTrpGammas})}) to be an arbitrarily small positive number, while $\gamma_-$ is less than a fixed negative number; the upshot is that the function $\delta_\Gamma(\Im(\sigma)-\eps)$ appearing in \citeAF{(\ref*{EqSpHiInvReg})} (with $\eps>0$ there arbitrary) can be taken to be less than any desired positive constant. That is, in the notation of \citeAF{(\ref*{EqSpHiNormb}) and (\ref*{EqSpHiInvReg})} (except for using the notation $\alpha_\sface$ instead of $\alpha_+$), we have for all $\sigma\in\R$, $|\sigma|\geq\frac{c}{2}$, the uniform boundedness of
  \begin{equation}
  \label{EqAdmPfHiRes}
    (\sigma\pa_\sigma)^j\wh{L_b}(\sigma)^{-1} \colon \bar H_{(\scop,|\sigma|^{-1});\bop^+}^{(\sfs_\scop-1;k+j),\sfr+\alpha_\sface+1,\sfb} \to |\sigma|^{-1+(j+1)\eps}\bar H_{(\scop,|\sigma|^{-1});\bop^+}^{(\sfs_\scop+j;k),\sfr+\alpha_\sface,\sfb}
  \end{equation}
  for any fixed $\eps>0$ and $j,k\in\N_0$. Here, $H_{(\scop,h);\bop}^{(\sfs;k),\sfr,\sfb}$ is a semiclassical scattering Sobolev space with additional $k$ orders of (non-se\-mi\-clas\-si\-cal) b-regularity. (For $\sfs\in\N_0$ and $\sfr,\sfb\in\R$ and $k=0$, this is the space $\rho^{-\sfr}h^{-\sfb}H_{\scop,h}^s$ where $H_{\scop,h}^s$ consists of functions on $X$ which are in $L^2$ together with up to $s$-fold derivatives along $h\pa_x$; for $k\in\N$, one also tests for up to $k$ orders of regularity with respect to $\la x\ra\pa_x$. See \citeAF{\S\ref*{SssMUK}} for details.) The symbol ``$+$'' in the subscript of $H_{(\scop,h);\bop^+}$ means that we test for $k$ orders of regularity (i.e., membership in $H_{\scop,h}$) under application of $\la x\ra\pa_x$ or multiplication by $h^{-1}$. Using \citeAF{Lemma~\ref*{LemmaMUetbFTb}}, we have
  \[
    \|u_{\rm hi}\|_{H_{\tbop;\bop}^{(\sfs;k),(\alpha_\sface,0)}(M_0)}^2 \sim \sum_{j=j_1+j_2\leq k} \int_{\R\setminus[-\frac{c}{2},\frac{c}{2}]} \|\sigma^{j_1}(\sigma\pa_\sigma)^{j_2}\wh{u_{\rm hi}}(\sigma)\|_{H_{(\scop,|\sigma|^{-1});\bop}^{(\sfs_\sigma;k-j),\sfs_\sigma+\alpha_\sface,\sfs_\sigma}(X)}^2\,\dd\sigma
  \]
  Since $\wh{u_{\rm hi}}(\sigma)=(1-\chi(|\sigma|))\wh{L_b}(\sigma)^{-1}\hat f(\sigma)$, we can bound this using~\eqref{EqAdmPfHiRes} (with $k-j=k-(j_1+j_2-j_3)$ and $j_2$ in place of $k$ and $j$, respectively) by
  \begin{align*}
    &\sum_{j=j_1+j_2+j_3\leq k} \int_{\R\setminus[-\frac{c}{2},\frac{c}{2}]} \Bigl\| |\sigma|^{j_1} \bigl( (\sigma\pa_\sigma)^{j_2}\wh{L_b}(\sigma)^{-1} \bigr) \bigl( (\sigma\pa_\sigma)^{j_3}\hat f(\sigma) \bigr) \Bigr\|_{H_{(\scop,|\sigma|^{-1});\bop}^{(\sfs_\sigma;k-j),\sfs_\sigma+\alpha_\sface,\sfs_\sigma}(X)}^2 \\
    &\qquad \lesssim \int_{\R\setminus[-\frac{c}{2},\frac{c}{2}]} \sum_{j_1+j_2+j_3\leq k} |\sigma|^{j_1-1+(j_2+1)\eps} \| (\sigma\pa_\sigma)^{j_3}\hat f(\sigma) \|_{H_{(\scop,|\sigma|^{-1});\bop^+}^{(\sfs_\sigma-1;k-j_1-j_3),\sfs_\sigma+\alpha_\sface+1,\sfs_\sigma}}^2 \\
    &\qquad \lesssim \int_{\R\setminus[-\frac{c}{2},\frac{c}{2}]} \sum_{j_1+j_2+j_3\leq k}\sum_{j_4=0}^{k-j_1-j_3} |\sigma|^{j_1-1+(j_2+1)\eps}|\sigma|^{j_4}\|(\sigma\pa_\sigma)^{j_3}\hat f(\sigma)\|_{H_{(\scop,|\sigma|^{-1});\bop}^{(\sfs_\sigma;k-j_1-j_3-j_4),\sfs_\sigma+\alpha_\sface+1,\sfs_\sigma}}^2\,\dd\sigma.
  \end{align*}
  With $k\in\N_0$ fixed, we take $\eps=\frac{1}{k+1}$. The total number of factors of $|\sigma|$ and $\sigma\pa_\sigma$ is then $\leq j_1-1+(k+1)\eps+j_4+j_3\leq k$, so this is bounded by
  \[
    \sum_{j=j_1+j_2\leq k} \|(\sigma\pa_\sigma)^{j_1}|\sigma|^{j_2}\hat f(\sigma)\|_{H_{(\scop,|\sigma|^{-1});\bop}^{(\sfs_\sigma-1;k-j),\sfs_\sigma+\alpha_\sface+1,\sfs_\sigma}}^2\,\dd\sigma \lesssim \|f\|_{H_{\tbop;\bop}^{(\sfs;k),(\alpha_\sface+1,0)}(M_0)}^2,
  \]
  where we used \citeAF{Lemma~\ref*{LemmaMUetbFTb}} for the last step.
\end{proof}

\subsubsection{Low-frequency estimate; combination}
\label{SssAdmPfLo}

We next study the low-energy part $u_{\rm lo}:=u-u_{\rm hi}$ of~\eqref{EqAdmIFT}, which is thus given by
\begin{equation}
\label{EqAdmPfLoDef}
  u_{\rm lo}(t_*,\cdot) = \frac{1}{2\pi} \int_{\gamma_-\cup\gamma_0\cup\gamma_+} e^{-i\sigma t_*} \chi(|\sigma|)\wh{L_b}(\sigma)^{-1}\hat f(\sigma)\,\dd\sigma.
\end{equation}
Let us use Proposition~\ref{PropAdmLoUnif} to define
\begin{equation}
\label{EqAdmPfLoTilde}
  \bigl( \hat u_{{\rm reg},1}(\sigma,\cdot),\ \hat b_1(\sigma),\ \hat\scal_1(\sigma) \bigr) := \wt{L_b}(\sigma)^{-1} \bigl( \hat f(\sigma,\cdot),\ 0,\ 0 \bigr),\quad |\sigma|\leq c.
\end{equation}
Thus, $\hat u_{{\rm reg},1}$ (as a distribution on $X^\circ$), $\hat b_1$, and $\hat\scal_1$ are holomorphic for $\Im\sigma>0$, continuous down to $[-c,c]\setminus\{0\}$, and uniformly bounded in the norm on the left-hand side of~\eqref{EqAdmLoUnifEst} by the norm of $\hat f(\sigma)$ in $\bar H_{\scbtop,|\sigma|}^{(\sfs-1;k),(\sfr+\alpha_\sface+1,\alpha_\sface+2,0)}(X)$. The first component of $\wt{L_b}(\sigma)(\hat u_{{\rm reg},1}(\sigma),\hat b_1(\sigma),\hat\scal_1(\sigma))=(\hat f(\sigma),0,0)$ (see~\eqref{EqAdmLoDef}) gives
\begin{equation}
\label{EqAdmPfLo}
  \hat u_{\rm lo}(\sigma) = \chi(|\sigma|)\bigl(\hat u_{{\rm reg},1}(\sigma) + \cF(\dot g_b^{\Ups,\aug})(\sigma)(\hat b_1(\sigma)) + \cF(h_{b,\rms 1}^{\leq 1,\aug})(\sigma)(\hat\scal_1(\sigma))\bigr)
\end{equation}
for $\sigma\neq 0$ with $\Im\sigma\geq 0$ and $|\sigma|\leq c$. We first study the regular part:

\begin{lemma}[Low-frequency estimate: regular part]
\label{LemmaAdmPfLoReg}
  For the spacetime 2-tensor
  \[
    u_{\rm reg}(t_*,\cdot):=\frac{1}{2\pi}\int_{\gamma_-\cup\gamma_0\cup\gamma_+} e^{-i\sigma t_*}\chi(|\sigma|)\hat u_{{\rm reg},1}(\sigma,\cdot)\,\dd\sigma,
  \]
  we have, for any fixed $k\in\N_0$, an estimate
  \[
    \|u_{\rm reg}\|_{H_{\tbop;\bop}^{(\sfs;k),(\alpha_\sface,0)}(M_0)} \leq C\|f\|_{H_{\tbop;\bop}^{(\sfs-1;k),(\alpha_\sface+2,0)}(\Omega)^{\bullet,-}}.
  \]
\end{lemma}

There is no 3b-derivative loss anymore; this only arose in Lemma~\ref{LemmaAdmPfHi} due to trapping.

\begin{proof}[Proof of Lemma~\usref{LemmaAdmPfLoReg}]
  By the properties of $\hat u_{{\rm reg},1}$ recorded after~\eqref{EqAdmPfLoTilde}, we can shift the integration contour down to $[-c,c]$. We then use Lemma~\ref{LemmaT3bFT}, including for spatial b-derivatives as well as temporal b-derivatives ($t_*\pa_{t_*}$, which becomes $-\sigma\pa_\sigma-1$ on the spectral side)---see \citeAF{Lemma~\ref*{LemmaMUetbFTb}}---as well as~\eqref{EqAdmLoUnifInvReg} to bound
  \begin{align*}
    \|u_{\rm reg}\|_{H_{\tbop;\bop}^{(\sfs;k),(\alpha_\sface,0)}}^2 &\sim \int_{-c}^c \sum_{j=0}^k \bigl\|(\sigma\pa_\sigma)^j\bigl(\chi(|\sigma|)\hat u_{{\rm reg},1}(\sigma,\cdot)\bigr)\bigr\|_{H_{(\scbtop,|\sigma|);\bop}^{(\sfs;k-j),(\sfs+\alpha_\sface,\alpha_\sface,0)}}^2\,\dd\sigma \\
      &\lesssim \int_{-c}^c \sum_{j=0}^k \sum_{j_1=0}^j \bigl\|(\sigma\pa_\sigma)^{j_1}\bigl(\chi(|\sigma|)\hat f(\sigma,\cdot)\bigr)\bigr\|_{H_{(\scbtop,|\sigma|);\bop}^{(\sfs-1;k-j+(j-j_1));(\sfs+\alpha_\sface+1,\alpha_\sface+2,0)}}^2\,\dd\sigma \\
      &\lesssim \int_{-c}^c \sum_{j=0}^k \|(\sigma\pa_\sigma)^j\hat f(\sigma,\cdot)\|_{H_{(\scbtop,|\sigma|);\bop}^{(\sfs-1;k-j);(\sfs+\alpha_\sface+1,\alpha_\sface+2,0)}}^2\,\dd\sigma \\
      &\lesssim \|f\|_{H_{\tbop;\bop}^{(\sfs-1;k),(\alpha_\sface+2,0)}}^2.\qedhere
  \end{align*}
\end{proof}

Next, we study the (singular at $\sigma=0$) second and third terms in~\eqref{EqAdmPfLo}. We first note:

\begin{lemma}[Coefficients of singular terms]
\label{LemmaAdmPfLoCoeff}
  For any fixed $k\in\N_0$, we have the estimates
  \[
    \int_{-c}^c \sum_{j=0}^k |(\sigma\pa_\sigma)^j \hat b_1(\sigma)|^2\,\dd\sigma,\ 
     \int_{-c}^c \sum_{j=0}^k |(\sigma\pa_\sigma)^j \hat\scal_1(\sigma)|^2\,\dd\sigma \leq C\|f\|_{H_{\tbop;\bop}^{(\sfs-1;k),(\alpha_\sface+2,0)}(\Omega)^{\bullet,-}}.
  \]
\end{lemma}
\begin{proof}
  This follows from the bounds for $\hat b_1$ in terms of $\hat f$ given by Proposition~\ref{PropAdmLoUnif} in the same way as in the proof of Lemma~\ref{LemmaAdmPfLoReg}; similarly for $\hat\scal_1$.
\end{proof}

Set
\begin{equation}
\label{EqAdmPfLodotghs1}
\begin{split}
  \dot g(t_*,\cdot) &:= \frac{1}{2\pi}\int_{\gamma_-\cup\gamma_0\cup\gamma_+} e^{-i\sigma t_*} \chi(|\sigma|) \cF(\dot g_b^{\Ups,\aug})(\sigma)(\hat b_1(\sigma))\,\dd\sigma, \\
  h_{\rms 1}(t_*,\cdot) &:= \frac{1}{2\pi}\int_{\gamma_-\cup\gamma_0\cup\gamma_+} e^{-i\sigma t_*} \chi(|\sigma|) \cF(h_{b,\rms 1}^{\leq 1,\aug})(\sigma)(\hat\scal_1(\sigma))\,\dd\sigma;
\end{split}
\end{equation}
then
\begin{equation}
\label{EqAdmPfLoDecomp}
  u = u_{\rm hi} + u_{\rm reg} + \dot g + h_{\rms 1}.
\end{equation}
While we do have $L^2$-bounds for $b_1:=\cF^{-1}(\chi(|\sigma|)\hat b_1(\sigma))$ by Lemma~\ref{LemmaAdmPfLoCoeff}, the $t_*$-convolution $\dot g_b^{\Ups,\aug}*b_1$ is not well-defined for general $b_1\in L^2(\R_{t_*})$. One must, in some way, use the holomorphicity of $\hat b_1$ in $\Im\sigma>0$, roughly corresponding to forward support in $t_*$, to make sense of this convolution. We do this using arguments inspired by those starting at \citeAF{(\ref*{EqA2AdmDecomp})} (which in turn go back to \cite[Proof of Theorem~3.22]{HintzGlueLocIII}).

We first recall that $\cF(h_{b,\rms 1}^{\leq 1,\aug})(\sigma)$ should be thought of as an improved (at $\tface\subset X_\scbtop$) version of $-\sigma^{-2}h_{b,\rms 1}+i\sigma^{-1}\breve h_{b,\rms 1}^1$ (as per the discussion starting with~\eqref{EqAdmLoNaive}), which has a double pole at $\sigma=0$; similarly for $\cF(\dot g_b^{\Ups,\aug})$, which should be thought of as having a single pole. Differentiation in $t_*$ removes these poles on the Fourier transform side; we record:

\begin{lemma}[Bounds for derivatives]
\label{LemmaAdmPfLoDer}
  For all $k\in\N_0$ and $\eps>0$, we have
  \begin{align}
  \label{EqAdmPfLoDer1}
    \pa_{t_*}\dot g &\in H_{\tbop;\bop}^{(\sfs;k),(-\frac12-\eps,0)}(M_0,|\dd g_b|;S^2\cT^*), \\
  \label{EqAdmPfLoDer2}
    \pa_{t_*}^2 h_{\rms 1} &\in H_{\tbop;\bop}^{(\sfs;k),(\frac12-\eps,0)}(M_0,|\dd g_b|;S^2\cT^*),
  \end{align}
  with norms bounded by $\|f\|_{H_{\tbop;\bop}^{(\sfs-1;k),(\alpha_\sface+2,0)}}(\Omega,|\dd g_b|;S^2\cT^*)^{\bullet,-}$.
\end{lemma}
\begin{proof}
  \pfstep{Control of $\pa_{t_*}\dot g$.} We have
  \begin{equation}
  \label{EqAdmPfLoDerCont}
    \pa_{t_*}\dot g = \frac{1}{2\pi}\int_{\gamma_-\cup\gamma_0\cup\gamma_+} e^{-i\sigma t_*}\chi(|\sigma|) \cF(\pa_{t_*}\dot g_b^{\Ups,\aug})(\sigma)(\hat b_1(\sigma))\,\dd\sigma.
  \end{equation}
  Now, $\pa_{t_*}$ maps $\dot g_b^{\Ups,\aug}(1,0)\in\cA^{1,1,((0,0),1+\eps_\ind)}(\Omega)$ (now on the radiative compactification $M$, with weights at $\scri^+$, $\iota^+$, and $\cK^+$, and support in $t_*\geq 1$)---which is a consequence of~\eqref{EqAdmLoImKerr}---to $\cA^{1,2,2}$ since $\pa_{t_*}\in\rho_{\iota^+}\rho_\cK\Vb(M)$ (which gives order $1$ gains in the $\iota^+$- and $\cK^+$-orders), and $\pa_{t_*}=t_*^{-1}t_*\pa_{t_*}$ is $\rho_{\iota^+}\rho_\cK$ times a b-normal vector field at $\cK^+$ (which thus annihilates the $t_*^0$ leading-order term), so the total $\cK^+$-index set and conormal remainder decay order are $((2,0),2+\eps_\ind)$. (One can also differentiate~\eqref{EqAdmLoImLot} to reach the same conclusion.) We also have $\pa_{t_*}\dot g_b^{\Ups,\aug}(0,\dot\bha)\in\cA^{\infty,2,2}\subset\cA^{1,2,2}$ (see~\eqref{EqAdmLoImKerr2}). This yields
  \begin{equation}
  \label{EqAdmPfPaGb}
    \pa_{t_*}\dot g_b^{\Ups,\aug} \in \Hb^{\infty,1-\eps,2-\eps,2-\eps}(\Omega)^{\bullet,-}\quad\forall\,\eps>0,
  \end{equation}
  where we use an unweighted b-density. Due to the integrability of this in $t_*$ (for each fixed spatial point $x\in\R^3$), the Fourier transform of this is continuous down to the real axis, and thus we can shift the part $\gamma_0$ of the contour in~\eqref{EqAdmPfLoDerCont} down to $[-\frac{c}{2},\frac{c}{2}]$, so
  \begin{equation}
  \label{EqAdmPfLoDerReal}
    \pa_{t_*}\dot g = \frac{1}{2\pi} \int_{-c}^c e^{-i\sigma t_*} \chi(|\sigma|) \cF(\pa_{t_*}\dot g_b^{\Ups,\aug})(\sigma)(\hat b_1(\sigma))\,\dd\sigma.
  \end{equation}

  We can use~\eqref{EqTFHbLo} to control the Fourier transform of~\eqref{EqAdmPfPaGb}; thus,
  \begin{equation}
  \label{EqAdmPfLoDerF1}
    \bigl(\cF(\pa_{t_*}\dot g_b^{\Ups,\aug})(\pm\sigma)\bigr)\big|_{\sigma\in[0,1]} \in \Hb^{\infty,\ 1-\eps,\ 1-2\eps,\ ((0,0),1-\eps)}(X_\scbtop;S^2\cT^*_X) \otimes (\C^4)^*.
  \end{equation}
  for all $\eps>0$ (and using an unweighted b-density). Mirroring the arguments after~\eqref{EqAdmLoImfHb}, this in turn implies uniform bounds for $\cF(\pa_{t_*}\dot g_b^{\Ups,\aug})(\sigma)$ (and its iterated $\sigma\pa_\sigma$-derivatives) in the space $H_{\bop,\sigma}^{\infty,(1-\eps,1-2\eps,0)}(X,\mu_\bop)$, with $\mu_\bop$ an unweighted b-density on $X$. Using~\citeAF{(\ref*{EqDResscbt1})}, this implies uniform (in $\sigma\in[0,1]$) bounds in $H_{(\scbtop,\sigma);\bop}^{(m;k),(\sfr,-\frac12-\eps,0)}(X)$ (now using $|\dd g_b|_X|$) for all fixed $m,k\in\N_0$ and $\eps>0$; the requirement on $\sfr$ here is that $\sfr<-\frac12$ at the zero section $\cR_{\rm out}$ of $\Tscbt^*_\scface X$. For the inverse Fourier transform of $\chi(|\sigma|)\cF(\pa_{t_*}\dot g_b^{\Ups,\aug})(\sigma)(\hat b_1(\sigma))$, this now implies (using Lemma~\ref{LemmaT3bFT} for the case $k=0$, and \citeAF{Lemma~\ref*{LemmaMUetbFTb}} for general $k$) the membership of~\eqref{EqAdmPfLoDerReal} in $\Htb^{(\sfs';k),(\alpha',0)}(M_0,|\dd g_b|)$ in view of Lemma~\ref{LemmaAdmPfLoCoeff}, provided $\alpha'<-\frac12$ and $\sfs'+\alpha'<-\frac12$ at $\cR_{\rm out}$; these conditions hold for $\alpha'=-\frac12-\eps$ and $\sfs'=\sfs$ (see also Remark~\ref{RmkAdmUpper}), so~\eqref{EqAdmPfLoDer1} follows.

  \pfstep{Control of $\pa_{t_*}^2 h_{\rms 1}$.} We use the description
  \[
    h_{b,\rms 1}^{\leq 1,\aug} = \chi_\cK h_{b,\rms 1}^{\leq 1} + h',\quad h'\in\cA^{1,1,-\eps}(\Omega)
  \]
  (with support in $t_*\geq 1$) that follows from~\eqref{EqAdmLoImLot}. Since $\pa_{t_*}^2 h_{b,\rms 1}^{\leq 1}=0$, we have
  \[
    \pa_{t_*}^2 h_{b,\rms 1}^{\leq 1,\aug} = 2\underbrace{(\pa_{t_*}\chi_\cK)}_{\in\cA^{\infty,1,\infty}}\underbrace{h_{b,\rms 1}}_{\in\cA^{*,2,*}} \!+\;\underbrace{(\pa_{t_*}^2\chi_\cK)}_{\in\cA^{\infty,2,\infty}}\underbrace{h_{b,\rms 1}^{\leq 1}}_{\in\cA^{*,1,*}} \!\!{}+ \pa_{t_*}^2 h'.
  \]
  We further split $h'=\chi_\cK h'+(1-\chi_\cK)h'$ and note that $\pa_{t_*}^2(\chi_\cK h')\in\cA^{\infty,3,2-\eps}$ to obtain the description
  \begin{equation}
  \label{EqAdmPfLoDer2Decomp}
    \pa_{t_*}^2 h_{b,\rms 1}^{\leq 1,\aug} = h_1 + \pa_{t_*}^2 h_2,\quad h_1 \in \cA^{\infty,3,2-\eps},\ h_2\in\cA^{1,1,\infty};
  \end{equation}
  a fortiori, $h_1\in\Hb^{\infty,\ \infty,\ 3-\eps,\ 2-\eps}(\Omega)^{\bullet,-}$ and $h_2\in\Hb^{\infty,\ 1-\eps,\ 1-\eps,\ \infty}(\Omega)^{\bullet,-}$ for all $\eps>0$. This is integrable in $t_*$, thus has continuous Fourier transform down to the real axis, and therefore
  \begin{equation}
  \label{EqAdmPfLoDer2Real}
    \pa_{t_*}^2 h_{\rms 1} = \frac{1}{2\pi} \int_{-c}^c e^{-i\sigma t_*}\chi(|\sigma|)\cF(\pa_{t_*}^2 h_{b,\rms 1}^{\leq 1,\aug})(\hat\scal_1(\sigma))\,\dd\sigma.
  \end{equation}

  Applying~\eqref{EqTFHbLo} to~\eqref{EqAdmPfLoDer2Decomp}, we have\footnote{If one only used $\pa_{t_*}^2 h_{b,\rms 1}^{\leq 1,\aug}\in\cA^{1,3,2-\eps}$, this membership would not follow from~\eqref{EqTFHbLo} since the $\scri^+$-order $1$ is too weak compared to the $\iota^+$-order $3$. This is why we instead use the careful splitting in~\eqref{EqAdmPfLoDer2Decomp}.}
  \begin{align}
    \bigl(\cF(\pa_{t_*}^2 h_{b,\rms 1}^{\leq 1,\aug})(\pm\sigma)\bigr)\big|_{\sigma\in[0,1]} &\in \Hb^{\infty,\ 3-\eps,\ 2-\eps,\ ((0,0),1-\eps)}(X_\scbtop) + \sigma^2 \Hb^{\infty,\ 1-\eps,\ -\eps,\ (0,0)}(X_\scbtop) \nonumber\\
  \label{EqAdmPfLoDerF2}
      &\subset \Hb^{\infty,\ 1-\eps,\ 2-\eps,\ ((0,0),1-\eps)}(X_\scbtop).
  \end{align}
  This gives uniform bounds in $H_{(\scbtop,\sigma);\bop}^{(m;k),(\sfr,\frac12-\eps,0)}(X,|\dd g_b|_X|)$ provided $\sfr<-\frac12$ at $\cR_{\rm out}$. We can then use Lemma~\ref{LemmaAdmPfLoCoeff} and \citeAF{Lemma~\ref*{LemmaMUetbFTb}} as before to deduce that~\eqref{EqAdmPfLoDer2Real} defines an element of $H_{\tbop;\bop}^{(\sfs';k),(\alpha',0)}$ provided $\alpha'<\frac12$ and $\sfs'+\alpha'<-\frac12$ at $\cR_{\rm out}$; these conditions hold for $\alpha'=\frac12-\eps$ and $\sfs'=\sfs$ (see again Remark~\ref{RmkAdmUpper}).
\end{proof}

Our goal is to obtain a bound for
\[
  u_{\rm sing} := \dot g+h_{\rms 1}
\]
in $t_*\geq 1$, which we will obtain by integrating $\pa_{t_*}^2(\dot g+h_{\rms 1})$ from $t_*<1$ towards the future.\footnote{Unlike in \cite[\S{3.4}]{HintzGlueLocIII}, we do not control $\dot g$ and $h_{\rms 1}$ separately here, which would be more delicate due to the non-product (in $t_*$ and spatial variables) nature of these terms.} Since $u=0$ for $t_*<1$, we first use~\eqref{EqAdmPfLoDecomp} to write
\begin{equation}
\label{EqAdmPfLo0}
  u_{\rm sing} = \dot g + h_{\rms 1} = -(u_{\rm hi}+u_{\rm reg}),\quad t_*<1.
\end{equation}
Thus, Lemmas~\ref{LemmaAdmPfHi} and \ref{LemmaAdmPfLoReg} give a norm bound for
\begin{equation}
\label{EqAdmPfLopa0}
  u_{\rm sing}|_{\{t_*<1\}} \in \bar H_{\tbop;\bop}^{(\sfs;k),(\alpha_\sface,0)}(\Omega_-),\quad\Omega_-:=\cl_{M_0}\{t_*\leq 1\},
\end{equation}
in terms of $\|f\|_{H_{\tbop;\bop}^{(\sfs;k),(\alpha_\sface+2,0)}}$. Using $\pa_{t_*}\in\rho_\sface\Vtb(M_0)$, this gives
\begin{equation}
\label{EqAdmPfLopa1}
  \pa_{t_*}u_{\rm sing}|_{\{t_*<1\}} \in \bar H_{\tbop;\bop}^{(\sfs-1;k),(\alpha_\sface+1,0)}(\Omega_-).
\end{equation}
But Lemma~\ref{LemmaAdmPfLoDer} gives the global membership
\begin{equation}
\label{EqAdmPfLopa2}
  \pa_{t_*}^2 u_{\rm sing} \in H_{\tbop;\bop}^{(\sfs-1;k),(\frac12-\eps,0)}(M_0),\quad\eps>0.
\end{equation}
We can therefore integrate~\eqref{EqAdmPfLopa2} in $t_*$ using~\eqref{EqAdmPfLopa1} as the ``initial condition;'' we use the following variant of \citeAF{Lemma~\ref*{LemmaA2AdmInt}} for this purpose:

\begin{lemma}[Integration of $\pa_{t_*}$ on 3b-Sobolev spaces]
\label{LemmaAdmPfInt}
  Let $\sfs\in\CI(\Stb^*M_0)$ be a $t_*$-translation-invariant 3b-differential order function. Let $k\in\N_0$, $\beta\in\R$, and $\ell>\frac12$. With $\Omega_-=\cl_{M_0}\{t_*\leq 1\}$ as above, suppose that $u|_{\Omega_-}\in\bar H_{\tbop;\bop}^{(\sfs;k),(\beta-1,-\ell)}(\Omega_-)$ and $\pa_{t_*}u\in H_{\tbop;\bop}^{(\sfs;k),(\beta,-\ell+1)}(M_0)$. (We use the product of $|\dd t_*|$ with any smooth weighted b-density on $X$ to define $L^2(M_0)$.) Then $u\in H_{\tbop;\bop}^{(\sfs;k),(\beta-1,-\ell)}(M_0)$.
\end{lemma}
\begin{proof}
  Write $f:=\pa_{t_*}u$. Let $\chi\in\CI(\R)$ be equal to $1$ on $[-\frac14,\infty)$ and equal to $0$ on $(-\infty,-\frac12]$. Then $u':=\chi(\frac{t_*}{r})u$ equals $u$ near $\{t_*\geq 1\}$, vanishes for $\frac{t_*}{r}\leq-\frac12$, and satisfies $\pa_{t_*}u'=\chi(\frac{t_*}{r})\pa_{t_*}u + t_*^{-1}\chi_1(\frac{t_*}{r})u=:f'$ where $\chi_1(x):=x\chi_0'(x)\in\CIdot([-\frac12,-\frac14])$. The a priori assumptions on $u$ imply then that $\pa_{t_*}u'\in H_{\tbop;\bop}^{(\sfs;k),(\beta,-\ell+1)}$ since $|t_*|^{-1}$ is a local defining function of $\sface$ near $\supp\chi_1(\frac{t_*}{r})$. Renaming $u',f'$ into $u,f$, we have thus shown that we may assume without loss that
  \[
    u = 0\ \ \text{for}\ \ t_*\leq-\frac{r}{2}.
  \]
  Furthermore, derivatives along $\pa_{t_*}$ and $\la x\ra\pa_x$ commute with $\pa_{t_*}$, and the identity $\pa_{t_*}(t_*\pa_{t_*}u)=t_*\pa_{t_*}f+f$ then shows that the claim for general $k\in\N_0$ follows from the case $k=0$, which we now treat.

  Suppose first that $\sfs=0$. Let $t=t_*+r$, then $t\geq\frac{r}{2}$ on $\supp u$, and thus $\rho_\sface=r^{-1}$ and $\rho_\cK=\frac{r}{t}$ are local defining functions of $\sface$ and $\cK^+\subset M_0$ on $\supp u$. Since $\pa_{t_*}$ commutes with multiplication by powers of $r$, we may assume $\beta=0$. Dropping angular variables from the notation, we then have $f(t,r)=\rho_\cK^{-\ell+1}f_0(t,r)$, $f_0\in L^2(|\dd t\,\dd r|)$, and thus for any fixed $\alpha\in(-\frac12,\ell-1)$ we can estimate the squared $\rho_\sface^{-1}\rho_\cK^{-\ell}L^2$-norm of $u$ by
  \begin{align*}
    &\int_{\bhm_0}^\infty \int_{\frac{r}{2}}^\infty \biggl| r^{\ell-1}t^{-\ell} \int_{\frac{r}{2}}^t r^{-\ell+1}s^{\ell-1}f_0(s,r)\,\dd s\biggr|^2\,\dd t\,\dd r \\
    &\qquad \leq\int_{\bhm_0}^\infty \int_{\frac{r}{2}}^\infty t^{-2\ell}\biggl( \int_{\frac{r}{2}}^t s^{2\alpha}\,\dd s\biggr) \biggl( \int_{\frac{r}{2}}^t s^{2\ell-2\alpha-1} |f_0(s,r)|^2\,\dd s\biggr)\,\dd t\,\dd r \\
    &\qquad \lesssim \int_{\frac{\bhm_0}{2}}^\infty \int_{\frac{r}{2}}^\infty s^{2\ell-2\alpha-1}|f_0(s,r)|^2 \biggl(\int_s^\infty t^{-2\ell+2\alpha+1}\,\dd t\biggr)\,\dd s\,\dd r \\
    &\qquad \lesssim \|f_0\|_{L^2}^2.
  \end{align*}
  The case of general $t_*$-translation-invariant orders $\sfs$ is treated exactly as in the proof of \citeAF{Lemma~\ref*{LemmaA2AdmInt}}.
\end{proof}

We therefore obtain the global membership
\begin{equation}
\label{EqAdmPfLopa12}
  \pa_{t_*}u_{\rm sing} \in H_{\tbop;\bop}^{(\sfs-1;k),(-\frac12-\eps,-1)}(M_0).
\end{equation}
(This is new information for $t_*\geq 1$. For $t_*<1$, we have the stronger~\eqref{EqAdmPfLopa1}.) We can now integrate~\eqref{EqAdmPfLopa12} with initial condition~\eqref{EqAdmPfLopa0}; Lemma~\ref{LemmaAdmPfInt} gives
\[
  u_{\rm sing} \in H_{\tbop;\bop}^{(\sfs-1;k),(-\frac32-\eps,-2)}(M_0)\quad\forall\,\eps>0.
\]
But recall that $u=u_{\rm hi}+u_{\rm reg}+u_{\rm sing}$ is supported in $t_*\geq 1$; recalling Lemmas~\ref{LemmaAdmPfHi} and \ref{LemmaAdmPfLoReg}, we have thus proved
\[
  u \in H_{\tbop;\bop}^{(\sfs-1;k),(-\frac32-\eps,-2)}(\Omega,|\dd g_b|;S^2\cT^*)^{\bullet,-}\quad\forall\,\eps>0.
\]
This proves~\eqref{EqAdm} (since $\alpha_\sface-\delta<-\frac32$) and thus finishes the proof of Theorem~\ref{ThmAdm}.

\section{Decay of metric perturbations on dynamical backgrounds}
\label{SD}

We now come to the heart of the paper. We will study forward problems for the linearization of the gauge-fixed Einstein equation around metrics $g=g_b+h$ that will arise in the nonlinear iteration scheme. The simplest version of the nonlinear forward map, in which the Kerr parameters are equal to the subextremal parameters $b_0$ of the metric we will ultimately study small perturbations of, is
\[
  P_0(h) = \Ric(g_{b_0}+h) - \delta_{g_{b_0},E^\cC}^* \Ups_{E^\Ups}(g_{b_0}+h,g_{b_0})
\]
in the notation of Definitions~\ref{Def1Gauge} and \ref{Def1Symm}; this is equal to $P(g_{b_0}+h,g_{b_0})$ in the notation of~\eqref{Eq1Ein}. \textit{Here, we fix the modifications $E^\cC$ and $E^\Ups$ as at the beginning of~\S\usref{SAdm}, so according to Lemma~\usref{LemmaWG0Pair} and Proposition~\usref{PropWGMode}, Theorem~\usref{ThmWCRec} with $v^\cC=\frac12$ and $C_0=100$, further $(1-e^\cC)(1-v^\cC)\gamma^\cC>100$, and such that Proposition~\usref{PropWEtf} holds.}

As usual in nonlinear iteration schemes, we need to show, roughly speaking, that the assumptions on the dynamical metric $g$ are consistent with the asymptotic behavior of linear forward solutions. Since general forward solutions are not even expected to decay in time (e.g., due to the presence of the generalized zero energy states $h_{b,\rms 1}^{\leq 1}(\dot\scal)$ and $\dot g_b^\Ups(\dot b)$), we must include additional arguments in the forward map that capture these contributions in an appropriate fashion (e.g., via working relative to an initially boosted Kerr metric and making the final Kerr parameters $b$ part of the arguments). There are many contributions to the late-time asymptotics of linear forward solutions that do not decay at the rate $t_*^{-4-\eps_\cK}$ in spatially compact regions (which is the rate motivated already in~\S\ref{SssINHeur}), and indeed all of the states constructed in Propositions~\ref{PropWG0Symm} and \ref{PropWG0Large} will feature (appropriately modulated in $t_*$) in the course of our asymptotic analysis. The pure gauge contributions that do not have $t_*^{-2-\eps_\cK}$-decay must moreover be eliminated using gauge modifications, as explained in~\S\ref{SssINElim}.

Rather than presenting, from the outset, the (rather ornate) nonlinear map that captures all of these not-sufficiently-fast-decaying contributions and gauge modifications, we shall largely proceed in a step-by-step (i.e., bottom-up) fashion---thereby also explaining the origin of the final nonlinear map in Definition~\ref{DefD6Aug5} (see also~\eqref{EqEfPMap} where all of its ingredients are collected in one place). Correspondingly, we will only use a limited amount of foresight as far as the asymptotics of metric perturbations $h$ and source terms $f$ at the boundary hypersurfaces $\scri^+$, $\iota^+$, and $\cK^+$ of $M$ (see Definition~\ref{DefKMfdRad}) are concerned.

We do encode the parameters $b=(\bhm,\bha)\approx b_0=(\bhm_0,\bha_0)$ (final black hole parameters) and $\scal\approx 0\in\scalspace_1$ (Lorentz boost) from the start; the background metric for our gauge condition will be
\begin{equation}
\label{EqDgPOU}
  g_{b_0,b,-\scal} := (1-\chi_\cK)\phi_{-\scal}^*g_{b_0}+\chi_\cK g_b,
\end{equation}
where we recall $\phi_{-\scal}=\phi_\scal^{-1}$ from~\eqref{EqKBoVscal} and \eqref{EqKBoMap}. (The minus sign is merely a matter of convention.) Moreover, we will, at least initially, write the dynamical spacetime metric as $g_{b_0,b,-\scal}+h$. Thus, we initially consider
\[
  P_0(h,\scal,b) := P(g,g^0) = \Ric(g) - \delta_{g^0,E^\cC}^*\Ups_{E^\Ups}(g,g^0),\quad g:=g^0+h,\ g^0:=g_{b_0,b,-\scal},
\]
where we recall $P(g,g^0)$ from~\eqref{Eq1Ein}. We first show (Corollary~\ref{CorDAdm} in~\S\ref{SsDAdm}) that the wave-type operator $2 L$ (relative to the metric $g$), where $L:=D_{(h,\scal,b)}P_0(\cdot,0,0)=L_{g,g^0}$ (in the notation of~\eqref{Eq1EinLin} and \eqref{EqIEin}), is admissible relative to $2 L_b$ (and likewise for more general linearized gauge-fixed Einstein operators), thus making the linear main result \citeAF{Theorem~\ref{ThmF}} available and giving weak decay (e.g., $t_*^{\frac32}$-growth in spatially compact sets) but full b-regularity for $u$ (Theorem~\ref{ThmDAdmReg}). As already explained in~\S\ref{SssINHeur}, the strategy for extracting stronger asymptotics is to rewrite the PDE for $u$ as
\begin{equation}
\label{EqDRewrite}
  L_b u = f: := f - \tilde L u,\quad \tilde L:=L-L_b;
\end{equation}
here $\tilde L$ has decaying coefficients (in particular, $\cO(t_*^{-2-\eps-\cK})$ in spatially compact sets), and the inversion of $L_b$ uncovers the late-time asymptotics of $u$ step-by-step. Control of solutions of $L_{b_0} u=f'$ is effected using several tools: the solution of model problems at $\scri^+$ (see~\S\ref{SsDScri}) and at $\iota^+$ (using material from~\S\ref{SsipInv}) on spacetime, and Fourier-based techniques, the heart of which is a detailed analysis of the resolvent $\wh{L_b}(\sigma)^{-1}$ at low energies for which we utilize the augmented operator $\wt{L_b}(\sigma)$ introduced in~\S\ref{SssAdmLoDef} as well as results (from~\S\ref{Ssiptf}) on model problems arising at the transition from zero to nonzero frequencies.

The following is a rough overview of how we extract increasingly precise asymptotics and decay for $u$.
\begin{enumerate}
\item \S\ref{SsDScri}: the Fourier transform is very precise only when acting on functions with good $\scri^+$-decay (cf.\ the conditions~\eqref{EqTFHbphgBetaScri}), so as a first step we produce a formal solution of~\eqref{EqDRewrite} at $\scri^+$ (Proposition~\ref{PropDScriFormal}). We also show how to recover sharp asymptotics at $\scri^+$ (Proposition~\ref{PropDScriPhg}) (after they may have gotten scrambled, e.g., when applying the inverse Fourier transform to the resolvent output).
\item \S\ref{SsD1Alm}: by passing to the Fourier transform in $t_*$ and inverting $\wh{L_{b_0}}(\sigma)$, we show that $|u|\lesssim t_*^{1+\eps}$ in spatially compact sets (see Proposition~\ref{PropD1Alm} for the full statement, including decay in the $\iota^+$-regime); this is an order $\frac12-\eps$ improvement over the initial $t_*^{\frac32}$-bound.
\item \S\ref{SsD2Boost}: plugging this improvement into the right-hand side of~\eqref{EqDRewrite}, the inversion of $L_{b_0}$ now produces a $t_* h_{b,\rms 1}(\dot\scal)$ leading-order term (Proposition~\ref{PropD2Boost}). As motivated in~\S\ref{SssINElim}, we shall counteract this term by boosting the initial Kerr black hole (\S\ref{SssD2BoostNo}); this introduces our first modification (or rather: augmentation) of $P_0$ (Definition~\ref{DefD2EinsteinAug}). (The operator $P_0$ is in fact \emph{insufficient} for this purpose, as the elimination of $t_* h_{b,\rms 1}$ via the adjustment of $\scal$ requires also metric patches in the transition region $\supp\dd\chi_\cK$.)
\item \S\ref{SsD3Alm}: having eliminated the asymptotic boost in the previous step, we now prove the almost-boundedness of $u$, so $|u|\lesssim t_*^\eps$ in spatially compact sets (see Proposition~\ref{PropD3Alm} for the full result).
\item \S\ref{SsD4Par}: we show that $u$ has mild $t_*$-decay up to finite-dimensional obstacles which encode changes of the final black hole parameters as well as a logarithmic plus constant modulation of its center-of-mass (Proposition~\ref{PropD4Par}). We thus make the final black hole parameters part of the set of arguments of an augmentation of $P_0$, and moreover introduce gauge modifications to eliminate the center-of-mass motion (Definition~\ref{DefD4EinsteinAug}).
\item \S\ref{SsD5Alm}: we argue that the obstruction to $t_*^{-1+\eps}$-decay of $u$ is given by $t_*^{-\alpha}$-shifts in the center-of-mass of the final black hole for $\alpha\in(0,1)$ (see~\S\ref{SssIEinLa} and Remark~\ref{RmkD5Orig} for the origin of such terms); we eliminate these shifts via further gauge modifications (which decay relative to those introduced in the previous step). The upshot is a bound $|u|\lesssim t_*^{-1+\eps}$ in the forward cone (Proposition~\ref{PropD5Alm}).
\item \S\ref{SsD6Better}: when establishing the bound in the previous step, much information on the asymptotics in spatially compact sets and at future timelike infinity (more precisely, at $\cK^+$ and $\iota^+$) has been discarded; roughly speaking, one can already establish partial polyhomogeneity at $\iota^+$ and $t_*^{-3+\eps}$-decay in spatially compact sets up to essentially explicit terms with less decay. We then prove (Theorem~\ref{ThmD6}) that one can in fact obtain an expansion into explicit terms up to $t_*^{-4-\eps_\cK}$-decay upon making further gauge modifications which now eliminate not only center-of-mass shifts but also other pure gauge solutions (featuring the zero energy states constructed in Proposition~\ref{PropWG0Large}, which do not have as clear a geometric interpretation). The full decay result, and thus the main result of the present section, is Corollary~\ref{CorD6Impr}.
\end{enumerate}

\begin{rmk}[Linear stability]
\label{RmkDLinStab}
  As a special case of the arguments presented in this section, we obtain a proof of the linear stability of subextremal Kerr with stronger decay (namely, $t_*^{-3}$) than existing results \cite{AnderssonBackdahlBlueMaKerr,HaefnerHintzVasyKerr,HaefnerHintzVasyKerrLarge}. See Corollary~\ref{CorD6LinStab}.
\end{rmk}

\begin{rmk}[Exponents appearing in the asymptotic expansion]
\label{RmkDAsyTerms}
  Our asymptotic analysis in this section will be very coarse as far as the bookkeeping of exponents $(\lambda,k)$ of terms $\sim t_*^{-\lambda}(\log t_*)^k$ (in spatially compact sets, or at punctured future timelike infinity) in the asymptotic expansion of $u$ are concerned. This is a strength in that our ability to prove strong decay for $u$ \emph{despite} imprecise bookkeeping means that our methods for eliminating poorly decaying (pure gauge) contributions are robust. In a different context, however, it is a weakness since the resulting over-parameterization of the number of asymptotic degrees of freedom (some of which are, in reality, not present) is not compatible with an efficient nonlinear \emph{forward} map that one would like to map the fast-decaying gravitational wave tail and various modification parameters to a source term $f$ to which the analysis in the present section applies (which in particular requires it to have sufficient decay in the forward cone). In~\S\ref{SEf} below, we shall thus post-process the asymptotics produced here into a more economical form.
\end{rmk}

We proceed to define the basic class of metric perturbations we will consider (until we make further refinements starting in~\S\ref{SsD5Alm} below). We first use \citeAF{Lemma~\ref*{LemmaSDGTime}}, with $\ell_\sscri$ and the Kerr parameters there taken to be equal to $\frac12$ and $b_0=(\bhm_0,\bha_0)$, in order to construct a function
\begin{equation}
\label{EqDftstar}
  \ft_*
\end{equation}
on $\{t_*\geq 1\}$ which is a time function for all sufficiently small perturbations of $g_{b_0}$ in the precise sense described there and discussed further below. (We recall that the construction gives $\ft_*=t_*(1-(\frac{t_*}{r})^{\frac12})$ for small $\frac{t_*}{r}$, i.e., near $\scri^+$, furthermore $0<\delta\leq\ft_*/t_*\leq 1$ for some $\delta>0$, and $\ft_*=c_0 t_*+c_1$ for $r\leq 2\bhm_0$ for some $c_0,c_1>0$; and $\ft_*\leq t_*$.) We then work on the domain
\begin{equation}
\label{EqDOmegaStar}
  \Omega_* := \cl_M \{ \ft_*\geq 1 \}.
\end{equation}
\emph{For the remainder of this section, all $L^2$-based spaces are defined with respect to an unweighted b-density on $M$, e.g., on $\Omega_*$ the density $t_*^{-1}\la x\ra^{-3}|\dd t_*\,\dd x|$.} Moreover, we shall require for the cutoff functions in Definition~\ref{DefKBoCutoff} that, moreover,
\begin{equation}
\label{EqDCutoffs}
  \supp\chi_\cK^\flat \subset \chi_\cK^{-1}(1) \subset \supp\chi_\cK \subset (\chi_\cK^\sharp)^{-1}(1) \subset \supp\chi_\cK^\sharp \subset \{ \ft_*\geq 10 \}.
\end{equation}

\begin{definition}[Basic metric perturbations and source terms]
\label{DefDMetBasic}
  Let $\cE_+\subset\C\times\N_0$ be an index set with
  \begin{equation}
  \label{EqDMetBasicEplus}
    (1,0)\in\cE_+,\quad
    \min\Re\cE_+\geq 1,\quad
    j\cE_+\subset\cE_+\ \forall\,j\in\N.
  \end{equation}
  Let $\ell_\sscri,\ell_+,\ell_\cK\in\R$. We then introduce:
  \begin{enumerate}
  \item{\rm (General metric perturbations.)} Let $\cE_\sscri^\cC$ be an index set satisfying the properties of Definition~\usref{DefExP} (so in particular $\min\Re\cE_\sscri^\cC>1+2\gamma^\Ups$). Then for $k\geq 5$,
    \begin{equation}
    \label{EqDMetBasicHb}
      \Hb^{k,\ \bigl(\la\cE_\sscri^\cC\ra,\ell_\sscri\bigr),\ (\cE_+,\ell_+),\ \ell_\cK}(\Omega_*)
    \end{equation}
    consists of all real symmetric 2-tensors $h$ such that (analogously to~\eqref{EqExPSpaceProj} and using the notation of Definition~\usref{DefExPComp})
    \begin{equation}
    \label{EqDMetBasicProj}
    \begin{alignedat}{2}
      \pi^\cC h &\in H_\bop^{k,(\cE_\sscri^\cC,\ell_\sscri),\ (\cE_+,\ell_+),\ \ell_\cK}, \\
      \pi_{0 1}h &\in H_\bop^{k,(\cE_{\sscri,0 1},\ell_\sscri),\ (\cE_+,\ell_+),\ \ell_\cK}, &\qquad
      \pi_{1 /}h &\in H_\bop^{k,(\cE_{\sscri,1 /},\ell_\sscri),\ (\cE_+,\ell_+),\ \ell_\cK}, \\
      \slpi_0 h &\in H_\bop^{k,(\slcE_{\sscri,0},\ell_\sscri),\ (\cE_+,\ell_+),\ \ell_\cK}, &\qquad
      \pi_{1 1}h &\in H_\bop^{k,(\cE_{\sscri,1 1},\ell_\sscri),\ (\cE_+,\ell_+),\ \ell_\cK}.
    \end{alignedat}
    \end{equation}
  \item\label{ItDMetBasicSrc}{\rm (Source terms.)} Define $\cF_{\sscri,1 1}$ by~\eqref{EqExFwSrc11}. For $k\geq 3$, we denote by
    \[
      \Hb^{k,\ \bigl(\la\cE_\sscri^\cC+1\ra',\ell_\sscri+1\bigr),\ (\cE_++2,\ell_++2),\ \ell_\cK}(\Omega_*)
    \]
    the space of all real symmetric 2-tensors $f$ such that (analogously to Definition~\usref{DefExFwSrc})
    \begin{align*}
      \pi^\cC f,\ \pi_{0 1}f,\ \pi_{1 /}f,\ \slpi_0 f &\in \Hb^{k,\ (\cE_\sscri^\cC+1,\ell_\sscri+1),\ (\cE_++2,\ell_++2),\ \ell_\cK}, \\
      \pi_{1 1}f &\in \Hb^{k,\ (\cF_{\sscri,1 1},\ell_\sscri+1),\ (\cE_++2,\ell_++2),\ \ell_\cK}.
    \end{align*}
  \item\label{ItDMetBasic2}{\rm ($\scri^+$-smooth metric perturbations.)} We define
  \[
    \Hb^{k,\ \bigl(\la(2,0)\ra_0,\ell_\sscri\bigr),\ (\cE_+,\ell_+),\ \cE_\cK}(\Omega_*)
  \]
  to consist of all real symmetric 2-tensors $h$ with $\pi^\cC h$, $\pi_{0 1}h$, $\pi_{1 /}h\in\Hb^{k,\ ((2,0),\ell_\sscri),\ (\cE_+,\ell_+),\ \cE_\cK}$ and $\slpi_0 h$, $\pi_{1 1}h\in\Hb^{k,\ ((1,0),\ell_\sscri),\ (\cE_+,\ell_+),\ \cE_\cK}$.
  \end{enumerate}
\end{definition}

Note that the second and third conditions in~\eqref{EqDMetBasicEplus} are completely analogous to Definition~\ref{DefExP}\eqref{ItExPE0}, except for the stronger lower bound on $\min\Re\cE_+$ (which mildly simplifies the bookkeeping, and will, in any case, be satisfied in our application). The requirement $(1,0)\in\cE_+$ ensures that smooth decaying perturbations are allowed as far as their behavior at $(\iota^+)^\circ$ is concerned. (We do allow for logarithmic terms at $\iota^+$ as well; terms of order $(1,1)$ will indeed arise, e.g., as metric patches when eliminating logarithmic center-of-mass motions.) Part~\eqref{ItDMetBasic2} of this definition is modeled on Definition~\ref{DefExPLead}.

When $\ell_\sscri<1$, the space~\eqref{EqDMetBasicHb} is equal to $\Hb^{k,\ell_\sscri,(\cE_+,\ell_+),\ell_\cK}$. The decay rates of remainder terms of partial polyhomogeneous expansions of metric perturbations ($h$) and source terms ($f$) will be captured using parameters
\begin{equation}
\label{EqDMetBasicEllEps}
  0 < \eps_\cK < \eps_+ < \min(\eps_\ind,\eps_\sscri) < 1.
\end{equation}
Metric perturbations will be required to have remainder terms with weights
\begin{equation}
\label{EqDMetBasicEll}
  3+\eps_\sscri,\quad
  3+\eps_+,\quad
  2+\eps_\cK.
\end{equation}
That is, we will study the linearization of the gauge-fixed Einstein operator around metrics that are the sum of a Kerr metric, or rather of $g_{b_0,b,-\scal}$ from~\eqref{EqDgPOU}, and a metric perturbation of class
\begin{equation}
\label{EqDMetBasich}
  h \in \Hb^{k,\ \bigl(\la\cE_\sscri^\cC\ra,3+\eps_\sscri\bigr),\ (\cE_+,3+\eps_+),\ 2+\eps_\cK}(\Omega_*)
\end{equation}
for large $k$.

\begin{rmk}[Source terms]
\label{RmkDSource}
  The decay rates we will ultimately require for source terms $f$ are the same as the conormal remainders of metric perturbations in Definition~\ref{DefDMetBasic} up to the usual shifts of $1$ and $2$ orders at $\scri^+$ and $\iota^+$, respectively (arising from $L_{b_0}\in\rho_\sscri\rho_+^2\Diffb^2$, similarly for linearizations around dynamical metrics), so $\rho_\sscri^{3+\eps_\sscri}$ and $\rho_+^{5+\eps_+}$ at $\scri^+$ and $\iota^+$, respectively. However, we will require $\rho_\cK^{4+\eps_\cK}$-decay of $f$ at $\cK^+$ due to the second order pole of the resolvent $\wh{L_b}(\sigma)^{-1}$ at $\sigma=0$, which (in a rough accounting) causes forward solutions to lose $2$ powers of $t_*$-decay. (See also the discussion following~\eqref{EqINHeurf}.) At intermediate steps of our arguments, the conditions on $f$ can (and will) be relaxed. This discrepancy between $h$ and $f$ will be bridged once we sharpen the class of metric perturbations, which will ultimately feature certain partial (non-polyhomogeneous) expansions plus remainders with the ``expected'' decay rates $3+\eps_+$ and $4+\eps_\cK$ at $\iota^+$ and $\cK^+$, respectively; see Definition~\ref{DefD6Aug5} for details.
\end{rmk}

\subsection{Structure of the dynamical linearized operator}
\label{SsDAdm}

The geometric and analytic requirements on dynamical metrics and wave-type operators in \citeAF{\S\ref*{SS}} are phrased in terms of the edge-3b-(co)tangent space, which we briefly recall. (This is a variant of Definition~\ref{DefExeb} and the discussion in~\S\ref{SssT3b}.) We work here on the domains
\[
  \Omega_{\frac12}:=\cl_{M_{\frac12}}\{t_*\geq 1\},\quad
  \Omega_{*,\frac12}:=\cl_{M_{\frac12}}\{\ft_*\geq 1\},
\]
where $M_{\frac12}$ is defined in Definition~\ref{DefKMfdRad2}. To wit, we write
\[
  \cV_\etbop(\Omega_{\frac12})
\]
for the space of all smooth vector fields on $\Omega_{\frac12}$ which are tangent to all boundary hypersurfaces ``at infinity,'' i.e., to $\Omega_{\frac12}\cap H$ for $H=\scri^+,\iota^+,\cK^+$, moreover tangent to each fiber of $\scri^+$ (see the discussion after~\eqref{EqKMfdCoordHalf}), and whose restrictions to a collar neighborhood of $\cK^+$ are smooth 3b-vector fields on $M_0$. Thus, they are edge-b-vector fields near $\scri^+$ and 3b-vector fields near $\cK^+$. In the region $\frac{r}{t_*}<2$ (so near $\cK^+$), such vector fields are smooth (in $r^{-1}$, $\frac{t_*}{r}$, and $\omega\in\Sph^2$) linear combinations of $r\pa_{t_*}$, $r\pa_r$, and $\pa_\omega$, while in $\frac{r}{t_*}>1$ (so near $\scri^+$) they are smooth (on $M_{\frac12}$) linear combinations of $x_\sscri\pa_{x_\sscri}$, $\rho_+\pa_{\rho_+}$, and $x_\sscri\pa_\omega$ in the coordinates $x_\sscri=\rho_\sscri^{\frac12}$, $\rho_+$, $\omega$ from~\eqref{EqKMfdCoordScriip}. An alternative description near $\scri^+$ arises from the expressions~\eqref{EqKNullComp}, which shows that elements of $\cV_\etbop$ are smooth linear combinations of $x_\sscri^{-2}\rho_+^{-1}\pa_0$, $\rho_+^{-1}\pa_1$, and $x_\sscri\pa_\omega$ near $\scri^+$. In particular,
\begin{equation}
\label{EqDAdmCoordetb}
  \pa_0\in x_\sscri^2\rho_+\Vetb=r^{-1}\Vetb,\quad
  \pa_1\in\rho_+\Vetb,\quad
  r^{-1}\pa_\omega\in x_\sscri\rho_+\Vetb.
\end{equation}
(This is completely analogous to an observation in the region $t_*\leq 1$ made in~\eqref{EqExebpa01}.) The associated classes of e3b-differential operators are denoted
\[
  \Diff_\etbop^m(\Omega_{\frac12}),
\]
and we can also consider versions weighted by powers of $x_\sscri$, $\rho_+$, and $\rho_\cK$ (defining functions of $\scri^+$, $\iota^+$, and $\cK^+$ on $\Omega_{\frac12}$).

We write $\Tetb_{\Omega_{\frac12}}M_{\frac12}\to\Omega_{\frac12}$ for the bundle (identified with the standard tangent bundle over the interior of $\Omega_{\frac12}$) for which the space of smooth sections is exactly $\cV_\etbop(\Omega_{\frac12})$, and $\Tetb^*_{\Omega_{\frac12}}M_{\frac12}$ for the dual bundle. We recall from \citeAF{Lemma~\ref*{LemmaCTebsc}} (or by duality from~\eqref{EqDAdmCoordetb}) that a local spanning set of the space $x_\sscri^{-2}\rho_+^{-2}\CI(\Omega_{\frac12};S^2\,\Tetb_{\Omega_{\frac12}}^* M_{\frac12})$ near $\scri^+$ is given by
\begin{equation}
\label{EqDAdmMete3b}
  x_\sscri^2(\dd x^0)^2,\ \dd x^0\otimes_s\dd x^1,\ x_\sscri\,\dd x_0\otimes_s r\,\dd\omega,\ x_\sscri^{-2}(\dd x^1)^2,\ x_\sscri^{-1}\,\dd x^1\otimes_s r\,\dd\omega,\ r\,\dd\omega\otimes_s r\,\dd\omega.
\end{equation}
This is in fact a global frame on $\Omega_{\frac12}$. (Away from $\sscri^+$ the factors of $x_\sscri$ can be dropped.) An inspection of~\eqref{EqKNullSchw} thus shows that
\[
  g_b \in x_\sscri^{-2}\rho_+^{-2}\CI(\Omega_{\frac12};S^2\,\Tetb^*_{\Omega_{\frac12}}M_{\frac12})
\]
when the angular momentum is $0$, and for general $b$ as a consequence of~\eqref{EqKMetDiff}; and indeed this is non-degenerate in that $g_b^{-1}\in x_\sscri^2\rho_+^2\CI(\Omega_{\frac12};S^2\,\Tetb_{\Omega_{\frac12}}M_{\frac12})$. (These observations were already made in \citeAF{Lemma~\ref*{LemmaTsKLMetric}(\ref*{ItTsKLMetrice3b})}.)

\begin{lemma}[Edge-3b-behavior of metric perturbations]
\label{LemmaDAdmMet}
  In the notation of Definition~\usref{DefDMetBasic} and recalling~\eqref{EqDMetBasicEllEps}, consider $h\in\cX^k:=\Hb^{k,\bigl(\la\cE_\sscri^\cC\ra,3+\eps_\sscri\bigr),\ (\cE_+,3+\eps_+),\ 2+\eps_\cK}(\Omega_*)$. Then
  \[
    h \in \tilde\sG_\bop^{k-3,(2\ell_\sscri,\ell_+,\ell_\cK)} := x_\sscri^{-2}\rho_+^{-2}\cC_\bop^{k-3,(2\ell_\sscri,\ell_+,\ell_\cK)}(\Omega_{*,\frac12};S^2\,\Tetb^*_{\Omega_{*,\frac12}}M_{\frac12})
  \]
  (where weights are powers of $x_\sscri$, $\rho_+$, and $\rho_\cK$), provided that\footnote{We use notation consistent with \citeAF{\S\ref*{SssSDW}} here. We caution the reader that the meaning of $\ell_\sscri$ here (referring to the $\scri^+$-decay rate of e3b-2-tensors) is different than that of $\ell_\sscri$ in Definition~\ref{DefExP} (where it refers to the $\scri^+$-decay rate of scattering 2-tensors, analogously to $3+\eps_\sscri$ in the present section). --- The e3b-perspective will not play any role outside of the present section, so this temporary clash of notation will not cause any confusion later on.}
  \begin{equation}
  \label{EqDAdmMetells}
    \ell_\sscri<\min(2\gamma^\Ups,\tfrac12),\quad
    \ell_+<1,\quad
    \ell_\cK\leq 2+\eps_\cK.
  \end{equation}
\end{lemma}
\begin{proof}
  Sobolev embedding (see \citeAF{Lemma~\ref*{LemmaMUCe3bSob}}) gives $h\in\cC_\bop^{k-3,(1,1-\eps,2+\eps_\cK)}(\Omega_*;S^2\cT^*)$ for all $\eps>0$ (using powers of $\rho_\sscri$, $\rho_+$, and $\rho_\cK$ for weights); moreover, we have $h_{0 0}\in\cC_\bop^{k-3,(1+2\gamma^\Ups,1-\eps,2+\eps_\cK)}$ by~\eqref{EqDMetBasicProj} since $h_{0 0}$ is a component of $\pi^\cC h$, and $\min\Re\cE_\sscri^\cC>1+2\gamma^\Ups$. On $\Omega_{*,\frac12}$ (and thus using powers of $x_\sscri$ for $\scri^+$-weights), this gives
  \[
    h \in \cC_\bop^{k-3,(2,\ 1-\eps,\ 2+\eps_\cK)}(\Omega_{*,\frac12};S^2\cT^*),\quad
    h_{0 0} \in \cC_\bop^{k-3,(2+4\gamma^\Ups,\ 1-\eps,\ 2+\eps_\cK)}(\Omega_{*,\frac12}).
  \]
  The claim now follows from~\eqref{EqDAdmMete3b}.
\end{proof}

As a consequence of (the proof of) Lemma~\ref{LemmaDAdmMet}, \citeAF{Lemma~\ref*{LemmaSDGTime} and Remark~\ref*{RmkSDGOtherKerr}} and Sobolev embedding imply that $\dd\ft_*$ is past timelike for all metrics $g_{b_0,b,-\scal}+h$ when $b$ is close to $b_0$, $\scal\in\scalspace_1$ is small, and the norm of $h$ in $\cX^3$ is sufficiently small. Moreover, such metrics $g_b+h$ are asymptotically Kerr metrics in the sense of \citeAF{Definition~\ref*{DefSDGMetric}}.

Note next that the computations in~\S\ref{SsExOp} of metric coefficients, and thus also of the Christoffel symbols, curvature components, the structure of covariant derivatives and, finally, of the linearized gauge-fixed Einstein operator in Proposition~\ref{PropExOpLin}, carry over with only notational changes from a neighborhood of $I^0\cup\scri^+$ to a neighborhood of $\scri^+\cup\iota^+$. (Indeed, the weights of the Kerr metric as an edge-b-metric near $\scri^+$ are the same at $I^0$ and $\iota^+$, namely $-2$ for both; and also the $I^0$- and $\iota^+$-weights of the basic differential operators~\eqref{EqKNullComp2} are the same.) We record this analogously to~\eqref{EqExOpLinbNormal}--\eqref{EqExOpLinebNormal}, but now using the second expressions from~\eqref{EqKNullComp} in~\eqref{EqExOpLinb}--\eqref{EqExOpLineb}.

\begin{prop}[Structure of $L_{g,g^0}$: forward cone]
\label{PropDAdmLin}
  Let $b=(\bhm,\bha)$ be close to $b_0=(\bhm_0,\bha_0)$, and let $\scal\in\scalspace_1$ be small. Let $g^0=g_{b_0,b,-\scal}$, or more generally
  \begin{subequations}
  \begin{align}
  \label{EqDAdmLinBg1}
    g^0&-g_{\bhm_0}\in\Hb^{\infty,\ \bigl(\la(2,0)\ra_0,\infty\bigr),\ \bigl((1,0),\infty\bigr),\ (0,0)}(\Omega_*), \\
  \label{EqDAdmLinBg2}
    g^0&=g_b\ \text{in a neighborhood of $\cK^+$}.
  \end{align}
  \end{subequations}
  Let $h\in\Hb^{k,(\la\cE_\sscri^\cC\ra,3+\eps_\sscri),\ (\cE_+,3+\eps_+),\ 2+\eps_\cK}(\Omega_*)$, with sufficiently small $\Hb^{3,(1,0,0)}$-norm,\footnote{The somewhat arbitrary weights here are chosen simply so that this smallness implies also that $h$ is small in $x_\sscri^{-2}\rho_+^{-2}L^\infty$ as a section of $S^2\,\Tetb^*$ and thus $g_{b_0}+h$ is Lorentzian.} and set
  \begin{equation}
  \label{EqDAdmLinKMet}
    g:=g_{b_0,b,-\scal}+h.
  \end{equation}
  Fix $\ell_\sscri$, $\ell_+$, and $\ell_\cK$ satisfying~\eqref{EqDAdmMetells}.
  \begin{enumerate}
  \item\label{ItDAdmLinScri}{\rm (Near $\scri^+$.)} Let us use the coordinates $\rho_\sscri=x_\sscri^2$, $\rho_+$, and $\omega$ from~\eqref{EqKMfdCoordScriip}. Then, for arbitrary $c>0$ and for $\frac{r}{t_*}\geq c$ (and correspondingly omitting weights at $\cK^+$, and writing \textnormal{``$\ebop$''} instead of \textnormal{``$\etbop$''}), the operator $L_{g,g^0}$ defined by~\eqref{Eq1EinLin} and~\eqref{Eq1Ein} satisfies
    \begin{subequations}
    \begin{align}
    \label{EqDAdmLinbNormal}
    \begin{split}
      L_{g,g^0} &\equiv \rho_\sscri\rho_+^2\Bigl[-(\rho_\sscri\pa_{\rho_\sscri}-(I+A_h))(\rho_\sscri\pa_{\rho_\sscri}-\rho_+\pa_{\rho_+}) + \rho_+^{-1}B_h \Bigr] \\
            &\quad\qquad \bmod (\rho_\sscri^2\rho_+^2\CI+H_\bop^{k-2,(\cE_\sscri^\cC,3+\eps_\sscri),\ (\cE_++2,\,5+\eps_+)})\Diffb^2,
    \end{split} \\
    \label{EqDAdmLinebNormal}
    \begin{split}
      2 L_{g,g^0} &\equiv \frac12 x_\sscri^2\rho_+^2\Bigl[-(x_\sscri\pa_{x_\sscri}-2(I+A_h))(x_\sscri\pa_{x_\sscri}-2\rho_+\pa_{\rho_+}) + 4\rho_+^{-1}B_h\Bigr] + r^{-2}\slDelta \\
            &\quad\qquad \bmod x_\sscri^2\rho_+^2(x_\sscri\CI+H_\bop^{k-2,(2\ell_\sscri,\ell_+)})\Diff_\ebop^2,
    \end{split}
    \end{align}
    \end{subequations}
    where $A_h$ and $B_h$ are given by~\eqref{EqExOpLinAB} in the splitting~\eqref{EqExOpLinSplit}. The operator $L_b$ has the same structure with $h=0$ (thus $A_h=A_0=\ubar S_{\ubar E^\Ups,\ubar E^\cC}$ in the notation of Lemma~\usref{LemmaWEOpMink} and $B_h=B_0=0$) and without the $\Hb^{k-2}$-errors.
  \item\label{ItDAdmLinK}{\rm (Near $\iota^+\cup\cK^+$.)} Fix a cutoff function $\chi_\sscri=\chi_\sscri(x_\sscri)$, equal to $1$ near $\scri^+$ and $0$ near $\cK^+$. Then
    \begin{subequations}
    \begin{align}
    \label{EqDAdmLb}
    \begin{split}
      &L_{g,g^0} - \Bigl[ L_b + \chi_\sscri \rho_\sscri\rho_+^2\bigl((A_h-A_0)(\rho_\sscri\pa_{\rho_\sscri}-\rho_+\pa_{\rho_+}) + \rho_+^{-1}B_h\bigr) \Bigr] \\
      &\qquad \in H_\bop^{k-2,(\cE^\cC_\sscri,3+\eps_\sscri),\ (\cE_++2,\,5+\eps_+),\ 2+\eps_\cK}\,\Diffb^2(\Omega_*;S^2\cT^*),
    \end{split} \\
    \label{EqDAdmLeb}
    \begin{split}
      &2 L_{g,g^0} - \Bigl[ 2 L_b + \chi_\sscri\frac12 x_\sscri^2\rho_+^2\bigl(2(A_h-A_0)(x_\sscri\pa_{x_\sscri}-2\rho_+\pa_{\rho_+}) + 4\rho_+^{-1}B_h\bigr) \Bigr] \\
      &\qquad \in x_\sscri^2\rho_+^2 H_\bop^{k-2,(2\ell_\sscri,\ell_+,\ell_\cK)}\Diff_\etbop^2(\Omega_*;S^2\cT^*).
    \end{split}
    \end{align}
    \end{subequations}
  \end{enumerate}
\end{prop}

The second summand in square brackets in~\eqref{EqDAdmLb} lies in $\Hb^{k-1,\ (\cE_\sscri^\tot,3+\eps_\sscri),\ (\cE_++2,\,5+\eps_+),\ \infty}\Diffb^1$. We separate it because it contains a leading-order term at $\scri^+$ (corresponding to the element $(1,0)\in\cE_\sscri^\tot$); we point out that it \emph{decays} towards $\iota^+$ relative to $\rho_\sscri\rho_+^2\Diffb$ by almost a full order.

\begin{rmk}[Compatibility across $\iota^+$]
\label{RmkDAdmComp}
  At this point, working relative to $g_{b_0,b,-\scal}$ is not particularly important. For example, the two metrics $g_{b_0}+h$ and $g_{b_0,b,0}=\chi_\cK g_b+(1-\chi_\cK)g_{b_0}+h$ differ near $(\iota^+)^\circ$ by $\chi_\cK(g_b-g_{b_0})\in\rho_+\CI$ (as sections of $S^2\cT^*$). Thus, the difference of $g_{b_0}$ and $g_b$ can be absorbed by a re-definition of $h$. (This is one place where the requirement $(1,0)\in\cE_+$ in Definition~\ref{DefDMetBasic} is convenient.) The same is true even if we replace $g_{b_0}$ by the pullback $\phi_{-\scal}^*g_{b_0}$ along a cut-off Lorentz boost by Lemma~\ref{LemmaKBo}. Note indeed that
  \[
    g_{b_0,b,-\scal}-g_{b_0,b,0}=(1-\chi_\cK)(\phi_{-\scal}^*g_{b_0}-g_{b_0})\in\Hb^{\infty,\ \bigl(\la(2,0)\ra_0,\infty\bigr),\ ((1,0),\infty),\ \infty}(\Omega_*)
  \]
  can be absorbed into $h$ (via a re-definition of $h$) for small $b$ and $\scal$.
\end{rmk}

\begin{proof}[Proof of Proposition~\usref{PropDAdmLin}]
  We already argued for~\eqref{EqDAdmLinbNormal}--\eqref{EqDAdmLinebNormal}. The final statement of part~\eqref{ItDAdmLinScri} is clear for $b=b_0$ (since $L_{b_0}=L_{g_{b_0},g_{b_0}}$). It holds for general $b$ just like the analogous calculations near $I^0\cap\scri^+$ (cf.\ Proposition~\ref{PropExOpLin} with $g=g^0=g_b$) do.

  Part~\eqref{ItDAdmLinScri} gives the structure~\eqref{EqDAdmLb}--\eqref{EqDAdmLeb} near $\scri^+$; in fact, we obtain~\eqref{EqDAdmLb}--\eqref{EqDAdmLeb} globally \emph{except} that the error terms a priori lie in the larger spaces
  \[
    \bigl(\rho_\sscri\rho_+^2\cdot\rho_\sscri\CI+\Hb^{k-2,(\cE_\sscri^\cC,3+\eps_\sscri),\ (\cE_++2,\,5+\eps_+)})\Diffb^2
  \]
  and $x_\sscri^2\rho_+^2(x_\sscri\CI+\Hb^{k-2,(2\ell_\sscri,\ell_+)})\Diff_\ebop^2$, respectively. Thus, only the contributions from the smooth part $g_{b_0,b,-\scal}$ of $g$ matter for the computation of the leading-order behavior (i.e., normal operator) at $\iota^+$. Away from $\cK^+$, we only need to use the even weaker information that $g_{b_0,b,-\scal}$ and $g^0$ are equal to the Minkowski metric $\ubar g$ modulo $r^{-1}\CI(\Omega_*;S^2\cT^*)$ to conclude that~\eqref{EqDAdmLb} and \eqref{EqDAdmLeb} hold with the Minkowskian operator $\ubar L$ (see~\eqref{EqWEOpMink}) in place of $L_b$ near $(\iota^+)^\circ$.

  It remains to study a neighborhood of $\cK^+$ where $g^0=g_b$ and $g=g_b+h$. We omit the weights at $\scri^+$ and the subscript ``$\eop$'' from the notation. The metric coefficients of $g$ differ from those of $g_b$ by elements of $\Hb^{k,(\cE_+,3+\eps_+),2+\eps_\cK}$, likewise for the coefficients of $g^{-1}-g_b^{-1}$ (which uses the fact that $j\cE_+\subset\cE_+$ for all $j\geq 2$ to handle nonlinear expressions, as in the proof of Lemma~\ref{LemmaExOpMet}). Therefore, the differences of Christoffel symbols (in Cartesian coordinates $t_*,x$) of $g$ and $g_b$ are of class $\Hb^{k-1,(\cE_++1,\,4+\eps_+),2+\eps_\cK}$ (since $\pa_{t_*}$, $\pa_x\in\rho_+\Vtb\subset\rho_+\Vb$ on $M$ near $\cK^+$), and the differences of curvature components are then of class $\Hb^{k-2,(\cE_++2,\,5+\eps_+),2+\eps_\cK}$. It then follows using~\eqref{Eq1LinEinGauged} that $L_{g,g^0}-L_b$ is of class $\Hb^{k-2,(\cE_++2,\,5+\eps_+),2+\eps_\cK}\Difftb^2$; this is, in fact, true for each summand of~\eqref{Eq1LinEinGauged} individually. A fortiori, this gives~\eqref{EqDAdmLeb}, and in view of $\Diff_\tbop\subset\Diff_\bop$ also~\eqref{EqDAdmLb}.
\end{proof}

\subsubsection{Forward mapping properties}
\label{SssDAdmFw}

Recall Definition~\ref{DefDMetBasic}. For $L_{g,g^0}$ defined by \eqref{Eq1EinLin} and \eqref{Eq1Ein}, we then have:

\begin{prop}[Forward mapping properties of $L_{g,g^0}$]
\label{PropDAdmFw}
  Let $\cE_\sscri^\cC$ and $\cE_+$ be as in Definition~\usref{DefDMetBasic}, and let $\ell_\sscri,\ell_+,\ell_\cK\in\R$. Let $g^0$ be as in~\eqref{EqDAdmLinBg1}--\eqref{EqDAdmLinBg2}, and let $h$ and $g$ be as in~\eqref{EqDAdmLinKMet}. Then, for $k\geq 5$,
  \[
    L_{g,g^0} \colon \Hb^{k,\ \bigl(\la\cE_\sscri^\cC\ra,\ell_\sscri\bigr),\ (\cE_+,\ell_+),\ \ell_\cK}(\Omega_*) \to \Hb^{k-2,\ \bigl(\la\cE_\sscri^\cC+1\ra',\ell_\sscri+1\bigr),\ (\cE_++2,\ell_++2),\ \ell_\cK}(\Omega_*).
  \]
  Furthermore, for all $\eps>0$,
  \begin{equation}
  \label{EqDAdmFw}
  \begin{split}
    L_{g,g^0}-L_b \colon &\Hb^{k,\ \bigl(\la\cE_\sscri^\cC\ra,\ell_\sscri\bigr),\ (\cE_+,\ell_+),\ \ell_\cK}(\Omega_*) \\
      &\quad \to \Hb^{k-2,\ \bigl(\la\cE_\sscri^\cC+1\ra',\ell_\sscri+1\bigr),\ (2\cE_++2,\,\ell_++3-\eps),\ \ell_\cK+2+\eps_\cK}(\Omega_*).
  \end{split}
  \end{equation}
\end{prop}
\begin{proof}
  The first part is completely analogous to Proposition~\ref{PropExFwL} (with the same proof, \emph{mutatis mutandis}). In the second part, only the improved asymptotics and decay at $\iota^+$ and $\cK^+$ require an argument; but these follow from~\eqref{EqDAdmLb} and the fact that the $\iota^+$-index set and decay rate of $A_h-A_0$ and $\rho_+^{-1}B_h$ are $\cE_+$ and $3+\eps_+>\min\Re\cE_+=1$.
\end{proof}

We next prove an analogue of Corollary~\ref{CorExFwN}.

\begin{cor}[Forward mapping properties of $P$]
\label{CorDAdmFwN}
  Let $g^0=g_{b_0,b,-\scal}$ where $|b-b_0|$ and $\scal\in\scalspace_1$ are small. Let $g=g^0+h$ where $h\in\Hb^{k,\ \bigl(\la\cE_\sscri^\cC\ra,\,3+\eps_\sscri\bigr),\ (\cE_+,\,3+\eps_+),\ 2+\eps_\cK}(\Omega_*)$ has small $\Hb^{3,(1,0,0)}$-norm. Then
  \begin{equation}
  \label{EqDAdmFwN}
    P_0(h,\scal,b) = P(g,g^0) = \Ric(g) - \delta_{g^0,E^\cC}^*\Ups_{E^\Ups}(g,g^0) \in \Hb^{\infty,\ \bigl(\la\cE_\sscri^\cC+1\ra',4+\eps_\sscri\bigr),\ (\cE_++2,\,5+\eps_+),\ 2+\eps_\cK}(\Omega_*).
  \end{equation}
\end{cor}
\begin{proof}
  Near $\scri^+$, where $\Ric(g^0)=0$, the same arguments as in the proof of Corollary~\ref{CorExFwN} (with $b,\bhm$ equal to $b_0,\bhm_0$ in present notation) apply. The only additional additive term of $P(g,g^0)$ not accounted for in~\eqref{EqExFwNMem} is
  \[
    P(g^0,g^0) = \Ric(g_{b_0,b,-\scal}),
  \]
  which is supported on $\supp\dd\chi_\cK$. Since $g_{b_0,b,-\scal}-\ubar g$ is a smooth section of $S^2\cT^*$ and of class $\rho_+\CI$ near $(\iota^+)^\circ$, we have $P(g^0,g^0)\in\rho_+^3\CI$ there. Since $(3,0)\subset\cE_++2$ by~\eqref{EqDMetBasicEplus}, this term lies in the space in~\eqref{EqDAdmFwN}.
\end{proof}

\subsubsection{Admissibility}
\label{SssDAdm}

The structure recorded in Proposition~\ref{PropDAdmLin} has important consequences for the forward inverse of $L_{g,g^0}$. Recall from Theorem~\ref{ThmAdm} that for fixed $\alpha_\sface\in(-\frac32,-\frac32+\eps_\ind)$, the operator $L_b$ is $2$-admissible with $\sface$-loss $\delta$, where $\delta>\alpha_\sface+\frac32\in(0,\eps_\ind)$ can be taken to be $<\eps_\ind$ (and thus $<1$). We then first record:

\begin{cor}[Admissibility]
\label{CorDAdm}
  Let $g^0=g_{b_0,b,-\scal}$ (or more generally as in~\eqref{EqDAdmLinBg1}--\eqref{EqDAdmLinBg2}). Then for all 2-tensors $h\in\cX^k:=\Hb^{k,\bigl(\la\cE_\sscri^\cC\ra,3+\eps_\sscri\bigr),\ (\cE_+,\,3+\eps_+),\ 2+\eps_\cK}(\Omega_*)$ with sufficiently small norm in $\cX^3$, the operator $L_{g,g^0}$, where $g=g_{b_0,b}+h$ as in~\eqref{EqDAdmLinKMet}, is \emph{admissible (relative to $L_b$) of class $((0;k-5),(2\ell_\sscri,\ell_+,\ell_\cK))$} in the sense of \citeAF{Definition~\ref*{DefSDWAdm}} for all $\ell_\sscri<\min(2\gamma^\Ups,\frac12)$, $\ell_+\in(\delta,1)$, and $\ell_\cK\in(2,2+\eps_\cK]$.
\end{cor}

\begin{rmk}[Regularity of $h$]
\label{RmkDAdmRegh}
  The requirements on $h$ in Corollary~\ref{CorDAdm} are much stronger than what is needed for the mere proof of admissibility; it would suffice that $h\in\Hb^{k,(\la\cE_\sscri^\cC\ra,\ell_\sscri),\ell_+,\ell_\cK}$ where $\ell_\sscri<\min(2\gamma^\Ups,\frac12)$, $0<\ell_+<1$, and $\ell_\cK>2$.
\end{rmk}

\begin{proof}[Proof of Corollary~\usref{CorDAdm}]
  By Sobolev embedding, the space on the right in~\eqref{EqDAdmLeb} is contained in the space $x_\sscri^2\rho_+^2\cC_\bop^{k-5,(2\ell_\sscri,\ell_+,\ell_\cK)}\Diff_\etbop^2$, which is the same as \citeAF{(\ref*{EqSDWAdmOp})} (with $(d_0;k)$ there replaced by $(0;k-5)$ in present notation), with the identification $\tilde p_1=(A_h-A_0)|_{\scri^+}$ and $p_0=4\rho_+^{-1}B_h|_{\scri^+}$; these lie in $\rho_+^{\ell_+}\cC_\bop^{k-4}(\sscri^+;\End(S^2\cT^*))$, as follows from their explicit expressions in~\eqref{EqExOpLinAB} and the fact that $\pa_1\in\rho_+\Vb$ near $\scri^+$ (cf.\ also the final statement of Proposition~\ref{PropExOpLin}). Note also that $p_1$ in the notation of \citeAF{Definition~\ref*{DefSDWAdm}(\ref*{ItSDWAdmSplit})} is equal to $A_0+(A_h-A_0)=A_h$, and thus the required bundle splitting in which $p_1$ is lower triangular and $p_0$ is strictly lower triangular is given by~\eqref{EqExOpLinSplit}. The weights $\ell_+\in(\delta,1)$ and $\ell_\cK>2$ satisfy the conditions in~\citeAF{(\ref*{EqSDWAdmWeights})}.
\end{proof}

The key consequence of Corollary~\ref{CorDAdm} is that the main result of \cite{HintzNonstat2} becomes applicable: forward solutions of $L_{g,g^0}u=f$ can be controlled with full b-regularity on $M$ and matching b-tame estimates. We state \citeAF{Theorem~\ref*{ThmF}}, in the form using unweighted b-densities as in \citeAF{(\ref*{EqFGammas})--(\ref*{EqFTame20})}; this is the starting point for the subsequent analysis in which we will prove stronger decay and asymptotics for forcing terms with more decay.

\begin{thm}[Regularity of forward solutions]
\label{ThmDAdmReg}
  Let $\beta_+\in(\frac12,\frac12+\eps_\ind)$ and $\beta_\sscri\in(\beta_+,1)$; fix $\delta>\beta_+-\frac12$. There exist $\eps_0,d\in\N$ such that the following holds. For all $k\in\N_0$ and
  \begin{enumerate}
  \item Kerr parameters $b$ with $|b-b_0|<\eps_0$,
  \item the background metric $g^0:=g_{b_0,b,-\scal}$ (see~\eqref{EqDgPOU}) for $|\scal|<\eps_0$,
  \item $h\in\cX^{k+d}:=\Hb^{k+d,\ \bigl(\la\cE_\sscri^\cC\ra,\,3+\eps_\sscri\bigr),\ (\cE_+,\,3+\eps_+),\ 2+\eps_\cK}(\Omega_*)$ with $\cX^d$-norm less than $\eps_0$,
  \end{enumerate}
  and for all source terms
  \begin{equation}
  \label{EqDAdmRegf}
    f\in\Hb^{k+d,\ \beta_\sscri+1,\ \beta_++2,\ \frac12}(\Omega_*;S^2\cT^*)^{\bullet,-},
  \end{equation}
  the unique forward solution $u$ of $L_{g,g^0}u=f$ (where $g:=g_{b_0,b,-\scal}+h$, and $L_{g,g^0}$ is given by~\eqref{Eq1EinLin}) satisfies
  \begin{equation}
  \label{EqDAdmRegu}
    u\in\Hb^{k,\ \beta_\sscri,\ \beta_+-\delta,\ -\frac32}(\Omega_*;S^2\cT^*)^{\bullet,-}
  \end{equation}
  together with the tame estimate
  \begin{equation}
  \label{EqDAdmRegTame}
    \|u\|_{\Hb^{k,\ \beta_\sscri,\ \beta_+-\delta,\ -2}} \leq C_k\Bigl( \|f\|_{\Hb^{k+d,\ \beta_\sscri+1,\beta_++2,0}} + \|h\|_{\cX^{k+d}} \|f\|_{\Hb^{d,\ \beta_\sscri+1,\ \beta_++2,\ 0}} \Bigr).
  \end{equation}
  Moreover, for any fixed $\eps>0$, the solution $u$ depends continuously on $b_0,b,\scal,h$ in the norm topology of $\Hb^{k-1,\ \beta_\sscri-\eps,\ \beta_+-\delta-\eps,\ -\frac32-\eps}(\Omega_*;S^2\cT^*)^{\bullet,-}$.
\end{thm}

Since the non-smooth contributions to the coefficients of $L_{g,g^0}$ come solely from $h$, the norm of $h$ in~\eqref{EqDAdmRegTame} is what takes the place of the norm on $P-P_0$ in the notation of \citeAF{(\ref*{EqFTame})} (up to increasing $d$ further by a constant amount).

\begin{rmk}[Tame estimates]
\label{RmkDAdmRegTame}
  Starting with the tame estimate~\eqref{EqDAdmRegTame} and re-writing
  \begin{equation}
  \label{EqDAdmRegTameRewrite}
    L_b u=f-(L_{g,g^0}-L_b)u,
  \end{equation}
  the additional source term $(L_{g,g^0}-L_b)u$ (which is controlled by~\eqref{EqDAdmFw}) satisfies tame estimates in weighted b-Sobolev spaces in terms of the norms of $u$ and $h$ separately and linearly as well as products of the norm of $u$ and a norm of $h$, with only one of the two norms being a high-regularity norm (cf.\ \citeAF{Lemma~\ref*{LemmaMTameMult}}). Since $L_b$ depends only on the finite-dimensional parameter $b$, the inversion of $L_b$ will preserve tame estimates. The upshot is that the tameness of the initial tame estimate~\eqref{EqDAdmRegTame} will propagate throughout all of our arguments; and as long as one uses the re-writing~\eqref{EqDAdmRegTameRewrite} only a fixed finite number of times $N$, the total b-regularity loss $N d$ is independent of the initial b-regularity order $k$ (and one starts with $k\geq (N+1)d$ then, say)---as required for an eventual application of a Nash--Moser iteration scheme (see \citeAF{\S\ref*{SsA1NM}}). Consequently, we shall omit explicit statements of tame estimates at intermediate steps of our argument, and work with infinite degrees of b-regularity for notational simplicity. (See Corollary~\ref{CorD6Impr} for the final, tame, estimate of the present section.)
\end{rmk}

\subsection{Asymptotics and formal solutions at \texorpdfstring{$\scri^+$}{null infinity}}
\label{SsDScri}

\emph{For the remainder of the present subsection, we work in the neighborhood $[0,c_\sscri]_{\rho_\sscri}\times[0,1]_{\rho_+}\times\Sph^2$ of $\scri^+\cap\{t_*\geq 0\}$ in the coordinates~\eqref{EqKMfdCoordScriip}; here $c_\sscri>0$ is some small positive number. We do not record any $\cK^+$-decay orders.} Let
\[
  L =: L_{g,g^0},
\]
where $g:=g^0+h$ and $g^0:=g_{b_0,b,-\scal}$ are as in Theorem~\ref{ThmDAdmReg}. We first prove the partial polyhomogeneity of solutions $u$ of $L u=f$ at $\scri^+$, without a loss in whatever a priori control at $\iota^+$ one assumes from the outset (e.g., $(\beta_+-\delta)$-decay as in~\eqref{EqDAdmRegu}):

\begin{prop}[Partial polyhomogeneity at $\scri^+$]
\label{PropDScriPhg}
  Let $\cE_\sscri^\cC,\cE_+$ be index sets as in Definition~\usref{DefDMetBasic}. Let $\ell_+<\ell_\sscri\leq 3+\eps_\sscri$ and $\ell_+\notin\Re\pi_1\cE_+$. Suppose $u$ and $f$ vanish for $\ft_*\leq 1$, and
  \begin{equation}
  \label{EqDScriPhgEq}
    L u = f \in \Hb^{\infty,\ \bigl(\la\cE_\sscri^\cC+1\ra',\ell_\sscri+1\bigr),\ (\cE_++2,\ell_++2)},\quad
    u\in\Hb^{\infty,\ \alpha_\sscri,\ (\cE_+,\ell_+)}\ \text{for some}\ \alpha_\sscri\in\R.
  \end{equation}
  Then, in fact,
  \[
    u\in\Hb^{\infty,\ \bigl(\la\cE_\sscri^\cC\ra,\ell_\sscri\bigr),\ (\cE_+,\ell_+)}.
  \]
\end{prop}

The key structural information is given by~\eqref{EqDAdmLinbNormal}, which we state in the form
\begin{equation}
\label{EqDScriOp}
\begin{split}
  &\rho_\sscri^{-1}\rho_+^{-2}L = L^0 + \tilde L, \\
  &\qquad L^0 = -(\rho_\sscri\pa_{\rho_\sscri}-(I+A_h))(\rho_\sscri\pa_{\rho_\sscri}-\rho_+\pa_{\rho_+}) + \rho_+^{-1}B_h, \\
  &\qquad \tilde L \in (\rho_\sscri\CI + \Hb^{k-2,(\cE_\sscri^\cC-1,\,2+\eps_\sscri),\ (\cE_+,\,3+\eps_+)})\Diffb^2.
\end{split}
\end{equation}

The proof proceeds in two steps. First we obtain the desired control at $\scri^+$ by integrating $L^0$ in the future direction, albeit with a loss of information at $\iota^+$:

\begin{lemma}[Precise control at $\scri^+$]
\label{LemmaDScriPhg1}
  Under the assumptions of Proposition~\usref{PropDScriPhg}, we have
  \[
    u \in \Hb^{\infty,\ \bigl(\la\cE_\sscri^\cC\ra,\ell_\sscri\bigr),\ \alpha_+}\quad\forall\,\alpha_+<\min\bigl(\alpha_\sscri,\ 1,\ \ell_+\bigr).
  \]
\end{lemma}
\begin{proof}
  Without loss, we may reduce $\alpha_\sscri$ in~\eqref{EqDScriPhgEq}, if necessary, so as to arrange $\alpha_\sscri<1$. Thus,
  \begin{equation}
  \label{EqDScriPhg1u}
    u\in\Hb^{\infty,\alpha_\sscri,\alpha_+}.
  \end{equation}

  \pfstep{Step~1. Sharper weight at $\scri^+$.} Given~\eqref{EqDScriPhg1u}, we have $\tilde L u\in\Hb^{\infty,\alpha_\sscri+\eta,\alpha_+}$, where we pick $\eta\in(0,2\gamma^\Ups]$ (which thus satisfies $\eta<\min\Re\cE_\sscri^\cC-1\leq 1$) such that $\alpha_\sscri+\eta<1$. Consider then the first component of
  \[
    L^0 u = \rho_\sscri^{-1}\rho_+^{-2}f - \tilde L u \in \Hb^{\infty,\alpha_\sscri+\eta,\alpha_+}
  \]
  in the splitting~\eqref{EqExOpLinSplit}. (Note that $\rho_\sscri^{-1}\rho_+^{-2}f\in\Hb^{\infty,\ 1-\eps,\ \alpha_+}$ for all $\eps>0$.) Integrating $\rho_\sscri\pa_{\rho_\sscri}-2(1-v^\cC)\gamma^\cC$ yields $(\rho_\sscri\pa_{\rho_\sscri}-\rho_+\pa_{\rho_+})u_{0 0}\in\Hb^{\infty,\alpha_\sscri+\eta,\alpha_+}$ by Lemma~\ref{LemmaTMIntFuchs}. Next, we use Lemma~\ref{LemmaTIntHyp}, the vanishing of $u$ in $\rho_+\geq 1$, and the fact that $\alpha_+<\alpha_\sscri+\eta$ to conclude that $u_{0 0}\in\Hb^{\infty,\alpha_\sscri+\eta,\alpha_+}$, which is an $\eta$-improvement at $\scri^+$ over~\eqref{EqDScriPhg1u}. Arguing similarly for the second component gives $u_{0 /}\in\Hb^{\infty,\alpha_\sscri+\eta,\alpha_+}$. For the third component $\sltr u$ of $u$, we put the contribution from $u_{0 0}$ (via $(A_h)_{3,1}$) on the right-hand side: this yields the equation
  \[
    \bigl(\rho_\sscri\pa_{\rho_\sscri} - (1+(1-e^\cC)(1-v^\cC)\gamma^\cC)\bigr)(\rho_\sscri\pa_{\rho_\sscri}-\rho_+\pa_{\rho_+})\sltr u \in \Hb^{\infty,\alpha_\sscri+\eta,\alpha_+},
  \]
  and thus also $\sltr u\in\Hb^{\infty,\alpha_\sscri+\eta,\alpha_+}$. Since the components of $A_h$ and $B_h$ lie in the space $\CI+\Hb^{\infty,(\cE_\sscri^\tot-1,2+\eps_\sscri),(\cE_+,3+\eps_+)}$ (and are thus, in particular, bounded conormal), completely analogous arguments apply for all subsequent components. (Recall here that all eigenvalues of $A_h$ are $\geq 1$, and thus no expansion terms arise at $\scri^+$ at this point since we are assuming $\alpha_\sscri+\eta<1$; and the assumptions $\alpha_+<\alpha_\sscri$ and $\alpha_+<1$ imply that Lemma~\ref{LemmaTIntHyp} is applicable indeed.) This gives~\eqref{EqDScriPhg1u} with $\alpha_\sscri+\eta$ in place of $\alpha_\sscri$.

  Iterating this argument, we obtain (for the original value of $\alpha_+$)
  \begin{equation}
  \label{EqDScriPhg1u2}
    u \in \bigcap_{\alpha_\sscri<1} \Hb^{\infty,\alpha_\sscri,\alpha_+}.
  \end{equation}

  \pfstep{Step~2. First expansion terms at $\scri^+$.} Given~\eqref{EqDScriPhg1u2}, we now have $\tilde L u\in\Hb^{\infty,1+\eta,\alpha_+}$ for all $\eta<2\gamma^\Ups$; we shall for now use this only for $\eta<(1-e^\Ups)\gamma^\Ups$ (the smallest nonzero eigenvalue of $A_h$). Integrating the first component of $L^0 u\in\Hb^{\infty,1+\eta,\alpha_+}$ gives $u_{0 0}\in\Hb^{\infty,1+\eta,\alpha_+}$, and similarly $u_{0 /}$, $\sltr u$, $u_{0 1}$, $u_{1 /}\in\Hb^{\infty,1+\eta,\alpha_+}$. (For the latter two, this uses $\eta<(1-e^\Ups)\gamma^\Ups$ and $\eta<\gamma^\Ups$, respectively.) The equation for $\slpi_0 u$ reads
  \[
    (\rho_\sscri\pa_{\rho_\sscri}-1)(\rho_\sscri\pa_{\rho_\sscri}-\rho_+\pa_{\rho_+})\slpi_0 u \in \Hb^{\infty,1+\eta,\alpha_+},
  \]
  and by Lemmas~\ref{LemmaTMIntFuchs} and \ref{LemmaTIntHyp}, its solution satisfies $\slpi_0 u\in\Hb^{\infty,((1,0),1+\eta),\alpha_+}$. Finally, in the equation for $u_{1 1}$, we put all off-diagonal contributions of $A_h$ and $\rho_+^{-1}B_h$ (in~\eqref{EqExOpLinAB}) on the right-hand side; those contributions arising from components other than $(A_h)_{7,6}$ lie in $\Hb^{\infty,1+\eta,\alpha_+}$, while the contribution from $(A_h)_{7,6}$ acting on $\slpi_0 u$ lies in $\Hb^{\infty,((1,0),1+\eta),\alpha_+}$. The equation for $u_{1 1}$ thus reads
  \begin{equation}
  \label{EqDScriPhg1u11Eq}
    \bigl(\rho_\sscri\pa_{\rho_\sscri}-(1+2\gamma^\Ups)\bigr)(\rho_\sscri\pa_{\rho_\sscri}-\rho_+\pa_{\rho_+})u_{1 1} \in \Hb^{\infty,((1,0),1+\eta),\alpha_+}.
  \end{equation}
  Its source term captures also the $(\dd x^1)^2$-component of $\rho_\sscri^{-1}\rho_+^{-2}f$. Again by Lemmas~\ref{LemmaTMIntFuchs} (and using $1+\eta<1+2\gamma^\Ups$) and \ref{LemmaTIntHyp}, this gives $u_{1 1}\in\Hb^{\infty,((1,0),1+\eta),\alpha_+}$. We have now shown that
  \begin{equation}
  \label{EqDScriPhg1u3}
    u \in \Hb^{\infty,\ (\la\cE_\sscri^\cC\ra,\alpha_\sscri),\ \alpha_+}\quad \forall\,\alpha_\sscri<1+(1-e^\Ups)\gamma^\Ups.
  \end{equation}

  \pfstep{Step~3. Full expansion.} Starting with~\eqref{EqDScriPhg1u3} for some value of $\alpha_\sscri$, we will prove the same membership for the value $\min(\ell_\sscri,\alpha_\sscri+2\gamma^\Ups)$. Since $u\in\Hb^{\infty,(\cE_\sscri^\tot,\alpha_\sscri),\alpha_+}$, we have $\tilde L u\in\Hb^{\infty,(\cE_\sscri^\cC,\alpha_\sscri+\eta),\alpha_+}$ where $\eta<\min\Re(\cE_\sscri^\cC-1)$; this uses that $(\cE_\sscri^\cC-1)+\cE_\sscri^\tot\subset\cE_\sscri^\cC$ (see Definition~\ref{DefExP}). Integration of the first three components of
  \[
    L^0 u \in \Hb^{\infty,\ \bigl(\cE_\sscri^\cC,\min(\ell_\sscri,\alpha_\sscri+\eta)\bigr),\ \alpha_+}
  \]
  (the weight $\ell_\sscri$ coming from $f$ in~\eqref{EqDScriPhgEq}). Successive integration of the first three components of this equation, and using $\ell_\sscri\leq 3+\eps_\sscri<1+(1-e^\cC)(1-v^\cC)\gamma^\cC$, yields the memberships
  \begin{equation}
  \label{EqDScriPhg1u4}
    u_{0 0},\ u_{0 /}, \sltr u \in \Hb^{\infty,\ \bigl(\cE_\sscri^\cC,\min(\ell_\sscri,\alpha_\sscri+\eta)\bigr),\ \alpha_+}.
  \end{equation}
  For the next three components, we note that the off-diagonal terms of $A_h$ and $\rho_+^{-1}B_h$ couple to~\eqref{EqDScriPhg1u4} via b-differential operators with coefficients of class $\CI+\Hb^{\infty,(\cE_\sscri^\tot-1,2+\eps_\sscri),(\cE_+,3+\eps_+)}$; moving these terms on the right-hand side thus shows that the equations for the fourth, fifth, and sixth component of $u$ read
  \[
    \bigl(\rho_\sscri\pa_{\rho_\sscri}-(1+\lambda)\bigr)(\rho_\sscri\pa_{\rho_\sscri}-\rho_+\pa_{\rho_+})v \in \Hb^{\infty,\ \bigl(\cE_\sscri^\cC,\min(\ell_\sscri,\alpha_\sscri+\eta)\bigr),\ \alpha_+},
  \]
  where $(v,\lambda)=(u_{0 1},(1-e^\Ups)\gamma^\Ups)$, $(\pi_{1 /}u,\gamma^\Ups)$, $(\slpi_0 u,0)$. (We use here that $\min\Re\cE_\sscri^\cC>1$ and thus $(2+\eps_\sscri)+\min\Re\cE_\sscri^\cC>\ell_\sscri$.) Integration yields, in the notation of~\eqref{EqExPOther},
  \begin{align*}
    u_{0 1} &\in \Hb^{\infty,\ \bigl(\cE_{\sscri,0 1},\min(\ell_\sscri,\alpha_\sscri+\eta)\bigr),\ \alpha_+}, \\
    u_{1 /} &\in \Hb^{\infty,\ \bigl(\cE_{\sscri,1 /},\min(\ell_\sscri,\alpha_\sscri+\eta)\bigr),\ \alpha_+}, \\
    \slpi_0 u &\in \Hb^{\infty,\ \bigl(\slcE_{\sscri,0},\min(\ell_\sscri,\alpha_\sscri+\eta)\bigr),\ \alpha_+}.
  \end{align*}
  In the equation for the final component $u_{1 1}$ of $u$, there is a term coupling $(A_h)_{7,6}=-\frac12\pa_1(r h^{\bar a\bar b})\in\Hb^{\infty,(\slcE_{\sscri,0}-1,2+\eps_\sscri),(\cE_+,3+\eps_+)}$ with $\slpi_0 u$, which produces a term in
  \[
    \Hb^{\infty,\ \bigl(2\slcE_{\sscri,0}-1,\,\min(\ell_\sscri,\alpha_\sscri+\eta)\bigr),\ \alpha_+} = \Hb^{\infty,\ \bigl((1,0)\cup\cE_\sscri^\cC,\min(\ell_\sscri,\alpha_\sscri+\eta)\bigr),\ \alpha_+}.
  \]
  The coupling of $(A_h)_{7,4}$ with $u_{0 1}$ moreover contributes the index set $(1+(1-e^\Ups)\gamma^\Ups,0)$ at $\scri^+$, so that altogether
  \[
    \bigl(\rho_\sscri\pa_{\rho_\sscri}-(1+2\gamma^\Ups)\bigr)(\rho_\sscri\pa_{\rho_\sscri}-\rho_+\pa_{\rho_+})u_{1 1} \in \Hb^{\infty,\ \bigl((1,0)\cup(1+(1-e^\Ups)\gamma^\Ups,0)\cup\cE_\sscri^\cC,\,\min(\ell_\sscri,\alpha_\sscri+\eta)\bigr),\ \alpha_+}.
  \]
  (This is similar to~\eqref{EqExPhgh11}.) Integrating this gives $u_{1 1}\in\Hb^{\infty,\ \bigl(\cE_{\sscri,1 1},\,\min(\ell_\sscri,\alpha_\sscri+\eta)\bigr),\ \alpha_+}$.

  We have now improved~\eqref{EqDScriPhg1u3} to
  \[
    u \in \Hb^{\infty,\ \bigl(\la\cE_\sscri^\cC\ra,\min(\alpha_\sscri+\eta,\ell_\sscri)\bigr),\ \alpha_+}
  \]
  for all $\eta<\min\Re(\cE_\sscri^\cC-1)$. Iterating this finitely many times proves the Lemma.
\end{proof}

\begin{proof}[Proof of Proposition~\usref{PropDScriPhg}]
  \pfstep{Step~1. Precise asymptotics at $\iota^+$, weaker remainder at $\scri^+$.} In order to recover the behavior of $u$ at $\iota^+$ assumed in~\eqref{EqDScriPhgEq}, we now integrate the hyperbolic vector field $\rho_\sscri\pa_{\rho_\sscri}-\rho_+\pa_{\rho_+}$ \emph{backwards} from $\rho_\sscri=c_\sscri>0$. While Lemma~\ref{LemmaTIntHyp} would seem to produce $\scri^+$-index sets for $u$ involving extended unions with $\cE_+$, Lemma~\ref{LemmaDScriPhg1} shows that the $\scri^+$-index set of $u$ cannot, in fact, contain any elements beyond the ones encoded in the membership in $\Hb^{\infty,(\la\cE_\sscri^\cC\ra,\ell_\sscri),[\cdots]}$. So suppose that for some $\alpha_\sscri<\ell_+$, we have already proved
  \begin{equation}
  \label{EqDScriPhgu}
    u \in \Hb^{\infty,\ (\la\cE_\sscri^\cC\ra,\alpha_\sscri),\ (\cE_+,\ell_+)}
  \end{equation}
  when $\alpha_\sscri>1$, or $u\in\Hb^{\infty,\alpha_\sscri,(\cE_+,\ell_+)}$ when $\alpha_\sscri<1$. Writing the equation for $u$ again as
  \[
    L^0 u = \rho_\sscri^{-1}\rho_+^{-2}f-\tilde L u,
  \]
  note that $\tilde L u\in\Hb^{\infty,(\cF,\alpha_\sscri+\eta),(\cE_+,\ell_+)}$ for some index set $\cF$ and all $\eta<2\gamma^\Ups$; the point, as usual, is the faster decay order of the remainder term at $\scri^+$ compared to that of $u$ itself. We can then integrate this equation component by component as in the proof of Lemma~\ref{LemmaDScriPhg1}, except in the application of Lemma~\ref{LemmaTIntHyp} we integrate from $\rho_\sscri=c_\sscri$ (instead of from $\rho_+=1$); this necessitates the relationship $\alpha_\sscri+\eta<\ell_+$ of weights, so upon decreasing $\eta$ if necessary to ensure this relationship, we conclude that~\eqref{EqDScriPhgu} holds with $\alpha_\sscri+\eta$ in place of $\alpha_\sscri$.

  \pfstep{Step~2. Precise asymptotics at $\iota^+$ and $\scri^+$.} Having now obtained
  \[
    u\in\Hb^{\infty,\ (\la\cE_\sscri^\cC\ra,\alpha_\sscri),\ (\cE_+,\ell_+)}
  \]
  for all $\alpha_\sscri<\ell_+$, we again integrate in the forward direction. We proceed as in the proof of Lemma~\ref{LemmaDScriPhg1}, thus increasing $\alpha_\sscri$ by a definite amount $\eta>0$ at each step. We comment on two aspects of the application of Lemma~\ref{LemmaTIntHyp} in each step:
  \begin{enumerate}
  \item The source term for the integration of $\rho_\sscri\pa_{\rho_\sscri}-\rho_+\pa_{\rho_+}$ (when improving the $\scri^+$-remainder weight of components of $u$) has remainder weight $\alpha_\sscri+\eta>\ell_+$, and hence the result of forward integration is indeed controlled by Lemma~\ref{LemmaTIntHyp}.
  \item Lemma~\ref{LemmaTIntHyp} produces, a priori, $\iota^+$-index sets for $u$ which are extended unions of $\cE_+$ and index sets of components of $u$; but in view of the a priori assumption that $u$ has index set $\cE_+$ at $\iota^+$, exponents other than those already contained in $\cE_+$ cannot, in fact, occur.
  \end{enumerate}
  After finitely many steps, we thus conclude that $u\in\Hb^{\infty,(\la\cE_\sscri^\cC\ra,\ell_\sscri),(\cE_+,\ell_+)}$, as desired.
\end{proof}

We next construct formal solutions to the equation $L_b u=f$ at $\scri^+$. (As motivated after~\eqref{EqDRewrite}, the Fourier transform in $t_*$ becomes effective afterwards for solving away the remaining source term.) \emph{We continue to work in $[0,c_\sscri]_{\rho_\sscri}\times[0,1]_{\rho_+}\times\Sph^2$ and omit weights at $\cK^+$ from the notation.}

\begin{prop}[Formal solutions of $L_b$ at $\scri^+$]
\label{PropDScriFormal}
  Let $\cE_\sscri^\cC$ be as in Definition~\usref{DefDMetBasic}, and let $\breve\cE_+\subset\C\times\N_0$ be an arbitrary index set. Let $\breve\ell_+<\ell_\sscri\leq 3+\eps_\sscri$ with $\breve\ell_+,\ell_\sscri\notin\Re\pi_1\cE_\sscri^\cC$ and $\breve\ell_+,\ell_\sscri\neq 1,1+(1-e^\Ups)\gamma^\Ups$. Let
  \[
    f \in \Hb^{\infty,\ \bigl(\la\cE_\sscri^\cC+1\ra',\ell_\sscri+1\bigr),\ (\breve\cE_++2,\breve\ell_++2)}
  \]
  be given, with $f=0$ for $t_*<1$. Then there exist an index set $\cE_+^\sharp$ depending only on $\cE_\sscri^\cC$ and $\breve\cE_+$, and a symmetric 2-tensor
  \[
    u \in \Hb^{\infty,\ \bigl(\la\cE_\sscri^\cC\ra,\ell_\sscri\bigr),\ (\cE_+^\sharp,\breve\ell_+)}
  \]
  vanishing on $\{t_*<1\}\cup\{\rho_\sscri>\frac12 c_\sscri\}$ such that
  \[
    f - L_b u \in \Hb^{\infty,\ \ell_\sscri+1,\ (\cE_+^\sharp+2,\breve\ell_++2)}(\Omega_*;S^2\cT^*).
  \]
  Concretely, recalling Definition~\usref{DefTMIndex}, put
  \begin{subequations}
  \begin{align}
  \label{EqDScriFormalInd1}
    \cE_+^{\sharp,(1)} &:= \breve\cE_+\extcup\{(1,0)\}\extcup\{(1+(1-e^\Ups)\gamma^\Ups,0)\}, \\
  \label{EqDScriFormalInd2}
    \cE_+^{\sharp,(j+1)} &:= \cE_+^{\sharp,(j)} \extcup \{(z_1^{(j)},k_1^{(j)}\} \extcup \cdots \extcup \{(z_N^{(j)},k_N^{(j)})\},\quad j\in\N,
  \end{align}
  \end{subequations}
  where the $z_i^{(j)}$ are pairwise distinct with
  \[
    \{z_1^{(j)},\ldots,z_N^{(j)}\} = \{ z\in\pi_1\cE_\sscri^\cC \colon j<\Re z\leq j+1 \},\quad
    k_i^{(j)} := k(\cE_\sscri^\cC,z_i^{(j)});
  \]
  then one can take $\cE_+^\sharp=\bigcup_{j=1}^\infty \cE_+^{\sharp,(j)}$. If, for some $C_0\in\R$, all $\rho_\sscri^{z+1}(\log\rho_\sscri)^k$-terms in the $\scri^+$-expansion of $f$ vanish for $\Re z\leq C_0$, then in these definitions of $\cE_+^{\sharp,(j)}$ for $j\in\N$ one can omit all of these $\{(z,k)\}$ in~\eqref{EqDScriFormalInd1}--\eqref{EqDScriFormalInd2}.
\end{prop}
\begin{proof}
   The key structure is a smooth coefficient version of~\eqref{EqDScriOp}, namely
  \begin{equation}
  \label{EqDScriFormalL}
  \begin{split}
    &\rho_\sscri^{-1}\rho_+^{-2}L_b = L^0 + \tilde L, \\
    &\qquad L^0=-(\rho_\sscri\pa_{\rho_\sscri}-(I+A_0))(\rho_\sscri\pa_{\rho_\sscri}-\rho_+\pa_{\rho_+}),\quad
    \tilde L\in\rho_\sscri\Diffb^2.
  \end{split}
  \end{equation}

  Let $\chi\in\CIc([0,\frac12 c_\sscri))$ be equal to $1$ near $0$; we write $\chi=\chi(\rho_\sscri)$. We can then write
  \begin{align*}
    &f = \chi\bigl(\rho_\sscri^2 f_{1 1}^{(2,0)} + \rho_\sscri^{2+(1-e^\Ups)\gamma^\Ups}f_{1 1}^{(2+(1-e^\Ups)\gamma^\Ups,0)}\bigr)\,(\dd x^1)^2 + \tilde f, \\
    &\qquad f_{1 1}^{(2,0)}\in\Hb^{\infty,(\breve\cE_++2,\breve\ell_++2)}(\Omega_*\cap\scri^+),\quad
    \tilde f\in\Hb^{\infty,\ (\cE_\sscri^\cC+1,\ell_\sscri+1),\ (\breve\cE_++2,\breve\ell_++2)}.
  \end{align*}

  \pfstep{Step~1. Solving away $f_{1 1}^{(2,0)}$.} For the $\rho_\sscri^1$-term of $u$, we make the ansatz $\chi\rho_\sscri u_{1 1}^{(1,0)}\,(\dd x^1)^2$. Recalling from~\eqref{EqExOpLinAB} the structure of $A_0$, the coefficient $u_{1 1}^{(1,0)}=u_{1 1}^{(1,0)}(\rho_+,\omega)$ then needs to solve
  \[
    -(1-(1+2\gamma^\Ups))(1-\rho_+\pa_{\rho_+})u_{1 1}^{(1,0)} = \rho_+^{-2}f_{1 1}^{(2,0)},
  \]
  i.e., $(\rho_+\pa_{\rho_+}-1)u_{1 1}^{(1,0)}=-\frac{1}{2\gamma^\Ups}\rho_+^{-2}f_{1 1}^{(2,0)}$. We integrate this with initial condition $0$ at $\rho_+=1$; by Lemma~\ref{LemmaTMIntFuchs2}, the solution satisfies
  \[
    u_{1 1}^{(1,0)} \in \Hb^{\infty,\ \bigl(\breve\cE_+\extcup\{(1,0)\},\breve\ell_+\bigr)}(\Omega_*\cap\scri^+).
  \]
  Upon replacing $f$ by $f-L_b(\chi\rho_\sscri u_{1 1}^{(1,0)})$, we have now reduced to the case $f_{1 1}^{(2,0)}=0$ (while the term $f_{1 1}^{(2+(1-e^\Ups)\gamma^\Ups,0)}$ is unchanged).

  \pfstep{Step~2. Solving away $f_{1 1}^{(2+(1-e^\Ups)\gamma^\Ups,0)}$.} In a similar vein to the construction in Step~1, we include in $u$ a term $\chi\rho_\sscri^{1+(1-e^\Ups)\gamma^\Ups}u_{1 1}^{(1+(1-e^\Ups)\gamma^\Ups,0)}\,(\dd x^1)^2$, which needs to solve
  \[
    -\bigl((1+(1-e^\Ups)\gamma^\Ups)-(1+2\gamma^\Ups)\bigr)(1+(1-e^\Ups)\gamma^\Ups-\rho_+\pa_{\rho_+})u_{1 1}^{(1+(1-e^\Ups)\gamma^\Ups,0)} = \rho_+^{-2}f_{1 1}^{(2+(1-e^\Ups)\gamma^\Ups,0)}.
  \]
  We control the solution with initial condition $0$ at $\rho_+=1$ using Lemma~\ref{LemmaTMIntFuchs2}, so
  \[
    u_{1 1}^{(1+(1-e^\Ups)\gamma^\Ups,0)} \in \Hb^{\infty,\ \bigl(\breve\cE_+\extcup\{(1+(1-e^\Ups)\gamma^\Ups,0)\},\breve\ell_+\bigr)}(\Omega_*\cap\scri^+).
  \]
  Upon replacing $f$ further by $f-L_b(\chi\rho_\sscri^{1+(1-e^\Ups)\gamma^\Ups}u_{1 1}^{(1+(1-e^\Ups)\gamma^\Ups,0)})$, we have now reduced to the case
  \begin{equation}
  \label{EqDScriFormalf}
    f \in \Hb^{\infty,\ (\cF_\sscri+1,\ell_\sscri+1),\ (\cF_++2,\breve\ell_++2)},
  \end{equation}
  where, at this point, $\cF_\sscri=\cE_\sscri^\cC$ (thus $\min\Re\cF_\sscri>1+2\gamma^\Ups$) and $\cF_+=\cE_+^{\sharp,(1)}$.

  \pfstep{Step~3. Solving away the remaining terms.} Given $f$ as in~\eqref{EqDScriFormalf}, consider $z\in\C$ such that
  \begin{equation}
  \label{EqDScriFormalExp}
    z\in\pi_1\cF_\sscri,\quad
    \Re z<\min\Re\cF_\sscri+1,\quad
    \Re z\leq \ell_\sscri.
  \end{equation}
  Set $k:=k(\cF_\sscri,z)$ and write the corresponding term in the expansion of $f$ as $\chi\rho_\sscri^{z+1}(\log\rho_\sscri)^k f^{(z+1,k)}$, where thus $f^{(z+1,k)}\in\Hb^{\infty,(\cF_++2,\breve\ell_++2)}(\Omega_*\cap\scri^+;S^2\cT^*)$. Define $u^{(z,k)}$ as the solution of
  \[
    -(z-(I+A_0))(z-\rho_+\pa_{\rho_+})u^{(z,k)} = \rho_+^{-2}f^{(z+1,k)},\quad u^{(z,k)}=u^{(z,k)}(t_*,\omega)=0\ \text{for}\ t_*\leq 1.
  \]
  Since $z$ does not lie in the spectrum of $I+A_0$, we can invert $z-(I+A_0)$ and thus find a (unique) solution $u^{(z,k)}$ of this ODE which, by Lemma~\ref{LemmaTMIntFuchs2}, satisfies
  \begin{equation}
  \label{EqDScriuzk}
    u^{(z,k)}\in\Hb^{\infty,\ \bigl(\cF_+\extcup\{(z,0)\},\breve\ell_+\bigr)}(\Omega_*\cap\scri^+).
  \end{equation}
  Moreover, one computes
  \begin{align*}
    &\rho_\sscri^{-1}\rho_+^{-2}\cdot\rho_\sscri^{z+1}(\log\rho_\sscri)^k f^{(z+1,k)} + (\rho_\sscri\pa_{\rho_\sscri}-(I+A_0))(\rho_\sscri\pa_{\rho_\sscri}-\rho_+\pa_{\rho_+})\bigl(\rho_\sscri^z(\log\rho_\sscri)^k u^{(z,k)}\bigr) \\
    &\qquad = \rho_\sscri^{-1}\rho_+^{-2}\sum_{k'=0}^{k-1}\rho_\sscri^{z+1}(\log\rho_\sscri)^{k'} f^{\prime(z+1,k')}
  \end{align*}
  for some (explicit) $f^{\prime(z+1,k')}\in\Hb^{\infty, \bigl((\cF_+\extcup\{(z,0)\})+2,\,\breve\ell_++2\bigr)}$. Upon replacing the source term $f$ by $f-L_b(\chi\rho_\sscri^z(\log\rho_\sscri)^k u^{(z,k)})$, the remaining error again satisfies~\eqref{EqDScriFormalf} but with $\cF_\sscri$ and $\cF_+$ replaced by $\cF_\sscri\setminus\{(z,k)\}$ and $\cF_+\extcup\{(z,0)\}$, respectively. (Note that the action of the term $\tilde L$ in~\eqref{EqDScriFormalL} produces a sum of terms of the form $\rho_\sscri^{z+1+j}(\log\rho_\sscri)^{k''}\cdot\Hb^{\infty,\bigl((\cF_+\extcup\{(z,0)\})+2,\,\breve\ell_++2\bigr)}$ for $j=1,2,3,\ldots$ and $k''=0,\ldots,k$.) Repeating this procedure for $(z,k-1)$, $\ldots$, $(z,0)$ ultimately replaces $\cF_\sscri$ and $\cF_+$ by $\cF_\sscri\setminus\{(z,j)\colon j=0,\ldots,k\}$ and $\cF_+\extcup\{(z,k)\}$, respectively.

  Once all terms of $f$ at $\scri^+$ corresponding to all exponents $z_1,\ldots,z_N$ satisfying~\eqref{EqDScriFormalExp} have been eliminated in this fashion, we have in effect replaced $\cF_\sscri$ and $\cF_+$ (which is $\cE_+^{\sharp,(j)}$ after the $j$-th iteration) by $\{(z,k)\in\cF_\sscri\colon\Re z\geq\min\Re\cF_\sscri+1\}$ and $\cF_+\extcup\{(z_1,k_1)\}\extcup\cdots\extcup\{(z_N,k_N)\}$, respectively, where $k_i=k(\cF_\sscri,z_i)$ (the latter index set thus being $\cE_+^{\sharp,(j+1)}$).
\end{proof}

\subsection{Step~1: almost linear bounds}
\label{SsD1Alm}

From this point onward, we consider metrics $g=g_{b_0,b,-\scal}+h$, where $|b-b_0|$ and $\scal\in\scalspace_1$ are small and $h\in\Hb^{\infty,\ \bigl(\la\cE_\sscri^\cC\ra,3+\eps_\sscri\bigr),\ (\cE_+,3+\eps_+),\ 2+\eps_\cK}(\Omega_*)$, as well as $g^0=g_{b_0,b,-\scal}$ as in Theorem~\ref{ThmDAdmReg}; and we write
\[
  L := L_{g,g^0}.
\]
We first prove a moderate improvement of the bound~\eqref{EqDAdmRegu} for $f$ with slightly more decay at $\iota^+$ and $\cK^+$ than required in~\eqref{EqDAdmRegf}: we will obtain almost $\frac12$ an order of additional $t_*$-decay. (Since it is the asymptotics as $t_*\to\infty$, i.e., at $\iota^+\cup\cK^+$, that is our focus now, we shall always require the strongest assumptions on $f$ at $\scri^+$ that we will ultimately need, even if at intermediate stages, such as this one, less decay would suffice.) The Fourier-based part of our arguments will naturally take place on the manifold
\begin{equation}
\label{EqD1AlmMp}
  M' = [\,\ol{\R_{t_*}}\times X;\{\pm\infty\}\times\pa X\,]
\end{equation}
introduced in~\eqref{EqKMfdSliceMfd} (which is contained in the manifold~\eqref{EqTFsfM} for $n=3$), and with boundary hypersurfaces denoted by $\scri^+$, $\iota=\iota^-\cup\iota^+$, and $\cK=\cK^+\cup\cK^-$; see Figure~\ref{FigD1AlmMp} (which is essentially identical to Figure~\ref{FigTFM} except for the range of $r$).

\begin{figure}[!ht]
\centering
\includegraphics{FigD1AlmMp-r}
\caption{The manifold $M'$ and its boundary hypersurfaces, together with some local coordinates (suppressing the spherical coordinates $\omega=\frac{x}{|x|}\in\Sph^2$); cf.\ Figure~\ref{FigIMwc}. The nomenclature $\iota^-$ is not accurate, as from the perspective of the Minkowski metric it contains points of spacelike infinity and past null infinity as well; we use it for symmetry reasons, and recall that we are in any case mainly interested in the region $t_*\geq 1$ here.}
\label{FigD1AlmMp}
\end{figure}

The Fourier transform, restricted to frequencies in $[-1,1]$, maps elements of weighted b-Sobolev spaces on $M'$ to similar spaces on $X_\scbtop^\pm=[\pm[0,1]_\sigma\times X;\{0\}\times\pa X]$ (which was introduced in~\eqref{EqTFXscbtpm}). Similarly to~\eqref{EqTFHbM}, we denote weighted b-Sobolev spaces on $M'$ by
\[
  \Hb^{k,\ \alpha_\sscri,\ \alpha_\iota,\ \alpha_\cK}(M') = \rho_\sscri^{\alpha_\sscri}\rho_\iota^{\alpha_\iota}\rho_\cK^{\alpha_\cK}\Hb^k(M'),
\]
where we always use an unweighted b-density (e.g., $|\frac{\dd t_*\,\dd x}{\la t_*\ra |x|^3}|$) to define the underlying $L^2$-space, and $\rho_\sscri$, $\rho_\iota$, and $\rho_\cK$ are defining functions of the lift of $\ol\R\times\pa X$, the lift of $\{\pm\infty\}\times\pa X$ (which has two connected components, namely future and past punctured timelike infinity), and the lift of $\{\pm\infty\}\times X$ (which also has two connected components, namely the future and past Kerr face). We also use the notation $\rho_+$ for a defining function of the lift of $\{+\infty\}\times\pa X$.

\begin{prop}[Almost linear bounds]
\label{PropD1Alm}
  Let $\alpha_\cK\in[\frac12,1)$, and let $\alpha_+>\alpha_\cK$. Then for $f\in\Hb^{\infty,\ \bigl(\la\cE_\sscri^\cC+1\ra',4+\eps_\sscri\bigr),\ \alpha_++2,\ \alpha_\cK}(\Omega_*)^{\bullet,-}$, the forward solution of $L u=f$ satisfies
  \begin{equation}
  \label{EqD1Alm}
    u \in \Hb^{\infty,\ \bigl(\la\cE_\sscri^\cC\ra,3+\eps_\sscri\bigr),\ \alpha_\cK-\eps,\ \alpha_\cK-2}(\Omega_*)^{\bullet,-}\quad\forall\,\eps>0.
  \end{equation}
  Moreover, in this space, $u$ depends continuously on the parameters $b_0,b,\scal,h$ involved in the definition of $L$.
\end{prop}

As we shall see in the proof, the decay rate of $u$ would be $\alpha_+$, provided $\alpha_+\in(\alpha_\cK,\alpha_\cK+\eps_\ind)$, if it were not for the singular terms of $\wh{L_b}(\sigma)^{-1}$ at $\sigma=0$ (which give rise to $\dot g$ and $h_{\rms 1}$ in~\eqref{EqD1Almu2} below). For now, rather rough estimates for these singular contributions suffice.

\begin{proof}[Proof of Proposition~\usref{PropD1Alm}]
  Theorem~\ref{ThmDAdmReg} gives $u\in\Hb^{\infty,(1-\eps,\frac12-\eps,-\frac32)}(\Omega_*;S^2\cT^*)^{\bullet,-}$ for all $\eps>0$. Proposition~\ref{PropDScriPhg} improves the $\scri^+$-asymptotics, so
  \begin{equation}
  \label{EqD1AlmIter}
    u\in\Hb^{\infty,\ \bigl(\la\cE_\sscri^\cC\ra,3+\eps_\sscri\bigr),\ \alpha'_\cK-\eps,\ \alpha'_\cK-2}(\Omega_*)^{\bullet,-}
  \end{equation}
  for $\alpha'_\cK=\frac12$ and all $\eps>0$. In order to set up an iterative scheme for improving this at $\iota^+$ and $\cK^+$, suppose we have established this for some (possibly larger value of) $\alpha'_\cK\in[\frac12,\alpha_\cK)$. (The arguments below go through unchanged also if $\alpha'_\cK>\frac12-\eps_\cK$. In light of the final part of Theorem~\ref{ThmDAdmReg}, this then implies the continuous dependence of $u$, in all spaces in which we estimate $u$ in the course of our arguments here, on the parameters $b_0,b,\scal,h$ involved in the definition of $L$.)

  By~\eqref{EqDAdmFw}, we have $(L-L_b)u\in\Hb^{\infty,\ (\la\cE_\sscri^\cC+1\ra',4+\eps_\sscri),\ \alpha'_\cK+3-\eps,\ \alpha_\cK'+\eps_\cK}$, and therefore
  \begin{equation}
  \label{EqD1AlmEq}
    L_b u = f' := f - (L-L_b)u \in \Hb^{\infty,\ \bigl(\la\cE_\sscri^\cC+1\ra',4+\eps_\sscri\bigr),\ \tilde\alpha_++2,\ \tilde\alpha_\cK}(\Omega_*)^{\bullet,-}
  \end{equation}
  where we may take $\tilde\alpha_+=\min(\alpha'_\cK+1-\eps,\alpha_+)$ and $\tilde\alpha_\cK=\min(\alpha'_\cK+\eps_\cK,\alpha_\cK)$. For bookkeeping purposes, it is convenient to relax the $\iota^+$-weight to satisfy
  \[
    \tilde\alpha_+ \in (\tilde\alpha_\cK,\tilde\alpha_\cK+\eps_\ind),\quad \tilde\alpha_+\leq\alpha_+<1.
  \]
  (This is less than $1$ and thus smaller than $\alpha'_\cK+1-\eps>\frac32-2\eps$. Note also that $\tilde\alpha_\cK<1$.) We control $u$ by solving~\eqref{EqD1AlmEq} in three steps.

  \pfstep{Step~1. Solution near $\scri^+$.} We use Proposition~\ref{PropDScriFormal} to find $u_\sscri\in\Hb^{\infty, (\la\cE_\sscri^\cC\ra,3+\eps_\sscri),\ \tilde\alpha_+,\ \infty}$ (with support near $\scri^+$ and vanishing for $t_*<1$) such that
  \begin{equation}
  \label{EqD1AlmFlat}
    u_\flat := u-u_\sscri \implies L_b u_\flat =: f_\flat \in \Hb^{\infty,\ 4+\eps_\sscri,\ \tilde\alpha_++2,\ \tilde\alpha_\cK}(\Omega;S^2\cT^*)^{\bullet,-}.
  \end{equation}
  (Carefully note that the explicitly constructed $u_\sscri$ inherits the $\iota^+$-decay rate from $f'$ in~\eqref{EqD1AlmEq}, which is stronger than the $\iota^+$-decay rate that we started out with in~\eqref{EqD1AlmIter}.)

  \pfstep{Step~2. Improved decay via spectral theory.} We invert $L_b$ in~\eqref{EqD1AlmFlat} using the Fourier transform. As in~\S\ref{SsAdmPf}, we split the solution into two parts: recalling the contours $\gamma_-$, $\gamma_0$, and $\gamma_+$ from~\eqref{EqAdmIFT} and the cutoff function $\chi=\chi(|\sigma|)\in\CIc([0,c))$, which is equal to $1$ on $[0,\frac{c}{2}]$, we set
  \begin{subequations}
  \begin{equation}
  \label{EqD1Almu}
  \begin{split}
    &u_\flat = u_{\rm lo} + u_{\rm hi}, \\
    &\qquad u_{\rm hi}(t_*,\cdot) := \frac{1}{2\pi}\int_\R e^{-i\sigma t_*}(1-\chi(|\sigma|))\wh{L_b}(\sigma)^{-1}\wh{f_\flat}(\sigma)\,\dd\sigma, \\
    &\qquad u_{\rm lo} := u_{\rm reg} + \dot g + h_{s 1},
  \end{split}
  \end{equation}
  where, recalling $\dot g_b^{\Ups,\aug}$ and $h_{b,\rms 1}^{\leq 1,\aug}$ from Lemma~\ref{LemmaAdmLoIm}, the low-energy pieces are defined by
  \begin{equation}
  \label{EqD1Almu2}
  \begin{alignedat}{2}
    &\bigl(\hat u_{{\rm reg},1}(\sigma),\ \hat b_1(\sigma),\ \hat\scal_1(\sigma)\bigr) := \wt{L_b}(\sigma)^{-1}\bigl(\wh{f_\flat}(\sigma,\cdot),\ 0,\ 0\bigr),\hspace{-22em}&& \\
    &&\qquad u_{\rm reg}(t_*,\cdot) &:= \frac{1}{2\pi}\int_{\gamma_-\cup\gamma_0\cup\gamma_+} e^{-i\sigma t_*}\chi(|\sigma|)\hat u_{{\rm reg},1}(\sigma,\cdot)\,\dd\sigma, \\
    &&\qquad \dot g(t_*,\cdot) &:= \frac{1}{2\pi}\int_{\gamma_-\cup\gamma_0\cup\gamma_+} e^{-i\sigma t_*}\chi(|\sigma|)\cF(\dot g_b^{\Ups,\aug})(\sigma)(\hat b_1(\sigma))\,\dd\sigma, \\
    &&\qquad h_{\rms 1}(t_*,\cdot) &:= \frac{1}{2\pi}\int_{\gamma_-\cup\gamma_0\cup\gamma_+} e^{-i\sigma t_*}\chi(|\sigma|)\cF(h_{b,\rms 1}^{\leq 1,\aug})(\sigma)(\hat\scal_1(\sigma))\,\dd\sigma;
  \end{alignedat}
  \end{equation}
  \end{subequations}
  these are the same expressions as used before in~\eqref{EqAdmPfHiDef}, \eqref{EqAdmPfLoTilde}, and \eqref{EqAdmPfLodotghs1}.

  As shown in \citeAF{Corollary~\ref*{CorDResHiLoc}}, the high-energy piece $u_{\rm hi}$ has arbitrary $\iota^+$- and $\cK^+$-decay (due to its $\sigma$-regularity), so in particular
  \begin{equation}
  \label{EqD1Almuhi}
    u_{\rm hi} \in \Hb^{\infty,\ 1-\eps,\ \tilde\alpha_+,\ \tilde\alpha_\cK}(M';S^2\cT^*)\quad\forall\,\eps>0.
  \end{equation}
  (Sharp decay at $\scri^+$ will be recovered in Step~3 below.) For the low-energy piece, we first observe that~\eqref{EqTFHbLo} gives
  \[
    \bigl(\wh{f_\flat}(\sigma)\bigr)\big|_{\sigma\in\pm[0,c]} \in \Hb^{\infty,\ \tilde\alpha_++2-\eps,\ \tilde\alpha_++1,\ \tilde\alpha_\cK-1}(X_\scbtop^\pm)
  \]
  in the notation of~\eqref{EqTFHbX}. Since $\tilde\alpha_+-\tilde\alpha_\cK\in(0,\eps_\ind)$ lies in the indicial gap of $\wh{L_b}(0)$, we can apply \citeAF{Proposition~\ref*{PropDResLo}} (applied to $\wt{L_b}(\sigma)$) to infer\footnote{The case $\tilde\alpha_\cK=\frac12$ follows directly from the conormal regularity of $\wt{L_b}(\sigma)^{-1}$ established there, while the case of general $\tilde\alpha_\cK$ is deduced from the case $\tilde\alpha_\cK=\frac12$ by conjugation by $|\sigma|^{\tilde\alpha_\cK-\frac12}$. This gives the stated memberships of $\hat b_1$ and $\hat\scal_1$.}
  \begin{equation}
  \label{EqD1AlmuregMem}
  \begin{split}
    \bigl(\hat u_{{\rm reg},1}(\sigma)\bigr)\big|_{\sigma\in\pm[0,c]} &\in \Hb^{\infty,\ 1-\eps,\ \tilde\alpha_+-1,\ \tilde\alpha_\cK-1}(X_\scbtop^\pm)\quad\forall\,\eps>0, \\
    \hat b_1 &\in \Hb^{\infty, \tilde\alpha_\cK-1}(\pm[0,c),|\tfrac{\dd\sigma}{\sigma}|;\C^4), \\
    \hat\scal_1 &\in \Hb^{\infty, \tilde\alpha_\cK-1}(\pm[0,c),|\tfrac{\dd\sigma}{\sigma}|;\scalspace_1).
  \end{split}
  \end{equation}
  (This is completely analogous to \citeAF{(\ref*{EqA2ureg})}, except we keep more precise track of the $\scface$-decay order here.) For $u_{\rm reg}$, we can shift the integration contour to $[-c,c]$ as in the proof of Lemma~\ref{LemmaAdmPfLoReg} (i.e., using the holomorphicity of $\hat u_{{\rm reg},1}$ in the upper half plane, uniform boundedness in the space of distributions down to the real axis, and continuity down to $\R\setminus\{0\}$) and then apply~\eqref{EqTFHbInvLo} to obtain
  \begin{equation}
  \label{EqD1Almureg}
    u_{\rm reg} \in \Hb^{\infty,\ \tilde\alpha_+-\eps,\ \tilde\alpha_+,\ \tilde\alpha_\cK}(M')\quad\forall\,\eps>0.
  \end{equation}

  We analyze $u_{\rm sing}:=\dot g+h_{\rms 1}$ similarly to the arguments in~\S\ref{SssAdmPfLo}. First of all, we recall from~\eqref{EqAdmPfLoDerF1} and \eqref{EqAdmPfLoDerF2} that
  \begin{equation}
  \label{EqD1AlmAug}
  \begin{split}
    \bigl(\cF(\pa_{t_*}\dot g_b^{\Ups,\aug})(\sigma)\bigr)\big|_{\sigma\in\pm[0,c]} &\in \Hb^{\infty,\ 1-\eps,\ 1-\eps,\ ((0,0),1-\eps)}(X_\scbtop^\pm), \\
    \bigl(\cF(\pa_{t_*}^2 h_{b,\rms 1}^{\leq 1,\aug})(\sigma)\bigr)\big|_{\sigma\in\pm[0,c]} &\in \Hb^{\infty,\ 1-\eps,\ 2-\eps,\ ((0,0),1-\eps)}(X_\scbtop^\pm)
  \end{split}
  \end{equation}
  for all $\eps>0$. Therefore,
  \begin{equation}
  \label{EqD1AlmSingFT}
  \begin{alignedat}{2}
    \cF(\pa_{t_*}\dot g) &= \chi(|\sigma|)\cF(\pa_{t_*}\dot g_b^{\Ups,\aug})(\hat b_1(\sigma)) &&\in \Hb^{\infty,\ 1-\eps,\ \tilde\alpha_\cK-\eps,\ \tilde\alpha_\cK-1}(X_\scbtop^\pm), \\
    \cF(\pa_{t_*}^2 h_{\rms 1}) &= \chi(|\sigma|)\cF(\pa_{t_*}^2 h_{b,\rms 1}^{\leq 1,\aug})(\hat\scal_1(\sigma)) && \in \Hb^{\infty,\ 1-\eps,\ \tilde\alpha_\cK+1-\eps,\ \tilde\alpha_\cK-1}(X_\scbtop^\pm).
  \end{alignedat}
  \end{equation}
  Upon multiplication of the first membership by $-i\sigma$ and adding the second membership, we obtain
  \begin{equation}
  \label{EqD1Almpa2using}
    \cF(\pa_{t_*}^2 u_{\rm sing}) \in \Hb^{\infty,\ 1-\eps,\ \tilde\alpha_\cK+1-\eps,\ \tilde\alpha_\cK}(X_\scbtop^\pm).
  \end{equation}
  Now, $u_{\rm sing}=-(u_{\rm hi}+u_{\rm reg})$ for $\ft_*<1$ due to the vanishing of $u$ there, so from~\eqref{EqD1Almuhi} and \eqref{EqD1Almureg}, and using $\pa_{t_*}\in\rho_\iota\rho_\cK\Vb(M')$ and $\tilde\alpha_+<1$, we obtain
  \begin{equation}
  \label{EqD1AlmPa2Ini}
    \pa_{t_*}u_{\rm sing}|_{\{\ft_*<1\}} \in \bar H_\bop^{\infty,\ \tilde\alpha_+-\eps,\ \tilde\alpha_++1,\ \tilde\alpha_\cK+1}(\Omega_-),\quad \Omega_-:=\cl_{M'}\{\ft_*<1\}.
  \end{equation}
  We now apply Lemma~\ref{LemmaD1IntFT} below (which uses~\eqref{EqD1Almpa2using}, and~\eqref{EqD1AlmPa2Ini} only on the even smaller domain denoted $\Omega_<$ there) to deduce that
  \[
    \pa_{t_*}u_{\rm sing} \in \Hb^{\infty,\ 1-\eps,\ \tilde\alpha_\cK+1-\eps,\ \tilde\alpha_\cK-1}(M');
  \]
  for\footnote{The $\iota^+$-weight $\tilde\alpha_\cK+1-\eps$ here is what we are after. If we only used $\pa_{t_*}^2 u_{\rm sing}\in\Hb^{\infty,\ 1-\eps,\ \tilde\alpha_\cK+2-\eps,\ \tilde\alpha_\cK+1}(M')$, which follows from~\eqref{EqD1Almpa2using} and~\eqref{EqTFHbInvLo}, then integration of $\pa_{t_*}$ using the less subtle Lemma~\ref{LemmaD1Intb} below would yield an $\iota^+$-decay rate of $\min(1-\eps,\tilde\alpha_\cK+1-\eps)=1-\eps$ only. (A further integration of $\pa_{t_*}$ would then give for $u_{\rm sing}$ itself only the $\iota^+$-order $-\eps$, rather than the desired $\tilde\alpha_\cK-\eps$.)} the $\iota$-order of this, we use that $\min(\tilde\alpha_++1,\tilde\alpha_\cK+1-\eps)=\tilde\alpha_\cK+1-\eps$. We integrate $\pa_{t_*}$ once more, now using as the initial condition the membership
  \[
    u_{\rm sing}|_{\{\ft_*<1\}} \in \bar H_\bop^{\infty,\ \tilde\alpha_+-\eps,\ \tilde\alpha_+,\ \tilde\alpha_\cK}(\Omega_-)
  \]
  (again obtained from~\eqref{EqD1Almuhi} and \eqref{EqD1Almureg}) and Lemma~\ref{LemmaD1Intb}\eqref{ItD1Intb1} below, in order to obtain
  \begin{equation}
  \label{EqD1Almusing}
    u_{\rm sing} \in \Hb^{\infty,\ \tilde\alpha_+-\eps,\ \tilde\alpha_\cK-\eps,\ \tilde\alpha_\cK-2}(M').
  \end{equation}
  (For the $\iota^+$-order here, we use that $\tilde\alpha_\cK-\eps<\tilde\alpha_+-\eps$.)

  Combining~\eqref{EqD1Almuhi}, \eqref{EqD1Almureg}, and~\eqref{EqD1Almusing} (and recalling the forward support property of $u$), we have now shown
  \begin{equation}
  \label{EqD1AlmuImpr}
    u = u_{\rm hi} + u_{\rm reg} + u_{\rm sing} \in \Hb^{\infty,\ \tilde\alpha_+-\eps,\ \tilde\alpha_\cK-\eps,\ \tilde\alpha_\cK-2}(\Omega_*;S^2\cT^*)^{\bullet,-}\quad\forall\,\eps>0.
  \end{equation}

  \pfstep{Step~3. Recovery of sharp decay at $\scri^+$.} We again apply Proposition~\ref{PropDScriPhg} to the equation $L u=f$, now with the improved information~\eqref{EqD1AlmuImpr} at $\iota^+$; this yields
  \[
    u \in \Hb^{\infty,\ \bigl(\la\cE_\sscri^\cC\ra,3+\eps_\sscri\bigr),\ \tilde\alpha_\cK-\eps,\ \tilde\alpha_\cK-2}(\Omega_*)^{\bullet,-}.
  \]
  This improves on~\eqref{EqD1AlmIter} by the amount $\min(\eps_\cK,\alpha_\cK-\tilde\alpha_\cK)$ at $\iota^+$ and $\cK^+$. After finitely many iterations, we thus obtain~\eqref{EqD1Alm}.
\end{proof}

To complete the proof, we need two results on the integration of $\pa_{t_*}$. The first result (stated in a more general form that we need later on) is as follows.

\begin{figure}[!ht]
\centering
\includegraphics{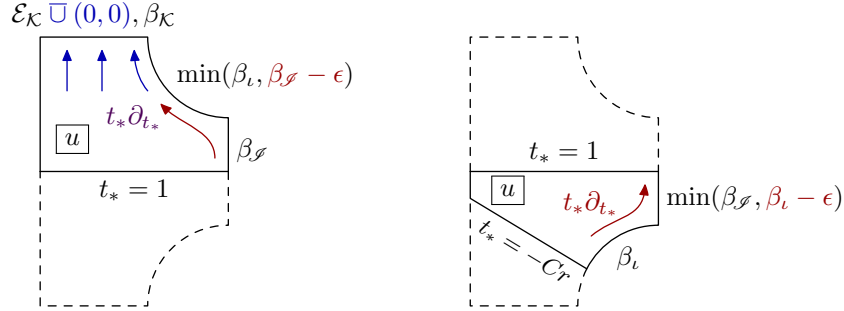}
\caption{\textit{On the left:} illustration of the two steps in the proof of Lemma~\ref{LemmaD1Intb}\eqref{ItD1Intb1}. In the first step, we integrate along a hyperbolic vector field, which transports $\scri^+$-decay to $\iota^+$. In the second step, we integrate along the b-normal vector field at $\cK^+$. The orders of $u$ are stated next to the respective boundary hypersurfaces. \textit{On the right}: illustration of the main content of Lemma~\ref{LemmaD1Intb}\eqref{ItD1Intb2}. Integration along a hyperbolic vector field transports $\iota$-decay to $\scri^+$.}
\label{FigD1AlmInt}
\end{figure}

\begin{lemma}[Integration of $\pa_{t_*}$ on b-Sobolev spaces]
\label{LemmaD1Intb}
  Let $k\in\N_0$ and $\beta_\sscri,\beta_\iota,\beta_\cK\in\R$, with $\beta_\cK\neq 0$. Let $\cE_\cK$ be an index set. Set $\Omega_-:=\cl_{M'}\{t_*\leq 1\}$ and $\Omega_<:=\cl_{M'}\{t_*\leq -C r\}$ where $C>0$ is fixed but arbitrary.
  \begin{enumerate}
  \item\label{ItD1Intb1}{\rm (Integration from $t_*=1$.)} Suppose that
    \[
      u|_{\Omega_-}\in\bar H_\bop^{k,\ \beta_\sscri,\ \beta_\iota,\ (\cE_\cK,\beta_\cK)}(\Omega_-),\quad
      \pa_{t_*}u\in\Hb^{k,\ \beta_\sscri,\ \beta_\iota+1,\ (\cE_\cK+1,\beta_\cK+1)}(M').
    \]
    Then, for all $\eps>0$,
    \begin{equation}
    \label{EqD1AlmInt}
      u \in \Hb^{k,\ \beta_\sscri,\ \min(\beta_\iota,\beta_\sscri-\eps),\ \bigl(\cE_\cK\extcup\{(0,0)\},\,\beta_\cK\bigr)}(M').
    \end{equation}
  \item\label{ItD1Intb2}{\rm (Integration from $t_*=-C r$.)} Suppose that\footnote{Since $\Omega_<\cap\scri^+=\emptyset$, the first weight (i.e., the $\scri^+$-weight) of $u|_{\Omega_<}$ is arbitrary.}
    \[
      u|_{\Omega_<}\in\bar H_\bop^{k,\ \infty,\ \beta_\iota,\ \beta_\cK}(\Omega_<),\quad
      (\pa_{t_*}u)|_{\Omega_-}\in\bar H_\bop^{k,\ \beta_\sscri,\ \beta_\iota+1,\ \beta_\cK+1}(\Omega_-).
    \]
    Then, for all $\eps>0$,
    \begin{equation}
    \label{EqD1Intb2}
      u|_{\Omega_-} \in \bar H_\bop^{k,\ \min(\beta_\sscri,\beta_\iota-\eps),\ \beta_\iota,\ \beta_\cK}(\Omega_-);
    \end{equation}
    similarly for\footnote{The $\cK^-$-order in the conclusion is inherited directly from the assumption. (The only nontrivial part of the result is the $\scri^+$-order in~\eqref{EqD1Intb2}.)} $(\cE_\cK[+1],\beta_\cK[+1])$ in place of $\beta_\cK[+1]$.
  \end{enumerate}
\end{lemma}
\begin{proof}
  \pfstep{Part~\eqref{ItD1Intb1}.} We first study the propagation along $\pa_{t_*}$ near $\scri^+\cap\iota^+$ in the coordinates $\rho_\sscri=\frac{t_*}{r}\in[0,c_\sscri)$ and $\rho_+=\frac{1}{t_*}\in[0,2)$ (cf.\ \eqref{EqKMfdCoordScriip}) in the region $t_*>0$. In these coordinates, $t_*\pa_{t_*}=\rho_\sscri\pa_{\rho_\sscri}-\rho_+\pa_{\rho_+}$ is a hyperbolic vector field, and the assumptions on $u$ read, upon restriction to this coordinate system,
  \begin{alignat*}{2}
    u|_{\{\rho_+>1\}} &\in &\rho_\sscri^{\beta_\sscri}&\bar H_\bop^k([0,c_\sscri)_{\rho_\sscri}\times[1,2)_{\rho_+}\times\Sph^2), \\
    (\rho_\sscri\pa_{\rho_\sscri}-\rho_+\pa_{\rho_+})u &\in &\rho_\sscri^{\beta_\sscri}\rho_\iota^{\beta_\iota}&\Hb^k([0,c_\sscri)_{\rho_\sscri}\times[0,2)_{\rho_+}\times\Sph^2).
  \end{alignat*}
  Lemma~\ref{LemmaTIntHyp} (with $\alpha=\beta_\sscri$ and $\beta\leq\beta_\iota$, $\beta<\beta_\sscri$) implies
  \begin{equation}
  \label{EqD1AlmInt1}
    u \in \rho_\sscri^{\beta_\sscri}\rho_\iota^{\beta'_\iota}\Hb^k,\quad \beta'_\iota:=\min(\beta_\iota,\beta_\sscri-\eps).
  \end{equation}
  Since these arguments are valid outside of any fixed neighborhood of $\cK^+$, we conclude that $u$ lies in $H_{\bop,\loc}^{k,\ \beta_\sscri,\ \beta'_\iota,\ 0}(M'\setminus\cK^+)$ (the third order, which refers to decay at $\cK^+$, being irrelevant here).

  We next switch to coordinates valid near $\iota^+\cap\cK^+$, namely $\rho_+=\frac{1}{r}\in[0,c_+)$ and $\rho_\cK=\frac{r}{t_*}\in[0,1)$ (see~\eqref{EqKMfdCoordipK}). In these, $t_*\pa_{t_*}=-\rho_\cK\pa_{\rho_\cK}$, and thus Lemma~\ref{LemmaTMIntFuchs} becomes applicable (with $z=0$): in view of
  \begin{alignat*}{2}
    u|_{\{\rho_\cK\geq\delta>0\}} &\in &\rho_\iota^{\beta'_\iota}\bar H_\bop^k&([\delta,1)_{\rho_\cK}\times[0,c_+)_{\rho_+}\times\Sph^2), \\
    \rho_\cK\pa_{\rho_\cK}u &\in\ &\Hb^{k,\ (\cE_\cK,\beta_\cK),\ \beta_\iota}&([0,1)_{\rho_\cK}\times[0,c_+)_{\rho_+}\times\Sph^2), 
  \end{alignat*}
  this yields
  \[
    u \in \Hb^{k,\ \bigl(\cE_\cK\extcup\{(0,0)\},\,\beta_\cK\bigr),\ \beta'_\iota}
  \]
  in this coordinate system. Together with~\eqref{EqD1AlmInt1}, this proves~\eqref{EqD1AlmInt}. (See Figure~\ref{FigD1AlmInt} for an illustration of the proof.)

  \pfstep{Part~\eqref{ItD1Intb2}.} We cover $\Omega_-\setminus\Omega_<$ using the coordinate chart $\rho_0=\frac{1}{t_*+2}$, $\rho_\sscri=\frac{t_*+2}{r}$ from~\eqref{EqKMfdCoordI0Scri}, so $-(t_*+2)\pa_{t_*}=\rho_0\pa_{\rho_0}-\rho_\sscri\pa_{\rho_\sscri}$ is a hyperbolic vector field. Omitting the weight at $\cK$ from the notation for the moment, we have $u|_{\Omega_<}\in\rho_\iota^{\beta_\iota}\bar H_\bop^k$ and $-(t_*+2)\pa_{t_*}u|_{\Omega_-}\in \rho_\sscri^{\beta_\sscri}\rho_\iota^{\beta_\iota}\bar H_\bop^k$; regarding the $\iota$-weight here, note that $-(t_*+2)^{-1}$ is a local defining function of $\iota^-$. Using Lemma~\ref{LemmaTIntHyp} thus gives
  \begin{equation}
  \label{EqD1IntFTOmegam}
    u|_{\Omega_-} \in \bar H_\bop^{k,\ \beta_\sscri',\ \beta_\iota,\ \beta_\cK}(\Omega_-)
  \end{equation}
  provided $\beta_\sscri'\leq\beta_\sscri$ and $\beta_\sscri'<\beta_\iota$. (We copied the $\cK^-$-weight $\beta_\cK$ from the a priori assumption on $u$.) This is the content of~\eqref{EqD1Intb2}.
\end{proof}

The second integration result is more subtle:

\begin{lemma}[Integration of $\pa_{t_*}$ on b-Sobolev spaces, with moment condition]
\label{LemmaD1IntFT}
  Let $\Omega_<:=\cl_{M'}\{t_*<-r\}$. Let $\beta_\sscri,\beta_\iota,\beta_\cK\in\R$ with $\beta_\iota>\beta_\sscri$ and $\beta_\cK\leq\beta_\iota-\beta_\sscri$, $\beta_\cK<1$, $\beta_\cK\neq 0$. Suppose that\footnote{Note that the $\scri^+$-order is irrelevant in the first assumption since $\Omega_<\cap\scri^+=\emptyset$.} $u|_{\Omega_<}\in\bar H_\bop^{k,\ \infty,\ \beta_\iota,\ \beta_\cK}(\Omega_<)$, and $\supp\cF u\subset[-1,1]\times X$ with
  \begin{equation}
  \label{EqD1IntFTAssm}
    \bigl(\cF(\pa_{t_*}u)(\sigma)\bigr)\big|_{\sigma\in\pm[0,1]} \in \Hb^{k+\ell,\ \beta_\sscri,\ (\beta_\iota+1)-1,\ \bigl((0,0),\,(\beta_\cK+1)-1\bigr)}(X_\scbtop^\pm),
  \end{equation}
  where $\ell=\ell(\beta_\sscri,\beta_\iota,\beta_\cK)$ is some ($k$-independent) integer; assume moreover that in the case $\beta_\cK>0$, the restriction of~\eqref{EqD1IntFTAssm} to $\zface$ is independent of the ``$\pm$'' sign. Then
  \[
    u \in \Hb^{k,\ \beta_\sscri,\ \beta_\iota-\eps,\ \bigl((0,0),\beta_\cK\bigr)}(M')\quad\forall\,\eps>0.
  \]
\end{lemma}

The membership~\eqref{EqD1IntFTAssm} implies $\pa_{t_*}u\in\Hb^{k,\ \beta_\sscri-\eps,\ \beta_\iota+1,\ \beta_\cK+1}(M')$ for all $\eps>0$ by~\eqref{EqTFHbInvLo}. This gives $u|_{\Omega_-}\in\bar H_\bop^{k,\ \beta_\sscri-\eps,\ \beta_\iota,\ \beta_\cK}(\Omega_-)$ by Lemma~\ref{LemmaD1Intb}\eqref{ItD1Intb2}, and thus Lemma~\ref{LemmaD1Intb}\eqref{ItD1Intb1} would at best yield the $\iota^+$-decay rate $\min(\beta_\iota,\beta_\sscri-\eps)=\beta_\sscri-\eps$ under the present assumptions: the $\scri^+$-decay rate of $\pa_{t_*}u$ gets inherited by $u$ and then transported to $\iota^+$. It is the ``moment condition'' on $\pa_{t_*}u$, encoded by the vanishing of $\cF(\pa_{t_*}u)$ at $\tface$ relative to $\scface\subset X_\scbtop^\pm$ (see~\eqref{EqD1IntFTMoment} below for the concrete consequence we will use), which means that the ``leading-order term'' of $u$ at $\scri^+$ obtained by integration of $\pa_{t_*}u$ vanishes at the future endpoints of $\scri^+$, thus leading to the stronger $\iota^+$-decay rate $\beta_\iota$ of $u$.

\begin{proof}[Proof of Lemma~\usref{LemmaD1IntFT}]
  Let $\chi=\chi(\frac{t_*}{r})$ be equal to $0$ on $(-\infty,-3]$ and $1$ on $[-2,\infty)$; our task is to estimate $\chi u$. Set $f:=\pa_{t_*}u$. Since $\pa_{t_*}\chi\in\rho_\sscri^\infty\rho_\iota\rho_\cK^\infty\CI(M')$, we have
  \begin{equation}
  \label{EqD1Intfp}
    \pa_{t_*}(\chi u) = f' + \chi f,\quad f' := [\pa_{t_*},\chi]u \in \Hb^{k,\ \infty,\ \beta_\iota+1,\ \infty}(M').
  \end{equation}

  Let $\chi_\scface\in\CIc([0,2))$ be equal to $1$ on $[0,1]$; we regard this as a function $\chi_\scface=\chi_\scface(\frac{\rho}{|\sigma|})$ on $X_\scbtop^\pm$ which thus equals $1$ near $\scface$ and $0$ near $\zface$. We then split
  \[
    \chi f = \chi f_1 + \chi f_2, \quad f_1 := \cF^{-1}\bigl( \chi_\scface \cF f \bigr), \ f_2 := \cF^{-1}\bigl( (1-\chi_\scface)\cF f \bigr).
  \]
  Since $(1-\chi_\scface)\cF f\in\Hb^{k+\ell,\ \infty,\ \beta_\iota,\ ((0,0),\beta_\cK)}(X_\scbtop^\pm)$, we can use \eqref{EqTFHbInvLo} to deduce that
  \begin{equation}
  \label{EqD1Intf2}
    f_2,\ \chi f_2 \in \Hb^{k,\ \beta_\iota+1-\eps,\ \beta_\iota+1,\ \beta_\cK+1}(M'),
  \end{equation}
  the membership for $\chi f_2$ following from that for $f_2$ since $\chi$ is smooth (thus, bounded and conormal) on $M'$.

  Consider now
  \[
    (\chi u)(t_*,x)=\int_{-\infty}^{t_*} \pa_{t_*}(\chi u(s,x))\,\dd s=\underbrace{\int_{-\infty}^{t_*} f'(s,x)\,\dd s}_{=:F'(t_*,x)} + \underbrace{\int_{-\infty}^{t_*}(\chi f_1)(s,x)\,\dd s}_{=:F_1(t_*,x)} + \underbrace{\int_{-\infty}^{t_*} (\chi f_2)(s,x)\,\dd s}_{=:F_2(t_*,x)}.
  \]
  Applying Lemma~\ref{LemmaD1Intb}\eqref{ItD1Intb2} with $C=3$ to the function $F'(t_*,x)$---which vanishes identically for $t_*\leq -3 r$---and recalling~\eqref{EqD1Intfp}, we obtain $F'|_{\Omega_-}\in\bar H_\bop^{k,\ \beta_\iota-\eps,\ \beta_\iota,\ \infty}(\Omega_-)$; integrating further in $t_*$ using Lemma~\ref{LemmaD1Intb}\eqref{ItD1Intb1} yields
  \begin{equation}
  \label{EqD1IntFTFp}
    F'\in\Hb^{k,\ \beta_\iota-\eps,\ \beta_\iota-2\eps,\ (0,0)}(M')
  \end{equation}
  for all $\eps>0$. We can similarly analyze the third integral, $F_2(t_*,x)$, using~\eqref{EqD1Intf2}: on $\Omega_-$, we have $F_2\in\bar H_\bop^{k,\ \beta_\iota-\eps,\ \beta_\iota,\ \infty}$, and on all of $M'$ we then get
  \begin{equation}
  \label{EqD1IntFTF2}
    F_2 \in H_\bop^{k,\ \beta_\iota-\eps,\ \beta_\iota-2\eps,\ ((0,0),\beta_\cK)}(M').
  \end{equation}

  The only subtle term is the second integral, $F_1$. Since $\chi_\scface=0$ near $\zface\subset X_\scbtop^\pm$, we have
  \[
    \cF f_1 = \chi_\scface\cF f \in \Hb^{k+\ell,\ \beta_\sscri,\ \beta_\iota,\ \infty}(X_\scbtop^\pm) \subset \rho^{\beta_\sscri}|\sigma|^{\beta_\iota-\beta_\sscri}\Hb^{k+\ell}\bigl(\pm[0,1]_\sigma\times X;|\tfrac{\dd\sigma}{\sigma}|\,\dd\mu_\bop\bigr),
  \]
  where $\dd\mu_\bop:=\frac{|\dd x|}{|x|^3}$ is an unweighted b-density on $X$. Since $\beta_\iota-\beta_\sscri>0$, this is continuous across $\sigma=0$ and vanishes at $\sigma=0$; we thus have
  \begin{equation}
  \label{EqD1IntFTMoment}
    f_1,\ \chi f_1 \in \rho^{\beta_\sscri}\la t_*\ra^{-(\beta_\iota-\beta_\sscri+1)}\Hb^k(\ol{\R_{t_*}}\times X),\quad
    \int_{-\infty}^\infty f_1(t_*,x)\,\dd t_*=0,
  \end{equation}
  with the membership for $\chi f_1$ again following from that for $f_1$ since $\chi$ is bounded and conormal on $\ol{\R_{t_*}}\times X$ (since it is smooth, and hence bounded conormal, on the blow-up $M'$). Upon integrating $\chi f_1$ from $t_*=-\infty$, we then obtain
  \[
    F_1|_{\{t_*\leq 1\}} \in \rho^{\beta_\sscri}\la t_*\ra^{-(\beta_\iota-\beta_\sscri)}\bar H_\bop^k([-\infty,1]\times X) = \bar H_\bop^{k,\ \beta_\sscri,\ \beta_\iota,\ \beta_\iota-\beta_\sscri}(\Omega_-).
  \]
  Using the moment condition~\eqref{EqD1IntFTMoment}, we rewrite $F_1(t_*,x)=\int_{-\infty}^{t_*} f_1(s,x)\,\dd s+\int_{-\infty}^{t_*}((\chi-1)f_1)(s,x)\,\dd s$ for $t_*\geq -1$, say, as
  \[
    F_1(t_*,x) = -\int_{-\infty}^{-t_*} f_1(-s,x)\,\dd s + \underbrace{\int_{-\infty}^{t_*} \bigl((\chi-1)f_1\bigr)(s,x)\,\dd s}_{=:F_3(t_*,x)}.
  \]
  The first integral lies in $\rho^{\beta_\sscri}\la t_*\ra^{-(\beta_\iota-\beta_\sscri)}\bar H_\bop^k([-1,\infty]\times X)$. To control $F_3$, we first note that~\eqref{EqD1IntFTMoment} implies $F_3\in\rho^{\beta_\sscri}\la t_*\ra^{-(\beta_\iota-\beta_\sscri)}\bar H_\bop^k([-\infty,1]_{t_*}\times X)$, so $F_3\in\bar H_\bop^{k,\ \infty,\ \beta_\iota,\ \beta_\iota-\beta_\sscri}(\Omega_<)$. Since $\pa_{t_*}F_3=(\chi-1)f_1\in\Hb^{k,\ \infty,\ \beta_\iota+1,\ \beta_\iota-\beta_\sscri+1}(M')$, we can (arguing similarly as for $F'$ above) integrate $\pa_{t_*}F_3$ further using Lemma~\ref{LemmaD1Intb}\eqref{ItD1Intb2} to get $F_3\in\bar H_\bop^{k,\ \beta_\iota-\eps,\ \beta_\iota,\ \beta_\iota-\beta_\sscri}(\Omega_-)$, and then Lemma~\ref{LemmaD1Intb}\eqref{ItD1Intb1} gives $F_3\in\Hb^{k,\ \beta_\iota-\eps,\ \beta_\iota-2\eps,\ \bigl((0,0),\beta_\iota-\beta_\sscri\bigr)}(M')$. Altogether then, we have shown that
  \[
    F_1 \in \Hb^{k,\ \beta_\sscri,\ \beta_\iota-2\eps,\ \bigl((0,0),\beta_\iota-\beta_\sscri\bigr)}(M')
  \]
  for all $\eps>0$. Adding this to~\eqref{EqD1IntFTFp} and \eqref{EqD1IntFTF2} yields $\chi u\in\Hb^{k,\ \beta_\sscri,\ \beta_\iota-\eps,\ \bigl((0,0),\,\min(\beta_\cK,\beta_\iota-\beta_\sscri)\bigr)}(M')$ for all $\eps>0$ and thus the claim.
\end{proof}

\subsection{Step~2: boost parameter changes}
\label{SsD2Boost}

\emph{We use the same notation as in~\S\usref{SsD1Alm}.} Having established almost-$t_*^1$-bounds for $u$ solving $L u=f$ with sufficiently decaying $f$, the right-hand side of the stationary equation $L_b u=f-(L-L_b)u$ now has $t_*^{-1-\eta}$-decay where $\eta>0$. The late-time asymptotics of $u$ one can extract from this involve, in particular, two $t_*$-integrals of this (cf.\ the arguments concerning $u_{\rm sing}=\dot g+h_{\rms 1}$ after~\eqref{EqD1Almpa2using}), which leads to a $t_*^1$ leading-order term plus a $t_*^{1-\eta}$ remainder. We will show that this leading-order term is given by $t_* h_{b,\rms 1}(\dot\scal)$ for some $\scal^{(-1)}\in\scalspace_1$; this has the interpretation of a metric perturbation due to an infinitesimal change of the linear momentum of the final black hole, as we shall discuss in detail in~\S\ref{SssD2BoostNo} (following the rough discussion in~\S\ref{SssINElim}).

\subsubsection{Boost parameter}
\label{SssD2BoostGet}

The precise result on the existence of a linear (in $t_*$) leading-order term is as follows.

\begin{prop}[Extraction of the boost parameter]
\label{PropD2Boost}
  Let $\alpha_\cK\in(1,1+\eps_\cK)$ and $\alpha_+>\alpha_\cK$. Then for $f\in\Hb^{\infty,\ \bigl(\la\cE_\sscri^\cC+1\ra',4+\eps_\sscri\bigr),\ \alpha_++2,\ \alpha_\cK}(\Omega_*)^{\bullet,-}$, the forward solution of $L u=f$ can be written as
  \begin{equation}
  \label{EqD2Boost}
    u = \chi_\cK h_{b,\rms 1}^{\leq 1}(\dot\scal) + \tilde u,\quad
    \dot\scal\in\scalspace_1,\ 
    \tilde u \in \Hb^{\infty,\ \bigl(\la\cE_\sscri^\cC\ra,3+\eps_\sscri\bigr),\ 1-\eps,\ \alpha_\cK-2}(\Omega_*)^{\bullet,-}\ \forall\,\eps>0,
  \end{equation}
  where we recall $h_{b,\rms 1}^{\leq 1}$ from~\eqref{EqWG0GenStateNot} and \eqref{EqWG0hs1breve}. Here $\dot\scal$ and $\tilde u$ depend continuously on $f$.
\end{prop}

Note that $\tilde u$ has sub-linear growth in $t_*$, and thus is of lower order (in the sense of decay as $t_*\to\infty$) than the leading-order term $t_* h_{b,\rms 1}(\dot\scal)$ of $u$. Unlike in~\eqref{EqD1Alm}, the $\iota^+$-decay rate is not $\alpha_\cK-\eps$ for all $\eps>0$ (which would be $>1$): it is limited by $1$ due to transport from $\scri^+$, and this limitation is also consistent with the expectation that $u$ should feature linearized Kerr metrics $\dot g_b^\Ups(\dot b)$, $\dot b=(\dot\bhm,\dot\bha)$, in its late-time asymptotics (which have $\iota^+$-order $1$ when $\dot\bhm\neq 0$).

As a preparation for the proof, we recall for the augmented operator $\wt{L_b}(\sigma)$ from~\eqref{EqAdmLoDef} the invertibility statement at zero energy from Proposition~\ref{PropAdmLoDef0}, and moreover record:

\begin{lemma}[Augmented operator near zero energy, I]
\label{LemmaD2wtLb}
  For $\sigma=|\sigma|\hat\sigma$ with $|\sigma|\leq c$ and $\hat\sigma\in e^{i[0,\pi]}$, the operator $\wt{L_b}(\sigma)-\wt{L_b}(0)$ is (uniformly in $|\sigma|$ and $\hat\sigma$) of class
  \begin{equation}
  \label{EqD2BoostwtLbDiff}
    \begin{pmatrix}
      \rho_\tface^2\rho_\zface\Diffb^1 & H_{\bop,|\sigma|}^{k,\ 3-\eps,\ 2+\eps_\ind-\eps,\ 1-\eps} & H_{\bop,|\sigma|}^{k,\ 3-\eps,\ 2+\eps_\ind-\eps,\ 1-\eps} \\ 0 & 0 & 0 \\ 0 & 0 & 0
    \end{pmatrix}
  \end{equation}
  on $X_\scbtop=[[0,1]_{|\sigma|}\times X;\{0\}\times\pa X]$ for all $\eps>0$; here we use the $H_{\bop,|\sigma|}$-norm defined in~\eqref{EqAdmLoImfHb2}. This is true also for all of its derivatives along $\sigma\pa_\sigma$.
\end{lemma}
\begin{proof}
  For the $(1,1)$-entry, this follows from~\eqref{EqAdmLoUnifDiff} since $\Diff_\scbtop\subset\Diff_\bop$; and $[\sigma\pa_\sigma,\cdot]$ maps $\rho_\tface^2\rho_\zface\Diffb^2$ (or, as a matter of fact, any other polynomially weighted version of $\Diffb^2$) into itself. For the $(1,2)$- and $(1,3)$-entries, we use Lemma~\ref{LemmaAdmLob}.
\end{proof}

\begin{proof}[Proof of Proposition~\usref{PropD2Boost}]
  Proposition~\ref{PropD1Alm} gives
  \[
    u \in \Hb^{\infty,\ \bigl(\la\cE_\sscri^\cC\ra,3+\eps_\sscri\bigr),\ 1-\eps,\ \alpha'_\cK-2}(\Omega_*)^{\bullet,-}
  \]
  for all $\eps>0$ and $\alpha'_\cK<1$. Therefore, using again~\eqref{EqDAdmFw},
  \[
    L_b u = f' := f - (L-L_b)u \in \Hb^{\infty,\ \bigl(\la\cE_\sscri^\cC+1\ra',4+\eps_\sscri\bigr),\ \tilde\alpha_++2,\ \tilde\alpha_\cK}(\Omega_*)^{\bullet,-},
  \]
  where one can take any $\tilde\alpha_+\leq\min((1-\eps)+1-\eps,\alpha_+)$, for any $\eps>0$, and $\tilde\alpha_\cK=\min(\alpha'_\cK+\eps_\cK,\alpha_\cK)=\alpha_\cK$. We may reduce $\alpha_+$ so that $\tilde\alpha_+=\alpha_+$ but still $\alpha_+>\alpha_\cK$, and we may moreover arrange for $\alpha_+\in(\alpha_\cK,1+\eps_\ind)$ (which implies $\alpha_+<\alpha_\cK+\eps_\ind$). In summary,
  \begin{equation}
  \label{EqD2BoostEq}
    L_b u = f' \in \Hb^{\infty,\ \bigl(\la\cE_\sscri^\cC+1\ra',4+\eps_\sscri\bigr),\ \alpha_++2,\ \alpha_\cK},\quad \alpha_+\in(\alpha_\cK,1+\eps_\ind).
  \end{equation}

  \pfstep{Step~1. Solution near $\scri^+$.} We begin the analysis of~\eqref{EqD2BoostEq} as in the proof of Proposition~\ref{PropD1Alm}. Proposition~\ref{PropDScriFormal} produces $u_\sscri\in\Hb^{\infty,\ \bigl(\la\cE_\sscri^\cC\ra,3+\eps_\sscri\bigr),\ \bigl((1,0),\alpha_+\bigr),\ \infty}$ with support disjoint from $\cK^+$ such that
  \begin{subequations}
  \begin{equation}
  \label{EqD2BoostRem30}
    u'_\flat := u - u_\sscri\implies L_b u'_\flat =: f'_\flat \in \Hb^{\infty,\ 4+\eps_\sscri,\ \bigl((3,0),\alpha_++2\bigr),\ \alpha_\cK}(\Omega;S^2\cT^*)^{\bullet,-}.
  \end{equation}
  Note that $f'_\flat$ does not decay at rate $\alpha_++2$ at $\iota^+$; rather, it has a $t_*^{-3}$ leading-order term there with coefficient
  \begin{equation}
  \label{EqD2BoostRem302}
    f_+=f_+(R,\omega):=(t_*^3 f'_\flat)_{\iota^+} \in \Hb^{\infty,\ 4+\eps_\sscri,\ \infty}(\iota^+)
  \end{equation}
  \end{subequations}
  whose support is disjoint from $\cK^+$. Thus, $f'_\flat-\chi_\iota t_*^{-3}f_+\in\Hb^{\infty,\ 4+\eps_\sscri,\ \alpha_++2,\ \alpha_\cK}$ where $\chi_\iota$ localizes near $\iota^+$.

  \pfstep{Step~2. Solution near $\iota^+$.} We solve away $t_*^{-3}f_+$ using Proposition~\ref{PropipGr} for $\lambda=1$ (for which $N_{\iota^+}(\ubar L,\lambda)^{-1}$ is defined by Proposition~\ref{PropipMero}\eqref{ItipMeroInv}); this produces
  \begin{equation}
  \label{EqD2Boost2u2}
    u_\iota\in\Hb^{\infty,\ \bigl(\la\cE_\sscri^\cC\ra,3+\eps_\sscri\bigr),\ 1-\eps,\ -\eps}(\Omega_*)^{\bullet,-}
  \end{equation}
  with support near $\iota^+\cup\cK^+$ (this membership following from~\eqref{EqipGru}, $\min\Re\cE_\cK\geq 0$, and $\lambda=1$) satisfying
  \[
    f'_{\iota,\flat} := \chi_\iota t_*^{-3}f_+ - L_b u_\iota \in \Hb^{\infty,\ (\cE_{\iota^+,\sscri}^\tot+2,\,4+\eps_\sscri),\ 3+\cE_\ind,\ 4+\eps_\ind-\eps}(\Omega_*)^{\bullet,-},
  \]
  the point being that this remainder has order $\geq 3+\eps_\ind>3$ at $\iota^+$; we thus only need to record that it has weight $\alpha_+$ for the present arguments. The $\scri^+$-index set of $f'_{\iota,\flat}$ is again nontrivial, \emph{but} it satisfies $\Re\cE_{\iota^+,\sscri}^\tot+2\geq 3$; Proposition~\ref{PropDScriFormal}---specifically, the last part for any $C_0<2$---thus produces $u'_\sscri\in\Hb^{\infty,\ \bigl(\la\cE_\sscri^\cC\ra,3+\eps_\sscri\bigr),\ \alpha_+,\ \infty}(\Omega_*)^{\bullet,-}$ with
  \[
    f'_{\iota,\flat} - L_b u'_\sscri \in \Hb^{\infty,\ 4+\eps_\sscri,\ \alpha_++2,\ 4+\eps_\ind-\eps}(\Omega_*)^{\bullet,-}.
  \]
  The difference $u_\flat:=u_\flat'-u_\iota-u'_\sscri$ thus solves an equation
  \[
    L_b u_\flat = (f'_\flat - \chi_\iota t_*^{-3}f_+) + (f'_{\iota,\flat} - L_b u'_\sscri) =: f_\flat \in \Hb^{\infty,\ 4+\eps_\sscri,\ \alpha_++2,\ \alpha_\cK}(\Omega_*)^{\bullet,-}.
  \]

  \pfstep{Step~3. Improved decay via spectral theory.} We split $u_\flat=u_{\rm lo}+u_{\rm hi}$ as in~\eqref{EqD1Almu}--\eqref{EqD1Almu2}. The high-energy piece again has arbitrary $t_*$-decay, so
  \[
    u_{\rm hi} \in \Hb^{\infty,\ 1-\eps,\alpha_+,\alpha_\cK}(M';S^2\cT^*)\quad\forall\,\eps>0.
  \]
  For the low-energy piece, we first use~\eqref{EqTFHbLo} to obtain
  \[
    \bigl(\wh{f_\flat}(\sigma)\bigr)\big|_{\sigma\in\pm[0,c]} \in \Hb^{\infty,\ \alpha_++2-\eps,\ \alpha_++1,\ ((0,0),\alpha_\cK-1)}(X_\scbtop^\pm)\quad\forall\,\eps>0.
  \]

  \pfsubstep{Step~3.1.}{Zero energy piece.} Unlike in~\S\ref{SsD1Alm}, we now need to control the contribution of the $\zface$-leading-order term $\wh{f_\flat}(0)\in\Hb^{\infty,\ \alpha_++1}(X;S^2\cT^*_X)$. To this effect, we use Proposition~\ref{PropAdmLoDef0} (and $\alpha_+-1\in(0,\eps_\ind)$) to determine
  \begin{equation}
  \label{EqD2Boost0Inv}
    \bigl(\hat u_{{\rm reg},1}(0),\ \hat b_1(0),\ \hat\scal_1(0)\bigr) := \wt{L_b}(0)^{-1}\bigl(\wh{f_\flat}(0,\cdot),\ 0,\ 0\bigr),
  \end{equation}
  so $\hat u_{{\rm reg},1}(0)\in\Hb^{\infty,\alpha_+-1}(X;S^2\cT^*_X)$, $\hat b_1(0)\in\C^4$, $\hat\scal_1(0)\in\scalspace_1$. Denote by $\pi_1\colon\sD'(X^\circ;S^2\cT^*_X)\oplus\C^4\oplus\scalspace_1\to\sD'(X^\circ;S^2\cT^*_X)$ the projection onto the first summand, and fix a smooth cutoff $\chi_\zface$ that is $1$ near $\zface$ and vanishes near $\scface$ (such as $\chi_\zface=\chi_\zface(\frac{\sigma}{\rho})\in\CIc((-2,2))$, equal to $1$ on $[-1,1]$). We shall argue that~\eqref{EqD2Boost0Inv} gives an approximate description of $\wt{L_b}(\sigma)^{-1}(\wh{f_\flat}(\sigma),0,0)$ near zero energy. To this end, we first estimate the first component of $(\wh{f_\flat}(\sigma),0,0)-\wt{L_b}(\sigma)\bigl(\chi_\zface\hat u_{{\rm reg},1}(0),\hat b_1(0),\hat\scal_1(0)\bigr)$ using the re-writing
  \begin{align}
  \label{EqD2Boost0InvErr1}
    \hat f_2(\sigma) &:= \wh{f_\flat}(\sigma) - \pi_1\wt{L_b}(\sigma)\bigl( \chi_\zface \hat u_{{\rm reg},1}(0),\ \hat b_1(0),\ \hat\scal_1(0) \bigr) \\
    & = (1-\chi_\zface)\wh{f_\flat}(\sigma) + \chi_\zface\bigl(\wh{f_\flat}(\sigma)-\wh{f_\flat}(0)\bigr) + \Bigl(\chi_\zface\wh{f_\flat}(0) - \pi_1\wt{L_b}(0)\bigl(\chi_\zface u_{{\rm reg},1}(0),\ \hat b_1(0),\ \hat\scal_1(0)\bigr)\Bigr) \nonumber\\
    & \quad\qquad - \pi_1\bigl(\wt{L_b}(\sigma)-\wt{L_b}(0)\bigr)(\chi_\zface\hat u_{{\rm reg},1}(0),\ \hat b_1(0),\ \hat\scal_1(0)\bigr). \nonumber
  \end{align}
  The first term vanishes in a neighborhood of $\zface$ and thus lies in $\Hb^{\infty,\ \alpha_++2-\eps,\ \alpha_++1,\ \infty}(X_\scbtop^\pm)$. The second summand vanishes at $\zface$ and thus lies in $\Hb^{\infty,\ \alpha_++2-\eps,\ \alpha_++1,\ \alpha_\cK-1}$ (recall here that $\alpha_\cK-1\in(0,1)$). The third term, by virtue of~\eqref{EqD2Boost0Inv}, equals
  \[
    -[\wh{L_b}(0),\chi_\zface]\hat u_{{\rm reg},1}(0) \in \rho_\scface^\infty\rho_\tface^2\rho_\zface^\infty\Diff_\bop^2(X_\scbtop^\pm) \Hb^{\infty,\alpha_+-1}(X) \subset \Hb^{\infty,\ \infty,\ \alpha_++1,\ \infty}(X_\scbtop^\pm).
  \]
  For the fourth term, we use Lemma~\ref{LemmaD2wtLb}. In view of $\chi_\zface\hat u_{{\rm reg},1}(0)\in\Hb^{\infty,\ \infty,\ \alpha_+-1,\ (0,0)}(X_\scbtop^\pm)$, it thus lies in
  \[
    \Hb^{\infty,\ \infty,\ \alpha_++1,\ (1,0)}(X_\scbtop^\pm) + \Hb^{\infty,\ 3-\eps,\ 2+\eps_\ind-\eps,\ 1-\eps}(X_\scbtop^\pm)
  \]
  for all $\eps>0$. Recalling from~\eqref{EqD2BoostEq} that $\alpha_+<1+\eps_\ind$, we have thus shown that
  \begin{subequations}
  \begin{equation}
  \label{EqD2Boost0InvErr12}
    \bigl(\hat f_2(\sigma)\bigr)\big|_{\sigma\in\pm[0,c]} \in \Hb^{\infty,\ \alpha_++2-\eps,\ \alpha_++1,\ \alpha_\cK-1}(X_\scbtop^\pm).
  \end{equation}

  Regarding the full output of $\wt{L_b}(\sigma)$, we note that
  \begin{equation}
  \label{EqD2Boost0InvErr}
    \bigl(\wh{f_\flat}(\sigma,\cdot),\ 0,\ 0\bigr) - \wt{L_b}(\sigma)\bigl(\chi_\zface\hat u_{{\rm reg},1}(0),\ \hat b_1(0),\ \hat\scal_1(0)\bigr) = \bigl(\hat f_2(\sigma,\cdot),\ 0,\ 0\bigr)
  \end{equation}
  \end{subequations}
  for sufficiently small $|\sigma|$ (depending only on the choice of $\chi_\zface$ and $f_1^*$, $f_2^*$ in~\eqref{EqAdmLofstar}), so for $|\sigma|\leq c$ if we reduce the value of $c$ (see~\eqref{EqAdmIFT}), as we may; this follows from the fact that $\supp((1-\chi_\zface)\hat u_{{\rm reg},1}(0))$ is disjoint from $\supp f_1^*$ and $\supp f_2^*$ when $|\sigma|$ is small.

  \pfsubstep{Step~3.2.}{Remaining piece.} In view of~\eqref{EqD2Boost0InvErr}, we have (for small $|\sigma|$)
  \begin{align*}
    &\bigl(\hat u_{{\rm reg},1}(\sigma),\ \hat b_1(\sigma),\ \hat\scal_1(\sigma)\bigr) := \wt{L_b}(\sigma)^{-1}\bigl(\wh{f_\flat}(\sigma,\cdot),\ 0,\ 0\bigr) \\
    &\qquad = \bigl(\chi_\zface\hat u_{{\rm reg},1}(0),\ \hat b_1(0),\ \hat\scal_1(0)\bigr) + \wt{L_b}(\sigma)^{-1}\bigl(\hat f_2(\sigma,\cdot),\ 0,\ 0\bigr).
  \end{align*}
  Using~\eqref{EqD2Boost0InvErr12} and \citeAF{Proposition~\ref*{PropDResLo}}, we can control the second summand completely analogously to~\eqref{EqD1AlmuregMem} (with $\alpha_+$ and $\alpha_\cK$ in place of $\tilde\alpha_+$ and $\tilde\alpha_\cK$, respectively). For the sum, we thus obtain
  \begin{equation}
  \label{EqD2BoosturegMem}
  \begin{split}
    \bigl(\hat u_{{\rm reg},1}(\sigma)\bigr)\big|_{\sigma\in\pm[0,c]} &\in \Hb^{\infty,\ 1-\eps,\ \alpha_+-1,\ ((0,0),\alpha_\cK-1)}(X_\scbtop^\pm)\quad\forall\,\eps>0, \\
    \hat b_1 &\in \Hb^{\infty,\ ((0,0),\alpha_\cK-1)}(\pm[0,c),|\tfrac{\dd\sigma}{\sigma}|;\C^4), \\
    \hat\scal_1 &\in \Hb^{\infty, ((0,0),\alpha_\cK-1)}(\pm[0,c),|\tfrac{\dd\sigma}{\sigma}|;\scalspace_1);
  \end{split}
  \end{equation}
  and the restrictions to $\zface$ (for $\hat u_{{\rm reg},1}$) and to $\sigma=0$ (for $\hat b_1$ and $\hat\scal_1$) do not depend on the choice of ``$\pm$'' sign: our construction shows that they are given by~\eqref{EqD2Boost0Inv}.

  We now mirror the arguments following~\eqref{EqD1AlmuregMem} (and recall the notation~\eqref{EqD1Almu2}). Using~\eqref{EqD2BoosturegMem} and~\eqref{EqTFHbInvLo}, we get
  \[
    u_{\rm reg} \in \Hb^{\infty,\ 1-\eps,\ \alpha_+,\ \alpha_\cK}(M')\quad\forall\,\eps>0.
  \]
  Next, the memberships~\eqref{EqD1AlmAug} and~\eqref{EqD2BoosturegMem} imply
  \begin{align*}
    \cF(\pa_{t_*}\dot g) &\in \Hb^{\infty,\ 1-\eps,\ 1-\eps,\ ((0,0),\alpha_\cK-1)}(X_\scbtop^\pm), \\
    \cF(\pa_{t_*}^2 h_{\rms 1}) &\in \Hb^{\infty,\ 1-\eps,\ 2-\eps,\ ((0,0),\alpha_\cK-1)}(X_\scbtop^\pm).
  \end{align*}
  Therefore, for $u_{\rm sing}=\dot g+h_{\rms 1}$ we have
  \begin{subequations}
  \begin{equation}
  \label{EqD2Boostpa2sing}
    \cF(\pa_{t_*}^2 u_{\rm sing}) \in \Hb^{\infty,\ 1-\eps,\ 2-\eps,\ ((0,0),\alpha_\cK-1)}(X_\scbtop^\pm);
  \end{equation}
  and for later use, we record that since $\cF(\pa_{t_*}^2\dot g)(\sigma)=-i\sigma\cF(\pa_{t_*}\dot g)(\sigma)$ vanishes at $\sigma=0$, we have
  \begin{equation}
  \label{EqD2Boostpa2sing2}
    \cF(\pa_{t_*}^2 u_{\rm sing})(0)=\cF(\pa_{t_*}^2 h_{\rms 1})(0)=\cF(\pa_{t_*}^2 h_{b,\rms 1}^{\leq 1,\aug})(\hat\scal_1(0));
  \end{equation}
  \end{subequations}
  we use here that $u_{\rm sing}=-(u_{\rm hi}+u_{\rm reg})$ restricts to $\Omega_-:=\cl_{M'}\{\ft_*<1\}$ as an element of the space $\bar H_\bop^{\infty,\ 1-\eps,\ \alpha_+,\ \alpha_\cK}$, so $\pa_{t_*}u_{\rm sing}|_{\Omega_-}\in\bar H_\bop^{\infty,\ 1-\eps,\ \alpha_++1,\ \alpha_\cK+1}$, Lemma~\ref{LemmaD1IntFT} (with $\beta_\sscri=1-\eps$, $\beta_\iota=2-\eps$, and $\beta_\cK=\alpha_\cK-1$) and~\eqref{EqD2Boostpa2sing} yield
  \begin{equation}
  \label{EqD2Boostpa1sing}
    \pa_{t_*}u_{\rm sing} \in \Hb^{\infty,\ 1-\eps,\ 2-\eps,\ ((0,0),\alpha_\cK-1)}(M').
  \end{equation}
  Integrating this using Lemma~\ref{LemmaD1Intb} yields
  \begin{equation}
  \label{EqD2Boostusing}
    u_{\rm sing} \in \Hb^{\infty,\ 1-\eps,\ 1-\eps,\ ((-1,0),\alpha_\cK-2)}(M').
  \end{equation}

  \pfsubstep{Step~3.3.}{Computation of the leading-order term.} We first present a simple but not fully rigorous argument to compute the $t_*^1$-term (corresponding to the element $(-1,0)$ of the $\cK^+$-index set of~\eqref{EqD2Boostusing}) $t_* u_{\cK^+}(x)$ of $u_{\rm sing}$, where $u_{\cK^+}=(t_*^{-1}u_{\rm sing})|_{\cK^+}\in\Hb^{\infty,2-\eps}(X)$. Namely, it arises from the most singular term of $\cF u(\sigma)$ at $\sigma=0$, which is given by $-\sigma^{-2}h_{b,\rms 1}(\hat\scal_1(0))$ (cf.\ \eqref{EqAdmLoNaive}), defined as the distributional limit as $\Im\sigma\searrow 0$. The inverse Fourier transform of this is $t_* H(t_*)h_{b,\rms 1}(\hat\scal_1(0))$, so $u_{\cK^+}=h_{b,\rms 1}(\hat\scal_1(0))$.

  A rigorous argument goes as follows. First, note that $u_{\cK^+}$ is the restriction to $\cK^+$ of $\pa_{t_*}u_{\rm sing}$ in~\eqref{EqD2Boostpa1sing}, which in turn is the total $t_*$-integral of $\pa_{t_*}^2 u_{\rm sing}$ (which in compact spatial regions has $\la t_*\ra^{-\alpha_\cK}$-decay by~\eqref{EqD2Boostpa2sing} and using that $\pa_{t_*}u_{\rm sing}$ vanishes at $t_*=-\infty$), i.e., it is given by~\eqref{EqD2Boostpa2sing2}. But by~\eqref{EqAdmLoImLot} and recalling $h_{b,\rms 1}^{\leq 1}=t_* h_{b,\rms 1}+\breve h_{b,\rms 1}^1$, the total integral of $\pa_{t_*}^2 h_{b,\rms 1}^{\leq 1,\aug}$ is equal to that of $\pa_{t_*}^2(\chi_\cK h_{b,\rms 1}^{\leq 1})$, which is equal to $\pa_{t_*}(\chi_\cK h_{b,\rms 1}^{\leq 1})|_{-\infty}^\infty=h_{b,\rms 1}$. Altogether, therefore, we find that
  \begin{equation}
  \label{EqD2BoostLot}
    u_{\rm sing} - \chi_\cK t_* h_{b,\rms 1}(\dot\scal) \in \Hb^{\infty,\ 1-\eps,\ 1-\eps,\ \alpha_\cK-2}(M'),\quad \dot\scal:=\hat\scal_1(0).
  \end{equation}
  We may replace $t_* h_{b,\rms 1}$ by $h_{b,\rms 1}^{\leq 1}$ up to an error $\chi_\cK\breve h_{b,\rms 1}^1(\dot\scal)$ of class $\chi_\cK\cA^1(X)\subset\Hb^{\infty,\ \infty,\ 1-\eps,\ -\eps}(M')$, which thus also lies in $\Hb^{\infty,\ 1-\eps,\ 1-\eps,\ \alpha_\cK-2}$. Adding~\eqref{EqD2BoostLot} and this to $u_{\rm hi}$ and $u_{\rm reg}$ yields the term $\tilde u$ in the decomposition~\eqref{EqD2Boost}, except for an imprecise $\scri^+$-decay order.

  \pfstep{Step~4. Recovery of sharp decay at $\scri^+$.} Near $\scri^+$, where $\chi_\cK$ vanishes, we have $L\tilde u=f$. We apply Proposition~\ref{PropDScriPhg} to this equation to deduce that $\tilde u$ has the precise $\scri^+$-decay claimed in~\eqref{EqD2Boost}. The proof is complete.
\end{proof}

\subsubsection{Undoing the boost}
\label{SssD2BoostNo}

We begin by explaining a particular way (compatible with the structure of gauge-fixed linearized Einstein operators) in which the leading-order term $\chi_\cK h_{b,\rms 1}^{\leq 1}(\dot\scal)$ in~\eqref{EqD2Boost} arises for forward solutions of $L$. To this end, let us first perform a computation for the linearization $L_{b_0}=L_{g_{b_0},g_{b_0}}=D_{g_{b_0}}\Ric+\delta_{g_{b_0},E^\cC}^*\delta_{g_{b_0},E^\Ups}\sfG_{g_{b_0}}$ around the Kerr metric $g_{b_0}$ that we are considering perturbations of. On the one hand, the linearization in $\scal$ of
\[
  \scalspace_1 \ni \scal \mapsto \Ric(g_{b_0,b_0,-\scal}) - \delta_{g_{b_0,b_0,-\scal},E^\cC}^* \Ups_{E^\Ups}(g_{b_0,b_0,-\scal},\,g_{b_0,b_0,-\scal}),\quad g_{b_0,b_0,-\scal} = (1-\chi_\cK)\phi_{-\scal}^*g_{b_0} + \chi_\cK g_{b_0},
\]
where we recalled the notation from~\eqref{EqDgPOU} as well as~\eqref{EqKBoVscal} and \eqref{EqKBoMap}, around $\scal=0$ maps
\begin{equation}
\label{EqD2BoostLin}
\begin{split}
  \dot\scal \mapsto -D_{g_{b_0}}\Ric\bigl( (1-\chi_\cK)\cL_{V(\dot\scal)}g_{b_0} \bigr) &= -D_{g_{b_0}}\Ric\Bigl((1-\chi_\cK)\delta_{g_{b_0}}^*\bigl((1-\chi_\cK^\flat)g_{b_0}\ubar g^{-1}\ubar\omega_{\rms 1}^{(0),\leq 1}(\dot\scal)\bigr)\Bigr) \\
    &= -D_{g_{b_0}}\Ric\Bigl((1-\chi_\cK)\delta_{g_{b_0}}^*\bigl(g_{b_0}\ubar g^{-1}\ubar\omega_{\rms 1}^{(0),\leq 1}(\dot\scal)\bigr)\Bigr) \\
    &= -D_{g_{b_0}}\Ric\Bigl( [\delta_{g_{b_0}}^*,\chi_\cK]\bigl(g_{b_0}\ubar g^{-1}\ubar\omega_{\rms 1}^{(0),\leq 1}(\dot\scal)\bigr)\Bigr)
\end{split}
\end{equation}
on $\Omega_*$; recall here that $\chi_\cK^\flat=0$ on $\supp(1-\chi_\cK)$, and $D_{g_{b_0}}\Ric\circ\delta_{g_{b_0}}^*=0$.

On the other hand, recalling that $h_{b_0,\rms 1}^{\leq 1}=\delta_{g_{b_0}}^*\omega_{b_0,\rms 1}^{(0),\leq 1}$ satisfies both the linearized Einstein equation and the linearized gauge condition (see Proposition~\ref{PropWG0Symm}\eqref{ItWG0SymmGrad}), we compute
\begin{equation}
\label{EqD2MetLin}
\begin{split}
  L_{b_0}\bigl(\chi_\cK h_{b_0,\rms 1}^{\leq 1}(\dot\scal)\bigr) &= D_{g_{b_0}}\Ric\bigl(\chi_\cK h_{b_0,\rms 1}^{\leq 1}(\dot\scal)\bigr) + \delta_{g_{b_0},E^\cC}^*[\delta_{g_{b_0},E^\Ups}\sfG_{g_{b_0}},\chi_\cK]h_{b_0,\rms 1}^{\leq 1}(\dot\scal) \\
    &= -D_{g_{b_0}}\Ric\Bigl( [\delta_{g_{b_0}}^*,\chi_\cK] \omega_{b_0,\rms 1}^{(0),\leq 1}(\dot\scal) \Bigr) + \delta_{g_{b_0},E^\cC}^*[\delta_{g_{b_0},E^\Ups}\sfG_{g_{b_0}},\chi_\cK]h_{b_0,\rms 1}^{\leq 1}(\dot\scal).
\end{split}
\end{equation}
Defining the (gauge-fixed Einstein) operator
\begin{subequations}
\begin{equation}
\label{EqD2tildeP}
\begin{split}
  \tilde P(h,\scal) &:= \Ric\Bigl( g_{b_0,b_0,-\scal} - [\delta_{g_{b_0}}^*,\chi_\cK]\bigl(\omega_{b_0,\rms 1}^{(0),\leq 1}(\scal)-g_{b_0}\ubar g^{-1}\ubar\omega_{\rms 1}^{(0),\leq 1}(\scal)\bigr) + h \Bigr) \\
    &\quad\quad - \delta_{g_{b_0,b_0,-\scal},E^\cC}^*\Bigl( \Ups_{E^\Ups}\bigl( g_{b_0,b_0,-\scal} + h, g_{b_0,b_0,-\scal} \bigr) - [\delta_{g_{b_0},E^\Ups}\sfG_{g_{b_0}},\chi_\cK]h_{b_0,\rms 1}^{\leq 1}(\scal) \Bigr),
\end{split}
\end{equation}
we can combine~\eqref{EqD2BoostLin} and \eqref{EqD2MetLin} into
\begin{equation}
\label{EqD2BoostRewrite}
  L_{b_0}\bigl(\chi_\cK h_{b_0,\rms 1}^{\leq 1}(\dot\scal)\bigr) = D_{(0,0)}\tilde P\bigl(\chi_\cK h_{b_0,\rms 1}^{\leq 1}(\dot\scal),0\bigr) = D_{(0,0)}\tilde P(0,\dot\scal).
\end{equation}
\end{subequations}
The interpretation of this equation is that the source term $D_{(0,0)}\tilde P(0,\dot\scal)$---which roughly speaking corresponds to a source term for gravitational waves arising from boosting of the ``initial'' metric $g_{b_0}$ along $\dot\scal$ and transitioning to the unboosted $g_{b_0}$ near $\cK^+$---generates a metric perturbation in which the final black hole is boosted along $-\dot\scal$, but without any initial boost. The crucial advantage of this re-interpretation of $\chi_\cK h_{b_0,\rms 1}^{\leq 1}$ is that while boosts are highly incompatible with the Kerr metric near $\cK^+$ (as they boost the black hole to move to a different point at future timelike infinity), they are well-behaved elsewhere (as discussed in particular in~\S\S\ref{SsKBo} and \ref{SsExBo}).

\begin{rmk}[Comparison with a naive attempt]
\label{RmkD2Naive}
  Since $h_{b,\rms 1}^{\leq 1}$ is a pure gauge perturbation, one may attempt to remove it by a simple gauge modification (but without changing the background metric from $g_{b_0}$ to $g_{b_0,b_0,-\scal}$). According to the computation~\eqref{EqD2MetLin}, the required gauge modification (i.e., the argument of $\delta_{g_{b_0},E^\cC}^*$) has good decay properties at $\iota^+\cap\supp\dd\chi_\cK$ (see~\eqref{EqD2CorrTheta} below); however, the metric perturbation $[\delta_{g_{b_0}}^*,\chi_\cK]\omega_{b_0,\rms 1}^{(0),\leq 1}=\dd\chi_\cK\otimes_s\omega_{b_0,\rms 1}^{(0),\leq 1}$ that this causes (in addition to the desired $\chi_\cK h_{b_0,\rms 1}^{\leq 1})$) is of size $\cO(1)$ at $\iota^+\cap\supp\dd\chi_\cK$, i.e., \emph{it does not decay}. In a nonlinear iteration scheme, this would mean that the metric in the next step would deviate from the Minkowski (or Kerr) metric to leading order at $\iota^+$, leading to a host of issues; e.g., $L_{b_0}$ would no longer be a good model on all of $\iota^+\cap\cK^+$. The issue, in a nutshell, is the poor decay of the gauge potential $\omega_{b_0,\rms 1}^{(0),\leq 1}$; this also makes it impossible to eliminate, say, $\tilde\chi h_{b_0,\rms 1}^{\leq 1}$ via a computation analogous to~\eqref{EqD2MetLin} when $\tilde\chi$ is a different cutoff function that equals $1$ near $\cK^+$, e.g., $\tilde\chi=\tilde\chi(t_*)$ which equals $1$ for large $t_*$. --- What the passage from $g_{b_0}$ to $g_{b_0,b_0,-\scal}$ does is to effectively avoid the need for any transition from an unboosted to a boosted frame altogether; when linking our present analysis to the initial value problem studied in Theorem~\ref{ThmExBoStab}, we may indeed start there with boosted \emph{initial} data, with the boost parameter ultimately chosen so as to avoid the emergence of nontrivial boosts in the late-time asymptotics.
\end{rmk}

Before proceeding, let us introduce notation for the terms appearing in~\eqref{EqD2tildeP}.

\begin{definition}[Correction terms \#1: boosts]
\label{DefD2Corr}
  For $\scal\in\scalspace_1$, and for the fixed cutoff $\chi_\cK\in\CI(M)$ from~\eqref{DefKBoCutoff} and~\eqref{EqDCutoffs}, we define
  \begin{align*}
    h^{(-1)}(\scal) &:= -[\delta_{g_{b_0}}^*,\chi_\cK]\bigl(\omega_{b_0,\rms 1}^{(0),\leq 1}(\scal) - g_{b_0}\ubar g^{-1}\ubar\omega_{\rms 1}^{(0),\leq 1}(\scal)\bigr), \\
    \vartheta^{(-1)}(\scal) &:= [\delta_{g_{b_0},E^\Ups}\sfG_{g_{b_0}},\chi_\cK]h_{b_0,\rms 1}^{\leq 1}(\scal).
  \end{align*}
\end{definition}

\begin{lemma}[Structural properties of correction terms]
\label{LemmaD2Corr}
  The tensor $h^{(-1)}(\scal)$ and the 1-form $\vartheta^{(-1)}(\scal)$ are supported on $\supp\dd\chi_\cK\subset\Omega$. Recalling that $\cE_\ind$ is an index set, independent of any parameters besides $\gamma^\Ups,e^\Ups,v^\cC,\gamma^\cC$ in Definitions~\usref{Def1Gauge} and \usref{Def1Symm}, with $\min\Re\cE_\ind\geq\eps_\ind$, we have
  \begin{align}
  \label{EqD2Corrh}
    h^{(-1)}(\scal)&\in\cA^{\infty,\ (1,1)\cup(1+\cE_\ind),\ \infty}, \\
  \label{EqD2CorrRich}
    D_{g^0}\Ric\bigl(h^{(-1)}(\scal)\bigr) &\in \cA^{\infty,\ (3,0)\cup(3+\cE_\ind),\ \infty}, \\
  \label{EqD2CorrTheta}
    \vartheta^{(-1)}(\scal) &\in \cA^{\infty,\ (2,0)\cup(2+\cE_\ind),\ \infty}
  \end{align}
  for any metric $g^0\in\CI(M;S^2\cT^*)$ that equals $\ubar g$ modulo $\rho_+\CI$. (Here $\cA=\cA(\Omega_*;\cT^*)^{\bullet,-}$ or $\cA(\Omega_*;S^2\cT^*)^{\bullet,-}$.)
\end{lemma}
\begin{proof}
  The operators $[\delta_{g_{b_0}}^*,\chi_\cK]=\dd\chi_\cK\otimes_s(\cdot)$ and $[\delta_{g_{b_0},E^\Ups}\sfG_{g_{b_0}},\chi_\cK]$ both lie in $\rho_\sscri^\infty\rho_+\rho_\cK^\infty\Diffb^0(M')$. Let us use the leading-order description in Proposition~\ref{PropWG0Symm}\eqref{ItWG0SymmStates} and the fact (from~\eqref{EqKMetDiff}) that $g_{b_0}\ubar g^{-1}\equiv I\bmod r^{-1}\CI=\rho_\sscri\rho_+\CI$ as an endomorphism of $\cT^*$: we then find (omitting $\scal$ from the notation) that on $\supp\dd\chi_\cK$ (and correspondingly only recording the weight at $\iota^+$),
  \[
    \underbrace{t_*\bigl(\omega_{b_0,\rms 1}^{(0)} - \ubar\omega_{\rms 1}^{(0)}\bigr)}_{\in\rho_+^{-1}\cA^{1+\cE_\ind}} + \underbrace{\breve\omega_{b_0,\rms 1}^{(0),1} - \breve{\ubar\omega}_{\rms 1}^{(0),1}}_{\cA^{(0,1)\cup\cE_\ind}} + \underbrace{(I-g_{b_0}\ubar g^{-1})}_{\in\rho_+\CI}\underbrace{\ubar\omega_{\rms 1}^{(0),\leq 1}}_{\in \rho_+^{-1}\CI} \in \cA^{(0,1)\cup\cE_\ind}.
  \]
  This implies~\eqref{EqD2Corrh}. Since $D_{g^0}\Ric\in\rho_+^2\Diffb^2$ near $\supp\dd\chi_\cK$, the membership~\eqref{EqD2CorrRich} is almost a direct consequence of~\eqref{EqD2Corrh}; we only need to show that the coefficient corresponding to the element $(3,1)$ of the index set $((1,1)\cup(1+\cE_\ind))+2$ (i.e., the $\rho_+^3\log\rho_+$-coefficient) vanishes. The full $(3,1)$-term can be computed using~\eqref{EgWG0SymmLog} to be a constant multiple of
  \begin{equation}
  \label{EqD2CorrNoLog}
  \begin{split}
    D_{\ubar g}\Ric\bigl([\ubar\delta^*,\chi_\cK]\log(\rho)\ubar\omega_{\rms 1}^{(0)}\bigr) &= -D_{\ubar g}\Ric\Bigl(\chi_\cK\ubar\delta^*\bigl(\log(\rho)\ubar\omega_{\rms 1}^{(0)}\bigr)\Bigr) \\
      &= -D_{\ubar g}\Ric\bigl( \chi_\cK [\ubar\delta^*,\log\rho]\ubar\omega_{\rms 1}^{(0)} \bigr) = -D_{\ubar g}\Ric\Bigl( \chi_\cK\frac{\dd\rho}{\rho}\otimes_s\ubar\omega_{\rms 1}^{(0)}\Bigr).
  \end{split}
  \end{equation}
  The left-hand side evidently vanishes outside of $\supp\dd\chi_\cK$, and the right hand side is the application of $D_{\ubar g}\Ric\in\rho_+^2\Diffb^2$ to an element of $\rho_+\CI$ (as a section of $\cT^*$), so this lies in $\rho_+^3\CI$ and thus does not have a logarithmic factor.

  For the membership~\eqref{EqD2CorrTheta}, we note that $h_{b_0,\rms 1}^{\leq 1}$ has index set $(1,0)\cup(1+\cE_\ind)$ at $\iota^+$, as follows from~\eqref{EqWG0hs1}--\eqref{EqWG0hs1breve}.
\end{proof}

Returning to the output of Proposition~\ref{PropD2Boost}, the key idea, explained already in~\S\ref{SssINElim}, for eliminating the term $\chi_\cK t_*h_{b,\rms 1}^{\leq 1}(\dot\scal)$ for solutions of general linearized operators $L_{g,g^0}$ (i.e., not just $L_{b_0}$) is as follows. The forward solution with source term $D_{(h,\scal)}\tilde P(0,\scal')$ generates an asymptotic boost with parameter $\dot\scal=\scal'$ in the sense of~\eqref{EqD2BoostRewrite} when $(h,\scal)=(0,0)$. For small $(h,\scal)$, the same is true with $\dot\scal\approx\scal'$, i.e., the (linear) map $\scal'\mapsto\dot\scal$ is a small perturbation of the identity and thus in particular surjective. (As already pointed out in~\S\ref{SssINElim}, this is conceptually very close to ideas used in \cite[Theorem~5.14]{HintzVasyKdSStability}.) See Proposition~\ref{PropD2Boost2} below for details.

\begin{definition}[Gauge-fixed Einstein operator, augmentation \#1]
\label{DefD2EinsteinAug}
  For small $\scal\in\scalspace_1$, for $b\in\R^4$ close to $b_0=(\bhm_0,\bha_0)$, and for $h\in\cX^\infty=\Hb^{\infty,\ \bigl(\la\cE_\sscri^\cC\ra,3+\eps_\sscri\bigr),\ (\cE_+,3+\eps_+),\ 2+\eps_\cK}(\Omega_*)$ with sufficiently small $\cX^d$-norm (for some large but fixed $d$, cf.\ Theorem~\usref{ThmDAdmReg}), set
  \begin{equation}
  \label{EqD2EinsteinAug}
  \begin{split}
    &P(h,\scal,b) := \Ric\bigl(g^0 + h^{(-1)}(\scal) + h\bigr) - \delta_{g^0,E^\cC}^*\bigl(\Ups_{E^\Ups}(g^0+h,g^0) - \vartheta^{(-1)}(\scal)\bigr), \\
    &\qquad g^0 := g_{b_0,b,-\scal} = (1-\chi_\cK)\phi_{-\scal}^*g_{b_0} + \chi_\cK g_b.
  \end{split}
  \end{equation}
  We denote its linearization in $h$ by
  \begin{equation}
  \label{EqD2EinsteinAugLin}
    L_{h,\scal,b} := D_{(h,\scal,b)}P(\cdot,0,0).
  \end{equation}
\end{definition}

\begin{rmk}[Gauge condition]
\label{RmkD2Gauge}
  We point out a minor subtlety in~\eqref{EqD2EinsteinAug}, which is that the spacetime metric $g^0+h^{(-1)}(\scal)+h$ is not the same as the metric $g^0+h$ appearing in the gauge condition. An alternative definition of $P$ as indicated in Remark~\ref{RmkD2Alt} would eliminate this discrepancy. Importantly, however, this discrepancy is captured by a finite-dimensional parameter (namely, $\scal$), and the gauge condition does concern the full gravitational wave tail $h$.
\end{rmk}

If we require\footnote{We do not require $(1,1)\in\cE_+$ for aesthetic reasons, as logarithmic terms at $\iota^+$ at order $(1,1)$ will only arise from $h^{(-1)}$ and other similarly explicit terms later on.} $\cE_+\supset 1+\cE_\ind$, then $h^{(-1)}(\scal)+h\in\bar H_\bop^{\infty,\ \bigl(\la\cE_\sscri^\cC\ra,3+\eps_\sscri\bigr),\ \bigl(\cE_+\cup(1,1),\,3+\eps_+\bigr),\ 2+\eps_\cK}(\Omega_*)$. For bookkeeping purposes, it is thus convenient to require
\begin{equation}
\label{EqD2EplusCond}
  j(\cE_+\cup(1,1))\subset\cE_+\cup(1,1)\quad\forall\, j\in\N.
\end{equation}
The term $h^{(-1)}(\scal)$ only affects the linearization of the Ricci curvature and thus contributes a term of class $\cA^{(\cE_+\cup(1,1))+2}\Diffb^2$ near $(\iota^+)^\circ$ (and which vanishes near $\scri^+$ and $\cK^+$). We conclude that the linearization $L_{h,\scal,b}$ has the same structure as in Proposition~\ref{PropDAdmLin} (with $\iota^+$-index set $\cE_+\cup(1,1)$). For small $h$, $\scal$, and $|b-b_0|$, the operator $L_{h,\scal,b}$ is thus admissible (i.e., Corollary~\ref{CorDAdm} holds verbatim), and hence Theorem~\ref{ThmDAdmReg} applies as well; and thus Propositions~\ref{PropD1Alm} and \ref{PropD2Boost} hold for $L=L_{h,\scal,b}$.

\begin{rmk}[Alternative definitions of $P$]
\label{RmkD2Alt}
  One could alternatively include $h^{(-1)}(\scal)$ also as a summand in the first argument of $\Ups_{E^\Ups}$ in~\eqref{EqD2EinsteinAug}, which would moreover necessitate adding to $\vartheta^{(-1)}(\scal)$ the counterterm $D_{g_{b_0},g_{b_0}}\Ups_{E^\Ups}(h^{(-1)}(\scal),0)$. The linearization of the resulting operator in $h$ would be equal to $L_{g,g^0}$ where $g=g^0+h^{(-1)}(\scal)+h$, and thus Theorem~\ref{ThmDAdmReg} would be immediately applicable as stated. While slightly cleaner in this sense, it would unnecessarily complicate the expression for $P$.
\end{rmk}

Now, if we are to consider $D_{(h,\scal,b)}P(0,\scal',0)$ as a source term designed to create an asymptotic term $t_*h_{b,\rms 1}(\dot\scal)$, we first need:

\begin{lemma}[Linearization of $P$ in the parameters, \#1]
\label{LemmaD2LinPar}
  For $\scal,b$, and $h$ as in Definition~\usref{DefD2EinsteinAug}, and recalling the cutoffs~\eqref{EqDCutoffs}, we can write
  \begin{equation}
  \label{EqD2LinParScal}
  \begin{split}
    &D_{(h,\scal,b)}P(0,\scal',0) = (1-\chi_\cK^\flat)\bigl(f^1_{h,\scal,b}(\scal') + \chi_\cK^\sharp f^2_{h,\scal,b}(\scal')\bigr), \\
    &\qquad f^1_{h,\scal,b}(\scal') \in \Hb^{\infty,\ \bigl(\la\cE_\sscri^\cC+1\ra',\,4+\eps_\sscri\bigr),\ (\cE_++3,\,6+\eps_+),\ 0}, \\
    &\qquad f^2_{h,\scal,b}(\scal') \in \hspace{5.5em}\Hb^{\infty,\ 0,\ (\tilde\cE_++2,\,6+\eps_+-\eps),\ 0}\quad\forall\,\eps>0,
  \end{split}
  \end{equation}
  on $\Omega_*$; here,
  \begin{equation}
  \label{EqD2LinPartildeE}
    \tilde\cE_+ := (1,0) \cup (1+\cE_\ind) \cup \Bigl(\cE_+ + \bigl((1,1)\cup(1+\cE_\ind)\bigr)\Bigr)
  \end{equation}
  where $\cE_\ind$ is an $\cE_+$-independent index set with $\min\Re\cE_\ind\geq\eps_\ind>0$. If $h$ vanishes for $\ft_*<1$, then so does~\eqref{EqD2LinParScal}.
\end{lemma}

\begin{rmk}[Splitting]
\label{RmkD2Splitting}
  The term $f^2_{h,\scal,b}$ has a prefactor $(1-\chi_\cK^\flat)\chi_\cK^\sharp$, which is equal to $1$ on the support of $\dd\chi_\cK$ but vanishes near $\scri^+$ and $\cK^+$. Its $\iota^+$-index sets is larger than the index set $\cE_++3$ of the term $f^1_{h,\scal,b}$; while the former includes $(3,0)$, the latter has $\min\Re\geq 4$. Since the element $(3,0)$ in the $\iota^+$-index set of source terms plays a mildly distinguished role (e.g., it will lead to logarithmic center-of-mass motions in Proposition~\ref{PropD4Par}), it is convenient to explicitly record the support properties of such order $(3,0)$ terms. Furthermore, if we included a term with $\iota^+$-index set $(3,0)$ in $f^1_{h,\scal,b}$, Proposition~\ref{PropDScriFormal} on formal solutions near $\scri^+$ would create error terms with index set $(3,1)$ at $\iota^+$ (which might lead to size $(\log t_*)^2$ center-of-mass motions); the splitting~\eqref{EqD2LinParScal} allows us to avoid having to face such terms.
\end{rmk}

\begin{proof}[Proof of Lemma~\usref{LemmaD2LinPar}]
  We set $g^0=g_{b_0,b,-\scal}$ as in~\eqref{EqD2EinsteinAug}. On the set $\chi_\cK^{-1}(1)$, $P(h,\scal,b)$ does not depend on $\scal$, and thus~\eqref{EqD2LinParScal} is supported in $\supp(1-\chi_\cK)$.

  \pfstep{Analysis near $(\iota^+)^\circ$.} We only record orders at $\iota^+$. The linearization $\dot g$ of $g=g^0+h^{(-1)}(\scal)+h$ in $\scal$ is a sum of terms of class $\rho_+\CI$ (from $g^0$, by~\eqref{EqKBo}) and $\cA^{(1,1)\cup(1+\cE_\ind)}$ (from $h^{(-1)}$, by~\eqref{EqD2Corrh}), and thus
  \begin{equation}
  \label{EqD2LinParDRic}
    D_g\Ric(\dot g) \in \rho_+^2\Bigl(\CI + \Hb^{\infty,\ \bigl((1,1)\cup\cE_+,\,3+\eps_+\bigr)}\Bigr)\Diffb^2(\dot g) \subset \Hb^{\infty,\bigl((3,1)\cup(\tilde\cE_++2),\,6+\eps_+-\eps\bigr)}\quad\forall\,\eps>0;
  \end{equation}
  but the logarithmic $(3,1)$-term is, in fact, absent by~\eqref{EqD2CorrRich} (with $D_g\Ric-D_{g^0}\Ric$ contributing only lower-order terms), so the index set is $\tilde\cE_++2$ simply.

  Consider next the linearization of
  \begin{equation}
  \label{EqD2LinParGauge}
    \delta_{g^0,E^\cC}^*\bigl(\Ups_{E^\Ups}(g^0+h,g^0)-\vartheta^{(-1)}(\scal)\bigr)
  \end{equation}
  in $\scal$. The operator $\delta_{g^0,E^\cC}^*\in\rho_+\Diffb^1$ is equal to $\delta_{\ubar g,E^\cC}^*\in\rho_+\Diffb^1$ to leading order at $(\iota^+)^\circ$, and its principal symbol is independent of the metric altogether; therefore, its dependence on $\scal$ arises only at the level $\rho_+^2\Diffb^0$. Computing $\Ups_{E^\Ups}(g^0+h,g^0)$ in Cartesian coordinates on $\R^4$, one finds that it gains a power of $\rho_+$ relative to $h$ (cf.\ \eqref{EqExOpLinEC} and \eqref{EqExOpLinCgg0}, where the role of $\iota^+$ is played by $I^0$), so $\Ups_{E^\Ups}(g^0+h,g^0)\in\Hb^{\infty,(\cE_++1,\,4+\eps_+)}$; and since $\rho_+\CI$-contributions to $g^0$ only enter at sub-leading order, the linearization of $\Ups_{E^\Ups}(g^0+h,g^0)$ in $\scal$ is of class $\Hb^{\infty,(\cE_++2,5+\eps_+)}$. Recalling~\eqref{EqD2CorrTheta}, the linearization of~\eqref{EqD2LinParGauge} in $\scal$ thus lies in $\Hb^{\infty,\bigl((3,0)\cup(3+\cE_\ind)\cup(\cE_++3),\,6+\eps_+\bigr)}$, which is consistent with the $\iota^+$-order of~\eqref{EqD2LinParScal}.

  \pfstep{Analysis near $\scri^+$.} In the set where $\chi_\cK=0$, we have
  \[
    P(h,\scal,b) = P(\phi_{-\scal}^*g_{b_0}+h, \phi_{-\scal}^*g_{b_0}),
  \]
  which lies in $\Hb^{\infty, \bigl(\la\cE_\sscri^\cC+1\ra',\,4+\eps_\sscri\bigr),\ (\cE_++2,\,5+\eps_+)}$ (omitting the $\cK^+$-order), and thus so does its linearization in $\scal$; in fact, this linearization has order $(\cE_++3,6+\eps_+)$ at $\iota^+$ since the restriction of $\phi_{-\scal}^*g_{b_0}\in\CI$ to $\iota^+$ is independent of $\scal$.
\end{proof}

In the stability problem, $h$ is not supported in $\Omega_*$---unless the evolving metric is equal to $g_{b_0}$ outside of $\Omega_*$, which can only happen when the initial data are equal to those of $g_{b_0}$ outside of a compact set (see also Remark~\ref{RmkExBoCpt}). But without the support condition on $h$, Propositions~\ref{PropD1Alm} and \ref{PropD2Boost} do not apply as stated to the source term $D_{(h,\scal,b)}P(0,\scal',0)$. One resolution is to use an extension of Theorem~\ref{ThmDAdmReg} to initial value problems with trivial data at $\ft_*=1$ (and then the subsequent asymptotic analysis is unaffected, except for the need to localize to large $t_*$). A solution that is more appropriate for our stability problem is instead to recall from Theorem~\ref{ThmExBoStab} the tensors $h_{-\scal}$ which near $\scri^+\cap\{\ft_*\leq 2\}$ solve $P(\phi_{-\scal}^*g_{b_0}+h_{-\scal},\phi_{-\scal}^*g_{b_0})=0$, with suitable initial data at the initial Cauchy hypersurface $\Sigma_\IVP$; here we work in the exterior domain~\eqref{EqExDomain} with $T$ large enough (but fixed) so that $\{\ft_*\leq 2\}$ is contained in it.\footnote{Strictly speaking, Theorem~\ref{ThmExBoStab} as stated only constructs $h_{-\scal}$ for large $r$. But standard finite-time solvability results for quasilinear wave equations on compact spacetime domains with spacelike boundary hypersurfaces produce $h_{-\scal}$ also for bounded $r$ and thus in the full region $\{t_\IVP\geq 0,\ \ft_*\leq 2\}$, provided initial data are given for it on all of $\Sigma_\IVP$; this is of course the case for the global stability problem we are studying. (See also the proof of Theorem~\ref{ThmSt}.)} With the aim of extending $h_{-\scal}$ to the forward cone, fix then
\begin{subequations}
\begin{equation}
\label{EqD2PFwdCutoff}
  \chi_-=\chi_-(\ft_*)\in\CI(\R),\quad \chi_-|_{(-\infty,\frac74]}=1,\ \chi_-|_{[2,\infty)}=0,
\end{equation}
and define
\begin{equation}
\label{EqD2PFwd}
  P'(h,\scal,b) := P(\chi_-h_{-\scal}+h,\scal,b).
\end{equation}
\end{subequations}
Since for $\ft_*\leq\frac74$ we have $P'(0,\scal,b)=P(\phi_{-\scal}^*g_{b_0}+h_{-\scal},\phi_{-\scal}^*g_{b_0})=0$, the task for the stability problem is indeed to find \emph{forward} solutions $h$, i.e., $\supp h\subset\{\ft_*\geq 1\}$, of the nonlinear equation $P'(h,\scal,b)=0$ (or more elaborate versions thereof). More generally, if $h$ vanishes for $\ft_*\leq c\in[1,\frac74]$, then so does $P'(h,\scal,b)$. The linearizations of $P'$ in $h$, $\scal$, and $b$ have the same structure as those of $P$ (and differ only on $\supp\chi_-\subset\{\ft_*\leq 2\}$)---including, crucially, their property of being supported in $\ft_*\geq 1$---and hence our arguments are unaffected by a switch from $L_{h,\scal,b}$ to $L'_{h,\scal,b}:=D_{(h,\scal,b)}P'(\cdot,0,0)$. \emph{For simplicity of notation, however, we shall for now work with $h$ vanishing in $\{\ft_*<1\}$, i.e.,
\begin{equation}
\label{EqD2hSupp}
  h \in \cX^\infty := \Hb^{\infty,\ \bigl(\la\cE_\sscri^\cC\ra,3+\eps_\sscri\bigr),\ (\cE_+,3+\eps_+),\ 2+\eps_\cK}(\Omega_*)^{\bullet,-},
\end{equation}
but do not include $\chi_- h_{-\scal}$ in the definition of the gauge-fixed Einstein operator;} we will restore $\chi_- h_{-\scal}$ only at the end (see Definition~\ref{DefD6Aug5} and Corollary~\ref{CorD6Impr}).

Another issue is that the $\iota^+$-decay of $f_{h,\scal,b}^2$ in~\eqref{EqD2LinParScal} is \emph{equal} to $3$ (due to $(1,0)\in\tilde\cE_+$), which is weaker than what is allowed in the statement Proposition~\ref{PropD2Boost}; the proof however does deal with $(3,0)$-terms at $\iota^+$ (cf.\ \eqref{EqD2BoostRem30}), so no further arguments are needed here:

\begin{prop}[Elimination of boosts]
\fakephantomsection
\label{PropD2Boost2}
  \begin{enumerate}
  \item\label{ItD2Boost2Asy}{\rm (Asymptotics.)} Let $\alpha_\cK\in(1,1+\eps_\cK)$ and $\alpha_+>\alpha_\cK$. Then for
    \begin{equation}
    \label{EqD2Boost2Asyf}
    \begin{split}
      &f=f_1+(1-\chi_\cK^\flat)f_2, \\
      &\quad f_1\in\Hb^{\infty,\ \bigl(\la\cE_\sscri^\cC+1\ra',\,4+\eps_\sscri\bigr),\ \alpha_++2,\ \alpha_\cK}(\Omega_*)^{\bullet,-}, \\
      &\quad f_2\in\Hb^{\infty,\ 4+\eps_\sscri,\ (3,0),\ 0}(\Omega_*;S^2\cT^*)^{\bullet,-},
    \end{split}
    \end{equation}
    the forward solution of $L_{h,\scal,b}u=f$ (where $L_{h,\scal,b}$ was defined in~\eqref{EqD2EinsteinAugLin}) can be written as
    \begin{equation}
    \label{EqD2Boost2}
      u = \chi_\cK h_{b,\rms 1}^{\leq 1}(\dot\scal(f)) + \tilde u,\quad
      \dot\scal(f)\in\scalspace_1,\ 
      \tilde u \in \Hb^{\infty,\ \bigl(\la\cE_\sscri^\cC\ra,3+\eps_\sscri\bigr),\ 1-\eps,\ \alpha_\cK-2}(\Omega_*)^{\bullet,-}\ \forall\,\eps>0.
    \end{equation}
    The terms $\dot\scal(f)$ and $\tilde u$ depend linearly and continuously on $f$.
  \item\label{ItD2Boost2Par}{\rm (Boost parameter.)} Define the linear map
    \begin{equation}
    \label{EqD2Boost2Par}
      \scal^{(-1)}_{h,b,\scal} \in \cL(\scalspace_1)
    \end{equation}
    to map $\scal'$ to the parameter $\dot\scal(f)$ in~\eqref{EqD2Boost2} for the source term $f=D_{(h,\scal,b)}P(0,\scal',0)$. Then $\scal_{h,b,\scal}^{(-1)}$ is invertible provided $h$, $b$, and $\scal$ are sufficiently small in $\cX^d$ (for some sufficiently large but fixed $d$), $\R^4$, and $\scalspace_1\cong\R^3$, respectively.
  \end{enumerate}
\end{prop}

For part~\eqref{ItD2Boost2Par}, the source terms one needs to consider are of the form~\eqref{EqD2Boost2Asyf} by Lemma~\ref{LemmaD2LinPar}.

\begin{proof}[Proof of Proposition~\usref{PropD2Boost2}]
  \pfstep{Part~\eqref{ItD2Boost2Asy}.} One solves away $f_1$ at $\scri^+$ using Proposition~\ref{PropDScriFormal} as usual; this leaves a remaining source term $f_{1,\flat}$ with orders $4+\eps_\sscri$ and $((3,0),\alpha_++2)$ at $\scri^+$ and $\iota^+$, respectively. The $(3,0)$-terms of $f_{1,\flat}$ and of $f_2$ at $\iota^+$ already arose in~\eqref{EqD2BoostRem30}--\eqref{EqD2BoostRem302}, including the condition that their supports are disjoint from $\cK^+$. Thus, the remaining arguments in the proof of Proposition~\ref{PropDScriFormal} apply without change. (Alternatively, one can solve away the leading-order term of $f_2$ directly using Proposition~\ref{PropipGr}, at which point the remaining source is of the class considered in Proposition~\ref{PropD2Boost}, so Proposition~\ref{PropD2Boost} applies directly; this point of view is useful in the context of Remark~\ref{RmkD2Boost2Gen} below.)

  \pfstep{Part~\eqref{ItD2Boost2Par}.} By~\eqref{EqD2BoostRewrite}, the map $\scal_{0,b_0,0}^{(-1)}$ is the identity. The continuous dependence statement in Theorem~\ref{ThmDAdmReg} implies the continuous dependence of $u$ in the space~\eqref{EqDAdmRegu} on the parameters $h,b,\scal$ (in a $k$-independent but sufficiently high-regularity space) and on the input $f$ in the space~\eqref{EqDAdmRegf}. This continuous dependence then persists throughout all subsequent arguments in Propositions~\ref{PropD1Alm} and \ref{PropD2Boost} as well as in the present proof, leading to the continuous dependence of $\dot\scal$ on $(h,b,\scal)$ when $\scal'$ ranges over the elements of a basis of $\scalspace_1\cong\R^3$. This implies that $\scal^{(-1)}_{h,b,\scal}$ is close to the identity map when $h,b,\scal$ are small, and thus invertible.
\end{proof}

\begin{rmk}[More general source terms]
\label{RmkD2Boost2Gen}
  One can relax the requirements on the form of $f$ in~\eqref{EqD2Boost2Asyf} to $f=f_1+f_2$ where $f_1$ is as in~\eqref{EqD2Boost2Asyf} and $f_2\in\Hb^{\infty,\ 4+\eps_\sscri,\ (3,0),\ 1+\eps_\cK}(\Omega_*)^{\bullet,-}$, thus allowing for possibly nontrivial but decaying behavior near $\cK^+$. To see that Proposition~\ref{PropD2Boost2} remains valid \emph{without any changes in the conclusions}, one simply solves away the $\rho_+^3$-leading-order term of $f_2$ using Proposition~\ref{PropipGr} (with $\lambda=1$, $\ell_\sscri=3+\eps_\sscri$, and $\ell_\cK=\eps_\cK-\eps>0$ for arbitrarily small $\eps>0$) and then applies Proposition~\ref{PropD2Boost} to control the forward solution produced by the remaining source term.
\end{rmk}

Given $f$ as in~\eqref{EqD2Boost2Asyf}, consider now the equation
\begin{subequations}
\begin{equation}
\label{EqD2Boost2Comb}
  D_{(h,\scal,b)}P\bigl(u,\ (\scal_{h,b,\scal}^{(-1)})^{-1}\dot\scal(f),\ 0\bigr) = f
\end{equation}
for $u$. Setting $\scal':=(\scal_{h,b,\scal}^{(-1)})^{-1}\dot\scal(f)$, this is equivalent to $L_{h,\scal,b}u=f-D_{(h,\scal,b)}P(0,\scal',0)$; the solution of this equation is given by~\eqref{EqD2Boost2} with boost parameter equal to $\dot\scal(f)-\scal_{h,b,\scal}^{(-1)}(\scal')=0$; therefore,
\begin{equation}
\label{EqD2Boost2CombSol}
  u \in \Hb^{\infty,\ \bigl(\la\cE_\sscri^\cC\ra,3+\eps_\sscri\bigr),\ 1-\eps,\ \alpha_\cK-2}(\Omega_*)^{\bullet,-}.
\end{equation}
\end{subequations}
Crucially, the $t_*$-growth (in compact spatial regions) of this is $t_*^{2-\alpha_\cK}=o(t_*)$; we have thus succeeded in eliminating the asymptotic Lorentz boost contribution to $u$. We describe the asymptotic behavior of this $u$ (for $f$ with the appropriate additional amount of decay) more precisely in what follows.

\subsection{Step~3: almost boundedness}
\label{SsD3Alm}

With the contribution $u_2$ in~\eqref{EqD2Boost2u2} being almost bounded (in fact, it is of size $\log t_*$ at $\cK^+$ due to the first term in~\eqref{EqipGr}), and with the linearly growing (in $t_*$) generalized zero energy state $h_{b,\rms 1}^{\leq 1}(\dot\scal)$ eliminated in~\eqref{EqD2Boost2Comb}--\eqref{EqD2Boost2CombSol}, the next threshold $t_*$-decay rate is $0$; there, black hole parameter changes and (logarithmic or stationary) shifts of the center-of-mass of the black hole will appear. Before getting to that point, we need to upgrade the $t_*^{1-\eps}$-bounds for forward solutions of $L_{h,\scal,b}$ (defined in~\eqref{EqD2EinsteinAugLin}) from~\eqref{EqD2Boost2CombSol} to $t_*^{+\eps}$-bounds:

\begin{prop}[Almost boundedness]
\label{PropD3Alm}
  Let $1<\alpha_\cK<\alpha_+<2$. Let $f$ be of the form\footnote{These conditions are compatible with Lemma~\ref{LemmaD2LinPar}, and indeed more permissive similarly to Remark~\ref{RmkD2Boost2Gen} (but with a stronger $\cK^+$-order of $f_2$ for compatibility with almost-boundedness of $u$).}
  \begin{equation}
  \label{EqD3Almf}
  \begin{split}
    &f=f_1+f_2, \\
    &\quad f_1\in\Hb^{\infty,\ \bigl(\la\cE_\sscri^\cC+1\ra',\,4+\eps_\sscri\bigr),\ \alpha_++2,\ \alpha_\cK}(\Omega_*)^{\bullet,-}, \\
    &\quad f_2\in\Hb^{\infty,\ 4+\eps_\sscri,\ \tilde\cE_++2,\ 2-\eps}(\Omega_*)^{\bullet,-}
  \end{split}
  \end{equation}
  for all $\eps>0$, where $\tilde\cE_+$ is given by~\eqref{EqD2LinPartildeE} (or any other index set with $\min\Re(\tilde\cE_+\setminus\{(1,0)\})>1$). Suppose that for the forward solution of $L_{h,\scal,b}u=f$, the leading-order term in~\eqref{EqD2Boost2} vanishes. Then
  \begin{equation}
  \label{EqD3Almu}
    u \in \Hb^{\infty,\ \bigl(\la\cE_\sscri^\cC\ra,\,3+\eps_\sscri\bigr),\ 1-\eps,\ \alpha_\cK-2}(\Omega_*)^{\bullet,-}\quad\forall\,\eps>0.
  \end{equation}
\end{prop}

The improvement over Proposition~\ref{PropD2Boost2} is that $\alpha_\cK$ can now be arbitrarily close to $2$. As follows from the proof below, the solution $u$ in fact has a partial expansion at $\iota^+$ (determined by the $\iota^+$-index set of $f_2$). We do not record it here since the $\rho_+^{1-\eps}$-decay recorded in~\eqref{EqD3Almu} is enough for the iterative argument for improving the $\cK^+$-decay at the present stage; as we shall see starting in~\S\ref{SsD6Better} below, we need more precise (partially polyhomogeneous) information on $u$ at $\iota^+$ only when we wish to extract terms in the partial expansion of $u$ at $\cK^+$ that are of size $\cO(t_*^{-1})$.

Given an arbitrary $f$ of the form~\eqref{EqD3Almf}, the assumptions of Proposition~\ref{PropD3Alm} are satisfied for $f-D_{(h,\scal,b)}P(0,(\scal_{h,b,\scal}^{(-1)})^{-1}\dot\scal(f),0)$ in place of $f$, i.e., for the solution~\eqref{EqD2Boost2CombSol} of~\eqref{EqD2Boost2Comb}. Note here that the two summands in~\eqref{EqD2LinParScal} have $\iota^+$-orders $\min(\cE_++3)\geq 4$ and $\tilde\cE_++2=(3,0)\cup(3+\cE_\ind)\cup(\cdots)$, where ``$\cdots$'' is an index set with minimal real part $\geq 4$. Thus, the requirement $\alpha_+<2$ ensures that merely conormal terms with $\rho_+^4$-decay (times possible logarithmic factors) can be absorbed into $f_1$ in~\eqref{EqD3Almf}.

We again use for $\wt{L_b}(\sigma)$ the zero energy invertibility (Proposition~\ref{PropAdmLoDef0}) and the following variant of Lemma~\ref{LemmaD2wtLb}:

\begin{lemma}[Augmented operator near zero energy, II]
\label{LemmaD3wtLb}
  The operator $(\wt{L_b}(\pm\varsigma)-\wt{L_b}(0))_{\varsigma\in[0,c]}$, as a single operator on $X_\scbtop$, is of class
  \[
  \begin{pmatrix}
    \rho_\tface^2\rho_\zface\Diffb^1 & \Hb^{\infty,\ 3-\eps,\ \cE_\ind+2,\ ((1,0),3+\eps_\ind-\eps)}(X_\scbtop) & \Hb^{\infty,\ 3-\eps,\ \cE_\ind+2,\ ((1,0),3+\eps_\ind-\eps)}(X_\scbtop) \\
    0 & 0 & 0 \\
    0 & 0 & 0
  \end{pmatrix}.
  \]
\end{lemma}
\begin{proof}
  We only need to argue for the memberships of the $(1,2)$- and $(1,3)$-entries; but these are immediate consequences of Lemma~\ref{LemmaAdmLob}.
\end{proof}

\begin{proof}[Proof of Proposition~\usref{PropD3Alm}]
  Suppose that
  \begin{equation}
  \label{EqD3AlmStart}
    u \in \Hb^{\infty,\ \bigl(\la\cE_\sscri^\cC\ra,3+\eps_\sscri\bigr),\ 1-\eps,\ \alpha'_\cK-2}(\Omega_*)^{\bullet,-}
  \end{equation}
  for all $\eps>0$ and for some $\alpha'_\cK\in(1,2)$; by Proposition~\ref{PropD2Boost2}, this holds for all $\alpha'_\cK<1+\eps_\ind$ with $\alpha'_\cK\leq\alpha_\cK$. Using~\eqref{EqDAdmFw} (which holds also for $L_{h,\scal,b}-L_b$ with additional logarithms in the $\iota^+$-index set) by the discussion after Definition~\ref{DefD2EinsteinAug}), this implies
  \begin{equation}
  \label{EqD3AlmOpDiff}
    (L_{h,\scal,b}-L_b)u \in \Hb^{\infty,\ \bigl(\la\cE_\sscri^\cC+1\ra',\,4+\eps_\sscri\bigr),\ \tilde\alpha_++2,\ \tilde\alpha_\cK}(\Omega_*)^{\bullet,-},
  \end{equation}
  where one can take any $\tilde\alpha_+\leq\min((1-\eps)+1-\eps,\alpha_+)=\alpha_+$ and $\tilde\alpha_\cK=\min(\alpha'_\cK+\eps_\cK,\alpha_\cK)$; to simplify notation, replace $\alpha'_\cK$ by $\min(\alpha'_\cK,\alpha_\cK-\eps_\cK)$, so $\tilde\alpha_\cK=\alpha'_\cK+\eps_\cK$. We then reduce $\alpha_+$ (if necessary) to arrange $\alpha_+<\tilde\alpha_\cK+\eps_\ind$. In summary,
  \begin{equation}
  \label{EqD3AlmEq}
  \begin{split}
    L_b u =: f' = f'_1 + f'_2,\quad
      &f'_1 \in \Hb^{\infty,\ \bigl(\la\cE_\sscri^\cC+1\ra',\,4+\eps_\sscri\bigr),\ \alpha_++2,\ \tilde\alpha_\cK}(\Omega_*)^{\bullet,-},\ \ f'_2:=f_2, \\
      &\ 1<\tilde\alpha_\cK=\alpha'_\cK+\eps_\cK<\alpha_+<\min(2,\tilde\alpha_\cK+\eps_\ind).
  \end{split}
  \end{equation}

  \pfstep{Step~1. Solution near $\scri^+$.} Applying Proposition~\ref{PropDScriFormal} to $f_1'$ (and with $\breve\cE_+=\emptyset$ and $\breve\ell_+=\alpha_+$) produces an index set $\cE_+^\sharp$, depending only on $\cE_\sscri^\cC$ (and with smallest elements $(1,0)$ and $(1+(1-e^\Ups)\gamma^\Ups,0)$) and
  \begin{subequations}
  \begin{equation}
  \label{EqD3Almu1scri}
    u_{1,\sscri}\in\Hb^{\infty,\ \bigl(\la\cE_\sscri^\cC\ra,3+\eps_\sscri\bigr),\ (\cE_+^\sharp,\alpha_+),\ \infty},
  \end{equation}
  with support contained in a neighborhood of $\scri^+$ (and in particular disjoint from $\cK^+$) such that
  \begin{equation}
  \label{EqD3Almu1scriErr}
    f'_{1,\flat} := f'_1 - L_b u_{1,\sscri} \in \Hb^{\infty,\ 4+\eps_\sscri,\ (\cE_+^\sharp+2,\alpha_++2),\ \tilde\alpha_\cK}
  \end{equation}
  \end{subequations}
  vanishes to high order at $\scri^+$. (The coefficients of the $\iota^+$-expansion of $f'_{1,\flat}$ have supports disjoint from $\cK^+$.) The remaining source term is $f'_{1,\flat}+f'_2$, which thus has $\iota^+$-index set $(\cE_+^\sharp\cup\tilde\cE_+)+2=(3,0)\cup(\cdots)$.

  \pfstep{Step~2. Solution near $\iota^+$.} The $\iota^+$-expansion of $f'_{1,\flat}+f_2'$ is a sum of terms
  \begin{equation}
  \label{EqD3AlmipTerms}
  \begin{split}
    &t_*^{-\lambda-2}(\log t_*)^k f_+^{(\lambda+2,k)}(R,\omega), \\
    &\quad (\lambda,k)\in\cE_+^\sharp\cup\tilde\cE_+,\quad \Re\lambda\leq\alpha_+<2, \\
    &\quad f_+^{(\lambda+2,k)}\in\Hb^{\infty,\ 4+\eps_\sscri,\ 2-\eps-(\lambda+2)}(\iota^+),\ \ \supp f_+^{(\lambda+2,k)}\cap(\iota^+\cap\cK^+)=\emptyset.
  \end{split}
  \end{equation}
  Note here that we express these terms using $t_*^{-1}=\rho_+\rho_\cK$ instead of $\rho_+$; thus $t_*^{-\lambda-2}f_+^{(\lambda+2,k)}\in\Hb^{\infty,\ 4+\eps_\sscri,\ (\lambda+2,0),\ 2-\eps}(\Omega_*;S^2\cT^*)$ has strong decay at $\cK^+$ (compatible with the error term~\eqref{EqipGrLb} for $\lambda=1$, which is the smallest $\lambda$ that occurs, and $\ell_\cK=1-\eps$).

  Denote by $c_0$ the minimal real part among all $\lambda$ for which $f_+^{(\lambda+2,k)}\neq 0$ for some $k$. We can solve away the terms~\eqref{EqD3AlmipTerms} corresponding to those $\lambda$ with $\Re\lambda\in[c_0,c_0+\eps_\ind)$ using Corollary~\ref{CoripGrQhom} (or Proposition~\ref{PropipGr} for $\lambda=1$ and thus $k=0$) using an element of $\Hb^{\infty,\ \bigl(\la\cE_{\iota^+,\sscri}^\cC\ra,\,3+\eps_\sscri\bigr),\ 1-\eps,\ -\eps}(\Omega_*)^{\bullet,-}$ (cf.\ \eqref{EqipGrQhomu} and use that $\Re\lambda\geq 1$ for all $\lambda$ appearing here). The remaining error terms are again a sum of terms of the form~\eqref{EqD3AlmipTerms}, but now the new value of $c_0$ is increased by at least $\eps_\ind$; moreover, the $\scri^+$-order of the remaining error terms is $(\cE_{\iota^+,\sscri}^\tot+2,\,4+\eps_\sscri)$ instead of merely $4+\eps_\sscri$, but this can be solved away using Proposition~\ref{PropDScriFormal} (for arbitrary $C_0<2$) by means of an element of $\Hb^{\infty,\ \bigl(\la\cE_\sscri^\cC\ra,\,3+\eps_\sscri\bigr),\ 1-\eps,\ \infty}$ that is, in fact, partially polyhomogeneous at $\iota^+$, with index set inherited from that of the aforementioned remaining error terms (so with $\iota^+$-order $\geq c_0+\eps_\ind$) and with conormal remainder of decay order $\alpha_+<2$. Repeating this procedure finitely many times eliminates the full $\iota^+$-expansion of $f'_{1,\flat}+(1-\chi_\cK^\flat)f'_2$. In summary, these arguments produce
  \begin{equation}
  \label{EqD3Almuip}
    u_\iota \in \Hb^{\infty,\ \bigl(\la\cE_\sscri^\cC\ra,3+\eps_\sscri\bigr),\ 1-\eps,\ -\eps}(\Omega_*)^{\bullet,-}
  \end{equation}
  such that
  \[
    f'_{1,\flat}+f'_2 - L_b u_\iota \in \Hb^{\infty,\ 4+\eps_\sscri,\ \alpha_++2,\ \tilde\alpha_\cK}
  \]
  has trivial $\iota^+$-index set.\footnote{These arguments in fact prove the partial polyhomogeneity of $u_\iota$; this is however not a piece of information we need to record at this stage.} Recalling~\eqref{EqD3AlmEq} and~\eqref{EqD3Almu1scri}--\eqref{EqD3Almu1scriErr}, this means that
  \begin{equation}
  \label{EqD3Almscriip}
  \begin{split}
    &u_\flat := u - u_{1,\sscri} - u_\iota \\
    &\qquad \implies L_b u_\flat = f'_{1,\flat} + f'_2 - L_b u_\iota =: f_\flat \in \Hb^{\infty,\ 4+\eps_\sscri,\ \alpha_++2,\ \tilde\alpha_\cK}(\Omega_*;S^2\cT^*)^{\bullet,-}.
  \end{split}
  \end{equation}
  Note that, a fortiori, both $u_{1,\sscri}$ and $u_\iota$ lie in the space $\Hb^{\infty,\ \bigl(\la\cE_\sscri^\cC\ra,3+\eps_\sscri\bigr),\ 1-\eps,\ -\eps}$, and thus in~\eqref{EqD3Almu}.

  \pfstep{Step~3. Improved decay via spectral theory.} We can now gainfully work on the spectral side as in Step~3 of the proof of Proposition~\ref{PropD2Boost}. As always, the high-energy piece $u_{\rm hi}(t_*):=\frac{1}{2\pi}\int_\R e^{-i\sigma t_*}(1-\chi(|\sigma|))\wh{L_b}(\sigma)^{-1}\wh{f_\flat}(\sigma)\,\dd\sigma$ of $u_\flat$ (cf.\ \eqref{EqD1Almu}) has arbitrary decay at $\iota^+$ and $\cK^+$, so in particular
  \begin{equation}
  \label{EqD3Almuhi}
    u_{\rm hi} \in \Hb^{\infty,\ 1-\eps,\ \alpha_+,\ \tilde\alpha_\cK}(M';S^2\cT^*).
  \end{equation}
  For the low-energy piece, we first use~\eqref{EqTFHbLo} to get
  \begin{equation}
  \label{EqD3Almfflat}
    \bigl(\wh{f_\flat}(\sigma)\bigr)\big|_{\sigma\in\pm[0,c]} \in \Hb^{\infty,\ \alpha_++2-\eps,\ \alpha_++1,\ ((0,0),\tilde\alpha_\cK-1)}(X_\scbtop^\pm)\quad\forall\,\eps>0;
  \end{equation}
  we then note that already the proof of Proposition~\ref{PropD1Alm} gives meaning to
  \[
    u_{\rm lo}(t_*,\cdot) = \frac{1}{2\pi}\int_{\gamma_-\cup\gamma_0\cup\gamma_+} e^{-i\sigma t_*}\chi(|\sigma|)\wh{L_b}(\sigma)^{-1}\wh{f_\flat}(\sigma)\,\dd\sigma
  \]
  as a tempered distribution (see~\eqref{EqD1Almureg} and \eqref{EqD1Almusing}), and thus to $\wh{u_{\rm lo}}(\sigma,\cdot)$ as a tempered distribution on $\R_\sigma$; a different perspective on this is that $\wh{u_{\rm lo}}(\sigma)$ is a distributional extension of $\chi(|\sigma|)\wh{L_b}(\sigma)^{-1}\wh{f_\flat}(\sigma)$ from $\R\setminus\{0\}$ to all of $\R$. The strategy for the remainder of the proof is to control $\wh{u_{\rm lo}}(\sigma)$ for $\sigma\in[-c,0)\cup(0,c]$ as precisely as needed to prove~\eqref{EqD3Almu}, with the distributional extension to $\sigma=0$ (where $\wh{u_{\rm lo}}(\sigma)$ does not lie in $L^1_\loc$).

  \pfsubstep{Step~3.1.}{Zero energy piece.}\label{ItD3Step31} We use Proposition~\ref{PropAdmLoDef0} to define
  \[
    \bigl(\hat u_{{\rm reg},1}(0),\ \hat b_1(0),\ \hat\scal_1(0)\bigr) := \wt{L_b}(0)^{-1}\bigl(\wh{f_\flat}(0),\ 0,\ 0\bigr).
  \]
  The assumed vanishing of the leading-order term in~\eqref{EqD2Boost2} implies, in view of the formula~\eqref{EqD2BoostLot} for it, that $\hat\scal_1(0)=0$. Moreover, since (by the definition~\eqref{EqAdmLoDef} and Lemma~\ref{LemmaAdmLob} as well as~\eqref{EqD3Almfflat})
  \begin{subequations}
  \begin{equation}
  \label{EqD3Almureg1Eq}
    \wh{L_b}(0)\hat u_{{\rm reg},1}(0) = \wh{f_\flat}(0) - \Bigl(\wh{L_b}(\sigma)\cF\bigl(\dot g_b^{\Ups,\aug}(\hat b_1(0))\bigr)(\sigma)\Bigr)\Big|_{\sigma=0} \in \Hb^{\infty,\ (\cE_\ind+2,\alpha_++1)}(X;S^2\cT^*_X)
  \end{equation}
  has a partial polyhomogeneous expansion, a normal operator argument yields
  \begin{equation}
  \label{EqD3Almureg1}
    \hat u_{{\rm reg},1}(0) \in \Hb^{\infty,\ (\cE_{\ind,+},\alpha_+-1)}(X;S^2\cT^*_X),
  \end{equation}
  \end{subequations}
  where $\cE_{\ind,+}:=\cE_\ind\extcup\{(z_1,0)\}\extcup\cdots\extcup\{(z_N,0)\}$, with $z_1,\ldots,z_N$ being all indicial roots of $\wh{L_b}(0)$ with $\Re z_j\in(0,\alpha_+-1]$; here we reduce $\alpha_+$ by an arbitrarily small amount if necessary to ensure that no indicial root has real part equal to $\alpha_+-1$. We then have
  \begin{equation}
  \label{EqD3AlmSolv0}
    \bigl(\wh{f_\flat}(\sigma),\ 0,\ 0\bigr) - \wt{L_b}(\sigma)\bigl(\chi_\zface\hat u_{{\rm reg},1}(0),\ \hat b_1(0),\ 0\bigr) =: \bigl(\hat f_2(\sigma),\ 0,\ 0\bigr)
  \end{equation}
  for small enough $|\sigma|$ as discussed after~\eqref{EqD2Boost0InvErr}, with $\hat f_2(\sigma)$ is given by~\eqref{EqD2Boost0InvErr1}; the first two summands on the right-hand side of~\eqref{EqD2Boost0InvErr1} are of class $\Hb^{\infty,\ \alpha_++2-\eps,\ \alpha_++1,\ \tilde\alpha_\cK-1}(X_\scbtop^\pm)$ (we call their sum $\hat f_{2,1}$), while the third summand is
  \[
    -[\wh{L_b}(0),\chi_\zface]\hat u_{{\rm reg},1}(0) \in \Hb^{\infty,\ \infty,\ (\cE_{\ind,+}+2,\alpha_++1),\ \infty}(X_\scbtop^\pm)
  \]
  by~\eqref{EqD3Almureg1}; and Lemma~\ref{LemmaD3wtLb} implies that the fourth summand satisfies
  \begin{align*}
    &-\pi_1(\wt{L_b}(\sigma)-\wt{L_b}(0))(\chi_\zface\hat u_{{\rm reg},1}(0),\ \hat b_1(0),\ 0) \\
    &\quad\qquad\in \Hb^{\infty,\ \infty,\ (\cE_{\ind,+}+2,\alpha_++1),\ (1,0)}(X_\scbtop^\pm) + \Hb^{\infty,\ 3-\eps,\ \cE_\ind+2,\ \bigl((1,0),3+\eps_\ind-\eps\bigr)}(X_\scbtop^\pm) \\
    &\quad\qquad \subset \Hb^{\infty,\ 3-\eps,\ (\cE_{\ind,+}+2,\alpha_++1),\ \bigl((1,0),3+\eps_\ind-\eps\bigr)}(X_\scbtop^\pm).
  \end{align*}
  We can thus write $\hat f_2=\hat f_{2,0}+\hat f_{2,1}$ where $\hat f_{2,0}$ captures the third and fourth summands just discussed, so
  \begin{equation}
  \label{EqD3Almf2}
  \begin{split}
    \hat f_{2,0} &\in \Hb^{\infty,\ 3-\eps,\ (\cE_{\ind,+}+2,\alpha_++1),\ \bigl((1,0),3+\eps_\ind-\eps\bigr)}(X_\scbtop^\pm) \\
      &\quad = |\sigma|\Hb^{\infty,\ 3-\eps,\ (\cE_{\ind,+}+1,\alpha_+),\ \bigl((0,0),2+\eps_\ind-\eps\bigr)}(X_\scbtop^\pm), \\
    \hat f_{2,1} &\in \Hb^{\infty,\ \alpha_++2-\eps,\ \alpha_++1,\ \tilde\alpha_\cK-1}(X_\scbtop^\pm),
  \end{split}
  \end{equation}
  (This means that via~\eqref{EqD3AlmSolv0} we have solved away the leading-order term of~\eqref{EqD3Almfflat} at $\zface$, but in doing so created some additional polyhomogeneous terms at $\tface$.)

  \pfsubstep{Step~3.2.}{Solving away the error at $\tface$.} The relative order of $\hat f_{2,0}$~\eqref{EqD3Almf2} at $\tface$ and $\zface$ is $\min\Re\cE_{\ind,+}+2-1=1+\min\Re\cE_\ind$ and thus less than $2$. Due to this lack of decay, a direct application of $\wh{L_b}(\sigma)^{-1}$ to $(\hat f_2(\sigma),\ 0,\ 0)$ by means of \citeAF{Proposition~\ref*{PropDResLo}} does not give precise control (i.e., without unacceptable losses in the $\zface$- and $\tface$-orders). Instead, we must first apply Proposition~\ref{PropiptfGr} to $|\sigma|^{-1}\hat f_{2,0}(\sigma)$ for each $(\alpha_0,k)\in\cE_{\ind,+}+1$ with $\Re\alpha_0<\alpha_+$ (note that $\Re\alpha_0\in(1,2)$ for all such $\alpha_0$) and with $\ell_\zface=2+\eps_\ind-\eps$ in order to solve away the leading-order terms of $|\sigma|^{-1}\hat f_2(\sigma)$ at $\tface$; we do this starting with the smallest values of $\Re\alpha_0$ and iteratively solve away the remainder terms~\eqref{EqiptfGrErr} until the minimal real part of the remaining error is $>\alpha_+$. Using only the rather coarse description~\eqref{EqiptfGru}, this produces
  \begin{subequations}
  \begin{equation}
  \label{EqD3Almu2}
  \begin{split}
    \hat u_2 &\in \bigcap_{\eps>0}|\sigma|\Hb^{\infty,\ 1-\eps,\ \min\Re\cE_{\ind,+}-1-\eps,\ \min\Re\cE_{\ind,+}-2-\eps}(X_\scbtop^\pm) \\
      &= \bigcap_{\eps>0}\Hb^{\infty,\ 1-\eps,\ \eps_\ind-\eps,\ -1+\eps_\ind-\eps}(X_\scbtop^\pm)
  \end{split}
  \end{equation}
  (which thus vanishes to positive order at $\tface$ and has an integrable singularity at $\zface$) such that
  \begin{equation}
  \label{EqD3Almu2Err}
    \bigl(\hat f_{2,0}(\sigma) - \wh{L_b}(\sigma)\hat u_2(\sigma)\bigr)\big|_{\sigma\in\pm[0,c]} \in |\sigma|\Hb^{\infty,\ 2-\eps,\ \alpha_+,\ \bigl((0,0),2+\eps_\ind-\eps\bigr)}(X_\scbtop^\pm).
  \end{equation}
  \end{subequations}
  Since $\tilde\alpha_\cK-1<1$, we can add to this the term $\hat f_{2,1}$ from~\eqref{EqD3Almf2} for a total of
  \begin{equation}
  \label{EqD3Almf3}
    \hat f_3(\sigma) := \hat f_2(\sigma) - \wh{L_b}(\sigma)\hat u_2(\sigma) \implies
    \bigl(\hat f_3(\sigma)\bigr)\big|_{\sigma\in\pm[0,c]} \in \Hb^{\infty,\ 2-\eps,\ \alpha_++1,\ \tilde\alpha_\cK-1}(X_\scbtop^\pm).
  \end{equation}
  This now has an acceptable relative decay rate at $\tface$ and $\zface$ of $(\alpha_++1)-(\tilde\alpha_\cK-1)\in(2,2+\eps_\ind)$ for the application of $\wt{L_b}(\sigma)^{-1}$.

  \pfsubstep{Step~3.3.}{Remaining piece; combination.} We apply $\wt{L_b}(\sigma)^{-1}$ to~\eqref{EqD3Almf3}; more precisely, we set
  \[
    \bigl(\hat u_3(\sigma),\ \hat b_3(\sigma),\ \hat\scal_3(\sigma)\bigr) := \wt{L_b}(\sigma)^{-1}\bigl(\hat f_3(\sigma),\ 0,\ 0\bigr),\quad \pm\sigma\in[0,c],
  \]
  which, by \citeAF{Proposition~\ref*{PropDResLo}}, satisfies
  \[
    \hat u_3 \in \Hb^{\infty,\ 1-\eps,\ \alpha_+-1,\ \tilde\alpha_\cK-1}(X_\scbtop^\pm),\quad
    \hat b_3 \in \Hb^{\infty,\ \tilde\alpha_\cK-1}(\pm[0,c];\C^4),\quad
    \hat\scal_3 \in \Hb^{\infty,\ \tilde\alpha_\cK-1}(\pm[0,c];\scalspace_1).
  \]
  Combining this with \eqref{EqD3Almureg1}--\eqref{EqD3AlmSolv0} and \eqref{EqD3Almu2} and recalling~\eqref{EqAdmLoDef} shows that\footnote{One reason for writing $\hat u_{\rm sing}$ in this fashion is that $\hat b,\hat\scal\in L^1$; another reason is that a single $t_*$-derivative of $\cF^{-1}\hat u_{\rm sing}$ can already be controlled using basic results on the inverse Fourier transform; see the discussion of~\eqref{EqD3Almpausing} below.}
  \begin{align}
    \wh{u_{\rm lo}}(\sigma) &= \hat u_{\rm reg}(\sigma) + \hat u_{\rm sing}(\sigma),\qquad \sigma\in\pm(0,c], \nonumber\\
    \hat u_{\rm reg}(\sigma) &:= \chi(|\sigma|)\bigl(\chi_\zface\hat u_{{\rm reg},1}(0) + \hat u_2(\sigma) + \hat u_3(\sigma)\bigr) \nonumber\\
      &\in \bigcap_{\eps>0}\Hb^{\infty,\ 1-\eps,\ \min(\alpha_+-1,\eps_\ind-\eps),\ -1+\eps_\ind-\eps}(X_\scbtop^\pm), \nonumber\\
  \label{EqD3Almusing}
    \hat u_{\rm sing}(\sigma) &:= \cF(\dot g_b^{\Ups,\aug})(\sigma)(\hat b(\sigma)) + \cF(\pa_{t_*}h_{b,\rms 1}^{\leq 1,\aug})(\sigma)(\hat\scal(\sigma)), \\
      &\qquad \hat b(\sigma) := \chi(|\sigma|)\bigl(\hat b_1(0)+\hat b_3(\sigma)\bigr)\in\Hb^{\infty,\ ((0,0),\tilde\alpha_\cK-1)}(\pm[0,c];\C^4), \nonumber\\
      &\qquad \hspace{-0.12em}\hat\scal(\sigma):=\chi(|\sigma|)i\sigma^{-1}\hat\scal_3(\sigma)\in\Hb^{\infty,\ \tilde\alpha_\cK-2}(\pm[0,c];\scalspace_1). \nonumber
  \end{align}
  Altogether, recalling~\eqref{EqD3Almscriip} and the subsequent discussion, as well as~\eqref{EqD3Almuhi}, we have
  \[
    u = u_{\rm lo} + u_\rem,\quad u_\rem:=u_{\rm hi}+u_{1,\sscri}+u_\iota \in \Hb^{\infty,\ \bigl(\la\cE_\sscri^\cC\ra,3+\eps_\sscri\bigr),\ 1-\eps,\ -\eps}(M';S^2\cT^*);
  \]
  and in fact $u_\rem$ decays (at rate $\tilde\alpha_\cK>1$) towards $\cK^-$. Since $\wh{u_{\rm lo}}$ is well-defined as a distribution on $\R$ and $\hat u_{\rm reg}$ as well (having an integrable singularity at $\sigma=0$), so is $\hat u_{\rm sing}$.\footnote{In other words, the construction of the singular piece of $\wh{u_{\rm lo}}$ in the proof of Proposition~\ref{PropD1Alm} via contour shifting yields an extension of $\hat u_{\rm sing}$ from $\R\setminus\{0\}$, which typically has a $|\sigma|^{-1}$-singularity from from $\cF(\dot g_b^{\Ups,\aug})(\sigma)(\hat b(0))$, to $\R$.} We can thus define $u_{\rm reg}$ and $u_{\rm sing}$ as the inverse Fourier transforms of $\hat u_{\rm reg}$ and $\hat u_{\rm sing}$; by~\eqref{EqTFHbInvLo},
  \[
    u_{\rm reg} \in \Hb^{\infty,\ 1-\eps,\ \min(\alpha_+,1+\eps_\ind-\eps),\ \eps_\ind-\eps}(M';S^2\cT^*)
  \]
  (which is consistent with~\eqref{EqD3Almu} as far as the $\iota^+$- and $\cK^+$-orders are concerned); for later use, we point out that this decays towards $\cK^-$.

  Now, since $u=u_{\rm reg}+u_{\rm sing}+u_\rem$ vanishes for $\ft_*<1$, we have
  \begin{equation}
  \label{EqD3AlmusingIni}
    u_{\rm sing} = -u_{\rm reg} - u_\rem \in \bar H_\bop^{\infty,\ 1-\eps,\ 1-\eps,\ \eps_\ind-\eps}(\Omega_-),\quad\Omega_-:=\cl_{M'}\{\ft_*<1\},
  \end{equation}
  where we use the decay of the two terms towards $\cK^-$. We shall control $u_{\rm sing}$ by integrating $\pa_{t_*}u_{\rm sing}$; for $\sigma\neq 0$, we have
  \begin{equation}
  \label{EqD3Almpausing}
  \begin{split}
    \cF(\pa_{t_*}u_{\rm sing})(\sigma) &= \cF(\pa_{t_*}\dot g_b^{\Ups,\aug})(\sigma)(\hat b(\sigma)) + \cF(\pa_{t_*}^2 h_{b,\rms 1}^{\leq 1,\aug})(\sigma)(\hat\scal(\sigma)) \\
      &\in \Hb^{\infty,\ 1-\eps,\ 1-\eps,\ \bigl((0,0),\tilde\alpha_\cK-1\bigr)}(X_\scbtop^\pm) + \Hb^{\infty,\ 1-\eps,\ 2-\eps,\ \tilde\alpha_\cK-2}(X_\scbtop^\pm) \\
      &\subset \Hb^{\infty,\ 1-\eps,\ 1-\eps,\ \tilde\alpha_\cK-2}(X_\scbtop^\pm),
  \end{split}
  \end{equation}
  where we used~\eqref{EqD1AlmAug} for the second line. This has an integrable singularity at $\sigma=0$, and thus the inverse Fourier transform of its unique $L^1_\loc$-extension differs from $\pa_{t_*}u_{\rm sing}$ by a term whose Fourier transform is supported at $\sigma=0$; the inverse Fourier transform of such a distribution is a polynomial in $t_*$ and does not decay at $\cK^-$ unless it vanishes. Therefore, we can use~\eqref{EqTFHbInvLo} to deduce
  \[
    \pa_{t_*}u_{\rm sing} \in \Hb^{\infty,\ 1-\eps,\ 2-\eps,\ \tilde\alpha_\cK-1}(M';S^2\cT^*).
  \]
  Applying Lemma~\ref{LemmaD1Intb}\eqref{ItD1Intb1} to this, with initial condition~\eqref{EqD3AlmusingIni}, yields\footnote{Unlike in previous steps, we only need to integrate \emph{once} to control $u_{\rm sing}$ globally; the reason is that the most singular term $\cF(h_{b,\rms 1}^{\leq 1,\aug})(\hat\scal(0))\sim|\sigma|^{-2}$ (which becomes integrable only after multiplication by $(-i\sigma)^2$) is absent.}
  \begin{equation}
  \label{EqD3Almusing2}
    u_{\rm sing} \in \Hb^{\infty,\ 1-\eps,\ 1-\eps,\ \tilde\alpha_\cK-2}(M';S^2\cT^*).
  \end{equation}
  We have thus shown that
  \[
    u = u_{\rm reg} + u_{\rm sing} + u_\rem \in \Hb^{\infty,\ 1-\eps,\ 1-\eps,\ \tilde\alpha_\cK-2}(\Omega_*;S^2\cT^*)^{\bullet,-}.
  \]

  \pfstep{Step~4. Recovery of sharp decay at $\scri^+$.} Applying Proposition~\ref{PropDScriPhg} to the equation $L u=f$ gives
  \[
    u \in \Hb^{\infty,\ \bigl(\la\cE_\sscri^\cC\ra,3+\eps_\sscri\bigr),\ 1-\eps,\ \tilde\alpha_\cK-2}(\Omega_*)^{\bullet,-}.
  \]
  This improves on~\eqref{EqD3AlmStart} in the $\cK^+$-order by a definite amount (see~\eqref{EqD3AlmEq}). After finitely many iterations, we thus obtain~\eqref{EqD3Almu}. This finishes the proof.
\end{proof}

\subsection{Step~4: black hole parameter changes; total/logarithmic center-of-mass motion}
\label{SsD4Par}

The only terms arising in the proof of Proposition~\ref{PropD3Alm} that do not decay at $\cK^+$ are $u_{\rm sing}$ (see~\eqref{EqD3Almusing} and \eqref{EqD3Almusing2}) as well as the contribution to $u_\iota$ (see~\eqref{EqD3Almuip}) arising from $(\lambda,k)=(1,0)$, which features a term $\log(t_*)h_{b,\rms 1}$ from~\eqref{EqipGr}, i.e., a logarithmic correction of the center-of-mass of the black hole. We proceed to show that for source terms with more than $t_*^{-2}$-decay at $\cK^+$, the term $u_{\rm sing}$ involves a constant center-of-mass shift as well as a correction of the black hole parameters.

\subsubsection{Extraction of parameters}
\label{SssD4ParEx}

Mirroring the procedure of~\S\ref{SsD2Boost}, we first prove the following analogue of Proposition~\ref{PropD2Boost}. The conditions we place on the source term~\eqref{EqD4Parf} below are still compatible with the boost term~\eqref{EqD2LinParScal}, and weakened compared to~\eqref{EqD3Almf} for compatibility with modification terms arising later on (cf.\ \eqref{EqD4LinParb} and \eqref{EqD4LinParCoM}).

\begin{prop}[Extraction of linearized black hole parameters and center-of-mass movement]
\label{PropD4Par}
  Let $\alpha_\cK\in(2,2+\eps_\cK)$, where we recall from~\eqref{EqDMetBasicEllEps} that $\eps_\cK\in(0,\eps_\ind)$. Let $f$ be of the form
  \begin{equation}
  \label{EqD4Parf}
  \begin{split}
    &f=f_1+f_2, \\
    &\quad f_1\in\bigcap_{\eps>0}\Hb^{\infty,\ \bigl(\la\cE_\sscri^\cC+1\ra',\,4+\eps_\sscri\bigr),\ 4-\eps,\ \alpha_\cK}(\Omega_*)^{\bullet,-}, \\
    &\quad f_2\in\Hb^{\infty,\ 4+\eps_\sscri,\ \tilde\cE_++2,\ \alpha_\cK}(\Omega_*;S^2\cT^*)^{\bullet,-},
  \end{split}
  \end{equation}
  where $\tilde\cE_+$ is the index set~\eqref{EqD2LinPartildeE} (or, in fact, any index set with $\min\Re(\tilde\cE_+\setminus\{(1,0)\})\geq\eps_\ind$). Suppose that for the forward solution of $L_{h,\scal,b}u=f$, the leading-order term in~\eqref{EqD2Boost2} vanishes. Then $u$ can be written as
  \begin{equation}
  \label{EqD4Paru}
  \begin{split}
    &u = \chi_\cK\dot g_b^\Ups(\dot b) + \chi_\cK h_{b,\rms 1}(\dot\scal_1^{(0)}) + \chi_\cK h_{b,\rms 1}^{\leq 4}(\dot\scal_1^{(0,1)}\log t_*) + \tilde u, \\
    &\quad \dot b\in\R^4,\quad \dot\scal_1^{(0)},\ \dot\scal_1^{(0,1)}\in\scalspace_1, \quad
      \tilde u \in \Hb^{\infty,\ \bigl(\la\cE_\sscri^\cC\ra,3+\eps_\sscri\bigr),\ 1-\eps,\ \alpha_\cK-2}(\Omega_*)^{\bullet,-},
  \end{split}
  \end{equation}
  where we recall $\dot g_b^\Ups$ from~\eqref{EqWEMode0KerrMem} and use the notation from Definition~\usref{DefipGrKerr}. Here $\dot b$, $\dot\scal_1^{(0)}$, and $\dot\scal_1^{(0,1)}$ depend continuously on $f$.
\end{prop}

Unlike in previous statements, we now stop (for purely aesthetic reasons) requiring the $\iota^+$-order of $f_1$ to be $>2+\alpha_\cK$. As we will see in later steps, the conormal decay order $4-\eps$ suffices to get all the way to $\cO(t_*^{-1+\eps})$-decay of the gravitational wave tail.

\begin{rmk}[Notation]
\label{RmkD4ParNot}
  The superscript ``$(0,1)$'' of $\dot\scal_1^{(0,1)}$ refers to the $\cK^+$-order $\rho_\cK^0(\log\rho_\cK)^1$ of the expansion term that this parameter belongs to. The subscript ``$1$'' refers to the fact that $\dot\scal_1^{(0,1)}$ lies in the space $\scalspace_1$.
\end{rmk}

\begin{proof}[Proof of Proposition~\usref{PropD4Par}]
  Proposition~\ref{PropD3Alm} establishes
  \[
    u \in \bigcap_{\eps>0}\Hb^{\infty,\ \bigl(\la\cE_\sscri^\cC\ra,3+\eps_\sscri\bigr),\ 1-\eps,\ -\eps}(\Omega_*)^{\bullet,-}.
  \]
  Using~\eqref{EqDAdmFw}, this gives $(L_{h,\scal,b}-L_b)u \in \Hb^{\infty,\ \bigl(\la\cE_\sscri^\cC+1\ra',\,4+\eps_\sscri\bigr),\ 4-\eps,\ 2+\eps_\cK-\eps}(\Omega_*)^{\bullet,-}$, and hence, in view of $\alpha_\cK<2+\eps_\cK$,
  \[
    L_b u =: f' = f'_1 + f'_2,\quad f'_1\in\Hb^{\infty,\ \bigl(\la\cE_\sscri^\cC+1\ra',\,4+\eps_\sscri\bigr),\ 4-\eps,\ \alpha_\cK}(\Omega_*)^{\bullet,-},\ \ f'_2:=f_2.
  \]

  \pfstep{Step~1. Solution near $\scri^+$.} By the same arguments as for~\eqref{EqD3Almu1scri}--\eqref{EqD3Almu1scriErr}, Proposition~\ref{PropDScriFormal} produces
  \begin{align}
  \label{EqD4u1scri}
    &u_{1,\sscri} \in \Hb^{\infty,\ \bigl(\la\cE_\sscri^\cC\ra,3+\eps_\sscri\bigr),\ (\cE_+^\sharp,2-\eps),\ \infty},\ \ \text{supported near $\scri^+$}, \\
    &\qquad f'_{1,\flat} := f'_1 - L_b u_{1,\sscri} \in \Hb^{\infty,\ 4+\eps_\sscri,\ (\cE_+^\sharp+2,4-\eps),\ \alpha_\cK}, \nonumber
  \end{align}
  where $\cE_+^\sharp$ depends only on $\cE_\sscri^\cC$ and is given by $(1,0)\cup(1+(1-e^\Ups)\gamma^\Ups,0)\cup(\cdots)$.

  \pfstep{Step~2. Solution near $\iota^+$.} The $\iota^+$-expansion of the remaining source term $f'_{1,\flat}+f'_2$ is a sum of terms~\eqref{EqD3AlmipTerms}, with $(\lambda,k)=(1,0)$ or $1+\eps_\ind\leq\Re\lambda<2$. We repeat the arguments in Step~2 of the proof of Proposition~\ref{PropD3Alm}, except we treat the $(3,0)$- (i.e., $\rho_+^3$-) term in the expansion of $f'_{1,\flat}+f'_2$ separately.

  \pfsubstep{Step~2.1.}{Leading-order term: $(\lambda,k)=(1,0)$.} The leading-order term at $\iota^+$ is $t_*^{-3}f_+^{(3,0)}(R,\omega)$ where $f_+^{(3,0)}\in\Hb^{\infty,\ 4+\eps_\sscri,\ \alpha_\cK-3}(\iota^+)$. We use Proposition~\ref{PropipGr} with $\lambda=1$ and $\ell_\cK=\alpha_\cK-1$ to obtain $\dot\scal_1^{(0,1)}\in\scalspace_1$ (which is $-\scal_1^{(-1)}$ in the notation of~\eqref{EqipGr}) and
  \begin{subequations}
  \begin{equation}
  \label{EqD4ip1}
  \begin{split}
    u_\iota^{(1,0)} &\in \Hb^{\infty,\ \bigl(\la\cE_{\iota^+,\sscri}^\cC\ra,3+\eps_\sscri\bigr),\ (1,0),\ \alpha_\cK} + \Hb^{\infty,\ \infty,\ (1,0)\cup(1+\cE_\ind),\ 1-\eps} \\
      &\subset \Hb^{\infty,\ \bigl(\la\cE_{\iota^+,\sscri}^\cC\ra,3+\eps_\sscri\bigr),\ (1,0)\cup(1+\cE_\ind),\ 1-\eps}
  \end{split}
  \end{equation}
  (capturing $\chi_\iota t_*^{-1}\tilde u$ in~\eqref{EqipGru} as well as all terms of $\chi_\cK u_{b,0}$ in~\eqref{EqipGr} except the first) such that
  \begin{align*}
    &\chi_\iota t_*^{-3}f_+^{(3,0)}(R,\omega) - L_b\Bigl( h_{b,\rms 1}^{\leq 4}(\dot\scal_1^{(0,1)}\log t_*) + u_\iota^{(1,0)} \Bigr) \\
    &\qquad \in \Hb^{\infty,\ \bigl(\cE_{\iota^+,\sscri}^\tot+2,\,4+\eps_\sscri\bigr),\ 3+\cE_\ind,\ \alpha_\cK}(\Omega_*;S^2\cT^*)^{\bullet,-}.
  \end{align*}
  We then use Proposition~\ref{PropDScriFormal} (with $\breve\cE_+=1+\cE_\ind$, $\breve\ell_+=2-\eps$, and arbitrary $C_0<2$) to solve this error term away at $\scri^+$ by means of
  \begin{equation}
  \label{EqD4ip1scri}
    u_\sscri^{(1,0)} \in \Hb^{\infty,\ \bigl(\la\cE_\sscri^\cC\ra,3+\eps_\sscri\bigr),\ (1+\cE_\ind,2-\eps),\ \infty}
  \end{equation}
  with support near $\scri^+$, so
  \begin{equation}
  \label{EqD4ip1Err}
  \begin{split}
    &\chi_\iota t_*^{-3}f_+^{(3,0)}(R,\omega) - L_b\Bigl( h_{b,\rms 1}^{\leq 4}(\dot\scal_1^{(0,1)}\log t_*) + u_\iota^{(1,0)} + u_\sscri^{(1,0)}\Bigr) \\
    &\qquad \in \Hb^{\infty,\ 4+\eps_\sscri,\ (3+\cE_\ind,4-\eps),\ \alpha_\cK}(\Omega_*;S^2\cT^*)^{\bullet,-}.
  \end{split}
  \end{equation}
  \end{subequations}

  \pfsubstep{Step~2.2.}{Lower-order terms: $\Re\lambda\in[1+\eps_\ind,2)$.} We now solve away the terms~\eqref{EqD3AlmipTerms} with $\Re\lambda\geq 1+\eps_\ind$ and $\alpha_\cK$ in place of $2-\eps$, as well as the terms of the $\iota^+$-expansion of~\eqref{EqD4ip1Err}, using Corollary~\ref{CoripGrQhom} as in the proof of Proposition~\ref{PropD3Alm} by means of elements of the space $\Hb^{\infty,\ \bigl(\la\cE_{\iota^+,\sscri}^\cC\ra,3+\eps_\sscri\bigr),\ 1+\eps_\ind-\eps,\ \eps_\ind-\eps}$, which however produces errors with $\scri^+$-order $(\cE_{\iota^+,\sscri}^\tot+2,\,4+\eps_\sscri)$ (but with improved decay at $\iota^+$) that we solve away using Proposition~\ref{PropDScriFormal} using elements of $\Hb^{\infty,\ \bigl(\la\cE_\sscri^\cC\ra,3+\eps_\sscri\bigr),\ 1+\eps_\ind-\eps, \infty}$ that, in fact, are partially polyhomogeneous at $\iota^+$; we can thus iteratively solve away all $\iota^+$-expansion terms using
  \[
    u_\iota^{\geq 1+\eps_\ind} \in \Hb^{\infty,\ \bigl(\la\cE_\sscri^\cC\ra,3+\eps_\sscri\bigr),\ 1+\eps_\ind-\eps,\ \eps_\ind-\eps}(\Omega_*)^{\bullet,-}.
  \]
  Adding this to~\eqref{EqD4ip1} and \eqref{EqD4ip1scri} produces
  \begin{subequations}
  \begin{equation}
  \label{EqD4ip}
    u_\iota \in \Hb^{\infty,\ \bigl(\la\cE_\sscri^\cC\ra,3+\eps_\sscri\bigr),\ 1-\eps,\ \eps_\ind-\eps}(\Omega_*)^{\bullet,-}
  \end{equation}
  such that
  \begin{equation}
  \label{EqD4ipErr}
  \begin{split}
    &\hspace{-7em} u_\flat := u - u_{1,\sscri} - h_{b,\rms 1}^{\leq 4}(\dot\scal_1^{(0,1)}\log t_*) - u_\iota \\
    \quad \implies L_b u_\flat &= f'_{1,\flat} + f'_2 - L_b\bigl(h_{b,\rms 1}^{\leq 4}(\dot\scal_1^{(0,1)}\log t_*)+u_\iota\bigr) \\
    \quad &=: f_\flat \in \Hb^{\infty,\ 4+\eps_\sscri,\ 4-\eps,\ \alpha_\cK}(\Omega_*;S^2\cT^*)^{\bullet,-}.
  \end{split}
  \end{equation}
  \end{subequations}

  \pfstep{Step~3. Improved decay via spectral theory.} We analyze the equation~\eqref{EqD4ipErr} using the Fourier transform in $t_*$. The high-energy piece $u_{\rm hi}$ of $u_\flat$ (see~\eqref{EqD1Almu}) has arbitrary decay as usual and thus satisfies
  \begin{equation}
  \label{EqD4uhi}
    u_{\rm hi} \in \Hb^{\infty,\ 1-\eps,\ 2-\eps,\ \alpha_\cK}(M';S^2\cT^*).
  \end{equation}
  For the low-energy piece, we use~\eqref{EqTFHbLo} to obtain
  \begin{equation}
  \label{EqD4fflat}
    \bigl(\wh{f_\flat}(\sigma)\bigr)\big|_{\sigma\in\pm[0,c]} \in \bigcap_{\eps>0}\Hb^{\infty,\ 4-\eps,\ 3-\eps,\ ((0,0),\alpha_\cK-1)}(X_\scbtop^\pm).
  \end{equation}

  \pfsubstep{Step~3.1.}{Zero energy piece.} We use Proposition~\ref{PropAdmLoDef0} as in Step~3.1 (but with $\alpha_+$ replaced by $2-\eps$) of the proof of Proposition~\ref{PropD3Alm} (see~\eqref{EqD3Almureg1}--\eqref{EqD3AlmSolv0}), thus producing
  \begin{equation}
  \label{EqD4u0}
  \begin{split}
    &\hat u_{{\rm reg},1}(0) \in \Hb^{\infty,\ (\cE_{\ind,+},2-\eps)}(X;S^2\cT^*_X),\quad \hat b_1(0)\in\R^4, \\
    &\qquad \bigl(\wh{f_\flat}(\sigma),\ 0,\ 0\bigr) - \wt{L_b}(\sigma)\bigl(\chi_\zface\hat u_{{\rm reg},1}(0),\ \hat b_1(0),\ 0\bigr) = \bigl(\hat f_{2,0}(\sigma) + \hat f_{2,1}(\sigma),\ 0,\ 0\bigr), \\
    &\qquad \qquad \hat f_{2,0} \in |\sigma|\Hb^{\infty,\ 3-\eps,\ (\cE_{\ind,+}+1,2-\eps),\ \bigl((0,0),2+\eps_\ind-\eps\bigr)}(X_\scbtop^\pm), \\
    &\qquad \qquad \hat f_{2,1} \in |\sigma|\Hb^{\infty,\ 4-\eps,\ 2-\eps,\ ((0,0),\alpha_\cK-2)}(X_\scbtop^\pm).
  \end{split}
  \end{equation}
  This is analogous to~\eqref{EqD3Almf2}, except now that $\alpha_\cK>2$, the term $\hat f_{2,1}$ still has a partial expansion at $\zface$. We recall that $\min\Re\cE_{\ind,+}\geq\eps_\ind$.

  \pfsubstep{Step~3.2.}{Solving away the error at $\tface$.} The arguments leading up to~\eqref{EqD3Almu2}--\eqref{EqD3Almu2Err} apply also in the current context and produce
  \begin{align}
  \label{EqD4u2}
    &\hat u_{2,0} \in \bigcap_{\eps>0} \Hb^{\infty,\ 1-\eps,\ \eps_\ind-\eps,\ -1+\eps_\ind-\eps}(X_\scbtop^\pm), \\
    &\qquad \bigl(\hat f_{2,0}(\sigma) - \wh{L_b}(\sigma)\hat u_{2,0}(\sigma)\bigr)\big|_{\sigma\in\pm[0,c]} \in |\sigma|\Hb^{\infty,\ 2-\eps,\ 2-\eps,\ \bigl((0,0),2+\eps_\ind-\eps\bigr)}(X_\scbtop^\pm). \nonumber
  \end{align}
  We denote the sum of this remaining error and $\hat f_{2,1}$ from~\eqref{EqD4u0} by
  \[
    \hat f_3 \in |\sigma|\Hb^{\infty,\ 2-\eps,\ 2-\eps,\ ((0,0),\alpha_\cK-2)}(X_\scbtop^\pm).
  \]
  This still has too little decay at $\tface$ relative to the $\zface$-order. We thus solve this away to leading order at $\tface$ using Proposition~\ref{PropiptfC} (with $\alpha=2-\eps$, $\ell_\zface=\alpha_\cK-2$, and $k_0=0$, and multiplied through by $|\sigma|$); this produces
  \begin{equation}
  \label{EqD4u3}
    \hat u_3 \in |\sigma|\Hb^{\infty,\ 1-\eps,\ -\eps,\ -1-\eps} = \Hb^{\infty,\ 1-\eps,\ 1-\eps,\ -\eps}(X_\scbtop^\pm)\quad\forall\,\eps>0
  \end{equation}
  such that
  \begin{equation}
  \label{EqD4Parf4}
    \bigl(\hat f_4(\sigma)\bigr)\big|_{\sigma\in\pm[0,c]} := \bigl(\sigma^{-1}\bigl(\hat f_3(\sigma)-\wh{L_b}(\sigma)\hat u_3(\sigma)\bigr)\bigr)\big|_{\sigma\in\pm[0,c]} \in \Hb^{\infty,\ 2-\eps,\ 2+\eps_\ind-\eps,\ ((0,0),\alpha_\cK-2)}(X_\scbtop^\pm).
  \end{equation}

  \pfstep{Step~4. The sub-leading zero energy piece.} This is a qualitatively new step, in which we will uncover the stationary center-of-mass shift. (The stationary Kerr parameter change is already encoded in $\hat b_1(0)$ in~\eqref{EqD4u0}.) To wit, note that $\hat f_4(\sigma)$ is of the same type as $\wh{f_\flat}(\sigma)$ in~\eqref{EqD4fflat}, but with $\alpha_\cK-2$ in place of $\alpha_\cK-1$ (and with weaker $\tface$- and $\scface$-orders, though this matters little here). Let us define\footnote{One should more precisely compute $\wt{L_b}(0)^{-1}(\hat f_4(\pm 0),0,0)$, which may depend on the choice of sign. We do not make this explicit in the notation, but we do argue for the validity of certain matching conditions later on where needed; see the discussion after~\eqref{EqD4Fpasing}.}
  \begin{equation}
  \label{EqD4Sub0}
    \bigl( \hat u_{{\rm reg},4}(0),\ \hat b_4(0),\ \hat\scal_4(0) \bigr) := \wt{L_b}(0)^{-1}\bigl(\hat f_4(0),\ 0,\ 0\bigr),
  \end{equation}
  so $\hat b_4(0)\in\C^4$, $\hat\scal_4(0)\in\scalspace_1$, and $\hat u_{{\rm reg},4}(0) \in \Hb^{\infty,\ \eps_\ind-\eps}(X;S^2\cT^*_X)$; and analogously to~\eqref{EqD3Almf2} (as far as the $\tface$- and $\zface$-orders are concerned, with $\alpha_+,\tilde\alpha_\cK$ there being $1+\eps_\ind-\eps,\alpha_\cK-1$ in present notation) we have
  \begin{align*}
    &\bigl(\hat f_4(\sigma),\ 0,\ 0\bigr) - \wt{L_b}(\sigma)\bigl(\chi_\zface\hat u_{{\rm reg},4}(0),\ \hat b_4(0),\ \hat\scal_4(0)\bigr) = \bigl(\hat f_{5,0}+\hat f_{5,1},\ 0,\ 0\bigr), \\
    &\qquad \hat f_{5,0} \in |\sigma|\Hb^{\infty,\ 2-\eps,\ 1+\eps_\ind-\eps,\ \bigl((0,0),2+\eps_\ind-\eps\bigr)}(X_\scbtop^\pm), \\
    &\qquad \hat f_{5,1} \in \Hb^{\infty,\ 2-\eps,\ 2+\eps_\ind-\eps,\ \alpha_\cK-2}(X_\scbtop^\pm).
  \end{align*}
  Since we have reached the most delicate term we are presently interested in, it suffices to record
  \begin{equation}
  \label{EqD4f5}
    \hat f_5 := \hat f_{5,0}+\hat f_{5,1} \in \Hb^{\infty,\ 2-\eps,\ 2+\eps_\ind-\eps,\ \alpha_\cK-2}(X_\scbtop^\pm),
  \end{equation}

  \pfstep{Step~5. Remaining piece; combination.} Note that~\eqref{EqD4f5} has relative $\tface$-order $(2+\eps_\ind-\eps)-(\alpha_\cK-2)\in(2,2+\eps_\ind)$ (since we can take $\eps>0$ arbitrarily small). We can thus use \citeAF{Proposition~\ref*{PropDResLo}} to obtain
  \begin{align}
  \label{EqD4u5}
    &\bigl(\hat u_5(\sigma),\ \hat b_5(\sigma),\ \hat\scal_5(\sigma)\bigr) := \wt{L_b}(\sigma)^{-1}\bigl(\hat f_5(\sigma),\ 0,\ 0\bigr), \\
    &\qquad \hat u_5 \in \Hb^{\infty,\ 1-\eps,\ \eps_\ind-\eps,\ \alpha_\cK-2}(X_\scbtop^\pm), \nonumber\\
    &\qquad \hat b_5\in\Hb^{\infty,\ \alpha_\cK-2}(\pm[0,c]_\sigma;\C^4),\ \ \hat\scal_5\in\Hb^{\infty,\ \alpha_\cK-2}(\pm[0,c]_\sigma;\scalspace_1). \nonumber
  \end{align}
  Altogether, we have
  \[
    u = u_{\rm sing} + u_{\rm log} + u_{\rm reg} + u_\rem,
  \]
  where (recalling~\eqref{EqD4u0}, \eqref{EqD4Sub0}, \eqref{EqD4u5} for $u_{\rm sing}$,~\eqref{EqD4ipErr} for $u_{\rm log}$, further \eqref{EqD4u0}, \eqref{EqD4u2}--\eqref{EqD4u3} for $u_{\rm reg}$, and \eqref{EqD4u1scri}, \eqref{EqD4ip} for $u_\rem$)
  \begin{align}
  \label{EqD4FTusing}
    (\cF u_{\rm sing})(\sigma) &:= \cF(\dot g_b^{\Ups,\aug})(\sigma)\bigl(\underbrace{\hat b_1(0) + \sigma\hat b_4(0) + \sigma\hat b_5(\sigma)}_{\in\Hb^{\infty,((0,0),\alpha_\cK-1)}}\bigr) + i\cF(\pa_{t_*}h_{b,\rms 1}^{\leq 1,\aug})(\sigma)\bigl(\underbrace{\hat\scal_4(0)+\hat\scal_5(\sigma)}_{\in\Hb^{\infty,((0,0),\alpha_\cK-2)}}\bigr), \\
    u_{\rm log} &:= \chi_\cK h_{b,\rms 1}^{\leq 4}(\dot\scal_1^{(0,1)}\log t_*), \nonumber\\
    (\cF u_{\rm reg})(\sigma) &:= \chi(|\sigma|) \Bigl( \chi_\zface\bigl(\hat u_{{\rm reg},1}(0) + \sigma\hat u_{{\rm reg},4}(0)\bigr) + \hat u_{2,0}(\sigma) + \hat u_3(\sigma) + \sigma\hat u_5(\sigma)\Bigr) \nonumber\\
      &\qquad \in \Hb^{\infty,\ 1-\eps,\ \eps_\ind-\eps,\ -1+\eps_\ind-\eps}(X_\scbtop^\pm), \nonumber\\
    u_\rem &:= u_{\rm hi} + u_{1,\sscri} + u_\iota \in \Hb^{\infty,\ 1-\eps,\ 1-\eps,\ \alpha_\cK}(M';S^2\cT^*). \nonumber
  \end{align}
  Extending $\cF u_{\rm reg}$ uniquely to an $L^1_\loc$-function of $\sigma\in[-c,c]$, we can use~\eqref{EqTFHbInvLo} to deduce
  \[
    u_{\rm reg} \in \Hb^{\infty,\ 1-\eps,\ 1+\eps_\ind-\eps,\ \eps_\ind-\eps} \subset \Hb^{\infty,\ 1-\eps,\ 1-\eps,\ \alpha_\cK-2}(M';S^2\cT^*).
  \]
  Recalling that $\cF u_{\rm sing}=\cF(u-u_{\rm log}-u_{\rm reg}-u_\rem)$ is a well-defined tempered distribution on $\R$, define $u_{\rm sing}$ as its inverse Fourier transform and note then that
  \begin{equation}
  \label{EqD4usingPast}
    u_{\rm sing} = -u_{\rm log}-u_{\rm reg}-u_\rem \in \bar H_\bop^{\infty,\ 1-\eps,\ 1-\eps,\ \alpha_\cK-2}(\Omega_-),\quad \Omega_-:=\cl_{M'}\{\ft_*<1\};
  \end{equation}
  and, analogously to~\eqref{EqD3Almpausing} (using~\eqref{EqD4FTusing} and \eqref{EqD1AlmAug}),
  \begin{equation}
  \label{EqD4Fpasing}
    \cF(\pa_{t_*}u_{\rm sing})(\sigma) \in \Hb^{\infty,\ 1-\eps,\ 1-\eps,\ ((0,0),\alpha_\cK-2)}(X_\scbtop^\pm).
  \end{equation}
  If the restrictions of this to $\zface\subset X_\scbtop^\pm$ did not match,\footnote{This would happen if and only if the value of $\hat\scal_4(0)$ in~\eqref{EqD4Sub0} depended on the sign of $\hat f_4(\pm 0)$.} then $\cF(\pa_{t_*}u_{\rm sing})$ would have a jump discontinuity across $\sigma=0$, and therefore $\pa_{t_*}u_{\rm sing}$ would be bounded \emph{from below} by $|t_*|^{-1}$ as $t_*\to-\infty$ in spatially compact sets; but this is incompatible with the $\cO(|t_*|^{-\alpha_\cK+1})=o(|t_*|^{-1})$ \emph{upper bound} that follows from~\eqref{EqD4usingPast}. Thus, we can use~\eqref{EqTFHbInvLo} to deduce that
  \[
    \pa_{t_*}u_{\rm sing} \in \Hb^{\infty,\ 1-\eps,\ 2-\eps,\ \alpha_\cK-1}(M';S^2\cT^*).
  \]
  Together with~\eqref{EqD4usingPast}, Lemma~\ref{LemmaD1Intb} then gives
  \[
    u_{\rm sing} \in \Hb^{\infty,\ 1-\eps,\ 1-\eps,\ ((0,0),\alpha_\cK-2)}(M';S^2\cT^*).
  \]
  If $u_{\rm log}$ and the $\cK^+$-leading-order term of $u_{\rm sing}$ vanished, we would get $u\in\Hb^{\infty,\ 1-\eps,\ 1-\eps,\ \alpha_\cK-2}$. To complete the proof of~\eqref{EqD4Paru}, it remains to characterize the leading-order term of $u_{\rm sing}$.

  \pfstep{Step~6. Leading-order term of $u_{\rm sing}$.} Recalling~\eqref{EqD4usingPast}, the constant term of $u_{\rm sing}$ as $t_*\to\infty$ is the total $t_*$-integral of $\pa_{t_*}u_{\rm sing}$, and hence equal to
  \[
    \cF(\pa_{t_*}u_{\rm sing})(0) = \cF(\pa_{t_*}\dot g_b^{\Ups,\aug})(0)(\hat b_1(0)) + \cF(\pa_{t_*}^2 h_{b,\rms 1}^{\leq 1,\aug})(0)(i\hat\scal_4(0)).
  \]
  But the total $t_*$-integrals of $\pa_{t_*}\dot g_b^{\Ups,\aug}$ and $\pa_{t_*}^2 h_{b,\rms 1}^{\leq 1,\aug}$ are $\dot g_b^\Ups$ and $h_{b,\rms 1}$, respectively, as follows from~\eqref{EqAdmLoImLot}. Therefore,
  \[
    u_{\rm sing} - \chi_\cK\bigl(\dot g_b^\Ups(\hat b_1(0)) - h_{b,\rms 1}(i\hat\scal_4(0))\bigr) \in \Hb^{\infty,\ 1-\eps,\ 1-\eps,\ \alpha_\cK-2}(M';S^2\cT^*).
  \]
  This identifies the parameters $\dot b$ and $\dot\scal_1^{(0)}$ as
  \begin{equation}
  \label{EqD4Params}
    \dot b = \hat b_1(0),\quad
    \hat\scal_1^{(0)} = i\hat\scal_4(0).
  \end{equation}
  The proof is complete.
\end{proof}

\subsubsection{Undoing the parameter changes and center-of-mass movements}
\label{SssD4ParNo}

Analogously to the discussion of the boosts starting with~\eqref{EqD2BoostLin}--\eqref{EqD2MetLin}, we first explain explain how to interpret and rewrite the terms in the expansion~\eqref{EqD4Paru} for $b=b_0$.

\pfstep{The term $\chi_\cK\dot g_{b_0}^\Ups(\dot b)$.} On the one hand, the linearization in $b$ of
\[
  b \mapsto \Ric(g_{b_0,b}) - \delta_{g_{b_0,b},E^\cC}^*\Ups_{E^\Ups}(g_{b_0,b},\,g_{b_0,b}),\quad g_{b_0,b}:=g_{b_0,b,0}=(1-\chi_\cK)g_{b_0}+\chi_\cK g_b,
\]
around $b=b_0$ maps
\[
  \dot b \mapsto D_{g_{b_0}}\Ric\bigl(\chi_\cK\dot g_{b_0}(\dot b)\bigr).
\]
On the other hand, recalling from Proposition~\ref{PropWEMode0Kerr} that $\dot g_{b_0}^\Ups(\dot b)=\dot g_{b_0}(\dot b)+\delta_{g_{b_0}}^*\dot\omega_{b_0}(\dot b)$ satisfies the linearized gauge condition (and of course the linearized Einstein equation), we compute
\begin{align*}
  L_{b_0}\bigl(\chi_\cK\dot g_{b_0}^\Ups(\dot b)\bigr) &= D_{g_{b_0}}\Ric\bigl(\chi_\cK\dot g_{b_0}(\dot b) + \chi_\cK\delta_{g_{b_0}}^*\dot\omega_{b_0}(\dot b)\bigr) + \delta_{g_{b_0}}^*[\delta_{g_{b_0},E^\Ups}\sfG_{g_{b_0}},\chi_\cK]\dot g_{b_0}^\Ups(\dot b) \\
    &= D_{g_{b_0}}\Ric\Bigl(\chi_\cK\dot g_{b_0}(\dot b) - [\delta_{g_{b_0}}^*,\chi_\cK]\dot\omega_{b_0}(\dot b)\Bigr) + \delta_{g_{b_0}}^*[\delta_{g_{b_0},E^\Ups}\sfG_{g_{b_0}},\chi_\cK]\dot g_{b_0}^\Ups(\dot b).
\end{align*}
Defining the gauge-fixed Einstein operator
\begin{subequations}
\begin{equation}
\label{EqD4ParNotildeP}
\begin{split}
  \tilde P(h,b) &:= \Ric\Bigl( g_{b_0,b}-[\delta_{g_{b_0}}^*,\chi_\cK]\dot\omega_{b_0}(b-b_0) + h \Bigr) \\
    &\qquad - \delta_{g_{b_0,b},E^\cC}^*\Bigl( \Ups_{E^\Ups}(g_{b_0,b}+h,g_{b_0,b}) - [\delta_{g_{b_0},E^\Ups}\sfG_{g_{b_0}},\chi_\cK]\dot g_{b_0}^\Ups(b-b_0)\Bigr),
\end{split}
\end{equation}
comparison of these two computations thus gives
\begin{equation}
\label{EqD4ParNoRewrite}
  L_{b_0}\bigl(\chi_\cK\dot g_{b_0}^\Ups(\dot b)\bigr) = D_{(0,b_0)}\tilde P\bigl(\chi_\cK\dot g_{b_0}^\Ups(\dot b),0\bigr) = D_{(0,b_0)}\tilde P(0,\dot b).
\end{equation}
\end{subequations}
This identity re-interprets the late-time contribution $\chi_\cK\dot g_b^\Ups(\dot b)$ to a solution of the linearized gauge-fixed Einstein equation as a change in the parameters $b$ of the final black hole. The mismatch between the correctly gauged linearized Kerr metric $\dot g_b^\Ups(\dot b)$ and the linearization of the presentation $g_b$ of the Kerr family we fixed in Definition~\ref{DefKMetcM} (prior to any gauge considerations) is accounted for by a mild modification $[\delta_{g_{b_0},E^\Ups}\sfG_{g_{b_0}},\chi_\cK]\dot g_{b_0}^\Ups(\dot b)=\cO(\rho_+^2)$ of the gauge condition in the transition region $\supp\dd\chi_\cK$ between the initial and final black hole background and an additional metric term $[\delta_{g_{b_0}}^*,\chi_\cK]\dot\omega_{b_0}=\dd\chi_\cK\otimes_s\dot\omega_{b_0}=\cO(\rho_+\log\rho_+)$ that is also supported there.

\begin{rmk}[Size of metric perturbations and gauge modifications]
\label{RmkD4ParSize}
  We recall here that metric perturbations of size $\cO(\rho_+)$ (up to a logarithm) are acceptable in Definition~\ref{DefDMetBasic}; and gauge modifications of size $\cO(\rho_+^2)$ (which $\delta_{g_{b_0},E^\cC}^*$ maps into $\cO(\rho_+^3)$) are compatible with our requirements, from Proposition~\ref{PropD2Boost2} onward, that source terms for $L_{b_0}$ have $\cO(\rho_+^3)$-decay at $\iota^+$.
\end{rmk}

\pfstep{The term $\chi_\cK h_{b,\rms 1}$.} We next turn to the constant center-of-mass shift in~\eqref{EqD4Paru}: we recall from~\S\ref{SssIEinElim} how on an exact Kerr spacetime it can be eliminated by a suitable gauge modification. To wit, since $h_{b_0,\rms 1}=\delta_{g_{b_0}}^*\omega_{b_0,\rms 1}^{(0)}$ (see~\eqref{EqWG0Symmomega} and \eqref{EqWG0hs1}) satisfies the linearized Einstein equation and gauge condition, we have (omitting the argument $\dot\scal_1^{(0)}\in\scalspace_1$ from the notation)
\begin{align*}
  L_{b_0}(\chi_\cK h_{b_0,\rms 1}) &= D_{g_{b_0}}\Ric\bigl(\chi_\cK\delta_{g_{b_0}}^*\omega_{b_0,\rms 1}^{(0)}\bigr) + \delta_{g_{b_0},E^\cC}^* [\delta_{g_{b_0},E^\Ups}\sfG_{g_{b_0}},\chi_\cK]h_{b_0,\rms 1} \\
    &= -D_{g_{b_0}}\Ric\bigl( [\delta_{g_{b_0}}^*,\chi_\cK]\omega_{b_0,\rms 1}^{(0)} \bigr) + \delta_{g_{b_0},E^\cC}^*[\delta_{g_{b_0},E^\Ups},\sfG_{g_{b_0}},\chi_\cK]h_{b_0,\rms 1};
\end{align*}
so the gauge modification $[\delta_{g_{b_0},E^\Ups}\sfG_{g_{b_0}},\chi_\cK]h_{b_0,\rms 1}$, which has support on $\supp\dd\chi_\cK$ and is of size $\cO(\rho_+^2)$, produces the asymptotic shift $\chi_\cK h_{b_0,\rms 1}$ up to a metric term $\dd\chi_\cK\otimes_s\omega_{b_0,\rms 1}^{(0)}$ with acceptable size $\cO(\rho_+)$. Ignoring the modifications made in~\eqref{EqD4ParNotildeP} to account for changes in the Kerr parameters, we are thus led to define
\begin{subequations}
\begin{equation}
\label{EqD4ParNotildeP2}
\begin{split}
  \tilde P(h,\scal_1^{(0)}) &:= \Ric\Bigl( g_{b_0} - [\delta_{g_{b_0}}^*,\chi_\cK]\omega_{b_0,\rms 1}^{(0)}(\scal_1^{(0)}) + h \Bigr) \\
    &\qquad - \delta_{g_{b_0},E^\cC}^*\Bigl(\Ups_{E^\Ups}(g_{b_0}+h,g_{b_0}) - [\delta_{g_{b_0},E^\Ups}\sfG_{g_{b_0}},\chi_\cK]h_{b_0,\rms 1}(\scal_1^{(0)})\Bigr);
\end{split}
\end{equation}
we then have
\begin{equation}
\label{EqD4ParNoRewrite2}
  L_{b_0}\bigl(\chi_\cK h_{b_0,\rms 1}(\scal_1^{(0)})\bigr) = D_{(0,0)}\tilde P\bigl(\chi_\cK h_{b_0,\rms 1}(\scal_1^{(0)}),0\bigr) = D_{(0,0)}\tilde P(0,\scal_1^{(0)}).
\end{equation}
\end{subequations}

\pfstep{The term $\chi_\cK h_{b,\rms 1}^{\leq 4}(\log t_*)$.} We can eliminate the logarithmic shift in~\eqref{EqD4Paru} in a similar fashion, though the computations are more involved; to a large degree, this was already explained following~\eqref{EqIEinElimDec}. First of all, we observe that
\begin{equation}
\label{EqD4ParNoDelvsH}
  h_{b_0,\rms 1}^{\leq 4}(a(t_*)) = \delta_{g_{b_0}}^*\omega_{b_0,\rms 1}^{(0),\leq 4}(a(t_*)) - \dd t_*\otimes_s\breve\omega_{b_0,\rms 1}^{(0),4}(a^{(5)}(t_*)),
\end{equation}
as follows from $\breve h_{b_0,\rms 1}^j=\wh{\delta_{g_{b_0}}^*}(0)\breve\omega_{b_0,\rms 1}^{(0),j}+\dd t_*\otimes_s\breve\omega_{b_0,\rms 1}^{(0),j-1}$ for $j=1,2,3,4$ (upon defining $\breve\omega_{b_0,\rms 1}^{(0),0}:=\omega_{b_0,\rms 1}^{(0)}$). For $a=\log t_*$ (times an element of $\scalspace_1$), we can thus replace $ h_{b_0,\rms 1}^{\leq 4}(\log t_*)$ by $\delta_{g_{b_0}}^*\omega_{b_0,\rms 1}^{(0),\leq 4}(\log t_*)$ up to an error $\dd t_*\otimes_s\breve\omega_{b_0,\rms 1}^{(0),4}(t_*^{-5})=\cO(\rho^{-4}t_*^{-5})=\cO(\rho_+\rho_\cK^5)$ with strong decay at $\cK^+$ (which can be absorbed into $\tilde u$ in~\eqref{EqD4Paru}). We then have
\begin{equation}
\label{EqD4ParRewrite3}
\begin{split}
  L_{b_0}\bigl(\chi_\cK\delta_{g_{b_0}}^*\omega_{b_0,\rms 1}^{(0),\leq 4}(a(t_*))\bigr) &= -D_{g_{b_0}}\Ric\Bigl([\delta_{g_{b_0}}^*,\chi_\cK]\omega_{b_0,\rms 1}^{(0),\leq 4}(a(t_*))\Bigr) \\
    &\quad \qquad + \delta_{g_{b_0},E^\cC}^*\delta_{g_{b_0},E^\Ups}\sfG_{g_{b_0}}\Bigl(\chi_\cK\delta_{g_{b_0}}^*\omega_{b_0,\rms 1}^{(0),\leq 4}(a(t_*))\Bigr),
\end{split}
\end{equation}
where the argument of $D_{g_{b_0}}\Ric$ is supported on $\supp\dd\chi_\cK$ and of size $\cO(\rho_+\rho^{-4}a^{(4)})=\cO(\rho_+)$ for $a=\log t_*$ (which thus has acceptable decay). In the second line, we further compute the gauge modification (i.e., the argument of $\delta_{g_{b_0},E^\cC}^*$) to be
\begin{subequations}
\begin{equation}
\label{EqD4ParNoLog1}
\begin{split}
  \delta_{g_{b_0},E^\Ups}\sfG_{g_{b_0}}\Bigl(\chi_\cK\delta_{g_{b_0}}^*\omega_{b_0,\rms 1}^{(0),\leq 4}(a(t_*))\Bigr) &= [\delta_{g_{b_0},E^\Ups}\sfG_{G_{b_0}},\chi_\cK] \Bigl( h_{b,\rms 1}^{\leq 4}(a(t_*)) + \dd t_*\otimes_s\breve\omega_{b_0,\rms 1}^{(0),4}(a^{(5)}(t_*))\Bigr) \\
    &\qquad + \frac12\chi_\cK\Box_{g_{b_0},E^\Ups}^\Ups \omega_{b_0,\rms 1}^{(0),\leq 4}(a(t_*));
\end{split}
\end{equation}
for $a=\log t_*$, the first term is of size $\cO(\rho_+^2)$ (with $h_{b,\rms 1}^{\leq 1}(\log t_*)$ only contributing terms of size $\cO(\rho_+^3\log\rho_+)$), while the second summand can be computed using the identity~\eqref{EqipGrKerrComp} (with $\Box_{g_{b_0},E^\Ups}^\Ups$ and $\omega$ in place of $L_b$ and $h$, and with $k=4$), so upon abbreviating $\Box^\Ups:=\Box_{g_{b_0},E^\Ups}^\Ups$,
\begin{equation}
\label{EqD4ParNoLog2}
\begin{split}
  \Box^\Ups\omega_{b_0,\rms 1}^{(0),\leq 4}(a(t_*)) &= a^{(5)}(t_*)\Bigl( [\Box^\Ups,t_*]\ftrans(0)\breve\omega_{b_0,\rms 1}^{(0),4} + \frac12[[\Box^\Ups,t_*],t_*]\breve\omega_{b_0,\rms 1}^{(0),3}\Bigr) \\
    &\qquad + a^{(6)}(t_*)\frac12[[\Box^\Ups,t_*],t_*]\breve\omega_{b_0,\rms 1}^{(0),4} \\
    &= \cO(\rho_+^2\rho_\cK^5)\quad\text{for}\ a=\log t_*.
\end{split}
\end{equation}
\end{subequations}
Altogether, the gauge modification is thus of the acceptable size $\cO(\rho_+^2\rho_\cK^5)$. (Upon acting on such a term with $\delta_{g_{b_0},E^\cC}^*$, this gives a $\cO(\rho_+^3\rho_\cK^5)$ source term. The decay at $\iota^+$ was already addressed in Remark~\ref{RmkD4ParSize}, and more than $4+\eps_\cK$ orders of $\cK^+$-decay are sufficient for all present and future purposes as well; see Remark~\ref{RmkDSource}.) We can thus eliminate the term $\chi_\cK\delta_{g_{b_0}}^*\omega_{b_0,\rms 1}^{(0),\leq 4}(\log t_*)$ much as in~\eqref{EqD4ParNotildeP2}--\eqref{EqD4ParNoRewrite2}.

The full expression for the gauge-fixed Einstein operator motivated by these considerations is given in Definition~\ref{DefD4EinsteinAug} below. First, we introduce notation for the metric and gauge modification terms arising above:

\begin{definition}[Correction terms \#2]
\label{DefD4Corr}
  For $\dot b\in\R^4$ and $\scal\in\scalspace_1$, and for the fixed cutoff $\chi_\cK$ from~\eqref{DefKBoCutoff} and~\eqref{EqDCutoffs}, we define
  \begin{align*}
    h^{(0)}(\dot b) &:= -[\delta_{g_{b_0}}^*,\chi_\cK]\dot\omega_{b_0}(\dot b), \\
    \vartheta^{(0)}(\dot b) &:= [\delta_{g_{b_0},E^\Ups}\sfG_{g_{b_0}},\chi_\cK]\dot g_{b_0}^\Ups(\dot b),
  \end{align*}
  further
  \begin{align*}
    h^{(0)}_{\rms 1}(\scal) &:= -[\delta_{g_{b_0}}^*,\chi_\cK]\omega_{b_0,\rms 1}^{(0)}(\scal), \\
    \vartheta^{(0)}_{\rms 1}(\scal) &:= [\delta_{g_{b_0},E^\Ups}\sfG_{g_{b_0}},\chi_\cK]h_{b_0,\rms 1}(\scal),
  \end{align*}
  and finally
  \begin{align}
    h^{(0,1)}_{\rms 1}(\scal) &:= -[\delta_{g_{b_0}}^*,\chi_\cK]\omega_{b_0,\rms 1}^{(0),\leq 4}(\scal\log t_*), \nonumber\\
  \label{EqD4CorrTheta01}
    \vartheta^{(0,1)}_{\rms 1}(\scal) &:= \delta_{g_{b_0},E^\Ups}\sfG_{g_{b_0}}\Bigl(\chi_\cK\delta_{g_{b_0}}^*\omega_{b_0,\rms 1}^{(0),\leq 4}(\scal\log t_*)\Bigr).
  \end{align}
\end{definition}

The analogue of Lemma~\ref{LemmaD2Corr} is:

\begin{lemma}[Structural properties of correction terms]
\label{LemmaD4Corr}
  For some index set $\cE_\ind$ (depending only on the parameters $\gamma^\Ups,e^\Ups,v^\cC,\gamma^\cC$) with $\min\Re\cE_\ind\geq\eps_\ind$, we have (writing $\cA=\cA(\Omega_*;\cT^*)^{\bullet,-}$ or $\cA(\Omega_*;S^2\cT^*)^{\bullet,-}$)
  \begin{align*}
    h^{(0)}(\dot b),\ h^{(0,1)}_{\rms 1}(\scal) &\in \cA^{\infty,\ (1,1)\cup(1+\cE_\ind),\ \infty}, \\
    h^{(0)}_{\rms 1}(\scal) &\in \cA^{\infty,\ (1,0)\cup(1+\cE_\ind),\ \infty},
  \end{align*}
  further
  \begin{equation}
  \label{EqD4CorrEin}
    D_{g^0}\Ric\bigl(h^{(0)}(\dot b)\bigr),\ D_{g^0}\Ric\bigl(h^{(0,1)}_{\rms 1}(\scal)\bigr) \in \cA^{\infty,\ (3,0)\cup(3+\cE_\ind),\ \infty}
  \end{equation}
  for any $g^0\in\CI(M;S^2\cT^*)$ that equals $\ubar g$ modulo $\rho_+\CI$; and finally
  \begin{align}
    \vartheta^{(0)}(\dot b) &\in \cA^{\infty,\ (2,0)\cup(2+\cE_\ind),\ \infty}, \nonumber\\
    \vartheta^{(0)}_{\rms 1}(\scal) &\in \cA^{\infty,\ (3,0)\cup(3+\cE_\ind),\ \infty}, \nonumber\\
  \label{EqD4CorrTheta101}
    \vartheta^{(0,1)}_{\rms 1}(\scal) &\in \cA^{\infty,\ (2,0)\cup(2+\cE_\ind),\ 5}.
  \end{align}
  The tensors $h^{(0)}$, $h_{\rms 1}^{(0)}$, $h_{\rms 1}^{(0,1)}$, $\vartheta^{(0)}$, and $\vartheta^{(0)}_{\rms 1}$ have supports in $\supp\dd\chi_\cK$ (and thus disjoint from $\scri^+$ and $\cK^+$), while $\supp\vartheta^{(0,1)}_{\rms 1}\subset\supp\chi_\cK$ (which is disjoint only from $\scri^+$).
\end{lemma}
\begin{proof}
  The commutators in Definition~\ref{DefD4Corr} are of class $\rho_\sscri^\infty\rho_+\rho_\cK^\infty\Diffb^0(M')$, which in view of $\dot\omega_{b_0}(\dot b)\in\cA^{(0,1)\cup\cE_\ind}(X;\cT^*_X)$ and $\dot g_{b_0}^\Ups(\dot b)\in\cA^{(1,0)\cup(1+\cE_\ind)}(X;S^2\cT^*_X)$ (from Proposition~\ref{PropWEMode0Kerr}) as well as $\omega_{b_0,\rms 1}^{(0)}(\scal)\in\cA^{(0,0)\cup\cE_\ind}$ and $h_{b_0,\rms 1}(\scal)\in\cA^{(2,0)\cup(2+\cE_\ind)}$ (from Proposition~\ref{PropWG0Symm}) yields the stated memberships of $h^{(0)}$, $h^{(0)}_{\rms 1}$, $\vartheta^{(0)}$, and $\vartheta^{(0)}_{\rms 1}$. Using moreover that $\breve\omega_{b_0,\rms 1}^{(0),j}\in\cA^{(-j,0)\cup(-j+\cE_\ind)}$, we control $h_{\rms 1}^{(0,1)}$. The claimed membership for $\vartheta^{(0,1)}_{\rms 1}$ follows by inspection of~\eqref{EqD4ParNoLog1}--\eqref{EqD4ParNoLog2}.

  Regarding~\eqref{EqD4CorrEin}, it follows from $D_{g^0}\Ric\in\rho_+^2\Diffb^2$ that the $\iota^+$-index set is contained in $(3,1)\cup(3+\cE_\ind)$. We thus only need to prove the vanishing of the logarithmic $(3,1)$-term. This is equal to a linear combination of the $(3,1)$-terms of
  \[
    D_{\ubar g}\Ric\bigl( [\ubar\delta^*,\chi_\cK]\log(\rho)\ubar\omega_{\rms l}^{(0)} \bigr)
  \]
  for $l=0$ (cf.\ \eqref{EqWEMode0KerrOmega}) and $l=1$ (from $\log(t_*)\omega_{b_0,\rms 1}^{(0)}(\scal)$)---which vanish by the computation~\eqref{EqD2CorrNoLog}. (Note that~\eqref{EqD2CorrNoLog} applies to $\ubar\omega_{\rms 0}^{(0)}$ as well without changes since this is a Killing 1-form for the Minkowski metric.)
\end{proof}

\begin{rmk}[$\cK^+$-order]
\label{RmkD4KOrder}
  We stress that the high $\cK^+$-order of $\vartheta_{\rms 1}^{(0,1)}(\scal)$ arises from the usage of a high-order Taylor-type expansion at $\cK^+$ of an approximate solution of the linear operator $L_{b_0}$ that has $\log(t_*)h_{b_0,\rms 1}(\scal)$ as its leading-order term. We could improve the order to any desired number $N+1\geq 5$ by invoking further terms $\breve\omega_{b_0,\rms 1}^{(j)}$, $j=5,\ldots,N$ (constructed using the same arguments as for Proposition~\ref{PropWG0Large}\eqref{ItWG0Larges1m1}). However, any order $>4$ (really $\geq 4+\eps_\ind-\eps$ for all $\eps>0$ to make the margin quantitative) suffices for our purposes since source terms for the linearized gauge-fixed Einstein equation with $4+\eps_\cK$ orders of vanishing at $\cK^+$ are good enough for the full asymptotic analysis needed to close a nonlinear iteration; see Corollary~\ref{CorD6Impr}. The same rationale will be used in the construction of modification terms in later sections.
\end{rmk}

The following extension of the operator from Definition~\ref{DefD2EinsteinAug} accounts for the additional expansion terms in~\eqref{EqD4Paru}:

\begin{definition}[Gauge-fixed Einstein operator, augmentation \#2]
\label{DefD4EinsteinAug}
  For symmetric 2-tensors $h\in\cX^\infty=\Hb^{\infty,\ \bigl(\la\cE_\sscri^\cC\ra,3+\eps_\sscri\bigr),\ (\cE_+,3+\eps_+),\ 2+\eps_\cK}(\Omega_*)$ with sufficiently small $\cX^d$-norm (for some large but fixed $d$), small $\scal,\scal_1^{(0,1)},\scal_1^{(0)}\in\scalspace_1$, and for $b\in\R^4$ close to $b_0=(\bhm_0,\bha_0)$, write
  \[
    \vecp := (\scal,b-b_0,\scal_1^{(0)},\scal_1^{(0,1)})
  \]
  and define
  \begin{equation}
  \label{EqD4EinsteinAug}
    P(h,\vecp) := \Ric\bigl( g^0 + h_\tot(\vecp) + h \bigr) - \delta_{g^0,E^\cC}^*\bigl( \Ups_{E^\Ups}(g^0+h,g^0) - \vartheta_{\rm tot}(\vecp) \bigr),
  \end{equation}
  where we recall Definitions~\usref{DefD2Corr} and \usref{DefD4Corr} and abbreviate
  \begin{align*}
    g^0 &:= g_{b_0,b,-\scal}, \\
    h_\tot(\vecp) &:= h^{(-1)}(\scal) + h^{(0)}(b-b_0) + h_{\rms 1}^{(0)}(\scal_1^{(0)}) + h_{\rms 1}^{(0,1)}(\scal_1^{(0,1)}), \\
    \vartheta_\tot(\vecp) &:= \vartheta^{(-1)}(\scal) + \vartheta^{(0)}(b-b_0) + \vartheta_{\rms 1}^{(0)}(\scal_1^{(0)}) + \vartheta_{\rms 1}^{(0,1)}(\scal_1^{(0,1)}).
  \end{align*}
  We denote its linearization in $h$ by
  \begin{equation}
  \label{EqD4EinsteinAugLin}
    L_{h,\vecp} := D_{(h,\vecp)}P(\cdot,\vec 0).
  \end{equation}
\end{definition}

Since $h_\tot(\vecp)$ is of the same class as $h^{(-1)}(\scal)$, the arguments following Definition~\ref{DefD2EinsteinAug} apply and imply that $L_{h,\vecp}$ is admissible, and all decay results we have proved thus far (in particular Propositions~\ref{PropD2Boost2} and~\ref{PropD4Par}) hold also for $L=L_{h,\vecp}$ when $\vecp$ is small.

Now, we wish to consider $D_{(h,\vecp)}P(0,\dot\vecp)$ as an additional source term for $L u=f$ that makes the boosts, Kerr parameter changes, and logarithmic or stationary center-of-mass motions vanish. Similarly to Lemma~\ref{LemmaD2LinPar}, we thus need to record:

\begin{lemma}[Linearization of $P$ in the parameters, \#2]
\label{LemmaD4LinPar}
  Let $h$ and $\vecp=(\scal,b-b_0,\scal_1^{(0)},\scal_1^{(0,1)})$ be as in Definition~\usref{DefD4EinsteinAug}; recall $\tilde\cE_+=(1,0)\cup(1+\cE_\ind)\cup\bigl(\cE_++((1,1)\cup(1+\cE_\ind))\bigr)$ from~\eqref{EqD2LinPartildeE}. Recalling the cutoffs~\eqref{EqDCutoffs}, we can write
  \begin{equation}
  \label{EqD4LinParScal}
  \begin{split}
    &D_{(h,\vecp)}P(0,(\scal',0,0,0)) = (1-\chi_\cK^\flat)\bigl(f^1_{h,\vecp}(\scal') + \chi_\cK^\sharp f^2_{h,\vecp}(\scal')\bigr), \\
    &\qquad f^1_{h,\vecp}(\scal') \in \Hb^{\infty,\ \bigl(\la\cE_\sscri^\cC+1\ra',\,4+\eps_\sscri\bigr),\ (\cE_++3,\,6+\eps_+),\ 0}, \\
    &\qquad f^2_{h,\vecp}(\scal') \in \Hb^{\infty,\ 0,\ (\tilde\cE_++2,\,6+\eps_+-\eps),\ 0}\quad\forall\,\eps>0,
  \end{split}
  \end{equation}
  as in~\eqref{EqD2LinParScal}, and
  \begin{equation}
  \label{EqD4LinParb}
    D_{(h,\vecp)}P(0,(0,b',0,0)) \in \chi_\cK^\sharp \Hb^{\infty,\ 0,\ (\tilde\cE_++2,\,6+\eps_+),\ 2+\eps_\cK};
  \end{equation}
  the $\cK^+$-order of this is equal to $\ell_\cK$ when the $\cK^+$-order of $h$ is equal to $\ell_\cK\in(0,4+\eps_\ind)$ instead of $2+\eps_\cK$. Furthermore,
  \begin{equation}
  \label{EqD4LinParCoM}
    D_{(h,\vecp)}P(0,(0,0,\scal_1^{(0)\prime},0)),\ D_{(h,\vecp)}P(0,(0,0,0,\scal_1^{(0,1)\prime})) \in \chi_\cK^\sharp \Hb^{\infty,\ 0,\ (\tilde\cE_++2,\,6+\eps_+-\eps),\ 5-\eps}.
  \end{equation}
\end{lemma}
\begin{proof}
  The linearization in $\scal$ can be computed exactly as in the proof of Lemma~\ref{LemmaD2LinPar}. For the linearization of $\Ric$, this uses that $h_\tot(\vecp)$ (appearing in~\eqref{EqD4EinsteinAug}) is of the same class as $h^{(-1)}(\scal)$ (appearing in~\eqref{EqD2EinsteinAug}), namely $\cA^{\infty,\ (1,1)\cup(1+\cE_\ind),\ \infty}$. The only additional term arises from the linearization of $\delta_{g^0,E^\cC}^*$ acting on the terms of $\vartheta_\tot(\vecp)$ besides $\vartheta^{(-1)}(\scal)$; but by Lemma~\ref{LemmaD4Corr}, these terms are of the same class $\cA^{\infty,\ (2,0)\cup(2+\cE_\ind),\ \infty}$ as $\vartheta^{(-1)}(\scal)$ (see~\eqref{EqD2CorrTheta}), and thus give rise to the same type of contribution (which is part of $f^1_{h,\scal,b}$ in the notation of~\eqref{EqD2LinParScal}.

  The linearization of $P$ in $b$ is supported in $\supp\chi_\cK$. We first work near $(\iota^+)^\circ$ and only record $\iota^+$-orders. Write $g=g^0+h_\tot(\vecp)+h$. Since $\chi_\cK\dot g_b(\dot b)\in\rho_+\CI$, we can use~\eqref{EqD2LinParDRic} to deduce that $D_g\Ric(\chi_\cK\dot g_b(\dot b))$ has $\iota^+$-order $((3,0)\cup(4,1)\cup(3+\cE_+),\,6+\eps_+)$, which is consistent with~\eqref{EqD4LinParb}. Similarly, one finds that $D_g\Ric(h^{(0)}(\dot b))$ (which has support in $\supp\dd\chi_\cK$) has order $((3,1)\cup(\tilde\cE_++2),6+\eps_+-\eps)$ for all $\eps>0$; but since the $(3,1)$-coefficient vanishes by~\eqref{EqD4CorrEin}, the index set is, in fact, $\tilde\cE_++2$. The linearization of the gauge term in $b$ can be analyzed exactly like the linearization in $\scal$ in the proof of Lemma~\ref{LemmaD2Corr}, except for one new term, namely, the linearization of $\delta_{g^0,E^\cC}^*\vartheta_\tot(\vecp)$ in $b$; this new term is the sum of the linearization of $\delta_{g^0,E^\cC}^*$ (which is of class $\rho_+^2\Diffb^0$) acting on $\vartheta_\tot(\vecp)$ (which has $\iota^+$-order $(2,0)\cup(2+\cE_\ind)$), and of $\delta_{g^0,E^\cC}^*\in\rho_+\Diffb^1$ on $\vartheta^{(0)}(\dot b)$---both summands having $\iota^+$-index set $(3,0)\cup(3+\cE_\ind)$.

  To control the linearization of $P$ in $b$ near $\cK^+$, we now work in the region where $\chi_\cK=1$, so
  \begin{align*}
    P(h,\vecp) &= \Ric(g_b+h) - \delta_{g_b,E^\cC}^*\bigl(\Ups_{E^\Ups}(g_b+h,g_b) - \vartheta_{\rms 1}^{(0,1)}(\scal_1^{(0,1)})\bigr) \\
      &= P(g_b+h,g_b) + \delta_{g_b,E^\cC}^*\vartheta_{\rms 1}^{(0,1)}(\scal_1^{(0,1)})
  \end{align*}
  in the notation of~\eqref{Eq1Ein}. We only record $\iota^+$- and $\cK^+$-orders now. By Corollary~\ref{CorDAdmFwN}, the first term has orders $(\cE_++2,5+\eps_+)$ and $2+\eps_\cK$ (or $\ell_\cK$ if that is the $\cK^+$-order of $h$); varying $b$ (which leaves $g_b|_{\iota^+}=\ubar g|_{\iota^+}$ unchanged) leads to changes of orders $(\cE_++3,6+\eps_+)$ and $2+\eps_\cK$ (or $\ell_\cK$). The linearization of the gauge term in $b$ has orders $(4,0)\cup(4+\cE_\ind)$ and $5$. Passing to $L^2$-based spaces gives~\eqref{EqD4LinParb}.

  The second tensor~\eqref{EqD4LinParCoM} is the sum of two terms. The first is $D_g\Ric(h_{\rms 1}^{(0,1)}(\scal_1^{(0,1)\prime}))$, which is supported on $\supp\dd\chi_\cK$; and the arguments involving~\eqref{EqD2LinParDRic} again apply. The second term is equal to $\delta_{g^0,E^\cC}^*\vartheta_{\rms 1}^{(0,1)}(\scal_1^{(0,1)\prime})\in\cA^{\infty,\ (3,0)\cup(3+\cE_\ind),\ 5}$. Similar (in fact, simpler) arguments apply also to the first term in~\eqref{EqD4LinParCoM}.
\end{proof}

Since the terms~\eqref{EqD4LinParScal}--\eqref{EqD4LinParCoM} are acceptable source terms $f$ in Proposition~\ref{PropD2Boost2}, a repetition of the arguments for the proof of Proposition~\ref{PropD2Boost2}\eqref{ItD2Boost2Par} produces a linear map $\scal_{h,\vecp}^{(-1)} \in \cL(\scalspace_1)$ that maps $\scal'$ to the boost parameter $\dot\scal$ for the solution of $L_{h,\vecp}u=D_{(h,\vecp)}P(0,(\scal',0,0,0))$ written as in~\eqref{EqD2Boost2}; and this map is invertible (for all sufficiently small $h\in\cX^d$ and $\vecp$) since it is the identity when $(h,\vecp)=(0,\vec 0)$. As already noted in~\S\ref{SssINElim}, we cannot, however, simply use $D_{(h,\vecp)}P(0,(0,b',0,0))$ (for suitable $b'$) to subsequently eliminate the first summand in~\eqref{EqD4Paru} (and similarly for the other terms): the reason is that when $(h,\vecp)\neq(0,\vec 0)$, one expects the source term $D_{(h,\vecp)}P(0,(0,b',0,0))$ to produce a metric perturbation featuring a \emph{non-zero}, albeit $\cO(\|h\|+|\vecp|)$-\emph{small} compared to $b'$, boost parameter. (Only in the exact Kerr case is there a clear separation of these source terms and the asymptotic parameters they eliminate, as, e.g., in~\eqref{EqD2BoostRewrite} and \eqref{EqD4ParNoRewrite}.) Thus, given a source term $f$, the correct choice of $\scal'$ depends (albeit weakly) on the choice of $b'$ etc., and indeed $\scal'$ is an affine-linear function of $(f,b',\ldots)$; and we must then show that for suitable initial choices of $b'$ etc,, and for the choice of $\scal'$ that is then forced upon us, we can eliminate the parameters $\dot b$ etc.\ in~\eqref{EqD4Paru} (which are then close to $b'$ etc.). We introduce some notation for such multi-step choices:

\begin{definition}[Modification parameters]
\label{DefD4ModPar}
  For $\vecp=(\scal,b-b_0,\scal^{(0)}_1,\scal^{(0,1)}_1)$, write $\vecp^{<0}:=\scal$ and $\vecp^{=0}:=(b-b_0,\scal^{(0)}_1,\scal^{(0,1)}_1)$. Similarly, given $\vecp'=(\scal',b',\scal^{(0)\prime}_1,\scal^{(0,1)\prime}_1)$, write
  \[
    \vecp^{\prime,<0} := \scal',\quad
    \vecp^{\prime,=0} := (b',\scal^{(0)\prime}_1,\scal^{(0,1)\prime}_1).
  \]
\end{definition}

Thus, $\vecp^{\prime,<0}$ collects all modification parameters required to eliminate asymptotic terms with $\cK^+$-decay orders $(\lambda,k)$ where $\Re\lambda<0$; the only such asymptotic term has $(\lambda,k)=(-1,0)$. On the other hand, the parameters in $\vecp^{\prime,=0}$ will serve to eliminate asymptotic terms with $\cK^+$-decay orders $(\lambda,k)$ where $\Re\lambda=0$; there are three such terms, two (namely, $\dot b$ and $\dot\scal_1^{(0)}$) for $(0,0)$ and one (namely, $\dot\scal_1^{(0,1)}$) for $(0,1)$.

\begin{lemma}[Elimination of $<0$ parameters]
\label{LemmaD4ElimLess}
  For small $h,\vecp$, and for all $f$ as in~\eqref{EqD4Parf} and $\vecp^{\prime,=0}$, there exists a unique
  \[
    \vecp_{h,\vecp}^{\prime,<0}(f,\vecp^{\prime,=0}) = \vecp_{h,\vecp}^{\prime,<0}(f,0) + o(1)\vecp^{\prime,=0}
  \]
  where $o(1)\to 0$ (in the space of linear maps on $\scalspace_1$) as $\|h\|_{\cX^d}+|\vecp|\to 0$, such that the asymptotic boost parameter (see~\eqref{EqD2Boost2}) for the forward solution of
  \begin{equation}
  \label{EqD4ElimLess}
    D_{(h,\vecp^{<0},\vecp^{=0})}P(u,0,0) = f-D_{(h,\vecp^{<0},\vecp^{=0})}P\bigl(0,\vecp_{h,\vecp}^{\prime,<0}(f,\vecp^{\prime,=0}),\vecp^{\prime,=0}\bigr)
  \end{equation}
  vanishes.
\end{lemma}
\begin{proof}
  This is a re-formulation of the discussion preceding Definition~\ref{DefD4ModPar}.
\end{proof}

Lemma~\ref{LemmaD4ElimLess} allows us to define:

\begin{definition}[Conditional asymptotic parameters]
\label{DefD4ModCond}
  Given $\vecp^{\prime,=0}$, write $\dot\vecp_{h,\vecp}^{=0}(f,\vecp^{\prime,=0})$ for the parameters $(\dot b,\dot\scal^{(0)}_1,\dot\scal^{(0,1)}_1)$ in~\eqref{EqD4Paru} for the forward solution of~\eqref{EqD4ElimLess}.
\end{definition}

Note that $\vecp_{h,\vecp}^{\prime,<0}$ and $\dot\vecp_{h,\vecp}^{=0}$ are linear in $(f,\vecp^{\prime,=0})$.

\begin{lemma}[Mapping properties for the $=0$ parameters]
\label{LemmaD4ModCond}
  For small $h,\vecp$, the map $\vecp^{\prime=0}\mapsto\dot{\vecp}_{h,\vecp}^{=0}(0,\vecp^{\prime,=0})$ is close to the identity map and thus invertible.
\end{lemma}
\begin{proof}
  The forward solution of $D_{(0,0,0)}P(u,0,0)=D_{(0,0,0)}P(0,0,\vecp^{\prime,=0})$ for $\vecp^{\prime,=0}=(b',\scal_1^{(0)\prime},\scal_1^{(0,1)\prime})$ is the sum of $\chi_\cK\dot g_{b_0}^\Ups(b')$ (by~\eqref{EqD4ParNoRewrite}, $\chi_\cK h_{b_0,\rms 1}(\scal^{(0)\prime}_1)$ (by~\eqref{EqD4ParNoRewrite2}), and $\chi_\cK\delta_{g_{b_0}}^*\omega_{b_0,\rms 1}^{(0),\leq 4}(\log(t_*)\scal_1^{(0,1)\prime})$ (by the arguments after~\eqref{EqD4ParRewrite3}). This implies that for $(h,\vecp)=(0,\vec 0)$, the parameter $\vecp_{0,\vec 0}^{\prime,<0}(f,\vecp^{\prime,=0})$ is independent of $\vecp^{\prime,=0}$, and that $\dot{\vecp}^{=0}_{0,\vec 0}(0,\vecp^{\prime,=0})=\vecp^{\prime,=0}$. The continuous dependence of forward solutions on $(h,\vecp)$ implies the claim.
\end{proof}

The upshot of our linear analysis thus far is:

\begin{cor}[Elimination of $=0$ parameters]
\label{CorD4ElimEqual}
  For small $h,\vecp$ (in the notation of Definition~\usref{DefD4EinsteinAug}), and for all $f$ as in~\eqref{EqD4Parf}, there exists a unique $\vecp_{h,\vecp}'(f)\in\scalspace_1\oplus\R^4\oplus\scalspace_1\oplus\scalspace_1$ such that the forward solution of
  \[
    D_{(h,\vecp)}P(u,0) = f - D_{(h,\vecp)}P\bigl(0,\vecp_{h,\vecp}'(f)\bigr)
  \]
  has neither asymptotic boost parameters nor the asymptotic parameters~\eqref{EqD4Paru}, and therefore
  \[
    u\in\Hb^{\infty,\ \bigl(\la\cE_\sscri^\cC\ra,3+\eps_\sscri\bigr),\ 1-\eps,\ \alpha_\cK-2}(\Omega_*)^{\bullet,-},
  \]
  where we recall that $\alpha_\cK\in(2,2+\eps_\cK)$. (This thus \emph{decays} at $\cK^+$.)
\end{cor}
\begin{proof}
  In the notation of Definition~\ref{DefD4ModCond}, we have
  \[
    \dot{\vecp}_{h,\vecp}^{=0}(f,\vecp^{\prime,=0}) = \dot{\vecp}_{h,\vecp}^{=0}(f,0) + \dot{\vecp}_{h,\vecp}^{=0}(0,\vecp^{\prime,=0}).
  \]
  By Lemma~\ref{LemmaD4ModCond}, there exists a unique value of $\vecp^{\prime,=0}$ for which this sum vanishes; defining $\vecp_{h,\vecp}^{\prime,=0}(f)$ to be equal to this value and then setting $\vecp_{h,\vecp}^{\prime,<0}(f):=\vecp_{h,\vecp}^{\prime,<0}(f,\vecp^{\prime,=0})=\vecp_{h,\vecp}^{\prime,<0}(f,\vecp_{h,\vecp}^{\prime,=0}(f))$ finishes the construction of $\vecp_{h,\vecp}'(f)=(\vecp_{h,\vecp}^{\prime,<0}(f),\vecp_{h,\vecp}^{\prime,=0}(f))$.
\end{proof}

The above arguments are closely related to \cite[Lemma~5.20]{HintzVasyKdSStability}, which concerns the elimination of non-decaying terms in the asymptotic expansion of solutions of the gauge-fixed Einstein equation on Kerr--de~Sitter backgrounds.

\subsection{Step~5: polynomially decaying center-of-mass motion; almost \texorpdfstring{$t_*^{-1}$}{1/t}-decay}
\label{SsD5Alm}

In the proof of Proposition~\ref{PropD4Par}, the terms that are \emph{not} of size $\cO(t_*^{-1})$ at $\cK^+$, besides the expansion terms in~\eqref{EqD4Paru}, arise in two places.
\begin{enumerate}
\myitem{ItD5AlmCoM}{i} They arise in Step~2.2 from $\iota^+$-source terms at order $t_*^{\lambda+2}(\log t_*)^k$ where $1<\Re\lambda<2$. More precisely, only the terms from~\eqref{EqipGrQhomu} arising from the first term in~\eqref{EqipGr} (which has a full order less $t_*$-decay than all subsequent terms) fail to decay like $t_*^{-1}$. These contributions are thus of the form $t_*^{-\lambda+1}(\log t_*)^k h_{b,\rms 1}$; they encode weakly decaying changes of the center-of-mass of the final black hole.
\myitem{ItD5AlmDec}{ii} They also arise in the term $u_{\rm sing}$ (see~\eqref{EqD4FTusing}), specifically from the second summand (the conormal remainder term of which is one order more singular than the first at $\sigma=0$); this leads to a term of the schematic form $\cO(t_*^{-\alpha_\cK+2})h_{b,\rms 1}$ upon inverse Fourier transforming. When the source $f$ has stronger decay at $\cK^+$, then one expects this term to inherit this stronger decay.
\end{enumerate}

Now, it appears that we cannot simply increase our requirements on the $\cK^+$-decay order of $f$ arbitrarily, since the requirements on $f$ must allow for the modification terms in Lemma~\ref{LemmaD4LinPar}; the delicate one is the linearization of $P$ in $b$ (see~\eqref{EqD4LinParb}), which gives a source term of the same size as the dynamical part $h$ of the metric $g_{b_0,b,-\scal}+h$ relative to which $L=L_{h,\vecp}$ in~\eqref{EqD4EinsteinAugLin} is a wave-type operator. For $h=\cO(t_*^{-2-\eps_\cK})$ (in spatially compact sets) without further structure at $\cK^+$, this would mean that we would expect $u$ to feature a (conormal but not polyhomogeneous) $\cO(t_*^{-\eps_\cK})$ center-of-mass motion, which we cannot eliminate with linear algebra arguments of the sort used in Corollary~\ref{CorD4ElimEqual} (cf.\ Remark~\ref{RmkINElimNon}). In order to get stronger decay of $u$, we must therefore make stronger demands on $h$. A simple solution is to require $4+\eps_\cK$ orders of decay of $h$ at $\cK^+$, i.e.,
\begin{equation}
\label{EqD5Almh}
  h \in \Hb^{\infty,\ \bigl(\la\cE_\sscri^\cC\ra,3+\eps_\sscri\bigr),\ (\cE_+,3+\eps_+),\ 4+\eps_\cK}(\Omega_*),
\end{equation}
instead of~\eqref{EqDMetBasich}. But we note here already that we will \emph{not} be able to recover this strong decay for solutions of $L u=f$. Instead, we shall find that $u$ has contributions of the schematic form $\chi_\cK h_{b,\rms 1}^{\leq 1}(\cO(t_*^{-2-\eps_\cK}))$, $\chi_\cK\dot g_b^\Ups(\cO(t_*^{-3-\eps_\cK}))$, and other contributions from zero energy states. These only decay at rate $t_*^{-2-\eps_\cK}$ (which is still consistent with~\eqref{EqDMetBasich} but not with~\eqref{EqD5Almh}). But note that if, say, we defined a gauge-fixed Einstein operator depending only on a metric perturbation~\eqref{EqD5Almh} and the black hole parameters $b$ as $\Ric(g_b+t_*^{-3-\eps_\cK}\dot g_b^\Ups+h)$ plus a gauge term, then since $\Ric(g_b+t_*^{-3-\eps_\cK}\dot g_b^\Ups)=\Ric(g_b)+D_{g_b}\Ric(t_*^{-3-\eps_\cK}\dot g_b^\Ups)=\cO(t_*^{-4-\eps_\cK})$, the linearization of $\Ric(g_b+t_*^{-3-\eps_\cK}\dot g_b^\Ups+h)$ in $b$ would be a sum of such a $\cO(t_*^{-4-\eps_\cK})$-term and another term $D_{g_b+t_*^{-3-\eps_\cK}\dot g_b^\Ups}\Ric(h)$, which thus inherits the $\rho_\cK^{4+\eps_\cK}$-decay from $h$. We expand on this in~\S\ref{SssD6Abs} when we extract these contributions to $u$ (and then make matching assumptions on $h$). \emph{For now, we use a hybrid perspective} that is forward compatible: we assume~\eqref{EqD5Almh}, but in our arguments for the improvement of decay \emph{only} use that the dynamical linearized gauge-fixed Einstein operator $L$ differs from its stationary model $L_b$ by terms with $t_*^{-2-\eps_\cK}$-decay at $\cK^+$---which is all we used in all of our arguments so far as well (based on~\eqref{EqDAdmFw}).

\bigskip

In order to argue, in a transparent fashion, for the possibility of eliminating pure gauge terms, it is useful to recall some of the index sets arising in our analysis. (We stress that the post-processing in~\S\ref{SEf} will remove the various redundancies in the index sets that we presently allow for.)
\begin{enumerate}
\myitem{ItD5Escri}{i} $\cE_\sscri^\cC$ is fixed according to Definition~\ref{DefExP} (and thus, in practice, is determined by the index set of the initial data via Theorem~\ref{ThmExPhg}).
\myitem{ItD5EUps}{ii} $\cE_\ind$ is a fixed index set with $\min\Re\cE_\ind\geq\eps_\ind$ that depends only on the parameters of the Minkowskian model $\ubar L$ (i.e., on the parameters in Definitions~\ref{Def1Gauge} and \ref{Def1Symm}).
\myitem{ItD5Eplus}{iii} $\cE_+$, with $(1,0)\in\cE_+$ and $\min\Re\cE_+\geq 1$, is a nonlinearly closed index set as in Definition~\ref{DefDMetBasic}; it is convenient to require
  \begin{equation}
  \label{EqD5EplusClosed}
    \cE_+ + \cE_\ind \subset \cE_+,\qquad j\bigl(\cE_+\cup(1,1)\bigr)\subset\cE_+\cup(1,1)\quad\forall\,j\in\N.
  \end{equation}
  (This implies the earlier conditions in~\eqref{EqD2EplusCond} and the paragraph preceding it.) Let
  \begin{equation}
  \label{EqD5Eplussharp}
    \cE_+^\sharp=(1,0)\cup(1+(1-e^\Ups)\gamma^\Ups,0)\cup(\cdots)
  \end{equation}
  be given by Proposition~\ref{PropDScriFormal} for $\breve\cE_+=\emptyset$, or more generally for $\breve\cE_+=\cE_++\cF$ where $\cF$ is some fixed index set with $\min\Re\cF>0$ (specified in the course of our arguments). We enlarge the index set defined in~\eqref{EqD2LinPartildeE} to
  \begin{equation}
  \label{EqD5TildeEplus}
    \tilde\cE_+ := (1,0) \cup (1+\cE_\ind) \cup \Bigl(\cE_+ + \bigl((1,1)\cup(1+\cE_\ind)\bigr)\Bigr) \cup \cE_+^\sharp.
  \end{equation}
\end{enumerate}

The correction terms in Lemma~\ref{LemmaD4LinPar} have $\iota^+$-index set contained in $\tilde\cE_++2$, and thus one expects to have a center-of-mass movement term as described in~\eqref{ItD5AlmCoM} above with $t_*^{-\lambda}(\log t_*)^k$-decay for each $(\lambda,k)\in\tilde\cE_+-1$ with $\Re\lambda>0$; indicial roots of the zero energy operator (which may produce further index sets at $\iota^+$) enlarge the set of $(\lambda,k)$ that we will encounter in practice to a possibly larger index set (with stationary or logarithmic motions already covered in Proposition~\ref{PropD4Par}). We will eliminate each one of them by means of a suitable gauge modification using the re-writing~\eqref{EqD4ParRewrite3}; analogously to the discussion preceding Definition~\ref{DefD4ModPar}, we must do this elimination one value of $\lambda$ (or rather $\Re\lambda$) at a time---but can treat all values of $k$ simultaneously---before uncovering the next term in the late-time asymptotics.

\begin{definition}[Correction terms \#3]
\label{DefD5Corr}
  For $\scal\in\scalspace_1$, $\lambda\in\C$ with $\Re\lambda>0$, and $k\in\N_0$, we define (using the notation of Definition~\usref{DefipGrKerr})
  \begin{subequations}
  \begin{alignat}{2}
  \label{EqD5Corrhs1}
    h_{\rms 1}^{(\lambda,k)} &:= -[\delta_{g_{b_0}}^*,\chi_\cK]\omega_{b_0,\rms 1}^{(0),\leq 4}\bigl(\scal t_*^{-\lambda}(\log t_*)^k\bigr) &&\in \cA^{\infty,\ (\lambda+1,k)\cup((\lambda+1,k)+\cE_\ind),\ \infty}, \\
  \label{EqD5Corrthetas1}
    \vartheta_{\rms 1}^{(\lambda,k)} &:= \delta_{g_{b_0},E^\Ups}\sfG_{g_{b_0}}\Bigl(\chi_\cK\delta_{g_{b_0}}^*\omega_{b_0,\rms 1}^{(0),\leq 4}\bigl(\scal t_*^{-\lambda}(\log t_*)^k\bigr) \Bigr) &&\in \cA^{\infty,\ (\lambda+2,k)\cup((\lambda+2,k)+\cE_\ind),\ \lambda+5-\eps}.
  \end{alignat}
  \end{subequations}
\end{definition}

The membership of $h_{\rms 1}^{(\lambda,k)}$ follows directly from its definition (and $[\delta_{g_{b_0}}^*,\chi_\cK]=\dd\chi_\cK\otimes_s(\cdot)\in\rho_\sscri^\infty\rho_+\rho_\cK^\infty\Diffb^0$), while that of $\vartheta_{\rms 1}^{(\lambda,k)}$ follows by inspection of~\eqref{EqD4ParNoLog1}--\eqref{EqD4ParNoLog2}. Our (rather wasteful) accounting of lower-order index sets here (i.e., the ``${}\!+\cE_\ind$'' terms) are a motivation for the first requirement in~\eqref{EqD5EplusClosed}.

\begin{definition}[Gauge-fixed Einstein operator, augmentation \#3]
\label{DefD5EinsteinAug}
  For symmetric $2$-tensors $h\in\cX^\infty=\Hb^{\infty,\ \bigl(\la\cE_\sscri^\cC\ra,3+\eps_\sscri\bigr),\ (\cE_+,3+\eps_+),\ 4+\eps_\cK}(\Omega_*)$ with sufficiently small $\cX^d$-norm, small $\vecp^{\leq 0}=(\scal,b-b_0,\scal_1^{(0)},\scal_1^{(0,1)})$, and small
  \[
    \vecp^{\in(0,1)} = \bigl(\scal_1^{(\lambda-1,k)}\bigr)_{\substack{ (\lambda,k)\in\tilde\cE'_{++} \\ 1<\Re\lambda<2 }},
  \]
  where $\tilde\cE'_{++}$ is an index set (defined in~\eqref{EqD5Eplusplus} below) with $\min\Re\tilde\cE'_{++}\geq 1+\eps_\ind>1$ and $\tilde\cE'_{++}+(\cE_\ind\cup\cE_+\cup\tilde\cE'_{++})\subset\tilde\cE'_{++}$, write $\vecp=(\vecp^{\leq 0},\vecp^{\in(0,1)})$ and $P(h,\vecp):=\Ric(g^0+h_\tot(\vecp)+h)-\delta_{g^0,E^\cC}^*\bigl(\Ups_{E^\Ups}(g^0+h,g^0)-\vartheta_\tot(\vecp)\bigr)$ as in~\eqref{EqD4EinsteinAug}, where $g^0 := g_{b_0,b,-\scal}$ as before, while we re-define for $\bullet=h,\vartheta$:
  \[
    \bullet_\tot(\vecp) := \bullet^{(-1)}(\scal) + \bullet^{(0)}(b-b_0) + \bullet_{\rms 1}^{(0)}(\scal_1^{(0)}) + \bullet_{\rms 1}^{(0,1)}(\scal_1^{(0,1)}) + \sum_{\substack{ (\lambda,k)\in\tilde\cE'_{++} \\ \Re\lambda<2 }} \bullet_{\rms 1}^{(\lambda-1,k)}(\scal_1^{(\lambda-1,k)}).
  \]
  We write $L_{h,\vecp}:=D_{(h,\vecp)}P(\cdot,0)$ for the linearization of $P$ in $h$.
\end{definition}

\begin{lemma}[Linearization of $P$ in the parameters, \#3]
\label{LemmaD5LinPar}
  The linearizations of $P$ in the components of $\vecp^{\leq 0}$ have the form~\eqref{EqD4LinParScal}--\eqref{EqD4LinParCoM} except with $\tilde\cE_++2$ replaced by $(\tilde\cE_+\cup\tilde\cE'_{++})+2$, and in particular $D_{(h,\vecp)}P(0,(0,b',0,\ldots,0))\in\chi_\cK^\sharp\Hb^{\infty,\ 0,\ \bigl((\tilde\cE_+\cup\tilde\cE'_{++})+2,\,6+\eps_+\bigr),\ 4+\eps_\cK}$ for $h$ satisfying~\eqref{EqD5Almh}. For $\Re\lambda>1$, we moreover have\footnote{By the conditions on $\tilde\cE'_{++}$ stated in Definition~\ref{DefD5EinsteinAug}, the $\iota^+$-index set of~\eqref{EqD5LinPar} is contained in $\tilde\cE_{++}+2$.}
  \begin{equation}
  \label{EqD5LinPar}
    D_{(h,\vecp)}P\bigl(0,(0,\ldots,\scal_1^{(\lambda-1,k)\prime},\ldots,0)\bigr) \in \chi_\cK^\sharp\Hb^{\infty,\ 0,\ \bigl( (\lambda+2,k)+((0,0)\cup\cE_\ind\cup\cE_+\cup\tilde\cE'_{++}),\,5+\eps_++\lambda-\eps\bigr),\ \lambda+4-\eps}.
  \end{equation}
  Furthermore,
  \begin{equation}
  \label{EqD5LinParDiff}
  \begin{split}
    &D_{(h,\vecp)}P\bigl(0,(0,\ldots,\scal_1^{(\lambda-1,k)\prime},\ldots,0)\bigr) - \chi_\iota \ubar L\Bigl( \chi_\cK\ubar\delta^*\ubar\omega_{\rms 1}^{(0),\leq 4}\bigl(\scal_1^{(\lambda-1,k)\prime}t_*^{-\lambda+1}(\log t_*)^k\bigr) \Bigr) \\
    &\qquad \in \chi_\cK^\sharp\Hb^{\infty,\ 0,\ \bigl((\lambda+2,k)+(\cE_\ind\cup\cE_+\cup\tilde\cE'_{++}),\,5+\eps_++\lambda-\eps\bigr),\ \lambda+4-\eps},
  \end{split}
  \end{equation}
  and this also holds if we instead subtract $\chi_\iota L_b\bigl(\chi_\cK\delta_{g_b}^*\omega_{b,\rms 1}^{(0),\leq 4}(\scal_1^{(\lambda-1,k)\prime}t_*^{-\lambda+1}(\log t_*)^k)\bigr)$.
\end{lemma}
\begin{proof}
  We only discuss (and thus only record) the $\iota^+$-index set. (Note that since the new modification terms have more decay than those previously discussed in Lemmas~\ref{LemmaD2LinPar} and \ref{LemmaD4LinPar}, the $\cK^+$-decay orders are the same as before, or in fact better except for the linearization in $b$.) The $\iota^+$-index set of $h_\tot(\vecp)$ is contained in $(1,1)\cup\cE_+\cup\tilde\cE'_{++}$ (using~\eqref{EqD5Corrhs1} for the new terms $h_{\rms 1}^{(\lambda-1,k)}$). The linearization $\dot g$ of $g=g^0+h_\tot(\vecp)+h\in(\CI+\Hb^{\infty,\ \bigl((1,1)\cup\cE_+\cup\tilde\cE'_{++},\,3+\eps_+\bigr)})(\Omega_*;S^2\cT^*)$ in $\scal_1^{(\lambda-1,k)}$ is $\dot g=h_{\rms 1}^{(\lambda-1,k)'}(\scal_1^{(\lambda-1,k)\prime})\in\cA^{(\lambda,k)\cup((\lambda,k)+\cE_\ind)}$, so (requiring $(1,1)\in\cE_\ind$)
  \begin{align*}
    D_g\Ric(\dot g) &\in \rho_+^2\Bigl(\CI+\Hb^{\infty,\ \bigl((1,1)\cup\cE_+\cup\tilde\cE'_{++},\,3+\eps_+\bigr)}\Bigr)\Diffb^2\bigl(\cA^{(\lambda,k)\cup((\lambda,k)+\cE_\ind)}\bigr) \\
      &\subset \Hb^{\infty,\ \bigl( (\lambda+2,k)\cup\bigl((\lambda+2,k)+(\cE_\ind\cup\cE_+\cup\tilde\cE'_{++})\bigr),\,5+\eps_++\lambda-\eps\bigr) }\quad\forall\,\eps>0.
  \end{align*}
  The linearization of the gauge term is $\delta_{g^0,E^\cC}^*\in\rho_+\Diffb^1$ acting on $\vartheta_{\rms 1}^{(\lambda-1,k)}$ and thus of class $\cA^{(\lambda+2,k)\cup((\lambda+2,k)+\cE_\ind)}$ by~\eqref{EqD5Corrthetas1}.

  For the proof of~\eqref{EqD5LinParDiff}, we note that $D_{(h,\vecp)}P(0,(0,\ldots,\scal_1^{(\lambda-1,k)\prime},\ldots,0))$ can be computed to leading order at $\iota^+$ by replacing all operators and tensors by their Minkowskian leading-order terms at $\iota^+$.
\end{proof}

\begin{rmk}[$\cK^+$-order]
\label{RmkD5KOrder}
  In the context of Remark~\ref{RmkD4KOrder}, we note that for $\Re\lambda>1$, the $\cK^+$-order of $\vartheta_{\rms 1}^{(\lambda-1,k)}$ (and thus of~\eqref{EqD5LinPar}) is $>5-\eps$ and thus larger than the order $4+\eps_\cK$ required for our source terms below and in the nonlinear analysis in~\S\S\ref{SEf}--\ref{SSt} below.
\end{rmk}

We can now state and prove the analogue of Corollary~\ref{CorD4ElimEqual} (which was based on Proposition~\ref{PropD4Par}):

\begin{prop}[Polynomially decaying center-of-mass motion and almost $t_*^{-1}$-decay]
\label{PropD5Alm}
  Let $h$ and $\vecp=(\vecp^{\leq 0},\vecp^{\in(0,1)})$ be as in Definition~\usref{DefD5EinsteinAug}, and let $\alpha_\cK\in(2,3)$. Let $f$ be of the form\footnote{The $\cK^+$-decay rate $4+\eps_\cK$ of $f_2$ is consistent with~\eqref{EqD5Almh} and the modification terms in Lemma~\ref{LemmaD5LinPar}.}
  \begin{equation}
  \label{EqD5Almf}
  \begin{split}
    &f = f_1 + f_2, \\
    &\quad f_1 \in \bigcap_{\eps>0} \Hb^{\infty,\ \bigl(\la\cE_\sscri^\cC+1\ra',\,4+\eps_\sscri\bigr),\ 4-\eps,\ \alpha_\cK}(\Omega_*)^{\bullet,-}, \\
    &\quad f_2 \in \Hb^{\infty,\ 4+\eps_\sscri,\ \tilde\cE_++2,\ 4+\eps_\cK}(\Omega_*;S^2\cT^*)^{\bullet,-}.
  \end{split}
  \end{equation}
  Then there exist $\vecp^{\prime,\leq 0}$ and $\vecp^{\prime,\in(0,1)}$ (unique when $\alpha_\cK-2>\Re(\lambda-1)$ for all $\lambda\in\pi_1\cE_+$ with $\Re\lambda<2$) such that the forward solution of
  \begin{subequations}
  \begin{equation}
  \label{EqD5AlmEq}
    L u := D_{(h,\vecp)}P\bigl(u,(0,0)\bigr) = f - D_{(h,\vecp)}P\bigl(0,(\vecp^{\prime,\leq 0},\vecp^{\prime,\in(0,1)})\bigr)
  \end{equation}
  satisfies
  \begin{equation}
  \label{EqD5Almu}
    u \in \Hb^{\infty,\ \bigl(\la\cE_\sscri^\cC\ra,\,3+\eps_\sscri\bigr),\ 1-\eps,\ \alpha_\cK-2}(\Omega_*)^{\bullet,-}.
  \end{equation}
  \end{subequations}
\end{prop}

\begin{rmk}[Structure of $f$ vs.\ modification terms]
\label{RmkD5AlmStruct}
  While the assumptions on $f$ in~\eqref{EqD5Almf} are motivated by the structure of the modification terms used in previous sections to eliminate non-decaying terms in the late-time asymptotics of $u$, we do \emph{not} make the assumptions flexible enough so that they would allow for the terms~\eqref{EqD5LinPar}; that is, we do \emph{not} require $\tilde\cE'_{++}\subset\cE_+$. The reason is that general terms of class~\eqref{EqD5LinPar} may produce center-of-mass shifts of size $t_*^{-\lambda+1}(\log t_*)^{k+1}$ (due to coincidences with indicial roots of $\wh{\ubar L}(0)$), i.e., with an additional factor of $\log t_*$, which one might try to undo using terms $h_{\rms 1}^{(\lambda-1,k+1)}$ and $\vartheta_{\rms 1}^{(\lambda-1,k+1)}$; but allowing general terms of class~\eqref{EqD5LinPar} with $k+1$ in place of $k$, one would then need terms with $(\lambda-1,k+2)$; and so on. It is the special nature of the correction terms in~\eqref{EqD5AlmEq} that prevents such an unbounded accumulation of factors of $\log t_*$. (See~\eqref{EqD5EqMod} and the subsequent discussion for more details.)
\end{rmk}

\begin{proof}[Proof of Proposition~\usref{PropD5Alm}]
  We set up an induction on $\alpha_\cK$ (in $\eps_\cK$-increments) as follows. For a given $\alpha_\cK\in(2,3)$, split the modification parameters into
  \begin{align*}
    \vecp^{\prime,\leq\alpha_\cK-2} &:= \Biggl(\scal',b',\scal_1^{(0)\prime},\scal^{(0,1)\prime}_1,(\scal_1^{(\lambda-1,k)\prime})_{\substack{(\lambda,k)\in\tilde\cE'_{++} \\ \Re(\lambda-1)\leq\alpha_\cK-2}}\,\Biggr), \\
    \vecp^{\prime,>\alpha_\cK-2} &:= \bigl(\scal_1^{(\lambda-1,k)\prime}\bigr)_{\substack{(\lambda,k)\in\tilde\cE'_{++} \\ \Re(\lambda-1)>\alpha_\cK-2}}\,.
  \end{align*}
  We shall then show that one can obtain~\eqref{EqD5Almu} using only $\vecp^{\prime,\leq\alpha_\cK-2}$; more precisely, for the purpose of making the inductive step it is important to allow for an arbitrary choice of $\vecp^{\prime,>\alpha_\cK-2}$ and find $\vecp^{\prime,\leq\alpha_\cK-2}$ (depending on the choice of $\vecp^{\prime,>\alpha_\cK-2}$) such that~\eqref{EqD5Almu} holds for the solution of~\eqref{EqD5AlmEq}. Note that by~\eqref{EqD5LinPar}, the modification parameters $\vecp^{\prime,>\alpha_\cK-2}$ lead to terms on the right-hand side of~\eqref{EqD5AlmEq} that are of class $\Hb^{\infty,\ \infty,\ \cE+2,\ 5}$ where $\cE$ is an index set with $\min\Re(\cE-1)>\alpha_\cK-2$ (so in particular $\min\Re(\cE+2)>3$ since $\alpha_\cK>2$). It is then convenient to make the inductive hypothesis even stronger and require that one can find $\vecp^{\prime,\leq\alpha_\cK-2}$ such that~\eqref{EqD5Almu} holds even in the case that
  \begin{equation}
  \label{EqD5AlmIndf2}
    \parbox{0.7\textwidth}{$f_2$ in~\eqref{EqD5Almf} has $\iota^+$-index set $(\tilde\cE_+\cup\cE)+2$ where $\cE$ is any index set with $\min\Re(\cE-1)>\alpha_\cK-2$.}
  \end{equation}
  Such $f_2$ can absorb the contribution of $\vecp^{\prime,>\alpha_\cK-2}$, and thus it suffices to consider $\vecp^{\prime,>\alpha_\cK-2}=\vec 0$.

  The base case is $\alpha_\cK\in(2,2+\eps_\cK)$, in which case this is the content of Corollary~\ref{CorD4ElimEqual} for $\vecp^{\prime,>\alpha_\cK-2}=\vecp^{\prime,\in(0,1)}=\vec 0$; more generally, the additional terms on the right-hand side of~\eqref{EqD5AlmEq} due to possibly non-zero choices of $\vecp^{\prime,\in(0,1)}$ are of the class~\eqref{EqD4Parf} (with the index set denoted $\tilde\cE_+$ there being a large enough index set so as to also allow for the terms~\eqref{EqD5LinPar} for all $(\lambda,k)\in\tilde\cE'_{++}$)  permitted as source terms in Corollary~\ref{CorD4ElimEqual}.

  \pfstep{Part~1. Extracting the next leading-order terms.} Let $\alpha'_\cK\in(2,3-\eps_\cK)$; we shall make the inductive step to $\alpha_\cK=\alpha'_\cK+\eps_\cK$. For arbitrary $\vecp^{\prime,>\alpha_\cK-2}$, or more generally for $f$ of the form~\eqref{EqD5Almf} with~\eqref{EqD5AlmIndf2} but setting $\vecp^{\prime,>\alpha_\cK-2}=0$, fix $\vecp^{\prime,\leq\alpha'_\cK-2}(f)$ such that~\eqref{EqD5Almu} holds with $\alpha'_\cK$ in place of $\alpha_\cK$; for now we set the remaining parameters
  \begin{equation}
  \label{EqD5AlmParamNew}
    \vecp^{\prime,\in(\alpha'_\cK-2,\alpha_\cK-2]}=(\scal_1^{(\lambda-1,k)\prime})_{ \substack{(\lambda,k)\in\tilde\cE'_{++} \\ \Re(\lambda-1)\in(\alpha'_\cK-2,\alpha_\cK-2] } }
  \end{equation}
  to be zero. Write $F_{\rm mod}^{\leq\alpha'_\cK-2}(f):=D_{(h,\vecp)}P\bigl(0,(\vecp^{\prime,\leq\alpha'_\cK-2}(f),0)\bigr)$; the inductive hypothesis thus states that the solution of $L u=f-F_{\rm mod}^{\leq\alpha'_\cK-2}(f)$ satisfies $u\in\Hb^{\infty,\ \bigl(\la\cE_\sscri^\cC\ra,\,3+\eps_\sscri\bigr),\ 1-\eps,\ \alpha'_\cK-2}$. Using~\eqref{EqDAdmFw} (which applies to $L-L_b$), we get
  \begin{align*}
    &L_b u = f'_1 + f'_2 - F_{\rm mod}^{\leq\alpha'_\cK-2}(f), \\
    &\qquad f'_1:=f_1-(L-L_b)u \in \Hb^{\infty,\ \bigl(\la\cE_\sscri^\cC+1\ra',\,4+\eps_\sscri\bigr),\ 4-\eps,\ \alpha_\cK},\ \ f'_2:=f_2.
  \end{align*}

  \pfsubstep{Step~1.1.}{Solution near $\scri^+$.} In view of~\eqref{EqD5TildeEplus}, Proposition~\ref{PropDScriFormal} (applied to $f'_1$) gives
  \begin{equation}
  \label{EqD5uscri}
  \begin{split}
    &u_{1,\sscri} \in \Hb^{\infty,\ \bigl(\la\cE_\sscri^\cC\ra,\,3+\eps_\sscri\bigr),\ (\tilde\cE_+,2-\eps),\ \infty},\ \text{supported near $\scri^+$}, \\
    &\qquad L_b(u-u_{1,\sscri}) =: f',\quad f' \in \Hb^{\infty,\ 4+\eps_\sscri,\ \bigl((\tilde\cE_+\cup\tilde\cE'_{++})+2,\,4-\eps\bigr),\ \alpha_\cK}(\Omega_*;S^2\cT^*)^{\bullet,-}.
  \end{split}
  \end{equation}
  The contribution $\tilde\cE'_{++}$ to the $\iota^+$-index set of $f'$ comes from $F_{\rm mod}^{\leq\alpha'_\cK-2}(f)$, specifically, the modification parameters $\scal_1^{(\lambda-1,k)\prime}$ for $\Re(\lambda-1)\leq\alpha'_\cK-2$; cf.\ \eqref{EqD5LinPar} (and the footnote preceding it).

  \pfsubstep{Step~1.2.}{Solving away the leading-order term at $\iota^+$.}\label{EqD5Step12} We repeat the arguments in~\eqref{EqD4ip1}--\eqref{EqD4ip1scri} to solve away the $\rho_+^3$-term of $f'$.\footnote{We are simplifying our bookkeeping in~\eqref{EqD5uscri}. In order to avoid collisions with indicial roots in the application of Proposition~\ref{PropipGr}, given only the membership of $f'$ in~\eqref{EqD5uscri}, one may need to reduce $\alpha_\cK$ by an arbitrarily small amount, though this is not necessary when $\alpha_\cK$ is sufficiently close to $3$.} The logarithmically divergent leading-order term is absent by our assumption that $u$ decays at $\cK^+$, so upon summing~\eqref{EqD4ip1} and \eqref{EqD4ip1scri}, we obtain
  \begin{equation}
  \label{EqD5uiota}
  \begin{split}
    &u_\iota \in \Hb^{\infty,\ \bigl(\la\cE_{\iota^+,\sscri}^\cC\ra,3+\eps_\sscri\bigr),\ \bigl((1,0)\cup(1+\cE_\ind),\,2-\eps\bigr),\ 1-\eps}, \\
    &\quad u_\flat := u-u_{1,\sscri}-u_\iota \implies L_b u_\flat =: f_\flat \in \Hb^{\infty,\ 4+\eps_\sscri,\ \bigl((\tilde\cE'_+\cup\tilde\cE'_{++})+2,\,4-\eps\bigr),\ \alpha_\cK}, \quad \tilde\cE'_+:=\tilde\cE_+\setminus\{(1,0)\}.
  \end{split}
  \end{equation}

  In previous proofs (e.g., Step~2.1 in the proof of Proposition~\ref{PropD4Par}), we continued solving away the partial expansion of the remaining error $f_\flat$ at $\iota^+$. We could do that here as well; the point would be that among the terms produced by Corollary~\ref{CoripGrQhom}, only the least-decaying one (the first term in~\eqref{EqipGr}, which describes a motion of the final black hole by an amount $\cO(t_*^{-\lambda+1})$ where $\lambda\in\pi_1(\tilde\cE'_+\cup\tilde\cE'_{++})$) would not lie in the space~\eqref{EqD5Almu}, while one might expect the assumption that $u$ has $\alpha'_\cK-2$ orders of $\cK^+$-decay to force those leading-order terms which do not have $\cO(t_*^{-(\alpha'_\cK-2)})$-decay to vanish. This is not necessarily correct, however: recall from the arguments leading up to~\eqref{EqD3Almu2} that the subsequent low-energy resolvent analysis (owing to Proposition~\ref{PropiptfGr}) would produce leading-order terms at $\zface\subset X_\scbtop^\pm$ with $\sigma^{\lambda-1}$-singularities for $\lambda\in\pi_1\cE_{\ind,+}$, which may include some $\lambda\in\pi_1(\tilde\cE_+'\cup\tilde\cE'_{++})-1$---which must therefore be combined with those from the prior $\iota^+$-analysis. ---  While this approach can thus certainly be pursued, we take a more efficient route here and pass \emph{directly} to the Fourier transform in equation~\eqref{EqD5uscri}. (The reason why we did not treat the term in Step~1.2 in this fashion is that on the spectral side it would arise via a non-integrable $\sigma^{-1}\log\sigma$-singularity; treating this term directly on spacetime thus avoids the need for arguments involving constructions of distributional extensions from $\R_\sigma\setminus\{0\}$ to $\R_\sigma$.)

  \pfsubstep{Step~1.3.}{Zero energy piece.} We express $u_\flat$ using the (inverse) Fourier transform as usual, with the high-energy piece (see~\eqref{EqD1Almu}) a fortiori lying in $\Hb^{\infty,\ 1-\eps,\ 2-\eps,\ \alpha_\cK}$ by \citeAF{Corollary~\ref*{CorDResHiLoc}}. At low energies, we now use Proposition~\ref{PropTFHbphg}\eqref{ItTFHbphgF} (with $\cE_\sscri=\emptyset$, $\cE_\iota=(\tilde\cE'_+\cup\tilde\cE'_{++})+2$, and $\beta_\sscri=4+\eps_\sscri$, $\beta_\iota=4-\eps$, and $\beta_\cK=\alpha_\cK$) to get
  \begin{equation}
  \label{EqD5AlmFT}
    \bigl(\wh{f_\flat}(\sigma)\bigr)\big|_{\sigma\in\pm[0,c]} \in \Hb^{\infty,\ 3-\eps,\ \bigl((\tilde\cE'_+\cup\tilde\cE'_{++})+1,\,3-\eps\bigr),\ ((0,0),\alpha_\cK-1)}(X_\scbtop^\pm)\quad\forall\,\eps>0.
  \end{equation}
  The restriction to $\zface$ is of class $\Hb^{\infty,\bigl((\tilde\cE'_+\cup\tilde\cE'_{++})+1,\,3-\eps\bigr)}(X;S^2\cT^*_X)$; since $\min\Re\tilde\cE'_++1>2$ and $\min\Re\tilde\cE'_{++}+1>2$, Proposition~\ref{PropAdmLoDef0} is applicable and gives
  \begin{equation}
  \label{EqD5Almureg1}
    \wt{L_b}(0)^{-1}(\wh{f_\flat}(0),0,0)=(\hat u_{{\rm reg},1}(0),0,0),\quad
    \hat u_{{\rm reg},1}(0)\in\Hb^{\infty,\ (\tilde\cE'_{+\sharp}-1,1-\eps)}(X;S^2\cT^*_X),
  \end{equation}
  where $\tilde\cE'_{+\sharp}\supset\tilde\cE'_+\cup\tilde\cE'_{++}$, with $\min\Re\tilde\cE'_{+\sharp}>1$, can be computed from $\tilde\cE'_+\cup\tilde\cE'_{++}$ and the indicial roots of $\wh{\ubar L}(0)$ with positive real parts (see Lemma~\ref{LemmaTMSolPhg} for the precise statement). Note here that the second and third components of the output of $\wt{L_b}(0)^{-1}$ (denoted $\hat b_1(0)$ and $\hat\scal_1(0)$ in earlier proofs, see, e.g., \eqref{EqD2Boost0Inv}), which encode the asymptotic boost and Kerr parameter changes in the asymptotic expansion of $u$, vanish (cf.\ \eqref{EqD2BoostLot} and \eqref{EqD4Params}) since we are assuming that $u$ decays at $\cK^+$. The remaining error term is
  \begin{equation}
  \label{EqD5Almf2}
  \begin{split}
    \bigl(\hat f_2(\sigma)\bigr)\big|_{\sigma\in\pm[0,c]} &:= \Bigl(\wh{f_\flat}(\sigma) - \wh{L_b}(\sigma)\bigl(\chi_\zface\hat u_{{\rm reg},1}(0)\bigr) \Bigr)\Big|_{\sigma\in\pm[0,c]} \\
      &\ \in |\sigma|\Hb^{\infty,\ 3-\eps,\ (\tilde\cE'_{+\sharp},2-\eps),\ ((0,0),\alpha_\cK-2)}(X_\scbtop^\pm).
  \end{split}
  \end{equation}

  \pfsubstep{Step~1.4.}{Solving away the error at $\tface$.}\label{ItD5Step14} We need to solve~\eqref{EqD5Almf2} away at $\tface$ until the $\tface$-decay of the remaining error is $>2$ (at which point $\wt{L_b}(0)^{-1}$ becomes applicable again). Consider thus a term in the $\tface$-expansion of $\sigma^{-1}\hat f_2(\sigma)$ on $X_\scbtop^\pm$,
  \begin{equation}
  \label{EqD5Almftf}
    \rho_\tface^\lambda(\log\rho_\tface)^k f^{(\lambda,k)}_{\tface,\pm}(\hat r,\omega),\quad f^{(\lambda,k)}_{\tface,\pm} \in \Hb^{\infty,\ 3-\eps,\ ((0,0),\alpha_\cK-2)}(\tface),\quad (\lambda,k)\in\tilde\cE'_{+\sharp},\ \Re\lambda<2,
  \end{equation}
  with the smallest value of $\Re\lambda$ among all non-vanishing terms; we recall here $\hat r=r|\sigma|$ and use the notation from~\S\ref{Ssiptf}. We apply Proposition~\ref{PropiptfGr} to it, with $\alpha_0=\lambda$, upon reducing $\alpha_\cK$ by an arbitrarily small amount if necessary to avoid collisions with indicial roots. (No such reduction is needed when $\alpha_\cK$ is sufficiently close to $3$.) We explicitly record only the most singular term
  \begin{subequations}
  \begin{equation}
  \label{EqD5Almutfsingpm}
    u_{\tface,\pm,{\rm sing}}^{(\lambda-1,k)} = \chi_\zface|\sigma|^{\lambda-2}(\log|\sigma|)^k|\sigma|\sigma^{-2}h_{b,\rms 1}(\dot\scal_{1,\pm}^{(\lambda-1,k)})
  \end{equation}
  arising from $k$-fold differentiation of~\eqref{EqiptfGruPrec2} in the parameter $\alpha_0=\lambda$ and note that the remaining terms are of class
  \begin{equation}
  \label{EqD5AlmutfRest}
    \tilde u_{\tface,\pm}^{(\lambda-1,k)} \in \Hb^{\infty,\ 1-\eps,\ \Re\lambda-2-\eps,\ \Re\lambda-2-\eps}(X_\scbtop^\pm)
  \end{equation}
  \end{subequations}
  which are more regular at $\zface$ than~\eqref{EqD5Almutfsingpm} by one order. (The only potential logarithmic terms from~\eqref{EqiptfGruPrec2} are part of~\eqref{EqD5AlmutfRest} since $-\lambda+2\neq 0$.) Subtracting the term $\wh{L_b}(\sigma)(u_{\tface,\pm,{\rm sing}}^{(\lambda-1,k)}+\tilde u_{\tface,\pm}^{(\lambda-1,k)})$ from $\hat f_2(\sigma)$ eliminates the term~\eqref{EqD5Almftf} at the price of an additional source term with $\tface$-index set $(\lambda,k+l)+\cE_\ind$ for $l=1$ or $l=0$ (depending on whether or not logarithmic terms arise from the terms $\fl^{-\alpha_0+2}$ in~\eqref{EqiptfGruPrec2}), which is still contained in $\tilde\cE'_{+\sharp}$ if we enlarge $\tilde\cE'_{+\sharp}$ appropriately. One then iterates this procedure.

  To explain how these terms contribute to the late-time asymptotics of $u$, consider the inverse Fourier transform of
  \begin{equation}
  \label{EqD5Almutfsing}
    \sigma\sum_\pm \chi_\zface H(\pm\sigma)u_{\tface,\pm,{\rm sing}}^{(\lambda-1,k)},
  \end{equation}
  the overall prefactor $\sigma$ undoing the division by $\sigma$ before~\eqref{EqD5Almftf}. We can write the total prefactor $|\sigma|^{\lambda-1}\sigma^{-1}(\log|\sigma|)^k$ as a linear combination of $(\sigma+i 0)^{\lambda-2}$ and $\sigma_-^{\lambda-2}$ when $k=0$ where $\sigma_-=\max(-\sigma,0)$, and for general $k$ of $(\sigma+i 0)^{\lambda-2}(\log(\sigma+i 0))^k$ and $\sigma_-^{\lambda-2}(\log\sigma_-)^j$, $j=0,\ldots,k$. The inverse Fourier transform of the former is a constant multiple of $H(t_*)t_*^{-\lambda+1}(\log t_*)^k$ and thus vanishes for negative $t_*$; the inverse Fourier transform of the latter vanishes for $t_*>0$ and is bounded \emph{from below} by a constant times $|t_*|^{-\lambda+1}(\log|t_*|)^j$ for $t_*\ll -1$ \emph{unless it vanishes}. As the arguments below will show, no other contributions to $u$ decay so poorly as $t_*\to-\infty$ that they could cancel these terms when $\Re(\lambda-1)\leq\alpha_\cK-2$, and thus these must, in fact, be absent since $u=0$ for $\ft_*<1$. (The same argument was already used in a related context after~\eqref{EqD4Fpasing}.) To simplify the notation, we shall thus suppress the ``$\sigma_-$''-terms. (For the terms with $\Re(\lambda-1)\geq\alpha_\cK-2$, one should include them, but they do not affect any arguments below.) Writing the ``${}\!+i 0$'' term of $\sigma\sum_\pm \chi_\zface H(\pm\sigma)u_{\tface,\pm,{\rm sing}}^{(\lambda-1,k)}$ as
  \[
    u_{\tface,{\rm sing}}^{(\lambda-1,k)} = \chi_\zface\,(\sigma+i 0)^{\lambda-2}(\log(\sigma+i 0))^k h_{b,\rms 1}(\dot\scal_1^{(\lambda-1,k)}),
  \]
  we thus have
  \begin{equation}
  \label{EqD5AlmutfsingFT}
    \cF^{-1}u_{\tface,{\rm sing}}^{(\lambda-1,k)}=c t_*^{-\lambda+1}(\log t_*)^k h_{b,\rms 1}(\dot\scal_1^{(\lambda-1,k)})+\cO(t_*^{-\infty}),\quad t_*\to\infty,
  \end{equation}
  in spatially compact sets; here $c=c(\lambda,k)$ is a numerical prefactor (cf.\ \cite[Exercise~2.6]{HintzMicro}). We now separate three cases.
  \begin{enumerate}
  \item If $\Re(\lambda-1)\leq\alpha'_\cK-2$, the assumption that $u=o(t_*^{-(\alpha'_\cK-2)})$ in spatially compact sets (in a pointwise sense, as follows from $u\in\Hb^{\infty,\ \cdots,\ \cdots,\ \alpha'_\cK-2}$ by Sobolev embedding) forces $\dot\scal_1^{(\lambda-1,k)}=0$: these are the decaying center-of-mass motions which we have already eliminated in the inductive hypothesis. Carefully note here that the set $\tilde\cE'_{+\sharp}$ may be strictly larger than $\tilde\cE'_{++}$, so this observation shows that the late-time asymptotics of $u$ cannot involve any $t_*^{-\lambda+1}(\log t_*)^k$-terms even for values of $(\lambda,k)$ that were not treated in the previous inductive step.
  \item The new terms in the asymptotic expansion of $u$ are the terms~\eqref{EqD5AlmutfsingFT} with $\Re(\lambda-1)\in(\alpha'_\cK-2,\alpha_\cK-2]$; we show in Part~2 of the proof how to eliminate them.
  \item If $\Re(\lambda-1)>\alpha_\cK-2$, we lump the terms~\eqref{EqD5Almutfsingpm} and~\eqref{EqD5AlmutfRest}, both multiplied by $\sigma$, together: their sum lies in $\Hb^{\infty,\ 1-\eps,\ \Re\lambda-1-\eps,\ \Re\lambda-2-\eps}\subset\Hb^{\infty,\ 1-\eps,\ -\eps,\ \alpha_\cK-3}(X_\scbtop^\pm)$ (and hence its inverse Fourier transform is of class $\Hb^{\infty,\ 1-\eps,\ 1-\eps,\ \alpha_\cK-2}(M';S^2\cT^*)$, which is contained in~\eqref{EqD5Almu}).
  \end{enumerate}

  Altogether, we have constructed
  \begin{equation}
  \label{EqD5utfsing}
    u_{\tface,{\rm sing}} := \sum_{\substack{(\lambda,k)\in\tilde\cE'_{+\sharp} \\ \Re(\lambda-1)\in(\alpha'_\cK-2,\alpha_\cK-2]}} u_{\tface,{\rm sing}}^{(\lambda-1,k)},\quad
    \tilde u_\tface \in \Hb^{\infty,\ 1-\eps,\ -\eps,\ \alpha_\cK-3}(X_\scbtop^\pm),
  \end{equation}
  where $\tilde u_\tface$ also absorbs the terms~\eqref{EqD5AlmutfRest} (multiplied by $\sigma$) for all $(\lambda,k)\in\tilde\cE'_{++}$, $\Re\lambda<2$, such that
  \begin{equation}
  \label{EqD5utff3}
    \hat f_3(\sigma) := \hat f_2(\sigma) - \wh{L_b}(\sigma)( u_{\tface,{\rm sing}} + \tilde u_\tface ) \in |\sigma|\Hb^{\infty,\ 2-\eps,\ 2-\eps,\ ((0,0),\alpha_\cK-2)}(X_\scbtop^\pm)
  \end{equation}
  does not have an expansion at $\tface$ anymore. This still has too little decay at $\tface$ relative to $\zface$; as in~\eqref{EqD4u3}, we thus apply Proposition~\ref{PropiptfC} to obtain
  \begin{equation}
  \label{EqD5utfu3}
    \hat u_3 \in |\sigma|\Hb^{\infty,\ 1-\eps,\ -\eps,\ -1-\eps} = \Hb^{\infty,\ 1-\eps,\ 1-\eps,\ -\eps}(X_\scbtop^\pm)\quad\forall\,\eps>0
  \end{equation}
  such that
  \begin{equation}
  \label{EqD5utff4}
    \bigl(\hat f_4(\sigma)\bigr)\big|_{\sigma\in\pm[0,c]} := \Bigl( \sigma^{-1}\bigl(\hat f_3(\sigma) - \wh{L_b}(\sigma)\hat u_3(\sigma)\bigr) \Bigr)\Big|_{\sigma\in\pm[0,c]} \in \Hb^{\infty,\ 2-\eps,\ 2+\eps_\ind-\eps,\ ((0,0),\alpha_\cK-2)}(X_\scbtop^\pm).
  \end{equation}

  \pfsubstep{Step~1.5.}{The sub-leading zero energy piece.} We next repeat the arguments around~\eqref{EqD4Sub0}, so
  \begin{equation}
  \label{EqD5ureg4}
    \bigl(\hat u_{{\rm reg},4}(0),\ \hat b_4(0),\ 0\bigr) := \wt{L_b}(0)^{-1}\bigl(\hat f_4(0),\ 0,\ 0\bigr)
  \end{equation}
  where $\hat u_{{\rm reg},4}(0)\in\Hb^{\infty,\ \eps_\ind-\eps}(X;S^2\cT^*_X)$; the third component vanishes since, by~\eqref{EqD4Params}, it encodes the constant asymptotic center-of-mass shift, which we have already eliminated. As in~\eqref{EqD4f5}, the remaining error is
  \begin{align}
    &\bigl( \hat f_4(\sigma),\ 0,\ 0 \bigr) - \wt{L_b}(\sigma)\bigl(\chi_\zface\hat u_{{\rm reg},4}(0),\ \hat b_4(0),\ 0\bigr) =: \bigl(\hat f_5(\sigma),\ 0,\ 0\bigr), \nonumber\\
  \label{EqD5f5}
    &\qquad \hat f_5 \in \Hb^{\infty,\ 2-\eps,\ 2+\eps_\ind-\eps,\ \alpha_\cK-2}(X_\scbtop^\pm).
  \end{align}

  \pfsubstep{Step~1.6.}{Solving away the error at $\tface$.} The orders of~\eqref{EqD5f5} at $\tface$ and $\zface$ differ by $2+\eps_\ind-\eps-(\alpha_\cK-2)$, which is less than $2$ when $\alpha_\cK\geq 2+\eps_\ind$. (This situation did not arise in Step~5 of the proof of Proposition~\ref{PropD4Par}.) Thus, we need to solve away $\hat f_5$ at $\tface$, which we do by repeatedly applying Proposition~\ref{PropiptfC} (with $k_0=1>\ell_\cK=\alpha_\cK-2$). This produces
  \begin{align*}
    &\hat u_5 \in \Hb^{\infty,\ 1-\eps,\ \eps_\ind-\eps,\ \eps_\ind-\eps-1}(X_\scbtop^\pm), \\
    &\qquad \hat f_6(\sigma) := \hat f_5(\sigma)-\wh{L_b}(\sigma)\hat u_5(\sigma) \implies \bigl(\hat f_6(\sigma)\bigr)\big|_{\sigma\in\pm[0,c]} \in \Hb^{\infty,\ 2-\eps,\ \alpha_\cK+\eps_\ind-\eps,\ \alpha_\cK-2}(X_\scbtop^\pm).
  \end{align*}

  \pfsubstep{Step~1.7.}{Remaining piece; combination.} As in~\eqref{EqD4u5}, we can now set
  \begin{align*}
    &\bigl(\hat u_6(\sigma),\ \hat b_6(\sigma),\ \hat\scal_6(\sigma) \bigr) := \wt{L_b}(\sigma)^{-1}\bigl(\hat f_6(\sigma),\ 0,\ 0\bigr), \\
    &\qquad \hat u_6 \in \Hb^{\infty,\ 1-\eps,\ \alpha_\cK-2+\eps_\ind-\eps,\ \alpha_\cK-2}(X_\scbtop^\pm),\ \ \hat b_6,\ \hat\scal_6 \in \Hb^{\infty,\ \alpha_\cK-2}(\pm[0,c]).
  \end{align*}
  Similarly to~\eqref{EqD4FTusing} (but with fewer poorly decaying terms), we then have
  \[
    u = u_{\rm sing} + u_{\rm exp} + u_{\rm reg} + u_\rem,
  \]
  where (using~\eqref{EqD1AlmAug} and recalling~\eqref{EqD5uscri}, \eqref{EqD5uiota}, and~\eqref{EqD5utfsing})
  \begin{align}
    (\cF u_{\rm sing})(\sigma) &:= i\cF(\pa_{t_*}\dot g_b^{\Ups,\aug})\bigl( \hat b_4(0) + \hat b_6(\sigma) \bigr) - \cF(\pa_{t_*}^2 h_{b,\rms 1}^{\leq 1,\aug})\bigl( \sigma^{-1}\hat\scal_6(\sigma) \bigr) \nonumber\\
      &\in \Hb^{\infty,\ 1-\eps,\ 1-\eps,\ \alpha_\cK-3}(X_\scbtop^\pm), \nonumber\\
    \cF u_{\rm exp} &:= u_{\tface,\rm sing}, \nonumber\\
  \label{EqD5ureg}
  \begin{split}
    (\cF u_{\rm reg})(\sigma) &:= \chi(|\sigma|) \Bigl( \chi_\zface\bigl(\hat u_{{\rm reg},1}(0)+\sigma\hat u_{{\rm reg},4}(0)\bigr) + \tilde u_\tface(\sigma) + \hat u_3(\sigma) + \sigma\hat u_5(\sigma) + \sigma\hat u_6(\sigma) \Bigr) \\
      &\in \Hb^{\infty,\ 1-\eps,\ -\eps,\ \alpha_\cK-3}(X_\scbtop^\pm),
  \end{split} \\
  \label{EqD5urem}
    u_\rem &:= u_{\rm hi} + u_{1,\sscri} + u_\iota \in \Hb^{\infty,\ 1-\eps,\ 1-\eps,\ 1-\eps}(M';S^2\cT^*).
  \end{align}
  Note that $\cF u_{\rm sing}$ is integrable in $\sigma$ (due to us having already eliminated the non-decaying boost, translation, and parameter change terms). Thus, \eqref{EqTFHbInvLo} gives
  \begin{equation}
  \label{EqD5SingRegMem}
    u_{\rm sing} \in \Hb^{\infty,\ 1-\eps,\ 2-\eps,\ \alpha_\cK-2}(M';S^2\cT^*), \quad
    u_{\rm reg} \in \Hb^{\infty,\ 1-\eps,\ 1-\eps,\ \alpha_\cK-2}(M';S^2\cT^*).
  \end{equation}
  (It is at this point that one can argue for the vanishing of the ``$\sigma_-$'' type contributions in the arguments following~\eqref{EqD5Almutfsing}, since $u_{\rm sing}+u_{\rm reg}+u_\rem=\cO(|t_*|^{-\alpha_\cK+2})$ as $t_*\to-\infty$.) On the other hand, $\cF u_{\rm exp}\in\Hb^{\infty,\ 2-\eps,\ \alpha'_\cK-1,\ \tilde\cE^{\prime,>\alpha'_\cK-1}_{+\sharp}-2}(X_\scbtop^\pm)$ (the $\scface$-order accommodating $h_{b,\rms 1}\in\cA^2(X;S^2\cT^*)\subset\Hb^{\infty,\ 2-\eps}$) where
  \[
    \tilde\cE^{\prime,>\alpha'_\cK-1}_{+\sharp}:=\{(\lambda,k)\in\tilde\cE'_{+\sharp}\colon 1<\Re\lambda<\alpha'_\cK-1\},
  \]
  so Proposition~\ref{PropTFHbphg}\eqref{ItTFHbphgFI} implies
  \[
    u_{\rm exp} \in \Hb^{\infty,\ 2-\eps,\ \alpha'_\cK,\ \tilde\cE^{\prime,>\alpha'_\cK-1}_{+\sharp}-1}(M';S^2\cT^*);
  \]
  and the $t_*^{-\lambda+1}(\log t_*)^k$-coefficient for $\Re(\lambda-1)\in(\alpha'_\cK-2,\alpha_\cK-2]$ is a non-zero multiple of $t_*^{-\lambda+1}(\log t_*)^k h_{b,\rms 1}(\dot\scal_1^{(\lambda-1,k)})$ by~\eqref{EqD5AlmutfsingFT}; so upon multiplying $\dot\scal_1^{(\lambda-1,k)}$ by a non-zero constant, we have (upon relaxing the $\iota^+$-order)
  \begin{equation}
  \label{EqD5uexpexp}
    u_{\rm exp} - \sum_{\substack{ (\lambda,k)\in\tilde\cE'_{+\sharp} \\ \Re(\lambda-1)\in(\alpha'_\cK-2,\alpha_\cK-2] }} \chi_\cK t_*^{-\lambda+1}(\log t_*)^k h_{b,\rms 1}(\dot\scal_1^{(\lambda-1,k)}) \in \Hb^{\infty,\ 2-\eps,\ 1-\eps,\ \alpha_\cK-2}(M';S^2\cT^*).
  \end{equation}
  We have now almost shown
  \begin{equation}
  \label{EqD5uFinal}
  \begin{split}
    &u = \sum_{ \substack{(\lambda,k)\in\tilde\cE'_{+\sharp} \\ \Re(\lambda-1)\in(\alpha'_\cK-2,\alpha_\cK-2] }} \chi_\cK t_*^{-\lambda+1}(\log t_*)^k h_{b,\rms 1}(\dot\scal_1^{(\lambda-1,k)}) + \tilde u, \\
    &\quad\hspace{3em} \dot\scal_1^{(\lambda-1,k)}=:\dot\scal_1^{(\lambda-1,k)}(f)\in\scalspace_1,\quad \tilde u\in\Hb^{\infty,\ \bigl(\la\cE_\sscri^\cC\ra,3+\eps_\sscri\bigr),\ 1-\eps,\ \alpha_\cK-2}(\Omega_*;S^2\cT^*)^{\bullet,-},
  \end{split}
  \end{equation}
  except thus far we have only obtained the $\scri^+$-order $1-\eps$; but this is improved to $(\la\cE_\sscri^\cC\ra,3+\eps_\sscri)$ by means of Proposition~\ref{PropDScriPhg} as usual. This establishes a partial expansion of $u$ into finitely many decaying center-of-mass motion terms, with a remainder the decays at the improved rate $\alpha_\cK-2$.

  \pfstep{Part~2. Eliminating the partial expansion.} In order to simplify the notation, let us first assume that the sum in~\eqref{EqD5uFinal} consists of a single term corresponding to some $(\lambda,k)\in\tilde\cE'_{+\sharp}$ with $\Re(\lambda-1)\in(\alpha'_\cK-2,\alpha_\cK-2]$. When $(\lambda,k)\in\tilde\cE_{++}$, we have already defined
  \begin{equation}
  \label{EqD5AlmMod}
    D_{(h,\vecp)}P(\scal_1^{(\lambda-1,k)\prime}):=D_{(h,\vecp)}P(0,(0,\ldots,\scal_1^{(\lambda-1,k)\prime},\ldots,0));
  \end{equation}
  and in general we define this (recalling Definition~\ref{DefD5Corr}) consistently to be given by the expression $D_g\Ric(h_{\rms 1}^{(\lambda-1,k)}(\scal_1^{(\lambda-1,k)\prime}))+\delta_{g^0,E^\cC}\vartheta_{\rms 1}^{(\lambda-1,k)}(\scal_1^{(\lambda-1,k)\prime})$ where $g=g^0+h_\tot(\vecp)+h$ in the notation of Definition~\ref{DefD5EinsteinAug}; note that~\eqref{EqD5LinPar}--\eqref{EqD5LinParDiff} still hold for~\eqref{EqD5AlmMod}. We henceforth abbreviate
  \[
    \scal := \scal_1^{(\lambda-1,k)\prime}.
  \]

  First of all, note that the inductive hypothesis produces parameters $\vecp^{\prime,\leq\alpha'_\cK-2}(\scal)$ such that the solution $v$ of $L v=-D_{(h,\vecp)}P(\scal)-D_{(h,\vecp)}P(0,(\vecp^{\prime,\leq\alpha'_\cK-2}(\scal),0))$ has $\cK^+$-order $\alpha'_\cK-2$ (and thus pointwise $o(t_*^{-(\alpha'_\cK-2)})$-decay in spatially compact sets) at $(\cK^+)^\circ$.

  \emph{Ideally}, one would then like to consider the ($\scal$-dependent) solution $u=u(\scal)$ of
  \begin{equation}
  \label{EqD5EqMod}
  \begin{split}
    L u(\scal) &= f - F_{\rm mod}^{\leq\alpha'_\cK-2}(f) - D_{(h,\vecp)}P(\scal) - D_{(h,\vecp)}P\bigl(0,(\vecp^{\prime,\leq\alpha'_\cK-2}(\scal),0)\bigr) \\
      &= f - D_{(h,\vecp)}P\bigl(0,(\vecp^{\prime,\leq\alpha'_\cK-2}(f),0)\bigr) - D_{(h,\vecp)}P(\scal) - D_{(h,\vecp)}P\bigl(0,(\vecp^{\prime,\leq\alpha'_\cK-2}(\scal),0)\bigr)
  \end{split}
  \end{equation}
  and choose $\scal\in\scalspace_1$ such that for $u(\scal)$, each term in the sum in~\eqref{EqD5uFinal} vanishes. Note that upon setting $f_\tot:=f-D_{(h,\vecp)}P(\scal)$, the right-hand side of~\eqref{EqD5EqMod} is $f_\tot-D_{(h,\vecp)}P(0,(\vecp^{\prime,\leq\alpha'_\cK-2}(f_\tot),0))$; thus the inductive hypothesis does apply to this equation, and hence $u(\scal)=o(t_*^{-\alpha'_\cK+2})$ at $(\cK^+)^\circ$. \emph{However} (and this makes Remark~\ref{RmkD5AlmStruct} more precise), the issue with the leading-order asymptotics of $u(\scal)$ is that the third term on the right-hand side of~\eqref{EqD5EqMod} features the element $(\lambda+2,k)$ in its $\iota^+$-index set; when $(\lambda,k)\notin\tilde\cE_{++}$, this was not allowed for in Part~1 (cf.\ \eqref{EqD5uscri}), and indeed for general source terms with $(\lambda+2,k)$ in their $\iota^+$-index sets, it may happen (in view of possible indicial root coincidences in~\eqref{EqD5Almureg1}) that one would get an additional $t_*^{-\lambda+1}(\log t_*)^{k+1}$-term, say, in~\eqref{EqD5uFinal}---which one would attempt to eliminate using a parameter $\scal_1^{(\lambda-1,k+1)\prime}$ at order $t_*^{-\lambda+1}(\log t_*)^{k+1}$, etc., leading to a never-ending proliferation of logarithms.

  It is thus crucial to exploit special properties of the term $-D_{(h,\vecp)}P(\scal)$ in~\eqref{EqD5EqMod}. When $h=0$ and $\vecp=0$, it sources the metric perturbation $-h_{b_0}^{(\lambda-1,k)}(\scal)$ where
  \begin{equation}
  \label{EqD5hb0Insert}
    h_{b_0}^{(\lambda-1,k)}(\scal):=\chi_\cK\delta_{g_{b_0}}^*\omega_{b_0,\rms 1}^{(0),\leq 4}(\scal t_*^{-\lambda+1}(\log t_*)^k)
  \end{equation}
  (cf.\ \eqref{EqD4ParRewrite3}), so it is natural to consider instead of~\eqref{EqD5EqMod} the equation for
  \[
    u_\Sigma(\scal) := u(\scal) + h_b^{(\lambda-1,k)}(\scal),
  \]
  which reads
  \begin{equation}
  \label{EqD5AlmGoodEqPrelim}
  \begin{split}
    L u_\Sigma(\scal) &= f - D_{(h,\vecp)}P\bigl(0,(\vecp^{\prime,\leq\alpha'_\cK-2}(f),0)\bigr) \\
      &\qquad - \bigl( D_{(h,\vecp)}P(\scal) - L h_b^{(\lambda-1,k)}(\scal) \bigr) - D_{(h,\vecp)}P\bigl(0,(\vecp^{\prime,\leq\alpha'_\cK-2}(\scal),0)\bigr).
  \end{split}
  \end{equation}
  The key point is that the first term on the second line now has more than $\alpha_\cK+1$ orders of decay at $\iota^+$, as follows by replacing $L$ by $L_b$ and using~\eqref{EqD5LinParDiff}; the $\cK^+$-order of the term $L h_b^{(\lambda-1,k)}(\scal)$, is equal to the minimum of that of $L_b h_b^{(\lambda-1,k)}(\scal)$ (which is $\Re\lambda+4-\eps$) and $(L-L_b)(h_b^{(\lambda-1,k)}(\scal))$ (which is $(\Re\lambda-1)+(2+\eps_\cK)>\alpha'_\cK-2+2+\eps_\cK=\alpha_\cK-2$), so $>\alpha_\cK-2$ and thus also acceptable. Note that the right-hand side of~\eqref{EqD5AlmGoodEqPrelim} is equal to
  \begin{equation}
  \label{EqD5AlmGoodEq}
    f_\tot-D_{(h,\vecp)}P\bigl(0,(\vecp^{\prime,\leq\alpha'_\cK-2}(f_\tot),0)\bigr),\quad f_\tot:=f-\bigl(D_{(h,\vecp)}P(\scal)-L(h_b^{(\lambda-1,k)}(\scal))\bigr);
  \end{equation}
  we use here that $\vecp^{\prime,\leq\alpha'_\cK-2}(L(h_b^{(\lambda-1,k)}(\scal)))=0$, which follows from the observation that $h_b^{(\lambda-1,k)}(\scal)$ does have the acceptable $o(t_*^{-\alpha'_\cK+2})$-decay at $(\cK^+)^\circ$.

  Thus, not only does the inductive hypothesis apply to this equation, but also the extraction of the leading-order terms of $u_\Sigma(\scal)$ leads to the expansion~\eqref{EqD5uFinal} \emph{with the same index set $\tilde\cE'_{+\sharp}$}, i.e., without a proliferation of factors of $\log t_*$. The total expansion of $u(\scal)=u_\Sigma(\scal)-h_b^{(\lambda-1,k)}(\scal)$, which, as we recall, we are (for notational simplicity) presently assuming to consist only of the $(\lambda,k)$-summand, is thus $\chi_\cK t_*^{-\lambda+1}(\log t_*)^k h_{b,\rms 1}(\scal_\tot)$ where
  \[
    \scal_\tot = \dot\scal_1^{(\lambda-1,k)}(f_\tot) - \scal = \dot\scal_1^{(\lambda-1,k)}(f) - \Bigl( \scal + \dot\scal_1^{(\lambda-1,k)}\bigl( D_{(h,\vecp)}P(\scal)-L(h_b^{(\lambda-1,k)}(\scal)) \bigr) \Bigr).
  \]
  But since $D_{(0,\vec 0)}P(\scal)-L_{b_0}(h_{b_0}^{(\lambda-1,k)}(\scal))=0$, the second summand is of size $o(1)|\scal|$ as $(h,\vecp)\to 0$ (in a fixed, sufficiently high-regularity, space). Since linear maps $I+o(1)\colon\scalspace_1\to\scalspace_1$ are invertible when $o(1)$ is sufficiently small, there thus exists a unique choice of $\scal\in\scalspace_1$ such that $\scal_\tot=0$. This finishes the construction of $\vecp^{\prime,\leq\alpha_\cK-2}=(\vecp^{\prime,\leq\alpha'_\cK-2},\scal_1^{(\lambda-1,k)\prime})$, $\scal_1^{(\lambda-1,k)\prime}:=\scal$.

  The arguments for general expansions~\eqref{EqD5uFinal} are the same except for mild notational complexities: instead of a single parameter $\scal_1^{(\lambda-1,k)\prime}$, one now finds unique
  \begin{equation}
  \label{EqD5AlmModParam}
    \vecp_\sharp^{\prime,\in(\alpha'_\cK-2,\alpha_\cK-2]} = (\scal_1^{(\lambda-1,k)\prime})_{ \substack{(\lambda,k)\in\tilde\cE'_{+\sharp} \\ \Re(\lambda-1)\in(\alpha'_\cK-2,\alpha_\cK-2]} }
  \end{equation}
  such that the expansion~\eqref{EqD5uFinal} of $u_\Sigma(\vecp_\sharp^{\prime,\in(\alpha'_\cK-2,\alpha_\cK-2]})-\sum h_b^{(\lambda-1,k)}(\scal_1^{(\lambda-1,k)\prime})$ is trivial, where $u_\Sigma$ solves the generalization
  \begin{align*}
    L u_\Sigma &= f_\tot - D_{(h,\vecp)}P\bigl(0,(\vecp^{\prime,\leq\alpha'_\cK-2}(f_\tot),0)\bigr), \\
    &\qquad f_\tot := f - \sum \Bigl(D_{(h,\vecp)}P(\scal_1^{(\lambda-1,k)\prime})-L\bigl(h_b^{(\lambda-1,k)}(\scal_1^{(\lambda-1,k)\prime})\bigr)\Bigr)
  \end{align*}
  of~\eqref{EqD5AlmGoodEqPrelim}. If we enlarge $\tilde\cE_{++}$ so that it contains
  \begin{equation}
  \label{EqD5AlmEplusplusExtra}
    \{(\lambda,k)\in\tilde\cE'_{+\sharp}\colon\Re(\lambda-1)\in(\alpha'_\cK-2,\alpha_\cK-2]\},
  \end{equation}
  this finishes the inductive step. It is critical to note that this enlargement can be done in a self-consistent manner (i.e., that ultimately a single choice of $\tilde\cE_{++}$ allows for a proof of~\eqref{EqD5AlmEq}--\eqref{EqD5Almu}): indeed, for the elimination of non-$o(t_*^{-\alpha'_\cK+2})$-terms, this enlargement (which only causes extra $\iota^+$-source of size $o(\rho_+^{\alpha_\cK'+1})$) has no effect in that it does \emph{not} force one to use any additional modification terms beyond~\eqref{EqD5AlmModParam} (with the index set $\tilde\cE_{+\sharp}'$ computed using $\tilde\cE'_{++}$ from the inductive hypothesis).
\end{proof}

The following procedure constructs an index set $\tilde\cE'_{++}$ satisfying all requirements above. Start with $\tilde\cE'_{++,(0)}=\emptyset$. Define then $\tilde\cE'_{+\sharp,(1)}$ to be the set $\tilde\cE'_{+\sharp}$ from~\eqref{EqD5Almureg1} for $\alpha'_\cK$ marginally above $2$ (so that no expansion term between $t_*^0$ and $t_*^{-\alpha'_\cK+2}$ exists) and $\alpha_\cK=2+\eps_\cK$, and enlarged at sub-leading orders as explained after~\eqref{EqD5AlmutfRest}. Define $\tilde\cE'_{++,(1)}$ to be the smallest index set containing
\[
  \bigl\{(\lambda,k)\in\tilde\cE'_{+\sharp,(1)} \colon \Re(\lambda-1)\in(\alpha'_\cK-2,\alpha_\cK-2] \bigr\}
\]
such that $\tilde\cE'_{++,(1)}+(\cE_\ind\cup\cE_+\cup\tilde\cE'_{++,(1)})\subset\tilde\cE'_{++,(1)}$. Consider then $\alpha'_\cK=2+\eps_\cK$ and $\alpha_\cK=\alpha'_\cK+\eps_\cK$, and perform the inductive step above with $\tilde\cE'_{++,(1)}$ in place of $\tilde\cE'_{++}$; this produces $\tilde\cE'_{+\sharp,(2)}$ in~\eqref{EqD5Almureg1} (and enlarged as after~\eqref{EqD5AlmutfRest}), and we then define $\tilde\cE'_{++,(2)}$ to be the smallest index set containing
\[
  \tilde\cE'_{++,(1)} \cup \bigl\{(\lambda,k)\in\tilde\cE'_{+\sharp,(2)} \colon \Re(\lambda-1)\in(\alpha'_\cK-2,\alpha_\cK-2] \bigr\}
\]
such that $\tilde\cE'_{++,(2)}+(\cE_\ind\cup\cE_+\cup\tilde\cE'_{++,(2)})\subset\tilde\cE'_{++,(2)}$. Repeating this process $N=\lceil\eps_\cK^{-1}\rceil+1<\infty$ many times (possibly with arbitrarily small reductions of the values of $\alpha_\cK$ to avoid collisions with indicial roots, as discussed at the beginning of Step~1.2 of the above proof) produces the desired set
\begin{equation}
\label{EqD5Eplusplus}
  \tilde\cE'_{++} := \tilde\cE'_{++,(N)}.
\end{equation}
This allows for a self-consistent elimination of the center-of-mass motions in Proposition~\ref{PropD5Alm} and satisfies $\tilde\cE'_{++}+(\cE_\ind\cup\cE_+\cup\tilde\cE'_{++})\subset\tilde\cE'_{++}$.

\begin{rmk}[Lower-order-only dependence of $\tilde\cE'_{++}$ from $\cE_+$]
\label{RmkD5EplusDep}
  Since $\min\Re\cE_+\geq 1$, the intersection of $\tilde\cE'_{++}$ with $\{\Re\lambda\in(1,2)\}$---which is the only part of $\tilde\cE'_{++}$ that is actually used in Definition~\ref{DefD5EinsteinAug} and Proposition~\ref{PropD5Alm}---is, in fact, independent of $\cE_+$.
\end{rmk}

\begin{rmk}[Origin of $t_*^{-\alpha}$ center-of-mass shifts]
\label{RmkD5Orig}
  The $t_*^{-\lambda+1}(\log t_*)^k$ center-of-mass shift term in the expansion~\eqref{EqD5uexpexp} arises from quasi-homogeneous $t_*^{-\lambda-1}$-terms in the $\iota^+$-expansion of the effective source term $f_\flat$ in~\eqref{EqD5uiota}; and these terms, in turn, can arise even for source terms $f$ in~\eqref{EqD5Almf} that vanish to high order at $\iota^+$, namely (via transport, cf.\ Figure~\ref{FigIDecay}) from the formal solution $u_{1,\sscri}$ at $\scri^+$ produced by Proposition~\ref{PropDScriFormal}, and thus from $r^{-\lambda-2}$-terms in the $\scri^+$-expansion of $f$ there (cf.\ the appearance of the $\iota^+$-index set $(z,0)$ in~\eqref{EqDScriuzk} appearing in the construction of the $\rho_\sscri^z$-term of the formal solution). Going further back, the relevant $f$ arises in the context of the stability problem from a source term of the type $P(\chi_- h_{-\scal},\scal,0)$ (cf.\ \eqref{EqD2PFwd}), say, and thus from $r^{-\lambda}$-terms in the $\scri^+$-expansion of the exterior solution $h_{-\scal}$. That expansion, finally, is determined by the expansion of the initial data of $h_{-\scal}$ (see~\eqref{EqExPhgData} and~\eqref{EqExBoIVPData}), and indeed the transport term of $L^0$ in~\eqref{EqExPhgScriKerr} is responsible for transporting $r^{-\lambda}$-terms of the polyhomogeneous expansion of the initial data at $r^{-1}=0$ from $I^0$ to $\scri^+$. Altogether, then, this explains how weakly decaying $r^{-\lambda}$-contributions to the initial data cause weakly decaying $t_*^{-\lambda+1}$-shifts of the center-of-mass of the black hole. --- The origin of the logarithmic shift in Proposition~\ref{PropD4Par} is slightly more subtle: it arises from $r^{-1}$-terms of the $(\dd x^1)^2$-component of metric perturbations (cf.\ also Remark~\ref{RmkipPResOver}), which exists due to its coupling to radiation, so it is typically present even when the initial data are Kerrian outside a compact set.
\end{rmk}

\subsection{Step~6: final gauge modifications; \texorpdfstring{$\iota^+$- and $\cK^+$-expansions, $t_*^{-4-\eps_\cK}$-remainder}{expansion at timelike infinity and t\textasciicircum(-4-eps)-remainder}}
\label{SsD6Better}

In the statement of Proposition~\ref{PropD5Alm}, we have put a considerable number of terms in the expansion of $u$ at $\iota^+$ and $\cK^+$ that were not needed for the proof into the remainder space~\eqref{EqD5Almu}: this includes partial expansions at $\iota^+$ (up to $\cO(\rho_+^{2-\eps})$-remainders) and partial expansions at $\cK^+$ produced by the inversion of $\iota^+$-, resp.\ $\tface$-normal operators using Proposition~\ref{PropipGr}, resp.\ Propositions~\ref{PropiptfGr} and \ref{PropiptfC}, as well as the decaying center-of-mass motion and black hole parameter changes arising from the singular terms of $\wh{L_b}(\sigma)^{-1}$. Let us first give a rough accounting of these terms.

\medskip

{\bf Expansion and decay at $\cK^+$.} Given the $\cO(t_*^{-1+\eps})$-decay of $u$ at $(\cK^+)^\circ$ that we have already established, one might expect from the decay of $(L-L_b)u=\cO(t_*^{-3-\eps_\cK+\eps})$ that one can already now prove that $u$ has a $\cO(t_*^{-3-\eps_\cK+\eps})$-remainder plus less-decaying pure gauge terms and the contributions of the singular part of $\wh{L_b}(\sigma)^{-1}$, which are $\cO(t_*^{-1-\eps_\cK+\eps})$ center-of-mass movements and $\cO(t_*^{-2-\eps_\cK+\eps})$ black hole parameter changes. Since in a nonlinear iteration scheme we need $u$ to have $\cO(t_*^{-2-\eps_\cK})$-decay (due to the $2$-admissibility of $L_b$, cf.\ Corollary~\ref{CorDAdm}), this indicates that at this point we are only slightly more than one power of $t_*$-decay at $\cK^+$ away from our goal; that is, we need to improve the decay of $u$ (after additional finite-dimensional gauge modifications) to $\cO(t_*^{-2-\eps_\cK})$.

Observe, though, that we need the forward map to map $u$ into a tensor $f$ with $\cO(t_*^{-4-\eps_\cK})$-decay. This means that we must show that the contributions to $u$ that are not of size $\cO(t_*^{-4-\eps_\cK})$ have spatial dependence given by large zero energy states (with, at worst, $\cO(t_*^{-2-\eps_\cK})$-modulations in time); recall here computations of the sort $L_b(h_{b,\rms 0}^{(0),\leq 3}(a(t_*)))=\cO(a^{(4)}(t_*))$ where $a=\cO(t_*^{-2-\eps_\cK})$, say, so $a^{(4)}$ decays more than fast enough for this to be an acceptably decaying (at $\cK^+$) source term in the next iteration step. In particular, it is \emph{not} necessary to show that these contributions are polyhomogeneous up to a $\cO(t_*^{-4-\eps_\cK})$-remainder term.

\medskip

{\bf Expansion and decay at $\iota^+$.} For the purpose of improving the decay of solutions of $L u=f$ in Proposition~\ref{PropD5Alm} beyond $t_*^{-1}$ by making further finite-dimensional gauge modifications, it is no longer sufficient to keep track of the ``structureless'' $\rho_+^{1-\eps}$-decay in~\eqref{EqD5Almu}; rather, we must now record the $\iota^+$-index set of $u$. Indeed, if we record a $\rho_+^{\beta_+}$-remainder of $u$ at $\iota^+$, with $\beta_+>1-\eps$, the right-hand side of the equation for $L_b u$ involves $(L-L_b)u$, which has a $\rho_+^{3+\beta_+}$-remainder (ignoring logarithmic factors). This, in turn, generically creates (roughly speaking) a non-polyhomogeneous $\cO(t_*^{-\beta_+})$ center-of-mass motion, which (due to the infinite-dimensional nature of weighted b-Sobolev spaces on $[1,\infty]_{t_*}$) we cannot eliminate using our linear algebra based approach (cf.\ Remark~\ref{RmkINElimNon}). It suffices to work with $\beta_+=2+\eps_+$, $\eps_+>\eps_\cK$, since terms of the schematic form $h_{b,\rms 1}^{\leq 1}(\cO(t_*^{-2-\eps_+}))$ are acceptable gravitational wave tails.

Notice, however, that a $\cO(\rho_+^{\beta_+})$-remainder of $u$ at $\iota^+$ would get mapped, via the (nonlinear) forward map (i.e., the gauge-fixed Einstein operator) into a $\cO(\rho_+^{2+\beta_+})$-remainder of the source term $f$ in the next step of a nonlinear iteration; thus we really need $\beta_+=3+\eps_+$ (so that $f$ creates $\cO(t_*^{-\beta_++1})=o(t_*^{-2-\eps_\cK})$ center-of-mass terms, among others) if we want to close a nonlinear iteration scheme.

Now, recording instead a partial expansion of $u$ at $\iota^+$ (modulo $\cO(\rho_+^{3+\eps_+})$ remainders), we expect further polyhomogeneous center-of-mass shifts; moreover, the sub-leading terms from the solutions of $\iota^+$- or $\tface$-model problems (via Proposition~\ref{PropipGr}, Corollary~\ref{CoripGrQhom}, and Proposition~\ref{PropiptfGr}) which we have thus far discarded due to their $\cO(t_*^{-1})$-decay---such as the $h_{b,\rms 0}^{(0),\leq 3}(t_*^{-1})$-term from~\eqref{EqipGr}---must now be dealt with. They are all pure gauge except for some terms with fast ($\cO(t_*^{-3})$) decay (namely, the terms involving $h_{b,\rms 2/\rmv 2}^{(-2),\leq 1}$ and $h_{b,\rms 3/\rmv 3}^{(-3)}$), and hence their elimination proceeds along very similar lines to the arguments in~\S\S\ref{SsD4Par} and \ref{SsD5Alm}; we spell this out below. We only need to eliminate terms that do not have size $\cO(t_*^{-2-\eps_\cK})$. (With $\eps_\cK>0$ so small that no term has decay between $t_*^{-2}$ and $t_*^{-2-\eps_\cK}$, these are the same as $o(t_*^{-2})$-terms.)

\subsubsection{Gauge modifications}
\label{SssD6Mod}

The key calculations for dealing with the expansion terms of Proposition~\ref{PropipGr} and Corollary~\ref{CoripGrQhom}, analogously to~\eqref{EqD4ParNoDelvsH}--\eqref{EqD4ParRewrite3}, are as follows.

\begin{enumerate}
\item{\rm (Rotations of the axis of rotation.)} Consider the second term in~\eqref{EqipGr}. For $a(t_*)=t_*^{-1}\vect$, $\vect\in\vectspace_1$, or $a(t_*)=t_*^{-\lambda}\vect$ with $\Re\lambda>1$, we have
  \[
    \chi_\cK h_{b_0,\rmv 1}^{\leq 3}(a(t_*)) = \chi_\cK\delta_{g_{b_0}}^*\omega_{b_0,\rmv 1}^{(-1),\leq 3}(a(t_*)) - \chi_\cK\,\dd t_*\otimes_s\breve\omega_{b_0,\rmv 1}^{(-1),3}(a^{(4)}(t_*)),
  \]
  similarly (via $k$-fold differentiation in $\lambda$) for $a(t_*)=t_*^{-\lambda}(\log t_*)^k\vect$. The second term on the right is of class $\cA^{\infty,\ (\lambda,0)\cup(\lambda+\cE_\ind),\ \lambda+4}$ and thus has acceptable $\iota^+$-decay (at least $1$ order) and $\cK^+$-decay (more than $4+\eps_\cK$ orders). Moreover,
  \begin{equation}
  \label{EqD6v1Corr}
  \begin{split}
    L_{b_0}\bigl(\chi_\cK \delta_{g_{b_0}}^*\omega_{b_0,\rmv 1}^{(-1),\leq 3}(a(t_*))\bigr) &= -D_{g_{b_0}}\Ric\Bigl([\delta_{g_{b_0}}^*,\chi_\cK]\omega_{b_0,\rmv 1}^{(-1),\leq 3}(a(t_*)) \Bigr) \\
      &\qquad + \delta_{g_{b_0},E^\cC}^*\delta_{g_{b_0},E^\Ups}\sfG_{g_{b_0}}\Bigl(\chi_\cK\delta_{g_{b_0}}^*\omega_{b_0,\rmv 1}^{(-1),\leq 3}(a(t_*))\Bigr);
  \end{split}
  \end{equation}
  the argument of $D_{g_{b_0}}\Ric$ has $\cO(\rho_+^{\lambda})$-decay at $\iota^+$, while that of $\delta_{g_{b_0},E^\cC}^*$ has $\cO(\rho_+^{\lambda+1})$-decay (as follows from the analogue of~\eqref{EqD4ParNoLog1}--\eqref{EqD4ParNoLog2}). --- Note for $\lambda=1$ that this means that undoing a $\cO(t_*^{-1})$-rotation of the axis of rotation of the black hole\footnote{We emphasize that this is \emph{different} from a $\cO(t_*^{-1})$-\emph{change} in the axis (or magnitude) of rotation, which would, e.g., include the possibility of settling down to Schwarzschild with time-dependent angular momentum parameter of size $t_*^{-1}$: this could \emph{not} be eliminated by pure gauge modifications, and it is thus important that such poorly decaying \emph{changes} of the axis of rotation do \emph{not} occur.} requires a $\cO(\rho_+^2)$-gauge modification, much like a $\cO(1)$-shift (which is thus stronger by a power of $t_*$) of the center-of-mass; the technical reason is that the relevant gauge potential $\omega_{b_0,\rmv 1}^{(-1)}=\cO(\rho^{-1})$ has $1$ order less spatial decay relative to $h_{b_0,\rmv 1}=\cO(\rho^2)$ than $\omega_{b_0,\rms 1}^{(0)}=\cO(1)$ relative to $h_{b_0,\rms 1}=\cO(\rho^2)$ (which in turn is still off by $1$ order compared to the relationship between gauge potentials that are unrelated to approximate symmetries and their symmetric gradients, such as those in Proposition~\ref{PropWG0Large}\eqref{ItWG0Larges01}--\eqref{ItWG0Largevl2}).
\item{\rm (Logarithmic time re-parameterization.)} Consider the third term in~\eqref{EqipGr}. Recall that $h_{b_0,\rms 0}^{(0)}=\delta_{g_{b_0}}^*\omega_{b_0,\rms 0}^{(0),\leq 1}$ (see~\eqref{EqWG0Larges00}) is pure gauge but with a gauge potential that grows (linearly) in time; so also here we need an exceptionally large gauge modification to eliminate such a term. Concretely, for $a(t_*)=t_*^{-\lambda}$, set $A(t_*)=\log t_*$ when $\lambda=1$ and $(-\lambda+1)^{-1}t_*^{-\lambda+1}$ when $\Re\lambda>1$; then
  \begin{equation}
  \label{EqD6s0Corr}
    \chi_\cK h_{b_0,\rms 0}^{(0),\leq 3}(a(t_*)) = \chi_\cK\delta_{g_{b_0}}^*\omega_{b_0,\rms 0}^{(0),\leq 4}(A(t_*)) - \chi_\cK\,\dd t_*\otimes_s \breve\omega_{b_0,\rms 0}^{(0),4}(a^{(4)}(t_*)).
  \end{equation}
  The analogue of~\eqref{EqD6v1Corr} is
  \begin{align*}
    L_{b_0}\bigl(\chi_\cK \delta_{g_{b_0}}^*\omega_{b_0,\rms 0}^{(0),\leq 4}(A(t_*))\bigr) &= -D_{g_{b_0}}\Ric\Bigl([\delta_{g_{b_0}}^*,\chi_\cK]\omega_{b_0,\rms 0}^{(0),\leq 4}(A(t_*)) \Bigr) \\
      &\qquad + \delta_{g_{b_0},E^\cC}^*\delta_{g_{b_0},E^\Ups}\sfG_{g_{b_0}}\Bigl(\chi_\cK\delta_{g_{b_0}}^*\omega_{b_0,\rms 0}^{(0),\leq 4}(A(t_*))\Bigr).
  \end{align*}
  For $A(t_*)=\log t_*$, the argument of $D_{g_{b_0}}\Ric$ is of size $\cO(\rho_+\log\rho_+)$ at $\iota^+$ (the logarithmic factor of which is already structurally familiar in the context of the elimination of the logarithmic center-of-mass motion as discussed after~\eqref{EqD4ParRewrite3} and in Lemma~\ref{LemmaD4Corr}), while that of $\delta_{g_{b_0},E^\cC}^*$ is of size $\cO(\rho_+^2)$.
\item{\rm (Non-exceptional terms.)} Among the other terms in~\eqref{EqipGr}, we only need to deal with those for now that correspond to indicial roots $-\mu$ of $\wh{\ubar L}(0)$ with $\Re\mu\in[0,1]$, which are those with $\mu=\lambda^\Ups_{\rms l,1}-1$ for $l=0,1$ and the terms involving $h_{b_0,\rms l}^{(-l+2)}$ for $l=2,3$ and $h_{b_0,\rmv l}^{(-l+2)}$ for $l=2$. The elimination of terms
  \[
    h_{b_0}^{(-\mu),\leq k}(a(t_*))=\delta_{g_{b_0}}^*\omega_{b_0}^{(-\mu-1),\leq k}(a(t_*))-\chi_\cK\,\dd t_*\otimes_s\breve\omega_{b_0}^{(-\mu-1),k}(a^{(k+1)}(t_*))
  \]
  then uses the analogue of~\eqref{EqD6v1Corr},
  \begin{equation}
  \label{EqD6CorrNonExc}
  \begin{split}
    L_{b_0}\bigl(\chi_\cK\delta_{g_{b_0}}^*\omega_{b_0}^{(-\mu-1),\leq k}(a(t_*))\bigr) &= -D_{g_{b_0}}\Ric\Bigl([\delta_{g_{b_0}}^*,\chi_\cK]\omega_{b_0}^{(-\mu-1),\leq k}(a(t_*))\Bigr) \\
      &\qquad + \delta_{g_{b_0},E^\cC}^*\delta_{g_{b_0},E^\Ups}\sfG_{g_{b_0}}\Bigl(\chi_\cK\delta_{g_{b_0}}^*\omega_{b_0}^{(-\mu-1),\leq k}(a(t_*))\Bigr).
  \end{split}
  \end{equation}
  The weakest-decaying $a(t_*)$ that arises for these terms is $a(t_*)=t_*^{-1-\mu}$, in which case the argument of $D_{g_{b_0}}\Ric$ has size $\cO(\rho_+^1)$, and the gauge modification (the argument of $\delta_{g_{b_0},E^\cC}^*$) is of size $\cO(\rho_+^2)$.
\end{enumerate}

Note that for $\lambda=1$, the only terms in~\eqref{EqipGr} that do not decay like $o(t_*^{-2})$ are those in the first line, further those in the second line for $l+j=1$, the $\rms l$ terms in the third line for $l=2,3$, and the $\rmv l$ term in the third line for $l=2$; we have thus discussed all of these above. We introduce correction terms for their elimination analogously to Definition~\ref{DefD5Corr}:

\begin{definition}[Correction terms \#4]
\label{DefD6Corr}
  Let $\scal\in\scalspace_l$, $\vect\in\vectspace_l$, and $\lambda\in\C$. We use the notation of Definition~\usref{DefipGrKerr} and recall the cutoffs~\eqref{EqDCutoffs}. We then define
  \begin{alignat*}{2}
    h_{\rmv 1}^{(\lambda,k)}(\vect) &:= -[\delta_{g_{b_0}}^*,\chi_\cK]\omega_{b_0,\rmv 1}^{(-1),\leq 3}\bigl(t_*^{-\lambda}(\log t_*)^k\vect\bigr) && \in \cA^{\infty,\ (\lambda,k)+((0,0)\cup\cE_\ind),\ \infty}, \\
    \vartheta_{\rmv 1}^{(\lambda,k)}(\vect) &:= \delta_{g_{b_0}}\sfG_{g_{b_0}}\Bigl(\chi_\cK\delta_{g_{b_0}}^*\omega_{b_0,\rmv 1}^{(0),\leq 3}\bigl(t_*^{-\lambda}(\log t_*)^k\vect\bigr)\Bigr) &&\in \cA^{\infty,\ (\lambda+1,k)+((0,0)\cup\cE_\ind),\ \lambda+4-\eps}.
  \end{alignat*}
  Furthermore, upon setting
  \begin{equation}
  \label{EqD6CorrA}
    A^{(\lambda,k)}(t_*):= \begin{cases} \frac{\dd^k}{\dd\lambda^k}\bigl((-\lambda+1)^{-1}t_*^{-\lambda+1}\bigr), & \lambda\neq 1, \\ \log t_*, & (\lambda,k)=(1,0) \end{cases}
  \end{equation}
  (which thus satisfies $\frac{\dd}{\dd t_*}A^{(\lambda,k)}(t_*)=t_*^{-\lambda}(\log t_*)^k$), we define
  \begin{alignat*}{2}
    h_{\rms 0}^{(0),(\lambda,k)} &:= -[\delta_{g_{b_0}}^*,\chi_\cK]\omega_{b_0,\rms 0}^{(0),\leq 4}(A^{(\lambda,k)}(t_*)) &&\in \cA^{\infty,\ (\lambda,k)+((0,0)\cup\cE_\ind),\ \infty}, \\
    \vartheta_{\rms 0}^{(0),(\lambda,k)} &:= \delta_{g_{b_0}}\sfG_{g_{b_0}}\Bigl(\chi_\cK\delta_{g_{b_0}}^*\omega_{b_0,\rms 0}^{(0),\leq 4}(A^{(\lambda,k)}(t_*))\Bigr) &&\in \cA^{\infty,\ (\lambda+1,k)+((0,0)\cup\cE_\ind),\ \lambda+4-\eps}
  \end{alignat*}
  for $\lambda\neq 1$, with the same definition also for $(\lambda,k)=(1,0)$, in which case the $\iota^+$-index set of $h_{\rms 0}^{(0),(1,0)}$ is $(1,1)+((0,0)\cup\cE_\ind)$. For $l=0,1$, we also define
  \begin{align*}
    h_{\rms l}^{(-\lambda^\Ups_{\rms l,1}+1),(\lambda,k)}(\scal) &:= -[\delta_{g_{b_0}}^*,\chi_\cK]\omega_{b_0,\rms l}^{(-\lambda^\Ups_{\rms l,1}),\leq 2}\bigl(t_*^{-\lambda}(\log t_*)^k\scal\bigr) \\
      &\in \cA^{\infty,\ (\lambda-\lambda^\Ups_{\rms l,1}+1,k) + ((0,0)\cup\cE_\ind),\ \infty}, \\
    \vartheta_{\rms l}^{(-\lambda^\Ups_{\rms l,1}+1),(\lambda,k)}(\scal) &:= \delta_{g_{b_0}}\sfG_{g_{b_0}}\Bigl(\chi_\cK\delta_{g_{b_0}}^*\omega_{b_0,\rms l}^{(-\lambda^\Ups_{\rms l,1}),\leq 2}\bigl(t_*^{-\lambda}(\log t_*)^k\scal\bigr)\Bigr) \\
      &\in \cA^{\infty,\ (\lambda-\lambda^\Ups_{\rms l,1}+2,k)+((0,0)\cup\cE_\ind),\ \lambda+3-\eps}.
  \end{align*}
  Lastly, for $l=2,3$ (in the scalar type sector) and $l=2$ (in the vector type sector), we set
  \begin{alignat*}{2}
    h_{\rms l}^{(-l+2),(\lambda,k)}(\scal) &:= -[\delta_{g_{b_0}}^*,\chi_\cK]\omega_{b_0,\rms l}^{(-l+1),\leq 5-l}\bigl(t_*^{-\lambda}(\log t_*)^k\scal\bigr) && \in \cA^{\infty,\,(\lambda-l+2,k)+((0,0)\cup\cE_\ind),\,\infty}, \\
    \vartheta_{\rms l}^{(-l+2),(\lambda,k)}(\scal) &:= \delta_{g_{b_0}}\sfG_{g_{b_0}}\Bigl(\chi_\cK\delta_{g_{b_0}}^*\omega_{b_0,\rms l}^{(-l+1),\leq 5-l}\bigl(t_*^{-\lambda}(\log t_*)^k\bigr)\Bigr) &&\in \cA^{\infty,\,(\lambda-l+3,k)+((0,0)\cup\cE_\ind),\,\lambda+6-l-\eps}; \\
    h_{\rmv l}^{(-l+1),(\lambda,k)}(\vect) &:= -[\delta_{g_{b_0}}^*,\chi_\cK]\omega_{b_0,\rmv l}^{(-l),\leq 4-l}\bigl(t_*^{-\lambda}(\log t_*)^k\vect\bigr) && \in \cA^{\infty,\,(\lambda-l+1,k)+((0,0)\cup\cE_\ind),\,\infty}, \\
    \vartheta_{\rmv l}^{(-l+1),(\lambda,k)}(\vect) &:= \delta_{g_{b_0}}\sfG_{g_{b_0}}\Bigl(\chi_\cK\delta_{g_{b_0}}^*\omega_{b_0,\rmv l}^{(-l),\leq 4-l}\bigl(t_*^{-\lambda}(\log t_*)^k\bigr)\Bigr) &&\in \cA^{\infty,\,(\lambda-l+2,k)+((0,0)\cup\cE_\ind),\,\lambda+5-l-\eps}.
  \end{alignat*}
\end{definition}

Note that we index these tensors according to the $t_*$-decay rate of the pure gauge term that they will serve to eliminate.\footnote{We could instead have chosen to index them according to the $\iota^+$-order of the source term that creates these pure gauge terms, which would be more in line with~\eqref{EqipGr} and also with the order of elimination in the proof of Theorem~\ref{ThmD6} below. But this would obscure the $\cK^+$-decay order, the improvement of which is our central aim.} Analogously to Lemma~\ref{LemmaD2Corr} (and also Lemma~\ref{LemmaD4Corr}), the $\rho_+\log\rho_+$-leading-order term of $h_{\rms 0}^{(0),(1,0)}$, which is given by $-[\ubar\delta^*,\chi_\cK]\ubar\omega_{\rms 0}^{(0)}$ (with $\ubar\omega_{\rms 0}^{(0)}=\pa_{t_*}^\flat$ a Killing 1-form on Minkowski space), is annihilated by $D_{g^0}\Ric$ for any metric $g^0$ that equals $\ubar g$ modulo $\rho_+\CI$, and thus
\begin{equation}
\label{EqD6Corrhs0}
  D_{g^0}\Ric(h_{\rms 0}^{(0),(1,0)}) \in \cA^{\infty,\ (3,0)\cup(3+\cE_\ind),\ \infty}.
\end{equation}
We use the tensors in Definitions~\ref{DefD2Corr}, \ref{DefD4Corr}, \ref{DefD5Corr}, and~\ref{DefD6Corr} to generalize Definition~\ref{DefD5EinsteinAug} as follows:

\begin{definition}[Gauge-fixed Einstein operator, augmentation \#4]
\label{DefD6EinsteinAug}
  Consider a symmetric 2-tensor $h\in\cX^\infty=\Hb^{\infty,\ \bigl(\la\cE_\sscri^\cC\ra,3+\eps_\sscri\bigr),\ (\cE_+,3+\eps_+),\ 2+\eps_\cK}(\Omega_*)$ with sufficiently small $\cX^d$-norm. Let $\vecp^{\leq 0}=(\scal,b-b_0,\scal_1^{(0)},\scal_1^{(0,1)})$ be small, and let
  \begin{equation}
  \label{EqD6EinsteinAugParam}
  \begin{split}
    \vecp^{\in(0,2]} &= \Bigl(\scal_1^{(\lambda-1,k)}, \quad \vect_1^{(1,0)},\ \vect_1^{(\lambda,k)}, \quad \scal_0^{(0),(1,0)},\ \scal_0^{(0),(\lambda,k)},\ \scal_2^{(0),(1,0)},\ \scal_2^{(0),(\lambda,k)}, \\
      &\qquad \bigl(\scal_l^{(-\lambda^\Ups_{\rms l,1}+1),(\lambda^\Ups_{\rms l,1},0)}\bigr)_{l=0,1},\ \bigl(\scal_l^{(-\lambda^\Ups_{\rms l,1}+1),(\lambda+\lambda^\Ups_{\rms l,1}-1,k)}\bigr)_{l=0,1}, \quad \scal_3^{(-1),(2,0)},\quad \vect_2^{(-1),(2,0)} \Bigr)
  \end{split}
  \end{equation}
  be small; here $(\lambda,k)\in\tilde\cE'_{++}$ for a suitable index set with $\min\Re\tilde\cE'_{++}\geq 1+\eps_\ind>1$ and $\tilde\cE'_{++}+(\cE_\ind\cup\cE_+\cup\tilde\cE'_{++})\subset\tilde\cE'_{++}$, and we include only those terms that have exponents $(\lambda+\mu,k)$ (where $\mu=0,\lambda^\Ups_{\rms 0,1}-1$, etc., and $\mu=0$ for $\vect_1^{(\lambda,k)}$) with $\lambda+\mu\leq 2$.\footnote{For the parameter $\scal_0^{(-\mu),(\lambda+\mu,k)}$, $\mu=\lambda^\Ups_{\rms 0,1}-1$, say, the sum $\lambda+\mu$ is the $t_*^{-1}$-power at which the corresponding pure gauge term arises in the $\cK^+$-expansion, while $\lambda$ is the $\iota_+$-decay rate of that term.} Write $\vecp=(\vecp^{\leq 0},\vecp^{\in(0,2]})$ and $P(h,\vecp):=\Ric(g^0+h_\tot(\vecp)+h)-\delta_{g^0,E^\cC}^*\bigl(\Ups_{E^\Ups}(g^0+h,g^0)-\vartheta_\tot(\vecp)\bigr)$ as in~\eqref{EqD4EinsteinAug}, where $g^0=g_{b_0,b,-\scal}$ as before, while we re-define for $\bullet=h,\vartheta$:
  \begin{equation}
  \label{EqD6EinsteinAugTot}
  \begin{split}
    \bullet_\tot(\vecp) &:= \bullet^{(-1)}(\scal) + \bullet^{(0)}(b-b_0) + \bullet_{\rms 1}^{(0)}(\scal_1^{(0)}) + \bullet_{\rms 1}^{(0,1)}(\scal_1^{(0,1)}) + \sum_{ \substack{ (\lambda,k)\in\tilde\cE'_{++} \\ \Re\lambda\leq 3 }} \bullet_{\rms 1}^{(\lambda-1,k)}(\scal_1^{(\lambda-1,k)}) \\
      &\qquad + \sum_{ \substack { \mu=\lambda^\Ups_{\rms l,1}-1: \\ l=0,1} }\sum_{ \substack{ (\lambda,k)\in\tilde\cE'_{++} \\ \Re(\lambda+\mu)\leq 2} } \bullet_{\rms l}^{(-\mu),(\lambda+\mu,k)}(\scal_l^{(-\mu),(\lambda+\mu)}) + \sum_{ \substack{ (\lambda,k)\in\tilde\cE'_{++} \\ \Re\lambda\leq 2 } } \bullet_{\rms 2}^{(0),(\lambda,k)}(\scal_2^{(0),(\lambda,k)}) \\
      &\qquad + \bullet_{\rms 3}^{(-1),(2,0)}(\scal_3^{(-1),(2,0)}) + \bullet_{\rmv 2}^{(-1),(2,0)}(\vect_2^{(-1),(2,0)}).
  \end{split}
  \end{equation}
  We write $D_{(h,\vecp)}P(\scal_0^{(0),(\lambda,k)\prime}):=D_{(h,\vecp)}P(0,(0,\ldots,\scal_0^{(0),(\lambda,k)\prime},\ldots,0))$ for the linearization of $P$ in the argument $\scal_0^{(0),(\lambda,k)}$ of $\vecp$; similarly for linearizations in (collections of) other arguments.
\end{definition}

The first parameter in each group (so $\vect_1^{(1,0)}$, $\scal_0^{(0),(1,0)}$, $\scal_2^{(0),(1,0)}$, $\scal_0^{(-\lambda^\Ups_{\rms 0,1}+1),(\lambda^\Ups_{\rms 0,1},0)}$, etc.) will be used to eliminate the least-decaying term in the late-time expansion of $u$ (so $t_*^{-1}h_{b,\rmv 1}$, $t_*^{-1}h_{b,\rms 0}^{(0)}$, $t_*^{-1}h_{b,\rms 2}^{(0)}$, $t_*^{-\lambda^\Ups_{\rms 0,1}}h_{b,\rms 0}^{(-\lambda^\Ups_{\rms 0,1}+1)}$, etc.) of the corresponding type. The generalization of Lemma~\ref{LemmaD5LinPar} reads:

\begin{lemma}[Linearization of $P$ in the parameters, \#4]
\label{LemmaD6LinPar}
  The linearizations of $P$ in the parameters $\vecp^{\leq 0}$ and $\scal_1^{(\lambda-1,k)}$ are as in Lemma~\usref{LemmaD5LinPar}. Moreover,
  \[
    D_{(h,\vecp)}P(\scal_0^{(0),(1,0)\prime}) \in \chi_\cK^\sharp\Hb^{\infty,\ 0,\ \bigl((3,0)+((0,0)\cup\cE_\ind\cup\cE_+\cup\tilde\cE'_{++}),\,6+\eps_+-\eps\bigr),\ 5-\eps},
  \]
  and for the linearization of $P$ in any other argument $\sfW_l^{(-\mu),(\lambda+\mu,k)}$ where $\sfW=\scal$ or $\vect$ (and also for $\vect_1^{(\lambda,k)}$, in which case $\mu=0$ in the memberships below), we have\footnote{The $\cK^+$-order is, more precisely, given by $\lambda+\mu+1+j-\eps$ where $j\in\N_0$ is the number of terms included in the gauge potentials in Definition~\ref{DefD6Corr} below the $t_*^{-\lambda}$-term, so, e.g., $j=3$ for $\scal_0^{(0),(\lambda,k)}$ and $j=2$ for $\scal_{\rms l}^{(-\lambda^\Ups_{\rms l,1}+1),(\lambda,k)}$.}
  \[
    D_{(h,\vecp)}P(\sfW_l^{(-\mu),(\lambda+\mu,k)\prime}) \in \chi_\cK^\sharp\Hb^{\infty,\ 0,\ \bigl((\lambda+2,k)+((0,0)\cup\cE_\ind\cup\cE_+\cup\tilde\cE'_{++}),\,5+\eps_++\lambda-\eps\bigr),\ 4+\eps_\ind-\eps}.
  \]
  Furthermore,
  \begin{equation}
  \label{EqD6LinParDiff}
  \begin{split}
    &D_{(h,\vecp)}P(\sfW_l^{(-\mu),(\lambda+\mu,k)\prime}) - \chi_\iota\ubar L\Bigl(\chi_\cK\ubar\delta^*\ubar\omega_{\rmw l}^{(-\mu-1),\leq j}\bigl(\sfW_l^{(-\mu),(\lambda+\mu,k)\prime}t_*^{-(\lambda+\mu)}(\log t_*)^k\bigr) \Bigr) \\
    &\qquad \in \chi_\cK^\sharp\Hb^{\infty,\ 0, \bigl((\lambda+2,k)+(\cE_\ind\cup\cE_+\cup\tilde\cE'_{++}),\,5+\eps_++\lambda-\eps\bigr),\ 4+\eps_\ind-\eps},
  \end{split}
  \end{equation}
  and this also holds if we use $L_b$, $\delta_{g_b}^*$, and $\omega_{b,\rmw l}^{(-\mu-1),\leq j}$ instead of $\ubar L$, $\ubar\delta^*$, and $\ubar\omega_{\rmw l}^{(-\mu-1),\leq j}$, with the exceptional case $D_{(h,\vecp)}P(\scal_0^{(0),(\lambda,k)\prime})-\chi_\iota\ubar L\bigl(\chi_\cK\ubar\delta^*\ubar\omega_{\rms 0}^{(0),\leq 4}(\scal_0^{(0),(\lambda,k)\prime}A^{(\lambda,k)}(t_*))\bigr)$ (which lies in the same space as~\eqref{EqD6LinParDiff}).
\end{lemma}
\begin{proof}
  The structure of the linearization of $P$ in the arguments described already in Lemma~\ref{LemmaD5LinPar} is unchanged since $h_\tot$ is of the same class (as far as index sets and decay of conormal remainder terms are concerned) as before. The arguments for the linearizations in the new parameters are identical to those in the proofs of Lemmas~\ref{LemmaD2LinPar}, \ref{LemmaD4LinPar}, and \ref{LemmaD5LinPar}.
\end{proof}

\subsubsection{Undoing pure gauge terms with less-than-\texorpdfstring{$t_*^{-2-\eps_\cK}$-}{t\textasciicircum(-2-eps)-}decay}
\label{SssD6Undo}

We now have all modification parameters in hand to prove optimal decay of forward solutions of $L u=f$. We recall the index sets from~\eqref{ItD5Escri}--\eqref{ItD5Eplus} on page~\pageref{ItD5Escri}; and we use unweighted b-densities. We moreover introduce the following notation:

\begin{definition}[Orders]
\label{DefD6Order}
  Fix a function $\chi_+\in\CI(\R)$ that vanishes on $(-\infty,1]$ and equals $1$ on $[2,\infty)$. Fix $\eps_\cK>0$. Given an index set $\cE$, a number $\alpha\in\R$, and a finite-dimensional vector space $V$, we define $\cE^{>2}:=\{(z,k)\in\cE\colon\Re z\geq 2+\eps_\cK\}$ and
  \[
    \dot H_\bop^{\infty,\,\scissors(\cE,\alpha)}([1,\infty]_{t_*};V) :=
    \begin{cases}
      \dot H_\bop^{\infty,\,(\cE^{>2}, \alpha)}([1,\infty];V), & \alpha<4+\eps_\cK, \\
      \Biggl\{ \sum_{ \substack{(\lambda,k)\in\cE^{>2} \\ \Re\lambda\leq 4+\eps_\cK} } \chi_+(t_*)c_{(\lambda,k)}t_*^{-\lambda}(\log t_*)^k \colon c_{(\lambda,k)}\in V \Biggr\}, & \alpha\geq 4+\eps_\cK.
    \end{cases}
  \]
  (These are spaces of $V$-valued functions on $\R_{t_*}$ with support in $[1,\infty)$.) When $\alpha\geq 4+\eps_\cK$, we also denote this space by $\dot H_\bop^{\infty,\scissors(\cE)}$ simply.
\end{definition}

The space $\dot H_\bop^{\infty,\,\scissors(\cE,\alpha)}$ thus records polyhomogeneous expansions up to conormal remainders with decay rate $\alpha$, but only allows for terms with at least $2+\eps_\cK$ orders of decay. Furthermore, when $\alpha\geq 4+\eps_\cK$, no conormal remainders are allowed, and in this case the space $\dot H_\bop^{\infty,\scissors(\cE,\alpha)}$ is finite-dimensional and consists only of finite polyhomogeneous expansions.

\begin{thm}[Sharp decay and partial polyhomogeneity of forward solutions]
\label{ThmD6}
  Let $h\in\cX^\infty:=\Hb^{\infty,\ \bigl(\la\cE_\sscri^\cC\ra,3+\eps_\sscri\bigr),\ (\cE_+,3+\eps_+),\ 4+\eps_\cK}(\Omega_*)^{\bullet,-}$ be small in $\cX^d$ for some large but fixed $d$, and let $\vecp=(\vecp^{\leq 0},\vecp^{\in(0,2]})$ (in the notation of Definition~\usref{DefD6EinsteinAug}) be small. Recall $0<\eps_\cK<\eps_+<\min(\eps_\ind,\eps_\sscri)$ from~\eqref{EqDMetBasicEllEps}. Let $f$ be of the form\footnote{Unlike in Proposition~\ref{PropD5Alm}, we now need to require more structure of $f_1$ at $\iota^+$. Recall from Remark~\ref{RmkD2Splitting} that the assumptions in~\eqref{EqD6f} are made specifically for compatibility with the modification terms arising from undoing boosts, cf.\ \eqref{EqD2LinParScal}.}
  \begin{equation}
  \label{EqD6f}
  \begin{split}
    &f = f_1 + f_2, \\
    &\quad f_1 \in \Hb^{\infty,\ \bigl(\la\cE_\sscri^\cC+1\ra',\,4+\eps_\sscri\bigr),\ (\cE_++3,\,5+\eps_+),\ 4+\eps_\cK}(\Omega_*)^{\bullet,-}, \\
    &\quad f_2 \in \Hb^{\infty,\ 4+\eps_\sscri,\ (\tilde\cE_++2,\,5+\eps_+),\ 4+\eps_\cK}(\Omega_*;S^2\cT^*)^{\bullet,-}.
  \end{split}
  \end{equation}
  Then there exists a unique $\vecp^\prime$ such that the forward solution of
  \[
    L u := D_{(h,\vecp)}P(u,0) = f - D_{(h,\vecp)}P(0,\vecp^\prime)
  \]
  takes the form
  \begin{subequations}
  \begin{equation}
  \label{EqD6u}
    u = \chi_\cK u_{\rm exp} + \tilde u,
  \end{equation}
  where (using the notation from Propositions~\usref{PropWG0Symm}, \usref{PropWG0Large}, \usref{PropWEMode0Kerr}, and~\usref{PropWE0} as well as Definition~\usref{DefipGrKerr}, and noting that $b=b_0+(b-b_0)$ is determined by our fixed choice of $b_0$ and the parameters $\vecp^{\leq 0}$)
  \begin{equation}
  \label{EqD6uexp}
  \begin{split}
    u_{\rm exp} &= h_{b,\rms 1}^{\leq 1}\bigl(\dot\scal_\rem(t_*)\bigr) + \dot g_b^\Ups\bigl(\dot b_\rem(t_*)\bigr) + h_{b,\rmv 1}^{\leq 3}\bigl(\dot\vect_1(t_*)\bigr) \\
      &\qquad + h_{b,\rms 0}^{(0),\leq 3}\bigl(\dot\scal_0^{(0)}(t_*)\bigr) + \sum_{ \substack{\mu=-\lambda^\Ups_{\rms l,l+j}+1 \\ 0\leq l\leq 3,\ j=0,1, \\ 1\leq l+j\leq 3} } h_{b,\rms l}^{(\mu),\leq 2}\bigl(\dot\scal_l^{(\mu)}(t_*)\bigr) \\
      &\qquad + \sum_{l=2}^5 h_{b,\rms l}^{(-l+2),\leq 5-l}\bigl(\dot\scal_l^{(-l+2)}(t_*)\bigr) + \sum_{l=2}^4 h_{b,\rmv l}^{(-l+1),\leq 4-l}\bigl(\dot\vect_l^{(-l+1)}(t_*)\bigr) \\
      &\qquad + h_{b,\rms 2}^{(-2),\leq 1}\bigl(\dot\scal_2^{(-2)}(t_*)\bigr) + h_{b,\rmv 2}^{(-2),\leq 1}\bigl(\dot\vect_2^{(-2)}(t_*)\bigr) + h_{b,\rms 3}^{(-3)}\bigl(\dot\scal_3^{(-3)}(t_*)\bigr) + h_{b,\rmv 3}^{(-3)}\bigl(\dot\vect_3^{(-3)}(t_*)\bigr),
  \end{split}
  \end{equation}
  with
  \begin{equation}
  \label{EqD6uexpComp}
  \begin{split}
    \dot\scal_\rem &\in \dot H_\bop^{\infty,\,2+\eps_\cK}([1,\infty]_{t_*};\scalspace_1) = t_*^{-2-\eps_\cK}\dot H_\bop^\infty([1,\infty],|\tfrac{\dd t_*}{t_*}|;\scalspace_1), \\
    \dot b_\rem &\in \dot H_\bop^{\infty,\,3+\eps_\cK}([1,\infty]_{t_*};\R^4), \\
    \dot\vect_1 &\in \dot H_\bop^{\infty,\ \scissors(\tilde\cE_\sharp,3+\eps_+)}([1,\infty];\vectspace_1\bigr), \\
    \dot\scal_0^{(0)} &\in \dot H_\bop^{\infty,\ \scissors(\tilde\cE_\sharp,3+\eps_+)}([1,\infty];\scalspace_0), \\
    \dot\scal_l^{(-\lambda^\Ups_{\rms l,l+j}+1)} &\in \dot H_\bop^{\infty,\ \scissors\bigl(\tilde\cE_\sharp+\lambda^\Ups_{\rms l,l+j}-1,\,3+\eps_++\lambda^\Ups_{\rms l,l+j}-1\bigr)}([1,\infty];\scalspace_l), \\
    \dot\scal_l^{(-l+2)} &\in \dot H_\bop^{\infty,\ \scissors(\tilde\cE_\sharp+l-2,\,3+\eps_++l-2)}([1,\infty];\scalspace_l), \\
    \dot\vect_l^{(-l+1)} &\in \dot H_\bop^{\infty,\ \scissors(\tilde\cE_\sharp+l-1,\,3+\eps_++l-1)}([1,\infty];\vectspace_l), \\
    \dot\scal_2^{(-2)} &\in \dot H_\bop^{\infty,\ \scissors(\tilde\cE_\sharp+2)}([1,\infty];\scalspace_2), \qquad
    \dot\vect_2^{(-2)} \in \dot H_\bop^{\infty,\ \scissors(\tilde\cE_\sharp+2)}([1,\infty];\vectspace_2), \\
    \dot\scal_3^{(-3)} &\in \dot H_\bop^{\infty,\ \scissors(\tilde\cE_\sharp+3)}([1,\infty];\scalspace_3), \qquad
    \dot\vect_3^{(-3)} \in \dot H_\bop^{\infty,\ \scissors(\tilde\cE_\sharp+3)}([1,\infty];\vectspace_3),
  \end{split}
  \end{equation}
  and
  \begin{equation}
  \label{EqD6tildeu}
    \tilde u \in \Hb^{\infty,\ \bigl(\la\cE_\sscri^\cC\ra,\,3+\eps_\sscri\bigr),\ (\tilde\cE_\sharp,\,3+\eps_+),\ 4+\eps_\cK}(\Omega_*)^{\bullet,-},
  \end{equation}
  \end{subequations}
  where $\tilde\cE_\sharp$ is an index set containing $(1,0)$ with $\min\Re(\tilde\cE_\sharp\setminus\{(1,0)\})>1$. Both the index set $\tilde\cE_\sharp$ and the index set $\tilde\cE'_{++}$ in Definition~\usref{DefD6EinsteinAug} are extended unions of index sets that depend only on $\ubar L$ and $\cE_\sscri^\cC$ with $\cE_++(z,l)$ for some $(z,l)\in\C\times\N_0$ with $\Re z\geq\eps_\ind$ that also only depend on $\ubar L$ and $\cE_\sscri^\cC$.
\end{thm}

\begin{rmk}[Discussion of $u_{\rm exp}$]
\fakephantomsection
\label{RmkD6uexp}
  \begin{enumerate}
  \item{\rm (Decay.)} Each term of $u_{\rm exp}$ has $\cK^+$-order $\geq 2+\eps_\cK$ (with equality for the first summand), so~\eqref{EqD6u} indeed states $\cO(t_*^{-2-\eps_\cK})$-decay of $u$ at $(\cK^+)^\circ$ (so in spatially compact regions). The terms comprising $u_{\rm exp}$ have spatial dependence given by pure gauge (all but $\dot g_b^\Ups(\dot b_\rem(t_*))$ and the final four terms) and physical zero energy modes of $L_b$. They are mapped by $L_b$ (and thus by $L$) into source terms with $\cK^+$-order $4+\eps_\cK$; this is discussed in more detail in~\S\ref{SssD6Abs} below.
  \item\label{ItD6uexpNum}{\rm (Number of summands for each term.)} It would suffice to use $h_{b,\rms 0}^{(0),\leq 1}$ in the expression for $u_{\rm exp}$, as the additional terms $\breve h_{b,\rms 0}^{(0),2}(\pa_{t_*}^2\dot\scal_0^{(0)})$ and $\breve h_{b,\rms 0}^{(0),3}(\pa_{t_*}^3\dot\scal_0^{(0)})$ have $\cK^+$-order $>4+\eps_\cK$ and thus could equivalently be put into $\tilde u$; similarly for some of the other terms. We choose not to make any such purely aesthetic optimizations here, however, and instead consistently use the same number of summands as in the construction of the modification terms; see, e.g.,~\eqref{EqD6s0Corr} and Definition~\ref{DefD6Corr} for the $\rms 0$ term, and see also~\eqref{EqipGr}.
  \item\label{ItD6uexpImpr}{\rm (Partial polyhomogeneity; possible improvements.)} The polyhomogeneous parts of the pure gauge terms (i.e., the terms in the expansions of $\dot\vect_1$, $\dot\scal_0^{(0)}$, $\dot\scal_l^{(-\lambda^\Ups_{\rms l,l+j}+1)}$, $\dot\scal_l^{(-l+2)}$, and $\dot\vect_l^{(-l+1)}$) could, in principle, be eliminated by means of further modification parameters. We do not pursue such a modest sharpening here, however, as our linear algebra based methods for eliminating pure gauge terms are not strong enough for the elimination of the conormal remainder terms (e.g., the ``structureless'' $\cO(t_*^{-2-\eps_\cK})$-center-of-mass motion) anyway.
  \item{\rm ($t_*^{-3}$-decay as a sharp boundary.)} Since the last four terms in~\eqref{EqD6uexp} are \emph{not} pure gauge, they cannot be eliminated; note that the $\rms 2$ and $\rmv 2$ terms have sharp $t_*^{-3}$-decay. The second term $\dot g_b^\Ups(\dot b_\rem(t_*))$ is not pure gauge either, but the limitation on its decay rate is due to the limited $\cK^+$-order of the source term assumed in~\eqref{EqD6f}. We stop at $2+\eps_\cK$ here since more than $2$ orders of decay of $u$ at $\cK^+$ suffice for the nonlinear iteration, but briefly discuss in~\S\ref{SsSt3} below how to push the decay rate of $u$ further. 
  \end{enumerate}
\end{rmk}

\begin{rmk}[Assumptions on $h$]
\label{RmkD6Assmh}
  Recalling the discussion around~\eqref{EqD5Almh}, we make the strong decay assumption on $h$ here solely to ensure that the linearization of $P$ in the black hole parameters $b$ has $4+\eps_\cK$ orders of $\cK^+$-decay, cf.\ Lemma~\ref{LemmaD5LinPar}. Apart from that, we only need $h$ to have $\cK^+$-weight $2+\eps_\cK$, and in particular only use that $L-L_b$ has $\cK^+$-decay order $2+\eps_\cK$. (We will explain in~\S\ref{SssD6Abs} below how incorporating the non-$\cO(t_*^{-4-\eps_\cK})$-terms in the expansion~\eqref{EqD6u} into the gauge-fixed Einstein operator gives rise to these two orders.) Similarly, recall from the discussion around~\eqref{EqD2hSupp} that the vanishing requirement on $h$ for $\ft_*<1$ is for notational convenience only.
\end{rmk}

\begin{rmk}[Index sets]
\label{RmkD6Index}
  A consequence of the final statement regarding the index sets (which extends the observation in Remark~\ref{RmkD5EplusDep}) is that one can self-consistently require the index set $\cE_+$ of the original metric perturbation $h$ to be large enough to contain $\tilde\cE_\sharp$. This can be regarded as (weak) evidence for the possibility of closing a nonlinear iteration scheme; however, it is crucial to develop a more concrete and constructive description of $\iota^+$-asymptotics in~\S\ref{SEf} (see also the discussion after Corollary~\ref{CorD6Impr} below).
\end{rmk}

We make a preliminary observation. Application of Proposition~\ref{PropipGr} with $\lambda=1$ will produce a term $t_*^{-1}h_{b,\rmv 1}(\dot\vect)$ (amongst other terms) that we will want to eliminate by means of $h_{\rmv 1}^{(1,0)}(\vect^{(1,0)\prime})$ and $\vartheta_{\rmv 1}^{(1,0)}(\vect^{(1,0)\prime})$ for a suitable $\vect^{(1,0)\prime}\approx\dot\vect$. Notice then that $D_{(h,\vecp)}P(\vect^{(1,0)\prime})$ is of size $\cO(\rho_+^3)$ at $\iota^+$, and hence of the same size as the modification terms that we used to eliminate the terms $t_*^0 h_{b,\rms 1}$, $t_*^0(\log t_*)h_{b,\rms 1}^{\leq 4}$, and $t_*^0\dot g_b^\Ups$ in~\S\ref{SssD4ParNo} (see Lemma~\ref{LemmaD4Corr}). Since in the proof of Proposition~\ref{PropD5Alm}, the polynomially decaying center-of-mass motions arose from increasingly decaying terms in the $\iota^+$-expansion of the effective right-hand side (by which we mean $f'$ in~\eqref{EqD5uscri}), we should now similarly introduce modification terms in the order determined by the order of decay of the $\iota^+$-expansion term of the effective right-hand side that generates the expansion term one wants to eliminate. In the concrete case of $t_*^{-1}h_{b,\rmv 1}(\dot\vect)$, this means that we will insert the modification term $-D_{(h,\vecp)}P(\vect^{(1,0)\prime})$ at the same time as the terms $-D_{(h,\vecp)}P(\scal_1^{(0,1)\prime})$ etc.\ from Corollary~\ref{CorD4ElimEqual}.

We also note that when eliminating terms that arise from an element $(\lambda+2,k)$ with $\Re\lambda>1$, say, in the $\iota^+$-index set of an effective right-hand side, we need to insert the modification term in the form $-\bigl(D_{(h,\vecp)}P(\vect_1^{(\lambda,k)\prime})-L(\chi_\cK\delta_{g_b}^*\omega_{b,\rmv 1}^{(-1),\leq 3}(t_*^{-\lambda}(\log t_*)^k\vect_1^{(\lambda,k)\prime}))\bigr)$ analogously to~\eqref{EqD5AlmGoodEqPrelim} in order to avoid the apparent runaway production of powers of $\log t_*$.

\begin{proof}[Proof of Theorem~\usref{ThmD6}]
  The proof has several parts:
  \begin{enumerate}
  \item In \textit{Part~1}, we sharpen and generalize Proposition~\ref{PropD5Alm} to allow for further modification terms that will be used in later steps, and to produce a partial expansion at $\iota^+$.
  \item In \textit{Part~2}, we explain how to push the decay of $u$ beyond $t_*^{-1}$ using some of the new modification parameters. Getting all the way to $\cO(t_*^{-2+\eps})$-decay is then a routine repetition of these arguments, similarly for getting beyond $t_*^{-2}$; this is briefly explained in \textit{Part~3}.
  \item In \textit{Part~4}, we give a more careful accounting of all terms that do not have $\cK^+$-order $4+\eps_\cK$, which will finish the proof.
  \end{enumerate}

  \pfstep{Notation.} We use two groupings of the components of the sought-after modification vector $\vecp'$ in the notation of Definition~\ref{DefD5EinsteinAug}. We begin with the grouping according to the $\iota^+$-decay order (recorded now in the \emph{sub}script) of the term which (as we are in the process of showing) they serve to eliminate, and thus set
  \begin{equation}
  \label{EqD6peq1}
  \begin{split}
    \vecp'_{=1} &:= \Bigl(\scal',b',\scal_1^{(0)\prime},\scal_1^{(0,1)\prime},\quad \vect_1^{(1,0)\prime},\ \scal_0^{(0),(1,0)\prime},\ \scal_2^{(0),(1,0)\prime}, \\
      &\quad\qquad \scal_0^{(-\lambda^\Ups_{\rms 0,1}+1),(\lambda^\Ups_{\rms 0,1},0)\prime},\ \scal_1^{(-\lambda^\Ups_{\rms 1,1}+1),(\lambda^\Ups_{\rms 1,1},0)\prime},\ \scal_3^{(-1),(2,0)\prime},\ \vect_2^{(-1),(2,0)\prime}\,\Bigr).
  \end{split}
  \end{equation}
  Recall here that the parameter $\scal_l^{(-\mu),(\lambda,k)}$ corresponds to a term in the $\cK^+$-expansion of $u$ that has $\iota^+$-index set starting with $(-\mu+\lambda,k)$, so the parameters $\vecp_{=1}$ indeed all arise at order $\rho_+$ (or $\rho_+\log\rho_+$) in the $\iota^+$-expansion of $u$ (equivalently, at order $\rho_+^{1+2}$ in the expansion of the effective source term). For $\alpha>1$, we set
  \begin{align*}
    \vecp'_{=\alpha} &:= \Bigl( \scal_1^{(\lambda-1,k)\prime},\ \vect_1^{(\lambda,k)\prime},\ \scal_0^{(0),(\lambda,k)\prime},\ \scal_2^{(0),(\lambda,k)\prime}, \\
      &\quad\qquad \scal_0^{(-\lambda^\Ups_{\rms 0,1}+1),(\lambda+\lambda^\Ups_{\rms 0,1}-1,k)\prime},\ \scal_1^{(-\lambda^\Ups_{\rms 1,1}+1),(\lambda+\lambda^\Ups_{\rms 1,1}-1,k)\prime} \colon (\lambda,k)\in\tilde\cE'_{++},\ \Re\lambda=\alpha \Bigr),
  \end{align*}
  and $\vecp'_{\in I}:=(\vecp'_\alpha\colon\alpha\in I\cap\Re\pi_1\tilde\cE'_{++})$ for $I\subset[1,3]$, as well as $\vecp'_{\leq\alpha}:=\vecp'_{\in[1,\alpha]}$. Recalling~\eqref{EqD6CorrA}, we moreover write
  \begin{align*}
    D_{(h,\vecp)}P(\vecp'_{=1}) &:= D_{(h,\vecp)}P(\vect_1^{(1,0)\prime}) + \cdots + D_{(h,\vecp)}P(\vect_2^{(-1),(2,0)\prime}), \\
    D_{(h,\vecp)}P(\vecp'_{=\alpha}) &:= \sum D_{(h,\vecp)}P(\scal_1^{(\lambda-1,k)\prime}) + \cdots + D_{(h,\vecp)}P(\scal_1^{(-\lambda^\Ups_{\rms 1,1}+1),(\lambda+\lambda_{\rms 1,1}-1,k)\prime}), \\
    h_b(\vecp'_{=\alpha}) &:= \sum_{ \substack{(\lambda,k)\in\tilde\cE'_{++} \\ \Re\lambda=\alpha} } \chi_\cK\delta_{g_b}^*\Biggl(\ \omega_{b,\rms 1}^{(0),\leq 4}\bigl(t_*^{-\lambda+1}(\log t_*)^k\scal_1^{(\lambda-1,k)\prime}\bigr) \\
      &\quad \hspace{7.8em} + \omega_{b,\rmv 1}^{(-1),\leq 3}\bigl(t_*^{-\lambda}(\log t_*)^k\vect_1^{(\lambda,k)\prime}\bigr) + \omega_{b,\rms 0}^{(0),\leq 4}\bigl(A^{(\lambda,k)}(t_*)\scal_0^{(0),(\lambda,k)\prime}\bigr) \\
      &\quad \hspace{7.8em} + \omega_{b,\rms 2}^{(-1),\leq 3}\bigl(t_*^{-\lambda}(\log t_*)^k\scal_2^{(0),(\lambda,k)\prime}\bigr) \\
      &\quad \hspace{7.8em} + \sum_{l=0}^1 \omega_{b,\rms l}^{(-\lambda^\Ups_{\rms l,1}),\leq 2}\bigl(t_*^{-\lambda-\lambda^\Ups_{\rms l,1}+1}(\log t_*)^k\scal_l^{(-\lambda^\Ups_{\rms l,1}+1),(\lambda+\lambda^\Ups_{\rms l,1}-1,k)\prime}\bigr)\ \Biggr).
  \end{align*}
  (Thus, $h_b(\vecp'_{=\alpha})$ generalizes~\eqref{EqD5hb0Insert}.) By~\eqref{EqD6LinParDiff}, the $\iota^+$-index set of $D_{(h,\vecp)}P(\vecp'_{=\alpha})-L(h_b(\vecp'_{=\alpha}))$ has $\min\Re\geq\alpha+2+\eps_\ind$. (The idea, as mentioned before, is that the source term $D_{(h,\vecp)}P(\vecp'_{=\alpha})$ should, approximately, create the metric perturbation $h_b(\vecp'_{=\alpha})$.) We define $h_b(\vecp'_{\in I})=\sum h_b(\vecp'_{=\alpha})$ where the sum is over all $\alpha\in I\cap\Re\pi_1\tilde\cE'_{++}$, similarly for $D_{(h,\vecp)}P(\vecp'_{\in I})$ etc.

  Now, the elimination of terms proceeds in increasing order of decay at $\cK^+$ (not at $\iota^+$), since we uncover the $\cK^+$-expansion step-by-step in $\eps_\cK$-increments, and need to eliminate these expansion terms before being able to extract the next ones. Thus, for $\alpha>0$, we will be able to prove $t_*^{-\alpha}$-decay of $u$ using only the modification parameters
  \begin{equation}
  \label{EqD6ParamDecayK}
    \vecp^{\prime,\leq\alpha}
  \end{equation}
  corresponding to expansion terms that do not decay like $o(t_*^{-\alpha})$; these are thus: $\vecp^{\prime,\leq 0}$; $\scal_1^{(\lambda-1,k)}$ for $\Re(\lambda-1)\leq\alpha$; further $\vect_1^{(1,0)}$, $\scal_0^{(0),(1,0)}$, and $\scal_2^{(0),(1,0)}$ when $1\leq\alpha$; $\vect_1^{(\lambda,k)}$ for $\Re\lambda\leq\alpha$; and the terms $\sfW_l^{(-\mu),(\lambda+\mu,k)}$ (where $\sfW=\scal,\vect$) in~\eqref{EqD6EinsteinAugParam} for $\Re(\lambda+\mu)\leq\alpha$. In particular, the remaining modification parameters,
  \[
    \vecp^{\prime,>\alpha},
  \]
  must remain free until the time that the terms they serve to eliminate are uncovered in the $\cK^+$-expansion. Carefully note, then, that if $\beta$ is the smallest real number such that the parameters $\vecp^{\prime,\leq\alpha}$ are a subset of the parameters $\vecp'_{\leq\beta}$ (e.g., for $\alpha=0$, one has $\beta=1$), the (typically non-empty) collection of parameters
  \[
    \vecp^{\prime,>\alpha}_{\leq\beta} := \vecp'_{\leq\beta}\setminus\vecp^{\prime,\leq\alpha}
  \]
  must remain free until the time that the terms they serve to eliminate are uncovered in the $\cK^+$-expansion.

  \pfstep{Part~1. Revisiting $\cO(t_*^{-1+\eps})$-decay; partial expansion at $\iota^+$.} We revisit the proof of Proposition~\ref{PropD5Alm}, with two new goals: first, we need to allow for the modification terms $\vecp'_{\in[1,2)}$ on the right-hand side (which includes more terms than used previously, but does not require any conceptually new arguments); secondly, we shall record the partial polyhomogeneous expansion of $u$ at $\iota^+$ up to a $\cO(\rho_+^{2-\eps})$ conormal remainder. The required modifications of the proof are rather minimal: the new modification terms have strong (i.e., $o(t_*^{-4-\eps_\cK})$) decay at $\cK^+$, the additional terms in $h_b(\vecp'_{=\alpha})$ for $\alpha\in(1,2)$ are of size $o(t_*^{-1})$, and since the proof of Proposition~\ref{PropD5Alm} did not use any $\iota^+$-expansion of $u$, we can extract this expansion at the very end.

  For this part, it suffices to assume that the $\iota^+$-order of $f_1$ is $4-\eps$ (as in~\eqref{EqD5Almf}).

  \pfsubstep{Step~1.1.}{Almost $t_*^{-1}$-decay.} We repeat the inductive procedure used in the proof of Proposition~\ref{PropD5Alm}, thus assuming we have obtained the $\cK^+$-order $\alpha'_\cK-2$ for $u$ where $\alpha'_\cK\in(2,3-\eps_\cK)$ and want to extract an expansion up to errors of order $\alpha_\cK-2$ where $\alpha_\cK:=\alpha'_\cK+\eps_\cK$. For the base case, when $\alpha'_\cK$ is close to $2$, consider arbitrary parameters $\vecp^{\prime,>0}$; these determine $\vecp^{\prime,\leq 0}=\vecp^{\prime,\leq 0}(f,\vecp^{\prime,>0})$ such that the ($\vecp^{\prime,>0}$-dependent) solution of
  \[
    L u = f - D_{(h,\vecp)}P(\vecp') = f - D_{(h,\vecp)}P(\vecp^{\prime,>0}) - D_{(h,\vecp)}P(\vecp^{\prime,\leq 0})
  \]
  does not feature any non-decaying terms (i.e., boosts, black hole parameter changes, and stationary or logarithmic center-of-mass motions); by Lemma~\ref{LemmaD6LinPar}, Corollary~\ref{CorD4ElimEqual} can indeed be applied to the source term $D_{(h,\vecp)}P(\vecp^{\prime,>0}_{=1})\in\chi_\cK^\sharp\Hb^{\infty,\ 0,\ \bigl((3,0)+((0,0)\cup\cE_\ind\cup\cE_+\cup\tilde\cE'_{++}),\,5+\eps_+\bigr),\ 4+\eps_\cK}$ and also to $D_{(h,\vecp)}P(\vecp^{\prime,>0}_{>1})$ which decays faster at $\iota^+$.

  The inductive hypothesis (cf.\ \eqref{EqD5AlmIndf2} but with $\alpha'_\cK$ in place of $\alpha_\cK$) is that for $f$ of the form~\eqref{EqD6f} but with $f_2$ having $\iota^+$-index set $(\tilde\cE_+\cup\cE)+2$ where $\cE$ is any index set with $\min\Re\cE>\alpha'_\cK-1$, and for any choice of (as-of-yet free) parameters $\vecp^{\prime,>\alpha'_\cK-2}$, we can find $\vecp^{\prime,\leq\alpha'_\cK-2}=\vecp^{\prime,\leq\alpha'_\cK-2}(f,\vecp^{\prime,>\alpha'_\cK-2})$ (which is a proper subset of the parameters $\vecp'_{\leq\alpha'_\cK-1}$, with some of the parameters in $\vecp'_{\leq\alpha'_\cK-1}$ being contained in $\vecp^{\prime,>\alpha'_\cK-2}$) such that the solution of $L u=f-D_{(h,\vecp)}P(\vecp')$ lies in the space $\Hb^{\infty,\ (\la\cE_\sscri^\cC\ra,3+\eps_\sscri),\ 1-\eps,\ \alpha'_\cK-2}$.

  For the inductive step (and thus now with $\min\Re\cE>\alpha_\cK-1$, which covers also the source terms $D_{(h,\vecp)}P(\vecp'_{>\alpha_\cK-1})$), we can extract the partial expansion~\eqref{EqD5uFinal} exactly as before. For its elimination, we will use arguments involving a generalization of~\eqref{EqD5AlmGoodEqPrelim}. Consider arbitrary parameters $\vecp^{\prime,>\alpha_\cK-2}$; these are presently free (i.e., not needed to push the decay of $u$ to $o(t_*^{-\alpha_\cK+2})$). These free parameters can be split into three groups.
  \begin{enumerate}
  \item $\vecp^{\prime,>\alpha_\cK-2}_{>\alpha_\cK-1}$: as mentioned before, the present assumptions on $f$ allow us to absorb possibly non-zero choices of $\vecp'_{>\alpha_\cK-1}$ into $f$. (These parameters contain the parameters $\scal_1^{(\lambda-1,k)}$ for $\Re(\lambda-1)>\alpha_\cK-2$ which we worked with in~\S\ref{SsD5Alm}.) We may thus reduce to the case
    \[
      \vecp^{\prime,>\alpha_\cK-2}_{>\alpha_\cK-1}=0.
    \]
  \item $\vecp^{\prime,>\alpha_\cK-2}_{\in(\alpha'_\cK-1,\alpha_\cK-1]}$: the corresponding modification terms are not of size $o(\rho_+^{\alpha_\cK+1})$ at $\iota^+$, and thus we must incorporate them into a generalized version of~\eqref{EqD5AlmGoodEqPrelim}. (Such parameters were not present in~\S\ref{SsD5Alm}.)
  \item $\vecp^{\prime,>\alpha_\cK-2}_{\leq\alpha'_\cK-1}$: the $\iota^+$-decay of the corresponding modification terms is weaker still. We will take the parameters $\vecp_{\rm induct,in}^{\prime,>\alpha'_\cK-2}$ with which we apply the inductive hypothesis to have all components $0$ except for the components $\vecp^{\prime,>\alpha_\cK-2}_{\leq\alpha'_\cK-1}$.
  \end{enumerate}
  The parameters we need to determine, given the arbitrary choice of $\vecp^{\prime,>\alpha_\cK-2}$, are $\vecp^{\prime,\in(\alpha'_\cK-2,\alpha_\cK-2]}$ and $\vecp^{\prime,\leq\alpha'_\cK-2}$. To this end, we consider the $\vecp^{\prime,\in(\alpha'_\cK-2,\alpha_\cK-2]}$-dependent equation
  \begin{equation}
  \label{EqD6AlmGoodEq}
  \begin{split}
    L\bigl(u_\Sigma(\vecp^{\prime,\in(\alpha'_\cK-2,\alpha_\cK-2]})\bigr) &= f - \Bigl( D_{(h,\vecp)}P(\vecp^{\prime,\in(\alpha'_\cK-2,\alpha_\cK-2]}) - L h_b(\vecp^{\prime,\in(\alpha'_\cK-2,\alpha_\cK-2]})\Bigr) \\
      &\qquad - \Bigl( D_{(h,\vecp)}P(\vecp^{\prime,>\alpha_\cK-2}_{\in(\alpha'_\cK-1,\alpha_\cK-1]}) - L h_b(\vecp^{\prime,>\alpha_\cK-2}_{\in(\alpha'_\cK-1,\alpha_\cK-1]})\Bigr) \\
      &\qquad - D_{(h,\vecp)}P(\vecp^{\prime,>\alpha'_\cK-2}_{\rm induct,in}) \\
      &\qquad - D_{(h,\vecp)}P(\vecp^{\prime,\leq\alpha'_\cK-2}_{\rm induct,out}),
  \end{split}
  \end{equation}
  where the parameters $\vecp^{\prime,\leq\alpha'_\cK-2}_{\rm induct,out}$ are produced by applying the inductive hypothesis to the source term given by the sum of the first two lines on the right and with free parameters taken to be $\vecp^{\prime,>\alpha'_\cK-2}_{\rm induct,in}$. Exactly as after~\eqref{EqD5AlmGoodEq}, the fact that the $\iota^+$-order of the first two modification terms on the right is $>\alpha_\cK-2$ (note for the first term that $\vecp^{\prime,\in(\alpha'_\cK-2,\alpha_\cK-2]}$ is a subset of $\vecp_{\in(\alpha'_\cK-1,\alpha_\cK-1]}$ since $2<\alpha'_\cK<\alpha_\cK<3$) implies that the aforementioned extraction of the next expansion terms can be applied to this equation, producing $\dot{\vecp}^{\in(\alpha'_\cK-2,\alpha_\cK-2]}$ (depending linearly on $f$ and $\vecp^{\prime,\in(\alpha'_\cK-2,\alpha_\cK-2]}$) such that (cf.\ \eqref{EqD5uFinal})
  \[
    u_\Sigma(\vecp^{\prime,\in(\alpha'_\cK-2,\alpha_\cK-2]}) = h_b(\dot{\vecp}^{\in(\alpha'_\cK-2,\alpha_\cK-2]}) + \tilde u_\Sigma,\quad \tilde u_\Sigma \in \Hb^{\infty,\ \bigl(\la\cE_\sscri^\cC\ra,3+\eps_\sscri\bigr),\ 1-\eps,\ \alpha_\cK-2}(\Omega_*;S^2\cT^*)^{\bullet,-}.
  \]
  Therefore, the total parameters $\dot{\vecp}_\tot^{\in(\alpha'_\cK-2,\alpha_\cK-2]}$ governing the new expansion terms of
  \begin{equation}
  \label{EqD6AlmGoodEqSol}
    u(\vecp^{\prime,\in(\alpha'_\cK-2,\alpha_\cK-2]}) := u_\Sigma(\vecp^{\prime,\in(\alpha'_\cK-2,\alpha_\cK-2]}) + h_b(\vecp^{\prime,\in(\alpha'_\cK-2,\alpha_\cK-2]}) + h_b(\vecp^{\prime,>\alpha_\cK-2}_{\in(\alpha'_\cK-1,\alpha_\cK-1]})
  \end{equation}
  modulo $\alpha_\cK-2$ orders of decay at $\cK^+$ (which in particular means that the third term in~\eqref{EqD6AlmGoodEqSol} does not enter here) are
  \begin{equation}
  \label{EqD6AlmTotalMod}
    \dot{\vecp}_\tot^{\in(\alpha'_\cK-2,\alpha_\cK-2]} = \dot{\vecp}^{\in(\alpha'_\cK-2,\alpha_\cK-2]} + \vecp^{\prime,\in(\alpha'_\cK-2,\alpha_\cK-2]}.
  \end{equation}
  But since the first modification term on the right of~\eqref{EqD6AlmGoodEq} vanishes for $h=0$ and $\vecp=0$, this is equal to a tuple depending linearly on $f$ plus a $o(1)$-perturbation (as $h,\vecp\to 0$) of the identity map applied to $\vecp^{\prime,\in(\alpha'_\cK-2,\alpha_\cK-2]}$, so there exists a unique choice of $\vecp^{\prime,\in(\alpha'_\cK-2,\alpha_\cK-2]}$ such that~\eqref{EqD6AlmTotalMod} vanishes. This finishes the inductive step.

  Taking $\alpha_\cK\nearrow 3$, we have now shown that for arbitrary choices of $\vecp^{\prime,\geq 1}$ there exist parameters $\vecp^{\prime,<1}$ such that, for $\vecp'=(\vecp^{\prime,<1},\vecp^{\prime,\geq 1})$,
  \begin{equation}
  \label{EqD6AlmSol}
    L u=f-D_{(h,\vecp)}P(\vecp') \implies u \in \Hb^{\infty,\ \bigl(\la\cE_\sscri^\cC\ra,3+\eps_\sscri\bigr),\ 1-\eps,\ 1-\eps}(\Omega_*;S^2\cT^*)^{\bullet,-}\quad\forall\,\eps>0.
  \end{equation}

  \pfsubstep{Step~1.2.}{Expansion at $\iota^+$.}\label{ItD6Step12} We revisit Steps~1.1--1.7 of the proof of Proposition~\ref{PropD5Alm} and now keep track of $\iota^+$-orders. The equation for $u$ in~\eqref{EqD6AlmSol} can be written as $L_b u=f-D_{(h,\vecp)}P(\vecp')-(L-L_b)u$, and thus, as in~\eqref{EqD5uscri} and \eqref{EqD5uiota}, one can construct $u_{1,\sscri}$ and $u_\iota$, with $\iota^+$-orders $(\tilde\cE_+,2-\eps)$ and $(1,0)\cup(1+\cE_\ind)\subset\tilde\cE_+$ (see~\eqref{EqD5TildeEplus}), respectively, such that for $u_\flat:=u-u_{1,\sscri}-u_\iota$, we have
  \[
    L_b u_\flat =: f_\flat \in \Hb^{\infty,\ 4+\eps_\sscri,\ \bigl((\tilde\cE'_+\cup\tilde\cE'_{++})+2,\,4-\eps\bigr),\ 3+\eps_\cK-\eps}.
  \]
  the index set $\tilde\cE'_{++}$ comes from $D_{(h,\vecp)}P(\vecp')$, and the $\cK^+$-order comes from $(L-L_b)u$ (using~\eqref{EqDAdmFw}). We analyze this on the Fourier transform side,\footnote{A conceptually somewhat cleaner approach for proving the polyhomogeneity of $u_\flat$ at $\iota^+$ would be to use a normal operator argument at $\iota^+$ involving the inversion of the $\iota^+$-normal operator family of $L_b$ (following the template of Lemma~\ref{LemmaTMSolPhg}). This would require proving suitable high-energy (i.e., large $|\Im\lambda|$) estimates in the context of Proposition~\ref{PropipNInv}; we leave it to the reader to pursue this approach.} beginning with the action of the zero energy operator inverse on $\wh{f_\flat}(0)$ as in~\eqref{EqD5Almureg1}; this produces a contribution
  \begin{equation}
  \label{EqD6Almureg1}
    \chi_\zface\hat u_{{\rm reg},1}(0)\in\Hb^{\infty,\ \infty,\ (\tilde\cE'_{+\sharp}-1,2-\eps),\ (0,0)}(X_\scbtop^\pm)
  \end{equation}
  to $\wh{L_b}(\sigma)^{-1}\wh{f_\flat}(\sigma)$ and leaves the error $\hat f_2\in|\sigma|\Hb^{\infty,\ 4-\eps,\ (\tilde\cE'_{+\sharp},2-\eps),\ ((0,0),1-\eps)}$ as in~\eqref{EqD5Almf2} (with $\alpha_\cK=3-\eps$). As after~\eqref{EqD5Almftf}, we apply Proposition~\ref{PropiptfGr}\eqref{ItiptfGrPrec} to its $\rho_\tface^\lambda(\log\rho_\tface)^k$-leading-order term at $\tface$, where $(\lambda,k)\in\tilde\cE'_{+\sharp}$ and $\Re\lambda\in(1,2)$. The most singular (at $\zface$) output~\eqref{EqD5Almutfsingpm} vanishes in view of the $\cK^+$-decay of~\eqref{EqD6AlmSol}; but we are presently more interested in the $\tface$-order, which is the same for all terms of~\eqref{EqiptfGruPrec} (or $k$-fold derivatives in $\alpha_0$ of terms of that form, evaluated at $\alpha_0=\lambda$), and indeed equal to $(\lambda-2,k)+((0,0)\cup\cE_\ind)$. Taking into account the overall factor of $|\sigma|$, these terms thus have the same $\tface$-index set $\tilde\cE'_{+\sharp}-1$ as the term~\eqref{EqD6Almureg1}.

  The remaining error is $\hat f_3$ as in~\eqref{EqD5utff3}, whose conormal term at $\tface$ is solved away using $\hat u_3$ in~\eqref{EqD5utfu3} (whose $\tface$-order $1-\eps$ corresponds, upon inverse Fourier transforming, to a presently acceptable $\iota^+$-order of $2-\eps$ by~\eqref{EqTFHbInvLo}). All further contributions to $\hat u$ in Steps~1.4--1.6 have $\tface$-orders at least $1-\eps$ (recalling that the outputs of these steps are multiplied by an overall factor of $\sigma$ in Step~1.7, due to the division by $\sigma$ in~\eqref{EqD5utff4}). Note that $u_\rem$ in~\eqref{EqD5urem} in fact has $\iota^+$-order $(\tilde\cE_+,2-\eps)$ (as follows from the above more careful accounting of $u_{1,\sscri}$ and $u_\iota$). Furthermore, we recall from~\eqref{EqD5SingRegMem} that the $\iota^+$-order of $u_{\rm sing}$ is $2-\eps$; and we can strengthen the membership of $u_{\rm reg}$ in~\eqref{EqD5SingRegMem} by noting that the $\tface$-order of $\cF u_{\rm reg}$ in~\eqref{EqD5ureg} is $(\tilde\cE'_{+\sharp}-1,1-\eps)$. We thus conclude from Proposition~\ref{PropTFHbphg}\eqref{ItTFHbphgFI} that the $\iota^+$-index set of $u$ is $\tilde\cE_{+\sharp}:=\tilde\cE_+\cup\tilde\cE'_{+\sharp}$, with conormal remainder of decay order $2-\eps$. We have thus sharpened~\eqref{EqD6AlmSol} to
  \begin{equation}
  \label{EqD6AlmSol2}
    u \in \Hb^{\infty,\ \bigl(\la\cE_\sscri^\cC\ra,\,3+\eps_\sscri\bigr),\ \bigl(\tilde\cE_{+\sharp},\,2-\eps\bigr),\ 1-\eps}(\Omega_*)^{\bullet,-}\quad\forall\,\eps>0.
  \end{equation}

  \pfstep{Part~2. Going beyond $t_*^{-1}$-decay.}\label{ItD6Step2} Our next goal is to extract the $t_*^{-1}(\log t_*)^k$-terms in the asymptotic expansion of $u$, and to use the parameters
  \begin{equation}
  \label{EqD6peq1Again}
    \vecp^{\prime,=1}=\Bigl((\scal_1^{(\lambda-1,k)})_{\Re\lambda=2},\ \vect_1^{(1,0)},\ \scal_0^{(0),(1,0)},\ \scal_2^{(0),(1,0)}\Bigr)
  \end{equation}
  to eliminate them.\footnote{The significance of crossing an integer decay rate is that the Taylor expansion at $\zface$ of the Fourier transform of the effective source term has one more term; so in addition to the sub-leading zero energy piece of~\eqref{EqD4Sub0}, we now have to keep track also of a sub-sub-leading zero energy piece.} We moreover keep more precise track of polyhomogeneous expansions at $\iota^+$ (and $\tface$ on the spectral side), in particular improving the conormal remainder term at $\iota^+$ from $2-\eps$ in~\eqref{EqD6AlmSol2} to $3-\eps$.\footnote{We could already have obtained this improvement in~\eqref{EqD6AlmSol2} if we had kept more precise track of $\iota^+$- and $\tface$-index sets in the course of the previous arguments. (That is, it is not the marginal improvement of the $\cK^+$-decay rate which we are about to discuss that facilitates the increased precision at $\iota^+$.)} On a technical note, it is not possible anymore to use the rather imprecise~\eqref{EqD1AlmAug} for this purpose, since only a $\tface$-decay rate of at least $2-\eps$ would suffice to get $3-\eps$ orders of decay at $\iota^+$ upon inverse Fourier transforming; while one could indeed use Proposition~\ref{PropTFHbphg} to control the Fourier transforms of $\pa_{t_*}\dot g_b^{\Ups,\aug}$ to the required level of precision, we instead switch to the augmentation $\wt{L_b'}(\sigma)$ defined in~\eqref{EqAdmLoSpec}, which due its definition fully on the spectral side is slightly easier to work with.

  \pfsubstep{Step~2.A.}{Extracting the $t_*^{-1}(\log t_*)^k$-terms.} We consider source terms $f=f_1+f_2$ of the form~\eqref{EqD6f} but with $f_2$ having $\iota^+$-index set $(\tilde\cE_+\cup\cE)+2$ where $\cE$ is any index set with $\min\Re>2$. (This in particular includes source terms $D_{(h,\vecp)}P(\vecp'_{>2})$, which are indeed the sole reason for introducing $\cE$.) There then exist (unique) parameters $\vecp^{\prime,<1}$ such that the solution of
  \[
    L u = f - D_{(h,\vecp)}P(\vecp^{\prime,<1})
  \]
  satisfies~\eqref{EqD6AlmSol2}. We rewrite this as
  \begin{equation}
  \label{EqD6Lbu}
    L_b u = \bigl(f_1 - (L-L_b)u\bigr) + f_2 - D_{(h,\vecp)}P(\vecp^{\prime,<1})
  \end{equation}
  All summands of $D_{(h,\vecp)}P(\vecp^{\prime,<1})$ other than $D_{(h,\vecp)}P(\scal')$ (which is used for undoing boosts) are of class $\chi_\cK^\sharp\Hb^{\infty,\ 0,\ ((\tilde\cE_+\cup\tilde\cE'_{++})+2,\,5+\eps_+),\ 4+\eps_\cK}$ and will be combined with $f_2$, as will the part of $D_{(h,\vecp)}P(\scal)$ denoted $(1-\chi_\cK^\flat)\chi_\cK^\sharp f^2_{\cdots}$ in~\eqref{EqD2LinParScal}; the part $(1-\chi_\cK)^\flat f^1_{\cdots}$ on the other hand is of the same form as $f_1$ in our present setup~\eqref{EqD6f}. Furthermore, we can control $(L-L_b)u$ using~\eqref{EqDAdmFw} albeit with $\cE_+$ enlarged to account for the $\iota^+$-index sets of $u$ and of $h_\tot(\vecp)$ in Definition~\ref{DefD6EinsteinAug}, i.e., with $\cE_+\cup(1,1)\cup\tilde\cE_{+\sharp}$ in place of $\cE_+$. Since the smallest element of $\cE_+$ and $\tilde\cE_{+\sharp}$ is $(1,0)$, one can self-consistently require $2(\cE_+\cup\tilde\cE_{+\sharp})+2\subset\cE_++3$ for $\cE_+$, and with the inclusion of $(1,1)$ one can self-consistently require
  \[
    2\bigl(\cE_+\cup(1,1)\cup\tilde\cE_{+\sharp}\bigr)+2\subset\cE_++(3,2),
  \]
  as we shall do. The $\cK^+$-order of $(L-L_b)u$ is $2+\eps_\cK$ more than that of $u$, so equal to
  \[
    \alpha_\cK:=3+\eps_\cK-\eps.
  \]
  The upshot is that
  \[
    L_b u \in \Hb^{\infty,\ \bigl(\la\cE_\sscri^\cC+1\ra',\,4+\eps_\sscri\bigr),\ \bigl(\cE_++(3,2),\,5-\eps\bigr),\ \alpha_\cK} + \Hb^{\infty,\ 4+\eps_\sscri,\ \bigl((\tilde\cE_+\cup\tilde\cE'_{++})+2,\,5+\eps_+\bigr),\ 4+\eps_\cK},
  \]
  where for ease of notation we include $\cE$ in $\tilde\cE_+$. As usual, the precise form of the $\iota^+$-index sets does not matter here; the main point is that $L_b u$ is partially polyhomogeneous at $\iota^+$ with remainder having $5-\eps$ orders of decay.

  \pfsubstep{Step~2.A.1.}{Solution near $\scri^+$.} We solve away the $\scri^+$-expansion of the first summand using Proposition~\ref{PropDScriFormal} with $\breve\cE_+=\cE_++(1,2)$; this produces an index set $\cE_+^\sharp=(1,0)\cup(1+(1-e^\Ups)\gamma^\Ups,0)\cup(\cdots)$ (which is larger than the one previously used, but agrees with the original choice in $\{\Re z<2\}$)---cf.\ the description following~\eqref{EqD5Eplussharp} where we now take advantage of the flexibility introduced there by taking $\cF=(1,2)$---and
  \begin{equation}
  \label{EqD6uscri}
  \begin{split}
    &u_{1,\sscri} \in \Hb^{\infty,\ \bigl(\la\cE_\sscri^\cC\ra,3+\eps_\sscri\bigr),\ (\cE_+^\sharp,3-\eps),\ \infty},\ \text{supported near $\scri^+$}, \\
    &\qquad L_b(u-u_{1,\sscri}) =: f',\quad f' \in \Hb^{\infty,\ 4+\eps_\sscri,\ \bigl((\tilde\cE_+\cup\tilde\cE'_{++})+2,\,5-\eps\bigr),\ \alpha_\cK}(\Omega_*;S^2\cT^*)^{\bullet,-}.
  \end{split}
  \end{equation}

  \pfsubstep{Step~2.A.2.}{Solving away the leading-order term at $\iota^+$.} Consider the $\rho_+^3$-leading-order term of $f'$ at $\iota^+$, which is given by $t_*^{-3}f_+^{(1,0)}$ where $f_+^{(1,0)}\in\Hb^{\infty,\ 4+\eps_\sscri,\ \eps_\cK-\eps}$ (recalling $\alpha_\cK-3=\eps_\cK-\eps$). We solve this away using Proposition~\ref{PropipGr} with $\lambda=1$; as in the previous steps (see~\eqref{EqD5uiota}), the logarithmically divergent first term in~\eqref{EqipGr} must vanish. But at the present step we can only regard those remaining terms as errors that have $\cK^+$-decay order $\alpha_\cK-2=1+\eps_\cK-\eps$. That is, we obtain
  \begin{align}
    &u_\iota^{(1,0)} = \chi_\cK u_{\iota,{\rm exp}}^{(1,0)} + \tilde u_\iota^{(1,0)}, \nonumber\\
  \label{EqD6uiotaexp}
    &\qquad u_{\iota,{\rm exp}}^{(1,0)} = h_{b,\rmv 1}^{\leq 3}(t_*^{-1}\dot\vect_1^{(1,0)}) + \dot\scal_0^{(0),(1,0)}h_{b,\rms 0}^{(0),\leq 3}(t_*^{-1}) + h_{b,\rms 2}^{(0),\leq 3}(t_*^{-1}\dot\scal_2^{(0),(1,0)}), \\
    &\qquad \tilde u_\iota^{(1,0)} \in \Hb^{\infty,\ \bigl(\la\cE_{\iota^+,\sscri}^\cC\ra,3+\eps_\sscri\bigr),\ (1,0)\cup(1+\cE_\ind),\ \alpha_\cK-2}(\Omega_*)^{\bullet,-}, \nonumber
  \end{align}
  where\footnote{Notation consistent with~\eqref{EqipGr} would be $\dot\vect_1^{(-1)}$, $\dot\scal_0^{(0)}$, and $\dot\scal_2^{(0)}$.} $\dot\vect_1^{(1,0)}\in\vectspace_1$, $\dot\scal_0^{(0),(1,0)}\in\scalspace_0$, and $\dot\scal_2^{(0),(1,0)}\in\scalspace_2$, such that
  \[
    \chi_\iota t_*^{-3}f_+^{(1,0)} - L_b u_\iota^{(1,0)} \in \Hb^{\infty,\ \bigl(\cE_{\iota^+,\sscri}^\tot+2,\,4+\eps_\sscri\bigr),\ 3+\cE_\ind,\ \alpha_\cK}.
  \]
  Similarly to~\eqref{EqD4ip1scri}, we solve this term away at $\scri^+$ using Proposition~\ref{PropDScriFormal} with $\breve\cE_+=1+\cE_\ind$ and arbitrary $C_0<2$, but now with $\breve\ell_+=3-\eps$ (since we wish to keep track of a more precise $\iota^+$-expansion of $u$). This thus produces a term with $\iota^+$-index set starting with $1+\cE_\ind$ and indeed
  \[
    u_\sscri^{(1,0)} \in \Hb^{\infty,\ \bigl(\la\cE_\sscri^\cC\ra,3+\eps_\sscri\bigr),\ (\tilde\cE'_+,3-\eps),\ \infty},\quad \tilde\cE'_+=\tilde\cE_+\setminus\{(1,0)\},
  \]
  such that
  \begin{equation}
  \label{EqD6uflat}
  \begin{split}
    &u_\flat := u - u_{1,\sscri} - u_\iota^{(1,0)} - u_\sscri^{(1,0)} \\
    &\qquad \implies L_b u_\flat =: f_\flat \in \Hb^{\infty,\ 4+\eps_\sscri,\ \bigl((\tilde\cE'_+\cup\tilde\cE'_{++})+2,\,5-\eps\bigr),\ \alpha_\cK}(\Omega_*;S^2\cT^*)^{\bullet,-}.
  \end{split}
  \end{equation}
  The gain compared to~\eqref{EqD6uscri} is the elimination of the element $(3,0)\in\tilde\cE_++2$ of $f'$ at $\iota^+$.

  \pfsubstep{Step~2.A.3.}{Zero energy piece.} The rest of the analysis is rather similar to that starting with~\eqref{EqD5AlmFT}, except the order in which one solves away error terms at $\zface$ and $\tface$ must be changed slightly owing to the particular relative decay rates that arise. We focus on the low-energy analysis, as the high-energy piece of $u_\flat$ has arbitrary decay as usual (recall here \citeAF{Proposition~\ref*{CorDResHiLoc}}). Since now $\alpha_\cK-1\in(2,3)$, the Taylor expansion of\footnote{The $\scface$-order was already rather immaterial in~\eqref{EqD5Almf2}, and we decide here to record only the order $2-\eps$, which is consistent with the best possible $\scface$-order $1-\eps$ that $\wh{L_b}(\sigma)^{-1}$ produces.}
  \[
    \bigl(\wh{f_\flat}(\sigma)\bigr)\big|_{\sigma\in\pm[0,c]} \in \Hb^{\infty,\ 2-\eps,\ \bigl((\tilde\cE'_+\cup\tilde\cE'_{++})+1,\,4-\eps\bigr),\ ((0,0),\alpha_\cK-1)}(X_\scbtop^\pm)
  \]
  at $\zface$ now has three instead of two terms; and we keep track of more terms at $\tface$ compared to~\eqref{EqD5AlmFT} due to our quest to get a more precise expansion of $u$ at $\iota^+$ upon inverse Fourier transforming. Analogously to~\eqref{EqD5Almureg1} but now using $\wt{L_b'}$, inversion of $\wt{L_b'}(0)$ produces
  \begin{equation}
  \label{EqD6ureg1}
  \begin{split}
    &\hat u_{{\rm reg},1}(0) \in \Hb^{\infty,\ (\tilde\cE'_{+\sharp}-1,2-\eps)}(X;S^2\cT^*_X)\ \text{such that} \\
    &\qquad \bigl(\hat f_2(\sigma)\bigr)\big|_{\sigma\in\pm[0,c]} := \Bigl(\wh{f_\flat}(\sigma) - \wh{L_b}(\sigma)\bigl(\chi_\zface\hat u_{{\rm reg},1}(0)\bigr)\Bigr)\Big|_{\sigma\in\pm[0,c]} \\
    &\qquad \hspace{7.1em} \in |\sigma|\Hb^{\infty,\ 2-\eps,\ (\tilde\cE'_{+\sharp},3-\eps),\ ((0,0),\alpha_\cK-2)}(X_\scbtop^\pm).
  \end{split}
  \end{equation}

  \pfsubstep{Step~2.A.4.}{Solving away the error at $\tface$.} Next, as around~\eqref{EqD5Almftf}, we solve away the $\tface$-expansion of $\sigma^{-1}\hat f_2(\sigma)$ until the $\tface$-order of the remaining error is $>2$ (so that the remaining error at $\zface\cong X$ has more than $2$ orders of decay at $\pa X$). The only terms not treated already in~\eqref{EqD5Almftf} are thus those corresponding to elements $(\lambda,k)\in\tilde\cE'_{+\sharp}$ with $\Re\lambda=2$. The output of Proposition~\ref{PropiptfGr} acting on the $\rho_\tface^\lambda(\log\rho_\tface)^k$-term of $\sigma^{-1}\hat f_2(\sigma)$, with $(\alpha_0,k):=(\lambda,k)\in\tilde\cE'_{+\sharp}$ and $\Re\lambda\leq 2$, does not feature the most singular first term of~\eqref{EqiptfGruPrec2} when $\Re\lambda<2$ (since we have already eliminated it), so these terms remain only for $\Re\lambda=2$ remain, and they are given by $u_{\tface,\pm,{\rm sing}}^{(\lambda-1,j)}=\chi_\zface|\sigma|^{\lambda-2}(\log|\sigma|)^j|\sigma|\sigma^{-2}h_{b,\rms 1}(\dot\scal_{1,\pm}^{(\lambda-1,k)})$ as in~\eqref{EqD5Almutfsingpm}, with $j=0,\ldots,k$ and also $j=k+1$ when $\lambda=2$ (since then $\fl^0(|\sigma|)=\log|\sigma|$ in~\eqref{EqiptfGruPrec2}). Moreover, with $\Re\lambda\in(1,2]$ still, all other terms in~\eqref{EqiptfGruPrec2} have $\zface$-order $\geq\Re\lambda-2$ (so order $\geq\Re\lambda-1>0$ after multiplying by $\sigma$, which translates into $o(t_*^{-1})$-decay upon inverse Fourier transforming). In summary, as in~\eqref{EqD5utfsing} but keeping more precise track of $\tface$-orders as in Part~1 of the current proof, we have constructed
  \[
    u_{\tface,{\rm sing}} = \sum_{ \substack{ (\lambda,k)\in\tilde\cE'_{+\sharp} \\ \Re\lambda=2 } } u_{\tface,{\rm sing}}^{(\lambda-1,k)},\quad
    \tilde u_\tface \in \Hb^{\infty,\ 1-\eps,\ \tilde\cE'_{+\sharp}-1,\ ((0,0),\alpha_\cK-2)}(X_\scbtop^\pm),
  \]
  where
  \[
    u_{\tface,{\rm sing}}^{(\lambda-1,k)}=
      \begin{cases}
        \chi_\zface\,(\log(\sigma+i 0))^{k+1}\,h_{b,\rms 1}(\dot\scal_1^{(1,k)}), & \lambda=2, \\
        \chi_\zface\,(\sigma+i 0)^{\lambda-2}(\log(\sigma+i 0))^k\,h_{b,\rms 1}(\dot\scal_1^{(\lambda-1,k)}), & \lambda\neq 2,
      \end{cases}
  \]
  such that
  \begin{align}
  \label{EqD6f3}
    &\hat f_3(\sigma) := \sigma^{-1}\Bigl(\hat f_2(\sigma)-\wh{L_b}(\sigma)(u_{\tface,{\rm sing}}+\tilde u_\tface)\Bigr) \in \Hb^{\infty,\ 2-\eps,\ (\tilde\cE^{\prime,>2}_{+\sharp},\,3-\eps),\ ((0,0),\alpha_\cK-2)}(X_\scbtop^\pm), \\
    &\hspace{22em} \tilde\cE^{\prime,>2}_{+\sharp} := \bigl\{ (z,k)\in\tilde\cE'_{+\sharp} \colon \Re z>2 \bigr\}. \nonumber
  \end{align}
  The inverse Fourier transform of $u_{\tface,{\rm sing}}^{(\lambda-1,k)}$ is a constant multiple of $t_*^{-\lambda+1}(\log t_*)^k h_{b,\rms 1}(\dot\scal_1^{(\lambda-1,k)})+\cO(t_*^{-\infty})$ in spatially compact sets.

  \pfsubstep{Step~2.A.5.}{The sub-leading zero energy piece.} Inverting $\wt{L_b'}(0)$ produces
  \begin{equation}
  \label{EqD6ureg3}
    \hat u_{{\rm reg},3}(0) \in \Hb^{\infty,\ (\tilde\cE''_{+\sharp},1-\eps)}(X;S^2\cT^*_X),\quad
    \hat b_3(0) \in \C^4,
  \end{equation}
  such that
  \begin{equation}
  \label{EqD6uregErr}
    \bigl( \hat f_3(\sigma),\ 0,\ 0 \bigr) - \wt{L_b'}(\sigma)\bigl(\chi_\zface\hat u_{{\rm reg},3}(0),\ \hat b_3(0),\ 0\bigr)
  \end{equation}
  vanishes at $\zface$. (The existence of the expansion of $\hat u_{{\rm reg},3}(0)$ follows from a normal operator argument as in~\eqref{EqD3Almureg1Eq}. The vanishing of the third component is equivalent to the absence of a stationary center-of-mass motion analogously to~\eqref{EqD5ureg4}.) In~\eqref{EqD6ureg3}, the index set $\tilde\cE''_{+\sharp}$ arises from Lemma~\ref{LemmaTMSolPhg} with $\cF=\tilde\cE^{\prime,>2}_{+\sharp}-2$ (with $\min\Re\cF\geq\eps_\ind$) and is thus typically larger than $\tilde\cE'_{+\sharp}$. Keeping in mind the overall factor of $\sigma$ from~\eqref{EqD6f3}, the contribution of $\sigma\hat u_{{\rm reg},3}(0)$ to the $\tface$-index set is $(\tilde\cE'_{+\sharp}-1)\cup(\tilde\cE''_{+\sharp}+1)$ and thus differs from $\tilde\cE'_{+\sharp}-1$ only in $\{\Re z>1\}$. We may arrange that this union is a subset of $\tilde\cE'_{+\sharp}-1$, and will do that below without further comment. Thus, dividing the first component of~\eqref{EqD6uregErr} by $\sigma$ produces
  \begin{equation}
  \label{EqD6f4}
    \hat f_4(\sigma) \in \Hb^{\infty,\ 2-\eps,\ (\tilde\cE^{\prime,>2}_{+\sharp}-1,\,2-\eps),\ ((0,0),\alpha_\cK-3)}(X_\scbtop^\pm).
  \end{equation}

  \pfsubstep{Step~2.A.6.}{Solving away the sub-leading error at $\tface$.} We next solve away the partial $\tface$-expansion of~\eqref{EqD6f4} using Proposition~\ref{PropiptfGr}. Since the output of Proposition~\ref{PropiptfGr} contributes with an overall factor of $\sigma^2$ to $\wh{u_\flat}(\sigma)$, even the most singular term in~\eqref{EqiptfGruPrec} contributes with a (vanishing) factor of at least $|\sigma|^{\eps_\ind-\eps}$ at $\zface$. Solving away also the conormal error term using Proposition~\ref{PropiptfC}, we thus obtain
  \begin{equation}
  \label{EqD6u4}
  \begin{split}
    &\hat u_4 \in \Hb^{\infty,\ 1-\eps,\ (\tilde\cE^{\prime,>2}_{+\sharp}-3,\,-\eps),\ \eps_\ind-\eps-2}(X_\scbtop^\pm), \\
    &\qquad \hat f_5(\sigma) := \hat f_4(\sigma) - \wh{L_b}(\sigma)\hat u_4(\sigma) \in \Hb^{\infty,\ 2-\eps,\ 2+\eps_\ind-\eps,\ ((0,0),\alpha_\cK-3)}(X_\scbtop^\pm).
  \end{split}
  \end{equation}

  \pfsubstep{Step~2.A.7.}{The sub-sub-leading zero energy piece.} We apply $\wt{L_b}(0)^{-1}$ one final time, now to $\hat f_5(0)$. This produces
  \[
    \hat u_{{\rm reg},5}(0) \in \Hb^{\infty,\ \eps_\ind-\eps}(X;S^2\cT^*_X),\quad
    \hat b_5(0)\in\C^4,\ \hat\scal_5(0)\in\scalspace_1
  \]
  such that the first component of $(\hat f_5(\sigma),0,0)-\wt{L_b'}(\sigma)(\chi_\zface\hat u_{{\rm reg},5}(0),\hat b_5(0),\hat\scal_5(0))$, divided by $\sigma$, satisfies
  \begin{equation}
  \label{EqD6f6}
    \hat f_6(\sigma) \in \Hb^{\infty,\ 2-\eps,\ 1+\eps_\ind-\eps,\ \alpha_\cK-4} = \Hb^{\infty,\ 2-\eps,\ 1+\eps_\ind-\eps,\ -1+\eps_\cK-\eps}(X_\scbtop^\pm).
  \end{equation}

  \pfsubstep{Step~2.A.8.}{Remaining piece; combination.} Since the relative order of~\eqref{EqD6f6} at $\tface$ and $\zface$ is $2+\eps_\ind-\eps_\cK\in(2,2+\eps_\ind)$, we can now use \citeAF{Proposition~\ref*{PropDResLo}} to determine
  \begin{align*}
    &\bigl(\hat u_6(\sigma),\ \hat b_6(\sigma),\ \hat\scal_6(\sigma)\bigr) := \wt{L_b'}(\sigma)^{-1}\bigl(\hat f_6(\sigma),\ 0,\ 0\bigr), \\
    &\qquad \hat u_6 \in \Hb^{\infty,\ 1-\eps,\ -1+\eps_\ind-\eps,\ \alpha_\cK-4}(X_\scbtop^\pm),\quad \hat b_6,\ \hat\scal_6 \in \Hb^{\infty,\,\alpha_\cK-4}(\pm[0,c]_\sigma).
  \end{align*}
  Altogether, recalling the tensors $\dot g_b^{\Ups,{\rm sing}}$ and $h_{b,\rms 1}^{\leq 1,{\rm sing}}$ from~\eqref{EqAdmLoErrtf1CorrErr} and~\eqref{EqAdmLoErrtf2CorrErr}, we have
  \begin{equation}
  \label{EqD6uflatFinal}
  \begin{split}
    \wh{u_\flat}(\sigma) &= \chi_\zface\bigl(\,\underbrace{\hat u_{{\rm reg},1}(0) + \sigma\hat u_{{\rm reg},3}(0) + \sigma^2\hat u_{{\rm reg},5}(0) + \sigma^3\hat u_6(\sigma)}_{\in\Hb^{\infty,\ 1-\eps,\ (\tilde\cE'_{+\sharp}-1,\,2-\eps),\ ((0,0),\alpha_\cK-1)}}\,\bigr) \\
      &\qquad + u_{\tface,{\rm sing}} + \underbrace{\tilde u_\tface + \sigma^2\hat u_4}_{\in \Hb^{\infty,\ 1-\eps,\ (\tilde\cE'_{+\sharp}-1,\,2-\eps),\ ((0,0),\alpha_\cK-2)} } \\
      &\qquad + \sigma\dot g_b^{\Ups,{\rm sing}}(\sigma)\bigl(\,\underbrace{\hat b_3(0) + \sigma\hat b_5(0) + \sigma^2\hat b_6(\sigma)}_{\in\Hb^{\infty,\,((0,0),\alpha_\cK-2)}}\,\bigr) + \sigma^2 h_{b,\rms 1}^{\leq 1,{\rm sing}}(\sigma)\bigl(\,\underbrace{\hat\scal_5(0) + \sigma\hat\scal_6(\sigma)}_{\in\Hb^{\infty,\,((0,0),\alpha_\cK-3)}}\,\bigr).
  \end{split}
  \end{equation}
  Recalling Remark~\ref{RmkAdmLoMem} and $\alpha_\cK=3+\eps_\cK-\eps\in(3,3+\eps_\ind)$, the third line lies in
  \begin{equation}
  \label{EqD6SingRes}
    \Hb^{\infty,\ 1-\eps,\ \bigl((1,0)\cup(1+\cE_\ind),\,\alpha_\cK-1\bigr),\ \bigl((0,0),\,1-\eps\bigr)} + \Hb^{\infty,\ 1-\eps,\ \bigl((2,0),\,\alpha_\cK-1\bigr),\ \bigl((0,0),\,\alpha_\cK-3\bigr)}.
  \end{equation}
  Proposition~\ref{PropTFHbphg} thus shows that the inverse Fourier transform of the sum of all terms on the right-hand side of~\eqref{EqD6uflatFinal} except for $u_{\tface,{\rm sing}}$ lies in $\Hb^{\infty,\ 1-\eps,\ \bigl(\tilde\cE'_{+\sharp},\ 3-\eps\bigr),\ \alpha_\cK-2}(M';S^2\cT^*)$. The partial polyhomogeneity of~\eqref{EqAdmLoMem} is used here to ensure that the conormal remainder of~\eqref{EqD6SingRes} has $\tface$-order at least $2-\eps$. Moreover, we can analyze potential mismatches at the two copies of $\zface$ inside of $X_\scbtop^\pm$ similarly to the arguments after~\eqref{EqD4Fpasing}: mismatches would lead to $|t_*|^{-1}$ lower bounds as $t_*\to-\infty$, which cannot be cancelled by $u_{\tface,{\rm sing}}$ or the other terms $u_{1,\sscri}$, $u_\iota^{(1,0)}$, and $u_\sscri^{(1,0)}$ (which vanish for $t_*\ll -1$) of $u$; this would yield a contradiction to the forward support property of $u$. The minor caveat, however, is that we have ignored the possibility of $\sigma_-^{\lambda-2}(\log\sigma_-)^j h_{b,\rms 1}(\scal_{1,-}^{(\lambda-1,j)\prime})$-terms in Step~2.A.4 (cf.\ Step~1.4 on page~\pageref{ItD5Step14}); only the terms with $\lambda=2$ and $j=0$ can cancel the mismatch. But this can simply be re-interpreted: the term $H(-\sigma)\scal_{1,-}^{(1,0)\prime}$ can be absorbed into the argument of $\sigma^2 h_{b,\rms 1}^{\leq 1,{\rm sing}}$ in~\eqref{EqD6uflatFinal}, and then both the ``$\sigma_-$''-terms and the mismatch must disappear by the forward support argument.

  Adding $u_{1,\sscri}+u_\iota^{(1,0)}+u_\sscri^{(1,0)}$ to obtain $u$ (see~\eqref{EqD6uflat}), we thus conclude that
  \begin{equation}
  \label{EqD6uFinal}
  \begin{split}
    u &= \chi_\cK \Biggl( h_{b,\rmv 1}^{\leq 3}(t_*^{-1}\dot\vect_1^{(1,0)}) + \dot\scal_0^{(0),(1,0)}h_{b,\rms 0}^{(0),\leq 3}(t_*^{-1}) + h_{b,\rms 2}^{(0),\leq 3}(t_*^{-1}\dot\scal_2^{(0),(1,0)}) \\
      &\quad \hspace{10em} + \sum_{ \substack{ (\lambda,k)\in\tilde\cE'_{+\sharp} \\ \Re\lambda=2 } } \chi_\cK t_*^{-\lambda+1}(\log t_*)^k h_{b,\rms 1}(\dot\scal_1^{(\lambda-1,k)}) \Biggr) + \tilde u, \\
    & \tilde u \in \Hb^{\infty,\ \bigl(\la\cE_\sscri^\cC\ra,3+\eps_\sscri\bigr),\ \bigl(\tilde\cE_{+\sharp},\,3-\eps\bigr),\ \alpha_\cK-2}(\Omega_*)^{\bullet,-},
  \end{split}
  \end{equation}
  where $\tilde\cE_{+\sharp}=(1,0)\cup\tilde\cE'_{+\sharp}$ (with the re-definition of $\tilde\cE'_{+\sharp}$ at higher orders that took place in the course of the above arguments). Recalling that $\alpha_\cK=3+\eps_\cK-\eps$, the error term $\tilde u$ thus has more-than-$t_*^{-1}$-decay at $\cK^+$; the failure of $u$ in~\eqref{EqD6AlmSol2} to decay faster than $t_*^{-1}$ is thus encoded in the finitely many (and finite-dimensional) parameters in~\eqref{EqD6uFinal}.

  \pfsubstep{Step~2.B.}{Eliminating the $t_*^{-1}(\log t_*)^k$-terms.} The novelty compared to Part~2 (or Proposition~\ref{PropD5Alm}) is that the $\iota^+$-decay rates of the terms in~\eqref{EqD6uFinal} and the modification terms required to eliminate them are no longer in lockstep. We shall consider an equation similar to~\eqref{EqD6AlmGoodEq}, now depending on the parameters $\vecp^{\prime,=1}$ (see~\eqref{EqD6peq1Again}). Note that $\vecp^{\prime,=1}$ is the union of $\vecp^{\prime,=1}_{=2}$ (the parameters $\scal_1^{(\lambda-1,k)\prime}$ for $\Re\lambda=2$) and $\vecp^{\prime,=1}_{<2}$ (the remaining parameters). Fixing arbitrary parameters $\vecp^{\prime,>1}$ (which will only be determined at later steps when we eliminate more decaying terms in the $\cK^+$-expansion), let us define
  \[
    \vecp^{\prime,\geq 1}_{\rm induct,in}(\vecp^{\prime,=1}_{<2})
  \]
  to have all parameters $0$ except for $\vecp^{\prime,>1}$ (not made explicit in the notation) and $\vecp^{\prime,=1}_{<2}$. Only the parameters $\vecp^{\prime,=1}_{=2}$ correspond to modification terms with $t_*^{-4}$ decay at $\iota^+$, which we need to exclude (for the purpose of preventing runaway powers of $\log t_*$). We then consider
  \begin{equation}
  \label{EqD6ElimGood}
  \begin{split}
    L\bigl( u_\Sigma(\vecp^{\prime,=1}) \bigr) &= f - \Bigl( D_{(h,\vecp)}P(\vecp^{\prime,=1}_{=2}) - L h_b(\vecp^{\prime,=1}_{=2}) \Bigr) \\
      &\qquad - D_{(h,\vecp)}P\bigl(\vecp^{\prime,\geq 1}_{\rm induct,in}(\vecp^{\prime,=1}_{<2})\bigr) - D_{(h,\vecp)}P(\vecp^{\prime,<1}_{\rm induct,out})
  \end{split}
  \end{equation}
  to which the analysis of Step~2.A applies. Here, the parameters $\vecp^{\prime,<1}_{\rm induct,out}$ (which depend on $f$ and $\vecp^{\prime,=1}$, and also on the arbitrary but fixed parameters $\vecp^{\prime,>1}$) are such that $u_\Sigma(\vecp^{\prime,=1})$ has the form~\eqref{EqD6uFinal}; such parameters indeed exist by Part~1. Thus, also
  \[
    u = u_\Sigma(\vecp^{\prime,=1}) + h_b(\vecp^{\prime,=1}_{=2})
  \]
  has an expansion of the form~\eqref{EqD6uFinal}; we claim that the coefficients
  \[
    \dot{\vecp}^{=1}=\bigl((\dot\scal_1^{(\lambda-1,k)})_{\Re\lambda=2},\,\dot\vect_1^{(1,0)},\,\dot\scal_0^{(0),(1,0)},\,\dot\scal_2^{(0),(1,0)}\bigr)
  \]
  in that expansion are the sum of a tuple depending linearly on $f$ plus a $o(1)$-perturbation (as $h,\vecp\to 0$) of $(\pm 1)$ times the identity map applied to $\vecp^{\prime,=1}$. For the coefficients $\dot{\vecp}^{=1}_{=2}=(\dot\scal_1^{(\lambda-1,k)})_{\Re\lambda=2}$, this follows (with the ``$+$'' sign) from the fact that for $(h,\vecp)=(0,\vec 0)$, the right-hand side of~\eqref{EqD6ElimGood} is independent of $\vecp^{\prime,=1}_{=2}$. For the remaining coefficients $\dot{\vecp}^{=1}_{<2}=(\dot\vect_1^{(1,0)},\dot\scal_0^{(0),(1,0)},\dot\scal_2^{(0),(1,0)})$, we similarly note that for $(h,\vecp)=(0,\vec 0)$, the contribution of $\vecp^{\prime,=1}_{<2}$ to the right-hand side of~\eqref{EqD6ElimGood} is $-L_{b_0}(h_{b_0}(\vecp^{\prime,=1}_{<2}))$ and thus contributes $-\vecp^{\prime,=1}_{<2}$ to $-\dot{\vecp}^{\prime,=1}_{<2}$.

  In summary, we have now shown that
  \begin{equation}
  \label{EqD6Part2}
    \text{\parbox{0.7\textwidth}{\it For all $\vecp^{\prime,>1}$, there exists a unique tuple $\vecp^{\prime,\leq 1}$ such that the solution of $L u=f-D_{(h,\vecp)}P(\vecp')$, $\vecp'=(\vecp^{\prime,\leq 1},\vecp^{\prime,>1})$ satisfies
      \[
        \hspace{-10em}u \in \Hb^{\infty,\ \bigl(\la\cE_\sscri^\cC\ra,\,3+\eps_\sscri\bigr),\ \bigl(\tilde\cE_{+\sharp},\,3-\eps\bigr),\ 1+\eps_\cK-\eps}(\Omega_*)^{\bullet,-}.
      \]
    }}
  \end{equation}

  \pfstep{Part~3. Reaching the $\cK^+$-order $2+\eps_\cK$.} Since further improvements of the $\cK^+$-order use the same arguments as before with only notational modifications, we shall not spell them out here. We only make the following observations that justify why we can reach the $\cK^+$-order $2+\eps_\cK\in(2,2+\eps_\ind)$ for $u$ by making use of the modification parameters $\vecp^{\prime,\leq 2}$.
  \begin{enumerate}
  \item Having established an expansion of $u$ at $\iota^+$ modulo conormal remainders with weight $3-\eps$ in~\eqref{EqD6Part2}, the term $(L-L_b)u$ in~\eqref{EqD6Lbu} has an expansion at $\iota^+$ up to weight $(3-\eps)+(3-\eps)>5+\eps_+$ by~\eqref{EqDAdmFw} (with $\ell_+=3-\eps$). Thus, the right-hand side of~\eqref{EqD6Lbu} has $\iota^+$-order $\bigl((\tilde\cE_+\cup\tilde\cE'_{++}\cup(\cE_++(1,2)))+2,\,5+\eps_+\bigr)$; the weight $5+\eps_+$ matches the weight of $f$ in~\eqref{EqD6f} and thus cannot be improved further. (We remark, though, that for the purpose of extracting asymptotics at $\cK^+$ up to order $\alpha_\cK-2$ where $\alpha_\cK=4+\eps_\cK$, it suffices to extract asymptotics at $\iota^+$ only up to order $\alpha_\cK-2+\eps$; for $\alpha_\cK=3-\eps$, this was shown in (the proof of) Proposition~\ref{PropD5Alm}.)
  \item The solution near $\scri^+$ in Step~2.A.1 is the same, except with $3+\eps_+$ and $5+\eps_+$ in place of $3-\eps$ and $5-\eps$ in~\eqref{EqD6uscri}.
  \item\label{ItD6iotaLot} In the inversion of the $\iota^+$-normal operator $N_{\iota^+}(\ubar L,1)$ in Step~2.A.2, we must record the expansion of $u_\iota^{(1,0)}$ up to remainders with $\cK^+$-order $2+\eps_\cK$. The crucial observation is that all terms in~\eqref{EqipGr}, with $\lambda=1$, that have $t_*$-order $\leq 2+\eps_\cK$ (so $\leq 2$)---which are the terms in the first line, the terms with $l+j=1$ in the second line, the scalar type terms with $l=2,3$, and the vector type terms with $l=2$ in the third line---are pure gauge and can be eliminated using the modification parameters $\scal_1^{(0,1)\prime}$, $\vect_1^{(1,0)}$, $\scal_0^{(1,0)}$, and $\scal_2^{(1,0)}$ (used already in Part~2) as well as $\scal_0^{(-\lambda^\Ups_{\rms l,1}+1),(\lambda^\Ups_{\rms l,1},0)}$, $l=0,1$, and $\scal_3^{(-1),(2,0)}$ and $\vect_2^{(-1),(2,0)}$.
  \item Step~2.A.3 is unchanged (except for the improvement of $\tface$-orders from $\bullet-\eps$ to $\bullet+\eps_\ind$). Step~2.A.4 is akin to solving away source terms whose $\iota^+$-orders are $>1$, and thus (similarly to the discussion in~\eqref{ItD6iotaLot} but with even more decay) the contributions from Proposition~\ref{PropiptfGr} that have order $\leq 1+\eps_\cK$ at $\zface$ are pure gauge. Thus, this step produces further terms in the asymptotic expansion at $\cK^+$ that can all be eliminated via gauge modifications as usual.
  \item The subsequent steps are all analogous. The only mild novelty occurs in the final accounting~\eqref{EqD6SingRes} of the terms in the third row of~\eqref{EqD6uflatFinal}: when $\alpha_\cK=4+\eps_\cK\in(4,4+\eps_\ind)$ and $\hat b\in\Hb^{\infty,\ ((0,0),\alpha_\cK-2)}(\pm[0,c]_\sigma;\C^4)$, Remark~\ref{RmkAdmLoMem} (or inspection of~\eqref{EqAdmLoErrtf1Corr}--\eqref{EqAdmLoErrtf1CorrErr}) implies that
    \begin{align*}
      \sigma\dot g_b^{\Ups,{\rm sing}}(\sigma)(\hat b(\sigma)) &\in \Hb^{\infty,\ 1-\eps,\ \bigl((1,1)\cup(1+\cE_\ind),\,\alpha_\cK-1-\eps\bigr),\ \bigl((0,0),1+\eps_\ind-\eps\bigr)} \\
        &\qquad + \chi_\zface \Hb^{\infty,\ \bigl((1,1),1+\eps_\ind-\eps\bigr)}(\pm[0,c];\C) h_{b,\rms 0}^{(0)},
    \end{align*}
    i.e., there is a logarithmic term at $\zface$. Its inverse Fourier transform is a multiple of $t_*^{-2}h_{b,\rms 0}^{(0)}$ and thus can be eliminated using the parameter $\scal_0^{(0),(2,0)}$ (which is included in~\eqref{EqD6EinsteinAugParam} if we require $\tilde\cE'_{++}\ni(2,0)$, as we may). Similarly, the singular $\sigma(\log|\sigma|)h_{b,\rms 1}^{\leq 1}$-term of $\sigma^2 h_{b,\rms 1}^{\leq 1,{\rm sing}}(\sigma)$ at $\zface$ contributes a multiple of $t_*^{-2}h_{b,\rms 1}$, which can be eliminated as before.
  \end{enumerate}

  The upshot is that for $f$ as in~\eqref{EqD6f},
  \begin{equation}
  \label{EqD6Part3}
    \text{\parbox{0.7\textwidth}{\it there exists a unique tuple $\vecp'=\vecp^{\prime,\leq 2}$ such that the solution of $L u=f-D_{(h,\vecp)}P(\vecp')$ satisfies
      \[
        \hspace{-10em}u \in \Hb^{\infty,\ \bigl(\la\cE_\sscri^\cC\ra,\,3+\eps_\sscri\bigr),\ \bigl(\tilde\cE_{+\sharp},\,3+\eps_+\bigr),\ 2+\eps_\cK}(\Omega_*)^{\bullet,-}.
      \]
    }}
  \end{equation}

  \pfstep{Part~4. Conormal expansion of $u$ modulo remainders with $\cK^+$-order $4+\eps_\cK$.}\label{ItD6Part4} Our final task is to show that $u$ in~\eqref{EqD6Part3} has a (only partially polyhomogeneous, and not entirely pure gauge) expansion at $\cK^+$ up to a remainder with $\cK^+$-order $4+\eps_\cK$. To this end, we go through the arguments in Parts~2 and 3, and now explicitly record all terms with $\cK^+$-orders $<4+\eps_\cK$.

  \pfsubstep{Step~4.1.}{Terms from the $\iota^+$-normal operator inversion at leading order.} As discussed in Part~3, we have~\eqref{EqD6uscri} but with $3-\eps$, $5-\eps$, and $\alpha_\cK$ replaced by $3+\eps_+$, $5+\eps_+$, and $4+\eps_\cK$, respectively. Regarding Step~2.A.2, we now apply Proposition~\ref{PropipGr} with $\lambda=1$ to $f_+^{(1,0)}\in\Hb^{\infty,\ 4+\eps_\sscri,\ 1+\eps_\cK}$ (where $1+\eps_\cK$ is now the $\cK^+$-order of $(t_*^3 f')|_{\iota^+}$); from the output~\eqref{EqipGr}, all terms with at most $t_*^{-2}$-decay in spatially compact sets vanish by~\eqref{EqD6Part3}; and the remaining terms are
  \begin{equation}
  \label{EqD6P4iota}
  \begin{split}
    &\sum_{ \substack{\mu=-\lambda^\Ups_{\rms l,l+j}+1 \\ (l,j)=(1,1),(2,0),(2,1),(3,0)} } h_{b,\rms l}^{(\mu),\leq 2}(t_*^{-1+\mu}\scal_l^{(\mu)}) \\
    &\hspace{3em} + \sum_{l=4}^5 h_{b,\rms l}^{(-l+2),\leq 5-l}(t_*^{-l+1}\scal_l^{(-l+2)}) + \sum_{l=3}^4 h_{b,\rmv l}^{(-l+1),\leq 4-l}(t_*^{-l}\vect_l^{(-l+1)}) \\
    &\hspace{3em} + h_{b,\rms 2}^{(-2),\leq 1}(t_*^{-3}\scal_2^{(-2)}) + h_{b,\rmv 2}^{(-2),\leq 1}(t_*^{-3}\vect_2^{(-2)}) + h_{b,\rms 3}^{(-3)}(t_*^{-4}\scal_3^{(-3)}) + h_{b,\rmv 3}^{(-3)}(t_*^{-4}\vect_3^{(-3)}).
  \end{split}
  \end{equation}

  \pfsubstep{Step~4.2.}{Polyhomogeneous terms from the first $\tface$-normal operator inversion.} We then pass to the spectral side; we still have~\eqref{EqD6ureg1} but with $\bullet-\eps$ replaced by $\bullet+\eps_+$, so now $\hat f_2\in|\sigma|\Hb^{\infty,\ 2-\eps,\ (\tilde\cE'_{+\sharp},3+\eps_+),\ ((0,0),2+\eps_\cK)}$, where we recall $\min\Re\tilde\cE'_{+\sharp}\geq 1+\eps_\ind>1$. As in Step~2.A.4, we use Proposition~\ref{PropiptfGr} (with $\ell_\zface=2+\eps_\cK$) to solve away the terms of the $\tface$-expansion of $\sigma^{-1}\hat f_2$ of order $\leq 2$. Let thus $(\lambda,k)\in\tilde\cE'_{+\sharp}$, $1<\Re\lambda\leq 2$. The terms in~\eqref{EqiptfGruPrec} that are singular at $\zface$ with $\zface$-order $\leq 0$ must vanish (as otherwise, upon putting back the factor of $\sigma$, they would contribute terms with $t_*^{-2}$-decay or less to~\eqref{EqD6Part3}); what remains are the analogues of~\eqref{EqD6P4iota} on the spectral side. Let us discuss only one exemplary such term, namely $\sigma\cdot\chi_\zface|\sigma|^{\lambda-2}\fl^{-\lambda+2}(\hat r)|\sigma|^2 h_{b,\rms 4}^{(-2),\leq 1}(\sigma,\scal_4^{(-2)})$, for two values of $\lambda$ and for $k=0$.
  \begin{enumerate}
  \item When $\lambda\neq 2$, we recall $\hat r=\frac{|\sigma|}{\rho}$ and expand $|\sigma|^{\lambda-2}\fl^{-\lambda+2}(\hat r)=(-\lambda+2)^{-1}(\rho^{\lambda-2}-|\sigma|^{\lambda-2})$; the singular (at $\zface$) term is thus a scalar multiple of $\chi_\zface \sigma|\sigma|^\lambda h_{b,\rms 4}^{(-2),\leq 1}(\sigma)$. The inverse Fourier transform is thus a multiple of $h_{b,\rms 4}^{(-2),\leq 1}(t_*^{-\lambda-2})$ up to errors that decay rapidly as $t_*\to\infty$ in spatially compact sets.
  \item When $\lambda=2$, we have $\fl^{-\lambda+2}(\hat r)=\log\hat r=\log|\sigma|-\log\rho$, and thus the singular term is $\chi_\zface\sigma|\sigma|^2(\log|\sigma|) h_{b,\rms 4}^{(-2),\leq 1}(\sigma,\scal_4^{(-2)})$. Now, $\sigma|\sigma|^2\log|\sigma|$ differs from $(\sigma+i 0)^3\log(\sigma+i 0)$ by a term $-i\pi\sigma|\sigma|^2 H(-\sigma)$ whose inverse Fourier transform has a lower bound by $|t_*|^{-4}$ as $t_*\to-\infty$, and this term cannot be cancelled by further terms that we describe in our analysis below, so it must be absent. The inverse Fourier transform of $\chi_\zface(\sigma+i 0)^3(\log(\sigma+i 0))h_{b,\rms 4}^{(-2),\leq 1}(\sigma,\scal_4^{(-2)})$ is a multiple of $h_{b,\rms 4}^{(-2),\leq 1}(t_*^{-4})$, up to rapidly decaying errors in spatially compact sets, for $t_*>1$.
  \end{enumerate}
  In both cases, the element $(\lambda,0)\in\tilde\cE'_{+\sharp}$ yields a contribution of $h_{b,\rms 4}^{(-2),\leq 1}(t_*^{-\lambda-2})$ to the asymptotics of $u$ at $\cK^+$. Similarly, say, the element $(\lambda,k)\in\tilde\cE'_{+\sharp}$ yields a contribution of $h_{b,\rms l}^{(-\mu)}(t_*^{-\lambda+\mu}(\log t_*)^k\scal_l^{(\mu)})$ for $\mu=-\lambda^\Ups_{\rms l,l+j}+1$ where $0\leq l\leq 3$, $j=0,1$, $1\leq l+j\leq 3$, provided $\Re(\lambda-\mu)>2$. In summary, we get additional terms of the form~\eqref{EqD6P4iota} except with arguments $t_*^{-\lambda+\mu}$, $t_*^{-\lambda}t_*^{-l+2}$, $t_*^{-\lambda}t_*^{-l+1}$, $t_*^{-\lambda}t_*^{-2}$ (twice), and $t_*^{-\lambda}t_*^{-3}$ (twice), respectively, and similar terms with factors of $(\log t_*)^k$. For $\Re\lambda=2$, we must also allow for $(l,j)=(0,1),(1,0)$ in the first sum of~\eqref{EqD6P4iota} (since $\Re(\lambda-\mu)\geq 2+\eps_\ind>2+\eps_\cK$ then). We finally remark that the second (remainder) term in~\eqref{EqiptfGruPrec} has $\zface$-remainder order $\ell_\zface=2+\eps_\cK$, which upon multiplication by $\sigma$ becomes $3+\eps_\cK$, and the inverse Fourier transform then produces the $\cK^+$-order $4+\eps_\cK$, which can thus be regarded as a remainder term for $u$.

  \pfsubstep{Step~4.3.}{Polyhomogeneous terms from the second $\tface$-normal operator inversion.} Upon subtracting the (grafted) solutions of the $\tface$-model problem from $\hat f_2$, we are left with the error~\eqref{EqD6f3} except with $\tface$- and $\zface$-remainder order $3+\eps_+$ and $2+\eps_\cK$ (rather than $3-\eps$ and $\alpha_\cK-2$), respectively. We then have~\eqref{EqD6ureg3} with remainder order $1+\eps_+$, and hence~\eqref{EqD6f4} now becomes $\hat f_4\in\Hb^{\infty,\ 2-\eps,\ (\tilde\cE^{\prime,>2}_{+\sharp}-1,2+\eps_+),\ ((0,0),1+\eps_\cK)}$ (the definition of which involves a total division by $\sigma^2$). We then solve away the $\tface$-expansion of $\hat f_4$ as in Step~4.2; this produces further polyhomogeneous terms (with more decay---essentially one order---than before). Unlike in Step~2.A.6, the conormal remainder term of $\hat f_4$ has more than two orders of $\tface$-decay, so we do not solve it away yet; the remainder is now $\hat f_5\in\Hb^{\infty,\ 2-\eps,\ 2+\eps_+,\ ((0,0),1+\eps_\cK)}$.

  \pfsubstep{Step~4.4.}{Conormal terms from the third $\tface$-normal operator inversion.} We can apply $\wt{L_b'}(0)^{-1}$ to $\hat f_5(0)$ (producing an element of $\Hb^{\infty,\ \eps_+}(X;S^2\cT^*_X)$, and elements of $\C^4$ and $\scalspace_1$), and are left to study $\wh{L_b}(\sigma)^{-1}$ acting on $\sigma$ times $\hat f_6\in\Hb^{\infty,\ 2-\eps,\ 1+\eps_+,\ ((0,0),\eps_\cK)}$ (which thus involves a total division by $\sigma^3$). The $\tface$-order of this being less than $2$, we now use Proposition~\ref{PropiptfC} repeatedly to solve away $\hat f_6$ at $\tface$ (each step improving the $\tface$-order by almost $\eps_\ind$). The most singular terms at $\zface$ arise here in the first step, where one applies Proposition~\ref{PropiptfC} with $\alpha=1+\eps_+$, $k_0=0$, and $\ell_\zface=\eps_\cK$; concretely, consider the terms in~\eqref{EqiptfCuPrec2} upon multiplication by the overall factor of $\sigma^3$. (Note that $u_b$ gets multiplied by $\chi_\zface$ in~\eqref{EqiptfCuPrec}, and $\sigma^3(1-\chi_\zface)u_b\in\sigma^3\Hb^{\infty,\ 1-\eps,\ \alpha-2,\ \infty}(X_\scbtop^\pm)$, with $\alpha=1+\eps_+$, has inverse Fourier transform in $\Hb^{\infty,\ 1-\eps,\ 3+\eps_+,\ \infty}(M';S^2\cT^*)$, which is a remainder term; thus, it suffices to study the inverse Fourier transform of $\sigma^3$ times~\eqref{EqiptfCuPrec2}.)
  \begin{enumerate}
  \myitem{ItID6tf31}{i} The contribution of the first term in~\eqref{EqiptfCuPrec2} is of class $\cF^{-1}(h_{b,\rms 1}^{\leq 4}(\Hb^{\infty, 1+\eps_+}(\pm[0,c];\scalspace_1)))$, so $h_{b,\rms 1}^{\leq 1}(\Hb^{\infty,\,2+\eps_+}([1,\infty]_{t_*};\scalspace_1))$, where we discard the terms from $\breve h_{b,\rms 1}^j$, $j=2,3,4$, since they come with $4+\eps_+>4+\eps_\cK$ orders of $\cK^+$-decay. This term can be absorbed into $\dot\scal_\rem$ in~\eqref{EqD6uexp} since $\eps_+>\eps_\cK$.
  \myitem{ItID6tf32}{ii} The second and third terms in~\eqref{EqiptfCuPrec2} contribute via $h_{b,\rmv 1}^{\leq 3}(\Hb^{\infty,3+\eps_+}([1,\infty]_{t_*};\vectspace_1))$ and $h_{b,\rms 0}^{(0),\leq 3}(\Hb^{\infty,3+\eps_+})$, respectively. (We do not need to record $\breve h_{b,\rmv 1}^j$ and $\breve h_{b,\rms 0}^{(0),j}$ terms for $j\geq 1$ since they come with $\cK^+$-decay order $4+\eps_+$, cf.\ Remark~\ref{RmkD6uexp}\eqref{ItD6uexpNum}.)
  \myitem{ItID6tf33}{iii} The other terms contribute at $t_*$-decay rates that are equal to $\eps_+$ plus a shift computed from the indicial root of $\wh{\ubar L}(0)$ they arise from; for example, the sum in the second line produces terms of class
    \[
      h_{b,\rms l}^{(\mu)}\bigl( \Hb^{\infty,\ 2+\eps_++\lambda^\Ups_{\rms l,l+j}}([1,\infty]_{t_*};\scalspace_l)\bigr)
    \]
    where $\mu=-\lambda^\Ups_{\rms l,l+j}+1$; the terms with $l+j\geq 2$ contribute remainder terms since then $2+\eps_++\lambda_{\rms l,l+j}^\Ups>4+\eps_+>4+\eps_\cK$, so only the cases $(l,j)=(0,1),(1,0)$ arise here.
  \end{enumerate}

  \pfsubstep{Step~4.5.}{Conormal terms from the inversion of $\wt{L_b'}(\sigma)$.} The remaining error is now $\hat f_7\in\Hb^{\infty,\ 2-\eps,\ 2+\eps_\ind-\eps,\ ((0,0),\eps_\cK)}$ (the definition of which involves a division by $\sigma^3$). We can solve this away at $\zface$ using $\wt{L_b'}(0)^{-1}$, with remaining error $\Hb^{\infty,\ 2-\eps,\ 2+\eps_\ind-\eps,\ \eps_\cK}$. To this error, finally, we apply the true (augmented) resolvent $\wt{L_b'}(\sigma)^{-1}$, which upon multiplication by $\sigma^3$ produces elements of $\Hb^{\infty,\ 1-\eps,\ 3+\eps_\ind-\eps,\ 3+\eps_\cK}$ (with inverse Fourier transform having $4+\eps_\cK$ orders at $\cK^+$) as well as the singular contribution
  \begin{equation}
  \label{EqD6PfSing}
  \begin{split}
    &\dot g_b^{\Ups,{\rm sing}}(\sigma)\bigl(\hat b(\sigma)\bigr) + h_{b,\rms 1}^{\leq 1,{\rm sing}}(\sigma)\bigl(\hat\scal(\sigma)\bigr), \\
    &\qquad \hat b \in \Hb^{\infty,\,3+\eps_\cK}(\pm[0,c]_\sigma;\C^4),\quad \hat\scal \in \Hb^{\infty,\,3+\eps_\cK}(\pm[0,c]_\sigma;\scalspace_1\otimes_\R\C)\bigr).
  \end{split}
  \end{equation}
  Consider the individual contributions from~\eqref{EqAdmLoErrtf1CorrErr}, expanded using~\eqref{EqAdmLoErrtf1Corr}, to the first summand. First of all, we can remove the cutoffs $\chi_\zface$ at the price of producing error terms whose inverse Fourier transforms have Schwartz decay at $\cK^+$ and order $4+\eps_\cK>3+\eps_+$ at $\iota^+$.
  \begin{itemize}
  \item The inverse Fourier transform of $i\sigma^{-1}\dot g_b^\Ups(\hat b(\sigma))$ (multiplied by a cutoff function $\chi(|\sigma|)\in\CIc((-c,c))$) is $\dot g_b^\Ups(\dot b_\rem)$ for $\dot b_\rem\in\Hb^{\infty,3+\eps_\cK}(\R_{t_*};\R^4)$. (In view of the forward support property of $u$, we may multiply $\dot b_\rem$ by a cutoff function in $t_*$ to make it have support in $[1,\infty]_{t_*}$.)
  \item $\cF^{-1}\bigl(\log(\hat r)\hat b(\sigma)h_{b,\rms 0}^{(0),\leq 3}(\sigma)\bigr)$ is the sum of two terms. The first is $\cF^{-1}\bigl(\log(|\sigma|)\hat b(\sigma)h_{b,\rms 0}^{(0),\leq 3}(\sigma)\bigr)$ and can (in light of $(3+\eps_\cK-\eps)+1>3+\eps_+$) be absorbed into the spacetime-$h_{b,\rms 0}^{(0)}(\Hb^{\infty,3+\eps_+})$-term mentioned in Step~4.4\eqref{ItID6tf32}. The second term is the inverse Fourier transform of $\log(\rho)\hat b(\sigma)h_{b,\rms 0}^{(0),\leq 3}(\sigma)\in\Hb^{\infty,\ -\eps,\ 3+\eps_\cK-\eps,\ 3+\eps_\cK}$, thus of class $\Hb^{\infty,\ -\eps,\ 4+\eps_\cK-\eps,\ 4+\eps_\cK}(M';S^2\cT^*)$; the $\iota$-order of this is $>3+\eps_+$, and the $\cK$-order is $4+\eps_\cK$ as required for remainder terms.
  \item The contribution from $|\sigma|^{\lambda^\Ups_{\rms 0,1}-1}$ in~\eqref{EqAdmLoErrtf1Corr} has total $|\sigma|$-weight $3+\eps_\cK+\lambda^\Ups_{\rms 0,1}-1\geq 3+\eps_\cK$ and thus contributes remainder terms upon inverse Fourier transforming.
  \item The contribution from $\dot g_{b,\tface}^{\Ups,{\rm corr}}(\hat b(\sigma))$ is of class $\Hb^{\infty,\ 1-\eps,\ 3+\eps_\cK-\eps,\ 3+\eps_\cK}(X_\scbtop^\pm)$ for all $\eps>0$ and thus has inverse Fourier transform of class $\Hb^{\infty,\ 1-\eps,\ 4+\eps_\cK-\eps,\ 4+\eps_\cK}(M')$, which is again an acceptable remainder term.
  \end{itemize}
  We similarly study the contributions from~\eqref{EqAdmLoErrtf2CorrErr}, expanded using~\eqref{EqAdmLoErrtf2Corr}, to the second summand of~\eqref{EqD6PfSing}:
  \begin{itemize}
  \item The inverse Fourier transform of $-\sigma^{-2}h_{b,\rms 1}(\hat\scal(\sigma))+i\sigma^{-1}\breve h_{b,\rms 1}^1(\hat\scal(\sigma))$ is equal to $h_{b,\rms 1}^{\leq 1}(\dot\scal_\rem)$ for $\dot\scal_\rem\in\Hb^{\infty,2+\eps_\cK}(\R_{t_*};\scalspace_1)$.
  \item The contribution from the part
    \[
      -\sigma^{-1}(\log|\sigma|)h_{b,\rms 1} + (\log|\sigma|) i\breve h_{b,\rms 1}^1
    \]
    of the first summand of~\eqref{EqAdmLoErrtf2Corr}, multiplied by $\hat\scal$, can be absorbed into the term $\dot\scal_\rem$ just discussed. Two terms remain. The first is the term $\log(\rho)\breve h_{b,\rms 1}^1(\Hb^{\infty,3+\eps_\cK})$ which is of class $\Hb^{\infty,\ 1-\eps,\ 4+\eps_\cK-\eps,\ 3+\eps_\cK}(X_\scbtop^\pm)$, and whose inverse Fourier transform is thus an acceptable remainder. The second type of term is of the form $(\log\hat r)\Hb^{\infty,3+\eps_\cK}(\pm[0,c]_\sigma)\cdot \sigma^j\breve h_{b,\rms 1}^{1+j}$, $j=1,2,3$, which is of class $\Hb^{\infty,\ -j-\eps,\ 3+\eps_\cK-\eps,\ 3+j+\eps_\cK-\eps}$ and thus contributes a remainder term upon inverse Fourier transforming.
  \item The contributions from the $|\sigma|^{\lambda^\Ups_{\rms 1,1}-1}$-term and of $h_{b,\rms 1,\tface}^{\leq 1,{\rm corr}}$ are acceptable remainder terms as before.
  \end{itemize}
  This completes the proof of Theorem~\ref{ThmD6}.
\end{proof}

\subsubsection{Absorbing the conormal expansion into the gauge-fixed Einstein operator}
\label{SssD6Abs}

--- \emph{We continue using the notation from~\S\usref{SssD6Undo}.} The decomposition~\eqref{EqD6u} suggests regarding the parameters~\eqref{EqD6uexpComp} of the partial polyhomogeneous expansion $u_{\rm exp}$ at $\cK^+$ as further arguments of an augmented version of the linearized gauge-fixed Einstein operator of Definition~\ref{DefD6EinsteinAug}, which would also enable us to separately record the strong decay of the remainder term $\tilde u$ in~\eqref{EqD6tildeu}. We moreover recall the discussion regarding the continuation of exterior solutions $h_{-\scal}$ around~\eqref{EqD2PFwdCutoff}--\eqref{EqD2PFwd}; for present purposes, it suffices to work in the set $\{\ft_*\geq 1\}$, so we assume that $h_{-\scal}$---defined for $\ft_*\leq 2$, with $\scri^+$-order $(\la\cE_\sscri^\cC\ra,3+\eps_\sscri)$, and depending smoothly on $\scal$ as in Theorem~\ref{ThmExBoStab}---solves
\begin{equation}
\label{EqD6AbsExt}
  \Ric(g_{b_0,b,-\scal}+h_{-\scal}) - \delta_{g_{b_0,b,-\scal},E^\cC}^*\Ups_{E^\Ups}(g_{b_0,b,-\scal}+h_{-\scal},\,g_{b_0,b,-\scal}) = 0\quad\text{on}\ \ft_*^{-1}([1,2]).
\end{equation}
We assume $h_{-\scal}$ to be small in some fixed high-regularity norm. (Since presently we are concerned with solutions of linearized equations with arbitrary source terms $f$, the reader may as well only consider the case that $h_{-\scal}=0$ for all $\scal$ in this section.) As in~\eqref{EqD2PFwdCutoff}, fix $\chi_-\in\CI(\R)$ to be equal to $1$ on $(-\infty,\frac74]$ and $0$ on $[2,\infty)$. This leads to:

\begin{definition}[Gauge-fixed Einstein operator, augmentation \#5]
\label{DefD6Aug5}
  Fix $h_{-\scal}$ as above. Let
  \begin{equation}
  \label{EqD6Aug5h}
    h\in\cX^\infty:=\Hb^{\infty,\ \bigl(\la\cE_\sscri^\cC\ra,3+\eps_\sscri\bigr),\ (\cE_+,3+\eps_+),\ 4+\eps_\cK}(\Omega_*)^{\bullet,-}
  \end{equation}
  be small in $\cX^d$ for some large but fixed $d$. Let $\vecp=(\vecp^{\leq 0},\vecp^{\in(0,2]})$ be as in Definition~\usref{DefD6EinsteinAug} and small, with $\vecp^{\leq 0}:=(\scal,b-b_0,\scal_1^{(0)},\scal_1^{(0,1)})$. Let
  \begin{subequations}
  \begin{equation}
  \label{EqD6Aug5q}
  \begin{split}
    \vecq &= \Bigl(\scal_\rem,\ b_\rem,\ \vect_1,\ \scal_0^{(0)},\ \bigl(\scal_l^{(-\lambda^\Ups_{\rms l,l+j}+1)}\bigr)_{ \substack{ 0\leq l\leq 3,\ j=0,1, \\ 1\leq l+j\leq 3 } }\,, \\
      &\quad\qquad \bigl(\scal_l^{(-l+2)}\bigr)_{2\leq l\leq 5},\ \bigl(\vect_l^{(-l+1)}\bigr)_{2\leq l\leq 4},\ \scal_2^{(-2)},\ \vect_2^{(-2)},\ \scal_3^{(-3)},\ \vect_3^{(-3)} \Bigr),
  \end{split}
  \end{equation}
  where $\scal_\rem\in\dot H_\bop^{\infty,\,2+\eps_\cK}([1,\infty]_{t_*};\scalspace_1)$, $b_\rem\in\dot H_\bop^{\infty,\,3+\eps_\cK}([1,\infty];\R^4)$, and (using Definition~\usref{DefD6Order})
  \begin{equation}
  \label{EqD6Aug5qCoeff}
    \sfW_l^{(-\mu)} \in \dot H_\bop^{\infty,\ \scissors(\cE_++\mu,3+\eps_++\mu)}([1,\infty]),\quad \sfW=\scal,\vect,
  \end{equation}
  \end{subequations}
  as well as $\vect_1\in\dot H_\bop^{\infty,\ \scissors(\cE_+,3+\eps_+)}([1,\infty];\vectspace_1)$; here $(\lambda,k)$ lies in an index set $\tilde\cE'_{++}$ with $\min\Re\tilde\cE'_{++}\geq 1+\eps_\ind>1$ as in Definition~\ref{DefD5EinsteinAug}. We assume the components of $\vecq$ to be small in their respective partially polyhomogeneous weighted $\Hb^d$-spaces. We then define
  \begin{equation}
  \label{EqD6Aug5Op}
  \begin{split}
    P(h,\vecp,\vecq) &:= \Ric\bigl(g^0+h_\tot(\vecp) + h_{b,\rm rem}(\vecq) + \chi_-h_{-\scal}+h\bigr) \\
      &\qquad - \delta_{g^0,E^\cC}^*\Bigl( \Ups_{E^\Ups}\bigl(g^0+h_{b,\rm rem}(\vecq)+\chi_- h_{-\scal}+h,g^0\bigr) - \vartheta_\tot(\vecp)\Bigr),
  \end{split}
  \end{equation}
  where $g^0=g_{b_0,b,-\scal}$, further $h_\tot(\vecp)$ and $\vartheta_\tot(\vecp)$ are given by~\eqref{EqD6EinsteinAugTot}, and
  \begin{equation}
  \label{EqD6Aug5hrem}
    h_{b,\rm rem}(\vecq) := \chi_\cK\Bigl( h_{b,\rms 1}^{\leq 1}\bigl(\scal_\rem(t_*)\bigr) + \dot g_b^\Ups\bigl( b_\rem(t_*)\bigr) + h_{b,\rmv 1}^{\leq 3}\bigl(\vect_1(t_*)\bigr) + \sum h_{b,\rmw l}^{(-\mu),\leq j}\bigl(\sfW_l^{(-\mu)}(t_*)\bigr)\Bigr),
  \end{equation}
  where the sum is over all choices of $(\bfw,\sfW)=(\rms,\scal),(\rmv,\vect)$ and $\mu,j$ appearing in~\eqref{EqD6uexp}.
\end{definition}

The idea is that the argument
\begin{equation}
\label{EqD6gtot}
  g:=g^0+h_\tot(\vecp)+h_{b,\rm rem}(\vecq)+\chi_- h_{-\scal}+h
\end{equation}
of $\Ric$ in~\eqref{EqD6Aug5Op} will, ultimately, be the full spacetime metric solving the initial value problem (with initial data as in~\eqref{EqExBoIVP}--\eqref{EqExBoIVPData}) globally: this defines the global future extension of the initial solution $g^0+h_{-\scal}=\phi_{-\scal}^*g_{b_0}+h_{-\scal}$ on $\{\ft_*\leq 1\}$. The support assumption on $h$ in~\eqref{EqD6Aug5h} is precisely what is needed to ensure that $g=g^0+h_{-\scal}$ for $\ft_*\leq 1$. (Recall from~\eqref{EqDCutoffs} that $h_\tot(\vecp)$ and $h_\rem(\vecq)$ vanish for $\ft_*\leq 10$.)

\begin{rmk}[Orders]
\fakephantomsection
\label{RmkD6Aug5Comm}
  \begin{enumerate}
  \item Observe that the $\cK^+$-order of~\eqref{EqD6Aug5h} is $4+\eps_\cK$, but due to the term $h_\rem(\vecq)$ we have $g-g_b\in\Hb^{\infty,\ ((1,1)\cup\cE_+,3+\eps_+),\ 2+\eps_\cK}$ away from $\scri^+$ (thus omitting the $\scri^+$-order)---with $\cK^+$-order thus equal to $2+\eps_\cK$ only---and therefore the linearization $L$ of $P$ in the first argument (for small $h,\vecp,\vecq$) has all the properties recorded in Proposition~\ref{PropDAdmFw}, Corollary~\ref{CorDAdm}, and Theorem~\ref{ThmDAdmReg}.
  \item As far as the index sets of the components of $\vecq$ are concerned, we use shifts of $\cE_+$ here much as in~\eqref{EqD6uexpComp} where instead $\tilde\cE_+$ is used; we can self-consistently arrange for $\cE_+\supset\tilde\cE_+$. The choice $\cE_+$ is the largest one such that the $\iota^+$-order of $h_\rem(\vecq)$ is $(\cE_+,3+\eps_+)$ and thus compatible with what we have throughout been allowing for the metric perturbation $h$ (as in~\eqref{EqD6Aug5h}).
  \end{enumerate}
\end{rmk}

Despite the weaker decay of $h_{b,\rm rem}(\vecq)$, a crucial observation (in the context of the discussion around~\eqref{EqD5Almh} and the linearization in $b$ discussed in Lemma~\ref{LemmaD5LinPar}) is that the linearization of $P$ in the final black hole parameters $b$ has $4+\eps_\cK$ orders of decay at $\cK^+$, as we show in Lemma~\ref{LemmaD6Aug5Linb} below, the reason being that the extra terms $h_{b,\rm rem}(\vecq)$ lie in the approximate nullspace of the ``linear model'' $L_b$ of $P$ at $\cK^+$. The linearizations of $P$ in all other parameters of $\vecp$ have the same structure as in Lemma~\ref{LemmaD6LinPar} (which refers back to Lemmas~\ref{LemmaD4LinPar} and \ref{LemmaD5LinPar}). Furthermore, since $\vecq$ and $h$ enter always as the sum $h_{b,\rm rem}(\vecq)+h$, the linearization of $P$ in $\vecq$ can be expressed in terms of the linearization in $h$, in the sense that
\begin{equation}
\label{EqD6Aug5Linq}
  D_{(h,\vecp,\vecq)}P(0,0,\vecq') = D_{(h,\vecp,\vecq)}P\bigl( h_{b,\rm rem}(\vecq'),0,0 \bigr).
\end{equation}
(As many times before, we use here that the right-hand side is a differential operator in the first argument and can thus be applied to any symmetric 2-tensor regardless of decay; so even though $h_{b,\rm rem}(\vecq')$ has weaker decay than what is allowed for in~\eqref{EqD6Aug5h}, the right-hand side of~\eqref{EqD6Aug5Linq} is well-defined.)

\begin{lemma}[Linearization in $b$]
\label{LemmaD6Aug5Linb}
  The linearization of $P$ in $b-b_0$ satisfies
  \[
    D_{(h,\vecp,\vecq)}P(b') \in \chi_\cK^\sharp\Hb^{\infty,\ 0,\ \bigl((\tilde\cE_+\cup\tilde\cE'_{++})+2,\,6+\eps_+\bigr),\ 4+\eps_\cK}.
  \]
\end{lemma}
\begin{proof}
  In view of Lemma~\ref{LemmaD5LinPar}, only the $\cK^+$-order requires justification. To this end, we note that on the neighborhood of $\cK^+$ where $\chi_\cK=1$, we have $g^0=g_b$ and thus (using~\eqref{EqD6Aug5Linq})
  \[
    P(h,\vecp,\vecq) = P(0,0,0) + \int_0^1 D_{(s h,s\vecp,0)}P(h,\vecp,0)\,\dd s + \int_0^1 D_{(h,\vecp,s\vecq)}P\bigl( h_{b,\rm rem}(\vecq),0,0 \bigr)\,\dd s.
  \]
  The first summand vanishes near $\cK^+$. The second summand has $\cK^+$-order $4+\eps_\cK$ since $h$ and $h_\tot(\vecp)$ do. The third term is the new one. Now, $D_{(h,\vecp,s\vecq)}P$ differs from $L_b$ by a b-differential operator with $\Hb^\infty$-coefficients that have $\cK^+$-order $2+\eps_\cK$ (since this is the size of the deviation of $g$ in~\eqref{EqD6gtot} from $g_b$), and this difference maps $h_{b,\rm rem}(\vecq)$ into a term with $\cK^+$-order $2(2+\eps_\cK)>4+\eps_\cK$. It then remains to show that
  \begin{equation}
  \label{EqD6Aug5Lb}
    L_b h_{b,\rm rem}(\vecq)\ \text{has $\cK^+$-order $4+\eps_\cK$.}
  \end{equation}
  We do this using formulas such as~\eqref{EqipGrKerrComp}, the point being that the right-hand side of~\eqref{EqipGrKerrComp} will involve derivatives of the arguments $\vecq$ of sufficiently high order so that they are of class $\dot H_\bop^{\infty,\,4+\eps_\cK}([1,\infty]_{t_*})$. Let us verify this for three exemplary terms. First, let $\dot\scal_\rem\in\dot H_\bop^{\infty,\,2+\eps_\cK}$ and consider
  \begin{align*}
    L_b h_{b,\rms 1}^{\leq 1}(\dot\scal_\rem(t_*)) &= \Bigl([L_b,t_*]\ftrans(0)\breve h_{b,\rms 1}^1\bigl(\dot\scal_\rem''(t_*)\bigr) + \frac12[[L_b,t_*],t_*]h_{b,\rms 1}\bigl(\dot\scal_\rem''(t_*)\bigr) \Bigr) \\
      &\qquad + \frac12[[L_b,t_*],t_*]\breve h_{b,\rms 1}^1\bigl(\dot\scal_\rem'''(t_*)\bigr).
  \end{align*}
  Since $\pa_{t_*}=t_*^{-1}\,t_*\pa_{t_*}$, every $t_*$-derivative increases the $t_*$-decay order by $1$ (at the expense of one b-derivative). Thus, $\dot\scal_\rem''\in\dot H_\bop^{\infty,\,4+\eps_\cK}$, likewise (a fortiori) for $\dot\scal_\rem'''$. Similarly, let $\dot b_\rem\in\dot H_\bop^{\infty,\,3+\eps_\cK}$; then
  \[
    L_b\dot g_b^\Ups\bigl(\dot b_\rem(t_*)\bigr) = [L_b,t_*]\ftrans(0)\dot g_b^\Ups\bigl(\dot b_\rem'(t_*)\bigr) + \frac12[[L_b,t_*],t_*]\dot g_b^\Ups\bigl(\dot b_\rem''(t_*)\bigr)
  \]
  has $\cK^+$-order $4+\eps_\cK$ since $\dot b_\rem'\in\dot H_\bop^{\infty,\,4+\eps_\cK}$. Lastly, for $2\leq l\leq 5$,
  \begin{align*}
    L_b h_{b,\rms l}^{(-l+2),\leq 5-l}\bigl(\dot\scal_l^{(-l+2)}(t_*)\bigr) &= [L_b,t_*]\ftrans(0)\breve h_{b,\rms l}^{(-l+2),5-l}\bigl((\dot\scal_l^{(-l+2)})^{(6-l)}(t_*)\bigr) \\
      &\qquad + \frac12[[L_b,t_*],t_*]\breve h_{b,\rms l}^{(-l+2),4-l}\bigl((\dot\scal_l^{(-l+2)})^{(7-l)}(t_*)\bigr) \\
      &\qquad + \frac12[[L_b,t_*],t_*]\breve h_{b,\rms l}^{(-l+2),5-l}\bigl((\dot\scal_l^{(-l+2)})^{(7-l)}(t_*)\bigr),
  \end{align*}
  with the convention $\breve h_{b,\rms l}^{(-l+2),0}=h_{b,\rms l}^{(-l+2)}$ and $\breve h_{b,\rms l}^{(-l+2),-1}=0$. The point now is that the order of $\geq 6-l$ many $t_*$-derivatives of $\dot\scal_l^{(-l+2)}(t_*)$ (for $\dot\scal_l^{(-l+2)}$ of the same class as $\scal_l^{(-l+2)}$ in~\eqref{EqD6Aug5qCoeff}) at $t_*^{-1}=0$ is $\textscissors(\cE_++l-2,3+\eps_++l-2)+6-l$, which (in view of $\min\Re\cE_+\geq 1$) is $\geq 5>4+\eps_\cK$, as required.
\end{proof}

We can now rephrase Theorem~\ref{ThmD6} in the following way (and recall that tame estimates follow from the preceding arguments, cf.\ Remark~\ref{RmkDAdmRegTame}):

\begin{cor}[Decay of forward solutions]
\label{CorD6Impr}
  Let $h$, $\vecp$, and $\vecq$ be as in Definition~\usref{DefD6Aug5}. Let\footnote{More generally, we can let $f$ be as in~\eqref{EqD6f}, but this flexibility will not be useful in the sequel.}
  \begin{equation}
  \label{EqD6Imprf}
    f\in\Hb^{\infty,\ \bigl(\la\cE_\sscri^\cC+1\ra',\,4+\eps_\sscri\bigr),\ 5+\eps_+,\ 4+\eps_\cK}(\Omega_*)^{\bullet,-}.
  \end{equation}
  Then there exist unique $\vecp'$ and $\vecq'$ such that the forward solution $h'$ of
  \[
    L h' := D_{(h,\vecp,\vecq)}P(h',0,0) = f - D_{(h,\vecp,\vecq)}P(0,\vecp',\vecq')
  \]
  satisfies
  \begin{equation}
  \label{EqD6Imprhp}
    h' \in \Hb^{\infty,\ \bigl(\la\cE_\sscri^\cC\ra,\,3+\eps_\sscri\bigr),\ (\tilde\cE_\sharp,\,3+\eps_+),\ 4+\eps_\cK}(\Omega_*)^{\bullet,-},
  \end{equation}
  where $\tilde\cE_\sharp$ (with $\min\Re(\tilde\cE_\sharp\setminus\{(1,0)\})>1$) is as in Theorem~\usref{ThmD6}. Moreover, we have tame estimates: for $d$ large enough, and for all $k\in\N_0$, we have
  \[
    \|h'\|_k + |\vecp'| + \|\vecq'\|_k \leq C_k \Bigl( \|f\|_{k+d} + \bigl( \|h\|_{k+d}+|\vecp| + \|\vecq\|_{k+d}\bigr) \|f\|_d \Bigr);
  \]
  here we write $\|h'\|_k$ for the $\Hb^{k,\ \bigl(\la\cE_\sscri^\cC\ra,\,3+\eps_\sscri\bigr),\ (\tilde\cE_\sharp,\,3+\eps_+),\ 4+\eps_\cK}(\Omega_*)^{\bullet,-}$-norm of $h'$, further $|\vecp'|$ for the norm of $\vecp'$ (which lies in a fixed finite-dimensional vector space that is the direct sum of copies of $\scalspace_l$ and $\vectspace_l$), and $\|\vecq'\|_k$ for the sum of the norms of the components of~\eqref{EqD6Aug5q} in partially polyhomogeneous $\Hb^d$-spaces (so of $\|\scal_{\rm rem}'\|_{\dot H_\bop^{d,\,2+\eps_\cK}([1,\infty];\scalspace_1)}$, etc.).
\end{cor}
\begin{proof}
  The statement of Theorem~\ref{ThmD6} holds for $D_{(h,\vecp,\vecq)}P(h',0,0)=f-D_{(h,\vecp,\vecq)}P(0,\vecp',0)$; recall from the above discussion that this is due to the sufficient ($4+\eps_\cK$ orders of) $\cK^+$-decay of the linearization of $P$ in the parameters $\vecp$ and the fact that $D_{(h,\vecp,\vecq)}P(h',0,0)$ can equivalently be expressed as $D_{(h+h_{b,\rm rem}(\vecq),\vecp)}P(h',0)$ in the notation of Definition~\ref{DefD6EinsteinAug} used in the statement of Theorem~\ref{ThmD6}, with $h+h_{b,\rm rem}(\vecq)$ having sufficient ($2+\eps_\cK$ orders of) $\cK^+$-decay for the iterative procedure for improving the decay of forward solutions to apply verbatim. Thus, for $\vecq'=0$, we obtain the description~\eqref{EqD6u}--\eqref{EqD6tildeu} for $h'=\chi_\cK u_{\rm exp}+\tilde u$. But then we simply note (using~\eqref{EqD6Aug5Linq}) that
  \[
    D_{(h,\vecp,\vecq)}P( \chi_\cK u_{\rm exp}, 0, 0 ) = D_{(h,\vecp,\vecq)}P (0,0,\vecq')
  \]
  where the parameters $\vecq'$ are given by~\eqref{EqD6uexpComp}.
\end{proof}

Corollary~\ref{CorD6Impr} succeeds in proving the \emph{same} decay at $\scri^+$, $\iota^+$, and $\cK^+$ for the gravitational wave tail $h'$ as what we assumed on the tail $h$ in~\eqref{EqD6Aug5h} around which we linearized. On the other hand, given small but \emph{arbitrary} $(h,\vecp,\vecq)$ in the spaces of Definition~\ref{DefD6Aug5}, the nonlinear error $P(h,\vecp,\vecq)$ is typically \emph{not} of the class of $f$ allowed in Corollary~\ref{CorD6Impr}. (For example, for general $\rho_+$-leading-order terms $h_\iota^{(1,0)}$ of $h$ at $\iota^+$, the $\rho_+^3$-leading-order term of $P(h,0,0)$, which is given by $\ubar L(t_*^{-1}h_\iota^{(1,0)})$, is typically nonzero.) A similar observation applies to the output of $D_{(h,\vecp,\vecq)}P$ acting on general $(h',\vecp',\vecq')$. One thus expects that a significantly smaller amount of information than $(h',\vecp',\vecq')$ should suffice to describe solutions of $D_{(h,\vecp,\vecq)}P(\cdots)=f$ where $f$ is of class~\eqref{EqD6Imprf}; likewise, then, triples $(h,\vecp,\vecq)$ for which $P(h,\vecp,\vecq)$ is of class~\eqref{EqD6Imprf} should be parameterizable using the same smaller amount of information. Implementing this idea is the content of~\S\ref{SEf}; the proof of nonlinear stability in~\S\ref{SSt} will then be straightforward.

In the special case $(h,\vecp,\vecq)=(0,0,0)$, Corollary~\ref{CorD6Impr} implies the linear stability of the subextremal Kerr black hole with metric $g_{b_0}$. To state this precisely, we use the notation $t_\IVP$ from Lemma~\ref{LemmaKMetIVP} and set $\Sigma_\IVP=t_\IVP^{-1}(0)\cap M\cong\ol{\R^3_x}\setminus\{r<\bhm_0\}$.

\begin{cor}[Linear stability]
\label{CorD6LinStab}
  Suppose we are given initial data $(\dot\gamma,\dot k)$ at $\Sigma_\IVP$ that satisfy the linearization of the constraint equations around the Kerr data $(\gamma_{b_0},k_{b_0})$ as well as $\dot\gamma,r\dot k\in\Hb^{\infty,(\cE_0,3+\eps_0)}(\Sigma_\IVP;S^2 T^*\Sigma_\IVP)$ where $\min\Re\cE_0>1$. Then there exist linearized Kerr parameters $\dot b$ and a symmetric 2-tensor
  \begin{subequations}
  \begin{equation}
  \label{EqD6LinStabh}
    \dot h\in\Hb^{\infty,\ (\cE_0,3+\eps_0),\ (\la\cE_\sscri^\cC\ra,3+\eps_\sscri),\ (\cE_\sharp,3+\eps_+),\ ((3,0),3+\eps_\cK)}(\{t_\IVP\geq 0\})
  \end{equation}
  (the index sets referring to $I^0$, $\scri^+$, $\iota^+$, and $\cK^+$) where $\cE_\sharp$ is an index set with $\min\Re(\cE_\sharp\setminus\{(1,0),(1,1)\})>1$, such that
  \begin{equation}
  \label{EqD6LinStab}
    \dot g := \chi_\cK\dot g_{b_0}(\dot b) + \dot h
  \end{equation}
  \end{subequations}
  solves the initial value problem for the linearized Einstein vacuum equation $D_{g_{b_0}}\Ric(\dot g)=0$ with initial data $(\dot\gamma,\dot k)$.
\end{cor}

Note that \eqref{EqD6LinStabh} asserts $t_*^{-3}$-decay to the linearized Kerr metric in spatially compact sets, and $t_*^{-1}\log t_*$-decay globally as $t_*\to\infty$. (The logarithmic factor is absent near $\cK^+$ and near $\scri^+$; see Remark~\ref{RmkStFurther}\eqref{ItStFurtherFull} in the nonlinear setting.) In spatially compact sets, this improves on the rates $t_*^{-\frac32+}$ and $t_*^{-1-\alpha}$, $\alpha\in(0,1)$, established in \cite{AnderssonBackdahlBlueMaKerr} and \cite{HaefnerHintzVasyKerr,HaefnerHintzVasyKerrLarge}.

\begin{proof}[Proof of Corollary~\usref{CorD6LinStab}]
  Since we do not use this result later, we only sketch its proof. The exterior linear stability result is easily proved using the methods of~\S\S\ref{SssIGen0}--\ref{SssIGenScri}, and the global linear stability problem can then be reduced, using a cutoff argument similar to~\eqref{EqINFwd}, to the solution of the linear equation $L h':=D_{(0,0,0)}P(h',0,0)=f-D_{(0,0,0)}P(0,\vecp',\vecq')$ where $f$ has $\scri^+$-order $(\la\cE_\sscri^\cC+1\ra',4+\eps_\sscri)$ and is supported near $\ft_*^{-1}([1,2])$, and we are free to choose $\vecp'$ and $\vecq'$. But Corollary~\ref{CorD6Impr} produces $\vecp'$ and $\vecq'$ such that $h'$ is of class~\eqref{EqD6Imprhp}. Note now that the linearization of the argument of $\Ric$ in~\eqref{EqD6Aug5Op} around $(h,\vecp,\vecq)=(0,0,0)$ in the direction $(h',\vecp',\vecq')$ can \emph{almost} be written in the form~\eqref{EqD6LinStab}; only the term $h_{b_0,\rem}(\vecq')$ requires some post-processing. But notice that all of its terms (see~\eqref{EqD6Aug5hrem}) other than $\chi_\cK\dot g_b^\Ups(b_\rem(t_*))$ and those involving $\scal_l^{(-l)}$ and $\vect_l^{(-l)}$ for $l=2,3$ can be written as pure gauge terms plus errors with $4+\eps_\cK$ orders of $\cK^+$-decay. Pure gauge terms, i.e., terms in the range of $\delta_{g_{b_0}}^*$, are annihilated by $D_{g_{b_0}}\Ric$ and can therefore simply be dropped near $\cK^+$. More globally, terms of the form $\chi_\cK\delta_{g_{b_0}}^*\omega$ can be absorbed into the tail $\dot h$ upon replacing them by $-\dd\chi_\cK\otimes_s\omega$; we use here that $D_{g_{b_0}}\Ric\circ\chi_\cK\delta_{g_{b_0}}^*=-D_{g_{b_0}}\Ric\circ[\delta_{g_{b_0}}^*,\chi_\cK]$. The $l=2$ terms have $t_*^{-3}$ decay, and $\dot g_b^\Ups(b_\rem(t_*))$ has $\cK^+$-order $3+\eps_\cK$, and hence the result follows.
\end{proof}

\section{Efficient parameterization of metric perturbations}
\label{SEf}

Given the discussion following Corollary~\ref{CorD6Impr}, our next task is to remove extraneous pieces from $(h,\vecp,\vecq)$ (see Definition~\ref{DefD6Aug5}) and write $(h,\vecp,\vecq)=\Phi(U)$ for a carefully chosen set of unknowns $U$ and a map $\Phi$ (with $\Phi(0)=(0,0,0)$). Roughly speaking, $U$ should encode some terms in the asymptotic expansion of $h'$ in Corollary~\ref{CorD6Impr} at $\scri^+$ and $\iota^+$ as well as some of the (gauge) modification parameters $\vecp$ and $\vecq$. There are two competing requirements.
\begin{enumerate}
\myitem{ItEfDec}{1} On the one hand, we need
  \[
    P(\Phi(U))\in\sF^\infty:=\Hb^{\infty,\ \bigl(\la\cE_\sscri^\cC+1\ra',4+\eps_\sscri\bigr),\ 5+\eps_+,\ 4+\eps_\cK}(\Omega_*)^{\bullet,-}
  \]
  for all small $U$; here $P$ is the gauge-fixed Einstein operator from Definition~\ref{DefD6Aug5}. While contributions to $h$ of class $\Hb^{\infty,\ \bigl(\la\cE_\sscri^\cC\ra,3+\eps_\sscri\bigr),\ 3+\eps_+,\ 4+\eps_\cK}$ are always acceptable in that $P(\cdot,0,0)$ maps them into $\sF^\infty$ (see Corollary~\ref{CorDAdmFwN}), this is not the case for, say, $\rho_+^1$-leading-order terms of $h$ at $\iota^+$. Roughly speaking, the terms in the $\iota^+$-expansion of $h$ must lie in the kernel of the $\iota^+$-normal operator, or interact with other contributions of $\vecp$, $\vecq$, and (nonlinear terms arising from) earlier terms in the $\iota^+$-expansion of $h$ to give contributions at $\iota^+$ with $5+\eps_+$ orders of decay.
\myitem{ItEfLin}{2} On the other hand, our choice of $U$ must still provide all the degrees of freedom that are necessary to ensure that for the solution of the forward problem $D_{\Phi(U)}P(h',\vecp',\vecq')=f$ produced by Corollary~\ref{CorD6Impr}, we can write $(h',\vecp',\vecq')=D_U\Phi(U')$ for a suitable $U'$. That is, the parameterization by $U$ is comprehensive enough to capture the behavior of solutions of the linearized problem.
\end{enumerate}

Roughly speaking, point~\eqref{ItEfDec} amounts to constructing formal solutions at $\iota^+\cup\cK^+$ ``backwards'' from infinity, whereas point~\eqref{ItEfLin} means that the data in this construction are exhaustive from the perspective of the asymptotics produced by the linearized forward problem.

Observe that the $\scri^+$-order $(\la\cE_\sscri^\cC+1\ra',4+\eps_\sscri)$ of the output of $P$ in Corollary~\ref{CorDAdmFwN} (and then also of $P$ in Definition~\ref{DefD6Aug5}) is already allowed for as the $\scri^+$-order of source terms in Corollary~\ref{CorD6Impr}, so it is \emph{not} necessary to encode the asymptotics of $h$ at $\scri^+$ in a similarly efficient manner.\footnote{This could, in principle, be done, since one can show that all terms in the $\scri^+$-expansion except for the radiation fields of $\pi_{0 1}h$, $\pi_{1 /}h$, $\slpi_0 h$, and $\pi_{1 1}h$ are determined by these radiation fields as well as the exterior solution $h_{-\scal}$ from~\eqref{EqD6AbsExt} (see Theorem~\ref{ThmExBoStab}) via a simple integration procedure. On the linear level, this follows from the arguments in the proof of Proposition~\ref{PropDScriPhg}; the nonlinear case is similar.} Our main work thus concerns the description of asymptotic expansions at $\iota^+$. Correspondingly, our main tools are results on the $\iota^+$-normal operator family $N_{\iota^+}(\ubar L,\lambda)$, $\lambda\in\C$, defined by~\eqref{EqipSpecFam2Expr}, from Proposition~\ref{PropipNInv} and~\S\ref{SssipP}.

We first discuss efficient parameterizations and how to use them for solving a nonlinear equation on partially polyhomogeneous spaces in the case of a simple ODE toy model in~\S\ref{SsEfT}. We then turn to the gauge-fixed Einstein equation: in~\S\ref{SsEfP}, we deal with point~\eqref{ItEfDec} above, and in~\S\ref{SsEfS} we revisit Corollary~\ref{CorD6Impr} to show that the parameterization we introduce in~\S\ref{SsEfP} is flexible enough to capture solutions of the linearized gauge-fixed Einstein equation; this deals with point~\eqref{ItEfLin} above.

\subsection{An ODE toy model}
\label{SsEfT}

We work with $\C^2$-valued functions on $[0,1]_\tau$ and consider the linear operator
\[
  L_0 := \tau\pa_\tau - \begin{pmatrix} 1 & 0 \\ 0 & 2 \end{pmatrix}.
\]
Its normal operator family $L_0(\lambda)=\lambda-\diag(1,2)$ is invertible at all $\lambda\in\C\setminus\{1,2\}$. Writing $e_1,e_2$ for the two standard basis vectors of $\C^2$, the spaces of (quasi)homogeneous solutions at the exceptional values of $\lambda$ are spanned by
\[
  \lambda=1:\ u_{(1,0)}(\tau):=\tau e_1;\quad
  \lambda=2:\ u_{(2,0)}(\tau):=\tau^2 e_2.
\]
For the nonlinearity, we take $N(u,u)$ where
\[
  N(u,v) := \begin{pmatrix} 0 \\ u_1 v_1 \end{pmatrix},\quad u=\begin{pmatrix} u_1 \\ u_2 \end{pmatrix},\ v=\begin{pmatrix} v_1 \\ v_2 \end{pmatrix}.
\]
We shall work with the function spaces
\[
  \Hb^{k,\,(\cE,\alpha)} := \dot H_\bop^{k,\,(\cE,\alpha)}\bigl([0,1]_\tau;|\tfrac{\dd\tau}{\tau}|\bigr)
\]
which (recalling Definition~\ref{DefTMphg}) consist of all functions of the form
\[
  u(\tau) = \chi(\tau)\sum_{ \substack{(z,k)\in\cE \\ \Re z\leq\alpha} } \tau^z(\log\tau)^k u_{(z,k)} + \tilde u(\tau),
\]
where $\chi\in\CIc([0,1))$ equals $1$ near $0$, further $u_{(z,k)}\in\C^2$, and finally $\tilde u(\tau)$ is a $\C^2$-valued function that vanishes for $\tau>1$ and satisfies $\int_0^1 \tau^{-2\alpha}|(\tau\pa_\tau)^j\tilde u(\tau)|^2\,\frac{\dd\tau}{\tau}<\infty$ for $j=0,\ldots,k$.

It is easy to show using the contraction mapping principle (and the fact that $\Hb^1$ is an algebra under pointwise multiplication) that if $\alpha\in(0,1)$, then for all sufficiently small $f\in\Hb^{0,\alpha}$, there exists a unique solution $u\in\Hb^{1,\alpha}$ of the equation $L_0 u-N(u)=f$. We shall henceforth ignore questions of regularity and work only with spaces $\Hb^{\infty,\,(\cE,\alpha)}$, and thus study
\begin{equation}
\label{EqEfTODE}
  P(u) := L_0 u - N(u) = f \in \Hb^{\infty,\alpha}.
\end{equation}
When $\alpha\in(1,2)$, the solution of~\eqref{EqEfTODE} lies in $\bigcap_{\eps>0}\Hb^{\infty,1-\eps}$ but generically does not inherit the $\alpha$ orders of decay from $f$; instead, $u$ will be the sum of $c u_{(1,0)}$ (for some generically non-zero $c\in\C$) and a remainder $\tilde u\in\Hb^{\infty,\alpha}$. When $\alpha\in(2,3)$, then the remainder $\tilde u\in\bigcap_{\eps>0}\Hb^{\infty,2-\eps}$ generically does not inherit the $\alpha$ orders of decay from $f$; instead, it will feature further (quasi-homogeneous) terms including a multiple of $u_{(2,0)}$ (and, as we will see below, a $\tau^2\log\tau$-term caused by a resonant nonlinear interaction).

These refined, partially polyhomogeneous, descriptions of the solution $u$ of~\eqref{EqEfTODE} can easily be established after the fact (i.e., starting with $u\in\bigcap_{\eps>0}\Hb^{\infty,1-\eps}$) by writing $L_0 u=f+N(u)$ and using basic regular-singular ODE analysis to uncover the expansion of $u$ step-by-step. What we wish to do is to describe a different approach in which the partial expansion of $u$ is encoded directly in an efficient manner. That is, our goal is to
\[
  \text{\parbox{0.8\textwidth}{\it solve~\eqref{EqEfTODE} \emph{directly} in the partially polyhomogeneous category---more precisely, in a space $\Hb^{\infty,\,(\cE,\alpha)}$ for $\alpha\in(2,3)$ and with $\min\Re\cE>0$.}}
\]
The reason for this particular goal is that attaining it will already showcase most of the ideas required in the Einstein case discussed in~\S\S\ref{SsEfP}--\ref{SsEfS} below.

\medskip

\pfstep{Step~1. Solving the linearized equation.} We first discuss the analogue of Corollary~\ref{CorD6Impr}. For now, we do not specify $\cE$, except we require $\min\Re\cE>0$. Let $u\in\Hb^{\infty,\,(\cE,\alpha)}$, and consider the linearized problem
\begin{equation}
\label{EqEfTLin}
  D_u P(u') = L_0 u' - 2 N(u,u') = f \in \Hb^{\infty,\alpha}.
\end{equation}
It is easy to show that the unique forward solution (i.e., vanishing for $\tau>1$) $u'$ is bounded as $\tau\searrow 0$. Since $u$ has a positive amount $\beta>0$ of decay (take any $\beta\in(0,\min\Re\cE)$), the term $N(u,u')$ has $\beta$ orders more decay than $u'$ itself, so re-writing~\eqref{EqEfTLin} as
\begin{equation}
\label{EqEfTAnalysis}
  L_0 u'=f+2 N(u,u'),
\end{equation}
we can use regular-singular ODE analysis to improve control of $u'$ by $\beta$ orders at a time. Since the first indicial root is $1$, we thus obtain $u'\in\Hb^{\infty,\,((1,0),1+\beta)}$. (The $\tau^1$-term is a multiple of the first indicial solution $u_{(1,0)}$, but we do not record this yet.) We can continue, now noting that $N(u,u')\in\Hb^{\infty,\ (\cE+1,\,\min(\alpha+1,1+2\beta))}$ is partially polyhomogeneous with more strongly decaying conormal remainder. If $2\in\pi_1(\cE+1)$, the fact that $2$ is an indicial root of $L_0$ may create logarithmic terms $\tau^2(\log\tau)^k u'_{(2,k)}$, $u'_{(2,k)}\in\C^2$. (Furthermore, a scalar multiple of the second indicial solution $u_{(2,0)}$ may arise here.) The upshot is that such arguments give
\begin{equation}
\label{EqEfTup}
  u' \in \Hb^{\infty,\,(\cE_\sharp,\alpha)}
\end{equation}
for some index set $\cE_\sharp$ depending only on $\cE$ that can be computed from the indicial roots $1$ and $2$ of $L_0$ and extended unions of shifts of $\cE$ by positive amounts. At this step, one could thus enlarge $\cE$, if necessary, such that one can guarantee $\cE_\sharp\subset\cE$.

\medskip

\pfstep{Step~2. Parameterizing approximate solutions of $P$.} While~\eqref{EqEfTup}, with $\cE_\sharp\subset\cE$, might suggest solving $P(u)=f$ using an iteration scheme on the space $\Hb^{\infty,\,(\cE,\alpha)}$, this is misguided: after all, the error $P(u)$ for $u\in\Hb^{\infty,\,(\cE,\alpha)}$ is (owing to the linear term $L_0$) typically only of class $\Hb^{\infty,\,(\cE,\alpha)}$ itself; and the solution of the linearized equation~\eqref{EqEfTLin} with forcing $f\in\Hb^{\infty,\,(\cE,\alpha)}$ (e.g., $f=-P(u)$) typically has a larger index set than~\eqref{EqEfTup} (and indeed larger than $\cE$ itself). That is, an iteration scheme set up in this naive fashion does not close.

The analysis of~\eqref{EqEfTup} shows that the asymptotics of solutions of the linearized equation~\eqref{EqEfTLin} include two degrees of freedom, corresponding to the two indicial solutions $u_{(1,0)}$ and $u_{(2,0)}$ of $L_0$. Therefore, \emph{we must include their scalar prefactors, denoted by $c_{(1,0)},c_{(2,0)}\in\C$, in our parameterization.} (More abstractly speaking, this means that we are using the space of indicial solutions of $L_0$, with (quasi)homogeneity $\leq\alpha$, in our parameterization.) For any $c_{(1,0)},c_{(2,0)}\in\C$ then, and for any $\tilde u\in\Hb^{\infty,\alpha}$ (which we also include in our parameterization), set
\[
  u_{[0]}:=c_{(1,0)}u_{(1,0)}+c_{(2,0)}u_{(2,0)}+\tilde u;
\]
we then have
\[
  P(u_{[0]}) = L_0 u_{[0]} - N(u_{[0]}) = L_0\tilde u - N(u_{[0]},u_{[0]}) \in \Hb^{\infty,\,\alpha} + \Hb^{\infty,\,((2,0),1+\alpha)} \subset \Hb^{\infty,\,((2,0),\alpha)}.
\]
While this decays better than the $\tau^1$-decay of the leading-order term of $u_{[0]}$, the nonlinear terms of $P$ (and if we replaced $L_0$ by an operator $L=L_0+\tilde L$ where $\tilde L\in\tau\Diffb^1$: the sub-leading linear terms of $L$) are responsible for the lack of decay relative to $\Hb^{\infty,\alpha}$. 

Now, adding, say, $\tau^\lambda u_0$ to $u_{[0]}$ where $\Re\lambda\in(1,2]$ would cause the output of $P$ to contain $(\lambda,0)$ in its index set \emph{unless} $\lambda=2$ and $u_0$ is a multiple of $u_{(2,0)}$; thus, the only correction we can make to $u_{[0]}$ without destroying the good decay of $P(\cdot)$ that we have already arranged must take place at the $\tau^2$-level. \emph{If} $L_0(2)$ were invertible (in which case $u_{(2,0)}$ would be absent), one could simply set $u_{[1]}:=u_{[0]}-\tau^2 L_0(2)^{-1}f_{[0],(2,0)}$ where
\[
  \tau^2 f_{[0],(2,0)},\quad f_{[0],(2,0)}\in\C^2,
\]
is the $\tau^2$-term of $P(u_{[0]})$. Nonlinear interactions of $u_{[0]}$ and this new term are of size $\tau^3$ and thus contained in the remainder space $\Hb^{\infty,\,\alpha}$ already. We would thus get $P(u_{[1]})\in\Hb^{\infty,\,\alpha}$ in this case. \emph{But} $L_0(2)$ is not invertible, and thus instead we need to resort to quasi-homogeneous correction terms. One can indeed solve
\begin{equation}
\label{EqEfTLog}
  L_0 v_{[1]} = -\tau^2 f_{[0],(2,0)},\quad v_{[1]}=\tau^2\log(\tau)v_{[1],(2,1)}+\tau^2 v_{[1],(2,0)}
\end{equation}
where $v_{[1],(2,1)}$ is a multiple of $\tau^{-2}u_{(2,0)}$ (uniquely determined by $f_{[0],(2,0)}$), while $\tau^2 v_{[1],(2,0)}$ is unique modulo the (quasi)homogeneous nullspace $\C u_{(2,0)}$ of $L_0$ at weight $\tau^2$; one can force uniqueness by requiring orthogonality of $v_{[1],(2,0)}$ to some element of $\C^2$ that is not orthogonal to $e_2=\tau^{-2}u_{(2,0)}$. (An elegant way to solve this equation uses the Mellin transform.) The indeterminacy of $v_{[1],(2,0)}$ is already accounted for by the parameter $c_{(2,0)}$. (Notice that in either case it is not necessary to compute $f_{[0],(2,0)}$ explicitly, though in this simple toy model one could.) Overall, one then sets $u_{[1]} := u_{[0]} + v_{[1]}$, which by the above prescription is completely determined by
\begin{subequations}
\begin{equation}
\label{EqEfTU}
  U := (c_{(1,0)},c_{(2,0)},\tilde u) \in \C \oplus \C \oplus \Hb^{\infty,\,\alpha}.
\end{equation}
We have thus defined
\begin{equation}
\label{EqEfTPhi}
  \Phi(U) := u_{[1]},
\end{equation}
\end{subequations}
and $P(\Phi(U))\in\Hb^{\infty,\alpha}$ indeed. We remark that $\Phi(U)\in\Hb^{\infty,\,((1,0)\cup(2,1),\alpha)}$.

\medskip

\pfstep{Step~3. Description of solutions of the linearized equation.} We wish to prove that given small $f\in\Hb^{\infty,\alpha}$, there exists $U$ of the form~\eqref{EqEfTU} such that $P(\Phi(U))=f$. A Newton-type iteration scheme for solving this requires us to be able to solve the linearized equation
\begin{equation}
\label{EqEfTLinPar}
  D_U(P\circ\Phi)(U')=f':=f-P(\Phi(U)) \in \Hb^{\infty,\alpha}
\end{equation}
for $U'$. Since
\[
  D_U(P\circ\Phi)(U') = D_u P \bigl( D_U\Phi(U') \bigr),\quad u:=\Phi(U),
\]
the solvability of~\eqref{EqEfTLinPar} is a nontrivial matter because it amounts to proving that the solution $u'\in\Hb^{\infty,\,(\cE_\sharp,\alpha)}$ of the linear equation $D_u P(u')=f'\in\Hb^{\infty,\,\alpha}$ (which is the same as~\eqref{EqEfTLin} up to relabeling $f'$ as $f$) from~\eqref{EqEfTup} is, in fact, of the very specific form
\begin{equation}
\label{EqEfTLinup}
  u' = D_U\Phi(U')
\end{equation}
for some (unique) $U'$.

We prove~\eqref{EqEfTLinup}, \emph{starting with~\eqref{EqEfTup}}, as follows. Consider again the equation
\[
  D_u P(u') = L_0 u' - 2 N(u,u') = f'.
\]
Consider the term $\tau^\lambda u_0$, $u_0\in\C^2$, with the least amount of decay in the expansion of $u'$, and suppose that $\Re\lambda\leq 1$. (We ignore the possibility of logarithmic factors at this stage; they can be easily handled.) Then $L_0(\lambda)u_0=0$, which implies that $u_0=0$ unless $\lambda=1$, in which case $\tau u_0=c_{(1,0)}u_{(1,0)}$ for some $c'_{(1,0)}\in\C$. We then note:
\begin{enumerate}[label=(\roman*)]
\item We have $D_U\Phi(c'_{(1,0)},0,0)\equiv c'_{(1,0)}u_{(1,0)}$ modulo terms with more (in fact, almost $2$ orders of) decay. Therefore, $u^{\prime,1}:=u'-D_U\Phi(c'_{(1,0)},0,0)$ has trivial $\tau^1$-coefficient and thus satisfies
  \begin{equation}
  \label{EqEfTupp}
    u^{\prime,1} \in \Hb^{\infty,\,(\cE_\sharp^{>1},\alpha)},\quad \cE_\sharp^{>1}:=\{(z,k)\in\cE_\sharp\colon\Re z>1\}.
  \end{equation}
  (We enlarge $\cE_\sharp$, if necessary, so that also $(2,1)\in\cE_\sharp$.)
\item The term $D_U\Phi(c'_{(1,0)},0,0)$ lies in the approximate nullspace of $D_u P$, since
  \begin{equation}
  \label{EqEfTDPDPhi}
  \begin{split}
    D_u P\bigl(D_U\Phi(c'_{(1,0)},0,0)\bigr) &= D_{\Phi(U)}P\bigl(D_U\Phi(c'_{(1,0)},0,0)\bigr) \\
      &= D_U(P\circ\Phi)(c'_{(1,0)},0,0) \in \Hb^{\infty,\alpha}.
  \end{split}
  \end{equation}
  This membership follows from the fact that, by construction, $P\circ\Phi$ takes values in $\Hb^{\infty,\alpha}$, and thus so does its linearization.
\end{enumerate}

In combination, we thus see that
\[
  D_u P(u^{\prime,1}) = f' - D_u P\bigl(D_U\Phi(c'_{(1,0)},0,0)\bigr) =: f^{\prime,1} \in \Hb^{\infty,\alpha}
\]
still has the same (strong) decay as $f'$ in~\eqref{EqEfTLinPar}; and $u^{\prime,1}$ has no $\tau^1$-leading-order term anymore, but rather is of size $o(\tau)$ as $\tau\to 0$. We continue the asymptotic analysis of this equation, keeping in mind the strong decay~\eqref{EqEfTupp} of $u^{\prime,1}$. We thus use $L_0 u^{\prime,1}=f^{\prime,1}+2 N(u,u^{\prime,1})$ and normal operator arguments for $L_0$ to extract the next term in the expansion of $u^{\prime,1}$; this must arise from the indicial root $2$, i.e., $u^{\prime,1}$ must be equal to $c'_{(2,0)}u_{(2,0)}$ plus a term with more decay still.\footnote{The $\tau^2\log(\tau)$-term of $u'$, cf.\ \eqref{EqEfTLog}, is absent in $u^{\prime,1}$: subtraction of $D_u\Phi(c'_{(1,0)},0,0)$ from $u'$ has (necessarily) eliminated it.} Mirroring the previous arguments, we then observe that $D_U\Phi(0,c'_{(2,0)},0)$ has the same structure and lies in the approximate kernel of $D_u P$; therefore,
\begin{align*}
  u^{\prime,2} := u^{\prime,1} - D_U\Phi(0,c'_{(2,0)},0) \implies &u^{\prime,2}\in\Hb^{\infty,\beta},\ \beta>2, \\
    &D_u P(u^{\prime,2})=f^{\prime,1}-D_u P\bigl(D_U\Phi(0,c'_{(2,0)},0)\bigr)=:f^{\prime,2}\in\Hb^{\infty,\alpha}.
\end{align*}
Using $L_0 u^{\prime,2}=f^{\prime,2}+2 N(u,u^{\prime,2})\in\Hb^{\infty,\alpha}$, the $o(\tau^2)$-decay of $u^{\prime,2}$, and the absence of indicial roots with real part in $(2,\alpha]$, this now implies
\[
  u^{\prime,2} \in \Hb^{\infty,\alpha},
\]
which is the remainder space of our data~\eqref{EqEfTU}. We have $u^{\prime,2}=D_U\Phi(0,0,u^{\prime,2})$, so $D_U(P\circ\Phi)(0,0,u^{\prime,2})=f^{\prime,2}$, and hence
\[
  U' := (c'_{(1,0)},c'_{(2,0)},u^{\prime,2}) \implies D_U(P\circ\Phi)(U') = f',
\]
as desired. This solves~\eqref{EqEfTLinPar}, which in combination with a nonlinear iteration scheme allows us to solve $P(\Phi(U))=f$ for small $f$.

\subsection{Parameterizing approximate solutions of the gauge-fixed Einstein operator}
\label{SsEfP}

Recall that our objective is to find a parameterization of $(h,\vecp,\vecq)$ in Definition~\ref{DefD6Aug5} that suffices to fully describe $\iota^+$-asymptotics as well as all (conormal remainder) terms with $\iota^+$-decay order $3+\eps_+$. We first present some motivation for our eventual choice.

The conormal remainder terms, which are the analogues of $\tilde u$ in~\eqref{EqEfTU}, are contributions of class $\Hb^{\infty,\ (\la\cE_\sscri^\cC\ra,3+\eps_\sscri),\ 3+\eps_+,\ 4+\eps_\cK}$ to $h$ in~\eqref{EqD6Aug5h} as well as the remainder terms of $\vecq$ in~\eqref{EqD6Aug5q}, i.e., with $\sfW_l^{(-\mu)}\in\dot H_\bop^{\infty,\,3+\eps_++\mu}$ instead of~\eqref{EqD6Aug5qCoeff}, unless $3+\eps_++\mu\geq 4+\eps_\cK$ in which case one drops this coefficient altogether; the latter scenario happens for all $\mu\geq 1$.

For the terms in the $\iota^+$-expansion, the basic idea is to use the solvability theory for the linear Minkowskian $\iota^+$-normal operator $\ubar L$ (defined in~\eqref{EqWEOpMink}) in Corollary~\ref{CoripPQhom} to iteratively compute what the terms in the $\iota^+$-expansion of a metric perturbation must be equal to. The indeterminacy of the term at order $t_*^{-\lambda}$ (where $\Re\lambda\leq 3+\eps_+<3+\eps_\sscri$) captured by the space $\Resspace_{\iota^+}^k(\ubar L,\cE_\sscri^\cC,\lambda)$ from Definition~\ref{DefipPRes} must be regarded as a further parameter (which we do concretely via~\eqref{EqipPResPar}); we recall here that this space is trivial for $\Re\lambda\leq 1$ unless $\lambda=1$. This is the analogue of the parameters $c_{(1,0)}$ and $c_{(2,0)}$ in~\eqref{EqEfTU}.

The terms produced by Corollary~\ref{CoripPQhom} (which applies also to elements of $\Resspace_{\iota^+}^k$) have an asymptotic expansion at $\iota^+\cap\cK^+$; since we must ensure a high order ($\geq 2+\eps_\cK$) of vanishing of dynamical metrics at $\cK^+$, we can thus determine the modification parameters required to ensure the triviality of the first few terms (as many as needed). For example, in order to eliminate a term $t_*^{-1}R^0\ubar h_{\rms 2}^{(0)}$ in the expansion of the $t_*^{-1}$-term $u^{(1,0)}$ of a metric perturbation $u$ at $\iota^+$, one must modify the equation satisfied by $u^{(1,0)}$ to include a source term that is a multiple of (the $t_*^{-3}$ leading-order term of) $\ubar L\bigl(\chi_\cK\ubar\delta^*\ubar\omega_{\rms 2}^{(-1),\leq 3}(t_*^{-1})\bigr)$; note that this tensor vanishes to high order at $R=0$ (by calculations such as~\eqref{EqipAsyCompGen2} or~\eqref{EqD4ParRewrite3}), and the argument of $\ubar L$ exactly reproduces the $t_*^{-1}R^0\ubar h_{\rms 2}^{(0)}$ leading-order term we wanted to eliminate. Moreover, the $t_*^{-3}$-term of this tensor matches the $t_*^{-3}$-leading-order term produced by the linearization of $P$ in the parameter $\scal_2^{(0),(1,0)}$ (see Definitions~\ref{DefD6Corr} and \ref{DefD6EinsteinAug}).

The only remaining independent terms are those that, in linearized theory, are related to the zero energy bound states of $L_b$ (as opposed to arising from the expansion at $\iota^+\cap\cK^+$ of solutions of $\iota^+$-model problems). These are the final black hole parameters $b$, the boost parameter $\scal$, and the center-of-mass shift $\scal_1^{(0)}$. (Recall Propositions~\ref{PropD2Boost} and \ref{PropD4Par} on how they arise in the linear theory.)

These considerations suggest:

\begin{definition}[Data space]
\label{DefEfD}
  Recall from~\eqref{EqKParam0} that $b_0=(\bhm_0,\bha_0)$ denotes subextremal Kerr parameters near which we work (cf.\ Definition~\ref{DefKMetcM}). Let $k\in\N$, and fix $0<\eps_\cK<\eps_+<\eps_\sscri$ as in~\eqref{EqDMetBasicEllEps}, and with $\eps_+<\eps_\ind$ where we recall $0<\eps_\ind\ll 1$ from~\eqref{EqWEIndEps}; and we moreover reduce $\eps_\ind$ further, if necessary, so as to ensure that $\lambda\in\spec_{\iota^+}^{<3+\eps_\ind}(\ubar L,\cE_\sscri^\cC)$ (see Definition~\usref{DefipPSpec}) implies $\Re\lambda\leq 3$. Fix moreover an index set $\cE_\sscri^\cC$ satisfying the properties in Definition~\ref{DefExP}. Recall the domain $\Omega_*=\cl_M\{\ft_*\geq 1\}$ from~\eqref{EqDOmegaStar}. We then define the space
  \[
    \fD^k(\Omega_*) = \fD^{k,\,\eps_\sscri,\,\eps_+,\,\eps_\cK}(\Omega_*)
  \]
  to consist of all tuples
  \begin{equation}
  \label{EqEfPU}
  \begin{split}
    U &= \Bigl(b-b_0,\scal,\scal_1^{(0)}, \ \bigl( u^0_\lambda, f_\lambda \bigr)_{\lambda\in\spec_{\iota^+}^{<3+\eps_+}(\ubar L,\cE_\sscri^\cC)}, \ \scal_\rem,b_\rem,\ \vect_{1,\rem}, \\
      &\quad\hspace{13em} \bigl(\scal_{l,\rem}^{(0)}\bigr)_{l=0,2},\ \bigl(\scal_{l,\rem}^{(-\lambda^\Ups_{\rms l,1}+1)}\bigr)_{l=0,1},\ \tilde h \Bigr)
  \end{split}
  \end{equation}
  where, recalling~\eqref{EqipPResFin}--\eqref{EqipPResFinHb},
  \begin{subequations}
  \begin{align}
  \label{EqEfPBH}
    &b=(\bhm,\bha) \in \R^4, \quad \scal,\ \scal_1^{(0)} \in \scalspace_1, \\
  \label{EqEfPScat}
    &u^0_\lambda \in \Resspace(\ubar L,\lambda),\quad f_\lambda\in\Poly^{K(\lambda)}\Bigl(\C_z;\Hb^{k,\ \bigl[\la\cE_\sscri^\cC+1\ra',\leq 4+\eps_\sscri\bigr]}\Bigr),
  \end{align}
  where $K(\lambda)\in\N_0$ is the unique integer with $(\lambda,K(\lambda))\in\Spec_{\iota^+}^{<3+\eps_+}(\ubar L,\cE_\sscri^\cC)$; moreover
  \begin{equation}
  \label{EqEfPRem}
  \begin{split}
    \scal_\rem &\in \dot H_\bop^{k,\,2+\eps_\cK}([1,\infty]_{t_*};\scalspace_1), \\
    b_\rem &\in \dot H_\bop^{k,\,3+\eps_\cK}([1,\infty];\R^4), \\
    \vect_{1,\rem} &\in \dot H_\bop^{k,\,3+\eps_+}([1,\infty];\vectspace_1), \\
    \scal_{l,\rem}^{(0)} &\in \dot H_\bop^{k,\,3+\eps_+}([1,\infty];\scalspace_l),\quad l=0,2, \\
    \scal_{l,\rem}^{(-\lambda^\Ups_{\rms l,1}+1)} &\in \dot H_\bop^{k,\,3+\eps_++(\lambda^\Ups_{\rms l,1}-1)}([1,\infty];\scalspace_l),\quad l=0,1;
  \end{split}
  \end{equation}
  and finally, recalling Definition~\ref{DefDMetBasic},
  \begin{equation}
  \label{EqEfPtildeh}
    \tilde h \in \Hb^{k,\ \bigl(\la\cE_\sscri^\cC\ra,3+\eps_\sscri\bigr),\ 3+\eps_+,\ 4+\eps_\cK}(\Omega_*)^{\bullet,-}.
  \end{equation}
  \end{subequations}
  (We use unweighted b-densities to define the underlying $L^2$-spaces.)
\end{definition}

From a tuple $U\in\fD^\infty(\Omega_*)$ as in~\eqref{EqEfPU}, we now construct $(h,\vecp,\vecq)$---and, in the process, also (in principle) the index sets\footnote{Note that in order to define the forward map $P(h,\vecp,\vecq)$ in~\eqref{EqD6Aug5Op}, one can use \emph{arbitrary} index sets $\cE_+$ and $\tilde\cE'_{++}$ (as long as $\min\Re(\cE_+\setminus\{(1,0)\})$, $\min\Re\tilde\cE'_{++}>1$).} $\cE_+$, with $\min\Re\cE_+\setminus\{(1,0)\}>1$, and $\tilde\cE'_{++}$, with $\min\Re\tilde\cE'_{++}>1$, appearing in Definitions~\ref{DefD6EinsteinAug} and \ref{DefD6Aug5} (though we shall not write them down explicitly here)---such that
\begin{equation}
\label{EqEfPGoal}
  P(h,\vecp,\vecq)\in\Hb^{\infty,\ \bigl(\la\cE_\sscri^\cC+1\ra',4+\eps_\sscri\bigr),\ 5+\eps_+,\ 4+\eps_\cK}(\Omega_*)^{\bullet,-},
\end{equation}
and where the construction of $(h,\vecp,\vecq)$ incorporates the parameters from $U$ in a ``natural'' way: the parameters $b-b_0,\scal,\scal_1^{(0)}$ and $\scal_\rem,b_\rem$ will be taken over directly, the other parameters with subscript ``$\rem$'' will be conormal remainder terms of the corresponding parameters without this subscript that comprise $\vecq$ in~\eqref{EqD6Aug5q}, and the parameters $(u^0_\lambda,f_\lambda)$ will contribute to the $\iota^+$-expansion of $h$ at order $t_*^{-\lambda}$.

For the convenience of the reader, we first collect the relevant notation and properties of $P$ and its arguments in one location.
\begin{enumerate}
\item $h\in\Hb^{\infty,\ \bigl(\la\cE_\sscri^\cC\ra,3+\eps_\sscri\bigr),\ (\cE_+,3+\eps_+),\ 4+\eps_\cK}$ (from~\eqref{EqD6Aug5h}). (For the map $P$ below to be defined, it suffices that $h$ has any positive rate of decay towards $\iota^+$ and $\cK^+$ and is small in a corresponding weighted b-Sobolev space with sufficiently high regularity.)
\item $\vecp$ (from Definition~\ref{DefD6EinsteinAug}) is a finite-dimensional parameter,
  \begin{equation}
  \label{EqEfPvecp}
  \begin{split}
    \vecp &= \Bigl( \scal, b-b_0, \scal_1^{(0)},\scal_1^{(0,1)},\ \ \scal_1^{(\lambda-1,k)},\ \ \vect_1^{(\tilde\lambda,k)},\ \scal_0^{(0),(\tilde\lambda,k)},\ \scal_2^{(0),(\tilde\lambda,k)}, \\
      &\quad \hspace{5em} \bigl(\scal_l^{(-\lambda^\Ups_{\rms l,1}+1),(\tilde\lambda+\lambda^\Ups_{\rms l,1}-1,k)}\bigr)_{l=0,1},\ \scal_3^{(-1),(2,0)},\ \vect_2^{(-1),(2,0)} \Bigr)
  \end{split}
  \end{equation}
  where $(\lambda,k)\in\tilde\cE'_{++}$ and $(\tilde\lambda,k)\in(1,0)\cup\tilde\cE'_{++}$, and we only include terms with exponents $(\lambda-1,k)$, resp.\ $(\tilde\lambda+\mu,k)$ with $\lambda-1\leq 2$, resp.\ $\tilde\lambda+\mu\leq 2$; here $\scal_l^{(\cdots)}\in\scalspace_l$ and $\vect_l^{(\cdots)}\in\vectspace_l$.
\item $\vecq$ (from~\eqref{EqD6Aug5q}) is given by
  \begin{equation}
  \label{EqEfPvecq}
  \begin{split}
    \vecq &= \Bigl(\scal_\rem,b_\rem,\vect_1,\scal_0^{(0)}, \bigl(\scal_l^{(-\lambda^\Ups_{\rms l,l+j}+1)}\bigr)_{ \substack{ 0\leq l\leq 3,\ j=0,1, \\ 1\leq l+j\leq 3 } }, \\
      &\quad \hspace{3em} \bigl(\scal_l^{(-l+2)}\bigr)_{2\leq l\leq 5},\ \bigl(\vect_l^{(-l+1)}\bigr)_{2\leq l\leq 4},\ \bigl(\scal_l^{(-l)},\vect_l^{(-l)}\bigr)_{l=2,3} \Bigr),
  \end{split}
  \end{equation}
  where $\scal_\rem$ and $b_\rem$ are as in~\eqref{EqEfPRem}, while, in the notation of Definition~\ref{DefD6Order}, $\vect_1\in\dot H_\bop^{\infty,\ \scissors(\cE_+,3+\eps_+)}([1,\infty]_{t_*})$ and $\sfW_l^{(-\mu)}\in\dot H_\bop^{\infty,\ \scissors(\cE_++\mu,3+\eps_++\mu)}([1,\infty]_{t_*})$ for $\sfW=\scal,\vect$.
\item Ignoring the initial solution $h_{-\scal}$ from~\eqref{EqD6AbsExt} for now (which in~\eqref{EqD6Aug5Op} enters only via terms supported in $\ft_*\leq 2$), we set
  \begin{equation}
  \label{EqEfPMap}
  \begin{split}
    P(h,\vecp,\vecq)&=\Ric\bigl(g^0+h_\tot(\vecp)+h_{b,\rem}(\vecq)+h\bigr) \\
      &\quad\qquad - \delta_{g^0,E^\cC}^*\Bigl(\Ups_{E^\Ups}\bigl(g^0+h_{b,\rem}(\vecq)+h,\,g^0\bigr) - \vartheta_\tot(\vecp)\Bigr),
  \end{split}
  \end{equation}
  where $g^0=g_{b_0,b,-\scal}=(1-\chi_\cK)\phi_{-\scal}^*g_{b_0}+\chi_\cK g_b$ (from~\eqref{EqDgPOU}, \eqref{EqKBoMap}, and Definition~\ref{DefKBoCutoff} and~\eqref{EqDCutoffs}), furthermore (using the notation from Definition~\ref{DefipGrKerr}, the large and generalized zero energy states from Propositions~\ref{PropWG0Symm}, \ref{PropWG0Large}, \ref{PropWEMode0Kerr}, and~\ref{PropWE0}, and the antiderivative $A^{(\lambda,k)}(t_*)$ of $t_*^{-\lambda}(\log t_*)^k$ from~\eqref{EqD6CorrA})
  \begin{align*}
    h_\tot(\vecp) &= \underbrace{-\dd\chi_\cK\otimes_s\bigl(\omega_{b_0,\rms 1}^{(0),\leq 1}(\scal)-g_{b_0}\ubar g^{-1}\ubar\omega_{\rms 1}^{(0),\leq 1}(\scal)\bigr)}_{h^{(-1)}(\scal)\ \text{(Def.~\ref{DefD2Corr})}}{} \underbrace{{}-\dd\chi_\cK\otimes_s\bigl(\dot\omega_{b_0}(b-b_0)+\omega_{b_0,\rms 1}^{(0)}(\scal_1^{(0)})\bigr)}_{h^{(0)}(b-b_0)+h_{\rms 1}^{(0)}(\scal_1^{(0)})\ \text{(Def.~\ref{DefD4Corr})}} \\
      &\quad \underbrace{{}- \dd\chi_\cK \otimes_s \biggl( \omega_{b_0,\rms 1}^{(0),\leq 4}\bigl(\log(t_*)\scal_1^{(0,1)}\bigr) + \sum_{(\lambda,k)} \omega_{b_0,\rms 1}^{(0),\leq 4}\bigl(t_*^{-\lambda+1}(\log t_*)^k\scal_1^{(\lambda,k)}\bigr)\biggr)}_{h_{\rms 1}^{(0,1)}(\scal_1^{(0,1)})\ \text{(Def.~\ref{DefD4Corr})}\ +\ \sum h_{\rms 1}^{(\lambda-1,k)}(\scal_1^{(\lambda-1,k)})\ \text{(Def.~\ref{DefD5Corr})}} \\
      &\quad \underbrace{{}- \dd\chi_\cK \otimes_s \sum_{(\tilde\lambda,k)}\omega_{b_0,\rmv 1}^{(-1),\leq 3}\bigl(t_*^{-\tilde\lambda}(\log t_*)^k\vect_1^{(\tilde\lambda,k)}\bigr)}_{ h_{\rmv 1}^{(\tilde\lambda,k)}(\vect_1^{(\tilde\lambda,k)})\ \text{(Def.~\ref{DefD6Corr})}} \\
      &\quad \underbrace{{}- \dd\chi_\cK \otimes_s \sum_{(\tilde\lambda,k)} \omega_{b_0,\rms 0}^{(0),\leq 4}\bigl(A^{(\lambda,k)}(t_*)\scal_0^{(0),(\tilde\lambda,k)}\bigr) }_{h_{\rms 0}^{(0),(\tilde\lambda,k)}(\scal_0^{(0),(\tilde\lambda,k)})\ \text{(Def.~\ref{DefD6Corr})}} \\
      &\quad  \underbrace{{}- \dd\chi_\cK \otimes_s \sum \omega_{b_0,\rmw l}^{(-\mu-1),\leq j}\bigl( t_*^{-\tilde\lambda}(\log t_*)^k\sfW_l^{(-\mu),(\tilde\lambda+\mu,k)}\bigr) }_{ h_{\rms 2}^{(0),(\tilde\lambda,k)},\ h_{\rms l}^{(-\lambda^\Ups_{\rms l,1}+1),(\tilde\lambda,k)},\ h_{\rms 3/\rmv 3}^{(-1),(2,0)}\ \text{(Def.~\ref{DefD6Corr})}},
  \end{align*}
  where the sum involving $\sfW$ is over the last four parts of~\eqref{EqEfPvecp} (starting with $\scal_2^{(0),(\tilde\lambda,k)}$), and $j$ (depending on $\mu$) is the largest number for which these tensors have been defined;\footnote{These are the same numbers appearing in~\eqref{EqipGr}, \eqref{EqD6uexp}, the definition of the modification terms, and other places; cf.\ Remark~\ref{RmkD6uexp}\eqref{ItD6uexpNum}.} and we recall that $\dot\omega_b\in\cA^{(1,1)\cup(1+\cE_\ind)}(X;\cT^*_X)$ has the form~\eqref{EqWEMode0KerrOmega}. Moreover,
  \begin{align*}
    \vartheta_\tot(\vecp) &= \underbrace{[\delta_{g_{b_0},E^\Ups}\sfG_{g_{b_0}},\chi_\cK]\bigl(h_{b_0,\rms 1}^{\leq 1}(\scal) + \dot g_{b_0}^\Ups(b-b_0) + h_{b_0,\rms 1}(\scal_1^{(0)})\bigr)}_{\vartheta^{(-1)}(\scal)\ \text{(Def.~\ref{DefD2Corr})}\ +\ \vartheta^{(0)}(b-b_0)+\vartheta_{\rms 1}^{(0)}(\scal_1^{(0)})\ \text{(Def.~\ref{DefD4Corr})}} \\
      &\quad + \underbrace{ \delta_{g_{b_0},E^\Ups}\sfG_{g_{b_0}}\Biggl(\chi_\cK \delta_{g_{b_0}}^*\biggl( \omega_{b_0,\rms 1}^{(0),\leq 4}\bigl(\log(t_*)\scal_1^{(0,1)}\bigr) + \sum_{(\lambda,k)} \omega_{b_0,\rms 1}^{(0),\leq 4}\bigl(t_*^{-\lambda+1}(\log t_*)^k\scal_1^{(\lambda,k)}\bigr)\biggr)\Biggr) }_{ \vartheta_{\rms 1}^{(0,1)}(\scal_1^{(0,1)})\ \text{(Def.~\ref{DefD4Corr})}\ +\ \sum \vartheta_{\rms 1}^{(\lambda-1,k)}(\scal_1^{(\lambda-1,k)})\ \text{(Def.~\ref{DefD5Corr})}} \\
      &\quad + \underbrace{ \delta_{g_{b_0},E^\Ups}\sfG_{g_{b_0}}\Biggl( \chi_\cK\delta_{g_{b_0}}^* \sum_{(\tilde\lambda,k)} \omega_{b_0,\rmv 1}^{(-1),\leq 3}\bigl(t_*^{-\tilde\lambda}(\log t_*)^k\vect_1^{(\tilde\lambda,k)}\bigr) \Biggr) }_{\vartheta_{\rmv 1}^{(\tilde\lambda,k)}(\vect_1^{(\tilde\lambda,k)})\ \text{(Def.~\ref{DefD6Corr})}} \\
      &\quad + \underbrace{ \delta_{g_{b_0},E^\Ups}\sfG_{g_{b_0}}\Biggl( \chi_\cK\delta_{g_{b_0}}^* \sum_{(\tilde\lambda,k)} \omega_{b_0,\rms 0}^{(0),\leq 4}\bigl(A^{(\lambda,k)}(t_*)\scal_0^{(0),(\tilde\lambda,k)}\bigr) \Biggr)}_{\vartheta_{\rms 0}^{(0),(\tilde\lambda,k)}(\scal_0^{(0),(\tilde\lambda,k)})\ \text{(Def.~\ref{DefD6Corr})}} \\
      &\quad + \underbrace{ \delta_{g_{b_0},E^\Ups}\sfG_{g_{b_0}}\Biggl( \chi_\cK\delta_{g_{b_0}}^*\sum \omega_{b_0,\rmw l}^{(-\mu-1),\leq j}\bigl(t_*^{-\tilde\lambda}(\log t_*)^k\sfW_l^{(-\mu),(\tilde\lambda+\mu,k)}\bigr) \Biggr) }_{\vartheta_{\rms 2}^{(0),(\tilde\lambda,k)},\ \vartheta_{\rms l}^{(-\lambda^\Ups_{\rms l,1}+1),(\tilde\lambda,k)},\ \vartheta_{\rms/\rmv 3}^{(-1),(2,0)}\ \text{(Def.~\ref{DefD6Corr})}},
  \end{align*}
  and finally (from~\eqref{EqD6Aug5hrem})
  \begin{equation}
  \label{EqEfhrem}
    h_{b,\rem}(\vecq) = \chi_\cK\Bigl(h_{b,\rms 1}^{\leq 1}(\scal_\rem)+\dot g_b^\Ups(b_\rem)+h_{b,\rmv 1}^{\leq 3}(\vect_1) + \sum h_{b,\rmw l}^{(-\mu),\leq j}(\sfW_l^{(-\mu)})\Bigr),
  \end{equation}
  where again $j$ is the largest number for which these tensors are defined. (We recall also that the superscript $-\mu-1$ of $\omega_{b_0,\cdots}^{(-\mu-1),\cdots}$ matches the size $\rho^{-\mu-1}=r^{\mu+1}$ of this 1-form as $\rho=r^{-1}\to 0$; similarly, $h_{b_0,\cdots}^{(-\mu),\cdots}$ is of size $\rho^{-\mu}$.)
\end{enumerate}

\subsubsection{Constructing metric perturbations and modification parameters from data}
\label{SssEfPC}

Besides the cutoff function $\chi_\cK$ ($1$ near $\cK^+$, $0$ near $\scri^+$ and $0$ for $\ft_*\leq 10$), we fix a cutoff function $\chi_\iota$ that equals $1$ near $\iota^+$ and $0$ near $\ft_*\leq 10$.

Given $U=(b-b_0,\scal,\scal_1^{(0)},(u^0_\lambda,f_\lambda),\scal_\rem,b_\rem,\vect_{1,\rem},(\scal_{l,\rem}^{(0)})_{l=0,2},(\scal_{l,\rem}^{(-\lambda^\Ups_{\rms l,1}+1)})_{l=0,1},\tilde h)\in\fD^\infty(\Omega_*)$ as in~\eqref{EqEfPU}, set\footnote{That is, $\vecq_{[0]}$ collects all $\bullet_\rem$ terms of $U$, while those components of~\eqref{EqEfPvecq} that are purely polyhomogeneous are set to $0$; they will be determined later from the $R\to 0$ asymptotics of solutions of $\iota^+$-model problems.}
\begin{align*}
  h_{[0]} &:= \chi_\iota\sum_{\lambda\in\spec_{\iota^+}^{<3+\eps_+}(\ubar L,\cE_\sscri^\cC)} u_\lambda + \tilde h, \quad u_\lambda:=\Res_{\iota^+}(\ubar L,\cE_\sscri^\cC,\lambda)(u^0_\lambda,f_\lambda), \\
  \vecp_{[0]} &:= \bigl( \scal, b-b_0, \scal_1^{(0)}, 0, 0, \ldots, 0 \bigr), \\
  \vecq_{[0]} &:= \Bigl( \scal_\rem, b_\rem, \vect_{1,\rem}, \scal_{0,\rem}^{(0)}, \bigl(\scal_{0,\rem}^{(-\lambda^\Ups_{\rms 0,1}+1)},\scal_{1,\rem}^{(-\lambda^\Ups_{\rms 1,1}+1)},0,\ldots,0\bigr), \\
    &\quad\hspace{12em} \bigl(\scal_{2,\rem}^{(0)},0,0,0\bigr),\ (0,0,0),\ (0,0,0,0) \Bigr).
\end{align*}
(This uses all components of $U$ in a ``natural'' fashion.) Recall from~\eqref{EqipPSpecEx} that $\spec_{\iota^+}^{<3+\eps_+}(\ubar L,\cE_\sscri^\cC)$ contains $1$ (with $u_1^0=0$ then, since $\Resspace(\ubar L,1)=\{0\}$, and the corresponding $u_1$ not having any factors of $\log t_*$), while all other elements have real parts $\geq 1+\eps_\ind$. We also recall that
\[
  \iota^+ = [0,\infty]_R \times \Sph^2_\omega,\quad R=\frac{r}{t_*}=\frac{|x|}{t_*},\ \omega=\frac{x}{|x|}.
\]

We have
\begin{equation}
\label{EqEfP0}
  P(h_{[0]},\vecp_{[0]},\vecq_{[0]})\in\Hb^{\infty,\ \bigl(\la\cE_\sscri^\cC+1\ra',4+\eps_\sscri\bigr),\ (3,0)\cup(3+\cF),\ 1-\eps}(\Omega_*)^{\bullet,-}
\end{equation}
for all $\eps>0$ and for some index set $\cF$ with $\min\Re\cF\geq\eps_\ind$. This follows from a variant of Corollary~\ref{CorDAdmFwN} with $\cE_+=(1,1)\cup(1+\cE_\ind)\cup\Spec_{\iota^+}^{<3+\eps_+}(\ubar L,\cE_\sscri^\cC)$ and the following observation regarding the absence of an element $(3,0)$ in the $\iota^+$-index set: since the arguments of $\Ric$ and the gauge 1-form in~\eqref{EqEfPMap} are equal to the Minkowski metric $\ubar g$ to leading order at $\iota^+$, one can compute the $\iota^+$-expansion of~\eqref{EqEfPMap} using (higher-order) linearizations around the Minkowski metric; thus, despite the presence of some metric terms of size $t_*^{-1}\log t_*$ (see Lemmas~\ref{LemmaD2Corr} and \ref{LemmaD4Corr}, and also~\eqref{EqD6Corrhs0}), they are annihilated to leading order by $D_{\ubar g}\Ric$.

The $\cK^+$-order $1-\eps$ of~\eqref{EqEfP0} is due to the fact that $h_{[0]}$ has this $\cK^+$-order; more precisely, $h_{[0]}$ is polyhomogeneous at $\cK^+$ with index set $(1,0)\cup(\cdots)$ determined by $\Spec_{\iota^+}^{<3+\eps_+}(\ubar L,\cE_\sscri^\cC)$ and the polyhomogeneous expansions of each $u_\lambda$ at $R=0$ as recorded in Corollary~\ref{CoripPQhom}\eqref{ItipPQhomExp}, and thus also~\eqref{EqEfP0} is of order $(\cF_\cK,4+\eps_\cK)$ for some index set $\cF_\cK=(1,0)\cup(\cdots)$. (As part of our construction, we will improve on $h_{[0]}$ by making it ultimately have $\cK^+$-order $4+\eps_\cK$. If $h_{[0]}$ had $\cK^+$-order $4+\eps_\cK$, then so would~\eqref{EqEfP0}.) Thus, we do \emph{not} have to invert any model problems at $\cK^+$ (e.g., the Kerr model $L_b$) to improve the $\cK^+$-order of~\eqref{EqEfP0}: this improvement will be an immediate consequence of improvements of the $\iota^+\cap\cK^+$-orders of the terms in the $\iota^+$-expansion of $h_{[0]}$. (See~\eqref{EqEfPCh1} below for a first such improvement.) We recall here that while $h_{b,\rem}(\vecq)$ has $\cK^+$-order $2+\eps_\cK$ only, the fact that $L_b h_{b,\rem}(\vecq)$ has $\cK^+$-order $4+\eps_\cK$ (see~\eqref{EqD6Aug5Lb}) implies that indeed so does $P(h,\vecp,\vecq)$ when $h$ has $\cK^+$-order $4+\eps_\cK$; cf.\ the proof of Lemma~\ref{LemmaD6Aug5Linb}.

\pfstep{Step~1. Determination of $t_*^{-1}$-terms at $\iota^+$.} Following the strategy outlined at the beginning of~\S\ref{SsEfP}, we begin by computing the $t_*^{-3}$-coefficient $f^{(3,0)}=f^{(3,0)}(R,\omega)$ of $P(h_{[0]},\vecp_{[0]},\vecq_{[0]})$ at $\iota^+$. This does not depend on $\vecq_{[0]}$ or $\tilde h$. Note that $\Ric(g^0)=0$ outside of $\supp\dd\chi_\cK$, so if we write $(t_*(g^0-\ubar g))|_{\iota^+}=:\tilde g^0=\tilde g^0(R,\omega)\in\CIc((\iota^+)^\circ;S^2\cT^*)$, we have
\begin{equation}
\label{EqEfPCt3}
  \bigl(t_*^3 D_{\ubar g}\Ric(t_*^{-1}\tilde g^0)\bigr)\big|_{\iota^+}\in\CIc((\iota^+)^\circ;S^2\cT^*),
\end{equation}
and this is the $t_*^{-3}$-coefficient of $\Ric(g^0)=\Ric(\ubar g+(g^0-\ubar g))$. The contributions from $\vecp_{[0]}$ are the $t_*^{-3}$-terms of $D_{\ubar g}\Ric\bigl( h^{(-1)}(\scal) + h^{(0)}(b-b_0) + h_{\rms 1}^{(0)}(\scal_1^{(0)}) \bigr)$ at $\iota^+$ as well as those of $\delta_{g^0,E^\cC}^*\vartheta_\tot(\vecp_{[0]})$, i.e., those of $\ubar\delta_{\ubar E^\cC}^*\bigl(\vartheta^{(-1)}(\scal)+\vartheta^{(0)}(b-b_0)+\vartheta_{\rms 1}^{(0)}(\scal_1^{(0)})\bigr)$, all of which are also of class $\CIc((\iota^+)^\circ;S^2\cT^*)$ (being supported in $\supp\dd\chi_\cK$). The final contribution is from the term $u_1\in t_*^{-1}\Hb^{\infty,\ (\la\cE_\sscri^\cC\ra,3+\eps_\sscri),\ (\cE_\cK,3+\eps_\cK)}(\iota^+)$ of $h_{[0]}$ in the notation of~\eqref{EqipPSpaces} (and Proposition~\ref{PropipNInv} for $\cE_\cK$), but this vanishes since $\ubar L u_1=t_*^{-3}N_{\iota^+}(\ubar L,1)(t_* u_1)=0$. We have thus proved that
\begin{equation}
\label{EqEfPcf3}
  f^{(3,0)} = f^{(3,0)}(\scal,b,\scal_1^{(0)}) \in \CIc((\iota^+)^\circ;S^2\cT^*).
\end{equation}
The two deficiencies which we proceed to rectify are that
\begin{enumerate}[label=(\roman*)]
\item $f^{(3,0)}$ does not vanish (unless $\vecp_{[0]}=0$), and
\item the $t_*^{-1}$-coefficient of the total metric perturbation does not vanish to order $4+\eps_\cK$ at $\cK^+$, the sole culprit at present being the term $u_1$ of $h_{[0]}$.
\end{enumerate}

First, we use Proposition~\ref{PropipNInv} with $\lambda=1$ (but give up some precision by only recording the orders needed for present purposes) to produce
\begin{equation}
\label{EqEfPch1}
  h^{(1,0)} = h^{(1,0)}(R,\omega) := -N_{\iota^+}(\ubar L,1)^{-1}f^{(3,0)} \in \cA^{(\la\cE_\sscri^\cC\ra,3+\eps_\sscri),\ (\cE_\cK,3+\eps_\cK)}(\iota^+).
\end{equation}
The $t_*^{-3}$-term of $P(u_1+\chi_\iota t_*^{-1}h^{(1,0)},\vecp_{[0]},\vecq_{[0]})$ then vanishes. Now, $\ubar L(u_1+t_*^{-1}h^{(1,0)})=0$ and Proposition~\ref{PropipNAsy} (with $\lambda=1$ and $\ell_\cK=3+\eps_\cK$) show that $t_*(u_1+t_*^{-1}h^{(1,0)})$ (which is homogeneous of degree $0$, i.e., in the coordinates $(t_*,R,\omega)$ it depends only on $(R,\omega)$) admits the expansion~\eqref{EqipNAsyu0}; in the notation of~\eqref{EqipNAsyu0} (multiplied by $t_*^{-1}$), we then note the following.
\begin{enumerate}
\item{\rm (Elimination of the $\ubar h_{\rms 1}^{\leq 4}$-term.)} The term\footnote{Recall Definition~\ref{DefipAsyForm} for the notation.} $\chi_\cK\ubar h_{\rms 1}^{\leq 4}(t_*^{-2}\scal_1^{(-1)})$ is, modulo a multiple of $\chi_\cK t_*^{-5}\breve{\ubar\omega}_{\rms 1}^{(0),4}$ (with $\breve{\ubar\omega}_{\rms 1}^{(0),4}=\cO(\rho^{-4})$), equal to $-\chi_\cK\ubar\delta^*\bigl(\ubar\omega_{\rms 1}^{(0),\leq 4}(\log(t_*)\scal_1^{(-1)})\bigr)$. (This follows from a variant of the identity~\eqref{EqD4ParNoDelvsH}; and the first two terms $-\chi_\cK\log(t_*)\ubar\delta^*\ubar\omega_{\rms 1}^{(0)}=0$ and $-\chi_\cK t_*^{-1}([\ubar\delta^*,t_*]\ubar\omega_{\rms 1}^{(0)}+\ubar\delta^*\breve{\ubar\omega}_{\rms 1}^{(0),1})=0$ in the expansion of $-\chi_\cK\ubar\delta^*\bigl(\ubar\omega_{\rms 1}^{(0),\leq 4}(\log(t_*)\scal_1^{(-1)})\bigr)$ vanish.) Furthermore, we have
  \begin{equation}
  \label{EqEfPCs1}
    \ubar L\Bigl( \chi_\cK\ubar\delta^*\bigl( \ubar\omega_{\rms 1}^{(0),\leq 4}\bigl(\log(t_*)\scal_1^{(-1)}\bigr) \bigr) \Bigr) = D_{\ubar g}\Ric\bigl(\ubar h_{\rms 1}^{(0,1)}(\scal_1^{(-1)})\bigr) - \ubar\delta_{\ubar E^\cC}^*\ubar\vartheta_{\rms 1}^{(0,1)}(\scal_1^{(-1)})
  \end{equation}
  if we denote by
  \begin{align*}
    \ubar h_{\rms 1}^{(0,1)}(\scal_1^{(-1)}) &:= -\dd\chi_\cK \otimes \ubar\omega_{\rms 1}^{(0),\leq 4}\bigl(\log(t_*)\scal_1^{(-1)}\bigr), \\
    \ubar\vartheta_{\rms 1}^{(0,1)}(\scal_1^{(-1)}) &:= \ubar\delta_{\ubar E^\Ups}\ul\sfG\Bigl(\chi_\cK \ubar\delta^* \ubar\omega_{\rms 1}^{(0),\leq 4}\bigl(\log(t_*)\scal_1^{(-1)}\bigr)\Bigr)
  \end{align*}
  the $\iota^+$-leading-order terms of $h_{\rms 1}^{(0,1)}(\scal_1^{(-1)})$ and $\vartheta_{\rms 1}^{(0,1)}(\scal_1^{(-1)})$, respectively. Reading the identity~\eqref{EqEfPCs1} from right to left thus means that changing the parameter $\scal_1^{(0,1)}$ of $\vecp_{[0]}$ from $0$ to $\scal_1^{(-1)}$ modifies $f^{(3,0)}$ in~\eqref{EqEfPcf3} by adding the source term~\eqref{EqEfPCs1}, thus modifying the $\ubar h_{\rms 1}^{\leq 4}(t_*^{-2})$-coefficient of $\bigl(t_*(u_1+t_*^{-1}h^{(1,0)})\bigr)\big|_{\iota^+}$ (for the $h^{(1,0)}$ defined as in~\eqref{EqEfPch1} but with respect to the new $f^{(3,0)}$) by subtracting the coefficient of the argument of $\ubar L$ in~\eqref{EqEfPCs1}---the total coefficient thus being equal to $0$. --- The conceptual content of this argument (in which the passage to Minkowskian operators and tensors is done merely to emphasize that these arguments only involve leading-order behavior at $\iota^+$) is thus that the modification term $\scal_1^{(0,1)}$ can be used to eliminate the logarithmic center-of-mass correction of the final black hole when solving the gauge-fixed Einstein equation \emph{backwards} from $\iota^+\cup\cK^+$ (which is thus the backwards analogue of why we introduced $\scal_1^{(0,1)}$ in~\S\ref{SssD4ParNo} in the first place); note that we now work entirely on the Minkowski background.
\item\label{ItPfPCs12}{\rm (Elimination of other terms that are not of size $t_*^{-1}R^1$.)} Such terms (arising from indicial roots of $\wh{\ubar L}(0)$ with real parts in $[-1,0]$) would be metric contributions of size $\rho_+\rho_\cK^2$ near $\cK^+$ and thus unacceptable. These are the terms $\ubar h_{\rmv 1}^{\leq 3}(t_*^{-2}\vect_1^{(-1)})$, $\scal_l^{(0)}\ubar h_{\rms l}^{(0),\leq 3}(t_*^{-1})$ ($l=0,2$), $\ubar h_{\rms l}^{(\mu),\leq 2}(t_*^{-1+\mu}\scal_l^{(\mu)})$ for $\mu=-\lambda^\Ups_{\rms l,1}+1$, $l=0,1$, and finally $\ubar h_{\rms 3}^{(-1),\leq 2}(t_*^{-2}\scal_3^{(-1)})$ and $\ubar h_{\rmv 2}^{(-1),\leq 2}(t_*^{-2}\vect_2^{(-1)})$ in~\eqref{EqipNAsyu0}. They can be eliminated in a completely analogous fashion by changing the parameters $\vect_1^{(1,0)}$, $\scal_l^{(0),(1,0)}$, $\scal_l^{(\mu),(1-\mu,0)}$, $\scal_3^{(-1),(2,0)}$, and $\vect_2^{(-1),(2,0)}$ of $\vecp_{[0]}$ from $0$ to $\vect_1^{(-1)}$, $\scal_1^{(0)}$, $\scal_l^{(\mu)}$, $\scal_3^{(-1)}$, and $\vect_2^{(-1)}$, respectively (see~\eqref{EqEfPvecp}).
\item{\rm (Elimination of the remaining terms.)} The remaining terms are of size $t_*^{-1}R^{1+\eps_\cK}$ (since $\eps_\cK<\eps_\ind$), i.e., $\rho_+\rho_\cK^{2+\eps_\cK}$ (in a spacetime-$L^2$-sense) near $\cK^+$ and thus need not be eliminated (cf.\ Theorem~\ref{ThmD6}). Instead, we can absorb them into $h_{b,\rem}$ by adjusting the corresponding coefficients of $\vecq$. (This was precisely the reason for the introduction of $\vecq$ in Definition~\ref{DefD6Aug5}.)
\end{enumerate}

In summary, we have thus shown that from the asymptotic expansion of the $t_*^{-1}$-term of $u_1-t_*^{-1}N_{\iota^+}(\ubar L,1)^{-1}f^{(3,0)}$ (for the original $f^{(3,0)}$ from~\eqref{EqEfPcf3}) at $\iota^+$ towards $\iota^+\cap\cK^+=\iota^+\cap R^{-1}(0)$ one can read off $\vecp'_{[1]}$ and $\vecq'_{[1]}$ such that, upon setting $\tilde h_1:=-t_*^{-1}N_{\iota^+}(\ubar L,1)^{-1}f_{\rm new}^{(3,0)}$ where $f_{\rm new}^{(3,0)}$ is the $t_*^{-3}$-coefficient of $P(h_{[0]},\vecp_{[0]}+\vecp'_{[1]},\vecq_{[0]}+\vecq'_{[1]})$, we have, for
\[
  h_{[1]} := h_{[0]} + \chi_\iota\tilde h_1,\quad
  \vecp_{[1]} := \vecp_{[0]} + \vecp'_{[1]},\quad
  \vecq_{[1]} := \vecq_{[0]} + \vecq'_{[1]},
\]
the following two properties:
\begin{enumerate}
\myitem{ItEfPCh1}{i}{\rm (Metric perturbation.)} The $t_*^{-1}$-coefficient of $h_{[1]}$ is of class
  \begin{equation}
  \label{EqEfPCh1}
    \Hb^{\infty,\ \bigl(\la\cE_\sscri^\cC\ra,3+\eps_\sscri\bigr),\ 3+\eps_\cK}(\iota^+)
  \end{equation}
  (and thus $t_*^{-1}$ times it vanishes to order $4+\eps_\cK$ at $\cK^+$).
\myitem{ItEfPCh2}{ii}{\rm (Error term.)} We have
  \begin{equation}
  \label{EqEfPCP1}
    P(h_{[1]},\vecp_{[1]},\vecq_{[1]})\in\Hb^{\infty,\ \bigl(\la\cE_\sscri^\cC+1\ra',4+\eps_\sscri\bigr),\ (\cF+2,5+\eps_+),\ (\cF_\cK,4+\eps_\cK)}(\Omega_*)^{\bullet,-}
  \end{equation}
  for some $\cF$ and $\cF_\cK$ with $\min\Re\cF,\min\Re\cF_\cK\geq 1+\eps_\ind$. (The $t_*^{-3}$-coefficient of this tensor at $\iota^+$ thus vanishes; this improves upon~\eqref{EqEfP0}. The improved order of decay at $\cK^+$ is inherited from the improved $\cK^+$-order of $h_{[1]}$ relative to $h_{[0]}$ due to point~\eqref{ItEfPCh1}.)
\end{enumerate}

\medskip

\pfstep{Step~2. Determination of lower-order terms at $\iota^+$.} We improve on~\eqref{EqEfPCP1} by iteratively eliminating the leading-order terms at $\iota^+$. This amounts to correcting $h_{[1]}$ at increasingly high orders at $\iota^+$ by adding solutions of $\iota^+$-model problems, while utilizing the modification parameters in $\vecp$, resp.\ $\vecq$ (to eliminate, resp.\ account for terms in the $\iota^+\cap\cK^+$-expansion of the solutions of $\iota^+$-model problems that have at most $t_*^{-2-\eps_\cK}$, resp.\ more than $t_*^{-2-\eps_\cK}$-decay) to ensure that all coefficients of the $\iota^+$-expansion of the corrected $h_{[1]}$ up to some $\iota^+$-order $\alpha_+<3+\eps_+$ have $4+\eps_\cK$ orders of decay at $\cK^+$. (This then automatically improves the $\cK^+$-order of~\eqref{EqEfPCP1} as well.)

More quantitatively, we set up an iterative procedure as follows. Let $i\in\N$ and $\alpha_+\in(1,3+\eps_+)$, and assume that we have constructed $\tilde h_\lambda$ for a finite set of $\lambda$ with $\lambda=1$ or $\Re\lambda\in(1,\alpha_+)$ which, as functions of $(t_*,R,\omega)$, are quasi-homogeneous of degree $-\lambda$ in $t_*$, such that for
\begin{equation}
\label{EqEfPChj}
  h_{[i]} = \chi_\iota\sum_{\lambda\in\spec_{\iota^+}^{<3+\eps_+}(\ubar L,\cE_\sscri^\cC)} u_\lambda + \tilde h + \sum_\lambda \tilde h_\lambda,
\end{equation}
every $t_*^{-\lambda}(\log t_*)^k$-term in the $\iota^+$-expansion of $h_{[i]}$ for $\Re\lambda<\alpha_+$ has $\cK^+$-order $4+\eps_\cK$. (Thus, only the $t_*^{-\lambda}(\log t_*)^k$-terms of $h_{[i]}$ with $\Re\lambda\geq\alpha_+$ may still have nontrivial $\cK^+$-orders; since each $u_\lambda$ is a sum of terms of the form $t_*^{-\lambda}(\log t_*)^k\Hb^{\infty,\ (\la\cE_\sscri^\cC\ra,3+\eps_\sscri),\ ((0,0),\eps_\ind)}(\iota^+)$ by Remark~\ref{RmkipPResAsy}, this means that the $\cK^+$-order of $h_{[i]}$ is at least $\alpha_+-\eps$ for all $\eps>0$. When $\alpha_+\geq 3$, there are no such nontrivial terms anymore,\footnote{Our iterative construction here improves the order of vanishing at $\iota^+$ by a definite amount $\eps_\ind>0$ at each step, and thus there is a universal finite upper bound on the $\iota^+$-index sets we encounter. We may thus take $\eps_+>0$ so small that no element in the $\iota^+$-index set of any $h_{[i]}$ has real part in $[3,3+\eps_+]$, which means that once $\alpha_+\geq 3$, we have already completed the construction of the sought-after $\iota^+$-expansion of $h_{[i]}$.} and thus the $\cK^+$-order of $h_{[i]}$ is then equal to the $\cK^+$-order $4+\eps_\cK$ of $\tilde h$.) Step~1 above produces~\eqref{EqEfPChj} for $i=1$, $\alpha_+=1+\eps_\ind$, and a single term $\tilde h_1$.

We furthermore assume that we have determined tuples $\vecp_{[i]}$ and $\vecq_{[i]}$ as in~\eqref{EqEfPvecp} and \eqref{EqEfPvecq}, where only the coefficients contributing at those $\iota^+$-orders that arise in~\eqref{EqEfPChj} and have real parts $<\alpha_+$ are possibly non-zero. (Cf.\ again the construction in Step~1 for $i=1$.) Assume finally that
\begin{equation}
\label{EqEfPCPj}
  P(h_{[i]},\vecp_{[i]},\vecq_{[i]})\in\Hb^{\infty,\ \bigl(\la\cE_\sscri^\cC+1\ra',\,4+\eps_\sscri\bigr),\ (\cF_{[i]}+2,\,5+\eps_+),\ \alpha'_\cK},\quad \min\Re\cF_{[i]}\geq\alpha_+,
\end{equation}
where $\alpha'_\cK=\alpha_+-\eps$, unless $\alpha_+\geq 3$ in which case $\alpha'_\cK=4+\eps_\cK$.\footnote{The discontinuous jump of $\alpha'_\cK$ from almost $3$ to $4+\eps_\cK$ when $\alpha_+$ crosses $3$ has a simple explanation. If $\alpha_+<3$ is close to $3$, then $h_{[i]}$ typically has an (as-of-yet untreated) $t_*^{-3}$-term $\chi_\iota t_*^{-3}h_{[i]}^{(3,0)}(R,\omega)$, with $h_{[i]}^{(3,0)}$ having an $R\to 0$ expansion including terms such as $\scal_0 R^0\ubar h_{\rms 0}^{(0)}$. The failure of~\eqref{EqEfPCPj} to have $\cK^+$-order $4+\eps_\cK$ is then due to the fact that $D_{g_b}\Ric(\chi_\iota \scal_0 t_*^{-3}\ubar h_{\rms 0}^{(0)})\neq o(t_*^{-3})$ at $\cK^+$. But as soon as one grafts $\ubar h_{\rms 0}^{(0)}$ correctly into spacetime, i.e., replaces $\scal_0 t_*^{-3}\ubar h_{\rms 0}^{(0)}$ here by $\scal_0 h_{b,\rms 0}^{(0),\leq 3}(t_*^{-3})$, the $\cK^+$-order of this contribution improves to $\geq 4+\eps_\cK$; and this replacement is precisely what putting $\scal_0$ into the $\scal_0^{(0)}$-component of $\vecq$ in~\eqref{EqEfPvecq} does (cf.\ \eqref{EqEfhrem}, where $\scal_0^{(0)}$ contributes to the final term). Note that merely conormal remainders are directly grafted correctly in~\eqref{EqEfhrem} (and in any case not constructed in our present generalized Taylor series analysis at $\iota^+$).} Here, $\cF_{[i]}$ is an index set that can in principle be computed explicitly from $\Spec_{\iota^+}^{<\alpha_+}(\ubar L,\cE_\sscri^\cC)$ and the $\iota^+$-index sets of the $\tilde h_\lambda$; and we have $\cF_{[i]}=\emptyset$ when $\alpha_+\geq 3$ (in which case, as mentioned before, we have completed the determination of the $\iota^+$-expansion of the metric perturbation up to order $3+\eps_+$ conormal remainders).

When $\alpha_+\geq 3$, we are done, so let us assume that
\[
  \alpha_+\leq 3+\eps_+-\eps_\ind.
\]
For the sake of notational simplicity, let us assume that there is only one $\lambda_0\in\pi_1\cF_{[i]}$ with $\Re\lambda_0\in[\alpha_+,\alpha_++\eps_\ind)$; and let $k:=k(\cF_{[i]},\lambda_0)$ (see Definition~\ref{DefTMIndex}\eqref{ItTMIndexMaxLog}). We distinguish two cases:
\begin{enumerate}
\myitem{ItEfPCRes}{a}{\rm (Resonant case.)} $\lambda_0\in\spec_{\iota^+}^{<3+\eps_+}(\ubar L,\cE_\sscri^\cC)$; denote by $K\in\N$ the unique integer with $(\lambda_0,K-1)\in\Spec_{\iota^+}^{<3+\eps_+}(\ubar L,\cE_\sscri^\cC)$.
\myitem{ItEfPCNonRes}{b}{\rm (Non-resonant case.)} $\lambda_0\notin\spec_{\iota^+}^{<3+\eps_+}(\ubar L,\cE_\sscri^\cC)$.
\end{enumerate}
Let us moreover assume that $\spec_{\iota^+}^{<3+\eps_+}(\ubar L,\cE_\sscri^\cC)\cap\{\Re\lambda\in[\alpha_+,\alpha_++\eps_\ind)\}$ consists of at most one element. (If there are several elements $\lambda_0$ in $\cF_{[i]}$ and/or $\spec_{\iota^+}^{<3+\eps_+}(\ubar L,\cE_\sscri^\cC)$ with these properties, one merely needs to treat each element separately in the arguments below and sum up all resulting corrections to $h_{[i]}$, $\vecp_{[i]}$, and $\vecq_{[i]}$. Indeed, for our leading-order, Minkowskian, analysis at $\iota^+$, each element can be treated independently, as their nonlinear interactions in the output of $P$ occur only at orders $2+2\alpha_+>2+\alpha_++\eps_\ind$.)

Our goal is to find $\tilde h_{\lambda_0}=\tilde h_{\lambda_0}(t_*,R,\omega)$ which is quasi-homogeneous at $\iota^+$ of degree $\lambda_0$ in $t_*^{-1}$, and $\vecp'_{[i+1]}$, $\vecq'_{[i+1]}$ such that $h_{[i+1]}:=h_{[i]}+\chi_\iota\tilde h_{\lambda_0}$ and $\vecp_{[i+1]}:=\vecp_{[i]}+\vecp'_{[i+1]}$, $\vecq_{[i+1]}:=\vecq_{[i]}+\vecq'_{[i+1]}$ satisfy the same properties as~\eqref{EqEfPChj}--\eqref{EqEfPCPj} above except with $\alpha_++\eps_\ind$ in place of $\alpha_+$.

Let us denote the quasi-homogeneous $t_*^{-\lambda_0-2}$-term in the $\iota^+$-expansion of~\eqref{EqEfPCPj} by
\begin{equation}
\label{EqEfPCfQhom}
  f^{(\lambda_0+2)} = \sum_{j=0}^J t_*^{-\lambda_0-2}(\log t_*)^j f_j(R,\omega).
\end{equation}
Similarly to~\eqref{EqEfPcf3}, we need to characterize the behavior of $f^{(j)}$ at $\iota^+\cap\cK^+=\{R=0\}$.

\begin{lemma}[Orders of $f_j$]
\label{LemmaEfPCfjOrder}
  We have $f_j \in \Hb^{\infty,\ \bigl(\la\cE_\sscri^\cC+1\ra',4+\eps_\sscri\bigr),\ 4+\eps_\ind-\eps-(\Re\lambda_0+2)}(\iota^+)$ for all $j=0,\ldots,J$ and for all $\eps>0$.
\end{lemma}
\begin{proof}
  By expanding $P(h_{[i]},\vecp_{[i]},\vecq_{[i]})$ in Taylor series at $\iota^+$, one finds that the contributions to $f^{(\lambda_0+2)}$ arise from the following terms.
  \begin{enumerate}
  \item The term $u_{\lambda_0}$ of~\eqref{EqEfPChj}. This is present only in the resonant case, but in any case contributes only linearly, and via $\ubar L u_{\lambda_0}=0$.
  \item Nonlinear interactions of terms in~\eqref{EqEfPChj} with $\Re\lambda<\lambda_0$. The individual factors already have strong decay at $\cK^+$ (namely, $\cK^+$-orders $4+\eps_\cK$), so their nonlinear interactions have (more than) $4+\eps_\cK$ orders of decay at $\cK^+$. There are further terms arising from nonlinear interactions of the different summands in the arguments of $\Ric$ and $\Ups_{E^\Ups}$ in~\eqref{EqEfPMap}, but all of their $\cK^+$-orders are the sums of the orders of the individual summands (at $\iota^+$-orders $<\alpha_+$) and thus $>4+\eps_\cK$.
  \item Linear contributions from lower-order terms in the $\iota^+$-expansion of the modification terms $h_\tot(\vecp_{[i]})$, $\vartheta_\tot(\vecp_{[i]})$, and $h_{b,\rem}(\vecq_{[i]})$. But by their construction these terms have $>4+\eps_\ind-\eps$ orders of $\cK^+$-decay for all $\eps>0$; see Remark~\ref{RmkD4KOrder} regarding $h_\tot(\vecp_{[i]})$ and $\vartheta_\tot(\vecp_{[i]})$, and regarding $h_{b,{\rm rem}}(\vecq_{[i]})$ we note that only the terms involving $\vect_1$ and $\sfW_l^{(-\mu)}$ in~\eqref{EqEfhrem} have potentially non-trivial polyhomogeneous expansions at $\iota^+$, each term of which contributes to the output of $L_b$ (and thus of $P$) at $\cK^+$-order $\geq\min\Re\textscissors(\cE_+)+4$ and $\geq\min\Re\textscissors(\cE_++\mu)+(j+1)$, cf.\ (the computations in the proof of) Lemma~\ref{LemmaD6Aug5Linb}, and thus at $\cK^+$-order $\geq 4+\eps_\ind-\eps$.
  \end{enumerate}
  The stated $\cK^+$-order of $f_j$ is then such that the $\cK^+$-order of each summand of~\eqref{EqEfPCfQhom} is $4+\eps_\ind-\eps$ for all $\eps>0$.

  Recall that we are assuming that the components $\scal_1^{(\lambda_0-1,k)}$ (for all $k\in\N_0$), $\vect_1^{(\lambda_0,k)}$, $\scal_l^{(0),(\lambda_0,k)}$ ($l=0,2$) and $\scal_l^{(-\lambda^\Ups_{\rms l,1}+1),(\lambda_0+\lambda_{\rms l,1}^\Ups-1,k)}$ ($l=0,1$) of $\vecp_{[i]}$, which through~\eqref{EqEfPMap} would contribute terms with $\iota^+$-order $((\lambda_0+2,k),\lambda_0+2+\eps_\ind)$, as well as the components of $\vecq_{[i]}$ that contribute at the same order (e.g., the $(\lambda_0,k)$-terms of $\scal_0^{(0)}$ and $\scal_2^{(0)}$ in~\eqref{EqEfPvecq} in the case $\Re\lambda_0>2+\eps_\cK$) vanish.\footnote{Recall that they will be used momentarily to arrange the improved $\cK^+$-order of the $t_*^{-\lambda_0}$-term of $h_{[i+1]}$ and of the $t_*^{-\lambda_0-2}$-term of $P(h_{[i+1]},\vecp_{[i+1]},\vecq_{[i+1]})$.}
\end{proof}

If one modifies $\vecp_{[i]}$ and $\vecq_{[i]}$ by making the (finitely many, finite-dimensional) components $\scal_1^{(\lambda_0-1,k)}$ etc.\ mentioned in the proof of Lemma~\ref{LemmaEfPCfjOrder} nonzero, then the order $J$ of~\eqref{EqEfPCfQhom} may increase (if one uses $k>J$), but the conclusion of Lemma~\ref{LemmaEfPCfjOrder} would remain valid: indeed, the additional contributions are given by $\ubar L\bigl(\chi_\cK\ubar\delta^*(\ubar\omega_{\rms 1}^{(0),\leq 4}(t_*^{-\lambda_0+1}(\log t_*)^k\scal_1^{(\lambda_0-1,k)}))\bigr)$ (cf.\ \eqref{EqEfPCs1}) etc., as these are the $\iota^+$-leading-order terms of
\begin{equation}
\label{EqEfPCAddl}
  D_{g_{b_0}}\Ric( h_{\rms 1}^{(\lambda_0-1,k)} ) + \delta_{g_{b_0},E^\cC}^* \vartheta_{\rms 1}^{(\lambda_0-1,k)} \in \cA^{\infty,\ \bigl((\lambda_0+2,k),\,\lambda_0+2+\eps_\ind\bigr),\ 4+\eps_\ind-\eps}(\Omega_*;S^2\cT^*)
\end{equation}
etc.; recall here Remarks~\ref{RmkD4KOrder} and \ref{RmkD5KOrder}.

With this in mind, we only discuss the resonant case~\eqref{ItEfPCRes}; the arguments are very similar to those in Step~1. (In the non-resonant case~\eqref{ItEfPCNonRes}, the arguments simplify.) We use Corollary~\ref{CoripPQhom} with $\ell_\cK=4+\eps_\ind-\eps-\Re\lambda_0$ (which lies in $(1,3+\eps_\ind)$) and $-f^{(\lambda_0+2)}$ in place of $f$; this produces
\[
  h^{(\lambda_0)} = \sum_{j=0}^{J+K} t_*^{-\lambda_0}(\log t_*)^j h_j(R,\omega),\quad h_j\in\Hb^{\infty,\ \bigl(\la\cE_\sscri^\cC\ra,3+\eps_\sscri\bigr),\ (\cE_\cK,\ell_\cK)}(\iota^+)
\]
such that $\ubar L h^{(\lambda_0)}=-f^{(\lambda_0+2)}$; and $h^{(\lambda_0)}$ has an asymptotic expansion as $R\to 0$ as described in Corollary~\ref{CoripPQhom}\eqref{ItipPQhomExp} modulo a remainder of class $\sum_{j=0}^{J+K} t_*^{-\lambda_0}(\log t_*)^j\Hb^{\infty,\ (\la\cE_\sscri^\cC\ra,3+\eps_\sscri),\ \ell_\cK}(\iota^+)$ (which thus has $\cK^+$-order $\ell_\cK+\Re\lambda_0-\eps=4+\eps_\ind-2\eps$ for all $\eps>0$, so $>4+\eps_\cK$). The (quasi-homogeneous) $t_*^{-\lambda_0-2}$-term of $P(h_{[i]}+\chi_\iota h^{(\lambda_0)},\vecp_{[i]},\vecq_{[i]})$, which is given by $f^{(\lambda_0+2)}+\ubar L h^{(\lambda_0)}$, then vanishes. Note that the (quasi-homogeneous) $t_*^{-\lambda_0}$-term of $h_{[i]}+\chi_\iota h^{(\lambda_0)}$ is equal to
\begin{equation}
\label{EqEfPCLot0}
  u_{\lambda_0} + h^{(\lambda_0)} = \sum_{j=0}^{J+K} t_*^{-\lambda_0}(\log t_*)^j h'_j(R,\omega).
\end{equation}
Recalling that $\ubar L(u_{\lambda_0}+h^{(\lambda_0)})=\ubar L h^{(\lambda_0)}=-f^{(\lambda_0+2)}$, Lemma~\ref{LemmaEfPCfjOrder}, and Remark~\ref{RmkipPResAsy}, this is of the same class and has the same type of expansion as $h^{(\lambda_0)}$ above. We proceed to explain how to use the modification parameters to improve the $\cK^+$-order of~\eqref{EqEfPCLot0} to $4+\eps_\cK$; the strategy, as before, is to use the coefficients of the $R\to 0$ expansion to read off the modification parameters that eliminate said coefficients.
\begin{enumerate}
\item{\rm (Elimination of the $\ubar h_{\rms 1}^{\leq 4}$-term.)} Consider a term
  \begin{equation}
  \label{EqEfPhs1}
    \chi_\cK\ubar h_{\rms 1}^{\leq 4}\bigl(t_*^{-\lambda_0-1}(\log t_*)^k\scal_1^{(\lambda_0-1,k)\prime}\bigr),\quad k=J+K,
  \end{equation}
  in the asymptotic expansion of~\eqref{EqEfPCLot0} (coming from~\eqref{EqipNAsyu0}). We can replace this term by $\frac{1}{(-\lambda_0+1)(-\lambda_0)}$ times $\chi_\cK\ubar\delta^*\omega_{\rms 1}^{(0),\leq 4}\bigl(t_*^{-\lambda_0+1}(\log t_*)^k\scal_1^{(\lambda_0-1,k)\prime}\bigr)$ modulo terms that have strong $\cK^+$-decay or that can be absorbed into a re-definition of the lower-order terms $\scal_1^{(\lambda_0-1,j)\prime}$, $j\leq k-1$. Now, the term $\ubar L\bigl(\chi_\cK\ubar\delta^*\ubar\omega_{\rms 1}^{(0),\leq 4}(t_*^{-\lambda_0+1}(\log t_*)^k\scal_1^{(\lambda_0-1,k)\prime})\bigr)$ is, to leading order at $\iota^+$, equal to~\eqref{EqEfPCAddl} (with argument $\scal_1^{(\lambda_0-1,k)\prime}$); cf.\ also \eqref{EqEfPCs1}. Thus, changing the parameter $\scal_1^{(\lambda_0-1,k)}$ of $\vecp_{[i]}$ from $0$ to $\frac{1}{(-\lambda_0+1)(-\lambda_0)}\scal_1^{(\lambda_0-1,k)\prime}$ modifies $f^{(\lambda_0+2)}$ by adding this term, thus modifying $h^{(\lambda_0)}$ by subtracting the argument $\chi_\cK\ubar\delta^*\ubar\omega_{\rms 1}^{(0),\leq 4}(t_*^{-\lambda_0+1}(\log t_*)^k\scal_1^{(\lambda_0-1,k)\prime})$ from it, and thereby eliminating the term~\eqref{EqEfPhs1}. --- We proceed in this fashion iteratively for $k=J+K-1,J+K-2,\ldots,0$, thus producing $\scal_1^{(\lambda_0-1,k)}$, $k=0,\ldots,K+J$. They have the property that the additional $t_*^{-\lambda_0-2}$-terms of $P(\cdots)$ arising from the corresponding modification parameters contribute to $f^{(\lambda_0+2)}$ in such a manner that one obtains $f_{\rm new}^{(\lambda_0+2)}$ with the property that for the solution $h^{(\lambda_0)}_{\rm new}$ of $\ubar L h^{(\lambda_0)}_{\rm new}=-f^{(\lambda_0+2)}_{\rm new}$, the sum $u_{\lambda_0}+h^{(\lambda_0)}_{\rm new}$ has trivial $\ubar h_{\rms 1}^{\leq 4}$-terms in its expansion at $R\to 0$.
\item{\rm (Elimination of other terms with limited decay.)} We only need to eliminate those polyhomogeneous terms in the asymptotic expansion of~\eqref{EqEfPCLot0} at $R=0$ that are of the form $R^\mu(\log R)^k$ where $\Re(\lambda_0+\mu)\leq 2+\eps_\cK$. For small enough $\eps_\cK>0$ and due to the discreteness of all index sets, this is equivalent to $\Re(\lambda_0+\mu)\leq 2$. Since $\Re\lambda_0>1$, these are at most those terms already discussed in point~\eqref{ItPfPCs12} on page~\pageref{ItPfPCs12}, so they are all pure gauge and can be eliminated in a completely analogous fashion. (As a concrete example, when $\lambda_0=2$, we eliminate in this manner the terms $\ubar h_{\rms l}^{(0),\leq 3}(t_*^{-2}(\log t_*)^k\scal_l^{(0)})$, $l=0,2$, coming from~\eqref{EqipNAsyu0} using $\scal_l^{(0),(2,k)}$.)
\item{\rm (Accounting of the remaining terms.)} Among the remaining terms in the $R\to 0$ expansion of~\eqref{EqEfPCLot0}, those at order $R^\mu(\log R)^k$ with $\Re(\lambda_0+\mu)>4+\eps_\cK$ do not need to be eliminated at all since they already have a sufficiently strong $\cK^+$-decay order. The remaining terms, which have $\Re(\lambda_0+\mu)\in[2+\eps_\cK,4+\eps_\cK]$, can be absorbed into $h_{b,\rem}$ by adjusting the corresponding coefficients of $\vecq_{[i]}$. (In the example of $\lambda_0=2$, we thus eliminate in this manner the term $\ubar h_{\rmv 1}^{\leq 3}(t_*^{-3}(\log t_*)^k\vect_1^{(-1)})$ coming from~\eqref{EqipNAsyu0} using the $t_*^{-3}(\log t_*)^k$-coefficient of $\vect_1$ in~\eqref{EqEfPvecq}, further $\ubar h_{\rms l}^{(\mu),\leq 2}(t_*^{-2+\mu}\scal_l^{(\mu)})$ for $\mu=-\lambda^\Ups_{\rms l,l+j}+1$ where $0\leq l\leq 2$, $j=0,1$, $l+j\in\{1,2\}$, using the $t_*^{-2+\mu}(\log t_*)^k$-coefficient of $\scal_l^{(-\lambda^\Ups_{\rms l,1}+1)}$, further $\ubar h_{\rmv l}^{(-l+1),\leq 4-l}(t_*^{-2-l+1}(\log t_*)^k\vect_l^{(-l+1)})$ for $l=2,3$ using the $t_*^{-2-l+1}(\log t_*)^k$-coefficient of $\vect_l^{(-l+1),(2,k)}$, and finally $\ubar h_{\rmw 2}^{(-2),\leq 1}(t_*^{-4}(\log t_*)^k\sfW_2^{(-2)})$, $(\rmw,\sfW)=(\rms,\scal),(\rmv,\vect)$ using the $t_*^{-4}(\log t_*)^k$-coefficient of $\sfW_l^{(-2),(2,k)}$.)
\end{enumerate}

Note that these arguments also determine the next term $(\lambda_0,J+K)$ in the index sets of the metric perturbation at $\iota^+$ as well as of the modification parameters $\vecp,\vecq$. In summary, we have now constructed $\vecp'_{[i+1]}$ and $\vecq'_{[i+1]}$ such that, upon taking $\tilde h_{\lambda_0}$ to be the tensor given by Corollary~\ref{CoripPQhom} for the right-hand side $-f_{\rm new}^{(\lambda_0+2)}$, where $f_{\rm new}^{(\lambda_0+2)}$ is the quasi-homogeneous $t_*^{\lambda_0-2}$-term of $P(h_{[i]},\vecp_{[i]}+\vecp'_{[i+1]},\vecq_{[i]}+\vecq'_{[i+1]})$, we have, upon defining
\[
  h_{[i+1]} := h_{[i]} + \chi_\iota\tilde h_{\lambda_0},\quad
  \vecp_{[i+1]} := \vecp_{[i]} + \vecp'_{[i+1]},\quad
  \vecq_{[i+1]} := \vecq_{[i]} + \vecq'_{[i+1]},
\]
the following two properties:
\begin{enumerate}[label=(\roman*)]
\item For all $\lambda$ with $\Re\lambda<\alpha_++\eps_\ind$, the $t_*^{-\lambda}(\log t_*)^k$-term in the $\iota^+$-expansion of $h_{[i+1]}$ has $\cK^+$-order $4+\eps_\cK$.
\item $P(h_{[i+1]},\vecp_{[i+1]},\vecq_{[i+1]})\in\Hb^{\infty,\ \bigl(\la\cE_\sscri^\cC+1\ra',4+\eps_\sscri\bigr),\ (\cF_{[i+1]},5+\eps_+),\ \alpha''_\cK}$ where $\min\Re\cF_{[i+1]}\geq\alpha_++\eps_\ind$, and $\alpha''_\cK=\alpha_++\eps_\ind-\eps$ unless $\alpha_++\eps_\ind>3$ in which case $\alpha''_\cK=4+\eps_\cK$.
\end{enumerate}

\pfstep{Summary.} After $I<\infty$ many steps---$I:=\lceil 3\eps_\ind^{-1}\rceil+1$ steps suffice---the above construction produces $(h,\vecp,\vecq):=(h_{[I]},\vecp_{[I]},\vecq_{[I]})$ such that $P(h,\vecp,\vecq)$ has orders $5+\eps_+$ and $4+\eps_\cK$ at $\iota^+$ and $\cK^+$, respectively, i.e., \eqref{EqEfPGoal} holds. We summarize the key properties of this construction as follows.

\begin{thm}[Efficient parameterization, I: construction of approximate solutions from data]
\label{ThmEfP}
  For fixed index sets $\cE_\sscri^\cC$ as in Definition~\usref{DefExP} and all sufficiently small $0<\eps_\cK<\eps_+<\eps_\sscri<1$, there exist an index set $\cE_+$, with $(1,0)\in\cE_+$ and $\min\Re\cE_+\setminus\{(1,0)\}>1$, and constants $\eps_0>0$ and $d\in\N$ such that there exists $d$ with the following property. There exists an $N$-fold differentiable (for any fixed $N$) nonlinear map $\Phi$, defined on $\{U\in\fD^\infty(\Omega_*)\colon\|U\|_{\fD^d}<\eps_0\}$ and extending continuously to
  \[
    \Phi \colon \fD^{k+d}(\Omega_*) \to \Hb^{k+2,\ \bigl(\la\cE_\sscri^\cC\ra,3+\eps_\sscri\bigr),\ (\cE_+,3+\eps_+),\ 4+\eps_\cK}(\Omega_*)^{\bullet,-} \oplus \fP \oplus \fQ^k,
  \]
  where $\fP$ and $\fQ^k$ are the spaces in which $\vecp$ and $\vecq$ in~\eqref{EqEfPvecp} and \eqref{EqEfPvecq} (with index sets $\tilde\cE'_{++}:=\cE_+\setminus\{(1,0)\}$) lie (with $k+2$ the regularity order of the conormal remainders of the components of~\eqref{EqEfPvecq}) such that
  \begin{equation}
  \label{EqEfPPPhi}
    P\bigl(\Phi(U)\bigr) \in \Hb^{k-2,\ \bigl(\la\cE_\sscri^\cC+1\ra',4+\eps_\sscri\bigr),\ 5+\eps_+,\ 4+\eps_\cK}(\Omega_*)^{\bullet,-}.
  \end{equation}
  Moreover, writing $U\in\fD^\infty(\Omega_*)$ as
  \[
    U=\Bigl(b-b_0,\scal,\scal_1^{(0)},\bigl(u^0_\lambda,f_\lambda\bigr)_{\lambda\in\spec_{\iota^+}^{<3+\eps_+}(\ubar L,\cE_\sscri^\cC)},\scal_\rem,b_\rem,\vect_{1,\rem},\bigl(\scal_{l,\rem}^{(0)}\bigr)_{l=0,2},\bigl(\scal_{l,\rem}^{(-\lambda^\Ups_{\rms l,1}+1)}\bigr)_{l=0,1},\tilde h\Bigr)
  \]
  and $u_\lambda:=\Res_{\iota^+}(\ubar L,\cE_\sscri^\cC,\lambda)(u^0_\lambda,f_\lambda)$ (using~\eqref{EqipPResPar}), the output $\Phi(U)=(h,\vecp,\vecq)$ has the following properties:
  \begin{enumerate}
  \item\label{ItEfPBH}{\rm (Respecting final black hole, boost, and translational parameters.)} The first three components $\scal,b-b_0,\scal_1^{(0)}$ of $\vecp$ (see~\eqref{EqEfPvecp}) are equal to the eponymous components of $U$.
  \item\label{ItEfPDec}{\rm (Respecting decaying error terms.)} The components $\scal_\rem,b_\rem$ of $\vecq$ (see~\eqref{EqEfPvecq}) are equal to the eponymous components of $U$, while the conormal remainder terms of $\vect_1$, $\scal^{(0)}_0$, $\scal^{(0)}_2$, and $\scal_l^{(-\lambda^\Ups_{\rms l,1}+1)}$, $l=0,1$, in~\eqref{EqEfPvecq} are equal to $\vect_{1,\rem}$, $\scal^{(0)}_{0,\rem}$, $\scal^{(0)}_{2,\rem}$, and $\scal_{l,\rem}^{(-\lambda^\Ups_{\rms l,1}+1)}$, respectively. Moreover, changing $\tilde h$ to $\tilde h+\tilde h'$ changes the component $h$ of the output of $\Phi$ to $h+\tilde h'$.
  \item\label{ItEfPInf}{\rm (Influence of $u_\lambda$.)} Suppose $U_1$ and $U_2\in\fD^\infty(\Omega_*)$ differ only in the component $(u_{\lambda_0}^0,f_{\lambda_0})$, $\lambda_0\in\spec_{\iota^+}^{3+\eps_+}(\ubar L,\cE_\sscri^\cC)$, by $(u^{\prime 0}_{\lambda_0},f'_{\lambda_0})$. Set
  \[
    (h',\vecp',\vecq'):=\Phi(U_1)-\Phi(U_2),\quad
    u'_\lambda:=\Res_{\iota^+}(\ubar L,\cE_\sscri^\cC,\lambda_0)(u^{\prime 0}_{\lambda_0},f'_{\lambda_0}).
  \]
  Then $h'$ has $\iota^+$-order $((\lambda_0,k),\lambda_0+\eps_\ind)$ for some $k$, and all components of $\vecp',\vecq'$ whose contributions to $g^0+h_\tot(\vecp)+h_{b,\rem}(\vecq)+h$ in~\eqref{EqEfPMap} have $\iota^+$-order less than $\lambda_0+2$ vanish; moreover, the quasi-homogeneous $t_*^{-\lambda_0}$-term $h^{\prime(\lambda_0)}$ of $h'$ at $\iota^+$ is equal to $u'_{\lambda_0}-\tilde u$ where $\tilde u$ is the sum of the $t_*^{-\lambda_0}$-terms of $h_\tot(\vecp')+h_{b,\rem}(\vecq')$.
  \end{enumerate}
\end{thm}

\begin{rmk}[Linearization in $U$]
\label{RmkEfPLin}
  If one linearizes~\eqref{EqEfPPPhi} in $U$, one obtains
  \begin{equation}
  \label{EqEfPLin}
    D_{\Phi(U)}P\bigl(D_U\Phi(U')\bigr) \in \Hb^{k-2,\ \bigl(\la\cE_\sscri^\cC+1\ra',4+\eps_\sscri\bigr),\ 5+\eps_+,\ 4+\eps_\cK}(\Omega_*)^{\bullet,-}.
  \end{equation}
  Writing $(h',\vecp',\vecq'):=D_U\Phi(U')$, the left-hand side is $D_{(h,\vecp,\vecq)}P(h',\vecp',\vecq')$, so $(h',\vecp',\vecq')$ is an approximate solution of the linearized operator. Linearized versions of parts~\eqref{ItEfPBH}--\eqref{ItEfPInf} then remain valid.
\end{rmk}

\begin{rmk}[Interpretation of part~\eqref{ItEfPInf}]
\label{RmkEfPInf}
  Theorem~\ref{ThmEfP}\eqref{ItEfPInf} asserts that each $(u_{\lambda_0}^0,f_{\lambda_0})$ contributes fully to the asymptotic expansion of the component $h$ of $\Phi(U)$. This contribution is not quite the addition of $u_{\lambda_0}$ to $h$ simply (plus corrections at lower orders than $t_*^{-\lambda_0}$ at $\iota^+$ to ensure the fast vanishing of the output of $P$), since typical terms $u_{\lambda_0}$ have a nontrivial expansion at $\cK^+$. Instead, an inspection of the construction shows that $\Phi$ utilizes the modification parameters at $\iota^+$-order $t_*^{-\lambda_0-2}$ to eliminate the $\cK^+$-expansion of $u_{\lambda_0}$ modulo errors with acceptable $4+\eps_\cK$ orders of $\cK^+$-decay. As a concrete example, and without logarithmic factors for notational simplicity, if the expansion of $u_{\lambda_0}$ only includes the single term $\chi_\cK\ubar h_{\rms 1}^{\leq 4}(t_*^{-\lambda_0-1}\scal)$ (cf.\ the first term of~\eqref{EqipNAsyu0}, times $t_*^{-\lambda_0}$) for $\Re\lambda_0>1$ or $\chi_\cK\ubar h_{\rms 2}^{(0),\leq 3}(t_*^{-\lambda_0}\scal_2^{(0)})$ for $\Re\lambda_0\geq 1$, then $\tilde u=\frac{1}{(-\lambda_0+1)(-\lambda_0)}\chi_\cK\ubar\delta^*\ubar\omega_{\rms 1}^{(0),\leq 4}(t_*^{-\lambda_0+1}\scal)$ or $\tilde u=\chi_\cK\ubar\delta^*\ubar\omega_{\rms 2}^{(-1),\leq 3}(t_*^{-\lambda_0}\scal_2^{(0)})$.
\end{rmk}

\begin{rmk}[Backward vs.\ forward solutions]
\label{RmkEfPSimple}
  The construction of the map $\Phi$, which amounts to solving the nonlinear gauge-fixed Einstein equation to sufficiently high order at $\iota^+$ and $\cK^+$, is vastly simpler than the proof of the existence of partial expansions and of the possibility of removing all terms in the $\cK^+$-expansion of solutions of linearized \emph{forward} solutions in Corollary~\ref{CorD6Impr}. In the present section we only needed to concern ourselves with the Minkowskian model problem at $\iota^+$ (and the elimination of asymptotic expansions at $\iota^+\cap\cK^+$ for such model problems), whereas in~\S\ref{SD} we had to work on fully dynamical spacetimes settling down to a Kerr spacetime at $\cK^+$. Notice also that while, say, Step~1 in the proof of Theorem~\ref{ThmEfP} shows how the parameters $b,\scal,\scal_1^{(0)}$, and $(u_1^0,f_1)$ uniquely determine the $t_*^{-1}$-term of $h$ as well as all modification parameters at that order, it is, conversely, not possible to extract the full $R\to 0$ expansion (up to $\cO(R^{3+\eps_\cK})$-remainders) of the $t_*^{-1}$-term of forward solutions of $L u=f$ in the context of Theorem~\ref{ThmD6} until one has proved strong $\cK^+$-decay of $u$. --- It may also be instructive to compare Theorem~\ref{ThmEfP} with what would be an analogue in the Kerr--de~Sitter setting of \cite{HintzVasyKdSStability,HintzPetersenVasyKdS}: the only parameter with nontrivial behavior at $\tau=e^{-t_*}=0$ is the final KdS black hole parameter $b$, while the gravitational wave tail $\wt g$ and the contributions from gauge modifications $\theta$ either already have the required exponential decay as $t_*\to\infty$ or are compactly supported. Thus, there is no need at all for a map like $\Phi$ above: the naively defined nonlinear forward map, with arguments $b,\wt g,\theta$, already maps into an acceptable $\Hb^{\infty,\alpha}$-space of source terms for the next linearized problem in the nonlinear iteration. (The precise formulation in \cite[\S{11.2}]{HintzVasyKdSStability} is only slightly more complicated.)
\end{rmk}

The map $\Phi$ in Theorem~\ref{ThmEfP} does not encode an initial solution of the Einstein equation in $\ft_*\leq 2$ that one may wish to extend to $\ft_*\geq 1$. Since this is only an issue for bounded $\ft_*$ (and thus not at $\iota^+$ and $\cK^+$), such a generalization is straightforward to implement. To wit, suppose that for small $\scal\in\scalspace_1$ we are given
\begin{subequations}
\begin{equation}
\label{EqEfPIniFam}
  h_\scal\in\bar H_\bop^{\infty,\ (\la\cE_\sscri^\cC\ra,3+\eps_\sscri)}(\{1\leq\ft_*\leq 2\}),\ \text{with smooth dependence on $\scal$},
\end{equation}
such that\footnote{This is a family of initial solutions of the gauge-fixed Einstein equation we wish to extend to the future; cf.\ Theorem~\ref{ThmExBoStab} and the discussion preceding Definition~\ref{DefD6Aug5}.}
\begin{equation}
\label{EqEfPIniEin}
  \Ric(\phi_\scal^*g_{b_0}+h_\scal) - \delta_{g_{b_0},E^\cC}^*\Ups_{E^\Ups}(\phi_\scal^*g_{b_0}+h_\scal,\phi_\scal^*g_{b_0}) = 0.
\end{equation}
\end{subequations}
Fix $\chi_-\in\CI(\R)$ to be equal to $1$ on $(-\infty,\frac74]$ and $0$ on $[2,\infty)$. We can then define
\begin{equation}
\label{EqEfPIni}
  \Psi(U) := \bigl(\chi_-(\ft_*)h_{-\scal},0,0\bigr) + \Phi(U)
\end{equation}
where we use the component $\scal$ of $U$ on the right-hand side. Note then that $P(\Psi(0))=\Ric(\phi_{-\scal}^*g_{b_0}+\chi_- h_{-\scal})-\delta_{g_{b_0},E^\cC}^*\Ups_{E^\Ups}(\phi_{-\scal}^*g_{b_0}+h_{-\scal},\phi_{-\scal}^*g_{b_0})$ is supported in $\supp\dd\chi_-$, and hence vanishes for $\ft_*\leq\frac74$ and for $\ft_*\geq 2$. More generally, for $h$ (the last component of $U$) with $\supp h\subset\{\ft_*\geq c\}$ for some $c\in[1,\frac74]$, we have $\supp P(\Psi(U))\subset\{\ft_*\geq c\}$ as well. Therefore:

\begin{lemma}[Properties of $\Psi$]
\label{LemmaEfPPsi}
  Theorem~\usref{ThmEfP} remain valid for $\Psi$ in place of $\Phi$.
\end{lemma}
\begin{proof}
  Since $\chi_-(\ft_*)h_{-\scal}=0$ for $\ft_*\geq 2$, we have $P(\Psi(U))=P(\Phi(U))$ for $\ft_*\geq 2$. Furthermore, the $\scri^+$-order of $P(\Psi(U))$ is still equal to $(\la\cE_\sscri^\cC+1\ra',4+\eps_\sscri)$: this only requires an argument near $\scri^+\cap\supp\dd\chi_-$, where however this follows from Corollary~\ref{CorExFwN}\eqref{ItExFwN1}. Finally, the observation about $\supp P(\Psi(U))$ above, with $c=1$, implies that $P(\Psi(U))$ has supported character at the initial hypersurface of $\Omega_*$.
\end{proof}

The map $\Psi$ generalizes $\Phi$ in that the trivial family $h_\scal=0$ satisfies~\eqref{EqEfPIniFam}--\eqref{EqEfPIni}, and in that case $\Psi=\Phi$.

\subsection{Description of solutions of the linearized equation}
\label{SsEfS}

Our remaining task is to show that the map $\Phi$ from Theorem~\ref{ThmEfP} is flexible enough to accommodate the asymptotics of forward solutions of the linearization of $P$ around $(h,\vecp,\vecq)=\Phi(U)$ produced by Corollary~\ref{CorD6Impr}. This is the content of the following result:

\begin{thm}[Efficient parameterization of linear forward solutions]
\label{ThmEfS}
  Let $U\in\fD^\infty$ be small in $\fD^d$ for some large but fixed $d$; use the notation of Theorem~\usref{ThmEfP} and set $(h,\vecp,\vecq):=\Phi(U)$. Suppose we are given
  \[
    f\in\Hb^{\infty,\ \bigl(\la\cE_\sscri^\cC+1\ra',4+\eps_\sscri\bigr),\ 5+\eps_+,\ 4+\eps_\cK}(\Omega_*)^{\bullet,-}.
  \]
  Then there exist $U'\in\fD^\infty$ such that
  \begin{equation}
  \label{EqEfSEq}
    D_{\Phi(U)}P\bigl(D_U\Phi(U')\bigr) = f.
  \end{equation}
  Moreover, we have tame estimates: for $d$ large enough, and for all $k\in\N_0$, we have
  \[
    \|U'\|_k \leq C_k\Bigl( \|f\|_{k+d} + \|U\|_{k+d}\|f\|_d \Bigr).
  \]
  The same remains true for $\Psi$ in place of $\Phi$, defined by~\eqref{EqEfPIni} with respect to a family of initial solutions $h_\scal$ as in~\eqref{EqEfPIniFam}.
\end{thm}

\textit{The arguments for $\Psi$ are identical to those for $\Phi$, and we shall only work with $\Phi$ below.}

Let us expand~\eqref{EqEfSEq} by writing $D_U\Phi(U')=:(h',\vecp',\vecq')$; then
\[
  D_{(h,\vecp,\vecq)}P(h',0,0) = f - D_{(h,\vecp,\vecq)}P(0,\vecp',\vecq').
\]
Thus, Theorem~\ref{ThmEfS} has \emph{almost} the same content as Corollary~\ref{CorD6Impr}, except for the crucial improvement that it provides a considerably sharper---and self-consistent for nonlinear purposes, in that $f=-\Phi(U)$ is an acceptable source term by~\eqref{EqEfPPPhi}---description of $(h',\vecp',\vecq')$.

\begin{rmk}[Non-uniqueness]
\label{RmkEfSNonUniq}
  Neither $D_{(h,\vecp,\vecq)}P$ nor $D_U\Phi$ are injective. For example, contributions to the component $b_\rem$ of $\vecq'$ of class $\Hb^{\infty,4+\eps_\cK}$ can alternatively be absorbed into a re-definition of $h$. What the proof below will show is that the \emph{particular} solution $(h',\vecp',\vecq')$ constructed by Corollary~\ref{CorD6Impr} (and for which we have tame estimates) is of the form $D_U\Phi(U')$, with a constructive procedure for the determination of $U'$ (which thus inherits the tame estimates from $(h',\vecp',\vecq')$).
\end{rmk}

\begin{proof}[Proof of Theorem~\usref{ThmEfS}]
  Let $h'\in\Hb^{\infty,\ (\la\cE_\sscri^\cC\ra,3+\eps_\sscri),\ (\tilde\cE_\sharp,3+\eps_+),\ 4+\eps_\cK}(\Omega_*)^{\bullet,-}$ and $\vecp',\vecq'$ be the tensors and parameters produced by Corollary~\ref{CorD6Impr}. Thus,
  \begin{equation}
  \label{EqEfSEqPf}
    D_{(h,\vecp,\vecq)}P(h',\vecp',\vecq') = f \in \Hb^{\infty,\ \bigl(\la\cE_\sscri^\cC+1\ra',4+\eps_\sscri\bigr),\ 5+\eps_+,\ 4+\eps_\cK}(\Omega_*)^{\bullet,-}.
  \end{equation}
  We remark that at this point, the index sets $\tilde\cE_\sharp$ and $\tilde\cE'_{++}$ used for the $\iota^+$-expansion of $h'$, for indexing the modification parameters $\vecp'$, and for the $t_*\to\infty$ expansions of the components of $\vecq'$ may be different than the index set $\cE_+$ of Theorem~\ref{ThmEfP}; part of our task is thus to prove that extraneous expansion terms and parameters (compared to what Theorem~\ref{ThmEfP} uses) must in fact vanish. The only properties we need to use are
  \begin{equation}
  \label{EqEfSIndexSets}
    (1,0)\in\cE_+,\,\tilde\cE_\sharp;\qquad \min\Re\bigl(\cE_+\setminus\{(1,0)\}\bigr),\ \min\Re\bigl(\tilde\cE_\sharp\setminus\{(1,0)\}\bigr),\ \min\Re\tilde\cE'_{++} \geq 1+\eps_\ind.
  \end{equation}

  We shall iteratively peel off the terms in the $\iota^+$-expansions of the linearizations of $g^0+h_\tot(\vecp)+h_{b,\rem}(\vecq)+h$, $g^0+h_{b,\rem}(\vecq)+h$, and $\vartheta_\tot(\vecp)$ in~\eqref{EqEfPMap}, ordered by the $\iota^+$-decay rate of their contributions to the outputs of
  \begin{equation}
  \label{EqEfSLinTerms}
    D_{(h,\vecp,\vecq)}P(h',0,0),\ D_{(h,\vecp,\vecq)}P(0,\vecp',0),\ D_{(h,\vecp,\vecq)}P(0,0,\vecq').
  \end{equation}
  In view of~\eqref{EqEfSIndexSets}, these three tensors have $\iota^+$-order
  \[
    (\cF+2,5+\eps_+);\qquad (1,0)\in\cF,\quad \min\Re\bigl(\cF\setminus\{(1,0)\}\bigr)\geq 1+\eps_{\rm ind}.
  \]
  (Recall here that the $t_*^{-1}\log t_*$-terms of $h_\tot(\vecp')$ do not cause $t_*^{-3}\log t_*$-terms in the output, as follows from~\eqref{EqD2CorrRich}, \eqref{EqD4CorrEin}, and \eqref{EqD6Corrhs0}.)

  As in~\S\ref{SssEfPC}, our arguments are based on (generalized) Taylor expansions at $\iota^+$. The strategy is to write the vanishing of what one should naively expect to be the leading-order term of $D_{(h,\vecp,\vecq)}P(h',\vecp',\vecq')$ at $\iota^+$ (i.e., the sum of the leading-order terms of~\eqref{EqEfSLinTerms}---so the $t_*^{-3}$-term in the first step) as an identity on the level of the Minkowskian normal operator $\ubar L$. We then use, essentially, normal operator arguments, relying on the (spectral) properties of $\ubar L$ and $N_{\iota^+}(\ubar L,\lambda)$ proved in~\S\ref{SsipInv}, to infer that the components of $h'$, $\vecp'$, and $\vecq'$ that enter at a given order must solve a linear equation (involving $\ubar L$ acting on quasi-homogeneous inputs), for which we have a full solvability and uniqueness theory (Corollary~\ref{CoripPQhom}, Definition~\ref{DefipPRes})---and thus these components must necessarily match those constructed in (a linearized version of) the proof of Theorem~\ref{ThmEfP} above.

  \pfstep{Step~1. Terms contributing to~\eqref{EqEfSLinTerms} at order $t_*^{-3}$.} These terms are:
  \begin{enumerate}
  \item The $t_*^{-1}$-term $t_*^{-1}h^{(1,0)}$ of $h$, where
    \begin{equation}
    \label{EqEfSh10}
      h^{(1,0)} \in \Hb^{\infty,\ \bigl(\la\cE_\sscri^\cC\ra,3+\eps_\sscri\bigr),\ 3+\eps_\cK}(\iota^+);
    \end{equation}
    this contributes via $\ubar L(t_*^{-1}h^{(1,0)})=t_*^{-3}N_{\iota^+}(\ubar L,1)h^{(1,0)}$.
  \item All $t_*^{-1}$- and $t_*^{-1}\log t_*$-terms of $D_\vecp h_\tot(\vecp')=h_\tot(\vecp')$. These are the terms $h^{(-1)}(\scal')$, $h^{(0)}(b')$, $h_{\rms 1}^{(0)}(\scal_1^{(0)\prime})$ (cf.\ the contributions to~\eqref{EqEfPcf3}), $h_{\rms 1}^{(0,1)}(\scal_1^{(0,1)\prime})$ (cf.\ the discussion around the identity~\eqref{EqEfPCs1}), as well as the terms $\ubar h_{\rmv 1}^{\leq 3}(t_*^{-2}\vect_1^{(1,0)\prime})$, $\ubar h_{\rms l}^{(0),\leq 3}(t_*^{-1}\scal_l^{(0),(1,0)\prime})$ ($l=0,2$), $\ubar h_{\rms l}^{(\mu),\leq 2}(t_*^{-1+\mu}\scal_l^{(\mu),(1-\mu,0)\prime})$ for $l=0,1$ and $\mu=-\lambda^\Ups_{\rms l,1}+1$, and $\ubar h_{\rms 3}^{(-1),\leq 2}(t_*^{-2}\scal_3^{(-1),(2,0)\prime})$ and $\ubar h_{\rmv 2}^{(-1),\leq 2}(t_*^{-2}\vect_2^{(-1),(2,0)\prime})$ as discussed in point~\eqref{ItPfPCs12} on page~\pageref{ItPfPCs12}. These contribute via $D_{\ubar g}\Ric(h_\tot(\vecp'))$. Similarly, the corresponding $\vartheta$-terms contribute via $\ubar\delta_{\ubar E^\cC}^*\vartheta_\tot(\vecp')$.
  \item All $t_*^{-1}$-terms of $D_\vecq h_{b,\rem}(\vecq')=h_{b,\rem}(\vecq')$, which like $t_*^{-1}h^{(1,0)}$ also contribute through $N_{\iota^+}(\ubar L,1)$. These are the leading-order terms of those partially polyhomogeneous terms among~\eqref{EqD6Aug5qCoeff} for which $\textscissors$ does not cut out any part of the index set (i.e., when $\min\Re(\cE_++\mu)>2$ there), so for example the $t_*^{-3}$-term $\scal_4^{(-2)\prime,(3)}$ of $\scal_4^{(-2)\prime}$, etc).
  \end{enumerate}
  For notational simplicity, let us suppose that among these terms, all terms other than possibly
  \begin{equation}
  \label{EqEfSLint3Terms}
    h^{(1,0)}=h^{(1,0)}(R,\omega),\quad b',\,\scal_1^{(0,1)\prime},\,\scal_2^{(0),(1,0)\prime}\ \text{(from $\vecp'$)}, \quad \scal_4^{(-2)\prime,(3)}\ \text{(from $\vecq'$)}
  \end{equation}
  vanish. (The arguments in the general case are completely analogous, and only involve more notation.)

  \pfsubstep{Step~1.1.}{Identity at order $t_*^{-3}$ from $(h',\vecp',\vecq')$.} The $t_*^{-3}$-term of~\eqref{EqEfSEqPf} vanishes, which means that that of
  \begin{align}
    D_{g^0}\Ric(\dot g^0) &+ \ubar L(t_*^{-1}h^{(1,0)}) \nonumber\\
      &+ D_{\ubar g}\Ric\Bigl( h^{(0)}(b') + h_{\rms 1}^{(0,1)}(\scal_1^{(0,1)\prime}) + h_{\rms 2}^{(0),(1,0)}(\scal_2^{(0),(1,0)\prime}) \Bigr) \nonumber\\
     &\qquad + \ubar\delta_{E^\cC}^* \Bigl(\vartheta^{(0)}(b') + \vartheta_{\rms 1}^{(0,1)}(\scal_1^{(0,1)\prime}) + \vartheta_{\rms 2}^{(0),(1,0)}(\scal_2^{(0),(1,0)\prime}) \Bigr) \nonumber\\
  \label{EqEfSLinTerms4}
     & + \ubar L \Bigl( \chi_\cK h_{b,\rms 4}^{(-2),\leq 1}\bigl(t_*^{-3}\scal_4^{(-2)\prime,(3)}\bigr) \Bigr)
  \end{align}
  vanishes; here
  \[
    \dot g^0 := \frac{\dd}{\dd s} g_{b_0,b+s b',-(\scal+s\scal')}\Big|_{s=0}
  \]
  is the linearization of $g^0=g_{b_0,b,-\scal}$ in the parameters $(b,\scal)$ (so for $\scal'=0$, this is $\chi_\cK\dot g_b(b')$); moreover, the second and third line collect the terms from $\vecp'$, and the fourth line collects the term from $\vecq'$; and we have already partially passed to the $\iota^+$-model metric $\ubar g$. Let us evaluate the $t_*^{-3}$-term of~\eqref{EqEfSLinTerms4}.
  \begin{enumerate}[label=(\alph*)]
  \item Similarly to~\eqref{EqEfPCt3}, the fact that $\Ric(g_{b_0,b,-\scal})$ is supported on $\supp\dd\chi_\cK$ for all $b,\scal$ implies that the first term has $t_*^{-3}$-coefficient of class $\CIc((\iota^+)^\circ;S^2\cT^*)$, and likewise for the two other terms involving $b'$ in lines 2 and 3; let us denote the sum of these $t_*^{-3}$-coefficients by $f^{(3,0)}_\cp(b')=f^{(3,0)}_\cp(b')(R,\omega)\in\CIc((\iota^+)^\circ;S^2\cT^*)$. An efficient characterization is that
    \begin{equation}
    \label{EqEfSf3Char}
      f^{(3,0)}_\cp(b')\ \text{is the $t_*^{-3}$-coefficient of $D_{(h,\vecp,\vecq)}P(b')$.}
    \end{equation}
    (If we allowed for $\scal'$, we would need to work instead with $f^{(3,0)}_\cp(\scal',b'):=D_{(h,\vecp,\vecq)}P(\scal',b')$.)
  \item The sum of terms involving $\scal_1^{(0,1)\prime}$ and $\scal_2^{(0),(1,0)\prime}$ has (by their definition) the same $t_*^{-3}$-leading-order term at $\iota^+$ as
    \begin{equation}
    \label{EqEfSLinLTerms}
      \ubar L \biggl( \chi_\cK \ubar\delta^* \Bigl( \ubar\omega_{\rms 1}^{(0),\leq 4}\bigl(\log(t_*)\scal_1^{(0,1)\prime}\bigr) + \ubar\omega_{\rms 2}^{(-1),\leq 3}\bigl(t_*^{-1}\scal_2^{(0),(1,0)\prime}\bigr) \Bigr)\biggr).
    \end{equation}
    Recalling the (type of) computation~\eqref{EqD4ParNoDelvsH}, the argument of $\ubar L$ here can be expanded as the sum of $\chi_\cK\ubar h_{\rms 1}^{\leq 4}(-t_*^{-2}\scal_1^{(0,1)\prime})+\chi_\cK\ubar h_{\rms 2}^{(0),\leq 3}(t_*^{-1}\scal_2^{(0),(1,0)\prime})$ and the terms $4!t_*^{-5}\chi_\cK\,\dd t_*\otimes_s\breve{\ubar\omega}_{\rms 1}^{(0),4}(\scal_1^{(0,1)\prime})$ and $4!t_*^{-5}\chi_\cK\,\dd t_*\otimes_s\breve{\ubar\omega}_{\rms 2}^{(-1),3}(\scal_2^{(0),(1,0)\prime})$; the point is that the argument of~\eqref{EqEfSLinLTerms} is homogeneous of degree\footnote{As already in the arguments~\S\ref{SssEfPC}, the $\rms 1$ term is exceptional since $\ubar\omega_{\rms 1}^{(0),\leq 1}$ is Killing on Minkowski space. The $\rms 0$ term (i.e., the third one in~\eqref{EqipNAsyu0}) is similarly exceptional since the gauge potential has a $\log t_*$ factor; this is handled in a similar way. All other terms, such as the $\rms 2$ term treated here, are not exceptional.} $-1$ in $t_*$ (as a function of $(t_*,R,\omega)$), and its $R\to 0$ expansion is of the form~\eqref{EqipNAsyu0} (for $\lambda=1$, and with only the first and scalar $l=2$ terms present here).
    \item The $t_*^{-1}$-leading-order term of the argument of~\eqref{EqEfSLinTerms4} similarly accounts for the $l=4$ term in~\eqref{EqipNAsyu0}.
  \end{enumerate}
  We therefore have the identity
  \begin{equation}
  \label{EqEfSLinLIdent}
  \begin{split}
    &\ubar L\bigl(t_*^{-1}h^{(1,0)}+h^{(1)\prime}\bigr) + t_*^{-3}f^{(3,0)}_\cp(b') = 0, \\
    &\quad h^{(1)\prime} := \chi_\cK\ubar\delta^*\Bigl(\ubar\omega_{\rms 1}^{(0),\leq 4}\bigl(\log(t_*)\scal_1^{(0,1)\prime}\bigr) + \ubar\omega_{\rms 2}^{(-1),\leq 3}\bigl(t_*^{-1}\scal_2^{(0),(1,0)\prime}\bigr)\Bigr) + \chi_\cK\ubar h_{\rms 4}^{(-2),\leq 1}\bigl(t_*^{-3}\scal_4^{(-2)\prime,(3)}\bigr),
  \end{split}
  \end{equation}
  of tensors that, in the coordinates $(t_*,R,\omega)$, are homogeneous of degree $3$ in $t_*^{-1}$; here $h^{(1)\prime}=h^{(1)\prime}(t_*,R,\omega)$ is homogeneous of degree $1$.

  \pfsubstep{Step~1.2.}{Identity at order $t_*^{-3}$ from $D_U\Phi(\cdots)$.} We compare these computations to the computation of $D_U\Phi(b',0,\ldots,0)$. Recall from Remark~\ref{RmkEfPLin} that $D_U\Phi(\cdots)$ produces formal solutions at $\iota^+$ (and $\cK^+$) of the linearization $D_{\Phi(U)}P$. At leading order at $\iota^+$, this amounts to the following: given $b'$, find modification parameters that enter at $\iota^+$-order $t_*^{-3}$---so in line with the notational simplification made above, parameters $\dot\scal_1^{(0,1)}$, $\dot\scal_2^{(0),(1,0)}$, and $\dot\scal_4^{(-2),(3)}$---and a homogeneous tensor $\tilde h_1\in t_*^{-1}\Hb^{\infty,\ (\la\cE_\sscri^\cC\ra,3+\eps_\sscri),\ 3+\eps_\cK}(\iota^+)$ (which thus has $\cK^+$-order $4+\eps_\cK$) such that
  \begin{equation}
  \label{EqEfSLinPSolv}
    D_{(h,\vecp,\vecq)}P\bigl(\tilde h_1,\ (b',\dot\scal_1^{(0,1)},\dot\scal_2^{(0),(1,0)}),\ (\dot\scal_4^{(-2),(3)}t_*^{-3})\bigr)
  \end{equation}
  (where we omit the components of the argument of $D_{(h,\vecp,\vecq)}P$ that are $0$) has vanishing $t_*^{-3}$-coefficient at $\iota^+$. (And then $D_U\Phi(b',0,\ldots,0)$ is equal to the parameters of $D_{(h,\vecp,\vecq)}P$ in~\eqref{EqEfSLinPSolv}, up to corrections that do not affect the $t_*^{-3}$-term of the output of $P$.) By the \emph{same} computations as above, this means that
  \begin{align*}
    &\ubar L\bigl(\tilde h_1 + \dot h^{(1)}\bigr) + t_*^{-3}f_\cp^{(3,0)}(b') = 0, \\
    &\quad \text{with}\ \dot h^{(1)}\ \text{defined like $h^{(1)\prime}$ in~\eqref{EqEfSLinLIdent} but with $\dot\scal_1^{(0,1)}$ etc.\ in place of $\scal_1^{(0,1)\prime}$}.
  \end{align*}
  Subtracting this identity from~\eqref{EqEfSLinLIdent}, we conclude that
  \begin{equation}
  \label{EqEfSLinu1Def}
    u_1 := t_*^{-1}h^{(1,0)} - \tilde h_1 + (h^{(1)\prime}-\dot h^{(1)}) \in \ker\ubar L;
  \end{equation}
  and this tensor is of class $t_*^{-1}\Hb^{\infty,\ (\la\cE_\sscri^\cC\ra,3+\eps_\sscri),\ (\cE_\cK,3+\eps_\cK)}(\iota^+)$, i.e., it defines an element
  \[
    u_1\in\Resspace_{\iota^+}^\infty(\ubar L,\cE_\sscri^\cC,1).
  \]
  Recalling~\eqref{EqipPResPar} and $\Resspace_{\iota^+}(\ubar L,1)=\{0\}$, we can moreover write $u_1=\Res_{\iota^+}(\ubar L,\cE_\sscri^\cC,1)(0,f'_1)$ for $f'_1\in\Hb^{\infty,\ [\la\cE_\sscri^\cC+1\ra',\leq 4+\eps_\sscri]}$.

  \pfsubstep{Step~1.3.}{Comparison; elimination of leading-order terms at $\iota^+$.} Re-arranging~\eqref{EqEfSLinu1Def}, we see that
  \[
    u_1 + \Bigl(\tilde h_1 - \bigl(h^{(1)\prime}-\dot h^{(1)}\bigr)\Bigr)  = t_*^{-1}h^{(1,0)}
  \]
  has $\cK^+$-order $4+\eps_\cK$ (recall~\eqref{EqEfSh10}). Recalling Theorem~\ref{ThmEfP}\eqref{ItEfPInf} (in whose notation $\tilde u=h^{(1)\prime}-\dot h^{(1)}$), we thus obtain
  \begin{align*}
    D_U\Phi(b',0,0,(0,f'_1),0,\ldots,0)&\equiv\bigl( \tilde h_1 + u_1 - (h^{(1)\prime}-\dot h^{(1)}),\ (b',\scal_1^{(0,1)\prime},\scal_2^{(0),(1,0)\prime}),\ (\scal_4^{(-2)\prime,(3)}t_*^{-3}) \bigr) \\
      &= \bigl( t_*^{-1}h^{(1,0)},\ (b',\scal_1^{(0,1)\prime},\scal_2^{(0),(1,0)\prime}),\ (\scal_4^{(-2)\prime,(3)}t_*^{-3}) \bigr)
  \end{align*}
  modulo lower-order terms that do not affect the $t_*^{-3}$-term of the output of $D_{(h,\vecp,\vecq)}P$. (Note here that $h^{(1,0)}$, $(\scal_1^{(0,1)\prime},\scal_2^{(0),(1,0)\prime})$, and $(\scal_4^{(-2)\prime,(3)}t_*^{-3})$ are uniquely determined, given $u_1$ and $b'$.)

  \medskip

  In summary, if we replace $(h',\vecp',\vecq')$ by $(h',\vecp',\vecq')-D_U\Phi(b',0,0,(0,f'_1),0,\ldots,0)$, the observation~\eqref{EqEfPLin} implies that we still have~\eqref{EqEfSEqPf}, albeit for a new $f$ (which however still has the decay orders stated there), and all components of the new $(h',\vecp',\vecq')$ that contribute to the linearizations of $g^0$, $h_\tot(\vecp)$, $h_{b,\rem}(\vecq)$, and $h$ at orders $t_*^{-1}\log t_*$ or $t_*^{-1}$, or to the linearization of $\vartheta_\tot(\vecp)$ at order $t_*^{-2}$, \emph{vanish}. (After all, this replacement effectively sets $t_*^{-1}h^{(1,0)}$, $(b',\scal_1^{(0,1)\prime},\scal_2^{(0),(1,0)\prime})$, and $\scal_4^{(-2)\prime,(3)}t_*^{-3}$ in~\eqref{EqEfSLint3Terms} to $0$.) The only change if one drops the restriction made in~\eqref{EqEfSLint3Terms} is that one needs to replace $(h',\vecp',\vecq')$ by $(h',\vecp',\vecq')-D_U\Phi(b',\scal',\scal_1^{(0)\prime},(0,f'_1),0,\ldots,0)$ (where $f'_1$ is determined as above). In full detail, then, for the new $(h',\vecp',\vecq')$, the components
  \[
    \scal',\ b',\ \scal_1^{(0)\prime},\ \scal_1^{(0,1)\prime},\ \vect_1^{(1,0)},\ \scal_l^{(0),(1,0)}\ (l=0,2),\ \scal_l^{(-\lambda^\Ups_{\rms l,1}+1),(\lambda^\Ups_{\rms l,1},0)}\ (l=0,1),\ \scal_3^{(-1),(2,0)\prime},\ \vect_2^{(-1),(2,0)\prime}
  \]
  in the notation of~\eqref{EqEfPvecp} (with primes attached) all vanish, and so do the polyhomogeneous leading-order terms of order $(1+\mu,0)$ of the components of~\eqref{EqEfPvecq} (with primes attached). Note in particular that now the linearization of $g^0$ (which depends linearly on $\scal'$ and $b'$) vanishes. Given this improvement, the tensors~\eqref{EqEfSLinTerms} each have $\iota^+$-order $(\cF+2,5+\eps_+)$ where now $\min\Re\cF\geq 1+\eps_\ind$.

  \pfstep{Step~2: lower-order terms.} We shall study the contributions from $h'$, $\vecp'$, and $\vecq'$ that have the least amount of decay at $\iota^+$. Let thus $\lambda_0$ with $\Re\lambda_0\in[1+\eps_\ind,3+\eps_+]$ be such that there are non-vanishing terms in the $\iota^+$-expansion of at least one of the terms $h_\tot(\vecp')$, $h_{b,\rem}(\vecq)$, and $h$ that are quasi-homogeneous in $t_*^{-1}$ (as functions of $(t_*,R,\omega)$) of order $\lambda_0$, or of the term $\vartheta_\tot(\vecp')$ of order $\lambda_0+1$; and no terms have order $\mu_0$, resp.\ $\mu_0+1$ with $\Re\mu_0<\Re\lambda_0$. Similarly to Step~1, we shall study the equation that follows from~\eqref{EqEfSEqPf} at (quasi-homogeneous) order $t_*^{-\lambda_0-2}$.

  For notational simplicity, let us consider the case that\footnote{The only change for general $\Re\lambda_0$ is that some terms that are part of $\vecp'$ below would need to be put into $\vecq'$ instead. For example, for $\lambda_0=3$, there is no more a term $\scal_2^{(0),(3,0)\prime}$ in $\vecp'$, as we do not need to eliminate $t_*^{-3}h_{b,\rms 2}^{(0)}$-contributions to the late-time asymptotics of what is denoted by $u$ in~\eqref{EqD6u} (as this has more than $2+\eps_\cK$ orders of $\cK^+$-decay); instead, such a contribution would be put into $u_{\rm exp}$ there, specifically, the $t_*^{-3}$-coefficient of $\dot\scal_2^{(0)}$. As we already saw in~\eqref{EqEfSLinTerms4} and \eqref{EqEfSLinLTerms}, the only (largely cosmetic) difference between treating contributions from $\vecp'$ and $\vecq'$ at $\iota^+$ is that the former contribute via $\ubar L(\chi_\cK\ubar\delta^*\ubar\omega(\cdots))$ and the latter via $\ubar L(\chi_\cK\ubar h(\cdots))$.} $\Re\lambda_0\leq 2$, and suppose that among these terms, all terms other than possibly
  \begin{align}
  \label{EqEfShl0}
    &h^{(\lambda_0)} = \sum_{j=0}^J t_*^{-\lambda_0}(\log t_*)^j h_j(R,\omega),\quad h_j\in\Hb^{\infty,\ \bigl(\la\cE_\sscri^\cC\ra,3+\eps_\sscri\bigr),\ 4+\eps_\cK-\Re\lambda_0-\eps}(\iota^+), \\
    &\scal_1^{(\lambda_0-1,j)\prime},\ \scal_2^{(0),(\lambda_0,j)\prime}\ \text{(from $\vecp'$)}, \nonumber\\
    &\scal_3^{(-1)\prime,(\lambda_0+1,j)}\ \text{(the $t_*^{-\lambda_0-1}(\log t_*)^j$-coefficient of $\scal_3^{(-1)\prime}$ from $\vecq'$)} \nonumber
  \end{align}
  vanish. The $\iota^+\cap\cK^+$-order of $h_j$ here is computed from the $\cK^+$-order $4+\eps_\cK$ of $h^{(\lambda_0)}$. As in Step~1.1, the vanishing of the $t_*^{-\lambda_0-2}$-term of $D_{(h,\vecp,\vecq)}P(h',\vecp',\vecq')$ at $\iota^+$ amounts to an identity involving the Minkowskian model operator $\ubar L$.\footnote{Note here that since all terms of $(h',\vecp',\vecq')$ that could contribute at lower orders have already been eliminated (i.e., they vanish), there are no non-linear terms involving these; and all terms of $(h',\vecp',\vecq')$ besides these and those we are considering presently contribute at higher orders of vanishing at $\iota^+$, and hence play no role at present.} To wit, the quasi-homogeneous $t_*^{-\lambda_0-2}$-term of
  \begin{align*}
    &\ubar L h^{(\lambda_0)} \\
    &+ D_{\ubar g}\Ric\biggl( \sum_j h_{\rms 1}^{(\lambda_0-1,j)}(\scal_1^{(\lambda_0-1,j)\prime}) + \sum_j h_{\rms 2}^{(0),(\lambda_0,j)}(\scal_2^{(0),(\lambda_0,j)\prime})\biggr) \\
    &\qquad + \ubar\delta_{\ubar E^\cC}^*\biggl(\sum_j\vartheta_{\rms 1}^{(\lambda_0-1,j)}(\scal_1^{(\lambda_0-1,j)\prime}) + \sum_j\vartheta_{\rms 2}^{(0),(\lambda_0,j)}(\scal_2^{(0),(\lambda_0,j)\prime}) \biggr) \\
    &+ \ubar L\biggl(\sum_j\chi_\cK h_{b,\rms 3}^{(-1),\leq 2}\bigl( t_*^{-\lambda_0-1}(\log t_*)^j\scal_3^{(-1)\prime,(\lambda_0+1,j)}\bigr)\biggr)
  \end{align*}
  vanishes. This is equivalent to the identity
  \begin{align}
  \label{EqEfSLinId2}
    &\ubar L\bigl( h^{(\lambda_0)} + h^{(\lambda_0)\prime} \bigr) = 0, \\
    &\qquad h^{(\lambda_0)\prime} := \chi_\cK\ubar\delta^*\biggl( \sum_j \ubar\omega_{\rms 1}^{(0),\leq 4}\bigl(t_*^{-\lambda_0+1}(\log t_*)^j\scal_1^{(\lambda_0-1,j)\prime}\bigr) + \sum_j \ubar\omega_{\rms 2}^{(-1),\leq 3}\bigl(t_*^{-\lambda_0}(\log t_*)^j\scal_2^{(0),(\lambda_0,j)\prime}\bigr) \biggr) \nonumber\\
    &\qquad \hspace{4em} + \sum_j \chi_\cK\ubar h_{\rms 3}^{(-1),\leq 2}\bigl(t_*^{-\lambda_0-1}(\log t_*)^j\scal_3^{(-1)\prime,(\lambda_0+1,j)}\bigr) \nonumber
  \end{align}
  of quasi-homogeneous tensors of degree $\lambda_0+2$ in $t_*^{-1}$ (in the coordinates $(t_*,R,\omega)$). Note that $h^{(\lambda_0)}+h^{(\lambda_0)\prime}$ is of class $\sum_j t_*^{-\lambda_0}(\log t_*)^j \Hb^{\infty,\ (\la\cE_\sscri^\cC\ra,3+\eps_\sscri),\ \beta_\cK}(\iota^+)$ for all $\beta_\cK<0$, or in fact with $\beta_\cK$ replaced by $(\cE_\cK,4+\eps_\ind-\Re\lambda_0-\eps)$ in the notation of Proposition~\ref{PropipNInv}, Corollary~\ref{CoripPQhom}, and Definition~\ref{DefipPRes}. From the equation~\eqref{EqEfSLinId2}, we thus conclude that
  \begin{equation}
  \label{EqEfSLinul0}
    u_{\lambda_0} := h^{(\lambda_0)} + h^{(\lambda_0)\prime} \in \Resspace_{\iota^+}^\infty(\ubar L,\cE_\sscri^\cC,\lambda_0).
  \end{equation}

  \pfsubstep{Case~A.}{$\lambda_0$ is not in the $\iota^+$-spectrum of $\ubar L$.} By definition (see also Lemma~\ref{LemmaipPRes}), the condition $\lambda_0\notin\spec_{\iota^+}^{<3+\eps_+}(\ubar L,\cE_\sscri^\cC)$ means that $\ubar L$ has \emph{no} nontrivial quasi-homogeneous solutions of class $\sum_j t_*^{-\lambda_0}(\log t_*)^j \Hb^{\infty,\ (\la\cE_\sscri^\cC\ra,3+\eps_\sscri),\ \beta_\cK}(\iota^+)$, i.e., the space in~\eqref{EqEfSLinul0} is trivial. Therefore,
  \begin{equation}
  \label{EqEfSLinId2Triv}
    u_{\lambda_0} = h^{(\lambda_0)}+h^{(\lambda_0)\prime}=0.
  \end{equation}
  Since $h^{(\lambda_0)\prime}$ encodes an asymptotic expansion as $R\to 0$ while $h^{(\lambda_0)}$ has a trivial expansion, both summands must vanish individually. In more detail, consider the order of vanishing of the coefficients of $t_*^{-\lambda_0}(\log t_*)^j$ as $R\to 0$ where $j$ is the largest logarithmic power appearing in $h^{(\lambda_0)}$ or $h^{(\lambda_0)\prime}$. The coefficient of $h^{(\lambda_0)}$ vanishes to order $4+\eps_\cK-\Re\lambda_0-\eps\geq 2+\eps_\cK-\eps$ as $R\to 0$ (as we are currently working with $\Re\lambda_0\leq 2$). On the other hand, the $t_*^{-\lambda_0}(\log t_*)^j$-coefficient of $h^{(\lambda_0)\prime}$ has an asymptotic expansion as $R\to 0$ starting with the terms $R\,\rho\ubar h_{\rms 1}^2(\scal_1^{(\lambda_0-1,j)\prime})$, $\ubar h_{\rms 2}^{(0)}(\scal_2^{(0),(\lambda_0,j)\prime})$, and $R\,\rho\ubar h_{\rms 3}^{(-1)}(\scal_3^{(-1)\prime,(\lambda_0+1,j)})$, which does \emph{not} vanish to order $2+\eps_\cK-\eps$ unless the parameters $\scal_1^{(\lambda_0-1,j)\prime}$, $\scal_2^{(0),(\lambda_0,j)\prime}$, and $\scal_3^{(-1)\prime,(\lambda_0+1,j)}$ vanish. Arguing inductively for decreasing $j$ shows that \emph{all} of these parameters vanish, and thus $h^{(\lambda_0)\prime}=0$; and then~\eqref{EqEfSLinId2Triv} implies $h^{(\lambda_0)}=0$, as claimed. --- We have thus shown that these terms in the $\iota^+$-expansion of the solution and of the modification parameters in Corollary~\ref{CorD6Impr} are extraneous (for the source terms of present interest, i.e., of class~\eqref{EqEfSEqPf}).

  \pfsubstep{Case~B.}{$\lambda_0$ is in the $\iota^+$-spectrum of $\ubar L$.} In this case, we rewrite~\eqref{EqEfSLinul0} using~\eqref{EqipPResPar} as
  \[
    h^{(\lambda_0)} = u_{\lambda_0} - h^{(\lambda_0)\prime},\quad u_{\lambda_0}=\Res_{\iota^+}(\ubar L,\cE_\sscri^\cC,\lambda_0)(u_{\lambda_0}^0,f_{\lambda_0}),
  \]
  for suitable $(u_{\lambda_0}^0,f_{\lambda_0})$ in the domain of~\eqref{EqipPResPar}. Consider now $D_U\Phi(0,0,0,(u_{\lambda_0}^0,f_{\lambda_0}),0,\ldots,0)$; if the argument of $D_U\Phi$ here were $(0,\ldots,0)$, the output would of course be $0$, so by Theorem~\ref{ThmEfP}\eqref{ItEfPInf} and Remark~\ref{RmkEfPLin} (and recalling our notationally restricted setting), we must have
  \begin{align*}
    &D_U\Phi(0,0,0,(u_{\lambda_0}^0,f_{\lambda_0}),0,\ldots,0) \\
    &\quad \equiv \Biggl( u_{\lambda_0}-h^{(\lambda_0)\prime},\ \bigl(\scal_1^{(\lambda_0-1,j)\prime},\scal_2^{(0),(\lambda_0,j)\prime}\bigr),\ \biggl(\sum_j t_*^{-\lambda_0-1}(\log t_*)^j\scal_3^{(-1)\prime,(\lambda_0+1,j)}\biggr) \Biggr)
  \end{align*}
  modulo lower-order terms that do not affect the $t_*^{-\lambda_0-2}$-term of the output of $D_{(h,\vecp,\vecq)}P$. Upon replacing $(h',\vecp',\vecq')$ by $(h',\vecp',\vecq')-D_U\Phi(0,0,0,(u_{\lambda_0}^0,f_{\lambda_0}),0,\ldots,0)$, we thus effectively set all of the tensors and parameters in~\eqref{EqEfShl0} to $0$. The upshot is that we have now reduced to the case that all terms of $h',\vecp',\vecq'$ that contribute to the output of $D_{(h,\vecp,\vecq)}P$ at $\iota^+$-order $\lambda_0+2$ vanish.

  \pfstep{Step~3. Conclusion.} By iterating Step~2, we eliminate all terms in the $\iota^+$-expansion of $h'$ as well as all the modification parameters $\vecp'$ and all parts of $\vecq'$ that contribute polyhomogeneous terms to the output of $D_{(h,\vecp,\vecq)}P$ at $\iota^+$ one by one by subtracting from $(h',\vecp',\vecq')$ the output of $D_U\Phi$ on suitable inputs. After finitely many steps, we have thus reduced to the case that
  \[
    h' \in \Hb^{\infty,\ \bigl(\la\cE_\sscri^\cC\ra,3+\eps_\sscri\bigr),\ 3+\eps_+,\ 4+\eps_\cK}(\Omega_*)^{\bullet,-}
  \]
  has a \emph{trivial} $\iota^+$-expansion, further $\vecp'=0$ \emph{vanishes identically}, and finally all polyhomogeneous terms of $\vecq'$ \emph{vanish}, so in the notation of~\eqref{EqEfPvecq} (with primes attached) and recalling Definition~\ref{DefD6Order}, we have
  \begin{align*}
    \scal'_\rem &\in \dot H_\bop^{\infty,\ 2+\eps_\cK}, \\
    b'_\rem &\in \dot H_\bop^{\infty,\ 3+\eps_\cK}, \\
    \vect_1' &\in \dot H_\bop^{\infty,\ 3+\eps_+}, \\
    \scal_0^{(0)\prime} &\in \dot H_\bop^{\infty,\ 3+\eps_+}, \\
    \scal_l^{(-\lambda^\Ups_{\rms l,1}+1)\prime} &\in \dot H_\bop^{\infty,\ 3+\eps_++(\lambda^\Ups_{\rms l,1}-1)},\quad l=0,1, \\
    \scal_l^{(-\lambda^\Ups_{\rms l,l+j}+1)\prime} &=0,\quad 1\leq l\leq 3,\ j=0,1,\ 2\leq l+j\leq 3, \\
    \scal_2^{(0)\prime} &\in \dot H_\bop^{\infty,\ 3+\eps_+}, \\
    \scal_l^{(-l+2)\prime} &= 0,\quad l=3,4,5, \\
    \vect_l^{(-l+1)\prime} &= 0,\quad l=2,3,4, \\
    \scal_l^{(-l)\prime} &= 0,\quad l=2,3, \\
    \vect_l^{(-l)\prime} &= 0,\quad l=2,3.
  \end{align*}
  Therefore, recalling~\eqref{EqEfPU}, we have
  \[
    (h',\vecp',\vecq') = (h',0,\vecq') = D_U\Phi\Bigl( 0,0,0,(0,0),\scal'_\rem,b'_\rem,\vect_1',(\scal_l^{(0)\prime})_{l=0,2},(\scal_l^{(-\lambda^\Ups_{\rms l,1}+1)\prime})_{l=0,1},h'\Bigr),
  \]
  as these conormal remainder terms simply contribute additively to the output of $\Phi$ and thus to that of $D_U\Phi$. The proof is complete.
\end{proof}

\section{Nonlinear stability}
\label{SSt}

We show in~\S\ref{SsSt} that the global nonlinear stability of the subextremal family of Kerr black holes can be proved by concatenating the exterior stability result (Theorem~\ref{ThmExBoStab}) with a Nash--Moser iteration based on Theorem~\ref{ThmEfS} for stability in the forward causal cone. Minor extensions and sharpenings of the nonlinear stability result, Theorem~\ref{ThmSt}, are briefly indicated in~\S\S\ref{SsSt3}--\ref{SsStPhg}.

\subsection{Precise statement and proof}
\label{SsSt}

To state the precise result, we recall the asymptotic boost diffeomorphism $\phi_\scal$ from~\eqref{EqKBoMap}, where $\scal\in\scalspace_1$ (see Definition~\ref{DefTY}) is small; and $\phi_\scal$ is the identity in $\{r\leq 5\bhm_0\}$. Recall also the cutoff function $\chi_\cK$ from Definition~\ref{DefKBoCutoff} which we can take to be of the form $\chi_\cK(t_*,r)=\chi_0(t_*)\chi_1(\frac{r}{t_*})$ where $\chi_0|_{[-\infty,T_0]}=0$, $\chi_0|_{[T_0+1,\infty)}=1$ for some large $T_0$, and $\chi_1|_{[0,\frac14]}=1$ and $\chi_1|_{[\frac12,\infty)}=0$. We will take $T_0$ such that $\chi_\cK$ vanishes for $\ft_*\leq 10$ where $\ft_*$ is the hyperboloidal time function from~\eqref{EqDftstar}. Moreover, we recall the compactified spacetime manifold $M$ from~\eqref{EqKMfdRadM} on which one can conveniently express partial polyhomogeneity (cf.\ Definition~\ref{DefTMphg}).

\begin{thm}[Nonlinear stability]
\label{ThmSt}
  Fix subextremal Kerr parameters $b_0=(\bhm_0,\bha_0)$, $|\bha_0|<\bhm_0$. Recalling the time function $t_\IVP$ from Lemma~\usref{LemmaKMetIVP}, let $\Sigma_\IVP=t_\IVP^{-1}(0)$ be a Cauchy hypersurface of the subset
  \[
    \Omega:=\{t_\IVP\geq 0,\ r\geq\bhm_0\}
  \]
  of the subextremal Kerr spacetime manifold $(M^\circ,g_{b_0})$; recall from Definition~\usref{DefKMetData} that the initial data of $g_{b_0}$ are denoted by $\gamma_{b_0}$ and $k_{b_0}$. Let $\eps_0>0$, and let $\cE_0\subset\C\times\N_0$ be an index set with $\min\Re\cE_0>1+\eps_0$ and $j\cE_0\subset\cE_0$ for all $j\in\N$. Then there exist $\eps>0$, $d\in\N_0$, as well as $0<\eps_\cK<\eps_+<\eps_\sscri<\eps_0$ and index sets $\cE_\sscri^\cC$ as in Definition~\usref{DefExP} and $\cE_+$ with $(1,0),(1,1)\in\cE_+$ and $\min\Re(\cE_+\setminus\{(1,0),(1,1)\})>1$ such that the following holds. Suppose we are given initial data $\gamma,k$ on $\Sigma_\IVP$ (i.e., $\gamma$ is a Riemannian metric on $\Sigma_\IVP$ and $k$ a symmetric 2-tensor) satisfying the constraint equations, the decay and regularity properties\footnote{As discussed in~\S\ref{SsExID}, one identifies $\Sigma_\IVP$ with the complement of a ball in $\R_x^3$; the meaning of the conditions~\eqref{EqStMem}--\eqref{EqStSmall} is then that they hold for each component in the frame $\pa_{x^i}$, $i=1,2,3$.}
  \begin{subequations}
  \begin{equation}
  \label{EqStMem}
  \begin{split}
    \gamma-\gamma_{b_0} &\in \Hb^{\infty,(\cE_0,3+\eps_0)}(\Sigma_\IVP), \\
    k-k_{b_0} &\in \Hb^{\infty,(\cE_0+1,4+\eps_0)}(\Sigma_\IVP),
  \end{split}
  \end{equation}
  and the smallness conditions
  \begin{equation}
  \label{EqStSmall}
    \|\gamma-\gamma_{b_0}\|_{\Hb^{d,(\cE_0,3+\eps_0)}(\Sigma_\IVP)},\ \|k-k_{b_0}\|_{\Hb^{d,(\cE_0+1,4+\eps_0)}(\Sigma_\IVP)} < \eps.
  \end{equation}
  \end{subequations}
  Then there exist $b=(\bhm,\bha)$ and $\scal\in\scalspace_1$, with $|b-b_0|$ and $|\scal|$ small, and a partially polyhomogeneous tensor
  \begin{equation}
  \label{EqSth}
    h \in \Hb^{\infty,\ (\cE_0,3+\eps_0),\ (\la\cE_\sscri^\cC\ra,3+\eps_\sscri),\ (\cE_+,3+\eps_+),\ 2+\eps_\cK} ( \phi_\scal(\Omega) ),
  \end{equation}
  where the orders refer to the boundary hypersurfaces $I^0$, $\scri^+$ (with the convention of Definitions~\usref{DefExP} and \usref{DefDMetBasic}), $\iota^+$, and $\cK^+$ (in this order) of $\phi_\scal(\Omega)$ inside of $M$, such that the spacetime metric
  \begin{equation}
  \label{EqStMetric}
    g = g_{b_0,b,-\scal} + h,\quad g_{b_0,b,-\scal} := (1-\chi_\cK)(\phi_\scal)_*g_{b_0}+\chi_\cK g_b,
  \end{equation}
  attains the initial data $(\phi_\scal)_*(\gamma,k)$ at $\phi_\scal(\Sigma_\IVP)$ and satisfies
  \[
    \Ric(g) = 0.
  \]
\end{thm}

The logarithmic term at $\iota^+$ (encoded by $(1,1)\in\cE_+$) is supported only on $\supp\dd\chi_\cK$ (see Remark~\ref{RmkStFurther}\eqref{ItStFurtherFull} below). Thus, $g$ settles down to the Kerr metric $g_b$ at the rates
\begin{enumerate}
\item $\cO(t_*^{-2-\eps_\cK})$ in spatially compact subsets of $\{r\geq\bhm_0\}$,
\item $\cO(r^{-1}(t_*/r)^{-2-\eps_\cK})$ in a narrow timelike cone $\frac{r}{t_*}\ll 1$ around the final black hole,
\item $\cO(t_*^{-1}\log t_*)$ as $t_*\to\infty$ in a region $\frac{r}{t_*}\in[c,C]$ for some $0<c<C<\infty$,
\item $\cO(r^{-1})$ in the region $\frac{r}{t_*}\geq C$.
\end{enumerate}
(For the structure of null infinity, it is important to stress that at $\scri^+$, the index set collection $\la\cE_\sscri^\cC\ra$ encodes different decay rates for different components of $h$ (see~\eqref{EqExPSpaceProj} and~\eqref{EqDMetBasicProj}).

\begin{figure}[!ht]
\centering
\includegraphics{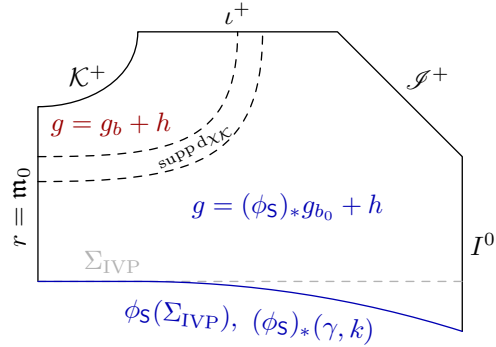}
\caption{Illustration of Theorem~\ref{ThmSt}. The initial data $(\gamma,k)$ are attained at a boosted version $\phi_\scal(\Sigma_\IVP)$ of $\Sigma_\IVP=t_\IVP^{-1}(0)$. The two dashed lines bound the set $\supp\dd\chi_\cK$ where the reference metric $g_{b_0,b,-\scal}$ transitions from $(\phi_\scal)_*g_{b_0}$ to $g_b$.}
\label{FigSt}
\end{figure}

See Figure~\ref{FigSt} for an illustration of Theorem~\ref{ThmSt}. Theorem~\ref{ThmISimple} follows from Theorem~\ref{ThmSt}; only the particular manifold on which the metric is defined is different here than in Theorem~\ref{ThmSt}, but this is easily fixed by pulling back the metric of Theorem~\ref{ThmSt} along a diffeomorphism of $M$ (generated by the flow of a smooth b-vector field) that maps $\phi_\scal(\Omega)$ here to $\Omega$ in Theorem~\ref{ThmISimple} and is the identity near $\scri^+\cup\iota^+\cup\cK^+$.

\begin{rmk}[Initial data, I: asymptotics]
\label{RmkStData}
  Theorem~\ref{ThmSt} remains valid if in the assumptions~\eqref{EqStMem}--\eqref{EqStSmall} and in the conclusion~\eqref{EqStMetric} we replace $b_0$ with parameters $b_1$ satisfying $|b_1-b_0|<\eps$; one merely needs to replace all occurrences of $b_0$ in the proof by $b_1$. (Recall that the particular choice $r=\bhm_0$ of the interior spacelike boundary hypersurface of $\Omega$ is inconsequential; when working with $b_1=(\bhm_1,\bha_1)$ instead of $b_0$, this choice is still acceptable as long as $\bhm_0\in(r_{\bhm_1,\bha_1}^-,r_{\bhm_1,\bha_1}^+)$ lies between the Cauchy and event horizon of $g_{b_1}$, which is true when $b_1$ is close enough to $b_0$.) This mild generalization allows for small changes of the ADM mass $\bhm_0$. Note that $\gamma$ in~\eqref{EqStMem} is allowed to deviate from $\gamma_{b_0}$ by terms of order $r^{-1-\delta}$ for any fixed $\delta\geq\eps_0>0$, so small changes of the angular momentum parameter $\bha_0$ (which enter at order $r^{-2}$, cf.\ \eqref{EqKMetDiff}) are likewise allowed, but the assumptions in~\eqref{EqStMem} allow for considerably more flexibility.
\end{rmk}

\begin{rmk}[Initial data, II: choice of Cauchy hypersurface]
\label{RmkStDataCauchy}
  The particular choice of Cauchy hypersurface is of little consequence. One can equally well work with initial data on a smooth spacelike hypersurface that is equal to, say, a level set of the Boyer--Lindquist time function $\ft$ (see~\eqref{EqKMetBLCoord}) near infinity and transversal to the future event horizon. We explain in~\S\ref{SssExIBL} how to produce from such a setup a local solution and from that, in turn, initial data that can be used as an input in Theorem~\ref{ThmSt} (which then produces a global extension of said local solution).
\end{rmk}

\begin{rmk}[Further properties of $g$]
\label{RmkStFurther}
  The proof below will express $g$ for $\ft_*\geq 2$ (where $\ft_*$ is the hyperboloidal time function from~\eqref{EqDftstar}) using Theorem~\ref{ThmEfP}. Therefore:
  \begin{enumerate}
  \item\label{ItStFurtherFull}{\rm (Full description in the forward cone.)} $g|_{\{\ft_*\geq 2\}}$ is the (restriction to this region of) the argument of $\Ric$ in~\eqref{EqEfPMap} and thus equal to
    \[
      g = g^0 + h_\tot(\vecp) + h_{b,\rem}(\vecq) + \tilde h
    \]
    for suitable $\tilde h\in\Hb^{\infty,\ (\la\cE_\sscri^\cC\ra,3+\eps_\sscri),\ (\cE'_+,3+\eps_+),\ 4+\eps_\cK}$ (recording only the orders at $\scri^+$, $\iota^+$, and $\cK^+$) and parameters $\vecp$ and $\vecq$ as recalled in~\eqref{EqEfPvecp}--\eqref{EqEfPvecq}; here we write $\cE'_+:=\cE_+\setminus\{(1,1)\}$ for an index set of the form used in~\S\ref{SEf}, namely $(1,0)\in\cE'_+$ and $\min\Re(\cE'_+\setminus\{(1,0)\})>1$. (In particular, the $t_*^{-1}\log t_*$-term of $h$ in~\eqref{EqSth} arises solely from $h_\tot(\vecp)$, and more specifically still from the components $\scal,b-b_0,\scal_1^{(0,1)}$, and $\scal_0^{(0),(1,0)}$ of~\eqref{EqEfPvecp}; and thus it is supported on $\supp\dd\chi_\cK$.)
  \item{\rm (Gauge condition.)} The first argument of $\Ups_{E^\Ups}$ in~\eqref{EqEfPMap} is equal to $g-h_\tot(\vecp)$. Globally, then, the gauge condition for $g$ reads
    \begin{equation}
    \label{EqStFurtherGauge}
      \Ups_{\chi_0 E^\Ups}\bigl(g-h_\tot(\vecp),\,g^0\bigr) - \vartheta_\tot(\vecp) = 0,
    \end{equation}
    where $\Ups_{\chi_0 E^\Ups}$ (with $\chi_0=1$ near $\scri^+$, and $\chi_0=0$ near $\Sigma_\IVP$ and thus near $\phi_\scal(\Sigma_\IVP)$) was defined in~\eqref{EqExBoGauge}. (Near $\phi_\scal(\Sigma_\IVP)$, this is the standard generalized wave map (or wave coordinate) gauge condition $\Ups_0\bigl((\phi_\scal)_*g_{b_0}+h,(\phi_\scal)_*g_{b_0}\bigr)=0$.)
  \end{enumerate}
\end{rmk}

\begin{proof}[Proof of Theorem~\usref{ThmSt}]
  \pfstep{Step~1. Parameters for constraint damping and gauge.} For the purpose of constraint damping, we use the modified symmetric gradient $\delta_{g^0,E^\cC}^*$ defined in Definition~\ref{Def1Symm} for the choice of $E^\cC$ given by the main result of \cite{HintzKerrCD} as recalled in Theorem~\ref{ThmWCRec} (with $v^\cC=\frac12$, $C_0=100$, and $(1-e^\cC)(1-v^\cC)\gamma^\cC>100$). In order to obtain improved decay rates of metric coefficients at $\scri^+$, we furthermore use, until further notice (specifically, outside of $\supp\chi_\cK$), a generalized harmonic gauge 1-form $\Ups_{\chi_0 E^\Ups}(g,g^0)$ as defined in Definition~\ref{Def1Gauge} and~\eqref{EqExBoGauge}, where for the parameters $E^\Ups=(\cd^\Ups,e^\Ups,\gamma^\Ups;\cd^\Ups_{\cH^+},\gamma^\Ups_{\cH^+})$ in Definition~\ref{Def1Gauge} we fix $\cd^\Ups$ and $\cd^\Ups_{\cH^+}$ as in Lemma~\ref{LemmaWG0Pair} and Proposition~\ref{PropWGMode}, and take $e^\Ups,\gamma^\Ups,\gamma^\Ups_{\cH^+}>0$ small enough so that the conclusions of Proposition~\ref{PropWGMode} (i.e., mode stability for the gauge potential wave operator $\Box_{g_{b_0},E^\Ups}^\Ups=2\delta_{g_{b_0},E^\Ups}\sfG_{g_{b_0}}\delta_{g_{b_0}}^*$) as well as of Proposition~\ref{PropWEtf} (concerning the $\tface$-admissibility of the spectral family of the linearized gauge-fixed Einstein operator) hold. We moreover require $\gamma^\Ups$ to be so small that $1+2\gamma^\Ups<\min\Re\cE_0$ (which, upon propagating polyhomogeneity to $\scri^+$, is motivated by Definition~\ref{DefExP}\eqref{ItExPEscri}).

  \medskip

  \pfstep{Step~2. Solution up to $\ft_*=2$.} We first use Proposition~\ref{PropExID} to produce Cauchy data $h_0,h_1\in\Hb^{\infty,(\cE_0,3+\eps_0)}(\Sigma_\IVP;S^2\cT^*_{\Sigma_\IVP})$ such that $\gamma$ and $k$ are the first and second fundamental form, respectively, of any Lorentzian metric $\hat g$ with $(\hat g-g_{b_0})|_{\Sigma_\IVP}=h_0$ and $(r\cL_{\pa_{t_\IVP}}(\hat g-g_{b_0}))|_{\Sigma_\IVP}=h_1$. We can then use Theorem~\ref{ThmExPhg} with these data, albeit with the gauge condition $\Ups_{\chi_0 E^\Ups}(g_{b_0}+h,g_{b_0})=0$ (see the discussion around~\eqref{EqExBoGauge}) to produce an index set $\cE_\sscri^\cC$ and an exterior solution $h\in\Hb^{\infty,\ (\cE_0,3+\eps_0),\ (\la\cE_\sscri^\cC\ra,3+\eps_\sscri)}(\Omega_{\ext,r_0})$, for $r_0$ large, of the initial value problem for~\eqref{EqExBoStabPDE}, i.e.,
  \[
    \left\{
    \begin{aligned}
      \Ric(g_{b_0}+h) - \delta_{g_{b_0},E^\cC}^*\Ups_{\chi_0 E^\Ups}(g_{b_0}+h,g_{b_0})&=0 && \text{in}\ \Omega_{\ext,r_0}, \\
      h&=h_0 && \text{on}\ \Sigma_{\IVP,r_0}, \\
      r\cL_{\pa_{t_\IVP}}h&=h_1 && \text{on}\ \Sigma_{\IVP,r_0};
    \end{aligned}
    \right.
  \]
  here we use the notation $\Omega_{\ext,r_0}$ and $\Sigma_{\IVP,r_0}$ from Lemma~\ref{LemmaEx0Dom}. Standard local well-posedness results for quasilinear wave equations produce also an interior solution $h$ on any fixed finite $t_\IVP$-interval, say on $t_\IVP^{-1}([0,1])$, and also for small negative times (restricted purely by domain of dependence considerations). Note that the metric $g_{b_0}+h$ has initial data $(\gamma,k)$ at $\Sigma_\IVP$.

  For all small $\scal\in\scalspace_1$, the hypersurface $\phi_\scal(\Sigma_\IVP)$ is contained in the set where $h$ is thus defined. We can therefore pull back the initial data from the boosted hypersurface $\phi_\scal(\Sigma_\IVP)$,
  \[
    h_{\scal,0} := (\phi_\scal^*h)|_{\Sigma_\IVP},\quad
    h_{\scal,1} := \bigl(r\cL_{\pa_{\IVP}}(\phi_\scal^*h)\bigr)\big|_{\Sigma_\IVP},
  \]
  as in~\eqref{EqExBoIVPData} (but now globally on $\Sigma_\IVP$); we still have $h_{\scal,0},h_{\scal,1}\in\Hb^{\infty,(\cE_0,3+\eps_0)}(\Sigma_\IVP)$. We then define $h_\scal$ to be the solution of the initial value problem
  \begin{equation}
  \label{EqStIVPPullback}
    \left\{
    \begin{aligned}
      \Ric(\phi_\scal^*g_{b_0}+h_\scal) - \delta_{\phi_\scal^*g_{b_0},E^\cC}^*\Ups_{\chi_0 E^\Ups}\bigl(\phi_\scal^*g_{b_0}+h_\scal,\,\phi_\scal^*g_{b_0})&=0, && \\
      h_\scal&=h_{\scal,0} && \text{on}\ \Sigma_{\IVP}, \\
      r\cL_{\pa_{t_\IVP}}h_\scal&=h_{\scal,1} && \text{on}\ \Sigma_{\IVP};
    \end{aligned}
    \right.
  \end{equation}
  this is described by Theorem~\ref{ThmExBoStab} (the proof of which applies without any changes near the full Cauchy hypersurface $\Sigma_\IVP$); in particular, we have $h_\scal=\phi_\scal^*h$ in a neighborhood of $\Sigma_\IVP$, and thus $\phi_\scal^*g_{b_0}+h_\scal=\phi_\scal^*(g_{b_0}+h)$ near $\Sigma_\IVP$. (The original initial data are attained by $\phi_\scal^*(g_{b_0}+h)$ at $\phi_\scal^{-1}(\Sigma_\IVP)$, in that these are $\phi_\scal^*(\gamma,k)$.) Moreover, $h_\scal\in\Hb^{\infty,\ (\cE_0,3+\eps_0),\ (\la\cE_\sscri^\cC\ra,3+\eps_\sscri)}$ again exists in a full neighborhood of $\Sigma_\IVP$ and up to some finite cross section of $\scri^+$; for globally small initial data, $h_\scal$ thus exists (and is small, cf.\ \eqref{EqExPhgSmall}) on
  \[
    \phi_\scal(\{ t_\IVP \geq 0 \}) \cap \{ \ft_*\leq 2 \},
  \]
  where we recall the hyperboloidal time function $\ft_*$ from~\eqref{EqDftstar}. This set contains $\ft_*^{-1}([1,2])$.

  \medskip
  \pfstep{Step~3. Solving the gauge-fixed Einstein equation in $\ft_*\geq 1$.} Given the family $h_\scal$ of ``initial'' solutions, we now fix $\chi_-\in\CI(\R)$ to be equal to $1$ on $(-\infty,\frac74]$ and $0$ on $[2,\infty)$. Let us use the notation of Definition~\ref{DefEfD} (which defines the space $\fD^\infty$ of unknowns $U=(b-b_0,\scal,\ldots)$ describing metric perturbations with support in $\ft_*\geq 1$ and gauge modifications) and Theorem~\ref{ThmEfP} (which produces a map $\Phi$ mapping $U$ that are small in $\fD^d$ to inputs $(h,\vecp,\vecq)$ of the nonlinear gauge-fixed Einstein operator $P$ defined in~\eqref{EqEfPMap}, and which moreover produces small constants $0<\eps_\cK<\eps_+$). We then recall from~\eqref{EqEfPIni} the definition
  \begin{equation}
  \label{EqStPsi}
    \Psi(U) := \bigl(\chi_-(\ft_*)h_{-\scal},0,0\bigr) + \Phi(U),
  \end{equation}
  where $\scal$ is the second component of the tuple $U$. Since $\Phi(0)=(0,0,0)$, we have $P(\Psi(0))=\Ric(g_{b_0}+\chi_-h_{\text{\bf 0}})-\delta_{g_{b_0},E^\cC}^*\Ups_{E^\Ups}(g_{b_0}+\chi_-h_{\text{\bf 0}},\,g_{b_0})$ where $h_{\text{\bf 0}}=h_\scal|_{\scal=0}$; this lies in $\Hb^{\infty,\ (\la\cE_\sscri^\cC+1\ra',4+\eps_\sscri)}$ and is small in any desired fixed high-regularity version of this space, and supported in $\ft_*^{-1}([\frac74,2])$. Since $h_{-\scal}$ solves the gauge-fixed Einstein equation for $\ft_*\leq 2$ relative to $\phi_{-\scal}^*g_{b_0}$ (this being the content of~\eqref{EqStIVPPullback} for $-\scal$ in place of $\scal$), and thus $\chi_-h_{-\scal}$ does so as well for $\ft_*\leq\frac74$, we moreover have
  \[
    P(\Psi(U)) = 0\ \text{for}\ \ft_*\leq c
  \]
  for all $U$ whose final component in~\eqref{EqEfPU}, which is the gravitational wave tail $\tilde h$ (that is always required to vanish for $\ft_*\leq 1$), vanishes for $\ft_*\leq c\in[\frac32,\frac74]$.

  We now set up a Nash--Moser iteration to solve $P(\Psi(U))=0$ for $U\in\fD^\infty$, and with the final component of $U$ supported in $\ft_*\geq\frac32$. We use the simple Nash--Moser iteration scheme of Saint Raymond \cite{SaintRaymondNashMoser} for the map $P\circ\Psi$. We check the assumptions of the main result of \cite{SaintRaymondNashMoser}:
  \begin{enumerate}
  \item By Theorem~\ref{ThmEfP}, the map $P\circ\Psi$ maps a neighborhood of $0$ in $\fD^\infty$ to a neighborhood of $0$ in $\bfB^\infty:=\Hb^{\infty,\ (\la\cE_\sscri^\cC+1\ra',4+\eps_\sscri),\ 5+\eps_+,\ 4+\eps_\cK}(\Omega_*)^{\bullet,-}$ (where we recall the notation $\Omega_*=\{\ft_*\geq 1\}$); and it does so in a $\cC^2$ manner.
  \item The tame estimates \cite[(1)]{SaintRaymondNashMoser} for the forward map $P\circ\Psi$ follow, as usual, from Moser-type estimates for products of functions in Sobolev spaces. The key input here is that $\Phi$ satisfies such tame estimates, which is a consequence of the fact that the construction of $\Phi$, besides nonlinear functions (e.g., products and reciprocals) applied to collections of Sobolev functions, only involves the inversion of model operators (all of them related to the Minkowskian model operator $\ubar L$) with smooth coefficients.
  \item For all small $U\in\fD^\infty$ and for all $f\in\bfB^\infty$, the linearized problem $D_U(P\circ\Psi)(U')=f$, which reads
    \[
      D_{\Psi(U)}P \bigl( D_U\Psi(U') \bigr) = f,
    \]
    has a solution as a consequence of Theorem~\ref{ThmEfS}; and the solution $U'$ satisfies tame estimates. Notice also that the support of the metric perturbation encoded by the summand $D_U\Phi(U')$ of $D_U\Psi(U')=D_\scal(h_{-\bullet})(\scal')+D_U\Phi(U')$ is contained in the union of $\supp\chi_\cK\subset\{\ft_*\geq 10\}$ and the support of the last component $\tilde h'$ of $U'$; so by finite speed of propagation for the linear wave-type operator $L$ in Corollary~\ref{CorD6Impr} (the output of which Theorem~\ref{ThmEfS} merely repackages), we conclude that if $f$ is supported in $\ft_*\geq c\in[\frac32,\frac74]$, then so is $\tilde h'$.
  \item We need to define smoothing operators for those components of $U$ that do not lie in finite-dimensional vector spaces. For the final six components $\scal_\rem,\ldots,\tilde h$ in~\eqref{EqEfPU}, which lie in weighted b-Sobolev spaces (with $\tilde h$ having also a partial polyhomogeneous expansion), such smoothing operators are essentially standard. Indeed, undoing the weight, and passing to logarithmic coordinates (after localizing to charts in the case of $\tilde h$), b-Sobolev spaces are isometric to standard Sobolev spaces on Euclidean space, and then the smoothing operators of \cite[Lemma~6.12 and Corollary~6.14]{HintzGlueLocII} can be used; these have the added benefit that they increase supports by very small amounts (which sum to any small number, say $\frac14$, over the entire iteration scheme). Smoothing operators on partially polyhomogeneous spaces (in which $\tilde h$ lies) are defined by smoothing each term in a polyhomogeneous expansion separately, and also smoothing the remainder term (that now lies in a standard weighted b-Sobolev space). Consider finally the components of the (finitely many) pairs $(u_\lambda^0,f_\lambda)$ in~\eqref{EqEfPU}: the first component $u_\lambda^0$ lies in a finite-dimensional space (see the discussion around~\eqref{EqipPResFin}) and thus needs no smoothing, while $f_\lambda$ is a polynomial (of bounded degree) with values in a space encoding polyhomogeneous expansions; so $f_\lambda$ can be smoothed as explained above.
  \end{enumerate}

  By the main result of \cite{SaintRaymondNashMoser}, then, we can find a small $U\in\fD^\infty$ such that $P(\Psi(U))=0$, provided $P(\Psi(0))$ is sufficiently small in $\bfB^\infty$ with respect to some fixed high-regularity norm---which is the case under the assumption~\eqref{EqStSmall} when $d$ there is sufficiently large and $\eps>0$ is sufficiently small. Since we are using smoothing operators with controlled and small increase in supports, the final component $\tilde h$ of $U$ automatically has support in $\ft_*\geq\frac32$ (or $\ft_*\geq\frac74-\delta$ for any fixed value of $\delta>0$).

  \pfstep{Step~4. Conclusion.} Having constructed $U\in\fD^\infty$ solving $P(\Psi(U))=0$, set $(h,\vecp,\vecq):=\Psi(U)$; this thus solves $P(\chi_-(\ft_*)h_{-\scal}+h,\vecp,\vecq)=0$ in the notation of~\eqref{EqEfPMap}, where the first two components of $\vecp=(b-b_0,\scal,\ldots)$ encode the final Kerr black hole parameters $b$ and the boost parameter $\scal$, and moreover $h$ is supported in $\ft_*\geq\frac32$. Define then the metric
  \[
    g :=
    \begin{cases}
      g_{b_0,b,-\scal} + \chi_-(\ft_*)h_{-\scal} + h_\tot(\vecp) + h_{b,\rem}(\vecq) + h, & \ft_*\geq 1, \\
      g_{b_0,b,-\scal} + h_{-\scal} = \phi_{-\scal}^*g_{b_0}+h_{-\scal}, & \ft_*\leq \frac32;
    \end{cases}
  \]
  the two definitions agree on the overlap of their domains of definition (where $\chi_-=1$ and $h_\tot=h_{b,\rem}=h=0$). As recalled from Theorem~\ref{ThmExBoStab} after~\eqref{EqStIVPPullback}, we have $g=\phi_{-\scal}^*(g_{b_0}+h)$ near $\Sigma_\IVP$, and its initial data at $\phi_{-\scal}^{-1}(\Sigma_\IVP)=\phi_\scal(\Sigma_\IVP)$ are $(\phi_\scal)_*(\gamma,k)$. Note also that since $h$ is defined for $\{t_\IVP\geq 0\}$, the metric $g$ is defined in $\phi_\scal(\{t_\IVP\geq 0\})$.

  Recalling $g^0=g_{b_0,b,-\scal}$, define the gauge 1-form
  \[
    \eta :=
    \begin{cases}
      \Ups_{E^\Ups}\bigl(g^0 + h_{b,\rem}(\vecq) + h,\,g^0\bigr) - \vartheta_\tot(\vecp), & \ft_*\geq 1, \\
      \Ups_{\chi_0 E^\Ups}( \phi_{-\scal}^*g_{b_0} + h_{-\scal},\,\phi_{-\scal}^*g_{b_0}), & \ft_*\leq\frac32,
    \end{cases}
  \]
  which is consistently defined on the overlap; and we have $\eta=0$ at $\phi_\scal(\Sigma_\IVP)$. The equation $P(h,\vecp,\vecq)=0$ reads
  \begin{equation}
  \label{EqStRic}
    \Ric(g) - \delta_{g^0,E^\cC}^*\eta = 0.
  \end{equation}
  The constraint equations for $(\phi_\scal)_*(\gamma,k)$ at $\phi_\scal(\Sigma_\IVP)$, which can be expressed using the Einstein tensor $\Ein(g)=\sfG_g\Ric(g)$ as $\Ein(g)(\nu,\cdot)=0$ where $\nu$ is the unit normal to $\phi_\scal(\Sigma_\IVP)$, then gives $(\sfG_g\delta_{g^0,E^\cC}^*\eta)(\nu,\cdot)=0$, which by a short calculation implies that also the normal derivative of $\eta$ vanishes at $\phi_\scal(\Sigma_\IVP)$. Applying the second Bianchi identity to~\eqref{EqStRic} gives the homogeneous wave equation $\delta_g\sfG_g\delta_{g^0,E^\cC}^*\eta=0$ for $\eta$; and thus $\eta=0$ and therefore also $\Ric(g)=0$. This proves the claims in Remark~\ref{RmkStFurther}. To deduce the less precise statement~\eqref{EqSth} from this, one only needs to inspect the decay rates and orders of the terms comprising $g$:
  \begin{enumerate}
  \item the weakest decay at $\iota^+$ arises from the terms $h^{(-1)}(\scal)$ (Definition~\ref{DefD2Corr} and Lemma~\ref{LemmaD2Corr}), $h^{(0)}(b-b_0)$, $h_{\rms 1}^{(0,1)}(\scal_1^{(0,1)})$ (Definition~\ref{DefD4Corr} and Lemma~\ref{LemmaD4Corr}), and $h_{\rms 0}^{(0),(1,0)}(\scal_0^{(0),(1,0)})$ (Definition~\ref{DefD6Corr}). They have a $t_*^{-1}\log t_*$ leading-order term at $\iota^+$;
  \item the weakest decay at $\cK^+$ arises from the term $\chi_\cK h_{b,\rms 1}^{\leq 1}(\scal_\rem)$, $\scal_\rem\in\Hb^{\infty,\,2+\eps_\cK}$, of $h_{b,\rem}(\vecq)$, which has order $2+\eps_\cK$ at $\cK^+$.
  \end{enumerate}
  This completes the proof of Theorem~\ref{ThmSt}.
\end{proof}

\subsection{Sharp \texorpdfstring{$t_*^{-3}$-}{inverse cubic }decay}
\label{SsSt3}

While we will not give complete arguments, we shall indicate how to prove the following strengthening of Theorem~\ref{ThmSt} for a more restrictive class of initial data:

\begin{thm}[Nonlinear stability with $t_*^{-3}$-decay in spatially compact sets]
\label{ThmSt3}
  We use the assumptions and notation of Theorem~\usref{ThmSt}, except instead of~\eqref{EqStMem}--\eqref{EqStSmall} we now require one more order of decay for the conormal remainder term, i.e.,
  \[
    \gamma-\gamma_{b_0},\ r(k-k_{b_0})\in\Hb^{\infty,(\cE_0,4+\eps_0)}(\Sigma_\IVP),\quad
    \|\gamma-\gamma_{b_0}\|_{\Hb^{d,(\cE_0,4+\eps_0)}},\ \|k-k_{b_0}\|_{\Hb^{d,(\cE_0+1,5+\eps_0)}} < \eps.
  \]
  Then the conclusions of Theorem~\usref{ThmSt} hold with~\eqref{EqSth} strengthened to
  \begin{equation}
  \label{EqSt3}
    h \in \Hb^{\infty,\ (\cE_0,4+\eps_0),\ (\la\cE_\sscri^\cC\ra,4+\eps_\sscri),\ (\cE_+,4+\eps_+),\ ((3,0),3+\eps_\cK)}(\phi_\scal(\Omega))
  \end{equation}
  (so the solution of the initial value problem is $g_{b_0,b,-\scal}+h$). More precisely, recalling the generalized zero energy states~\eqref{EqWE0s2}, there exist $\scal_2\in\scalspace_2$ and $\vect_2\in\vectspace_2$ such that
  \begin{equation}
  \label{EqSt3Asy}
    h - t_*^{-3}\chi_\cK \bigl( h_{b,\rms 2}^{(-2)}(\scal_2) + h_{b,\rmv 2}^{(-2)}(\vect_2) \bigr)
  \end{equation}
  has order $3+\eps_\cK$ at $\cK^+$ (and the same orders as~\eqref{EqSt3} at the other boundary hypersurfaces), so in particular has $t_*^{-3-\eps_\cK}$-decay in spatially compact regions.
\end{thm}

\begin{rmk}[The question of sharpness]
\label{RmkSt3Sharp}
  Since $h_{b,\rms 2}^{(-2)}$ and $h_{b,\rmv 2}^{(-2)}$ are \emph{not} pure gauge, their contributions to the late-time asymptotics of $h$ (which is $g-g_b$ on the set $\chi_\cK^{-1}(1)$) cannot be eliminated by a gauge modification. While this suggests that $t_*^{-3}$ is generically the sharp decay rate for the decay of the metric coefficients, one can draw this conclusion only if $\scal_2,\vect_2\neq 0$ generically; we do not attempt in this paper to determine whether or not this is so.
\end{rmk}

\begin{proof}[Sketch of the proof of Theorem~\usref{ThmSt3}]
  Consider the output of Theorem~\ref{ThmD6}, specifically the contributions to $u_{\rm exp}$ in~\eqref{EqD6uexp}--\eqref{EqD6uexpComp} that do \emph{not} decay at a rate $>3$ at $\cK^+$; there are three types of such contributions:
  \begin{enumerate}
  \item the residual center-of-mass motion encoded by $\dot\scal_\rem$ in~\eqref{EqD6uexpComp};
  \item the $t_*^{-3}$-coefficients in the polyhomogeneous of expansions $\dot\scal_2^{(-2)}(t_*)$ and $\dot\vect_2^{(-2)}(t_*)$ at $t_*^{-1}=0$ (recalling that $\tilde\cE_\sharp=(1,0)\cup(\cdots)$);
  \item $t_*^{-z}(\log t_*)^k$-coefficients, with $\Re z\leq 3$, in the polyhomogeneous expansions of $\dot\vect_1(t_*)$, $\dot\scal_l^{(-\lambda^\Ups_{\rms l,l+j}+1)}(t_*)$ (for $l+j\leq 2$), $\dot\scal_l^{(-l+2)}(t_*)$ (for $l\leq 4$), and $\dot\vect_l^{(-l+1)}(t_*)$ (for $l\leq 3$).
  \end{enumerate}
  The third class of terms can be eliminated using gauge modifications, as was already noted in Remark~\ref{RmkD6uexp}\eqref{ItD6uexpImpr}. The second class cannot be eliminated, and is precisely what appears in~\eqref{EqSt3Asy}. For the term $h_{b,\rms 1}^{\leq 1}(\dot\scal_\rem)$ however, finite-dimensional gauge modifications do not suffice.

  \medskip

  The simple remedy is to increase the decay rate of the conormal remainder terms of metric perturbations $h$ at $\scri^+$, $\iota^+$, and $\cK^+$ by a full order, so $4+\eps_\sscri$, $4+\eps_+$, and $5+\eps_\cK$, respectively; cf.\ \eqref{EqSt3}. The key task is to show that an analogue of Corollary~\ref{CorD6Impr} (with the decay orders of $f$ in~\eqref{EqD6Imprf} strengthened to $5+\eps_\sscri$, $6+\eps_+$, and $5+\eps_\cK$) remains valid in this setting. This requires two additions to the arguments presented there.

  \pfstep{Addition~1: gauge modifications.} We need to ensure that the gauge modifications comprising $\vartheta_\tot(\vecp)$ have more than $5$ orders of $\cK^+$-decay. In view of the observation in Remark~\ref{RmkD4KOrder}, this can be arranged by including one further term in the definition of the gauge modification terms. For example, one now needs to set $\vartheta_{\rms 1}^{(0,1)}(\scal):=\delta_{g_{b_0},E^\Ups}\sfG_{g_{b_0}}\bigl(\chi_\cK\delta_{g_{b_0}}^*\omega_{b_0,\rms 1}^{(0),\leq 5}(\scal\log t_*)\bigr)$, which by comparison with~\eqref{EqD4CorrTheta01} includes an additional term $4! t_*^{-5}\breve\omega_{b_0,\rms 1}^{(0),5}(\scal)$; similarly for $\vartheta_{\rms 1}^{(\lambda,k)}$ in Definition~\ref{DefD5Corr}, and the gauge modifications in Definition~\ref{DefD6Corr}. --- In order to implement this, we need to extend the result Proposition~\ref{PropWG0Large} to produce the relevant 1-forms $\breve\omega_{b_0,\rms 1}^{(0),5}$ etc. This, in fact, requires no additional effort at all since the construction is exactly the same as that of $\breve\omega_{b_0,\rms 1}^{(0),4}$ etc.; in fact, we only stopped at $\breve\omega_{b_0,\rms 1}^{(0),4}$ there since that was all that we needed for the proof of nonlinear stability above.
  
  \pfstep{Addition~2: refined expansions at $\tface\cap\zface$ and $\iota^+\cap\cK^+$.} Consider the solution of the $\iota^+$-model problem in Step~1.2 of the proof of Proposition~\ref{PropD5Alm} (page~\pageref{EqD5Step12}), as grafted into the Kerr spacetime via Proposition~\ref{PropipGr} with $\lambda=1$. It is then crucial that we allow $\ell_\cK=4+\eps_\cK$ (or arbitrary $\ell_\cK<4+\eps_\ind$) in Proposition~\ref{PropipGr} in order to ensure that the remainder term~\eqref{EqipGrLb} has $\cK^+$-order $5+\eps_\cK>5$ (consistent with our strengthened requirements on $f$). This requires two ingredients.
  \begin{enumerate}[label=(\roman*)]
  \item In the context of Proposition~\ref{PropipNAsy}, we need to expand $u_0$ in~\eqref{EqipNAsyu0} to one order further, thus capturing the polyhomogeneous expansion modulo $R^{4+\eps_\ind-\eps}$ for all $\eps>0$. Besides using the higher-order terms $\breve{\ubar h}_{\rms 1}^5$ etc.\ that one obtains as a by-product of the improvements in the \emph{Addition~1} above, and also using higher-order versions of the tensors in Proposition~\ref{PropWE0} (i.e., $\breve h_{b,\rms 3/\rmv 2}^{(-2),2}$ and $\breve h_{b,\rms 3/\rmv 3}^{(-3),1}$) this requires including the contribution of further indicial solutions of $\wh{\ubar L}(0)$ that correspond to indicial roots $\lambda$ with $\Re\lambda\in[-4,-3)$. (See Lemma~\ref{LemmaWEInd} for a complete list of those.) 
  \item We must moreover show that each indicial solution corresponding to these new indicial roots is, in fact, the leading-order term at $\rho=r^{-1}=0$ of a large zero energy state for the linear Kerr model operator $L_b$ (see~\eqref{EqWEOp}) for the purpose of grafting (Proposition~\ref{PropipGr}) and thus ensuring that the contributions of these terms in the $\cK^+$-expansion of solutions of the linearized gauge-fixed Einstein equation are annihilated by $L_b$ (leading to an analogue of~\eqref{EqD6Aug5Lb} with $5+\eps_\cK$ in place of $4+\eps_\cK$). For the indicial roots corresponding to pure gauge solutions, this follows easily from the construction of the associated gauge potentials (analogously to Proposition~\ref{PropWG0Large}). The only exceptional indicial root is $-4$, for which one needs to construct large zero energy states $h_{b,\rms 4}^{(-4)}$ and $h_{b,\rmv 4}^{(-4)}$ analogously to Proposition~\ref{PropWE0}.
  \end{enumerate}
  The same ingredients serve to push the expansions (at $\ztface=\tface\cap\zface$) and grafting (into $X_\scbtop^\pm$) of solutions of $\tface$-model problems discussed in~\S\ref{Ssiptf} one order further.

  \medskip
  Upon completing these refined and additional constructions, the proofs of Theorem~\ref{ThmD6} and Corollary~\ref{CorD6Impr}---and then also of Theorems~\ref{ThmEfP}, \ref{ThmEfS}, and the nonlinear stability itself---apply with only minor changes. The only additional work consists of eliminating the pure gauge contributions at orders $\leq 3$ and continuing the extraction of $\iota^+$- and $\cK^+$-expansions one order further; in the context of Part~4 of the proof of Theorem~\ref{ThmD6} (page~\pageref{ItD6Part4}), this amounts to Taylor expanding the low-energy resolvent to one additional order. The decay rate of the conormal remainder terms in~\eqref{EqD6uexpComp} is then correspondingly one order higher; in particular, the residual center-of-mass motion term $\dot\scal_\rem\in\dot H_\bop^{\infty,\,3+\eps_\cK}$ then has more than $3$ orders of decay.
\end{proof}

\subsection{Further remarks on asymptotics}
\label{SsStPhg}

There are several interesting future avenues for post-processing Theorem~\ref{ThmSt} (i.e., taking it as a black box and extracting further information). First:

\begin{conj}[Full polyhomogeneity of the spacetime metric]
\label{ConjStPhg}
   Suppose the initial data $\gamma$ and $k$ in Theorem~\usref{ThmSt} are polyhomogeneous, i.e., we have $\gamma-\gamma_{b_0}\in\cA^{\cE_0}(\Sigma_\IVP)$ and $k-k_{b_0}\in\cA^{\cE_0+1}(\Sigma_\IVP)$ (instead of merely~\eqref{EqStMem}) where we recall $\min\Re\cE_0>1$. Then the spacetime metric $g$ produced by (the proof of) Theorem~\usref{ThmSt} is polyhomogeneous on $\phi_\scal(\Omega)\subset M$, i.e., we have
   \[
     h \in \cA^{\cE_0,\ \la\cE_\sscri^\cC\ra,\ \cE_+,\ \cE_\cK}(\phi_\scal(\Omega))
   \]
   (instead of merely~\eqref{EqSth}) where $\cE_\cK$ is an index set with $\min\Re\cE_\cK\geq 3$, and in fact $\min\Re(\cE_\cK\setminus\{(3,0)\})>3$.
\end{conj}

The order $3$ decay of $h$ at $\cK^+$ is taken from Theorem~\ref{ThmSt3}.

Let us give supporting evidence for this conjecture. First of all, away from $\iota^+\cup\cK^+$, it was proved for a different gauge in \cite[\S{7}]{HintzVasyMink4}. For the gauge used here, one only needs to extend the arguments from \S\ref{SsExPhg} mildly: when pushing past the $\scri^+$-order $1+(1-e^\cC)(1-v^\cC)\gamma^\cC$, one needs to record further functional degrees of freedom encoded by the radiation fields of the components $\pi^\cC h$, but otherwise the arguments are unaffected.

Near $\iota^+\cup\cK^+$ on the other hand, one expresses the nonlinear PDE $P(h,\vecp,\vecq)=0$, $(h,\vecp,\vecq)=\Psi(U)$ (in the notation of~\eqref{EqStPsi}), similarly to Corollaries~\ref{CorDAdmFwN} and \ref{CorExFwN} as an integral of its linearizations; more precisely, one writes $0=P(0,\vecp,0)+\int_0^1 D_{(s h,\vecp,s\vecq)}P(h,0,\vecq)\,\dd s$. The term $P(0,\vecp,0)$ is polyhomogeneous. The linearization of $P$ in the first and third components can be written as $L_b$---modulo error terms with $3-\eps$ and $2+\eps_\cK$ orders of decay (as b-operators) at $\iota^+$ and $\cK^+$, respectively---acting on $h+h_{b,\rem}(\vecq)$. (Recall that this observation, made precise by Proposition~\ref{PropDAdmFw}, was the key for the successive improvements of decay in~\S\ref{SD}.) Upon inverting $L_b$ on the spectral side, one thus expects to be able to extract the polyhomogeneous expansion of the delicate (as in: not automatically polyhomogeneous) part $h+h_{b,\rem}(\vecq)$ of the metric perturbation (see~\eqref{EqEfPMap}) step-by-step in $\eps_\cK$-increments. The key task is to show that the output of $\wh{L_b}(\sigma)^{-1}$ has increasingly long partial polyhomogeneous expansions at $\tface$ and $\zface$ (corresponding, via the inverse Fourier transform and Proposition~\ref{PropTFHbphg}\eqref{ItTFHbphgFI}, to increasingly long expansions at $\iota^+$ and $\cK^+$) when the input does. This has been shown in considerable generality by Sussman \cite{SussmanResolventPhg}. We expect his arguments to carry over to the present setting as well; in particular, this requires handling the singularity of the resolvent at $\sigma=0$ (or working with the augmentation $\wt{L_b}(\sigma)$ that has a regular resolvent).

\bigskip

Secondly, by adapting the arguments from \cite[\S{8}]{HintzVasyMink4} to the present gauge, one expects to be able to derive a formula for the Bondi mass $M_B(g;u)$ of $g$ \cite{BondivdBMetznerGravity}; this requires computing Christoffel symbols and curvature components to higher order than needed for our stability argument (compare Lemmas~\ref{LemmaExOpChr} and \ref{LemmaExOpRiem} with the expressions in \cite[\S{A.2}]{HintzVasyMink4}). An interesting aspect that arises in such an analysis of the metric $g=g_{b_0,b,-\scal}+h$ from~\eqref{EqStMetric} is the following. First of all, near $\scri^+$, we have $g=(\phi_\scal)_*g_{b_0}+h$. Therefore, $\phi_\scal^*g=g_{b_0}+\phi_\scal^*h$ there; so if we use the level sets of the optical function $u=t_*$ to fiber $\scri^+$, the Bondi mass $M_B(\phi_\scal^*g;t_*)$ of $\phi_\scal^*g$ tends to the initial ADM mass $\bhm_0$ as $t_*\to-\infty$. By contrast, the Bondi mass $M_B(g;t_*)$ of $g$ itself (but \emph{still with respect to} $t_*$ rather than $(\phi_\scal)_*t_*$) tends, as $t_*\to-\infty$, to
\begin{equation}
\label{EqStMassBoosted}
  \frac{1}{4\pi} \bhm_0 \int_0^{2\pi}\int_0^\pi \frac{1}{(\cosh\chi+\cos\theta\,\sinh\chi)^3}\,\sin\theta\,\dd\theta\,\dd\phi = \bhm_0 \cosh\chi
\end{equation}
where $\chi=\frac12\|\scal\|_{L^\infty}$ is the boost parameter (cf.\ the $(\dd t_*)^2$-component of $\phi_\scal^*g_{b_0}$ as read off from~\eqref{EqKBoh11}--\eqref{EqKBo}); the factor $\frac12$ here comes from~\eqref{EqKBoVscal}.\footnote{The initial Bondi linear momentum of $\phi_\scal^*g$ on the other hand is zero, whereas that of $g$ is $(0,0,\bhm\sinh\chi)$ if the boost is in the $z$-direction; the Bondi 4-momenta are thus $(\bhm,0,0,0)$ and $(\bhm\cosh\chi,0,0,\bhm\sinh\chi)$, respectively, which are simply boosted relative to another. So the Bondi mass is really the $0$-component of the Bondi $4$-momentum.}

Similarly then, since for the metric $\phi_\scal^*g$ the final black hole with parameters $b=(\bhm,\bha)$ is boosted, one expects the final Bondi mass $M_B(\phi_\scal^*g;+\infty)$ to be equal to $\bhm\cosh\chi$. (On the other hand, for the metric $g$ itself, the final black hole is not boosted and thus the final Bondi mass should be $M_B(g;+\infty)=\bhm$.) Thus:

\begin{conj}[Bondi mass]
\label{ConjMass}
  In the notation of Theorem~\usref{ThmSt}, and with $\chi=\frac12\|\scal\|_{L^\infty}$, the Bondi mass $M_B(\phi_\scal^*g,t_*)$ of $\phi_\scal^*g$ (which thus has initial data $(\gamma,k)$ at $\Sigma_\IVP$, with ADM mass $\bhm_0$ and vanishing linear momentum), defined with respect to the level sets of $t_*$ on $\scri^+$, satisfies
  \[
    \lim_{t_*\to-\infty} M_B(\phi_\scal^*g;t_*) = \bhm_0,\quad
    \lim_{t_*\to+\infty} M_B(\phi_\scal^*g;t_*) = \bhm\cosh\chi.
  \]
  It satisfies the Bondi mass loss formula
  \begin{equation}
  \label{EqMassLoss}
    \frac{\dd}{\dd u}M_B(\phi_\scal^*g;u)=-\frac{1}{32\pi}\int_{\scri^+\cap\{t_*=u\}} |r\pa_{t_*}\slpi_0(\phi_\scal^*h)|_\slg^2\,\dd\slg.
  \end{equation}
\end{conj}

A consequence would be the inequality
\begin{equation}
\label{EqMassIneq}
  \bhm\cosh\chi \leq \bhm_0
\end{equation}
between the Bondi masses (``energies'') of the final and initial black holes, consistently with the interpretation of the right-hand side of~\eqref{EqMassLoss} as energy carried away by gravitational radiation. (One can similarly conjecture that $M_B(g;t_*)\to\bhm_0\cosh\chi$, resp.\ $\bhm$ as $t_*\to-\infty$, resp.\ $t_*\to+\infty$, again with an accompanying mass loss formula; note, however, that the conclusion $\bhm\leq\bhm_0\cosh\chi$ one could draw from this would be weaker than~\eqref{EqMassIneq}.)

We leave the investigation of Conjecture~\ref{ConjMass}, and related statements concerning the full Bondi 4-momentum and notions of angular momentum, to future work.

\clearpage
\bibliographystyle{alphaurl}


\begin{thebibliography}{GMGCH05}

\bibitem[AAG23a]{AngelopoulosAretakisGajicKerr}
Yannis Angelopoulos, Stefanos Aretakis, and Dejan Gajic.
\newblock Late-time tails and mode coupling of linear waves on {K}err
  spacetimes.
\newblock {\em Advances in Mathematics}, 417:108939, 2023.
\newblock \href {https://doi.org/https://doi.org/10.1016/j.aim.2023.108939}
  {\path{doi:https://doi.org/10.1016/j.aim.2023.108939}}.

\bibitem[AAG23b]{AngelopoulosAretakisGajicRNPrice}
Yannis Angelopoulos, Stefanos Aretakis, and Dejan Gajic.
\newblock Price's law and precise late-time asymptotics for subextremal
  {R}eissner-{N}ordstr\"om black holes.
\newblock {\em Ann. Henri Poincar\'e}, 24(9):3215--3287, 2023.
\newblock \href {https://doi.org/10.1007/s00023-023-01328-8}
  {\path{doi:10.1007/s00023-023-01328-8}}.

\bibitem[AB15a]{AnderssonBlueHiddenKerr}
Lars Andersson and Pieter Blue.
\newblock Hidden symmetries and decay for the wave equation on the {K}err
  spacetime.
\newblock {\em Ann. of Math. (2)}, 182(3):787--853, 2015.
\newblock \href {https://doi.org/10.4007/annals.2015.182.3.1}
  {\path{doi:10.4007/annals.2015.182.3.1}}.

\bibitem[AB15b]{AnderssonBlueMaxwellKerr}
Lars Andersson and Pieter Blue.
\newblock Uniform energy bound and asymptotics for the {M}axwell field on a
  slowly rotating {K}err black hole exterior.
\newblock {\em J. Hyperbolic Differ. Equ.}, 12(4):689--743, 2015.
\newblock \href {https://doi.org/10.1142/S0219891615500204}
  {\path{doi:10.1142/S0219891615500204}}.

\bibitem[ABBM25]{AnderssonBackdahlBlueMaKerr}
Lars Andersson, Thomas B\"ackdahl, Pieter Blue, and Siyuan Ma.
\newblock Stability for linearized gravity on the {K}err spacetime.
\newblock {\em Ann. PDE}, 11(1):Paper No. 11, 161, 2025.
\newblock \href {https://doi.org/10.1007/s40818-024-00193-w}
  {\path{doi:10.1007/s40818-024-00193-w}}.

\bibitem[AHW24]{AnderssonHaefnerWhitingMode}
Lars Andersson, Dietrich H{\"a}fner, and Bernard~F. Whiting.
\newblock Mode analysis for the linearized {E}instein equations on the {K}err
  metric: the large $a$ case.
\newblock {\em J.~Eur.~Math.~Soc., online first}, 2024.
\newblock \href {https://doi.org/10.4171/jems/1544}
  {\path{doi:10.4171/jems/1544}}.

\bibitem[AKU26a]{AngelopoulosKehleUngerModuli}
Yannis Angelopoulos, Christoph Kehle, and Ryan Unger.
\newblock The moduli space of dynamical spherically symmetric black hole
  spacetimes and the extremal threshold.
\newblock {\em Preprint, arXiv:2603.10378}, 2026.

\bibitem[AKU26b]{AngelopoulosKehleUngerXRNStab}
Yannis Angelopoulos, Christoph Kehle, and Ryan Unger.
\newblock Nonlinear stability of extremal {R}eissner-{N}ordstr\"om black holes
  in spherical symmetry.
\newblock volume~14, pages Paper No. e3, 84, 2026.
\newblock \href {https://doi.org/10.1017/fmp.2025.10017}
  {\path{doi:10.1017/fmp.2025.10017}}.

\bibitem[AMPW17]{AnderssonMaPaganiniWhitingModeStab}
Lars Andersson, Siyuan Ma, Claudio Paganini, and Bernard~F. Whiting.
\newblock Mode stability on the real axis.
\newblock {\em J. Math. Phys.}, 58(7):072501, 19, 2017.
\newblock \href {https://doi.org/10.1063/1.4991656}
  {\path{doi:10.1063/1.4991656}}.

\bibitem[And05]{AndersonStabilityEvenDS}
Michael~T. Anderson.
\newblock Existence and stability of even-dimensional asymptotically de
  {S}itter spaces.
\newblock {\em Ann. Henri Poincar\'e}, 6(5):801--820, 2005.
\newblock \href {https://doi.org/10.1007/s00023-005-0224-x}
  {\path{doi:10.1007/s00023-005-0224-x}}.

\bibitem[Are12]{AretakisExtremalKerr}
Stefanos Aretakis.
\newblock Decay of axisymmetric solutions of the wave equation on extreme
  {K}err backgrounds.
\newblock {\em J. Funct. Anal.}, 263(9):2770--2831, 2012.
\newblock \href {https://doi.org/10.1016/j.jfa.2012.08.015}
  {\path{doi:10.1016/j.jfa.2012.08.015}}.

\bibitem[Bac91]{BachelotSchwarzschildScattering}
Alain Bachelot.
\newblock Gravitational scattering of electromagnetic field by {Schwarzschild}
  black-hole.
\newblock {\em Annales de l'I.H.P. Physique th\'eorique}, 54(3):261--320, 1991.
\newblock URL: \url{http://www.numdam.org/item/AIHPA_1991__54_3_261_0/}.

\bibitem[BdC25]{BenomioTdCMaxwell}
Gabriele Benomio and Rita~Teixeira da~Costa.
\newblock {T}he {M}axwell equations on full sub-extremal and extremal {K}err
  spacetimes.
\newblock {\em Preprint, arXiv:2512.08917}, 2025.

\bibitem[BDGR24]{BaskinDollGellRedmanKG}
Dean Baskin, Moritz Doll, and Jesse Gell-Redman.
\newblock The {K}lein-{G}ordon equation on asymptotically {M}inkowski
  spacetimes: causal propagators.
\newblock {\em Preprint, arXiv:2409.01134}, 2024.

\bibitem[Ben24]{BenomioSchwarzschildLinear}
Gabriele Benomio.
\newblock A new gauge for gravitational perturbations of {K}err spacetimes
  {II}: the linear stability of {S}chwarzschild revisited.
\newblock {\em Arch. Ration. Mech. Anal.}, 248(5):Paper No. 92, 70, 2024.
\newblock \href {https://doi.org/10.1007/s00205-024-02036-1}
  {\path{doi:10.1007/s00205-024-02036-1}}.

\bibitem[Ben25]{BenomioFormalismI}
Gabriele Benomio.
\newblock A new gauge for gravitational perturbations of {K}err spacetimes {I}:
  {T}he linearised theory.
\newblock volume~26, pages 1573--1731, 2025.
\newblock \href {https://doi.org/10.1007/s00023-024-01472-9}
  {\path{doi:10.1007/s00023-024-01472-9}}.

\bibitem[BFHR99]{BrodbeckFrittelliHubnerReulaSCP}
Othmar Brodbeck, Simonetta Frittelli, Peter H\"ubner, and Oscar~A. Reula.
\newblock Einstein's equations with asymptotically stable constraint
  propagation.
\newblock {\em J. Math. Phys.}, 40(2):909--923, 1999.
\newblock \href {https://doi.org/10.1063/1.532694}
  {\path{doi:10.1063/1.532694}}.

\bibitem[BH08]{BonyHaefnerDecay}
Jean-Fran{\c{c}}ois Bony and Dietrich H{\"a}fner.
\newblock Decay and non-decay of the local energy for the wave equation on the
  de {S}itter--{S}chwarzschild metric.
\newblock {\em Communications in Mathematical Physics}, 282(3):697--719, 2008.
\newblock \href {https://doi.org/10.1007/s00220-008-0553-y}
  {\path{doi:10.1007/s00220-008-0553-y}}.

\bibitem[BL67]{BoyerLindquistKerr}
Robert~H. Boyer and Richard~W. Lindquist.
\newblock Maximal analytic extension of the {K}err metric.
\newblock {\em J. Mathematical Phys.}, 8:265--281, 1967.
\newblock \href {https://doi.org/10.1063/1.1705193}
  {\path{doi:10.1063/1.1705193}}.

\bibitem[BMB93]{BachelotMotetBachelotSchwarzschild}
Alain Bachelot and Agn\`es Motet-Bachelot.
\newblock Les r\'esonances d'un trou noir de {S}chwarzschild.
\newblock {\em Ann. Inst. H. Poincar\'e{} Phys. Th\'eor.}, 59(1):3--68, 1993.
\newblock URL: \url{http://www.numdam.org/item?id=AIHPA_1993__59_1_3_0}.

\bibitem[BS03]{BlueSofferSchwarzschildDecay}
Pieter Blue and Avy Soffer.
\newblock {S}emilinear wave equations on the {S}chwarzschild manifold. {I}.
  {L}ocal decay estimates.
\newblock {\em Adv. Differential Equations}, 8(5):595--614, 2003.
\newblock URL: \url{https://projecteuclid.org:443/euclid.ade/1355926842}.

\bibitem[BS05]{BlueSofferSchwarzschildSpin2}
P.~Blue and A.~Soffer.
\newblock The wave equation on the {S}chwarzschild metric. {II}. {L}ocal decay
  for the spin-2 {R}egge-{W}heeler equation.
\newblock {\em J. Math. Phys.}, 46(1):012502, 9, 2005.
\newblock \href {https://doi.org/10.1063/1.1824211}
  {\path{doi:10.1063/1.1824211}}.

\bibitem[BvdBM62]{BondivdBMetznerGravity}
H.~Bondi, M.~G.~J. van~der Burg, and A.~W.~K. Metzner.
\newblock Gravitational waves in general relativity. {VII}. {W}aves from
  axi-symmetric isolated systems.
\newblock {\em Proc. Roy. Soc. London Ser. A}, 269:21--52, 1962.
\newblock \href {https://doi.org/10.1098/rspa.1962.0161}
  {\path{doi:10.1098/rspa.1962.0161}}.

\bibitem[BVW15]{BaskinVasyWunschRadMink}
Dean Baskin, Andr\'as Vasy, and Jared Wunsch.
\newblock Asymptotics of radiation fields in asymptotically {M}inkowski space.
\newblock {\em Amer. J. Math.}, 137(5):1293--1364, 2015.
\newblock \href {https://doi.org/10.1353/ajm.2015.0033}
  {\path{doi:10.1353/ajm.2015.0033}}.

\bibitem[BVW18]{BaskinVasyWunschRadMink2}
Dean Baskin, Andr\'as Vasy, and Jared Wunsch.
\newblock Asymptotics of scalar waves on long-range asymptotically {M}inkowski
  spaces.
\newblock {\em Adv. Math.}, 328:160--216, 2018.
\newblock \href {https://doi.org/10.1016/j.aim.2018.01.012}
  {\path{doi:10.1016/j.aim.2018.01.012}}.

\bibitem[BZ09]{BieriZipserStability}
Lydia Bieri and Nina Zipser.
\newblock {\em Extensions of the stability theorem of the {M}inkowski space in
  general relativity}, volume~45 of {\em AMS/IP Studies in Advanced
  Mathematics}.
\newblock American Mathematical Society, Providence, RI; International Press,
  Cambridge, MA, 2009.
\newblock \href {https://doi.org/10.1090/amsip/045}
  {\path{doi:10.1090/amsip/045}}.

\bibitem[Car68]{CarterHamiltonJacobiEinstein}
Brandon Carter.
\newblock Hamilton--{J}acobi and {S}chr\"odinger separable solutions of
  {E}instein's equations.
\newblock {\em Comm. Math. Phys.}, 10:280--310, 1968.
\newblock URL: \url{http://projecteuclid.org/euclid.cmp/1103841118}.

\bibitem[CBG69]{ChoquetBruhatGerochMGHD}
Yvonne Choquet-Bruhat and Robert Geroch.
\newblock Global aspects of the {C}auchy problem in general relativity.
\newblock {\em Comm. Math. Phys.}, 14:329--335, 1969.
\newblock URL: \url{http://projecteuclid.org/euclid.cmp/1103841822}.

\bibitem[CD02]{ChruscielDelaySimple}
Piotr~T. Chru\'sciel and Erwann Delay.
\newblock Existence of non-trivial, vacuum, asymptotically simple spacetimes.
\newblock {\em Classical Quantum Gravity}, 19(9):L71--L79, 2002.
\newblock \href {https://doi.org/10.1088/0264-9381/19/9/101}
  {\path{doi:10.1088/0264-9381/19/9/101}}.

\bibitem[CD03]{ChruscielDelayMapping}
Piotr~T. Chru\'sciel and Erwann Delay.
\newblock On mapping properties of the general relativistic constraints
  operator in weighted function spaces, with applications.
\newblock {\em M\'em. Soc. Math. Fr. (N.S.)}, (94):vi+103, 2003.
\newblock \href {https://doi.org/10.24033/msmf.407}
  {\path{doi:10.24033/msmf.407}}.

\bibitem[Cha98]{ChandrasekharBlackHoles}
Subrahmanyan Chandrasekhar.
\newblock {\em The mathematical theory of black holes}.
\newblock Oxford Classic Texts in the Physical Sciences. The Clarendon Press,
  Oxford University Press, New York, 1998.
\newblock Reprint of the 1992 edition.

\bibitem[Chr86]{ChristodoulouGlobalSolutionsSmallData}
Demetrios Christodoulou.
\newblock Global solutions of nonlinear hyperbolic equations for small initial
  data.
\newblock {\em Comm. Pure Appl. Math.}, 39(2):267--282, 1986.
\newblock \href {https://doi.org/10.1002/cpa.3160390205}
  {\path{doi:10.1002/cpa.3160390205}}.

\bibitem[Chr02]{ChristodoulouNoPeeling}
Demetrious Christodoulou.
\newblock {{T}he {G}lobal {I}nitial {V}alue {P}roblem in {G}eneral
  {R}elativity}.
\newblock In V.~G. {Gurzadyan}, R.~T. {Jantzen}, and R.~{Ruffini}, editors,
  {\em The Ninth Marcel Grossmann Meeting}, pages 44--54, December 2002.
\newblock \href {https://doi.org/10.1142/9789812777386_0004}
  {\path{doi:10.1142/9789812777386_0004}}.

\bibitem[Cic23]{CicortasScatteringdS}
Serban Cicortas.
\newblock {S}cattering for the {W}ave {E}quation on de {S}itter {S}pace in
  {A}ll {E}ven {S}patial {D}imensions.
\newblock {\em Preprint, arXiv:2309.07342}, 2023.

\bibitem[CK93]{ChristodoulouKlainermanStability}
Demetrios Christodoulou and Sergiu Klainerman.
\newblock {\em The global nonlinear stability of the {M}inkowski space},
  volume~41 of {\em Princeton Mathematical Series}.
\newblock Princeton University Press, Princeton, NJ, 1993.

\bibitem[CK24]{ChenKlainermanHorizon}
Xuantao Chen and Sergiu Klainerman.
\newblock Regularity of the {F}uture {E}vent {H}orizon in {P}erturbations of
  {K}err.
\newblock {\em Preprint, arXiv:2409.05700}, 2024.

\bibitem[CK25]{ChenKlainermanConstraints}
Xuantao Chen and Sergiu Klainerman.
\newblock Solving the constraint equation for general free data.
\newblock {\em arXiv preprint arXiv:2512.22704}, 2025.

\bibitem[COM81]{ChristodoulouOMurchadhaBoost}
D.~Christodoulou and N.~\'O{}~Murchadha.
\newblock The boost problem in general relativity.
\newblock {\em Comm. Math. Phys.}, 80(2):271--300, 1981.
\newblock URL: \url{http://projecteuclid.org/euclid.cmp/1103919879}.

\bibitem[Cor00]{CorvinoScalar}
Justin Corvino.
\newblock Scalar curvature deformation and a gluing construction for the
  {E}instein constraint equations.
\newblock {\em Comm. Math. Phys.}, 214(1):137--189, 2000.
\newblock \href {https://doi.org/10.1007/PL00005533}
  {\path{doi:10.1007/PL00005533}}.

\bibitem[CS06]{CorvinoSchoenAsymptotics}
Justin Corvino and Richard~M. Schoen.
\newblock On the asymptotics for the vacuum {E}instein constraint equations.
\newblock {\em J. Differential Geom.}, 73(2):185--217, 2006.
\newblock URL: \url{http://projecteuclid.org/euclid.jdg/1146169910}.

\bibitem[Den82]{DenckerPolarization}
Nils Dencker.
\newblock On the propagation of polarization sets for systems of real principal
  type.
\newblock {\em J. Functional Analysis}, 46(3):351--372, 1982.
\newblock \href {https://doi.org/10.1016/0022-1236(82)90051-9}
  {\path{doi:10.1016/0022-1236(82)90051-9}}.

\bibitem[DH72]{DuistermaatHormanderFIO2}
Johannes~J. Duistermaat and Lars H{\"o}rmander.
\newblock Fourier integral operators. {II}.
\newblock {\em Acta Math.}, 128(3-4):183--269, 1972.
\newblock \href {https://doi.org/10.1007/BF02392165}
  {\path{doi:10.1007/BF02392165}}.

\bibitem[DHR19a]{DafermosHolzegelRodnianskiTeukolsky}
Mihalis Dafermos, Gustav Holzegel, and Igor Rodnianski.
\newblock Boundedness and decay for the {T}eukolsky equation on {K}err
  spacetimes {I}: {T}he case {$|a|\ll M$}.
\newblock {\em Ann. PDE}, 5(1):Paper No. 2, 118, 2019.
\newblock \href {https://doi.org/10.1007/s40818-018-0058-8}
  {\path{doi:10.1007/s40818-018-0058-8}}.

\bibitem[DHR19b]{DafermosHolzegelRodnianskiSchwarzschildStability}
Mihalis Dafermos, Gustav Holzegel, and Igor Rodnianski.
\newblock The linear stability of the {S}chwarzschild solution to gravitational
  perturbations.
\newblock {\em Acta Math.}, 222(1):1--214, 2019.
\newblock \href {https://doi.org/10.4310/ACTA.2019.v222.n1.a1}
  {\path{doi:10.4310/ACTA.2019.v222.n1.a1}}.

\bibitem[DHRT]{DafermosHolzegelRodnianskiTaylorQuasilinear}
Mihalis Dafermos, Gustav Holzegel, Igor Rodnianski, and Martin Taylor.
\newblock {Q}uasilinear wave equations on asymptotically flat spacetimes with
  applications to {K}err black holes.
\newblock {\em Preprint, arXiv:2212.14093. To appear in \textit{Analysis \&
  PDE}}.

\bibitem[DHRT21]{DafermosHolzegelRodnianskiTaylorSchwarzschild}
Mihalis Dafermos, Gustav Holzegel, Igor Rodnianski, and Martin Taylor.
\newblock The nonlinear stability of the {S}chwarzschild solution to
  gravitational perturbations.
\newblock {\em Preprint, arXiv:2104.08222}, 2021.

\bibitem[DHRT24]{DafermosHolzegelRodnianskiTaylorQuasilinear2}
Mihalis Dafermos, Gustav Holzegel, Igor Rodnianski, and Martin Taylor.
\newblock Quasilinear wave equations on kerr black holes in the full
  subextremal range $| a|< m$.
\newblock {\em arXiv preprint arXiv:2410.03639}, 2024.

\bibitem[DL25]{DafermosLukKerrCauchyHorI}
Mihalis Dafermos and Jonathan Luk.
\newblock The interior of dynamical vacuum black holes {I}: {T}he
  {$C^0$}-stability of the {K}err {C}auchy horizon.
\newblock {\em Ann. of Math. (2)}, 202(2):309--630, 2025.
\newblock \href {https://doi.org/10.4007/annals.2025.202.2.1}
  {\path{doi:10.4007/annals.2025.202.2.1}}.

\bibitem[DR09]{DafermosRodnianskiRedShift}
Mihalis Dafermos and Igor Rodnianski.
\newblock The red-shift effect and radiation decay on black hole spacetimes.
\newblock {\em Comm. Pure Appl. Math.}, 62(7):859--919, 2009.
\newblock \href {https://doi.org/10.1002/cpa.20281}
  {\path{doi:10.1002/cpa.20281}}.

\bibitem[DRSR16]{DafermosRodnianskiShlapentokhRothmanDecay}
Mihalis Dafermos, Igor Rodnianski, and Yakov Shlapentokh-Rothman.
\newblock Decay for solutions of the wave equation on {K}err exterior
  spacetimes {III}: {T}he full subextremal case {$|a|<M$}.
\newblock {\em Ann. of Math. (2)}, 183(3):787--913, 2016.
\newblock \href {https://doi.org/10.4007/annals.2016.183.3.2}
  {\path{doi:10.4007/annals.2016.183.3.2}}.

\bibitem[DSS11]{DonningerSchlagSofferPrice}
Roland Donninger, Wilhelm Schlag, and Avy Soffer.
\newblock A proof of {P}rice's law on {S}chwarzschild black hole manifolds for
  all angular momenta.
\newblock {\em Adv. Math.}, 226(1):484--540, 2011.
\newblock \href {https://doi.org/10.1016/j.aim.2010.06.026}
  {\path{doi:10.1016/j.aim.2010.06.026}}.

\bibitem[DSS12]{DonningerSchlagSofferSchwarzschild}
Roland Donninger, Wilhelm Schlag, and Avy Soffer.
\newblock On pointwise decay of linear waves on a {S}chwarzschild black hole
  background.
\newblock {\em Comm. Math. Phys.}, 309(1):51--86, 2012.
\newblock \href {https://doi.org/10.1007/s00220-011-1393-8}
  {\path{doi:10.1007/s00220-011-1393-8}}.

\bibitem[Dya11a]{DyatlovQNMExtended}
Semyon Dyatlov.
\newblock Exponential energy decay for {K}err--de {S}itter black holes beyond
  event horizons.
\newblock {\em Math. Res. Lett.}, 18(5):1023--1035, 2011.
\newblock \href {https://doi.org/10.4310/MRL.2011.v18.n5.a19}
  {\path{doi:10.4310/MRL.2011.v18.n5.a19}}.

\bibitem[Dya11b]{DyatlovQNM}
Semyon Dyatlov.
\newblock Quasi-normal modes and exponential energy decay for the {K}err-de
  {S}itter black hole.
\newblock {\em Comm. Math. Phys.}, 306(1):119--163, 2011.
\newblock \href {https://doi.org/10.1007/s00220-011-1286-x}
  {\path{doi:10.1007/s00220-011-1286-x}}.

\bibitem[Dya12]{DyatlovAsymptoticDistribution}
Semyon Dyatlov.
\newblock Asymptotic distribution of quasi-normal modes for {K}err--de {S}itter
  black holes.
\newblock {\em Annales Henri Poincar{\'e}}, 13(5):1101--1166, 2012.
\newblock \href {https://doi.org/10.1007/s00023-012-0159-y}
  {\path{doi:10.1007/s00023-012-0159-y}}.

\bibitem[Dya15]{DyatlovWaveAsymptotics}
Semyon Dyatlov.
\newblock Asymptotics of linear waves and resonances with applications to black
  holes.
\newblock {\em Comm. Math. Phys.}, 335(3):1445--1485, 2015.
\newblock \href {https://doi.org/10.1007/s00220-014-2255-y}
  {\path{doi:10.1007/s00220-014-2255-y}}.

\bibitem[DZ19]{DyatlovZworskiBook}
Semyon Dyatlov and Maciej Zworski.
\newblock {\em Mathematical theory of scattering resonances}, volume 200 of
  {\em Graduate Studies in Mathematics}.
\newblock American Mathematical Society, Providence, RI, 2019.
\newblock \href {https://doi.org/10.1090/gsm/200} {\path{doi:10.1090/gsm/200}}.

\bibitem[Fan26a]{FangKdSLinear}
Allen~Juntao Fang.
\newblock Linear stability of the slowly-rotating {K}err--de {S}itter family.
\newblock {\em Ann. PDE}, 12(1):Paper No. 15, 201, 2026.
\newblock \href {https://doi.org/10.1007/s40818-026-00236-4}
  {\path{doi:10.1007/s40818-026-00236-4}}.

\bibitem[Fan26b]{FangKdS}
Allen~Juntao Fang.
\newblock Nonlinear stability of the slowly-rotating {K}err--de {S}itter
  family.
\newblock {\em Ann. PDE}, 12(1):Paper No. 5, 79, 2026.
\newblock \href {https://doi.org/10.1007/s40818-025-00227-x}
  {\path{doi:10.1007/s40818-025-00227-x}}.

\bibitem[FB52]{ChoquetBruhatLocalEinstein}
Yvonne Four\`es-Bruhat.
\newblock Th\'eor\`eme d'existence pour certains syst\`emes d'\'equations aux
  d\'eriv\'ees partielles non lin\'eaires.
\newblock {\em Acta Math.}, 88:141--225, 1952.
\newblock \href {https://doi.org/10.1007/BF02392131}
  {\path{doi:10.1007/BF02392131}}.

\bibitem[FJS21]{FajmanJoudiouxSmuleviciEinsteinVlasov}
David Fajman, J{\'e}r{\'e}mie Joudioux, and Jacques Smulevici.
\newblock {T}he {S}tability of the {M}inkowski space for the
  {E}instein-{V}lasov system.
\newblock {\em Anal. PDE}, 14(2):425--531, 2021.
\newblock \href {https://doi.org/10.2140/apde.2021.14.425}
  {\path{doi:10.2140/apde.2021.14.425}}.

\bibitem[Fri86a]{FriedrichDeSitterPastSimple}
Helmut Friedrich.
\newblock Existence and structure of past asymptotically simple solutions of
  {E}instein's field equations with positive cosmological constant.
\newblock {\em J. Geom. Phys.}, 3(1):101--117, 1986.
\newblock \href {https://doi.org/10.1016/0393-0440(86)90004-5}
  {\path{doi:10.1016/0393-0440(86)90004-5}}.

\bibitem[Fri86b]{FriedrichStability}
Helmut Friedrich.
\newblock On the existence of {$n$}-geodesically complete or future complete
  solutions of {E}instein's field equations with smooth asymptotic structure.
\newblock {\em Comm. Math. Phys.}, 107(4):587--609, 1986.
\newblock \href {https://doi.org/10.1007/BF01205488}
  {\path{doi:10.1007/BF01205488}}.

\bibitem[Fri98]{FriedrichSpaceConformal}
Helmut Friedrich.
\newblock Gravitational fields near space-like and null infinity.
\newblock {\em J. Geom. Phys.}, 24(2):83--163, 1998.
\newblock \href {https://doi.org/10.1016/S0393-0440(97)82168-7}
  {\path{doi:10.1016/S0393-0440(97)82168-7}}.

\bibitem[Fri04]{FriedrichSmoothScriReview}
Helmut Friedrich.
\newblock Smoothness at null infinity and the structure of initial data.
\newblock In {\em The {E}instein equations and the large scale behavior of
  gravitational fields}, pages 121--203. Birkh\"auser, Basel, 2004.

\bibitem[FS24]{FournodavlosSchlueExpanding}
Grigorios Fournodavlos and Volker Schlue.
\newblock Stability of the expanding region of {K}err de {S}itter spacetimes.
\newblock {\em Preprint, arXiv:2408.02596}, 2024.

\bibitem[FST25]{FangSzeftelTouatiBHData}
Allen~Juntao Fang, J{\'e}r{\'e}mie Szeftel, and Arthur Touati.
\newblock Spacelike initial data for black hole stability.
\newblock {\em Communications in Mathematical Physics}, 406(10):235, Sep 2025.
\newblock \href {https://doi.org/10.1007/s00220-025-05416-0}
  {\path{doi:10.1007/s00220-025-05416-0}}.

\bibitem[Gaj23a]{GajicXKerrInstab}
Dejan Gajic.
\newblock Azimuthal instabilities on extremal {K}err.
\newblock {\em Preprint, arXiv:2302.06636}, 2023.

\bibitem[Gaj23b]{GajicInverseSquare}
Dejan Gajic.
\newblock Late-time asymptotics for geometric wave equations with
  inverse-square potentials.
\newblock {\em J. Funct. Anal.}, 285(7):Paper No. 110058, 114, 2023.
\newblock \href {https://doi.org/10.1016/j.jfa.2023.110058}
  {\path{doi:10.1016/j.jfa.2023.110058}}.

\bibitem[GH08]{GuillarmouHassellResI}
Colin Guillarmou and Andrew Hassell.
\newblock Resolvent at low energy and {R}iesz transform for {S}chr\"{o}dinger
  operators on asymptotically conic manifolds. {I}.
\newblock {\em Math. Ann.}, 341(4):859--896, 2008.
\newblock \href {https://doi.org/10.1007/s00208-008-0216-5}
  {\path{doi:10.1007/s00208-008-0216-5}}.

\bibitem[Gio20a]{GiorgiRNLinearSmall}
Elena Giorgi.
\newblock The linear stability of {R}eissner-{N}ordstr\"om spacetime for small
  charge.
\newblock {\em Ann. PDE}, 6(2):Paper No. 8, 145, 2020.
\newblock \href {https://doi.org/10.1007/s40818-020-00082-y}
  {\path{doi:10.1007/s40818-020-00082-y}}.

\bibitem[Gio20b]{GiorgiRNLinear}
Elena Giorgi.
\newblock The linear stability of {R}eissner-{N}ordstr\"om spacetime: the full
  subextremal range {$|Q|<M$}.
\newblock {\em Comm. Math. Phys.}, 380(3):1313--1360, 2020.
\newblock \href {https://doi.org/10.1007/s00220-020-03893-z}
  {\path{doi:10.1007/s00220-020-03893-z}}.

\bibitem[GKS20]{GiorgiKlainermanSzeftelFormalism}
Elena Giorgi, Sergiu Klainerman, and J{\'e}r{\'e}mie Szeftel.
\newblock A general formalism for the stability of {K}err.
\newblock {\em Preprint, arXiv:2002.02740}, 2020.

\bibitem[GKS24]{GiorgiKlainermanSzeftelStability}
Elena Giorgi, Sergiu Klainerman, and J\'er\'emie Szeftel.
\newblock Wave equations estimates and the nonlinear stability of slowly
  rotating {K}err black holes.
\newblock {\em Pure Appl. Math. Q.}, 20(7):2865--3849, 2024.
\newblock \href {https://doi.org/10.4310/pamq.241128023033}
  {\path{doi:10.4310/pamq.241128023033}}.

\bibitem[GL91]{GrahamLeeConformalEinstein}
C.~Robin Graham and John~M. Lee.
\newblock Einstein metrics with prescribed conformal infinity on the ball.
\newblock {\em Adv. Math.}, 87(2):186--225, 1991.
\newblock \href {https://doi.org/10.1016/0001-8708(91)90071-E}
  {\path{doi:10.1016/0001-8708(91)90071-E}}.

\bibitem[GMGCH05]{GundlachCalabreseHinderMartinConstraintDamping}
Carsten Gundlach, Jose~M. Martin-Garcia, Gioel Calabrese, and Ian Hinder.
\newblock {Constraint damping in the Z4 formulation and harmonic gauge}.
\newblock {\em Class. Quant. Grav.}, 22:3767--3774, 2005.
\newblock \href {http://arxiv.org/abs/gr-qc/0504114}
  {\path{arXiv:gr-qc/0504114}}, \href
  {https://doi.org/10.1088/0264-9381/22/17/025}
  {\path{doi:10.1088/0264-9381/22/17/025}}.

\bibitem[Gui05]{GuillarmouMeromorphic}
Colin Guillarmou.
\newblock {Meromorphic properties of the resolvent on asymptotically hyperbolic
  manifolds}.
\newblock {\em Duke Mathematical Journal}, 129(1):1--37, 2005.
\newblock \href {https://doi.org/10.1215/S0012-7094-04-12911-2}
  {\path{doi:10.1215/S0012-7094-04-12911-2}}.

\bibitem[GW24]{GajicWarnickKerrQNM}
Dejan Gajic and Claude~M. Warnick.
\newblock Quasinormal modes on {K}err spacetimes.
\newblock {\em Preprint, arXiv:2407.04098}, 2024.

\bibitem[Ham82]{HamiltonNashMoser}
Richard~S. Hamilton.
\newblock The inverse function theorem of {N}ash and {M}oser.
\newblock {\em Bull. Amer. Math. Soc. (N.S.)}, 7(1):65--222, 1982.
\newblock \href {https://doi.org/10.1090/S0273-0979-1982-15004-2}
  {\path{doi:10.1090/S0273-0979-1982-15004-2}}.

\bibitem[He26]{HeLinearKerrNewman}
Lili He.
\newblock The linear stability of weakly charged and slowly rotating
  {K}err-{N}ewman family of charged black holes.
\newblock {\em Ann. PDE}, 12(1):Paper No. 3, 277, 2026.
\newblock \href {https://doi.org/10.1007/s40818-025-00219-x}
  {\path{doi:10.1007/s40818-025-00219-x}}.

\bibitem[HHV21]{HaefnerHintzVasyKerr}
Dietrich H{\"a}fner, Peter Hintz, and Andr{\'a}s Vasy.
\newblock Linear stability of slowly rotating {K}err black holes.
\newblock {\em Inventiones mathematicae}, 223:1227--1406, 2021.
\newblock \href {https://doi.org/10.1007/s00222-020-01002-4}
  {\path{doi:10.1007/s00222-020-01002-4}}.

\bibitem[HHV24]{HaefnerHintzVasyKerrErratum}
Dietrich H{\"a}fner, Peter Hintz, and Andr{\'a}s Vasy.
\newblock Correction to: Linear stability of slowly rotating kerr black holes.
\newblock {\em Inventiones mathematicae}, 236(1):477--481, Apr 2024.
\newblock \href {https://doi.org/10.1007/s00222-024-01240-w}
  {\path{doi:10.1007/s00222-024-01240-w}}.

\bibitem[HHV25]{HaefnerHintzVasyKerrLarge}
Dietrich H\"afner, Peter Hintz, and Andr\'as Vasy.
\newblock Linear stability of {K}err black holes in the full subextremal range.
\newblock {\em Preprint, arXiv:2506.21183}, 2025.

\bibitem[Hig87]{HiguchiSpherical}
Atsushi Higuchi.
\newblock Symmetric tensor spherical harmonics on the {$N$}-sphere and their
  application to the de {S}itter group {${\rm SO}(N,1)$}.
\newblock {\em J. Math. Phys.}, 28(7):1553--1566, 1987.
\newblock \href {https://doi.org/10.1063/1.527513}
  {\path{doi:10.1063/1.527513}}.

\bibitem[Hin17]{HintzPsdoInner}
Peter Hintz.
\newblock Resonance expansions for tensor-valued waves on asymptotically
  {K}err--de {S}itter spaces.
\newblock {\em J. Spectr. Theory}, 7:519--557, 2017.
\newblock \href {https://doi.org/10.4171/JST/171} {\path{doi:10.4171/JST/171}}.

\bibitem[Hin18]{HintzKNdSStability}
Peter Hintz.
\newblock {N}on-linear {S}tability of the {K}err--{N}ewman--de {S}itter
  {F}amily of {C}harged {B}lack {H}oles.
\newblock {\em Annals of PDE}, 4(1):11, Apr 2018.
\newblock \href {https://doi.org/10.1007/s40818-018-0047-y}
  {\path{doi:10.1007/s40818-018-0047-y}}.

\bibitem[Hin22a]{HintzPrice}
Peter Hintz.
\newblock {A} sharp version of {P}rice's law for wave decay on asymptotically
  flat spacetimes.
\newblock {\em Communications in Mathematical Physics}, 389:491--542, 2022.
\newblock \href {https://doi.org/10.1007/s00220-021-04276-8}
  {\path{doi:10.1007/s00220-021-04276-8}}.

\bibitem[Hin22b]{HintzConicPowers}
Peter Hintz.
\newblock Resolvents and complex powers of semiclassical cone operators.
\newblock {\em Mathematische Nachrichten}, 295(10):1990--2035, 2022.
\newblock \href {https://doi.org/10.1002/mana.202100004}
  {\path{doi:10.1002/mana.202100004}}.

\bibitem[Hin23a]{HintzMink4Gauge}
Peter Hintz.
\newblock Exterior stability of {M}inkowski space in generalized harmonic
  gauge.
\newblock {\em Archive for Rational Mechanics and Analysis}, 247(99), 2023.
\newblock \href {https://doi.org/10.1007/s00205-023-01931-3}
  {\path{doi:10.1007/s00205-023-01931-3}}.

\bibitem[Hin23b]{HintzGlueLocI}
Peter Hintz.
\newblock Gluing small black holes along timelike geodesics {I}: formal
  solution.
\newblock {\em Preprint, arXiv:2306.07409}, 2023.

\bibitem[Hin23c]{HintzNonstat}
Peter Hintz.
\newblock Linear waves on non-stationary asymptotically flat spacetimes. {I}.
\newblock {\em Preprint, arXiv:2302.14647}, 2023.

\bibitem[Hin23d]{Hintz3b}
Peter Hintz.
\newblock Microlocal analysis of operators with asymptotic translation- and
  dilation-invariances.
\newblock {\em Preprint, arXiv:2302.13803}, 2023.

\bibitem[Hin24a]{HintzGlueLocII}
Peter Hintz.
\newblock Gluing small black holes along timelike geodesics {II}: uniform
  analysis on glued spacetimes.
\newblock {\em Preprint, arXiv:2408.06712}, 2024.

\bibitem[Hin24b]{HintzGlueLocIII}
Peter Hintz.
\newblock Gluing small black holes along timelike geodesics {III}: construction
  of true solutions and extreme mass ratio mergers.
\newblock {\em Preprint, arXiv:2408.06715}, 2024.

\bibitem[Hin24c]{HintzConicProp}
Peter Hintz.
\newblock Semiclassical propagation through cone points.
\newblock {\em Analysis \& PDE}, 17(10):3477--3550, 2024.
\newblock \href {https://doi.org/10.2140/apde.2024.17.3477}
  {\path{doi:10.2140/apde.2024.17.3477}}.

\bibitem[Hin25a]{HintzMicro}
Peter Hintz.
\newblock {\em {A}n {I}ntroduction to {M}icrolocal {A}nalysis}.
\newblock Graduate Texts in Mathematics, No. 304. Springer-Verlag, Cham, 2025.
\newblock \href {https://doi.org/10.1007/978-3-031-90706-7}
  {\path{doi:10.1007/978-3-031-90706-7}}.

\bibitem[Hin25b]{HintzKdSMS}
Peter Hintz.
\newblock Mode stability and shallow quasinormal modes of {K}err--de {S}itter
  black holes away from extremality.
\newblock {\em J. Eur. Math. Soc. (JEMS)}, 27(12):4891--4996, 2025.
\newblock \href {https://doi.org/10.4171/jems/1463}
  {\path{doi:10.4171/jems/1463}}.

\bibitem[Hin25c]{HintzUnDet}
Peter Hintz.
\newblock Underdetermined-elliptic {PDE} on asymptotically {E}uclidean
  manifolds, and generalizations.
\newblock {\em Math. Res. Lett.}, 32(3):789--831, 2025.
\newblock \href {https://doi.org/10.4310/mrl.250729160443}
  {\path{doi:10.4310/mrl.250729160443}}.

\bibitem[Hin26a]{HintzKerrCD}
Peter Hintz.
\newblock {C}onstraint damping on subextremal {K}err spacetimes.
\newblock {\em Preprint, arXiv:2606.27658v1}, 2026.

\bibitem[Hin26b]{HintzHorizons}
Peter Hintz.
\newblock Horizons of some asymptotically stationary space-times.
\newblock {\em Duke Math. J.}, 175(7):1345--1361, 2026.
\newblock \href {https://doi.org/10.1215/00127094-2025-0058}
  {\path{doi:10.1215/00127094-2025-0058}}.

\bibitem[Hin26c]{HintzNonstat2}
Peter Hintz.
\newblock ({N}on-)linear waves on asymptotically flat spacetimes. {II}:
  trapping, bound states, nonlinear applications.
\newblock {\em Preprint, arXiv:2606.28008v1}, 2026.

\bibitem[Hin26d]{HintzScaledBddGeo}
Peter Hintz.
\newblock Pseudodifferential operators on manifolds with scaled bounded
  geometry.
\newblock {\em Commun. Am. Math. Soc.}, 6:153--243, 2026.
\newblock \href {https://doi.org/10.1090/cams/58} {\path{doi:10.1090/cams/58}}.

\bibitem[HK16]{HungKellerAxial}
Pei-Ken Hung and Jordan Keller.
\newblock Linear stability of {S}chwarzschild spacetime subject to axial
  perturbations.
\newblock {\em Preprint, arXiv:1610.08547}, 2016.

\bibitem[HK23]{HolzegelKauffmannKerrFirstOrder}
Gustav Holzegel and Christopher Kauffman.
\newblock The wave equation on the {S}chwarzschild spacetime with small
  non-decaying first-order terms.
\newblock {\em J. Hyperbolic Differ. Equ.}, 20(4):825--834, 2023.
\newblock \href {https://doi.org/10.1142/S0219891623500273}
  {\path{doi:10.1142/S0219891623500273}}.

\bibitem[HK26]{HeKlainermanKerr}
Lili He and Sergiu Klainerman.
\newblock A physical space derivation of {M}orawetz-{E}nergy estimates in
  {K}err spacetimes with large angular momentum.
\newblock {\em Preprint, arXiv:2607.08958}, 2026.

\bibitem[HKW20]{HungKellerWangSchwarzschild}
Pei-Ken Hung, Jordan Keller, and Mu-Tao Wang.
\newblock Linear stability of {S}chwarzschild spacetime: decay of metric
  coefficients.
\newblock {\em J. Differential Geom.}, 116(3):481--541, 2020.
\newblock \href {https://doi.org/10.4310/jdg/1606964416}
  {\path{doi:10.4310/jdg/1606964416}}.

\bibitem[HMV08]{HassellMelroseVasySymbolicOrderZero}
Andrew Hassell, Richard Melrose, and Andr\'as Vasy.
\newblock Microlocal propagation near radial points and scattering for symbolic
  potentials of order zero.
\newblock {\em Anal. PDE}, 1(2):127--196, 2008.
\newblock \href {https://doi.org/10.2140/apde.2008.1.127}
  {\path{doi:10.2140/apde.2008.1.127}}.

\bibitem[H{\"o}r07]{HormanderAnalysisPDE3}
Lars H{\"o}rmander.
\newblock {\em The analysis of linear partial differential operators. {III}}.
\newblock Classics in Mathematics. Springer, Berlin, 2007.
\newblock Pseudo-differential operators, Reprint of the 1994 edition.
\newblock \href {https://doi.org/10.1007/978-3-540-49938-1}
  {\path{doi:10.1007/978-3-540-49938-1}}.

\bibitem[HPS77]{HirschPughShubInvariantManifolds}
M.~W. Hirsch, C.~C. Pugh, and M.~Shub.
\newblock {\em Invariant manifolds}, volume Vol. 583 of {\em Lecture Notes in
  Mathematics}.
\newblock Springer-Verlag, Berlin-New York, 1977.

\bibitem[HPV25]{HintzPetersenVasyKdS}
Peter Hintz, Oliver Petersen, and Andr\'as Vasy.
\newblock {C}onditional non-linear stability of {K}err--de~{S}itter spacetimes:
  the full subextremal range.
\newblock {\em Preprint, arXiv:2508.06620}, 2025.

\bibitem[Hun18]{HungSchwarzschildOdd}
Pei-Ken Hung.
\newblock The {L}inear {S}tability of the {S}chwarzschild {S}pacetime in the
  {H}armonic {G}auge: {O}dd {P}art.
\newblock 2018.
\newblock Thesis (Ph.D.)--Columbia University.
\newblock URL:
  \url{http://gateway.proquest.com/openurl?url_ver=Z39.88-2004&rft_val_fmt=info:ofi/fmt:kev:mtx:dissertation&res_dat=xri:pqm&rft_dat=xri:pqdiss:10791286}.

\bibitem[Hun19]{HungSchwarzschildEven}
Pei-Ken Hung.
\newblock {T}he linear stability of the {S}chwarzschild spacetime in the
  harmonic gauge: even part.
\newblock {\em Preprint, arXiv:1909.06733}, 2019.

\bibitem[HV15]{HaberVasyPropagation}
Nick Haber and Andr\'as Vasy.
\newblock Propagation of singularities around a {L}agrangian submanifold of
  radial points.
\newblock volume 143, pages 679--726. 2015.
\newblock \href {https://doi.org/10.24033/bsmf.2702}
  {\path{doi:10.24033/bsmf.2702}}.

\bibitem[HV18a]{HintzVasyKdsFormResonances}
Peter Hintz and Andr{\'a}s Vasy.
\newblock {A}symptotics for the wave equation on differential forms on
  {K}err--de {S}itter space.
\newblock {\em Journal of Differential Geometry}, 110(2):221--279, 2018.
\newblock \href {https://doi.org/10.4310/jdg/1538791244}
  {\path{doi:10.4310/jdg/1538791244}}.

\bibitem[HV18b]{HintzVasyKdSStability}
Peter Hintz and Andr{\'a}s Vasy.
\newblock {T}he global non-linear stability of the {K}err--de {S}itter family
  of black holes.
\newblock {\em Acta mathematica}, 220:1--206, 2018.
\newblock \href {https://doi.org/10.4310/acta.2018.v220.n1.a1}
  {\path{doi:10.4310/acta.2018.v220.n1.a1}}.

\bibitem[HV20]{HintzVasyMink4}
Peter Hintz and Andr{\'a}s Vasy.
\newblock {S}tability of {M}inkowski space and polyhomogeneity of the metric.
\newblock {\em Annals of PDE}, 6(2), 2020.
\newblock \href {https://doi.org/10.1007/s40818-020-0077-0}
  {\path{doi:10.1007/s40818-020-0077-0}}.

\bibitem[HV24]{HintzVasyKdSCosm}
Peter Hintz and Andr{\'a}s Vasy.
\newblock Stability of the expanding region of {K}err--de~{S}itter spacetimes
  and smoothness at the conformal boundary.
\newblock {\em Preprint, arXiv:2409.15460}, 2024.

\bibitem[HV26]{HintzVasyScrieb}
Peter Hintz and Andr\'as Vasy.
\newblock Microlocal analysis near null infinity in asymptotically flat
  spacetimes.
\newblock {\em Anal. PDE}, 19(1):1--106, 2026.
\newblock \href {https://doi.org/10.2140/apde.2026.19.1}
  {\path{doi:10.2140/apde.2026.19.1}}.

\bibitem[IP22]{IonescuPausaderKG}
Alexandru~D. Ionescu and Beno\^{i}t Pausader.
\newblock {\em The {E}instein--{K}lein--{G}ordon {C}oupled {S}ystem: {G}lobal
  {S}tability of the {M}inkowski {S}olution}, volume 213 of {\em Annals of
  Mathematics Studies}.
\newblock Princeton University Press, Princeton, NJ, 2022.
\newblock \href {https://doi.org/10.1515/9780691233031}
  {\path{doi:10.1515/9780691233031}}.

\bibitem[JK79]{JensenKatoResolvent}
Arne Jensen and Tosio Kato.
\newblock Spectral properties of {S}chr\"odinger operators and time-decay of
  the wave functions.
\newblock {\em Duke Math. J.}, 46(3):583--611, 1979.
\newblock URL: \url{http://projecteuclid.org/euclid.dmj/1077313577}.

\bibitem[Joh19]{JohnsonSchwarzschild}
Thomas~William Johnson.
\newblock The linear stability of the {S}chwarzschild solution to gravitational
  perturbations in the generalised wave gauge.
\newblock {\em Ann. PDE}, 5(2):Paper No. 13, 92, 2019.
\newblock \href {https://doi.org/10.1007/s40818-019-0069-0}
  {\path{doi:10.1007/s40818-019-0069-0}}.

\bibitem[Keh22a]{KehrbergerScri2}
Leonhard M.~A. Kehrberger.
\newblock The case against smooth null infinity {II}: a logarithmically
  modified {P}rice's law.
\newblock {\em Adv. Theor. Math. Phys.}, 26(10):3633--3676, 2022.
\newblock \href {https://doi.org/10.4310/atmp.2022.v26.n10.a6}
  {\path{doi:10.4310/atmp.2022.v26.n10.a6}}.

\bibitem[Keh22b]{KehrbergerScri3}
Leonhard M.~A. Kehrberger.
\newblock The case against smooth null infinity {III}: {E}arly-time asymptotics
  for higher {$\ell $}-modes of linear waves on a {S}chwarzschild background.
\newblock {\em Ann. PDE}, 8(2):Paper No. 12, 117, 2022.
\newblock \href {https://doi.org/10.1007/s40818-022-00129-2}
  {\path{doi:10.1007/s40818-022-00129-2}}.

\bibitem[Keh22c]{KehrbergerScri1}
Lionor M.~A. Kehrberger.
\newblock The case against smooth null infinity {I}: heuristics and
  counter-examples.
\newblock {\em Annales Henri Poincar{\'e}}, 23(3):829--921, Mar 2022.
\newblock \href {https://doi.org/10.1007/s00023-021-01108-2}
  {\path{doi:10.1007/s00023-021-01108-2}}.

\bibitem[Keh24]{KehrbergerScri4}
Leonhard Kehrberger.
\newblock The case against smooth null infinity {IV}: {L}inearized gravity
  around {S}chwarzschild---an overview.
\newblock {\em Philos. Trans. Roy. Soc. A}, 382(2267):Paper No. 20230039, 27,
  2024.
\newblock \href {https://doi.org/10.1098/rsta.2023.0039}
  {\path{doi:10.1098/rsta.2023.0039}}.

\bibitem[Kei18]{KeirWeak}
Joseph Keir.
\newblock {T}he weak null condition and global existence using the p-weighted
  energy method.
\newblock {\em Preprint, arXiv:1808.09982}, 2018.

\bibitem[Ker63]{KerrKerr}
Roy~P. Kerr.
\newblock Gravitational field of a spinning mass as an example of algebraically
  special metrics.
\newblock {\em Phys. Rev. Lett.}, 11:237--238, 1963.
\newblock \href {https://doi.org/10.1103/PhysRevLett.11.237}
  {\path{doi:10.1103/PhysRevLett.11.237}}.

\bibitem[KK25]{KadarKehrbergerPhgScatter}
Istvan Kadar and Lionor Kehrberger.
\newblock {S}cattering, {P}olyhomogeneity and {A}symptotics for {Q}uasilinear
  {W}ave {E}quations {F}rom {P}ast to {F}uture {N}ull {I}nfinity.
\newblock {\em Preprint, arXiv:2501.09814}, 2025.

\bibitem[Kla86]{KlainermanNullCondition}
Sergiu Klainerman.
\newblock The null condition and global existence to nonlinear wave equations.
\newblock In {\em Nonlinear systems of partial differential equations in
  applied mathematics, {P}art 1 ({S}anta {F}e, {N}.{M}., 1984)}, Lectures in
  Appl. Math., pages 293--326. Amer. Math. Soc., Providence, RI, 1986.

\bibitem[KN03]{KlainermanNicoloEvolution}
Sergiu Klainerman and Francesco Nicol\`o.
\newblock {\em The evolution problem in general relativity}, volume~25 of {\em
  Progress in Mathematical Physics}.
\newblock Birkh\"auser Boston, Inc., Boston, MA, 2003.
\newblock \href {https://doi.org/10.1007/978-1-4612-2084-8}
  {\path{doi:10.1007/978-1-4612-2084-8}}.

\bibitem[KS21]{KlainermanSzeftelPolarized}
Sergiu Klainerman and J{\'e}r{\'e}mie Szeftel.
\newblock {\em {G}lobal {N}onlinear {S}tability of {S}chwarzschild {S}pacetime
  under {P}olarized {P}erturbations}, volume 210 of {\em Annals of Mathematics
  Studies}.
\newblock Princeton University Press, Princeton, NJ, 2021.
\newblock \href {https://doi.org/10.1515/9780691218526}
  {\path{doi:10.1515/9780691218526}}.

\bibitem[KS22a]{KlainermanSzeftelGCM1}
Sergiu Klainerman and J\'er\'emie Szeftel.
\newblock Construction of {GCM} spheres in perturbations of {K}err.
\newblock {\em Ann. PDE}, 8(2):Paper No. 17, 153, 2022.
\newblock \href {https://doi.org/10.1007/s40818-022-00131-8}
  {\path{doi:10.1007/s40818-022-00131-8}}.

\bibitem[KS22b]{KlainermanSzeftelGCM2}
Sergiu Klainerman and J\'er\'emie Szeftel.
\newblock Effective results on uniformization and intrinsic {GCM} spheres in
  perturbations of {K}err.
\newblock {\em Ann. PDE}, 8(2):Paper No. 18, 89, 2022.
\newblock With an appendix by Camillo De Lellis.
\newblock \href {https://doi.org/10.1007/s40818-022-00132-7}
  {\path{doi:10.1007/s40818-022-00132-7}}.

\bibitem[KS23]{KlainermanSzeftelKerr}
Sergiu Klainerman and J\'er\'emie Szeftel.
\newblock Kerr stability for small angular momentum.
\newblock {\em Pure Appl. Math. Q.}, 19(3):791--1678, 2023.
\newblock \href {https://doi.org/10.4310/pamq.2023.v19.n3.a1}
  {\path{doi:10.4310/pamq.2023.v19.n3.a1}}.

\bibitem[KU22]{KehleUngerThirdLaw}
Christoph Kehle and Ryan Unger.
\newblock Gravitational collapse to extremal black holes and the third law of
  black hole thermodynamics.
\newblock {\em Preprint, arXiv:2211.15742}, 2022.

\bibitem[KU24]{KehleUngerExtremalCritical}
Christoph Kehle and Ryan Unger.
\newblock Extremal black hole formation as a critical phenomenon.
\newblock {\em Preprint, arXiv:2402.10190}, 2024.

\bibitem[Lei26]{LeimbacherdS}
Maurus Leimbacher.
\newblock {S}tability of de~{S}itter space and expansion at the conformal
  boundary.
\newblock {\em Preprint, arXiv:2605.03481}, 2026.

\bibitem[Li25]{LiInternalCorners}
Zhenhao Li.
\newblock Internal waves in aquariums with characteristic corners.
\newblock {\em Pure Appl. Anal.}, 7(2):445--534, 2025.
\newblock \href {https://doi.org/10.2140/paa.2025.7.445}
  {\path{doi:10.2140/paa.2025.7.445}}.

\bibitem[LM16]{LeFlochMaEinsteinMassive}
Philippe~G. LeFloch and Yue Ma.
\newblock The global nonlinear stability of {M}inkowski space for
  self-gravitating massive fields.
\newblock {\em Comm. Math. Phys.}, 346(2):603--665, 2016.
\newblock \href {https://doi.org/10.1007/s00220-015-2549-8}
  {\path{doi:10.1007/s00220-015-2549-8}}.

\bibitem[LO24]{LukOhTwoTails}
Jonathan Luk and Sung-Jin Oh.
\newblock Late time tail of waves on dynamic asymptotically flat spacetimes of
  odd space dimensions.
\newblock {\em Preprint, arXiv:2404.02220}, 2024.

\bibitem[Loo22a]{LooiDecayCubic}
Shi-Zhuo Looi.
\newblock Decay rates for cubic and higher order nonlinear wave equations on
  asymptotically flat spacetimes.
\newblock {\em Preprint, arXiv:2207.10280}, 2022.

\bibitem[Loo22b]{LooiDecayQuasilinearImproved}
Shi-Zhuo Looi.
\newblock Improved decay for quasilinear wave equations close to asymptotically
  flat spacetimes including black hole spacetimes.
\newblock {\em Preprint, arXiv:2208.05439}, 2022.

\bibitem[Loo22c]{LooiDecayEnergyCritical}
Shi-Zhuo Looi.
\newblock Pointwise decay for the energy-critical nonlinear wave equation.
\newblock {\em Preprint, arXiv:2205.13197}, 2022.

\bibitem[LR03]{LindbladRodnianskiWeakNull}
Hans Lindblad and Igor Rodnianski.
\newblock The weak null condition for {E}instein's equations.
\newblock {\em C. R. Math. Acad. Sci. Paris}, 336(11):901--906, 2003.
\newblock \href {https://doi.org/10.1016/S1631-073X(03)00231-0}
  {\path{doi:10.1016/S1631-073X(03)00231-0}}.

\bibitem[LR05]{LindbladRodnianskiGlobalExistence}
Hans Lindblad and Igor Rodnianski.
\newblock Global existence for the {E}instein vacuum equations in wave
  coordinates.
\newblock {\em Comm. Math. Phys.}, 256(1):43--110, 2005.
\newblock \href {https://doi.org/10.1007/s00220-004-1281-6}
  {\path{doi:10.1007/s00220-004-1281-6}}.

\bibitem[LR10]{LindbladRodnianskiGlobalStability}
Hans Lindblad and Igor Rodnianski.
\newblock The global stability of {M}inkowski space-time in harmonic gauge.
\newblock {\em Ann. of Math. (2)}, 171(3):1401--1477, 2010.
\newblock \href {https://doi.org/10.4007/annals.2010.171.1401}
  {\path{doi:10.4007/annals.2010.171.1401}}.

\bibitem[LT18]{LindbladTohaneanuSchwarzschildQuasi}
Hans Lindblad and Mihai Tohaneanu.
\newblock Global existence for quasilinear wave equations close to
  {S}chwarzschild.
\newblock {\em Comm. Partial Differential Equations}, 43(6):893--944, 2018.
\newblock \href {https://doi.org/10.1080/03605302.2018.1476529}
  {\path{doi:10.1080/03605302.2018.1476529}}.

\bibitem[LT20a]{LindbladTaylorVlasov}
Hans Lindblad and Martin Taylor.
\newblock {G}lobal {S}tability of {M}inkowski {S}pace for the
  {E}instein--{V}lasov {S}ystem in the {H}armonic {G}auge.
\newblock {\em Archive for Rational Mechanics and Analysis}, 235(1):517--633,
  2020.
\newblock \href {https://doi.org/10.1007/s00205-019-01425-1}
  {\path{doi:10.1007/s00205-019-01425-1}}.

\bibitem[LT20b]{LindbladTohaneanuKerrQuasi}
Hans Lindblad and Mihai Tohaneanu.
\newblock {A} {L}ocal {E}nergy {E}stimate for {W}ave {E}quations on {M}etrics
  {A}symptotically {C}lose to {K}err.
\newblock {\em Annales Henri Poincar{\'e}}, 21(11):3659--3726, Nov 2020.
\newblock \href {https://doi.org/10.1007/s00023-020-00950-0}
  {\path{doi:10.1007/s00023-020-00950-0}}.

\bibitem[LX24]{LooiXiongSemilinearAsymp}
Shi-Zhuo Looi and Haoren Xiong.
\newblock Asymptotic expansions for semilinear waves on asymptotically flat
  spacetimes.
\newblock {\em Preprint, arXiv:2407.08997}, 2024.

\bibitem[Ma20]{MaMaxwellKerr}
Siyuan Ma.
\newblock {U}niform energy bound and {M}orawetz estimate for extreme components
  of spin fields in the exterior of a slowly rotating {K}err black hole {I}:
  {M}axwell field.
\newblock {\em Annales Henri Poincar{\'e}}, 21(3):815--863, Mar 2020.
\newblock \href {https://doi.org/10.1007/s00023-020-00884-7}
  {\path{doi:10.1007/s00023-020-00884-7}}.

\bibitem[Ma22]{MaMaxwellAlmost}
Siyuan Ma.
\newblock Almost {P}rice's law in {S}chwarzschild and decay estimates in {K}err
  for {M}axwell field.
\newblock {\em J. Differential Equations}, 339:1--89, 2022.
\newblock \href {https://doi.org/10.1016/j.jde.2022.08.021}
  {\path{doi:10.1016/j.jde.2022.08.021}}.

\bibitem[Mar83]{MarckParallelNull}
Jean-Alain Marck.
\newblock Parallel-tetrad on null geodesics in {K}err and {K}err-~{N}ewman
  space-time.
\newblock {\em Phys. Lett. A}, 97(4):140--142, 1983.
\newblock \href {https://doi.org/10.1016/0375-9601(83)90197-4}
  {\path{doi:10.1016/0375-9601(83)90197-4}}.

\bibitem[Mav23]{MavrogiannisSdSMorawetz}
Georgios Mavrogiannis.
\newblock Morawetz estimates without relative degeneration and exponential
  decay on {S}chwarzschild--de {S}itter spacetimes.
\newblock {\em Ann. Henri Poincar\'e}, 24(9):3113--3152, 2023.
\newblock \href {https://doi.org/10.1007/s00023-023-01293-2}
  {\path{doi:10.1007/s00023-023-01293-2}}.

\bibitem[Mav24]{MavrogiannisSdSQuasi}
Georgios Mavrogiannis.
\newblock Quasilinear wave equations on {S}chwarzschild--de {S}itter.
\newblock {\em Comm. Partial Differential Equations}, 49(1-2):38--87, 2024.
\newblock \href {https://doi.org/10.1080/03605302.2023.2295035}
  {\path{doi:10.1080/03605302.2023.2295035}}.

\bibitem[Mav25a]{MavrogiannisKdSMorawetz}
Georgios Mavrogiannis.
\newblock {B}oundedness and {M}orawetz estimates on subextremal {K}err de
  {S}itter.
\newblock {\em Preprint, arXiv:2503.22077}, 2025.

\bibitem[Mav25b]{MavrogiannisKdSDecay}
Georgios Mavrogiannis.
\newblock {R}elatively non-degenerate integrated decay estimates on subextremal
  {K}err de {S}itter.
\newblock {\em Preprint, arXiv:2503.22090}, 2025.

\bibitem[Maz91]{MazzeoEdge}
Rafe~R. Mazzeo.
\newblock {Elliptic theory of differential edge operators I}.
\newblock {\em Communications in Partial Differential Equations},
  16(10):1615--1664, 1991.
\newblock \href {https://doi.org/10.1080/03605309108820815}
  {\path{doi:10.1080/03605309108820815}}.

\bibitem[Mel81]{MelroseTransformation}
Richard~B. Melrose.
\newblock Transformation of boundary problems.
\newblock {\em Acta Math.}, 147(3-4):149--236, 1981.
\newblock \href {https://doi.org/10.1007/BF02392873}
  {\path{doi:10.1007/BF02392873}}.

\bibitem[Mel93]{MelroseAPS}
Richard~B. Melrose.
\newblock {\em The {A}tiyah-{P}atodi-{S}inger index theorem}, volume~4 of {\em
  Research Notes in Mathematics}.
\newblock A K Peters, Ltd., Wellesley, MA, 1993.
\newblock \href {https://doi.org/10.1016/0377-0257(93)80040-i}
  {\path{doi:10.1016/0377-0257(93)80040-i}}.

\bibitem[Mel94]{MelroseEuclideanSpectralTheory}
Richard~B. Melrose.
\newblock Spectral and scattering theory for the {L}aplacian on asymptotically
  {E}uclidian spaces.
\newblock In {\em Spectral and scattering theory ({S}anda, 1992)}, volume 161
  of {\em Lecture Notes in Pure and Appl. Math.}, pages 85--130. Dekker, New
  York, 1994.

\bibitem[Mel96]{MelroseDiffOnMwc}
Richard~B. Melrose.
\newblock Differential analysis on manifolds with corners.
\newblock {\em Book, in preparation, available online}, 1996.
\newblock URL: \url{https://math.mit.edu/~rbm/daomwcf.ps}.

\bibitem[Mil23]{MilletTeukolskyDecay}
Pascal Millet.
\newblock {O}ptimal decay for solutions of the {T}eukolsky equation on the
  {K}err metric for the full subextremal range $|a|<{M}$.
\newblock {\em Preprint, arXiv:2302.06946}, 2023.

\bibitem[MM83]{MelroseMendozaB}
Richard~B. Melrose and Gerardo Mendoza.
\newblock {\em Elliptic operators of totally characteristic type}.
\newblock Mathematical Sciences Research Institute, 1983.

\bibitem[MM87]{MazzeoMelroseHyp}
Rafe~R. Mazzeo and Richard~B. Melrose.
\newblock Meromorphic extension of the resolvent on complete spaces with
  asymptotically constant negative curvature.
\newblock {\em J. Funct. Anal.}, 75(2):260--310, 1987.
\newblock \href {https://doi.org/10.1016/0022-1236(87)90097-8}
  {\path{doi:10.1016/0022-1236(87)90097-8}}.

\bibitem[MMTT10]{MarzuolaMetcalfeTataruTohaneanuStrichartz}
Jeremy Marzuola, Jason Metcalfe, Daniel Tataru, and Mihai Tohaneanu.
\newblock Strichartz estimates on {S}chwarzschild black hole backgrounds.
\newblock {\em Comm. Math. Phys.}, 293(1):37--83, 2010.
\newblock \href {https://doi.org/10.1007/s00220-009-0940-z}
  {\path{doi:10.1007/s00220-009-0940-z}}.

\bibitem[Mor24]{MorganDecay}
Katrina Morgan.
\newblock The effect of metric behavior at spatial infinity on pointwise wave
  decay in the asymptotically flat stationary setting.
\newblock {\em Amer. J. Math.}, 146(1):47--105, 2024.
\newblock \href {https://doi.org/10.1353/ajm.2024.a917539}
  {\path{doi:10.1353/ajm.2024.a917539}}.

\bibitem[MS24]{MaSzeftelEnergyKerr}
Siyuan Ma and J{\'e}r{\'e}mie Szeftel.
\newblock {E}nergy-{M}orawetz estimates for the wave equation in perturbations
  of {K}err.
\newblock {\em Preprint, arXiv:2410.02341}, 2024.

\bibitem[MS26]{MaSzeftelTeukolsky}
Siyuan Ma and J{\'e}r{\'e}mie Szeftel.
\newblock {E}nergy-{M}orawetz estimates for {T}eukolsky equations in
  perturbations of {K}err.
\newblock {\em Preprint, arXiv:2603.23437}, 2026.

\bibitem[MSBV14a]{MelroseSaBarretoVasyResolvent}
Richard Melrose, Ant\^onio S\'a{}~Barreto, and Andr\'as Vasy.
\newblock Analytic continuation and semiclassical resolvent estimates on
  asymptotically hyperbolic spaces.
\newblock {\em Comm. Partial Differential Equations}, 39(3):452--511, 2014.
\newblock \href {https://doi.org/10.1080/03605302.2013.866957}
  {\path{doi:10.1080/03605302.2013.866957}}.

\bibitem[MSBV14b]{MelroseSaBarretoVasySdS}
Richard Melrose, Ant\^onio S\'a{}~Barreto, and Andr\'as Vasy.
\newblock Asymptotics of solutions of the wave equation on de
  {S}itter-{S}chwarzschild space.
\newblock {\em Comm. Partial Differential Equations}, 39(3):512--529, 2014.
\newblock \href {https://doi.org/10.1080/03605302.2013.866958}
  {\path{doi:10.1080/03605302.2013.866958}}.

\bibitem[MTT12]{MetcalfeTataruTohaneanuPriceNonstationary}
Jason Metcalfe, Daniel Tataru, and Mihai Tohaneanu.
\newblock Price's law on nonstationary space-times.
\newblock {\em Adv. Math.}, 230(3):995--1028, 2012.
\newblock \href {https://doi.org/10.1016/j.aim.2012.03.010}
  {\path{doi:10.1016/j.aim.2012.03.010}}.

\bibitem[MTT17]{MetcalfeTataruTohaneanuMaxwellKerr}
Jason Metcalfe, Daniel Tataru, and Mihai Tohaneanu.
\newblock Pointwise decay for the {M}axwell field on black hole space-times.
\newblock {\em Adv. Math.}, 316:53--93, 2017.
\newblock \href {https://doi.org/10.1016/j.aim.2017.05.024}
  {\path{doi:10.1016/j.aim.2017.05.024}}.

\bibitem[MW23]{MorganWunschPrice}
Katrina Morgan and Jared Wunsch.
\newblock Generalized {P}rice's law on fractional-order asymptotically flat
  stationary spacetimes.
\newblock {\em Math. Res. Lett.}, 30(3):865--911, 2023.
\newblock \href {https://doi.org/10.4310/mrl.2023.v30.n3.a10}
  {\path{doi:10.4310/mrl.2023.v30.n3.a10}}.

\bibitem[MZ22]{MaZhangSchwarzschildDiracSharp}
Siyuan Ma and Lin Zhang.
\newblock Sharp decay estimates for massless {D}irac fields on a
  {S}chwarzschild background.
\newblock {\em Journal of Functional Analysis}, 282(6):109375, 2022.
\newblock \href {https://doi.org/10.1016/j.jfa.2021.109375}
  {\path{doi:10.1016/j.jfa.2021.109375}}.

\bibitem[MZ23]{MaZhangTeukolsky}
Siyuan Ma and Lin Zhang.
\newblock {S}harp decay for {T}eukolsky equation in {K}err spacetimes.
\newblock {\em Communications in Mathematical Physics}, Feb 2023.
\newblock \href {https://doi.org/10.1007/s00220-023-04640-w}
  {\path{doi:10.1007/s00220-023-04640-w}}.

\bibitem[N{\"u}t25]{NuetziStability}
Andrea N{\"u}tzi.
\newblock {P}erturbations of {M}inkowski spacetime with regular conformal
  compactification.
\newblock {\em Preprint, arXiv:2510.01964}, 2025.

\bibitem[PB04]{PriceBurkoLaw}
Richard~H. Price and Lior~M. Burko.
\newblock Late time tails from momentarily stationary, compact initial data in
  {S}chwarzschild spacetimes.
\newblock {\em Phys. Rev. D (3)}, 70(8):084039, 6, 2004.
\newblock \href {https://doi.org/10.1103/PhysRevD.70.084039}
  {\path{doi:10.1103/PhysRevD.70.084039}}.

\bibitem[Pen65]{PenroseAsymptotics}
Roger Penrose.
\newblock Zero rest-mass fields including gravitation: asymptotic behaviour.
\newblock In {\em Proceedings of the Royal Society of London A: Mathematical,
  Physical and Engineering Sciences}, volume 284, pages 159--203. The Royal
  Society, 1965.

\bibitem[Pen74]{PenroseSCC}
Roger Penrose.
\newblock Gravitational collapse.
\newblock In C.~Dewitt-Morette, editor, {\em Gravitational Radiation and
  Gravitational Collapse}, volume~64, pages 82--91. IAU Symposium, Springer,
  1974.

\bibitem[Pri72a]{PriceLawI}
Richard~H. Price.
\newblock Nonspherical perturbations of relativistic gravitational collapse.
  {I}. {S}calar and gravitational perturbations.
\newblock {\em Phys. Rev. D (3)}, 5:2419--2438, 1972.
\newblock \href {https://doi.org/10.1103/PhysRevD.5.2419}
  {\path{doi:10.1103/PhysRevD.5.2419}}.

\bibitem[Pri72b]{PriceLawII}
Richard~H. Price.
\newblock Nonspherical perturbations of relativistic gravitational collapse.
  {II}. {I}nteger-spin, zero-rest-mass fields.
\newblock {\em Phys. Rev. D (3)}, 5:2439--2454, 1972.
\newblock \href {https://doi.org/10.1103/PhysRevD.5.2439}
  {\path{doi:10.1103/PhysRevD.5.2439}}.

\bibitem[PV21]{PetersenVasySubextremal}
Oliver~Lindblad Petersen and Andr{\'a}s Vasy.
\newblock Wave equations in the {K}err--de {S}itter spacetime: the full
  subextremal range.
\newblock {\em Preprint, arXiv:2112.0135}, 2021.
\newblock URL: \url{https://arxiv.org/abs/2112.0135}.

\bibitem[Rin08]{RingstromEinsteinScalarStability}
Hans Ringstr{\"o}m.
\newblock Future stability of the {E}instein--non-linear scalar field system.
\newblock {\em Inventiones mathematicae}, 173(1):123--208, 2008.
\newblock \href {https://doi.org/10.1007/s00222-008-0117-y}
  {\path{doi:10.1007/s00222-008-0117-y}}.

\bibitem[RW57]{ReggeWheelerSchwarzschild}
Tullio Regge and John~A. Wheeler.
\newblock Stability of a {S}chwarzschild singularity.
\newblock {\em Phys. Rev. (2)}, 108:1063--1069, 1957.

\bibitem[Sbi15]{SbierskiBeams}
Jan Sbierski.
\newblock Characterisation of the energy of {G}aussian beams on {L}orentzian
  manifolds: with applications to black hole spacetimes.
\newblock {\em Anal. PDE}, 8(6):1379--1420, 2015.
\newblock \href {https://doi.org/10.2140/apde.2015.8.1379}
  {\path{doi:10.2140/apde.2015.8.1379}}.

\bibitem[Sbi16]{SbierskiMGHD}
Jan Sbierski.
\newblock On the existence of a maximal {C}auchy development for the {E}instein
  equations: a dezornification.
\newblock {\em Ann. Henri Poincar\'e}, 17(2):301--329, 2016.
\newblock \href {https://doi.org/10.1007/s00023-015-0401-5}
  {\path{doi:10.1007/s00023-015-0401-5}}.

\bibitem[SBW16]{SaBarretoWangResolvent}
Ant\^onio S\'a{}~Barreto and Yiran Wang.
\newblock The semiclassical resolvent on conformally compact manifolds with
  variable curvature at infinity.
\newblock {\em Comm. Partial Differential Equations}, 41(8):1230--1302, 2016.
\newblock \href {https://doi.org/10.1080/03605302.2016.1190377}
  {\path{doi:10.1080/03605302.2016.1190377}}.

\bibitem[SBZ97]{SaBarretoZworskiResonances}
Ant{\^o}nio S{\'a}~Barreto and Maciej Zworski.
\newblock Distribution of resonances for spherical black holes.
\newblock {\em Mathematical Research Letters}, 4:103--122, 1997.
\newblock URL: \url{https://dx.doi.org/10.4310/MRL.1997.v4.n1.a10}.

\bibitem[She22]{ShenMinkExtStab}
Dawei Shen.
\newblock Stability of {M}inkowski spacetime in exterior regions.
\newblock {\em Preprint, arXiv:2211.15230}, 2022.

\bibitem[She23]{ShenGCMKerr}
Dawei Shen.
\newblock Construction of {GCM} hypersurfaces in perturbations of {K}err.
\newblock {\em Ann. PDE}, 9(1):Paper No. 11, 112, 2023.
\newblock \href {https://doi.org/10.1007/s40818-023-00152-x}
  {\path{doi:10.1007/s40818-023-00152-x}}.

\bibitem[She24]{ShenMinkBorderline}
Dawei Shen.
\newblock {E}xterior stability of {M}inkowski spacetime with borderline decay.
\newblock {\em Preprint, arXiv:2405.00735}, 2024.

\bibitem[SR89]{SaintRaymondNashMoser}
Xavier Saint~Raymond.
\newblock A simple {N}ash-{M}oser implicit function theorem.
\newblock {\em Enseign. Math. (2)}, 35(3-4):217--226, 1989.

\bibitem[SR15]{ShlapentokhRothmanModeStability}
Yakov Shlapentokh-Rothman.
\newblock Quantitative mode stability for the wave equation on the {K}err
  spacetime.
\newblock {\em Ann. Henri Poincar\'e}, 16(1):289--345, 2015.
\newblock \href {https://doi.org/10.1007/s00023-014-0315-7}
  {\path{doi:10.1007/s00023-014-0315-7}}.

\bibitem[SRdC20]{ShlapentokhRothmanTeixeiradCTeukolskyI}
Yakov Shlapentokh-Rothman and Rita~Teixeira da~Costa.
\newblock {B}oundedness and decay for the {T}eukolsky equation on {K}err in the
  full subextremal range $|a|<{M}$: frequency space analysis.
\newblock {\em Preprint, arXiv:2007.07211}, 2020.

\bibitem[SRdC23]{ShlapentokhRothmanTeixeiradCTeukolskyII}
Yakov Shlapentokh-Rothman and Rita~Teixeira da~Costa.
\newblock {B}oundedness and decay for the {T}eukolsky equation on {K}err in the
  full subextremal range $|a|<{M}$: physical space analysis.
\newblock {\em Preprint, arXiv:2302.08916}, 2023.

\bibitem[SRT25]{ShlapentokhRothmanTohaneanuC1Kerr}
Yakov Shlapentokh-Rothman and Mihai Tohaneanu.
\newblock {B}oundedness for the wave equation on ${C}^1$ stationary
  axisymmetric perturbations of {K}err.
\newblock {\em Preprint, arXiv:2511.08751}, 2025.

\bibitem[ST15]{SterbenzTataruMaxwellSchwarzschild}
Jacob Sterbenz and Daniel Tataru.
\newblock Local energy decay for {M}axwell fields {P}art {I}: {S}pherically
  symmetric black-hole backgrounds.
\newblock {\em Int. Math. Res. Not.}, (11):3298--3342, 2015.

\bibitem[Stu24]{StuckerKerrQNM}
Thomas Stucker.
\newblock Quasinormal modes for the {K}err black hole.
\newblock {\em Preprint, arXiv:2407.04612}, 2024.

\bibitem[Sus24]{SussmanResolventPhg}
Ethan Sussman.
\newblock Complete asymptotic analysis of low energy scattering for
  {S}chr\"odinger operators with a short-range potential.
\newblock {\em Preprint, arXiv:2411.04220}, 2024.

\bibitem[Sus26]{SussmanKG}
Ethan Sussman.
\newblock Massive wave propagation near null infinity.
\newblock {\em Ann. PDE}, 12(1):Paper No. 9, 127, 2026.
\newblock \href {https://doi.org/10.1007/s40818-026-00232-8}
  {\path{doi:10.1007/s40818-026-00232-8}}.

\bibitem[Tat13]{TataruDecayAsympFlat}
Daniel Tataru.
\newblock Local decay of waves on asymptotically flat stationary space-times.
\newblock {\em Amer. J. Math.}, 135(2):361--401, 2013.
\newblock \href {https://doi.org/10.1353/ajm.2013.0012}
  {\path{doi:10.1353/ajm.2013.0012}}.

\bibitem[Tay17]{TaylorEinsteinVlasov}
Martin Taylor.
\newblock The global nonlinear stability of {M}inkowski space for the massless
  {E}instein-{V}lasov system.
\newblock {\em Ann. PDE}, 3(1):Paper No. 9, 177, 2017.
\newblock \href {https://doi.org/10.1007/s40818-017-0026-8}
  {\path{doi:10.1007/s40818-017-0026-8}}.

\bibitem[TdC20]{TeixeiradCModes}
Rita Teixeira~da Costa.
\newblock Mode stability for the {T}eukolsky equation on extremal and
  subextremal {K}err spacetimes.
\newblock {\em Comm. Math. Phys.}, 378(1):705--781, 2020.
\newblock \href {https://doi.org/10.1007/s00220-020-03796-z}
  {\path{doi:10.1007/s00220-020-03796-z}}.

\bibitem[Teu72]{TeukolskySeparation}
Saul~A. Teukolsky.
\newblock {R}otating black holes: {S}eparable wave equations for gravitational
  and electromagnetic perturbations.
\newblock {\em Physical Review Letters}, 29(16):1114, 1972.

\bibitem[Toh12]{TohaneanuKerrStrichartz}
Mihai Tohaneanu.
\newblock Strichartz estimates on {K}err black hole backgrounds.
\newblock {\em Trans. Amer. Math. Soc.}, 364(2):689--702, 2012.
\newblock \href {https://doi.org/10.1090/S0002-9947-2011-05405-X}
  {\path{doi:10.1090/S0002-9947-2011-05405-X}}.

\bibitem[TT11]{TataruTohaneanuKerrLocalEnergy}
Daniel Tataru and Mihai Tohaneanu.
\newblock A local energy estimate on {K}err black hole backgrounds.
\newblock {\em Int. Math. Res. Not. IMRN}, (2):248--292, 2011.
\newblock \href {https://doi.org/10.1093/imrn/rnq069}
  {\path{doi:10.1093/imrn/rnq069}}.

\bibitem[Vas00]{VasyThreeBody}
Andr\'as Vasy.
\newblock Propagation of singularities in three-body scattering.
\newblock {\em Ast\'erisque}, (262):vi+151, 2000.

\bibitem[Vas13]{VasyMicroKerrdS}
Andr{\'a}s Vasy.
\newblock Microlocal analysis of asymptotically hyperbolic and {K}err--de
  {S}itter spaces (with an appendix by {S}emyon {D}yatlov).
\newblock {\em Invent. Math.}, 194(2):381--513, 2013.
\newblock \href {https://doi.org/10.1007/s00222-012-0446-8}
  {\path{doi:10.1007/s00222-012-0446-8}}.

\bibitem[Vas14]{VasyMinkDSHypRelation}
Andr\'as Vasy.
\newblock Resolvents, {P}oisson operators and scattering matrices on
  asymptotically hyperbolic and de {S}itter spaces.
\newblock {\em J. Spectr. Theory}, 4(4):643--673, 2014.
\newblock \href {https://doi.org/10.4171/JST/82} {\path{doi:10.4171/JST/82}}.

\bibitem[Vas21]{VasyLowEnergyLag}
Andr{\'a}s Vasy.
\newblock {R}esolvent near zero energy on {R}iemannian scattering
  (asymptotically conic) spaces, a {L}agrangian approach.
\newblock {\em Communications in Partial Differential Equations},
  46(5):823--863, 2021.
\newblock \href {https://doi.org/10.1080/03605302.2020.1857401}
  {\path{doi:10.1080/03605302.2020.1857401}}.

\bibitem[Vis70]{VishveshwaraSchwarzschild}
C.~V. Vishveshwara.
\newblock {S}tability of the {S}chwarzschild {M}etric.
\newblock {\em Phys. Rev. D}, 1:2870--2879, May 1970.

\bibitem[Wal78]{WaldKerrPerturbation}
Robert~M. Wald.
\newblock Construction of solutions of gravitational, electromagnetic, or other
  perturbation equations from solutions of decoupled equations.
\newblock {\em Phys. Rev. Lett.}, 41(4):203--206, 1978.
\newblock \href {https://doi.org/10.1103/PhysRevLett.41.203}
  {\path{doi:10.1103/PhysRevLett.41.203}}.

\bibitem[Wan10]{WangThesis}
Fang Wang.
\newblock {\em Radiation field for vacuum {E}instein equation}.
\newblock PhD thesis, Massachusetts Institute of Technology, 2010.

\bibitem[Wan13]{WangRadiation}
Fang Wang.
\newblock {R}adiation field for {E}instein vacuum equations with spacial
  dimension $n\geq 4$.
\newblock {\em Preprint, arXiv:1304.0407}, 2013.

\bibitem[Wan20]{WangEinsteinKleinGordon}
Qian Wang.
\newblock An intrinsic hyperboloid approach for {E}instein {K}lein-{G}ordon
  equations.
\newblock {\em J. Differential Geom.}, 115(1):27--109, 2020.
\newblock \href {https://doi.org/10.4310/jdg/1586224841}
  {\path{doi:10.4310/jdg/1586224841}}.

\bibitem[Whi89]{WhitingKerrModeStability}
Bernard~F. Whiting.
\newblock Mode stability of the {K}err black hole.
\newblock {\em J. Math. Phys.}, 30(6):1301--1305, 1989.
\newblock \href {https://doi.org/10.1063/1.528308}
  {\path{doi:10.1063/1.528308}}.

\bibitem[WZ11]{WunschZworskiNormHypResolvent}
Jared Wunsch and Maciej Zworski.
\newblock Resolvent estimates for normally hyperbolic trapped sets.
\newblock {\em Ann. Henri Poincar\'e}, 12(7):1349--1385, 2011.
\newblock \href {https://doi.org/10.1007/s00023-011-0108-1}
  {\path{doi:10.1007/s00023-011-0108-1}}.

\bibitem[Zer70]{ZerilliPotential}
Frank~J. Zerilli.
\newblock {E}ffective {P}otential for {E}ven-{P}arity {R}egge--{W}heeler
  {G}ravitational {P}erturbation {E}quations.
\newblock {\em Phys. Rev. Lett.}, 24:737--738, Mar 1970.

\bibitem[Zwo16]{ZworskiRevisitVasy}
Maciej Zworski.
\newblock Resonances for asymptotically hyperbolic manifolds: {V}asy's method
  revisited.
\newblock {\em J. Spectr. Theory}, 6(4):1087--1114, 2016.
\newblock \href {https://doi.org/10.4171/JST/153} {\path{doi:10.4171/JST/153}}.

\end{thebibliography}

\end{document}